\documentclass{dune}
\pdfoutput=1

\usepackage[pdftex,bookmarks,hidelinks]{hyperref}

\graphicspath{ {graphics/} }

\newif\ifdp
\newif\ifsp

\def\expshort{DUNE\xspace}

\def\explong{The Deep Underground Neutrino Experiment\xspace}

\def\thedocsubtitle{Deep Underground Neutrino Experiment (DUNE)} 
\def\tdrtitle{Technical Design Report}
\def\voltitleexec{Introduction to DUNE\xspace}
\def\volnumberexec{I}

\def\voltitlephysics{DUNE Physics\xspace}
\def\volnumberphysics{II}

\def\physchtools{Volume~\volnumberphysics{}, \voltitlephysics{}, Chapter~4\xspace}
\def\physchlbl{Volume~\volnumberphysics{}, \voltitlephysics{}, Chapter~5\xspace}
\def\physchndk{Volume~\volnumberphysics{}, \voltitlephysics{}, Chapter~6\xspace}
\def\physchsnb{Volume~\volnumberphysics{}, \voltitlephysics{}, Chapter~7\xspace}

\def\voltitletc{DUNE Far Detector Technical Coordination\xspace}
\def\volnumbertc{III}

\def\tcchjpo{Volume~\volnumbertc{}, \voltitletc{}, Chapter~4\xspace}
\def\tcchfac{Volume~\volnumbertc{}, \voltitletc{}, Chapter~5\xspace}

\def\tcchqa{Volume~\volnumbertc{}, \voltitletc{}, Chapter~9\xspace}
\def\tcchesh{Volume~\volnumbertc{}, \voltitletc{}, Chapter~10\xspace}
\def\tcchappx{Volume~\volnumbertc{}, \voltitletc{}, Chapter~11\xspace}

\def\voltitlesp{The DUNE Far Detector Single-Phase Technology\xspace}
\def\volnumbersp{IV}

\def\voltitledp{The DUNE Far Detector Dual-Phase Technology\xspace}
\def\volnumberdp{V}

\def\dpchpds{Volume~\volnumberdp{}, \voltitledp{}, Chapter~5\xspace}

\def\dpchcisc{Volume~\volnumberdp{}, \voltitledp{}, Chapter~6\xspace}

\newcommand{\refsec}[2]{Volume~\csname volnumber#1\endcsname \xspace Section~#2}
\newcommand{\refch}[2]{Volume~\csname volnumber#1\endcsname \xspace Chapter~#2}
\newcommand{\refinch}[2]{#2 in Volume~\csname volnumber#1\endcsname \xspace}

\newcommand{\bigo}[1]{\ensuremath{\mathcal{O}(#1)}}

\newcommand{\numu}{\ensuremath{\nu_\mu}\xspace}
\newcommand{\nue}{\ensuremath{\nu_e}\xspace}

\def\argon40{${}^{40}$Ar}       
\def\Ar39{$^{39}$Ar}
\def\Cl40{$^{40}$Cl}
\def\K40{$^{40}$K}
\def\B8{$^{8}$B}
\newcommand\isotope[2]{\textsuperscript{#2}#1} 

\def\fdfiducialmass{\SI{40}{\kt}\xspace}
\def\driftvelocity{\SI{1.6}{\milli\meter/\micro\second}\xspace} 

\def\larmass{\SI{17.5}{\kt}\xspace} 

\def\cryostatht{\SI{17.8}{\meter}\xspace} 
\def\cryostatlen{\SI{65.8}{\meter}\xspace} 
\def\cryostatwdth{\SI{18.9}{\meter}\xspace} 

\def\nominalmodsize{\SI{10}{kt}\xspace} 
\def\dunelifetime{\SI{20}{years}\xspace} 
 
\def\cooldown{cool-down\xspace} 

\def\spmaxfield{\SI{500}{\volt/\centi\meter}\xspace} 
\def\spactivelarmass{\SI{10}{\kt}\xspace} 
\def\spmaxdrift{\SI{3.5}{\m}\xspace}
\def\tpcheight{\SI{12.0}{\meter}\xspace} 
\def\sptpclen{\SI{58.2}{\meter}\xspace} 
\def\spfcmodlen{\SI{3.5}{\m}} 
\def\planespace{\SI{4.8}{\milli\meter}\xspace}
\def\sptargetdriftvolt{$-\SI{180}{\kilo\volt}$\xspace} 
\def\sptargetdriftvoltpos{\SI{180}{\kilo\volt}\xspace} 
\def\coldbox{cold box\xspace} 
\def\Coldbox{Cold box\xspace} 

\def\spreadout{\SI{5.4}{\ms}\xspace}

\def\snbtime{\SI{100}{\s}\xspace}
\def\snbpretime{\SI{10}{\s}\xspace}

\def\daqpower{\SI[inter-unit-product =$\cdot$]{500}{\kilo\volt\ampere}\xspace}
\def\surfdaqpower{\SI[inter-unit-product =$\cdot$]{50}{\kilo\volt\ampere}\xspace}

\def\cucracks{\SI{60}{racks}\xspace}
\def\daqracks{\SI{56}{racks}\xspace}
\def\surfdaqracks{\SI{8}{racks}\xspace}

\def\offsitepbpy{\SI{30}{\PB/\year}\xspace}

\newcommand{\efield}{E field\xspace}

\newcommand{\rms}{RMS\xspace} 
\newcommand{\threed}{3D\xspace}
\newcommand{\twod}{2D\xspace}
\newcommand{\fdth}{feedthrough\xspace} 
\newcommand{\phel}{photoelectron\xspace} 
\newcommand{\frfour}{FR-4\xspace} 

\def\mindriftfield{\SI{250}{\volt/\cm}\xspace}
\def\mindriftfieldgoal{\SI{500}{\volt/\cm}\xspace}
\def\elecnoisefe{ \SI{1000}{e$^-$}\xspace}

\def\apacollwireangle{$\SI{0}{^\circ}$\xspace}
\def\apainducwireangle{$\pm\SI{35.7}{^\circ}$\xspace}
\def\uvpitch{\SI{4.7}{\milli\meter}\xspace}
\def\xgpitch{\SI{4.8}{\milli\meter}\xspace}
\def\wirepitchtol{$\pm$\SI{0.5}{\milli\meter}\xspace}

\def\fepeaktime{\SI{1}{\micro\second}\xspace}

\def\cathodemegohm{\SI{1}{\mega\ohm/square}\xspace}
\def\cathodegigohm{\SI{1}{\giga\ohm/square}\xspace}
\def\localefield{\SI{30}{\kV/\cm}\xspace}

\newcommand{\lsim}{{\;\raise0.3ex\hbox{$<$\kern-0.75em\raise-1.1ex\hbox{$\sim$}}\;}}
\newcommand{\gsim}{{\;\raise0.3ex\hbox{$>$\kern-0.75em\raise-1.1ex\hbox{$\sim$}}\;}}
\newcommand{\beq}{\begin{equation}}
\newcommand{\eeq}{\end{equation}}
\newcommand{\bea}{\begin{eqnarray}}
\newcommand{\eea}{\end{eqnarray}}

\mathchardef\minus="002D

\newcommand{\startpduneiispinstall}{March 2021\xspace}
\newcommand{\startpduneiidpinstall}{March 2022\xspace}
\newcommand{\sdlwavailable}{April 2022\xspace}
\newcommand{\cucbenocc}{October 2022\xspace}
\newcommand{\accesscuccountrm}{April  2023\xspace}
\newcommand{\accesstopfirstcryo}{January 2024\xspace}
\newcommand{\startfirsttpcinstall}{August 2024\xspace}
\newcommand{\accesstopsecondcryo}{January 2025\xspace}
\newcommand{\firsttpcinstallend}{May 2025\xspace}
\newcommand{\startsecondtpcinstall}{August 2025\xspace}
\newcommand{\secondtpcinstallend}{May 2026\xspace}

\newcommand{\rrt}[1]{}

\newcommand{\kamland}{KamLAND\xspace} 
\newcommand{\microboone}{MicroBooNE\xspace} 
\newcommand{\minerva}{MINERvA\xspace} 
\newcommand{\nova}{NOvA\xspace} 

\newcommand{\lartpc}{LArTPC\xspace}

\newcommand{\fnal}{Fermilab\xspace} 
\newcommand{\surf}{SURF\xspace}

\newcommand{\detmodule}{detector module\xspace}
\newcommand{\dual}{DP\xspace}

\newcommand{\single}{SP\xspace}

\newcommand{\spmod}{SP detector module\xspace}
\newcommand{\lar}{LAr\xspace}
\newcommand{\lntwo}{LN$_2$\xspace}  

\DeclareSIUnit \s {\second}
\DeclareSIUnit \MB {\mega\byte}
\DeclareSIUnit \GB {\giga\byte}
\DeclareSIUnit \TB {\tera\byte}
\DeclareSIUnit \PB {\peta\byte}
\DeclareSIUnit \Mbps {\mega\bit/\s}
\DeclareSIUnit \Gbps {\giga\bit/\s}
\DeclareSIUnit \Tbps {\tera\bit/\s}
\DeclareSIUnit \Pbps {\peta\bit/\s}
\DeclareSIUnit \kton {\kilo\tonne} 
\DeclareSIUnit \kt {\kilo\tonne}
\DeclareSIUnit \Mt {\mega\tonne}
\DeclareSIUnit \eV {\electronvolt}
\DeclareSIUnit \keV {\kilo\electronvolt}
\DeclareSIUnit \MeV {\mega\electronvolt}
\DeclareSIUnit \GeV {\giga\electronvolt}
\DeclareSIUnit \m {\meter}
\DeclareSIUnit \cm {\centi\meter}
\DeclareSIUnit \in {\inchcommand}
\DeclareSIUnit \km {\kilo\meter}
\DeclareSIUnit \kV {\kilo\volt}
\DeclareSIUnit \kW {\kilo\watt}
\DeclareSIUnit \MW {\mega\watt}
\DeclareSIUnit \MHz {\mega\hertz}
\DeclareSIUnit \mrad {\milli\radian}
\DeclareSIUnit \year {year}
\DeclareSIUnit \POT {POT}
\DeclareSIUnit \sig {$\sigma$}
\DeclareSIUnit\parsec{pc}
\DeclareSIUnit\lightyear{ly}
\DeclareSIUnit\foot{ft}
\DeclareSIUnit\ft{ft}
\DeclareSIUnit \ppb{ppb}
\DeclareSIUnit \ppt{ppt}
\DeclareSIUnit \samples{S}

\usepackage[toc]{glossaries}
\makeglossaries

\newcommand{\dshort}[1]{\glsentrytext{#1}}  
\newcommand{\dshorts}[1]{\glsentryshortpl{#1}}  
\newcommand{\dlong}[1]{\glsentrylong{#1}}  

\newcommand{\dfirst}[1]{\glsfirst{#1}\glsunset{#1}}
\newcommand{\dfirsts}[1]{\glsfirstplural{#1}\glsunset{#1}}

\newcommand{\dword}[1]{\gls{#1}}
\newcommand{\dwords}[1]{\glspl{#1}}
\newcommand{\Dword}[1]{\Gls{#1}}
\newcommand{\Dwords}[1]{\Glspl{#1}}

\newcommand{\newduneword}[3]{
    \newglossaryentry{#1}{
        text={#2},
        long={#2},
        name={\glsentrylong{#1}},
        first={\glsentryname{#1}},
        firstplural={\glsentrylong{#1}\glspluralsuffix},
        description={#3},
        sort={#2}
    }
}

\newcommand{\newduneabbrev}[4]{
  \newglossaryentry{#1}{
    text={#2},
    long={#3},
    shortplural={{#2}\glspluralsuffix},
    longplural={{#3}\glspluralsuffix{}},
    name={\glsentrylong{#1}{} (\glsentrytext{#1}{})},
    first={#3 (#2)},
    firstplural={#3\glspluralsuffix{} (\glsentrytext{#1}\glspluralsuffix{})},
    description={#4},
    sort={#2}
  }
}

\newcommand{\newduneabbrevs}[5]{
  \newglossaryentry{#1}{
    text={#2},
    long={#3},
    plural={#4},
    shortplural={{#2}\glspluralsuffix},
    longplural={#4},
    name={\glsentrylong{#1}{} (\glsentrytext{#1}{})},
    first={#3 (#2)},
    firstplural={#4 (\glsentrytext{#1}\glspluralsuffix{})},
    description={#5},
    sort={#2}    
  }
}

\newduneword{dword}{DUNE Word}{A term in the DUNE lexicon}

\newduneword{nasa}{NASA}{U.S. National Aereonautics and Space Administration}

\newduneabbrev{nd}{ND}{near detector}{Refers to the detector(s) 
 installed close to the neutrino source at \fnal }

\newduneabbrev{fd}{FD}{far detector}{The \SI{70}{kt} total (\fdfiducialmass fiducial) mass \gls{lartpc} DUNE detector, composed of four \larmass total (\nominalmodsize fiducial) mass modules,  
  to be installed at the far site at \surf in
  Lead, SD, USA}

\newduneabbrev{sp}{SP}{single-phase}{Distinguishes one of the DUNE far detector technologies by the fact that it operates using argon in its liquid phase only}

\newduneabbrev{dp}{DP}{dual-phase}{Distinguishes one of the DUNE far detector technologies by the fact that it operates using argon 
 in both gas and liquid phases}

\newduneabbrev{pds}{PD system}{photon detection system}{The detector 
  subsystem sensitive to light produced in the \lar }

\newduneabbrev{hvs}{HVS}{high voltage system}{The detector 
  subsystem that provides the \gls{tpc} drift field}

\newduneabbrev{tpc}{TPC}{time projection chamber}{A type of particle detector that uses an \efield together with a sensitive volume of gas or liquid, e.g., \gls{lar}, to perform a \threed reconstruction of a particle trajectory or interaction. The activity is recorded by digitizing the waveforms of current
  induced on the anode as the distribution of ionization charge passes by
  or is collected on the electrode} 

\newduneabbrev{lartpc}{LArTPC}{liquid argon time-projection chamber}{A \gls{tpc} filled with liquid argon; 
the basis for the \gls{dune} \gls{fd} modules} 

\newduneabbrevs{apa}{APA}{anode plane assembly}{anode plane assemblies}{A unit of the \single
  detector module containing the elements sensitive to ionization in the \lar. 
  It contains two faces each of three planes of wires, and interfaces to the cold
  electronics and photon detection system} 

\newduneabbrev{awg}{AWG}{American wire gauge} {U.S. standard set of non-ferrous wire conductor sizes}

\newduneabbrev{ufer}{Ufer}{concrete encased electrode} {U.S. National Electrical Code grounding method refered to as Concrete Encased Electrode}

\newduneabbrev{cro}{CRO}{charge readout}{The system for detecting
  ionization charge distributions in a \dual detector module}

\newduneabbrev{lro}{LRO}{light readout}{The system for detecting
  scintillation photons in a \dual  detector module}

\newduneabbrev{shv}{SHV}{safe high voltage}{Type of bayonet mount
connector used on coaxial cables that has additional insulation 
compared to standard BNC and MHV connectors that makes it safer
for handling \gls{hv} by preventing accidental contact with the
live wire connector in an unmated connector or plug}

\newduneabbrev{fe}{FE}{front-end}{The front-end refers a point that is
  ``upstream'' of the data flow for a particular subsystem. 
  For example the \gls{sp} front-end electronics is where the cold electronics
  meet the sense wires of the TPC and the front-end \gls{daq} is where the
  \gls{daq} meets the output of the electronics}

\newduneabbrev{daqrou}{DAQ RU}{DAQ readout unit}{The first element in the data flow of the \gls{daq}}

\newduneabbrev{cots}{COTS}{commercial off-the-shelf}{Items, typically hardware such as 
computers, that may be purchased whole, without any custom design or fabrication and 
thus at normal consumer prices and availability}

\newduneabbrev{i2c}{I2C}{Inter-Integrated Circuit}{I$^2$C or I2C is a synchronous, 
multi-master, multi-slave, packet switched, single-ended, serial computer bus widely used 
for attaching lower-speed peripheral ICs to processors and microcontrollers in short-distance, 
intra-board communication} 

\newduneabbrev{spi}{SPI}{Serial Peripheral Interface}{The Serial Peripheral Interface is a 
synchronous serial communication interface specification used for short distance 
communication, primarily in embedded systems}

\newduneabbrev{miso}{MISO}{master in slave out}{The Master In Slave Out is a logic
signal on the \gls{spi} bus on which the data from the slave are transmitted once
a request from the master is received} 

\newduneabbrev{mosi}{MOSI}{master out slave in}{The Master Out Slave In is a logic
signal on the \gls{spi} bus on which the data from the master is transmitted} 

\newduneabbrev{uart}{UART}{Universal Asynchrous Receiver/Transmitter}{A universal 
asynchronous receiver-transmitter is a computer hardware device for asynchronous 
serial communication in which the data format and transmission speeds are configurable}

\newduneword{cr}{CR}{Capacitance-Resistance} 

\newduneword{dc}{DC}{direct coupling} 

\newduneword{ac}{AC}{capacitive coupling}  

\newduneabbrev{pll}{PLL}{Phase-Locked Loop}{A control system that generates an
output signal whose phase is related to the phase of an input signal}  

\newduneword{fifo}{FIFO}{First-In-First-Out} 

\newduneword{tsmc}{TSMC}{Taiwan Semiconductor Manufacturing Company}

\newduneword{saci}{SACI}{\gls{slac} \gls{asic} Control Interface}

\newduneword{om3}{OM3}{Type of multi-mode fiber optic cable, typically capable of \SI{10}{Gbps} data transmission at lengths up to \SI{300}{m}}

\newduneword{om4}{OM4}{Type of multi-mode fiber optic cable, typically capable of \SI{10}{Gbps} data transmission at lengths up to \SI{550}{m}}

\newduneword{qfp}{QFP}{Quad Flat Package} 

\newduneabbrev{ams}{AMS}{analog and mixed signal}{Verilog-AMS is a derivative of the Verilog hardware description language that includes analog and mixed-signal extensions (AMS) in order to define the behavior of analog and mixed-signal systems}

\newduneabbrev{hepa}{HEPA}{High Efficiency Particulate Air}{The High Efficiency Particulate Air filters are a type of air filter that remove \num{99.97}\% of particles that have a size greater han or equal to \SI{0.3}{$\mu$m}}  

\newduneabbrev{uvm}{UVM}{universal verification methodology}{The Universal Verification Methodology is a standardized methodology for verifying integrated circuit designs}   

\newduneword{lhc}{LHC}{Large Hadron Collider}

\newduneabbrev{lsb}{LSB}{least significant bit}{The bit with the lowest numerical value in a binary number}

\newduneabbrev{ldo}{LDO}{low-dropout regulator}{A low-dropout or LDO regulator is a \gls{dc} linear voltage regulator that can regulate the output voltage even when the supply voltage is very close to the output voltage}

\newduneabbrev{adc}{ADC}{analog-to-digital converter}{A sampling of a voltage
  resulting in a discrete integer count corresponding in some way to
  the input}

\newduneabbrev{inl}{INL}{integral non-linearity}{A commonly used measure of performance in \glspl{adc}. It is the deviation between the ideal input threshold value and the measured threshold level of a certain output code}

\newduneabbrev{dnl}{DNL}{differential non-linearity}{A commonly used measure of performance in \glspl{adc}. The DNL error is defined as the difference between an actual step width and the ideal value of one \gls{lsb}}

\newduneword{pnp}{PNP}{Type of bipolar junction transistor consistning of a
layer of N-doped semiconductor sandwiched between two layers of P-doped material}

\newduneword{spice}{SPICE}{SPICE
(``Simulation Program with Integrated Circuit Emphasis'') is a general-purpose, 
open-source analog electronic circuit simulator. It is a program used in integrated 
circuit and board-level design to check the integrity of circuit designs and to 
predict circuit behavior}

\newduneabbrev{daq}{DAQ}{data acquisition}{The data acquisition system
  accepts data from the detector \gls{fe} electronics, buffers
  the data, performs a \gls{trigdecision}, builds events from the selected
  data and delivers the result to the offline \gls{diskbuffer}}

\newduneword{detmodule}{detector module}{The entire DUNE far detector is
  segmented into four modules, each with a nominal \SI{10}{\kton}
  fiducial mass}

\newduneword{detunit}{detector unit}{A 
portion of a \gls{detmodule} may be further partitioned into a number of similar parts.   For example the \gls{sp} \gls{tpc} 
is made up of \gls{apa}  units (and other elements)}

\newduneword{diskbuffer}{secondary DAQ buffer}{A secondary
  \dshort{daq} buffer holds a small subset of the full rate as
  selected by a \gls{trigcommand}. 
  This buffer also marks the interface with the DUNE Offline}

\newduneabbrev{om}{OM}{online monitoring}{Processes that run inside
  the \gls{daq} on data ``in flight,'' specifically before landing on the
  offline disk buffer, and that provide feedback on the operation of
  the \gls{daq} itself and the general health of the data it is marshalling}

\newduneabbrev{dqm}{DQM}{data quality monitoring}{Analysis of the raw
  data to monitor the integrity of the data and the performance of the
  detectors and their electronics. This type of monitoring may be
  performed in real time, within the \gls{daq} system, or in later
  stages of processing, using disk files as input}

\newduneword{dumpbuffer}{DAQ dump buffer}{This \gls{daq} buffer
  accepts a high-rate data stream, in aggregate, from an associated
  portion of a \gls{detmodule} sufficient to collect all data likely relevant to
  a potential \gls{snb}}

\newduneabbrev{etl}{ETL}{external trigger logic}{Trigger processing
  that consumes \gls{detmodule} level \gls{trignote} information
  and other global sources of trigger input and emits
  \gls{trigcommand} information back to the \glspl{mtl}}
\newduneabbrev{daqeti}{ETI}{external trigger interface}{Interface between \glspl{mtl} and external source and sinks of relevant trigger information}

\newduneword{trignote}{trigger notification}{Information provided by
  \gls{mtl} to \gls{etl} about \gls{trigdecision} 
  processing}

\newduneword{trigprimitive}{trigger primitive}{Information derived by
  the \gls{daq} \gls{fe} hardware that describes a region of space (e.g.,
  one or several neighboring channels) and time (e.g., a contiguous set
  of \gls{adc} sample ticks) associated with some activity}

\newduneword{externtrigger}{external trigger candidate}{Information
  provided to the \gls{mtl} about events external to a
  \gls{detmodule} so that it may be considered in forming
  \glspl{trigcommand}}

\newduneabbrev{daqoob}{OOB dispatcher}{out-of-band trigger command
  dispatcher}{This component is responsible for dispatching a \gls{snb} dump
  command to all \glspl{daqfer} in the \gls{detmodule}}

\newduneabbrev{mtl}{MTL}{module trigger logic}{Trigger processing
  that consumes \gls{detunit} level \gls{trigcommand} information
  and emits \glspl{trigcommand}. 
  It provides the \gls{etl} with \glspl{trignote} and receives back any
  \glspl{externtrigger}}

\newduneword{octant}{octant}{Any of the eight parts into which 4$\pi$
  is divided by three mutually perpendicular axes. 
  In particular in referencing the value for the mixing angle
  $\theta_{23}$}

\newduneword{trigcandidate}{trigger candidate}{Summary information derived
  from the full data stream and representing a contribution toward
  forming a \gls{trigdecision}}

\newduneword{trigcommand}{trigger command}{Information derived from
  one or more \glspl{trigcandidate}  that directs elements of the
  \gls{detmodule} to read out a portion of the data stream}

\newduneabbrev{tcm}{TCM}{trigger command message}{A message flowing
  down the trigger hierarchy from global to local context.  Also see \gls{tpm}}

\newduneabbrev{mlt}{MLT}{module level trigger}{The \gls{daq} component responsible for producing a \gls{trigdecision} that will be used to command the readout of a detector module}

\newduneword{trigdecision}{trigger decision}{The process by which
  \glspl{trigcandidate} are converted into \glspl{trigcommand}}

\newduneabbrev{tpm}{TPM}{trigger primitive message}{A message flowing
  up the trigger hierarchy from local to global context.  Also see \gls{tcm}}

\newduneabbrev{ipc}{IPC}{inter-process communication}{A system for software elements to exchange information between threads, local processes or across a data network.  An IPC system is typically specified in terms of protocols  composed of message types and their associated data schema}

\newduneword{daqdispre}{discovery and presence}{As used in the context of the \gls{ipc}, a system that provides mechanisms for a node on a communication network to learn of the existence of peers and their identity (discovery) as well as determine if they are currently operational or have become unresponsive (presence)}

\newduneabbrev{pubsub}{PUB/SUB}{publish-subscribe communication pattern}{An \gls{ipc} communication pattern where one element, the publisher, sends data to all connected elements, the subscribers.  Each subscriber may connect to multiple publishers.  A variant is PUB/SUB with topics where a subscriber may register an identifier, the topic, to limit the information received to just an associated subset}

\newduneabbrev{eb}{EB}{event builder}{A software agent that executes \glspl{trigcommand}  for one  \gls{detmodule} by reading out the requested data}

\newduneabbrev{daqdfo}{DFO}{data flow orchestrator}{The process by which trigger commands are executed in parallel and asynchronous manner by the back-end output subsystem of the \gls{daq}}

\newduneabbrev{daqubi}{UBI}{upstream DAQ buffer interface}{The process which provides read-only access to data residing in the upstream \gls{daq} buffers to processes on the network}

\newduneabbrev{cob}{COB}{cluster on board}{An ATCA motherboard housing four RCEs}

\newduneabbrev{rce}{RCE}{reconfigurable computing element}{Data processor located outside of the cryostat on a \gls{cob} that contains \gls{fpga}, RAM and \gls{ssd} resources, responsible for buffering data, producing trigger primitives, responding to triggered requests for data and synching \gls{snb} dumps}

\newduneabbrev{bow}{BOW}{Bump On Wire}{A working name for the front-end readout computing elements used in the nominal \gls{daq} design to interface the \dual  crates to the \gls{daq} front-end computers}

\newduneabbrev{atca}{ATCA}{Advanced Telecommunications Computing
  Architecture}{An advanced computer architecture specification developed for the telecommunications, military, and aerospace industries that incorporates the latest trends in  high-speed interconnect technologies, next-generation processors, and improved reliability, availability and serviceability} 

\newduneabbrev{utca}{$\mu$TCA}{Micro Telecommunications Computing Architecture}{The computer architecture specification followed by the crates that house charge and light readout electronics in the \gls{dpmod}} 

\newduneabbrev{udp}{UDP}{user datagram protocol}{A simple,
  connectionless Internet protocol that supports data integrity
  checksums, requires no handshaking, and does not guarantee packet delivery}

\newduneabbrev{amc}{AMC}{advanced mezzanine card}{Holds digitizing
  electronics and lives in \gls{utca} crates}

\newduneabbrev{rf}{RF}{radio frequency}{Electromagnetic emissions
  that are within the (radio) frequency band of sensitivity of the detector
  electronics}

\newduneabbrev{fpga}{FPGA}{field programmable gate array}{An
integrated circuit technology that allows the hardware to be reconfigured to
execute different algorithms after its manufacture and deployment}

\newduneabbrev{fmc}{FMC}{FPGA mezzanine card}{Boards holding \glspl{fpga} and other integrated circuitry that attach to a motherboard}

\newduneabbrev{felix}{FELIX}{Front-End Link eXchange}{A
  high-throughput interface between \gls{fe} and trigger electronics
  and the standard PCIe computer bus}

\newduneword{daqpart}{DAQ partition}{A cohesive and
 coherent collection of \gls{daq} hardware and software working together to trigger and read out some portion of one detector module; it consists of an integral number of \glspl{daqfrag}. 
 Multiple \gls{daq} partitions may operate simultaneously, but each instance operates independently}

\newduneabbrev{fec}{DAQ FEC}{DAQ front-end computer}{The portion of one
  \gls{daqpart} that hosts the \gls{daqdr}, \gls{daqbuf} and
  \gls{daqds}.  It hosts the \gls{daqfer} and corresponding portion of the \gls{daqbuf}}

\newduneword{daqfrag}{DAQ front-end fragment}{The portion of one
  \gls{daqpart} relating to a single \gls{fec} and corresponding to an
  integral number of \glspl{detunit}.  See also \gls{datafrag}}

\newduneword{datafrag}{data fragment}{A block of data read out from a single \gls{daqfrag} that
span a contiguous period of time as requested by a \gls{trigcommand}}

\newduneabbrev{daqfer}{FER}{DAQ front-end readout}{The portion of a
  \gls{daqfrag} that accepts data from the detector electronics and
  provides it to the \gls{fec}}

\newduneabbrev{daqdr}{DDR}{DAQ data receiver}{The portion of the
  \gls{daqfrag} that accepts data from the \gls{daqfer}, emits
  trigger candidates produced from the input trigger primitives, and
  forwards the full data stream to the \gls{daqbuf}}

\newduneword{daqbuf}{DAQ primary buffer}{The portion
  of the \gls{daqfrag} that accepts full data stream from the
  corresponding \gls{detunit} and retains it sufficiently long for it
  to be available to produce a \gls{datafrag}}

\newduneword{daqds}{data selector}{The portion of the \gls{daqfrag}
  that accepts \glspl{trigcommand} and returns the corresponding
  \gls{datafrag}.  Not to be confused with \gls{daqdsn}}

\newduneword{daqdsn}{data selection}{The process of forming a trigger decision for selecting a subset of detector data for output by the \gls{daq} from the content of the detector data itself.  Not to be confused with \gls{daqds}}

\newduneabbrev{daqros}{DAQ RO}{DAQ readout subsystem}{The subsystem of the \gls{daq} for accepting and buffering data input from detector electronics}

\newduneabbrev{daqdss}{DAQ DS}{DAQ data selection subsystem}{The subsystem of the \gls{daq} responsible for forming a trigger decision based on a portion of the input data stream.  The majority subset of the \gls{daqtrs}}

\newduneabbrev{daqtrs}{DAQ TS}{DAQ trigger subsystem}{The subsystem of the \gls{daq} responsible for forming a trigger decision}

\newduneabbrev{daqbes}{DAQ BE}{DAQ back-end subsystem}{The portion of the \gls{daq} that is generally toward its output end.  It is responsible for accepting and executing trigger commands and marshaling the data they address to output storage buffers}

\newduneabbrev{daqtss}{DAQ TSS}{DAQ timing and synchronization subsystem}{The portion of the \gls{daq} that provides for timing and synchronization to various components}

\newduneabbrev{femb}{FEMB}{front-end mother board}{Refers a unit of
  the \gls{sp} \gls{ce} that contains the \gls{fe} amplifier
  and \gls{adc} \glspl{asic} covering 128 channels}

\newduneword{asic}{ASIC}{application-specific integrated circuit}

\newduneword{lv}{LV}{low voltage}

\newduneabbrev{iceberg}{ICEBERG}{ICEBERG R\&D cryostat and electronics}{Integrated Cryostat and Electronics Built for Experimental Research Goals:
a new double-walled cryostat built and installed at \gls{fnal} 
for liquid argon detector R\&D and for testing of DUNE detector components}

\newduneword{coldadc}{ColdADC}{A newly developed 16-channels \gls{asic} providing analog to digital conversion}

\newduneword{coldata}{COLDATA}{A 64-channel control and communications \gls{asic}}

\newduneword{cryo}{CRYO}{Integrated ASIC including \gls{fe} circuitry providing signal amplification and pulse shaping, analog to digital conversion, and control and communication functionalities for 64 channels}

\newduneword{larasic}{LArASIC}{A 16-channel \gls{fe} \gls{asic} that provides signal amplification and pulse shaping}

\newduneword{cmos}{CMOS}{Complementary metal-oxide-semiconductor}

\newduneabbrev{enc}{ENC}{equivalent noise charge}{The equivalent noise charge is the input charge that corresponds to a 
$\gls{snr}=1$}

\newduneword{sar}{SAR}{successive approximation register}

\newduneword{protodune}{ProtoDUNE}{Either of the two DUNE prototype detectors constructed at \gls{cern}. 
  One prototype implements \gls{sp} technology and the other \gls{dp}}
  
\newduneword{protodune2}{ProtoDUNE-2}{The second run of a \gls{protodune} detector}

\newduneword{pdsp}{ProtoDUNE-SP}{The \gls{sp} \gls{protodune} detector at \gls{cern}}

\newduneword{pddp}{ProtoDUNE-DP}{The \gls{dp} \gls{protodune} detector at \gls{cern}}

\newduneword{wa105}{WA105 DP demonstrator}{The \SI[product-units=power]{3x1x1}{m} WA105 \gls{dp} prototype detector at \gls{cern}}

\newduneword{rawevent}{DAQ event block}{The unit of data output by the
  \gls{daq}.  
  It contains trigger and detector data spanning a unique, contiguous
  time period and a subset of the detector channels}

\newduneabbrev{ssd}{SSD}{solid-state disk}{Any storage device that
  may provide sufficient write throughput to receive, both collectively and
  distributed, the sustained full rate of data from a \gls{detmodule}
  for many seconds}
\newduneabbrev{nvme}{NVMe}{Non-volatile memory express}{A specification for an interface to storage media attached via PCIe}

\newduneabbrev{hlt}{HLT}{high-level trigger}{This is actually a filter applied to data that has been triggered and aggregated in order to further reduce or characterize it}

\newduneabbrev{pid}{PID}{particle ID}{Particle identification}

\newduneword{readout window}{readout window}{A fixed, atomic and
  continuous period of time over which data from a \gls{detmodule}, in
  whole or in part, is recorded. 
  This period may differ based on the trigger that initiated the
  readout}

\newduneabbrev{zs}{ZS}{zero-suppression}{Used to delete some portion of a
  data stream that does not significantly deviate from zero or
  intrinsic noise levels. 
  It may be applied at different granularity from per-channel to per
  \gls{detunit}}

\newduneabbrev{rc}{RC}{run control}{The system for configuring,
  starting and terminating the \gls{daq}}

\newduneword{r-c}{RC}{resistive-capacitive (circuit)}

\newduneabbrev{daqccm}{CCM}{DAQ control, configuration and monitoring subsystem}{A system for controlling, configuring and monitoring other systems in particular those that make up the \gls{daq} where the CCM encompasses \gls{rc}}

\newduneword{daqrun}{DAQ run}{A period of time over which relevant data taking conditions and \gls{daq} configuration are asserted to be unchanged. 
  Multiple \gls{daq} runs may occur simultaneously when multiple \glspl{daqpart} are active. 
  This term should not be confused with DUNE experiment or beam ``runs'' that typically span many \gls{daq} runs}
\newduneword{daqrunnum}{DAQ run number}{A monotonically increasing count that uniquely and globally identifies a \gls{daqrun}}

\newduneabbrev{snb}{SNB}{supernova neutrino burst}{A prompt 
  increase in the flux of low-energy neutrinos emitted in the first few seconds of a core-collapse supernova.  It can also refer to a trigger command type that may be due to this phenomenon,
  or detector conditions that mimic its interaction signature}

\newduneabbrev{snble}{SNB/LE}{supernova neutrino burst and low
  energy}{Supernova neutrino burst and low-energy physics program}

\newduneabbrev{snews}{SNEWS}{SuperNova Early Warning System}{A global
  supernova neutrino burst trigger formed by a coincidence of \gls{snb} 
  triggers collected from participating experiments}

\newduneabbrev{pps}{1PPS signal}{one-pulse-per-second signal}{An
  electrical signal with a fast rise time and that arrives in real
  time with a precise period of one second}

\newduneabbrev{sls}{SLS}{spill location system}{A system residing at
  the DUNE far detector site that provides information, possibly
  predictive, indicating periods of time when neutrinos are being
  produced by the \fnal Main Injector beam spills}

\newduneabbrev{wib}{WIB}{warm interface board}{Digital electronics
  situated just outside the \gls{sp} cryostat that receives digital data
  from the \glspl{femb} over cold copper connections and sends it to the \gls{rce}
  \gls{fe} readout hardware}

\newduneabbrev{gps}{GPS}{Global Positioning System}{A satellite-based system that provides a highly accurate \gls{pps} that may be used to synchronize clocks and determine location}

\newduneabbrev{ntp}{NTP}{Network Time Protocol}{A networking protocol that allows synchronizing of clocks to within a few \si{\milli\second} of a time standard on a local network and within a few tens of \si{\milli\second} over the Internet} 

\newduneabbrev{ptproto}{PTP}{Precision Time Protocol}{A networking protocol that allows synchronizing of clocks to within a few \si{\micro\second} of a time standard on a local network} 

\newduneabbrev{irig}{IRIG}{inter-range instrumentation group}{A standards body that defined a time-code standard for transferring timing information}

\newduneabbrev{nic}{NIC}{network interface controller}{Hardware for controlling the interface to a communication network.  Typically, one that obeys the Ethernet protocol}

\newduneabbrev{wiec}{WIEC}{warm interface electronics crate}{Crates mounted on the signal flanges that contain the \glspl{wib}}

\newduneabbrev{ptc}{PTC}{power and timing card}{Cards that provide further processing and distribution of the signals entering and exiting the \gls{sp} cryostat}

\newduneabbrev{ptb}{PTB}{power and timing backplane}{Backplane used to connect the \gls{wib}s and the \gls{ptc}s on the \gls{wiec}. Also connects the \gls{ce} flange on the cryostat penetration}

\newduneabbrev{sipm}{SiPM}{silicon photomultiplier}{A solid-state
  avalanche photodiode sensitive to single \phel signals}

\newduneabbrev{cisc}{CISC}{cryogenic instrumentation and slow controls}{Includes equipment to monitor all detector  components and  \gls{lar} quality and behavior, and provides a control system for many of the detector components}

\newduneword{fte}{FTE}{full-time equivalent. A unit of labor
  for the project. One year of work from one person}

\newduneword{art}{art}{A software framework implementing an
  event-based execution paradigm} 

\newduneabbrev{sam}{SAM}{sequential
  access via metadata}{A data-handling system to store and retrieve
  files and associated metadata, including a complete record of the
  processing that has used the files}

\newduneword{artdaq}{artdaq}{A data acquisition toolkit for data transfer, aggregation and processing}

\newduneword{beamline}{beamline}{A sequence of control and monitoring devices used for the formation of a directed collection of particles}
\newduneabbrev{cdr}{CDR}{conceptual design report}{A formal project
  document 
   that describes the experiment
  at a conceptual level}

\newduneabbrev{cf}{CF}{conventional facilities}{Pertaining to
  construction and operation of buildings and conventional infrastructure, and for \gls{lbnf-dune}, CF includes the excavation caverns}

\newduneabbrev{cp}{CP}{charge parity}{Product of charge and parity
  transformations}

\newduneabbrev{cpt}{CPT}{charge, parity, and time reversal symmetry}{product of charge, parity
  and time-reversal transformations}

\newduneabbrev{cpv}{CPV}{charge-parity symmetry violation}{Lack of
  symmetry in a system before and after charge and parity
  transformations are applied. 
  For CP symmetry to hold,  a particle turns into its
 corresponding antiparticle under a charge transformation, and a parity
transformation inverts its space coordinates, i.e., 
produces the mirror image}

\newduneword{doe}{DOE}{U.S. Department of Energy}

\newduneabbrev{fra}{FRA}{Fermi Research Alliance}{A joint partnership of the University of Chicago and the Universities Research Association (URA) that manages and operates Fermilab on behalf of the \gls{doe}}

\newduneabbrev{dune}{DUNE}{Deep Underground Neutrino Experiment}{A leading-edge, international experiment for neutrino science and proton decay studies}

\newduneabbrev{esh}{ES\&H}{environment, safety and health}{A discipline and specialty that studies and implements practical aspects of environmental protection and safety at work} 

\newduneabbrev{ppe}{PPE}{personnel protective equipment}{Equipment worn to minimize exposure to hazards that cause serious workplace injuries and illnesses}

\newduneabbrev{odh}{ODH}{oxygen deficiency hazard}{a hazard that occurs when inert gases such as nitrogen, helium, or argon displace room air and thus reduce the percentage of oxygen below the level required for human life}

\newduneabbrev{feshm}{FESHM}{Fermilab Environment, Safety and Health Manual}{The document that contains Fermilab's policies and procedures designed to manage environment, safety, and health in all its programs}

\newduneabbrev{fscf}{FSCF}{far site conventional facilities}{The
  \gls{cf} at the DUNE far detector site, \surf}
  
\newduneabbrev{nscf}{NSCF}{near site conventional facilities}{The
  \gls{cf} at the DUNE near detector site, \fnal}

\newduneabbrevs{gut}{GUT}{grand unified theory}{grand unified theories}{A class of theories that unifies the electro-weak and strong forces}

\newduneabbrev{lar}{LAr}{liquid argon}{Argon in its liquid phase; it is a cryogenic liquid with a boiling point of $\SI{-90}{^\circ{C}}$ (\SI{87}{K}) and density of \SI{1.4}{g/ml}}

\newduneabbrev{lbl}{LBL}{long-baseline}{Refers to the distance between the 
  neutrino source  and the \gls{fd}.  It can also refer to the distance between the near and far detectors. 
  The ``long'' designation is an approximate and relative distinction. For DUNE, this distance  (between \gls{fnal} and \gls{surf}) is approximately \SI{1300}{km}}

\newduneabbrev{lbnf}{LBNF}{Long-Baseline Neutrino Facility}{The
  organizational entity responsible for developing the neutrino beam, the cryostats
  and cryogenics systems, and the conventional facilities for DUNE}
  
\newduneabbrev{lbnf-dune}{LBNF/DUNE}{LBNF and DUNE project}{The overall global project, including \gls{lbnf} and \gls{dune}}

\newduneabbrev{lbnc}{LBNC}{Long-Baseline Neutrino Committee}{The committee, composed of internationally prominent scientists with relevant expertise, charged by the \gls{fnal} director to review the scientific, technical, and managerial progress, plans and decisions associated with \gls{dune}}

\newduneabbrev{ncg}{NCG}{Neutrino Cost Group}{A group of internationally prominent scientists with relevant experience that is charged by the \gls{fnal} director to review the cost, schedule, and associated risks for the \gls{dune} experiment}

\newduneabbrev{mh}{MH}{mass hierarchy}{Describes the separation
  between the mass squared differences related to the solar and
  atmospheric neutrino problems}

\newduneabbrev{mi}{MI}{Fermilab Main Injector}{An accelerator at
  \fnal that provides a beam of high-energy protons that upon
  striking a target produce secondaries that decay to provide the
  neutrinos directed toward the DUNE far detector}

\newduneabbrev{pot}{POT}{protons on target}{Typically used as a unit
  of normalization for the number of protons striking the neutrino
  production target}

\newduneabbrev{qa}{QA}{quality assurance}{The set of actions taken to provide confidence that quality requirements are fulfilled, and to detect and correct poor results}

\newduneabbrev{qc}{QC}{quality control}{An aggregate of activities (such as design analysis and inspection for defects) performed to ensure adequate quality in manufactured products}

\newduneabbrev{sm}{SM}{standard model}{Refers to a theory describing
  the interaction of elementary particles}

\newduneabbrev{tdr}{TDR}{technical design report}{A formal project
  document 
  that describes the experiment at a technical level}

\newduneabbrev{prelimdr}{PDR}{preliminary design report}{A formal project
  document 
  that describes the experiment at a preliminary design level}

\newduneabbrev{tp}{IDR}{interim design report}{An intermediate
milestone on the path to a full \gls{tdr}} 

\newduneabbrev{ckm}{CKM matrix}{Cabibbo-Kobayashi-Maskawa
  matrix}{Refers to the matrix describing the mixing between mass and
  weak eigenstates of quarks}

\newduneabbrev{cl}{CL}{confidence level}{Refers to a probability
  used to determine the value of a random variable given its
  distribution}

\newduneabbrev{pmns}{PMNS}{Pontecorvo-Maki-Nakagawa-Sakata}{A type of matrix that describes the mixing between mass and weak eigenstates of
  the neutrino}

\newduneabbrevs{cpa}{CPA}{cathode plane assembly}{cathode plane assemblies}{The component of the \single detector module that provides the drift HV cathode}

\newduneabbrev{fc}{FC}{field cage}{The component of a \gls{lartpc} that contains and shapes the applied \efield}

\newduneword{cpafc}{CPA/FC}{A pair of \gls{cpa} panels and the top and bottom \gls{fc} portions that attach to the pair; an intermediate assembly for installation into the \gls{spmod} }

\newduneabbrev{topfc}{top FC}{top field cage}{The horizontal portions of the \gls{sp} \gls{fc}   on the top of the \gls{tpc}}

\newduneabbrev{botfc}{bottom FC}{bottom field cage}{The horizontal portions of the \gls{sp} \gls{fc} on the bottom of the \gls{tpc}}

\newduneabbrev{ewfc}{endwall FC}{endwall field cage}{The vertical portions of the \gls{sp} \gls{fc} near the wall}

\newduneabbrev{gp}{GP}{ground plane}{An electrode held electrically neutral relative to Earth ground voltage; it is mounted on the \gls{fc} in a \gls{spmod} to protect the cryostat wall}

  \newduneword{gg}{ground grid}{An electrode held electrically neutral relative to Earth ground voltage; it is installed between the cathode and the \glspl{pd} in a \gls{dpmod} to protect the \glspl{pmt}, maintaining high transparency to light}

\newduneabbrev{alara}{ALARA}{as low as reasonably
  achievable}{Typically used with regard management of radiation
  exposure but may be used more generally. It means making every
  reasonable effort to maintain e.g., exposures, to as far below the
  limits as practical, consistent with the purpose for that the
  activity is undertaken}

\newduneabbrev{ecal}{ECAL}{electromagnetic calorimeter}{A detector
  component that measures energy deposition of traversing particles (in the near detector conceptual design)}

\newduneabbrev{hv}{HV}{high voltage}{Generally describes a voltage
  applied to drive the motion of free electrons through some media, e.g., LAr}

\newduneword{spmod}{SP module}{single-phase DUNE \gls{fd} module}
\newduneword{dpmod}{DP module}{dual-phase DUNE \gls{fd} module}

\newduneabbrev{tcoord}{TC}{technical coordinator}{A member of the \gls{dune} management team responsible for organizing the technical aspects of the project effort; is head of \gls{tc}}

\newduneabbrev{rcoord}{RC}{resource coordinator}{A member of the \gls{dune} management team responsible for coordinating the financial resources of the project effort}

\newduneword{tc}{technical coordination}{The DUNE organization responsible for overall integration 
of the detector elements and successful execution of the detector
construction project; areas of responsibility include 
general project oversight, systems engineering, \gls{qa} 
and safety}

\newduneabbrev{exb}{EB}{executive board}{The highest level DUNE
  decision-making body for the collaboration}

\newduneabbrev{tb}{TB}{technical board}{The DUNE organization responsible for
  evaluating technical decisions}

\newduneabbrev{rrb}{RRB}{Resources Review Board}{A part of \gls{dune}'s international project governance structure, composed of representatives of all funding agencies that sponsor the project, and of  \gls{fnal} management, established to provide coordination among funding partners and oversight of \gls{dune}}

\newduneabbrev{inc}{INC}{International Neutrino Council}{A highest-level international advisory body to the U.S. \gls{doe} and the  \gls{fnal} directorate on matters related to the  \gls{lbnf} and the  \gls{pip2} projects. This council is composed of representatives from the international funding agencies and  \gls{cern} that make major contributions the infrastructure}

\newduneabbrev{cc}{CC}{charged current}{Refers to an interaction
  between elementary particles where a charged weak force carrier
  ($W^+$ or $W^-$) is exchanged}

\newduneabbrev{dis}{DIS}{deep inelastic scattering}{Refers to the 
  interaction of an elementary charged particle with a nucleus in an
  energy range where the interaction can be modeled as taking place with
  individual nucleons}

\newduneabbrev{fsi}{FSI}{final-state interactions}{Refers to
  interactions between elementary or composite particles subsequent to
  the initial, fundamental particle interaction, such as may occur as
  the products exit a nucleus}

\newduneword{geant4}{Geant4}{A
  software toolkit for the simulation of the passage of particles
  through matter using \gls{mc} methods}

\newduneabbrev{genie}{GENIE}{Generates Events for Neutrino Interaction
  Experiments}{Software providing an object-oriented neutrino
  interaction simulation resulting in kinematics of the products of
  the interaction}

\newduneabbrev{mc}{MC}{Monte Carlo}{Refers to a method of numerical
  integration that entails the statistical sampling of the integrand
  function. 
  Forms the basis for some types of detector and physics simulations}

\newduneabbrev{qe}{QE}{quasi-elastic}{Refers to interaction between
  elementary particles and a nucleus in an energy range where the
  interaction can be modeled as occurring between constituent quarks
  of one nucleon and resulting in no bulk recoil of the resulting
  nucleus}

\newduneabbrev{mou}{MoU}{memorandum of understanding}{A document
  summarizing an agreement between two or more parties}

\newduneabbrev{pip2}{PIP-II}{Proton Improvement Plan II}{A \gls{fnal} project for
  improving the protons on target delivered delivered by the \gls{lbnf} neutrino production beam. 
  This is version two of this plan and it is planned to be followed by a PIP-III}
  
\newduneabbrev{sdsta}{SDSTA}{South Dakota Science and Technology
  Authority}{The legal entity that manages \gls{surf}, in Lead, S.D}
  
\newduneabbrev{sdsd}{SDSD}{Fermilab South Dakota Services Division}{A Fermilab division responsible providing host laboratory functions at SURF in South Dakota}

\newduneabbrev{firus}{FIRUS}{Facility Information Reporting Utility System}
 {The safety system at \surf}

\newduneabbrev{bsi}{BSI}{building and site infrastructure}
 {The work package for outfitting of the \gls{lbnf} underground infrastructure}

\newduneabbrev{wbs}{WBS}{work breakdown structure}{An organizational
  project management tool by which the tasks to be performed are
  partitioned in a hierarchical manner}

\newduneabbrev{br}{BR}{branching ratio}{A fractional probability for a
  decay of a composite particle to occur into some specified set or
  sets of products}
\newduneword{bsm}{BSM}{beyond the standard model}

\newduneabbrev{dm}{DM}{dark matter}{The term given to the unknown
  matter or force that explains measurements of galaxy motion 
  that are otherwise inconsistent with the amount of mass associated
  with the observed amount of photon production}
  
  \newduneabbrev{bdm}{BDM}{boosted dark matter}{A new model that describes a relativistic dark matter particle boosted by the annihilation of heavier dark matter participles in the galactic center or the sun}

\newduneabbrev{cern}{CERN}{European Organization for Nuclear
Research}{The leading particle physics laboratory in Europe and home to the ProtoDUNEs. (In French, the Organisation Europ\'{e}enne pour la Recherche Nucl\'{e}aire, derived from Conseil Europ\'{e}en pour la Recherche Nucl\'{e}aire}

\newduneabbrev{dsnb}{DSNB}{diffuse supernova neutrino background}{The
  term describing the pervasive, constant flux of neutrinos due to all
  past supernova neutrino bursts}

\newduneabbrev{espp}{ESPP}{European Strategy for Particle Physics}{The
cornerstone of Europe's
decision-making process for the long-term future of the
field. Mandated by the \gls{cern} Council, it is formed through a broad
consultation of the grass-roots particle physics community, it
actively solicits the opinions of physicists from around the world,
and it is developed in close coordination with similar processes in
the USA and Japan in order to ensure coordination between regions and
optimal use of resources globally}

\newduneabbrev{gar}{GAr}{gaseous argon}{argon in its gas phase}
\newduneabbrev{gartpc}{GArTPC}{gaseous argon time-projection chamber}{A \gls{tpc} filled with gaseous argon; a possible technology choice for the \gls{nd}}

\newduneabbrev{globes}{GLoBES}{General Long-Baseline Experiment
  Simulator}{A software package for simulating energy spectra of
  neutrino flux, interactions, and energy spectra measured after application of some
  model of a detector response)}

\newduneabbrev{snowglobes}{SNOwGLoBES}{SuperNova
Observatories with GLoBES} {From the official description~\cite{snowglobes}: 
SNOwGLoBES is public software for computing interaction rates and distributions of observed quantities for \gls{snb} neutrinos in common detector materials} 

\newduneword{l/e}{L/E}{length-to-energy ratio}
\newduneword{lri}{LRI}{long-range interactions}

\newduneabbrev{nc}{NC}{neutral current}{Refers to an interaction
  between elementary particles where a neutrally charged weak force carrier
  ($Z^0$) is exchanged}

\newduneabbrev{nh}{NH}{normal hierarchy}{Refers to the neutrino mass
  eigenstate ordering whereby the sign of the mass squared difference
  associated with the atmospheric neutrino problem is positive}

\newduneabbrev{ih}{IH}{inverted hierarchy}{Refers to the neutrino mass
  eigenstate ordering whereby the sign of the mass squared difference
  associated with the atmospheric neutrino problem is negative}

\newduneabbrev{no}{NO}{normal ordering}{Refers to the neutrino mass
  eigenstate ordering whereby the sign of the mass squared difference
  associated with the atmospheric neutrino problem is positive}

\newduneabbrev{io}{IO}{inverted ordering}{Refers to the neutrino mass
  eigenstate ordering whereby the sign of the mass squared difference
  associated with the atmospheric neutrino problem is negative}

\newduneabbrev{msw}{MSW}{Mikheyev-Smirnov-Wolfenstein effect}{Explains
  the oscillatory behavior of neutrinos produced inside the sun as
  they traverse the solar matter}

\newduneabbrev{nsi}{NSI}{nonstandard interaction}{A general class of
  theory of elementary particles other than the Standard Model}

\newduneabbrev{pfive}{P5}{Particle Physics Project Prioritization
Panel}{The Particle Physics Project Prioritization Panel (P5) was a
subpanel of the High Energy Physics Advisory Panel (HEPAP). It completed
its Report, a ten-year strategic plan for high energy physics in the
U.S., in 2014. This report included a recommendation that ``host a world-leading neutrino
program that will have an optimized set of short- and long-baseline neutrino oscillation experiments, and its long-term focus
is a reformulated venture referred to here as the Long Baseline
Neutrino Facility (LBNF)''}

\newduneabbrev{sme}{SME}{standard-model extension}{an effective field theory that contains the \gls{sm}, general relativity, and all possible operators that break Lorentz symmetry (Wikipedia)}

\newduneabbrev{susy}{SUSY}{supersymmetry}{Theoretical symmetry between a fermion and a boson}

\newduneabbrev{wimp}{WIMP}{weakly-interacting massive particle}{A
  hypothesized particle that may be a component of dark matter}

\newduneabbrev{ce}{CE}{cold electronics}{Analog and digital readout electronics that operate at cryogenic temperatures}

\newduneabbrev{crp}{CRP}{charge-readout plane}{In the \gls{dp} technology, a  collection of
  electrodes in a planar arrangement placed at a particular voltage
  relative to some applied \efield such that drifting electrons
  may be collected and their number and time may be measured}

\newduneabbrev{dram}{DRAM}{dynamic random access memory}{A computer memory technology}

\newduneabbrev{fnal}{Fermilab}
{Fermi National Accelerator Laboratory}{U.S. national laboratory in Batavia, IL. It is the laboratory that hosts \gls{dune} and serves as its near site}

\newduneabbrev{bnl}{BNL}{Brookhaven National Laboratory}{US national laboratory in Upton, NY}

\newduneabbrev{slac}{SLAC}{SLAC National Accelerator Laboratory}{US national laboratory in Menlo Park, CA}

\newduneabbrev{lbnl}{LBNL}{Lawrence Berkeley National Laboratory}{US national laboratory in Berkeley, CA}

\newduneabbrev{anl}{ANL}{Argonne National Laboratory}{US national laboratory in Lemont, IL}

\newduneabbrev{lanl}{LANL}{Los Alamos National Laboratory}{US national laboratory in Los Alamos, NM}

\newduneabbrev{fs}{FS}{full stream}{Relates to a data stream that has not undergone selection, compression or other form of reduction}

\newduneabbrev{lem}{LEM}{large electron multiplier}{A micro-pattern detector suitable for use in ultra-pure argon vapor; LEMs consist of copper-clad PCB boards with sub-millimeter-size holes through which electrons undergo amplification}

\newduneabbrev{lng}{LNG}{liquefied natural gas}{Pertaining to natural gas in its liquid phase}

\newduneabbrev{mip}{MIP}{minimum ionizing particle}{Refers to a
  particle traversing some medium such that the particle's mean energy loss is  
  near the minimum}

\newduneabbrev{pd}{PD}{photon detector}{The detector
  elements involved in measurement of the number and arrival times of
  optical photons produced in a detector module} 

\newduneabbrev{pmt}{PMT}{photomultiplier tube}{A device that makes use
  of the photoelectric effect to produce an electrical signal from the
  arrival of optical photons}

\newduneabbrev{ppm}{ppm}{parts per million}{A concentration equal to one part in $10^{-6}$}
\newduneabbrev{ppb}{ppb}{parts per billion}{A concentration equal to one part in $10^{-9}$}
\newduneabbrev{ppt}{ppt}{parts per trillion}{A concentration equal to one part in $10^{-12}$}

\newduneword{rio}{RIO}{reconfigurable input output}

\newduneabbrev{s/n}{S/N}{signal-to-noise}{signal-to-noise ratio}
\newduneword{snr}{\mbox{S/N}}{signal-to-noise ratio}

\newduneword{ssp}{SSP}{SiPM signal processor}

\newduneabbrev{sbn}{SBN}{Short-Baseline Neutrino}{A \gls{fnal} program consisting of three collaborations, \gls{microboone}, \gls{sbnd}, and \gls{icarus}, to perform sensitive searches for $\nue$ appearance and $\numu$ disappearance in the Booster Neutrino Beam}

\newduneword{stt}{STT}{straw tube tracker}

\newduneword{wire board}{wire board}{At the head end of the APA in the \single TPC, stacks of electronics boards referred to as ``wire boards'' are arrayed to anchor the wires.  They also provide the connection between the wires and the cold electronics} 

\newduneabbrev{wls}{WLS}{wavelength-shifting}{A material or process by
  which incident photons are absorbed by a material and photons are
  emitted at a different, typically longer, wavelength}
  
\newduneabbrev{tpb}{TPB}{tetra-phenyl butadiene}{A 
\gls{wls} material}

\newduneabbrev{ptp}{PTP}{p-terphenyl}{A 
\gls{wls} material}

\newduneabbrev{sft}{SFT}{signal feedthrough}{A cryostat penetration allowing for the passage of cables or other extended parts}
\newduneabbrev{sftchimney}{SFT chimney}{signal feedthrough chimney}{In the \dual technology, a volume above the cryostat penetration used for a signal feedthrough}

\newduneabbrev{catiroc}{CATIROC}{charge and time integrated readout chip}{A complete read-out chip manufactured in AustriaMicroSystem designed to read arrays of 16 photomultipliers}

\newduneabbrev{wr}{WR}{White Rabbit}{A component of the timing system that forwards clock signal and time-of-day reference data to the master timing unit}

\newduneabbrev{mch}{MCH}{MicroTCA Carrier Hub}{An network switching device}

\newduneabbrev{wrmch}{WR-MCH}{White Rabbit \gls{utca} Carrier Hub}{A card mounted in \gls{utca} crate that recieves time syncronization information and trigger data packets over \gls{wr} network and disributes them to the \gls{amc} over \gls{utca} backplane} 

\newduneabbrev{wrtsn}{WR-TSN}{White Rabbit TimeStamping Node}{A unit on the \gls{wr} network that timestamps the trigger signals and sends out trigger data packets to \gls{wrmch}}

\newduneword{cmp}{CMP}{configuration management plan}
\newduneword{qap}{QAP}{quality assurance plan} 
\newduneword{ieshp}{IESHP}{integrated environmental, safety and health plan}
\newduneword{dmp}{DMP}{data management plan} 
\newduneword{qam}{QAM}{quality assurance manager} 

\newduneabbrev{dss}{DSS}{detector support system}{The system used to support a \gls{sp} \gls{detmodule} within its cryostat}

\newduneabbrev{ddss}{DDSS}{DUNE detector safety system}{The system used to manage key aspects of detector safety}

\newduneabbrev{lc}{LC}{logistics center}{A facility where \gls{lbnf} and \gls{dune} components will be received and transhipped to \gls{surf}}

\newduneabbrev{tco}{TCO}{temporary construction opening}{An opening in the side of a cryostat through which detector elements are brought into the cryostat; utilized during construction and installation}

\newduneabbrev{surf}{SURF}{Sanford Underground Research Facility}{The laboratory in South Dakota where the \gls{lbnf} \gls{fscf} will be constructed and the \gls{dune} \gls{fd} will be installed and operated}

\newduneabbrev{sit}{SIT}{surface installation team}{An organizational unit responsible for logistics and integration in South Dakota}

\newduneabbrev{uit}{UIT}{underground installation team}{An organizational unit responsible for installation in the underground area at the \gls{surf} site}

\newduneabbrev{cmgc}{CMGC}{construction manager/general contractor}{The organizational unit responsible for management of the construction of conventional facilities at the underground area at the \surf site}

\newduneword{cdrev}{conceptual design review}{A project management device by which a conceptual design is reviewed} 
\newduneword{pdr}{preliminary design review}{A project management device by which an early design is reviewed} 
\newduneword{fdr}{final design review}{A project management device by which a final design is reviewed}
\newduneword{prr}{production readiness review}{A project management device by which the production readiness is reviewed}
\newduneword{irr}{installation readiness review}{A project management device by which the plan for installation is reviewed}
\newduneword{orr}{operational readiness review}{A project management device by which the operational readiness is reviewed}
\newduneword{ppr}{production progress review}{A project management device by which the progress of production is reviewed} 
\newduneabbrev{edms}{EDMS}{engineering document management system}{A computerized document management system developed and supported at \gls{cern} in which some DUNE documents, drawings and engineering models are managed}
\newduneabbrev{ecr}{ECR}{engineering change request}{The first step in the change control process in which a proposed change is described}
\newduneabbrev{docdb}{DocDB}{Document DataBase}{A computerized document management system developed and supported at \gls{fnal} in which virtually all LBNF and most DUNE documents are managed}

\newduneword{wrgm}{WR grandmaster}{White Rabbit grandmaster}

\newduneabbrev{larsoft}{LArSoft}{Liquid Argon Software}{A shared base of physics software across \lartpc experiments}
\newduneword{nova}{NOvA}{The \nova off-axis neutrino oscillation experiment at \gls{fnal}}
\newduneword{minerva}{MINERvA}{The \minerva neutrino cross sections experiment at  \gls{fnal}}
\newduneword{microboone}{MicroBooNE}{The \lartpc-based \microboone neutrino oscillation experiment at  \gls{fnal}}
\newduneword{sbnd}{SBND}{The Short-Baseline Near Detector experiment at  \gls{fnal}}
\newduneabbrev{nexo}{nEXO}{Enriched Xenon Observatory}{Experiment at Lawrence Livermore National Laboratory (U.S. national lab in Livermore, CA)searching for new physics with neutrinoless double-beta decay}
\newduneword{argoneut}{ArgoNeuT}{The ArgoNeuT test-beam experiment and \gls{lartpc} prototype at  \gls{fnal}}
\newduneword{icarus}{ICARUS}{A neutrino experiment that was located at the Laboratori Nazionali del Gran Sasso (LNGS) in Italy, then refurbished at \gls{cern} for re-use in the same neutrino beam from \gls{fnal} used by the MiniBooNE, \gls{microboone} and \gls{sbnd} experiments. The ICARUS detector is being reassembled at \gls{fnal}}
\newduneword{atlas}{ATLAS}{One of two general-purpose detectors at the \gls{lhc}. It investigates a wide range of physics, from the search for the Higgs boson to extra dimensions and particles that could make up \gls{dm}}

\newduneword{lbne}{LBNE}{Long Baseline Neutrino Experiment (a terminated US project that was reformulated in 2014 under the auspices of the new \gls{dune} collaboration, an internationally coordinated and internationally funded program, with \gls{fnal} as host)}

\newduneabbrev{lbno}{LBNO}{Long Baseline Neutrino Observatory} {A terminated European project that, during its six-year duration, assessed the feasibility of a next-generation deep underground neutrino observatory in Europe)}

\newduneword{wirecell}{Wire-Cell}{A tomographic automated \threed neutrino event reconstruction method for \lartpc{}s}
\newduneabbrev{wct}{WCT}{Wire-Cell Toolkit}{A software toolkit with data flow processing components for \lartpc noise and signal simulation, noise filtering, signal processing, and tomographic \threed ionization activity imaging}
\newduneword{ftslite}{F-FTS-lite}{Light-weight version of the \fnal File Transfer system used for rapid data transfers out of the online systems}
\newduneabbrev{fts}{FTS}{File Transfer System}{A file transfer system developed at \fnal to catalog and move data to permanent storage}

\newduneword{35t}{35 ton prototype}{A prototype cryostat and \gls{sp} detector built at \fnal before the \gls{protodune} detectors}

\newduneabbrev{mcr}{MCR}{main communications room}{Space at the \gls{fd} site for cyber infrastructure}

\newduneabbrev{cuc}{CUC}{central utility cavern}{The utility cavern at the 4850L of \gls{surf} located between the two detector caverns. It contains utilities such as central cryogenics and other systems, and the underground data center and control room}

\newduneabbrev{cfd}{CFD}{computational fluid dynamics}{High performance computer-assisted modeling of fluid dynamical systems}
\newduneword{vuv}{VUV}{vacuum ultra-violet}
\newduneword{tallbo}{TallBo}{A cylindrical cryostat at \gls{fnal} primarily used for developing scintillation light collection technologies for \gls{lartpc} detectors}

\newduneword{root}{ROOT}{A modular scientific software toolkit. It provides all the functionalities needed to deal with big data processing, statistical analysis, visualisation and storage. It is mainly written in C++ but integrated with other languages such as Python and R}

\newduneabbrev{eos}{EOS}{EOS}{The XRootD-based distributed file system developed by CERN}
\newduneabbrev{ehn1}{EHN1}{Experiment Hall North One}{Location at CERN of the ProtoDUNE experiments}
\newduneword{led}{LED}{Light-emitting diode}
\newduneabbrev{rtd}{RTD}{resistance temperature detector}{A temperature sensor consisting of a material with an accurate and reproducible resistance/temperature relationship}
\newduneword{swc}{SWC}{Software \& Computing}
\newduneabbrev{las}{LAS}{LEM-anode Sandwich}{In the \dual technology, a \gls{lem} and its corresponding anode are mounted together in a module called a LEM-anode sandwich}

\newduneword{roi}{ROI}{region of interest}
\newduneabbrev{hpc}{HPC}{high-performance computing}{high-performance computing facilities; generally computing facilities emphasizing parallel computing with aggregate power of more than a teraflop}

\newduneword{comfund}{common fund}{The shared resources of the collaboration}
\newduneabbrev{ims}{IMS}{integrated master schedule}{A project management device consisting of linked tasks and milestones}

\newduneword{hvdb}{HVDB}{HV divider board}
\newduneword{sas}{SAS}{Another term for the materials airlock; a pass-through chamber used to ensure safe transfer of materials into a clean room, avoiding contamination in both directions}

\newduneabbrev{fea}{FEA}{finite element analysis}{Simulation of a physical phenomenon using the numerical technique called Finite Element Method (FEM), a numerical method for solving problems of engineering and mathematical physics}

\newduneword{fss}{FSS}{field shaping strips}
\newduneword{lvds}{LVDS}{low-voltage differential signaling}

\newduneword{esd}{ESD}{electrostatic discharge}

\newduneabbrev{rp}{RP}{resistive panel}{Resistive panels form the constant potential surfaces for a \gls{spmod} \gls{cpa}; they are composed of a thin layer of carbon-impregnated Kapton and laminated to both sides of a \frfour sheet}

\newduneword{uhmwpe}{UHMWPE}{ultra-high molecular weight polyethylene}

\newduneword{cts}{CTS}{Cryogenic Test System}
\newduneword{plc}{PLC}{programmable logic controller}

\newduneword{mppc}{MPPC}{\SI{6}{mm}$\times$\SI{6}{mm} Multi-Pixel Photon Counters produced by Hamamatsu\texttrademark{} Photonics K.K}

\newduneabbrev{sfp}{SFP}{small form-factor pluggable}{a particular standard for optical transceivers}

\newduneabbrev{minipod}{MiniPOD}{miniature parallel optical device}{a family of types of multi-channel optical transceivers}

\newduneword{ccc}{CCC}{configuration change command}
\newduneword{act}{ACT}{activation time stamp}
\newduneword{lcm}{LCM}{light calibration module}
\newduneword{lpm}{LPM}{light pulser module}
\newduneword{dac}{DAC}{digital-to-analog converter}
\newduneword{arapuca}{ARAPUCA}{A \gls{pds} design that consists of a light trap that captures wavelength-shifted photons inside boxes with highly reflective internal surfaces until they are eventually detected by \gls{sipm} detectors or are lost}
\newduneword{sarapu}{S-ARAPUCA}{Standard \gls{arapuca} design with different \gls{wls} coatings on both faces of the dichroic filter window(s) of the cell}
\newduneword{xarapu}{X-ARAPUCA}{Extended \gls{arapuca} design with \gls{wls} coating on only the external face of the dichroic filter window(s) but with a \gls{wls} doped plate inside the cell}
\newduneword{feb}{FEB}{front-end board}

\newduneabbrev{lsnd}{LSND}{Liquid Scintilator Neutrino Detector}{A scintillation detector and associated experiment located at Los Alamos National Laboratory}

\newduneabbrev{cvn}{CVN}{convolutional visual network}{An algorithm for identifying neutrino interactions based on their topology and without the need for detailed reconstruction algorithms}

\newduneword{pandora}{Pandora}{The Pandora multi-algorithm approach to pattern recognition} 

\newduneabbrev{pma}{PMA}{Projection Matching Algorithm}{A reconstruction algorithm that combines \twod reconstructed objects to form a \threed representation}
\newduneabbrev{bdt}{BDT}{boosted decision tree}{A method of multivariate analysis}
\newduneabbrev{cnn}{CNN}{convolutional neural network}{A deep learning technique most commonly applied to analyzing visual imagery}
\newduneword{pdg}{PDG}{Particle Data Group}

\newduneword{pci}{PCI}{Peripheral Component Interconnect}

\newduneword{labview}{LabVIEW}{Laboratory Virtual Instrument Engineering Workbench is a system-design platform and development environment for a visual programming language from National Instruments}

\newduneword{pcb}{PCB}{printed circuit board}

\newduneword{crio}{cRIO}{Compact Reconfigurable Input Output}

\newduneword{dcs}{DCS}{Distributed Communications System}

\newduneword{opc-ua}{OPC-UA}{OPC  Unified Architecture is a machine to machine communication protocol for industrial automation developed by the OPC Foundation. OPC stands for Object Linking and Embedding for Process Control}

\newduneword{cabangle}{Cabibbo angle}{A quark mixing parameter that governs the coupling of up quarks to strange quarks}
\newduneword{valor}{VALOR}{A neutrino oscillation fitting framework that is used by \gls{t2k}; the name stands for VALencia-Oxford-Rutherford, the original three institutions that developed it}
\newduneword{cafana}{CAFAna}{Common Analysis File Analysis}
\newduneabbrev{pca}{PCA}{principal component analysis}{A statistical procedure that uses an orthogonal transformation to convert a set of observations of possibly correlated variables into a set of values of linearly uncorrelated variables called principal components (Wikipedia)}
\newduneword{numi}{NuMI}{a set of facilities at \fnal, collectively called ``Neutrinos at the Main Injector.''  The NuMI neutrino beamline target system converts an intense proton beam into a focused neutrino beam}
\newduneword{gibuu}{GiBUU}{Giessen Boltzmann-Uehling-Uhlenback Project; a unified theory and transport framework in the MeV and GeV energy regimes for elementary reactions on nuclei }
\newduneabbrev{rpa}{RPA}{random phase approximation} {an approximation method commonly used for describing the dynamic linear electronic response of electron systems (Wikipedia)}
\newduneword{t2k}{T2K}{T2K (Tokai to Kamioka) is a long-baseline neutrino experiment in Japan studying neutrino oscillations}
\newduneword{mptdet}{MPT detector}{multipurpose tracking detector}

\newduneword{lariat}{LArIAT}{The repurposed ArgoNeuT \gls{lartpc}, modified for use in a charged particle beam, dedicated to the calibration and precise characterization of the output response of these detectors}

\newduneword{captain}{CAPTAIN}{Experimental program sited at \gls{lanl} that is designed to make measurements of scientific importance to \gls{lbl} neutrino physics and physics topics that will be explored by large underground detectors}

\newduneword{dayabay}{Daya Bay}{a neutrino-oscillation experiment in Daya Bay, China, designed to measure the mixing angle $\Theta_{13}$  using antineutrinos produced by the reactors of the Daya Bay and Ling Ao nuclear power plants}

\newduneword{nuwro}{NuWro}{neutrino interaction generator}

\newduneabbrev{neut}{NEUT}{neutrino interaction generator}{A neutrino interaction simulation program library for the studies of atmospheric accelerator neutrinos}

\newduneword{minos}{MINOS}{A long-baseline neutrino experiment, with a near detector at \gls{fnal} and a far detector in the Soudan mine in Minnesota, designed to observe the phenomena of neutrino oscillations (ended data runs in 2012)}

\newduneabbrev{efig}{EFIG}{Experimental Facilities Interface Group}{The body responsible for the required high-level coordination between the \gls{lbnf} and \gls{dune} projects}
\newduneword{ashriver}{Ash River}{The Ash River, Minnesota, USA \gls{nova} experiment far site, used as an assembly test site for \gls{dune}} 

\newduneword{ipd}{project integration director}{Responsible for integration and installation of \gls{lbnf} and \gls{dune} deliverables in South Dakota. Manages the \gls{integoff}}

\newduneabbrev{jpo}{JPO}{Joint Project Office}{The framework through which team members from the LBNF project office, \gls{integoff}, and DUNE \gls{tc} work together to provide coherence in project support functions across the global enterprise. 
Its functions include global project configuration and integration, installation planning and coordination, scheduling, safety assurance, technical review planning and oversight, development of partner agreements, and financial reporting}

\newduneword{ifbeam}{IFbeam}{Database that stores beamline information 
indexed by timestamp}

\newduneabbrev{marley}{MARLEY}{Model of Argon Reaction Low Energy
Yields}{Developed at UC Davis, MARLEY is the first realistic model of
neutrino electron interactions on argon for enegies less than \SI{50}{MeV}. This includes the energy range important for \gls{snb}
neutrinos and also solar 8--boron neutrinos}

\newduneabbrev{es}{ES}{elastic scattering}{Events in which a neutrino
elastically scatters off of another particle}

\newduneabbrev{cno}{CNO}{carbon nitrogen oxygen}{The CNO cycle (for carbon-nitrogen-oxygen) is one of the two known sets of fusion reactions by which stars convert
hydrogen to helium, the other being the proton-proton chain reaction
(pp-chain reaction). In the CNO cycle, four protons fuse, using
carbon, nitrogen, and oxygen isotopes as catalysts, to produce one
alpha particle, two positrons and two electron neutrinos}

\newduneabbrev{sdwf}{SDWF}{South Dakota Warehouse Facility}{Warehousing operations in South Dakota responsible for receiving LBNF and DUNE goods and coordinating shipments to the Ross shaft at \gls{surf}}

\newduneabbrev{wms}{WMS}{warehouse management system}{Commercial software package used to track shipments and interface to freight forwarders. This includes a database for shipping}

\newduneabbrev{dcdb}{DCDB}{DUNE construction database}{Database used by DUNE to track the history and testing of all parts of each \gls{detmodule}}

\newduneabbrev{aup}{AUP}{acceptance for use and possession}{Required for beneficial occupancy of the underground areas at SURF for LBNF and DUNE}

\newduneabbrev{bms}{BMS}{building management system}{Part of the safety system at \gls{surf} that includes the fire and life safety system}
\newduneabbrev{fls}{FLS}{fire and life safety system}{Part of the safety system at \gls{surf}}

\newduneabbrev{sno}{SNO}{Sudbury Neutrino Observatory}{The Sudbury
Neutrino Observatory was a detector built 6800 feet under ground, in
INCO's Creighton mine near Sudbury, Ontario, Canada. SNO was a
heavy-water Cherenkov detector designed to detect neutrinos produced
by fusion reactions in the sun}

\newduneword{sk}{Super-Kamiokande}{Experiment sited in the Kamioka-mine, Hida-city, Gifu, Japan that uses a large water Cherenkov detector to study neutrino properties through the observation of solar neutrinos, atmospheric neutrinos and man-made neutrinos}

\newduneabbrev{id}{ID}{inner diameter}{Inner diameter of a tube}

\newduneabbrev{od}{OD}{outer diameter}{Outer diameter of a tube}

\newduneabbrev{rms}{RMS}{root mean square}{The square root of the arithmetic mean of the squares of a set of values, used as a measure of the typical magnitude of a set of numbers, regardless of their sign}

\newduneabbrev{orc}{ORC}{operational readiness clearance}{Final safety approval prior to the start of operation}

\newduneabbrev{gsc}{GSC group}{global safety coordination group}{DUNE group that evaluates applicable codes and standards, including international code equivalency, for the design, assembly, and installation of the \gls{fd}}

\newduneabbrev{ha}{HA}{hazard analysis}{A first step in a process to assess risk; the result of hazard analysis is the identification of the hazards present for a task or process}
\newduneword{har}{HAR}{hazard analysis report}

\newduneabbrev{tap}{TAP}{trip action plan}{A document required for any trip by a worker to the underground area at \gls{surf}, per that site's access control program; 
it describes the work to be accomplished during the trip} 

\newduneword{em}{EM}{emergency management}
\newduneword{ert}{ERT}{emergency response team}

\newduneabbrev{ndk}{NDK}{nucleon decay}{The hypothetical, baryon number violating decay of a proton or a bound neutron into lighter particles}

\newduneabbrev{emi}{EMI}{electromagnetic interference}{Disturbance generated by an external source that affects an electrical circuit by electromagnetic induction, electrostatic coupling, or conduction}

\newduneabbrev{pe}{PE}{photoelectron}{An electron ejected from the surface of a material by the photoelectric effect}

\newduneabbrev{spe}{SPE}{single photoelectron}{A single photoelectron}

\newduneabbrev{fwhm}{FWHM}{full width at half maximum}{Width of a distribution measured between those points at which the distribution is equal to half of its maximum amplitude}

\newduneabbrev{gdml}{GDML}{geometry description markup language}{An application-indepedent, geometry-description format based on XML}

\newduneabbrev{xml}{XML}{extensible markup language}{A markup language that defines a set of rules for encoding documents in a format that is both human-readable and machine-readable}

\newduneabbrev{crt}{CRT}{cosmic ray tagger}{Detector external to the TPC designed to tag TPC-traversing cosmic ray particles}

\newduneabbrev{sn}{SN}{supernova}{Event that occurs upon the death of certain types of stars}

\newduneabbrev{wg}{WG}{working group}{A group of persons working together to achieve specified goals}

\newduneabbrev{ctsf}{CTSF}{coating, testing and storage facility}{A facility where the the \dual photon detectors will be coated, tested, and stored}

\newduneword{rucio}{Rucio}{Data management system originally developed
by \gls{atlas} but now open-source and shared across HEP}
\newduneabbrev{doma}{DOMA}{data organization, management, and
access}{data organization, management, and access efforts through the
HEP Software Foundation}

\newduneabbrev{hsf}{HSC}{HEP Software Foundation Collaboration}{A foundation that facilitates cooperation and common efforts in high energy physics software and computing internationally}

\newduneabbrev{wlcg}{WLCG}{Worldwide LHC Computing Grid}{Worldwide LHC
Computing Grid}
\newduneabbrev{osg}{OSG}{Open Science Grid}{Open Science Grid}
\newduneabbrev{sci}{SCI}{Scientific Computing Infrastructure}{Proposed
extension of the infrastructure component of \gls{wlcg} to other
experiments}
\newduneabbrev{csc}{CSC}{computing and software consortium}{DUNE
computing and software consortium}

\newduneword{dirac}{DIRAC}{Computing workflow management designed for
LHCb and now used by many HEP experiments}

\newduneword{frp}{FRP}{fiber-reinforced plastic}
\newduneabbrev{hdpe}{HDPE}{high-density polyethylene}{High-density polyethylene plastic}
\newduneword{hvps}{HVPS}{\gls{hv} power supply}
\newduneword{aisi}{AISI}{American Iron and Steel Institute}
\newduneword{ific}{IFIC}{Instituto de Fisica Corpuscular (in Valencia, Spain)}
\newduneabbrev{rsds}{RSDS}{radioactive source deployment system}{Proposed calibration system based on the deployment of
radioactive sources inside the \gls{dune} cryostat}
\newduneword{2p2h}{2p2h}{two particle, two hole}
\newduneabbrev{duneprism}{DUNE-PRISM}{\gls{dune} Precision Reaction-Independent Spectrum Measurement}{a mobile near detector that can perform measurements over a range of angles off-axis from the neutrino beam direction in order to sample many different neutrino energy distributions}
\newduneword{arcube}{ArgonCube}{The name of the core part of the \gls{dune} \gls{nd}, a \gls{lartpc}}

\newduneabbrev{citf}{CITF}{cryogenic instrumentation test facility}{A facility at \fnal with small ($<\,\SI{1}{ton}$) to intermediate ($\sim\,\SI{1}{ton}$) volumes of instrumented, purified TPC-grade \lar, used for testing devices intended for use in \gls{dune}}

\newduneabbrev{3dst}{3DST}{3D scintillator tracker}{The core part of the \threed projection scintillator tracker spectrometer in the near detector conceptual design}
\newduneabbrev{3dsts}{3DST-S}{3D scintillator tracker spectrometer}{The \threed projection scintillator tracker spectrometer  in the near detector conceptual design}
\newduneabbrev{mpd}{MPD}{multi-purpose detector}{A component of the near detector conceptual design; it is a magnetized system consisting of a \gls{hpgtpc} and a surrounding \gls{ecal}}
\newduneabbrev{hpg}{HPG}{high-pressure gas}{gas at high pressure to be used in a \gls{hpgtpc}} 
\newduneabbrev{hpgtpc}{HPgTPC}{high-pressure gaseous argon TPC}{A \gls{tpc} filled with gaseous argon; a possible component of the \gls{dune} \gls{nd}}

\newduneword{src}{SRC}{short-range correlated nucleon-nucleon interactions}
\newduneword{larpix}{LArPix}{ \gls{asic} pixelated charge readout for a \gls{tpc} }
\newduneword{arclt}{ArCLight}{a light detector \gls{arcube} effort}
\newduneword{fhc}{FHC}{forward horn current ($\numu$ mode)}
\newduneword{rhc}{RHC}{reverse horn current ($\overline{\nu}_{\mu}$ mode)}
\newduneword{mwpc}{MWPC}{multi-wire proportional chamber}
\newduneword{na61}{NA61}{CERN hadron production experiment}
\newduneword{pdnd}{ProtoDUNE-ND}{a prototype \gls{dune} \gls{nd}}
\newduneword{ccqe}{CCQE}{charged current quasielastic interaction} 
\newduneabbrev{roc}{ROC}{readout chamber}{readout chamber for gaseous argon \gls{tpc}}
\newduneabbrev{iroc}{IROC}{inner readout chamber}{inner (radial) readout chamber for gaseous argon \gls{tpc}}
\newduneabbrev{oroc}{OROC}{outer readout chamber}{outer (radial) readout chamber for gaseous argon \gls{tpc}}

\newduneword{lux}{LUX}{Large Underground Xenon (LUX) dark matter detector at \gls{surf} }

\newduneword{mjdemo}{Majorana Demonstrator}{Experiment sited at \gls{surf} that  seeks to determine whether neutrinos are their own antiparticles}

\newduneword{lz}{LZ}{Experiment sited at \gls{surf} that  seeks to detect faint interactions between galactic dark matter and regular matter}

\newduneword{mu2e}{Mu2e}{An experiment sited at \gls{fnal} that searches for charged-lepton flavor violation and seeks to discover physics beyond the \gls{sm}}

\newduneword{pdsp2}{ProtoDUNE-SP-2}{A second test run in the singe-phase
ProtoDUNE test stand at CERN, acting as a validation of the final
single-phase detector design}

\newduneword{osha}{OSHA}{Occupational Safety and Health Administration (USA Department of Labor) formed by the Occupational Safety and Health Act of 1970}
\newduneabbrev{pns}{PNS}{pulsed neutron source}{Calibration system based
on neutron capture gamma showers spread out in the whole detector}

\newduneabbrev{fv}{FV}{fiducial volume}{The detector volume within the \gls{tpc} 
that is selected for physics analysis through cuts on reconstructed event position}

\newduneword{p6}{P6}{framework used to plan and status the resource-loaded schedule of activities associated with the USA contributions to \gls{lbnf} and \gls{dune} }
\newduneabbrev{evms}{EVMS}{earned value management system}{Earned Value Management is a systematic approach to the integration and measurement of cost, schedule, and technical (scope) accomplishments on a project or task. It provides both the government and contractors the ability to examine detailed schedule information, critical program and technical milestones, and cost data (text from the US DOE); the EVMS is a system that implements this approach}

\newduneword{core}{CORE}{CORE contributions are in either monetary units or labor hours. They can be technical components for the facility or experiment and the effort of the staff needed to produce, install, and test them;  major facilities for the experiment; or other products and services relevant for the completion of the facility or experiment} 

\newduneabbrev{ahj}{AHJ}{Authority Having Jurisdiction}{An organization, office, or individual responsible for enforcing the requirements of a code or standard, or for approving equipment, materials, an installation, or a procedure (OSHA)}
\newduneword{cte}{CTE}{coefficient of thermal expansion}

\newduneabbrev{opc}{OPC}{open platform communications}{Open platform communications is a series of standards and specifications for industrial telecommunication} 
\newduneword{scada}{SCADA}{supervisory control and data acquisition}
\newduneword{ln}{LN$_2$}{liquid nitrogen}
\newduneabbrev{lapd}{LAPD}{Liquid Argon Purity Demonstrator}{Cryostat at Fermilab for long-term studies requiring a large volume of argon}

\newduneabbrev{pab}{PAB}{Proton Assembly Building}{Home of several \gls{lar} facilities at Fermilab}
\newduneword{hep}{HEP}{high energy physics}
\newduneword{sc}{SC}{scientific computing}  
\newduneword{cms}{CMS}{Compact Muon Solenoid experiment at CERN}
\newduneword{alice}{ALICE}{A Large Ion Collider Experiment, at CERN}
\newduneword{gpib}{GPIB}{general purpose interface bus}

\newduneabbrev{pfparticle}{PFParticle}{particle flow particle}{Each of the individual reconstructed particles in the hierarchy (or particle flow) describing the reconstructed event interaction}

\newduneabbrev{mcparticle}{MCParticle}{Monte Carlo Particle}{Individual true simulated particle}
\newduneword{au}{AU}{astronomical unit}
\newduneword{nufit}{NuFIT 4.0}{The NuFIT 4.0 global fit to neutrino oscillation data}

\newduneabbrev{sgft}{SGFT}{term}{add def (DP install)}
\newduneword{uhv}{UHV}{ultra high vacuum}
\newduneword{lps}{LPS}{laser positioning system}

\newduneword{unicamp}{UNICAMP}{University of Campinas, Sao Paulo, Brazil}
 
\newduneabbrev{fbk}{FBK}{Fondazione Bruno Kessler}{FBK is a research non-profit entity in Trento, Italy that partners in the development of technology with applications in various fields including High Energy Physics}

\newduneword{fft}{FFT}{fast Fourier transform}
\newduneabbrev{enob}{ENOB}{effective number of bits}{The effective number of bits is a measure of the dynamic range of an \gls{adc} and its associated circuitry. The resolution of an \gls{adc} is specified by the number of bits used to represent the analog value, in principle giving 2N signal levels for an N-bit signal. However, all real \gls{adc} circuits introduce noise and distortion. ENOB specifies the resolution of an ideal \gls{adc} circuit that would have the same resolution as the circuit under consideration}
\newduneabbrev{sndr}{SNDR}{signal to noise and distortion ratio}{Also known as SINAD. Ratio of the \gls{rms} signal amplitude to the mean value of the root-sum-square of all other spectral components, including harmonics, but excluding \gls{dc} levels. It is a good indication of the overall dynamic performance of an \gls{adc} because it includes all components which make up noise and distortion}
\newduneabbrev{sfdr}{SFDR}{spurious free dynamic range}{Spurious free dynamic range is the ratio of the \gls{rms} value of the signal to the \gls{rms} value of the worst spurious signal regardless of where it falls in the frequency spectrum. The worst spur may or may not be a harmonic of the original signal}
\newduneabbrev{thd}{THD}{total harmonic distortion}{Total harmonic distortion is the ratio of the \gls{rms} value of the fundamental signal to the mean value of the root-sum-square of its harmonics} 
\newduneword{tvs}{TVS}{transient voltage suppression}

\newduneword{riskprob}{risk probabilities}{The risk probability, after taking into account the planned mitigation activities, is ranked as 
L (low $<\,$\SI{10}{\%}), 
M (medium \SIrange{10}{25}{\%}), or 
H (high $>\,$\SI{25}{\%}). 
The cost and schedule impacts are ranked as 
L (cost increase $<\,$\SI{5}{\%}, schedule delay $<\,$\num{2} months), 
M (\SIrange{5}{25}{\%} and 2--6 months, respectively) and 
H ($>\,$\SI{20}{\%} and $>\,$2 months, respectively)}

\newduneabbrev{lbls}{LBLS}{laser beam location system}
{Auxiliary calibration system providing an independent location measurement of the ionization laser beams direction}

\newduneabbrev{lsst}{LSST}{Large Synoptic Survey Telescope}{8.4 m telescope with 3.2G-pixel camera that will start taking data in 2023}
\newduneabbrev{ska}{SKA}{Square Kilometer Array}{International radio telescope array planned to start data-taking in 2027}
\newduneabbrev{hyperk}{HyperK}{Hyper Kamiokande}{260 kt water Cerenkov neutrino detector to begin construction at Kamiokande in 2020}
\newduneword{lhcb}{LHCb}{LHC experiment dedicated to forward physics}
\newduneword{belleii}{Belle II}{B-factory experiment now running at KEK}

 \newduneabbrev{ldm}{LDM}{light-mass dark matter}{Refers to dark matter particles with mass values much lower than the electroweak scale, specifically below the 1~GeV level}
 
\newduneabbrev{bnv}{BNV}{baryon-number violating}{Describing an interaction where \gls{baryonnumber} is not conserved}

\newduneword{bugey}{Bugey}{Neutrino experiment that operated at the Bugey nuclear power plant in France}

\newduneword{minosplus}{MINOS$+$}{The successor to the \gls{minos} experiment, utilizing the same detectors and beam line, but operating at higher beam energy tune than \gls{minos}, parasitic with \gls{nova}}

\newduneword{baryonnumber}{baryon number}{A quantity expressing the total number of baryons in a system minus the number of antibaryons}

\newduneword{np04}{NP04}{CERN North Area hadron beamline used for the \gls{sp} test beam run}

\newduneword{ua1}{UA1}{UA1 (Underground Area 1) was a particle detector at \gls{cern}'s  Super Proton Synchrotron (SPS). It ran from 1981 until 1990, when the SPS was used as a proton-antiproton collider, searching for traces of W and Z particles in collisions. (CERN) The UA1 dipole magnet was reused in the NOMAD experiment and currently provides the magnetic field for the \gls{t2k} ND280 detector}

\newduneword{ssc}{SSC}{The Superconducting Super Collider was to be a huge underground ring complex beneath the area near Waxahachie, Texas, USA, that would have been the world’s most energetic particle accelerator. It was begun in 1990, but canceled by the U.S. Congress in 1993 (scientificamerican.com Oct 2013)}

\newduneword{daphne}{DAPHNE}{Detector electronics for Acquiring PHotons from NEutrinos is a custom-developed warm front-end waveform digitizing electronics module derived from the readout system developed at Fermilab for the Mu2e experiment}
 
\newduneword{nersc}{NERSC}{National Energy Research Computing Facility at \gls{lbnl}}

  \newduneword{integoff}{integration office}{The office that incorporates the onsite team responsible for coordinating integration and installation activities at SURF}

\newduneabbrev{sma}{SMA}{SubMiniature version A}{Connector interface for coaxial cables
with a screw-type coupling mechanism}

\newduneword{kloe}{KLOE}{KLOE is a $e^+ e^-$ collider detector spectrometer operated at DAFNE, the $\phi$-meson factory at Frascati, Rome. In DUNE it will consist of a \SI{26}{cm} Pb+scintillating fiber ECAL surrounding a cylindrical open detector region that is  \SI{4.00}{m} in diameter and \SI{4.30}{m} long. The ECAL and detector region are embedded in a \SI{0.6}{T} magnetic field created by a \SI{4.86}{m} diameter superconducting coil and a \SI{475}{tonne} iron yoke}

\newduneword{ro}{review office}{An office within the \gls{integoff} that organizes reviews }

\newduneabbrev{doecd}{CD}{critical decision}{The U.S. DOE's Order 413.3B outlines a series of staged project approvals, each of which is referred to as a critical decision (CD)}

\newduneabbrev{lbnfspac}{LBNF SPAC}{LBNF Strategic Project Advisory Committee}{A committee charged by the host laboratory director to provide expert, independent advice on significant issues and strategies related to LBNF project organization, management, and risks}

\newduneabbrev{sand}{SAND}{System for on-Axis Neutrino Detection}{The beam monitor component of the near detector that remains on-axis at all times and serves as a dedicated neutrino spectrum monitor}

\newduneword{4850l}{4850L}{The depth in feet (1480 m) of the top of the cryostats underground at SURF; used more generally to refer to the DUNE underground area. Called the ``4850 level'' or ``4850L''}

\hypersetup{
    pdftitle={\expshort TDR \thedocsubtitle},
    pdfauthor={\expshort Collaboration},
    final=true,
    colorlinks=false,
    linktocpage=true,
    linkbordercolor=blue,
    citebordercolor=green,
    urlbordercolor=magenta,
    filecolor=black,
    pdfpagemode=UseOutlines,
    pdfborderstyle={/S/U},  
}

\sptrue
\renewcommand\thedoctitle{\voltitlesp} 
\newcommand\thevolumenumber{\volnumbersp} 

\begin{document}

\pagestyle{titlepage}
\includepdf[pages={-}]{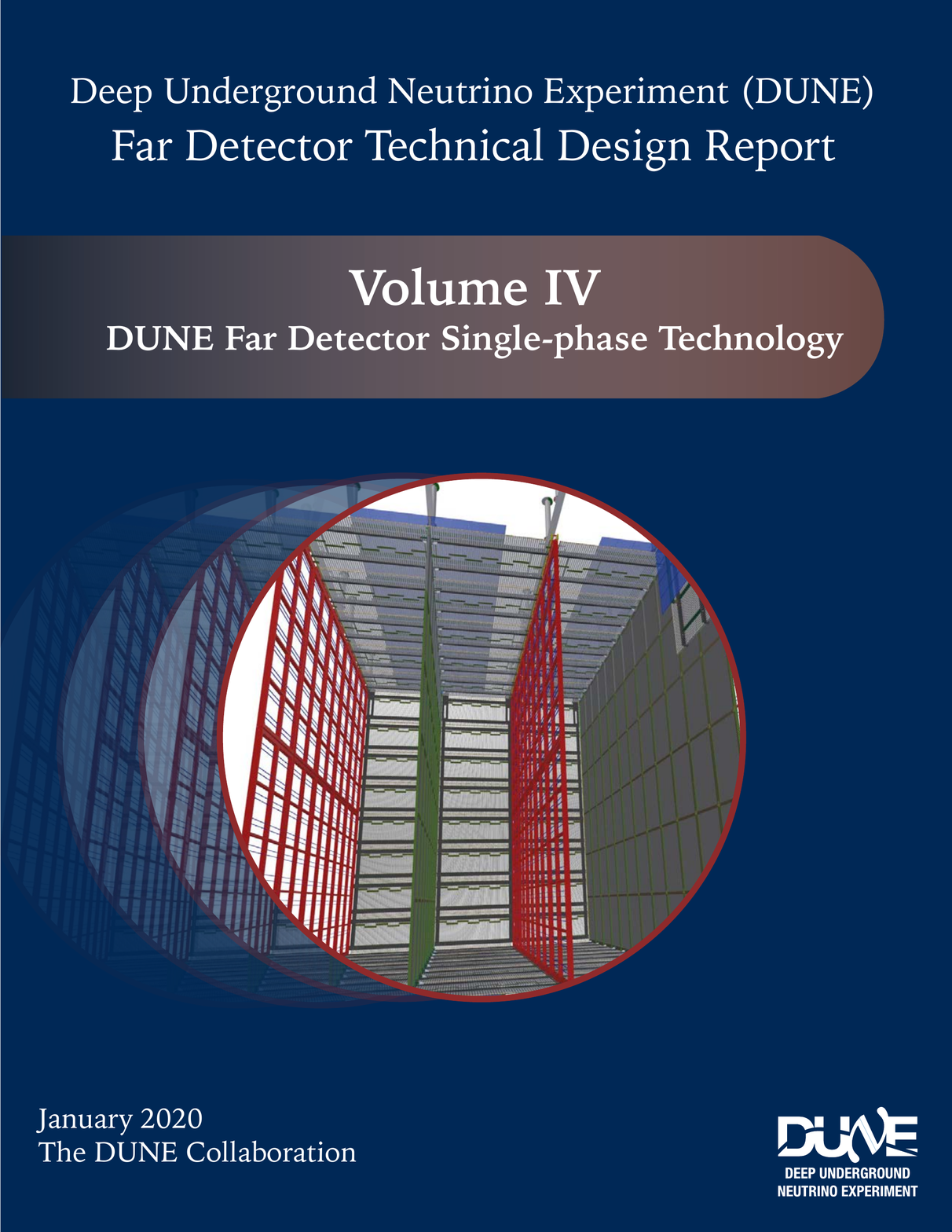}
\cleardoublepage

\cleardoublepage
\vspace*{16cm} 
  {\small  This document was prepared by the DUNE collaboration using the resources of the Fermi National Accelerator Laboratory (Fermilab), a U.S. Department of Energy, Office of Science, HEP User Facility. Fermilab is managed by Fermi Research Alliance, LLC (FRA), acting under Contract No. DE-AC02-07CH11359.
  
The DUNE collaboration also acknowledges the international, national, and regional funding agencies supporting the institutions who have contributed to completing this Technical Design Report.  
  }
\includepdf[pages={-}]{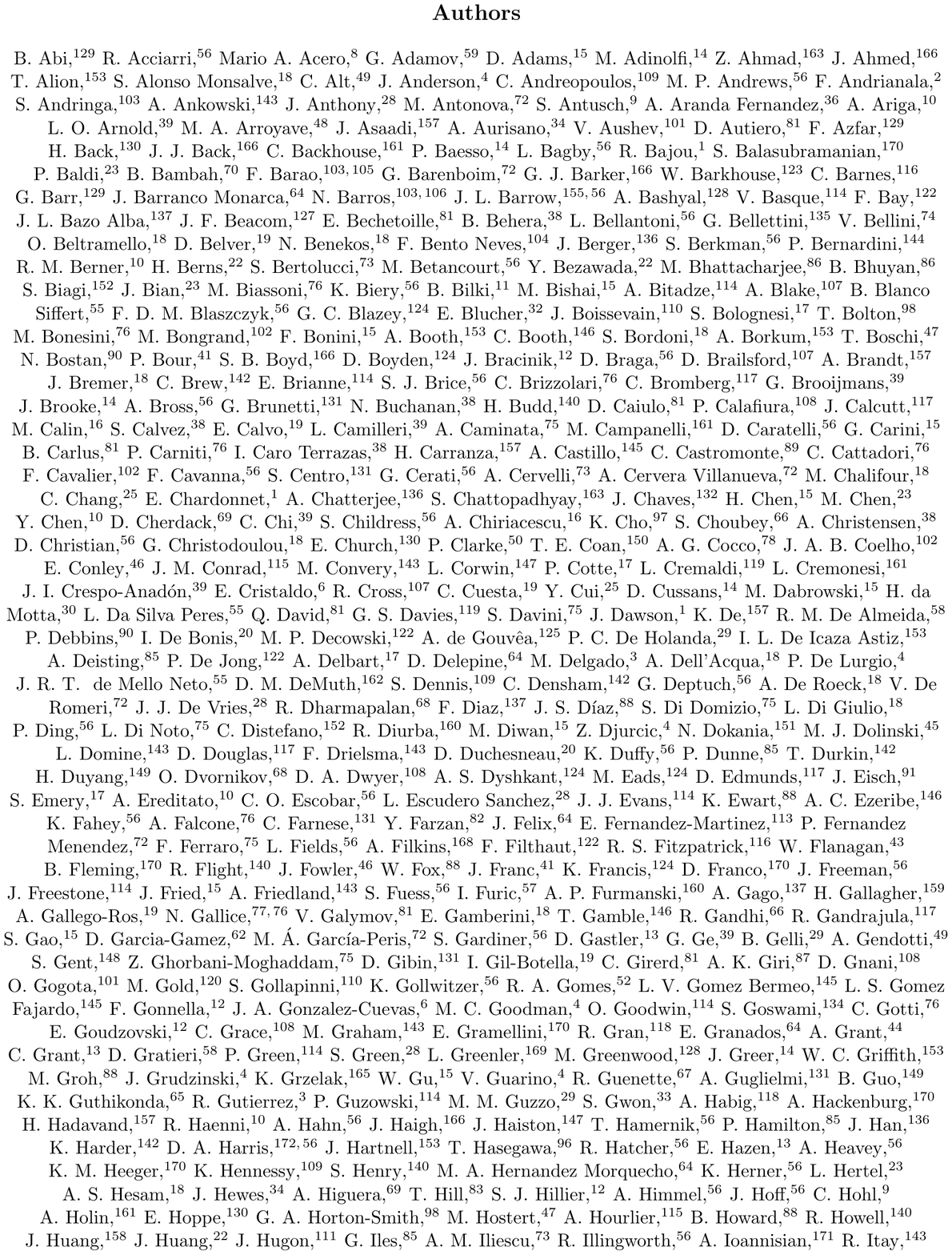}

\renewcommand{\familydefault}{\sfdefault}
\renewcommand{\thepage}{\roman{page}}
\setcounter{page}{0}

\pagestyle{plain}

\textsf{\tableofcontents}

\textsf{\listoffigures}

\textsf{\listoftables}
  \vspace{4mm}
  \addcontentsline{toc}{chapter}{A Roadmap of the DUNE Technical Design Report}

\iffinal\else
\textsf{\listoftodos}
\clearpage
\fi

\renewcommand{\thepage}{\arabic{page}}
\setcounter{page}{1}

\pagestyle{fancy}

\renewcommand{\chaptermark}[1]{%
\markboth{Chapter \thechapter:\ #1}{}}
\fancyhead{}
\fancyhead[RO,LE]{\textsf{\footnotesize \thechapter--\thepage}}
\fancyhead[LO,RE]{\textsf{\footnotesize \leftmark}}

\fancyfoot{}
\fancyfoot[RO]{\textsf{\footnotesize The DUNE Technical Design Report}}
\fancyfoot[LO]{\textsf{\footnotesize \thedoctitle}}
\fancypagestyle{plain}{}

\renewcommand{\headrule}{\vspace{-4mm}\color[gray]{0.5}{\rule{\headwidth}{0.5pt}}}



\cleardoublepage
\chapter*{A Roadmap of the DUNE Technical Design Report}

The \dword{dune} \dword{fd} \dword{tdr} describes the proposed physics program,  
detector designs, and management structures and procedures at the technical design stage.  

The TDR is composed of five volumes, as follows:

\begin{itemize}
\item Volume~\volnumberexec{} (\voltitleexec{}) provides an overview of all of DUNE for science policy professionals.

\item Volume~\volnumberphysics{} (\voltitlephysics{}) describes the DUNE physics program.

\item Volume~\volnumbertc{} (\voltitletc{}) outlines DUNE management structures, methodologies, procedures, requirements, and risks. 

\item Volume~\volnumbersp{} (\voltitlesp{}) and Volume~\volnumberdp{} (\voltitledp{}) describe the two \dword{fd} \dword{lartpc} technologies.

\end{itemize}

The text includes terms that hyperlink to definitions in a volume-specific glossary. These terms  appear underlined in some online browsers, if enabled in the browser's settings.

\cleardoublepage

\cleardoublepage

\chapter{Executive Summary}
\label{ch:fdsp-execsum}

\section{Introduction}
\label{sec:fdsp-exec-introduction}

The overriding physics goals of the \dword{dune} are the search for leptonic \dword{cp} violation, the search for nucleon decay as a signature of a Grand Unified Theory underlying the Standard Model, and the observation of  \dwords{snb} from supernovae. Central to achieving this physics program is the construction of a detector that combines the many-kiloton fiducial mass necessary for rare event searches with sub-centimeter spatial resolution in its ability to image those events, allowing us to identify the signatures of the physics processes we seek among the numerous backgrounds. The \dword{sp} \dword{lartpc}~\cite{Rubbia:1977zz} allows us to achieve these dual goals, providing a way to read out with sub-centimeter granularity the patterns of ionization in $\SI{10}{kt}$ volumes of \dword{lar} resulting from the \si{\mega\electronvolt}-scale interactions of solar and \dword{snb} neutrinos up to the \si{\giga\electronvolt}-scale interactions of neutrinos from the \dword{lbnf} beam.

To search for leptonic \dword{cp} violation, we must study \nue appearance in the \dword{lbnf} \numu beam. This requires the ability to separate electromagnetic activity induced by \dword{cc} \nue interactions from similar activity arising from photons, such as photons from $\pi^{0}$ decay. Two signatures allow this: photon showers are typically preceded by a gap prior to conversion, characterized by the \SI{18}{cm} conversion length in \dword{lar}; and the initial part of a photon shower, where an electron-positron pair is produced, has twice the $\mathrm{d}E/\mathrm{d}x$ of the initial part of an electron-induced shower. To search for nucleon decay, where the primary channel of interest is $p\rightarrow K^{+}\overline{\nu}$, we must identify kaon tracks as short as a few centimeters. It is also vital to accurately fiducialize these nucleon-decay events to suppress cosmic-muon-induced backgrounds, and here the detection of argon-scintillation photons is important for determining the time of the event. Detecting a \dword{snb} poses different challenges: those of dealing with a high data-rate and maintaining the high detector up-time required to ensure we do not miss one of these rare events. The signature of a \dword{snb} is a collection of MeV-energy electron tracks a few centimeters in length from \dword{cc} $\nu_{e}$ interactions, spread over the entire detector volume. To fully reconstruct a \dword{snb}, the entire detector must be read out, a data-rate of up to $\SI{2}{\tera\byte/\second}$, for \SIrange{30}{100}{s}, including a $\sim\!\SI{4}{s}$ pre-trigger window.

In this Executive Summary, we give an overview of the basic operating principles of a \dword{sp} \dword{lartpc}, followed by a description of the \dword{dune} implementation. We then discuss each of the subsystems separately, connecting the high-level design requirements and decisions to the overriding physics goals of \dword{dune}.

\section{The Single-Phase Liquid Argon Time-Projection Chamber}
\label{sec:fdsp-exec-splar}

\begin{dunefigure}[The single-phase (SP) \dshort{lartpc} operating principle]{fig:LArTPC}
{The general operating principle of the single-phase liquid argon time-projection chamber.}
\includegraphics[trim={5cm 0 5cm 0},clip,width=0.8\textwidth]{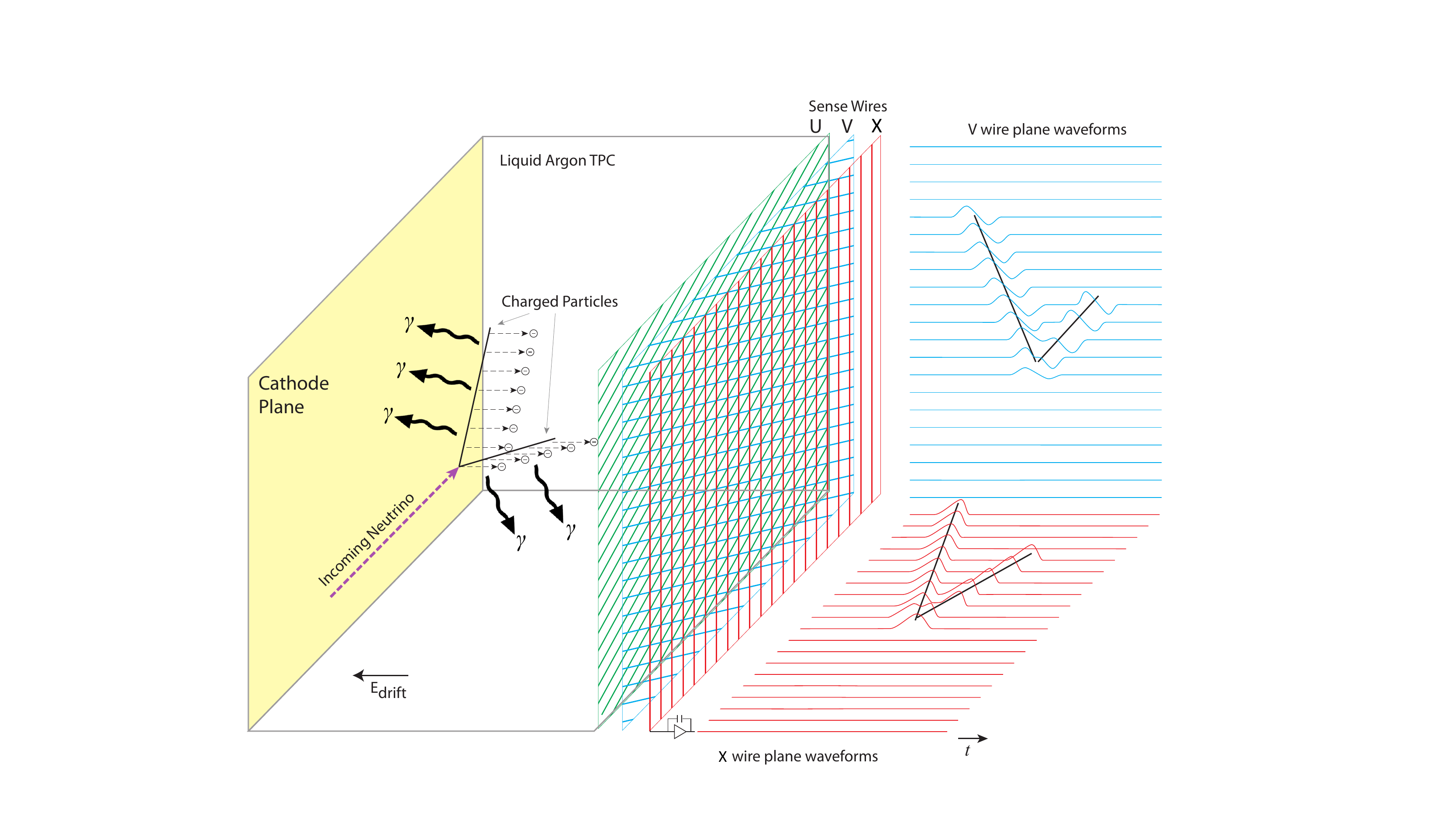}
\end{dunefigure}

Figure~\ref{fig:LArTPC} shows a schematic of the general operating principle of a \dword{sp} \dword{lartpc}, as has been previously demonstrated by
ICARUS~\cite{Icarus-T600},
\dword{argoneut}~\cite{Anderson:2012vc}, \dword{microboone}~\cite{microboone}, \dword{lariat}~\cite{Cavanna:2014iqa}, and \dword{protodune}~\cite{Abi:2017aow}. A large volume of \dword{lar} is subjected to a strong electric field of a few hundred volts per centimeter. Charged particles passing through the detector ionize the argon atoms, and the ionization electrons drift in the \efield to the anode wall on a timescale of milliseconds. This anode consists of layers of active wires forming a grid. The relative voltage between the layers is chosen to ensure all but the final layer are transparent to the drifting electrons, and these first layers produce bipolar induction signals as the electrons pass through them. The final layer collects the drifting electrons, resulting in a monopolar signal.

\dword{lar} is also an excellent scintillator, emitting \dword{vuv} light at a wavelength of \SI{127}{\nano\meter}. This prompt scintillation light, which crosses the detector on a timescale of nanoseconds, is shifted into the visible and collected by \dword{pd} devices. The \dword{pd}s can provide a $t_{0}$ determination for events, telling us when the ionization electrons begin to drift. Relative to this $t_{0}$, the time at which the ionization electrons reach the anode allows reconstruction of the event topology along the drift direction, which is crucial to fiducialize nucleon-decay events and to apply drift corrections to the ionization charge.

The pattern of current observed on the grid of anode wires provides the information for reconstruction in the two coordinates perpendicular to the drift direction. A closer spacing of the wires, therefore, results in better spatial resolution, but, in addition to increasing the cost of the readout electronics due to the additional wire channels, a closer spacing worsens the \dword{s/n} of the ionization measurement because the same amount of ionization charge is now divided over more channels. \dword{s/n} is an important consideration because the measurement of the ionization collected is a direct measurement of the $\mathrm{d}E/\mathrm{d}x$ of the charged particles, which is what allows us to perform both calorimetry and particle identification.

\section{The \dshort{dune} Single-Phase Far Detector Module}
\label{sec:fdsp-exec-dunefd}

\begin{dunefigure}[A \nominalmodsize DUNE far detector SP module]{fig:DUNESchematic}
{A \nominalmodsize \dword{dune} \dword{fd} \dword{spmod}, showing the alternating \sptpclen{} long (into the page), \tpcheight{} high anode (A) and cathode (C) planes, as well as the \dword{fc} that surrounds the drift regions between the anode and cathode planes. On the right-hand cathode plane, the foremost portion of the \dword{fc} is shown in its undeployed (folded) state.}
\includegraphics[width=0.65\textwidth]{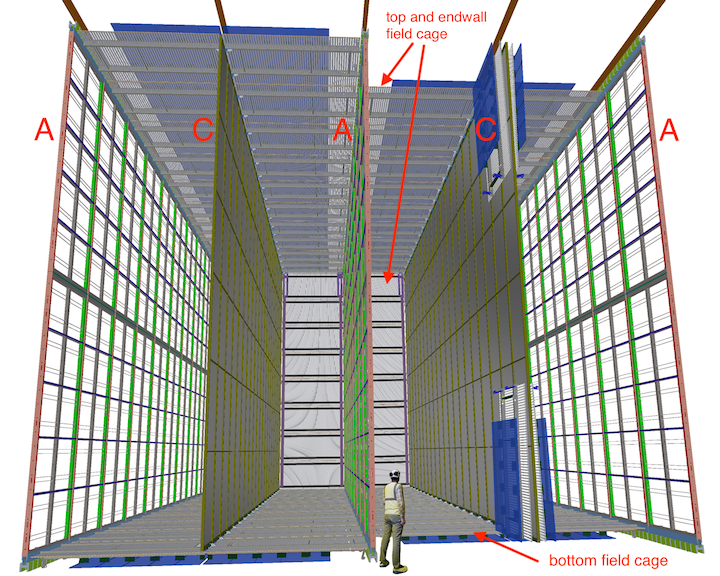}
\end{dunefigure}

The DUNE \dword{sp} \dword{lartpc} consists of four modules of \SI{10}{\kilo\tonne} fiducial mass (\SI{17.5}{\kilo\tonne} total mass), contributing to the full \SI{40}{\kilo\tonne} \dword{fd} fiducial mass. Figure~\ref{fig:DUNESchematic} shows a \SI{10}{\kilo\tonne} module, and the key parameters of a \dword{sp} module are listed in Table~\ref{tab:sp-key-parameters}. Inside a cryostat of outer dimensions 
\cryostatlen$\times$\cryostatht$\times$\cryostatwdth
 (shown in Figure~\ref{fig:Cryostat}), four $\SI{3.5}{\meter}$ drift volumes are created between five alternating anode and cathode walls, each wall having dimensions of $\SI{58}{\meter}\times \SI{12}{\meter}$.

\begin{dunetable}
[Key parameters for a \nominalmodsize  \dshort{fd} \dshort{spmod}]
{p{0.65\textwidth}p{0.25\textwidth}}
{tab:sp-key-parameters}
{Key parameters for a \nominalmodsize  \dshort{fd} \dshort{spmod}.}
Item & Quantity   \\ \toprowrule
TPC size & \tpcheight$\times \SI{14.0}{\meter} \times$\sptpclen \\ \colhline
Nominal fiducial mass & \spactivelarmass \\ \colhline
\dshort{apa} size & $\SI{6}{\meter}\times\SI{2.3}{\meter}$ \\ \colhline
\dshort{cpa} size & $\SI{1.2}{\meter}\times\SI{4}{\meter}$ \\ \colhline
Number of \dshort{apa}s & 150 \\ \colhline
Number of \dshort{cpa}s & 300 \\ \colhline
Number of \dshort{xarapu} \dshort{pd} bars & 1500 \\ \colhline
\dshort{xarapu} \dshort{pd} bar size & $\SI{209}{\cm}\times\SI{12}{cm}\times\SI{2}{\cm}$ \\ \colhline
Design voltage & \sptargetdriftvolt \\ \colhline
Design drift field & \spmaxfield \\ \colhline
Drift length & \spmaxdrift \\ \colhline
Drift speed & $\SI{1.6}{\mm/\micro\second}$ \\
\end{dunetable}

\begin{dunefigure}[A far detector (\dshort{fd}) cryostat]{fig:Cryostat}{A  \cryostatlen (L) by \cryostatwdth (W) by \cryostatht{} (H) outer-dimension cryostat that houses a \nominalmodsize \dword{fd} module. A mezzanine (light blue) installed \SI{2.3}{m} above the cryostat supports both  detector and cryogenics instrumentation. At lower left, between the \lar recirculation pumps (green) installed on the cavern floor,  the figure of a person indicates the scale.}
\includegraphics[width=0.8\textwidth]{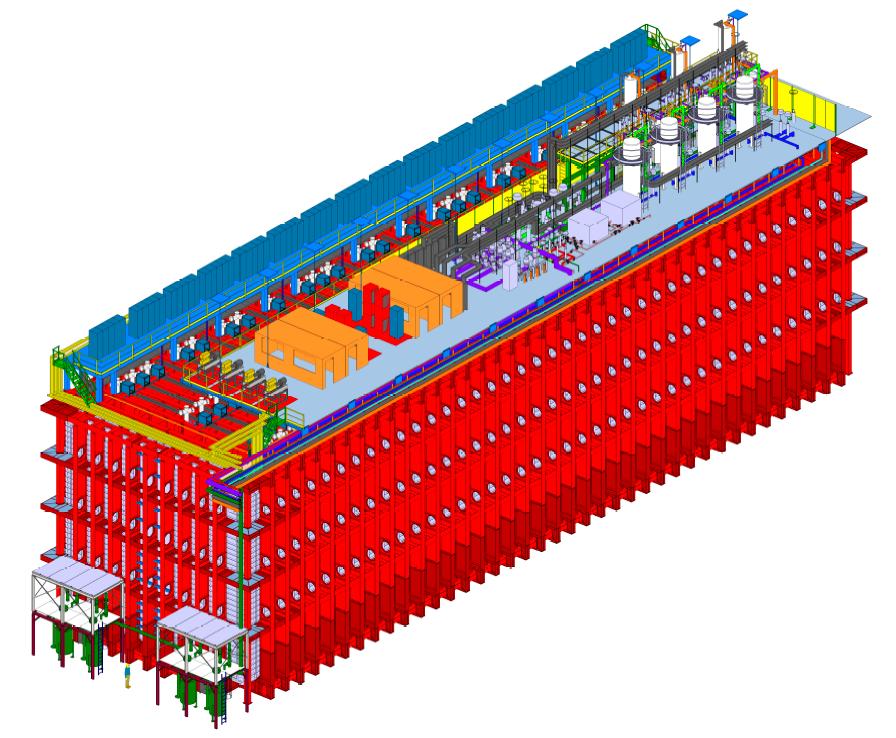}
\end{dunefigure}

The \dword{fd} is located underground, at the \SI{4850}{\foot} level of the \dword{surf} in South Dakota. The detector is \SI{1300}{\km} from the source of the \dword{lbnf} neutrino beam at \dword{fnal}; this baseline provides the matter effects necessary for \dword{dune} to determine the neutrino mass hierarchy. The \dword{surf} underground campus is shown in Figure~\ref{fig:SURFCaverns}. The four \nominalmodsize \dword{fd} modules will be located in the two main caverns, which are each \SI{144.5}{\meter} long, \SI{19.8}{\meter} wide and 
\SI{28.0}{\meter}high. Each cavern houses two \nominalmodsize modules, one either side of the central access drift. Between the two caverns is the \dword{cuc}, a \SI{190}{\meter} long, \SI{19.3}{\meter} wide, 
\SI{11.0}{\meter} high cavern in which many of the utilities and the upstream \dword{daq} reside.

\begin{dunefigure}[The underground layout of the \dshort{surf} laboratory]{fig:SURFCaverns}{The underground layout of the \dword{surf} laboratory. The two main caverns each hold two \nominalmodsize modules, one either side of the central access drift. The \dword{cuc} houses utilities and the upstream \dword{daq}.}
\includegraphics[width=\textwidth]{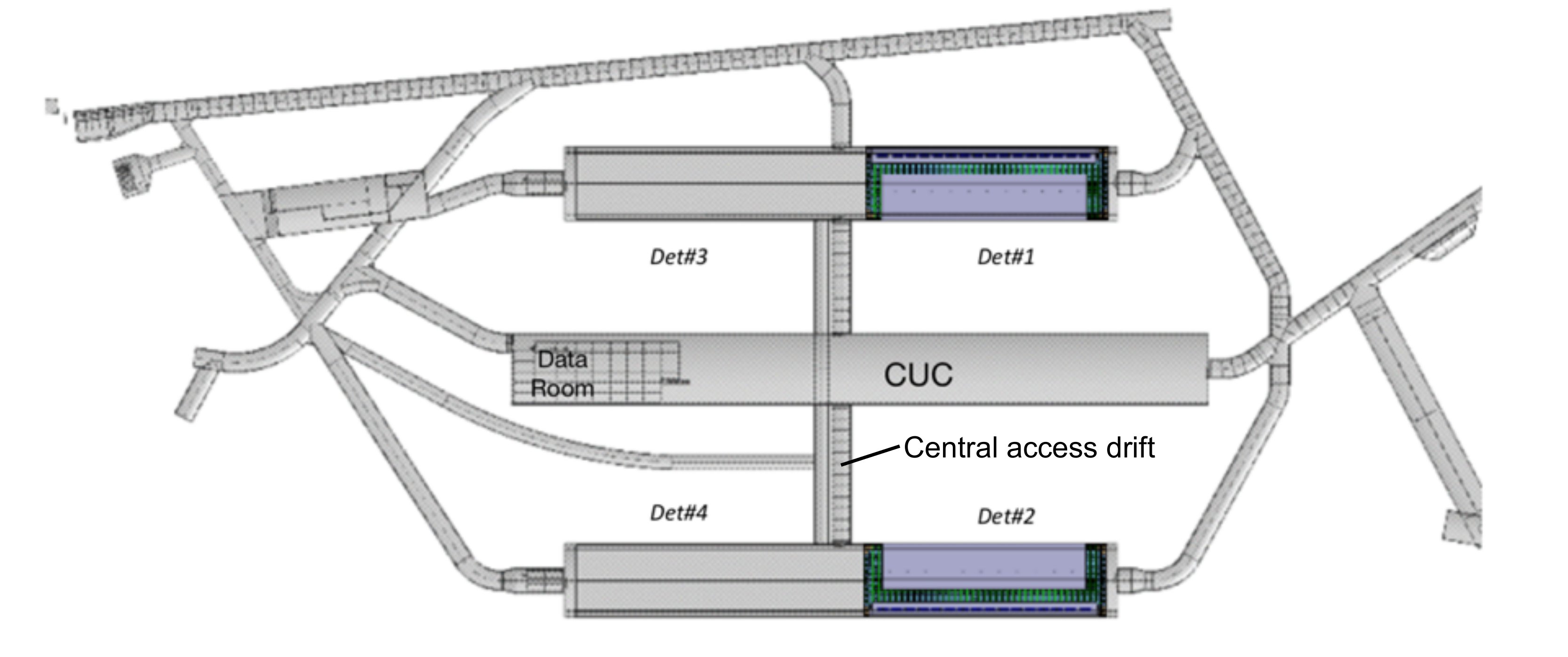}
\end{dunefigure}

Each cathode wall in a module is called a \dword{cpa} array. The \dword{cpa} is the $\SI{1.2}{\meter}\times\SI{4}{\meter}$ panel from which the \dword{cpa} arrays are formed; each \dword{cpa} array contains $150$ \dword{cpa}s. The \dword{cpa} arrays are held at $-\SI{180}{\kilo\volt}$. With the anode walls held close to ground, this results in a uniform \SI{500}{\volt/\centi\meter} \efield across the drift volume. A \dword{fc} surrounds the remaining open sides of the \dword{tpc}, ensuring the field is uniform to better than 1\% throughout the active volume. A typical minimum ionizing particle passing through the argon produces around $60k$ ionization electrons per centimeter, which drift towards the anodes at around $\SI{1.6}{\mm/\micro\second}$; the time taken to drift the full distance from cathode to anode would therefore be around $\SI{2.2}{\milli\second}$.

The anode walls are each made up of 50 \dword{apa} units that are $\SI{6}{\meter}\times\SI{2.3}{\meter}$ in dimension. As shown in Figure~\ref{fig:APAStack}, the \dword{apa}s hang vertically; each anode wall is two \dword{apa}s high and 25 \dword{apa}s wide. The \dword{apa}s are two-sided, with three active wire layers and an additional shielding layer, also called a grid layer, wrapped around them. The wire spacing on the layers is $\sim\!\SI{5}{\mm}$. The collection layer is called the $X$-layer; the induction layer immediately next to that is called the $V$-layer; the next induction layer is the $U$-layer; and the shielding layer is the $G$-layer. $X$-layer and $G$-layer wires are vertical; the $U$- and $V$-layer wires are at $\pm\SI{35.7}{\degree}$ to the vertical.

\begin{dunefigure}[A stack of two anode plane assemblies (APAs)]{fig:APAStack}
{Left: two \dword{apa}s linked together to form one unit of an \dword{apa} wall. \dword{pd} bars can be seen installed across the width of the \dword{apa}s. Right: a zoom into the top and bottom ends of the \dword{apa} stack showing the readout electronics, and the center of the stack where the \dword{apa}s are connected together.}
\includegraphics[width=.8\textwidth]{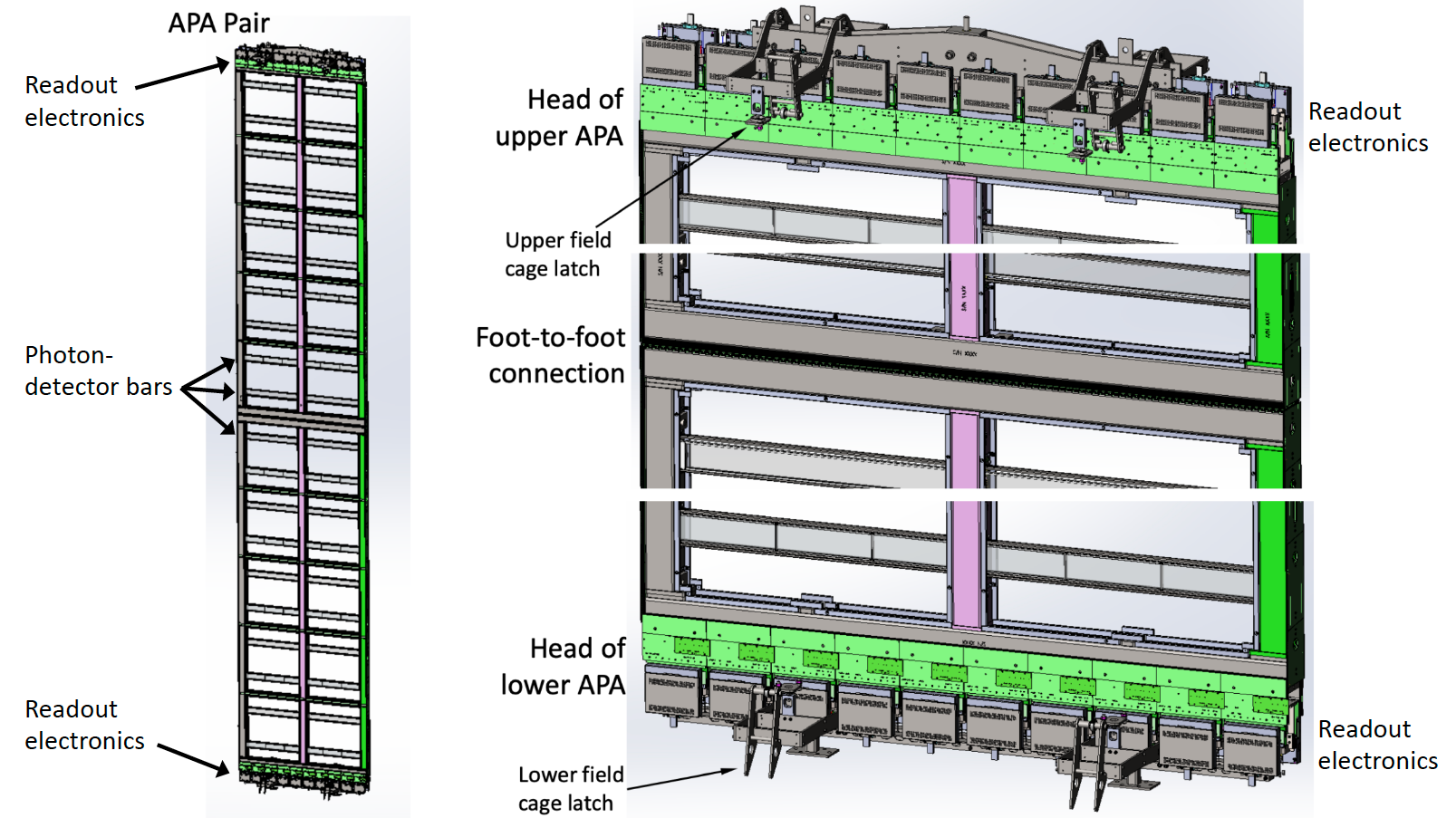}
\end{dunefigure}

Readout electronics, called \dword{ce}, are attached to the top end of the top \dword{apa} and the bottom end of the bottom \dword{apa}. These \dword{fe} electronics benefit from the low \dword{lar} temperature through the reduction of thermal noise. The front-end electronics shape, amplify, and digitize the signals from the induction and collection wires thanks to a series of three different types of \dword{asic} through which all signals pass.
Cables from the \dword{ce} pass through feedthroughs on the roof of the cryostat; cables from the motherboards on the bottom \dword{apa} pass through the inside of the hollow \dword{apa} frames up to the top.

Once signals from \dword{apa}s leave the cryostat through feedthroughs, they are passed to warm interface boards that put the signals onto \SI{10}{Gbps} optical fibers, ten fibers per \dword{apa}, which carry the signals to the upstream \dword{daq} system located in the \dword{cuc}. Each \SI{10}{\kilo\tonne} module has its own, independent \dword{daq} system, built around the \dword{felix} system, developed by \dword{cern}, which is responsible for triggering, buffering, and shipping data out to permanent storage above ground; when triggered, each \SI{10}{\kilo\tonne} module will provide data at a rate of up to \SI{2}{\tera\byte/\second}. This separation of \dword{daq} systems allows each module to run as an independent detector to minimize any chance of a complete \dword{fd} outage. Modules can, however, provide the others with a supernova trigger signal. The \dword{daq} system also provides the detector clock. A \dword{gps} \dword{pps} is used to time-stamp events, both to allow matching to the beam window and to allow time-stamping of supernova triggers. Within a \SI{10}{\kilo\tonne} module a $\SI{62.5}{\mega\hertz}$ master clock keeps all detector components synchronized to within \SI{1}{\nano\second}.

In addition to the ionization, charged particles passing through the argon produce approximately 24,000 scintillation photons per \si{\mega\electronvolt}. These photons are collected by devices called X-Arapucas, which are mounted in the \dword{apa}s, in between the two sets of wire layers, as shown in Figure~\ref{fig:APAStack}. There are ten X-Arapucas on each \dword{apa}, which are bars running the full \SI{2.3}{\meter} width of the \dword{apa}. The X-Arapuca bars consist of layers of dichroic filter and wavelength-shifter that shift the \dword{vuv} scintillation light into the visible and trap these visible photons, transporting them to \dword{sipm} devices. The signals from these \dword{sipm}s are sent along cables that pass through the hollow \dword{apa} frames, up to feedthroughs in the cryostat roof. The signals are then sent along \SI{10}{Gbps} optical fibers, one fiber per \dword{apa} (ten X-Arapuca bars), to the \dword{daq} system where the \dword{pd} and \dword{apa}-wire data-streams are merged.

\section{The Liquid Argon}
\label{sec:fdsp-exec-liquidargon}

The primary requirement of the \dword{lar} is its purity. Electronegative contaminants such as oxygen or water absorb ionization electrons as they drift. Nitrogen contaminants quench scintillation photons.

The target purity from electronegative contaminants in the argon is $<\!\!100$\,ppt (parts per trillion) O$_{2}$ equivalent, which is enough to ensure a $>\!\!\SI{3}{\milli\second}$ ionization-electron lifetime at the nominal \SI{500}{\volt/\centi\meter} drift voltage. This target electron lifetime means that, for a charged particle traveling near a \dword{cpa} array, there is 48\% attenuation of the ionization by the time it reaches the anode, which ensures that we achieve \dword{s/n} ratios of $S/N>5$ for the induction planes and $S/N>10$ for the collection planes, which are necessary to perform pattern recognition and two-track separation. We have an additional requirement for electronegative impurities released into the argon by detector components of $<\!30$\,ppt, to ensure such sources of contamination are negligible compared to the contamination inherent in the argon. Data from \dword{protodune} has shown that we can exceed our target argon purity, with electron lifetimes in excess of \SI{6}{\milli\second} achieved.

Nitrogen contamination must be $<\!25$\,ppm (parts per million). This is necessary to ensure we achieve our requirement of at least 0.5 photoelectrons per MeV detected for events in all parts of the detector, which in turn ensures, through the timing requirements discussed in Section~\ref{sec:fdsp-exec-pds}, that we can fiducialize nucleon decay events throughout the detector.

Fundamental to maintaining argon purity is the constant flow of argon through the purification system. It is, therefore, important to understand the fluid dynamics of the argon flow within the detector to ensure there are no dead regions where argon can become trapped. This fluid dynamics also informs the placement of purity, temperature, and level monitors.

\section{Photon Detection System}
\label{sec:fdsp-exec-pds}

Compared to the ionization electrons, which can take milliseconds to drift across the drift volume, the scintillation photons are fast, arriving at the \dword{pd}s nanoseconds after production. This scintillation light provides a $t_{0}$ for each event. By comparing the arrival time of ionization at the anode with this $t_{0}$, reconstruction in the drift direction becomes possible. A \SI{1}{\micro\second} requirement on the timing resolution of the \dword{pd} system enables $\sim\!\SI{1}{\mm}$ position resolution for \SI{10}{\mega\electronvolt} \dword{snb} events. The \dword{pd} $t_{0}$ is also vital in fiducializing nucleon-decay events, which allows us to reject cosmic-muon-induced background events that will occur near the edges of the detector modules. We must be able to do this throughout the entire active volume with $>\!99\%$ efficiency, leading to a requirement of at least 0.5 photoelectrons per MeV detected for events in all parts of the detector. These requirements are discussed later in Chapter~\ref{ch:fdsp-pd}.

\dword{pd} modules, shown in Figure~\ref{fig:PDModules}, are 
$\SI{209}{\cm}\times\SI{12}{\cm}\times\SI{2}{\cm}$ bars, ten of which are mounted in each \dword{apa} between the wire layers. Each bar contains 24 X-Arapuca\footnote{An arapuca is a South American bird trap, the name used here in analogy to the way the X-Arapuca devices trap photons.} cells, grouped into four supercells. An X-Arapuca cell is shown in Figure~\ref{fig:ArapucaCell}. The outer layers are dichroic filters transparent to the \SI{127}{\nano\meter} scintillation light. Between these filters is a \dword{wls} plate, which converts the UV photons into the visible spectrum (\SI{430}{\nano\meter}); one WLS plate runs the full length of each supercell.
Visible photons emitted inside the \dword{wls} plate at an angle to the surface greater than the critical angle reach \dword{sipm}s at the edges of the plates. Visible photons that escape the \dword{wls} plates are reflected off the dichroic filters, which have an optical cutoff, reflecting photons with wavelengths more than \SI{400}{\nano\meter} back into the \dword{wls} plates.

\begin{dunefigure}[Photon detector (PD) modules, mounted in an APA]{fig:PDModules}
{Left: an \dword{xarapu} \dword{pd} module. The 48 \dwords{sipm} that detect the light from the 24 cells are along the long edges of the module. Right: \dword{xarapu} \dword{pd} modules mounted inside an \dword{apa}.}
\includegraphics[width=0.49\textwidth]{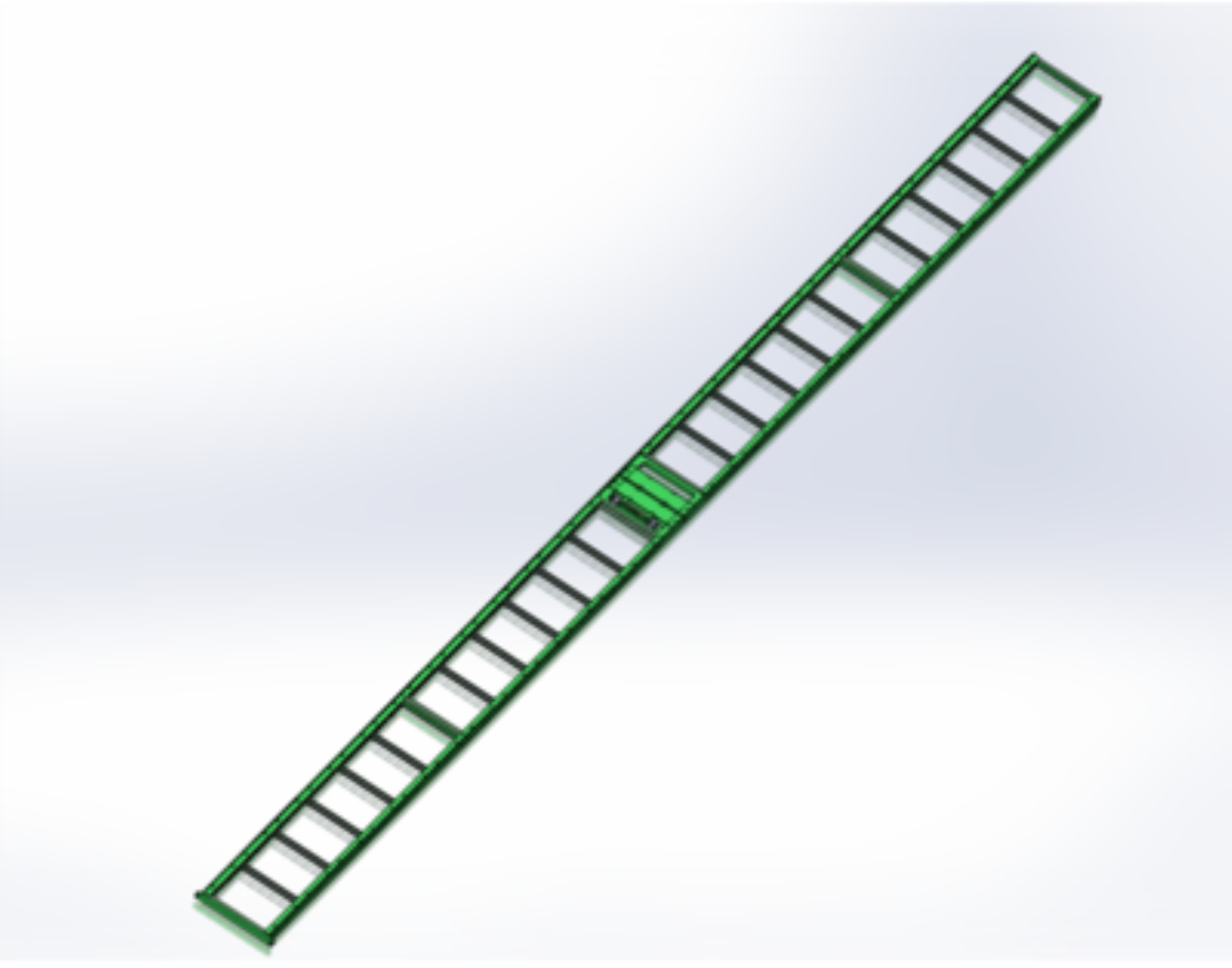}
\includegraphics[width=0.49\textwidth]{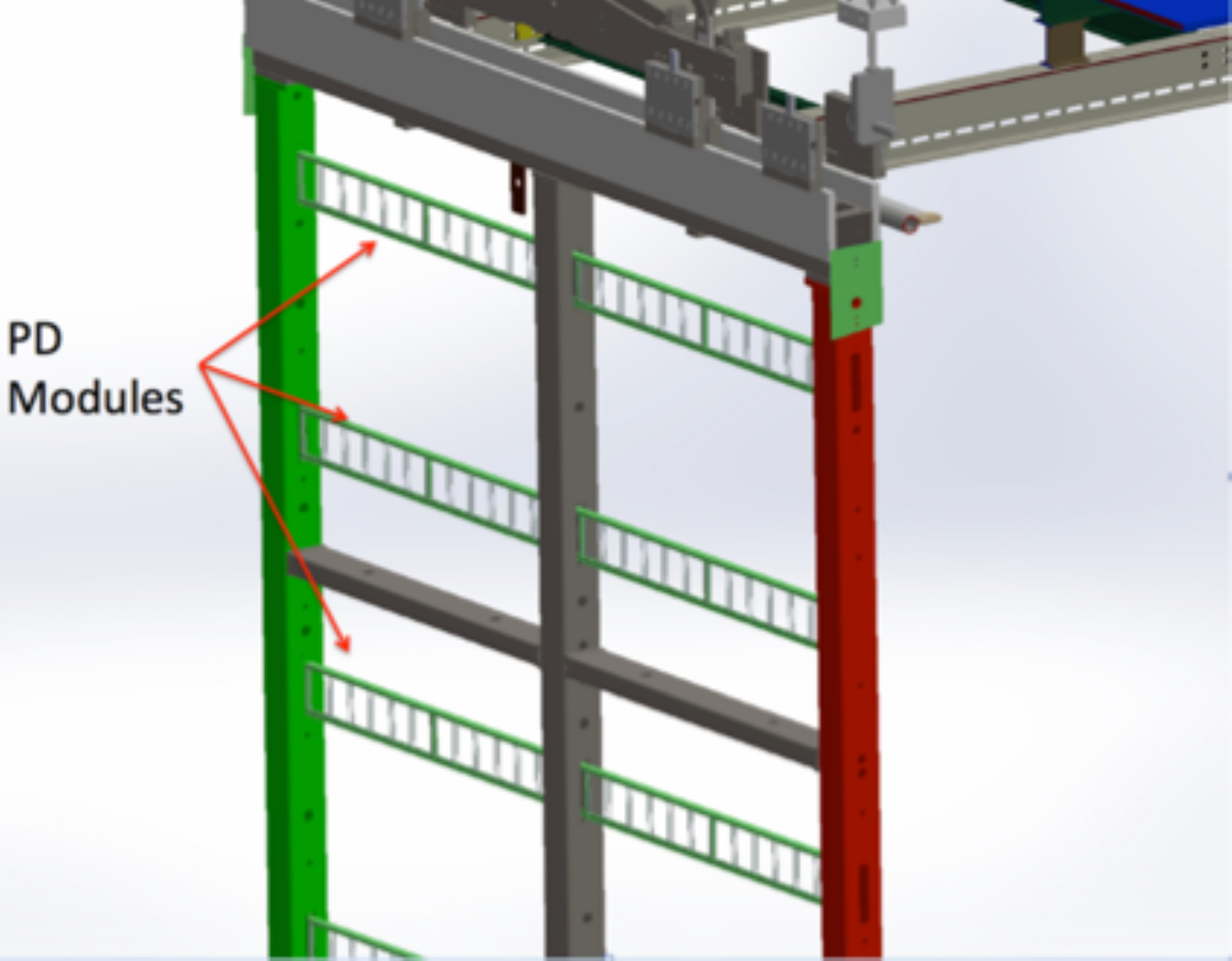}
\end{dunefigure}

\begin{dunefigure}[An \dshort{xarapu} \dshort{pd} cell]{fig:ArapucaCell}
{Left: an X-Arapuca cell. Right: an exploded view of the X-Arapuca cell, where the blue sheet is the wavelength-shifting plate and the yellow sheets the dichroic filters.}
\includegraphics[width=\textwidth]{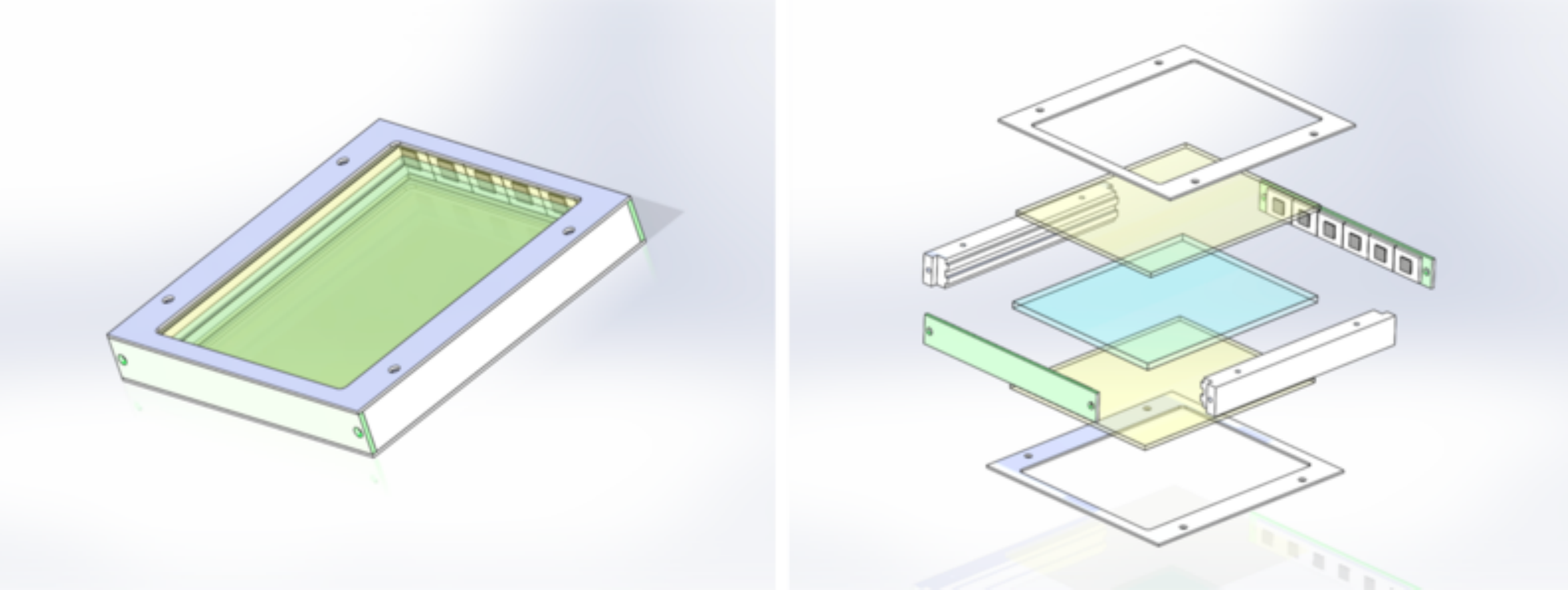}
\end{dunefigure}

The 48 \dword{sipm}s on each X-Arapuca supercell are ganged together and the signals are collected by front-end electronics, mounted on the supercell. The design of the front-end electronics is inspired by the system used for the Mu2e cosmic-ray tagger~\cite{bib:mu2e_tdr}, which uses commercial ultrasound \dword{asic}s. The front-end electronics define the \SI{1}{\micro\second} timing resolution of the \dword{pd} system.

\section{High Voltage, Cathode Planes and Field Cage}
\label{sec:fdsp-exec-hv}

The design voltage at which the \dword{dune} \dword{tpc} will operate is $-\SI{180}{\kilo\volt}$, corresponding to \SI{500}{\volt/\cm} across each drift volume. This voltage is a trade off. A higher voltage results in more charge collected, and hence better \dword{s/n} ratio, better calorimetry, and lower detection thresholds, as well as less saturation of free charge at the point of ionization. A higher voltage, however, also reduces the amount of scintillation light produced and requires more space between the \dword{cpa}s and the cryostat walls to prevent discharges, reducing the fiducial volume. The \dword{protodune} experience shows that we can achieve this design voltage; nevertheless, from \dword{microboone}, we also know that a drift voltage of \SI{250}{\volt/\cm} achieves an adequate \dword{s/n} ratio.

The \dword{hv} is supplied to the \dword{cpa} arrays. Each \dword{cpa} array (two per \SI{10}{\kilo\tonne} module) has its own independent high voltage supply. These commercial high voltage devices will supply a current of \SI{0.16}{\milli\ampere} at $-\SI{180}{\kilo\volt}$. The voltage is delivered, via $\sim\!\SI{30}{\meter}$ length commercial cables, through a series of few-\si{\mega\ohm} filtering resistors that act as low-pass filters to reduce noise and thereby satisfy the ripple-voltage requirement of $<\!\SI{0.9}{\milli\volt}$ on the \dword{cpa} array, which corresponds to a requirement of $<\!100\,e{-}$ of noise injected into the \dword{tpc} by the high-voltage system. The supply unit monitors the voltage and current every \SI{300}{\milli\second}; toroids mounted on the cables are sensitive to much faster changes in current and enable responses to current changes on a timescale of \SIrange{0.1}{10}{\micro\second}.

The high voltage passes into the cryostat through a feedthrough based on the ICARUS design~\cite{Icarus-T600}, the stainless steel conductor of which mates with the \dword{cpa} array via a spring-loaded feedthrough.
When at $-\SI{180}{\kilo\volt}$, each \dword{cpa} array stores \SI{400}{\joule} of energy, so the \dword{cpa}s must have at least $\SI{1}{\mega\ohm/\cm^{2}}$ resistance to prevent damage if the field is quenched. The \dword{cpa}, an example of which from \dword{protodune} is shown in Figure~\ref{fig:CPA}, is a $\SI{1.2}{\meter}\times\SI{4}{\meter}$ planar unit, each side of which is a \SI{3}{\mm} thick FR-4 sheet, onto which is laminated a thin layer of carbon-impregnated Kapton that forms the resistive cathode plane.

\begin{dunefigure}[A \dshort{pdsp} cathode plane assembly (CPA)]{fig:CPA}
{A \dshort{pdsp}  cathode plane assembly. The black surface is the carbon-impregnated Kapton resistive cathode plane.}
\includegraphics[width=0.35\textwidth]{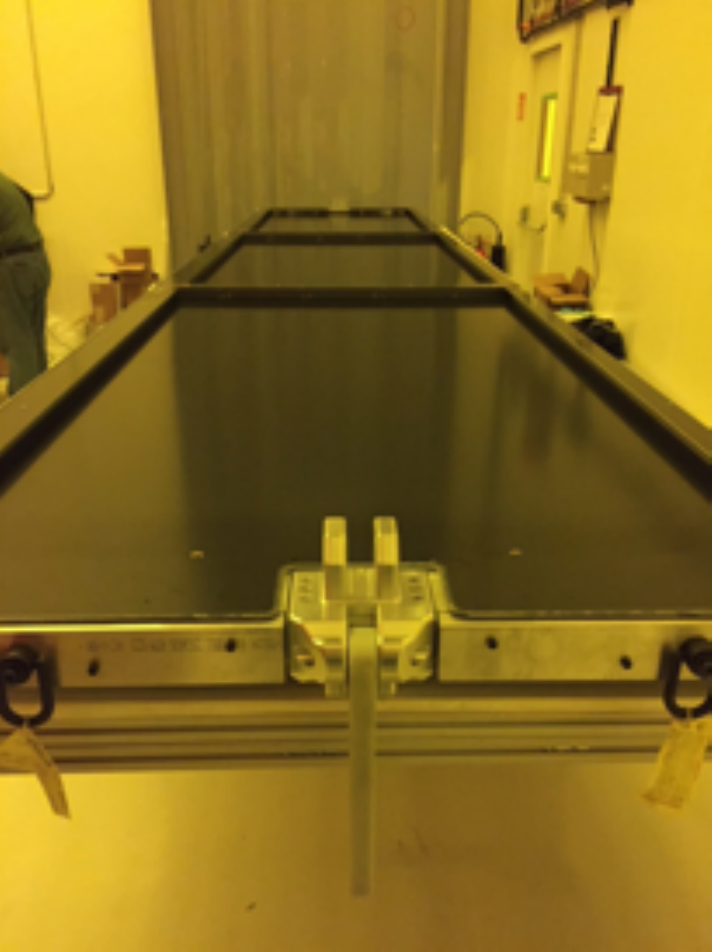}
\end{dunefigure}

The field must be uniform throughout the active \dword{tpc} volume to within 1\%, and this is achieved by a \dword{fc} that surrounds the drift volumes. The \dword{fc} is built from field-shaping aluminum profiles, terminated with \SI{6}{\mm} thick ultra-high molecular-weight polyethylene caps (see Figure~\ref{fig:FieldCage}). All surfaces on these profiles must be smooth to keep local fields below \SI{30}{\kilo\volt/\cm}, a requirement that reduces the possibility of voltage breakdowns in the argon; the shape of the profiles leads to a maximum local field near the surface of the \dword{fc} of $\sim\!\SI{12}{\kilo\volt/\cm}$. The aluminum profiles are connected together via a resistive divider chain; between each profile, two \SI{5}{\giga\ohm} resistors, arranged in parallel, provide a \SI{2.5}{\giga\ohm} resistance to create a nominal \SI{3}{\kilo\volt} drop.

\begin{dunefigure}[A section of the field cage (FC)]{fig:FieldCage}
{A section of the field cage, showing the extruded aluminum field-shaping profiles, with white polyethylene caps on the ends to prevent discharges.}
\includegraphics[width=0.5\textwidth]{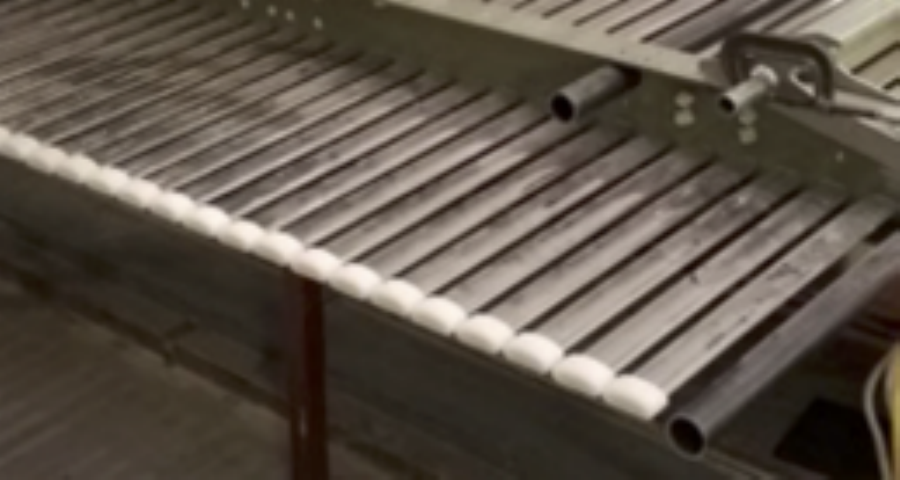}
\end{dunefigure}

\section{Anode Planes}
\label{sec:fdsp-exec-apas}

The \dword{apa}s are $\SI{6}{\meter}\times\SI{2.3}{\meter}$ planes that form the three anode walls of the \dword{tpc}. An \dword{apa} is shown in Figure~\ref{fig:APA}. In the \dword{fd}, the \dword{apa}s are mounted in pairs, in portrait orientation, one above another, with the head end of the top \dword{apa} at the top of the detector and the head end of the bottom \dword{apa} at the bottom of the detector.

\begin{dunefigure}[Schematic and photo of an APA]{fig:APA}
{Top: a schematic of an anode plane assembly. In black is the steel \dword{apa} frame. The green and pink areas indicate the directions of the induction wire layers. The blue area indicates the directions of the induction and shielding (grid) wire layers. The blue boxes at the right-hand end  are the \dword{ce}. Bottom: a \dword{protodune} \dword{apa} in a wire-winding machine. The right-hand end of the \dword{apa} as shown in this picture is the head end, onto which the \dword{ce} are mounted.}
\includegraphics[width=0.7\textwidth]{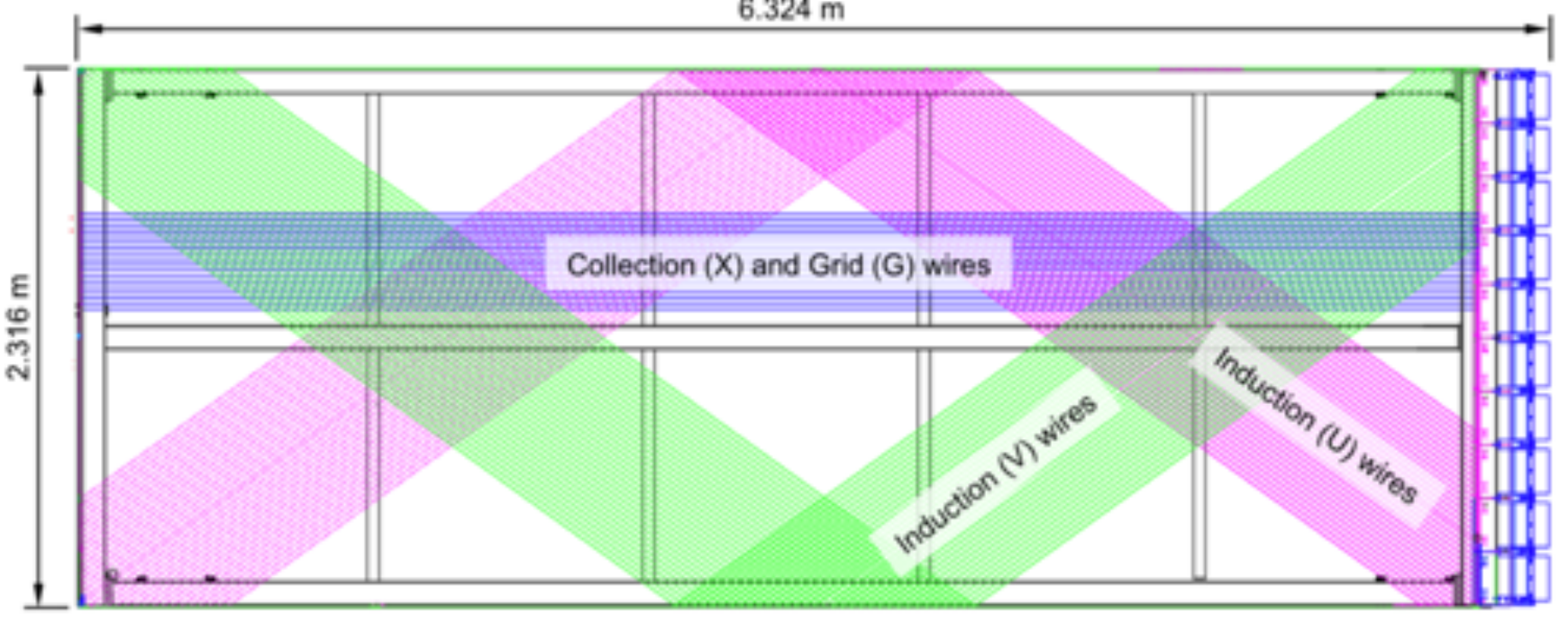}
\includegraphics[width=0.6\textwidth]{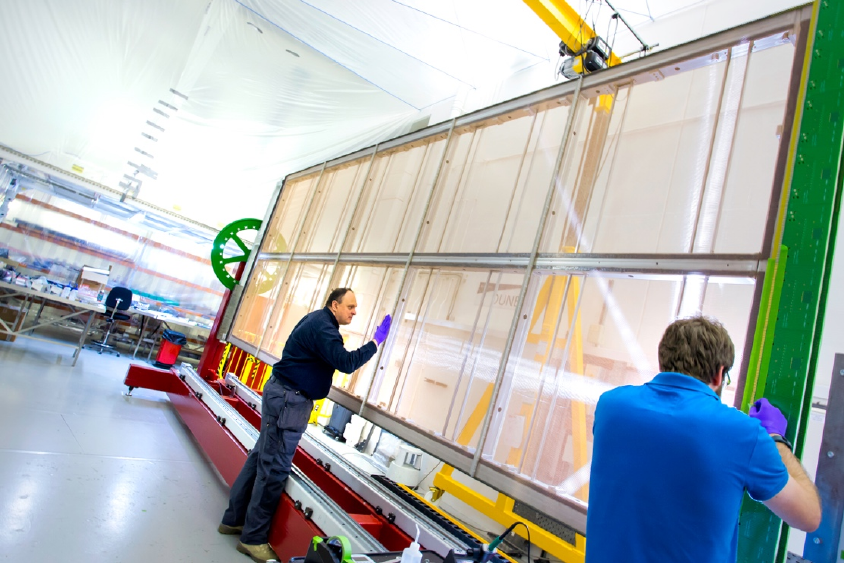}
\end{dunefigure}

The basic building block of the \dword{apa} is the steel frame that can be seen outlined in Figure~\ref{fig:APA}, consisting of three long steel bars, a head shaft onto which the \dword{ce} are mounted, a foot shaft, and four thinner cross-braces. The two outer long sections are $4\,\mathrm{inch}\times 4\,\mathrm{inch}$ square-profile steel tubes through which run the \dword{pd} cables and the \dword{ce} cables from the bottom \dword{apa} of a pair. The \dword{pd}s are mounted into the \dword{apa}s, after production, through slots in these long sections.

Mounted directly onto both sides of the \dword{apa} frame is a grounding mesh, which ensures any ionization produced inside the \dword{apa} cannot cause signals on the active wire layers. The four wire layers, consisting of \SI{152}{\micro\meter} diameter copper-beryllium wire, are wound around the grounding mesh. The inside layer is the collection layer, called the $X$-layer, the 960 wires of which run parallel to the long axis of the \dword{apa}. Next are the two induction layers, the $U$- and $V$-layers, each with 800 wires at $\pm\SI{35.7}{\degree}$ to the long axis. Finally, the uninstrumented shielding layer, the $G$-layer, has 960 wires running parallel to the $X$-layer wires; this layer shields the three active layers from long-range induction effects. The wire spacing on each layer is \SI{4.79}{\mm} for the $X$ and $G$ layers and \SI{4.67}{\mm} for the $U$ and $V$ layers; the inter-plane spacing is \SI{4.75}{\mm}. The wire spacing on each plane defines the spatial resolution of the \dword{apa}; it is wide enough to keep readout costs low and \dword{s/n} high, but small enough to enable reconstruction of short tracks such as few-\si{\cm} kaon tracks from proton-decay events. The tolerance both on the wire spacing in the plane and on the plane-to-plane spacing is \SI{0.5}{\mm}; this is most important in the plane-to-plane direction where the spacing ensures that the induction planes remain transparent to the drifting charge.

The wires are soldered to printed circuit boards located around the four sides of the \dword{apa}. These boards, shown in Figure~\ref{fig:GeometryBoardsAndCombs}, are called geometry boards since they define the wire spacing in all dimensions; they consist only of pads and traces: no active components. At the head end, these boards lie flat in the plane of the \dword{apa}, and the wires are terminated onto these boards for readout. On the remaining three sides, the boards sit on the sides of the \dword{apa}, perpendicular to the wire planes, and control the wrapping of the wires around the \dword{apa}. These wrap boards have insulating pins on their edges, around which the wires are wrapped, to set the wire spacing. At the head end, additional active boards are installed after all wires are wound: $G$-bias boards provide the necessary capacitance to the $G$-layer and a resistor to provide the bias voltage; $CR$-boards provide the interface between the $X$ and $U$ layers and the \dword{ce}, resistors providing the bias voltages and capacitors providing DC blocking. Relative to the ground, the four wire layers are biased to \SI{820}{\volt} ($X$-layer), \SI{0}{\volt} ($V$-layer), \SI{-370}{\volt} ($U$-layer), and \SI{-665}{\volt} ($G$-layer). To maintain the wire spacing across the \dword{apa}, wire-support combs, also shown in Figure~\ref{fig:GeometryBoardsAndCombs}, run along the four cross-braces across the short dimension of the \dword{apa}.

\begin{dunefigure}[Geometry boards and wire-support combs on an APA]{fig:GeometryBoardsAndCombs}
{Left: $V$-layer geometry boards, showing the head-end boards face-on and the wrap boards along the bottom. Back plastic insulating pins are visible on the edges of the wrap boards. The $V$-layer wires can be seen running diagonally, and the $X$-layer wires, horizontal in this picture, are visible behind those.  Right: wire-support combs, showing all four layers of wires.}
\includegraphics[width=\textwidth]{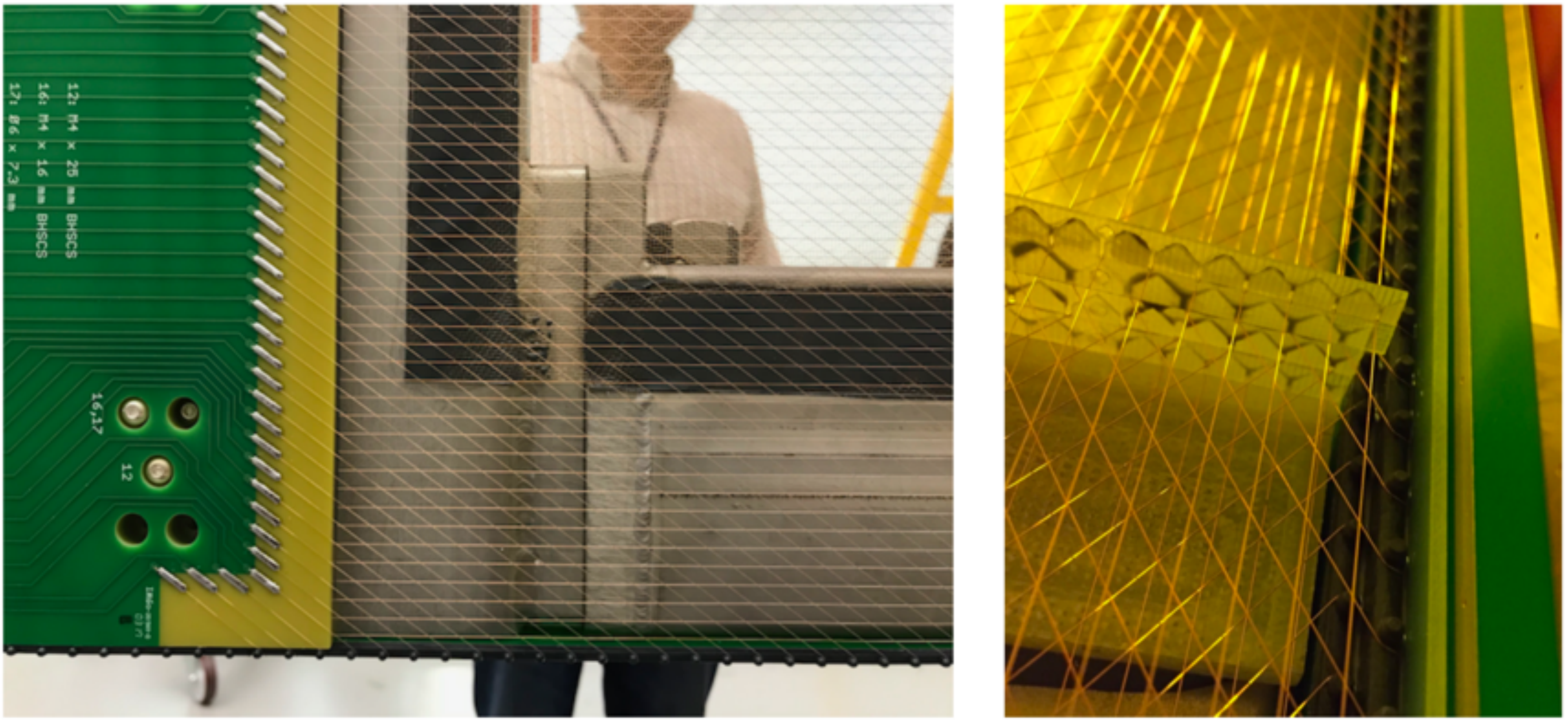}
\end{dunefigure}

\section{Electronics}
\label{sec:fdsp-exec-electronics}

The job of the readout electronics is to send out of the cryostat digitized waveforms from the \dword{apa} wires. To enable us to look at low-energy particles, we aim to keep noise to below $1000\,e^{-}$ per channel, which should be compared to the $20k$--$30k\,e^{-}$ per channel collected from a minimum-ionizing particle traveling parallel to the wire plane and perpendicular to the wire orientation. For large signals, we require a linear response up to $500k\,e^{-}$, which ensures that fewer than 10\% of beam events experience saturation. This can be achieved using 12\, \dword{adc} bits. In addition, the electronics are designed with a front-end peaking time of \SI{1}{\micro\second}, which matches the time for the electrons to drift between wires planes on the \dword{apa}; this then leads to a design sampling frequency of \SI{2}{\mega\hertz} to satisfy the Nyquist criterion.

The digitization electronics are mounted on the head ends of the \dword{apa}s in the \dword{lar} and are therefore referred to as \dword{ce}. The low, \SI{87}{\kelvin} temperatures reduce thermal noise. Figure~\ref{fig:ElectronicsBlockDiagram} shows a block diagram of the \dword{femb}s mounted on the \dword{apa}s. Each \dword{apa} is instrumented with 20 \dword{femb}s, each of which takes the signals from 40 $U$-layer wires, 40 $V$-layer wires, and 48 $X$-layer wires. The signals pass through a series of three \dword{asic}s. The first \dword{asic}, the front-end \dword{asic}, shapes and amplifies the signals. The next \dword{asic}, the \dword{adc} \dword{asic}, performs the analogue-to-digital conversion. Finally, a \dword{coldata} \dword{asic} merges the data streams from the preceding \dword{asic}s for transmission to the outside world; this \dword{coldata} \dword{asic} also controls the front-end motherboard and facilitates communications between the motherboard and the outside world.

\begin{dunefigure}[A block diagram of the APA readout electronics]{fig:ElectronicsBlockDiagram}
{Left: an \dword{apa} with 20 \dword{femb}s installed on the head end. Right: a block diagram of the readout electronics mounted on the APAs.}
\includegraphics[width=0.8\textwidth]{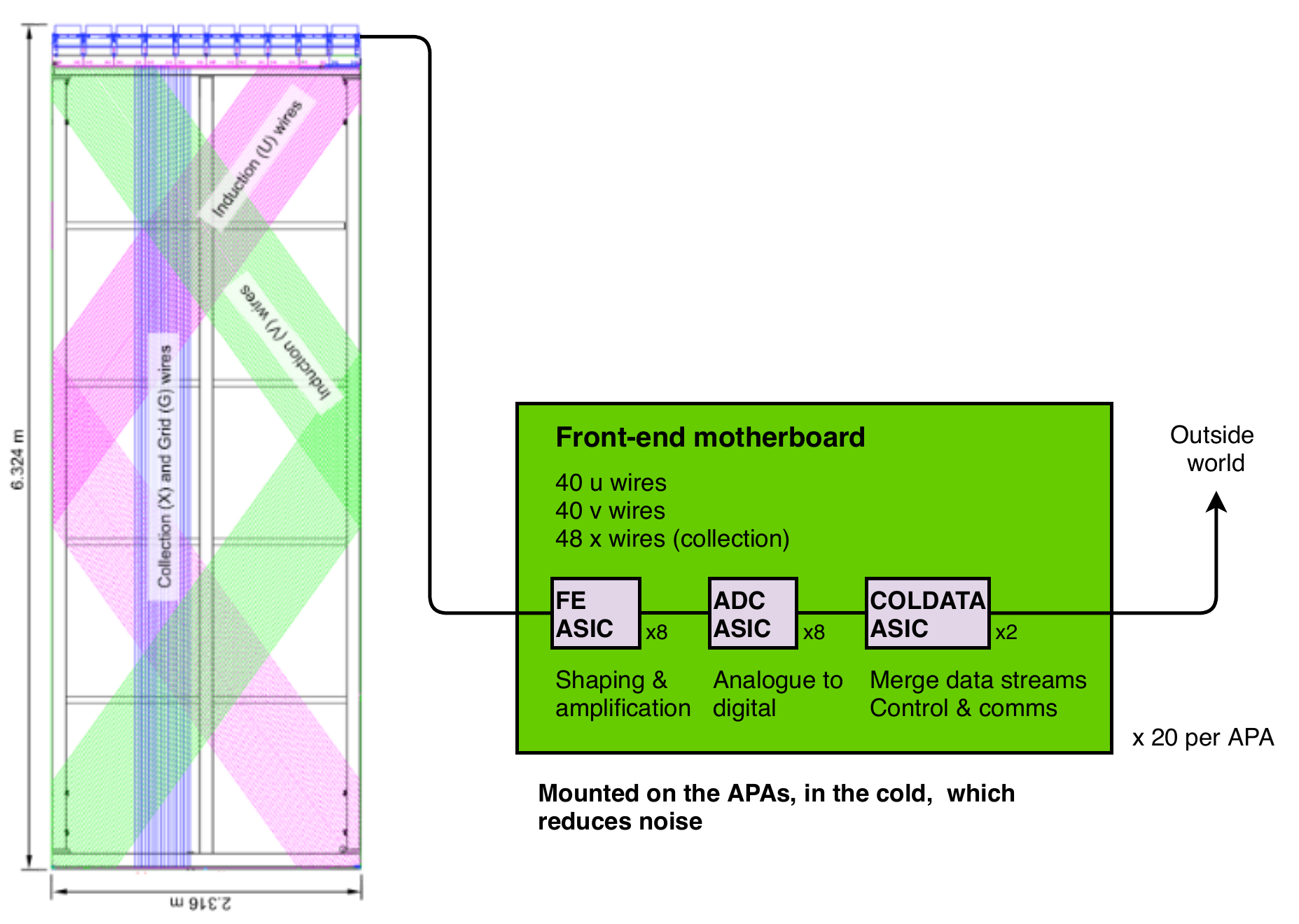}
\end{dunefigure}

The data passes out of the cryostat through feedthroughs in the roof. Mounted directly to each feedthrough is a \dword{wiec}. Each \dword{wiec} contains five \dword{wib}s, each of which processes the signals from five \dword{femb}s. A \dword{wiec} also contains a \dword{ptc} that provides the fiber interface to the timing system, fanning out timing and control systems, as well as the low-voltage power, to the \dword{wib}s via a \dword{ptb}.

\section{Data Acquisition}
\label{sec:fdsp-exec-daq}

The \dword{daq} is divided between an upstream section, located underground in the \dword{cuc}, and a downstream \dword{daqbes} above ground at \dword{surf}. All trigger decisions are made underground, and the data buffered underground until the \dword{daqbes} indicates it is ready to receive data, in order to minimize the rate of data flowing to the surface. An end-goal of the \dword{daq} is to achieve a data-rate to tape of no more than \SI{30}{\peta\byte/\year}.

The \dword{daq} architecture is based on the \dword{felix} system designed at \dword{cern} and used for the LHC
experiments. The 150 \dword{apa}s from each \SI{10}{\kilo\tonne} module are processed by 75 \dword{daqrou}; each \dword{daqrou} contains one \dword{felix} board. The \dword{pd}s from the module will have a lower data-rate since the \dword{pd} electronics, unlike that of the \dword{tpc}, perform zero-suppression; therefore the \dword{pd}s of a module will be processed by six to eight additional \dword{daqrou}s. The \dword{daq} will be partitionable: it will be possible to run multiple instances of the \dword{daq} simultaneously so that the majority of the detector can be taking physics data whilst other \dword{daq} instances are running test runs for development or special runs such as calibration runs. A key philosophy is that all the primary \dword{dune} physics goals can be achieved using only the \dword{tpc} as the trigger; information from the \dword{pd}s can then further enhance the trigger.

There will be two basic triggers operating. Beam, cosmic and nucleon decay events will be triggered using the localized high-energy trigger. This will trigger on localized regions of visible activity, for example in a single \dword{apa}, with a $>99\%$ trigger efficiency at \SI{100}{\mega\electronvolt} and a trigger threshold as low as \SI{10}{\mega\electronvolt}. A localized high-energy trigger will open a readout window of \SI{5.4}{\ms}, enough to read out the full TPC drift around an event. For \dword{snb}s, we will use an extended low-energy trigger. This will look for coincident regions of low-energy deposits, below \SI{10}{\mega\electronvolt}, across an entire module and in a \SI{10}{\second} period. An extended high-energy trigger will open a readout window of \SI{100}{\second} to capture a full \dword{snb}. The upstream \dword{daq} identifies per-channel regions of interest and forms them into trigger primitives. These are then formed into trigger candidates that contain information from an entire module; on these trigger candidates, trigger decisions are made. Once a trigger decision has been made, this will be communicated to the surface, and the data buffered underground until the \dword{daqbes} indicates it is ready to receive data.

The \dword{daq} must also provide the system clock that keeps the detector components synchronized and provides the timestamp for all data. The timestamp derives from a \dword{gps} \dword{pps} that is fed into the \dword{daq} with \SI{1}{\micro\second} precision, adequate to timestamp beam and supernova events. To provide the finer synchronization between detector components, a \SI{10}{\mega\hertz} reference clock drives the module's \SI{62.5}{\mega\hertz} master clock, which is fanned out to all detector components, providing an overall synchronization to a precision of \SI{1}{\nano\second}.

\section{Calibration}
\label{sec:fdsp-exec-calibration}

The challenge of calibrating the \dword{dune} \dword{fd} is to control the response of a huge cryogenic detector over a period of decades, a challenge amplified by the detector's location deep underground and therefore shielded from the cosmic muons that were typically used as standard candles by previous \dword{lartpc}s.

To achieve our $\mathcal{O}(\si{\giga\electronvolt})$ oscillation and nucleon decay physics goals, we must know our fiducial volume to 1--2\% and have a similar understanding of the vertex position resolution; understand the \nue event rate to 2\%; and control our lepton and hadron energy scales to 1\% and 3\%, respectively. At the $\mathcal{O}(\si{\mega\electronvolt})$ scale our physics requirements are driven by our goal of identifying, and measuring the spectral structure of, a \dword{snb}; here, we must achieve a 20--30\% energy resolution, understand our event timing to the \SI{1}{\micro\second} level, and measure our trigger efficiency and levels of radiological background. These are all high-level calibration requirements, but the underlying detector parameters that we are characterizing are parameters such as the energy deposited per unit length ($\mathrm{d}E/\mathrm{d}x$), ionization electron drift-lifetimes, scintillation light yield and detection efficiency, \efield maps, timing precision, \dword{tpc} alignment, and the behavior (noise, gain, cross-talk, linearity, etc.) of electronics channels.

The tools available to us for calibration include the \dword{lbnf} beam, atmospheric neutrinos, atmospheric muons, radiological backgrounds, and dedicated calibration devices that will be installed in the detector. At the lowest energies, we have deployable neutron sources and intrinsic radioactive sources; in particular the natural $^{39}$Ar component of the \dword{lar} with its \SI{565}{\kilo\electronvolt} end-point can, given its pervasive nature across the detector, be used to measure the spatial and temporal variations in electron lifetime. The possibility of deploying radioactive sources is also being explored. In the \SIrange{10}{100}{\mega\electronvolt} energy range we will use Michel electrons, photons from $\pi^{0}$ decay, stopping protons and both stopping and through-going muons. We will also have built-in lasers, purity monitors and thermometers, and the ability to inject charge into the readout electronics. Finally, data from the \dword{protodune} detectors will be invaluable in understanding the response and particle-identification capabilities of the \dword{fd}.

Once the first \SI{10}{\kilo\tonne} module is switched on, there will be a period of years before \dword{lbnf} beam sources are available for calibration --- and even then the statistics will be limited. In this time, cosmic muons will be available, but the low rate of these means that it will take months to years to build up the necessary statistics for calibration. The inclusive cosmic muon rate for each \SI{10}{\kilo\tonne} module is $1.3\times 10^{6}$ per year. However, for calibrations such as \dword{apa} alignment, the typical rate of useful muons is 3000--4000 per APA per year. For energy-scale calibrations, stopping cosmic muons are the most relevant and here the rate is 11000 per \SI{10}{\kilo\tonne} module per year. Therefore the earliest calibrations will come from dedicated calibration hardware systems and intrinsic radiological sources.

A \SI{266}{\nano\meter} laser will be used to ionize the argon, and this can be used to map the \efield and to make early measurements of \dword{apa} alignment. The laser system will be used throughout the lifetime of the detector to measure the gradual changes in the \efield map as positive ions accumulate and flow around the detector.
An externally deployed pulsed neutron source provides a triggered, well defined energy deposition from neutron capture in argon which is an important component of signal processes for \dword{snb} and \dword{lbl} physics. A radioactive source deployment system, which is complementary to the pulsed neutron system, can provide at known locations inside the detector a source of gamma rays in the same energy range as \dword{snb} and solar neutrino physics

Over time, the \dword{fd} calibration program will evolve as statistics from the cosmic rays and the \dword{lbnf} beam amass and add to the information gained from the calibration hardware systems. These numerous calibration tools will work alongside the detector monitoring system, the computational fluid dynamics models of the argon flow, and \dword{protodune} data to give us a detailed understanding of the \dword{fd} response across the \dword{dune} physics program.

\section{Installation}

A major challenge in building the \dword{dune} \dword{sp} modules is transporting all the detector and infrastructure components down the \SI{1500}{\meter} Ross shaft, to the detector caverns. To aid the planning of the installation phase, installation tests will be performed at the \dword{nova} \dword{fd} site in \dword{ashriver}, Minnesota, USA. These tests will allow us to develop our procedures, train the installation workers, and develop our labor planning through time and motion studies.

Once the module's cryostat has been installed, a \dword{tco} is left open at one end through which the detector components are installed. A cleanroom is built around the \dword{tco} to prevent any contamination entering the cryostat during installation. The \dword{dss} is then installed into the cryostat, ready to receive the \dword{tpc} components. 

Inside the cryostat, the various monitor devices (temperature, purity, argon level) are installed at the end furthest from the \dword{tco}. The far end of the \dword{fc} is then installed. Rows of \dword{apa}s and \dword{cpa}s, along with the top and bottom \dword{fc} sections, are then installed and cabled, working from the far end of the detector towards the \dword{tco}. The integration of the \dword{pd}s and \dword{ce} with the \dword{apa} happens in the cleanroom immediately outside the \dword{tco}. Finally, the second \dword{fc} end-wall is installed across the \dword{tco}, along with the monitoring devices at the \dword{tco} end. The \dword{tco} can then be closed up and the cryostat is ready to purge and fill with \dword{lar}. The warm electronics and \dword{daq} are installed in parallel with the \dword{tpc} installation.

Throughout the installation process, safety is the paramount consideration: safety both of personnel and of the detector components. Once the detectors are taking data, safety is still the priority with the \dword{ddss} monitoring for argon level drops, water leaks and smoke. A detailed detector and cavern grounding scheme has been developed that not only guards against ground loops, but also ensures that any power faults are safely shunted to the facility ground.

Throughout the project, \dword{qa} and \dword{qc} are written into all processes. Most detector components are constructed off-site at collaborating institutions; strict \dword{qc} procedures will be followed at all production sites to ensure that components are working within specifications before delivery to \dword{surf}. Underground at \dword{surf} integrated detector components are tested in the cleanroom to ensure functionality, before passing them through the \dword{tco} for installation. Finally, \dword{qc} is performed on all integrated components inside the cryostat, in particular to ensure that all connections have been made through to the \dword{cuc}.

\section{Schedule and Milestones}
\label{sec:fdsp-exec-sched}

A set of key milestones and dates  have been defined for planning purposes in the development of the \dword{tdr}.  The dates will be finalized once the international project baseline has been defined.  Table~\ref{tab:sp-key-dates} shows the key dates and milestones (colored rows) and indicates the way that detector consortia will add subsystem-specific milestones based on these dates (no background color). A more detailed schedule for the detector installation is discussed in Chapter~\ref{ch:sp-install}.
 
\begin{dunetable}
[Key milestones and dates]
{p{0.65\textwidth}p{0.25\textwidth}}
{tab:sp-key-dates}
{(Sample subsystem) construction schedule milestones leading to commissioning of the first two  FD modules. Key DUNE dates and milestones, defined for planning purposes in this TDR, are shown in orange.  Dates will be finalized following establishment of the international project baseline.}   
Milestone & Date (Month YYYY)   \\ \toprowrule
Technology Decision Dates &   April 2020   \\ \colhline
Final Design Review Dates &   June 2020   \\ \colhline
Start of module 0 component production for \dshort{protodune2} & August 2020  \\ \colhline
End of module 0 component production for \dshort{protodune2} & January 2021  \\ \colhline
\rowcolor{dunepeach} Start of \dshort{pdsp}-II installation& \startpduneiispinstall      \\ \colhline
\rowcolor{dunepeach} Start of \dshort{pddp}-II installation& \startpduneiidpinstall      \\ \colhline
\rowcolor{dunepeach}South Dakota Logistics Warehouse available& \sdlwavailable      \\ \colhline
 \dword{prr} dates &  September 2022    \\ \colhline
\rowcolor{dunepeach}Beneficial occupancy of cavern 1 and \dword{cuc}& \cucbenocc      \\ \colhline
Start procurement of (subsystem) hardware & December 2022 \\ \colhline
\rowcolor{dunepeach} \dshort{cuc} counting room accessible& \accesscuccountrm      \\ \colhline
\rowcolor{dunepeach}Top of \dshort{detmodule} \#1 cryostat accessible& \accesstopfirstcryo      \\ \colhline
\rowcolor{dunepeach}Start of \dshort{detmodule} \#1 \dshort{tpc} installation& \startfirsttpcinstall      \\ \colhline
\rowcolor{dunepeach}Top of \dshort{detmodule} \#2 cryostat accessible& \accesstopsecondcryo      \\ \colhline
\rowcolor{dunepeach}End of \dshort{detmodule} \#1 \dshort{tpc} installation& \firsttpcinstallend      \\ \colhline
 \rowcolor{dunepeach}Start of \dshort{detmodule} \#2 \dshort{tpc} installation& \startsecondtpcinstall      \\ \colhline
\rowcolor{dunepeach}End of \dshort{detmodule} \#2 \dshort{tpc} installation& \secondtpcinstallend      \\  \colhline
Full  (subsystem) commissioned and integrated into remote operations & July 2026 \\ 
\end{dunetable}

\section{Conclusion}
\label{sec:fdsp-exec-conclusion}

This executive summary has provided an overview of the design of the \SI{10}{\kilo\tonne} \dword{sp} \dword{lartpc} modules of the \dword{dune} \dword{fd}, explaining how key design choices have been made to ensure we can achieve our primary physics goals of searching for leptonic $CP$ violation, nucleon decay and neutrinos from supernova bursts. The chapters that follow go into significantly more detail about the design of the \dword{sp} \dword{fd} modules. In addition to describing the design and requirements, these chapters include details on the construction, integration and installation procedures, the \dword{qa} and \dword{qc} processes that have been developed to ensure that the detector will function for a period of decades, and the overall project management structure. The chapters also describe how the design has been validated and informed by \dword{protodune}.

\cleardoublepage

\chapter{Anode Plane Assemblies}
\label{ch:fdsp-apa}

\section{Anode Plane Assembly (APA) Overview}
\label{sec:fdsp-apa-intro}

The \dwords{apa}, or wire planes, are the \dword{dune} \dword{sp} module elements used to sense, through both signal induction and direct collection, the ionization electrons created when charged particles traverse the \dword{lar} volume inside the \dword{spmod}.   All elements of the \dword{dune} physics program depend on a high performing system of \dword{apa}s and their associated readout \dword{ce}.  

Volume~\volnumberphysics{} of this \dword{tdr}, \voltitlephysics, 
describes the simulations that rigorously establish the requirements for achieving the needed performance.  Here we summarize some of the \dword{apa} capabilities required for the key elements of neutrino \dword{cpv} and associated long-baseline oscillation physics, \dword{ndk}, and intra-galactic \dword{snb} searches.  As a multipurpose detector accessing physics from MeV to multi-GeV scales, the \dword{dune} \dword{lartpc} cannot be optimized for a narrow range of interaction signatures in the manner of noble liquid \dwords{tpc} dedicated to direct \dword{dm} or neutrino-less double beta decay searches.  The \dword{apa}s must collect ionization charge in a way that preserves the spatial and energy profiles of ionization events that range from few hundred keV point-like depositions (from low energy electrons and neutrons created in \dword{snb} neutrino interactions) to the double-kinked $K\rightarrow\mu\rightarrow{e}$ decay chain with its combination of highly- and minimum-ionizing particles (HIPs and MIPs) that is a key signature in proton decay searches.  The \dword{apa}s must record enough hits on tracks within a few cm of a neutrino interaction vertex to differentiate the 1 \dword{mip} $dE/dx$ signature of a $\nu_e$-induced electron from the 2 \dword{mip} signature of a $\nu$ neutral current photon conversion to enable the $\nu_\mu-\nu_e$ separation demanded for \dword{cpv} physics; and they must provide the pattern recognition and calorimetry for multi-GeV neutrino interaction products  spread over cubic meters of the detector needed for the precision neutrino energy estimates that allow separation of \dword{cpv} effects from those related to matter effects. 
 
Anode planes in the \dword{apa} must be well-shielded from possible high voltage breakdown events in the \dword{detmodule}.  The \dword{apa} wire spacing and orientations must maximize pattern recognition capabilities and \dword{s/n} in a cost-effective manner.  The \dword{apa} wires must maintain their positions to a level that is small compared to the wire spacing so that energy estimators based on range and multiple Coulomb scattering remain reliable over two decades of operation.  The wires must hold their tension, lest microphonic oscillations develop that degrade \dword{s/n} or anode plane field distortions arise that inhibit the transmission of drifting electrons through the induction planes to the collection plane.  Any wire break would destroy \dwords{fv}, so the \dword{apa} design must both minimize the possibility of this occurrence and contain the extent of any damage that would ensue should it happen.  An \dword{apa} implementation that meets all these goals follows in the remainder of this chapter, along with a summary of significant validations achieved through dedicated simulations and \dword{pdsp} construction and operations.

\begin{dunefigure}[A \nominalmodsize DUNE far detector SP module]{fig:DUNESchematic1ch1}
{A \nominalmodsize \dword{dune} \dword{fd} \dword{spmod}, showing the alternating \sptpclen{} long (into the page), \tpcheight{} high anode (A) and cathode (C) planes, as well as the \dfirst{fc} that surrounds the drift regions between the anode and cathode planes. On the right-hand cathode plane, the foremost portion of the \dword{fc} is shown in its undeployed (folded) state.}
\includegraphics[width=0.65\textwidth]{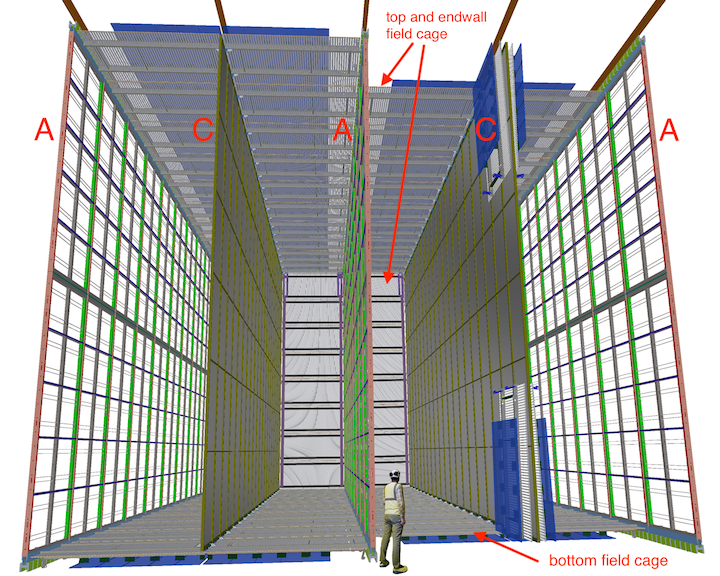}
\end{dunefigure}

To facilitate fabrication and installation underground, the anode design is modular, with \dword{apa}s tiled together to form the readout system for a \nominalmodsize \dword{detmodule}. A single \dword{apa} is \SI{6}{m} high by \SI{2.3}{m} wide, but two of them are connected vertically, and twenty-five of these vertical stacks are linked together to define a \tpcheight 
tall by \sptpclen 
long mostly-active readout plane.  As described below, the planes are active on both sides, so three such wire readout arrays (each one \tpcheight$\,\times\,$\sptpclen) 
are interleaved with two \dword{hv} surfaces to define four \spmaxdrift 
wide drift regions inside each \dword{spmod}, as Figure~\ref{fig:DUNESchematic1ch1} shows in the detector schematic views. Each \dword{sp} \nominalmodsize module, therefore, will contain 150 \dword{apa}s.

Each \dword{apa} frame is covered by more than \num{2500} sense wires laid in three planes  oriented at angles to each other: a vertical collection plane, $X$, and two induction planes at \apainducwireangle to the vertical, $U$ and $V$. Having three planes allows multi-dimensional reconstruction of particle tracks even when the particle propagates parallel to one of the wire plane directions.  An additional \num{960} wires that are not read out make up an outer shielding plane, $G$, to improve signal shapes on the $U$ induction channels.  The angled wires are wrapped around the frame from one side to the other, allowing all channels to be read out from one end of the \dword{apa} only (the top or bottom), thereby minimizing the dead regions between neighboring \dword{apa}s. Signals induced or collected on the wires are transferred through soldered connections to wire termination boards mounted at the end of the \dword{apa} frame that in turn connect to \dword{fe} readout \dword{ce} sitting in the \dword{lar}.  Figures~\ref{fig:tpc_apa1} and \ref{fig:sp-apa-head-xsec} illustrate the layout of the wires on an \dword{apa}, showing how they wrap around the frame and terminate on wire boards at the head end where readout \dword{ce} are mounted.

The \dword{apa}s are a critical interface point between the various detector subsystems within the \dword{spmod}.  As already mentioned, the \dword{tpc} readout \dword{ce} mount directly to the \dword{apa} frames.  The \dwords{pd} for detecting scintillation light produced in the \dword{lar} are also housed inside the frames, sandwiched between the wires on the two sides, requiring careful coordination in frame design as well as requiring transparency for the \dword{apa} structures.  In addition, the electric \dfirst{fc} panels connect directly to the edges of the \dword{apa} frames.  Finally, the \dword{apa}s must support routing cables for both the \dword{tpc} electronics and the \dword{pd} systems. All these considerations are important to the design, fabrication, and installation planning of the \dword{apa}s.

\begin{dunefigure}[Illustration of the APA wire layout]{fig:tpc_apa1}
{Illustration of the \dword{dune} \dword{apa} wire wrapping scheme showing small portions of the wires from the three signal planes ($U,V,X$). The fourth wire plane ($G$) above these three, and parallel to $X$, is present to improve the pulse shape on the $U$ plane signals. The \dword{tpc} electronics boxes, shown in blue on the right, mount directly to the frame and process signals from both the collection and induction channels. The \dword{apa} is shown turned on its side in a horizontal orientation.} 
\includegraphics[width=\textwidth,trim = 0mm 10mm 0mm 0mm,clip]{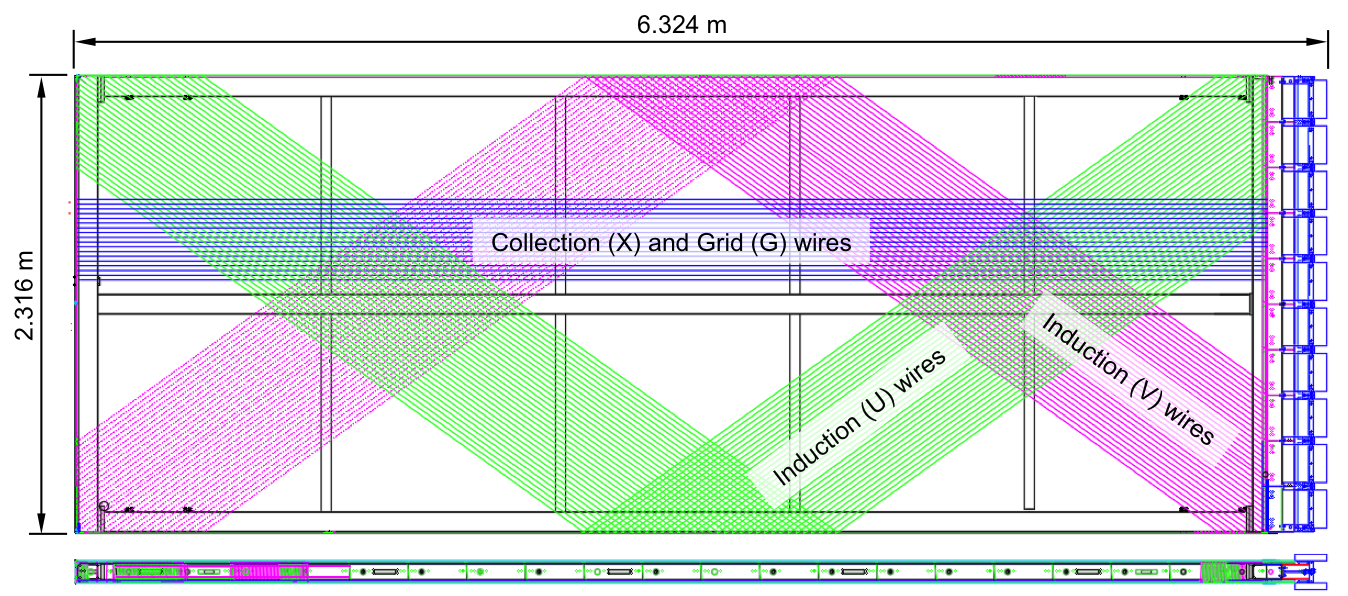} 
\end{dunefigure} 

\begin{dunefigure}[Cross section view of the head end and wire layers of an APA]{fig:sp-apa-head-xsec}
{Cross section view of an \dword{apa} frame near the head end showing the layers of wires ($G,U,V,X$) on both sides of the frame that terminate on wire boards, which connect to \dword{tpc} readout \dword{ce} through a capacitor-resistor chain on the \dword{cr} boards and a connector adapter board.} 
\includegraphics[width=1\textwidth]{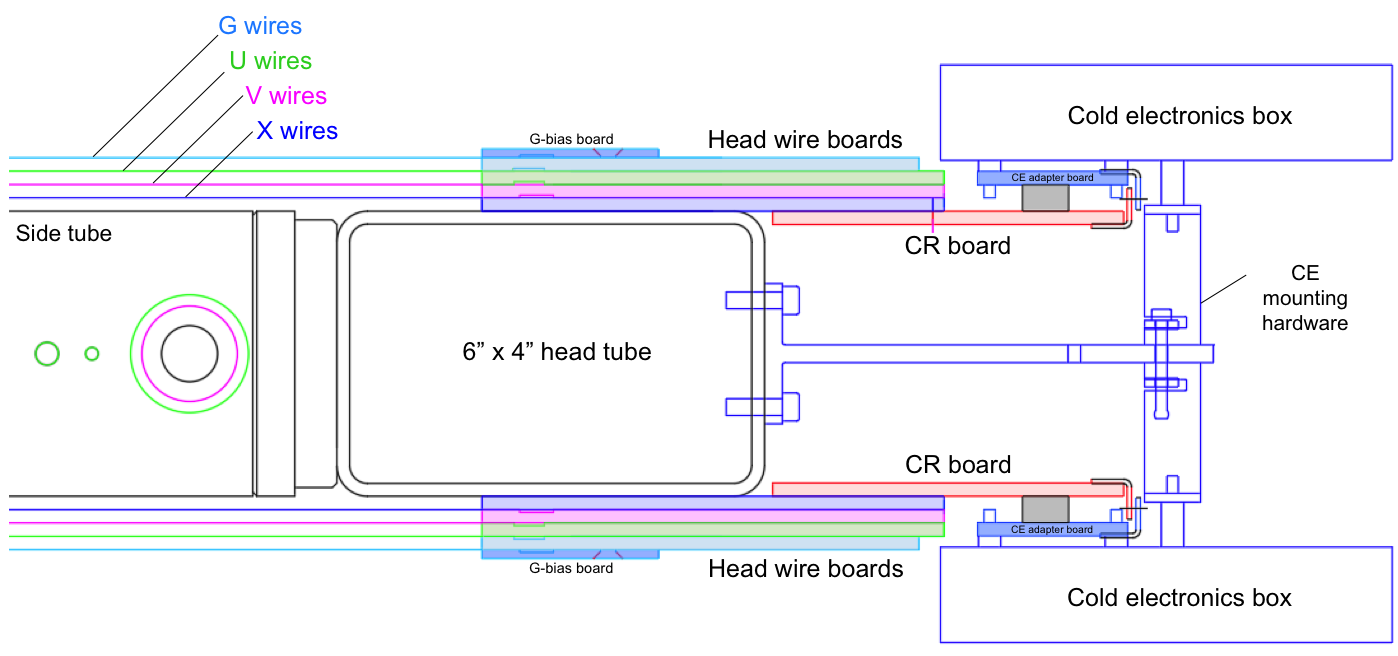} 
\end{dunefigure}

The \dword{apa} consortium 
oversees the design, construction, testing, and installation of the \dword{apa}s. Several \dword{apa} production sites will be set up in both the US and the UK with each nation producing  half of the \dword{apa}s needed for the 
\dwords{spmod}.  Production site setup is anticipated to begin in 2020, with \dword{apa} fabrication for the first \nominalmodsize \dword{spmod} running from 2021--2023.  

The Physical Sciences Laboratory (PSL) at the University of Wisconsin and the Daresbury Laboratory in the UK have recently produced full-scale \dword{apa}s for the \dword{pdsp} project at \dword{cern}. Figure~\ref{fig:apa-photo} shows a completed \dword{apa} produced at PSL just before shipment to \dword{cern}. 
This effort has greatly informed the design and production planning for the \dword{dune} \dwords{detmodule}, and \dword{pdsp} running has provided valuable validation for many fundamental aspects of the  \dword{apa} design. 

The remainder of this chapter is laid out as follows.  In Section~\ref{sec:fdsp-apa-design}, we present an overview of the design of the \dword{apa}s, focusing on the key design parameters and their connection to the physics requirements of \dword{dune}.  In Section~\ref{sec:fdsp-apa-qa}, we discuss quality assurance for the design with an emphasis on lessons learned from \dword{pdsp} construction and operations and a summary of remaining prototyping efforts being planned before the start of production next year. Section~\ref{sec:fdsp-apa-intfc} summarizes three important interfaces to the \dword{apa}s: \dword{tpc} cold electronics (\dword{ce}), photon detectors (\dword{pd}), and the cable routing for both systems.  In Section~\ref{sec:fdsp-apa-prod}, we detail the production plan for fabricating the large number of \dword{apa}s needed for the experiment including a description of the main construction sites being developed in the US and UK by the \dword{apa} consortium.  Section~\ref{sec:fdsp-apa-transport} describes some requirements for handling the large and delicate \dword{apa}s throughout construction and presents the design for a custom transport system for delivery to the far detector site for installation. Section~\ref{sec:fdsp-apa-safety} reviews the safety considerations for \dword{apa} construction and handling. Finally,  Section~\ref{sec:fdsp-apa-org} summarizes the organization of the \dword{apa} consortium that is responsible for building the \dword{apa}s and provides the high-level cost, schedule, and risk summary tables for the project. 

\begin{dunefigure}[A completed APA for \dshort{pdsp}]{fig:apa-photo}
{Completed \dshort{pdsp} \dshort{apa} ready for shipment to \dword{cern}.}
\includegraphics[width=1.0\textwidth,trim=20mm 80mm 0mm 60mm,clip]{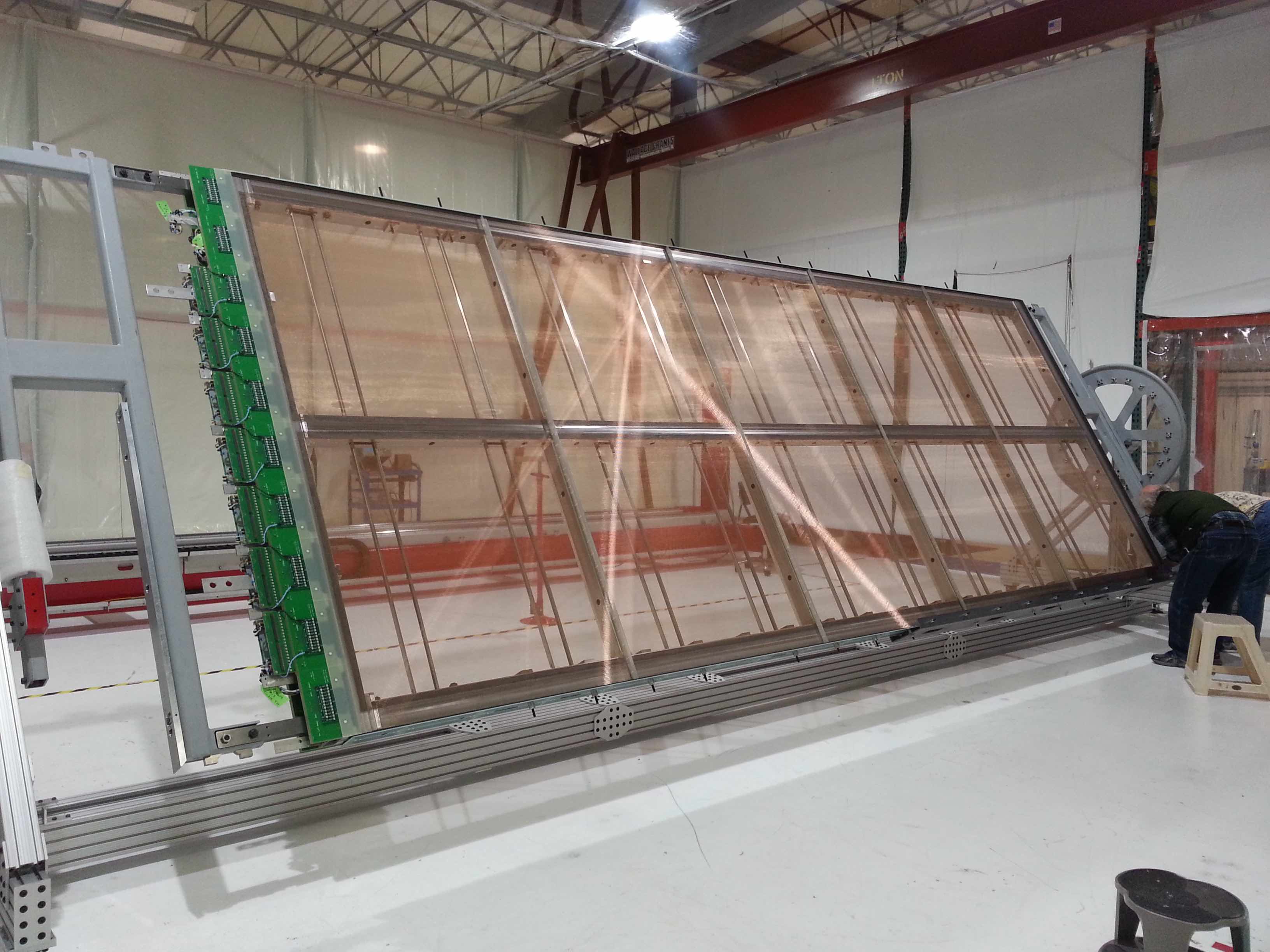}
\end{dunefigure}

\section{Design}
\label{sec:fdsp-apa-design}

The physics performance of the \dword{spmod} is a function of many intertwined detector parameters including argon purity, drift distance, \efield, wire pitch, wire length, and noise levels in the readout \dword{ce}.  Energy deposits from \dwords{mip} originating anywhere inside the active volume of the detector should be identifiable with near \num{100}\% efficiency.  This requirement constrains aspects of the \dword{apa} design, specifically, the limits on wire pitch, wire length, and choice of wire material.  This section details the design of an individual \dword{apa}. We begin with an overview of the key fundamental parameters of the \dword{apa}s and their connection to the physics requirements of the \dword{dune} experiment.

\subsection{APA Design Parameters}
\label{sec:fdsp-apa-design-overview}

Each \dword{apa} is \SI{6}{m} high, \SI{2.3}{m} wide, and \SI{15}{cm} thick.  The underlying support frame is made from stainless steel hollow tube sections that are precisely machined and bolted together. A fine, conducting mesh covers the rectangular openings in the frame on both sides to define a uniform electrical \dword{gp} behind the wires. The four layers of sense and shielding wires at varying angles relative to each other completely cover the frame. The wires terminate on boards that anchor them as well as provide the electrical connection to the \dword{tpc} readout \dword{ce}. Starting from the outermost wire layer, there is first an uninstrumented shielding (grid) plane (strung vertically, $G$), followed by two induction planes (strung at \apainducwireangle to the vertical, $U,V$), and finally the collection plane (vertical, $X$). All wire layers span the full height of the \dword{apa} frame. The two planes of induction wires wrap in a helical fashion around the long edges and over both sides of the \dword{apa}. Figures~\ref{fig:tpc_apa1} and \ref{fig:sp-apa-head-xsec} illustrate the layout of the wire layers.  Below, we summarize the key design parameters and the considerations driving the main design choices for the \dword{apa}s.  A tabulated summary of \dword{apa} specific requirements is also provided in Table~\ref{tab:specs:SP-APA}.

\begin{footnotesize}
\begin{longtable}{p{0.12\textwidth}p{0.18\textwidth}p{0.17\textwidth}p{0.25\textwidth}p{0.16\textwidth}}
\caption{APA specifications \fixmehl{ref \texttt{tab:spec:SP-APA}}} \\
  \rowcolor{dunesky}
       Label & Description  & Specification \newline (Goal) & Rationale & Validation \\  \colhline

  \newtag{SP-FD-6}{ spec:apa-gaps }  & Gaps between APAs   &  $<\,\SI{15}{mm}$ between APAs on same support beam; $<\,\SI{30}{mm}$ between APAs on different support beams &  Maintains fiducial volume.  Simplified contruction. &  ProtoDUNE \\ \colhline

  \newtag{SP-FD-7}{ spec:misalignment-field-uniformity }  & Drift field uniformity due to component alignment  &  $<\,1\,$\% throughout volume &  Maintains APA, CPA,  FC orientation and shape. &  ProtoDUNE \\ \colhline

  \newtag{SP-FD-8}{ spec:apa-wire-angles }  & APA wire angles  &  \SI{0}{\degree} for collection wires, $\pm\,$\SI{35.7}{\degree} for induction wires &  Minimize inter-APA dead space. &  Engineering calculation \\ \colhline

  \newtag{SP-FD-9}{ spec:apa-wire-spacing }  & APA wire spacing  &  \SI{4.669}{mm} for U,V; \SI{4.790}{mm} for X,G &  Enables 100\% efficient MIP detection, \SI{1.5}{cm} $yz$ vertex resolution. &  Simulation \\ \colhline

  \newtag{SP-FD-10}{ spec:apa-wire-pos-tolerance }  & APA wire position tolerance  &  $\pm\,\SI{0.5}{mm}$ &  Interplane electron transparency; $dE/dx$, range, and MCS calibration. &  ProtoDUNE and simulation \\ \colhline

  \newtag{SP-APA-1}{ spec:apa-unit-size }  & APA unit size  &  \SI{6.0}{m} tall $\times$ \SI{2.3}{m} wide &  Maximum size allowed for fabrication, transportation, and installation.  &  ProtoDUNE-SP  \\ \colhline

  \newtag{SP-APA-2}{ spec:apa-active-area }  & Active area  &  Maximize total active area. &  Maximize area for data collection  &  ProtoDUNE-SP  \\ \colhline

  \newtag{SP-APA-3}{ spec:apa-wire-tension }  & Wire tension  &  \SI{6}{N} $\pm$ \SI{1}{N} &  Prevent contact beween wires and minimize  break risk &  ProtoDUNE-SP \\ \colhline

  \newtag{SP-APA-4}{ spec:apa-bias-voltage }  & Wire plane bias voltages  &  The setup, including boards, must hold 150\% of max operating voltage. &  Headroom in case adjustments needed &  E-field simulation sets wire bias voltages. ProtoDUNE-SP confirms performance. \\ \colhline

  \newtag{SP-APA-5}{ spec:apa-frame-planarity }  & Frame planarity (twist limit)  &  $<$\SI{5}{mm} &  APA transparency.  Ensures wire plane spacing change of $<$0.5 mm.  &  ProtoDUNE-SP \\ \colhline

  \newtag{SP-APA-6}{ spec:apa-bad-channels }  & Missing/unreadable channels  &  $<$1\%, with a goal of $<$0.5\% &  Reconstruction efficiency &  ProtoDUNE-SP \\ \colhline

\label{tab:specs:SP-APA}
\end{longtable}
\end{footnotesize}

\begin{itemize}
\item \textbf{\dword{apa} size:} The size of the \dword{apa}s is chosen for fabrication purposes, compatibility with over-the-road shipping, and for eventual transport to the 4,850L at \dword{surf} and installation into 
a cryostat. The dimensions are also chosen so that an integral number of electronic readout channels and boards instrument the full area of the \dword{apa}.

\item \textbf{Detector active area:} \dword{apa}s should be sensitive over most of the full area of an \dword{apa} frame, with any dead regions between \dword{apa}s due to frame elements, wire boards, electronics, or cabling kept to a minimum.  The wrapped style shown in Figure~\ref{fig:tpc_apa1} allows all channels to be read out at the top of the \dword{apa}, eliminating the dead space between \dword{apa}s that would otherwise be created by electronics and associated cabling. In addition, in the design of the \dword{spmod}, a central row of 
\dword{apa}s is flanked by drift-field regions on either side (Figure~\ref{fig:DUNESchematic1ch1}), 
and the wrapped design allows the induction plane wires to sense drifted ionization that originates from either side of the \dword{apa}.  This double-sided feature is also effective for the \dword{apa}s located against the cryostat walls where the drift field is on only one side; the grid layer facing the wall effectively blocks any ionization generated outside the \dword{tpc} from drifting in to the wires on that side of the \dword{apa}.        

\item \textbf{Wire angles:} The $X$ wires run vertically to provide optimal reconstruction of beam-induced particle tracks, which are predominantly forward (in the beam direction). The angle of the induction planes on the \dword{apa}, \apainducwireangle, was chosen to ensure that each induction wire only crosses a given collection wire once, reducing the ambiguities that the reconstruction must address.  Simulation studies (see next item) show that this configuration performs similarly to an optimal 45$^\circ$ wire angle for the primary \dword{dune} physics channels.  The design angle of the induction wires, coupled with their pitch, also satisfies the requirement 
of using an integer multiple of electronics boards to read out one \dword{apa}.

\item \textbf{Wire pitch:} The wire spacing, \xgpitch for $(X,G)$ and \uvpitch for $(U,V)$, combined with key parameters for other \dword{tpc} systems 
can achieve the required performance for energy deposits by \dwords{mip} while providing good tracking resolution and good granularity for particle identification. The \single requirement that it be possible to determine the fiducial volume to \num{1}\% implies a vertex resolution of \SI{1.5}{cm} along each coordinate direction. The $\sim$\SI{4.7}{mm} wire pitch achieves this for the $y$ and $z$ coordinates.  The resolution on $x$, the drift coordinate, will be better than in the $y$--$z$ plane because of the combination of drift velocity and electronics sampling rate.  Finally, as already mentioned, the total number of wires on an \dword{apa} will match the granularity of the electronics boards (each \dword{femb} can read out \num{128} wires, mixed between the $U,V,X$ planes). This determines the exact wire spacings of \xgpitch on the collection plane and \uvpitch on the induction planes.  To achieve the reconstruction precision required (e.g., for $dE/dx$ reconstruction accuracy and multiple Coulomb scattering determination), the tolerance on the wire pitch is \wirepitchtol.

In 2017, the \dword{dune} \dword{fd} task force, using a full \dword{fd} simulation and reconstruction chain, performed detector optimization studies to quantify the impact of design choices, including wire pitch and wire angle, on \dword{dune} physics performance.  The results indicated that reducing wire spacing (to \SI{3}{mm}) or changing wire angle (to \num{45}$^\circ$) would not significantly affect the performance for the main physics goals of \dword{dune}, including $\nu_\mu $ to $\nu_e$ oscillations and \dword{cpv} sensitivity.  
A key low-level metric for oscillation physics is the ability to distinguish electrons versus photons in the detector because photon induced showers can fake electron showers and create \dword{nc} generated backgrounds in the $\nu_e$ \dword{cc} event sample.  Two important handles for reducing this contamination are (1)~the visible gap between the vertex of the neutrino interaction and the start of a photon shower, and 
(2)~the accordance of the energy density at the start of the shower with one \dword{mip} instead of two.

\begin{dunefigure}[Electron-photon separation dependence on wire pitch and angle]{fig:e-gamma}
{Summary of electron--photon separation performance studies from the \dword{dune} \dword{fd} task force. (a) $e$--$\gamma$ separation by $dE/dx$ for the nominal wire spacing and angle (\SI{4.7}{mm}/$37.5^\circ$) compared to \SI{3}{mm} spacing or 45$^\circ$ induction wire angles. (b) Electron signal selection efficiency versus photon (background) rejection for the different detector configurations. The \SI{3}{mm} wire pitch and 45$^\circ$ wire angle have similar effects, so the 45$^\circ$ curve is partly obscured by the \SI{3}{mm} curve.}
(a)
\includegraphics[height=0.24\textheight]{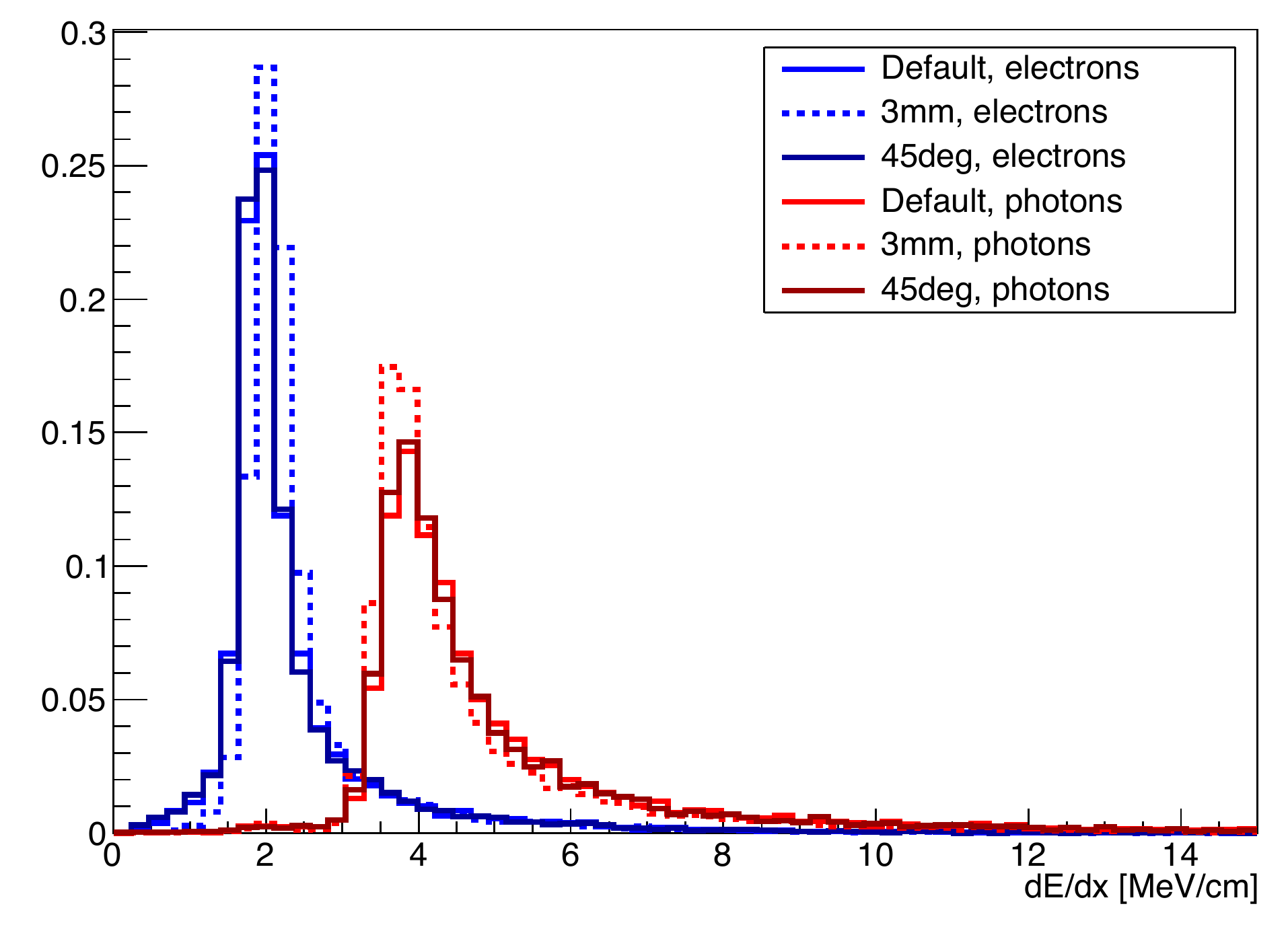} \qquad
(b)
\includegraphics[height=0.24\textheight]{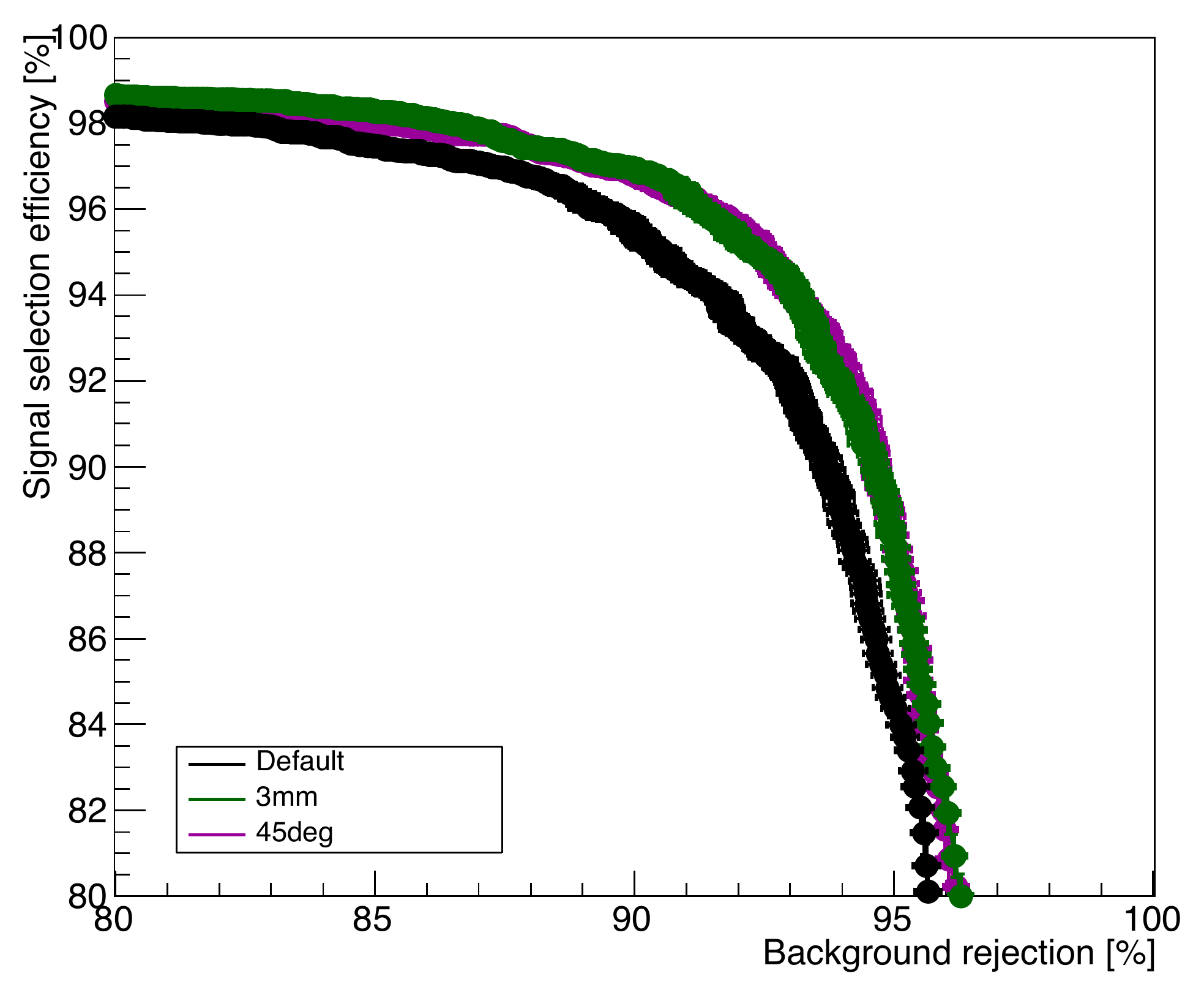} 
\end{dunefigure}

A detector spatial resolution much smaller than the radiation length for photons (\SI{0.47}{cm} vs. \SI{14}{cm}) allows the gap between the neutrino interaction vertex and a photon conversion point to be easily visible, and  Figure~\ref{fig:e-gamma}(a) shows the reconstructed ionization energy loss density ($dE/dx$) in the first centimeters of electron and photon showers, illustrating the separation between the single \dword{mip} signal from electrons and the double \dword{mip} signal when photons pair-produce an $e^+e^-$.  In the figure, the ($dE/dx$) separation for electrons and photons is compared for finer wire pitch (\SI{3}{mm}) and optimal wire angle ($45^\circ$). The final electron signal selection efficiency is also shown as a function of the background rejection rate for different wire configurations in Figure~\ref{fig:e-gamma}(b). At a signal efficiency of \num{90}\,\%, for example, the background rejection can be improved by about \num{1}\,\% using either \SI{3}{mm} spacing or 45$^\circ$ wire angles for the induction planes.  This slight improvement in background rejection with more dense hit information or more optimal wire angles is not surprising, but the effect on high-level physics sensitivities from these changes is very small. The conclusions of the \dword{fd} task force, therefore, were that the introduction of ambiguities into the reconstruction by increasing the wire angles is not a good trade off, and the increase in cost incurred by decreasing the wire pitch (and, therefore, increasing the number of readout channels) is not justified.

\item \textbf{Wire plane transparency and signal shapes:}  The ordering of the layers, starting from the active detector region, is $G$-$U$-$V$-$X$, followed by the grounding mesh. The operating voltages of the \dword{apa} layers are listed in Table~\ref{tab:bias}.  These were calculated by COMSOL software 
in order to maintain a \num{100}\% ionization electron transparency as they travel through the grid and induction wire planes. Figure~\ref{fig:apa-fields} shows the field simulation and expected signal shapes for the bias voltages listed in the table.  When operated at these voltages, the drifting ionization follows trajectories around the grid and induction wires, terminating on a collection plane wire. The grid and induction layers are completely transparent to drifting ionization, and the collection plane is completely opaque.  The grid layer is present for pulse-shaping and not connected to the electronics readout; it effectively shields the first induction plane from the drifting charge and removes a long leading edge from the signals on that layer.  These operating conditions were confirmed by a set of dedicated runs in \dword{pdsp} taken with various bias voltage settings during spring 2019 (see Sec.~\ref{sec:fdsp-apa-qa-protodune-ops} for a detailed discussion).

\begin{dunetable}[APA wire plane nominal bias voltages]{lr}{tab:bias}
{Baseline bias voltages for \dword{apa} wire layers for a 100\% ionization electron transparency as they travel through the grid and induction wire planes. These values were calculated by COMSOL software 
and confirmed by analytical calculations based on the conformal representation theory as well as dedicated data from \dword{pdsp}.} 
\textbf{Anode Plane} & \textbf{Bias Voltage} \\ \toprowrule
$G$ - Grid & \SI{-665}{V} \\ \colhline
$U$ - Induction & \SI{-370}{V{}} \\ \colhline
$V$ - Induction & \SI{0}{V} \\ \colhline
$X$ - Collection & \SI{820}{V} \\ \colhline
Grounding Mesh & \SI{0}{V} \\ 
\end{dunetable}

\begin{dunefigure}[Field line simulation around wire planes]{fig:apa-fields}
{Field lines (a) and resulting signal shapes on the APA induction and collection wires (b) according to a 2D electric field simulation.  The bi-polar nature of the induced signals on the $U$ and $V$ wires together with the uni-polar collection signals on $Y$ are clearly illustrated.}
a) \includegraphics[width=0.42\textwidth]{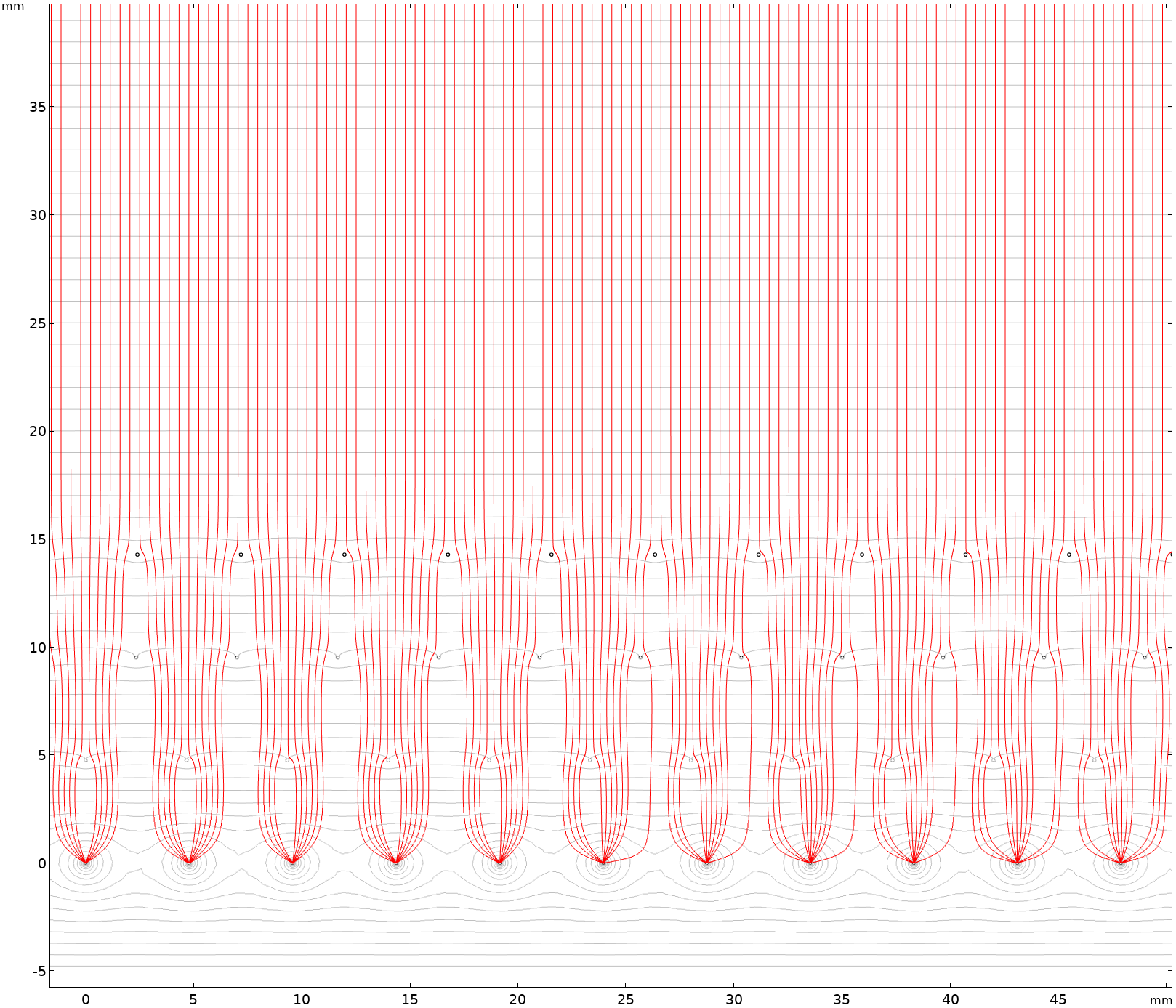}
b) \includegraphics[width=0.5\textwidth]{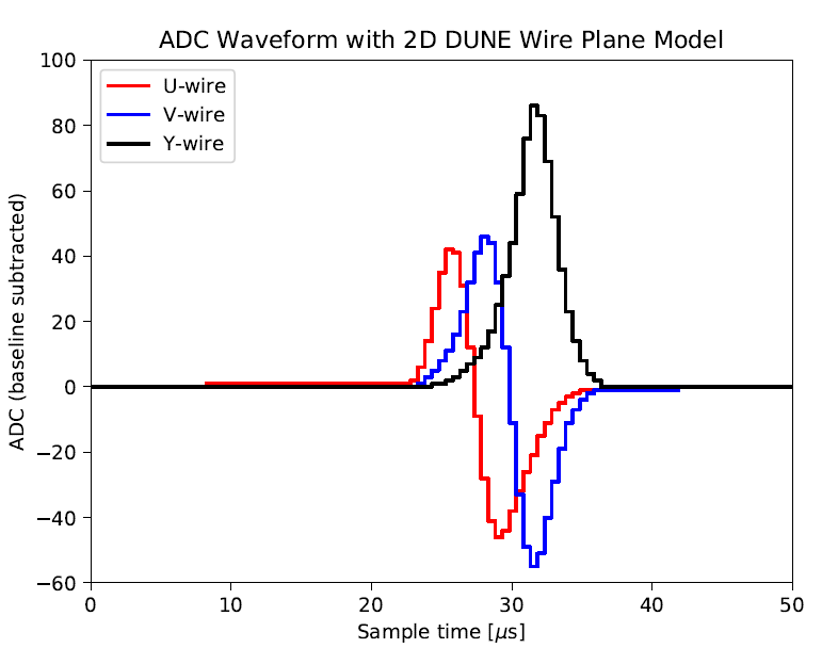}
\end{dunefigure}

\item \textbf{Wire type and tension:}  The wire selected for the \dword{apa}s is \SI{152}{$\mu$m} beryllium (\num{1.9}\%) copper wire, 
chosen for its mechanical and electrical properties, ease of soldering, and cost.  The tension on the wires, combined with intermediate support combs on the \dword{apa} frame cross beams (described in Section~\ref{sec:combs}), ensure that the wires are held taut in place with minimal sag.  Wire sag can affect the precision of reconstruction, as well as the transparency of the \dword{tpc} wire planes.  The tension must be low enough that when the wires are cooled, which increases their tension due to thermal contraction, they stay safely below the break load of the beryllium copper wire.  A tension of $6\pm$\SI{1}{N} is the baseline for \dword{dune}, to be confirmed after \dword{pdsp} analysis is completed.  
See Section~\ref{sec:fdsp-apa-wires} for more details about the \dword{apa} wires.

\end{itemize}

Table~\ref{tab:apaparameters} summarizes some of the principal design parameters for the \dword{spmod} 
\dwords{apa}.

\begin{dunetable}[APA design parameters]{lr}{tab:apaparameters}
{\dword{apa} design parameters}   
Parameter & Value  \\ \toprowrule
Active height & \SI{5.984}{m} \\ \colhline
Active width & \SI{2.300}{m} \\ \colhline
Wire pitch ($U,V$) & \uvpitch \\ \colhline
Wire pitch ($X,G$) & \xgpitch \\ \colhline
Wire pitch tolerance & \wirepitchtol \\ \colhline
Wire plane spacing & \planespace \\ \colhline
Wire plane spacing tolerance & $\pm$\SI{0.5}{mm} \\ \colhline
Wire Angle (w.r.t. vertical) ($U,V$) & \apainducwireangle{} \\ \colhline
Wire Angle (w.r.t. vertical) ($X,G$) & \apacollwireangle \\ \colhline
Number of wires / \dword{apa} & 960 ($X$), 960 ($G$), 800 ($U$), 800 ($V$) \\ \colhline
Number of electronic channels / \dword{apa} & 2560 \\ \colhline
Wire material & beryllium copper \\ \colhline
Wire diameter & 152 $\mu$m \\ 
\end{dunetable}

\subsection{Support Frames}
\label{sec:fdsp-apa-frames}

The \dword{apa} frames are built of rectangular hollow section (RHS) stainless steel tubes.  Figure~\ref{fig:apa-frame-full} shows three long tubes, a foot tube, a head tube, and eight cross-piece ribs that bolt together to create the \SI{6.0}{m} tall by \SI{2.3}{m} wide frame. All hollow sections are \SI{10.2}{cm} (\SI{4}{in}) deep with varying widths depending on their role. 

\begin{dunefigure}[Bare APA frame drawing]{fig:apa-frame-full}
{A \dword{dune} \dword{apa} frame showing the \num{13} separate stainless steel tube sections that bolt together to form a complete frame.  The long tubes and foot tube are a \num{10.2}$\times$\SI{10.2}{cm} (\num{4}$\times$\SI{4}{inch}) cross section, the head tube is \num{10.2}$\times$\SI{15.2}{cm} (\num{4}$\times$\SI{6}{inch}), and the ribs are \num{10.2}$\times$\SI{5.1}{cm} (\num{4}$\times$\SI{2}{inch}). 
}
\includegraphics[width=1\textwidth]{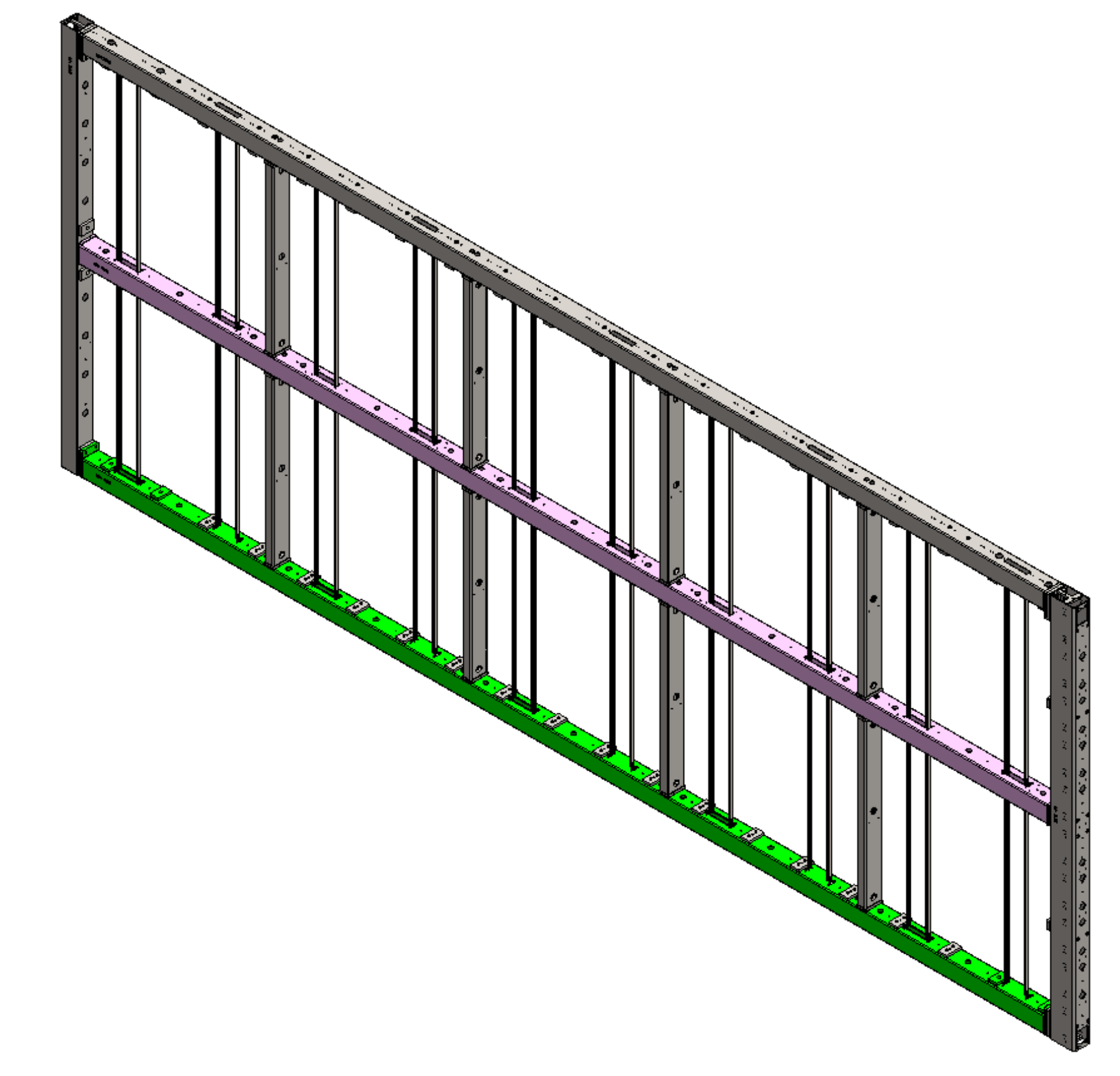}
\end{dunefigure}

\begin{dunefigure}[APA frame construction  details]{fig:apa-frame-details}
{\dword{apa} frame construction details. Top: The corner joint between the foot tube and the side tube. Middle: The joint between the side tube and a rib. Bottom: The joint between the head tube and the side tube.}

\includegraphics[height=0.31\textheight,trim=0mm 0mm 0mm 0mm,clip]{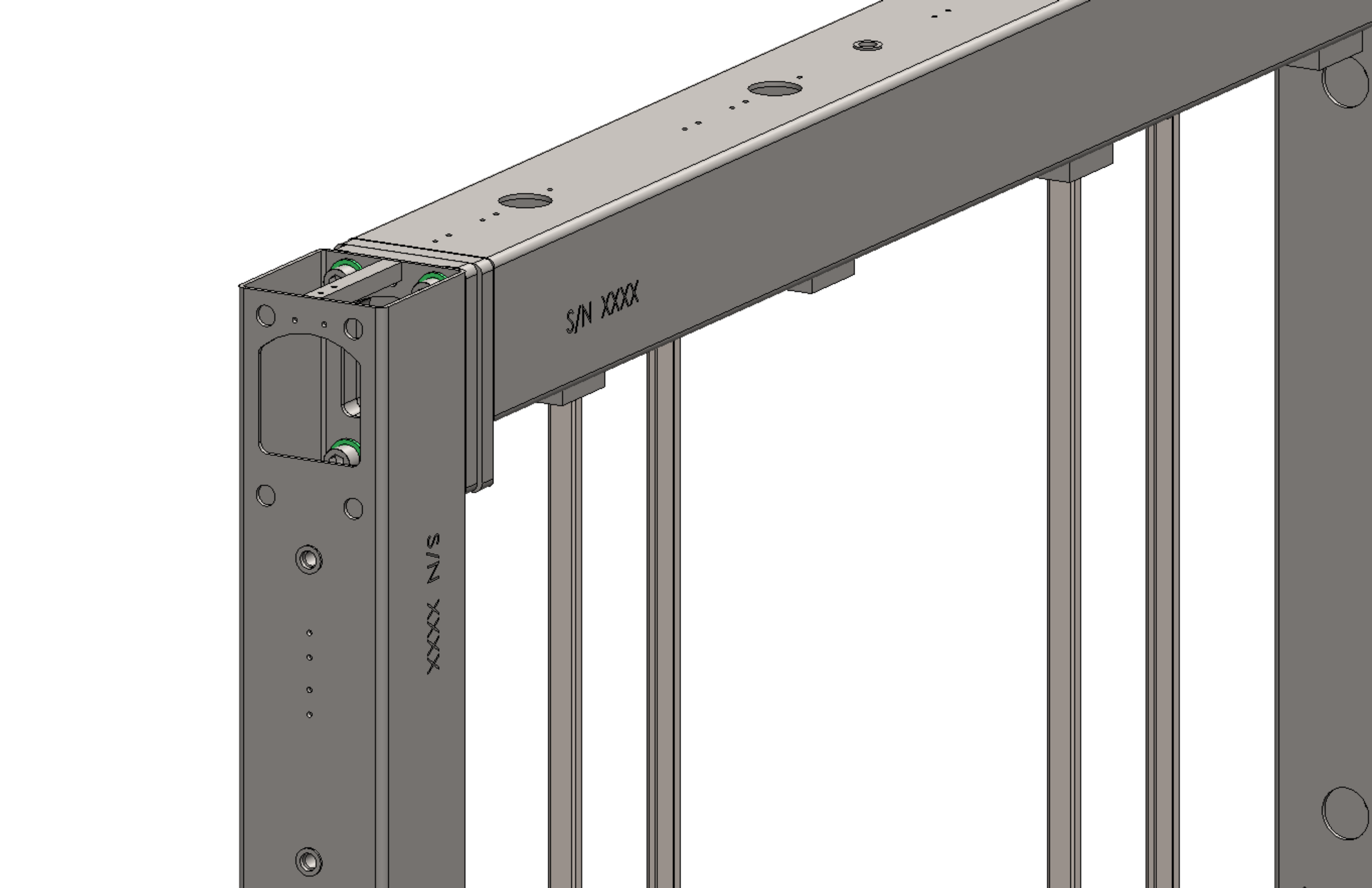}
\includegraphics[height=0.28\textheight,trim=0mm 0mm 0mm 0mm,clip]{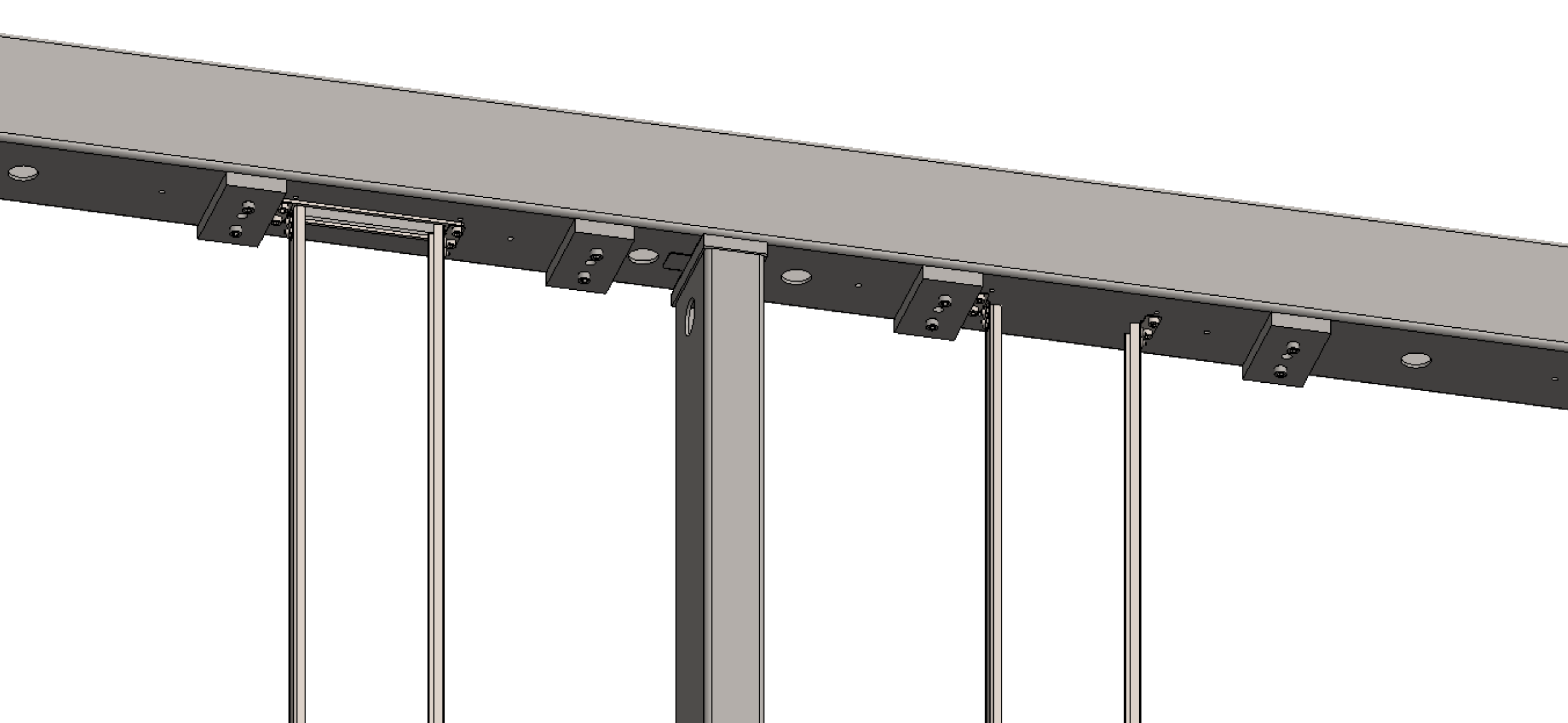}
\includegraphics[height=0.32\textheight]{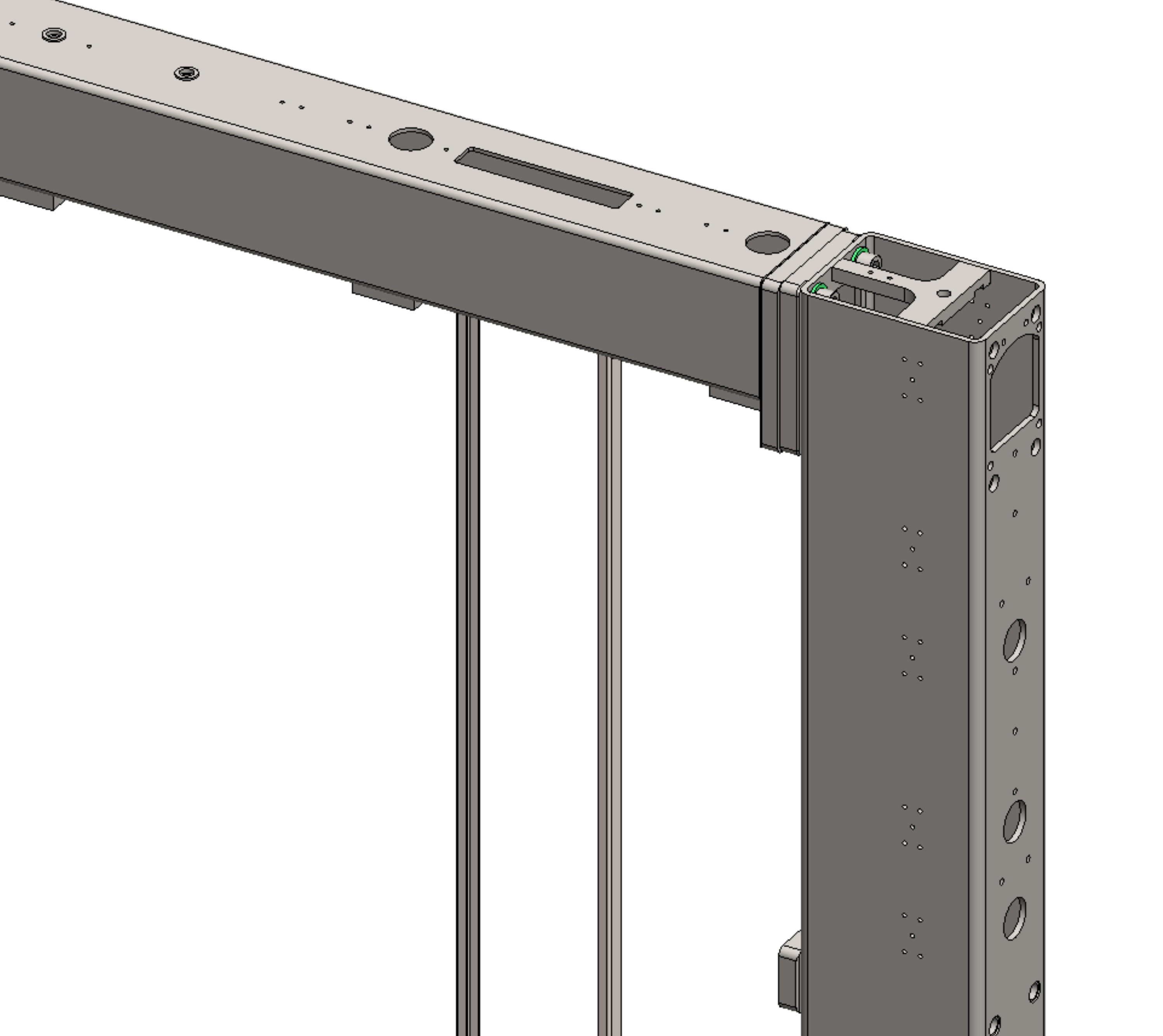}
\end{dunefigure}

The head and foot tubes are bolted to the side and center pieces via abutment flanges welded to the tubes. In production, the pieces can be individually machined to help achieve the flatness and shape tolerances.  During final assembly, shims are used to create a flat, rectangular frame of the specified dimensions.  The central cross pieces are similarly attached to the side pieces.  Figure~\ref{fig:apa-frame-details} shows models of the different joints.

The \dword{apa} frames also house the \dword{pds} (Chapter~\ref{ch:fdsp-pd}).  
Rectangular slots are machined in the outer frame tubes and guide rails are used to slide in \dword{pd} elements from the edges. 
(See Section~\ref{sec:fdsp-apa-intfc} for more details on interfacing with the \dword{pds}.)   

In a \dword{fd} \dword{spmod}, pairs of \dword{apa} frames will be mechanically connected to form a \tpcheight 
tall structure with electronics for \dword{tpc} readout at both the top and bottom of this two-frame assembly and \dwords{pd} installed throughout.  The \dword{apa} frame design, therefore, must support cable routing to the top of the detector from both the bottom \dword{apa} readout electronics and the \dwords{pd} mounted throughout both \dword{apa}s.  Section~\ref{sec:fdsp-apa-intfc} discusses the interfaces.

\subsection{Grounding Mesh}
\label{sec:fdsp-apa-mesh}

Beneath the layers of sense wires, the conducting surface should be uniform to evenly terminate the \efield and improve the uniformity of field lines around the wire planes. A fine woven mesh that is \num{85}\% optically transparent is used to allow scintillation photons to pass through to the \dwords{pd} mounted inside the frame.  The mesh also shields the \dword{apa} wires from spurious electrical signals from other parts of the \dword{apa} or the \dword{pd} system.  

In the \dword{pdsp} \dword{apa}s, the mesh was installed in four long sheets, along the length of the left- and right-hand halves of each side of the \dword{apa} and epoxied directly to the frame. This approach to mesh installation was found to be slow and cumbersome.  For the \dword{dune} mass production, a modular window-frame design is being developed, where mesh is pre-stretched over smaller sub-frames that can be clipped into each gap between cross beams in the full \dword{apa} frame.   This improves the reliability of the installed mesh (more uniform tension across the mesh) and allows much easier installation on the \dword{apa} frame. The mesh will be woven of conducting 304 stainless steel \SI{89}{\um} wire. It will be mounted on 304 stainless steel \SI{20}{mm}$\times$\SI{10}{mm} box section frames,  stretched over the frame with jigs and pneumatic actuators built for the purpose, and TIG welded around the top surface and again around the side surfaces. Five different panel designs are needed to match the openings in the \dword{apa} frames: two for the foot end, two for the head end, and one for the central panels that are all the same. There are 20 panels per \dword{apa}. Stainless steel brackets will be fixed to the inner window sections of the \dword{apa} frame and the panels will be secured into position using steel fasteners. The design ensures good electrical contact between the mesh and the frame. A full-scale \dword{apa} (\dword{apa}-07) has been built at Daresbury Lab for \dword{ce} testing at \dword{cern} using the mesh panel design. Figure~\ref{fig:tpc-apa-mesh} shows images of the mesh design and the prototypes built for \dword{apa} 7.

\begin{dunefigure}[Photos of APA grounding mesh]{fig:tpc-apa-mesh}
{\dword{apa} grounding mesh construction and installation. a) The mesh panel stretching jig, b) mesh being welded to the support frame, c) model showing the mesh sub-frame (in dark gray) fitting into the \dword{apa} frame (green), and d) photo of an installed mesh panel in \dword{apa} 7.}
\mbox{a) \includegraphics[height=0.23\textheight]{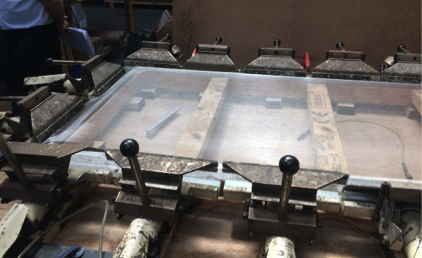} \hspace{0.0mm}
b) \includegraphics[height=0.23\textheight]{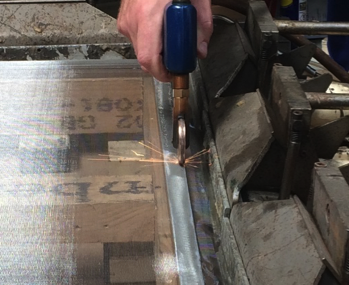}} \\
\vspace{3mm}
\hspace{0.4mm}
\mbox{c) \includegraphics[height=0.23\textheight]{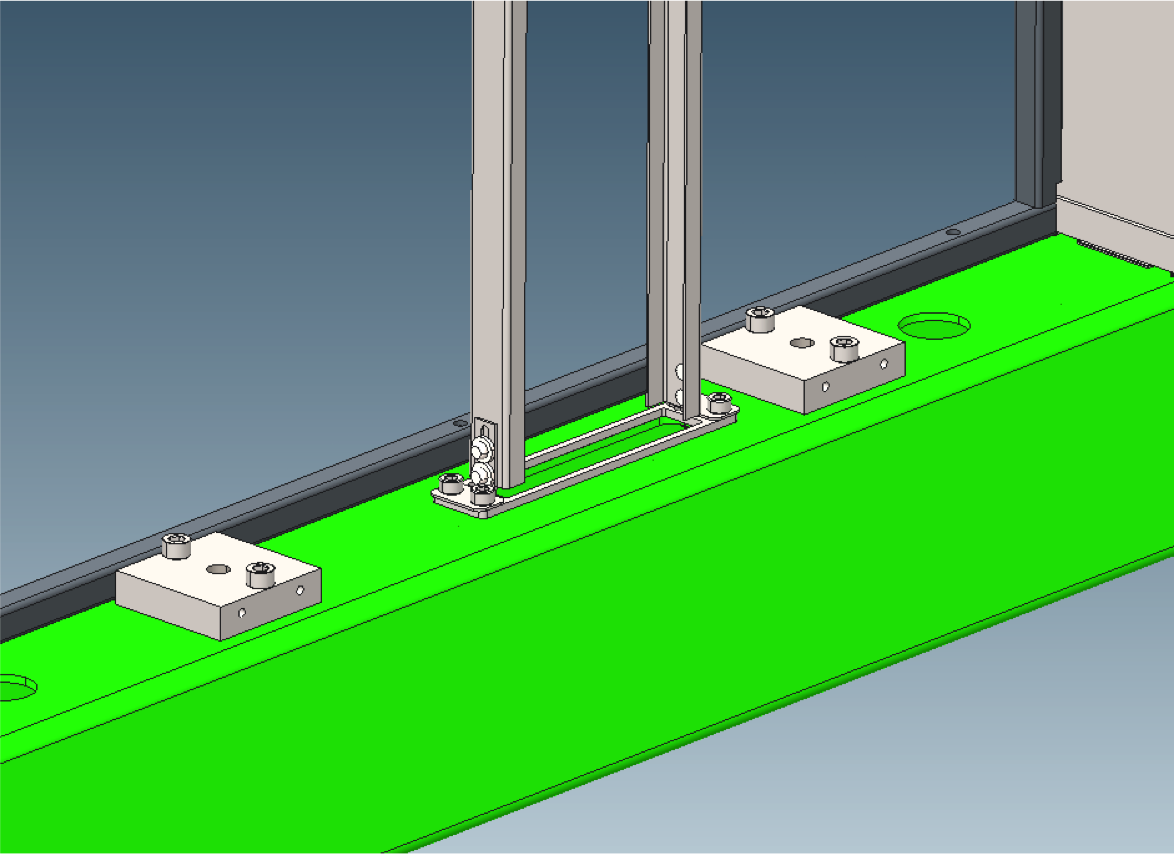}
d) \includegraphics[height=0.23\textheight]{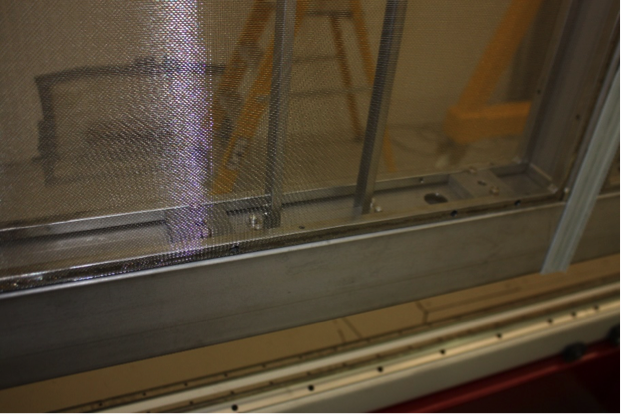}}
\end{dunefigure}

\subsection{Wires}
\label{sec:fdsp-apa-wires}

The \SI{152}{$\mu$m} (\SI{.006}{in}) diameter beryllium copper (CuBe) wire chosen for use in the \dword{apa}s is known for its high durability and yield strength. It is composed of \num{98}\,\% copper, \num{1.9}\,\% beryllium, and a negligible amount of other elements. Each \dword{apa} contains a total of \SI{23.4}{km} of wire.  

The key properties for its use in the \dword{apa}s are low resistivity, high tensile or yield strength, and a coefficient of thermal expansion suitable for use with the \dword{apa}'s stainless steel frame (see Table~\ref{tab:wire} for a summary of properties).  Tensile strength of the wire describes the wire-breaking stress.  The yield strength is the stress at which the wire starts to take a permanent (inelastic) deformation and is the important limit for this case. The wire spools purchased from Little Falls Alloys\footnote{Little Falls Alloys\texttrademark, \url{http://www.lfa-wire.com/}} for use on \dword{pdsp} were measured to have tensile strength higher than \SI{1380}{MPa} and yield strength more than \SI{1100}{MPa} (\SI{19.4}{N} for \SI{152}{$\mu$m} diameter wire).  The stress while in use is approximately \SI{336}{MPa} (\SI{6}{N}), leaving a comfortable margin.

The \dword{cte} describes how a material expands or contracts with changes in temperature.  The \dwords{cte} of CuBe alloy and \num{304} stainless steel are very similar.  Integrated down to \SI{87}{K}, they are \SI{2.7}{mm/m} for stainless steel and \SI{2.9}{mm/m} for CuBe. The wire contracts slightly more than the frame, so for a wire starting at \SI{6}{N} at room temperature the tension increases to around \SI{6.5}{N} when everything reaches \lar temperature.  

The rate of change in wire tension during \cooldown is also important.  In the worst case, the wire cools quickly to \SI{87}{K} before any significant cooling of the much larger frame.  In the limiting case with complete contraction of the wire and none in the frame, the tension would peak around \SI{11.7}{N}, which is still well under the \SI{19}{N} yield tension. In practice, however, the cooling will be done gradually to avoid this tension spike as well as other thermal shocks to the detectors.

\begin{dunetable}[Beryllium copper (CuBe) wire properties]{lr}{tab:wire}{Summary of properties of the beryllium copper wire used on the \dword{apa}s.}
Parameter & Value \\ \toprowrule
Resistivity & 7.68 $\mu\Omega$-cm $@$ 20$^{\circ}$ C \\ \colhline
Resistance & 4.4 $\Omega$/m $@$ 20$^{\circ}$ C \\ \colhline
Tensile strength (from property sheets)  & \SI{1436}{MPa} / \SI{25.8}{N} for \SI{152}{$\mu$m} wire \\ \colhline
CTE of beryllium copper integrated to \SI{87}{K}  & \SI{2.9e-3}{m/m} \\ \colhline
CTE of stainless steel integrated to \SI{87}{K}  & \SI{2.7e-3}{m/m} \\
\end{dunetable}

\subsection{Wire Boards and Anchoring Elements}
\label{sec:fdsp-apa-boards}

To guide and secure the \num{3520} wires on an \dword{apa}, stacks of custom \frfour circuit boards attach all along the outside edges of the frame, as shown in the engineering drawings in Figure~\ref{fig:apa-wire-boards}.  There are \num{337} total circuit boards on each \dword{apa} (\num{50550} in an \dword{spmod} with \num{150} \dword{apa}s), where this number includes 204 wire boards ($X/V/U/G = 30/72/72/30$), 72 cover boards to protect the wire solder pads and traces on the top layer of wire boards, 20 capacitive-resistance (\dword{cr}) boards, 20 adapter boards to connect the \dword{cr}s to the \dword{ce}, 20 $G$-layer bias boards, and one \dword{shv} board to distribute bias voltages to the planes.  Figure~\ref{fig:sp-apa-head-xsec} shows the positions of the boards at the head of the \dword{apa} and the path connecting \dword{tpc} wires to the \dword{ce}.  

\begin{dunefigure}[Wire carrier board layout on the APA frames]{fig:apa-wire-boards}
{Engineering drawings that illustrate the layering of the wire carrier boards that are secured along the perimeter of the \dword{apa} steel frames. Left: The full set of $V$-layer boards.  Right: Detail showing the stack of four boards at the head end of the \dword{apa} (bottom to top: $X,V,U,G$).}
\includegraphics[width=0.48\textwidth,trim = 12mm 0mm 5mm 0mm,clip]{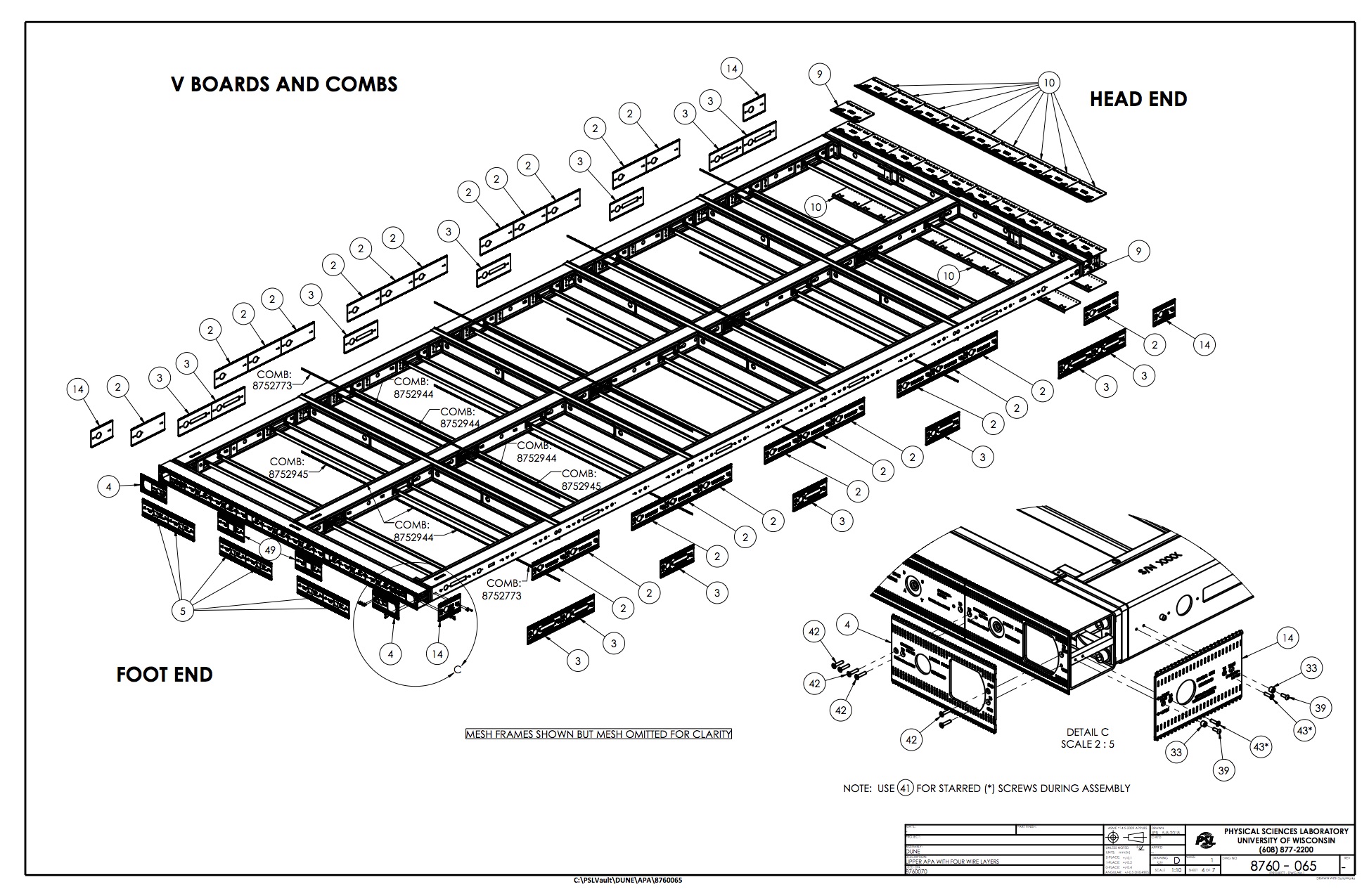}
\includegraphics[width=0.48\textwidth,trim = 12mm 0mm 5mm 0mm,clip]{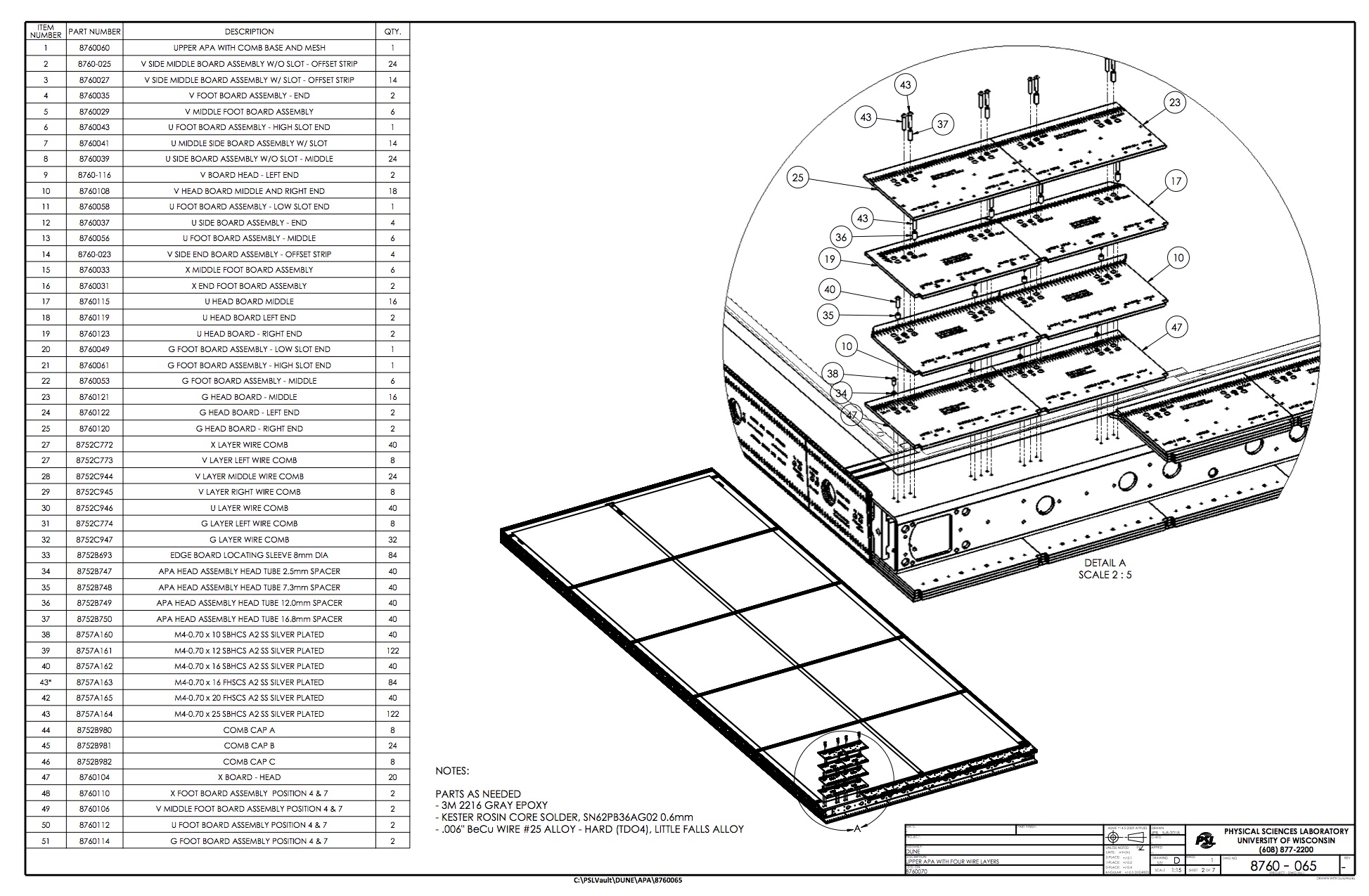}
\end{dunefigure}

\subsubsection{Head Wire Boards}
\label{sec:fdsp-apa-headboards}

All \dword{apa} wires terminate on wire boards stacked along the electronics end of the \dword{apa} frame.  The board stack at the head end is shown in an engineering drawing in the right panel of Figure~\ref{fig:apa-wire-boards}. A photograph showing the head boards and $G$-bias boards on one of the completed \dword{pdsp} \dword{apa}s is shown in Figure~\ref{fig:tpc_apa_electronics_connectiondiagram}. Attaching the wire boards begins with the $X$-plane (innermost). Once the $X$-plane wires are strung on both sides of the \dword{apa} frame, they are soldered and epoxied to their wire boards and trimmed. Next, the $V$-plane boards are epoxied in place and the $V$ wires installed, followed by the $U$-plane boards and wires, and finally the $G$-plane boards and wires. The wire plane spacing of \planespace 
is set by the thickness of these wire boards.   

\begin{dunefigure}[Wire board stack at the head end of an \dshort{apa}]{fig:tpc_apa_electronics_connectiondiagram}{The wire board stack at the head end of an \dword{apa}. The four wire boards within a stack can be seen on both the top and bottom sides of the \dword{apa}.  Also visible are the T-shaped brackets that will hold the \dword{ce} boxes when electronics are installed.   
}
\includegraphics[width=0.6\textwidth, trim=0mm 0mm 0mm 15mm, clip]{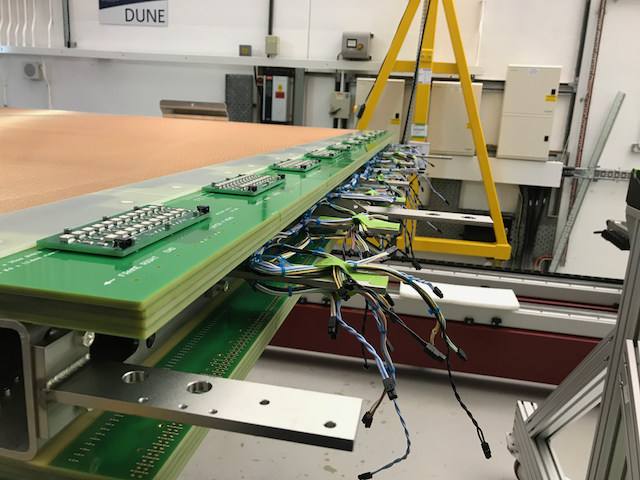}
\end{dunefigure} 

Mill-Max\footnote{Mill-Max\texttrademark{}, \url{https://www.mill-max.com/}} pins and sockets provide electrical connections between circuit boards within a stack. They are pressed into the circuit boards and are not repairable if damaged. To minimize the possibility of damaged pins, the boards are designed so that the first wire board attached to the frame has only sockets. All boards attached subsequently contain pins that plug into previously mounted boards. This process eliminates exposure of any pins to possible damage during winding, soldering, or trimming.

The $X$, $U$ and $V$ layers of wires are connected to the \dword{ce} (housed in boxes mounted on the \dword{apa}) either directly ($V$) or through DC-blocking capacitors ($U,X$).  Ten stacks of wire boards are installed across the width of each side along the head of the \dword{apa}.  The $X$-layer board in each stack has room for \num{48} wires, the $V$-layer has 40 wires, the $U$-layer \num{40} wires, and the $G$-layer \num{48} wires.  Each board stack, therefore, has \num{176} wires but only \num{128} signal channels since the $G$ wires are not read out. With a total of \num{20} stacks per \dword{apa}, this results in \num{2560} signal channels per \dword{apa} and a total of \num{3520} wires starting at the top of the \dword{apa} and ending at the bottom. Many of the capacitors and resistors that in principle could be on these wire boards are instead placed on the attached CR (capacitive-resistance) boards (Section~\ref{sec:crboards}) to improve their accessibility in case of component failure. 

\subsubsection{Side and Foot Wire Boards}

The boards along the sides and foot of the \dword{apa} have notches, pins, and other location features to hold wires in the correct position as they wrap around the edge from one side of the \dword{apa} to the other.  

The edge boards need a number of hole or slot features to provide access to the underlying frame (see Figure~\ref{fig:tpc_apa_sideboardmodel} for examples).  In order that these openings not be covered by wires, the sections of wire that would go over the openings are replaced by traces on the boards.  After the wires are wrapped, the wires over the opening are soldered to pads at the ends of the traces, and the section of wire between the pads is snipped out.  These traces can be easily and economically added to the boards by the many commercial fabricators who make circuit boards. 

\begin{dunefigure}[APA side boards showing traces that connect wires around openings]{fig:tpc_apa_sideboardmodel}
{Side boards with traces that connect wires around openings.  The wires are wound straight over the openings, then soldered to pads at the ends of the traces. The wire sections between the pads are then trimmed away.}
\includegraphics[height=0.28\textheight]{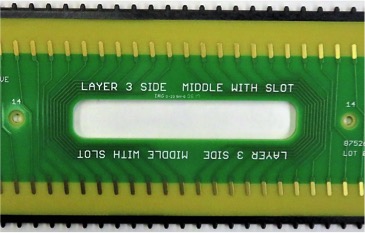} \quad
\includegraphics[height=0.28\textheight,trim=0mm 0mm 0mm 25mm,clip]{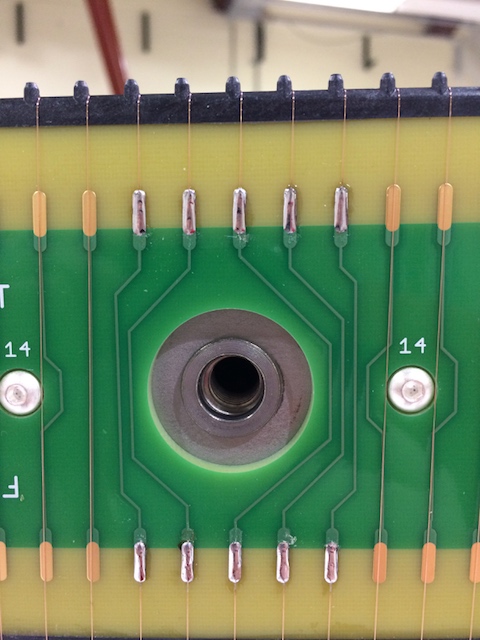}
\end{dunefigure}

The placement of the angled wires are fixed by teeth that are part of an injected molded strip glued to the edge of the \frfour boards.  The polymer used for the strips is Vectra e130i (a trade name for 30$\%$ glass filled liquid crystal polymer, or LCP). It retains its strength at cryogenic temperature and has a \dword{cte} similar enough to \frfour that differential contraction is not a problem.  The wires make a partial wrap around the pin as they change direction from the face of the \dword{apa} to the edge.

\subsubsection{Capacitive-Resistive (CR) Boards}
\label{sec:crboards}

The \dword{cr} boards carry a bias resistor and a DC-blocking capacitor for each wire in the $X$ and $U$-planes. These boards are attached to the board stacks after fabrication of all wire planes.   Electrical connections to the board stack are made through Mill-Max pins that plug into the wire boards. Connections from the \dword{cr} boards to the \dword{ce} are made through a pair of \num{96}-pin Samtec\footnote{Samtec\texttrademark \url{https://www.samtec.com/}} connectors.

Surface-mount bias resistors on the \dword{cr} boards have resistance of \SI{50}{\mega\ohm} and are constructed with a thick film on a ceramic substrate. Rated for \SI{2.0}{kV} operation, the resistors measure \SI{3.0 x 6.1}{mm} (\SI{0.12 x 0.24}{in}). The selected DC-blocking capacitors have capacitance of \SI{3.9}{nF} and are rated for \SI{2.0}{kV} operation. Measuring \SI{5.6 x 6.4}{mm} (\SI{0.22 x 0.25}{in}) across and
\SI{2.5}{mm} (\SI{0.10}{in}) high,  
the capacitors feature flexible terminals to comply with \dword{pcb} expansion and contraction. They are designed to withstand \num{1000} thermal cycles between the extremes of the operating temperature range. Tolerance is also \num{5}\,\%.

In addition to the bias and DC-blocking capacitors for all $X$ and $U$-plane wires, the \dword{cr} boards include two R-C filters for the bias voltages\footnote{The $V$-plane does not carry a bias voltage, so does not require these components.}. The resistors are of the same type used for wire biasing except with a resistance of \SI{5}{\mega\ohm}, consisting of two \SI{10}{\mega\ohm} resistors connected in parallel. Wire plane bias filter capacitors are \SI{39}{nF}, consisting of ten \SI{3.9}{nF} surface-mount capacitors connected in parallel. They are the same capacitors as those used for DC blocking.

The selected capacitors were designed by the manufacturer to withstand repeated temperature excursions over a wide range. Their mechanically compliant terminal structure accommodates \dword{cte} mismatches. The resistors use a thick-film technology that is also tolerant of wide temperature excursions.  Capacitors and resistors were qualified for \dword{pdsp} by testing samples repeatedly at room temperature and at \num{-190}\,$^\circ$C.  Performance criteria were measured across five thermal cycles, and no measurable changes were observed. During the production of \num{140} \dword{cr} boards, more than \num{10000} units of each component were tested at room temperature, at \dword{lar} temperature, and again at room temperature. No failures or measurable changes in performance were observed.

\subsubsection{Support Combs}
\label{sec:combs}

Support combs are glued at four points along each side of the \dword{apa}, along the four cross beams. These combs maintain the wire and plane spacing along the length of the \dword{apa}. A dedicated jig is used to install the combs and also provides the alignment and pressure as the glue dries. The glue used is the Gray epoxy \num{2216} described below. Before the jig can be removed and production can continue, an eight-hour cure time is required after comb installation on each side of the \dword{apa}.  Figure~\ref{fig:tpc_apa_sideboardphoto} shows a detail of the wire support combs on a \dword{pdsp} \dword{apa}.

\begin{dunefigure}[APA side boards on the frame]{fig:tpc_apa_sideboardphoto}
{Left: \dword{apa} corner where end boards meet side boards.  The injection molded teeth that guide the $U$ and $V$ wires around the edge are visible at the bottom. Right: The wire support combs.}
\includegraphics[height=0.32\textheight]{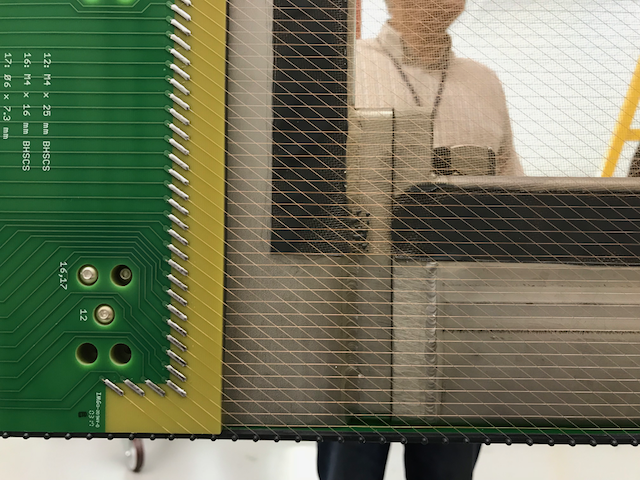} \quad
\includegraphics[height=0.32\textheight]{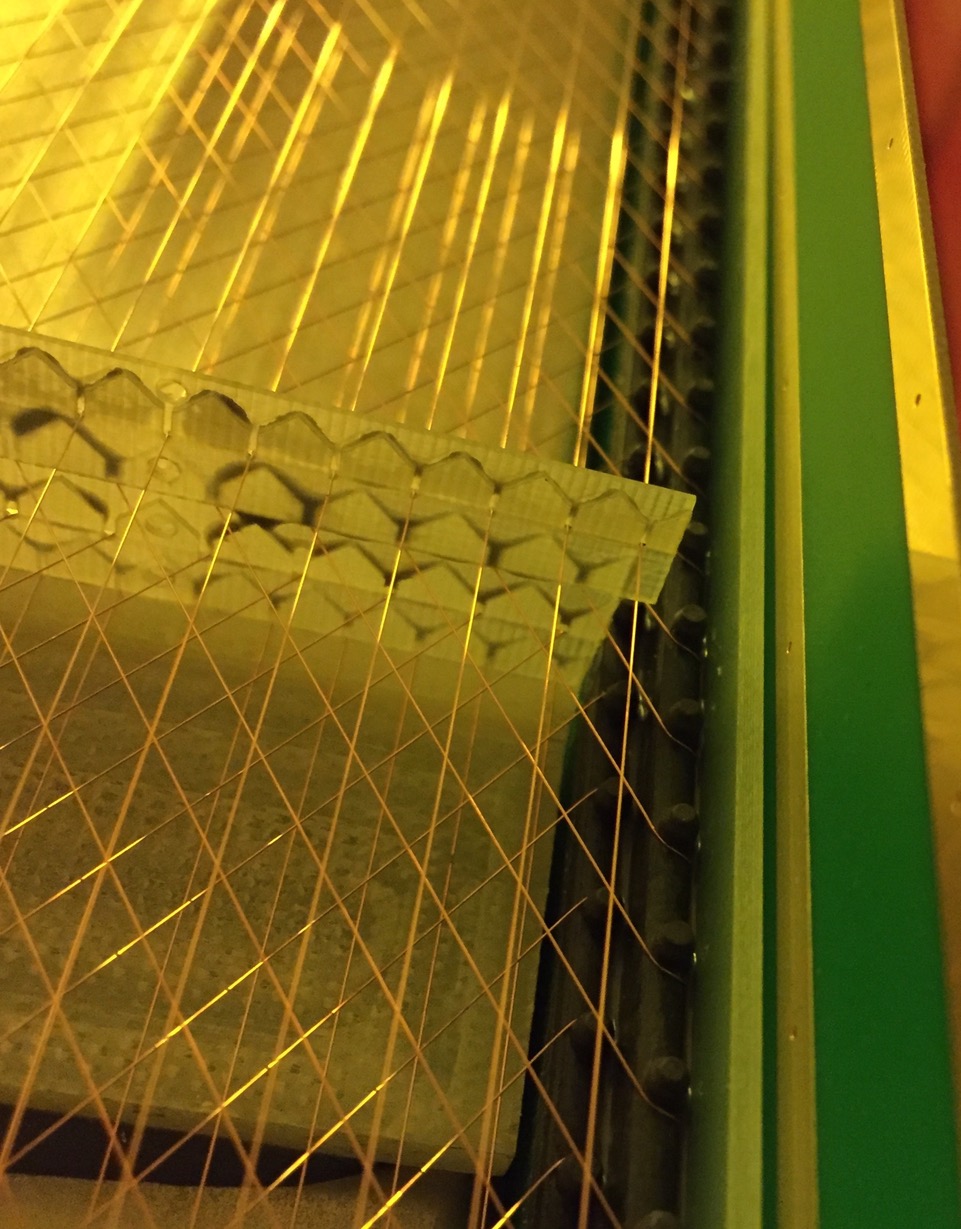}
\end{dunefigure}

\subsubsection{Solder and Epoxy}
\label{sec:glue-solder}

The ends of the wires are soldered to pads on the edges of the wire boards.  Solder provides both an electrical connection and a physical anchor to the wire pads. A 62$\%$ tin, 36$\%$ lead, and 2$\%$ silver solder was chosen.  A eutectic mix (63/37) is the best of the straight tin-lead solders, but the 2$\%$ added silver gives better creep resistance.  The solder contains a no-clean flux and does not need to be removed after soldering. Most of it is encapsulated when subsequent boards are epoxied in place.  At room temperatures and below, the flux is not conductive or corrosive.

Once a wire layer is complete, the next layer of boards is glued on; this glue provides an additional physical anchor. Gray epoxy \num{2216} by 3M\footnote{3M\texttrademark ~\url{https://www.3m.com/}} is used for the glue.  It is strong and widely used (therefore much data is available), and it retains good properties at cryogenic temperatures.

\subsection{The APA Pair} 
\label{sec:fdsp-apa-intfc-apa}

In an \dword{spmod}, pairs of \SI{6}{m} tall \dword{apa} frames are mechanically connected at their ends to form a \SI{12}{m} tall readout surface.  Figure~\ref{fig:apa-doublet} shows a connected pair (turned on its side) with dimensions.  The \dword{tpc} readout electronics require that the individual \dword{apa} frames be electrically isolated.   The left panel of Figure~\ref{fig:apa-connections} shows the design for mechanically connecting \dword{apa}s while maintaining electrical isolation.  The two \dword{apa}s are connected through a stainless steel link that is attached to both frames with a special shoulder screw.  The steel part of the link is electrically insulated from the frames using a G10 panel.  The links connect to the side tubes with a special shoulder screw that screws into plates welded to the frame.  

\begin{dunefigure}[APA pair diagram with dimensions]{fig:apa-doublet}
{Diagram of an \dword{apa} pair, with connected bottom and top \dword{apa}. The dimensions of the \dword{apa} pair, including the accompanying cold electronics and mechanical supports (the yoke), are indicated.}
\includegraphics[width=1\textwidth]{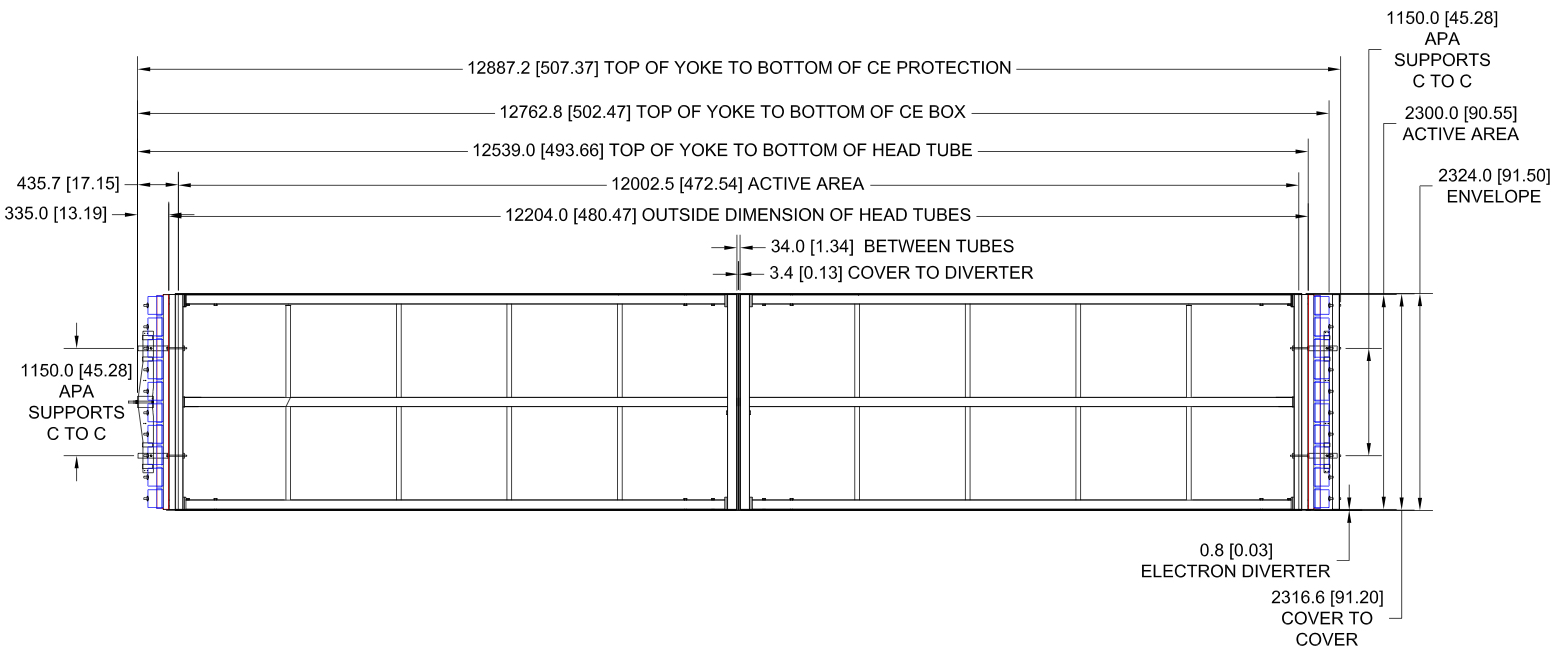} 
\end{dunefigure}

\begin{dunefigure}[APA-to-APA connections]{fig:apa-connections}
{Design for the \dword{apa}-to-\dword{apa} connections.  Left: For the vertical connection there are two steel links joining the upper APA to the lower APA; one link connected to one APA is shown here.  The steel part of the link is electrically insulated from the frames.  Right: Along adjoining vertical edges, two pins keep neighboring \dword{apa}s in plane. Two side tubes before engagement with the screw and insulating sleeve installed are shown at the top, and the engaged side tubes are shown below.}  
\includegraphics[height=0.29\textheight]{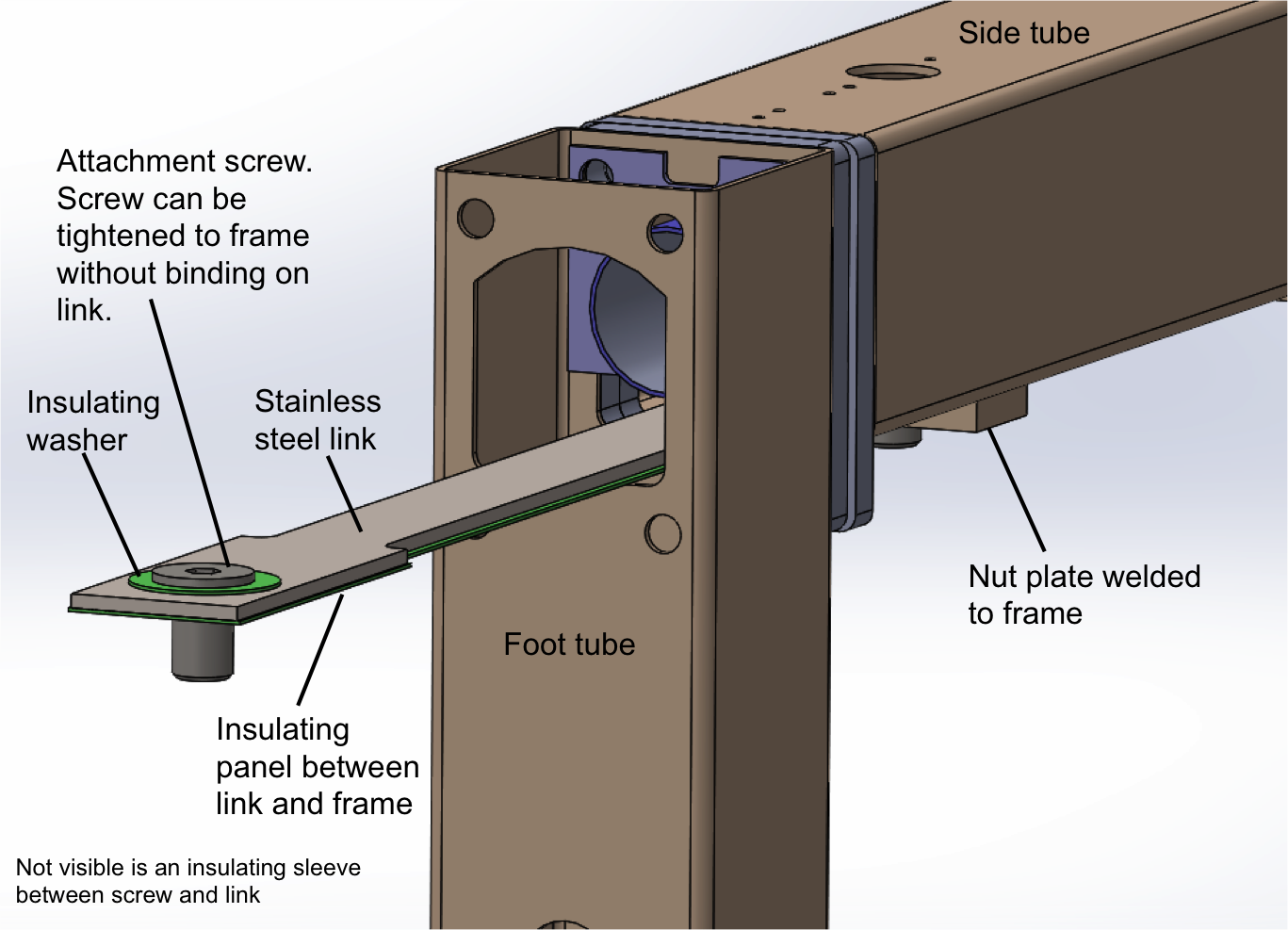} \qquad
\includegraphics[height=0.3\textheight]{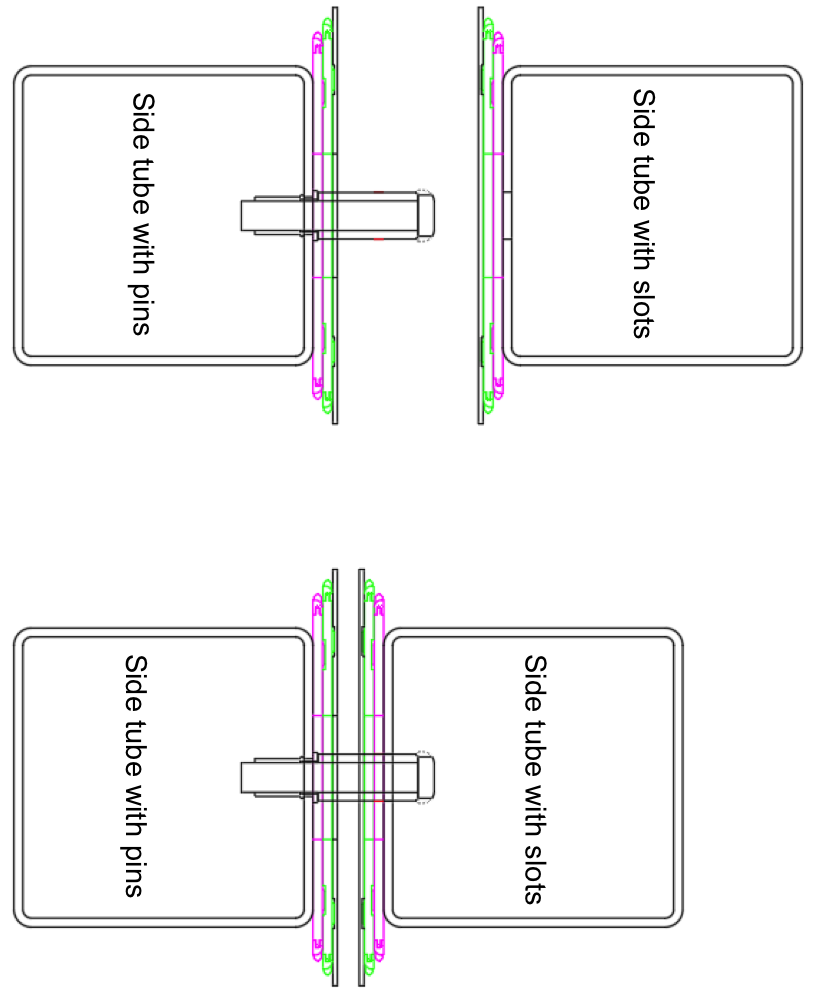}
\end{dunefigure}

The \dword{apa} yoke, shown in Figure~\ref{fig:apa-yoke}, is a bolted stainless steel structural assembly with a central support point and a pair of pins to connect to the load.  Two T-shaped brackets, referred to as the structural tees, mount to the head tube of the top \dword{apa} and provide the mating pin holes to connect the yoke to the \dword{apa}.  The center support point consists of a M20 stainless steel bolt, oversize washers, and a PEEK washer for electrical isolation.   The yoke is mounted to the top \dword{apa} before an \dword{apa} pair 
is assembled.  To move into the cryostat, the pair 
is hung from two trolleys that connect to the structural tees. Once in final position, the load of the \dword{apa} pair 
can be transferred from the trolleys to the \dword{dss} in the cryostat through the center support point of the yoke.  To accomplish this, the M20 bolt and washer assembly is inserted from the bottom of the yoke and the threaded end of the bolt connects to the \dword{dss}.

\begin{dunefigure}[Yoke that connects an APA pair 
to the DSS]{fig:apa-yoke} 
{The yoke at the top of an \dword{apa} pair 
that provides connection to the \dword{dss}.}
\includegraphics[width=0.95\textwidth]{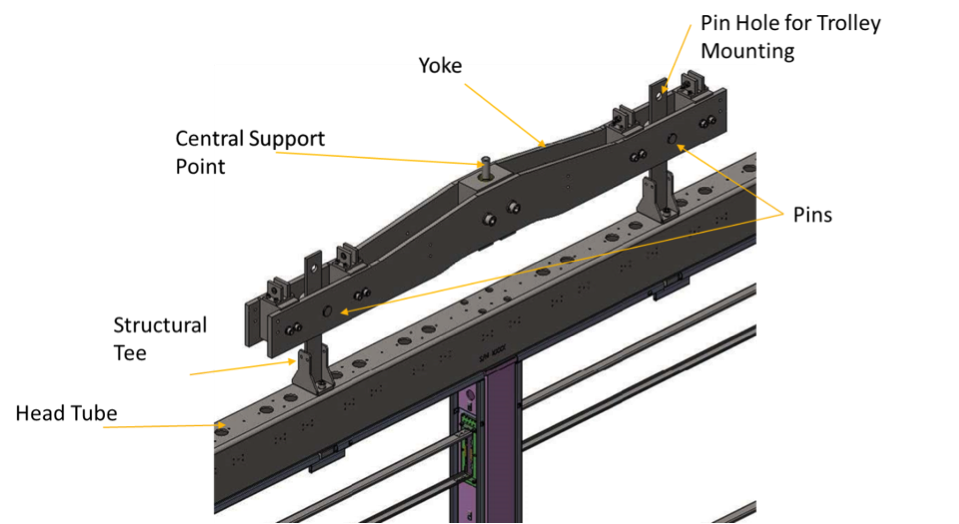} 
\end{dunefigure}

Adjacent \dword{apa} pairs 
are kept in plane with each other by simple insertion pins at the top and bottom of the side tubes.  The pins are made up of a screw and an insulating sleeve to ensure electrical isolation, and each pin engages a slot in the adjacent \dword{apa} pair 
side tubes. The right panel of Figure~\ref{fig:apa-connections} shows a schematic of the side pin connectors before and after insertion.  

Once installed in the detector, a physical gap of \SI{12}{mm} exists along this vertical connection between all adjacent \dword{apa}s at room temperature. Since the \dword{apa}s are suspended under the stainless steel \dword{dss} beams, which contract similarly to that of the \dword{apa} frames, the gaps between most adjacent \dword{apa}s stay about the same (\SI{12}{mm}) in the cold.  The \dword{dss} beams, however, are segmented at \SI{6.4}{m} length, and each beam segment is independently supported by two \dword{dss} feedthroughs, one of which does not allow lateral movement.  As a result, the gaps between \dword{dss} beams open up in the cold by another \SI{17}{mm},  
making the physical gaps \SI{29}{mm} as shown in Figure~\ref{fig:sp-apa-gaps}.  The actual gap between the \dword{apa}s active width \SI{28}{mm} is approximately \SI{16}{mm} wider than the physical gap (\SI{45}{mm}) in the two scenarios described above.

\begin{dunefigure}[APA-to-APA gaps]{fig:sp-apa-gaps}
{Illustration of the gap width between \dword{apa}s}  
\includegraphics[width=0.8\textwidth]{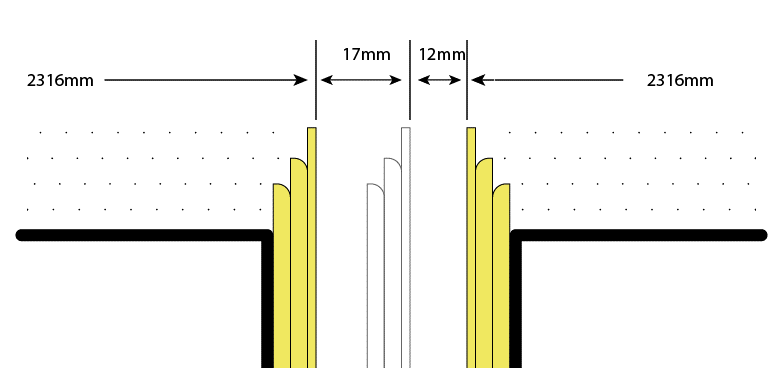} 
\end{dunefigure}

To minimize the loss of signal charge over the gaps between \dword{apa}s in \dword{pdsp}, special electrodes (electron diverters) were installed along the vertical gaps to nudge the incoming electrons into the active regions of the \dword{apa}s. The data from \dword{pdsp} are being used to study the impact of using electron diverters and determine the need for them in \dword{dune}.  See Section~\ref{sec:fdsp-apa-qa} for a discussion of the \dword{pdsp} data analysis (\dword{pdsp} had some gaps with electron diverters installed and some without, enabling comparisons of the tracking and calorimetry performance in the two cases). 

\subsection{APA Structural Analysis}

The \dword{apa}s will be subjected to a variety of load conditions throughout construction, installation, and operation of the experiment, so it is important to analyze carefully and confirm the design of the mechanical components.  A structural and safety analysis was performed to confirm the strength of the \dword{apa} frame, the \dword{apa} yoke, and the \dword{apa}-to-\dword{apa} link.  The full report can be found at~\cite{bib:cernedms2100877}. 
As noted, the way the \dword{apa} frame is supported and loaded changes during the construction and transport of the \dword{apa}. Twenty different load cases were checked.  These load cases cover the handling of the bare frame, the \dword{apa} during wiring, the fully integrated \dword{apa}, and the \dword{apa} pair.  The analysis covered the loaded \dword{apa} pair in the installed warm and dry state as well as spatial and transient thermal gradients that might be encountered during cool down.

The masses of components mounted on the frame were determined from the material densities and the geometry defined in the \threed models.  Loads from the supported masses and the APA wire load were applied to the frame in the analysis to replicate the way loads are applied to the actual frame.  The analysis was performed in accordance with the standard building code for large steel structures, the \dwords{asic} Specification for Structural Steel Buildings (AISC document 360-10). For stainless steel structures, the \dword{asic}  publication Design Guide 27: Structural Stainless Steel was also applied.  The analysis was performed using the Load and Resistance Factor Design method (LRFD).

Per LRFD, a load factor of 1.4 was applied to all loads and to the self-weight of the \dword{apa}  frame.  The factored loads were used to calculate the required strength or stress.  Strength reduction factors were assigned to the strength or stress rating of the component or material.  The strength reduction factors determined the allowable strengths and the design was considered to meet code as long as the allowable strength of the material or component is greater than the required strength as determined by the factored loading.

In order to evaluate the structure, a \dword{fea}  model of the \dword{apa}  frame was built in SolidWorks Simulation\footnote{\url{https://www.solidworks.com/}}.   For each load case, proper constraints were defined and factored loads were applied.  The stress in the frame members were directly extracted from the model.  Also extracted from the model were the forces and moments acting on the welded and bolted joints. These forces and moments and methods from code were used to determine the required strength.  The allowable strength was also determined using methods from the code.
For transportation cases, the analysis was used to determine that the maximum shock or g-load the \dword{apa}  frame can tolerate is 4g (\SI{39.2}{m/s^2}).  This value has been incorporated into the requirements for the design of the transport frame.

Two thermal cases were run. In one case, a steady state thermal gradient of \SI{17}{K/m} was applied to the frame in addition to the installed state loading.   The second thermal case was a transient case.  In this case, the fastest cool down rate the \dword{apa}  frame can tolerate without over stressing the wires and wire solder/epoxy joints was estimated.  The wire cools faster than the frame and the \cooldown rate is limited by a 75~$^\circ$C allowable differential temperature between the frame and the wire.  The estimation of the differential was done using a conservative method that is described in the section that presents the results for case 20 in the \dword{apa} analysis document.  The allowable cool down rate of the ambient environment is 70~$^\circ$C/hr.

In addition to the frame, the yoke was also analyzed for strength.  These components are not subjected to multiple load states and see their maximum loads when in the installed state.  The yoke was analyzed using \dword{fea} to check stress and buckling of the side plates.  The \dword{apa} -to-\dword{apa}  link was checked using methods for pinned connections defined in code.

Results for the frame analysis show the frame members are most heavily stressed in the transportation cases.  This is expected because the g-load was increased until strength limits were reached.  Here the ratio of allowable to required strength is 1.1 for both the beam structural members and for the welds and 1.5 for the bolts.
The results for the yoke analysis show that the allowable stress over the required strength for the yoke plates is 2.2.  The ratio of the load that will cause buckling to the applied load is 33.

The structural analysis clearly shows the \dword{apa}  frame members, welds, and bolts are strong enough to carry the loads.

\section{Quality Assurance}
\label{sec:fdsp-apa-qa}

The most important and complete information for assuring the quality of the \dword{apa} design, components, materials, and construction methods comes from the construction and operation of \dword{pdsp}.  We have learned much about the design and fabrication procedures that has informed the detailed design and plans for the DUNE \dword{apa} construction project. \dword{pdsp} included six full-scale \dword{dune} \dword{apa}s instrumenting two drift regions around a central cathode.  Four of the \dword{pdsp} \dword{apa}s were constructed in the USA at the University of Wisconsin-PSL, and two were made at Daresbury Laboratory in the UK. All were shipped to \dword{cern}, integrated with \dwords{pd} and \dword{ce}, and tested in a \coldbox prior to installation into the \dword{pdsp} cryostat.  Figure~\ref{fig:sp-apa-pd-photo} shows one of the drift regions in the fully constructed \dword{pdsp} detector.

\begin{dunefigure}[A drift region in ProtoDUNE-SP with installed APAs]{fig:sp-apa-pd-photo}
{One of the two drift regions in the \dword{pdsp} detector at \dword{cern} showing the three installed \dword{apa}s on the left.}
\includegraphics[width=0.9\textwidth]{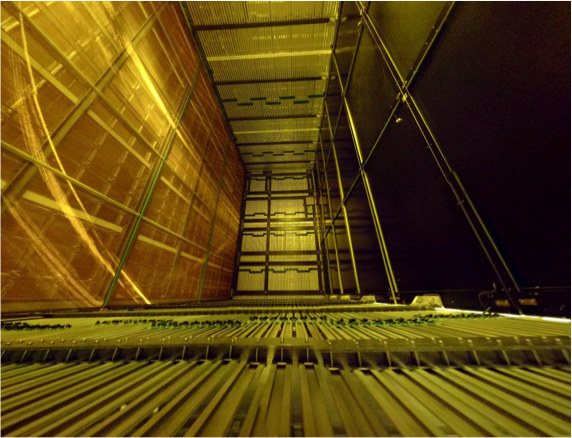}
\end{dunefigure}

\subsection{Results from ProtoDUNE-SP Construction}
\label{sec:fdsp-apa-qa-protodune-const}

A thorough set of \dword{qc} tests were performed and documented throughout the fabrication of the \dword{pdsp} \dword{apa}s.  The positive outcomes give great confidence in the quality of the overall \dword{apa} design  and construction techniques.  Here we summarize some of the testing that was done for \dword{pdsp} and the results.

After each wire layer was applied to an \dword{apa}, electrical continuity between the head and foot boards was checked for each wire.  This test is most useful for the $U$ and $V$ layers, where metal traces on the side boards can be damaged during construction. All boards were visually inspected as construction proceeded.

Channels were checked for isolation.  In the beginning, wire-to-wire isolation was measured over a long period of time, and no problems arose.  In the end, each wire was individually hipot tested (a dielectric withstand test) at \SI{1}{kV}. No failures were ever seen. However, leakage currents were seen to be highly dependent on relative humidity.  The surface of the epoxy has some affinity for moisture in the air and provides a measurable leakage path when relative humidity exceeds about \SI{60}{\%}. Tests have confirmed that in a dry environment, such as the \dword{lar} cryostat, these leakage currents disappear.

Wire tension was measured for all wires at production.  Figure~\ref{fig:sp-apa-pd-tension-prod} displays the measured tensions for wires on the instrumented wire planes ($X, U, V$) for the six \dword{pdsp} \dword{apa}s, four constructed at PSL in the US and two at Daresbury Laboratory in the UK.  
In total, \SI{4.4}{\%} of the \num{14972} wires considered for this analysis had a tension below \SI{4}{N}, and \SI{22.5}{\%} were above \SI{6}{N}. 
A wire which has a tension higher than specification should not impact the physics in any meaningful way. Wires with tension lower than specification could move slightly out of position and impact detector function primarily through modifying the local \efield. \efield modification can lead to the number of ionization electrons being incorrectly reconstructed in the deconvolution process or alter the transparency so that less than \SI{100}{\%} of the ionization electrons reach the collection plane. Because these processes change the amount of reconstructed charge, they would alter the reconstruction of the energy deposited by charged particles near these wires. A further complication from very low-tension wires might be an increase in noise level, introduced by wire vibrations, which can lead to vortex shedding.  Each of these impacts is expected to be quite small, but to confirm, cosmic muon tracks in \dword{pdsp} data are now being used to test if differences in response can be seen on wires with particularly low tension.  The target tension for \dword{dune} \dword{apa}s has already been increased to \SI{6}{N}, and these \dword{pdsp} studies will quantitatively inform a minimum tension requirement, but no challenges in meeting specifications are foreseen based on current knowledge from \dword{pdsp} construction.

\begin{dunefigure}[ProtoDUNE-SP wire tensions as measured during production]{fig:sp-apa-pd-tension-prod}
{Distributions of wire tensions in the \dword{pdsp} \dword{apa}s for wires longer than \SI{70}{cm}, as measured during production at PSL and Daresbury. For the $X$-plane, every wire has the same length (\SI{598.39}{cm}), and so every wire is included.  
The histograms for the six \dword{apa}s are stacked. 
}
\includegraphics[height=0.28\textheight,trim=0mm 0mm 0mm 0mm,clip]{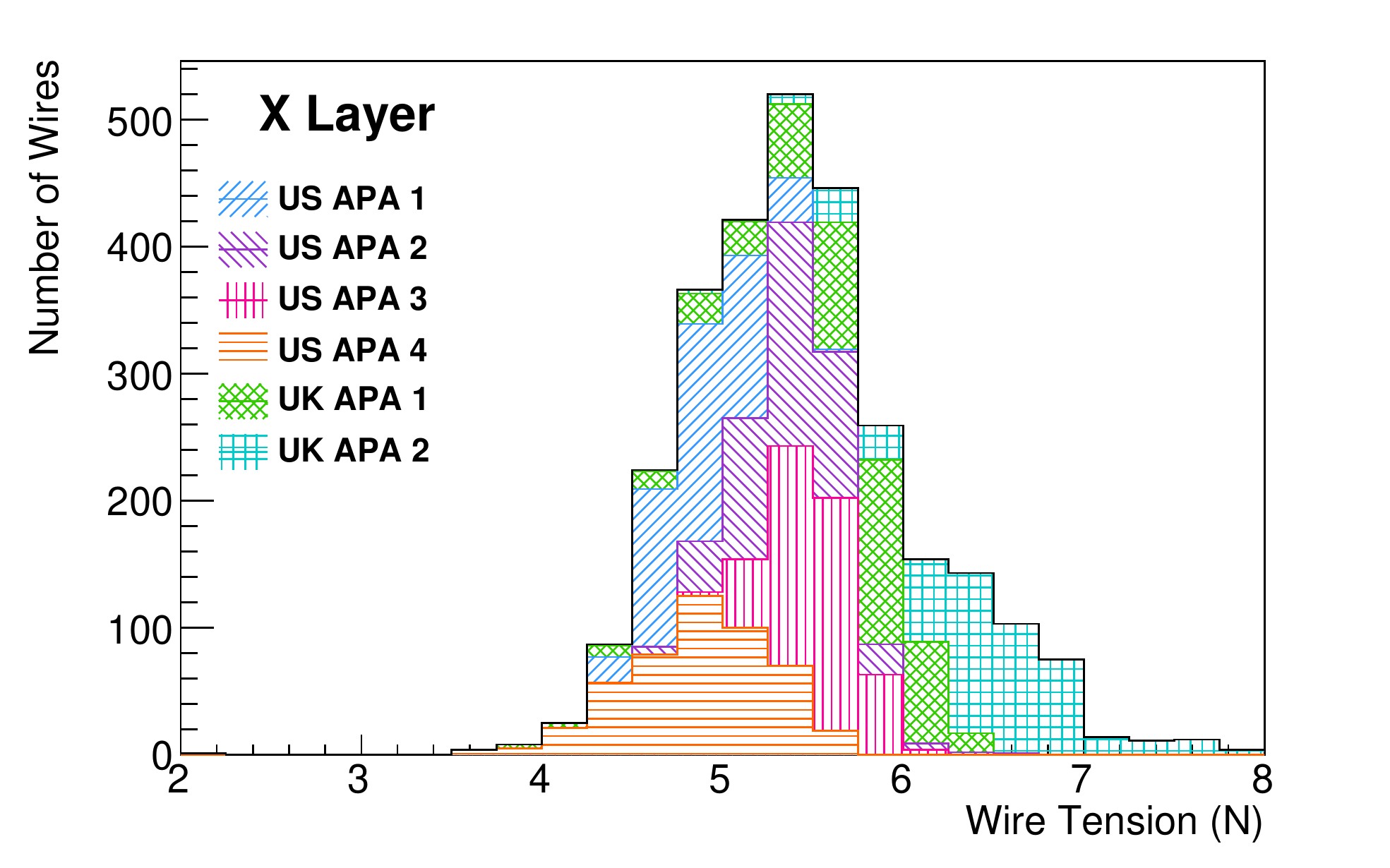}
\includegraphics[height=0.28\textheight,trim=0mm 0mm 0mm 0mm,clip]{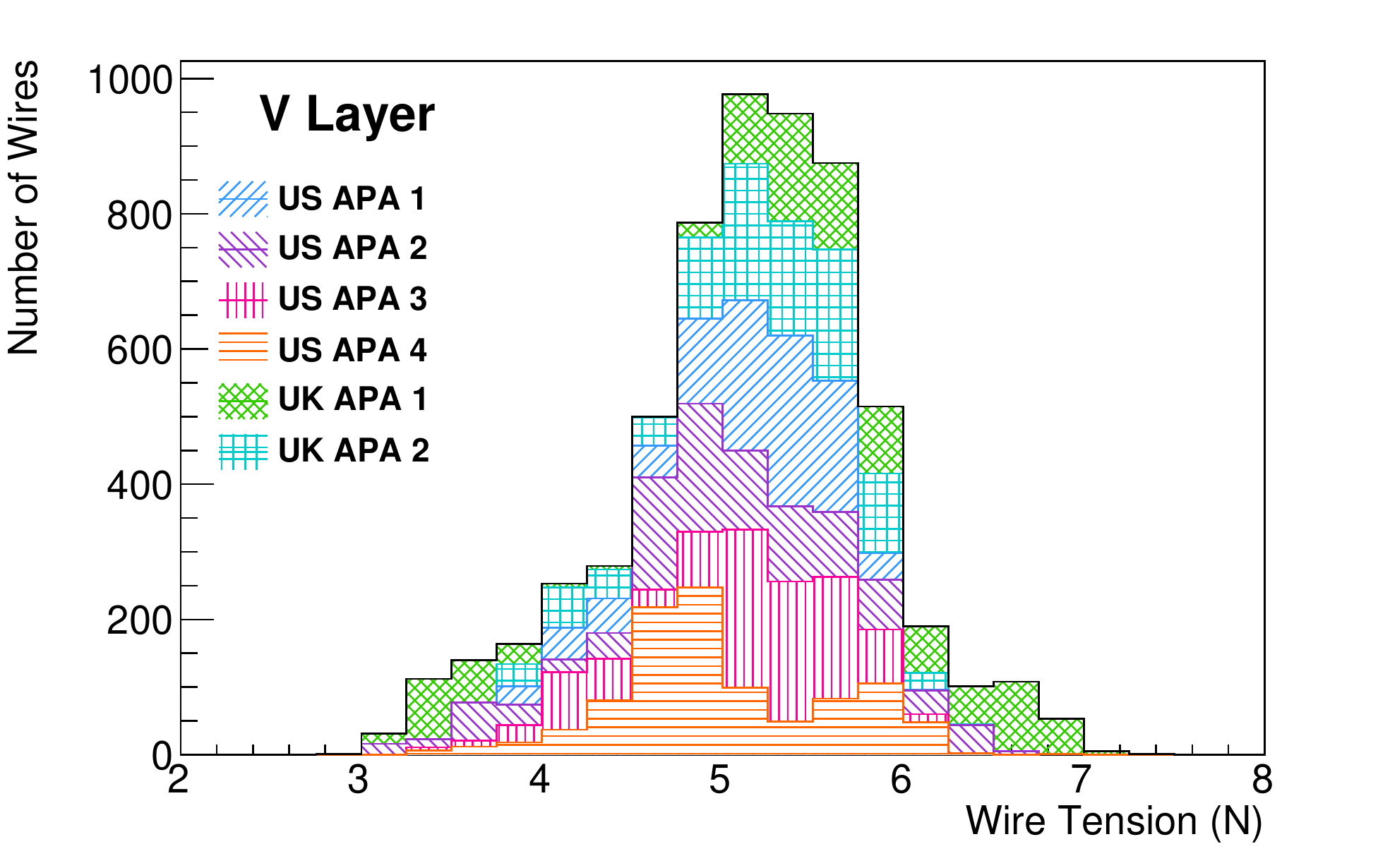}
\includegraphics[height=0.28\textheight,trim=0mm 0mm 0mm 0mm,clip]{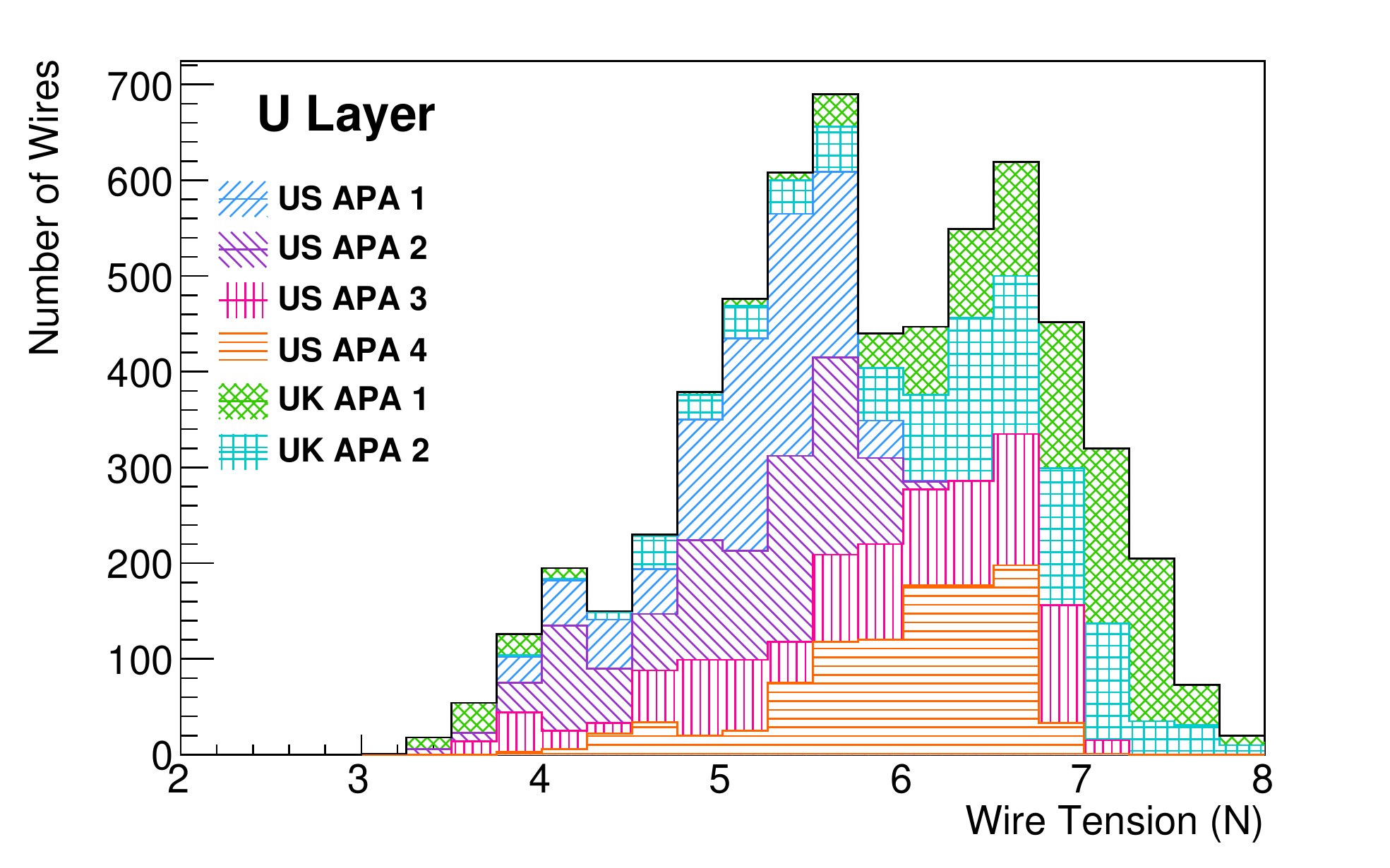}
\end{dunefigure}

Wire tension measurements were also performed for a subset of wires on each \dword{apa} after arriving at \dword{cern}. Figure~\ref{fig:sp-apa-pd-tension-cern} shows the comparison of tension values measured at \dword{cern} versus at the production site for the selected subset of wires from each wire plane. Based on the traveler documents provided by the production sites, wires having outlier tension values were selected from each \dword{apa} for re-measurement at \dword{cern}. In addition, a set of randomly selected wires from each plane was measured. In total, for six \dword{apa}s, $\sim$1500 wires had their tension re-measured at \dword{cern}. Measurements took place in the clean room with \dword{apa}s hanging vertically, the first time the tensions were sampled in this orientation. Tension measurements were performed by using the laser-photodiode based method, the same as at the production sites. 

\begin{dunefigure}[ProtoDUNE-SP wire tension measurement plots]{fig:sp-apa-pd-tension-cern}
{Comparison of wire tensions upon arrival at \dword{cern} versus at the production sites for a sample of wires on each of the \dword{pdsp} \dword{apa}s.}
\mbox{\includegraphics[height=0.23\textheight]{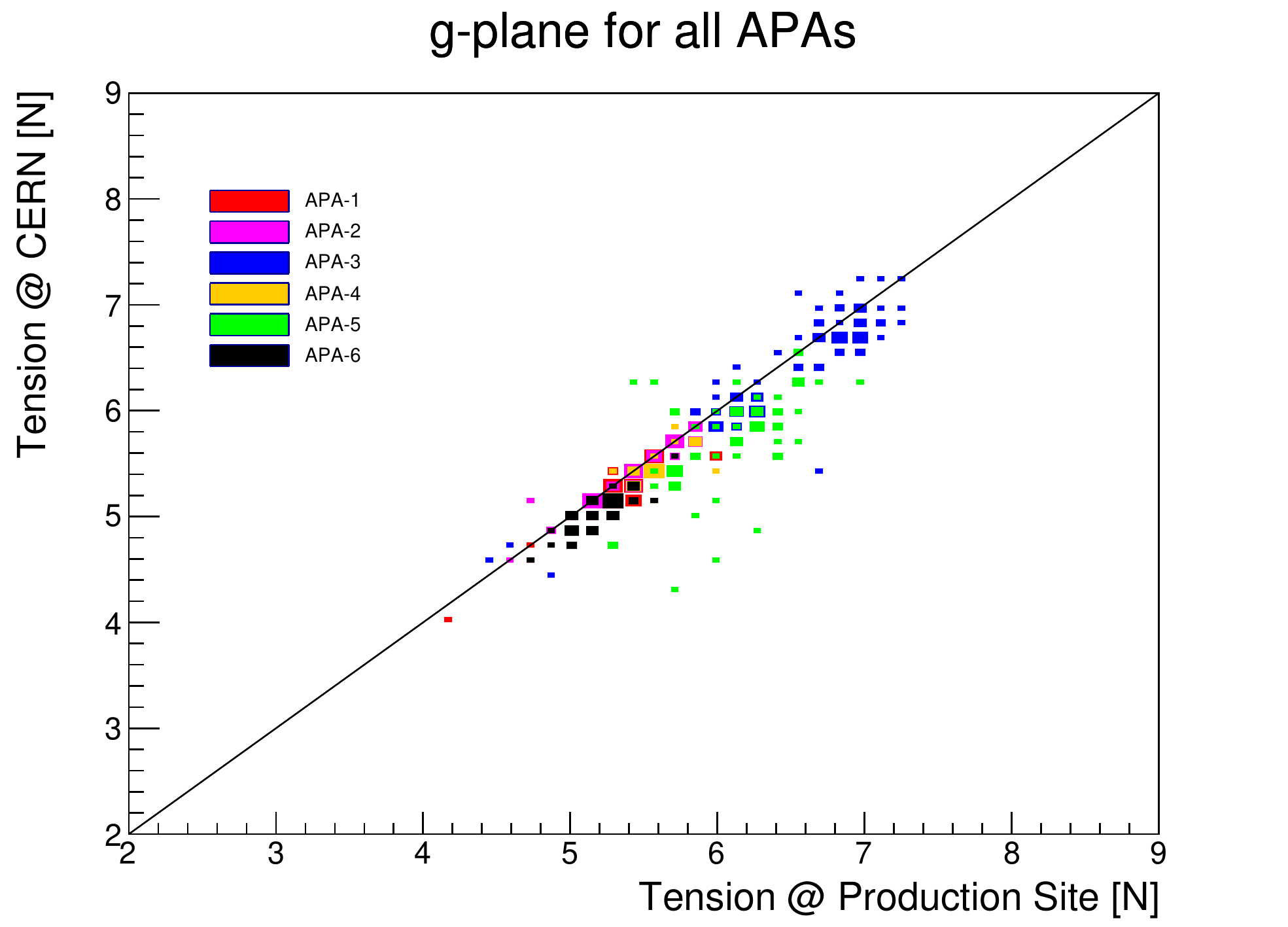} 
\includegraphics[height=0.23\textheight]{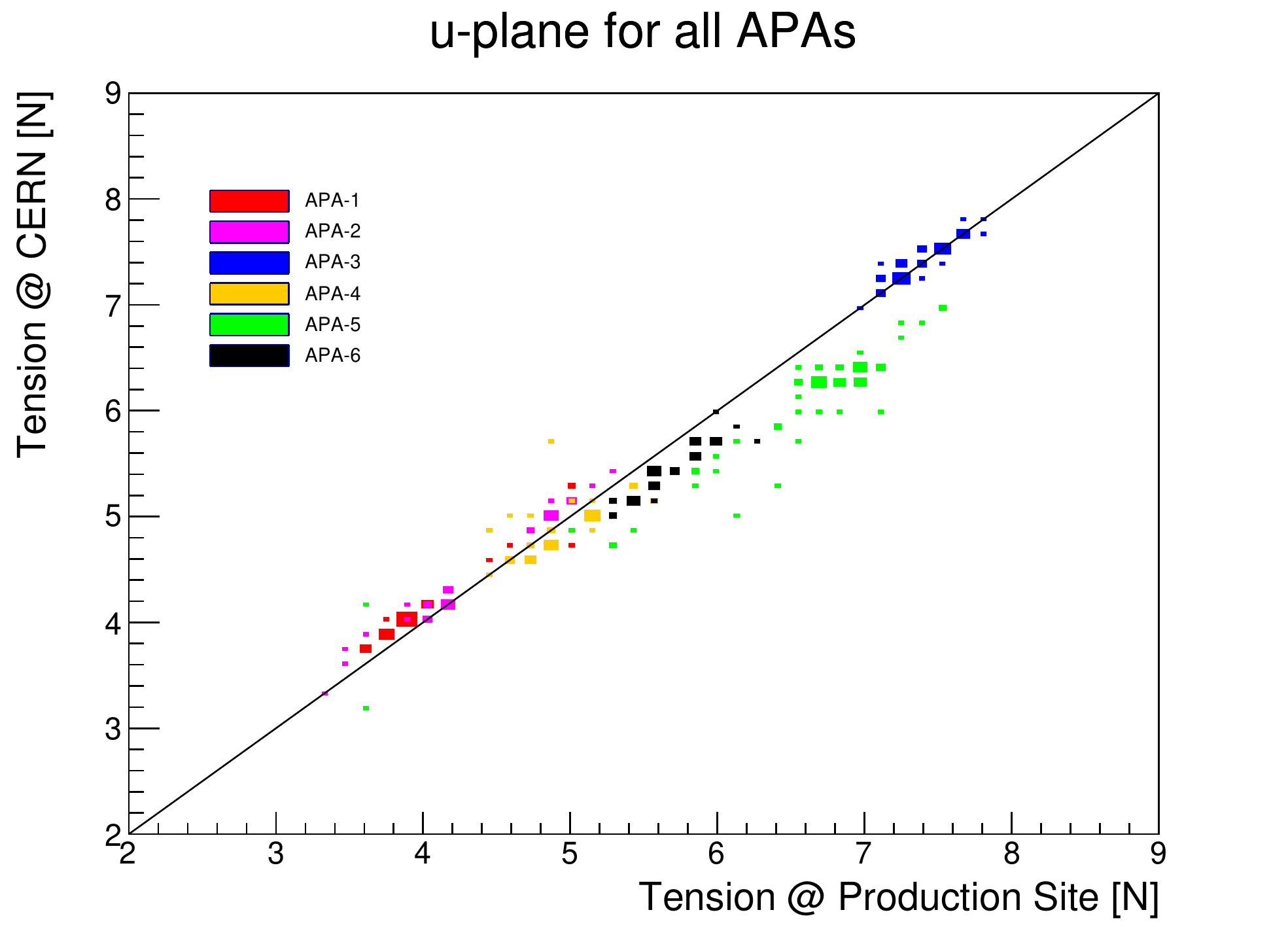}} \\
\vspace{3mm}
\mbox{\includegraphics[height=0.23\textheight]{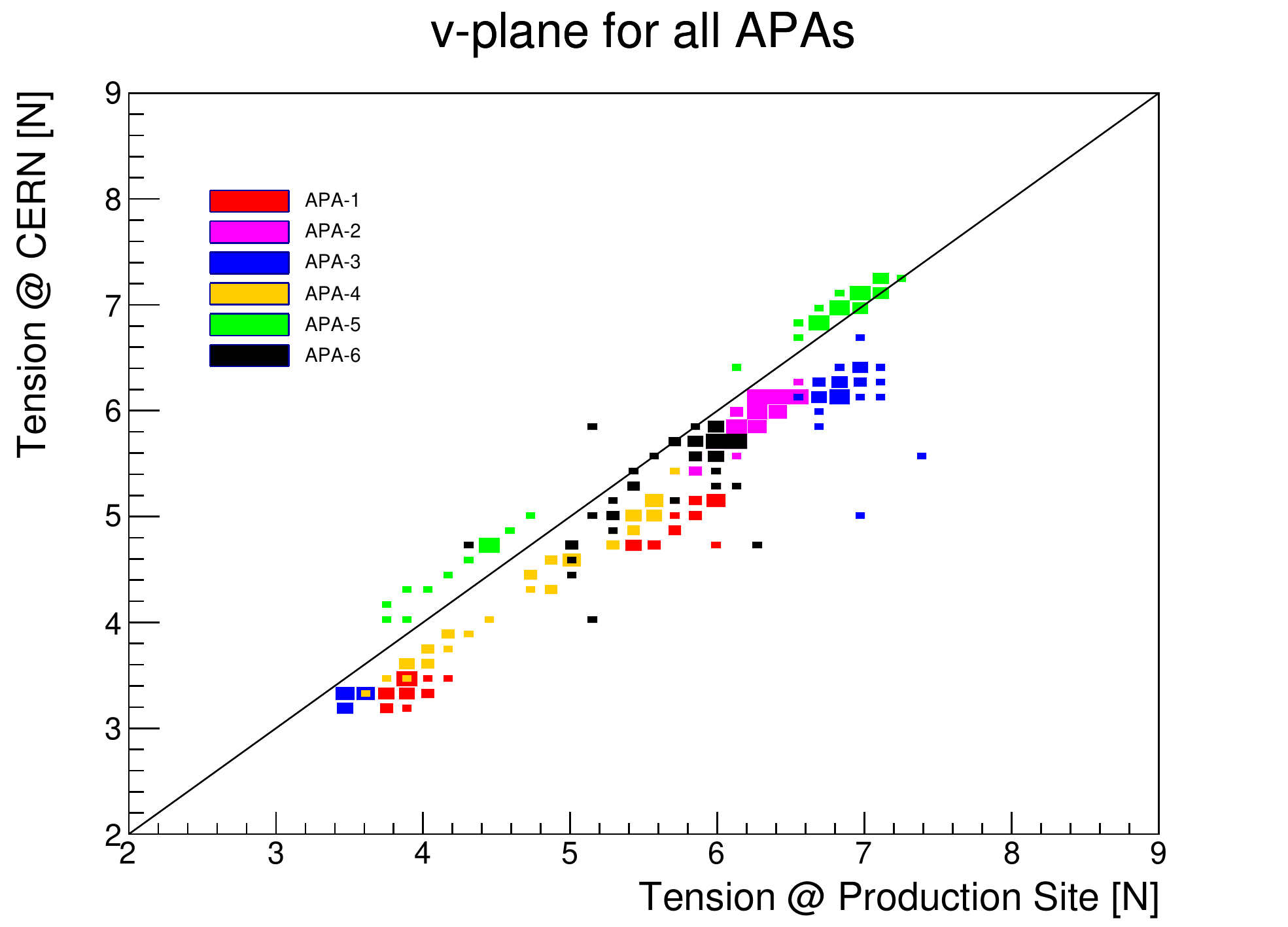} 
\includegraphics[height=0.23\textheight]{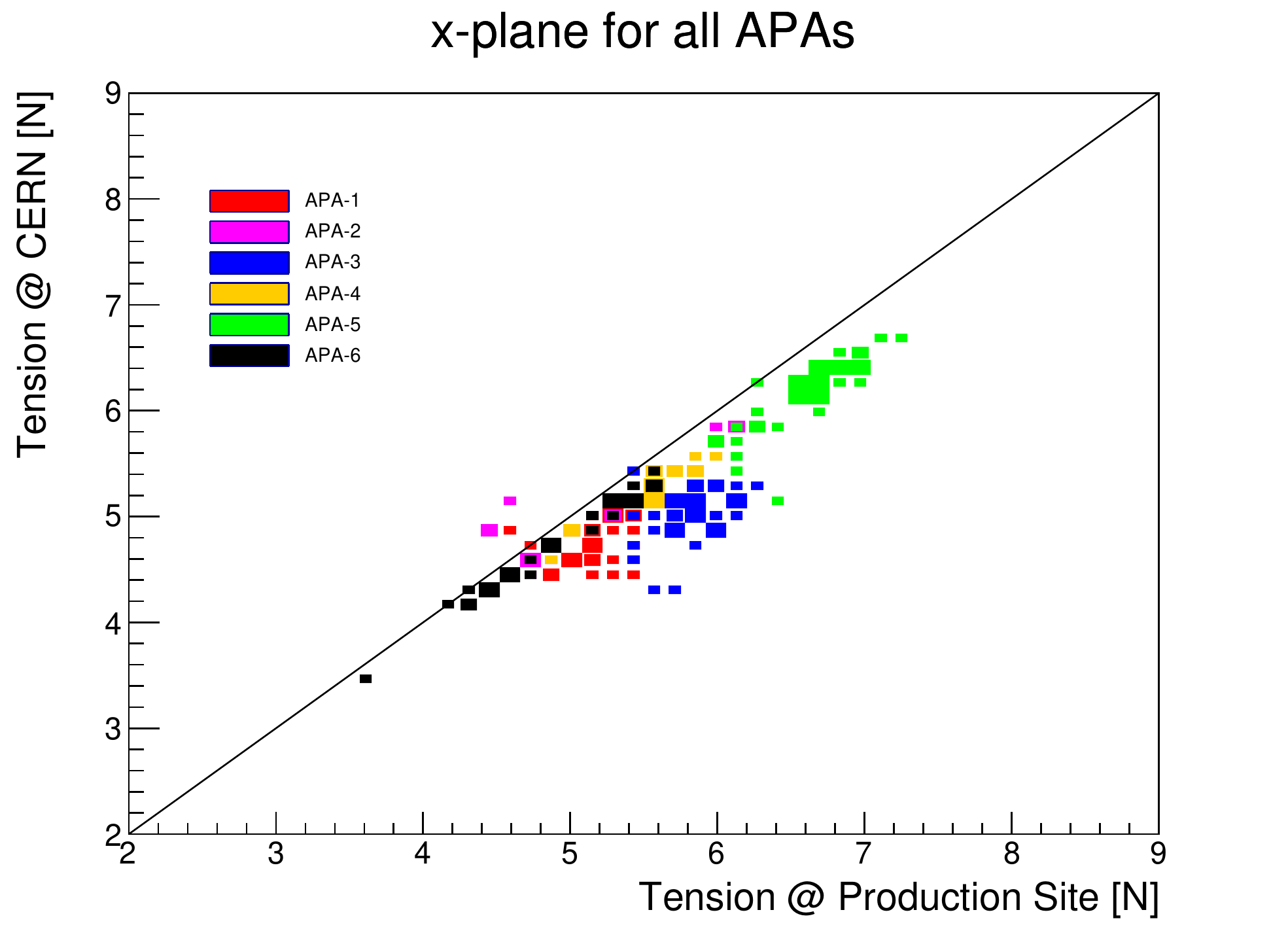}}
\end{dunefigure}

Finally, to test if a cold cycle had any effect on the wire tension, samples of wires were measured again after the \coldbox tests at \dword{cern}. This is the only tension data we have after a cold-cycle for \dword{pdsp} \dword{apa}s. Figure~\ref{fig:sp-apa-pd-tension-cold} presents the results, showing no significant change in the resonant frequency of the wires, indicating cold cycle does not have a significant effect on wire tension.

\begin{dunefigure}[ProtoDUNE-SP wire tension before and after cold tests]{fig:sp-apa-pd-tension-cold}
{Comparison of wire tensions after the \coldbox test versus before at \dword{cern} for a sample of wires on each of the \dword{pdsp} \dword{apa}s.}
\includegraphics[height=0.3\textheight]{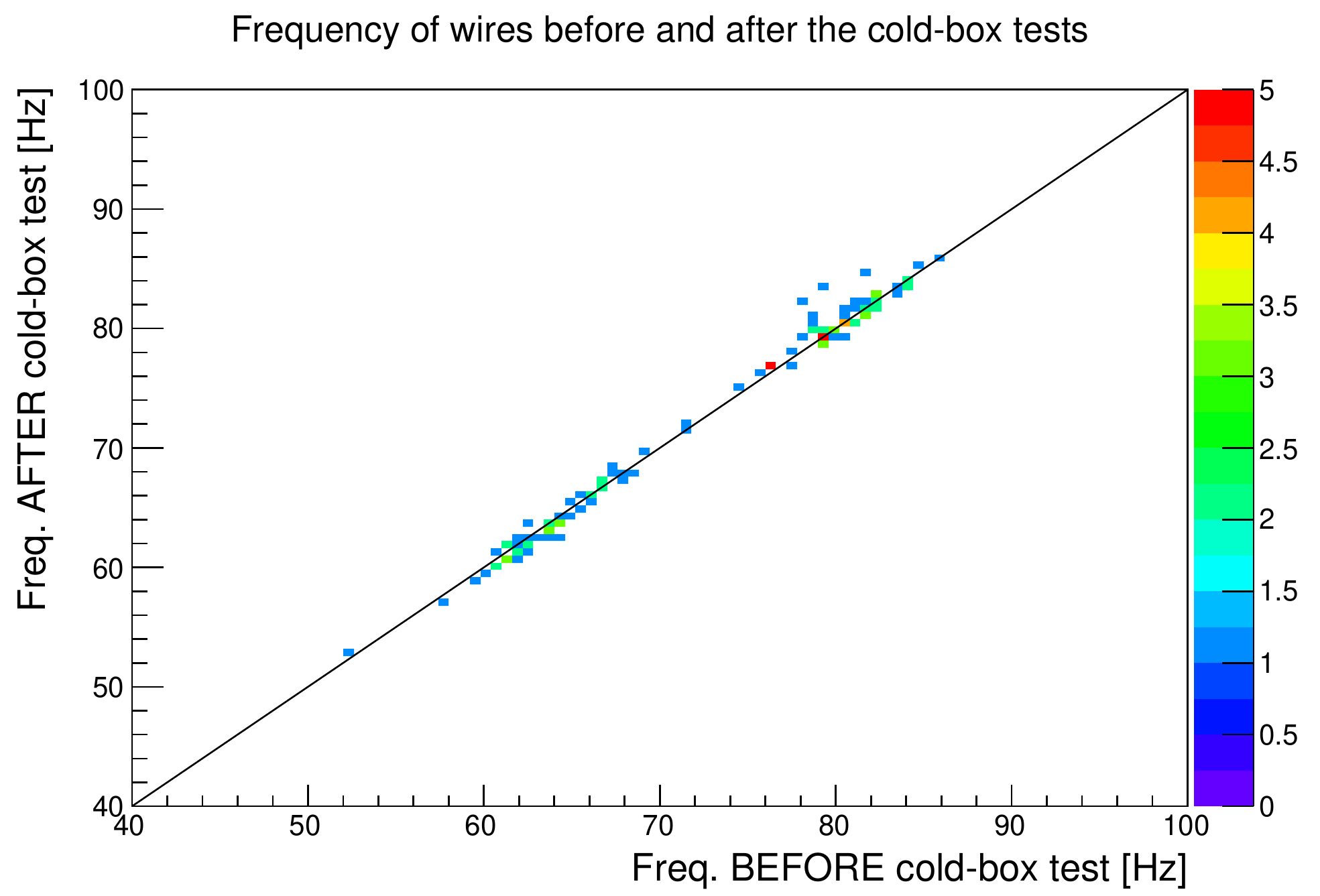} 
\end{dunefigure}

\subsection{Results from ProtoDUNE-SP Operation}
\label{sec:fdsp-apa-qa-protodune-ops}

Several useful analyses for evaluating the \dword{apa} design have been carried out including monitoring the number of non-responsive or disconnected channels in the detector, studying the impact of the electron diverters on reconstruction and calorimetry, and measuring the change in electron transparency with wire bias voltage.  The status of these studies is presented below.

\subsubsection{Disconnected Channels}
\label{sec:fdsp-apa-qa-protodune-ops-dead-channels}

\dword{apa} channels with a ``broken connection'' can be identified in \dword{pdsp} data by comparing channels that do not record hits during detector runs against channels that do respond to the internal calibration pulser system on the \dword{femb}s.  If pulser signals are seen on a channel with no hits, this most likely points to a mechanical failure in the wire path to the electronics.  The failure could be, for example, at a bad solder connection, a damaged trace on a wire board, or a faulty connection between a wire, \dword{cr}, and \dword{ce} adapter boards. Studies have been done using data throughout the \dword{pdsp} run, looking for channels non-responsive to ionization. Note that this analysis is insensitive to the $X$-plane wires that face the cryostat walls since no ionization arrives at those wires.

The results show a very low count of permanently disconnected channels in the \dword{pdsp} \dword{apa}s (28 channels out of 12,480 channels facing the drift volume). In addition, we identified 21 channels that are intermittently not responsive, most probably due to \dword{apa} problems. This is summarized in tables~\ref{tab:deadchan1} and \ref{tab:deadchan2}.  
The fractions of disconnected and intermittent channels are low, 0.22\% and 0.17\%, respectively. 

\begin{dunetable}[Disconnected channel counts in ProtoDUNE-SP]{lcccccc}{tab:deadchan1}{Summary of disconnected channels per plane in \dword{pdsp} due to mechanical failures in the \dword{apa}s.}
 & $U$-plane & $V$-plane & $X$-plane & Total Channels & ~~~Rate~~~ & Total \\ \toprowrule
Disconnected & 16 & 8 & 4 & \multirow{2}{*}{\num{12480}} & 0.22\% & \multirow{2}{*}{0.39\%} \\
Intermittent & 7 & 7 & 7 &  & 0.17\% & \\
\end{dunetable}

\begin{dunetable}[Dead channel counts in ProtoDUNE-SP]
{lcccccc}
{tab:deadchan2}
{Summary of disconnected channels per \dword{apa} in \dword{pdsp} due to mechanical failures in the \dword{apa}s.}
& APA 1 & APA 2 & APA 3 & APA 4 & APA 5 & APA 6 \\ \toprowrule
Disconnected & 4 & 5 & 8 & 3 & 1 & 7  \\
Intermittent & 10 & 0 & 1 & 4 & 3 & 3  \\
\end{dunetable}

So far, analysis of data throughout the \dword{pdsp} run shows no evidence of increasing numbers of disconnected or intermittent channels.

\subsubsection{Effect of Electron Diverters on Charge Collection}
\label{sec:fdsp-apa-qa-protodune-ops-electron-diverters-charge}

Active strip-electrode electron diverters were installed in \dword{pdsp} between \dword{apa}s~1 and~2 (ED12) 
and between \dword{apa}s~2 and~3 (ED23), which are both on the beam-right side of \dword{pdsp} for 
the 2018-2019 run.  The two inter-\dword{apa} gaps on the beam-left side did not have electron diverters in them. ED12 developed an electrical short early in the run, and as a consequence, both ED12 and ED23 were left unpowered for the beam run and all but a small number of test runs after the beam run.  A voltage divider on the electron diverter \dword{hv} distribution board provided a path to ground, and so the electron diverter strips were effectively grounded.  Since they protrude into the drift volume in front of the \dword{apa}s, the grounded diverters collect nearby drifting charge instead of diverting it towards the active apertures of the \dword{apa}s, 
leading to broken tracks with charge loss in the gaps.  When powered properly, charge is primarily displaced away from the gap, and tracks that are more isochronous provide good measurements of the charge arrival time delays due to the longer drift paths of diverted charge. Figure~\ref{fig:sp-apa-diverterevd100} shows the collection-plane view of the readout of \dword{apa}s~3 and~2 for a test run in which ED23 was powered at its nominal voltage. Figure~\ref{fig:sp-apa-nodiverterevd} shows the collection-plane view of a track crossing the drift volumes read out by \dword{apa}s~6 and~5, which do not have an electron diverter installed between them.  Timing and spatial distortions in the absence of diverters appear minimal.

The impact of charge distortions can be seen in Figure~\ref{fig:sp-apa-qc-diverterdqdx}, which shows the average $dQ/dx$ distributions for \dword{pdsp} run 5924, which has ED12 at ground voltage, ED23 at nominal voltage, and no diverters on the beam-left side of the detector between \dword{apa}s~4, 5, and~6.  Pronounced drops in the charge collected near ED12 (grounded diverter) are seen, while much smaller distortions are seen elsewhere.  Run 5924 was taken while the grid plane in \dword{apa}~3 was charging up, resulting in artifacts in the $dQ/dx$ measurements with a period of three wires.  \dword{apa}~2 has an artifact from an \dword{asic} with a slightly different gain reading out channels near the boundary with \dword{apa}~1, causing even and odd channels to be offset.

\begin{dunefigure}[\dshort{pdsp} event display; impact of a grounded electron diverter]{fig:sp-apa-diverterevd100}{Collection-plane charge signals in ProtoDUNE-SP for a single readout window in \dword{apa}s~3 (left) and~2 (right) for a test run in which ED23 was powered at its nominal voltage.  The horizontal axis is wire number, arranged spatially along the beam direction, and the vertical axis is readout time.  The event is run 5924, event 275.}
    \includegraphics[width=\textwidth,trim=85mm 0mm 85mm 0mm,clip]{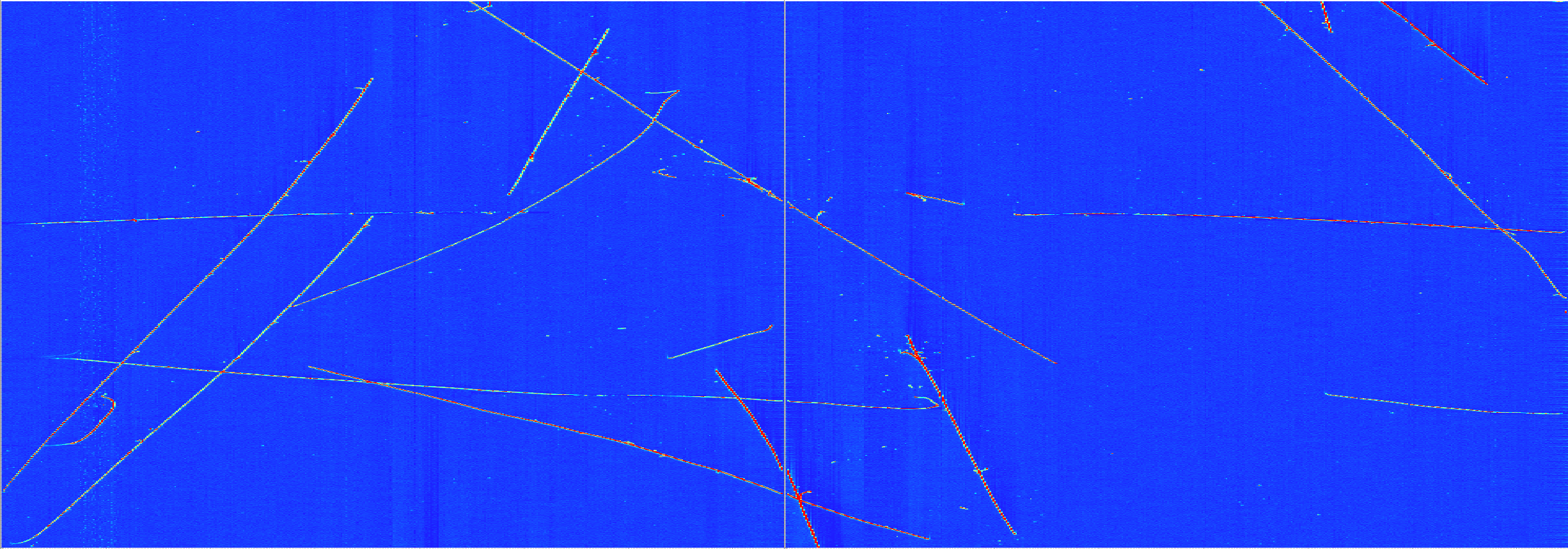}
\end{dunefigure}

\begin{dunefigure}[\dshort{pdsp} event display; track crossing gap without electron diverter]{fig:sp-apa-nodiverterevd}{Collection-plane event display for \dword{apa}s~6 (left) and~4 (right). No electron diverter was installed between these two \dword{apa}s.  The event is run 5439, event 13.}
    \includegraphics[width=\textwidth,trim=85mm 0mm 85mm 0mm,clip]{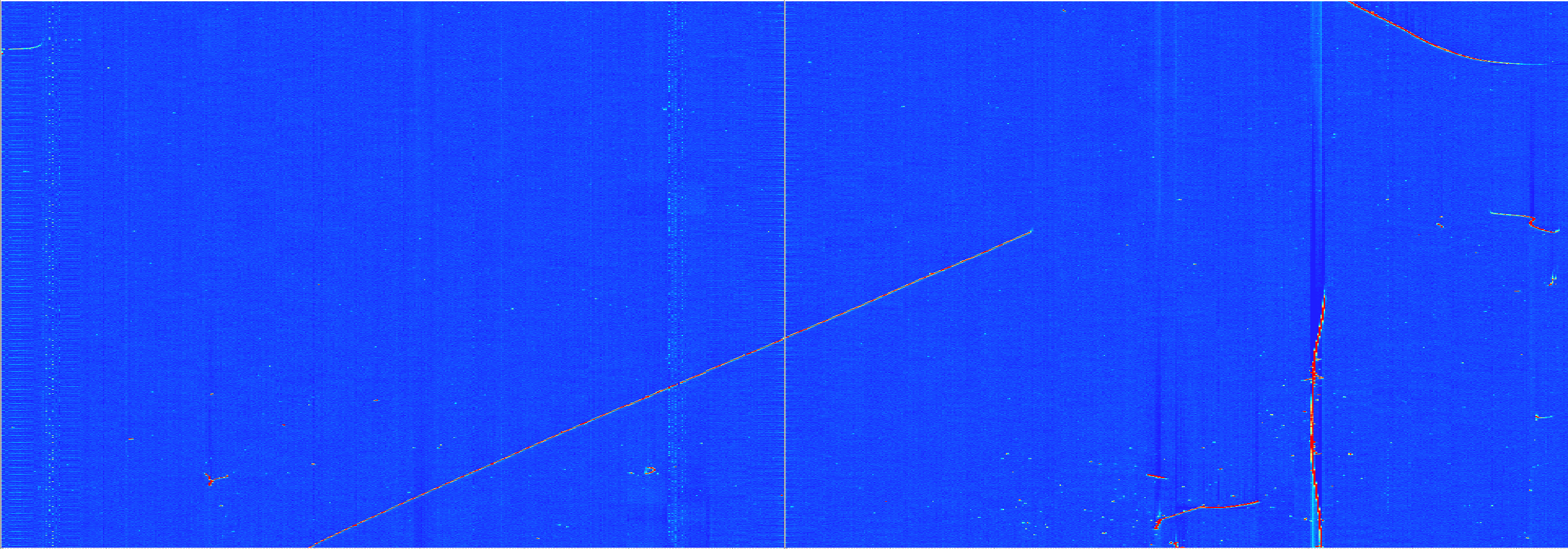}
\end{dunefigure}

\begin{dunefigure}[$dQ/dx$ distributions for \dshort{pdsp} with different diverter conditions]
{fig:sp-apa-qc-diverterdqdx}{The $dQ/dx$ distributions as a function of the collection wire number zoomed in near the gaps, using cosmic ray muons in \dword{pdsp} run 5924. The electron diverters are only instrumented for the gaps at the beam right side ($x<0$). The electron diverter between \dword{apa}~2 and \dword{apa}~3 was running at the nominal voltage while the electron diverter between \dword{apa}1 and \dword{apa}~2 was turned off. }
\includegraphics[width=0.48\textwidth]{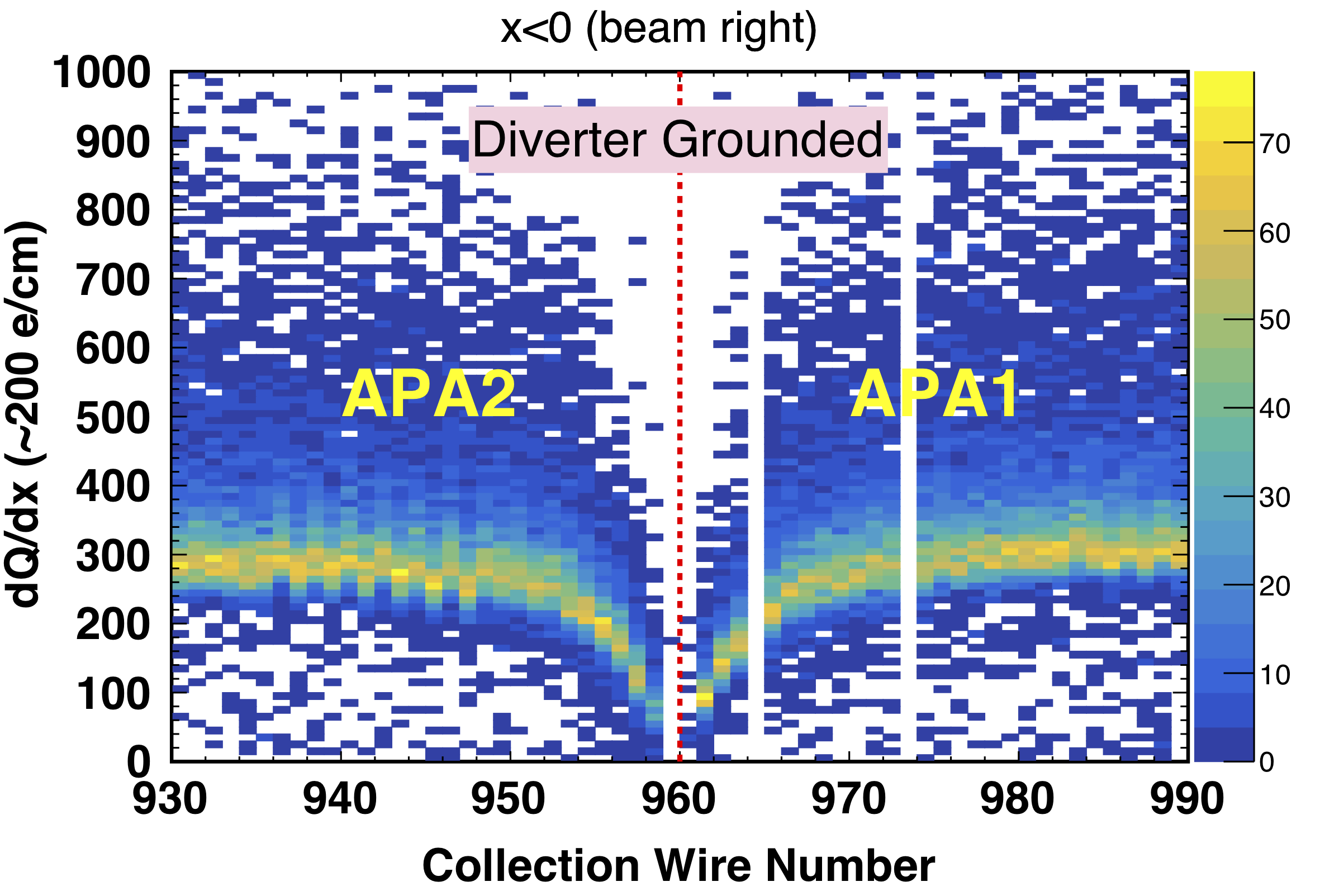}
\includegraphics[width=0.48\textwidth]{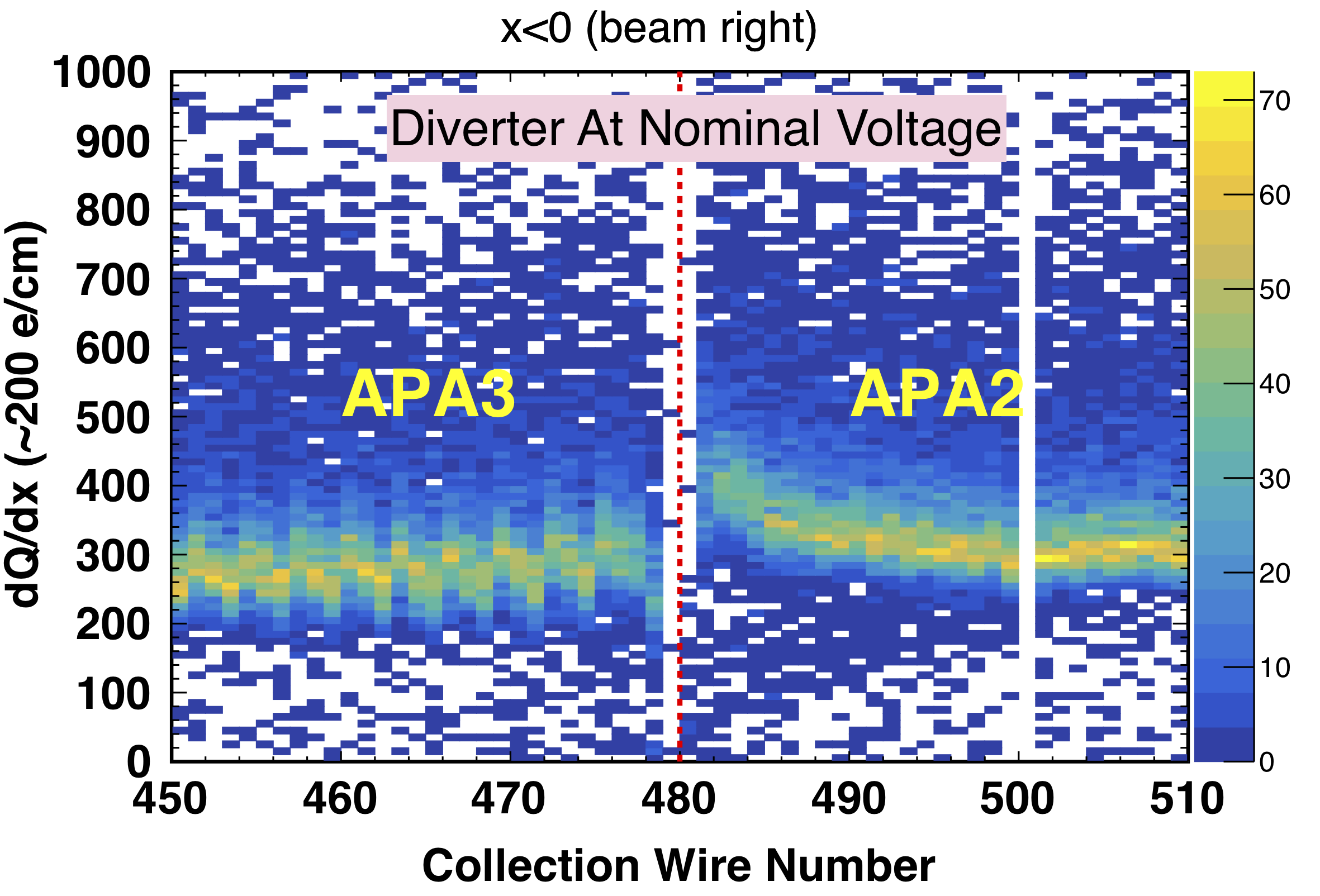}
\includegraphics[width=0.48\textwidth]{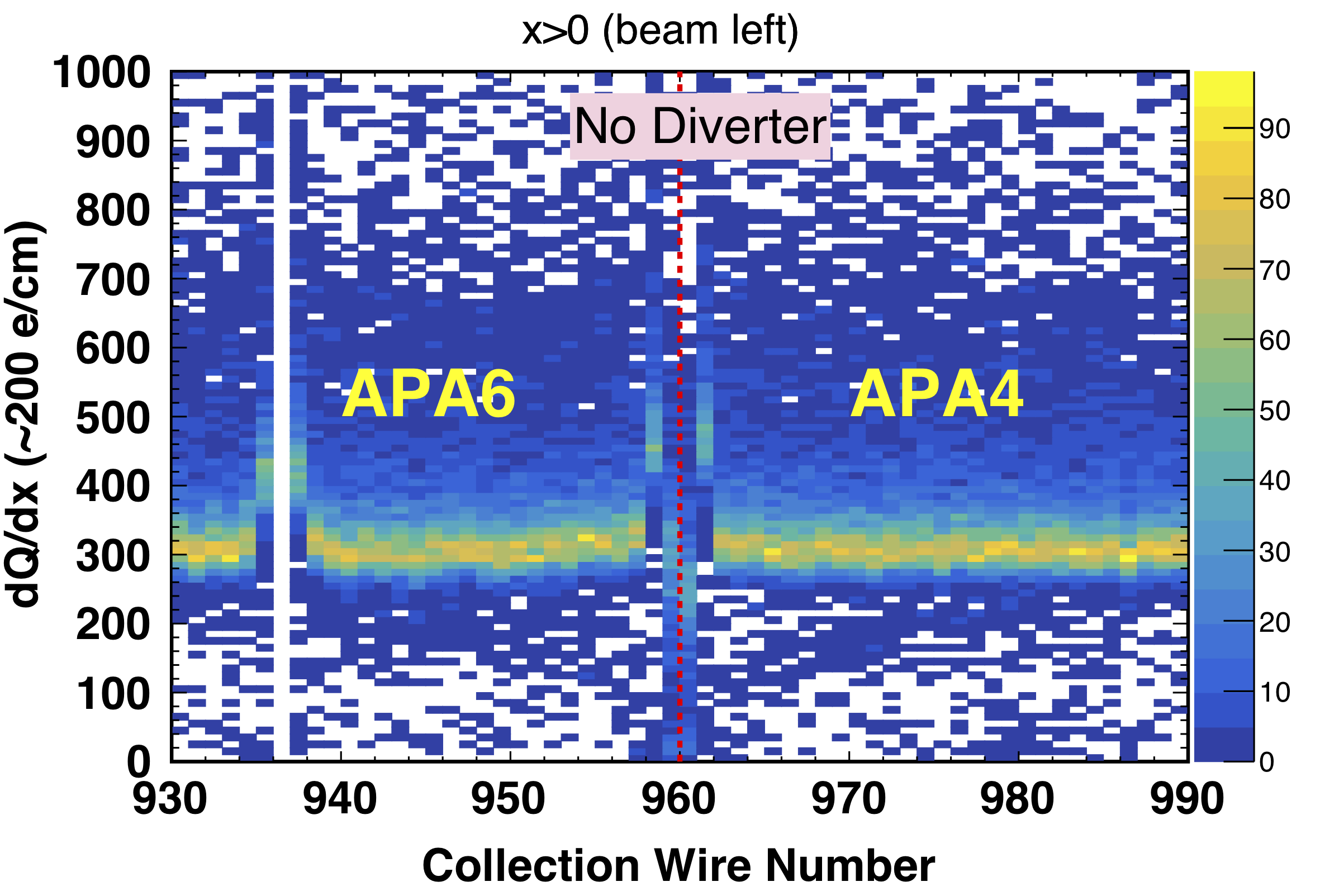}
\includegraphics[width=0.48\textwidth]{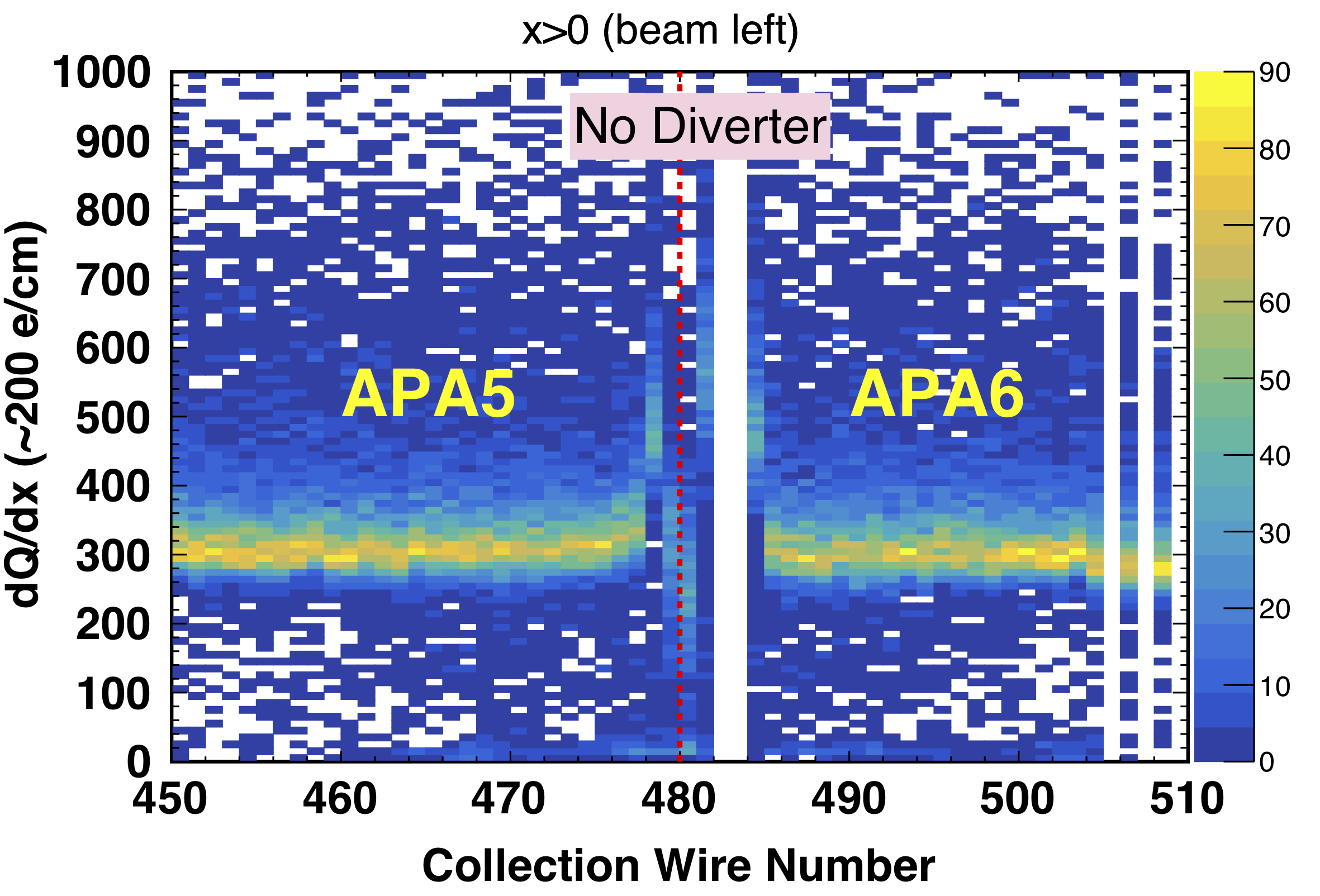}
\end{dunefigure}

\subsubsection{Effect of Wire Support Combs on Charge Collection}
\label{sec:fdsp-apa-qa-protodune-ops-combs-charge}

Inclusive distributions of charge deposition on each channel can be made with \dword{pdsp} data using the cosmic-ray tracks.  Tracks that cross from the cathode to the anode have unambiguous times even without association with \dwords{pd}, and thus distance-dependent corrections to the lifetime can be made.  The reconstruction of tracks in three dimensions makes use of the charge deposited in each of the three wire planes.  Maps of the median $dQ/dx$ response have been made for each plane in each \dword{apa} in the $(y,z)$ plane, the plane in which the \dword{apa} resides.  The granularity of these maps is the wire spacing, in both dimensions, and so the charge response of small segments of wires is measured.  These maps are projected onto the $U$, $V$, $y$, and $z$ coordinate axes in order to visualize more easily the impacts of localized detector inhomogeneities.

The wire-support combs are approximately evenly spaced in the $y$ coordinate.  In order to investigate the impact of the wire combs on charge collection and induction signals, the average of the median binned $dQ/dx$ values as a function of $y$ is shown for $U$, $V$, and collection-plane ($Z$) wires in Figure~\ref{fig:sp-apa-pd-comb-charge-impact}.  \dword{apa}~6, which is in the middle of the detector and thus is minimally affected by features on the neighboring field cages, is chosen so the effects of the combs are most visible, though similar effects are seen in all six \dword{apa}s in \dword{pdsp}.  Localized dips of the order of 2\% in the average signals can be seen at the locations of the combs in the $U$ and $V$ views, while the collection-plane channels show smaller dips and other features.  Charge is expected to divert around the dielectric combs after they charge up, and if the diversion is purely in the vertical direction, then the impact on the collection-plane response is expected to be suppressed.  The induction-plane response may be understood as the result of the dielectric comb locally polarizing in the field of the drifting charge, thus modifying the \efield at the wires.  This analysis well exhibits the uniformity of the response of the \dword{pdsp} \dword{apa}s as well as the level of detail that can be extracted from \dword{tpc} data for the precise calibration of the \dwords{spmod}.

\begin{dunefigure}[Average charge deposition on tracks vs. height]{fig:sp-apa-pd-comb-charge-impact}
{Average $dQ/dx$ on the $U$, $V$, and collection-plane ($Z$) wires in \dword{apa}~6 as a function of the height $y$ from the bottom of the \dword{pdsp} detector. }
\includegraphics[width=0.32\textwidth]{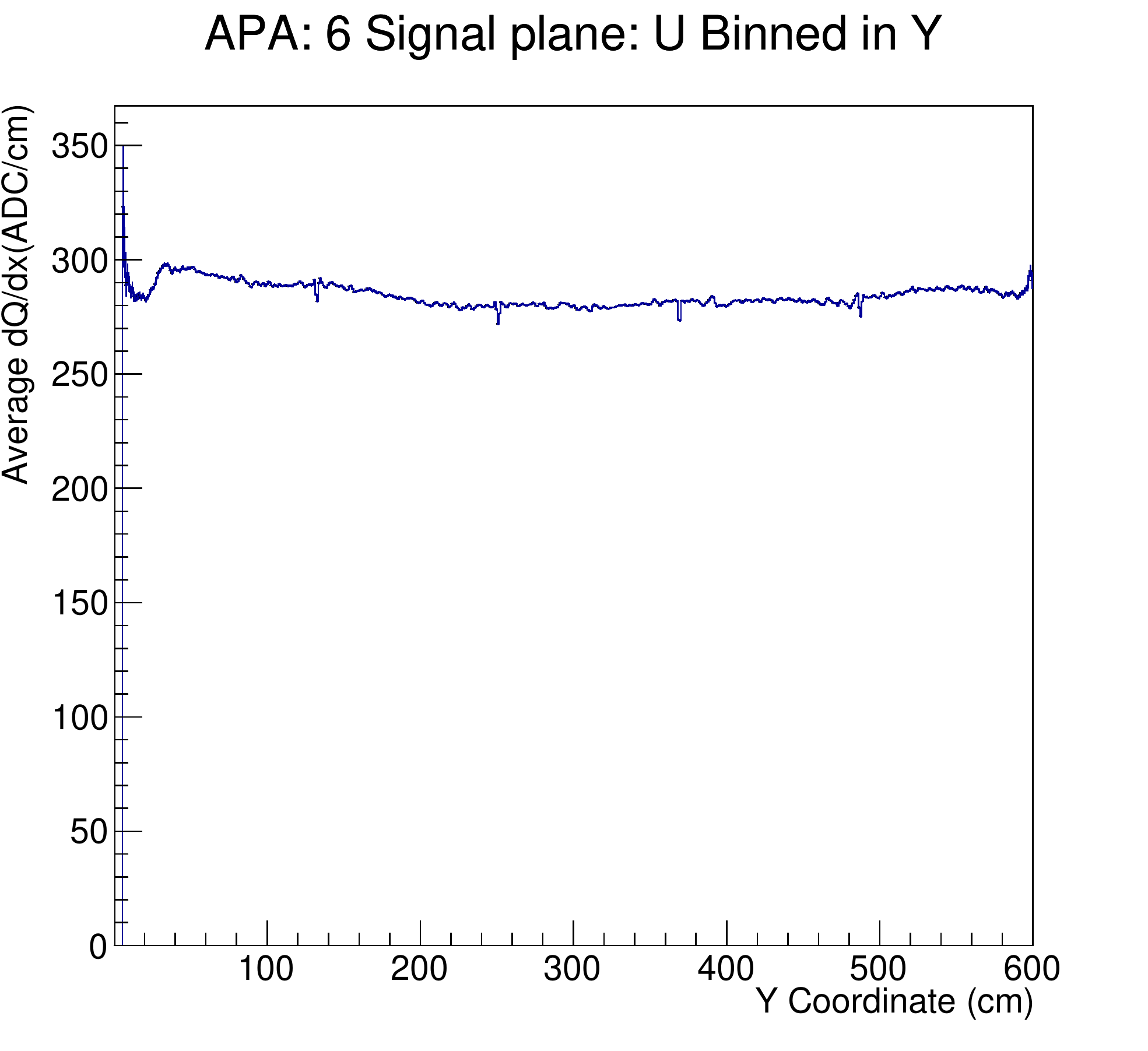}
\includegraphics[width=0.32\textwidth]{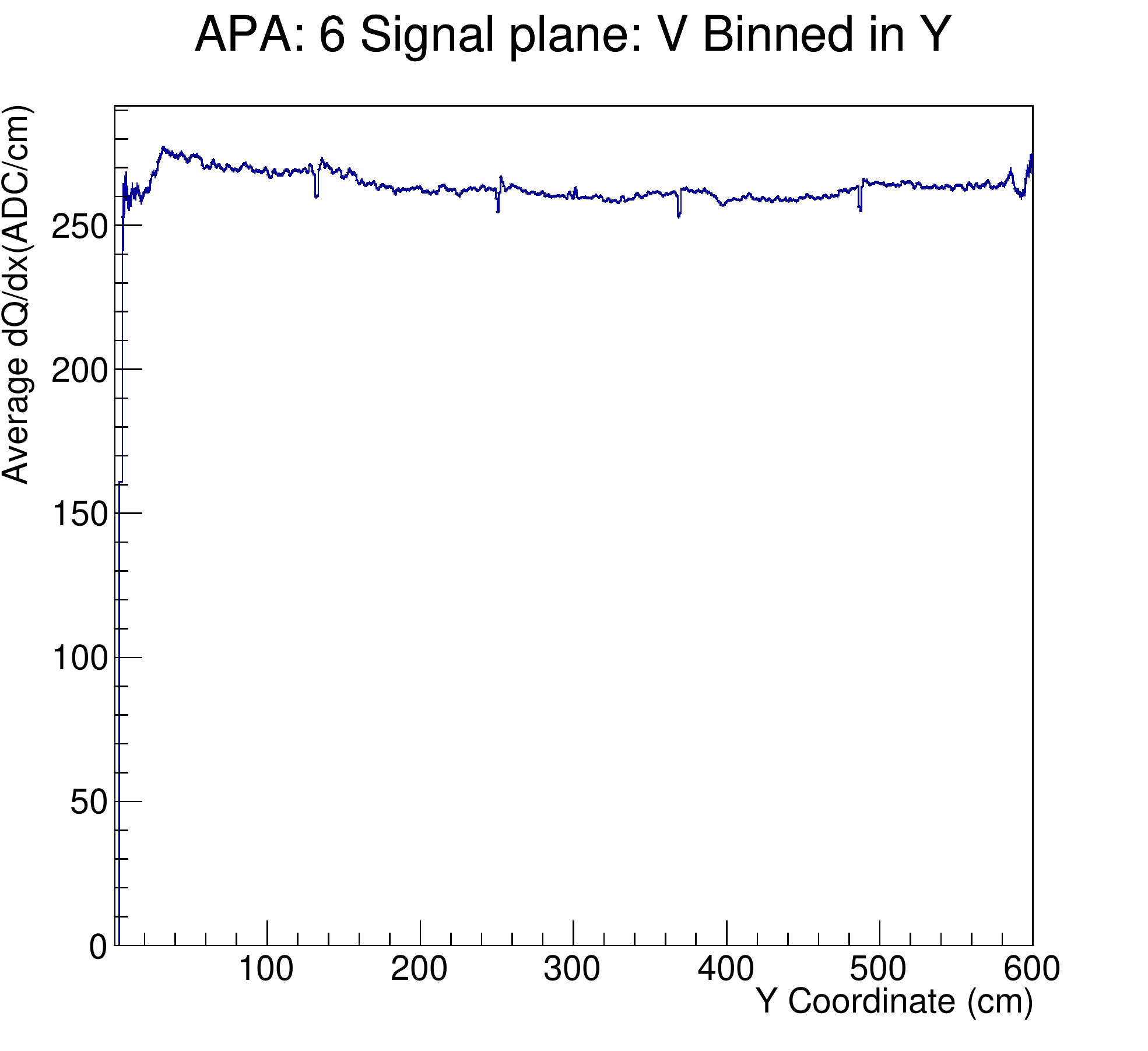}
\includegraphics[width=0.32\textwidth]{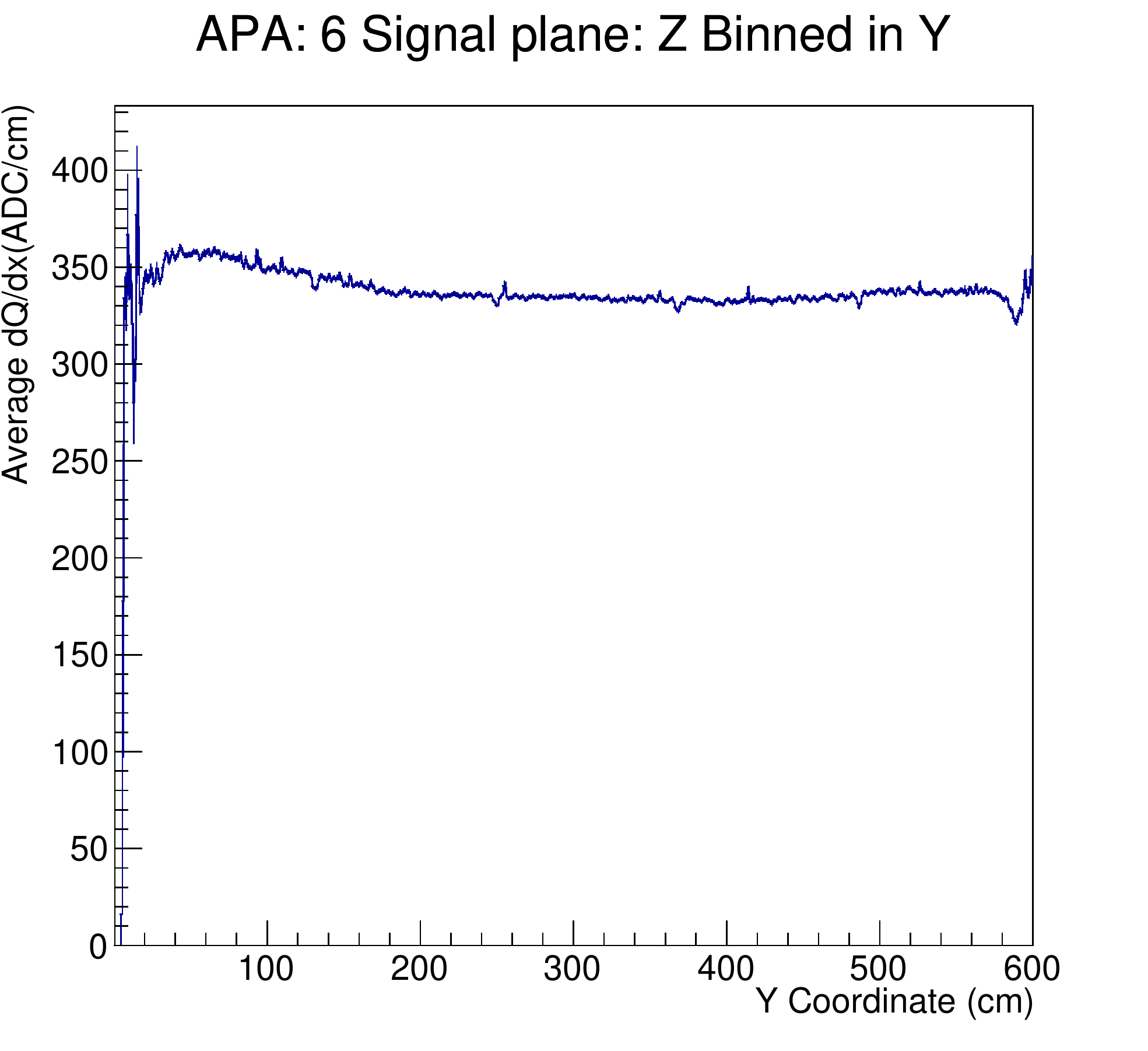}   
\end{dunefigure}

\subsubsection{Wire Bias Voltage Scans and Electron Transparency}
\label{sec:fdsp-apa-qa-protodune-ops-bias-scans}

A set of dedicated runs were taken at \dword{pdsp} in order to confirm the bias voltage settings calculated by the COMSOL software 
and presented in Section~\ref{sec:fdsp-apa-design-overview}. In particular, the bias voltages in the $G$ (grid), (induction) $U$, and (collection) $X$ wire plane were uniformly reduced from 5\% to 30\% relative to the nominal settings. For each wire plane, the transparency condition depends on the ratio of the \efield before and after the wire plane. Therefore, in the situation of uniform reduction of the bias voltages, some ionization electrons are expected to be collected by the grid plane, leading to a loss of ionization electrons collected by the $X$ wires. Figure~\ref{fig:protodune-bias-voltage-scan} shows the results from each of six \dword{apa}s in \dword{pdsp}. The ratio ``R'', ranging from 0.7 (30\% reduction) to 0.95 (5\% reduction), represents the different bias voltage settings used in these runs. ``T'' represents the transparency of the ionization electrons, which is proportional to the number of ionization electrons collected by the $X$ wire plane. As a result of the significant space charge effect in \dword{pdsp}, the sources of ionization electrons (presumably dominated by cosmic muons) are different for different \dword{apa}s. To facilitate the comparison among different \dword{apa}s, the transparency at each bias voltage setting is normalized by the transparency at the highest bias voltage setting (R=0.95). Except for \dword{apa}~3, all \dword{apa}s show a similar trend in the change of transparency. The spread represents the uncertainty in calculating the transparency. The grid plane of \dword{apa}~3 was found to be disconnected since December 2018, which led to incorrect bias voltage settings in these runs. This explained the abnormal behavior in its transparency data. Two sets of predictions (COMSOL vs. Garfield) are compared with the \dword{pdsp} data. The ranges of R in these predictions are different from that of the \dword{pdsp} data, since these two predictions were obtained prior to the \dword{pdsp} data taking. The COMSOL prediction is clearly confirmed by the \dword{pdsp} data, which also validates the nominal bias voltage settings listed in Section~\ref{sec:fdsp-apa-design-overview}. The incorrect prediction from the Garfield simulation is attributed to inaccurate \efield calculations near the boundary of the wires (\SI{152}{$\mu$m} diameter), which is much smaller than the wire pitch ($\sim$\SI{4.79}{mm}). 

\begin{dunefigure}[ProtoDUNE bias voltage scan data]{fig:protodune-bias-voltage-scan}
{The transparency results from the bias voltage scan in \dword{pdsp}. "R", the ratio to the nominal bias voltages, represents different bias voltage settings. "T" represents the transparency of the ionization electrons, which is proportional to the number of ionization electrons collected by the $X$ wire plane. The prediction of COMSOL (Garfield) is confirmed (refuted) by the \dword{pdsp} data. The abnormal behavior of \dword{apa}~3 is a result of incorrect bias voltage settings. See Section~\ref{sec:fdsp-apa-qa-g-plane} for more discussion.}
\includegraphics[width=0.75\textwidth]{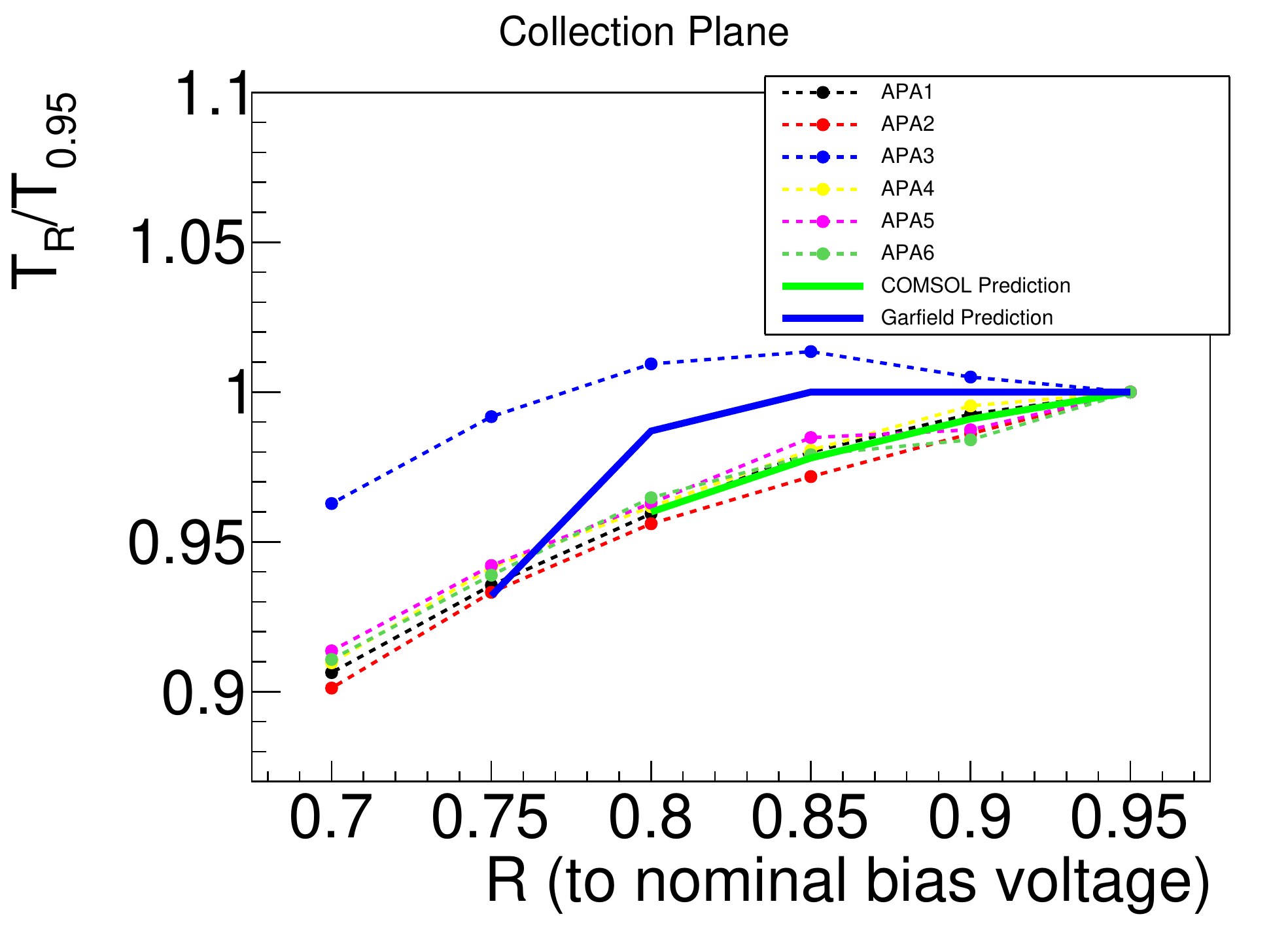}   
\end{dunefigure}

\subsubsection{Abnormal Behavior of G-plane on APA 3}
\label{sec:fdsp-apa-qa-g-plane}

Dedicated studies of $dQ/dx$, the recorded ionization charge per unit path length from cosmic muon tracks, have been performed for each \dword{apa}. For runs immediately after periods when the cathode \dword{hv} was off for an extended length of time, of the order of a few days, the average of the $dQ/dx$ distribution on \dword{apa}~3 collection and induction planes was found to be systematically lower than for the other \dword{apa}s.  The $dQ/dx$ would then slowly increase with time.  Detailed investigations showed that this behavior is explained by the assumption that the $G$-plane on \dword{apa}~3 is not connected to a proper reference voltage. When the cathode \dword{hv} is turned on after a long off period, the $G$-plane, initially at a floating potential close to ground, slowly charges up towards a negative \dword{hv},  re-establishing transparency for the ionization electrons towards the signal planes. It takes about 100 hours for the $G$-plane to reach a negative potential close to the nominal value that allows full transparency.

We are presently evaluating more accessible locations for the connection of the bias \dword{hv} cables from the cryostat feedthroughs to the \dword{apa}s, to minimize connection problems with the \dword{shv} connectors. In addition, during installation, we will include as part of the standard checkout procedure either a direct confirmation of the bias connection between a wire plane and its bias input on the feedthrough flange, or an indirect measurement of the connection by recording the charging current in the bias line when increasing the bias voltage to its nominal value.  The construction and integration tests with a pre-production \dword{apa}, described below in Sec.~\ref{sec:fdsp-apa-qa-prototyping}, will fully test any changes to the \dword{shv} system.

\subsection{Final Design Prototyping and Test Assemblages}
\label{sec:fdsp-apa-qa-prototyping}

To confirm modifications made to the \dword{apa} design and production process since \dword{pdsp} and to work through the multi-\dword{apa} assembly procedures, several prototypes are planned for 2020.

A seventh \dword{pdsp}-like \dword{apa} was completed at Daresbury Laboratory by utilizing an upgraded winding machine with the new interface arm design (see Section~\ref{sec:fdsp-apa-prod-tooling}). This \dword{apa} was shipped to \dword{cern} in March 2019 for a test of the \dword{ce} in the \coldbox, expected to be performed in 2020. In addition, work is in progress to implement a new winding head on the \dword{apa} winding machines, with automatic tension feedback and control on the wires. These same upgrades will be implemented on the winding machine at PSL in 2020. 

A top and bottom version of the new supporting \dword{apa} frame design were built in spring 2019 at PSL and shipped to \dword{ashriver}.  A full test of the \dword{apa} pair assembly procedure was successfully completed in early October 2019 (see Figure~\ref{fig:ashriver-integration}). The procedure of routing the \dword{ce} cables along the side tubes of the \dword{apa} pair was also successfully tested. In addition, a preliminary test of the installation of \dword{pds} prototype cables inside the \dword{apa} frames and the mating of cable connections between the lower and upper \dword{apa} was performed.  See Section~\ref{sec:fdsp-apa-intfc-cables}  for more information on cable routing in the \dword{apa} frames.

Also planned is the construction of a pre-production \dword{apa} for an integration test with the \dword{ce} and \dword{pds} systems at \dword{cern}, which will fully test all interface aspects. This test will inform the final design review of the \dword{apa} system in May 2020. 

In addition, three fully wound \dword{apa}s with pre-installed \dword{pds} cables, will be built by the end of 2020 for deployment in \dword{pdsp}-II, replacing the detectors of one of the drift volumes.  This will allow a test of all \dword{apa} components, including the larger size frames and geometry boards and a final tuning of the winding machines. The three \dword{apa}s will be shipped to \dword{cern}, integrated with \dword{ce} boxes and \dword{pds} detectors, and tested in the \coldbox before installation in \dword{pdsp}.  The pre-production \dword{apa} mentioned above could serve as one of the three if no design modifications are required.  These final prototyping activities will serve to test all critical aspects of the \dword{apa} design before starting \dword{dune} \dword{apa} production in 2020.  

\begin{dunefigure}[Ash River APA pair integration tests]{fig:ashriver-integration}
{\dword{apa} pair assembly and integration tests at \dword{ashriver}.}
\includegraphics[height=0.5\textheight]{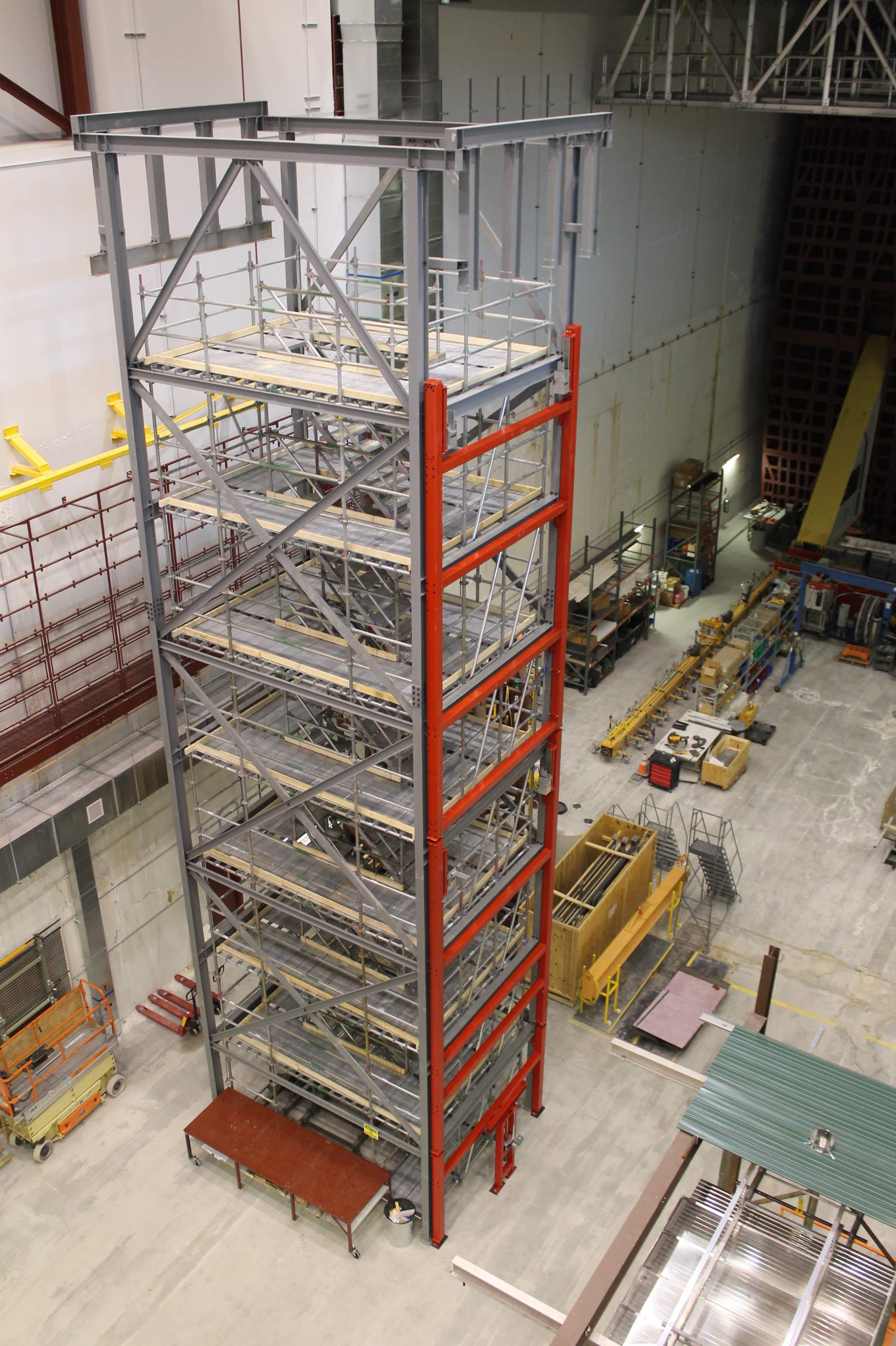} \quad
\includegraphics[height=0.5\textheight]{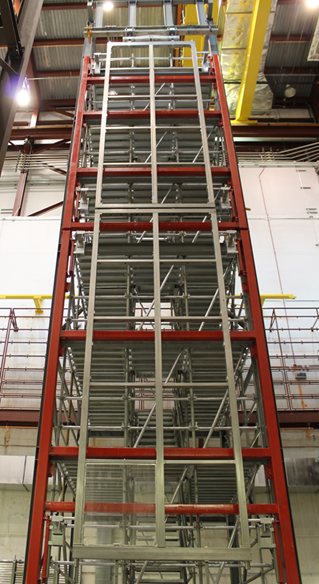} 
\end{dunefigure}
\section{Interfaces}
\label{sec:fdsp-apa-intfc}

The interfaces between the \dword{apa} consortium and other detector consortia, facilities, and working groups covers a wide range of activities. Table~\ref{tab:apa_interface_docdb} lists the interface control documents under development. In the following sections, we elaborate slightly on the interfaces with the \dword{tpc} readout electronics and the \dword{pds}, as well as the cable routing plan for both systems.  Other important interfaces are to the \dword{tpc} \dword{hv} system (the \dword{fc}) and the \dword{dss} inside the \dword{dune} cryostats.  

\begin{dunetable}
[APA interfaces]
{p{0.4\textwidth}p{0.2\textwidth}}
{tab:apa_interface_docdb}
{\dword{apa} Interface Links. 
}
Interfacing System & Linked Reference \\ \toprowrule
TPC electronics & \citedocdb{6670} \\ \colhline 
Photon detector system & \citedocdb{6667} \\ \colhline
Drift high voltage system & \citedocdb{6673} \\ \colhline
DAQ & \citedocdb{6676} \\ \colhline
Slow controls and cryogenics & \citedocdb{6679} \\ \colhline
Integration facility & \citedocdb{7021} \\ \colhline
Facility interfaces 
& \citedocdb{6967} \\ \colhline
Installation & \citedocdb{6994} \\ \colhline
Calibration & \citedocdb{7048} \\ \colhline
Software computing & \citedocdb{7102} \\ \colhline
Physics & \citedocdb{7075} \\
\end{dunetable}

\subsection{TPC Cold Electronics}
\label{sec:fdsp-apa-intfc-elec}

The \dword{tpc} readout electronics (\dword{ce}) are directly mounted to the \dword{apa} and thus immersed in \dword{lar} to reduce the input capacitance and inherent electronic noise.  With the wire-wrapped design, all \num{2560} wires to be read out (recall \num{960} are $G$-plane wires used for charge shielding only and are not read out) terminate on wire boards that stack along one end (the head) of the \dword{apa} frame.  The \num{2560} channels are read out by \num{20} \dword{fe} motherboards (\num{128} channels per board), each of which includes eight \num{16}-channel \dword{fe} \dwords{asic}, eight \num{16}-channel \dword{adc} \dwords{asic}, \dword{lv} regulators, and input signal protection circuits.  Figure~\ref{fig:apa_ce} shows a \dword{pdsp} \dword{apa} during integration at \dword{cern} with the \dword{tpc} electronics partially installed and a cable tray mounted above. 

\begin{dunefigure}[APA interface with TPC electronics]{fig:apa_ce}
{The head region of an \dword{apa} frame during installation at \dword{pdsp}.  On the left the head wire boards, \dword{cr} boards, and yoke are clearly visible. On the right, five of the 20 \dword{ce} boxes have been installed.}
\includegraphics[height=0.2\textheight]{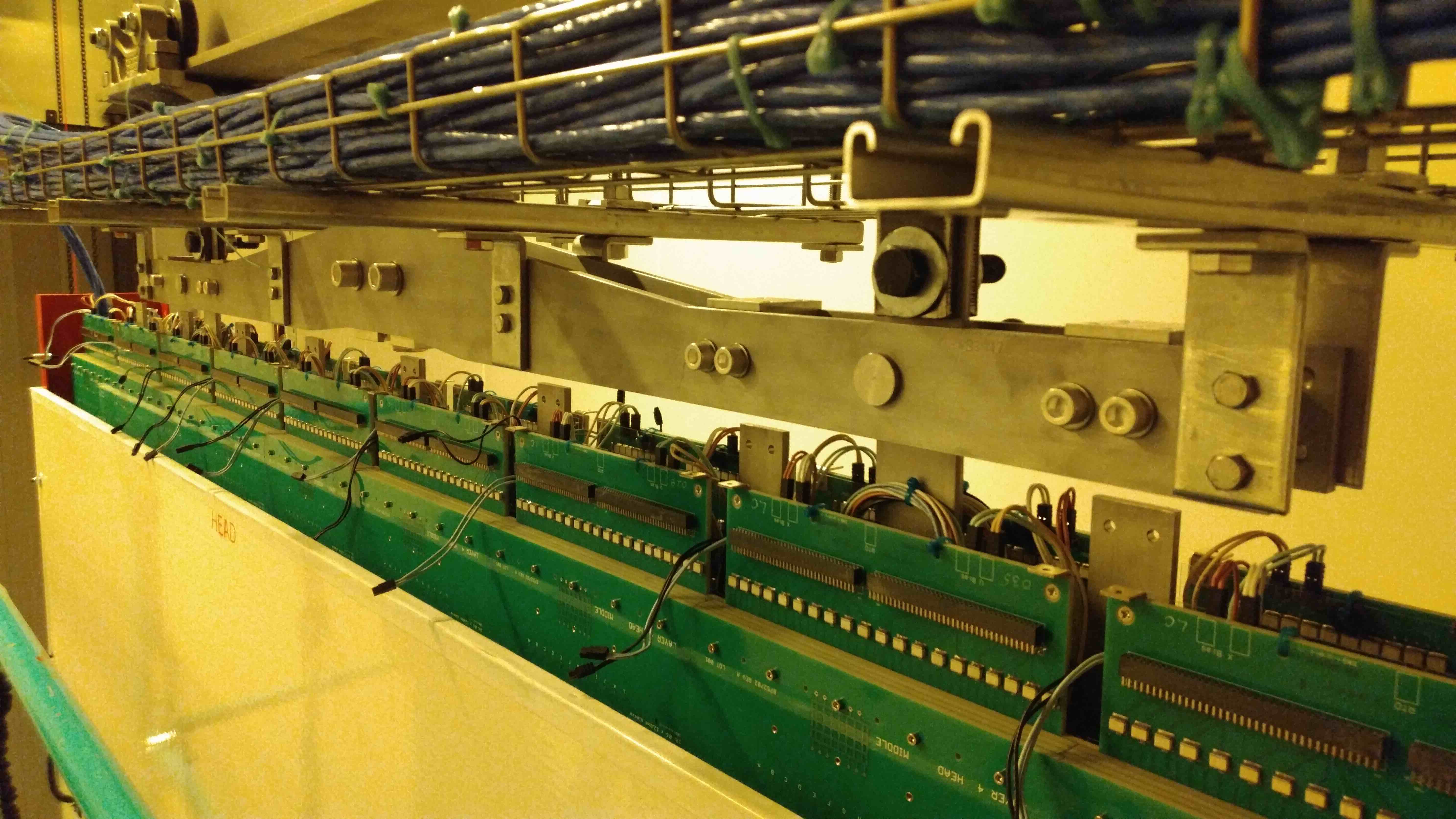}
\includegraphics[height=0.2\textheight]{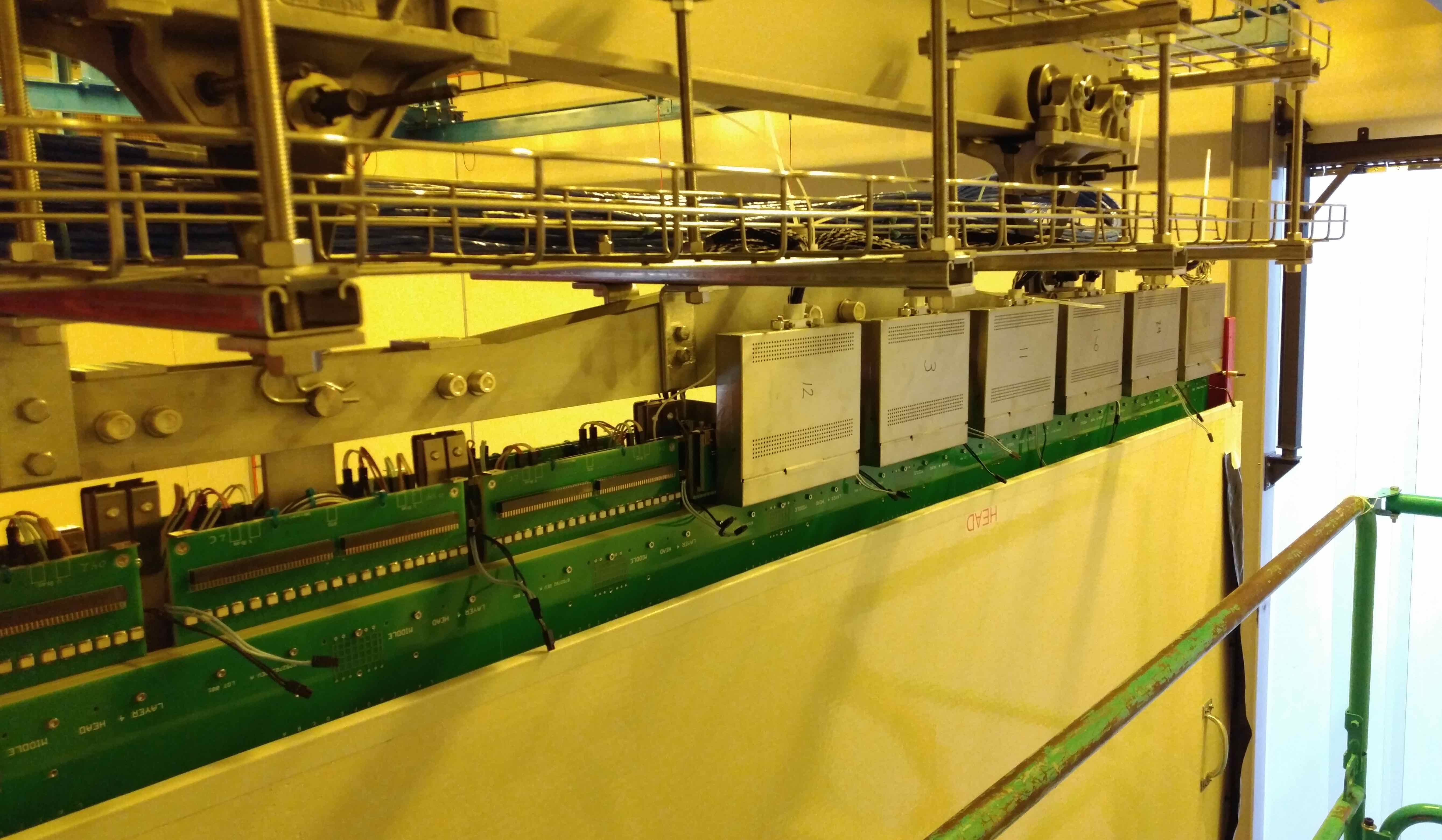}
\end{dunefigure}

The mechanical interface includes the support of the \num{20} \dword{ce} boxes, each housing a \num{128} channel \dword{femb}.  They are the gray vertically oriented boxes  on the right in Figure~\ref{fig:apa_ce}. 

The electrical interface covers the choice of wire-bias voltages to the four wire planes so that \num{100}\% transparency can be achieved for drifting ionization electrons, cable connection for the wire bias voltages from the cryostat feedthroughs to the \dword{cr} boards, interface boards connecting \dword{cr} boards and \dword{ce} boxes, filtering of wire-bias voltages through \dword{cr} boards to suppress potential electronic noise, and an overall grounding scheme and electrical isolation scheme for each \dword{apa}. The last item is particularly important to achieve the required low electronic noise levels.  See Section~\ref{sec:fdsp-tpcelec-design} 
for information on all parts of the \dword{ce} system.

\subsection{Photon Detection System}
\label{sec:fdsp-apa-intfc-pds}

The \dword{pds} is integrated into the \dword{apa} frame to form a single unit for detecting both ionization charge and scintillation light.  The \dword{apa} frame design must also accommodate cables for the \dwords{pd}.  
Individual \dword{pd} units are inserted through \num{10} slots machined in the side steel tubes of the frame and supported by rails mounted in the \dword{apa}. Figures~\ref{fig:apa-frame-full} and \ref{fig:apa-frame-details} show examples of these features in the frame. Figure~\ref{fig:apa-pd} shows a \dword{pd} module being inserted into a slot in the frame and mating with an electrical connector mounted along the center tube in the \dword{apa}.

\begin{dunefigure}[APA interface with PDs in ProtoDUNE-SP]{fig:apa-pd}
{Top: A \dword{pd} module in \dword{pdsp} being inserted into a slot in the frame. Bottom: The \dword{pd} unit mating with an electrical connector mounted along the center tube in the \dword{apa}.

}
\includegraphics[height=0.4\textheight,trim=0mm 0mm 0mm 0mm,clip]{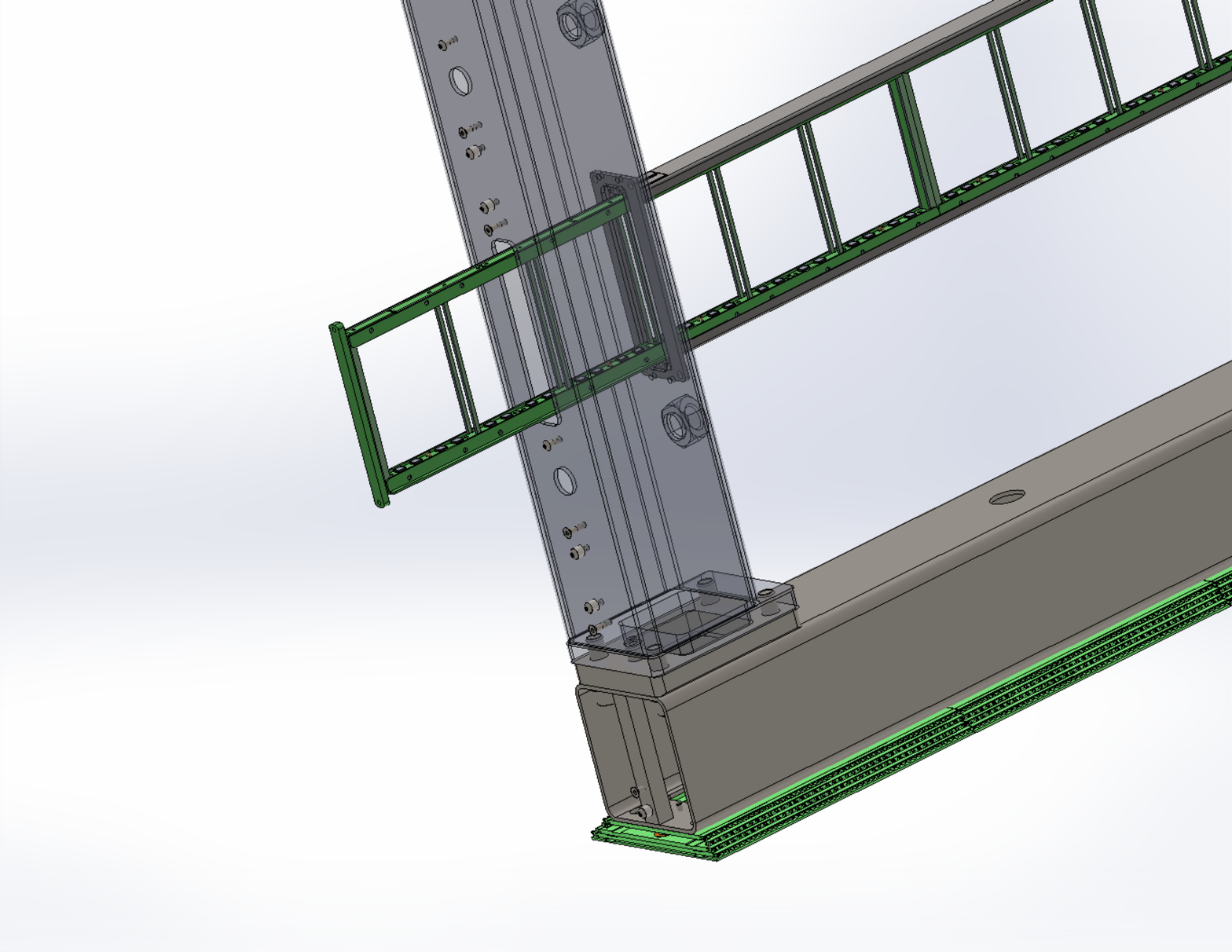}\\
\vspace{2mm}
\includegraphics[height=0.4\textheight,trim=0mm 0mm 0mm 0mm,clip]{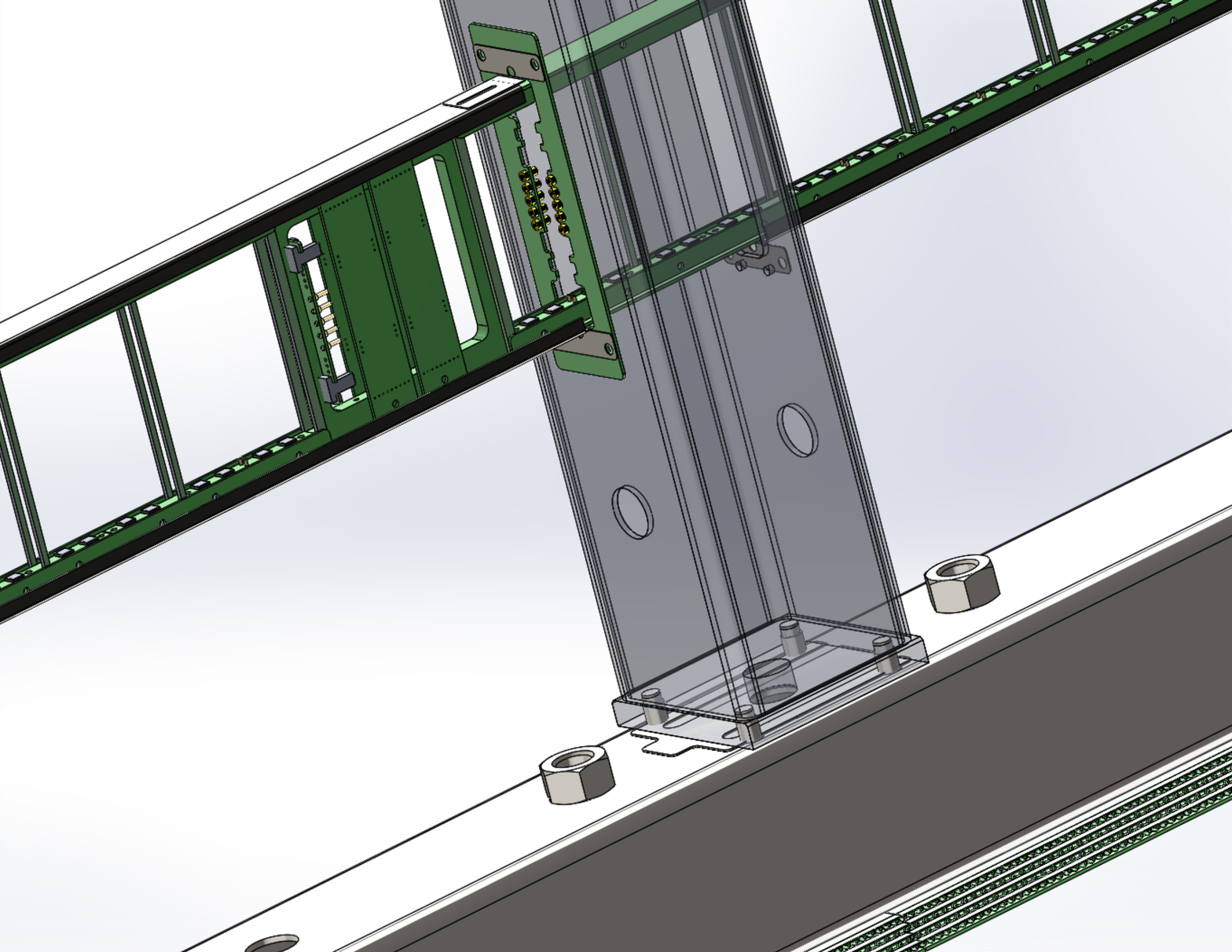}
\end{dunefigure}

The interface between the \dword{pds} and \dword{apa}s involves the mechanical hardware design and production, cable routing, and integration, installation, and testing procedures. The electrical interface includes a grounding scheme and electrical insulation. The strict requirements on noise from the \dword{ce} means the electrical interface must be defined together with the \dword{sp} \dword{tpc} electronics consortium. 

For more information on the \dword{pds}, see Chapter~\ref{ch:fdsp-pd} 

\subsection{Cable Routing}
\label{sec:fdsp-apa-intfc-cables}

Cable routing schemes for both the \dword{tpc} electronics and \dword{pds} must be integrated into the design of the \dword{apa}s.   The \dword{ce} signal and power cables must be routed so that the head end of the lower \dword{apa} in the two-\dword{apa} assembly can be reached. \dword{ce} cables, therefore, will be routed inside the two side beams of the \dword{apa} frames. Figure~\ref{fig:apa-cable-tube} depicts such a cable routing scheme.  The \dword{ce} cables at the lower end of the lower \dword{apa} are formed into two bundles, each about \SI{50}{mm} in diameter. Installation of the cables through the side tubes of the two stacked \dword{apa}s is done by pulling them through a large, smooth conduit placed inside each of the side tubes.  To fully accommodate the cables, the \dword{apa} frame hollow tube sections were enlarged relative to the \dword{pd} design from \SIrange{7.6}{10.2}{cm} (\SI{3}{in} to \SI{4}{in}) deep. Prototyping of this solution was carried out at PSL during summer 2018, as shown in the right photograph in Figure~\ref{fig:apa-cable-tube}.

\begin{dunefigure}[TPC electronics cable routing in the APAs]{fig:apa-cable-tube}
{Cable routing scheme. Left: Conduit for the \dword{ce} cables protruding from the end of the long side tubes of the \dword{apa} frame.  Right: One of the \dword{ce} cable bundles being pulled through conduit of equal length to the two stacked \dword{apa}s.  The cable bundle is wrapped with a protective cover of braided PET plastic.  }
\includegraphics[height=0.3\textheight,trim=5mm 0mm 0mm 7mm,clip]{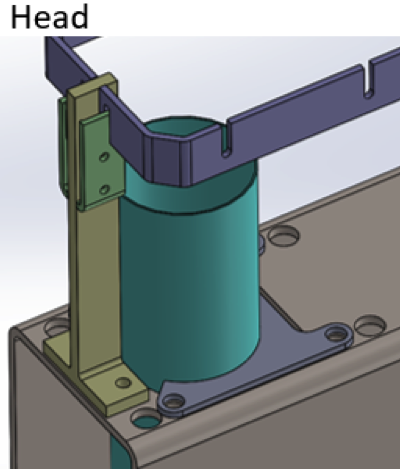} \quad
\includegraphics[height=0.3\textheight,trim=5mm 0mm 0mm 7mm,clip]{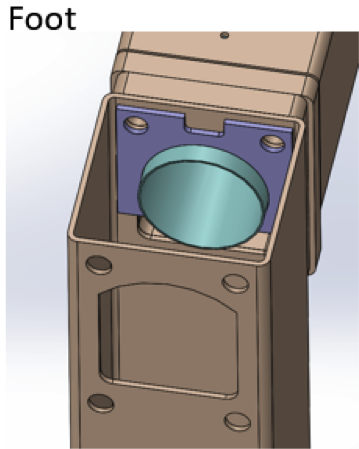} \quad
\includegraphics[height=0.3\textheight]{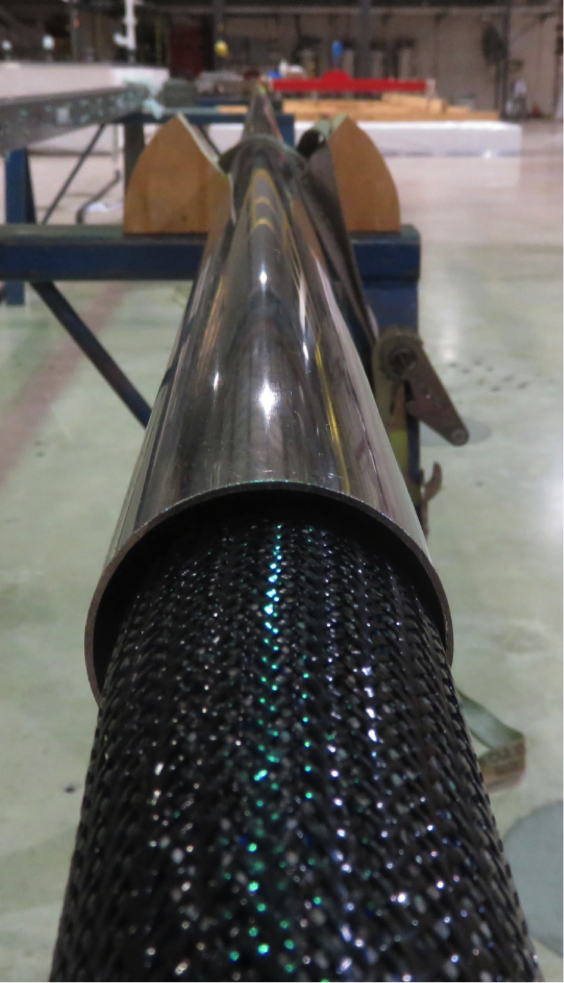}
\end{dunefigure}

The concept being developed for the cables of the  \dwords{pd} is depicted in Figure~\ref{fig:apa-pds-cables}.  The cables run along the outside of the central tube in the \dword{apa} frame, joining together into a bundle of five cables by the time they reach the top of the frame.  Cables from the bottom \dword{apa} in a stack are fed through the foot tubes to the upper \dword{apa} and ran along the outside tubes on either side.  In this way, all \dword{pd} cables make it to the head of the upper \dword{apa}.  

\begin{dunefigure}[APA-to-APA connection and cable routing]{fig:apa-pds-cables}
{A concept for \dword{pds} cable routing (shown horizontal). Top: The bottom \dword{apa}.  The \dwords{pd} are the ten transparent pieces spanning the frame -- two between each set of ribs.  They connect to their cables at the center tube.  The cables run up either side of the center tube (outside the tube) joining with others and forming two bundles of five cables by the time they reach the foot tube at the right end of this image. Middle: The top \dword{apa}. The two five-cable bundles from the lower \dword{apa} continue to the head tube of the upper \dword{apa} (at the right in this image) where they go through the head tube.  The cables from the \dwords{pd} in the upper \dword{apa} run up the outside of the center tube and form bundles which also go through the head tube. Bottom: Detail showing the five \dword{pds} cables gathered together at the foot tube (the top) of the bottom \dword{apa}.}
\includegraphics[width=0.78\textwidth]{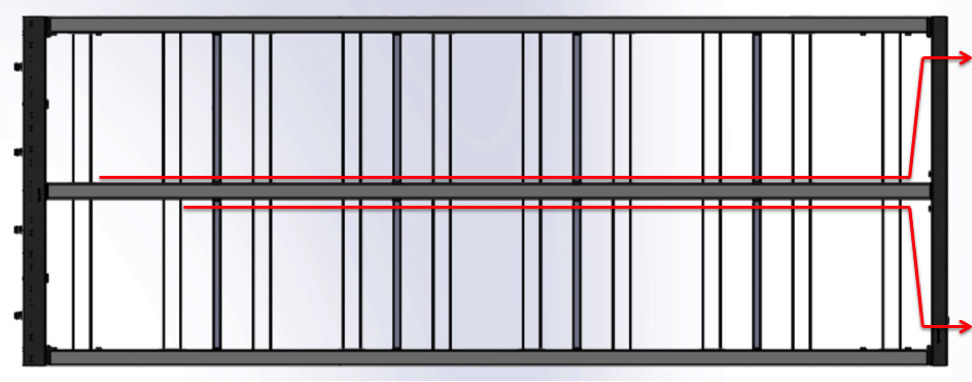}
\hspace*{4mm}\includegraphics[width=0.8\textwidth]{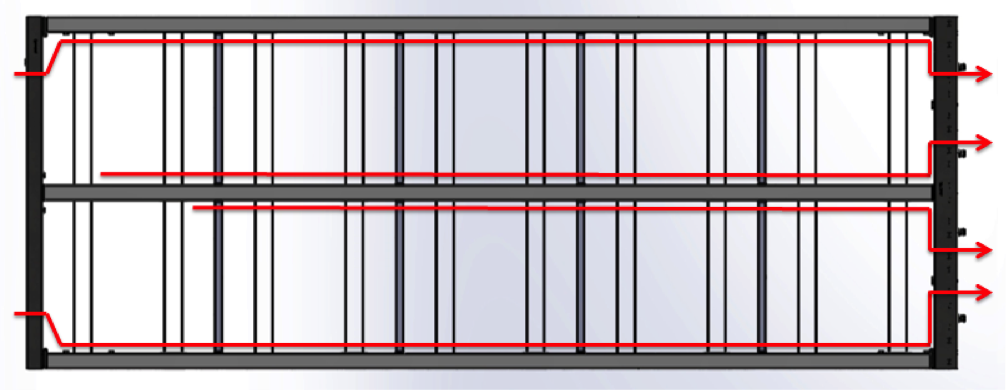}
\vspace{5mm} \\
\hspace*{-10mm}\includegraphics[width=0.9\textwidth]{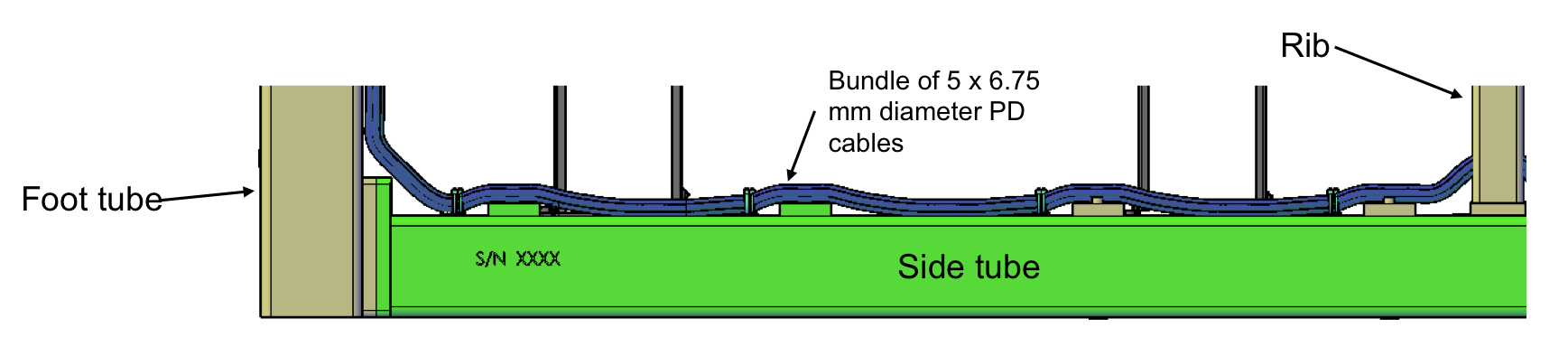}
\end{dunefigure}

\section{Production Plan}
\label{sec:fdsp-apa-prod}

The \dword{apa} consortium oversees the design, construction, and testing of the \dword{dune} \dword{spmod} \dword{apa}s. Production sites are being planned in the USA and UK. This approach allows the consortium to produce \dword{apa}s at the rate required to meet overall construction milestones and, at the same time, reduce risk to the project if any location has problems that slow the production pace.

The starting point for the \dword{apa} production plan for \dword{dune} \dwords{spmod} is the experience and lessons learned from \dword{pdsp} construction. For \dword{pdsp}, the \dword{apa}s have been constructed on single production lines set up at PSL in the USA and at Daresbury Laboratory in the UK.  The plan now is to construct \dword{apa}s for \dword{dune} at US and UK collaborating institutions with ten total production lines, four in the UK and six in the US.  

A production line is centered around a wire winding robot, or winder, that enables continuous wrapping of wire on a \SI{6}{m} long frame (see figures ~\ref{fig:winder} and \ref{fig:winder-photos}). 
Two process carts are needed to support the \dword{apa} during  board epoxy installation and \dword{qc} checks, among other construction processes. A means of lifting the \dword{apa} in and out of the winder is also required. A gantry-style crane was used for \dword{pdsp} construction.

The fabrication of an \dword{apa} is a  three-stage process requiring 
about \num{50} eight-hour shifts to complete, with a mix of engineering, technical, and scientific personnel.   The first stage, estimated at about four shifts, is a preparation stage in which  \dword{pds} cables and rails, wire mesh panels, comb bases, $X$-plane wire boards, and tension test boards are installed on the bare \dword{apa} frame. In the second and longest stage, lasting \num{38}--\num{40} shifts, the \dword{apa} occupies a winding machine.  All the wires are strung and attached in this stage, and tension and electrical tests of each channel are performed.
The third and final stage, requiring an estimated \num{8} shifts, is completed in a process cart and involves the installation of wire harnesses, G-bias boards, and cover boards. 
Protection panels are then installed over the wire planes and the completed \dword{apa} is transferred to a transport frame (see Section ~\ref{sec:fdsp-apa-transport}).   During \dword{pdsp} construction, we were able to complete an \dword{apa} in \num{64} shifts, on average. Several improvements to the process and tooling have been developed since then to reduce this to the maximum allowed \num{50} shifts. 

The approximately \num{40} shifts that an \dword{apa} spends in the winding machine combined with 
the total number of winders determines the overall pace of production since the pre- and post-winding stages can be done in parallel with winding.  
The overall production model assumes that the \dword{apa} production sites run one shift per day, that all winding machines are operated in parallel, and that two weeks per year are devoted to maintaining equipment.  The work plan at production sites further assumes a steady supply of the necessary hardware for \dword{apa} wiring, such as completed frames, grounding mesh panels, and wire boards.  Detailed planning is underway within the \dword{apa} consortium 
for collaborating institutions to help  source and test components and ensure their on-time delivery to production sites.        

Having several \dword{apa} production sites in two different countries presents quality assurance and quality control (\dshort{qa}/\dshort{qc}) challenges. 
A key requirement is that every \dword{apa} be the same, regardless of where it was constructed. To achieve this goal, we are building on \dword{pdsp} experience where six identical \dword{apa}s were built, four in the US and two in the UK. The same tooling, fabrication drawings, assembly, and test procedures were used at each location, and identical acceptance criteria were established at both sites.  This uniform approach to construction 
is necessary, and the \dword{apa} consortium is developing the necessary management structure to ensure that each production line follows the agreed-upon approach to achieve \dword{apa} performance requirements.

\subsection{Assembly Procedures and Tooling}
\label{sec:fdsp-apa-prod-tooling}

The central piece of equipment used in \dword{apa} production is the custom-designed wire winder machine, shown schematically in Figure~\ref{fig:winder} and in use in Figure~\ref{fig:winder-photos}.  An important centerpiece of the winder machine is the wiring head.  The head releases wire as motors move it up and down and across the frame, controlling the tension in the wire as it is laid.  The head then positions the wire at solder connection points for soldering by hand. The fully automated motion of the winder head is controlled by software, which is written in the widely used numerical control G programming language.  The winder also includes a built-in vision system to assist operators during winding, which is currently used at winding start-up to find a locator pin on the wire boards.  

\begin{dunefigure}[Winding machine schematic showing ongoing development]{fig:winder}
{Schematic of the custom-designed \dword{apa} wiring machine.  This shows the updated version with upper and lower supports and the spherical bearings for rotating the \dword{apa} on the winder.}
\includegraphics[width=0.95\textwidth]{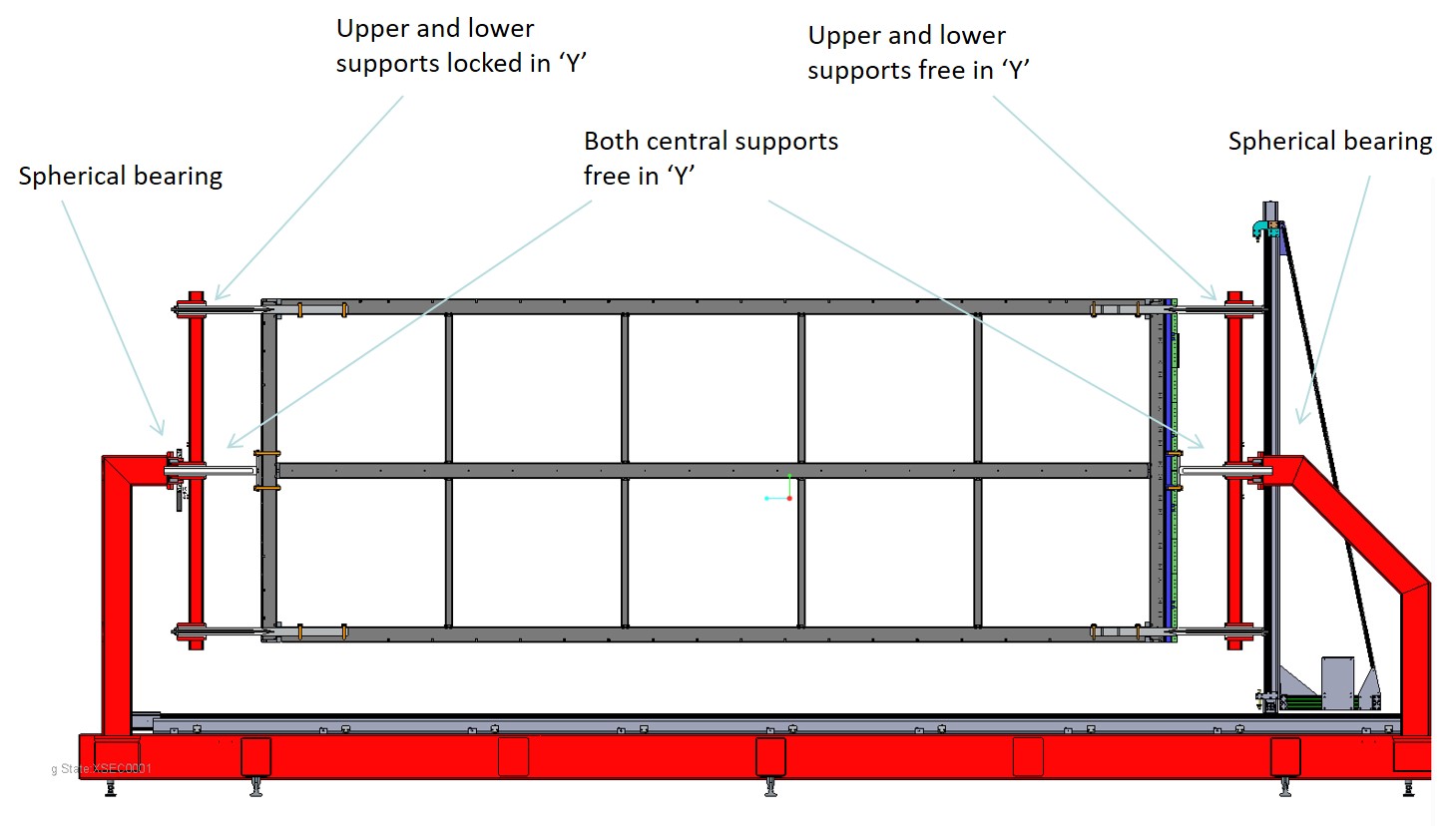} 
\end{dunefigure}

\begin{dunefigure}[APA wire winding machine]{fig:winder-photos}
{Left: Partly wired \dword{pdsp} \dword{apa} on the winding machine at Daresbury Lab, UK. Right: Partly wired \dword{pdsp} \dword{apa} on the winding machine during wire tension measurements at PSL.}
\includegraphics[height=0.3\textheight,trim=25mm 0mm 4mm 0mm,clip]{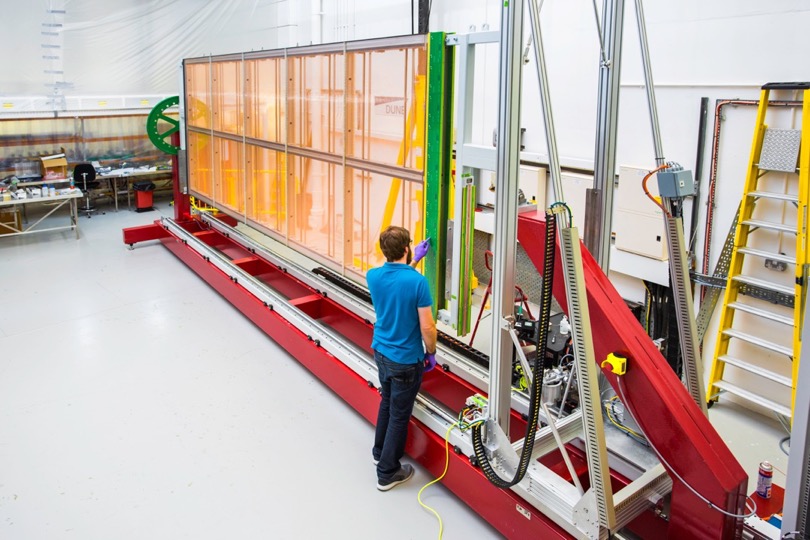}
\includegraphics[height=0.3\textheight,trim=200mm 0mm 30mm 0mm,clip]{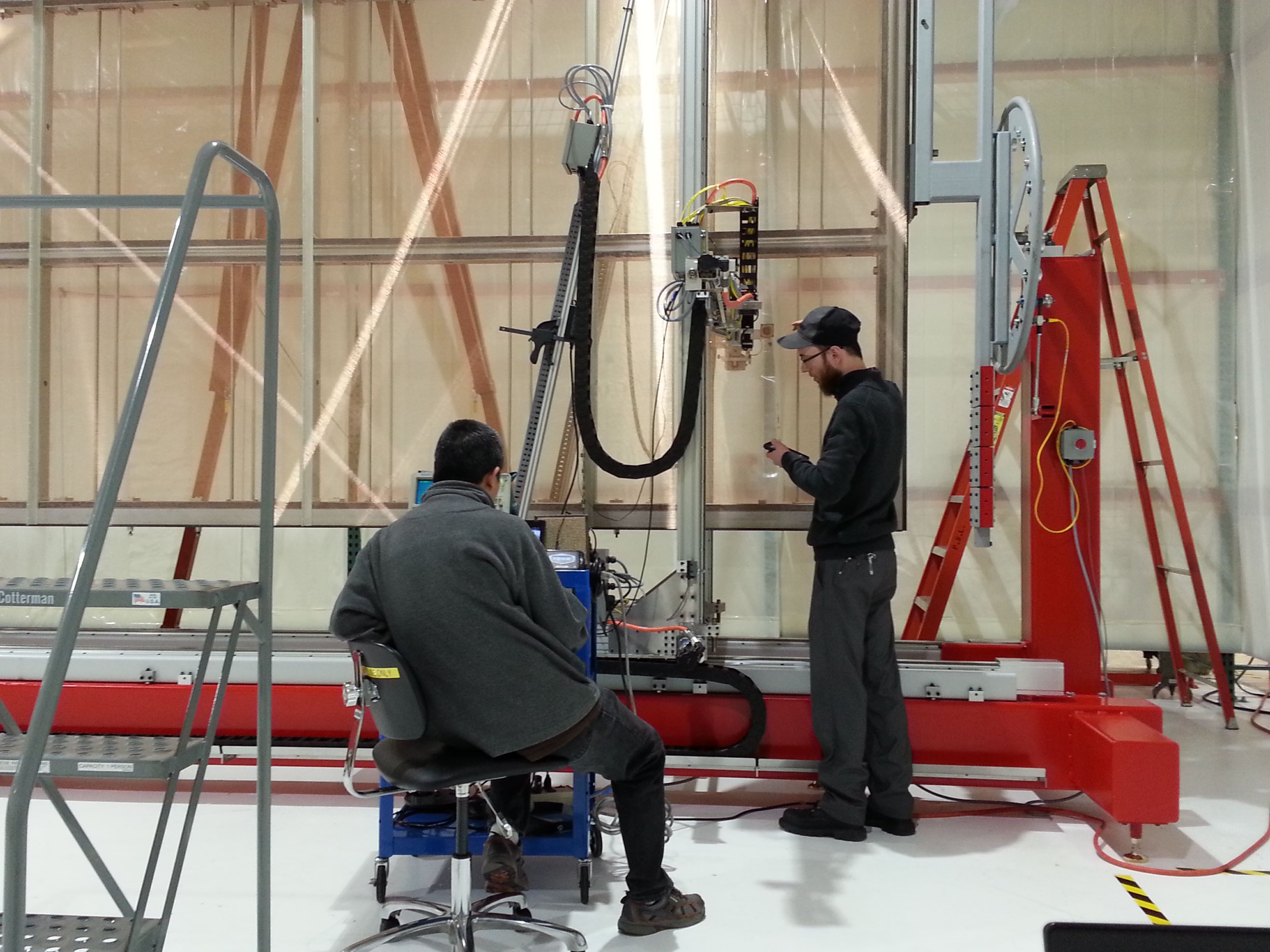}
\end{dunefigure}

In the scheme used for wiring the \dword{pdsp} \dword{apa}s, an \dword{apa} moved on and off the winder machine several times for wiring, soldering, and testing. 
 Several design changes were developed in 2018--2019 to enable the \dword{apa} to remain on the winding machine throughout the wiring process. The design concept allows the winder head to pass from one side to the other nearly continuously. The interface frames at either end have been replaced by retractable linear guided shafts. These can be withdrawn to allow the winding head to pass around the frame over the full height of the frame. These shafts have conical ends and are in shafts fixed to the internal frame tube to provide guides to location. This design change does not alter the design of the frame itself, but it does allow for rotation in the winding machine. It is now possible to carry out board installation, gluing, and soldering all while on the winding machine. This eliminates any transfer of the \dword{apa} to the process cart for the entire production operation, making it an inherently safer and faster production method.
 The upgraded design has been implemented on the winding machine at Daresbury, which has been used to build a new prototype, \dword{apa}~7 (Figure~\ref{fig:winder-upgrade-photos}). All winding, board installation, gluing, soldering and testing operations are being carried out in the winding machine. \dword{apa}~7 also incorporates the pre-built grounding mesh sub-frames, another new feature 
 that saves significant time in production.  

\begin{dunefigure}[The upgraded APA wire winding machine]{fig:winder-upgrade-photos}
{Left: Upgraded winding machine with new interface arm design being used to wire APA-07. Fitted mesh panels are also shown installed. Right: The V-layer soldering process at the head end of APA-07. Soldering can now be done with the \dword{apa} in the winding machine.}
\includegraphics[height=0.25\textheight,trim=10mm 0mm 0mm 0mm,clip]{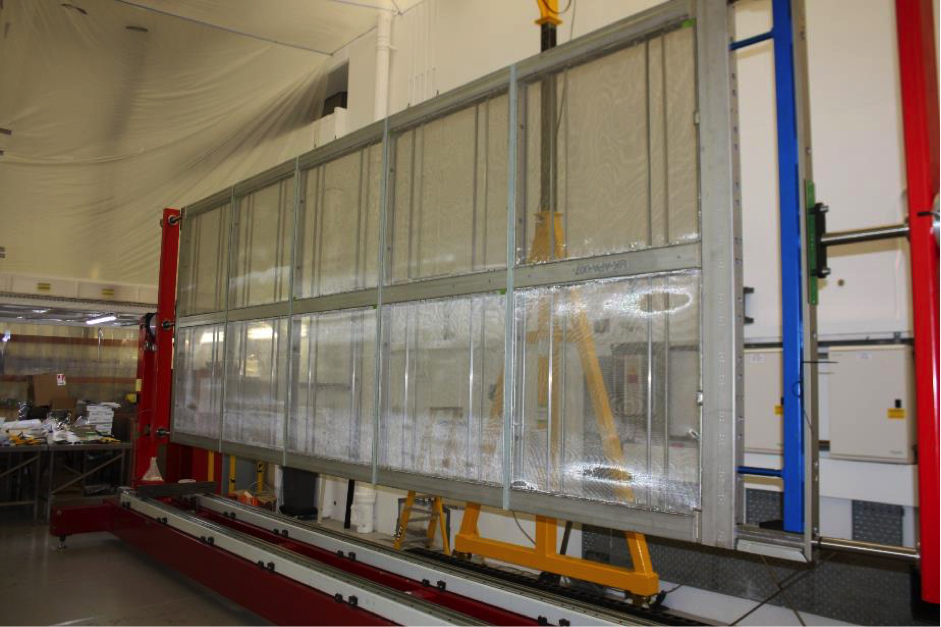} 
\includegraphics[height=0.25\textheight,trim=0mm 0mm 10mm 0mm,clip]{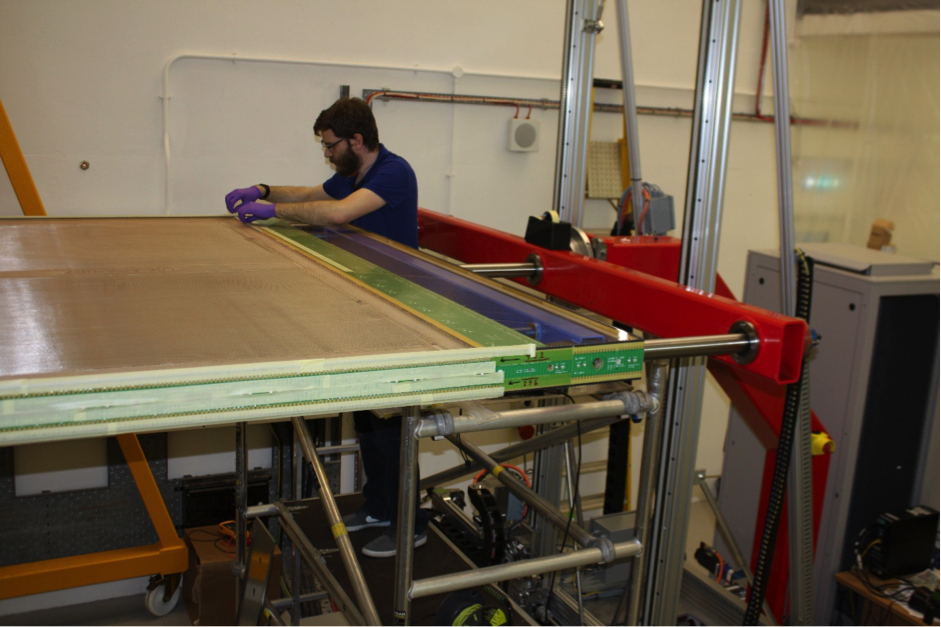}
\end{dunefigure}

The wiring head has also been updated. The upgraded design offers real-time tension feedback and control, which will save time in wiring and produce better tension uniformity across wires.  A prototype of the new head has been constructed and is undergoing extensive commissioning and qualification.   

An important element in the long-term use of the winders 
will be maintenance.  
During \dword{pdsp} construction winding machine problems traceable to a lack of routine maintenance occurred from time to time, shutting the production line down until repair or maintenance was performed. We will formulate a routine and preventive maintenance plan that minimizes winder downtime during \dword{apa} production for the \dword{spmod}.

The large process carts are important to the 
flow of activities during production 
(Figure~\ref{fig:apa-process-cart}). The process carts are used to hold \dword{apa}s during wiring preparations,  for \dword{qc} checks after wiring, and to safely move \dword{apa}s around within the assembly facility. Process carts are fitted with specialized 360$^\circ$ rotating casters that allow the cart, loaded with a fully assembled \dword{apa}, to maneuver corners while moving the large frames between preparation, assembly, and packing/shipping areas.

\begin{dunefigure}[APA on a process cart during construction]{fig:apa-process-cart}
{A \dword{pdsp} \dword{apa} being moved around the PSL production facility on a process cart.}
\includegraphics[height=0.35\textheight,trim=50mm 10mm 70mm 250mm,clip]{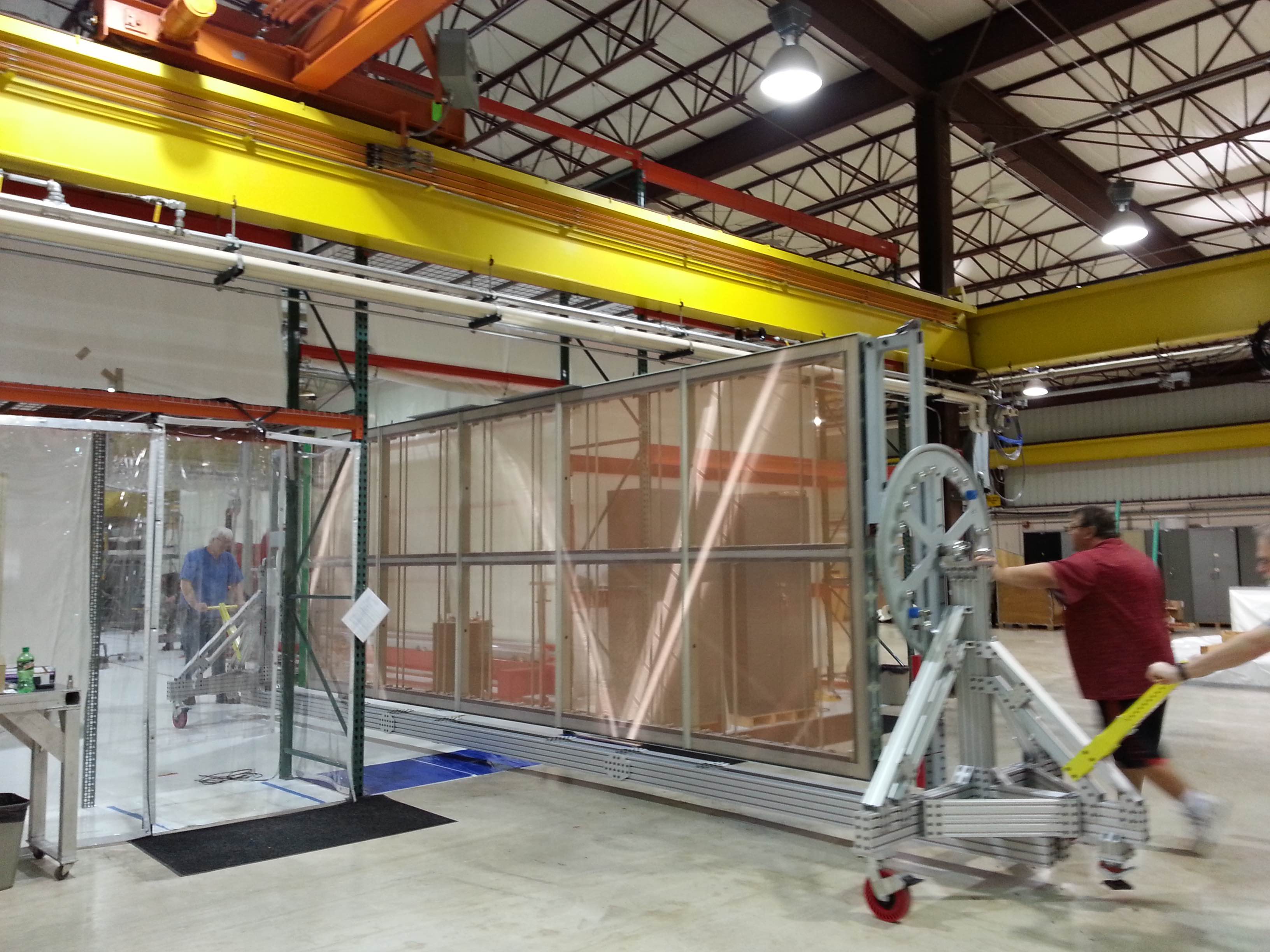}
\end{dunefigure}

\subsection{APA Production Sites}
\label{sec:fdsp-apa-prod-facility}

Multiple \dword{apa} production lines spread over several sites in the USA and the UK will provide some margin on the production schedule and provide backup in the event that technical problems occur at any particular site. 

The space requirements for each production line are driven by the size of the \dword{apa} frames and the winding robot used to build them. The approximate dimensions of a class \num{100000} clean space needed to house winder operations and associated tooling is \SI{175}{m$^2$}. The estimated requirement for inventory, work in progress, and completed \dword{apa}s is about \SI{600}{m$^2$}. Each facility  also needs temporary access to shipping and crating space of about \SI{200}{m$^2$}. Floor layouts at each institution are being developed, with current layouts shown in Figure~\ref{fig:factories}. Adequate space is available at each site, and the institutions have offered commitments for space for this purpose. 

At Daresbury Laboratory in the UK, the existing single production line used for \dword{pdsp} construction will be expanded to four.  The Inner Hall on the Daresbury site has been identified as an area that is large enough to be used for \dword{dune} \dword{apa} construction. It has good access and crane coverage throughout. Daresbury Laboratory management has agreed that the area is available, and a working environment that meets \dword{dune}'s safety standards is now being prepared, starting with clearing the current area of existing facilities, obsolete cranes, and ancillary equipment. The renovation of a plant room is also in progress, so that it can be used for storage and as a shipping area. The production area is designed to hold four winding machines and associated process equipment and tooling.  A production site design review of the Daresbury facility is planned for January 2020, and a production site readiness review is anticipated for June 2020, followed by the start of \dword{apa} production for \dword{dune} \dword{detmodule} \#1 in August 2020.

\begin{dunefigure}[Layouts of APA production sites]{fig:factories}
{Developing concepts for production site layouts at Daresbury Lab (top left), University of Chicago (top right), and Yale University (bottom left), and the existing APA production area at PSL (bottom right).}
\includegraphics[height=0.23\textheight]{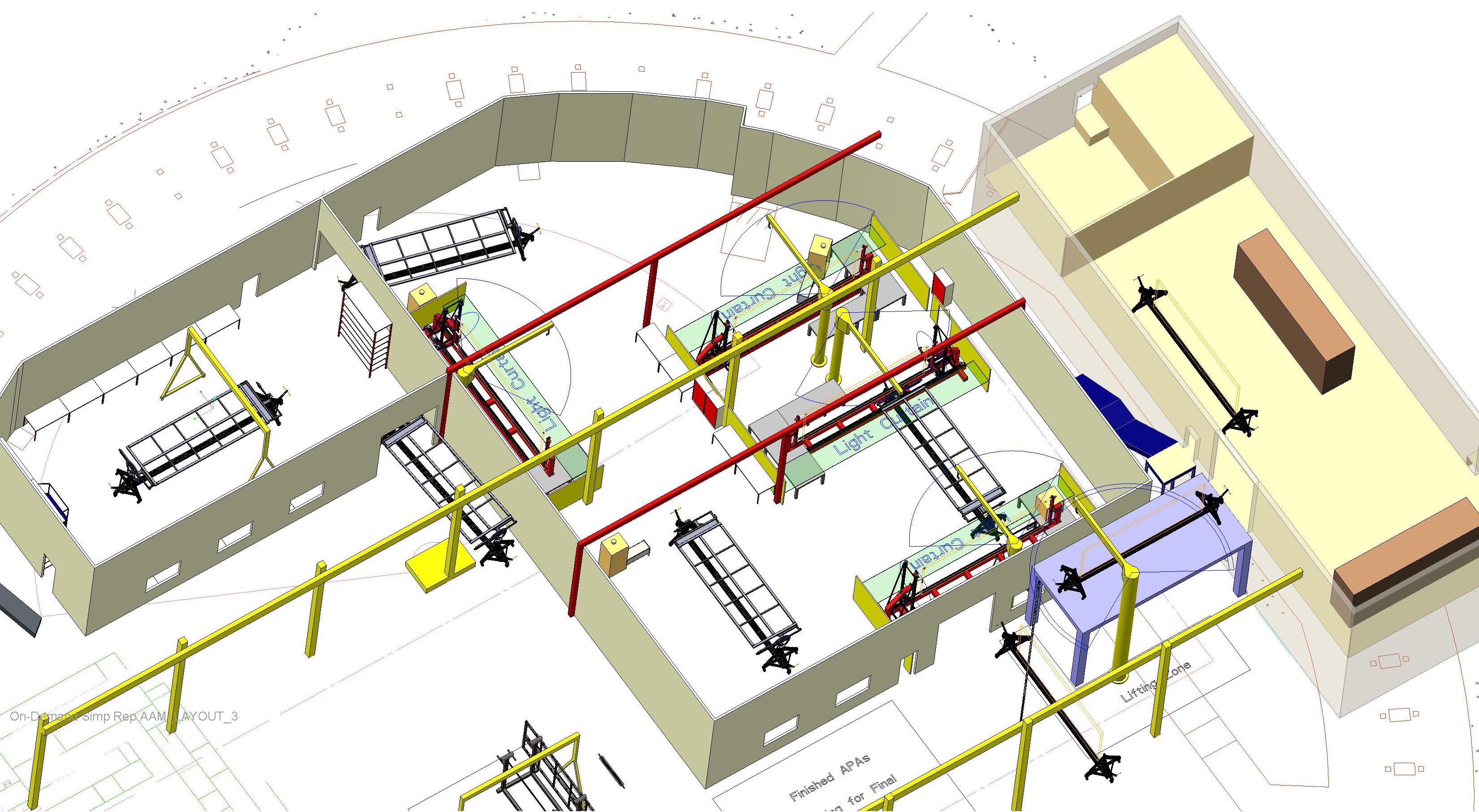} 
\includegraphics[height=0.23\textheight]{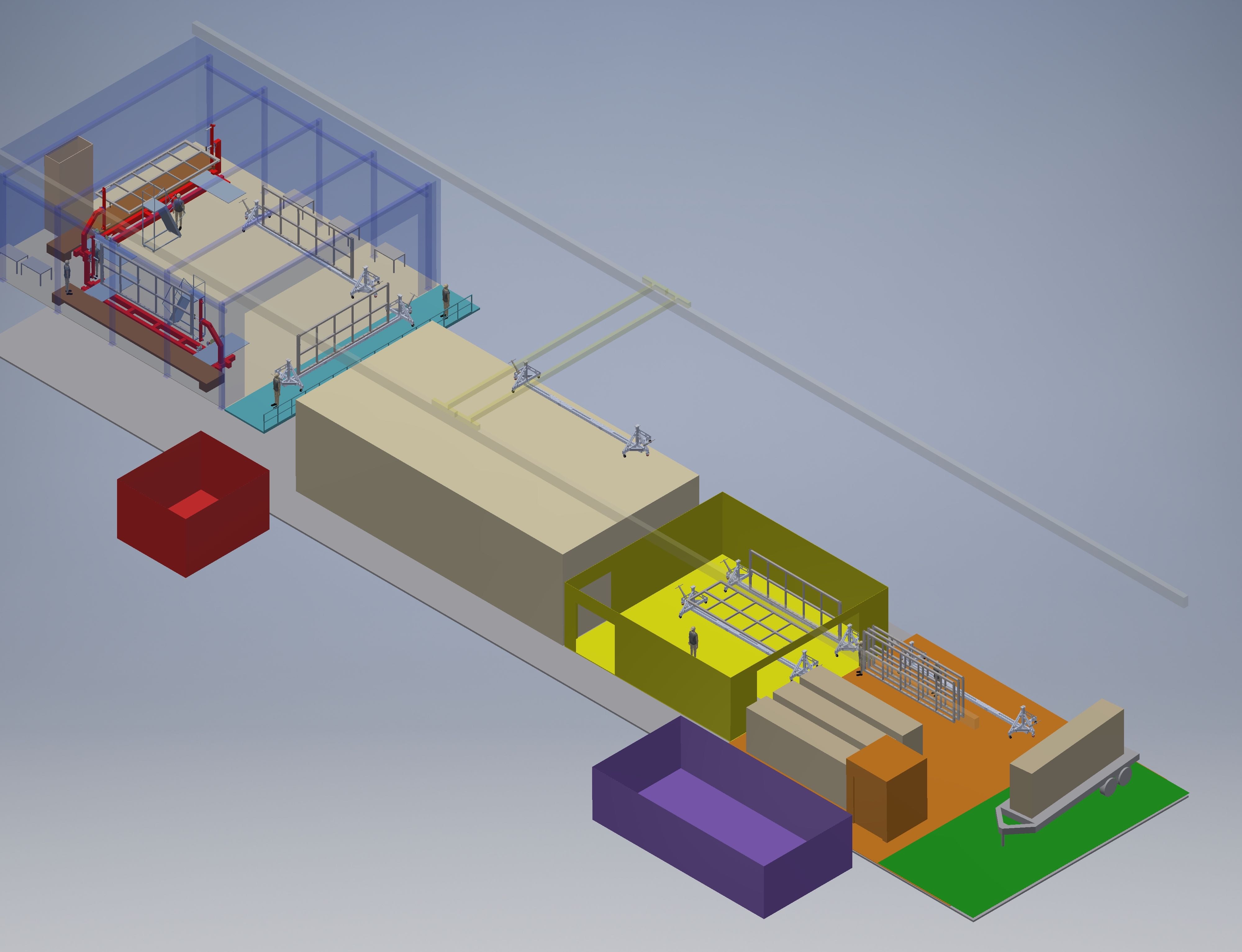} \\
\vspace{1mm}
\includegraphics[height=0.225\textheight]{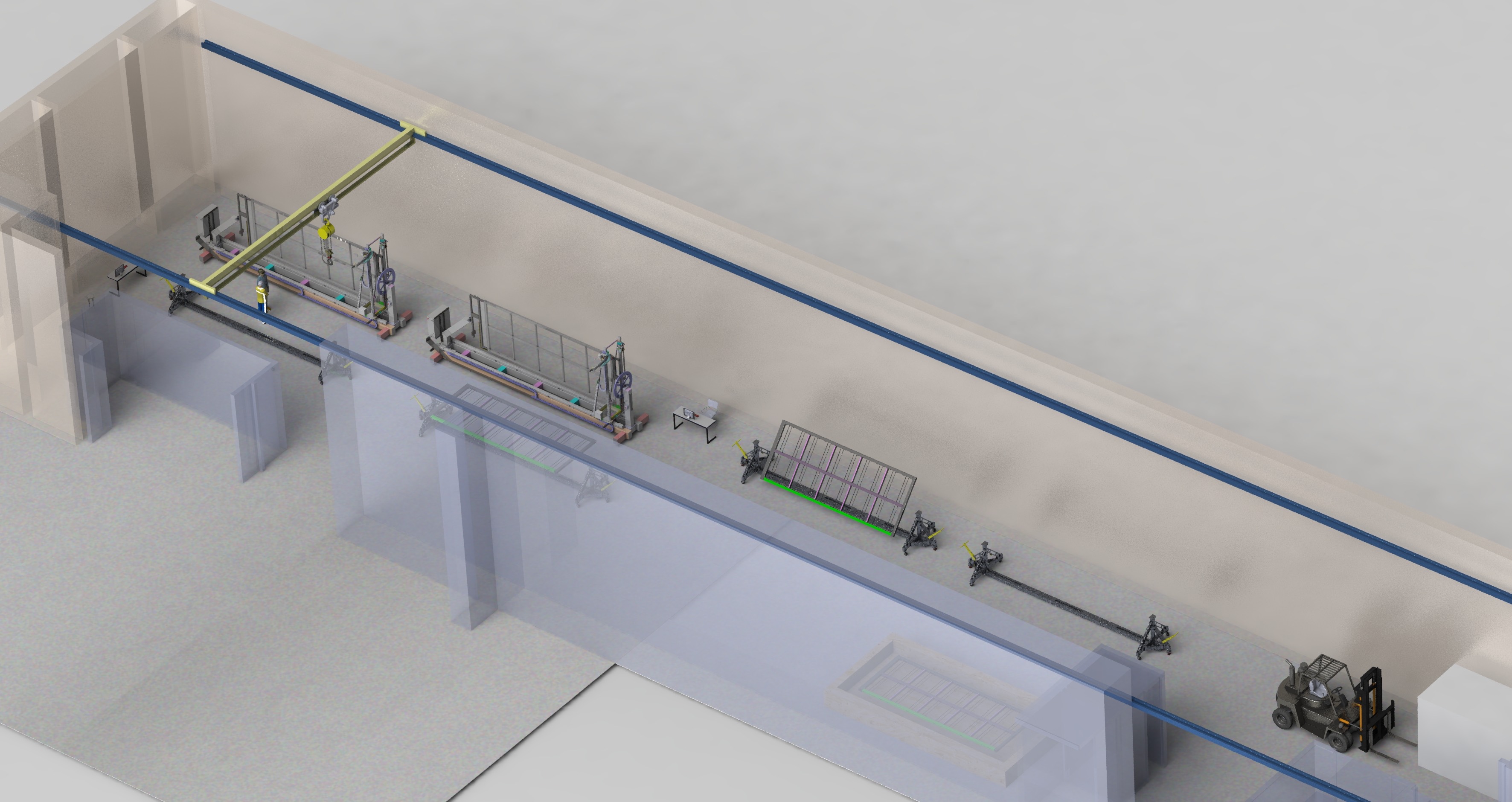}
\includegraphics[height=0.225\textheight]{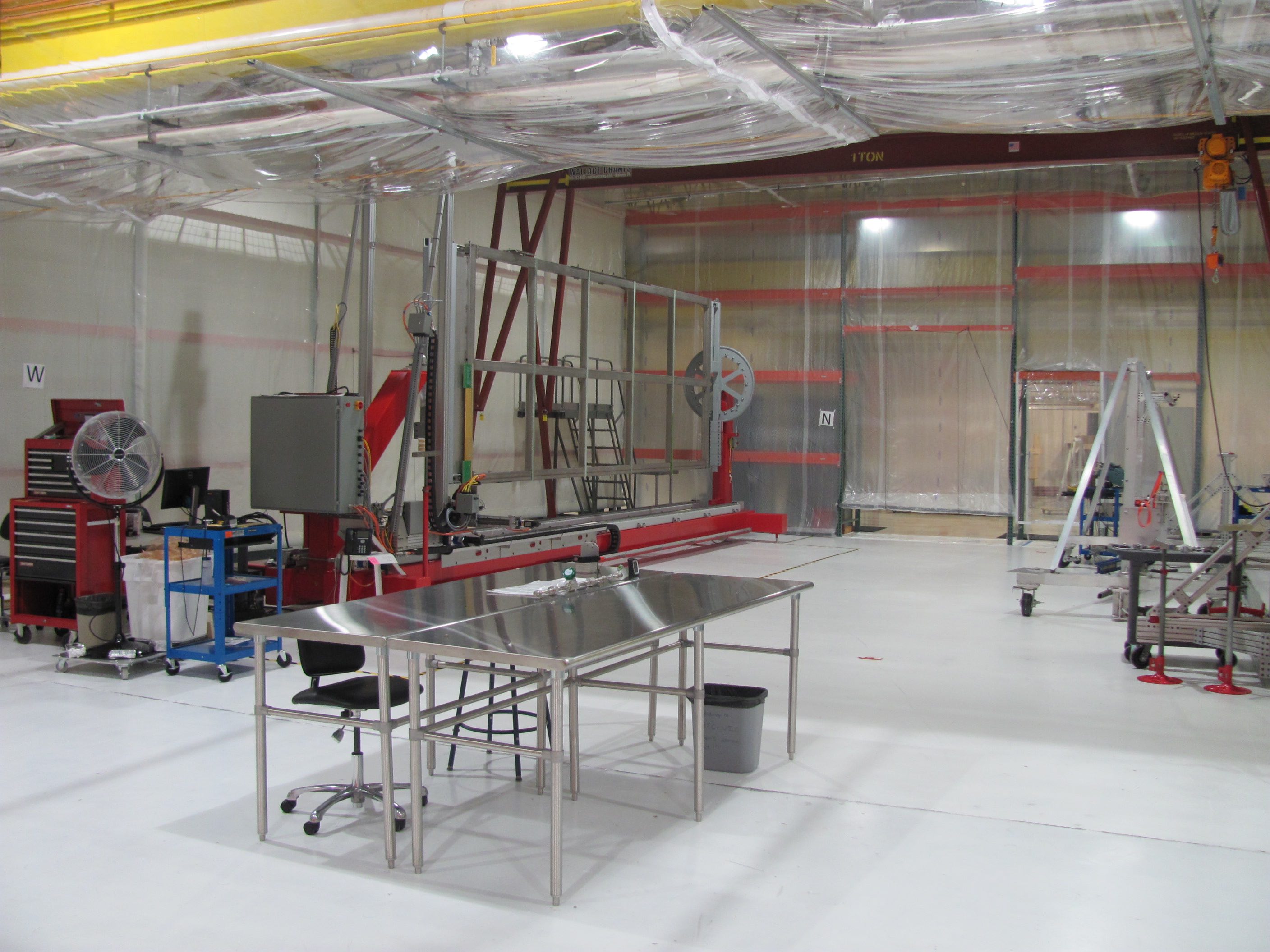} 
\end{dunefigure}

In the USA, there will be six total production lines at three sites: two at the University of Chicago, two at Yale University, and two at the University of Wisconsin's PSL, including the existing winder where the construction for \dword{pdsp} was carried out. 

The \dword{apa} production site at the University of Chicago will be housed in the Accelerator Building on the campus in Hyde Park.  The building has hosted the assembly of large apparatuses for numerous experiments over the course of its history and features an extensive high bay with an overhead crane, an indoor truck bay, clean laboratory spaces, a professional machine shop, and proximity to faculty and staff offices.  Winding will be done inside a clean room installed on the first floor-level mezzanine, where there is \SI{234}{m^2} of floor space above the machine shop.  A \SI{2}{ton} capacity bridge crane will be installed inside this clean room to move \dword{apa}s 
between the two winders and the process carts that will be located here.  \dword{apa}s will enter and exit the mezzanine by way of a loading deck external to the cleanroom.  Preparation of \dword{apa} frames, including mesh installation, will be done inside a second clean room on the basement level floor of the high bay.  Ample space, roughly \SI{170}{m^2}, between this clean room and the truck bay allows for simultaneous receiving of bare frames or other larger items, hoisting of \dword{apa}s to and from the mezzanine, and packaging of completed \dword{apa}s for outbound shipment.  When needed, additional off-site storage will be available for holding excess inventory and completed \dword{apa}s before they are transported to South Dakota.

Yale's Wright Laboratory will host another of the USA-based \dword{apa} production sites in a recently renewed area named ``The Vault'' where the nuclear accelerator operated previously.  The Vault is approximately \SI{720}{m^2} of total floor space and it satisfies all the safety and space requirements for an \dword{apa} production site. 
Indeed, the area, which is planned to be completely transformed into a cleanroom, can easily host two winders and four processing carts and has sufficient space for crating the 
\dword{apa}s for shipment and receiving and stocking all the material, e.g., bare frames, electronics boards, and mesh panels. A large high bay door at one end offers direct road access, allowing trucks to back inside the room where a \SI{10}{ton} crane operates all along the length.  Moreover, Wright laboratory has good support infrastructure, including cleanrooms and modern mechanical and prototyping workshops that are directly connected to the Vault. Faculty, researchers and postdoc offices are located upstairs in the same building. 

The Physical Sciences Laboratory (PSL) Rowe Technology Center has up to \SI{1850}{m^2} (\SI{20000}{ft^2}) total space available for continued \dword{dune} activities.  A clean work area that houses the existing winding machine used for \dword{pdsp} is already in place and will be used for \dword{dune} \dword{apa} construction. A second \dword{apa} production line using the updated winder design will assembled in 2020, and the existing winder will be upgraded.  PSL will host other major activities as well, including the assembly of bare \dword{apa} frames for wiring in the USA, production of \dword{cr} boards, and fabrication of \dword{apa} pair linkage and installation hardware.

Development work relevant for local planning at each site is rapidly advancing.  Figure~\ref{fig:factories} shows current conceptual layouts for the future production setups at Daresbury, Chicago, and Yale and a photograph of the existing \dword{apa} production facility at PSL-Wisconsin.  Production Site Design Reviews of the Chicago and Yale facilities are planned for early in 2020. 

\subsection{Material Supply}  
\label{sec:fdsp-apa-prod-supply}

Ensuring the reliable supply of raw materials and parts to each \dword{apa} production site is critical to keeping \dword{apa} production on schedule through the years of construction. Here the consortium institutions are pivotal in taking responsibility for delivery of \dword{apa} sub-elements. Supplier institutions will be responsible for sourcing, inspecting, cleaning, testing, \dword{qa}, and delivery of hardware to each production site.  In particular, the critical activities to supply production sites with the minimum needed \dword{apa} components for assembly include:

\begin{itemize}

\item {\bf Frame construction:} There will be separate sources of frames in the USA and the UK. The institutions responsible will rely on many lessons learned from \dword{pdsp}. The effort requires specialized resources and skills, including a large assembly area, certified welding capability, large-scale metrology tools and experience, and large-scale tooling and crane support. We are considering two approaches for sourcing: one is to outsource to an industrial supplier; the other is to procure all the major machined and welded components and then assemble and survey in-house. Material suppliers have been identified and used with good results on \dword{pdsp}.

\item {\bf Grounding mesh supply:} The modular grounding mesh frame design allows the mesh screens to be produced outside of the \dword{apa} production sites and supplied for \dword{apa} construction.  Suitable vendors to supply the needed units (20 mesh frames per \dword{apa}) will be identified in both the USA and UK.   

\item {\bf Wire wrapping board assembly:} Multiple consortium institutions will take on the responsibility of supplying the tens of thousands of wire-wrapping boards required for each \dword{sp} \dword{detmodule}. The side and foot boards with electrical traces are procured from suppliers and a separately bonded tooth strip is installed to provide wire placement support. 
The institutions responsible for boards will work with several vendors to reduce risk and ensure quality.

\item {\bf Wire procurement:} 
Approximately \SI{24}{km} of wire is required for 
each unit. During \dword{pdsp} construction, 
an excellent supplier 
worked with us to provide high-quality wire wound onto spools that we provide. These spools are then used directly on the winder head with no additional handling or re-spooling required. Wire samples from each spool are strength-tested before use.

\item {\bf Comb procurement:} Each institution will either work with our existing comb supplier or find other suppliers who can meet our requirements. The \dword{pdsp} supplier has been very reliable.

\end{itemize}

\subsection{Quality Control in APA Production}
\label{sec:fdsp-apa-prod-qc}

\Dword{qc} testing is a critical element of \dword{apa} production.  All \dword{qc} procedures are being developed by the consortium and will be implemented identically at all production sites in order to ensure a uniform quality product as well as uniform available data from all locations.  Important \dword{qc} checks are performed both at the level of components, before they can be used on an \dword{apa}, as well as on the completed \dword{apa}s, to ensure quality of the final product before leaving the production sites.  In addition, a 10\% sample of the completed \dword{apa}s produced at each of the production sites each year will be cold cycled in a cryogenic test facility available at PSL.

\subsubsection{APA Frame Acceptance Tests} 

Each \dword{apa} support frame must meet geometrical tolerances in order to produce a final \dword{apa} that meets requirements for physics. In particular, the wire plane-to-plane spacing must be within the specified tolerance of $\pm$\SI{0.5}{mm} (see Sec.~\ref{sec:fdsp-apa-design-overview}).  Flatness of the support frame, therefore, is a key feature and is defined as the minimum distance between two parallel planes that contain all the points on the surface of the \dword{apa}.  Although 
the frame could be distorted out of plane in several ways, the most likely causes are: 
(1)~a curve in the long side tubes causing the frame to bow out of plane, (2)~a twist in the frame from one end to the other, or (3)~a fold down the center-line (if the ends of the ribs are not adequately square).

As detailed in Section~\ref{sec:fdsp-apa-frames}, \dword{apa} frames are constructed of 13 separate rectangular hollow steel sections.  Before machining, a selection procedure is followed to choose the sections of the steel most suited to achieving the geometrical tolerances.  After assembly, a laser survey is performed on the bare frames before they can be delivered to an \dword{apa} production site. Three sets of data are compiled into a map that shows the amount of bow, twist, and fold in the frame. A visual file is also created for each \dword{apa} from measured data. 

A study was performed to determine the tolerances on the three distortions characterized above and is documented in~\cite{bib:docdb1300}.  It was determined that a \SI{0.5}{mm} change in the final wire plane spacing could result from:
\begin{enumerate}
\item An \SI{11}{mm} out-of-flatness caused by curved long side tubes.
\item A \SI{6}{mm} out-of-flatness due to a twist in the frame.  This is assumed to be evenly distributed between each of the 5 cells of the \dword{apa} with $\sim$\SI{1.2}{mm} out-of-flatness per cell.
\item A \SI{1.2}{mm} out-of-flatness due to a fold down the middle of the \dword{apa}.
\end{enumerate}

The bow, twist, and fold extracted from the survey data will be compared against these allowable amounts before the support frame is used to build an \dword{apa}.  Later, during \dword{apa} wiring at the production sites, a final frame survey will be completed after all electrical components have been installed, and the as-built plane-to-plane separations will be measured to verify that the distance between adjacent wire planes meets the tolerances.  

Another check performed at the \dword{apa} production site before the frame is transferred to a winder will confirm sufficient electrical contact between the mesh sub-panels and the \dword{apa} support frame.  A resistance measurement is taken immediately after mesh panel installation for all \num{20} panels before wiring begins.

\subsubsection{Material Supply Inspections}

All components require inspection and \dword{qc} checks before use on an \dword{apa}.  Most of these tests will be performed at locations other than the \dword{apa} production sites by institutions within the consortium before the hardware is shipped for use in \dword{apa} construction. This distributed model for component production and \dword{qc} is key to enabling the efficient assembly of \dword{apa}s at the production sites.   The critical path components are the support frames (one per \dword{apa}), grounding mesh panels (20 per \dword{apa}), and wire carrier boards (204 per \dword{apa}). Section~\ref{sec:fdsp-apa-org} provides details about which consortium institutions in the US and UK will be responsible for each of these work packages.  

\subsubsection{Wire Tension Measurements and Channel Continuity and Isolation Checks}
\label{sec:fdsp-apa-prod-qa-tension}

The tension of every wire will be measured during production to ensure wires have a low probability of breaking or moving excessively in the detector.  Every channel on the completed \dword{apa}s will also be tested for continuity across the \dword{apa} and isolation from other channels.  The plan is to perform all tests at once, using the methods described in this section.  As will be described in Section~\ref{sec:fdsp-apa-prod-coldtest}, it is also planned that 10$\%$ of the \dword{apa}s will be shipped to PSL for a cold test, where the full \dword{apa} will be brought to \dword{ln} temperature. Following the cold test, the wire tensions and continuity will be remeasured.  Finally, for this 10$\%$ sample of \dword{apa}s, a measurement of the wire plane spacing will be performed using a Faro arm that can precisely record the position of each wire plane in space. This checks that the \dword{qc} on the flatness of the support frames remains sufficient.  

Wire tensions will be measured after each new plane of wires is installed on an \dword{apa}. The optimal target tension has been set at \SI{6}{N} based on \dword{pdsp} experience.  \Dword{pdsp} data, where the tensions did have substantial variation, is also being used to study the effects of varying tensions and finalize the allowed range of values.

The technique  
used to measure tensions for the \dword{apa}s of \dword{pdsp} was based on a laser and a photodiode system~\cite{Acciarri:2016ugk}. In this method, the laser shines on an individual wire and its reflection is captured by the photodiode. An oscillation is produced in the measured voltage when a vibration is induced on the wire, such as by manually plucking it. This oscillation is dominated by the fundamental mode of the wire, which is set by the wire's tension. Since the length and density of the wire are known, the measured fundamental frequency can be converted into a tension value. The method works very well, but due to the necessity of aligning the laser and exciting and measuring wires individually, this technique can take tens of seconds per wire. Given the large number of wires per \dword{detmodule}, development of a faster technique represents a major opportunity for the full \dword{dune} construction.

A technique that can reduce the overall time required to measure the tension of every wire is currently being developed~\cite{Garcia-Gamez:2018frz}. In this method, DC and AC voltages are applied on the neighboring wires of a wire under test. A sine wave of the same frequency as that of the applied AC voltage is measured from the tested wire, since it is capacitively coupled to its neighbors. The amplitude of the sine wave exhibits a resonant behavior when the frequency of the AC voltage corresponds to the fundamental frequency of the wire. Thus, a frequency sweep of the AC voltage can be performed to determine at which frequency there is a resonance, from which the wire tension can be obtained. As electrical signals can be injected and measured in several wires simultaneously, this technique has the potential of measuring the tension of many wires at once.

A wire tension measurement device based on the electrical method is being developed within the context of the \dword{dune} \dword{apa}s. While the underlying principle of the electrical method has been demonstrated, its technical implementation requires consideration. The wire pitch of the \dword{apa}s requires summed input voltages on the order of \SI{500}{V} to reasonably discern resonances against noise. The head boards, cf.\ Section~\ref{sec:fdsp-apa-headboards}, have been designed to withstand temporarily such large differential voltages across neighboring channels. Additionally, the components of the \dword{cr} boards or of the \dword{ce} would interfere with the method and need to be absent.

The exact specifications of the measurement system are being finalized. It is planned to connect to one of the twenty head board stacks at a time. Within a given stack, the device is projected to inject and read out signals by groups of sixteen wires simultaneously. The device could be used to measure the tension of any wire layer at any stage of the production process, in particular after the winding of a wire layer or after all the wires are wound. The designs of the winder machine and of the \dword{apa} protection panels have clearance provisions for the usage of such a measurement device.

The measurement system design is a combination of a commercial \dword{fpga} board and a custom printed circuit board for analog signal processing. An \dword{fpga} board is used as it can produce a square wave at any frequency that is expected to be encountered while measuring a wire's fundamental frequency, i.e.,\ below \SI{5}{kHz}. In addition, the \dword{fpga} board can be used for digital signal processing of the readout signal. The analog circuitry would act as a bridge between the \dword{fpga} and the \dword{apa} wires. It is needed to filter the square wave into a sine wave, to amplify that sine wave before sending it to the wires and to digitize the readout signal before sending it to the \dword{fpga}. The analog board is also needed to provide electrical connections to the head boards. With such a design, it is expected that the concurrent tension measurement of eight wires would take on the order of ten seconds.

A prototype of the measurement device has been built. The main difference between the prototype and the planned design is that the former is restricted to three wires instead of sixteen: a single readout wire and two stimulus wires. The prototype has been employed on a test bench in which wires with the same physical properties as those that will be used in the \dword{apa}s have been wound. The wires were wound according to these wire parameters, which are similar to those of the \dword{apa}s: \SI{6}{N} tension, \SI{6}{m} length, \SI{4.7}{mm} pitch. The applied voltages were \SI{400}{V} DC and \SI{26}{V} for the AC peak amplitude. The results obtained are shown in Figure~\ref{fig:electrical-tension}. The expected resonant frequency is \SI{16.1}{Hz}. The observed resonant frequency is obtained from the raw data by offline data analysis using numerical algorithms that can be implemented directly in the \dword{fpga}. The value obtained is \SI{16.0}{Hz}, corresponding to a tension value of \SI{5.9}{N}, which is within a few percent of the physical value.

\begin{dunefigure}[Observed resonant frequency in electrical wire tension method]{fig:electrical-tension}
{Amplitude of the readout signal as a function of the stimulus frequency, as used in the electrical wire-tension method. The vertical line is located at the observed resonant frequency. The raw digitized signal values corresponding to the first data point of the main plot are shown in the inset plot.}
\includegraphics[height=0.33\textheight]{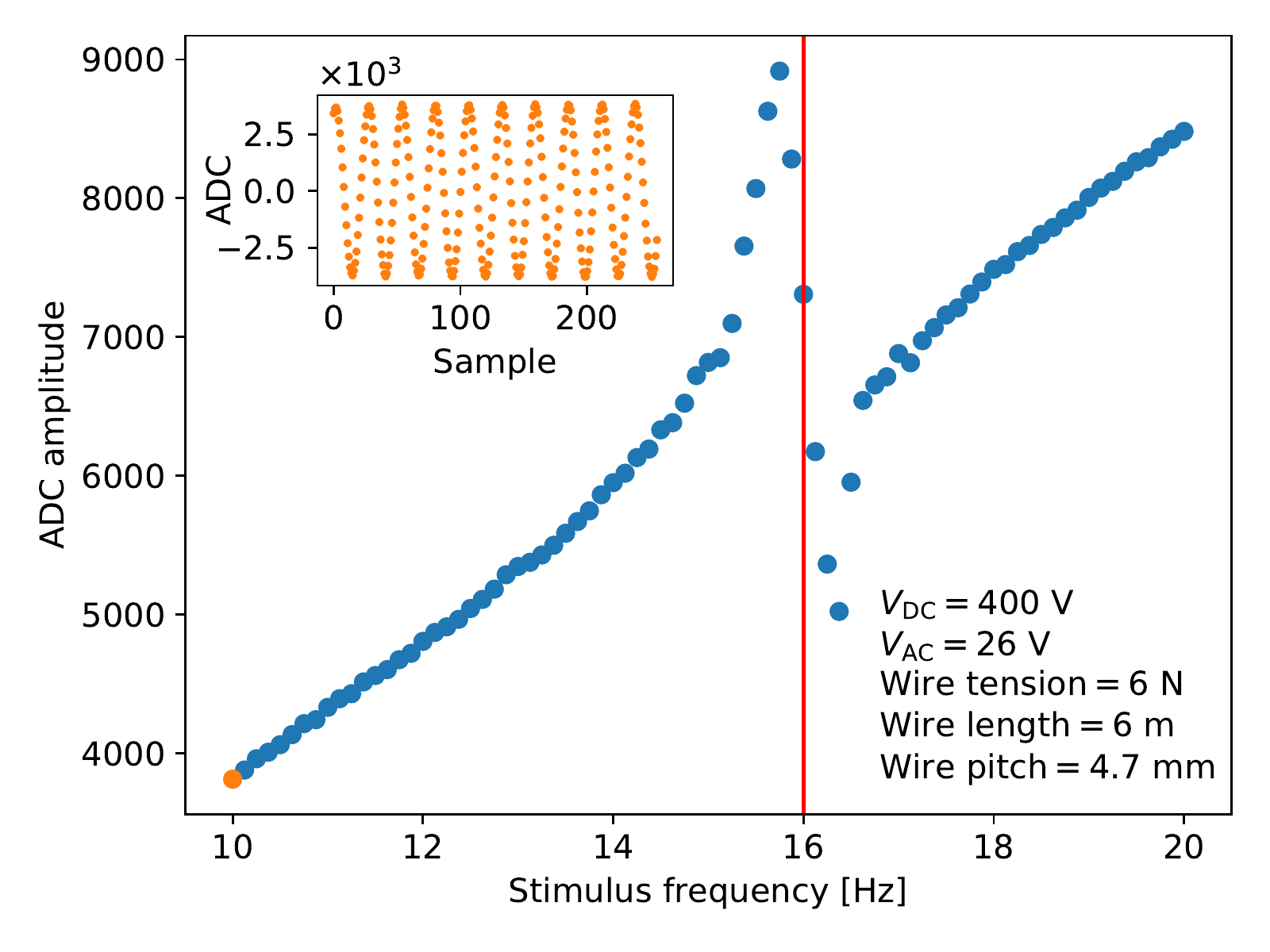}
\end{dunefigure}

In this test bench setup, no wire support combs, cf.\ Section~\ref{sec:combs}, are present. Their presence shortens the wire length that needs to be considered in this method, resulting in several higher resonant frequencies per wire. A similar effect happens for the wire channels that wrap around the \dword{apa} frame. Although they are a succession of wire segments electrically connected, the segments are mechanically independent and can have different tension values. Several resonant frequencies can be present per readout channel, possibly corresponding to different tension values.

In addition to measuring tension values, the measurement device is envisioned to be able to test wires for electrical isolation and electrical continuity, given the flexibility of the \dword{fpga} and provisions put in place in the design of the analog circuitry. Injecting a signal in a readout channel and detecting it in a different channel would indicate that these channels are not electrically isolated, for example, due to a solder bridge. The electrical continuity could be tested by sending a pulse down a channel and measuring the time it takes to travel through the wire and back to the measurement device.  If the measured time is shorter than expected, this could indicate cold solder joints, for example.

A final review of the electrical tension measuring system design will take place in spring 2020. Once completed, mobile \dword{apa} test stands will be built for each of the \dword{apa} production sites, the \dword{sdwf}, and \dword{surf}.  The introduction of the electrical testing methods for \dword{apa}s presents a fantastic opportunity for more efficient \dword{apa} fabrication and more flexible testing during the integration and installation phases.

\subsubsection{Cold Testing of APAs}
\label{sec:fdsp-apa-prod-coldtest}

The six \dword{apa}s produced for \dword{pdsp} have demonstrated clearly that the \dword{apa} design, materials, and fabrication methods are sufficiently robust to operate at \dword{lar} temperature.  No damage or change in performance due to cold have been identified during \dword{pdsp} running.  Nevertheless, over a five year construction effort, it is prudent to cold cycle a sample of the \dword{apa}s produced to ensure steady fabrication quality.  A cold testing facility sized for \dword{dune} \dword{apa}s exists at PSL and can be used for such tests. Throughout the construction project, it is anticipated that 10\% of the produced \dword{apa}s will be shipped to PSL for cold cycling.  This amounts to about 1 APA per year per production site during the project.  It is planned that all APAs will still be cold tested during integration at SURF and before installation in the DUNE cryostats.

\subsubsection{Documentation} 
\label{sec:fdsp-apa-prod-doc}

Each \dword{apa} is delivered with a traveler document in which specific assembly information is gathered, initially by hand on a paper copy, then entered into an electronic version for longer term storage.  The traveler database contains a detailed log of the production of each \dword{apa}, including where and when the \dword{apa} was built and the origin of all parts used in its construction. 

Assembly issues that arise during the construction of an \dword{apa} are gathered in an issue log for each \dword{apa}, and separate short reports provide details of what caused the occurrence, how the issue was immediately resolved, and what measures should be taken in the future to ensure the specific issue has a reduced risk of occurring.  

\section{Handling and Transport to SURF} 
\label{sec:fdsp-apa-transport}

Completed \dword{apa}s are shipped from the \dword{apa} production sites to 
the \dword{sdwf} in South Dakota. As they are transported to the \dword{4850l}, they are integrated 
with the \dword{tpc} \dword{fe} electronics and \dwords{pd} followed by installation in the cryostat. 
Extensive \dword{qc} testing will be performed before installation to ensure the fully integrated \dword{apa}s function properly.  
Installation activities at \dword{surf} are described in Chapter~\ref{ch:sp-install}. 

\subsection{APA Handling}
\label{sec:fdsp-apa-transport-handling}

The handling of the \dword{apa}s  must 
ensure their safety.  Several lifting and handling fixtures will be employed for transferring and manipulating the \dword{apa}s during fabrication, integration, and installation.  At the production sites a fixture called the edge lift kit will be used to transfer the \dword{apa} to and from the process cart and the winder, as well as to the transport containers.  The lift kit is shown schematically in Figure~\ref{fig:apa-edge-lift}.  It is essential that the fixture connect to the \dword{apa} along an outer edge because after wires are attached to the support frame, it can no longer be grabbed anywhere on the front or back face of the frame. 

\begin{dunefigure}[APA edge lifting fixture]{fig:apa-edge-lift}
{A custom lifting fixture is used to pick up an \dword{apa} from the long edge and safely handle it during the various construction steps at the production sites.}  
\includegraphics[width=0.8\textwidth]{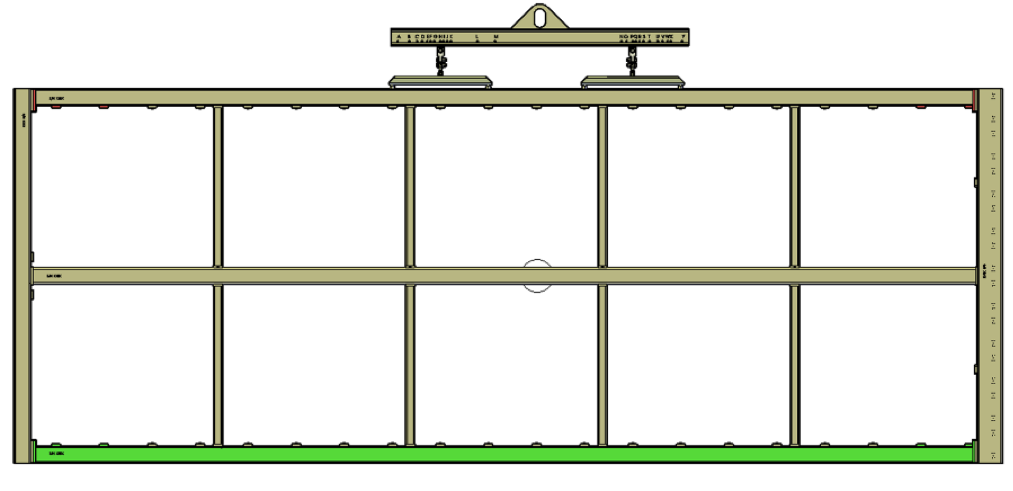} 
\end{dunefigure}

\subsection{APA Transport Frame and Shipping Strategy}
\label{sec:fdsp-apa-transport-container}

The transport packaging for the \dword{apa}s is designed to safely transport them from the production sites to the \dword{sdwf}. 
The design of the packaging is shown in Figure~\ref{fig:apa-transport-frame}. Light rigid metalized foam protective panels are attached via clamps affixed to the \dword{apa} frames and provide the primary protection for the wire planes. Pairs of \dword{apa}s (one upper and one lower in an \dword{apa} pair) are loaded onto welded structural steel transport frames at the factory. The \dword{apa} frames are bolted to mounts on the transport frames that incorporate shock-attenuating coil springs designed to reduce possible accelerations on the \dword{apa} frames to less than $4g$. The \dword{apa}s and transport frames will be instrumented with accelerometers to 
find out if the \dword{apa}s were subject to shocks above their specifications. Removable side frames, made from aluminum, are then bolted to the transport frames providing a structure around the \dword{apa}s, and this whole structure is then 
sealed in plastic sheeting. 

The packaged transport frames from the US sites will be covered in wooden panels, loaded on custom pallets, and shipped via truck from the \dword{apa} factories to the \dword{sdwf}. 
The packaged transport frames from the UK will be packed, in pairs, inside wooden crates for shipping. They then will be trucked to the nearby port in Liverpool, transported by ship to the port of Baltimore, and then shipped by truck to the \dword{sdwf}. 
\dword{apa}s may be stored for three years or longer at the \dword{sdwf} -- an \dword{apa} crate cannot arrive at \dword{surf} until it is required underground. 

\begin{dunefigure}[APA transport frame]{fig:apa-transport-frame}
{The current design of the APA shipping frame (maroon) and removable side frames (green) with two APAs covered with protective panels (shown in grey and tan). The external wooden packaging is not shown in this view.}  
\includegraphics[width=0.9\textwidth]{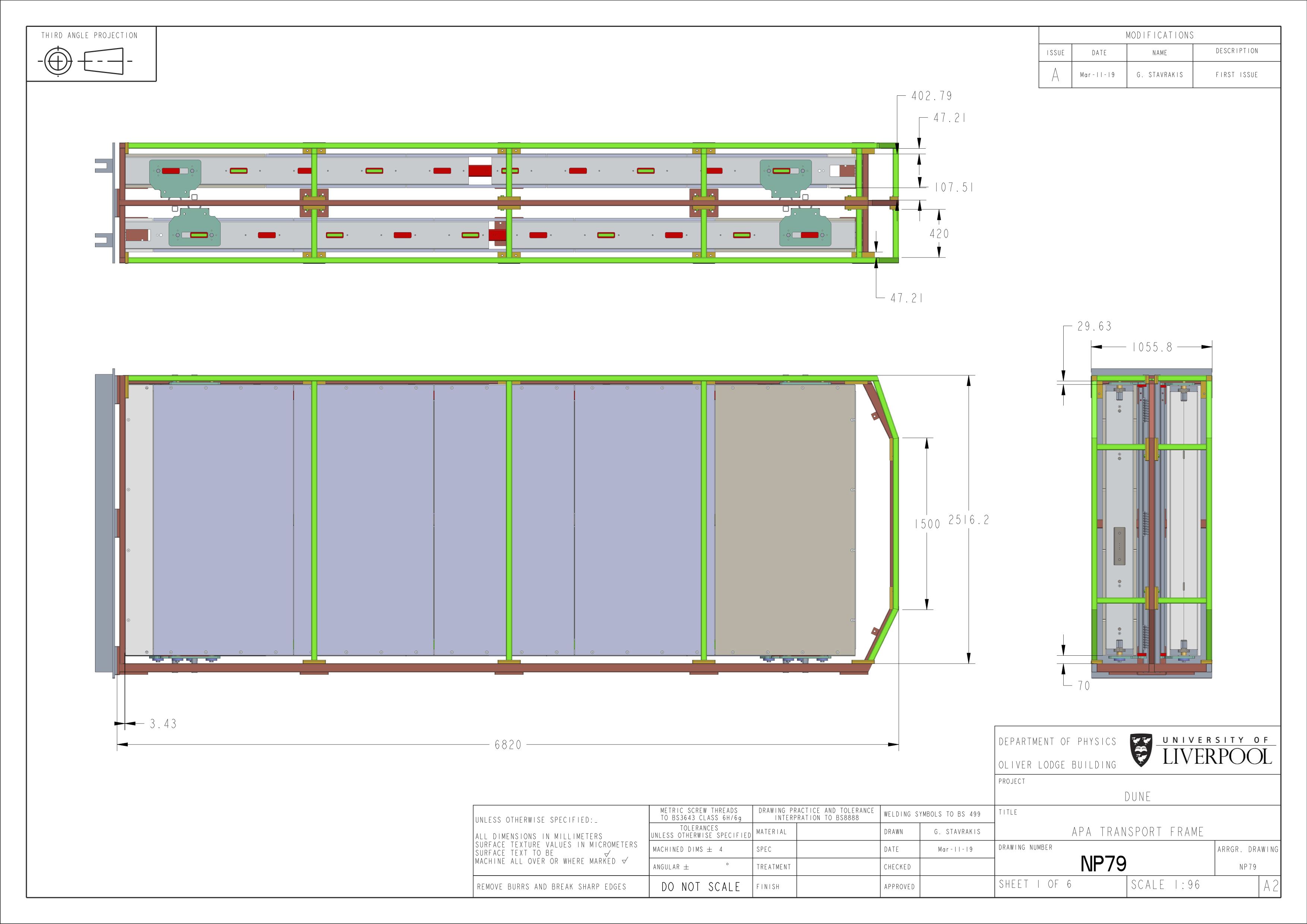} 
\end{dunefigure}

The size of the packaging and rigging hardware is constrained by the Ross headframe dimensions and over-the-road shipping requirements in the US. The design of the protective panels and the side frames allow for temporarily removing a portion of the shipping packaging and protective panels to access the \dword{apa} head boards for wire tension, isolation, and continuity tests after shipment and after transport underground. 

When a crate is required underground, 
it will be stripped of its wooden crating and 
 transported via Conestoga-type trailer to the headframe area. Near the headframe, the crates will be moved by forklift onto a cart on a rail system and rolled into the headframe. The inside portion of the headframe will have rigging gear attached to hard points on both short ends of the crate. 
 The crate's upper end will be attached 
 to the hoist below the cage and will be used to lift the crate from horizontal to vertical and pull it into the shaft. The shipping frame is designed to clear the headframe during this operation. The other end (lower) will be used to attach a horizontal tugger that will control the crate as it is pulled into the shaft station (Figure~\ref{fig:apa-transport-shaft}). When in the shaft, fixtures on the sides of the crate will engage wooden guides in the shaft to keep the crate from swinging or rotating while being lowered down the shaft. This operation is consistent with standard slung-load transport procedures at \dword{surf}. When the crate arrives underground, it will be pulled out of the shaft by reversing the shaft rigging operation; it will land on the opposite long edge of the crate that was used on the surface. The crate is placed on a transport cart and pulled down the drift to the cavern. When in the cavern, the \dword{apa}s will be uncrated, rotated to vertical by the cavern crane, mounted on a vertically oriented cart, tested, and stored temporarily (a few weeks) in the cavern adjacent to the clean room prior to final integration and installation. The transport frames and carts have been designed to be stable in each of these configurations. 

\begin{dunefigure}[APA loading into the mine shaft]{fig:apa-transport-shaft}
{Motion study of loading an \dword{apa} frame into the shaft.}  
\includegraphics[height=0.9\textheight,trim = 0mm 0mm 0mm 0mm, clip]{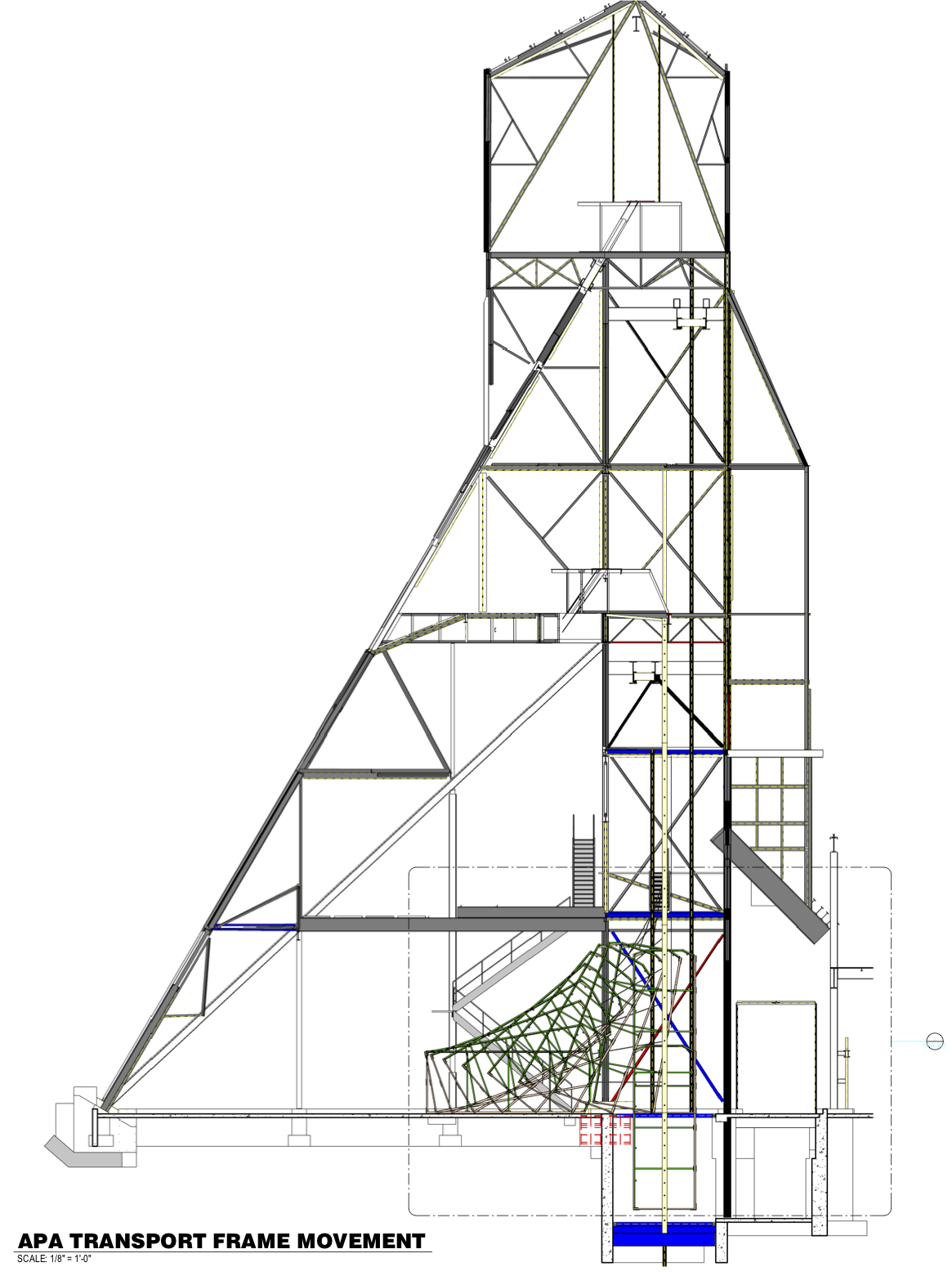} 
\end{dunefigure}

\subsection{APA Quality Control During Integration and Installation} 
\label{sec:fdsp-apa-transport-qc}

All active detector components are shipped to the \dword{sdwf} before final transport to \dword{surf}. 
After unpacking an \dword{apa} (underground at \dword{surf}), a visual inspection will be performed and wire continuity and tension measurements will be made. 
Tension values will be recorded in the database and compared with the original tension measurements performed at the production sites, as was done for \dword{pdsp} and shown in Figure~\ref{fig:sp-apa-pd-tension-cern}. Definite guidance for the acceptable tension values will be available to inform decisions on the quality of the \dword{apa}. Clear pass/fail criteria 
will be provided as well as clear procedures to deal with individual wires lying outside the acceptable values. 
This guidance will be informed also by the \dword{pdsp} experience. 
In addition, a continuity test and a leakage current test is performed on all channels and the data recorded in the database.

When all tests are successful, 
the \dword{apa} can be prepared 
for integration with the other components. 
This step is critical for ensuring high performance of the integrated \dword{apa}s. The procedures for \dword{apa} transport to the \dword{4850l} at \dword{surf}, integration with the \dword{pds} and \dword{ce}, and the schedule for testing the integrated \dword{apa} are addressed in 
Chapter~\ref{ch:sp-install}. \dword{apa} consortium personnel will play direct and key roles throughout the integration and installation activities.

\section{Safety Considerations}
\label{sec:fdsp-apa-safety}

The \dword{lbnf-dune} is committed to ensuring a safe work environment for workers at all institutions and facilities, from \dword{apa} fabrication to installation. The project utilizes the concept of an Integrated Safety Management System (ISMS) as an organized process whereby work is planned, performed, assessed, and systematically improved to promote the safe conduct of work. The \dword{lbnf-dune} Integrated Environment, Safety and Health Management Plan \cite{bib:docdb291} 
contains details on \dword{lbnf-dune} integrated safety management systems. This work planning and \dword{ha} program utilizes detailed work plan documents, hazard analysis reports, equipment documentation, safety data sheets, \dword{ppe}, and job task training to minimize work place hazards. 

Prior to  \dword{apa} production, applying the experience of \dword{pdsp}, the project will coordinate with fabrication partner facilities to develop work planning documents, and equipment documentation, such as the Interlock Safety System for \dword{apa} winding machines to implement an automated protection against personnel touching the winding arm while the system is in operation. Additionally, the project will work with the local institutions' \dword{esh} coordinators to ensure that \dword{esh} requirements within the home institution's \dword{esh} Manual address the hazards of the work activities occurring at the facility. Common job hazard analyses may be shared across multiple fabrication facilities.   

Handling of the large but delicate frames is a challenge. Procedures committed to the safety of personnel and equipment will be developed for all phases of construction, including frame assembly, wiring, transport, and integration, and installation in the cryostat. This documentation will continue to be developed through the \dword{ashriver} trial assembly process, which maps out the step-by-step procedures and brings together the documentation needed for approving the work plan to be applied at the far site. 

As is \dword{fnal}'s practice, all personnel have the right to stop work for any safety issues.

\section{Organization and Management}
\label{sec:fdsp-apa-org}

Coordination of the groups participating in the \dword{dune} \dword{apa} consortium is critical to successfully executing the large-scale multi-year construction project that is needed to produce high-quality \dword{apa}s for the \dword{dune} \dwords{spmod}.   The \dword{apa} consortium comprises \num{21} institutions, of which \num{14} are in the USA and \num{7} in the UK (see Table~\ref{tab:apa-institutions}). The consortium is organized along the main deliverables, which are the final design of the \dword{apa} and the \dword{apa} production and installation procedures (see Figure~\ref{fig:apa-consortium-structure}).  The two main centers of \dword{apa} construction are in the USA and the UK, so usually the leaders of a working group are chosen to represent the main stakeholders to ensure that common procedures and tooling are developed.  We plan to produce half of the \dword{dune} \dword{apa}s in the USA and half in the UK. 

\begin{dunefigure}[APA consortium organizational chart]{fig:apa-consortium-structure}
{\dword{apa} consortium organizational chart.}
\includegraphics[width=0.7\textwidth,trim=0mm 0mm 0mm 0mm,clip]{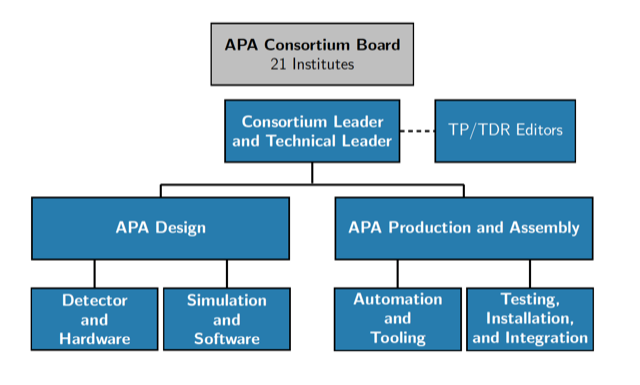}
\end{dunefigure}
\begin{dunetable}
[APA consortium institutions]
{ll}
{tab:apa-institutions}
{Current APA consortium institutions and countries.}
Institution & \bfseries{Country} \\ \toprowrule
University of Cambridge     &  UK       \\ \colhline
Daresbury Laboratory - Science and Technology Facilities Council & UK \\ \colhline
Lancaster University & UK \\ \colhline
University of Liverpool & UK \\ \colhline
University of Manchester & UK \\ \colhline
University of Sheffield & UK \\ \colhline
University of Sussex & UK \\ \colhline
Brookhaven National Laboratory & USA \\ \colhline
University of Chicago & USA \\ \colhline
Colorado State University & USA \\ \colhline
Harvard University & USA \\ \colhline
University of Houston & USA \\ \colhline
University of Iowa & USA \\ \colhline
University of Mississippi & USA \\ \colhline
Northern Illinois University & USA \\ \colhline
Syracuse University & USA \\ \colhline
University of Texas at Arlington & USA \\ \colhline
Tufts University & USA \\ \colhline
College of William \& Mary & USA \\ \colhline
University of Wisconsin-Madison, Physical Sciences Laboratory & USA \\ \colhline
Yale University & USA \\
\end{dunetable}

The university groups and \dword{bnl} are responsible for validating  the design, while engineering and the production set up is being developed at PSL 
(USA) and Daresbury Laboratory (UK), where the \dword{apa}s for \dword{pdsp} have been built. 
In addition to PSL and Daresbury Laboratory, the University of Chicago and Yale University are developing detailed plans for the layouts, activities, and schedule at each site.

In addition to the \dword{apa} production sites, a successful production effort will require significant and sustained contributions from university groups throughout the production process.  Table~\ref{tab:apa-institution-roles} and Figure~\ref{fig:apa-construction-org} list the main work packages that are part of the overall \dword{apa} construction process and the institutions in the USA and UK who are taking the leading roles in each effort.  The tasks range from the production of bare support frames, to the assembly and testing of many thousands of wire boards, to the procurement of the custom transport crates for shipping the completed \dword{apa}s.  The on-time supply of materials to each of the \dword{apa} production sites will be imperative to maintaining the production schedule, and detailed plans are being developed for the execution of the project in both countries.

\begin{dunefigure}[APA construction organizational chart]{fig:apa-construction-org}
{\dword{apa} construction project organizational chart.}
\includegraphics[width=1\textwidth,trim=0mm 0mm 0mm 0mm,clip]{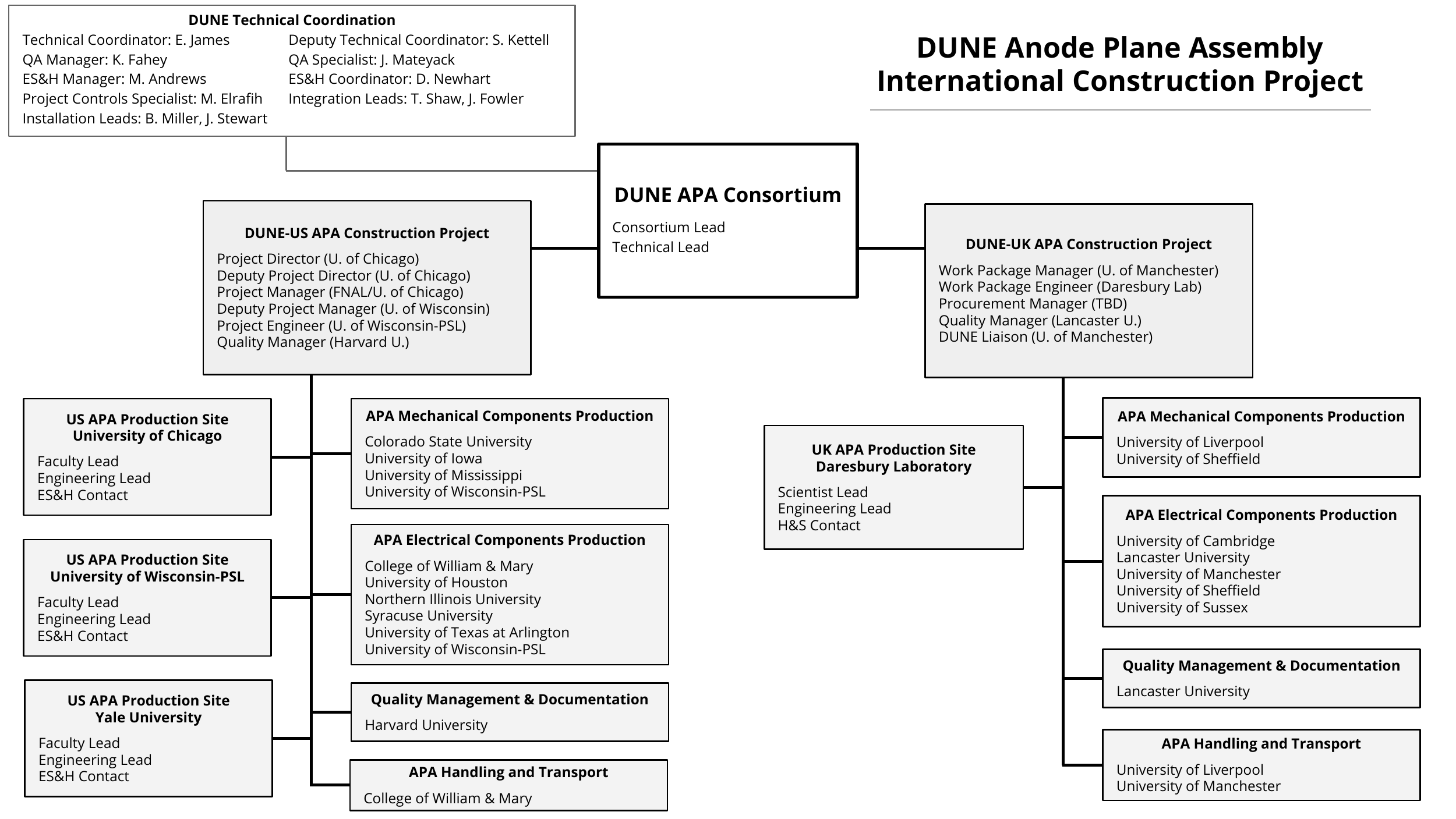}
\end{dunefigure}

\begin{dunetable}
[APA production institutional responsibilities]
{ll}
{tab:apa-institution-roles}
{Institutional responsibilities for \dword{apa} production in the UK and USA.}
APA Construction Work Packages & \multicolumn{1}{c}{\bfseries{Institutions}} \\ \toprowrule
\rowcolor{lightgray} \multicolumn{2}{c}{Production in the UK} \\ \colhline
APA production site  &  Daresbury Laboratory \\ \colhline
$U/V$-plane wire boards & University of Cambridge, University of Sussex \\ \colhline
$X/G$-plane wire boards & Lancaster University, University of Sheffield \\ \colhline
G-bias boards & University of Manchester \\ \colhline
CR boards & University of Manchester \\ \colhline
Cold electronics adapter boards & University of Sheffield \\ \colhline
Grounding mesh frames & University of Sheffield \\ \colhline
APA frames & University of Liverpool \\ \colhline
APA transport crates & University of Liverpool, University of Manchester \\ \colhline
Yokes and structural tees & University of Liverpool  \\ \colhline
QA/QC management & Lancaster University \\ \colhline
\rowcolor{lightgray} \multicolumn{2}{c}{Production in the US} \\ \colhline
APA production site & University of Wisconsin-PSL \\ \colhline
APA production site & University of Chicago  \\ \colhline
APA production site & Yale University  \\ \colhline 
U/V-plane wire boards & College of William \& Mary  \\ \colhline
X-plane wire boards & University of Texas at Arlington  \\ \colhline
G-plane wire boards & University of Houston  \\ \colhline
G-bias boards & Syracuse University  \\ \colhline
CR boards & University of Wisconsin-PSL \\ \colhline
Cold electronics adapter boards & Northern Illinois University  \\ \colhline
Grounding mesh frames & University of Chicago  \\ \colhline
APA frames & University of Iowa, University of Wisconsin-PSL \\ \colhline
APA transport crates & College of William \& Mary  \\ \colhline
Yokes and structural tees & University of Wisconsin-PSL  \\ \colhline
\dword{ce} interface hardware & Colorado State University, University of Wisconsin-PSL  \\ \colhline
QA/QC management, wire tension & Harvard University  \\ 
\end{dunetable}
\section{Schedule and Risks}
\label{sec:fdsp-apa-cost-sched}

\subsection{Schedule}

A schedule for key design and production readiness reviews leading up to the start of \dword{apa} production is provided in Table~\ref{tab:apa-reviews}. The high-level milestones for the final design and construction of the \dword{dune} \dword{apa}s between 2019 and 2026 are given in Table~\ref{tab:apa-milestones}.
\begin{dunetable}
[APA consortium review schedule]
{p{0.7\textwidth}p{0.18\textwidth}}
{tab:apa-reviews}
{Planned review schedule for the APA design and production preparations.}   
Review & \bfseries{Date}    \\ \toprowrule
\dword{apa} Electrical Preliminary Design Review & November 2019 \\ \colhline
\dword{apa} Production Site Design Internal Review -- UK & January 2020  \\ \colhline
\dword{apa} Transport Frame Preliminary Design Review & April 2020 \\ \colhline
Wire Tension Measurement Internal Review & March 2020 \\ \colhline
\dword{apa} Production Site Design Internal Review -- USA & April 2020  \\ \colhline
\dword{apa} Final Design Review & May 2020 \\ \colhline
Production Site Readiness Review  - UK & June 2020 \\ \colhline
Production Sites Readiness Review  -- U. of Wisconsin-PSL &  November 2020    \\ \colhline
Production Sites Readiness Review  -- U. of Chicago \& Yale U. &  April 2021    \\ 
\end{dunetable}

\begin{dunetable}
[APA consortium schedule]
{p{0.7\textwidth}p{0.18\textwidth}}
{tab:apa-milestones}
{Schedule milestones for the production and installation of \dwords{apa} for two \dword{sp} \dword{dune} far detector modules.}   
Milestone & \bfseries{Date}    \\ \toprowrule
Final report on the necessity and design of electron diverters & August 2019  \\ \colhline
Completion of \dword{apa} pair 
frame assembly \& cabling at \dword{ashriver} & October 2019 \\ \colhline

Decision on the wire tension measurement method & April 2020    \\ \colhline
Completion of winding machine modifications and commissioning & April 2020 \\ \colhline
Start of \dword{apa} Components Production -- UK & June 2020 \\ \colhline
Start of APA production for \dword{pdsp}-II & July 2020 \\ \colhline
Start of \dword{apa} Production for DUNE -- UK & August 2020 \\ \colhline 
Completion of \dword{apa} integration test with CE and PDS at CERN & September 2020 \\ \colhline
Start of \dword{apa} Components Production -- USA & November 2020 \\ \colhline
End of APA production for \dword{pdsp}-II & December 2020 \\ \colhline
Start of \dword{apa} Production for DUNE -- U. of Wisconsin-PSL & January 2021  \\ \colhline
\rowcolor{dunepeach} Start of \dword{pdsp}-II installation& \startpduneiispinstall \\ \colhline
Start of \dword{apa} Production for DUNE -- U. of Chicago \& Yale U. & June 2021  \\ \colhline
\rowcolor{dunepeach}South Dakota Logistics Warehouse available& \sdlwavailable \\ \colhline
\rowcolor{dunepeach}Beneficial occupancy of cavern 1 and \dword{cuc}& \cucbenocc \\ \colhline
\rowcolor{dunepeach} \dword{cuc} counting room accessible& \accesscuccountrm  \\ \colhline
End of \dword{apa} Production - \dword{detmodule} \#1  & September 2023 \\ \colhline
Start of \dword{apa} Production - \dword{detmodule} \#2  & October 2023 \\ \colhline
\rowcolor{dunepeach}Top of \dword{detmodule} \#1 cryostat accessible& \accesstopfirstcryo \\ \colhline
\rowcolor{dunepeach}Start of \dword{detmodule} \#1 TPC installation& \startfirsttpcinstall \\ \colhline
\rowcolor{dunepeach}Top of \dword{detmodule} \#2 accessible& \accesstopsecondcryo \\ \colhline
\rowcolor{dunepeach}End of \dword{detmodule} \#1 TPC installation& \firsttpcinstallend \\ \colhline
\rowcolor{dunepeach}Start of \dword{detmodule} \#2 TPC installation& \startsecondtpcinstall \\ \colhline
End of \dword{apa} Production - \dword{detmodule} \#2  & April 2026  \\ \colhline
\rowcolor{dunepeach}End of \dword{detmodule} \#2 TPC installation& \secondtpcinstallend \\
\end{dunetable}

Analysis of the \dword{pdsp} data will inform the decision on the electron diverters. 
Additional design considerations that cannot be directly tested through \dword{pdsp}, like the \dword{apa} pair assembly and related cabling issues require the full test with cabling of an \dword{apa} pair frame assembly, planned to be performed at \dword{ashriver}.  Also planned is the construction of a pre-production \dword{apa} for an integration test with \dword{ce} and \dword{pds}s at \dword{cern} in spring 2020, which will fully test all interface aspects. This test will inform the final design review of the \dword{apa} system in May 2020.

Design reviews of \dword{apa} production sites in the UK and USA, to validate the layout of the production lines, are planned for early 2020, together with the finalization of the winding machine modifications.  Production site readiness reviews are planned in June 2020 in the UK and in December 2020 in the USA.

Production of three \dword{apa}s for a final test in \dword{pdsp2} is foreseen in the second half of 2020. The pre-production \dword{apa} used for the integration test at \dword{cern} in spring 2020 could be used for installation in \dword{pdsp2}, if no additional modifications are required.

Dates are also provided in Table~\ref{tab:apa-milestones} for the start and end of \dword{apa} production for \dwords{detmodule} \#1 and \#2. Steady-state production rates are 24 \dword{apa}s/year at Daresbury Laboratory with four production lines, 12 \dword{apa}s/year both at Yale and Chicago, each with two production lines, and six \dword{apa}s/year at PSL, with one production line. The production time for \dword{detmodule} \#1 takes into account a gradual start-up of the production lines, and the different start dates and number of production lines in the UK and USA. The end of \dword{apa} production for \dword{detmodule} \#1 happens comfortably ten months before the start of installation. In the UK, with four assembly lines, in order to meet the installation date for detector module \#2 the \dword{apa} production time would need to be reduced by about seven months. This could be achieved by a reduction of the \dword{apa} assembly time, an opportunity mentioned in Section~\ref{sec:fdsp-apa-cost-sched-risks}, and, if necessary, by increasing the number of working shifts per week.
%

\subsection{Risks}
\label{sec:fdsp-apa-cost-sched-risks}

Risks have been identified for the finalization of the \dword{apa} design and the prototyping phase, for the setup of the production sites, for the production of \dword{apa}s, and for installation at \dword{surf}.  Risks are summarized in Table~\ref{tab:risks:SP-FD-APA}.  For each risk source, we describe a brief mitigation strategy as well as an estimation of the probability of occurring (P) and the impact that risk would have on costs (C) and on schedule (S).  These are each indicated as Low (L), Medium (M), or High (H) probability of impact. One opportunity is also listed and uses the same probability and impact indicators.

\begin{footnotesize}
\begin{longtable}{P{0.18\textwidth}P{0.20\textwidth}P{0.32\textwidth}P{0.02\textwidth}P{0.02\textwidth}P{0.02\textwidth}} 
\caption[APA risks]{APA risks (P=probability, C=cost, S=schedule) The risk probability, after taking into account the planned mitigation activities, is ranked as 
L (low $<\,$\SI{10}{\%}), 
M (medium \SIrange{10}{25}{\%}), or 
H (high $>\,$\SI{25}{\%}). 
The cost and schedule impacts are ranked as 
L (cost increase $<\,$\SI{5}{\%}, schedule delay $<\,$\num{2} months), 
M (\SIrange{5}{25}{\%} and 2--6 months, respectively) and 
H ($>\,$\SI{20}{\%} and $>\,$2 months, respectively). \fixmehl{ref \texttt{tab:risks:DP-FD-CISC}}} \\
\rowcolor{dunesky}
ID & Risk & Mitigation & P & C & S  \\  \colhline
RT-SP-APA-01 & Loss of key personnel & Implement succession planning and formal project documentation & L & L & M \\  \colhline
RT-SP-APA-02 & Delay in finalisation of APA frame design & Close oversight on prototypes and interface issues & L & L & M \\  \colhline
RT-SP-APA-03 & One additional pre-production APA may be necessary & Close oversight on approval of designs, commissioning of tooling and assembly procedures & L & L & L \\  \colhline
RT-SP-APA-04 & APA winder construction takes longer than planned & Detailed plan to stand up new winding machines at each facility & M & L & M \\  \colhline
RT-SP-APA-05 & Poor quality of APA frames and/or inaccuracy in the machining of holes and slots & Clearly specified requirements and seek out backup vendors & L & L & M \\  \colhline
RT-SP-APA-06 & Insufficient scientific manpower at APA assembly factories & Get institutional commitments for requests of necessary personnel in research grants & M & M & L \\  \colhline
RT-SP-APA-07 & APA production quality does not meet requirements & Close oversight on assembly procedures & L & M & M \\  \colhline
RT-SP-APA-08 & Materials shortage at factory & Develop and execute a supply chain management plan & M & L & L \\  \colhline
RT-SP-APA-09 & Failure of a winding machine - Drive chain parts failure & Regular maintenance and availability of spare parts & L & L & L \\  \colhline
RT-SP-APA-10 & APA assembly takes longer time than planned  & Estimates based on protoDUNE. Formal training of every tech/operator & L & M & M \\  \colhline
RT-SP-APA-11 & Loss of one APA due to an accident & Define handling procedures supported by engineering notes & M & L & L \\  \colhline
RT-SP-APA-12 & APA transport box inadequate & Construction and test of prototype transport boxes & L & L & M \\  \colhline
RO-SP-APA-01 & Reduction of the APA assembly time & Improvements in the winding head and wire tension mesurements & M & M & M \\  
 &  &  &  &  &  \\  \colhline

\label{tab:risks:SP-FD-APA}
\end{longtable}
\end{footnotesize}

Risks with medium or greater probability and/or medium or greater impact are discussed in more detail below:

\begin{itemize}
\item RT-SP-APA-01, Loss of key personnel:
\begin{itemize} 
\item \textit{Description}: If loss of key personnel happens, it will cause delays as knowledge is lost and new team members will need to come up to speed.
\item \textit{Mitigation}: Implement succession planning and formal project documentation at all stages. All key tasks to be shared between multiple people, including production site management.
\item \textit{Probability and impact}: While the post-mitigation probability is low, below 10\%, if the risk is realized, the impact on the schedule could range from a couple of months to a half-year.
\end{itemize}

\item RT-SP-APA-02, Delay in finalization of \dword{apa} frame design:
\begin{itemize} 
\item \textit{Description}: If problems are encountered with the \dword{apa} pair 
frame assembly and cabling tests at \dword{ashriver}, or with the integration test of the pre-production \dword{apa} at \dword{cern}, this will delay the finalization of the \dword{apa} frame design.
\item \textit{Mitigation}: Oversight of the \dword{apa} Consortium on the schedule of components procurement for the \dword{ashriver} tests and close coordination with the \dword{ashriver} team. Close coordination with \dword{ce} and \dword{pds}s consortia on all interface issues, to be formalized in the interface documents.
\item \textit{Probability and impact}: On the basis of the work done up to now we believe that the probability of this risk is low. However, if materialized, it would imply a delay in the start of \dword{apa} production from a couple of months to a half-year.
\end{itemize}

\item RT-SP-APA-04, \dword{apa} winders construction takes longer than planned:
\begin{itemize} 
\item \textit{Description}: If the construction of the winding machines takes longer than planned, the schedule for \dword{apa} production will be delayed, and additional labor for winders production will be needed. We plan the construction of four additional winders in the USA and the modification of winder, presently at PSL, as well as the construction of three additional winders in the UK, in addition to the modification and relocation of the winder at Daresbury Lab. The estimated time for the production of the additional winders is approximately one year, both in the UK and the USA.
\item \textit{Mitigation}: Get commitments from the relevant institutions for the necessary resources for winder production, both for space and skilled manpower availability. Develop and execute a detailed plan to set up new winding machines at each production site. This plan will include contingencies in the event that technical problems cause schedule delays.  
\item \textit{Probability and impact}: Winders are complex machines, and we estimate a medium probability for this risk, of less than 25\%. The impact on the schedule is also medium, with possible delays up to a half-year.
\end{itemize}

\item RT-SP-APA-05, Poor quality of \dword{apa} frames and/or inaccuracy in the machining of holes and slots:
\begin{itemize} 
\item \textit{Description}: \dword{apa} frames are constructed from structural stainless steel tubing. The quality of the material provided by the vendor may change with time and be outside the required tolerances. Problems with \dword{qa} during machining of holes and slots may result in unusable products. If this happens, it may delay the supply of frames of sufficient quality, which would delay the \dword{apa} construction schedule.
\item \textit{Mitigation}: All requirements must be clearly specified in the purchase contracts. We will establish a well managed relationship with a vendor to provide the stainless steel tubing and the machining for the components of an \dword{apa} frame. In addition, through our prototyping efforts, we will seek out at least one solid backup vendor for material supply and machining in both the USA and UK.
\item \textit{Probability and impact}: While the post-mitigation probability is low, below 10\%, if the risk is realized, the impact on the schedule is medium, since finding a new vendor may take up to a half-year.
\end{itemize}

\item RT-SP-APA-06, Insufficient scientific manpower at \dword{apa} assembly factories:
\begin{itemize} 
\item \textit{Description}: 
For US production, if it is not possible to recruit scientific resources, costed professional manpower is needed and costs will increase. This risk does not apply to the UK production since the required scientific staff is costed and awarded on project.
\item \textit{Mitigation}: Proactively contact institutions and get their commitments for inclusion of the necessary personnel in their research grants.
\item \textit{Probability and impact}: This is a medium probability risk; we estimate a 50\% probability that 50\% of the US scientific resources may be missing. The cost impact is also medium, up to 20\%.
\end{itemize}

\item RT-SP-APA-07, \dword{apa} production quality does not meet requirements:
\begin{itemize} 
\item \textit{Description}: If wire planes are outside the required tolerances, they will need to be reworked, and the \dword{apa} production schedule will be affected. A point of concern is to stay within limits for the tension of the wires. 
\item \textit{Mitigation}: The overall quality of each constructed \dword{apa} will be ensured by following detailed procedures for every step of the assembly process (e.g., mesh installation, board placement and gluing, soldering, wire winding, etc.).  These procedures already exist from our \dword{pdsp} work and are in the process of being modified to the final design of the DUNE \dword{fd} \dword{apa}s. For critical steps, an operator and quality control representative will record information in travelers for each \dword{apa}. 
\item \textit{Probability and impact}: Given the \dword{pdsp} experience and the steps outlined in the mitigation strategy, we can keep this risk probability low, below 10\%. If realized, we assume a maximum impact on cost and schedule of 20\%, corresponding to a medium impact.
\end{itemize}

\item RT-SP-APA-08, Materials shortage at an \dword{apa} production site:
\begin{itemize} 
\item \textit{Description}: A material shortage at an \dword{apa} production site would delay production.
\item \textit{Mitigation}: As part of our comprehensive production strategy, we are in the process of developing and executing a supply chain management plan. This plan will include the details of material source, delivery logistics, critical milestones, and personnel resources required to meet \dword{apa} production site needs for efficient \dword{apa} production. All suppliers (vendors, laboratories, academic institutions) will be included in the implementation of the supply chain plan. A key part of this plan will be the establishment of supplier metrics that will be gathered and reported to DUNE management by the \dword{apa} production manager. These metrics will serve as an early warning of potential problems and trigger mitigation efforts early in the cycle. 
\item \textit{Probability and impact}: Even with mitigation, this is a realistic risk with an estimated probability of up to 25\%. Delays on the schedule would probably not exceed a couple of months, making the impact low.
\end{itemize}

\item RT-SP-APA-10, \dword{apa} assembly takes longer than planned:
\begin{itemize} 
\item \textit{Description}: If the labor for \dword{apa} assembly is underestimated, it will correspondingly lengthen the time to produce \dword{apa}s. We estimate an upper limit on the additional required labor of 10\%. 
\item \textit{Mitigation}: \dword{apa} assembly time estimates have been based on \dword{pdsp} experience and improvements to the winding machine. Formal training of every technician/operator of the winding machine should maintain a high production efficiency. 
\item \textit{Probability and impact}: We believe that this risk probability is low, below 10\%, but even a 10\% increase in the required labor would have a medium impact on both cost and schedule.
\end{itemize}

\item RT-SP-APA-11, Loss of an \dword{apa} due to an accident:
\begin{itemize} 
\item \textit{Description}: If during \dword{apa} assembly or integration/installation an accident happens, this may  cause the destruction of an \dword{apa}. We already plan to build two spare \dword{apa}s for each detector module. In addition, we assume a probability of up to 10\% to lose an additional \dword{apa} during assembly or integration/installation. 
\item \textit{Mitigation}: We define procedures to handle \dword{apa}s at all stage of fabrication, integration/installation, together with associated engineering notes for all modes of handling. 
Wire planes are delicate, and once damaged they would not be repairable. 
\item \textit{Probability and impact}: This is a marginally medium risk, with low impact both on cost and schedule.
\end{itemize}

\item RT-SP-APA-12, \dword{apa} transport box inadequate:
\begin{itemize} 
\item \textit{Description}:  
If the transport box will not provide enough mechanical protection for the safe transportation of the \dword{apa}s or if the size of the box is inadequate for transfer to the underground location, it will impact the schedule.
\item \textit{Mitigation}: Construct and test prototype transport boxes. Test all handling steps at \dword{ashriver} and \dword{surf}. Coordinate closely with the team at \dword{surf} and \dword{apa} Consortium oversight of transport boxes.
\item \textit{Probability and impact}: Given the mitigation steps, we estimate a probability for this risk of less than 10\%. In case of redesign, it may have a medium impact on the schedule.
\end{itemize}

\item RO-SP-APA-01, Reduction of the \dword{apa} assembly time:
\begin{itemize} 
\item \textit{Description}: If the new winding head will provides better uniformity in wire tension, it will reduce the time necessary for re-tensioning of the wires. If the new electrical method for wire tension measurements work as planned, it will reduce substantially the time required for wire tension measurements. A saving of up to 10\% in \dword{apa} assembly time will be possible with these improvements. 
\item \textit{Opportunity}: This is an opportunity that the \dword{apa} Consortium is actively pursuing with ongoing testing of a new winding head design and development of the electronic tension measurement method. The \dword{apa} boards have been already redesigned to allow for electronic tension measurements. 
\item \textit{Probability and impact}: Given the preliminary results obtained up to now, we estimate a medium probability for this opportunity. A saving of 10\% in the \dword{apa} assembly time could be realized, corresponding to a medium impact for both cost and schedule.
\end{itemize}

\end{itemize}

\cleardoublepage

\chapter{High Voltage}
\label{ch:sp-hv}
\section{High Voltage System Overview}
\label{sec:fdsp-hv-ov}

\subsection{Introduction and Scope}
\label{sec:fdsp-hv-intro}

A \dword{lartpc} requires an equipotential cathode plane at \dword{hv} and a precisely regulated interior electric field (\efield{}) to drive 
electrons from particle interactions to sensor planes.  To achieve this, the \dword{dune}  \dword{sp} \dword{tpc} consists of 
\begin{itemize}
\item vertical cathode planes, called \dword{cpa} arrays, held at \dword{hv};
\item vertical anode planes, called \dword{apa} arrays, described in Chapter~\ref{ch:fdsp-apa}; and
\item formed sets of conductors at graded voltages surrounding 
 the drift volumes to ensure uniformity of the \efield; the conductors are collectively called the \dword{fc}.
\end{itemize}

The \single \dword{tpc} configuration is shown in Figure~\ref{fig:dune_sp_fd}.
The  drift fields transport the ionization electrons 
towards the \dword{apa}s at the sides and center.

\begin{dunefigure}[SP module schematic with one unit of top and bottom FC modules]
{fig:dune_sp_fd}
{A schematic of a \dword{spmod} showing the three \dword{apa} arrays (at the far left and right and in the center, all of which span the entire \sptpclen \dword{detmodule} length) and the two \dword{cpa} arrays, occupying the intermediate second and fourth positions. The top and bottom \dword{fc} modules are shown with \dwords{gp} in blue. 
On the right, the front top and bottom \dword{fc} modules are shown folded up against the \dword{cpa} panels to which they connect, as they are positioned for shipping and insertion into the cryostat.  The \dword{cpa}s, \dword{apa}s, and \dword{fc} together define the four drift volumes of the \dword{spmod}. The sizes and quantities of the \dword{fc} and \dword{cpa}-array components are listed in Tables~\ref{tab:cpaparts} and~\ref{tab:fcparts} and represented in this image.}
\includegraphics[width=0.95\textwidth]{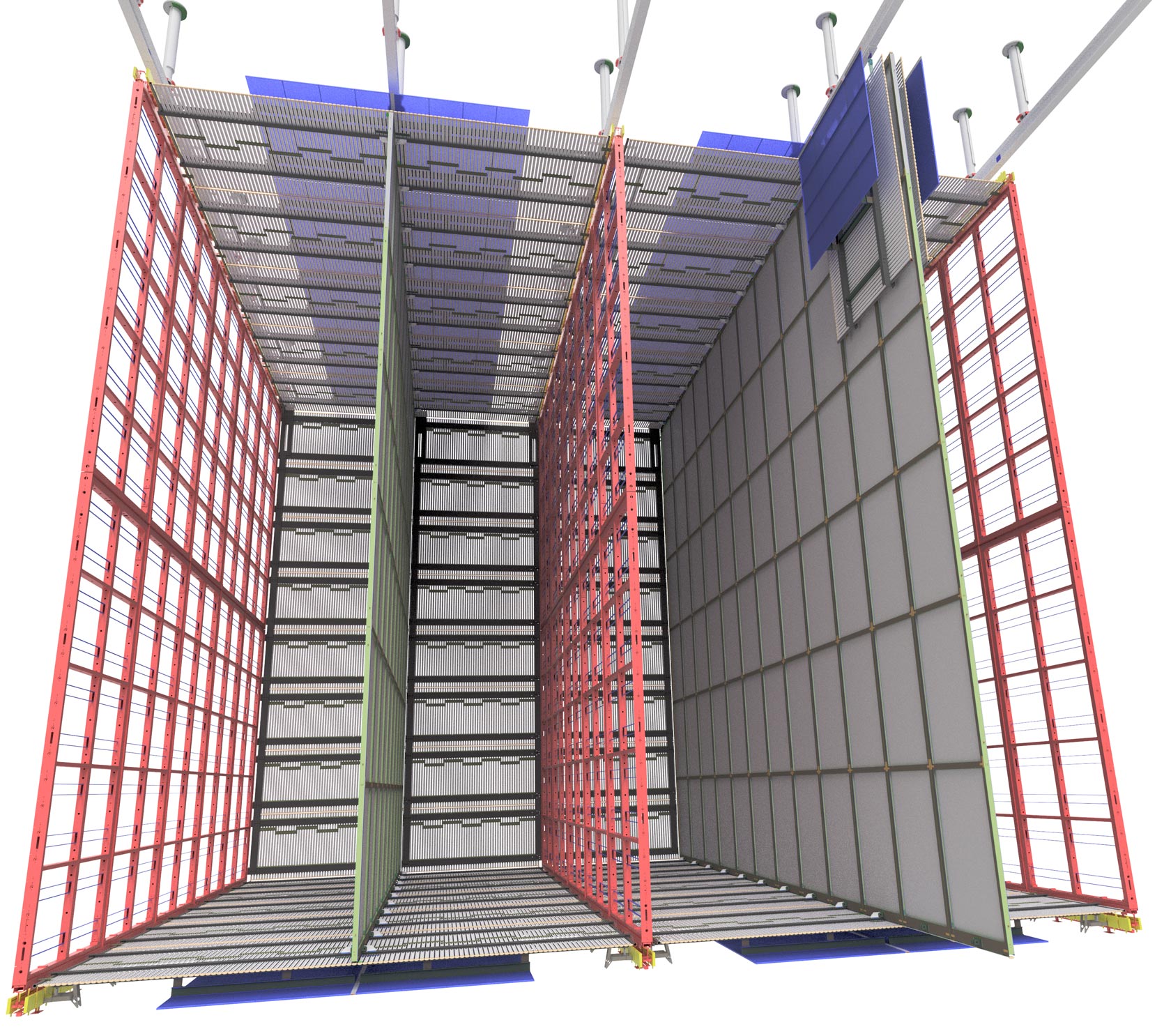}
\end{dunefigure}


The scope of the \single \dword{hv} system, provided by the DUNE \dword{hvs} consortium, includes the selection and procurement of materials for, and the fabrication, testing, delivery, and installation of systems to generate, distribute, and regulate the voltages that
create a stable and precise \efield{} within a \dword{spmod}. 

The \dword{hv} system consists of components both exterior and interior to the cryostat. The voltage generated at the \dword{hv} power supplies passes through the cables, filters, and the \dword{hv} \fdth into the cryostat. From the point of delivery into the cryostat, components that form part of the \dword{tpc} structure further distribute the voltage. The internal \dword{hv} components in fact form a large fraction of the total internal structures of the \dword{tpc} itself, and  
 effectively bound the  fiducial volume of the 
 \dword{detmodule}. 

The \dword{sp} \dword{hv} system consists of
\begin{itemize}
\item \dword{hv} power supplies, cables, filters, and feedthrough;
\item \dword{cpa} array;
\item \dword{topfc}, \dword{botfc}, and \dwords{gp}; and
\item \dword{ewfc}.
\end{itemize}

The system operates at the full range of voltages, 
$-$\sptargetdriftvoltpos to ground, inside the \dword{tpc} volume. 

The  \dword{sp} and  \dword{dp} modules will implement similar designs for some
 of the \dword{hv} system components, 
 in particular, aspects of the \dword{fc} and its supporting beams. This chapter describes the  \dword{sp} versions. 
 
\subsection{Design Specifications}
\label{sec:fdsp-hv-des-consid}

The working principle of the \dword{lartpc} relies on the application of a very uniform strong \efield in ultra-pure \dword{lar}.  A number of detector performance parameters benefit from such an \efield in ways that directly support the core components of the DUNE physics program.  Some of these are examined in detail in Volume~\volnumberphysics{}, \voltitlephysics{}.  Here we present a qualitative description of \efield impacts on physics to set context.

Since free electron drift velocity in \dword{lar} is a function of \efield, a uniform \efield leads to a simple time versus position mapping along the drift direction, enabling precise and efficient \threed reconstruction.  This allows, for example, the establishment of a well defined fiducial volume for beam neutrino events reconstructed in the \dword{fd}.  Since a neutrino \dword{cpv} measurement or neutrino \dword{mh} test at root consists of the comparison of normalized spectra for electron and muon neutrino and antineutrinos interactions in the fiducial volume of the \dword{fd} as projected from the \dword{nd}, fiducial volume characterization is critical.   

The optimal \efield range at which to operate the \dword{lartpc} is a trade-off  of detector performances that improve with increasing field against others that degrade. For instance, spectral information is necessary to separate \dword{cp} and \dword{mh} effects, necessitating efficient tracking and shower reconstruction and good energy resolution. To accomplish this, higher E field strength is generally better; more free charge is created at the ionization points, as electron-ion recombination decreases at higher fields, improving \dword{s/n} and calorimetry.

Drift times are reduced, resulting in less free electron capture from residual electronegative impurities, and hence better \dword{s/n}, even under less than optimal purity conditions.  Spatial resolution improves, as free electron diffusion (proportional to the square root of the drift time) lessens. 

Higher free charge production and lower electron capture allows for lower detection thresholds for components of electromagnetic showers, improving shower energy reconstruction. Lower detection thresholds also lead to higher detection efficiency for MeV-scale electron, photon, and neutron signatures of low-energy $\nu_e$ interactions from \dword{snb} events.  

The electron-ion recombination more strongly affects highly ionizing particles, usually protons.  With decreased recombination, less saturation of free charge production occurs,  leading to better particle identification and more precise energy measurements.  Lower recombination particularly aids in proton-kaon separation by $dE/dx$, a key component of a search for $p\rightarrow K^+ \nu$ baryon decay events. 

However, the \efield should not be raised beyond certain limits. For example, while free charge production increases with \efield, scintillation photon production decreases, resulting in fewer photons available for triggering and determination of $t_0$. Two-track separation can degrade if the drift velocity is increased while keeping the anode wire separation and electronic wave form sampling frequency fixed. The distance between the \dword{tpc} boundaries and the cryostat walls might  need to be increased for very high \efield{}s to prevent electrostatic discharge. This would in turn reduce the fraction of \dword{lar} in the \dword{fv}. The impacts of the first two effects are modest, and all effects are subsidiary to technical challenges in the delivery of high voltage to the cryostat and the maintenance of highly stable  \dword{hv}  surfaces for multiple decades of operation. These challenges require development of non-commercial cryogenic \dword{hv} feedthroughs, \dword{hv} ripple-repression through custom \dword{hv} \dword{r-c} circuits, careful construction and deployment of \dword{hv} cables, redundant \dword{hv} connections, high-precision monitoring, and best practices at all stages of design, installation, and operation.


To the best of present common knowledge, the response and stability of a \dword{lartpc} to \dword{hv} is strongly dependent on many boundary conditions that are not fully related to the  \dword{hv} design. There are for instance hints and tests that suggest that gas bubble formation as well as residual dust circulating in the \dword{lar} are  primary sources of \dword{hv} instability. Insulator charging up can also affect \dword{hv} performance in the long term.  
Finally, because we found no information on applying \SI{-180}{\kV} in an \dword{lar} detector, our approach to designing the \dword{hv} system relied heavily on past experience, applying in addition sufficient safety margins from previous designs. \dword{pdsp} has provided experience and understanding of \dword{hv} behavior, giving us confidence that the upgraded design documented in this \dword{tdr} is appropriate for underground long-term operation.  

Two decades of design and operational experience that began with \dword{icarus} have established that a \SI{500}{V/cm} field is an appropriate trade-off value that can be realistically achieved through utilization of cost-effective design and construction methods. In practice, achieving this design goal has been challenging as the drift distance has been progressively increased to the 
\SI{3.5}{m} foreseen for the \dword{spmod}, and overall detector optimization has proved to be important. For example, \dword{microboone} operates  at \SI{273}{V/cm} (lower than its nominal value of \SI{500}{V/cm}) and is able to operate well by exploiting its very high argon purity, (characterized by an electron lifetime in excess of 15 ms), as well as an excellent \dword{s/n} ratio from the \dword{fe} \dword{ce}.  
\dword{microboone} (and a number of other noble liquid \dwords{tpc}) compensated for electrostatic instability problems by achieving higher purity, and \dword{dune} might well operate in this mode during its run. 

In \dword{dune}, the minimum requirement of the drift \efield has been set to \mindriftfield, with a goal 
of \mindriftfieldgoal for long-term stable operation. With good free electron lifetime (>\SI{10}{ms}), and the electronics \dword{s/n} demonstrated in \dword{pdsp}, experience shows that \dword{dune} will be able to operate above \SI{250}{V/cm}. The advantage of running at higher \efield is that the lower electron-ion recombination rate and the higher electron drift velocity can compensate for any lower purity conditions that could arise during the planned operation period.

 Running \dword{pdsp} at a higher \dword{hv} value (as allowed by \dword{hv} cables and filtering systems) is under consideration to gain better understanding of the \dword{hv} stability issues.

Positive \dword{protodune} experience (see  Section~\ref{sec:fdsp-hv-protodune-lessons}) indicates that the \SI{500}{V/cm} \efield goal is within reach. This goal, combined with high \dword{lar} purity and a large \dword{s/n} ratio, will allow  a wide range of possible operating points to optimize detector performance for maximum physics potential over decades of stable conditions and very high live-time. 
The specification minimum of \SI{250}{V/cm} will provide adequate detector performance, assuming achievable purity and electronics parameters. 

The \dword{hv} system is designed to meet the physics requirements of the \dword{dune} experiment,  both physical  (e.g., \efield{}s that allow robust event reconstruction) and operational (e.g., avoiding over-complication that could affect 
the time available for collecting neutrino events). 
The important requirements and specifications for the \dword{hv} system are given in Table~\ref{tab:specs:SP-HV}. 

\begin{footnotesize}
\begin{longtable}{p{0.12\textwidth}p{0.18\textwidth}p{0.17\textwidth}p{0.25\textwidth}p{0.16\textwidth}}
\caption{HV specifications \fixmehl{ref \texttt{tab:spec:SP-HV}}} \\
  \rowcolor{dunesky}
       Label & Description  & Specification \newline (Goal) & Rationale & Validation \\  \colhline

   \newtag{SP-FD-1}{ spec:min-drift-field }  & Minimum drift field  &  $>$\,\SI{250}{ V/cm} \newline ($>\,\SI{500}{ V/cm}$) &  Lessens impacts of $e^-$-Ar recombination, $e^-$ lifetime, $e^-$ diffusion and space charge. &  ProtoDUNE \\ \colhline

  \newtag{SP-FD-11}{ spec:hvs-field-uniformity }  & Drift field uniformity due to HVS  &  $<\,\SI{1}{\%}$ throughout volume &  High reconstruction efficiency. &  ProtoDUNE and simulation \\ \colhline

  \newtag{SP-FD-12}{ spec:hv-ps-ripple }  & Cathode HV power supply ripple contribution to system noise  &  $<\,\SI{100}e^-$ &  Maximize live time; maintain high S/N. &  Engineering calculation, in situ measurement,   ProtoDUNE \\ \colhline
    
   \newtag{SP-FD-17}{ spec:cathode-resistivity }  & Cathode resistivity  &  $>\,\SI{1}{\mega\ohm/square}$ \newline ($>\,\SI{1}{\giga\ohm/square}$) &  Detector damage prevention. &  ProtoDUNE \\ \colhline

  \newtag{SP-FD-24}{ spec:local-e-fields }  & Local electric fields  &  $<\,\SI{30}{kV/cm}$ &  Maximize live time; maintain high S/N. &  ProtoDUNE \\ \colhline
    
   \newtag{SP-FD-29}{ spec:dp-det-uptime }  & Detector uptime  &  $>\,$98\% \newline ($>\,$99\%) &  Meet physics goals in timely fashion. &  ProtoDUNE \\ \colhline
    
   \newtag{SP-FD-30}{ spec:dp-det-mod-uptime }  & Individual detector module uptime  &  $>\,$90\% \newline ($>\,$95\%) &  Meet physics goals in timely fashion. &  ProtoDUNE \\ \colhline

  \newtag{SP-HV-1}{ spec:power-supply-stability }  & Maximize power supply stability  &  $>\,\SI{95}{\%}$ uptime &  Collect data over long period with high uptime. &  ProtoDUNE \\ \colhline
    
   \newtag{SP-HV-2}{ spec:hv-connection-redundancy }  & Provide redundancy in all \dshort{hv} connections.  &  Two-fold \newline (Four-fold) &  Avoid interrupting data collection or causing accesses to the interior of the detector. &  Assembly QC \\ \colhline

\label{tab:specs:SP-HV}
\end{longtable}
\end{footnotesize}  

We note that specification SP-FD-1 is discussed in the text above Table~\ref{tab:specs:SP-HV}. SP-FD-2  is met in case of stable \dword{hv} operation (lessons from \dword{pdsp}). The noise contribution from \dword{hv} instabilities is unclear and under investigation with \dword{pdsp}. The remaining requirements specific to the \dword{hv} system, summarized here, are all addressed and referred to in the remainder of this chapter.

\begin{itemize}
\item SP-FD-11: Non-uniformity could be due to defects in resistor chains; muon and laser calibrations will mitigate this effect. [Section~\ref{sec:fdsp-hv-des-fc}] 
\item SP-FD-12: Lessons learned from \dword{pdsp} demonstrate that the present filtering scheme is adequate. [\ref{sec:fdsp-hv-des-hvps}, \ref{sec:fdsp-hv-des-fc-profiles}] 
\item SP-FD-17: The \dword{cpa} design is based on few \si{\mega\ohm}/square resistivity surfaces. Such surfaces have been demonstrated in \dword{pdsp} to be adequate to prevent fast discharges that could potentially damage \dword{ce} and the cryostat (no event was ever recorded). Underground operation will allow higher resistivity, thus further slowing down the potential release of stored energy. [\ref{sec:fdsp-hv-cpa-arrays}, \ref{sec:fdsp-hv-des-fc-profiles}] 
\item SP-FD-24 is met by calculation in \dword{pdsp}. In the present design, the \efield in the critical region between \dword{fc} and \dword{gp} is further reduced. [\ref{sec:fdsp-hv-des-fc-profiles}, \ref{sec:fdsp-hv-design-interconnect}] 
\item SP-FD-29, 30: These uptime requirements are already met in \dword{pdsp}; the much lower ionization density in underground operation and optimization in the design (\dword{fc} to \dword{gp} distance) will ensure meeting the requirement even in the case of the much wider detector surface. 
\item SP-HV-1: The \dword{hv} distribution and filtering has been tested in \dword{pdsp}; the design of these items will be revised to minimize long-term degradation and maintenance requirements.  [\ref{sec:fdsp-hv-des-fc-profiles}] 
\item SP-HV-2: Two-fold redundant connections to the \dword{cpa} are foreseen. The \dword{hv} feedthrough and its connection to the \dword{cpa} is designed in such a way that it could be extracted and replaced even with the detector filled with \dword{lar} (based on \dword{icarus} experience). [\ref{sec:fdsp-hv-cpa-arrays}, \ref{sec:fdsp-hv-design-interconnect}] 
\end{itemize}
\subsection{Design Overview}
\label{sec:fdsp-hv-des}

\subsubsection{Cathode Plane Assembly (CPA) Arrays}
\label{sec:fdsp-hv-des-cpa}

\dword{cpa} arrays are made up of adjacent resistive cathode panels, secured in frames and connected by an \dword{hv} bus. \dword{hv} cups are mounted at both ends to receive input from the power supply.

Two \dword{cpa} arrays span the length and height of the \dword{spmod}, as shown in Figure~\ref{fig:dune_sp_fd}. 
Each array is assembled from a set of \num{25} adjacent full-height \dword{cpa} planes, 
each of which consists of two adjacent full-height panels. 
Each panel consists of three stacked units, approximately \SI{4}{\m} in $y$ (height) by \SI{1.2}{\meter} in the $z$-coordinate (parallel to beam). 
A unit consists of two 
vertically stacked \dwords{rp} framed by  \frfour\footnote{NEMA grade designation for flame-retardant glass-reinforced epoxy laminate material, multiple vendors, National Electrical Manufacturers Association\texttrademark{},  \url{https://www.nema.org/pages/default.aspx}.} members. 
The \dword{hv} cathode components are listed in Table~\ref{tab:cpaparts} and will hereafter be referred to by their names as defined in this table.

\begin{dunetable}
[HV cathode components]
{p{0.4\textwidth}p{0.12\textwidth}
p{0.12\textwidth}p{0.32\textwidth}}
{tab:cpaparts}
{\dword{hv} cathode components} 
Component and Quantity &  Length (z) & Height (y) & Per \dword{spmod} \\ \toprowrule
\dword{cpa} array (2 per \dword{spmod}) & \SI{58}{\meter} & \SI{12}{\meter} & 2  \\ \colhline
\dword{cpa} plane (25 per \dword{cpa} array)  & \SI{2.3}{\meter}  &\SI{12}{\meter} & 50  \\ \colhline
\dword{cpa} panel (2 per \dword{cpa} plane)  & \SI{1.2}{\meter}   & \SI{12}{\meter} & 100  \\ \colhline
\dword{cpa} unit (3 per \dword{cpa} panel)  & \SI{1.2}{\meter}  & \SI{4}{\meter} & 300 \\ \colhline
\dword{rp} (2 per \dword{cpa} unit)  & \SI{1.2}{\meter}  & \SI{2}{\meter} & 600 \\
\end{dunetable}
The \dwords{rp} are made of a highly resistive material. 
An  installation rail supports the \dword{cpa} panels from above through a single mechanical link. 

The cathode bias is provided by an external \dword{hv} power supply through an \dword{hv} \fdth connecting to the \dword{cpa} array 
inside the cryostat. 
 
\subsubsection{Field Cage}
\label{subsec:fdsp-hv-des-fc}

In the \dword{spmod}, an \dword{fc} covers the top, bottom, and endwalls of all the drift volumes, thus providing the necessary
boundary conditions to ensure a uniform \efield, unaffected by the presence of the cryostat walls. 
The \dword{fc} is made of adjacent extruded aluminum open profiles (electrodes) running perpendicular to the drift field and set at increasing potentials along the \spmaxdrift drift distance from the \dword{cpa} \dword{hv} (\SI{-180}{kV}) to ground potential at the \dword{apa} sensor arrays. 

The \dword{fc} modules come in two distinct types: the identical top and bottom modules, which are assembled to run the full length of the \dword{detmodule}, and the \dword{ewfc} modules, 
which are assembled to complete the detector at either end. 
The profiles in both types of modules are supported by \dword{frp}\footnote{Fiber-reinforced plastic, a composite material made of a polymer matrix reinforced with fibers, many vendors.} (fiber-reinforced plastic) structural beams.  

The \dword{topfc} and \dword{botfc} modules extend nominally  \SI{2.3}{\meter} in $z$ and \SI{3.5}{\meter} in $x$; the top and bottom of the \dword{spmod} each requires 25 modules lengthwise in $z$ and four across in $x$.  The \dword{ewfc} modules are \SI{3.5}{\meter} wide by \SI{1.5}{\meter} in high; each endwall requires four adjacent stacks, eight units high. A \dword{gp} consisting of modular 
perforated stainless steel sheets 
runs along the outside surface of each of the 
\dword{topfc} and \dword{botfc}, with a \SI{30}{\centi\meter} clearance. The \dword{ewfc} modules do not require a \dword{gp} because the distance to the cryostat wall is sufficient, approximately \SI{2}{\meter}.

To provide a linear voltage gradient within each drift volume, 
a chain of resistive divider boards connects the adjacent pairs of aluminum profiles along each \dword{fc} module. 

Table~\ref{tab:fcparts} lists the \dword{fc} components.
\begin{dunetable}
[HV field cage components]
{p{0.37\textwidth}p{0.13\textwidth}p{0.07\textwidth}p{0.07\textwidth}p{0.07\textwidth}
p{0.1\textwidth}p{0.08\textwidth}}
{tab:fcparts}{\dword{hv} field cage components}
Component & Count & Length (z) & Width (x) & Height (y) & Submodules & Grand Total \\ \toprowrule
\Dword{topfc} modules & 100 (4$\times$25) & 2.3 m & 3.5 m & - & - & 100 \\ \colhline
\Dword{botfc} modules & 100 (4$\times$25) & 2.3 m & 3.5 m & - & - & 100 \\ \colhline
Profiles per module \\(all top and bottom module types) & 57 & 2.3 m & - & - & - & 11400 \\ \colhline
\Dword{gp} modules per top or bottom \\ \dword{fc} module & 5 & 2.3 m & 0.7 m & - & - & 1000 \\ \colhline
\Dword{ewfc} plane & 2 & - & 14.4 m & 12 m & 4 & 2 \\ \colhline
\Dword{ewfc} modules per \dword{ewfc} & 32 & - & 3.5 m & 1.5 m & - & 64 \\ \colhline
Profiles per \dword{ewfc} module & 57 & - & - & 1.5 m & - & 3648 \\

\end{dunetable}

\subsubsection{Electrical Considerations}
\label{sec:fdsp-hv-des-elec}

As shown in Figure~\ref{fig:dune_sp_fd}, the outer \dword{apa} arrays face the cryostat walls, and the \dword{cpa} arrays are installed between the \dword{apa} arrays in two of the three interior positions (A-C-A-C-A).
In this configuration, as opposed to C-A-C-A-C,  most of the cathode plane surfaces are far away from the grounded cryostat walls, reducing electrostatic breakdown risks and decreasing the total energy stored in the \efield to \SI{800}{J}.

\begin{dunefigure}[Electric field distribution in the TPC]
{fig:e-field-distribution}
{A simplified cross sectional view of an outer drift volume of the \dword{tpc} showing the distribution of the static \efield (in V/m).  Since the electrostatic potential energy is proportional to $E^2$, most of the energy is stored between the \dword{fc} modules and their facing \dwords{gp}.   }
\centering
\includegraphics[width=0.7\textwidth]{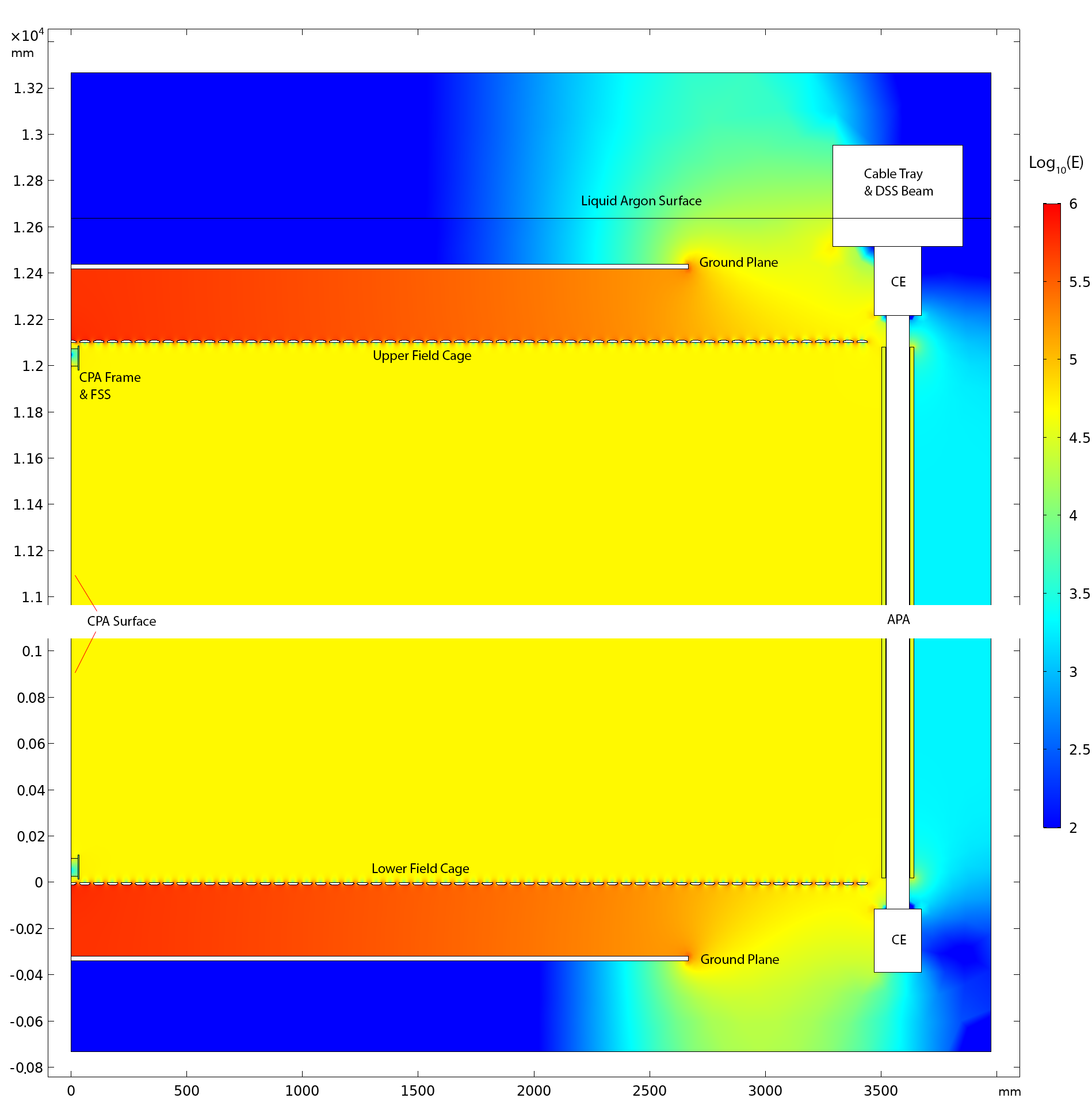} 
\end{dunefigure}

Figure~\ref{fig:e-field-distribution} maps out the \efield strength over a cross section of a drift volume.  
The energy is stored mostly in the high \efield{} region between the \dword{fc} and the facing \dwords{gp}.  In the case of an unexpected \dword{hv} breakdown, the entire \SI{400}{J} associated with one \dword{cpa} array could be discharged to ground,
potentially causing physical damage.
Given the difficulty of predicting the distribution of energy along a discharge path, we treat the possibility of discharged energy, conservatively, as a risk to the TPC components and the cryostat membrane. 

Previous large \dwords{lartpc} (e.g., \dword{icarus} and \dword{microboone}) have used continuous stainless steel tubes as their \dword{fc} electrodes;
however, a continuous electrode in a DUNE \dword{detmodule} would need to be at least \SI{140}{\m} long. This would increase the stored energy in each electrode and, in turn, increase the risk of damage in the case of a discharge. 

Subdividing the \dword{fc} into electrically isolated modules limits the stored energy in each \dword{fc} module, thereby minimizing the risk of damage. Each \dword{fc} module must have its own voltage divider network to create a linear voltage gradient. Dividing the \dword{fc} into mechanically and electrically independent modules also eases the construction and assembly of the \dword{fc} and greatly restricts the extent of drift field distortion caused by a resistor failure on the divider chain of a \dword{fc} module.

An \dword{hv} discharge onto a metallic cathode could cause the electrical potential of the entire cathode surface to swing from its nominal bias (e.g., $-$\sptargetdriftvoltpos) to \SI{0}{V} in a few nanoseconds, inducing a large current into the analog \dword{fe} amplifiers connected to the sensing wires on the \dword{apa}s (mostly to the first induction wire plane channels). 
An internal study\cite{bib:docdb1320} has shown that with a metallic cathode structure, an \dword{hv} discharge could swing the outer wire plane by nearly \SI{100}{V} and inject \SI{0.9}{A} current into the input of the \dword{fe} amplifiers connected to the first induction plane, possibly overwhelming the internal electrostatic discharge (\dword{esd}) protection in the \dword{fe} \dwords{asic}.  

On the other hand, a highly resistive cathode structure can significantly delay the change in its potential distribution in a discharge event due to its large distributed \dword{r-c} time constant. Such a delay reduces both the current flowing through the discharge path and the current induced on the anode readout amplifiers.  The upper limit in the cathode surface resistivity is 
determined by the voltage drop between the center and the edges of the cathode array driven by the ionization current flowing to the cathode.  For example, a surface resistivity of \SI{1}{\giga\ohm}/square  will have a voltage drop less than \SI{1}{V} from the $^{39}$Ar ionization flux at the underground site. Figure~\ref{fig:cpa-frame-discharge} illustrates the two main 
benefits in such a design in an event of \dword{hv} discharge at the edge of the cathode: (1) reducing the rate of transfer of the stored energy in the cathode plane to reduce the risk of damage to the \dword{hvs} and cryostat membrane; and (2) slowing down the change in cathode voltage distribution that capacitively injects charge into the readout electronics.
With a surface resistivity of \SI{1}{\giga\ohm}/square on the entire cathode, the time constant of a discharge is on the order of a few seconds. An \dword{hv} discharge on the edge of the cathode would inject a maximum current of only about 50\,$\mu$A into the \dword{fe} \dwords{asic}, avoiding damage.

\begin{dunefigure}[Simulated CPA discharge event]
{fig:cpa-frame-discharge}
{Simulated discharge event on a highly resistive cathode surface with a surface resistivity of \SI{1}{\giga\ohm}/square. Top:  stored energy on the cathode as a function of elapsed time from an \dword{hv} discharge. 0.2 second after the discharge, only about 15\% of the stored energy contributes to the discharge. Bottom: voltage distribution on a section of the cathode (2.3\,m\,$\times$\,12\,m) 0.2\,s after the discharge at the upper right edge.  Due to the long time constant of the cathode, most of the surface area remains at the \SI{-180}{\kV} operating potential. Only the region close to the discharge site shifts positively toward 0V. Charge injection to the wire readout electronics, proportional to $dV/dt$ averaged over the cathode area facing an \dword{apa}, is therefore greatly suppressed. }
\centering
\includegraphics[width=0.7\textwidth,trim=2mm 2mm 2mm 2mm,clip]{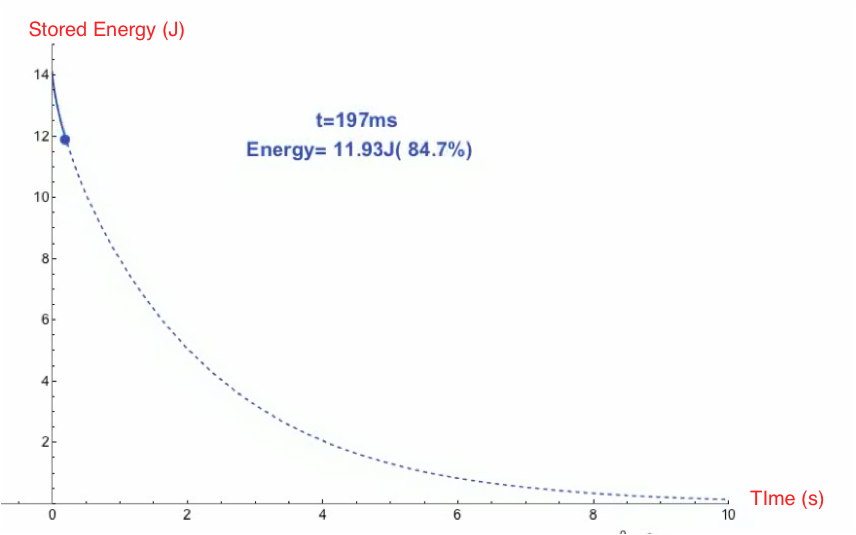} \\ \vspace{20pt}    
\includegraphics[width=0.6\textwidth,trim=2mm 2mm 2mm 2mm,clip]{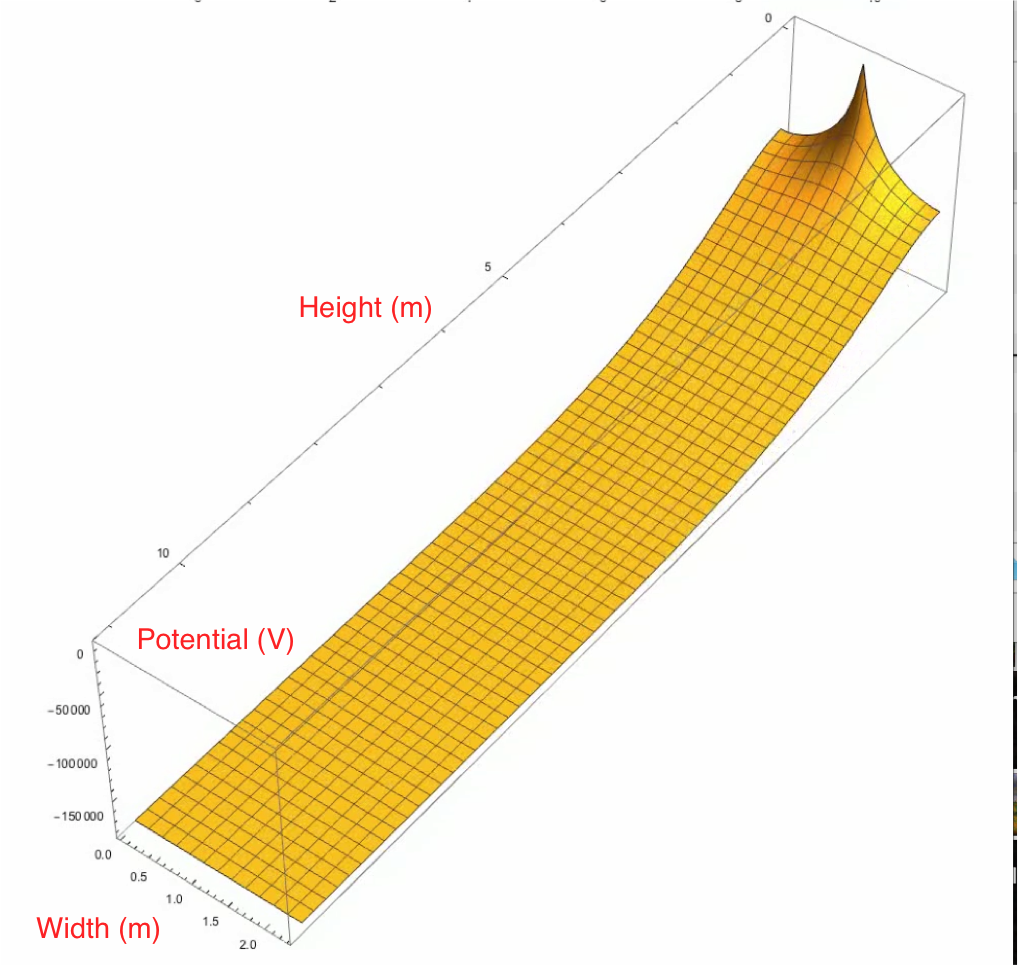}
\end{dunefigure}

\subsubsection{Structural Considerations}
\label{sec:fdsp-hv-des-des-sc}

The frames around the \dword{cpa} panels and the frames supporting the \dword{fc} aluminum profiles  
are made from materials with similar thermal expansion coefficients, minimizing issues of differential thermal expansion. The \dword{fc} frames 
are restrained at only one location.  The \dword{cpa}s and \dword{apa}s support the \dword{topfc} and \dword{botfc} modules, whereas installation rails above the \dword{cpa}s and \dword{apa}s support the \dword{ewfc} modules. 

All structural members of the \dword{cpa}s and \dwords{fc} are made of either \frfour or \dword{frp} with very similar coefficients of thermal expansion (\dword{cte}). However, the structures supporting the \dword{cpa}s and \dwords{fc} are made of stainless steel, with a \dword{cte} about 50\% greater.  To accommodate the mismatch in the \dwords{cte}, small expansion gaps are added between \dword{cpa}s at installation time. These gaps are set during installation between \dword{cpa} panels by adjusting the distance between the \dword{cpa} hanger bars and between \dword{cpa} planes at the top of the \dword{tpc} on the \dword{cpa} beam; these \SI{3}{mm} gaps, 49 of them in total, will disappear once the \dword{tpc} is submerged in \dword{lar}. 
 
\subsubsection{Design Validation}
\label{sec:fdsp-hv-des-des-val}

Successful \dword{protodune} running and extensive testing has 
validated the mechanical and electrical properties of materials selected for the \dword{hv} system.  These are fully documented in references~\cite{bib:docdb2338, bib:docdb1504, bib:docdb1601}. More details follow in Section~\ref{sec:fdsp-hv-protodune}.

Issues identified in earlier testing form the basis of an ongoing R\&D program. 

Operations experience from \dword{pdsp} is summarized in Section \ref{sec:fdsp-hv-protodune}. It revealed some instabilities in the \dword{hvs} operations.  Design changes (see Section \ref{sec:fdsp-hv-des-fc-gp}) have been introduced to the top and bottom \dword{fc} assemblies to further decrease the overall \efield between the profiles and the \dwords{gp}.

\subsection{HV System Safety}
\label{fdsp-hv-design-safety}

Safety is central to the design of the \dword{hv} system and is the highest priority concern in all phases: fabrication, installation, and operations. Documentation of assembly, testing, transport, and installation procedures is in progress and systematically catalogued. 
Particular attention was paid to these procedures in the design and construction of \dword{pdsp}, with the explicit understanding that they be applicable to the \dword{spmod}. The most critical procedures are also noted in the current \dword{hv} risk assessment.

The structural and electrical designs for the \dword{spmod} \dword{hv} are closely modeled on designs that were vetted and validated in the \dword{pdsp} construction. 
Prior to \dword{pdsp}, a full-voltage and full-scale \dword{hv} feedthrough, power supply, filtering, and monitoring system were tested at \dword{fnal}, along with the \dword{hv} connection cup and arm,  
after completing full safety reviews. 
These devices worked as designed and were used in \dword{pdsp}. They will be reproduced for the \dword{spmod},  except for specific optimizations described in this chapter. 

At full operating voltage, the \dword{fc} stores a substantial amount of energy.
As discussed in Section~\ref{sec:fdsp-hv-des-elec}, the \dword{cpa} is designed to limit the power dissipated during a power supply trip or other failure that unexpectedly drops the \dword{hv}.
Its design has succeeded in tests at full voltage over \num{2}\,m$^2$ surfaces and at larger scale in \dword{pdsp}.  

Integral to the \dword{pdsp} and \dword{spmod} design is the concept of pre-assembled modular panels of field-shaping conductors with individual voltage divider boards. The structural design and installation procedures used in \dword{pdsp} were selected to be compatible with use at the \dword{fd} site and were vetted by project engineers, engineering design review teams, and safety engineers at the \dword{cern}. Any revisions to these designs based on lessons learned in \dword{pdsp}  installation and operations will be reviewed both within the project and by \dword{fnal} \dword{esh} personnel. The safety features of the overall design are on solid footing. 

\section {HV Power Supply and Feedthrough}
\label{sec:fdsp-hv-des-hvps}

The \dword{hv} delivery system consists of
\begin{itemize}
\item two power supplies,
\item \dword{hv} cables,
\item filter resistors, and
\item \dword{hv} feedthrough into the cryostat.
\end{itemize}

For \dword{hv} delivery, two power supplies generate the voltage, one for each \dword{cpa} array. 
This separated setup accommodates any necessary different running voltages between the two \dword{cpa} arrays.
The cryostat design has two feedthrough ports for each \dword{cpa} array, one at each end of the cryostat. Correspondingly, two \dword{hv} receiving cups are mounted on the \dword{cpa} array frame. The spare downstream port provides redundancy against any failure of the primary \dword{hv} delivery system. In addition, the \dword{hv} feedthrough is designed to be extracted and replaced in case of misbehavior.

Each \dword{cpa} array separates and services two adjacent drift volumes, 
presenting a net resistance of \SI{1.14}{\giga\ohm} to each power supply. At the nominal \SI{180}{kV} cathode voltage, each power supply must provide \SI{0.16}{mA}. 
The power supply model planned for the \dword{spmod} is similar to that used on \dword{pdsp}.\footnote{Heinzinger, PNC HP300000 \dword{hv} power supply, Heinzinger\texttrademark{} Power Supplies, \url{http://www.heinzinger.com/}.}  
The \dword{hv} cables are commercially available models compatible with the selected power supplies. 

Filter resistors are placed between the power supply and the feedthrough.  Along with the cables, these resistors reduce the discharge impact by partitioning the stored energy in the system.  The resistors and cables together also serve as a low-pass filter reducing 
the \SI{30}{kHz} voltage ripple on the output of the power supply.  With filtering, such supplies have been used successfully in other \dword{lartpc} experiments, such as \dword{microboone} and \dword{icarus}. Figure~\ref{fig:ps_filter_ft_schematic} shows the \dword{hv} supply circuit.

\begin{dunefigure}[Power supply photos and schematic of HV delivery system to the cryostat]  
{fig:ps_filter_ft_schematic}
{Left: Photo of \SI{300}{kV} and \SI{200}{kV} power supplies. 
Right:  A schematic showing the \dword{hv} delivery system to the cryostat. 
One of the two filter resistors sits near the power supply; the other sits near the feedthrough.}
\begin{minipage}{\textwidth}
  \centering
 $\vcenter{\hbox{\includegraphics[width=0.25\textwidth]{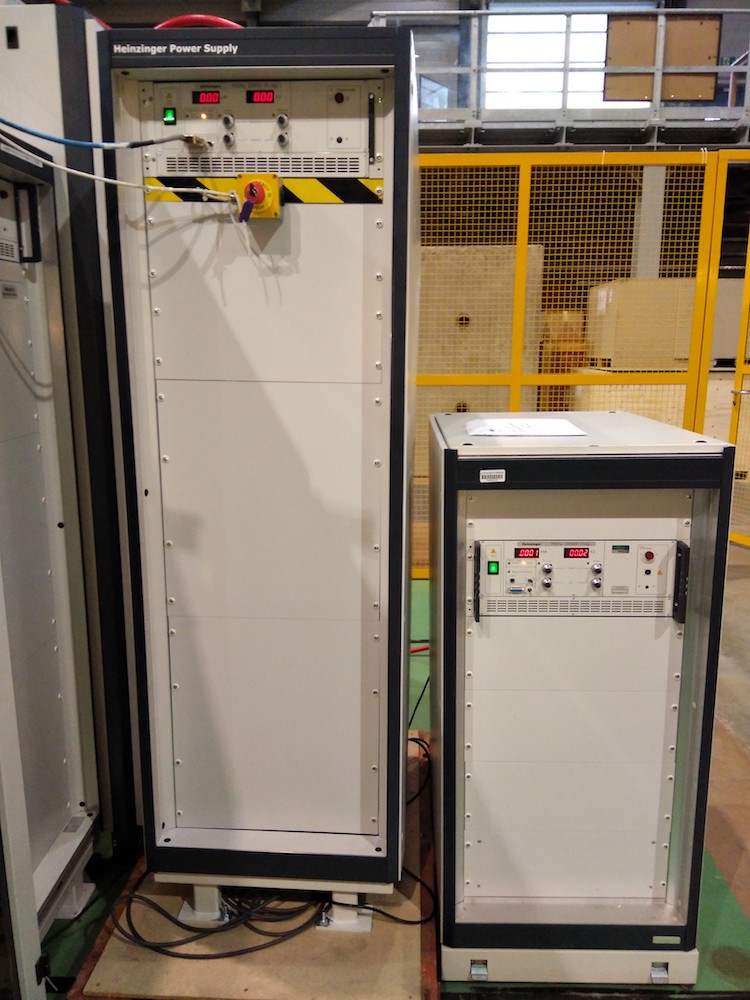}}}$
 \hspace*{0.001\textwidth}  $\vcenter{\hbox{\includegraphics[width=0.7\textwidth]{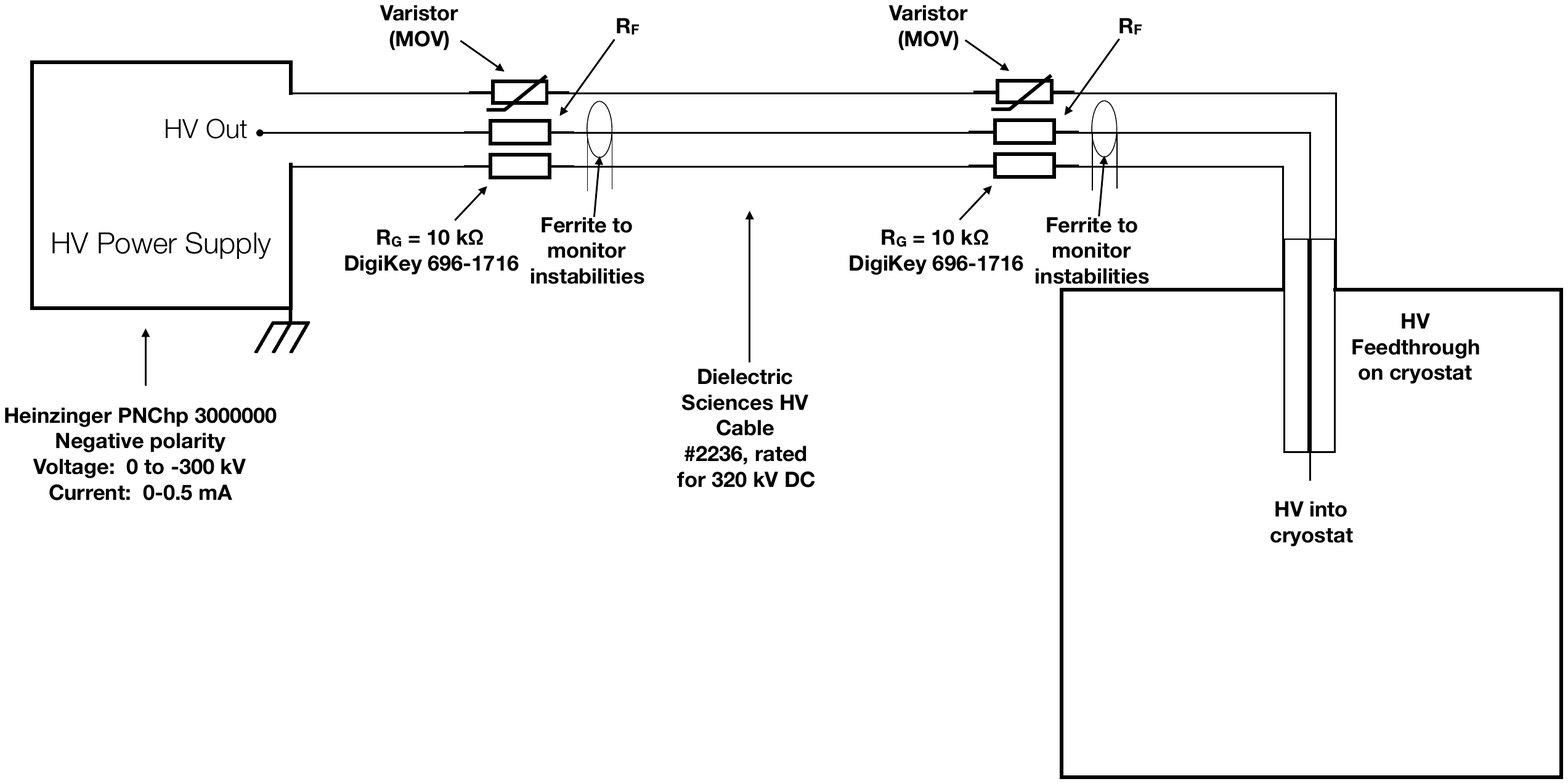}}}$
\end{minipage}
\end{dunefigure}
The requirement  
on low electronics noise sets the upper limit of residual voltage ripple on the cathode to be \SI{0.9}{mV}. 

Typically, commercial supplies specify a ripple variation limit of 
\SI{.001}{\%} around an absolute precision in nominal voltage of $\pm\,$\SI{50}{mV}.
Assuming cable lengths of \SI{30}{m} and \SI{3}{m} between the filters themselves and between the filter and \fdth, respectively, calculations and experience confirm that resistances as low as a few \si{\mega\ohm} yield the required noise reduction. 

The 
filter resistors are of a cylindrical design. 
Each end of a \dword{hv} resistor is electrically connected to a cable receptacle. 
The resistor 
must withstand a large over-power condition.  A cylindrical insulator is placed around the resistor.

The \dword{hv} feedthrough 
is based on the successful ICARUS design \cite{Icarus-T600}, 
which was adapted for \dword{pdsp}.  The voltage is transmitted by a stainless steel center conductor.  On the warm side of the cryostat, this conductor mates with a cable end.  Inside the cryostat, the end of the center conductor has a spring-loaded tip that 
contacts a receptacle cup mounted on the cathode, delivering \dword{hv} to the \dword{fc}.  The center conductor of the \fdth is surrounded by ultra-high molecular weight polyethylene (\dword{uhmwpe}), an insulator. This is illustrated in Figure~\ref{fig:filterAndFeedthrough}.

\begin{dunefigure}[Photo and drawing of a HV  feedthrough]{fig:filterAndFeedthrough}
{Photograph and drawing of a \dword{hv} \fdth{}. Photograph shows \dword{pdsp} installation. The distance from the cup to the top surface is approximately \SI{1.3}{\meter}. } 
\begin{minipage}{\textwidth}
  \centering
 $\vcenter{\hbox{\includegraphics[width=0.16\textwidth]{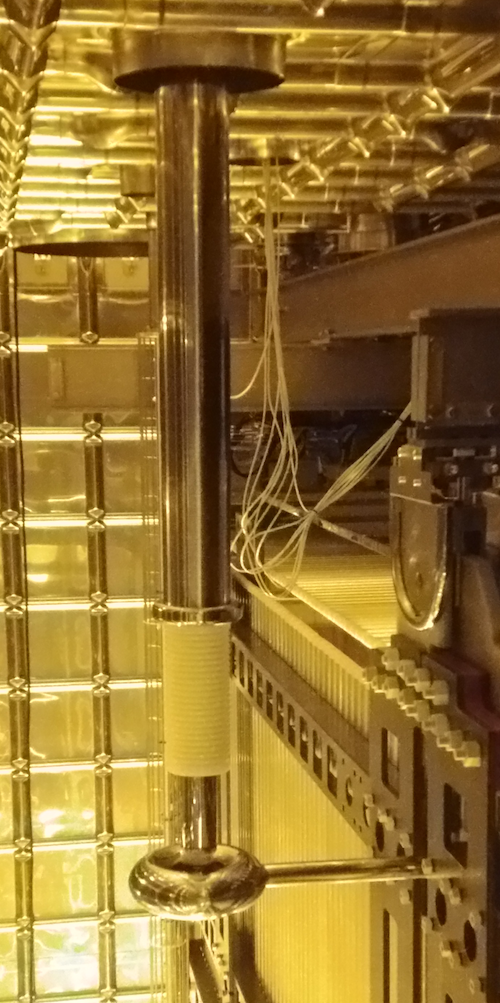}}}$
 \hspace*{0.001\textwidth}  $\vcenter{\hbox{\includegraphics[width=0.82\textwidth]{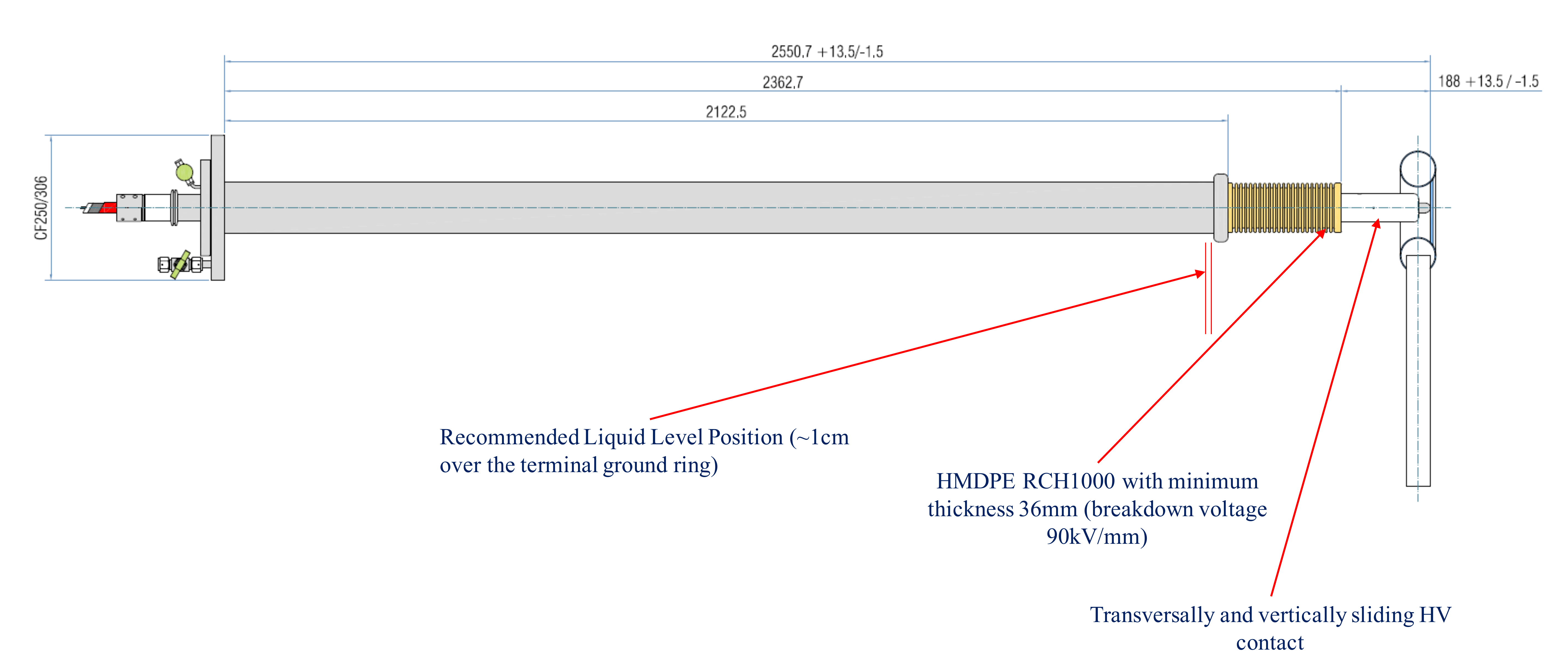}}}$
\end{minipage}
\end{dunefigure}

On a feedthrough, to a first approximation, the operating voltage upper bound is set by the maximum \efield{}. This \efield{} can be reduced by increasing the insulator radius.  For the target voltage, the feedthrough uses a \dword{uhmwpe} cylinder of approximately \SI{15}{cm} diameter.  In the gas space and into at least \SI{15}{\centi\meter} of the liquid, the insulator is surrounded by a tight-fitting stainless steel ground tube.  A 
Conflat industry-standard flange is welded onto the ground tube for attachment to the cryostat.

Outside the cryostat, the \dword{hv} power supply and cable-mounted toroids monitor the \dword{hv}.    The power supplies 
have capabilities down to tens of \si{\nano\ampere} in current read-back 
and are able to sample the current and voltage every \SI{300}{\ms}.  The cable-mounted toroids are sensitive to fast changes in current;  
the polarity of a toroid's signal  
indicates the location of the current-drawing feature as either upstream or downstream of it.  Experience from the DUNE \dword{35t} installation suggests that sensitivities to changing currents 
are on a timescale between \SIrange{0.1}{10}{\micro\s}.

Inside the cryostat, pick-off points near the anode monitor the current  
in each resistor chain.  Additionally, the voltage of the \dwords{gp} above and below each drift region can diagnose problems via a high-value resistor connecting the \dword{gp} to the cryostat.  In the DUNE \dword{35t}, such instrumentation provided useful information on \dword{hv} stability and locations of 
any stray charge flows. 

Both commercial and custom \dword{hv} components must be rated for sufficient voltage and must satisfy tests to meet the specifications 
summarized in Section~\ref{sec:fdsp-hv-des-consid}.  Section~\ref{sec:fdsp-hv-transport-QC} provides further details on these tests.

The resistances in the filters, in combination with the capacitances between the \dword{hv} system and the cathode,
 determine the attenuation of the tens-of-\si{\kilo\hertz} ripple from the power supply.  The filters  
are designed such that the ripple is reduced to an acceptable level when installed in the complete system, thus satisfying specification 
SP-FD-12 that the power supply ripple be negligible. 

\section{CPA Arrays}
\label{sec:fdsp-hv-cpa-arrays}

Two vertical, planar \dword{cpa} arrays held at \dword{hv} each provide constant-potential surfaces at \SI{-180}{\kV}. 
Each \dword{cpa} array also distributes \dword{hv} to the first profile on the top and bottom \dword{fc} and to the \dwords{ewfc}. The configuration of the \dword{cpa} arrays is described in Section~\ref{sec:fdsp-hv-des-cpa}.

\Dwords{rp} form the constant-potential surfaces of each \dword{cpa} unit. The \dwords{rp} are  composed of a thin layer of carbon-impregnated Kapton\footnote{DuPont\texttrademark{}, Kapton\textsuperscript{\textregistered} polymide film,  E. I. du Pont de Nemours and Company,  \url{http://www.dupont.com/}.} laminated to both sides of a \SI{3}{\milli\meter} thick \frfour sheet of \SI{1.2 x 2}{\meter} size.  

A \dword{cpa} array receives its \dword{hv} via the feedthrough that makes contact with the \dword{hv} bus mounted on the \dword{cpa} frame through a cup assembly attached to the frame, as shown in Figure~\ref{fig:cup_cpa}. 
One cup assembly attaches to each end of the two \dword{cpa} arrays, for a total of four. Details on the electrical connections are in Section~\ref{sec:fdsp-hv-design-interconnect}.

\begin{dunefigure}[HV input connection to CPA array]{fig:cup_cpa}{\dword{hv} input cup connection to \dword{cpa} array.}
\includegraphics[width=0.6\textwidth]{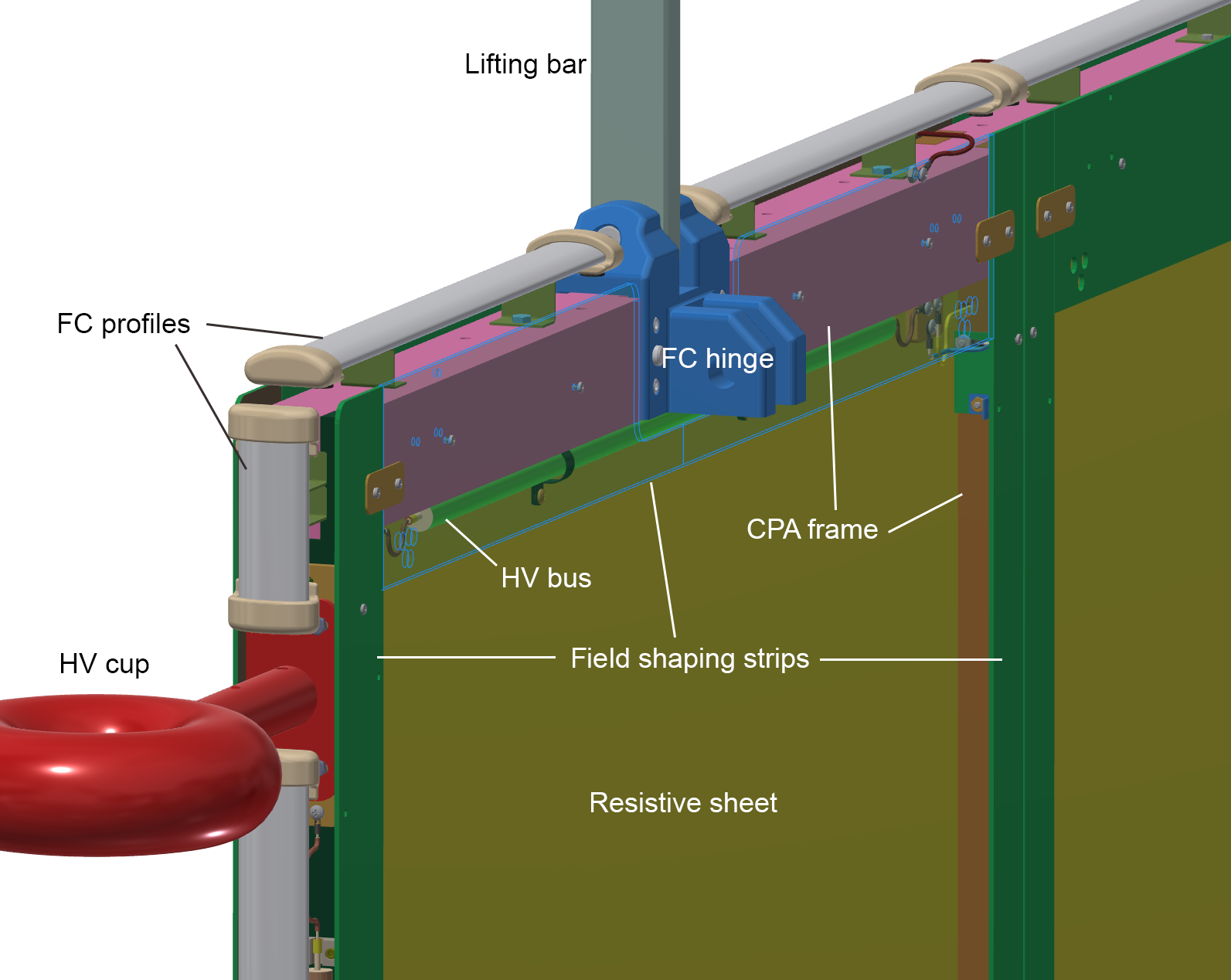} 
\end{dunefigure}

In accordance with specification SP-FD-17, 
the surface resistivity of the \dwords{rp} is required to be greater than \cathodemegohm to provide for slow reduction of accumulated charge in the event of a discharge.  Given the anticipated higher stored energy at the \dword{fd} 
relative to the prototypes, we set a goal of \cathodegigohm to further  protect against potential discharges.  
 
To maintain the position and flatness of the cathode, 
\SI{6}{cm} thick \frfour frames are placed at \SI{1.2}{m} intervals between the \dword{cpa} panels. This design ensures the cathode distortion caused by a small pressure differential (up to \SI{1}{Pa}) across the cathode surface from the convective flow of the \dword{lar} is less than \SI{1}{cm}, meeting the specification of less than \SI{3}{cm}, which causes a field uncertainty of 1\%.

The \dword{cpa} frames are required to support, in addition to the \dword{hv} components, the \dword{topfc} and \dword{botfc} units attached to both sides of the \dword{cpa} plane. The arrangement and deployment of these components is identical to that in \dword{pdsp}.  Figure~\ref{fig:cpa_panel-complete} shows a completed \dword{pdsp} \dword{cpa} panel on the production table ready for lifting into vertical position. 

\begin{dunefigure}[Completed ProtoDUNE-SP CPA panel on production table]{fig:cpa_panel-complete}{Completed \SI{6}{m} long \dword{pdsp} \dword{cpa}  panel on production table.  A \dword{cpa} plane is made up of two panels side-by-side.}
\includegraphics[width=0.7\textwidth]{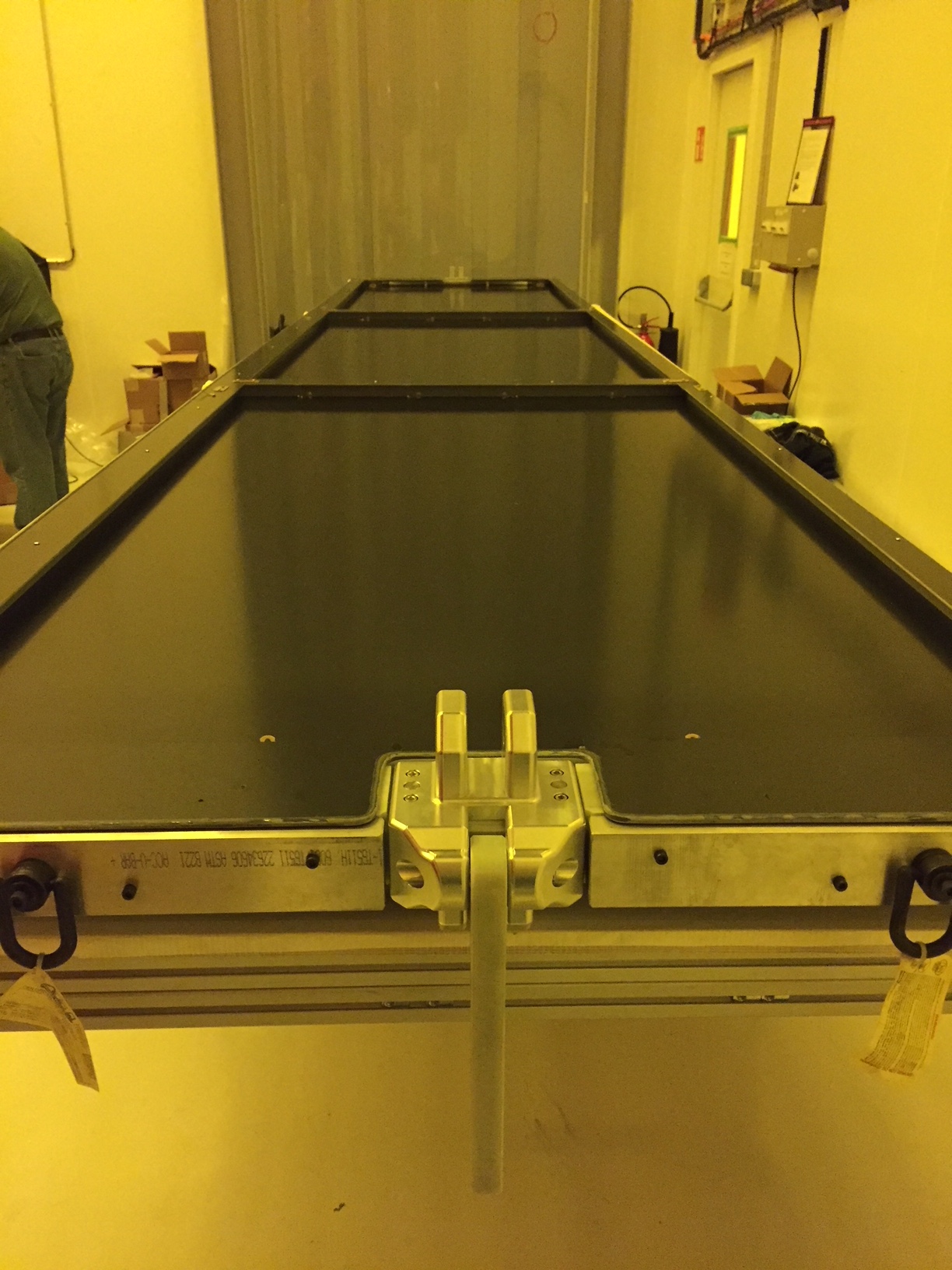}
\end{dunefigure}

 Since \frfour is a good insulator at cryogenic temperatures with a dielectric constant different from that of \dword{lar}, the presence of the \dword{cpa} panel frames causes a local \efield distortion that can become pronounced if the frame surface becomes charged 
from ionization in the \dword{tpc}.  To minimize this distortion, resistive field-shaping strips (\dword{fss}) are placed on the frame and biased at a different potential.  Figure~\ref{fig:fss_concept} illustrates the drift field uniformity improvement with these strips.

\begin{dunefigure}[Benefit of field-shaping strip concept]{fig:fss_concept}{A comparison of three cathode cross sections to illustrate the benefit of the \dword{fss}. Both equipotential lines (horizontal) and \efield{} lines (vertical) are shown.  The amplitude of the \efield{} is shown as color contours. Each color contour is a 10\% step of the nominal drift field.  The gray rectangles represent the frame and the resistive sheet in each case. Left: a conductive/resistive frame similar to that of \dword{icarus} or \dword{sbnd}; Middle: an insulating frame with the insulating surfaces charged to an equilibrium state; Right: an insulating frame covered with \dword{fss} (purple) and biased at the optimum potential. }
\includegraphics[width=0.95\textwidth]{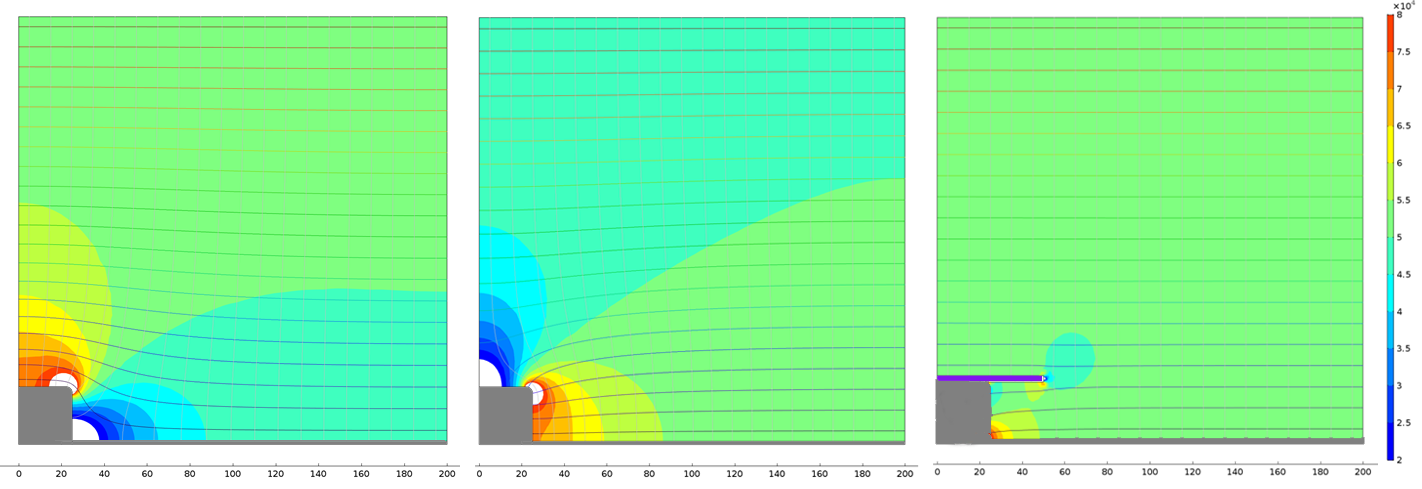} 
\end{dunefigure}

Other \dword{hv} components of the \dword{cpa} arrays include  edge aluminum profiles (to act as the first elements of the \dword{fc}) and cable segments forming the \dword{hv} bus. 

There are at least two instances of electrical connections on the \dword{cpa} array and between the \dword{cpa} array and other \dword{hv} system components (top, bottom, and \dwords{ewfc}), and four connections between \dwords{rp} in a \dword{cpa} unit.  Each of the different types of electrical connections on the \dword{cpa} were tested in a \dword{ln} tank at \dword{bnl}~\cite{bib:docdb2338}  and in \dword{pdsp}. 
No failures occurred at either \dword{bnl} or \dword{pdsp}. The \dword{hv} connection from the \dword{hv} power supply is a closed loop around the \dword{cpa} that can sustain at least one broken connection without loss of the cathode \dword{hv}.  This ensures compliance with requirement~\ref{ spec:hv-connection-redundancy }.

Visual inspection of these items during the assembly process is done to ensure that no accidental sharp points or edges have been introduced. The surface resistivity of the \dword{cpa} \dwords{rp} and the \dword{fss} are checked multiple times during assembly, first when the resistive panels and strips are received and after assembly into \dword{cpa} units on the table.  Coated parts that do not meet the minimum surface resistivity requirement are replaced.  This ensures that requirement SP-FD-17 is satisfied. No discharges were observed in \dword{protodune}, so no additional cryogenic tests are planned for the \dword{cpa}s for DUNE.

\section{Field Cage}
\label{sec:fdsp-hv-des-fc}

The \dword{fc} is introduced in Section~\ref{subsec:fdsp-hv-des-fc}. Its function, basic characteristics, and components are described there. 
The \dword{fc} 
is designed to 
meet the system specifications listed in Section~\ref{sec:fdsp-hv-des-consid}. 
To allow the system to reach the design \efield{} uniformity 
(specification SP-FD-11), 
all components other than the aluminum profiles, \dwords{gp}, and electronic divider boards are made of insulating \dword{frp} and \frfour materials, and the end of each profile is covered with a \dword{uhmwpe} end cap. \\

All voltage divider boards provide redundancy for establishing the profile-to-profile potential differences with only minor distortions to the \efield in case of failure of any individual part, and two redundant boards provide the connection from the \dwords{fc} modules to the \dword{cpa} 
(specification \ref{ spec:hv-connection-redundancy }).  
The aluminum profiles are attached to \dword{frp} pultruded structural elements, including I-beams and box beams.  
Pultruded \dword{frp} material is non-conductive and strong enough to withstand the \dword{fc} loads  in the temperature range of \SI{-150}{C} and \SI{23}{C}, as certified by vendors. Testing of the \dword{frp} joints were conducted at \dword{ln} temperatures~\cite{bib:docdb1504}. 
The material was stronger at these temperatures than at room temperature, 
providing confidence in the material behavior at \dword{lar}  temperature. The \dword{frp} material meets class A standards for fire and smoke development established by the International Building Code characterized by ASTM E84.\footnote{ASTM E84-20, Standard Test Method for Surface Burning Characteristics of Building Materials, ASTM International, West Conshohocken, PA, 2020, \url{https://www.astm.org}.}

The top and bottom \dwords{fc} 
are supported by the \dword{cpa} and \dword{apa} arrays. The \dword{ewfc} modules, 
\SI{1.5}{\m} tall by \SI{3.5}{\m} long, are stacked eight units high (\SI{12}{\m}) and are supported by the installation rails above the \dword{cpa} and \dword{apa} arrays.

\subsection{Field Cage Profiles}
\label{sec:fdsp-hv-des-fc-profiles}

The \dword{fc} consists of modules of extruded aluminum  field-shaping  
profiles, as listed in Table~\ref{tab:fcparts}. The shape of these 
profiles is chosen to minimize the strength of the \efield{} between a given profile and its neighbors and between a profile and
other surrounding parts, including the \dword{gp}. For example, with a \SI{30}{\cm} separation between the \dword{fc} and the \dword{gp}, the maximum \efield{} on the profiles surface is under \SI{10}{\kilo\volt\per\centi\meter} over the straight sections of the profiles at \SI{-180}{\kV} bias (Figure~\ref{fig:profile-e-field}).

\begin{dunefigure}
[\efield{} map and equipotential contours of profiles at \SI{-180}{\kV}]
{fig:profile-e-field}
{\efield{} map (color) and equipotential contours of an array of the \dword{fc} profiles biased up to \SI{-180}{\kV} and a ground clearance of \SI{30}{\cm}.} 
\includegraphics[width=0.8\textwidth]{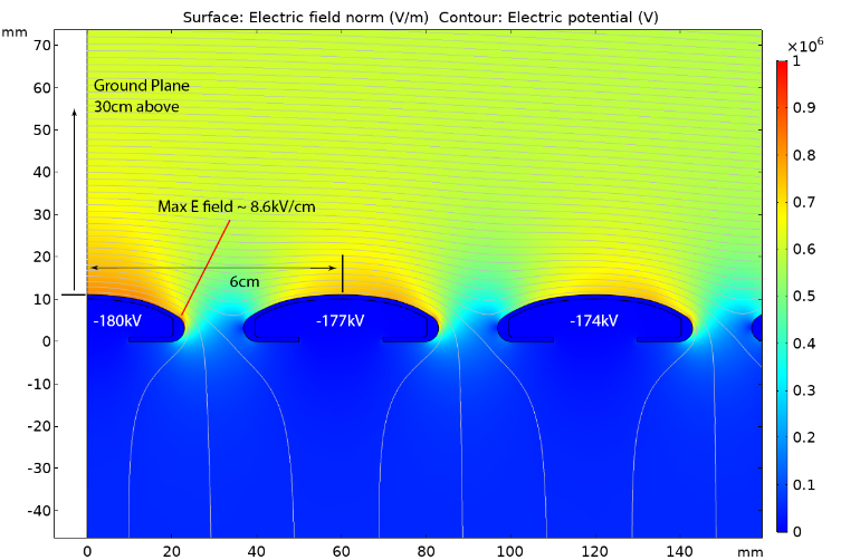}
\end{dunefigure}

The profile ends have higher surface \efield{}, especially those at the corners of the \dword{fc}. To prevent high voltage breakdown in the \dword{lar}, the ends of the profiles are encapsulated by custom \dword{uhmwpe} caps.  These caps are designed and experimentally verified to withstand the full voltage across their \SI{6}{\milli\m} thickness. 

The profiles and their end caps have been carefully modeled to ensure the resulting \efield{}
 does not approach \SI{30}{\kV}/{cm}~\cite{Blatter:2014wua} (specification SP-FD-24). This design concept was validated in a small-scale test setup at \dword{cern} before it was adapted for the \dword{spmod}.  
These features are designed to avoid sparking and thus to draw very small stable currents, 
which should produce a consistent load on the power supply 
(specifications~\ref{ spec:hv-ps-ripple }, \ref{ spec:cathode-resistivity }, 
and~\ref{ spec:power-supply-stability }. 

\subsection{Ground Planes}
\label{sec:fdsp-hv-des-fc-gp}

For safe and stable operation of the \dword{lar} cryogenics system, the cryostat requires a small fraction of its volume to be filled with gaseous argon. This small volume is commonly referred to as the ullage. To optimize use of the \dword{lar} in the cryostat, we will place the upper \dword{fc}, which forms the top boundary of the \dword{tpc}, just below the liquid surface.

The ullage contains many grounded 
conducting components with sharp features, near which the \efield could easily exceed the breakdown strength of gaseous argon if directly exposed to the upper \dword{fc}. 
To shield the high \efield from entering the gas ullage and thereby prevent such breakdowns, 
\dfirsts{gp}, 
in the form of tiled, perforated stainless steel sheet panels, are mounted on the outside surface of the 
\dword{topfc} module with a \SI{30}{cm} clearance. While critical over the region near the cathode, the need for such shielding diminishes toward the \dword{apa} end of the \dword{fc} due to the lower voltages on the \dword{fc} profiles in that region. 
The 30\,cm \dword{fc}-\dword{gp} distance represents a 50\% increase over the value used in \dword{pdsp}, to further reduce the maximum \efield in the \dword{tpc} and thus the possibility of discharges. The 20\,cm distance in \dword{pdsp} was due to an early DUNE design, where 20\,cm was the maximum possible distance that could maintain the \dword{gp} below the liquid level. With the current cryostat and \dword{spmod} design, more space is available, allowing an increase in the \dword{fc}-to-\dword{gp} distance. 
 
In addition to the increase in \dword{fc} to \dword{gp} clearance, we are also eliminating most of the insulating standoffs used in \dword{pdsp} that support \dword{gp} tiles from the \dword{fc} I-beams, in particular, those near the \dword{cpa} end of the \dword{fc}.  These standoffs  are deemed at risk of aiding discharges by providing a short path from \dword{fc} to \dword{gp} along corresponding straight edges.  Figure~\ref{fig:new-fc} illustrates the new configuration. Figure~\ref{fig:new-fc_mockup} shows a test stand built to demonstrate the coupling between \dword{fc} and \dword{gp}, with the standoffs near the cathode end removed. Tests in \dword{ashriver} are confirming that no changes in the assembly or deploying procedures are needed and that the mechanical stability of the full system is unaffected. An upcoming review will examine the design changes and related tests and calculations. Final validation of the complete \dword{hv} design for the \dword{spmod} will be performed in future \dword{protodune} running.

\begin{dunefigure}
[Current baseline FC+GP module design; changes relative to ProtoDUNE-SP]
{fig:new-fc}
{Comparison between the \dword{fd} \dword{topfc} module (top) and the \dword{pdsp} counterpart (bottom).  The changes to the \dword{cpa} side standoffs in the \dword{fd} version are highlighted in red circles.  The increase in the \dword{fc} to \dword{gp} separation is also shown here.} 
\includegraphics[width=0.9\textwidth]{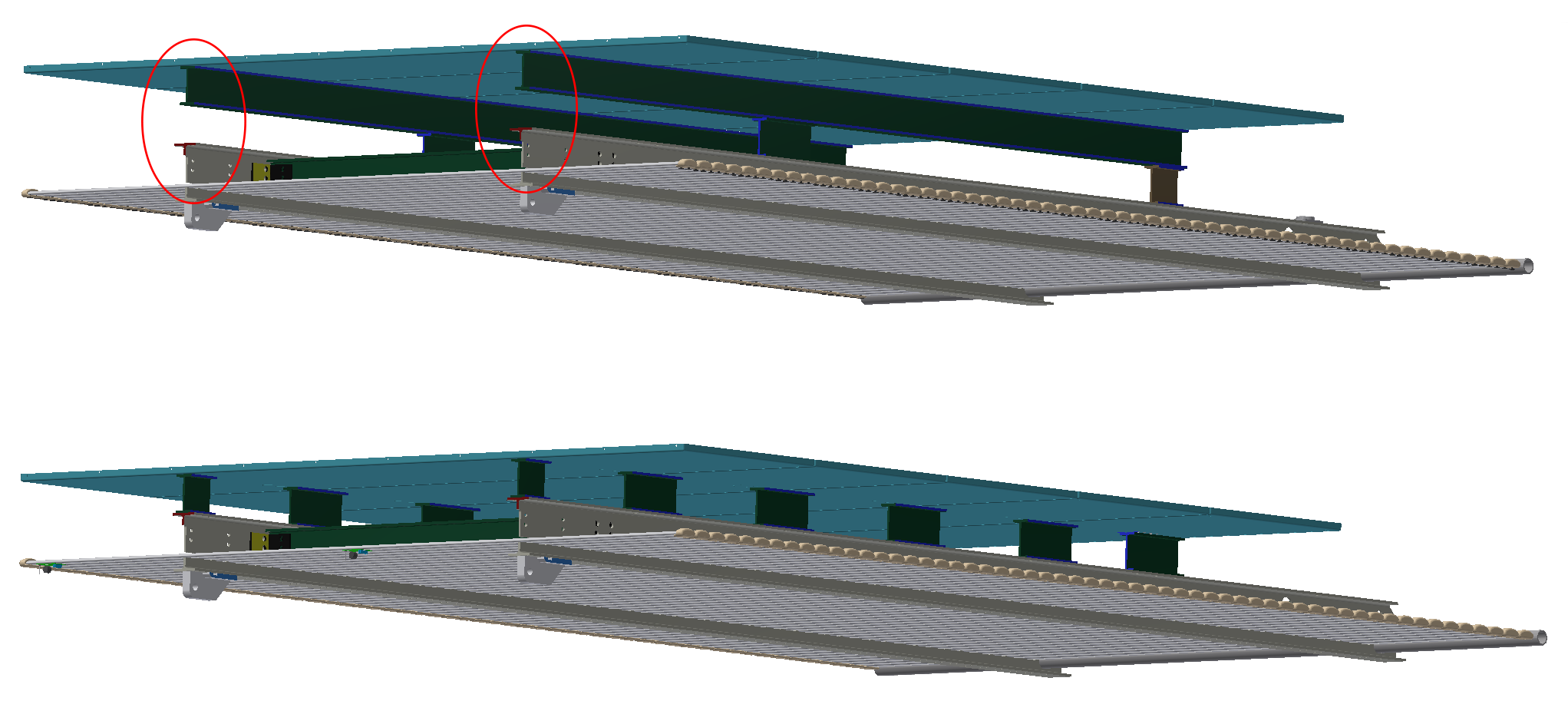}
\end{dunefigure}

\begin{dunefigure}
[Coupling between \dshort{fc} and \dshort{gp}]
{fig:new-fc_mockup}
{Photos of a test module demonstrating the coupling between the  \dword{fc} and \dword{gp}, with the standoffs near the cathode end (towards the right) removed.} 
\includegraphics[width=0.48\textwidth]{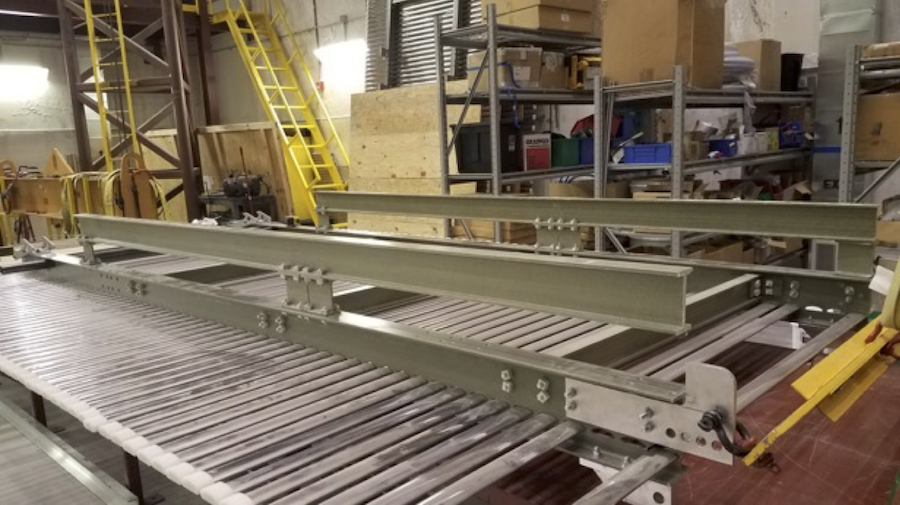}
\includegraphics[width=0.48\textwidth]{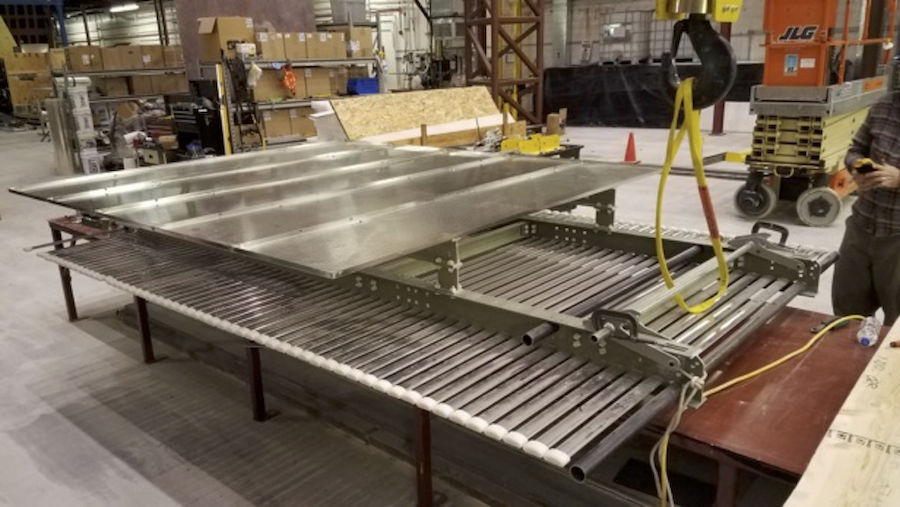}
\end{dunefigure}

On the bottom of the cryostat a similar set of \dwords{gp} 
will protect against 
breakdown in the liquid near cryogenic pipings and other sensors with sharp features. The same clearance will be used. No \dwords{gp} are planned beyond the two \dwords{ewfc} since there is sufficient clearance in those regions.

\subsection{Maximum Field Distortions}
\label{sec:fdsp-hv-des-fc-mfd}

The \dwords{fc} are designed to produce a uniform \efield with understood characteristics.
The largest known \efield distortion in the \dword{tpc} occurs around a large gap in the \dword{fc} between the endwall module and its neighboring top and bottom modules. This gap is necessary to allow the top and bottom modules to rotate past the \dword{ewfc} during deployment.  Figure~\ref{fig:fc-distortion} illustrates the extent of the distortion in this limiting scenario. 
In the \dword{spmod}, a total \dword{lar} mass of \SI{160}{kg} along these four edges of the \dword{tpc} suffers $>\,\SI{5}{\%}$ \efield distortions.  If the non-uniformity is not accounted for in reconstruction, this will result in uncertainties in $dE/dx$ in these regions exceeding 1\%. 

\begin{dunefigure}[\efield at edge of field cage]
{fig:fc-distortion}
{\efield at a corner between the bottom and endwall \dword{fc} modules, showing effects of a \SI{7}{cm} gap. Left: the extent of \num{5}\% \efield{} non-uniformity boundary (black surface, contains less than \SI{10}{kg} of \dword{lar}) and \num{10}\% non-uniformity boundary (white surface, contains $\sim$\SI{6}{kg} of \dword{lar}) inside the \dword{tpc}'s active volume. The inset is a view from the CAD model.  Right: electron drift lines originating from the cathode surface.}
\includegraphics[width=0.9\textwidth]{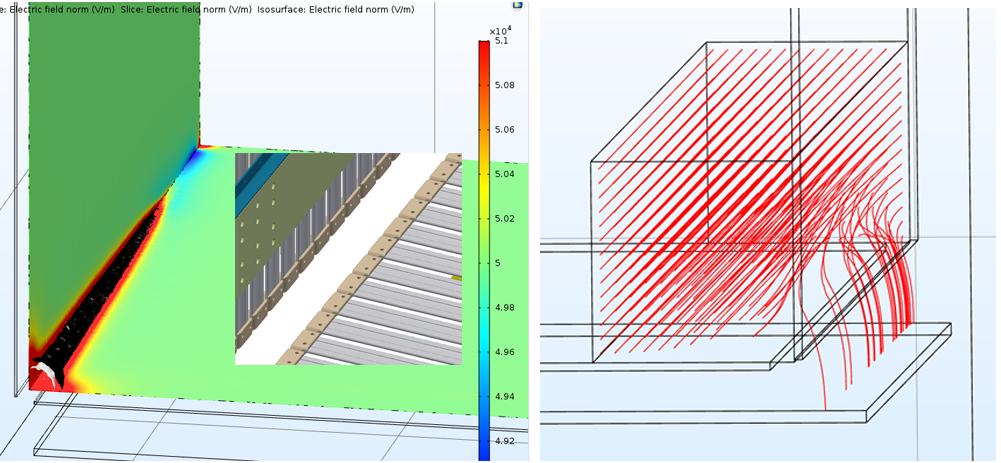}
\end{dunefigure}

\subsection{Top and Bottom Field Cage Modules}
\label{sec:fdsp-hv-des-fc-tbmods}

The \dword{topfc} and \dword{botfc} module dimensions are listed in Table~\ref{tab:fcparts}. The length, \spfcmodlen{}, is set by the length of the two \SI{15.2}{\cm} (\SI{6}\,in) \dword{frp} I-beams that form the primary support structure of the modules. The I-beams are connected to each other by three  \SI{7.6}{\cm} (\SI{3}\,in) \dword{frp} cross beams. The connections between the longitudinal and cross I-beams are made with L-shaped \dword{frp} braces that are attached to the I-beams with \dword{frp} spacer tubes, and secured with \dword{frp} threaded rods, \dword{frp} hex-head nuts, and custom-machined \frfour washer plates.

The \SI{2.3}{\m} module width corresponds to the length of the aluminum profiles, including the UHMW polyethylene end caps. Profiles are secured to the \dword{frp} frame using custom-machined double-holed stainless steel slip nuts that are slid into and electrically in direct contact with the aluminum profiles such that they straddle the webbing of the \SI{15}{\cm} I-beams, and are held in place with screws that penetrate the I-beam flanges. The profile offset with respect to the \dword{frp} frame is different for modules closest to the \dwords{ewfc}, 
and modules in the center of the active volume.

Five \dwords{gp} are connected to the outside (i.e., the non-drift side) of each \dword{topfc} and \dword{botfc} module. The \dwords{gp} are positioned $\sim$\SI{30}{\cm} away from the profiles, and begin at the \dword{cpa} end of the module, leaving the last 14 profiles (\SI{88}{\cm}) on the \dword{apa} end of the module exposed. Between the \dwords{gp} and the \SI{15}{\cm} I-beams standoffs made of short sections of \SI{10.2}{\cm} (4\,in)  \dword{frp} I-beams are connected with \dword{frp} threaded rods and slip nuts. The electrical connection between the \dwords{gp} is made with copper strips.

The connections between the top and bottom modules and the \dword{cpa}s are made with aluminum hinges, \SI{2.54}{\cm} (1\,in) in thickness, that allow the modules to be folded in on the \dword{cpa} during installation. The hinges are electrically connected to the second profile from the \dword{cpa}. The connections to the \dword{apa}s are made with stainless steel latches that are engaged once the top and bottom \dword{fc} modules are unfolded and fully extended toward the \dword{apa}.

The voltage drop between adjacent profiles is established by voltage divider boards screwed into the drift-volume side of the profiles. A custom-machined nut plate 
is inserted into the open slot of each profile and twisted \SI{90}{\degree} 
to lock into position. Two additional boards to connect the modules to the \dword{cpa}s 
screw into the last profile on the \dword{cpa} end of the module. This system is 
more fully described in Section~\ref{sec:fdsp-hv-design-interconnect}. A fully assembled module is pictured in Figure~\ref{fig:tbfc1-2}.

\begin{dunefigure}[Fully assembled \dshort{topfc} module with \dshort{gp}]{fig:tbfc1-2}{A fully assembled \dfirst{topfc} module with \dfirst{gp} is shown.} 
\includegraphics[width=0.45\textwidth]{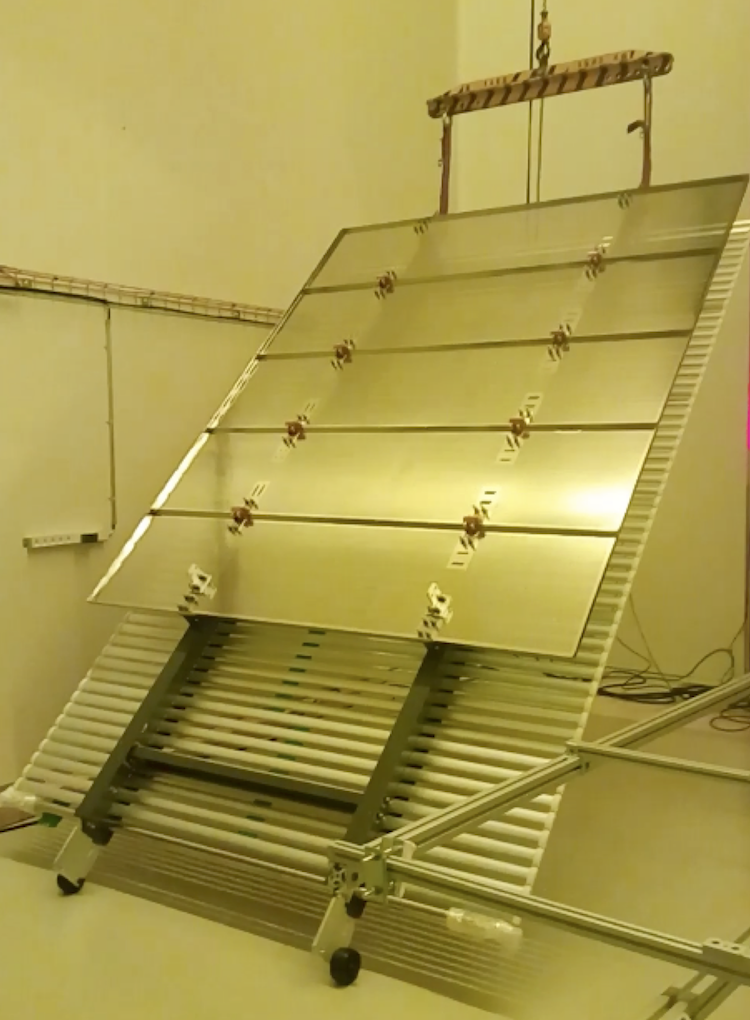}
\end{dunefigure}

\subsection{Endwall Field Cages (EWFC)}
\label{sec:fdsp-hv-des-fc-ewfc}

Each of the four drift volumes has two \dwords{ewfc}, one on each end. Each \dword{ewfc} is in turn composed of eight \dword{ewfc} modules.  The two endwalls are identical in construction, and are installed with an 180$^\circ$ rotation front to back.
Figure~\ref{fig:fc_endwall_panels} illustrates the layout for the topmost 
and the other panels, respectively.

Each \dword{ewfc} module is constructed of two \dword{frp} box beams, each \SI{3.5}{\m} long, as shown in \ref{fig:fc_endwall_panels}. 
The box beam design also incorporates cutouts on the outside face to minimize charge build up. Box beams are connected using \SI{1.27}{\cm} (\num{0.5}\,in) thick \dword{frp} plates. The plates are connected to the box beams using a shear pin and bolt arrangement. The inside plates facing the active volume are connected using special stainless steel slip nuts and stainless steel bolts. The field-shaping profiles are connected to the top box beam using stainless steel slip nuts, an \dword{frp} angle, and two screws each that pass through matching holes in the wings of the aluminum profiles. At the bottom box beam, the profiles are pulled against another \dword{frp} angle with a single screw and a slip nut that is held in place by friction.

\begin{dunefigure}[Endwall FC panels]{fig:fc_endwall_panels}{Top: Uppermost module of the \dword{ewfc}. The two G10 hanger plates connect the \dword{ewfc} to the \dword{dss} beams above the \dword{apa}s and \dword{cpa}s. Bottom: regular \dword{ewfc} module. Seven such modules stack vertically with the top module to form the \SI{12}{m} total height.}
\includegraphics[width=0.8\textwidth]{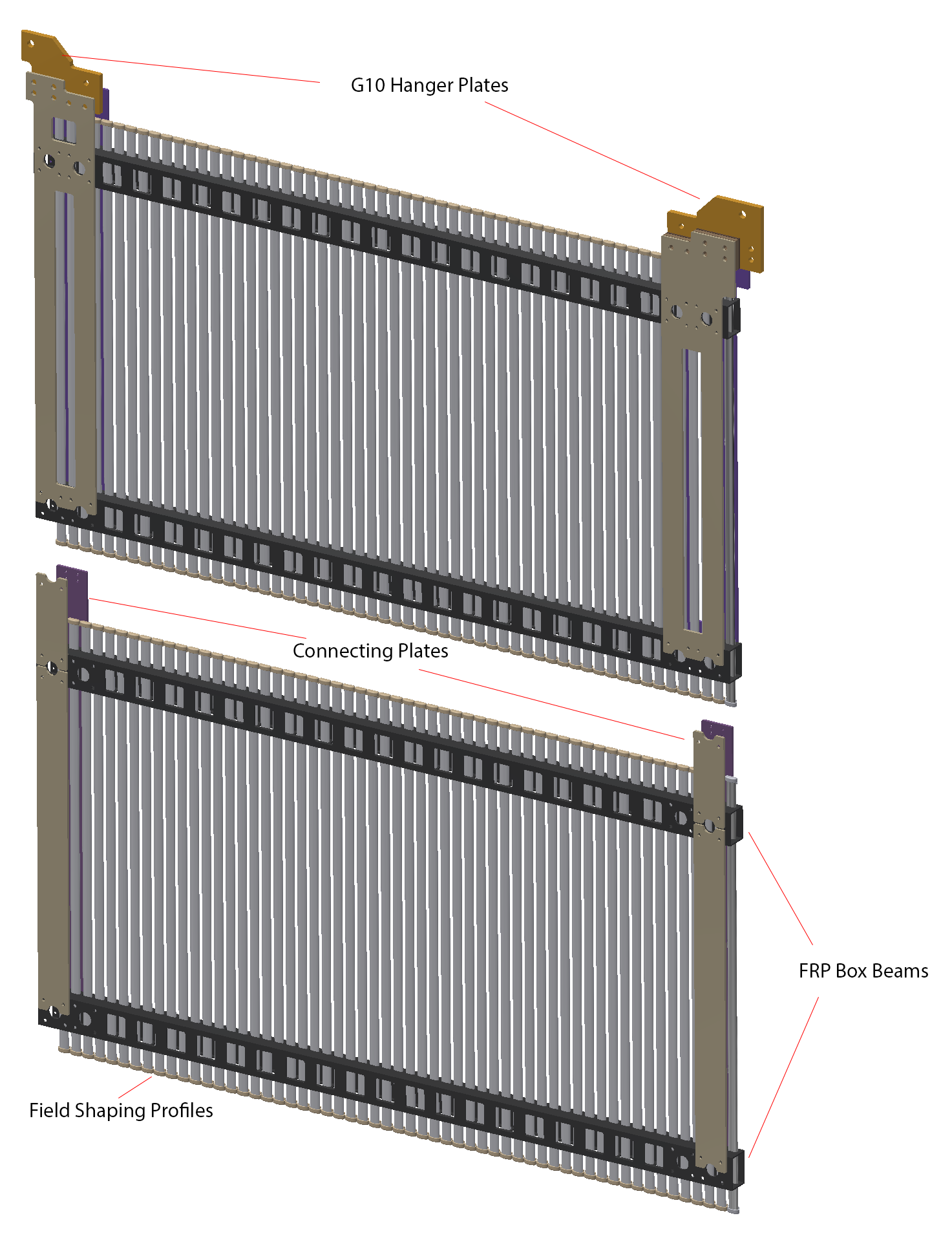}
\end{dunefigure}

\subsection{Voltage Divider Boards}
\label{sec:fdsp-hv-des-fc-vdb}

A resistive divider chain interconnects all the metal profiles of each \dword{fc} module to provide a linear voltage gradient between the cathode and anode planes.

The resistive divider chain is 
 a chain of resistor divider boards each with 
eight resistive stages in series. 
 Each stage (corresponding to \SI{6}{cm} gap between \dword{fc} profiles) consists of two \SI{5}{\giga\ohm} resistors in parallel yielding a parallel resistance of \SI{2.5}{\giga\ohm} per stage to hold a nominal voltage difference of \SI{3}{kV}. Each stage is protected against high voltage discharge transients by transient/surge absorbers (varistors). To achieve the desired clamping voltage, three varistors (with \SI{1.8}{kV} clamping voltage) are wired in series and placed in parallel with the associated resistors. A schematic of the resistor divider board is shown in Figure~\ref{fig:ResDivBoa}; an illustration of the resistor divider board used in \dword{pdsp} is shown as well.
These boards will be identical to the ones successfully mounted in the \dword{pdsp} \dword{fc}. 

\begin{dunefigure}[ProtoDUNE-SP HV resistor divider board]{fig:ResDivBoa}
  {Left: A \dword{pdsp} resistor divider board. Right: Schematic diagram of resistor divider board}
  \includegraphics[width=0.48\textwidth]{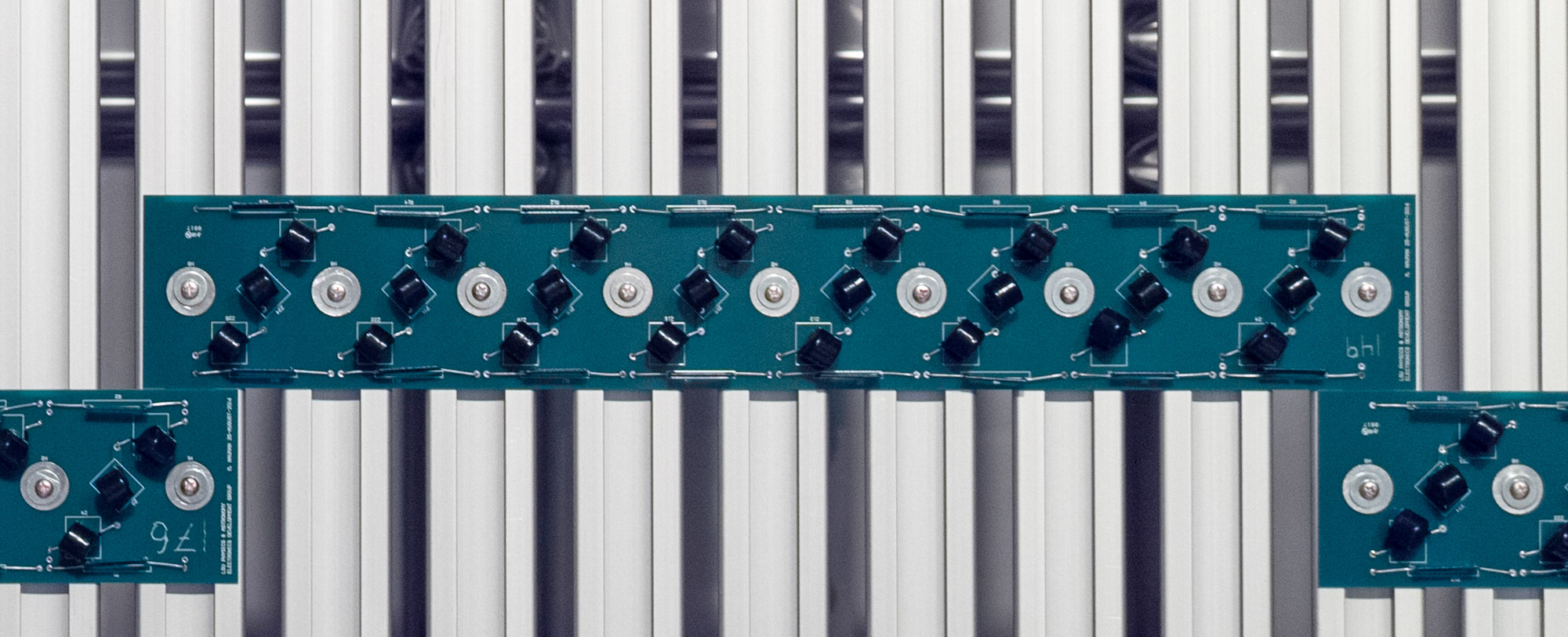}
  \includegraphics[width=0.48\textwidth]{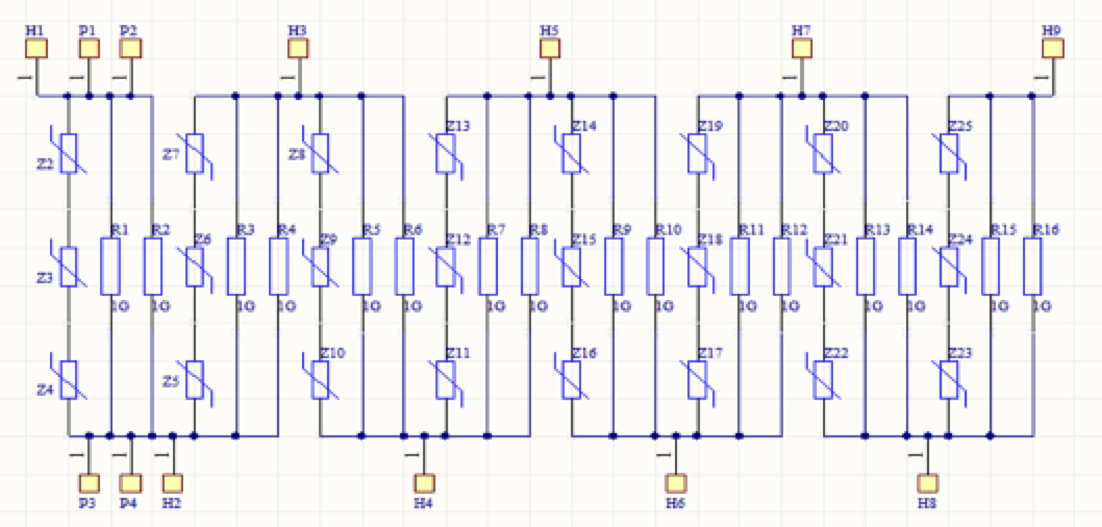}
\end{dunefigure}

The current drawn by each divider chain is about $1.2~\mu A$ at the nominal \efield{} of \SI{500}{V/cm}. A total of 132 resistive divider 
boards are connected in parallel to each \dword{cpa} array for a total of about $158~\mu A$, well within the capability of the selected \dword{hv} power supply.

There are about 30,000 resistors used on the \dwords{fc} in an \dword{spmod}. A resistor failure is a possible risk to the \dword{tpc}.  
An open resistor on the divider chain, the most common failure mode, would approximately double the voltage across the remaining resistor to \SI{6}{kV}.  This larger voltage would force the three varistors in parallel to that resistor into conduction mode, resulting in a voltage drop of roughly \SI{5}{kV} (\SI{1.7}{kV} $\times$ \num{3}), while the rest of the divider chain remains linear, with a slightly lower voltage gradient. 
Because the damage to the divider would be local to one module, its impact to the \dword{tpc} drift field is limited to region near this module, a benefit of the modular \dword{fc} design.
An example of a simulated \efield{} distortion that would be caused by a failed resistor is shown in Figure~\ref{fig:fc-broken-resistor}. 

\begin{dunefigure}[\efield distortion from broken voltage divider path]{fig:fc-broken-resistor}{Simulated \efield{} distortion from one broken resistor in the middle of the voltage divider chain on one \dword{botfc} module.The benefit of the redundancy scheme is emphasized by the limited extent of the \efield distortions. Left: Extent of \efield{} non-uniformity in the active volume of the \dword{tpc}. The green planes mark the boundaries of the active volume inside the \dword{fc}. The partial contour surfaces represent the volume boundaries where \efield{} exceeds 5\% (dark red, contains less than \SI{100}{kg} of \dword{lar}) and 10\% (dark blue, contains less than \SI{20}{kg} of \dword{lar}) of the nominal drift field. The units are \si{\volt\per\m} in the legend. Right: electron drift lines connecting the \dword{cpa} to \dword{apa} in a 
\dword{botfc} corner.  The maximum distortion to the field line is about \SI{5}{cm} for electrons starting at mid-drift at the bottom edge of the active volume.}
\includegraphics[width=0.9\textwidth]{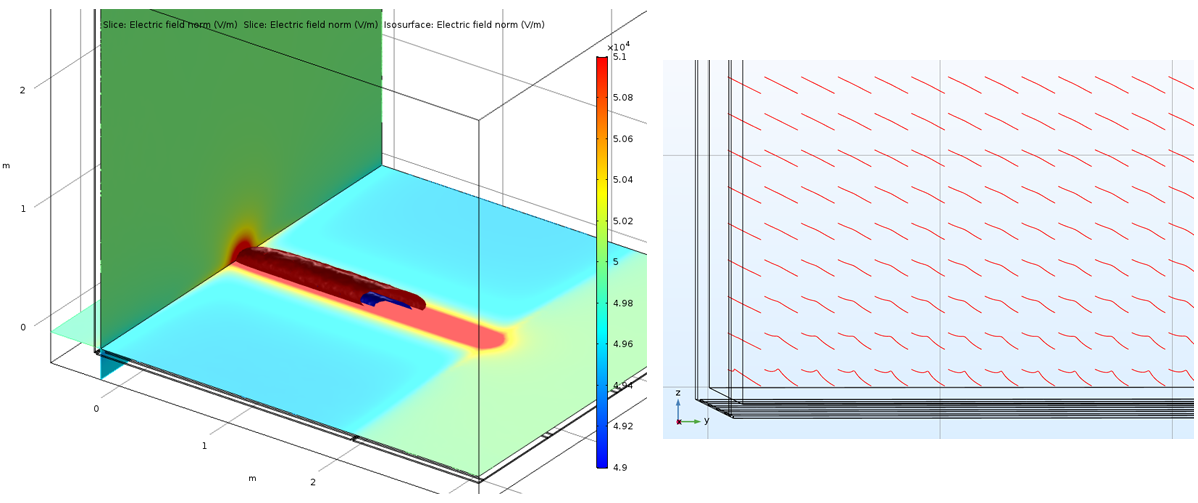}
\end{dunefigure}
The effect of the non-uniformity in resistor values can also be scaled from this study.  A 2\% change in a resistor value (1\% change from the 2R in parallel) would give about 1.5\% of the distortion from a broken resistor, i.e. less than 1\,mm of transverse distortion in track position, with no noticeable drift field amplitude change inside the active volume.
 
\section{Electrical Interconnections} 
\label{sec:fdsp-hv-design-interconnect}

Electrical interconnections are needed among the \dword{hv} delivery system, \dword{cpa} panels, \dword{fc} modules, and termination
boards on the \dword{apa} modules, as well as between resistive dividers and
the field-forming elements on the \dword{cpa}s and \dwords{fc}.  
Redundancy is
needed to avoid single points of failure. 
Some connections must be
insulated in order to avoid creating a discharge path that might
circumvent the discharge mitigation provided by the resistive \dword{cpa}
surface and \dword{fc} partitioning.  Certain connections must be
flexible in order to allow for \dword{fc} deployment, thermal
contraction, and motion between separately supported \dword{cpa}s components.  Figure~\ref{fig:fdsp-hv-design-interconnect-concept} shows a high-level
overview of the interconnections between the \dword{hv}, \dword{cpa}, and \dword{fc} modules.

\begin{dunefigure}[HV interconnection topology]{fig:fdsp-hv-design-interconnect-concept}
  {High-level topology of the \dword{hv} interconnections for one CPA array and adjacent field cages. Each pair of adjacent CPA panels is connected to two top field cage modules and two bottom field cage modules. A high voltage bus supplies the CPA panels at the top and bottom, and also supplies the endwall field cage modules. All field cages are terminated at the \dword{apa}s (not shown).}
  \includegraphics[width=0.7\textwidth]{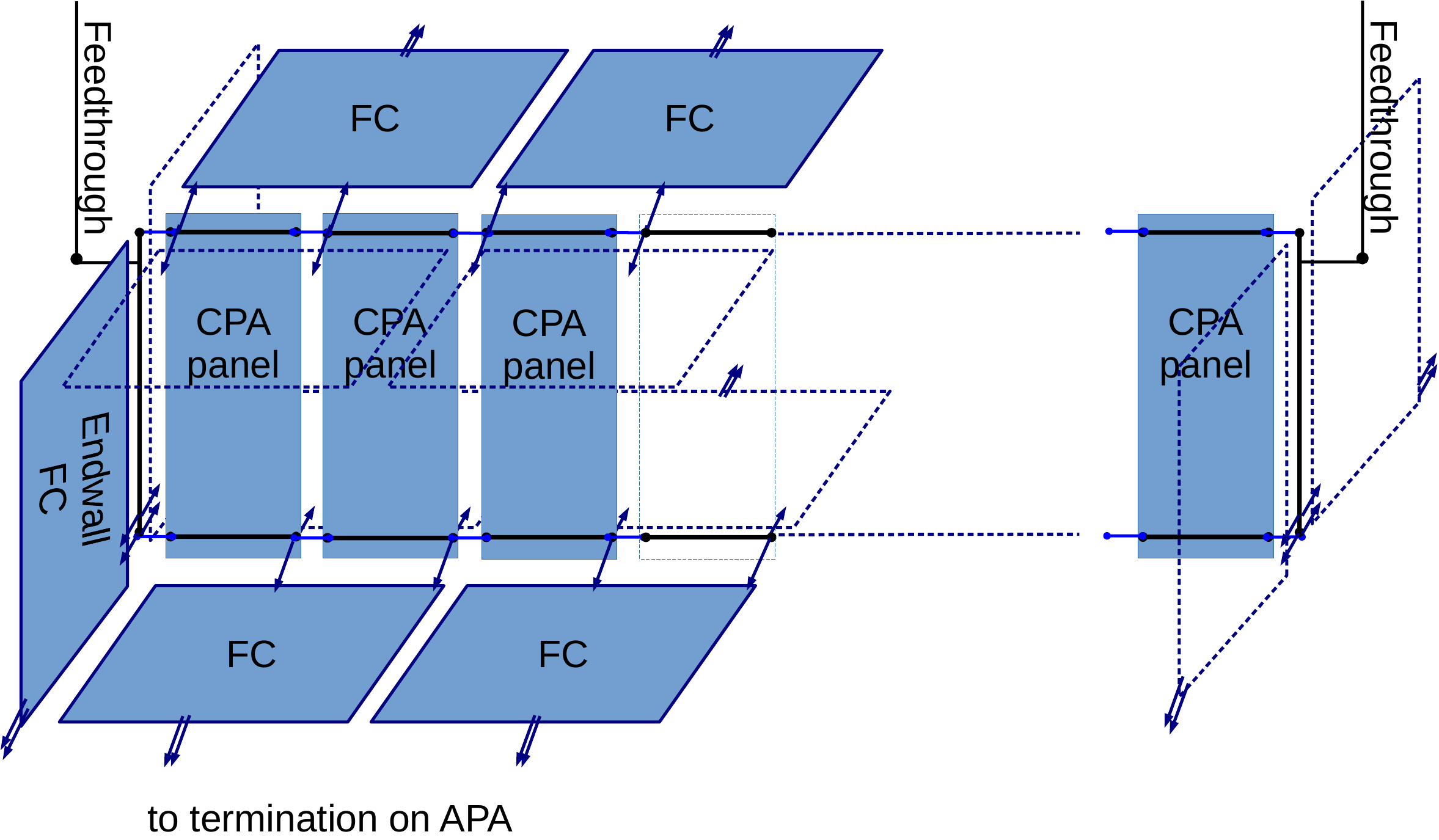}
\end{dunefigure}

High voltage feedthroughs connect to cups mounted on the \dword{cpa} frame
that attach to an \dword{hv} bus running through the \dword{cpa}s.  \Dword{hv} bus connections
between \dword{cpa} panels are made by flexible wires through holes in the
\dword{cpa} frame. The \dword{hv} bus is a loop that mitigates any risk of a single
failure point; the feedthrough at each end of each \dword{cpa} panel mitigate
risk of a double-break failure.  Voltage dividers on each \dword{cpa} panel
bias the \dword{fss} and the resistive dividers on the top
and bottom \dwords{fc}.  The \dword{cpa}-to-\dword{fc} connections are made using
flexible wire to accommodate \dword{fc} deployment.  To further
increase redundancy, two \dword{cpa} panels connect to each top or bottom
field cage, and two connections are also made to each \dword{ewfc}. Resistor divider boards attach directly to the interior side of
the \dword{fc} profiles with screws.   A redundant pair of flexible wires
connects a circuit board on the last profile of each \dword{fc} to a
bias-and-monitoring board mounted on the corresponding \dword{apa}.

Short sections of flexible wire at the ends of each \dword{hv} bus segment
attach to screws in brass tabs on the \dword{cpa} resistive panels (\dword{cpa} \dwords{rp}).
Vertical \dword{hv} bus segments on the outer ends of each \dword{cpa} plane connect
the top and bottom \dword{hv} buses to complete the loop.  Solid wire is used
to connect resistive panels within a \dword{cpa} panel.

Each \dword{fc} module is as electrically independent as possible in order to
mitigate discharge.  However, only the bottom module of each endwall
can make connections to the \dword{hv} bus and \dword{apa}, so each endwall module
is connected to its upper neighbor at its first and last profile
using metal strips.

Each \dword{fc} divider chain connects to an \dword{fc} termination board in parallel to a grounded fail-safe circuit at the \dword{apa} end.  The \dword{fc} termination boards are mounted on the top of the upper \dword{apa}s and bottom of lower \dword{apa}s.  Each board provides a default termination resistance, and an SHV cable connection to the outside of the cryostat, via the \dword{ce} signal feedthrough flange, through which we can either supply a different termination voltage to the \dword{fc} or monitor the current flowing through the divider chain.

All flexible wires have ring or spade terminals and are secured by
screws in brass tabs.  Spring washers are used with every electrical
screw connection in order to maintain good electrical contact with
motion and changes of temperature.

Table \ref{tab:sp-hv-interconnects} summarizes the interconnections required 
for the \dword{hv} system.

\begin{dunetable}
[HV system interconnections]
{p{0.35\linewidth}p{0.62\linewidth}}
{tab:sp-hv-interconnects}
{\dword{hv} system interconnections}   
 Connection & Method \\ \toprowrule
 \dword{hv} cup to \dword{hv} bus & wire to screw in \dword{hv} cup mount on \dword{cpa} frame \\ \colhline
 \dword{hv} bus between \dword{cpa} panels & wire between screws in brass tabs \\ \colhline
 \dword{hv} bus to \dword{fss} & wire to circuit board mounted on \dword{fss} \\ \colhline
 \dword{fss} to \dword{topfc} and \dword{botfc} & wire to circuit board on first \dword{fc} profile, two per \dword{fc} module \\ \colhline
 \dword{hv} bus to endwall \dword{fc} & wire to circuit board mounted on first \dword{fc} profile, two per endwall \\ \colhline
 \dword{fc} divider circuit boards & directly attached to profiles using screws and SS slip nuts \\ \colhline
 \dword{fc} to bias and monitoring termination & redundant wires from board mounted on last \dword{fc} profile \\ \colhline
 \dword{hv} bus to \dword{cpa} panels & brass tab on \dword{cpa} resistive panel \\ \colhline
 \dword{cpa} \dword{rp} interconnections & solid wire between screws in brass tabs \\ \colhline
 Endwall \dword{fc} module interconnections & metal strips, first and last profiles only
 \\ 
\end{dunetable}

The redundancy in electrical connections described above meets requirement~\ref{ spec:hv-connection-redundancy }. 
The \dword{hv} bus and interconnections are all made in low field regions in order to meet requirement \ref{ spec:local-e-fields }  
The \dword{hv} bus cable is rated at the full cathode \dword{hv} such that even in case of a rapid discharge of the \dword{hv} system no current can flow to the cathode or \dword{fc} except at the intended contact points, preserving the ability of the resistive cathode and \dwords{fc} to meet requirement \ref{ spec:cathode-resistivity }. 

\section{ProtoDUNE-SP High Voltage Experience}
\label{sec:fdsp-hv-protodune}

\dword{pdsp} \cite{Abi:2017aow} is a prototype for an \dword{spmod}. 
Approximately one twentieth the size of a \dword{spmod}, this detector implements an A-C-A configuration with one \dword{cpa} array that bisects the \dword{tpc} and two \dword{apa} arrays, one along each side. 
The \dword{cpa} array consists of 
six \dword{cpa} panels, each \SI{1.2}{m} wide by \SI{6.0}{m} high (half-height relative to an \dword{spmod}), 
and is positioned \SI{359}{cm} away from each \dword{apa} array, matching the maximum drift distance of an \dword{spmod}.

Six top and six bottom \dword{fc} modules connect the horizontal edges of the \dword{cpa} and \dword{apa} arrays, and four 
\dwords{ewfc} connect the vertical edges (two per drift volume). One of the drift volumes is pictured in Figure~\ref{fig:protodune_sp_hv}. 
Each \dword{ewfc} comprises four endwall modules (half-height relative to a \dword{spmod}).
A Heinzinger $-$\SI{300}{kV} \SI{0.5}{mA} \dword{hv} power supply delivers voltage to the cathode.
Two \dword{hv} filters in series between the power supply and \dword{hv} feedthrough filter out high-frequency fluctuations upstream of the cathode.

\begin{dunefigure}[Photo of ProtoDUNE-SP drift volume with HV components]
{fig:protodune_sp_hv}
{One of the two drift volumes of \dword{pdsp}. The \dword{fc} modules shown enclose the drift volume between the \dword{cpa} array (at the center of the image) and the \dword{apa} array (upper right). The \dwords{ewfc} are oriented vertically; the top and bottom units are horizontal. The staggered printed circuit boards connecting the \dword{ewfc} profiles are the voltage divider boards. 
}
\includegraphics[width=0.8\textwidth]{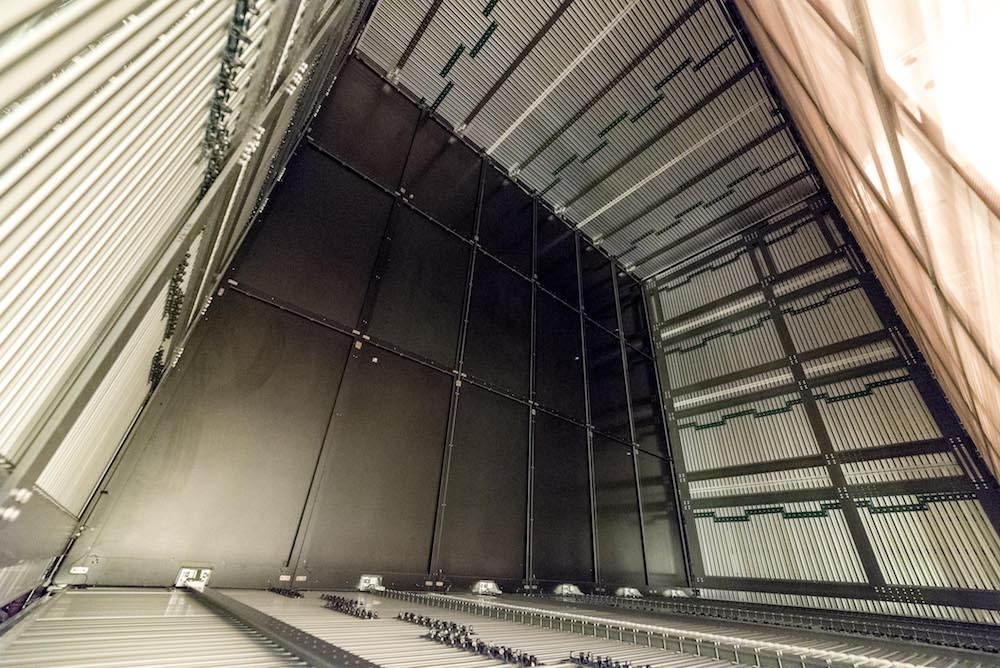}
\end{dunefigure}

\subsection{Summary of HV Construction}
\label{sec:fdsp-hv-protodune-summary}

The \dword{pdsp} \dword{hv} components underwent 
various levels of pre-assembly offsite prior to transport and final assembly in the \dword{pdsp} cleanroom adjacent to the cryostat.

Parts for the top and bottom \dword{fc} frames were procured and test fit at Stony Brook University before being shipped to \dword{cern} for module assembly in a cleanroom about \SI{5}{km} away from the detector hall.
Fully assembled modules were transported individually to the detector hall for storage until installation. \dword{cern} provided the \dwords{gp} for the top and bottom \dwords{fc} as well as the field shaping profiles for all \dwords{fc}.

Louisiana State University (LSU) provided all the voltage divider boards, then procured and test fit the \dword{ewfc} frame parts before shipping them fully assembled to \dword{cern}.
These profiles and the voltage divider boards were installed in the same \dword{cern} cleanroom facility as the other \dword{fc} components.

Argonne National Laboratory shipped the \dword{cpa} material to the detector hall as single pre-assembled resistive panels held in a \frfour frame; i.e., as \dword{cpa} units (Table~\ref{tab:cpaparts}). 
In the cleanroom adjacent to the \dword{pdsp} cryostat, three \dword{cpa} units were mechanically and electrically connected to produce a \dword{cpa} panel. 
The \dword{cpa} panels (one of which is pictured in Figure~\ref{fig:cpa_panel-complete}) were first assembled horizontally and then lifted and rotated to a vertical orientation where they were paired to make a \SI{6.0 x 2.3}{m} \dword{cpa} plane.

At this point, two top and two bottom \dword{fc} modules were brought to the cleanroom to be lifted, rotated to vertical, and attached to the \dword{cpa} plane. 
To fit through the \dword{tco}, the top \dwords{fc} were suspended from their support at the top of the \dword{cpa} plane to hang vertically, and the bottom \dwords{fc} were folded up and temporarily attached to their top \dword{fc} counterparts.
The resulting \dword{cpa}-\dword{fc} assemblies were rolled onto the central bridge beam inside the cryostat and 
deployed. 

Also in the \dword{pdsp} cleanroom, sets of four pre-assembled \dwords{ewfc} 
were each assembled into one \dword{ewfc} plane. 
Although not a component of the \dword{spmod} design, 
the beam plug was installed onto its corresponding module 
before the beam-right, upstream \dword{ewfc} was built.
An electric hoist lifted the top 
module to a height at which the next 
module could be wheeled underneath and connected via \dword{frp} plates.
The hoist then raised the pair, and the procedure would continue in this way until the \dword{ewfc} was four 
modules tall.
The load of the assembled endwall was then transferred to a trolley on a transport beam, which allowed it to be pushed into the cryostat onto the appropriate bridge beam.

The \dword{tpc} components of \dword{pdsp} were installed first for the drift space to the right of the delivered beam (beam-right), and then for the beam-left drift space.
The \dword{apa}s and the \dword{cpa} array were locked into position along their respective bridge beams, and then the bridge beams were locked into their positions along the drift direction.
Next, the two \dwords{ewfc} were moved and rotated into their upstream and downstream positions to bridge the gap between the vertical edges of the corresponding \dword{apa} and \dword{cpa}.
The \dwords{ewfc} loads were transferred onto the \dword{apa} and \dword{cpa}  bridge beams, which freed the intermediate bridge beam for top and bottom \dword{fc} deployment.
Two mechanical hoists were used to lower (raise) the bottom (top) \dword{fc} to bridge the gap between the horizontal edges of the \dword{apa}s and \dword{cpa}s.
Finally, the \dword{hv} cup was connected on the downstream \dword{cpa}, and the \dword{hv} feedthrough was lowered through the cryostat penetration to make contact with the cup.

\subsection{HV Commissioning and Beam Time Operation}
\label{sec:fdsp-hv-commissioning-operation}
During \cooldown and \dword{lar} filling, a power supply was used to supply $-$\SI{1}{kV} to the cathode and monitor the current draw of the system.
As the system cooled from room temperature to \dword{lar} temperature, the resistance increased by $\sim$10\%, consistent with expectations.
Once the \dword{lar} level had exceeded the height of the top \dwords{gp}, the voltage was ramped up to the nominal voltage.

The initial week of \dword{hv} operations showed no signs of any anomalous instabilities. Over the following weeks, the \dword{hv} power supply showed signs of instabilities that affected the quality of the \dword{hv} provided to the cathode plane. Replacement of the power supply  midway through the run resulted in higher stability of the warm side of the \dword{hv} system. The original power supply was sent to Heinzinger for inspection. The malfunctioning was confirmed to be due to unexpected excessive moisture that had accumulated in the \dword{hv} cable socket.

In addition, 
two types of instabilities emerged in the cold side of the \dword{hv} system. The first type was the so-called current blips, during which the system draws a small excessive current that persists for no more than a few seconds. The magnitude of the excess current during such events increased over the subsequent three weeks from 1\% to 20\%. The second type of instability, labeled ``current streamers,'' 
exhibited persistent excessive current draw from the \dword{hv} power supply with accompanying excessive current detected on a \dword{gp} and on the beam plug. These two types of instabilities were experienced periodically throughout the duration of the \dword{pdsp} beam run. The frequency of both types increased over time after the system was powered on, until a steady state of about ten current blips/day and one current streamer  every four hours was reached. These effects are consistent with a slow charging-up process of the insulating components of the \dwords{fc} supports, which then experience partial discharges that are recorded as \dword{hv} instabilities. This process appears to restart after every long \dword{hv}-off period. 

In addition, these processes seem to be enhanced by the \dword{lar} bulk high purity, which allows the electric current to develop. 
At low purity electronegative impurities act as quenchers, blocking the development of the leakage current. Despite the presence of two types of instabilities, the \dword{hv} system was able to consistently achieve >95\% uptime during the beam runs. The downtime was the result of short manual interventions to quench a current streamer (Figure \ref{fig:protoDUNE-SP_beamRunSum_HV}).

In some cases, 
mostly outside of the beam run period, we turned off the \dword{hv} system momentarily to allow the \dword{hv} system components to discharge. This is reflected as larger dips in the uptime plot. During moments when the rest of the subsystems (including the beam) were stable, the moving 12-hour \dword{hv} uptime fluctuated between 96\% and 98\%.

\begin{dunefigure}[ProtoDUNE-SP HV performance during the test beam run]
{fig:protoDUNE-SP_beamRunSum_HV}
{The performance of the \dword{hv} system across the test beam period, September-November 2018. The top panel shows the drift field delivered to the \dword{tpc}, the middle panel indicates \dword{hv} cuts during periods when the system is not nominal (some periods not visible due to their short timescale), and the bottom panel shows the moving 12-hour uptime of the \dword{hv} system based on these \dword{hv} cuts.}
\includegraphics[width=\textwidth]{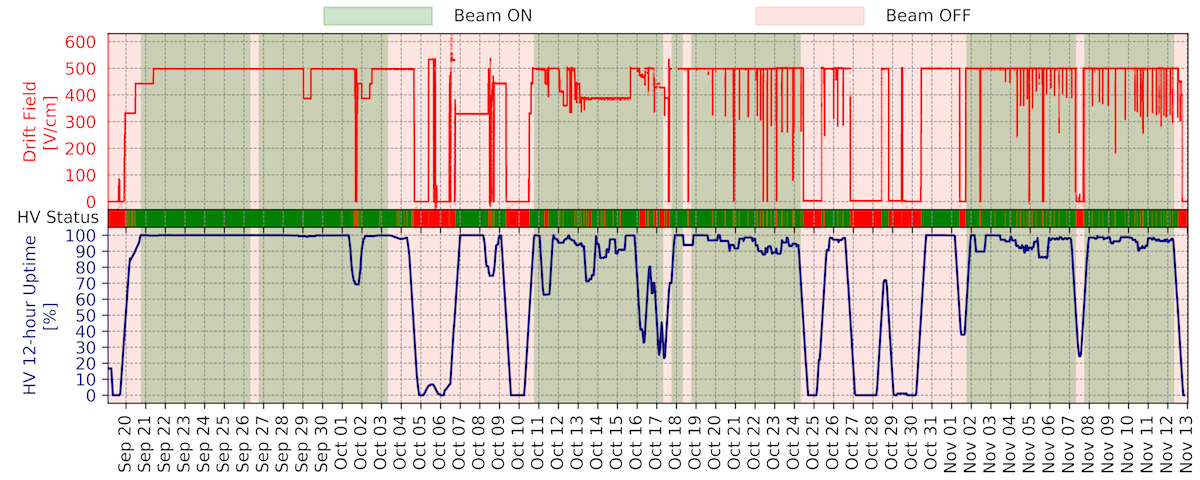}
\end{dunefigure}

The up-time during the week starting October 11th (Oct 11 in Figure \ref{fig:protoDUNE-SP_beamRunSum_HV}) is lower than the subsequent three beam-on weeks because the current streamers were addressed differently in these two periods. In the beginning, they were left to develop until they quenched themselves or until the \dword{hv} was manually ramped down. The \dword{hv} was brought back up when the current draw returned to nominal, according to the \dword{fc} resistance value. Automated controls to quench the current streamers were then successfully implemented in an auto-recovery mode. These helped significantly to increase the up-time, by optimizing the ramping down and up of the \dword{hv} power supply voltage, which was performed in less than four minutes (Figure \ref{fig:HV_autorecovery}).

\begin{dunefigure}[ProtoDUNE-SP HV autorecovery procedure]
{fig:HV_autorecovery}
{Example of the \dword{hv} automatic recovery procedure developed to detect and quench the current streamers: whenever an excess sustained current from the \dword{hv} PS is detected (obtained by continuously monitoring the total detector resistance experienced by the PS), the \dword{hv} delivered by the PS is lowered in discrete steps. At each step the total resistance is checked again, and if it agrees with the nominal detector resistance the \dword{hv} is ramped up again to its nominal value; otherwise the \dword{hv} is lowered to the next step.}
\includegraphics[width=0.5\textwidth]{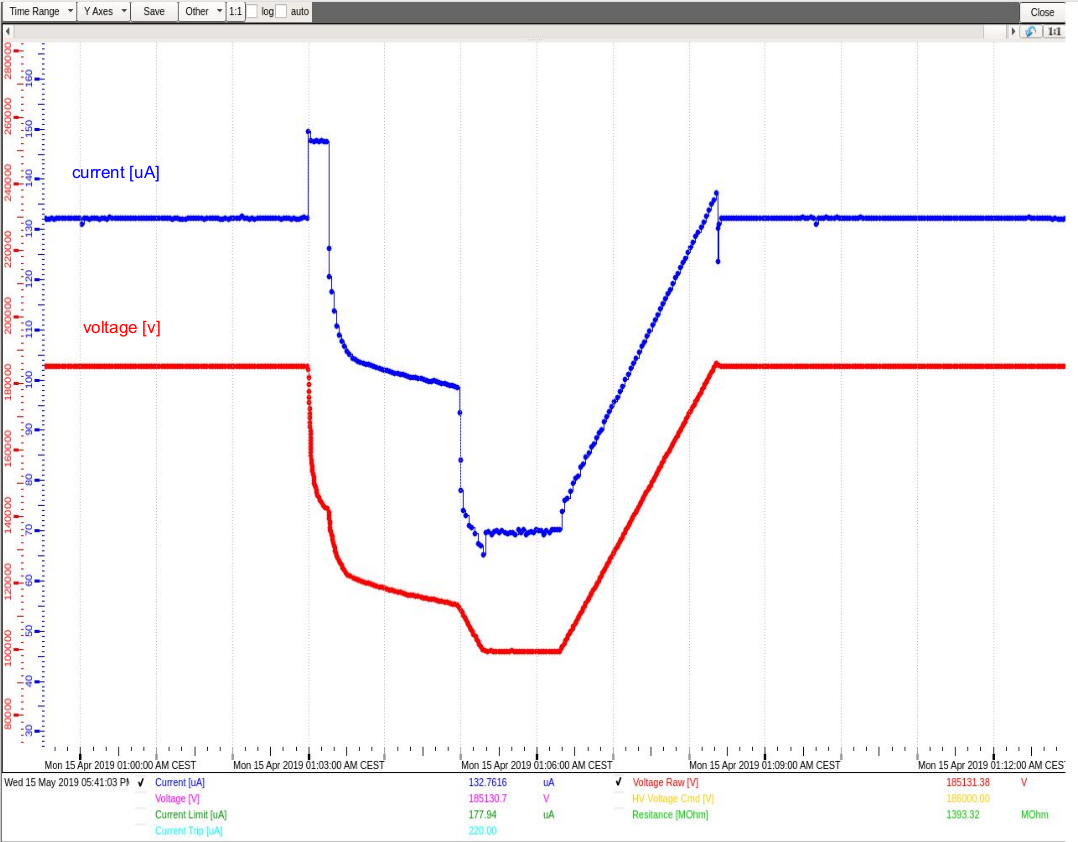}
\end{dunefigure}

\subsection{Post-beam Stability Runs with Cosmic Rays}
\label{sec:fdsp-hv-long-term-operation}

During the 2018 beam run periods, priority was given to operating the \dword{pdsp} detector with maximal up-time in order to collect as much beam data as possible at the nominal \dword{hv} conditions. Therefore, investigating the long-term behavior of the \dword{hv} instabilities and understanding their origin became goals of the long-term operation of \dword{pdsp} in 2019.

As mentioned above, it appears that the current streamer effect is a charging-up process with its frequency increasing with time after a long \dword{hv}-off period. This behavior has been repeatedly observed and confirmed in 2019. The current streamer rate stabilized at 4-6 per day, and the location was essentially always on the same single Ground Plane (GP\#6) out of the 12 monitored \dwords{gp}. Their rate and location were approximately independent of the \dword{hv} applied on the CPA in the $\SI{90}{kV}$ to $\SI{180}{kV}$ range.

More recently, after a change of the \dword{lar} re-circulation pump (April 2019), the detector was operated for several months in very stable cryogenic conditions and with very high and stable \dword{lar} purity (as measured by purity monitors and cosmic rays). During this period, the \dword{hv} system was set and operated at the nominal value of 180 kV at the \dword{cpa}  for several weeks without interruption. A significant evolution in the behavior of the \dword{hv} system was observed. 

To better understand the current streamers phenomenon, the \dword{hv} system was operated for about fifty days without the auto-recovery script, and the current streamers were left to evolve naturally. They typically lasted 6 to 12 hours, exhibiting steady current and voltage drawn from the \dword{hv} power supply and they eventually self-quenched without any intervention. The repetition rate was highly reduced to about one current streamer every 10-14 days; this rate can be compared to the 4-6 per day in the previous periods with auto-recovery on.

The auto-recovery script was then re-enabled and the current streamer rate stabilized at about one in every 20 hours; in addition, the intensity of the current streamer on the \dword{gp} was reduced with respect to the previous periods. As in the previous runs, the current streamers occurred always on the same \dword{gp} (GP\#6) with a small leakage current on the beam plug hose, which is close to GP\#6. 

This behavior is a further indication that the current streamers are in fact a slow discharge process of charged-up insulating materials present in the high-field region outside of the \dword{fc}. The auto-recovery mode does not allow a full discharge, so the charging up is faster, and the streamer repetition rate is shorter.

The \dword{lar} purity loss experienced at the end of July, 2019, was accompanied by the complete disappearance of any \dword{hv} instabilities. 
They gradually reappeared when the electron lifetime again exceeded 200 microseconds, and their intensity constantly increased as purity improved. This behavior replicated that observed after the initial filling, and supports the hypothesis that the \dword{hv} instabilities are enhanced by the absence of electronegative impurities in high-purity \dword{lar}.

The effects of the current streamers on the \dword{fe} electronic noise and the \dword{pd} background rate have been investigated. We have not observed any effect of the current streamers on the \dword{fe} electronics. On the other hand, recent analysis of the data collected by the \dword{pds} during active current streamers has indicated a high single photon rate on the upper upstream part of the \dword{tpc}. This is consistent with the activities recorded on GP\#6, which is located exactly at this upper upstream area. The analysis of the photon detection data is in progress with the main goal of narrowing down the position of current streamers and the localization of it, if possible (Figure~\ref{fig:PD_activity_on_streamers})
Visual inspection of this location when the detector is emptied will be required to further understand the \dword{hv} instability issues.

\begin{dunefigure}[Photon detector activity on streamers]
{fig:PD_activity_on_streamers}
{Preliminary analysis of the single photon activity rate in coincidence with a current streamer as a function of the position of the \dwords{pd} in the \dword{apa}s (beam right site). The rate clearly decreases proportionally to the distance of the \dword{pd} from the supposed location of the current streamer. More refined analysis is ongoing (including the beam left \dwords{pd}) to better locate the light source.}
\includegraphics[width=0.75\textwidth]{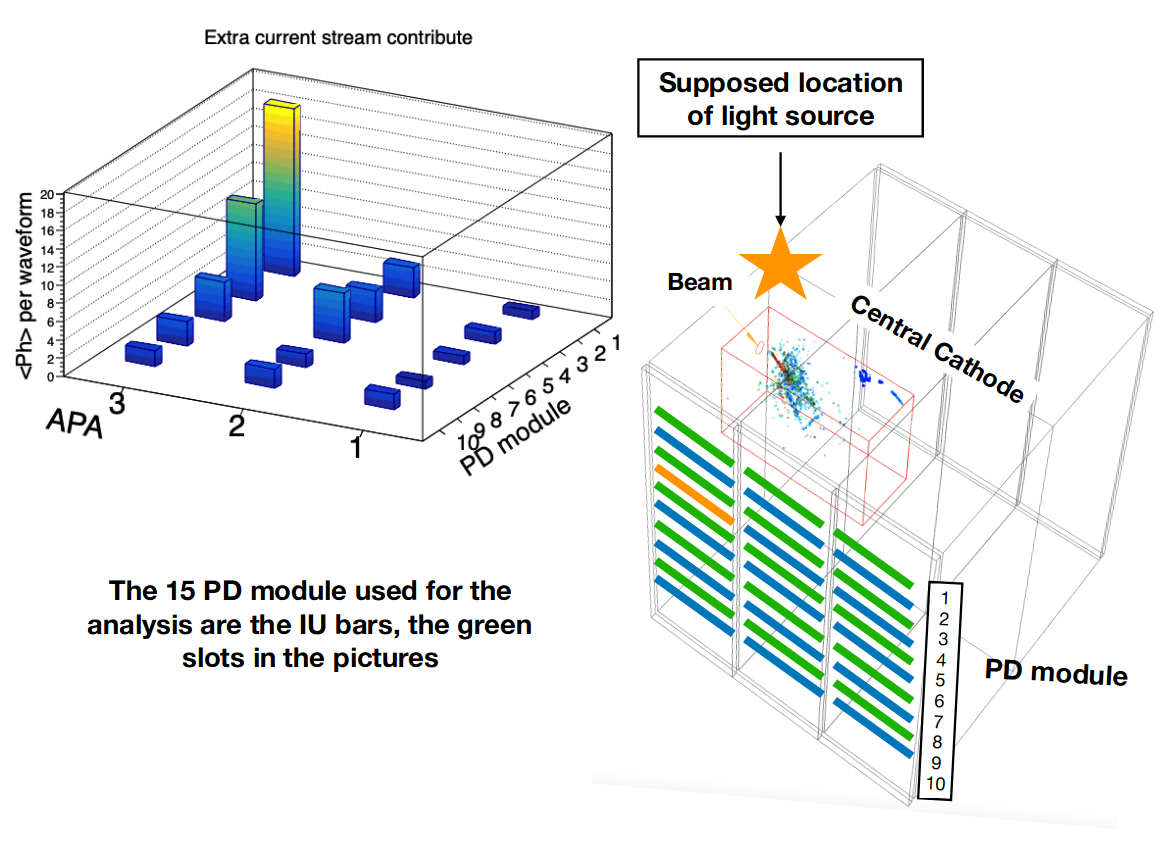}
\end{dunefigure}

It is planned to continue monitoring the \dword{hv} behavior, and in particular, before the end of the run, we plan to increase the \dword{hv} above the nominal value to possibly enhance anomalous effects in the \dword{hv} chain. Furthermore, the possible role of macroscopic impurities (metallic or insulating dust) circulation within the \dword{lar} is still to be understood, and specific running conditions will be implemented in \dword{pdsp} to better investigate this issue as well. 

Although 
we do not yet have 
precise knowledge of the origin of the current streamers, it is certainly safe to state that, in a time scale of nearly one year, we did not observe any degradation of the \dword{hv} system performance. On the contrary, the up-time has constantly improved (now it is above 99\%), and the instability rate and intensity have decreased. 

\subsection{Lessons from ProtoDUNE}
\label{sec:fdsp-hv-protodune-lessons}

The \dword{pdsp} \dword{hv} experience was, in general, very encouraging, having demonstrated 
an ability to operate the \dword{tpc} with a drift field of \spmaxfield{}. 
However, throughout the run, the system experienced various instabilities, 
discussed above. 
Systematic study of these instabilities continues.

\subsubsection{Design}
\label{sec:fdsp-hv-protodune-lessons-design}

The success of \dword{pdsp} validated the general design of the DUNE \dword{hv} system, but various opportunities for improvement during its construction and operation appeared. 
 In particular, we chose the following:

\begin{itemize}
\item adopt a ``pot-style'' filter resistor design (with input and output cables on the same end) 
to prevent leaks from causing interventions for refilling;
\item raise the \dword{hv} feedthrough cable insert to be above the cold insulation space, if space allows (to allow removing the cable while preventing moisture from entering and freezing on the walls, which could affect electrical contact);
\item add toroid signals 
to the feedthrough; and
\item improve stability by increasing the distance between the \dwords{gp} and field-shaping profiles and eliminating direct paths for potential surface currents.
\end{itemize}

The instrumented \dwords{gp} on the top and bottom \dwords{fc} proved invaluable for collecting information during moments of instability.
A dedicated \dword{daq} read out the signals from the \dword{gp} monitoring system, the beam plug current monitor, and the power supply at a rate of \SI{20}{kHz} on a trigger provided useful information for diagnosing the \dword{hv} behavior inside the \dword{tpc}.
This system was not operated continuously due to correspondingly large data disk storage requirements.
Toroid signals from the \dword{hv} filters were also helpful in localizing sources of instability, specifically for distinguishing issues on the warm side from issues inside the \dword{tpc}.


\subsubsection{Production, Handling and Quality Control}
\label{sec:fdsp-hv-protodune-lessons-prod}

The production and handling of \dword{hv} components must be approached with 
great care to avoid scratching and potentially compromising the electrical components. 
Part production should be carried out to avoid introducing sharp edges wherever possible.
The corners of the \dword{gp} panels had to be smoothed after some buckling was introduced during the pressing process, and a number of support hinges and clevises had sharp features removed by polishing.
The aluminum field-shaping profiles are particularly prone to scratches and must be packaged and handled so as to avoid direct contact with other profiles and materials.
Kapton strips were used to separate the profiles from the \dword{frp} of the \dword{fc} frames as they were being inserted to protect against scratching or removal of the profile coating.
Any scratches found in the \dword{frp} beams were covered with epoxy to prevent fibers from escaping into the \dword{lar}.

\Dword{qc} tests were conducted on \dword{hv} modules and individual components at every step: 
part procurement, production, integration, and installation.  For example, checklist forms were completed for component parts of detector modules as production proceeded.  Also, during the production process, documented procedures included \dword{qc} steps with checklist forms.  Printed copies of the checklists completed in the procurement and production stages were included as travelers in shipping crates.  
To ensure that nothing was compromised during transport, \dword{qc} tests were repeated on individual components and assembled pieces after shipping. 
Resistance between steps on the voltage divider boards was measured and verified to be within specification both after their production at LSU and after they were shipped to \dword{cern}.
Once the voltage divider boards were mounted onto an assembled \dword{fc} module, the resistance between adjacent profiles was measured to verify sound electrical connection.
In a similar way, \dword{qc} checks of connections between \dword{cpa} modules and between \dword{cpa} and \dword{fc} modules were performed after installation.

\dword{qc} tests on the \dword{hv} components of \dword{pdsp} required many measurements to be made with several different test devices.  Extrapolating these measurements to the scale of DUNE will require development of dedicated tools so that the \dword{qc} process can be made more efficient and optimal at each step.
For example, devices to measure the resistivity of \dword{cpa} coated resistive panels and field shaping strips will be provided to each of the designated production factories.  Also, designing a rig that can latch onto the \dword{fc} modules in such a way to make contact with all electrodes and control their voltages independently would allow for an automated loop across all steps.
Such dedicated equipment and automated procedures will be required en route to a full \dword{spmod}.

\subsubsection{Assembly and Installation}
\label{sec:fdsp-hv-protodune-lessons-assy}

The \dword{pdsp} experience allowed for a realistic estimation of the time involved to produce various \dword{hv} components for an \dword{spmod}.
The time involved for \dword{pdsp} was approximately as follows:
\begin{itemize}
\item \dword{fc} module assembly: 1.5 days/module with 2 workers,
\item \dword{ewfc} module assembly: 1.5 days/module with 2 workers, 
\item \dword{cpa} (2-panel) plane: 2 days/plane with 4 workers,
\item \dword{cpa} plane + \dword{fc} integration: 1 day/assembly with 4 workers,
\item \dword{ewfc} frame assembly: 7 days/module with 2 workers,
\item \dword{ewfc} final assembly: 4 hours/wall with 4-6 workers.
\end{itemize}
These estimates include time needed to perform the required \dword{qc} tests at each stage of the assembly and installation.

The \dword{pdsp} installation sequence had the beam-right drift volume deployed before the beam-left.
As anticipated from calculation and testing, asymmetries in the weight distribution before the beam-left drift was deployed produced temporary misalignments that propagated throughout the entire detector until the final left drift deployment, which corrected them.
The process of connecting individual endwall modules to build an endwall exposed another alignment issue.
The first endwall was 
significantly bowed initially.
A tool was built to adjust the angle between adjacent modules, which straightened out the wall.
The tool was also used while connecting modules for the remaining three endwalls, and no significant bowing was observed.

\subsection{Future R\&D}
\label{sec:fdsp-hv-protodune-RD}

The present \dword{hv} system design derives closely from the one in operation in \dword{pdsp}.  Operation of this detector in 2019 has allowed us to gain further confidence in depth concerning the long term stability and reliability of the \dword{hv} system under nominal conditions.  The present R\&D program, which will not extend beyond early 2020, has the goal to further improve, if required, the reliability of the system.   
In the R\&D program we plan to  

\begin{itemize}
\item evaluate the charge and discharge behavior of the \dword{uhmwpe} caps on the end of the profiles compared to metallic capped profiles.  The goal is to check if the end caps contribute to \dword{hv} instability. 

\item compare the high voltage stability of a new version of end wall profiles to the \dword{protodune} version.  The new version bends the top and bottom of the end wall profiles 90 degrees towards the ends of the top or bottom profiles, reducing the gap between field cage components at the two detector module ends and lowering the \efield{}s on the surfaces of the \dword{uhmwpe} caps and profiles near the profile ends.
\item evaluate resistive versus metallic caps.  If the \dword{uhmwpe} caps are 
problematic, find an alternative solution to maintain separated \dword{fc} modules.
\item study the surface-charging behavior of the \dword{fc} insulation structures.  Evaluation of general insulator performance for \dwords{lartpc}, including charge-up effects and geometry, remains an outstanding task.  In this test, the goal is to find out if any geometrical feature or surface treatment can reduce \dword{hv} instability.
\item evaluate higher-resistivity Kapton films.  The goal is to check the feasibility of increasing the surface resistivity of the cathode plane up to 1~G$\Omega$/square.  The task includes verifying the lamination quality on \frfour sheet and production availability.
\item perform further simulation of  \dword{hvs} discharge behavior. Although modeling other \dword{fc} designs and DUNE itself will take considerable effort, 
understanding the source of instabilities or exposing any design weaknesses would be worthwhile. 
\end{itemize}

\section{Interfaces }
\label{sec:fdsp-hv-intfc}

The \dword{hvs} has the largest surface area on the \dword{tpc} and interfaces with many other systems.  Table~\ref{tab:HVinterfaces} summarizes the interfaces with other consortia, highlights the key elements, and provides the links to the existing interface documents.

The two most important mechanical interfaces are with the \dword{dss} and the \dword{apa}.  The entire weight of the \dword{cpa}s, \dword{ewfc} and half the weight of the top and bottom \dword{fc} are supported by rails provided by the \dword{dss}.  The other half of the top and bottom \dword{fc} weight is transferred to the \dword{apa}s through latches mounted on the \dword{apa}s. All \dword{cpa}s and most of the \dword{fc} modules are also transported along the DSS rails to their final positions. The \dword{dss} rails ultimately determine the final locations of the \dword{cpa}s and \dwords{fc} on the \dword{tpc}.

Electrically, since the \dword{apa}s are at the detector ground, all \dword{hvs} field cage termination and fail-safe circuits are connected to the \dword{apa}s.  All cables used for the \dword{fc} termination pass through the \dword{apa} frame, to connect to the \dword{shv} cables provided by the \dword{ce} through the \dword{ce} signal flanges.  The \dword{tpc} electronics consortium also provides \dword{hvs} the \dword{fc} termination power supplies.

\begin{dunetable}
[HV system interfaces]
{p{0.25\textwidth}p{0.5\textwidth}l}
{tab:HVinterfaces}
{\dword{hv} system interface links }   
Interfacing System & Description & Linked Reference \\ \toprowrule
\dword{dss}  &  Support, positioning, and alignment of all \dword{cpa}, \dword{fc} modules inside the cryostat both warm and cold & \citedocdb{16766}  
\\ \colhline
\dword{apa} & \dword{fc} support (top, bottom, and end wall) on \dword{apa} frames; Mounting of FC termination filter boards and \dword{fc} fail-safe terminations; 
& \citedocdb{6673} 
\\ \colhline
\dword{ce} & \dword{fc} termination wire connectors on CE feedthrough flange, \dword{fc} termination wires routed with CE cables & \citedocdb{6739} 
 \\ \colhline
\dword{pds} & Mounting of PD calibration flash diffusers and routing of their fibers to \dword{cpa}s; Possible \dword{tpc} coated reflector foil on \dword{cpa}s. & \citedocdb{6721} 
 \\ \colhline
facility & Locations and specifications of the \dword{hv} \fdth ports; gas and \dword{lar} flow velocities and patterns. & \citedocdb{6985}  
\\ \colhline
calibration & \dword{fc} openings for the calibration laser heads & \citedocdb{7066}
\\ \colhline
\dword{cisc} & \dword{hv} vs. \dword{lar} level interlock, sensor locations in high field regions, cold/warm camera coverage, \dword{hv} signal monitoring, etc. & \citedocdb{6787} 
 \\ \colhline
\dword{sdwf} & Storage buffer, inspections/tests, repackage for underground delivery & \citedocdb{7039} 
 \\ \colhline
physics & Requirements: range of operating drift field, uniformity of the drift field; Supply detector geometry and \efield{} map. & \citedocdb{7093} 
 \\ 
\end{dunetable}

\section{Production and Assembly }
\label{sec:fdsp-hv-prod-assy}

\subsection{Power Supplies and Feedthrough}
\label{sec:fdsp-hv-supplies-feedthrough}

We plan to buy commercial power supplies through, among other vendors, Heinzinger. 
The \dword{hv} cable is commercially available.

The power supply is tested extensively along with the controls and monitoring software.  Features to be included in the software are
\begin{itemize}
\item the ability to ramp, or change, the voltage, set the ramp rate, and pause the ramp. 
In previous installations, the ramp rate was typically between \SIrange{60}{120}{V/s}.
\item an input for a user-defined current limit.  This parameter is the electric current (I) value at which the supply reduces the voltage output to stay below the current limit.  The current-limiting is done in hardware.
\item an input for a trip threshold.  At this current reading, the program would reduce the voltage output through software.  In previous experiments, the trip function in software would set the output to \SI{0}{kV}.
\end{itemize}
Additionally, the software must
record the current and voltage read-back values with a user-defined frequency, as well as any irregular current or voltage events.

The \dword{hv} feedthrough and filters are custom devices. As for \dword{pdsp}, the feedthrough  designs are made by collaborators and fabricated by an external company or major laboratory.
 Raw materials such as stainless steel, \dword{uhmwpe} rods, and flanges are readily available and are machined to make a feedthrough. Similarly, the resistors, steel or aluminum, and insulator material for the filters are readily available. The feedthrough and filters require testing before being delivered to the \dword{sdwf}. 
 
\subsection{Cathode Plane Assembly}
\label{sec:fdsp-hv-prod-cpa}
 The component parts of the \dword{cpa} array will be mainly produced by commercial companies except for specific items that are more efficiently produced by university collaborators.  Parts will be packaged into kits, each to contain the parts for a single \dword{cpa} panel
(three \dword{cpa} units). The parts in each kit are

\begin{itemize}
\item manufactured \frfour \dword{rp} frames, 
\item carbon-impregnated Kapton-coated \dwords{rp} and \dword{fss},
\item \dword{hv} cable segments and wire jumpers making up the \dword{cpa} \dword{hv} bus and \dword{rp} interconnects,
\item resistor boards connecting the \dwords{rp} to \dword{fss} (for raising the \dword{rp} \dword{hv} by \SI{1.5}{\kV}),
\item machined brass tabs for connecting \dwords{rp}, \dword{hv} bus, and \dword{fss}, and
\item top, bottom, and exterior edge profiles and associated connection hardware.
\end{itemize}
The kits are sent to the production factories, the locations of which will be determined later.  
The 
\dword{cpa} construction unit for installation into the \dword{spmod} at the \dword{surf} is a pair of \dword{cpa} panels called a \dword{cpa} plane. The production factories thus ship partially-assembled \dword{cpa} panels to \dword{surf} where panel assembly is completed and two panels are paired in the underground cleanroom to form a \dword{cpa} plane. During production, some storage (up to one month's installation rate) of \dword{cpa} shipping crates can occur at the \dword{sdwf} while waiting for movement into the \dword{surf} cleanroom.  No unpacking of crates is needed at the \dword{sdwf}; only visual inspection will be done to determine if any damage occurred during shipping.

The most basic element of the \dword{cpa} 
is an \dword{rp} mounted in a machined slot in the top, bottom and sides of \frfour frames.  
There are three different \dword{rp} types: an upper, which has as its top frame the \dword{cpa} mounting bracket and \dword{topfc} hinge, a middle, and a lower, which has as its bottom frame a \dword{botfc} hinge.  
Pairs of \dwords{rp} are bolted together and pinned to form \dword{cpa} units of size \SI{1.2}{\m} $\times$ \SI{4}{\m} for shipment. Three types of pairings are constructed to make a full six-\dword{rp}, \SI{12}{\m} tall \dword{cpa} panel: (1) an upper and a middle, (2) two middle, and (3) a middle and a lower.

The order in the shipping crate from top to bottom is: middle-and-lower, middle-and-middle, and upper-and-middle.   Two \dword{cpa} panels are shipped together in one crate; they are paired at \dword{surf} to form one \dword{cpa} plane.  The \dword{spmod} requires 100 upper, 100 lower, and 400 middle \dwords{rp} to make up the 100 \dword{cpa} panels (50 \dword{cpa} planes) of the \dword{tpc}.

In addition to the frames and \dwords{rp}, 
\dword{fss} are mounted on the exposed sides of the \frfour frames, aluminum profiles are attached to the exterior edges of the upper and lower \dwords{rp}, 
and cables are attached to the \dwords{rp} to form segments of the \dword{hv} bus.  

The \dword{cpa} units are assembled horizontally on a smooth, flat, highly stable table 
to ensure flatness and straightness of the entire panel before units are pinned together. There is one table per factory with up to three factories making \dword{cpa}s.

Figure~\ref{fig:12m-cpa} shows a  \SI{6}{\m} \dword{pdsp} \dword{cpa} panel (rear) and a \SI{12}{\m} \dword{pdsp} \dword{cpa} panel (foreground) at \dword{ashriver} Laboratory in Minnesota, USA.

\begin{dunefigure}[CPA mockup panels at Ash River]{fig:12m-cpa}{A \SI{12}{\m} DUNE-SP \dword{cpa} mock-up panel (foreground) and a 
half-height \SI{6}{\m} \dword{pdsp} panel mock-up (rear) at \dword{ashriver}, Minnesota.}  
\includegraphics[width=0.5\textwidth]{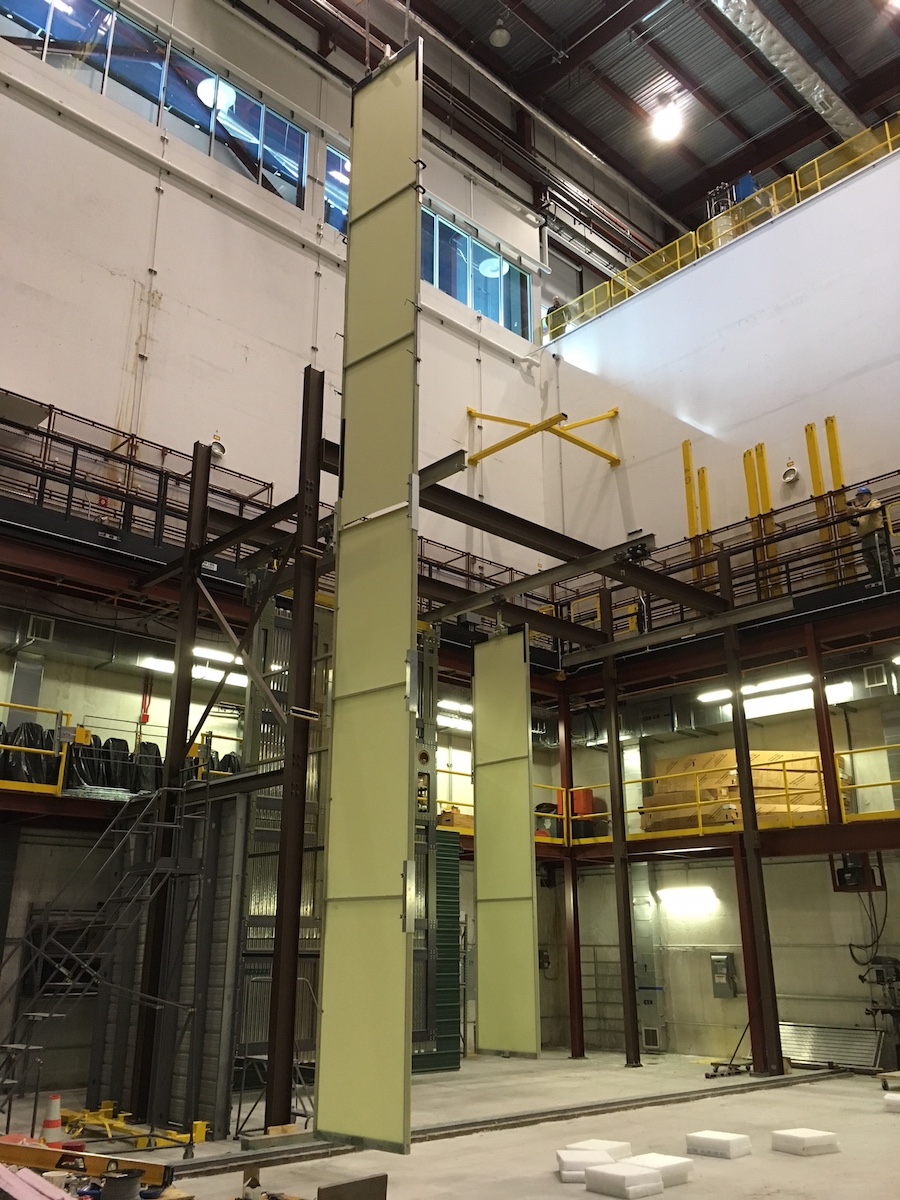}
\end{dunefigure}

\subsection{Field Cages}
\label{sec:fdsp-hv-prod-fc}

\subsubsection{Top and Bottom Field Cages}
\label{sec:fdsp-hv-prod-fc-tb}

Firms that specialize in the machining of fiberglass components for electrical applications will produce the \dword{frp} and \frfour components of the top and \dwords{botfc}, as was successfully done for \dword{pdsp}. 
All the machined edges except the small circular holes are to be coated with translucent epoxy. The stainless steel and aluminum components will be produced in university and commercial machine shops. University groups will likely fabricate the voltage divider boards and \dword{fc} and \dword{cpa} connection boards.

The \dword{frp} frame assembly process consists primarily of fastening together \dword{frp} I-beams with \dword{frp} threaded rods and hex nuts,  and securing them with a limited and specified torque to avoid damage to the threads. Detailed views of this procecure 
are shown in Figure~\ref{fig:tbfc3}.

\begin{dunefigure}[Top and bottom FC module frame assembly procedure]{fig:tbfc3}{The figure shows the procedure for connecting the cross beams to the main I-beams for the \dword{topfc}. Left: A display of the components of each connection, which (from top to bottom) are the threaded rods, the spacer tubes, washer plates, the hexagonal nuts, and an L-shaped \dword{frp} brace. An intermediate stage (middle) and final stage (right) of the assembly are also shown.}
\includegraphics[width=0.70\textwidth]{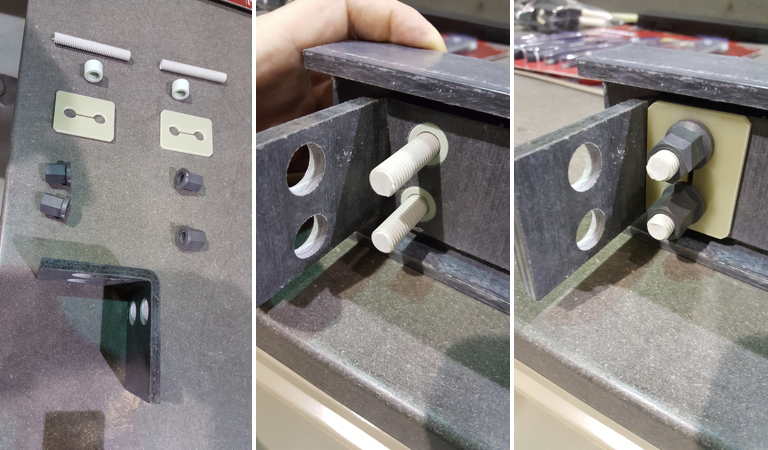}
\end{dunefigure}

Prior to sliding each profile into the \dword{frp} frame, the holes 
are covered with Kapton tape to avoid damage to the profile coating. An end cap is attached to each profile using plastic rivets, and then the profiles are aligned against an alignment fixture running the length of the \dword{fc}. After securing each profile to the frame, the tension in the mounting screws is adjusted to remove any angular deflection in the extended portion of the profile.

The \dwords{gp} are attached to the \SI{10}{\cm} stand-off I-beam sections with threaded rods and a machined plate. The copper strips are connected to adjacent modules at the same locations. Care must be taken to avoid bending the corners of the \dwords{gp} toward the profiles, particularly on the \dword{cpa} side of the module.

\subsubsection{Endwall Field Cages}
\label{sec:fdsp-hv-prod-fc-endw}

For the \dwords{ewfc}, all \dword{frp} plates are commercially cut to shape by water jet, as are the cutouts in the \dword{frp} box beams. 
Holes that accommodate G10 bushings are reamed in a machine shop. \dword{frp} frames are pre-assembled to ensure proper alignment of all \dword{frp} parts and 
holes (the profiles are not inserted at this stage). The \dword{frp} modules are hung off of each other by means of interconnecting \dword{frp} plates to ensure accurate alignment.

Next, parts are labeled, and the frames are taken apart. All components are cleaned by pressure washing or ultrasonic bath. All cut \dword{frp} surfaces are then coated with polyurethane, which contains the same main ingredient as the \dword{frp} resin, allowing it to bond well to the \dword{frp} fibers. Final panels are constructed from cleaned and inspected parts. Since assembly requires access to both sides of a module,
a dedicated assembly table has been manufactured that allows convenient module rotation. 

Figure~\ref{fig:endwall_assy_rot_table} shows a partially assembled \dword{ewfc} \dword{frp} frame on the assembly table.
\begin{dunefigure}[Endwall FC assembly table]{fig:endwall_assy_rot_table}{Assembly table with partially assembled \dword{ewfc} module. Box beams, cross beams, and slots for mounting of aluminum profiles are visible.}
 \includegraphics[width=0.8\textwidth]{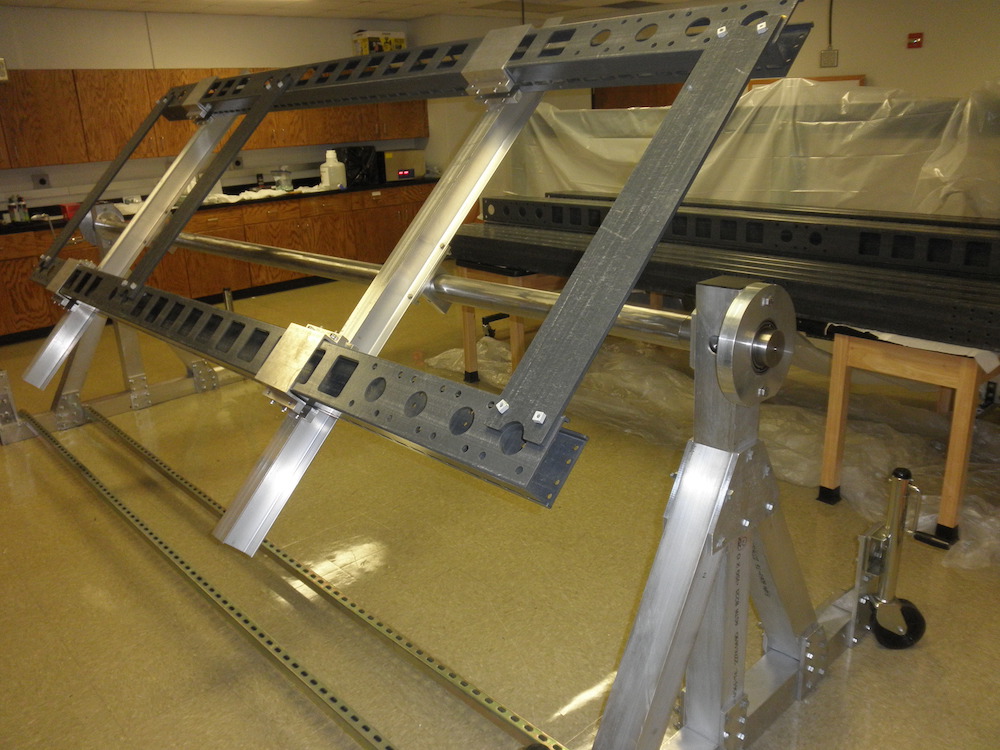}
 \end{dunefigure}
The \dword{frp} box beams are sandwiched between \SI{1.27}{\cm} (\num{0.5}\,in) thick \dword{frp} panels that are held on one side by means of G10 bushings and rods with square nuts.
On the other side M10 stainless steel bolts, which are clearly visible in Figure~\ref{fig:endwall_assy_detail},  
engage with large slip nuts that are inserted into the aluminum profiles. The profiles 
are pulled towards a \SI{2.5}{\cm} thick \dword{frp} plate located 
on the inside of the box beam.

\begin{dunefigure}[Hanging endwall FC frames] 
{fig:endwall_assy_detail}{
Top and center \dword{ewfc} module frames hanging. }
\includegraphics[width=0.7\textwidth]{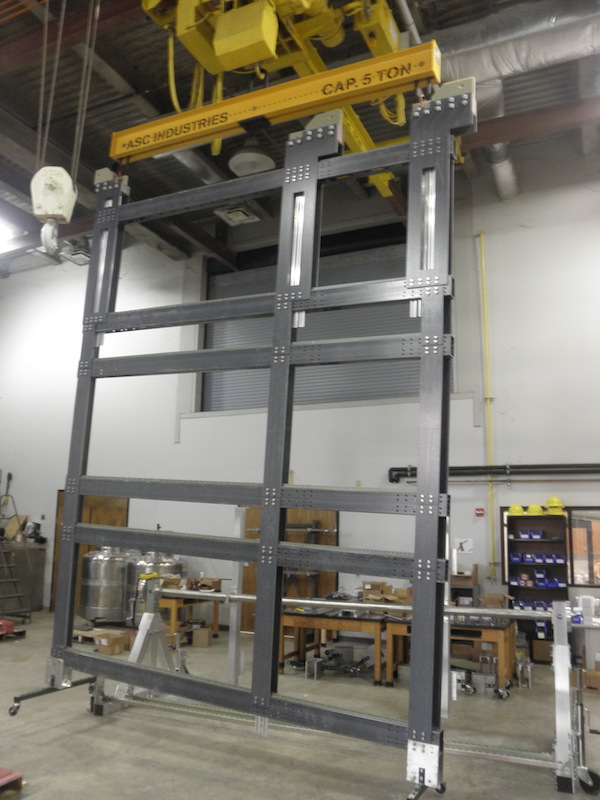}
\end{dunefigure}

Aluminum profiles are inserted into the cutouts of the box beams and attached with screws and stainless steel slip nuts to L-shaped \dword{frp} brackets that are mounted on the \dword{frp} box beams. Small changes in part sizes will help to simplify the assembly  procedure with respect to the one used for \dword{pdsp}. Currently, we expect that pre-assembly of the \dword{fc} endwall frames will no longer be required. The full modules will be assembled at the factory (LSU), and then complete \dword{ewfc} panels will be shipped to the \dword{sdwf}.

\subsection{Electrical Interconnections}
\label{sec:fdsp-hv-prod-interconnect}

All electrical fasteners and wires used on the \dword{cpa} arrays and \dword{fc} are produced
to specification by commercial vendors and packaged with the \dword{cpa} or \dword{fc} modules.  
As discussed in Sections~\ref{sec:fdsp-hv-prod-cpa} and~\ref{sec:fdsp-hv-prod-fc}), 
this includes. e.g., the \dword{hv} cable segments, wire jumpers, and machined brass
tabs.

University shops will produce and test circuit boards for 
\dword{hv} interconnections according to the same design used for \dword{pdsp}.  The \dword{fc} voltage dividers were produced for \dword{pdsp} at LSU, and the boards for \dword{cpa} frame bias and \dword{cpa}-\dword{fc} connections were produced at Kansas State University (KSU).
Both institutions have created custom test apparatuses for verifying proper operation of the boards at full voltage and over-voltage conditions, keeping the boards free of solder flux and flux-remover.  These institutions may scale up production and testing by the required order of magnitude for the \dword{spmod} or share this work with other institutions, whichever best meets the needs of the project. 

\subsection{Production Safety}
\label{sec:fdsp-hv-prod-safety}

Production of the \dword{fc} panels and resistor-divider boards will involve collaboration technical, scientific, and student labor and  does not present unusual industrial hazards. The \dword{hvs} consortium will work closely with each production site to ensure that procedures meet both \dword{fnal} and institutional requirements for safe procedures, \dword{ppe}, environmental protection, trained materials handling, and training. The vast majority of production part fabrication will be carried out commercially and shipping will be contracted through approved commercial shipping companies. Prior to approving a site as a production venue, it will be visited and reviewed by an external safety panel to ensure best practices are in place and maintained. 

Testing of the \dword{hv} feedthrough will be done in a closed cryostat to avoid exposure to high voltage and  to assure the nominal voltage is functional.  The power supply is grounded to the cryostat as a further safety measure. Tests for the \dword{pdsp} \dword{hv} feedthrough were done at \dword{cern} after a safety electrical and cryogenic review mainly focusing on the grounding of the whole test stand (power supply, cable, and cryostat where the feedthrough was tested) as well as interlocks. A safety document (PPSPS) was created, reviewed, and approved for this test. Similar testing and documentation will be done for the \dword{spmod}.

\section{Quality Control, Transport, and Installation}
\label{sec:fdsp-hv-transport}

The \dword{hvs} consortium has developed a comprehensive quality control (\dword{qc})
plan for the production, shipping, and installation of the \dword{spmod} \dword{hv} components. 
It is based partly on \dword{qc} procedures developed and implemented on \dword{pdsp} and on the \dword{nova} experiment's successful use of barcode tagging for identifying and tracking detector components.  Inventory tagging and tracking each component is crucial. Documentation in the form of printed checklists is maintained~\cite{bib:docdb10452}. 

Travelers have been replaced by a system of tags with bar codes attached to the units, which key to electronic \dword{qc} data. The tags will be large and brightly colored enough to be seen from both ends of the cryostat.
A particularly suitable choice is to use 
bright yellow cattle tags, plastic tags of about 10-12 square inches ($\sim\,\SI{70}{cm^{2}}$) on which a unique QR\footnote{Quick Response Code, The QR\texttrademark{} code system was invented in 1994 by the Japanese company Denso Wave. \url{https://www.qrcode.com/en/index.html}.} or bar code can be printed; they can be purchased very inexpensively in quantities of hundreds or thousands.

Scanned tags are removed after completion of electronic checklist forms linked to the tag's bar code.  At the end of \dword{tpc} installation, all \dword{qc} data for components at a particular location in the detector are stored electronically and linked to that location.

\subsection{Quality Control}
\label{sec:fdsp-hv-transport-QC}

Power supply devices used in an \dword{spmod} will be tested before installation.  Output voltages and currents will be checked on a known load. 

The feedthrough and filters will be tested at the same time, with the selected power supply.  The feedthrough must be verified to hold the required voltage in \dword{tpc}-quality \dword{lar} ($\tau\geq$\SI{1.6}{ms}) for several days.  The ground tube submersion and \efield{} environment of the test setup will be comparable to the real \dword{fc} setup or more challenging (e.g., the test liquid level can be lower than that in the \dword{spmod} but not higher).  Additionally, the feedthrough must be leak-tight to satisfy cryogenics requirements.

The \dword{qc} tests concerning the voltage divider boards are as follows: All individual resistors and
 varistors are submitted to a warm and cold (87 K) current-voltage measurement. This forms the basis for selecting components that meet specifications: all
 electrical components must pass visual inspection for mechanical damage; all measurement values (resistance, clamping voltage) must be within 2$\,\sigma$ of the mean for entire sample both in warm and cold tests.

The \dword{qc} process for mechanical components starts at the production factories by attaching a cattle tag with a unique code to each production element.  A file linked to each code contains the individual measurements and properties contained in the \dword{qc} checklists for that element.  The following is an example of how this system will be implemented for the \dword{cpa} components:

\begin{enumerate}
\item During assembly, \dword{qc} checklists are filled out electronically using a smart phone or tablet.  Once a \dword{cpa} unit 
is completely assembled and all checklists are complete, a coded temporary cattle tag is attached and scanned, linking the checklist information to the code on the tag. (The \dword{cpa} unit's individual parts are not tagged separately.)
\item 
A shipping crate will contain six 
\dword{cpa} units, each with its removable coded tag, plus any included hardware packages, each with a coded sticker.
\item A coded label on the shipping crate (paper sticker) will 
identify the contents of the crate (six codes + codes of hardware packages). The code on the label is used only for shipping purposes and for inventory purposes.  
\item In the \dword{surf} cleanroom, the first \dword{cpa} panel is assembled. A coded tag is attached to the \dword{cpa} panel and scanned.  Then the three individual \dword{cpa} unit tags are scanned and removed, linking them to the \dword{cpa} panel code.
\item The same procedure is followed for the second \dword{cpa} panel from the crate.  Each \dword{cpa} panel now has a single tag attached to it.
\item The 
\dword{cpa} panels are then combined into a \dword{cpa} plane, and a single coded tag is attached to the \dword{cpa} plane and scanned. 
The two individual \dword{cpa} panel tags are then scanned, linking their codes to the that of the \dword{cpa} plane. 
\item Top and bottom \dword{fc} modules 
are attached to both sides of the \dword{cpa} plane, and  a single coded tag is placed on this \dword{cpafc} assembly identifying the codes of each of the four \dword{fc} modules 
and the code of the \dword{cpa} planes;  
these five tags are removed after scanning.  
\item When moving the panel into the cryostat the code of the position tag on the \dword{dss} is scanned as well as the tag on the \dword{cpafc} assembly, and then both tags 
are removed.
\end{enumerate}
At this point, 
a sequence of linked codes associated with \dword{qc} checklists 
identify which \dword{cpa} and \dword{fc} modules 
are mounted in 
which \dword{dss} positions, and no tagging material remains in the cryostat.  A similar sequence is anticipated for the production of the top and bottom \dword{fc} units up to step 6; the endwalls are done separately but similarly.  
At the completion of installation in the cryostat and before top and bottom \dword{fc} deployment, visual inspection will confirm the absence of any tags.

\subsection{Transport and Handling}
\label{sec:fdsp-hv-transport-transport}

The \dword{hvs} consortium has studied 
options for 
transportation from \dword{hvs} production sites 
to the \dword{sdwf} 
and packaging of the shipped elements. 
We found that using reusable underground 
crates and returning them to the factories when empty is less expensive than using inexpensive, disposable crates for shipment from the factories to the \dword{sdwf},  
even with the extra shipment costs. 

We have identified a vendor that 
produces honeycombed PVC sheets of varying thicknesses that can be formed into crates. These 
can be loaded at the production sites, 
shipped to the \dword{sdwf}, and sent underground at \dword{surf}.  
We will require 50 shipments of crates containing two \dword{cpa} panels each to complete the \dword{spmod}.  
The reusable underground crate scheme requires only 20 crates to make the 50 shipments. Similar reductions are obtained for the top and bottom \dword{fc} modules.

Crates would be available at each factory at the start of production. 
As production proceeds, individual assembly units are bagged and sealed inside them.
When a full shipment of crates is ready at a factory, crates are sent by flatbed truck from the factory to the \dword{sdwf}.  The full crates are stored at the \dword{sdwf} until they can be received at SURF.  Some components may require \dword{qc} and/or minor assembly procedures to be done at the \dword{sdwf} before shipping to \dword{surf}.

At \dword{surf}, the crates are lowered into the staging area outside the cleanroom where they are unpacked. The assembly units are removed from their bags and taken into the cleanroom for installation. Only cleaned assembly units are allowed into the cleanroom; the crate is restricted to the staging area only. The empty crate is returned to the \dword{sdwf} and then sent back to a production factory for reloading.

\subsection{Safety during Handling} 
\label{sec:fdsp-hv-transport-safety}

In the current installation scenario, no assembly activities are foreseen at the \dword{sdwf} site for any components of the \dword{hv} system. Only visual inspection of the \dword{hvs} modules crate condition will be performed to verify the integrity after shipping. 
No disruption in installation should occur in the event of shipping damage since there is a one-month storage period at the \dword{sdwf} and two week's installation storage underground at \dword{surf}.  The \dword{hvs} consortium will coordinate procedures for underground handling with \dword{tc}.

A detailed Gantt chart on the production and installation schedule for the \dword{hvs} of the first \dword{spmod} is shown in Figure~\ref{fig:tasks}.

The installation activities are described in Chapter~\ref{ch:sp-install}.

\section{Organization and Management}
\label{sec:fdsp-hv-org}


\subsection{Institutional Responsibilities}
\label{sec:fdsp-hv-org-consortium}
The \dword{hvs} consortium 
includes all the institutions that have participated in the design, construction, and assembly of the \dword{hv} systems for both \dword{pdsp}  and \dword{pddp}. They are listed in 
Table~\ref{tab:instit}. 
The consortium  currently comprises several USA institutions and \dword{cern}, 
the only non-USA participant.

As it has been 
for \dword{protodune}, \dword{cern} is heavily committed to a significant role in the  \dword{fd} in terms of funding, personnel, 
 and the provision of infrastructure for R\&D and detector optimization. Moreover, \dword{cern} will be responsible for a significant fraction of subsystem deliverables; as such  \dword{cern} is actively in search of additional European institutions to attract into the consortium. 
 
At present, in the \dword{hv} current consortium organization, each institution is naturally assuming the same responsibilities that it assumed for 
\dword{pdsp} and \dword{pddp}. The 
consortium organizational structure includes a scientific lead (from \dword{cern}), a technical lead (from BNL), 
and an \dword{hvs} design and integration lead (from \dword{anl}). 

The successful experience gained with the \dword{pdsp} detector has demonstrated that the present \dword{hvs} consortium organization and the number of institutions are appropriate for the construction of the \dword{hv} system of the \dword{spmod}. Funding and the predominant participation of USA institutions are presently open issues that would benefit from more international participation. 

 The consortium is organized into working groups (``WG'' below) addressing the design and  R\&D phases of development, and the hardware production and installation.

\begin{itemize}
\item WG1: Design optimization for \dword{spmod} and \dword{dpmod}; assembly, system integration, detector simulation, physics requirements for monitoring and calibrations; 
\item WG2: R\&D activities, R\&D facilities; 
\item WG3: \dword{sp}-\dword{cpa}: Procurement of resistive panels, frame strips, electrical connections of planes; assembly, \dword{qc} at all stages, and shipment of these parts; \item WG4: \dword{dp} cathode and \dword{gp}:  material procurement; construction, assembly, shipment to \dword{sdwf} 
\dword{qa}, \dword{qc}; 
\item WG5:  modules: \dword{sp}-top/bottom-\dword{fc} module, \dword{sp}-endwall modules, \dword{dp}-\dword{fc} modules: procurement of mechanical and electrical components, assembly and shipping to \dword{sdwf}; and  
\item WG6: \dword{hv} supply and filtering, \dword{hv} power supply and cable procurement, R\&D tests, filtering and receptacle design and tests. 
\end{itemize}

Taking advantage of identified synergies, some activities of the \dword{sp} and \dword{dp} working groups are merged: \dword{hv} feedthrough, voltage dividers, aluminum profiles, \dword{frp} beams, and assembly infrastructure.

\begin{dunetable}
[HVS consortium institutions]
{ll}
{tab:instit}
{Institutions participating in the \dword{hvs} consortium}   
Institution & Country \\ \toprowrule

\dword{cern} & Switzerland \\ \colhline
Argonne National Lab& USA \\ \colhline
Brookhaven National Lab& USA \\ \colhline
University of California Berkeley / LBNL & USA \\ \colhline
 University of California Davis& USA \\ \colhline
\dword{fnal} & USA \\ \colhline
University of Houston & USA \\ \colhline
Kansas State University& USA \\ \colhline
Louisiana State University & USA \\ \colhline
 SUNY Stony Brook & USA \\ \colhline
 University of Texas Arlington & USA \\ \colhline
 Virginia Tech& USA \\ \colhline 
College of William and Mary& USA \\ 
\end{dunetable}
\subsection{Risks}
\label{sec:fdsp-hv-org-risk}

Table~\ref{tab:risks:SP-FD-HV} presents a summary of the risk items identified for the \dword{hv} system of  the \dword{fd} \dword{spmod}.

\begin{footnotesize}
\begin{longtable}{P{0.18\textwidth}P{0.20\textwidth}P{0.32\textwidth}P{0.02\textwidth}P{0.02\textwidth}P{0.02\textwidth}} 
\caption[HV risks]{HV risks (P=probability, C=cost, S=schedule) The risk probability, after taking into account the planned mitigation activities, is ranked as 
L (low $<\,$\SI{10}{\%}), 
M (medium \SIrange{10}{25}{\%}), or 
H (high $>\,$\SI{25}{\%}). 
The cost and schedule impacts are ranked as 
L (cost increase $<\,$\SI{5}{\%}, schedule delay $<\,$\num{2} months), 
M (\SIrange{5}{25}{\%} and 2--6 months, respectively) and 
H ($>\,$\SI{20}{\%} and $>\,$2 months, respectively).  \fixmehl{ref \texttt{tab:risks:SP-FD-HV}}} \\
\rowcolor{dunesky}
ID & Risk & Mitigation & P & C & S  \\  \colhline
RT-SP-HV-01 & Open circuit on the field cage divider chain & Component selection and cold tests. Varistor protection. & L & L & L \\  \colhline
RT-SP-HV-02 & Damage to the resistive Kapton film on CPA & Careful visual inspection of panel surfaces.  Replace panel if scratches are deep and long  & L & L & L \\  \colhline
RT-SP-HV-03 & Sole source for Kapton resistive surface; and may go out of production & Another potential source of resistive Kapton identified. Possible early purchase if single source. & M & L & L \\  \colhline
RT-SP-HV-04 & Detector components are damaged during shipment to the far site  & Spare parts at  LW. FC/CPA modules can be swapped and replaced from factories in a few days. & L & L & L \\  \colhline
RT-SP-HV-05 & Damages (scratches, bending) to aluminum profiles of Field Cage modules & Require sufficent spare profiles for substitution. Alternate: local coating with epoxy resin. & L & L & L \\  \colhline
RT-SP-HV-06 & Electric field uniformity is not adequate for muon momentum reconstruction  & Redundant components; rigorous screening. Structure based on CFD. Calibration can map E-field. & L & L & L \\  \colhline
RT-SP-HV-07 & Electric field is below goal during stable operations & Improve the protoDUNE SP HVS design to reduce surface E-field and eliminate exterior insulators. & M & L & L \\  \colhline
RT-SP-HV-08 & Damage to CE in event of discharge  & HVS was designed to reduce discharge to a safe level. Higher resistivity cathode could optimize. & L & L & L \\  \colhline
RT-SP-HV-09 & Free hanging frames can swing in the fluid flow  & Designed for flow using fluid model; Deformation can be calibrated by lasers or cosmic rays. & L & L & L \\  \colhline
RT-SP-HV-10 & FRP/ Polyethene/ laminated Kapton component lifetime is less than expected & Positive experience in other detectors. Gain experience with LAr TPC's; exchangeable feedthrough. & L & L & L \\  \colhline
RT-SP-HV-11 & International funding level for SP HVS too low & Cost reduction through design optimization. Effort to increase international collaboration. & M & M & M \\  \colhline
RT-SP-HV-12 & Underground installation is more labor intensive or slower than expected & SWF contingency, full-scale trial before installation. Estimates based on ProtoDUNE experience. & L & L & L \\  \colhline

\label{tab:risks:SP-FD-HV}
\end{longtable}
\end{footnotesize}

The first five risks refer to the construction and operation phases; risks 6 through 12  apply to the installation  and/or  detector operation phase. 

Most of the cited risks have already been addressed during the construction, commissioning, and operation of \dword{pdsp}. 
None have caused significant problems, with the partial exception of risk 9. 
Risk 9 requires an accurate analysis of collected muon data (this activity is in progress), and the disentangling of space charge effects.

Given the much larger detector scale and the more complex underground installation environment, the listed risks still 
apply to the \dwords{detmodule}. However, the positive experience gained with \dword{protodune} justifies the low risk probabilities assigned to most of the items.  To better justify these statements, brief explanations are given below, together with the identified mitigation actions.

Risk 1: An open circuit on the \dword{fc} could occur if a resistor in the conventional voltage dividers were to fail in the open condition, which 
could result in \dword{hv} discharges across the open circuit gap. Mitigation: Perform stringent component selection and cryogenic testing. Use parallel resistor chains to provide redundancy. Varistors, capable of withstanding several thousand amperes of current impulses, have been added in parallel with the resistor chains to protect them from large current surges. Check resistances 
several times during \dword{fc} fabrication and assembly phases, including once after the \dword{fc} deployment.   

Risk 2: Limited, local scratches could 
occur from accidental contacts during module assembly or installation. No mitigation is required if a scratch is limited in size. For larger scratches that can induce delamination, the mitigation is to replace the panel with a spare. 

Risk 3: About 12 rolls of resistive Kapton are needed (\SI{4}{ft} wide, \SI{300}{m} per roll)  for the \dword{cpa} panels of one \dword{spmod}.  The cost is 20k\$ per roll from the only vendor available up to now. Mitigation: Recently another source of resistive Kapton has become available and is being investigated. An early purchase is also under consideration in case of a single source condition.

Risk 4: Poor shipping techniques could cause damage to delicate components (e.g., broken \dword{cpa} panels, bent or heavily scratched aluminum profiles) that would cause the modules to fail \dword{qc} tests. 
If significant repairs to detector components are needed, they may require replacements.   Mitigation: Plan for an adequate number of spare elements and implement 
a documented \dword{qa} program for shipment packing with detailed review of shipping procedures, shipping containers, and testing in crates after arrival.

Risk 5: The surface of the aluminum profiles is very delicate and deep scratches could locally increase the \efield to close to the critical field of \SI{30}{\kV}/cm. The surface can be damaged during transport or manipulation in the assembly area. In case of significant damage, it cannot be repaired due to its conductive coating. For mitigation, ensure the availability of sufficient spare profiles 
on-site to allow last-minute substitutions. Alternatively, use a local coating with epoxy resin. 

Risk 6: Unexpected changes in \dword{fc} resistor values, cathode/\dword{fc} non-planarity or movement, and surface or space charge buildup can distort the \efield. As a consequence, the momentum of non-contained muons,  measured by estimating the multiple scattering rate for the observed track segments, could be incorrectly estimated, thereby degrading 
the momentum resolution for non-contained muons.  ${\nu}_{\mu}$ disappearance analyses and three-flavor fits could be affected 
leading to feed-down of high-energy neutrino backgrounds to low-energy reconstructed categories.  Mitigation: Consider addition of a laser calibration 
if calibration with cosmic crossing muons is not sufficient. 

Risk 7:  In \dword{protodune}, \dword{hv} instabilities appeared as current streams occurring at intervals of several hours and localized on  a specific \dword{fc} module. These required a several-minute ramp-down of the \dword{hv} from the nominal \SI{-180}{kV} to a lower value, typically \SI{-140}{kV}; see Section~\ref{sec:fdsp-hv-protodune}.
Investigations are underway to characterize and mitigate this risk. Recently, understanding of this risk has significantly progressed thanks to the 
long-term operation of \dword{pdsp} in 2019 while exposed only to cosmic rays. 
There are strong hints 
that the instabilities are due to charging-up processes on insulators. In addition, 
the current streamers are confirmed to have been localized mainly on 
one \dword{gp} (of 12). No degradation of the detector performance due to \dword{hv} instabilities has been observed. 
Moreover, these instabilities appear to decrease gradually in rate and intensity. At present, the detector down-time due to these instabilities is less than 1\%.  

\dword{protodune} long-term operation is also indicating that \dword{lar} purity does not  play a significant role in the onset of the \dword{hv} instability (for free electron lifetime above $\sim\,$\SI{1}{ms}).  The impact of the \dword{hv} instabilities on \dword{apa}/\dword{ce} and \dwords{pd} is also under investigation and at the moment appears to be negligible.

The mitigation for risk 7 involves improvement at the design level to increase as much as possible the distance between the \dword{fc} and \dwords{gp} and to avoid high-field regions by smoothing all electrodes exposed to \dword{hv}.  
\dword{pddp}, with a comparable design, will help determine 
the validity of these improvements.

Risk 8: A sudden discharge on the \dword{hv} system would inject charge to the \dword{fe} \dwords{asic}, overwhelming the protection circuits and causing permanent damage. Mitigation: Key aspects of the \dword{hvs} design were aimed at reducing the charge injection to a safe level for the \dword{ce}, such as segmenting the \dword{fc} and making the cathode planes resistive.  We are still searching for higher-resistivity material on the cathode to increase the safety factor. 

Risk 9:  Each cathode is made of lightweight, non-porous material with an area of \SI{58x12}{m} that could move under the  convectional flow of the \dword{lar}.  Mitigation: The \dword{cpa} structure is designed to withstand pressure from \dword{lar} flow based on fluid model predictions. Static deformation can be calibrated by lasers or cosmic rays.

Risk 10: Aging of insulator components in \dword{lar} could 
pose a problem,  but  experience in \dword{atlas} (Kapton \& PCB, $20+$ years), \dword{icarus} (G10, feedthrough $4+$ years; feedthrough exchangeable) is trending favorably. Mitigation: 
Continue to gain experience with \dwords{lartpc}. Make feedthrough exchangeable.

Risk 11: Current costing suggests that international funding could be 
insufficient. Mitigation: Implement cost reduction through design optimization and scaling. Make efforts to include more international 
institutions. 

Risk 12: Underground installation is more labor-intensive or slower than expected. Mitigation: Add labor contingency. Carry out full-scale installation trials at the  \dword{ashriver} site prior to installation. The estimates are based on \dword{protodune} experience. With the present knowledge, the \dword{hv} system is not on the critical path for installation.

\subsection{High-level Schedule}
\label{sec:fdsp-hv-org-cs}

Table~\ref{tab:HVsched} lists the most high-level milestones for the design, testing, production, and installation of the \dword{spmod} \dword{hvs}. Dates in this tentative schedule are based on the assumed start of installation of the first \dword{spmod} at \dword{surf}. The dates for the \dword{hvs} production of a second \dword{spmod} are 
included as a reference.

 The production scenario 
 for the schedule presented in Table~\ref{tab:HVsched} assumes 
 two factory sites for the \dword{cpa} construction, two for the top/bottom \dword{fc} modules and one for \dword{ewfc} modules. Given the present starting date 
 for  the first \dword{spmod} installation, this assumption is fully compatible with the time available after the operation of the \dword{protodune2} prototype.
 A more detailed schedule for production and installation of the first \dword{spmod} is found in 
 Figure~\ref{fig:tasks}.

\begin{dunetable}
[HVS consortium schedule]
{p{0.65\textwidth}p{0.25\textwidth}}
{tab:HVsched}
{High level Milestones and Schedule for the production of the \dword{hvs} of the \dword{spmod}}   
Milestone & Date (Month YYYY)   \\ \toprowrule
Technology Decision &      \\ \colhline
CPA/FC/Endwall 60\% Design Review & June 2019 \\ \colhline
CPA/FC/Endwall Mod 0 (for tests at \dshort{ashriver}) & June 2019 - June 2020 \\ \colhline
Final Design Review & June 2020     \\ \colhline
Start of module 0 component production for  \dshort{protodune2} & June 2020     \\ \colhline
End of module 0 component production for  \dshort{protodune2} &  March 2021    \\ \colhline
\rowcolor{dunepeach} Start of  \dshort{protodune2} installation& \startpduneiispinstall      \\ \colhline
\rowcolor{dunepeach} Start of  \dshort{protodune2} installation& \startpduneiidpinstall      \\ \colhline
\rowcolor{dunepeach}South Dakota Logistics Warehouse available& \sdlwavailable      \\ \colhline
\rowcolor{dunepeach}Beneficial occupancy of cavern 1 and \dshort{cuc}& \cucbenocc      \\ \colhline
\rowcolor{dunepeach}  \dshort{cuc} counting room accessible& \accesscuccountrm      \\ \colhline
 Top/Bottom FC  \dshort{prr} & July 2023    \\  \colhline
 Start of Top/Bottom FC production  & September 2023     \\ \colhline
CPA  \dshort{prr} &  October 2023    \\ \colhline
Start of  \dshort{cpa} production  & December 2023     \\ \colhline
\rowcolor{dunepeach}Top of  \dshort{detmodule} \#1 cryostat accessible& \accesstopfirstcryo      \\ \colhline
 Endwall FC  \dshort{prr} & February 2024     \\  \colhline
Start of Endwall FC production  & April 2024     \\ \colhline
End of  \dshort{cpa} production Detector \#1 & August 2024     \\ \colhline
End of Top/Bottom FC production Detector \#1 & August 2024  \\ \colhline
End of Endwall FC production  Detector \#1 & August 2024  \\ \colhline
\rowcolor{dunepeach}Start of  \dshort{detmodule} \#1 TPC installation& \startfirsttpcinstall      \\ \colhline
\rowcolor{dunepeach}Start of  \dshort{detmodule} \#1 TPC installation& \startfirsttpcinstall      \\ \colhline
\rowcolor{dunepeach}Top of  \dshort{detmodule} \#2 accessible& \accesstopsecondcryo      \\ \colhline
\rowcolor{dunepeach}End of  \dshort{detmodule} \#1 TPC installation& \firsttpcinstallend      \\ \colhline
 \rowcolor{dunepeach}Start of  \dshort{detmodule} \#2 TPC installation& \startsecondtpcinstall      \\ \colhline
End of  \dshort{cpa} production Detector \#2 & September 2025  \\ \colhline
End of Top/Bottom FC production Detector \#2  &  October 2025   \\ \colhline
End of Endwall FC production Detector \#2  & January 2026  \\  \colhline
\rowcolor{dunepeach}End of  \dshort{detmodule} \#2 TPC installation& \secondtpcinstallend      \\ 
\end{dunetable}

\begin{dunefigure}[HVS production and installation schedule for first SP detector module]{fig:tasks}{
Gantt chart providing a detailed view of the production and installation schedule for the \dword{hvs} for the first \dword{spmod}.}
\includegraphics[width=0.95\textwidth]{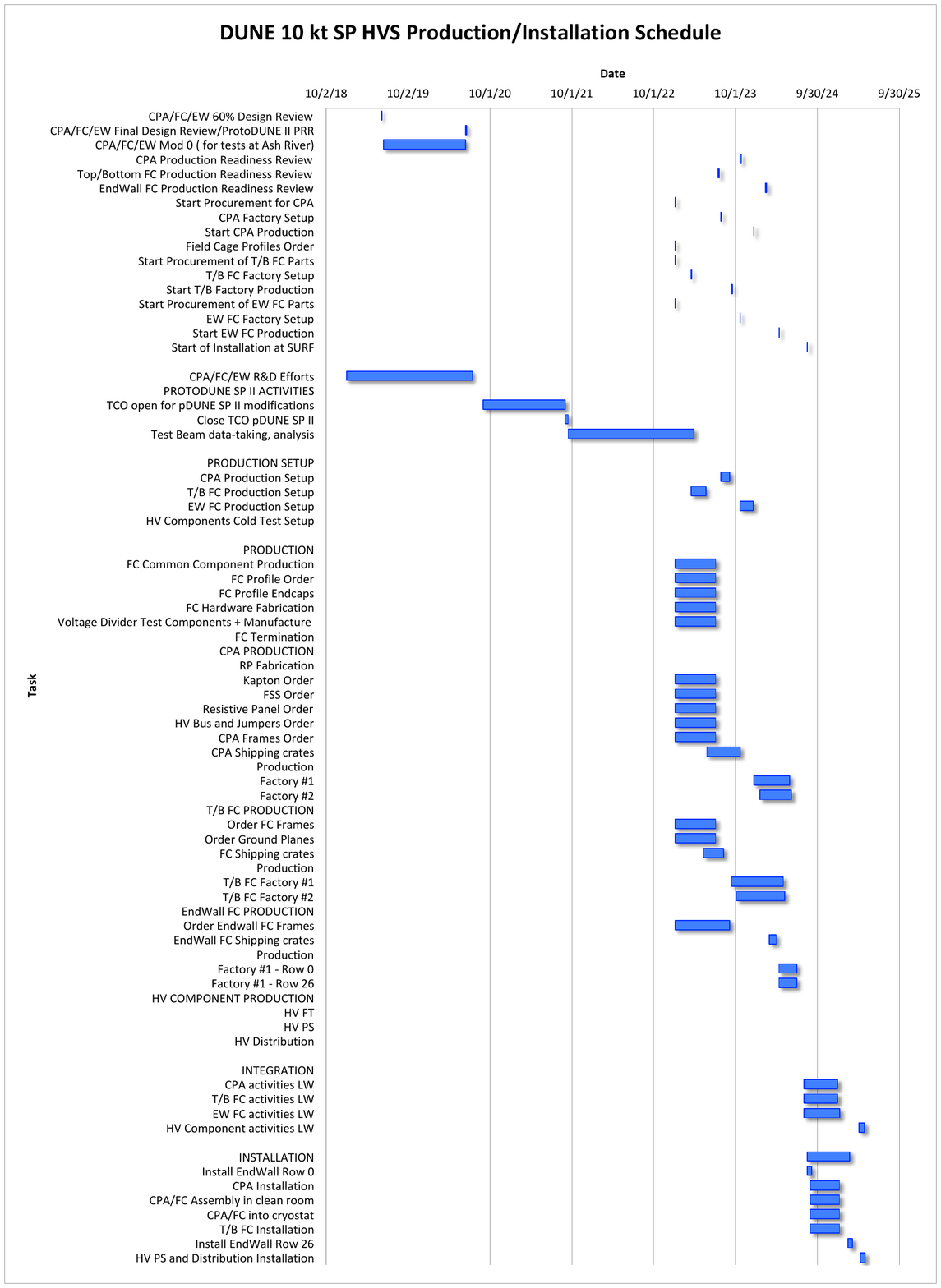}
\end{dunefigure}


\section{Appendix: Alternatives}
\label{sec:fdsp-hv-app-alt}

\subsection{Optical Reflectors on CPA}
\label{sec:fdsp-hv-app-alt-opt}

Since the \dwords{pd} in the current \dword{tpc} design are installed only on the \dword{apa} side of the drift volume and have low coverage, their responses to ionization inside the \dword{tpc} are highly dependent on drift distance and severely biased toward the \dword{apa}.  In order to improve the uniformity of response along the drift direction, the \dword{pd} consortium has proposed adding reflector foils coated with \dword{wls} to convert the UV photons arriving at the cathode into visible photons and bounce them back to the \dwords{pd} inside the \dword{apa}s.  Simulations have shown that addition of the reflectors  significantly improves the uniformity of response.

Implementing this concept, however, could dramatically alter the current \dword{cpa} characteristics and design.  The \dword{hvs} consortium has 
developed several concepts to accommodate the reflectors with minimal change to the 
\dword{cpa} design.  The main issue 
is the conductivity of the reflector foil versus the highly resistive nature of the \dword{cpa}.  To improve the light output, it would be best to 
cover as much of the cathode surfaces as possible, but 
large area coverage with conductive, (e.g., aluminum-coated) reflectors could short-circuit the resistive cathode and render it ineffective in slowing down the energy transfer during a potential \dword{hv} breakdown.  On the other hand, 
reflector foils made of insulating material would intercept the ionization charges drifting toward the cathode and become charged. This would alter the drift field uniformity and, worst yet, could result in random breakdown through the foil.

A design concept that is fairly simple to implement 
is depicted in Figure~\ref{fig:reflectorOnCPA}.  A 3M Vikuiti\footnote{Vikuiti\texttrademark is a light enhancement film produced by the 3M Company,  \url{http://multimedia.3m.com/mws/media/419882O/vikuititm-rear-projection-displays-brochure.pdf}.} 
reflector foil or equivalent is laminated onto a thin \frfour backing sheet to maintain thermal expansion compatibility
with the resistive \dword{cpa} panel, which also has an \frfour core. The reflector foil assembly is perforated at regular intervals to allow 
collection of electrons through the holes to the \dword{rp}  surface, minimizing the voltage build-up from charging of the non-perforated surfaces.  Several such foil assemblies are then tiled onto the existing \dwords{rp} with screws.

In order to advance the \dword{cpa} design while providing the option of adding the reflector foils at a later time, the \dword{hvs} consortium will design 
a hole pattern 
on the 
\dwords{rp} that could be used for mounting of reflector foils or panels, or left unused without negative consequences. In the meantime, \dword{hvs} and \dword{pd} consortia are conducting joint R\&D to evaluate a few design concepts and material choices.

\begin{dunefigure}[Concept to attach reflector to a CPA panel]{fig:reflectorOnCPA}{A concept to attach reflector foils to a \dword{cpa} panel. (Credit: BNL)}
\includegraphics[width=0.9\textwidth]{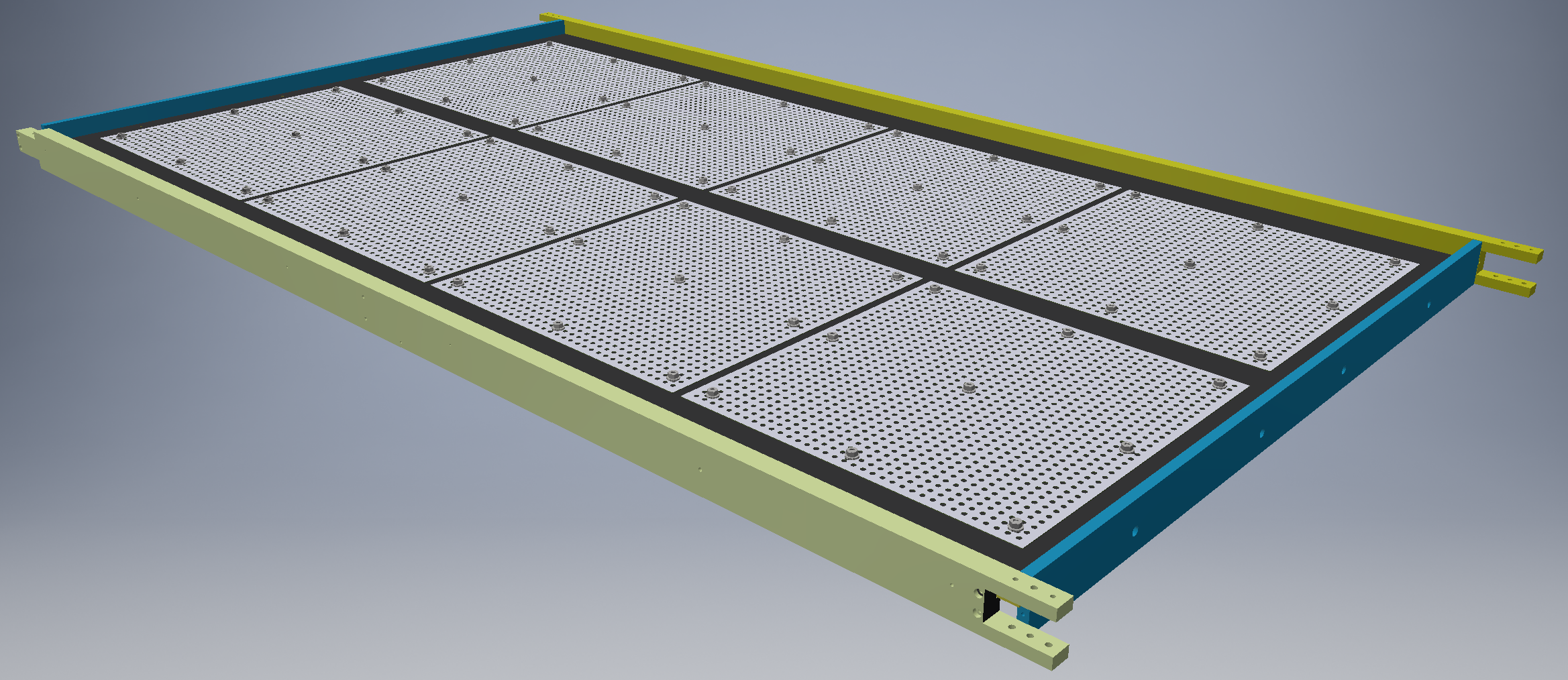}
\end{dunefigure}

\subsection{Calibration Laser Penetrations}
\label{sec:fdsp-hv-app-alt-cal}

The calibration consortium is developing requirements for calibrating the \efield.  One existing technique is to use UV laser beams to ionize the \dword{lar} and generate straight tracks along known trajectories.  Because the \dword{fc} surrounds the \dword{tpc} active volume, we can either shoot through the gaps between the \dword{fc} profiles (as in \dword{microboone}) or make openings in the \dword{fc} for the laser heads to pass through (as in \dword{sbnd}). Figure~\ref{fig:SBND_laser}  shows the design of a corner of the \dword{sbnd} \dword{tpc} with a \dword{fc}  opening and a calibration laser head through the opening.  Implementing such openings is straightforward if the openings are at the \dword{fc}  module boundaries.  Doing so through the interior surface of a \dword{fc} panel is more complicated but still simpler than the beam plug we designed for \dword{pdsp}.  There will be some minor drift field distortion around the openings.  Preliminary \dword{fea} studies have shown the field distortion to be negligible. 

\begin{dunefigure}[SBND laser arrangement]{fig:SBND_laser}{SBND field cage opening to allow a calibration laser head to pass through. (Credit: BNL)}
\includegraphics[width=0.7\textwidth]{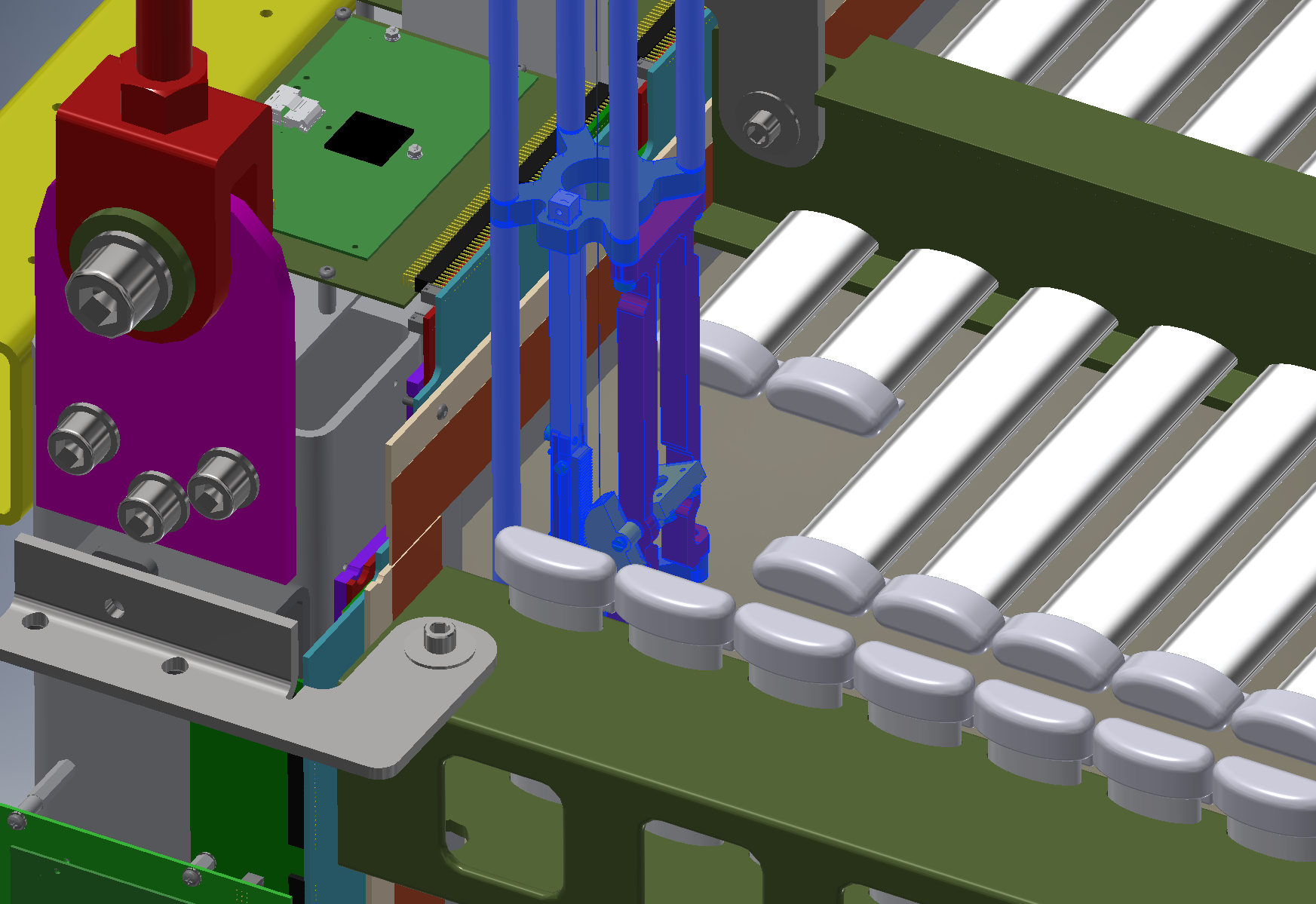}
\end{dunefigure}

\cleardoublepage

\chapter{TPC Electronics}
\label{ch:sp-tpcelec}

\section{System Overview}
\label{sec:fdsp-tpcelec-overview}

The \dword{tpc} electronics encompass the hardware systems necessary to 
amplify, digitize, and transmit the \dword{tpc} ionization charge 
signals out of a \dword{dune} \dword{sp} \dword{lartpc}. This 
includes the cryogenic \dword{fe} electronics (amplifiers, digitizers, 
digital controllers), power and data cabling and their cryostat 
feedthroughs, external (non-cryogenic) digital control electronics and 
power supplies, in addition to the system providing the bias voltage
to the \dwords{apa}.  The \dword{tpc} electronics as presented here does not 
include the electronics associated with the detection 
and recording of \dword{lar} scintillation photons, nor the \dword{daq} 
computing systems needed to capture and record these data.

The main difference between the \dword{dune} \dword{sp} \dword{detmodule} 
and previous experiments or prototypes using \dword{lar} technology is
that for the first time all the signal processing for the readout of the
wires of the \dword{apa}s takes place inside the \dword{lar}, in boards that 
are directly mounted on the \dword{apa}. This approach to the \dword{tpc}
readout was tested for the first time in the \dword{dune} \dword{35t},
and extensively tested in the \dword{pdsp} prototype. It has also 
been adopted by the \dword{sbnd} experiment. 
The \dword{tpc} \dword{fe} readout components immersed in the \dword{lar} are also referred 
to as the \dword{ce}. 

The \dword{fe}  electronics are mounted inside the \dword{lar} to exploit the fact that 
charge carrier mobility in silicon is higher, and thermal fluctuations are lower,  
at \dword{lar} temperature than at room temperature. For \dword{cmos} 
electronics, this results in substantially higher gain and lower noise 
at \dword{lar} temperature than at room temperature~\cite{DeGeronimo:2011zz}.
Mounting the \dword{fe} electronics on the \dword{apa} frames also minimizes 
the input capacitance, which further contributes to the noise reduction.  
Furthermore, placing the digitizing and multiplexing electronics inside 
the cryostat reduces the total number of penetrations into the cryostat 
and minimizes the number of cables coming out of it.

As the full \dword{tpc} electronics chain for the \dword{spmod} includes 
many components on the warm side of the cryostat as well, the \dword{dune} 
consortium designated to develop this system is formally called 
the \dword{dune} \dword{sp} \dword{tpc} electronics consortium.

This overview section starts with a review of the considerations that
have led to the proposed design for the \dword{dune} \dword{sp} detector, 
then discusses how the detector requirements follow from the physics goals 
of the experiment. The reader will find a detailed description of all the 
\dword{tpc} electronics detector components in Section \ref{sec:fdsp-tpcelec-design},
including a discussion of how the lessons learned from the construction,
integration, installation and commissioning of \dword{pdsp} have informed
the design of the \dword{dune} \dword{sp} module and how the early data
from \dword{pdsp} validate this design. The description of the detector
design is then followed by discussions 
of the \dword{qa} program and the plans for production and assembly, 
and for integration, installation and commissioning, in 
Sections~\ref{sec:fdsp-tpcelec-qa}--\ref{sec:fdsp-tpcelec-integration}. 
Section~\ref{sec:fdsp-tpcelec-interfaces} discusses the interfaces with 
detector components provided by other consortia, with \dword{tc}, and with the physics group. 
Sections~\ref{sec:fdsp-tpcelec-safety}--\ref{sec:fdsp-tpcelec-management} 
conclude the chapter with plans for addressing safety issues and risks during the
construction, installation, and operation of the detector, and 
an outline of the organization of the \dword{tpc} electronics consortium, 
with a timeline for the \dword{detmodule} construction and an estimate
of the resources required.

\subsection{Introduction}
\label{sec:fdsp-tpcelec-overview-intro}

In the \dword{dune} \dword{spmod} a \dword{mip} deposits on average between
\SI{20}{k}{e$^-$} and \SI{30}{k}{e$^-$} on each collection wire, assuming a drift \efield
of \spmaxfield and an electron lifetime of \SI{6}{ms}, as discussed in
Chapter~\ref{ch:fdsp-execsum}, and assuming full transparency during the 
electron transport through the grid plane of the \dword{apa} and its two planes of induction
wires, as discussed in Chapter~\ref{ch:fdsp-apa}. The larger of the two numbers 
is for \dwords{mip} close to the anode plane, and the smaller
takes into account the electron capture by electronegative impurities during the electron
drift for tracks close to the cathode plane. 

The \dword{dune} \dword{sp} \dword{tpc} is a unit-gain device where the 
electrical signal is produced by the drift of the charges near the wires, 
in contrast to signal production in gaseous wire 
chambers, where the \efield is strong enough to provide additional
ionization and signal multiplication. The signal induced in the \dword{dune}
\dword{spmod} wires is bipolar on the induction wires, negative when the
electrons drift toward the wires, and positive when they drift away from
the wires. On the collection wires signals are unipolar (negative).
The signal duration is of the order of microseconds, and tends to be narrower
for the collection plane due to the enhancement of the weighting field for
the collection wires. Due to the lack of amplification of electrons inside 
\dword{lar}, low noise is essential for the \dword{ce} to reliably extract the 
ionization electron signal from both the collection and induction wire planes. 

The reduction in noise level obtained with the \dword{ce} 
greatly extends the reach of the \dword{dune} 
physics program. It allows measurement of smaller charge deposits, which 
mitigates  the risks of inability to reach the desired drift field 
or a lower electron lifetime than desired due to electronegative impurities.
For example, given an electron lifetime of \SI{3}{ms} and a drift \efield
of \SI{0.25}{kV/cm}, the charge deposited in the collection wires from a 
\dword{mip} close to the cathode plane is reduced to \SI{10}{k}{e$^-$}.
The exact 
minimal \dword{s/n} required for pattern
recognition depends on the tracking algorithms and the offline signal 
processing. 
We use the
minimal requirement of a total \dword{enc} less than 1000 e$^-$, consistent 
with a \dword{s/n} of at least 10 on the collection wires, even in the
pessimistic case where the electron lifetime and the \efield just meet 
the required design values discussed in Chapter~\ref{ch:sp-hv}. Considering
the difference in signal amplitudes between collection and induction wires
and the bipolar shape of the signal on the latter, this requirement corresponds 
to a \dword{s/n} of at least 5 on the induction wires. This asymmetric requirement 
for the minimal \dword{s/n} on the collection and induction wires was first 
adopted by the \dword{sbnd} experiment~\cite{bib:sbnddoc1921}.

The goal is to keep the total noise level as low as possible. For example, an 
increase in the \dword{s/n} above 15 allows the observation of MeV-scale photons, 
as recently demonstrated by \dword{argoneut}~\cite{Acciarri:2018myr}. This enables  
reconstruction of both photons released during de-excitation of the nucleus and 
part of the energy transferred to final-state neutrons. Low noise is also crucial
for the baseline oscillation analysis described in \physchlbl. The event classification is based
on a \dword{cvn} that uses as inputs three images of the neutrino interactions, 
one for each of the three readout views, using the reconstructed hits on the 
individual wire planes. This approach relies on low noise levels. Decreasing 
the noise level also increases the reach of low-energy physics measurements 
like those associated with stellar core-collapse \dword{snb}. 
Finally, a low noise level opens up the possibility of using $\mathrm{{}^{39}Ar}$ 
beta decays to calibrate the \dword{dune} \dword{spmod}~\cite{MICROBOONE-NOTE-1050-PUB}.
Instead of zero suppression, the \dword{dune} \dword{daq} system uses lossless 
data compression, as discussed in Section~\ref{sec:daq:requirements}, that
becomes more efficient as the noise level is reduced. Therefore, the noise level 
also affects the bandwidth requirements for the \dword{daq} system,
discussed in Chapter~\ref{ch:daq};
these bandwidth requirements can be a limiting factor for low-energy
physics signals, particularly those of astrophysical origin.

\begin{dunefigure}
[The reference architecture for the TPC electronics]
{fig:ce-scheme}
{The reference architecture for the \dword{tpc} electronics. The basic unit is the
128-channel \dword{femb}. The scheme includes also the \dwords{sipm} used for the 
readout of the \dwords{pd}, as discussed in Chapter~\ref{ch:fdsp-pd}.}
\includegraphics[width=0.99\linewidth]{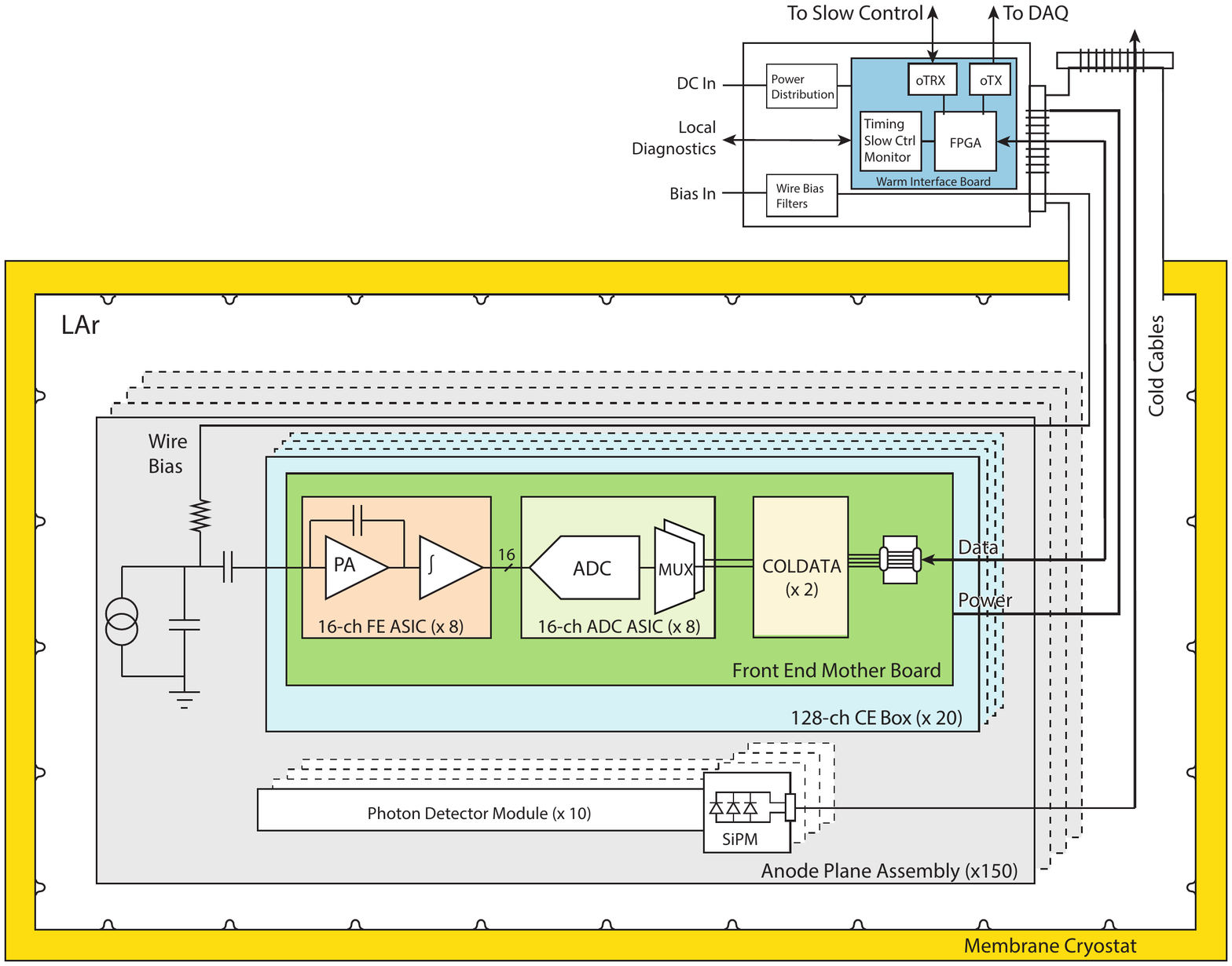}
\end{dunefigure}

To retain maximum flexibility in optimizing reconstruction algorithms after 
the \dword{dune} data is collected, the \dword{tpc} electronics are designed 
to produce a digital record representing the waveform of the current produced 
by charge collection and induction on the anode wires. Each anode wire signal is 
input to a charge-sensitive amplifier, followed by a pulse-shaping circuit and 
an \dword{adc}. To minimize the number of cables and cryostat penetrations, 
the \dwords{adc} as well as the amplifier/shapers are located in the \dword{lar}, 
and digitized data from many wires merge onto a much smaller set of high-speed 
serial links. The \dword{tpc} signal processing is implemented in \dwords{asic} 
using \dword{cmos} technology. The \dword{tpc} is continuously read out, resulting 
in a digitized \dword{adc} sample from each \dword{apa} channel (wire). The 
\dwords{asic} used for the readout of the \num{2560} wires of each \dword{apa} 
are mounted on \dfirsts{femb}, as shown in Figure~\ref{fig:ce-scheme}. These are
connected to \dwords{wib} located outside of the cryostat via the \dword{ce} signal 
cable flange located at the \dword{ce} \fdth at the top of the cryostat.
The \dwords{wib} are installed, together with \dfirsts{ptc} that distribute
the power and the clock and control signals, in a \dword{wiec} that is
mounted on the signal flange. From the \dwords{wib} the data is sent to 
the \dword{daq} back-end on an optical fiber network, as discussed in 
Chapter~\ref{ch:daq}. 

\subsection{Requirements and Specification}
\label{sec:fdsp-tpcelec-overview-requirements}

A number of specifications are imposed on the \dword{tpc} electronics in addition to the 
noise requirement (\dword{enc}~$<\SI{1000}{e^-}$). Some of them, labeled as 
SP-FD in Table~\ref{tab:specs:SP-ELEC}, are derived from \dword{dune}'s 
overall physics goals. The rest, labeled as SP-ELEC, are engineering specifications 
derived from the design choices for the \dword{ce}.

\begin{footnotesize}
\begin{longtable}{p{0.12\textwidth}p{0.18\textwidth}p{0.17\textwidth}p{0.25\textwidth}p{0.16\textwidth}}
\caption{TPC electronics specifications \fixmehl{ref \texttt{tab:spec:SP-ELEC}}} \\
  \rowcolor{dunesky}
       Label & Description  & Specification \newline (Goal) & Rationale & Validation \\  \colhline

  \newtag{SP-FD-2}{ spec:system-noise }  & System noise  &  $<\,\SI{1000}\,e^-$ &  Provides $>$5:1 S/N on induction planes for  pattern recognition and two-track separation. &  ProtoDUNE and simulation \\ \colhline

  \newtag{SP-FD-13}{ spec:fe-peak-time }  & Front-end peaking time  &  \SI{1}{\micro\second} &  Vertex resolution; optimized for \SI{5}{mm} wire spacing. &  ProtoDUNE and simulation \\ \colhline
    
   \newtag{SP-FD-14}{ spec:sp-signal-saturation }  & Signal saturation level  &  \num{500000} $e^-$ \newline (Adjustable so as to see saturation in less than \SI{10}{\%} of beam-produced events) &  Maintain calorimetric performance for multi-proton final state. &  Simulation \\ \colhline

  \newtag{SP-FD-19}{ spec:adc-sampling-freq }  & ADC sampling frequency  &  $\sim\,\SI{2}{\mega\hertz}$ &  Match \SI{1}{\micro\second} shaping time. &  Nyquist requirement and design choice \\ \colhline

  \newtag{SP-FD-20}{ spec:adc-number-of-bits }  & Number of ADC bits  &  \num{12} bits &  ADC noise contribution negligible (low end); match signal saturation specification (high end). &  Engineering calculation and design choice \\ \colhline

  \newtag{SP-FD-21}{ spec:ce-power-consumption }  & Cold electronics power consumption   &  $<\,\SI{50}{ mW/channel} $ &  No bubbles in LAr to reduce HV discharge risk. &  Bench test \\ \colhline

  \newtag{SP-FD-25}{ spec:non-fe-noise }  & Non-FE noise contributions  &  $<<\,\SI{1000}\,e^- $ &  High S/N for high reconstruction efficiency. &  Engineering calculation and ProtoDUNE \\ \colhline

  \newtag{SP-FD-28}{ spec:dead-channels }  & Dead channels  &  $<\,\SI{1}{\%}$ &  Minimize the degradation in physics performance over the $>\,20$-year detector operation. &  ProtoDUNE and bench tests \\ \colhline

  \newtag{SP-ELEC-1}{ spec:num-FE-baselines }  & Number of baselines in the front-end amplifier  &  \num{2} &  Use a single type of amplifier for both induction and collection wires &  ProtoDUNE \\ \colhline
    
   \newtag{SP-ELEC-2}{ spec:gain-FE-amplifier }  & Gain of the front-end amplifier  &  $\sim\SI{20}{mV/fC}$ \newline (Adjustable in the range \SIrange{5}{25}{mV/fC}) &  The gain of the FE amplifier is obtained from the maximum charge to be observed without saturation and from the operating voltage of the amplifier, that depends on the technology choice. &   \\ \colhline
    
   \newtag{SP-ELEC-3}{ spec:syncronization-CE }  & System synchronization  &  \SI{50}{ns} \newline (\SI{10}{ns}) &  The dispersion of the sampling times on different wires of the APA should be much smaller than the sampling time (500 ns) and give a negligible contribution to the hit resolution. &   \\ \colhline

  \newtag{SP-ELEC-4}{ spec:num-channels-FEMB }  & Number of channels per front-end motherboard  &  \num{128} &  The total number of wires on one side of an APA, 1,280, must be an integer multiple of the number of channels on the FEMBs. &  Design \\ \colhline
    
   \newtag{SP-ELEC-5}{ spec:FEMB-data-link }  & Number of links between the FEMB and the WIB  &  \num{4} at \SI{1.28}{Gbps} \newline (\num{2} at \SI{2.56}{Gbps}) &  Balance between reducing the number of links and reliability and power issues when increasing the data transmission speed. &  ProtoDUNE, Laboratory measurements on bit error rates \\ \colhline

  \newtag{SP-ELEC-6}{ spec:number-FEMB-per-WIB }  & Number of FEMBs per WIB  &  \num{4} &  The total number of FEMB per WIB is a balance between the complexity of the boards, the mechanics inside the WIEC, and the required processing power of the FPGA on the WIB.  &  ProtoDUNE, Design \\ \colhline

  \newtag{SP-ELEC-7}{ spec:WIB-data-link }  & Data transmission speed between the WIB and the DAQ backend  &  \SI{10}{Gbps} &  Balance between cost and reduction of the number of optical fiber links for each WIB. &  ProtoDUNE, Laboratory measurements on bit error rates \\ \colhline

  \newtag{SP-ELEC-8}{ spec:cold-cables-xsec }  & Maximum diameter of conduit enclosing the cold cables while they are routed through the APA frame  &  \SI{6.35}{cm} (2.5") &  Avoid the need for further changes to the APA frame and for routing the cables along the cryostat walls &  Tests on APA frame prototypes \\ \colhline

\label{tab:specs:SP-ELEC}
\end{longtable}
\end{footnotesize}

\begin{itemize}
\item SP-FD-13: The \dword{fe} peaking time must be in the range \numrange{1}{3}\,\si{\micro\second} 
to match the time required for the drifting charges to travel from one plane of anode
wires to the next, which corresponds to the typical duration of the signal observed
on the wires. The planes of anode wires are separated by \SI{4.75}{mm} 
(see Chapter~\ref{ch:fdsp-apa}), and the drift velocity for the \efield{}s 
considered for \dword{dune} is in the range \SIrange{1.2}{1.6}{mm/$\mu$s}
(\SIrange{1.4}{2.1}{mm/$\mu$s} for the gaps between the \dword{apa} wire planes).
A \dword{fe} peaking time  similar to the typical signal duration improves the 
detector's two-track resolution.  

\item SP-FD-14: The system must have a linear response up to an impulse input of 
at least \num{500000}\,$e^{-}$.  This corresponds roughly to the largest 
ionization signals expected. These occur in events where multiple protons are produced 
in the primary event vertex, in particular, when the trajectories of one 
or more of the protons are parallel to the wire, 
leading to collection of charge over a long path length within a short time.

\item SP-FD-19:  The \dword{adc} sampling frequency must be \SI{{\sim}2}{MHz},
This value is chosen to match a \dword{fe} shaping time of \SI{1}{\micro\second} 
(approximate Nyquist condition) while minimizing the data rate.

\item SP-FD-20: The \dword{adc} must digitize the charge deposited on the wires 
with 12~bits of precision.  The lower end of the \dword{adc} dynamic 
range is driven by the requirement that the digitization not contribute 
to the total electronics noise, as defined by requirement SP-FD-25. The upper end
is defined by SP-FD-14. Combining this with SP-FD-02 on the total electronics noise 
results in the need for 12~bits digitization. 

\item SP-FD-21: Preliminary studies indicate that the power dissipated by the 
electronics located in the \dword{lar} should be less than \SI{50}{mW/channel}. 
Lower power dissipation is desirable because the mass of the power cables scales 
with power. Ongoing studies focus on whether the amount of power dissipated by 
the electronics should be minimized further because of potential complications from 
argon boiling. 

\item SP-FD-25: 
The components of the readout chain, including the \dword{adc}
and the bias voltage supplies, together must not contribute significantly to
the overall noise. The \dword{adc} specifications for non-linearity and 
noise will depend on the gain of the \dword{fe}. Like in the case of
the requirement on the noise caused by voltage ripples on the cathode
(SP-FD-12) discussed in Section~\ref{sec:fdsp-hv-des-consid} we are
aiming to keep all sources of noise other than the \dword{fe}
amplifier below \SI{100}{$e^-$}.

\item SP-FD-28: The fraction of non-functioning channels over 
\dword{dune}'s nominal \dunelifetime lifetime must not exceed 1\%. 
Ongoing studies will quantify the effect of failures
in the \dword{tpc} and electronics, including
single wire failures, and failures of groups of
\num{16}, \num{64}, or \num{128} channels.

\item SP-ELEC-1: The \dword{fe} must have an adjustable baseline such that a single
amplifier can process both the bipolar signal from the induction wires and the mostly 
unipolar signal from the collection wires. 

\item SP-ELEC-2: The \dword{fe} must have a gain that allows using the entire
voltage range provided by the chosen chip fabrication technology and operating
voltage without saturation for physics signals up to those specified in
SP-FD-14. Multiple gain settings could be made available to allow for optimization
of the detector performance.

\item SP-ELEC-3: The dispersion of the sampling times on the different wires of
all the \dword{apa}s in one \dword{dune} \dword{sp} detector module should
be less than \SI{50}{ns}. This value is much smaller than the 
time difference between two subsequent samples on the same wire
as defined by the sampling frequency (SP-FD-19) such that it gives
a negligible contribution to the single hit resolution (assuming
a drift velocity of \SI{1.6}{mm/$\mu$s}, the requirement of \SI{50}{ns}
corresponds to a contribution to the single hit resolution of \SI{80}{$\mu$m}).

\item SP-ELEC-4: The readout electronics for the \dword{apa} wires must be organized
into \dwords{femb} containing 128 channels. This number is a sub-multiple
of the number of wires on an \dword{apa} and is determined
by geometrical considerations, e.g., the number, size, and form
factor of the \dword{cr} boards introduced in Section~\ref{sec:fdsp-tpcelec-design-bias}.

\item SP-ELEC-5: The data from the \dwords{femb} must be transmitted to the
\dwords{wib} on a maximum of four links per board, each with a maximum
speed of \SI{1.28}{Gbps}, to minimize
the number of connections on the cryostat penetrations. This 
requires data transmission at high speeds, which 
increases the power consumption inside the \dword{lar}.
A reduction in the number of links per \dword{femb} to two
(with a link speed of \SI{2.56}{Gbps}) will be investigated.

\item SP-ELEC-6: Each \dword{wib} must read out four \dwords{femb}. This number
is chosen to balance the complexity of the boards, the mechanics
of the \dword{wiec} that houses the \dwords{wib}, and the 
required processing power in the \dword{fpga} inside the
\dword{wib}.

\item SP-ELEC-7: Each \dword{wib} must transmit data to the \dword{daq}
back end on optical links at a speed of $\sim\!\SI{10}{Gbps}$. This speed 
is a compromise between the cost of optical transmitters and
receivers and the complexity of the readout fiber plant.

\item SP-ELEC-8: All the cables required to provide the low-voltage power
and the control and readout for the \dwords{femb} mounted on
the bottom \dword{apa}, plus the bias voltage cables for
the same \dword{apa}, must fit inside two conduits with a
diameter of \SI{6.35}{cm} (2.5 inch) that are inserted in the frame of
the \dword{apa}, as discussed in Section~\ref{sec:fdsp-tpcelec-interfaces-apa}. 
\end{itemize}

\subsection{Design}
\label{sec:fdsp-tpcelec-overview-design}

The reference design of the \dword{tpc} electronics detector components is
based on the specifications presented in Section~\ref{sec:fdsp-tpcelec-overview-requirements}. 
Each individual \dword{apa} has \num{2560} channels read out by \num{20} 
\dfirsts{femb}, with each \dword{femb} enabling digitized wire readout 
from \num{128} channels. One cable bundle connects each \dword{femb} to
the outside of the cryostat via a \dword{ce} signal cable flange located 
at the \dword{ce} \fdth at the top of the cryostat, where a single flange 
services each \dword{apa}, as shown in Figure~\ref{fig:connections}. 
Two \dword{ce} signal flanges are on each \fdth and together account for 
all electronics channels associated with a pair of \dword{apa}s (upper 
and lower, vertically arranged). Each cable bundle contains wires for 
low-voltage (\dword{lv}) power, high-speed data readout, and clock or 
digital-control signal distribution. Eight separate cables carry the 
\dword{tpc} wire bias voltages from the signal flange to the \dword{apa} 
wire bias boards, in addition to the bias voltages for the field cage 
termination electrodes and for the electron diverters. An additional 
flange on the top of each \fdth services the \dword{pds} cables associated 
with the \dword{apa} pair. Low-voltage power supplies and bias-voltage
power supplies are located on the top of the cryostat. 

\begin{dunefigure}
[Connections between the signal flanges and APA]
{fig:connections}
{Connections between the signal flanges and \dword{apa}. The lower 
\dword{apa} shares the \dword{pd} flange with the 
upper \dword{apa} but has a separate TPC readout flange.}
\includegraphics[width=0.7\textwidth]{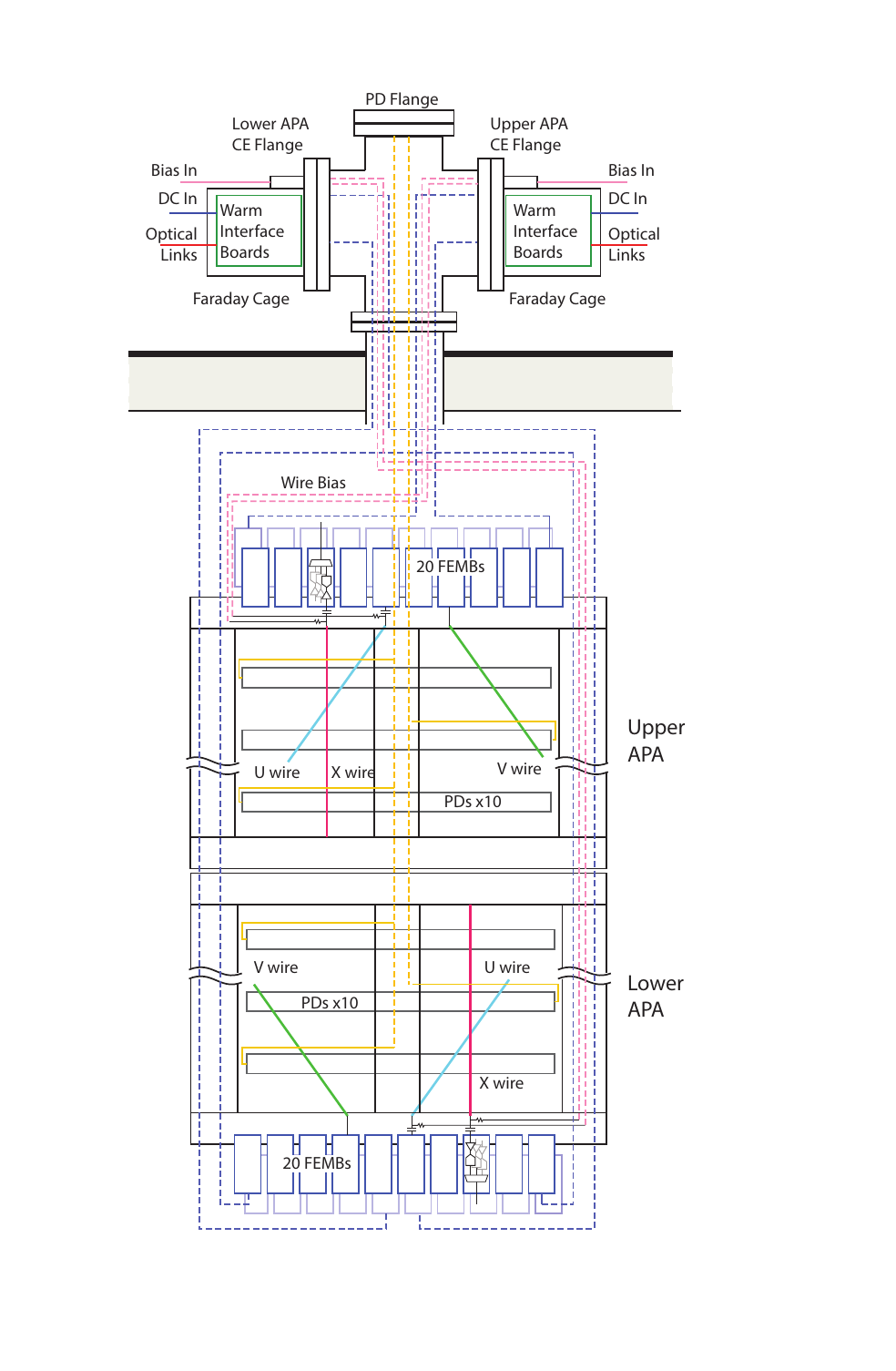}
\end{dunefigure}

The reference design for the \dword{ce} calls for three 
types of custom \dwords{asic} inside  the \dword{lar}:
\begin{itemize}
\item{a \num{16}-channel \dword{fe} \dword{asic} for amplification 
and pulse shaping (referred to as \dword{larasic});}
\item{a \num{16}-channel \num{12}-bit \dword{adc} \dword{asic} 
operating at \SI{{\sim}2}{MHz} (referred to as \dword{coldadc}); and}
\item{a \num{64}-channel control and communications \dword{asic} 
(referred to as \dword{coldata}).}
\end{itemize}

The \dword{tpc} electronics detector components required for one \dword{apa} are: 
\begin{itemize}
\item{\dwords{femb}, on which the \dwords{asic} are mounted, and 
which are installed on the \dword{apa}s;}
\item{cables for the data, clock, and control signals; \dword{lv} 
power; and wire bias voltages between the \dword{apa} and the 
signal flanges (cold cables);}
\item{signal flanges with a \dword{ce} \fdth to pass the data, clock, 
and control signals; \dword{lv} power; and \dword{apa} wire bias 
voltages between the inside and outside of the cryostat; and 
the corresponding cryostat penetrations and spool pieces;}
\item{\dwords{wiec} mounted on the signal flanges 
containing the \dwords{wib} and a \dword{ptc} for further processing
and distribution of the signals entering and exiting the cryostat;
low voltage power and clock and control signals are transmitted
from the \dwords{ptc} to the \dwords{wib} on 
the \dword{ptb}.}
\item{cables for \dword{lv} power and wire bias voltages between 
the signal flange and external power supplies (warm cables); and}
\item{\dword{lv} power supplies for the \dword{ce} and bias-voltage 
power supplies for the \dword{apa}s.}
\end{itemize}

The number of channels (wires) connected to each of these
components is given in Table~\ref{tab:elecNums}.

\begin{dunetable}
[TPC electronics components and quantities for a single APA]
{llr}
{tab:elecNums}
{TPC electronics components and quantities for a single \dword{apa} of the DUNE \dword{spmod}.}
\textbf{Element} &\textbf{Quantity} & \textbf{Channels per element}\\ \toprowrule
Front-end mother board (\dword{femb}) & \num{20} per \dword{apa} & \num{128} \\ \colhline
\dword{fe} \dword{asic} chip & \num{8} per \dword{femb} & \num{16} \\ \colhline
\dword{adc} \dword{asic} chip & \num{8} per \dword{femb} & \num{16} \\ \colhline
\dword{coldata} \dword{asic} chip & \num{2} per \dword{femb} & \num{64} \\ \colhline
Cold cable bundle & \num{1} per \dword{femb} & \num{128} \\ \colhline
Signal flange & \num{1} per \dword{apa} & \num{2560} \\ \colhline
\dword{ce} \fdth & \num{1} per \dword{apa} pair & \num{5120} \\ \colhline
Warm interface board (\dword{wib}) & \num{5} per \dword{apa} & \num{512} \\ \colhline
Warm interface electronics crate (\dword{wiec}) & \num{1} per \dword{apa} & \num{2560} \\ \colhline
Power and timing card (\dword{ptc}) & \num{1} per \dword{apa} & \num{2560} \\ \colhline
Power and timing backplane (\dword{ptb}) & \num{1} per \dword{apa} & \num{2560} \\
\end{dunetable}

The electronics located inside the cryostat cannot be replaced or repaired after the
cryostat has been filled with \dword{lar}. Successful operation of the readout electronics 
in \dword{lar} for the \dunelifetime of \dword{dune} operation imposes technological 
choices for the \dword{spmod} \dwords{asic}, 
and specific constraints on commercial components that are installed
inside the \dword{lar}. While the higher charge carrier 
mobility~\cite{Hairapetian1989} at \dword{lar} temperature than at room
temperature is central to improving the performance of the  \dword{ce}, it also leads
to the hot-carrier effect~\cite{Hot-electron}, which limits the lifetime of \dwords{asic}.
In n-type \dword{cmos} transistors, the carriers (electrons)
can acquire enough kinetic energy to ionize silicon in the active channel. This
charge can become trapped and lead to effects (including threshold shifts)
similar to those caused by radiation damage, which can cause \dword{cmos}
circuits to age much more quickly at \dword{lar} temperature, 
reducing performance and potentially causing failure. To mitigate the hot carrier effect,
the maximum \efield in transistor channels must be lower than 
that which could be used reliably at room temperature. The reduction of
the maximum \efield is achieved by operating the \dwords{asic} at a
reduced bias voltage and by increasing by $\sim$50\% the length 
of the transistors' channels. Another drawback of integrated circuits 
operated at \dword{lar} temperature is that the spread of the transistor 
properties becomes larger, making it more difficult to rely on transistor 
matching for circuit design. We must carefully test any commercial 
circuits used in the \dword{lar} to ensure they will perform well for 
the expected experiment lifetime. Reliability studies for \dword{tpc}
electronics designs under consideration are discussed in Section~\ref{sec:fdsp-tpcelec-qa-reliability}.

\section{System Design}
\label{sec:fdsp-tpcelec-design}

In order to achieve the lowest possible overall noise in the readout of 
the \dword{apa} wire planes, all possible sources of noise need to be kept to a
minimum. This requires minimizing the noise sources in each
of the components of the readout chain of the \dword{apa} wires, such as
the \dword{fe} amplifier noise. It also requires that all system aspects
are taken into account, including avoiding channeling noise inside
the cryostat through ground connections and through the readout
chain of other detector components, like the \dword{pds}, the \dword{hvs},
or the cryogenic instrumentation. 

In this section we describe the overall
system design of the \dword{tpc} electronics, starting in
Section~\ref{sec:fdsp-tpcelec-design-grounding} with a description of the
grounding and shielding scheme adopted in the \dword{dune} \dword{spmod}
to minimize the overall noise in the detector, followed in 
Section~\ref{sec:fdsp-tpcelec-design-bias} by a discussion of the bias
voltage distribution system. Later, we describe in
Section~\ref{sec:fdsp-tpcelec-design-femb} the \dwords{femb}, including
the design of the \dwords{asic} that are being considered for
use in \dword{dune}. In Section~\ref{sec:fdsp-tpcelec-design-infrastructure}
we discuss the infrastructure for the \dword{ce} inside the cryostat,
which includes the \coldbox{}es that shield the \dwords{femb}, the
cold cables, and the cable trays. Then in 
Sections~\ref{sec:fdsp-tpcelec-design-ft}-\ref{sec:fdsp-tpcelec-design-services} we discuss 
the infrastructure on the top of the cryostat, including the
feedthroughs, the \dwords{wiec}, the timing distribution and
synchronization system, and the services that provide the low
voltage power and the bias voltage to the \dword{tpc} electronics. 
The design presented here is very similar to the one used for
\dword{pdsp}. The results obtained with this detector are 
discussed in Section~\ref{sec:fdsp-tpcelec-overview-pdune}.
Then in Section~\ref{sec:fdsp-tpcelec-overview-lessons} we discuss
how the lessons learned from the construction, installation,
commissioning, and operation are informing the final design
of the \dword{dune} \dword{sp} detector module. Later in
Section~\ref{sec:fdsp-tpcelec-overview-remaining} we conclude
with a discussion of the design maturity and of
the remaining prototype activities that are required prior to
the beginning of the detector construction. Other aspects of system
design pertaining to the interfaces with other detector components, including
their grounding, are discussed in Section~\ref{sec:fdsp-tpcelec-interfaces}.

\subsection{Grounding and Shielding}
\label{sec:fdsp-tpcelec-design-grounding}

The overall approach to minimizing the system noise in the  \dword{spmod} 
relies on enclosing the sensitive wire planes in a nearly hermetic 
Faraday cage, and then carefully controlling currents flowing into or 
out of that protected volume through the unavoidable penetrations 
needed to build a working detector. Done carefully, this can result 
in avoiding all unwanted disturbances that result in detector noise. 
Such disturbances could either be induced on the signal wires by 
changing currents flowing inside the cryostat or even on the cryostat 
walls as, for instance, a temperature sensing circuit that acts as a 
receiving antenna on the outside of the cryostat and a transmitting 
antenna on the interior of the cryostat. In addition, unwanted signals 
might be injected into the electronics either in the cold or just 
outside the cryostat by direct conduction along unavoidable power 
or signal connections to other devices. This approach to minimizing
the detector noise by using appropriate grounding and shielding procedures
is discussed in detail in~\cite{radekaNoise}. 
It results in the 
following set of requirements that need to be respected during the
design and the construction of the  \dword{spmod}:
\begin{itemize}
\item{The \dword{apa} frame shall be connected to the common of
all the \dword{fe} \dwords{asic};}
\item{All electrical connections (low voltage power, bias voltage,
clock, control, and data readout) from one \dword{apa} shall lead to a 
single \dword{sft};}
\item{All \dword{apa}s shall be insulated from each other;}
\item{The common of the \dword{fe} \dword{asic} and the rest of the 
\dword{tpc} readout electronics shall be connected to
the common plane of the \dword{femb};}
\item{The return leads of the \dword{apa} power line and any shield
for the clock, control, and data readout shall be connected
to the common plane of the \dword{femb} at one end and
to the flange of the \dword{sft} at the other end; this shall be 
the only connection of the \dword{apa} frame to the cryostat;}
\item{The mechanical suspension from the frame of the \dword{apa}
to the cryostat shall be insulated;}
\item{The last stage of the sense wire and grid bias filters shall be
connected to the common of all the \dword{fe} \dwords{asic} and therefore
to the \dword{apa} frame.}
\item{Similarly the last element of the \dword{fc} divider chain shall
be connected with an appropriate termination to the \dword{apa}
frame, as discussed in Section~\ref{sec:fdsp-hv-design-interconnect}.}
\end{itemize}

These requirements have been 
followed already for the construction
of the \dword{pdsp} prototype. As discussed in Section~\ref{sec:fdsp-tpcelec-overview-pdune},
the initial results from the online monitoring and the
analysis of \dword{pdsp} indicate that the system
noise requirements for the \dword{dune} \dword{spmod}
can be met.

To minimize system noise, the \dword{tpc} electronics cables for each \dword{apa} 
enter the cryostat through a single \dword{ce} flange, as shown in 
Figure~\ref{fig:connections}. This creates, for grounding purposes, 
an integrated unit consisting of an \dword{apa} frame, \dword{femb}
ground for all \num{20} \dword{ce} modules, a \dword{tpc} flange, and 
warm interface electronics. The input amplifiers on the 
\dword{fe} \dwords{asic} have their ground terminals connected to 
the \dword{apa} frame. All power-return leads and cable shields are 
connected to both the ground plane of the \dword{femb} and to the 
\dword{tpc} signal flange.

The only location where this integrated unit makes electrical contact 
with the cryostat, which defines the detector ground and acts as a 
Faraday cage, is at a single point on the \dword{ce} \fdth board in 
the \dword{tpc} signal flange where the cables exit the cryostat. 
Mechanical suspension of the \dword{apa}s is accomplished using 
insulated supports. To avoid structural ground loops, the \dword{apa} 
frames described in Chapter~\ref{ch:fdsp-apa} are insulated from 
each other.

Filtering circuits for the \dword{apa} wire-bias voltages are 
locally referenced to the ground plane of the \dwords{femb} 
through low-impedance electrical connections. This approach 
ensures a ground-return path in close proximity to the 
bias-voltage and signal paths. The close proximity of the 
current paths minimizes the size of potential loops to further 
suppress noise pickup.

Signals associated with the \dword{pds}, described in 
Chapter~\ref{ch:fdsp-pd}, are carried directly on shielded, 
twisted-pair cables to the signal \fdth. The cable shields 
are connected to the cryostat at the \dword{pd} flange 
shown in Figure~\ref{fig:connections}, and to the \dword{pcb} 
shield layer on the \dwords{pd}. The cable shields have no 
electrical connection to the \dword{apa} frame or the \dword{tpc}
electronics.

Further aspects of the \dword{dune} grounding scheme are 
discussed in \tcchfac and in~\cite{DUNE:GroundingRules}.

\subsection{Distribution of Bias Voltages}
\label{sec:fdsp-tpcelec-design-bias}

Each side of an \dword{apa} includes four wire layers, as 
described in Section~\ref{sec:fdsp-apa-design}. Electrons passing 
through the wire grid must drift unimpeded until they reach 
the $X$-plane collection layer. The nominal bias voltages, chosen 
to result in this electrically transparent configuration, 
are given in Section~\ref{sec:fdsp-apa-design}. 

The filtering of wire bias voltages and the \dword{ac} coupling 
of wire signals passing onto the charge amplifier circuits is 
done on \dword{cr} boards that plug in between the \dword{apa} 
wire-board stacks and \dwords{femb}. The \dword{cr} boards
have already been described in Section~\ref{sec:crboards};
here we focus on the rationale for the choice of the 
resistance and capacitor values and their impact on the
wire signals. Each \dword{cr} board includes single \dword{r-c} filters 
for the $X$- and $U$-plane wire bias voltages, while the $V$-plane 
wires have a floating voltage. 
In addition, each board 
has \num{48} pairs of bias resistors and \dword{ac} coupling 
capacitors for $X$-plane wires, and \num{40} pairs for the $U$-plane 
wires. The coupling capacitors block \dword{dc} levels while passing \dword{ac} 
signals to the \dwords{femb}. On the \dwords{femb},
clamping diodes limit the input voltage received at the amplifier
circuits to between $\SI{1.8}{V}+U_D$ and $\SI{0}{V}-U_D$, where $U_D$
is the threshold voltage of the diode, approximately \SI{0.7}{V} at \dword{lar} temperature.
The amplifier circuit has a \SI{22}{nF} coupling capacitor at the
input to avoid leakage current from the protection clamping diodes.
Tests of the protection mechanism have been performed by discharging
\SI{4.7}{nF} capacitors holding a voltage of \SI{1}{kV} (\SI{2.35}{mJ} of
stored energy). The diodes survived more than 250 discharges at \dword{ln}
temperature. A schematic diagram of the \dword{apa} wire bias 
subsystem, identical to the one used in \dword{pdsp}, appears
in Figure~\ref{fig:CR-board}.

\begin{dunefigure}
[APA wire bias schematic diagram, including the CR board]
{fig:CR-board}
{\dword{dune} \dword{apa} wire bias schematic diagram including the \dword{cr} board.}
\includegraphics[width=0.9\linewidth]{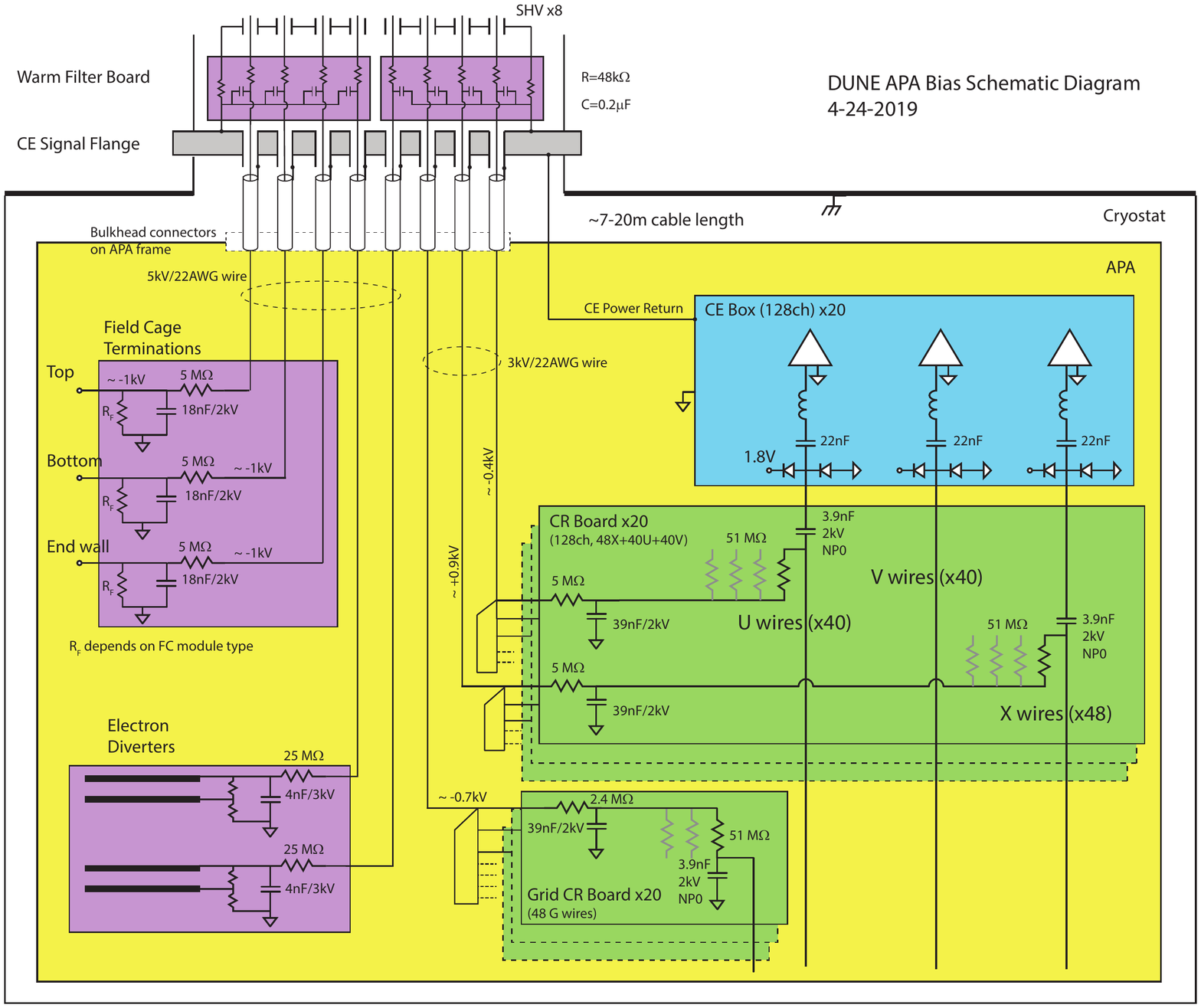}
\end{dunefigure}

Bias resistance values should be at least \SI{20}{\mega\ohm} to 
maintain negligible noise contributions. The higher value helps 
achieve a longer time constant for the high-pass coupling networks.
Time constants should be at least \num{25} times the electron 
drift time so that the undershoot in the digitized waveform
is small and easily correctable. 
As discussed in Section~\ref{sec:crboards},
the bias resistance value is \SI{51}{\mega\ohm}, while the 
\dword{dc}-blocking capacitors on each wire have a value of
\SI{3.9}{nF}. This gives a time constant of \SI{0.2}{s} that
is much larger than the drift time for electrons from tracks
passing near the cathode ($\sim\SI{2.3}{ms}$).

The bias-voltage filters are \dword{r-c} low-pass networks. Resistance 
values should be much smaller than the bias resistances to control 
cross-talk between wires and limit the voltage drop if any of the 
wires becomes shorted to the \dword{apa} frame. As discussed
in Section~\ref{sec:crboards}, these resistors have a resistance of
\SI{5}{\mega\ohm}, while the bias filter capacitors have a
capacitance of \SI{39}{nF}.

\subsection{Front-End Motherboard}
\label{sec:fdsp-tpcelec-design-femb}

Each \dword{apa} is instrumented with \num{20} \dwords{femb}.
The \dwords{femb} plug into the \dword{apa} \dword{cr} boards, 
making the connections from the wires to the charge amplifier 
circuits as short as possible. Each \dword{femb} receives signals 
from \num{40} $U$ wires, \num{40} $V$ wires, and \num{48} $X$ wires.
The reference \dword{femb} design contains eight \num{16}-channel 
\dword{larasic} chips, eight \num{16}-channel 
\dword{coldadc} \dwords{asic}, and two \dword{coldata} control and 
communication \dwords{asic} (see Figure~\ref{fig:ce-scheme}).
The \dword{femb} also contains regulators that produce the voltages 
required by the \dwords{asic} and filter those voltages, and 
a micro-electromechanical system oscillator that provides
a \SI{40}{MHz} reference to the \dword{coldata} \dword{pll}. The \dword{larasic} 
inputs are protected by an external series inductor and two 
diodes as well as the internal diode protection in the chip.

The \dword{pdsp} version of the \dword{femb} (which uses a single 
\dword{fpga} on a mezzanine card instead of two \dword{coldata} 
\dwords{asic}) is shown in Figure~\ref{fig:femb}. In the rest of
this section we describe the \dwords{asic} that will be installed
on the \dwords{femb} and discuss the procedure that will be 
followed to choose the \dword{asic} design 
to implement in the 
\dword{spmod}. 
In addition to
describing \dword{larasic}, \dword{coldadc}, and \dword{coldata},
we also discuss two alternative solutions, one based on a 
\dword{cots} \dword{adc}, and one where the functionality of the
three-\dword{asic} is implemented in a single chip, \dword{cryo}.

\begin{dunefigure}
[The complete FEMB assembly as used in ProtoDUNE-SP]
{fig:femb}
{The complete \dword{femb} assembly as used in the \dword{pdsp} 
detector. The cable shown is the high-speed data, clock, and control cable.}
\includegraphics[width=0.6\linewidth]{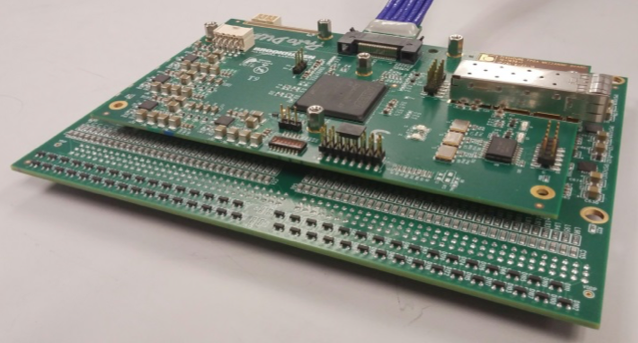}
\end{dunefigure}

The functionality of the \dword{femb} for \dword{dune} will be
almost identical to that of the \dword{femb} used in \dword{pdsp}.
The design will change slightly to accommodate the new \dwords{asic},
which will also entail changing the connections to the \dword{wib},
and changing the number of voltage regulators. In addition, the
connector for the control and data cold cables will be replaced
to address an issue observed in \dword{pdsp} that will be discussed
in Section~\ref{sec:fdsp-tpcelec-overview-lessons}. The new design,
shown in Figure~\ref{fig:femb-connector}, adds 
wings to the \dword{pcb} soldered to the cold cable, with 
standoffs to ensure the planarity of the connector
to the \dword{femb}, and a cutout in the \dword{pcb} to preclude 
any stresses introduced by height variations.

\begin{dunefigure}
[Modified design of the cold data cable and of the FEMB PCB]
{fig:femb-connector}
{Modified design of the cold data cable and of the \dword{femb} \dword{pcb}.}
\includegraphics[width=0.9\linewidth]{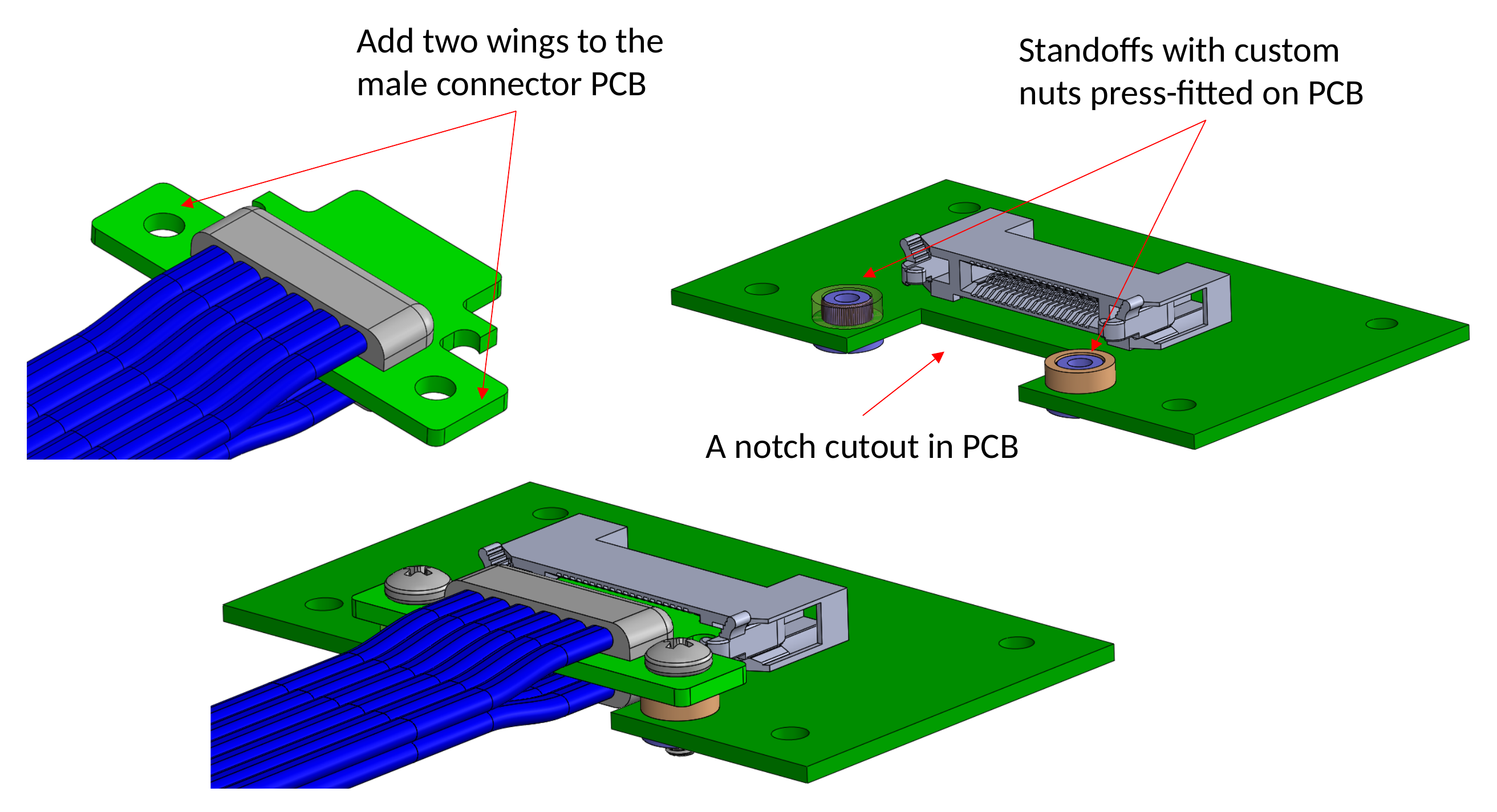}
\end{dunefigure}

All the discrete components mounted on the \dword{femb} have been
characterized for operation in \dword{lar}. In some cases (resistors,
capacitors, diodes) the components used on the \dword{pdsp} \dword{femb}
belong to the same family of components already used for other boards
operating in cryogenic environment, namely the boards used for the 
\dword{atlas} accordion \dword{lar} calorimeter, providing relevant information
on the lifetime of these components, which is discussed later in 
Section~\ref{sec:fdsp-tpcelec-qa-reliability}. There we also discuss
procedures for the measurement of the lifetime of discrete
components that have been adopted in recent years to demonstrate
that the \dword{tpc} electronics can survive in \dword{lar}. 
These types of measurements have been performed already for 
other neutrino experiments using the \dword{lar} \dword{tpc}
technology, while for the micro-mechanical oscillator we 
rely on characterizations performed by \dword{nasa}~\cite{nasa_mems}.

In the case of custom \dwords{asic}, appropriate steps must be taken prior 
to starting the layout of the chips. Both \dword{coldata} and 
\dword{coldadc} are implemented in the \dword{tsmc} \SI{65}{nm} \dword{cmos} 
process~\cite{TSMC65}. The designs were done using cold transistor models 
produced by Logix Consulting\footnote{Logix\texttrademark{} Consulting, http://www.lgx.com/.}.  Logix made measurements of 
 \dword{tsmc} \SI{65}{nm} transistors (supplied by \dword{fnal}) at \dword{ln} 
temperature and extracted and provided to the design teams 
\dword{spice}~\cite{Nagel:1973qq} 
models valid at \dword{ln} temperature.  These models were used in 
analog simulations of \dword{coldata} and \dword{coldadc} subcircuits.  
In order to eliminate the risk of accelerated aging due to the hot-carrier
effect~\cite{Hot-electron}, no transistor with a channel length
less than \SI{90}{nm} was used in either \dword{asic} design.
A special library of standard cells using \SI{90}{nm} channel-length 
transistors was developed by members of the University
of Pennsylvania and \dword{fnal} groups. Timing parameters were
developed for this standard cell library using the Cadence Liberate
tool\footnote{
\href{https://www.cadence.com/content/cadence-www/global/en_US/home/tools/custom-ic-analog-rf-design/library-characterization/liberate-characterization.html}{Cadence Liberate\texttrademark{}.}
} 
and the Logix \dword{spice} models. Most of the digital logic
used in \dword{coldadc} and \dword{coldata} was synthesized 
from Verilog code using this standard cell library and the Cadence Innovus tool\footnote{
\href{https://www.cadence.com/content/cadence-www/global/en_US/home/tools/digital-design-and-signoff/hierarchical-design-and-floorplanning/innovus-implementation-system.html}{Cadence Innovus\texttrademark{}}.}.
Innovus was also used for the layout of the synthesized logic.
The design of the \dword{cryo} \dword{asic} and of \dword{larasic}
are implemented in the \dword{tsmc} \SI{130}{nm} and \SI{180}{nm} 
\dword{cmos} process~\cite{TSMC130,TSMC180}, respectively. 
In the case of \dword{larasic}, 
the design uses models that were obtained by extrapolating the
parameters of the models provided by \dword{tsmc}, which are 
generally valid in the \SIrange{230}{400}{K}. In the case of the \dword{cryo}
\dword{asic}, cold transistor models were based on data taken at \dword{slac} with
\dword{tsmc}-produced \SI{130}{nm} transistors.

\subsubsection{Front-End \dshort{asic}}
\label{sec:fdsp-tpcelec-design-femb-fe}

\dword{larasic}~\cite{DeGeronimo:2011zz} receives 
current signals from the \dword{tpc} sense wires and provides a way to 
amplify and shape the signals for downstream signal digitization. 
\dword{larasic} has \num{16} channels and is implemented 
using the \dword{tsmc} \SI{180}{nm} \dword{cmos} process~\cite{TSMC180}. It 
integrates a band-gap reference to generate all the internal bias 
voltages and currents. This guarantees high stability of the operating 
point over a wide range of temperatures, including cryogenic temperatures. 
The channel schematic of \dword{larasic} is shown in 
Figure~\ref{fig:feasic1}. 

\begin{dunefigure}
[FE ASIC channel schematic]
{fig:feasic1}
{Channel schematic of \dword{larasic}, which includes a 
dual-stage charge amplifier and a \num{5}$^{th}$ order semi-Gaussian 
shaper with complex conjugate poles. Circuits in red circles are 
programmable to allow different gain and peaking time settings.}
\includegraphics[width=0.99\linewidth]{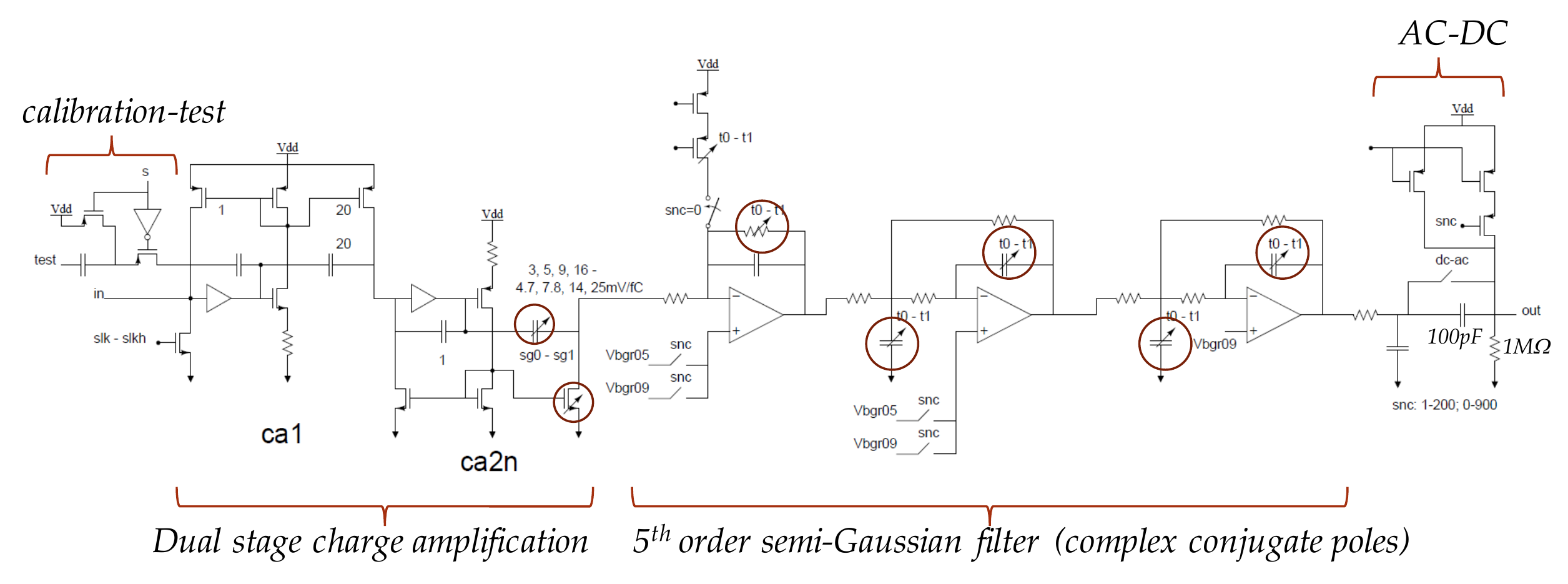}
\end{dunefigure}

Each \dword{larasic} channel has a dual-stage charge amplifier 
and a \num{5}$^{th}$ order semi-Gaussian shaper as an anti-aliasing 
filter for the \dword{tpc} signals. It has programmable gain 
selectable from one of \num{4.7}, \num{7.8}, \num{14}, and \SI{25}{mV/fC}
(corresponding to full-scale charge of \num{1.9e6}, \num{1.1e6}, \num{625e3}, 
and \SI{350e3}{$e^-$}), programmable peaking time selectable from one of 
\num{0.5}, \num{1}, \num{2}, and \SI{3}{$\mu$s}, and programmable 
baseline for operation with either the collection ($\sim$\SI{200}{mV}) 
or the induction ($\sim$\SI{900}{mV}) wires. All these parameters
can be set only at the \dword{asic} level, i.e., they affect the
behavior of \num{16} readout channels.
The design of
\dword{larasic} has been optimized for the capacitive
loads expected in the case of the \dword{dune} detector
(i.e., in the range \SIrange{170}{210}{pF}).
Each channel has an 
option to enable the output monitor to probe the analog signal, and 
an option to enable a high-performance output driver that can be 
used to drive a long cable. 

Each \dword{larasic} channel has a built-in charge calibration 
capacitor that can be enabled or disabled through a dedicated register. 
Measurements of the injection capacitance have been performed using an 
external precisely calibrated capacitor. These measurements show that
the calibration capacitance is extremely stable against temperature variations, 
changing from \SI{184}{fF} at room temperature to 
\SI{183}{fF} at \SI{77}{K}. This result and the measured stability of 
the peaking time demonstrate the high stability of the passive 
components as a function of temperature. Channel-to-channel and 
chip-to-chip variation in the calibration capacitor are typically 
less than \num{1}\%. The variations of the calibration capacitors
could be characterized prior to the beginning of \dword{dune}
data taking, using the \dword{qc} process, discussed in
Section~\ref{sec:fdsp-tpcelec-production-qc}.

Shared among the \num{16} channels in \dword{larasic} are 
the digital interface, programming registers, a temperature monitor, 
and a band-gap reference monitor. It is also possible to enable \dword{ac} 
coupling as mitigation of baseline variations induced by vibrations
of the \dword{apa} wire, a programmable input bias current 
selectable from one of \num{0.1}, \num{0.5}, \num{1}, or \SI{5}{nA}, 
as well as a programmable pulse generator with a \num{6}-bit 
\dword{dac} for calibration. The possibility of configuring various
parameters controlling the \dword{fe} amplifier (gain, peaking time,
baseline) has allowed \dword{pdsp} to reduce the impact of the
saturation effect discussed in Section~\ref{sec:fdsp-tpcelec-overview-lessons}, 
at the cost of a reduction in dynamic range for the collection wires.

The power dissipation of \dword{larasic} is about \SI{5.5}{mW} 
per channel at \SI{1.8}{V} supply voltage when the output buffer is
disabled (the output buffer is required only for transmitting analog
signals over long distances; it is not needed when \dword{larasic}
is mounted close to the \dword{adc} on the \dword{femb}).
The \dword{asic} is packaged in a commercial, fully encapsulated 
plastic \num{80} pin \dword{qfp}. Figure~\ref{fig:feasic2} shows the 
response of \dword{larasic} for all gains and peaking times 
and both baselines. Note that the gain is independent of the peaking 
time; the same amount of charge, in the impulse approximation, produces 
the same peak voltage signal regardless of the peaking time.

\begin{dunefigure}
[FE ASIC response and layout]
{fig:feasic2}
{Response of \dword{larasic} for four gains, four peaking times, 
and both baseline values (left; the time distance between the positive
and negative pulse for the induction wires has been exaggerated
for clarity reasons); layout of \num{16}-channel
\dword{larasic} version P3, where revisions with reference to version 
P2 are highlighted in yellow boxes (right).}
\includegraphics[width=0.53\linewidth]{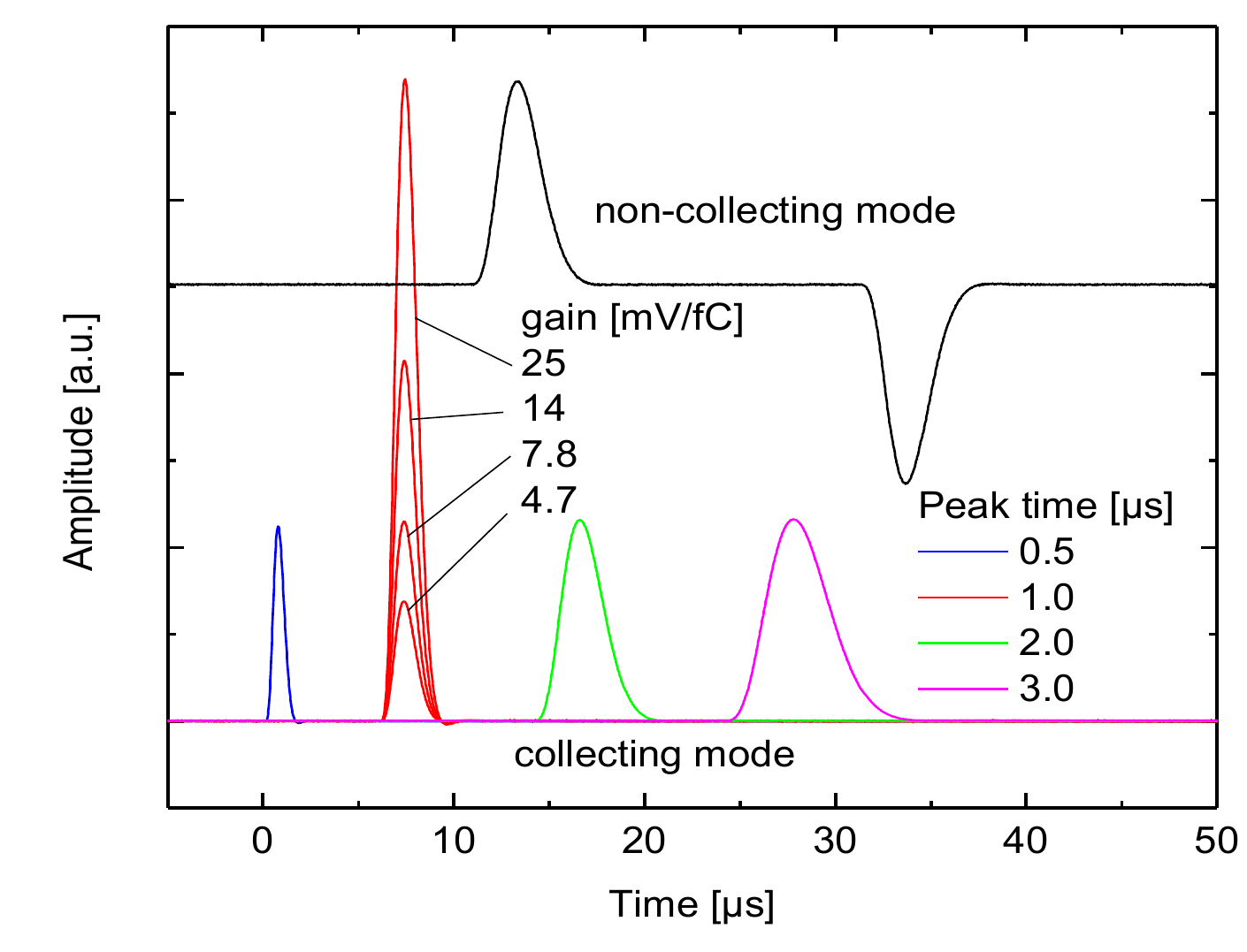}
\hspace{6mm}
\includegraphics[width=0.42\linewidth]{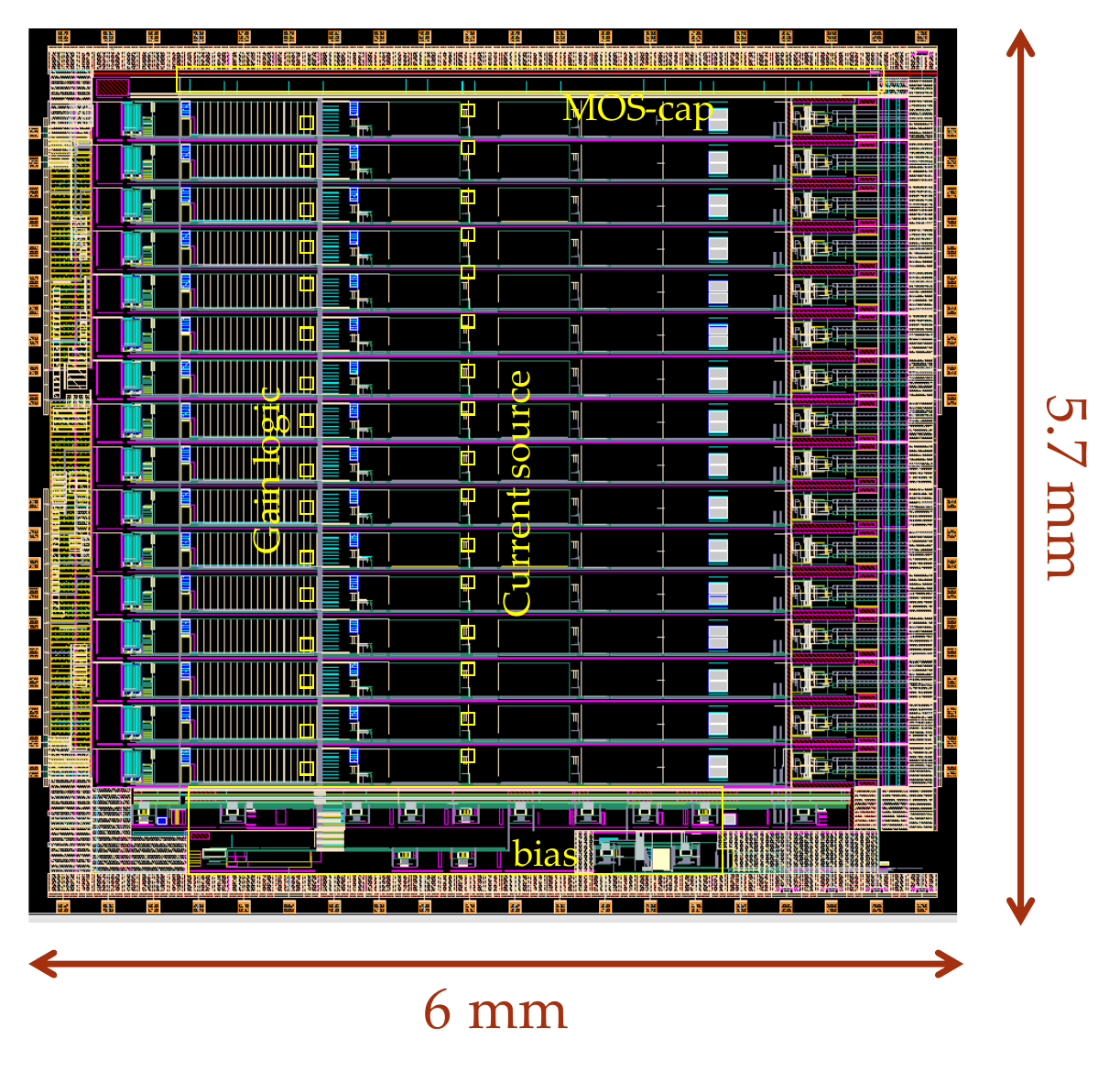}
\end{dunefigure}

Prototype version P2 \dword{larasic} chips have been evaluated and 
characterized at room temperature and \dword{ln} temperature (\SI{77}{K}). 
\num{960} P2 chips, totaling \num{15,360} channels, 
have been used to instrument six \dword{pdsp} \dword{apa}s successfully. 
Excessive stress in the package of \dword{larasic} at cryogenic 
temperature causes \dword{fe} channels to have a non-uniform baseline in 
collection mode, while the baseline \dword{dc} voltage in induction mode 
is uniform. A new prototype, version P3, was fabricated in March 2018 
to address this issue by making \dword{dc} circuits for the collection mode 
similar to the induction mode. At the same time, the default gain
setting was changed to \SI{14}{mV/fC}. The layout of P3
\dword{larasic} is also shown in Figure \ref{fig:feasic2}, with modifications 
highlighted in yellow boxes. The P3 \dword{larasic} chips were 
received and evaluated in September 2018. We have verified that with
the new design the \dword{fe} channels have a uniform baseline when
operated in the collection mode, and that \SI{14}{mV/fC} is the new 
default gain setting.

P3 \dword{larasic} will be further evaluated on \dwords{femb} 
in various integration test stands for performance studies, including 
the \num{40}\% \dword{apa} at \dword{bnl}, the \dword{iceberg} \dword{tpc} 
at \dword{fnal} and the seventh \dword{protodune} \dword{apa} 
in the \coldbox at \dword{cern}. Analysis of the \dword{pdsp} data
has highlighted a saturation problem in the design of the P2 \dword{larasic}
that we have observed also in bench tests of the P3 version. This problem,
discussed in detail in Section~\ref{sec:fdsp-tpcelec-overview-lessons},
will be addressed in the design of the next version of \dword{larasic}, P4,
for which we are also planning to implement a single-ended-to-differential
converter as an interface to the recently developed \dword{coldadc}.
The plan for solving the saturation problem in \dword{larasic} is discussed 
in Section~\ref{sec:fdsp-tpcelec-overview-remaining}.

\subsubsection{\dshort{coldadc} \dshort{asic}}
\label{sec:fdsp-tpcelec-design-femb-adc}

\dword{coldadc} is a low-noise \dword{adc} \dword{asic} designed to digitize
\num{16} input channels at a rate of $\sim\SI{2}{MHz}$, as required for the
\dword{dune} \dword{spmod}. \dword{coldadc} was designed to operate with 
an external \SI{64}{MHz} clock and an external \SI{2}{MHz} digitization clock.
The \SI{2}{MHz} clock is aligned on the rising edge of one of
the \SI{64}{MHz} transitions, as discussed in Section~\ref{sec:fdsp-tpcelec-design-femb-coldata}.
For the remainder of this section we assume that the main clock is operating
at \SI{64}{MHz}, but in the \dword{dune} \dword{spmod} this external clock will operate at \SI{62.5}{MHz}
as discussed in Section~\ref{sec:daq:design-timing}, and the waveforms from
the \dword{apa}s will be digitized every \SI{512}{ns}.
\dword{coldadc} is implemented in the \dword{tsmc}
\SI{65}{nm} \dword{cmos} technology and has been designed by a team of engineers
from \dword{lbnl}, \dword{bnl}, and \dword{fnal}.  The \dword{asic} uses a conservative,
industry-standard design including digital calibration.  Each \dword{coldadc}
receives \num{16} voltage outputs from a single \dword{larasic} chip.  The voltages
are buffered, multiplexed by \num{8}, and input to two \num{15}-stage pipelined \dwords{adc}
operating at \SI{16}{MHz}. The \SI{16}{MHz} clock is generated internally in
\dword{coldadc} and shares its rising edge with the \SI{2}{MHz} clock. 
The \dword{adc} uses the well known pipelined architecture
with redundancy~\cite{PipelinedADC}.  Digital logic is used to correct non-linearity
introduced by non-ideal amplifier gain and offsets in each pipeline
stage~\cite{CalibrationCorrection}, and an automatic calibration procedure is
implemented to determine the constants used in this logic.  The \dword{adc} produces
\num{16}-bit output which is expected to be truncated to \num{12} bits.

The \dword{adc} is highly programmable to optimize performance at different
temperatures.  Many circuit blocks can be bypassed, allowing the performance 
of the core digitization engine to be evaluated separately from the ancillary 
circuits. A block diagram of the chip is shown in Figure~\ref{fig:COLDADC_Block_Diagram}. 
Each of the major blocks is described below.

\begin{dunefigure}
[ColdADC block diagram]
{fig:COLDADC_Block_Diagram}
{\dword{coldadc} block diagram.}
\includegraphics[width=0.8\linewidth]{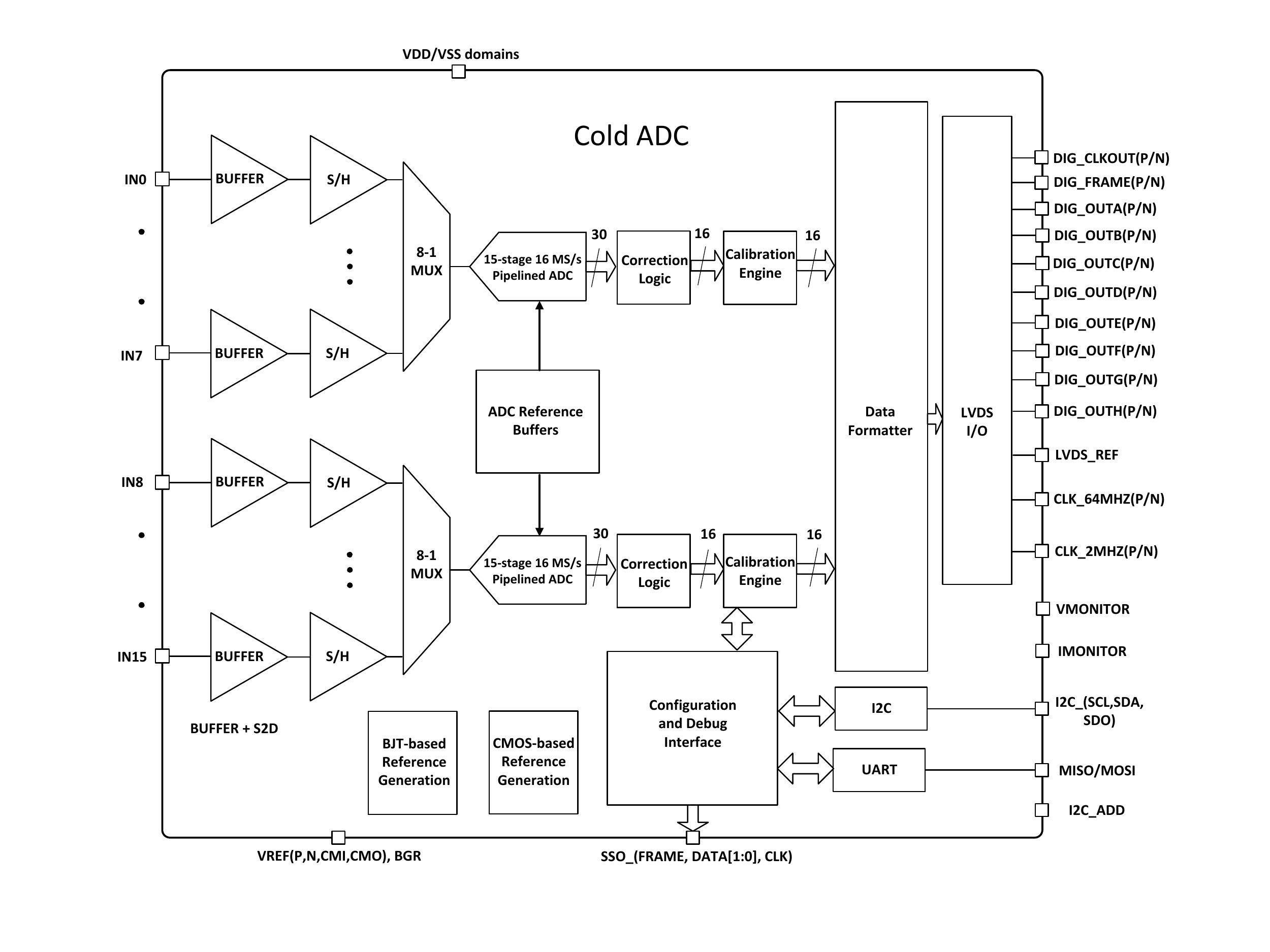}
\end{dunefigure}

All required reference voltages and currents are generated on-chip by programmable
circuit blocks. Independently adjustable bias voltage levels and currents are provided for the
input buffers, sample-and-hold amplifiers, \dwords{adc}, and \dword{adc}
reference buffers.
The most accurate reference voltage circuit is a band-gap reference
based on a \dword{pnp} transistor.  However, measurements made at \dword{bnl} and
\dword{lbnl} of a large \dword{pnp} transistor indicate that the foundry-provided \dword{spice} model
does not adequately describe the device operation at \lar temperature.  Thus,
a \dword{cmos}-based voltage reference has also been included in \dword{coldadc}.
As discussed below, bench tests of \dword{coldadc} prototypes show that both reference blocks 
perform well and meet requirements.

\dword{coldadc} has four possible ways to interface with \dword{larasic}.  It
can accept either single-ended inputs (provided by existing \dword{larasic}
chips) or differential inputs (foreseen for the future \dword{larasic} P4 upgrade).
In either case, it is also possible to bypass the input buffers and apply the inputs directly
to the sample-and-hold amplifiers. The role of the input buffers is to present a well defined
and easy-to-drive load to \dword{larasic}.  The sample-and-hold amplifiers are separated
into two groups of eight. They sample the waveform at the rising edge of the
(\SI{2}{MHz}) sampling clock. The \SI{16}{MHz} clock is then
used to clock an \num{8}-to-\num{1} multiplexer that presents eight samples in 
turn to one of the two \dword{adc} pipelines.

\begin{dunefigure}
[Circuit blocks in each ADC pipeline stage]
{fig:pipelinestage}
{Circuit blocks in each \dword{adc} pipeline stage. MUX selects one of three values
as the digitized output of the current stage and presents it to the ADD circuit, which 
adds it to the result calculated by previous pipeline stages.
SHA is a sample-and-hold amplifier, and
ADSC and DASC are low resolution 1.5 bit analog-to-digital and digital-to-analog 
subconverters, respectively.}
\includegraphics[width=0.9\linewidth]{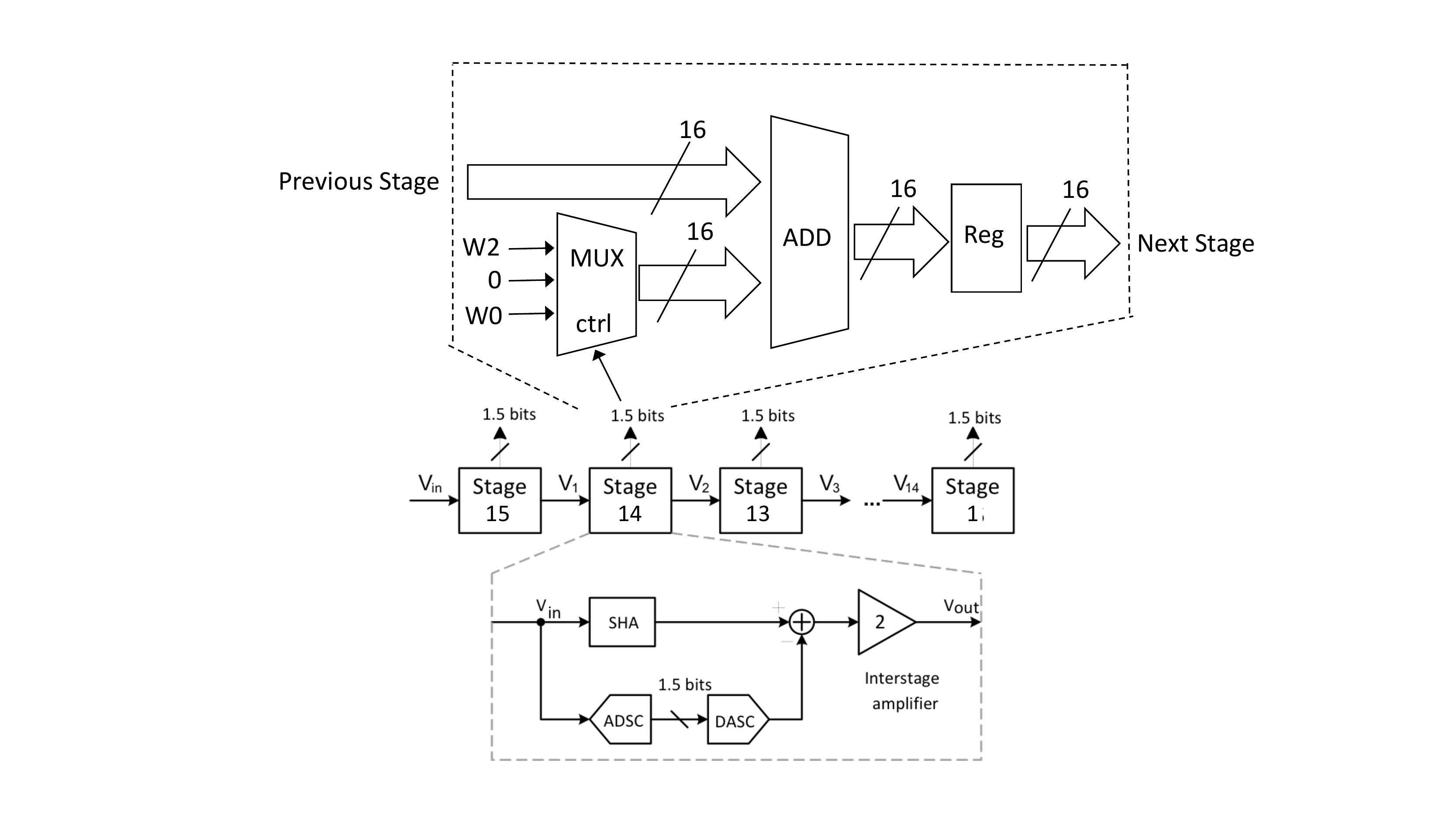}
\end{dunefigure}

A block diagram of an \dword{adc} pipeline is shown in
Figure~\ref{fig:pipelinestage}. Each of the \num{15} stages contains a low-resolution
\num{1.5}-bit analog-to-digital subconverter containing two comparators, a
\num{1.5}-bit digital-to-analog subconverter that produces a voltage based on
the two comparator outputs, an analog subtractor, a sample-and-hold amplifier,
and a gain stage (with a nominal gain of two). The transfer function of
each stage is identical and is shown in Figure~\ref{fig:ADCXFERFN} along with
the nominal ``weights'' (W$_0$ and W$_2$) that are added to form the output of the
pipeline. Each pipeline stage makes a three-level coarse decision based on the
analog input voltage, selects one of three digital weights to be added to the
results of previous stages, and passes a voltage to the next stage that is
proportional to the difference between the input voltage and the voltage
corresponding to the digital output of the stage.  Because the stages are
weighted by a factor of two, but have three possible digital results, there is
redundancy between stages that makes the final result independent of errors in
the comparator thresholds (up to $\pm V_r/4$ where the stage range is
$[-V_r,V_r]$).  An ``error'' in the output of one stage is corrected in subsequent
stages (usually the next stage).  In order to take advantage of this redundancy
provided by the pipelined architecture it is necessary to include at least one
``extra'' stage in the pipeline. 

\begin{dunefigure}
[ADC stage transfer function]
{fig:ADCXFERFN}
{\num{1.5}-bit stage transfer function and digital output. The voltage 
range of the \dword{adc} as a whole, and of each individual stage 
is $[-V_r,V_r]$.  Note that the voltage passed to the subsequent 
stage will not exceed the stage range even if a comparator threshold 
is wrong by up to $V_r/4$.}
\includegraphics[width=0.90\linewidth]{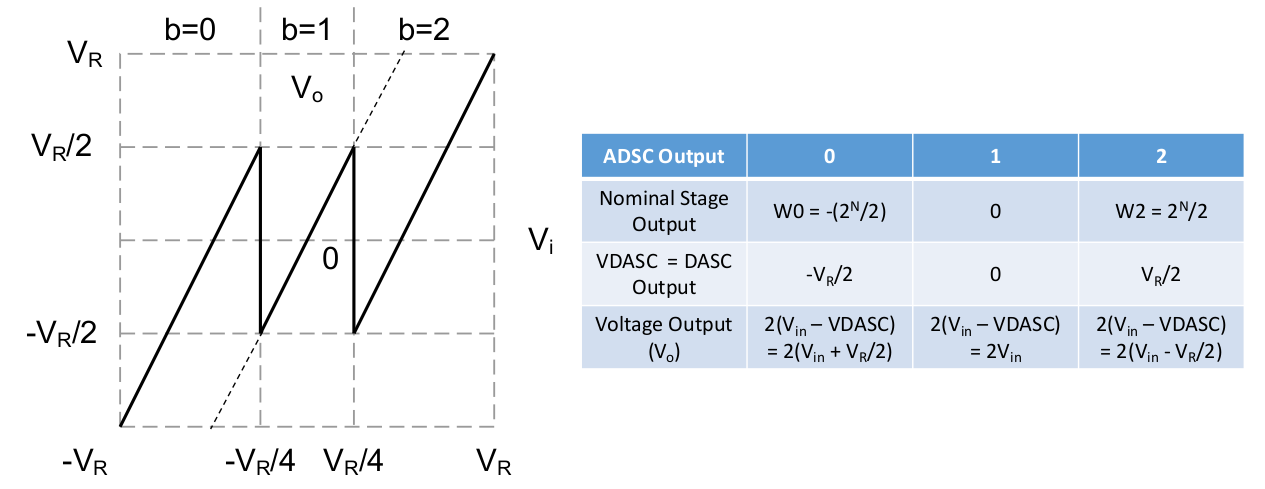}
\end{dunefigure}

The calibration logic allows the correction of errors caused by imperfections
in the voltage that are passed from one stage to the next.  These imperfections
arise from errors in each stage corresponding to $\pm V_r/2$ from the resistive
dividers and non-ideal effects in the gain and offset of the interstage amplifiers.
The calibration procedure relies on the fact that the required precision is easily
satisfied by the last stages of the pipeline.
The number of stages to be calibrated (maximum seven) is set by a programmable
register.  An iterative calibration procedure is used.  Starting with the least
significant stage to be calibrated, the input to the stage is set to the threshold
levels of $\pm V_r/4$ and the normal comparator outputs are overridden
and forced first to \num{1} and then to \num{0}.  The lower stages of the \dword{adc} digitize
the analog value output from the stage being calibrated and the difference between
the \dword{adc} output when the comparator is forced to \num{1} and the \dword{adc}
output when the comparator is forced to \num{0} is calculated.  These two differences
(W$_0$ expressed as a negative number and W$_2$ expressed as a positive number) are
stored and used as two of the three possible digital outputs of the stage being
calibrated (the third possible output being \num{0}).  This procedure is then repeated
for the next most significant pipeline stage until stage \num{15} has been calibrated.

The number of \dword{adc} bits that are useful depends on the effective
noise of the various subcircuits of the \dword{adc}.  The noise of the first few
pipeline stages (associated with the most significant bits) contributes more
heavily than subsequent stages.  For this reason, the first stages are designed
to be larger, lower noise, and to require more power than later stages. The capacitance
is reduced by a factor of two, relative to that of the sample-and-hold amplifier, for each
of the first three stages, and then kept constant. The total
effective noise expected is \SI{\sim130}{$\mu$V} \dword{rms}.  This is similar to the
quantization error of an ideal \num{12}-bit \dword{adc} with a voltage range of
\SI{1.5}{V} (slightly larger than the output range of \dword{larasic}, \SIrange{0.2}{1.6}{V})
for which the bin width is \SI{\sim366}{$\mu$V} and the quantization
error is \SI{\sim106}{$\mu$V}.

In normal operation, each pipelined \dword{adc} passes a \num{16}-bit result to the
data formatter on the rising edge of the \SI{16}{MHz} clock.  The data formatter
separates the two \num{16}-bit words into eight \num{4}-bit nibbles and serializes the
nibbles for output (most significant bit first) at \SI{64}{MHz}.   An output
clock and a frame marker are also generated.  The frame marker indicates the most
significant bit in each nibble of the first of eight channels digitized by one
of the \dword{adc} pipelines in each \SI{2}{MHz} sample period.  The output data
is generated on the falling edge of the output clock and is latched by the
\dword{coldata} \dword{asic} using the rising edge of the same clock.  

A second
mode of operation is included for debugging purposes.  In this mode, \num{2}-bit raw
stage results from each of the \num{15} stages of one of the two pipelines are formatted
into the most significant \num{15} bits of two \num{16}-bit words, broken into nibbles, and
output in the same manner as normal data.

Ten differential output drivers are used for the \SI{64}{MHz} output clock, frame
marker, and \dword{adc} data. The output drivers source and sink a current whose
value can be digitally controlled. The minimum current is \SI{165}{$\mu$A},
which corresponds to approximately \SI{3}{mV} peak-to-peak with \SI{100}{$\Omega$}
termination. Seven additional levels spaced by \SI{275}{$\mu$A} can be selected.
The maximum current is \SI{2.07}{mA}, about \num{2/3} of the \dword{lvds} standard of
\SI{3.5}{mA}.

The operation of \dword{coldadc} is controlled by a number of \num{8}-bit registers.
These registers can be written to and read back using either an \dword{i2c}
interface~\cite{bib:I2C} or a \dword{uart}. \dword{coldata} will use the \dword{i2c} interface. The
\dword{uart} is included in the first \dword{coldadc} prototype to facilitate chip testing
and for risk mitigation.

\dword{coldadc} was received at the end of January 2019.  Bench tests were performed at
\dword{bnl}, \dword{fnal}, and \dword{lbnl}. These tests used \dword{adc} chips mounted directly
on printed circuit boards, and were done at both room temperature and cryogenic temperature.
The tests concentrated first on functionality and later on performance.
A small number of problems were found during bench testing and will be described below. These problems
will not prevent system tests from being done with prototype \dword{coldadc} chips.

Both control interfaces (\dword{i2c} and \dword{uart}) operate as designed. All of the digital control
bits can be written and read. The \dword{lvds} I/O operates as designed and the drive current of the
\dword{lvds} can be selected as designed. The \dword{adc} pipeline functions as
designed, as does the data formatter. The automatic calibration logic does not work, but the pipelines 
can be calibrated off-chip using register-controlled debugging modes to force all of the steps of the 
calibration procedure. The sample-and-hold amplifiers and the multiplexer that connects the sample-and-hold 
outputs to the \dword{adc} pipelines operate correctly. Both the \dword{cmos} reference generation 
block and the band-gap reference block operate as designed, although a minor error in
a digital-to-analog converter in the band-gap reference block means that it must operate with the 
(nominally \SI{2.3}{V}) analog voltage set to \SI{2.7}{V}. Another error was discovered in the input 
buffer block. Level shifters intended to translate control bits in the \SI{1.2}{V} domain to the 
\SI{2.5}{V} domain were omitted. As a result, the \SI{1.2}{V} digital supply must be set to 
\SI{2.1}{V}. All of these design errors (including the auto-calibration failure) have been understood
and are easily corrected. Bench tests have proven that the \dword{coldadc} prototypes can be run at the 
required elevated voltage settings for many days without damage to the chips.

Performance measurements of \dword{coldadc} have also been done. The performance of many of 
the sub-circuits have been measured separately as well as the performance of the entire \dword{adc}.  
Here we present two measurements made at \dword{ln} temperature.

\begin{dunefigure}
[Static linearity of ColdADC]
{fig:ADCStaticLinearity}
{\dshort{dnl} (top) and \dshort{inl} distributions as a function of \dword{adc} code for \dword{coldadc}.}
\includegraphics[width=0.9\linewidth]{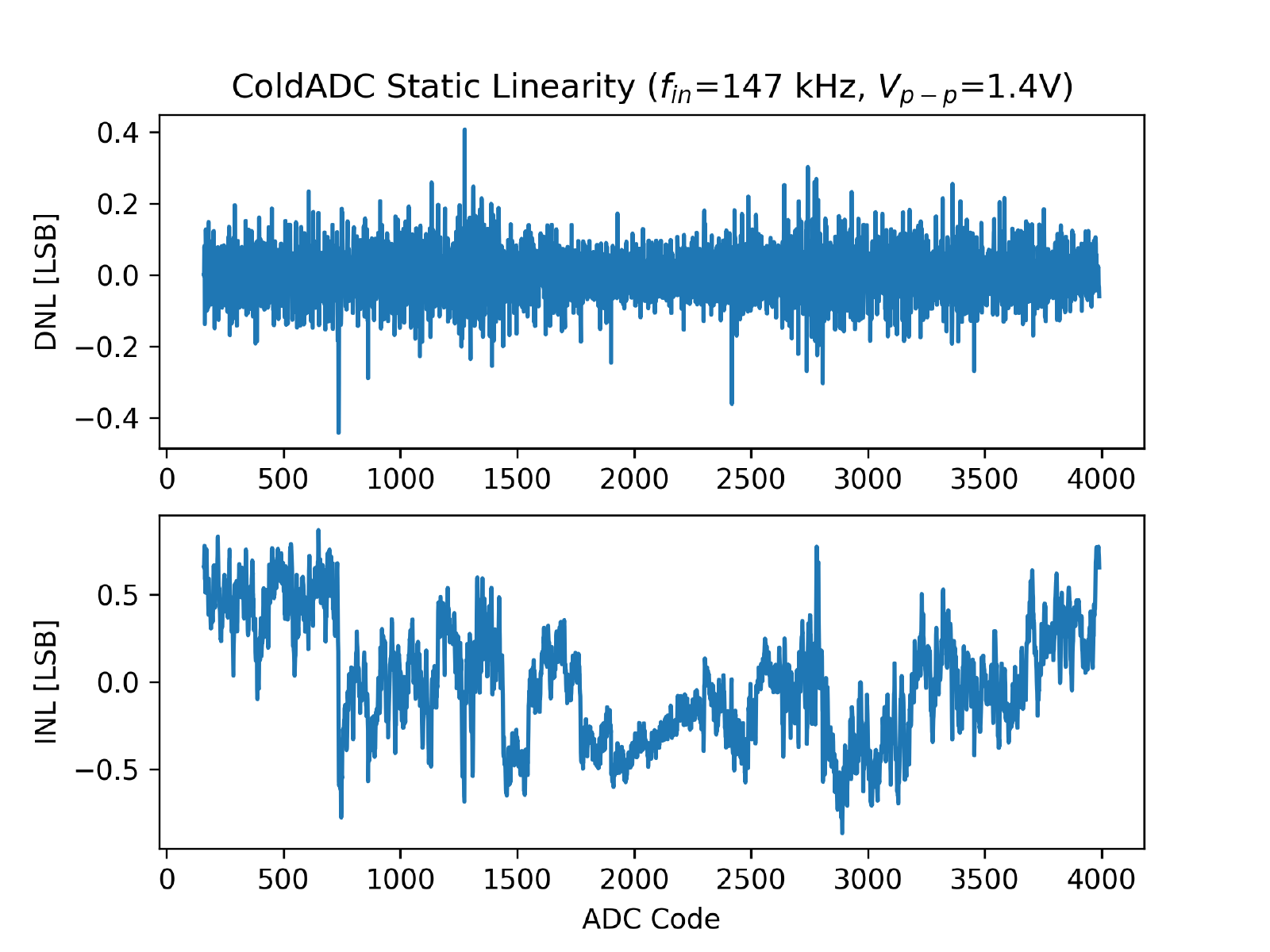}
\end{dunefigure}

The static linearity of the pipeline \dword{adc} was measured using a filtered sine wave connected to
the test inputs of \dword{coldadc}. The measured histogram of \dword{adc} codes was fitted to the 
probability density function for a sine wave. The calculation of the residuals to the fit yields the \dword{dnl} 
as a function of \dword{adc} code; the integral of \dword{dnl} is the \dword{inl}. These two
distributions are shown in Figure~\ref{fig:ADCStaticLinearity}, which was obtained using a sine 
wave of amplitude of \SI{1.4}{V} peak-to-peak (matching the \dword{larasic} dynamic range) and 
the nominal reference voltage settings (corresponding to a \SI{1.5}{V} dynamic range).

Dynamic linearity was also measured using a filtered sine wave. In this case, \dword{adc} codes were
collected for an integer number of sine wave cycles and a \dword{fft} was performed on the data. 
The \dword{sndr}, \dword{enob}, \dword{sfdr}, and the \dword{thd} were extracted from the \dword{fft}.
An example of the \dword{fft} is shown in Figure~\ref{fig:ADCDynamicLinearity}, which was obtained 
using a sine wave of amplitude \SI{1.5}{V} (matching the full range of the \dword{adc}).
The extracted \dword{enob} is over \num{11}, despite the non-linearity evident in 
Figure~\ref{fig:ADCStaticLinearity}, because the \dword{adc} noise is very low.
The dominant source of non-linearity has been demonstrated to be insufficient 
open-loop gain of the operational amplifier used in each pipeline stage. 
The design has already been modified to address this deficiency.

\begin{dunefigure}
[Dynamic Linearity]
{fig:ADCDynamicLinearity}
{Fourier transform of \dword{adc} codes collected with a coherently sampled sine wave input to a 
single-ended input buffer.}
\includegraphics[width=0.9\linewidth]{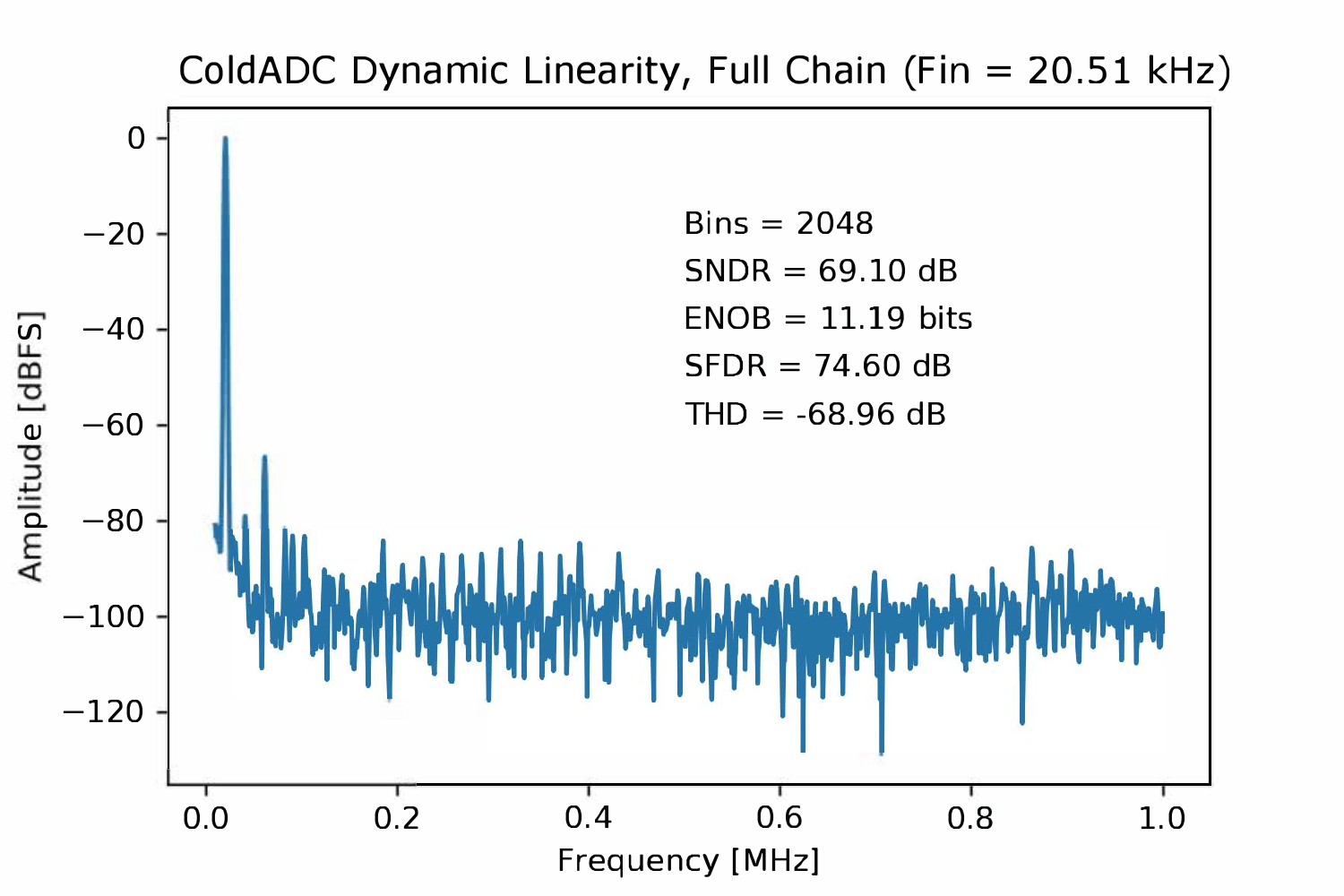}
\end{dunefigure}

\subsubsection{COLDATA \dshort{asic}}
\label{sec:fdsp-tpcelec-design-femb-coldata}

The \dword{coldata} \dword{asic} was designed by engineers from \dword{fnal} 
and Southern Methodist University. It is responsible for all communications 
between the \dwords{femb} and the electronics located outside the cryostat. 
Each \dword{femb} contains two \dword{coldata} chips. \dword{coldata} receives 
command-and-control information from a \dword{wib}. Each \dword{coldata} provides 
clocks to four \dword{coldadc}s and relays commands to four \dword{larasic}s
and four \dword{coldadc}s to set operating modes and 
initiate calibration procedures.  Each \dword{coldata} receives data from four 
\dword{coldadc}s, merges the data streams, provides 8b/10b encoding, serializes 
the data, and transmits the data to the warm electronics over two \SI{1.28}{Gbps} 
links.  These links are driven by line drivers with programmable pre-emphasis. 
Figure~\ref{fig:coldata_block_diagram} is a block diagram of \dword{coldata}.

\begin{dunefigure}
[ColdDATA block diagram]
{fig:coldata_block_diagram}
{\dword{coldata} block diagram.}
\includegraphics[width=0.99\linewidth]{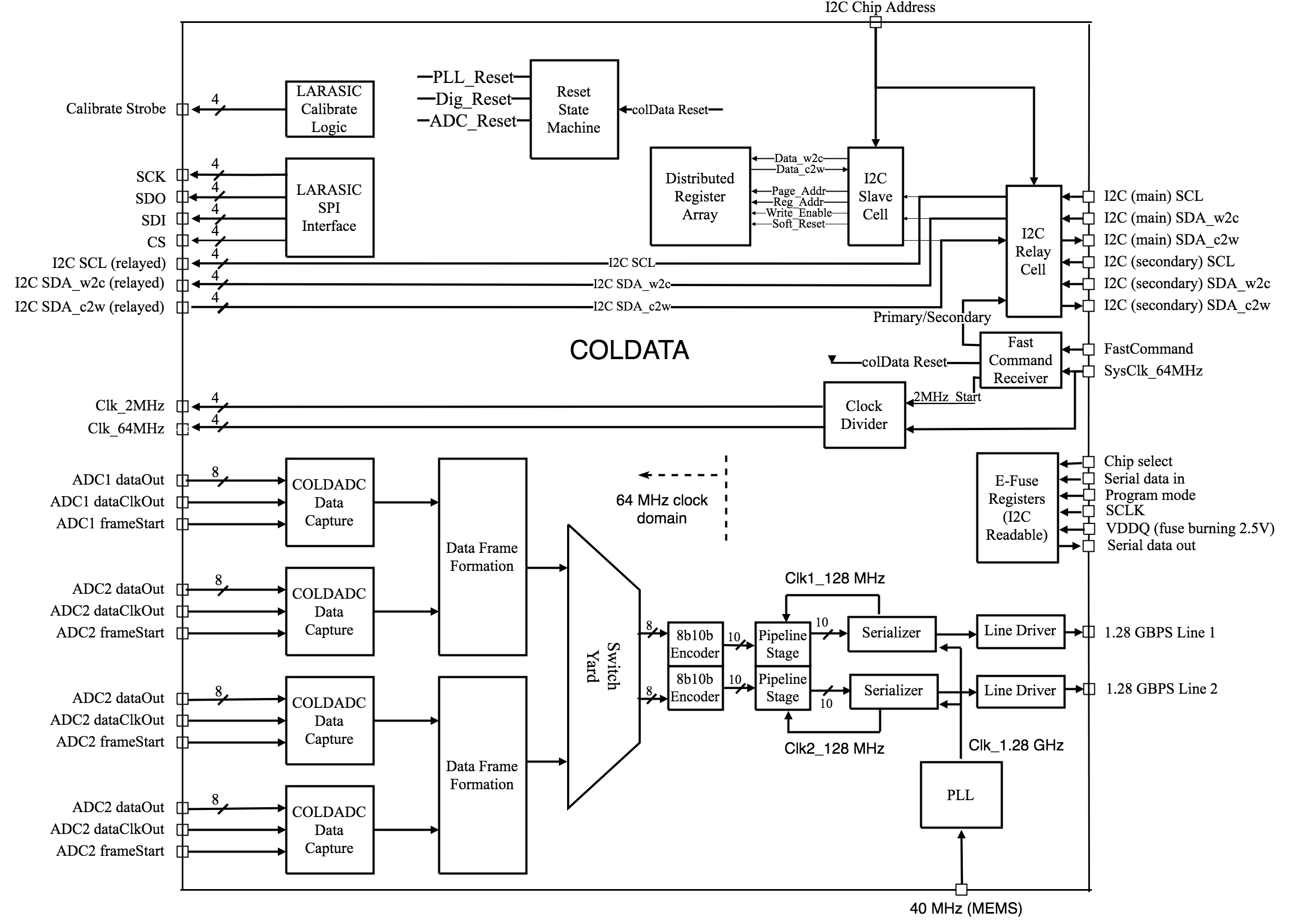}
\end{dunefigure}

The commands for the control of all the \dwords{asic} on a \dword{femb} are sent 
from a \dword{wib}  
using an \dword{i2c}-like protocol~\cite{bib:I2C}. The protocol used
in \dword{coldata} differs from the standard \dword{i2c} one.
Because of the long cables required between the \dword{wiec} and the 
\dwords{femb}, \dword{coldata} uses \dword{lv} differential pairs for both 
the \dword{i2c} clock and data. Separate point-to-point links are used for
data sent from warm-to-cold and for data sent from cold-to-warm.
In order to reduce the number of cables required, only one of the two 
\dword{coldata} chips on an \dword{femb} has its main \dword{i2c} interface 
directly connected to a \dword{wib}. That \dword{coldata} chip relays \dword{i2c} 
commands and data to the secondary \dword{coldata} chip and relays \dword{i2c} 
responses from the secondary \dword{coldata} to the \dword{wib}. 
Each  \dword{coldata} also relays \dword{i2c} commands and data sent from the 
\dword{wib} to one of the four \dword{coldadc} chips, and it relays data back to
the \dword{wib} from one of the  four \dword{coldadc} chips. 
The links on the \dword{femb} between \dword{coldata} 
chips and \dword{coldadc} chips use single-ended (\SI{2.25}{V}) \dword{cmos} 
signals.

The controls intended for the \dword{larasic} chips are interpreted 
inside \dword{coldata} and transmitted to the appropriate \dword{asic} using 
a \dword{spi}-like interface that uses single-ended (\SI{1.8}{V}) \dword{cmos} 
signals. The configuration registers in \dword{larasic} are configured to be 
loaded as a single-shift register. As data is shifted into \dword{larasic} on 
the \dword{mosi} line, bits from the other end of the shift register are shifted 
out on the \dword{miso} line. It is thus only possible to read \dword{larasic} 
configuration registers while writing new configuration data.

In addition to the configuration commands, \dword{coldata} 
receives a master clock and a fast command signal on a \dword{lv} 
differential pair from
the \dword{wib}. Currently the master clock is \SI{64}{MHz}, but it will be
changed to \SI{62.5}{MHz} to simplify the overall \dword{dune} \dword{spmod} 
synchronization, as already discussed in the case of \dword{coldadc}. The clock 
used for sampling the \dword{adc} is created inside
\dword{coldata} by dividing the master clock by \num{32}. The relative phase
of the \num{2} and the \SI{64}{MHz} clocks is set by an appropriate fast
command sent from the \dword{wib}.  Both the master clock and
the \dword{adc} sampling clocks are passed from \dword{coldata} to the four
\dword{coldadc} chips that it controls. Depending on the master clock frequency
the \dword{adc} will convert input data every \num{500} or \SI{512}{ns}, 
corresponding to a frequency of \num{2} or \SI{1.95}{MHz}. Signals that must be
executed at a known time use the fast command line. \dword{coldata} 
uses the falling edge of the master clock to sample fast command bits as shown 
in Figure~\ref{fig:coldata_fast_command_timing}. All legal fast commands 
are \dword{dc} balanced. An ``alert'' pattern is used to establish the 8-bit 
fast-command word boundary. An ``idle'' pattern is used when no command is being 
sent. Four commands are defined: ``Edge,'' which moves the rising edge of the 
\dword{adc} sampling clock to coincide with the next rising edge of the 
master MHz clock; ``Sync,'' which zeros the \num{8}-bit timestamp that is incremented 
on the rising edge of each \dword{adc} sampling clock; ``Reset,'' which resets 
\dword{coldata}; and ``Act,'' the function of which is determined by an \num{8}-bit 
register that is programmed using the \dword{i2c} interface.  

\begin{dunefigure}
[ColdDATA fast command timing]
{fig:coldata_fast_command_timing}
{Fast command timing: the leading edges of the fast command and of the master 
clock are equal time when produced on the \dword{wib}. The fast-command bits 
are captured by \dword{coldata} on the falling edge of the master clock and 
shifted into a register on the next positive edge.}
\includegraphics[width=0.4\linewidth]{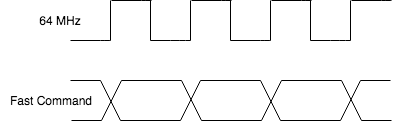}
\end{dunefigure}

\dword{coldata} receives digitized waveform data from 
four \dword{coldadc} \dwords{asic}. Each \dword{adc} presents its data on eight 
serial streams operating in parallel. Data from the \dwords{adc} is captured 
using the \dword{adc} ``dataClkOut'' signal  
(one per \dword{adc}) and the start of a 
sample period is indicated by the ``frameStart signal'' (one per \dword{adc}). 
Each \dword{adc} digitizes \num{16} channels of information and puts out \num{16} bits 
of data per channel. Information from two \dwords{adc} are merged by a Data 
Frame Formation block.  
The Data Frame Formation circuitry converts the two 
groups of sixteen \num{16}-bit words into one of three types of data frame. For normal 
data taking, either a \num{12}-bit format or a \num{14}-bit \dword{adc} format can be 
selected, discarding either the four or two lowest order bits. When the 
\num{12}-bit format is selected, a data frame consists of an 8b/10b command 
character (K28.2) and an \num{8}-bit time stamp, followed by \num{48} bytes of \dword{adc} 
data and two bytes of parity information. When the \num{14}-bit format is selected, 
a data frame consists of an 8b/10b command character (K28.3) and an \num{8}-bit time 
stamp, followed by \num{56} bytes of \dword{adc} data and two bytes of parity 
information. Two debugging frame formats are also defined. When the ``Frame-12 Test'' 
format is selected, a data frame consists of an 8b/10b command character 
(K28.0) and an \num{8}-bit time stamp, followed by \num{48} bytes of predefined data 
and two bytes of parity information. The final format is used when the 
\dwords{coldadc} are read out in debug mode. In this case, \num{30} bits of raw 
pipeline stage data are read out from one of the two pipelined \dwords{adc} 
in each \dword{coldadc} \dword{asic} and passed from \dword{coldadc} to 
\dword{coldata} using two \num{16}-bit frames. When the ``Frame-15'' format is 
selected, a \dword{coldata} output data frame consists of an 8b/10b command 
character (K28.6) and an \num{8}-bit time stamp, followed by \num{60} bytes of \dword{adc} 
data (\num{30} bytes from each \dword{coldadc}). No parity information is generated 
when this format is selected. This is to ensure that at least one idle 
character (K28.1) will be sent between each ``Frame-15.'' A series of 8b/10b 
command characters (K28.5) is sent at the end of each frame of \num{12}-bit or 
\num{14}-bit data to ensure synchronization of the high-speed links.

The serializers and output drivers operate asynchronously in a separate clock domain that
is not related to the master clock signal received from the \dword{wib}. Instead
they use clocks derived from a \SI{40}{MHz} micro-electromechanical
system oscillator on the \dword{femb}. A single \dword{pll} generates a 
\SI{1.28}{GHz} clock for both serializers and output drivers. The \num{10}-bit
serializers are implemented using two \num{5}:\num{1} multiplexers (clocked at \SI{128}{MHz}) 
followed by a single \num{2}:\num{1} multiplexer (clocked at \SI{640}{MHz}). Each serializer 
derives the \SI{640}{MHz} and \SI{128}{MHz} clock from the \SI{1.28}{GHz} 
clock provided by the \dword{pll} and provides its \SI{128}{MHz} clock to 
the Data Frame Formation block, which uses it at the output stage of a 
clock-domain-crossing \dword{fifo}. A link synchronization sequence of 8b/10b 
command characters (K28.5) is used when the link is reset to establish the 
boundary between \num{10}-bit ``words.'' Idle characters (K28.1) are inserted 
by the Data Frame Formation block when no data is ready for serialization 
(between data frames). The \SI{1.28}{Gbps} output drivers include programmable 
pre-emphasis. The pre-emphasis is achieved using a combination of a voltage 
mode circuit at the input to the current mode driver and current mode 
pre-emphasis integrated into the driver circuit. Measurements were made 
of the insertion loss (``S parameters'') as a function of frequency using
\SI{25}{m} and \SI{35}{m} lengths of the twinax cable identical to the 
cable used in \dword{pdsp}, and the output driver circuit including 
pre-emphasis was simulated using a \dword{spice} model based on these 
measurements. The \dword{pll} and serializer circuits used in \dword{coldata}
were included in the first partial prototype (CDP1) test chip that was 
produced in fall 2017 and shown to work as designed. The measured eye 
diagram after \SI{25}{m} of twinax cable immersed in \dword{lar} using 
a commercial equalizer on the receiving end is shown in Figure~\ref{fig:128Gbpseyesim}.
The pre-emphasis circuit has been added to the current mode driver, which 
was verified in CDP1 and can be disabled if desired. 

A conservative estimation of the power consumption of \dword{coldata},
that is dominated by the power required for the \dword{lvds} transmitters
and receivers, amounts to \SI{195}{mW} for each \num{64}-channel \dword{asic}.

Prototype \dword{coldata} chips were received in July 2019 and the first 
round of room and \dword{ln}\  temperature bench tests has been completed.  
All of the \dword{i2c} control paths have been verified. \dword{coldata} 
registers can be written and read using either the \dword{lvds} interface 
or the \dword{cmos} interface. \dword{coldadc} registers can be written and 
read using the \dword{i2c} relay. Fast commands are interpreted as designed; 
the 2 MHz clock phase can be controlled, and the various ``Act'' commands 
are executed as intended.  The \dword{pll} locks and the link speed is 
correct. Bench tests of \dword{coldata} were completed in December 
2019. Data integrity was verified more completely using test equipment 
capable of checking for link errors in test periods of days.  
The \dword{larasic} control path will also be more completely verified and
later system tests with packaged \dwords{asic} will be performed.

\begin{dunefigure}
[ColdDATA output eye diagram]
{fig:128Gbpseyesim}
{Eye diagram after \SI{25}{m} of \dword{pdsp} twinax at \dword{ln}
temperature for the \dword{coldata} \SI{1.28}{Gbps} output link.}  
\includegraphics[width=0.8\linewidth]{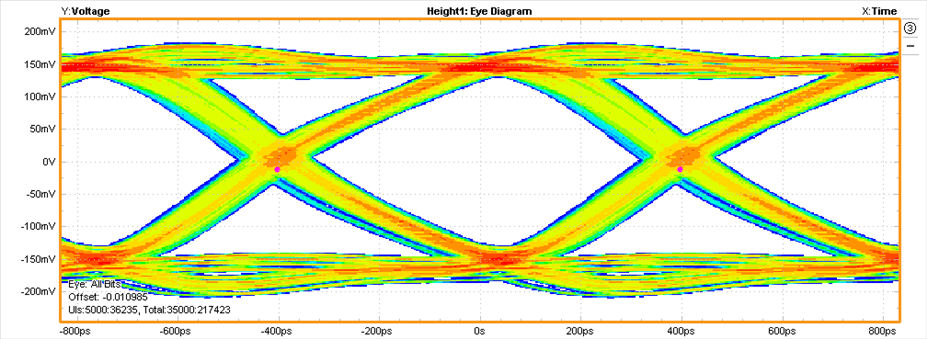}
\end{dunefigure}

\subsubsection{Alternative \dshort{asic} Solutions}
\label{sec:fdsp-tpcelec-design-asic-alternatives}

\subsubsubsection{Commercial Off-the-shelf \dshort{adc} Option}
\label{sec:fdsp-tpcelec-design-femb-alt-cots}

The \dword{sbnd} collaboration has been exploring the \dword{cots} \dword{adc} option 
for the \dword{tpc} readout electronics development since spring
2017~\cite{Chen:2018zic}. After a market survey, a few candidate \dwords{adc} 
using the \dword{sar} architecture were identified that would continue
to operate correctly when immersed in \dword{ln}. Starting in
July 2017, a lifetime study plan was developed to evaluate a \dword{cots} 
\dword{adc} option in two different phases: exploratory and validation. The 
lifetime study focused on the Analog Devices AD7274\footnote{AnalogDevices,
  AD7274\texttrademark{}, \url{https://www.analog.com/en/products/ad7274.html}.},
implemented in \dword{tsmc} \SI{350}{nm} \dword{cmos} technology, and has
demonstrated better performance in cryogenic operation compared to other candidates.

During the exploratory phase, fresh samples of the \dword{cots} \dword{adc} 
AD7274 were stressed with higher than nominal operation voltage, e.g.~\SI{5}{V},
while power consumption (drawn current) was monitored continuously. 
Periodically, the sample would be operated at nominal voltage (setting the power
supply input, V$_{DD}$, at \SI{2.5}{V}, and the voltage reference input, 
V$_{REF}$, at \SI{1.8}{V}) for a performance characterization test, where 
both the \dword{dnl} and \dword{inl} were monitored and analyzed in addition 
to the current. Stress test results were used to extrapolate the 
lifetime of the \dword{cots} \dword{adc}. The relation between the
\dword{cmos} transistor lifetime $\tau$ and the drain-source voltage $V_{ds}$, 
$\log\tau\propto1/V_{ds}$, is based on the creation of interface states by hot 
electrons and has been studies in the past extensively~\cite{Li:CELAr}.
The linear extrapolation of $\log\tau\propto1/V_{ds}$ is also used in industry (e.g. IBM) 
for accelerated stress testing. It was determined that a current drop 
of \num{1}\% on V$_{DD}$ would be used as the degradation criterion for the lifetime 
study. Following the development of this criterion, more devices were tested later
to validate what was learned in the exploratory phase.

The lifetime projection of the AD7274 \dword{adc} from the stress 
test with V$_{DD} >$ \SI{5}{V} is shown in Figure~\ref{fig:cotsadc-lifetime}. 
With the AD7274 operating at \SI{2.5}{V}, which is lower than the nominal 
\SI{3.6}{V} for the \SI{350}{nm} \dword{cmos} technology, the projected lifetime 
is more than than \num{1e6} years.

\begin{dunefigure}
[Lifetime projection of the COTS ADC]
{fig:cotsadc-lifetime}
{Lifetime projection of the \dword{cots} \dword{adc} AD7274 from the stress test 
with V$_{DD} >$ \SI{5}{V}. The current drop of 1\% on V$_{DD}$ is used as 
the degradation criterion. With nominal operation voltage of \SI{3.6}{V} for the 
\SI{350}{nm} \dword{cmos} technology, the lifetime is projected to be more 
than 100 years. For \dword{sbnd} and the \dword{dune} \dshort{fd}, the AD7274 will be
operated at \SI{2.5}{V} to add an additional margin; the expected lifetime is more 
than \num{1e6} years.}
\includegraphics[width=0.8\linewidth]{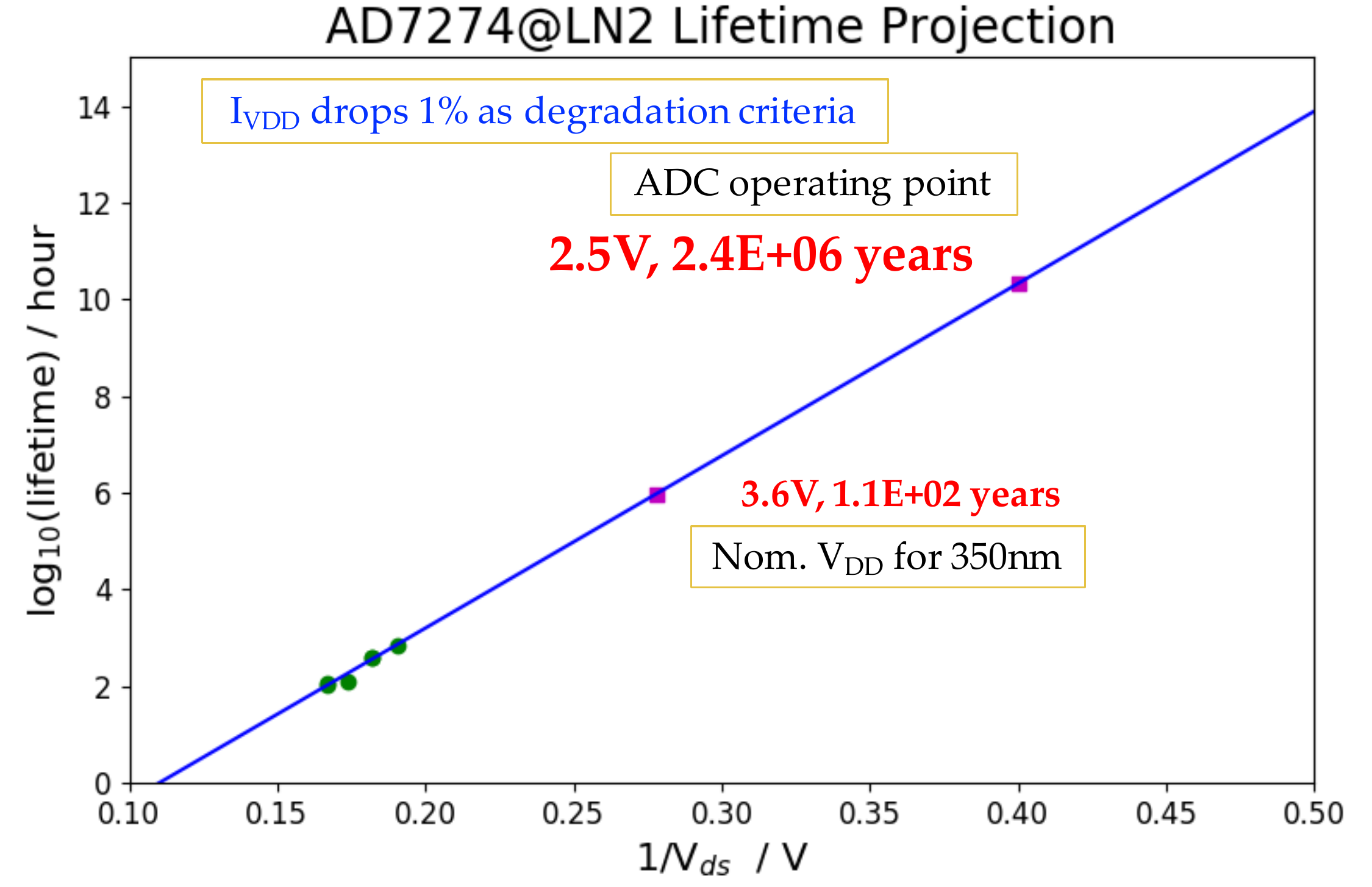}
\end{dunefigure}

Based on the lifetime study of AD7274, a \dword{femb} with the \dword{cots} 
\dword{adc} was developed and characterized for the \dword{sbnd} experiment. The 
integration test was carried out with 40\% \dword{apa} at \dword{bnl} and 
showed satisfactory noise performance as seen in Figure~\ref{fig:cotsadc-fembenc}.
The noise measurements obtained with the  40\% \dword{apa} at \dword{bnl} indicate
that the AD7274 gives a negligible contribution to the overall system noise, as
expected given that the \dword{adc} has an \dword{enob}
of \num{11.4}. The \dword{cots} \dword{adc} AD7274 serves as a backup solution for the 
\dword{spmod} \dword{tpc} readout electronics system. The current 
plan is to evaluate this \dword{adc} in the small \dword{tpc} installed in
\dword{iceberg} at \dword{fnal}. Ten \dwords{femb} with the \dword{cots} 
\dword{adc} have been fabricated and will be used to instrument the \dword{iceberg} 
\dword{tpc} for system integration tests in spring 2020. The main drawback of
the AD7274 \dword{adc} is that it is a single-channel chip, complicating the 
assembly of the \dwords{femb}.

\begin{dunefigure}
[Noise measurement of FEMBs with COTS ADC]
{fig:cotsadc-fembenc}
{The noise measurement of \dwords{femb} with \dword{cots} \dwords{adc} 
mounted on the \num{40}\% APA at \dword{bnl}. A picture of the 
\dword{femb} is shown in the top left corner. The induction plane 
(\SI{4}{m} wire length) has an \dword{enc} level of $\sim\SI{400}{e^-}$ with \SI{1}{$\mu$s} 
peaking time, while the collection plane (\SI{2.8}{m} wire length) has a noise level
of $\sim\SI{330}{e^-}$ with \SI{1}{$\mu$s} peaking time.}
\includegraphics[width=0.99\linewidth]{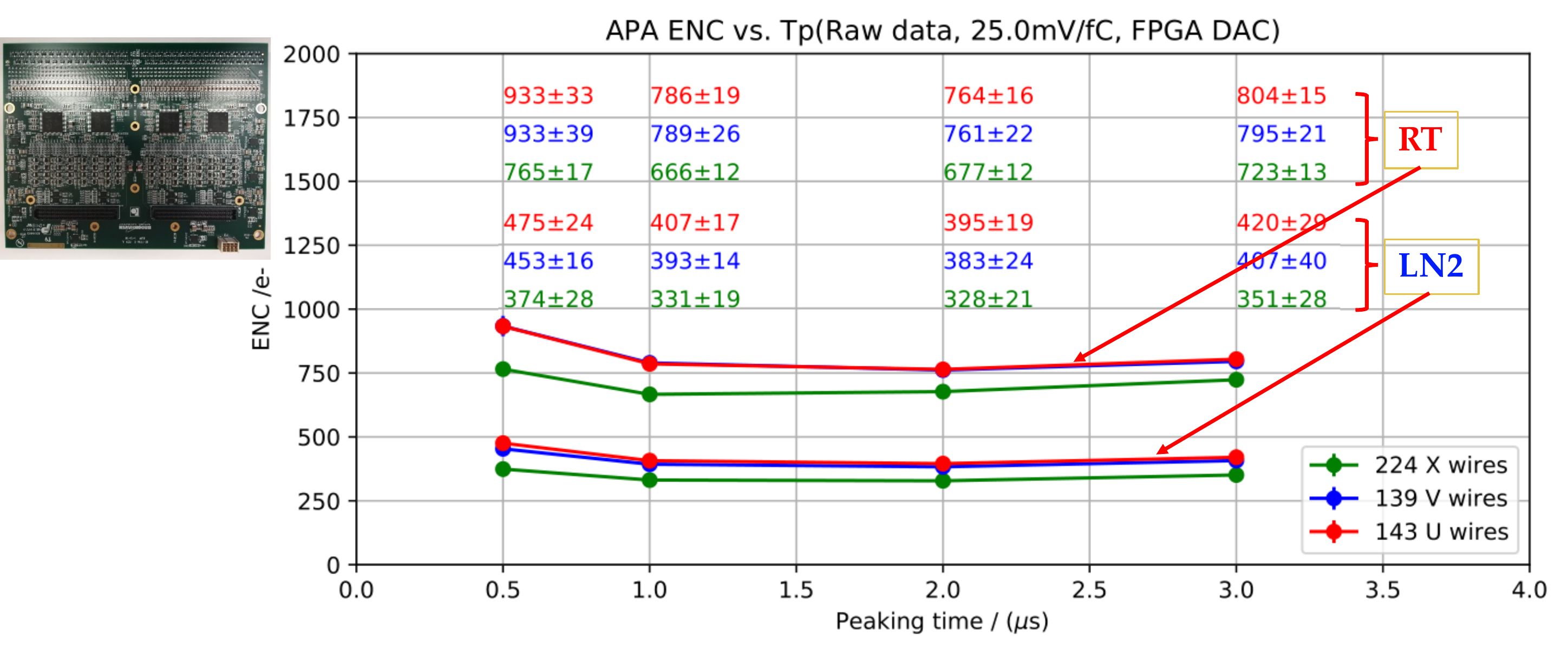}
\end{dunefigure}

\subsubsubsection{CRYO Option}
\label{sec:fdsp-tpcelec-design-femb-alt-cryo}

The \dword{slac} \dword{cryo} \dword{asic} differs from the reference three-chip 
design by combining the functions of an analog pre-amplifier, \dword{adc}, and 
data serialization along with transmission for \num{64} wire channels into a single 
chip. It is based on a design developed for the \dword{nexo} experiment~\cite{nEXO} 
and differs from it only in the design of the pre-amplifier, which is modified for 
the higher capacitance of the \dword{dune} \dword{spmod} wires compared to the short
strips of \dword{nexo}. The \dwords{femb} constructed using this chip would use only 
two \dwords{asic}, compared to the \num{18} (eight \dword{larasic}s, eight \dword{coldadc}s, and 
two \dword{coldata}s) needed in the reference design. This drastic reduction in 
part count may significantly improve \dword{femb} reliability, reduce power 
(\SI{40}{mW} per channel), and reduce costs related to production and testing.

Figure~\ref{fig:cryo-schematic} shows the overall architecture of the 
\dword{cryo} \dword{asic}, which is implemented in \SI{130}{nm} \dword{cmos}. 
It comprises two identical 32-channel blocks (banks) and a common section
providing biasing voltages and currents, as well as the controls signals, the clocks 
generation, and the configuration of the registers.

\begin{dunefigure}
[Overall architecture of the CRYO ASIC]
{fig:cryo-schematic}
{Overall architecture of the \dword{cryo} \dword{asic}.}
\includegraphics[width=0.9\textwidth]{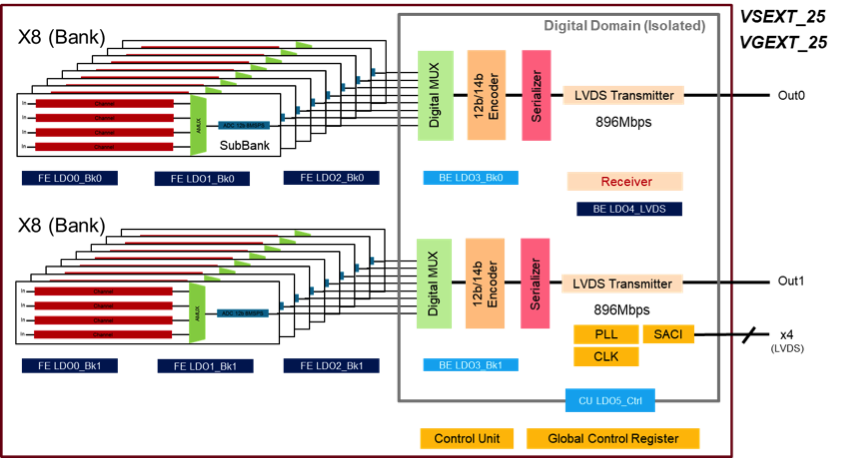}
\end{dunefigure}

The current signal from each wire is amplified using a pre-amplifier with pole-zero
cancellation~\cite{DeGeronimo:2011zz} and an anti-alias fifth-order 
Bessel filter (Figure~\ref{fig:cryo-FE}). Provisions are also made for injection 
of test pulses. Gain and peaking time are adjustable to values similar to 
those of the reference design. 
The four programmable gain settings of 6X, 3X, 1.5X, and 1X correspond to full-scale signals of 
$\SI{3.2e5}{e^{-}}$, $\SI{6.4e5}{e^{-}}$, $\SI{1.28e6}{e^{-}}$, and $\SI{1.92e6}{e^{-}}$. A 
filter with a Bessel shape has been chosen because of its flat group delay characteristic 
that minimizes waveform distortion as well as provides noise shaping performance similar 
to more classic semi-Gaussian shaper implementations. The four programmable peaking times 
of the filter are $\SI{0.6}{{\mu}s}$, $\SI{1.2}{{\mu}s}$, $\SI{2.4}{{\mu}s}$, 
and $\SI{3.6}{{\mu}s}$, corresponding to filter bandwidths equivalent to the ones used in 
the reference solution. Similarly to the reference design with three \dwords{asic}, each
channel can be configured, independently from the other channels, to have a 
baseline for operation consistent with either the collection or the induction wires. 
The outputs of the \dword{fe} amplifiers can be connected, one-at-a-time, 
to an analog monitor to probe the analog signal.

\begin{dunefigure}
[CRYO FE section architecture and typical front-end response]
{fig:cryo-FE}
{\dword{cryo} front-end section architecture (left); typical response of the \dword{cryo} front-end (right).}
\includegraphics[width=.54\textwidth]{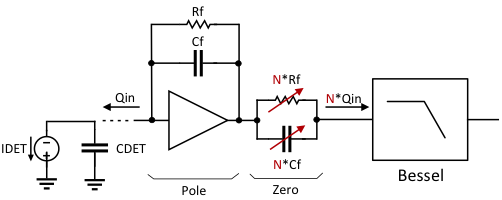}
\includegraphics[width=.45\textwidth,trim={0 5.5cm 6.5cm 3cm},clip]{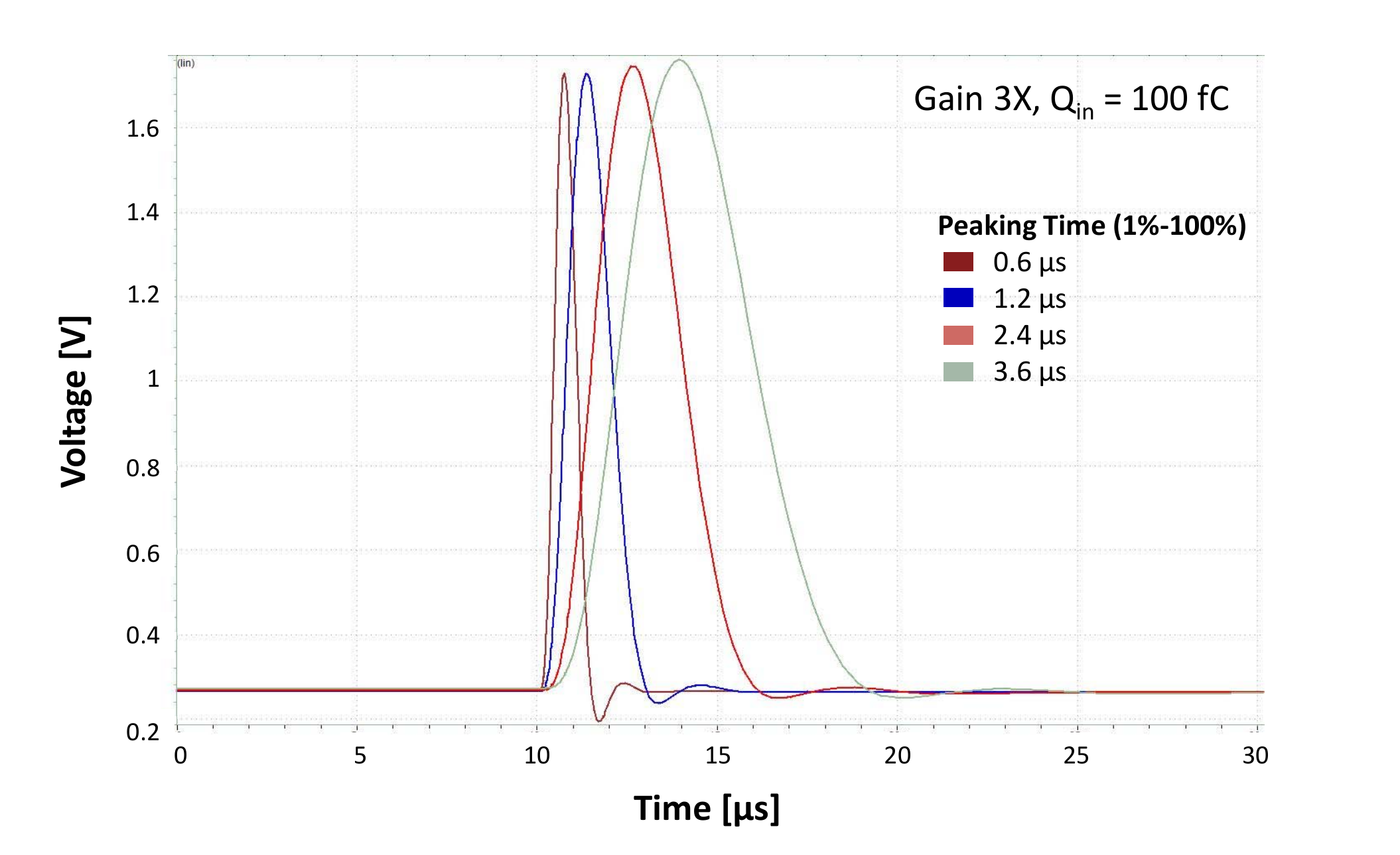}
\end{dunefigure}

Four input channels are multiplexed onto a single fully differential 12-bit \SI{8}{MSPS} 
\dword{adc}. Signals from the four channels are concurrently sampled onto a sample-and-hold 
stage. An \dword{adc} driver after the multiplexer performs the single-ended to differential 
conversion. The \dword{adc} has a pure \dword{sar} architecture (Figure~\ref{fig:cryo-ADC}) 
with a split-cap \dword{dac} based on V$_{cm}$ switching~\cite{5482529}, and has the option to be 
calibrated for offset compensation. External signals can be routed to the input of each single 
\dword{adc} allowing standalone characterization.

\begin{dunefigure}
[CRYO ADC architecture]
{fig:cryo-ADC}
{\dword{cryo} \dword{adc} architecture.}
\includegraphics[width=.9\textwidth]{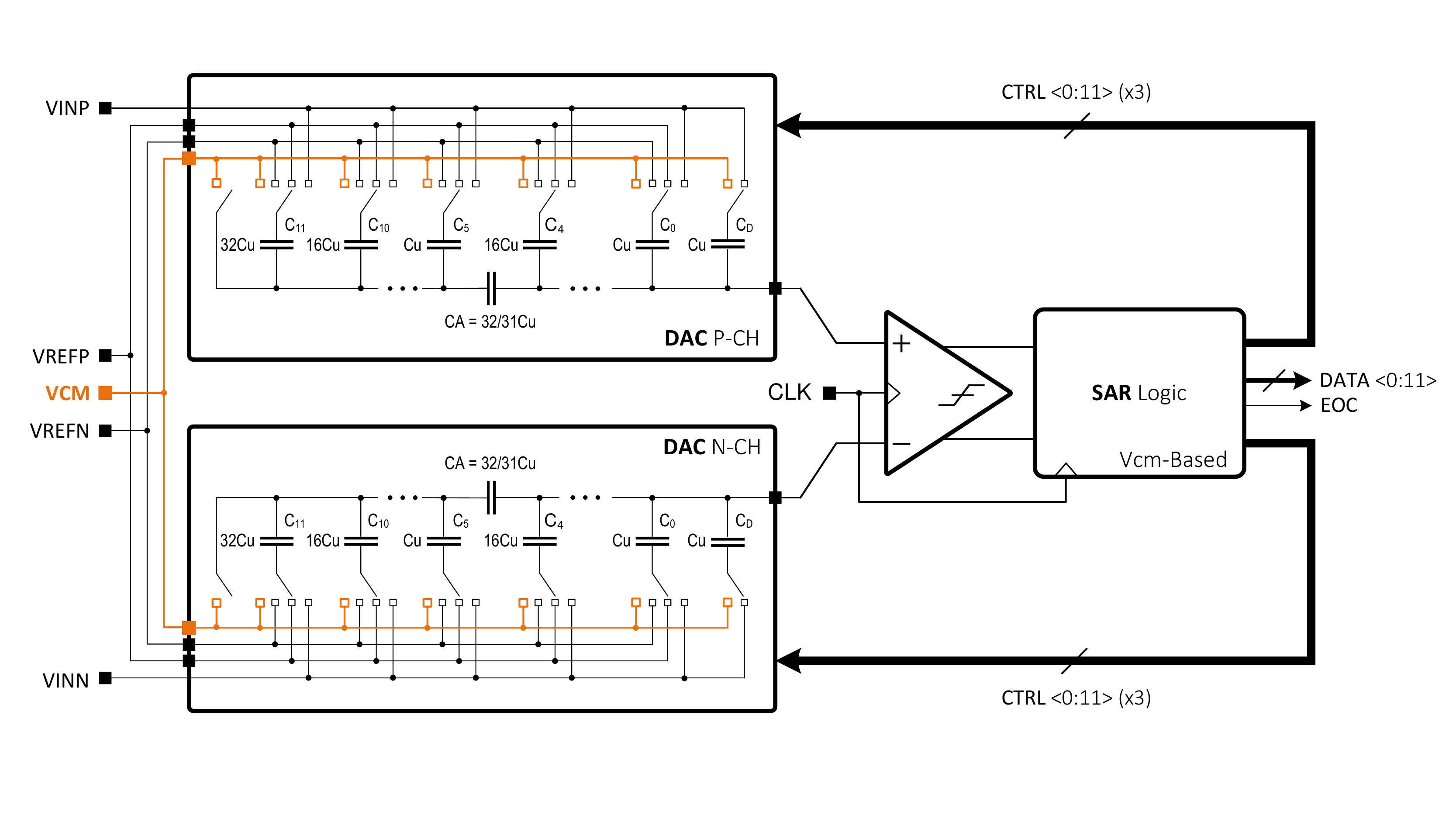}
\end{dunefigure}

The data serialization and transmission block uses a custom 12b/14b encoder, so 32 channels 
of 12-bit \SI{2}{MSPS} data can be transmitted with a digital bandwidth of only \SI{896}{Mbps}, 
which is significantly lower than the required bandwidth of the reference design (\SI{1.28}{Gbps}).

One key concern with mixed-signal \dwords{asic} is the possibility of interference from the 
digital side causing noise on the very sensitive pre-amplifier. To avoid this interference, 
the \dword{cryo} design uses well established techniques for isolating the substrate;
these are described in the literature~\cite{yeh} and have been successfully used in previous 
\dwords{asic}. Furthermore, power domains of the various sections of the \dwords{asic} are 
isolated using multiple internal \dwords{ldo}.

For reliability purposes the analog section of the \dword{asic} using thick oxide devices 
is biased at \SI{2}{V} ($\num{20}\%$ less than nominal voltage) and does not use minimum length devices.
The digital section of the \dword{asic} uses core devices biased at \SI{1}{V} (again $\num{20}\%$ 
less than nominal voltage).

The infrastructure requirements for a \dword{cryo} \dword{asic}-based system are similar 
to those of the reference option. However, in most cases, somewhat fewer resources are 
needed; for instance:
\begin{itemize}
\item A single voltage is needed for the power supply. This is used to generate the two 
supply voltages using internal voltage regulators.
\item The warm interface is different. \dword{cryo} operates synchronously
with a \SI{56}{MHz} clock, does not require a fast command, and uses the 
\dword{saci} protocol~\cite{SACI} for configuration rather than \dword{i2c}.
\end{itemize}

Simulation-based studies have been performed: using the $\SI{1.2}{{\mu}s}$ peaking time and an 
input capacitance of \SI{220}{pF} (close to that expected in the \dword{spmod}), the noise 
level is approximately $\SI{500}{e^{-}}$, similar to that expected with the reference 
\dword{larasic} design in \dword{lar} with the same input capacitance.

The first iteration of the \dword{cryo} \dword{asic} design (see Figure~\ref{fig:cryo-photos}) 
was submitted to MOSIS for fabrication in November 2018. The first prototypes were 
delivered at the end of January 2019.

\begin{dunefigure}
[Photos of CRYO ASIC prototype]
{fig:cryo-photos}
{Photo of the prototype \dword{cryo} cold board (left); zoomed-in photo of \dword{cryo} \dwords{asic} (right).}
\includegraphics[width=.3\textwidth]{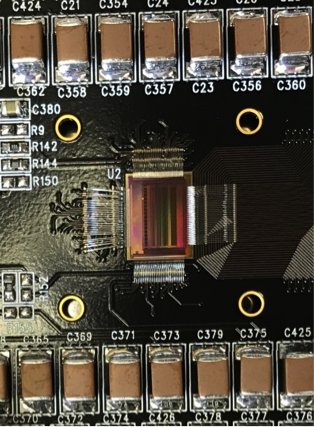}
\hspace{1cm}
\includegraphics[width=.4075\textwidth]{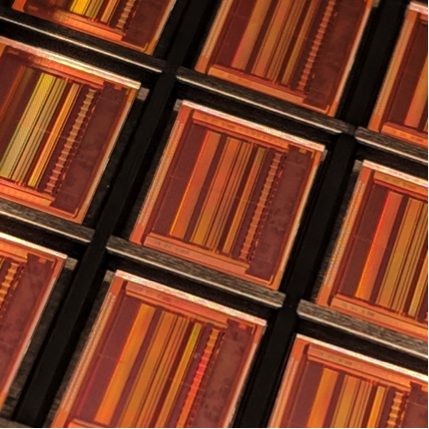}
\end{dunefigure}

The prototypes are under test in an existing test stand at \dword{slac} using the \dword{cts}
described in Section~\ref{sec:fdsp-tpcelec-qa-initial}. Subsequent system tests are planned 
using the facilities described in Section~\ref{sec:fdsp-tpcelec-qa-facilities}.

The first prototype of the \dword{asic} is functional at both room temperature and \dword{ln} 
temperature. In particular, all the key blocks have been verified. Configuration of all the 
64 channel registers (13 bits each) and the 17 (16-bit) global registers has been 
verified. Optimization of the register values is ongoing at both room and cold temperature.
Initialization procedures for the \dword{asic} power-up have been established. 
Operation of the on-chip \dwords{ldo} has been verified and expected supply levels are 
stable against changes in temperature. The analog monitor can be used to spy on 
the output of the amplifier for injected pulses on the \dword{fe} channels, prior
to the digitization of these signals by the internal \dwords{adc}, as shown in
Figure~\ref{fig:cryo-FEresponse}.

\begin{dunefigure}
[CRYO ASIC FE response at liquid nitrogen temperature]
{fig:cryo-FEresponse}
{\dword{cryo} \dword{asic} front-end response at liquid nitrogen temperature, presented at 
the analog monitor and acquired with an external \SI{50}{MSPS} \dword{adc}.}
\includegraphics[width=.65\textwidth]{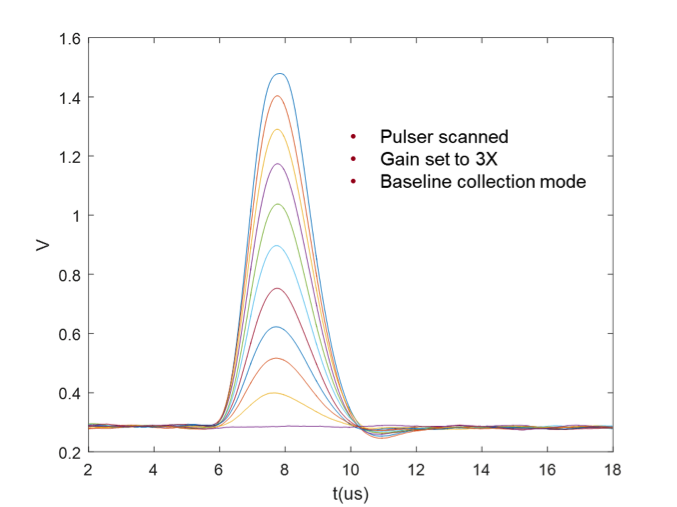}
\end{dunefigure}

Encoded data are transmitted and correctly decoded in the external \dword{fpga}. 
Figure~\ref{fig:cryo-analogmon} shows an example of a pulse injected in a channel 
visible on both the analog monitor as well as in the data acquired by the \dword{asic}. 
Data are acquired at \dword{ln} temperature at the nominal $\sim$\SI{2}{MSPS} rate.

\begin{dunefigure}
[Example of a pulse injected in a CRYO ASIC channel]
{fig:cryo-analogmon}
{Example of a pulse injected in a \dword{cryo} \dword{asic} channel, visible on both the analog monitor and in the output data.}
\includegraphics[width=.99\textwidth]{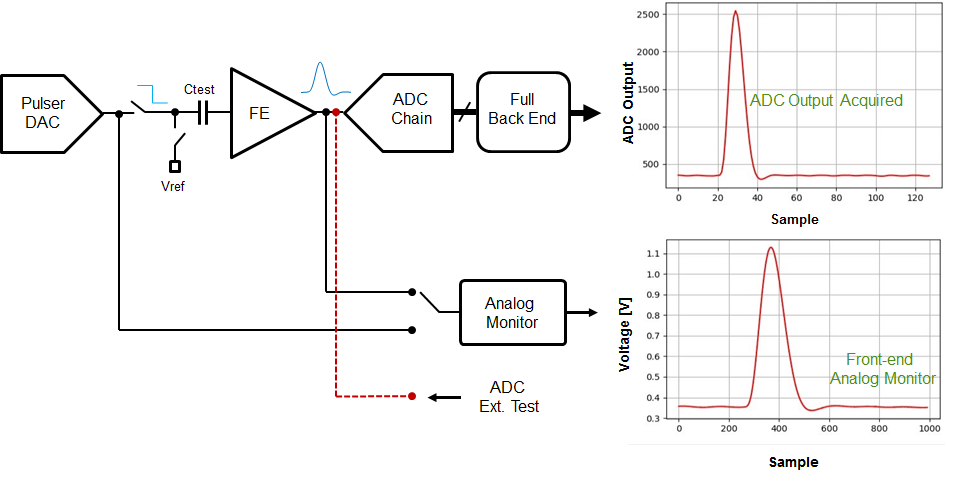}
\end{dunefigure}

From the functional point of view, a single unexpected behavior has been identified 
in the digital multiplexer that is used at the input of the encoders. The latches
at the input of the multiplexer show poor driving capability, resulting in the presence
of a ghost from a previously multiplexed channel. The effect is not present on the 
first 12 channels of each block which show expected behavior. The effect has been
replicated in simulation and a trivial fix has been implemented for 
the next version of the \dword{asic}.

Initial results on the performance of the \dword{adc} block of \dword{cryo} have
been obtained by directly injecting a linear voltage ramp (generated by an external
\num{20}-bit \dword{dac}) into the \dword{adc}. The distributions of the \dword{dnl}
and \dword{inl} obtained from these measurements are shown in Figure~\ref{fig:cryo-dnlinl}.
The maximum deviations of the \dword{dnl} and \dword{inl} from the reference
signal are \num{0.74} and \num{1.27} \dword{adc} counts, respectively, within
a usable dynamic range of $\sim\num{3000}$ \dword{adc} counts. From these
distributions, the values of \SI{65.75}{dB} and of \num{10.63} 
are estimated for the \dword{sndr} and \dword{enob}, respectively. These results indicate
that, from a static point of view, the \dword{adc} block of \dword{cryo}
meets the required performance for the \dword{dune} \dword{tpc} readout.
Further work is ongoing to characterize the dynamic response of the
\dword{adc} and to determine the overall linearity and noise performance
of the entire readout chain including the \dword{fe} amplifier.

\begin{dunefigure}
[Differential and integral non-linearities for the CRYO ADC]
{fig:cryo-dnlinl}
{Distribution of the \dword{dnl} (left) and \dword{inl} (right) for the
\dword{adc} block of \dword{cryo}.}
\includegraphics[width=.495\textwidth]{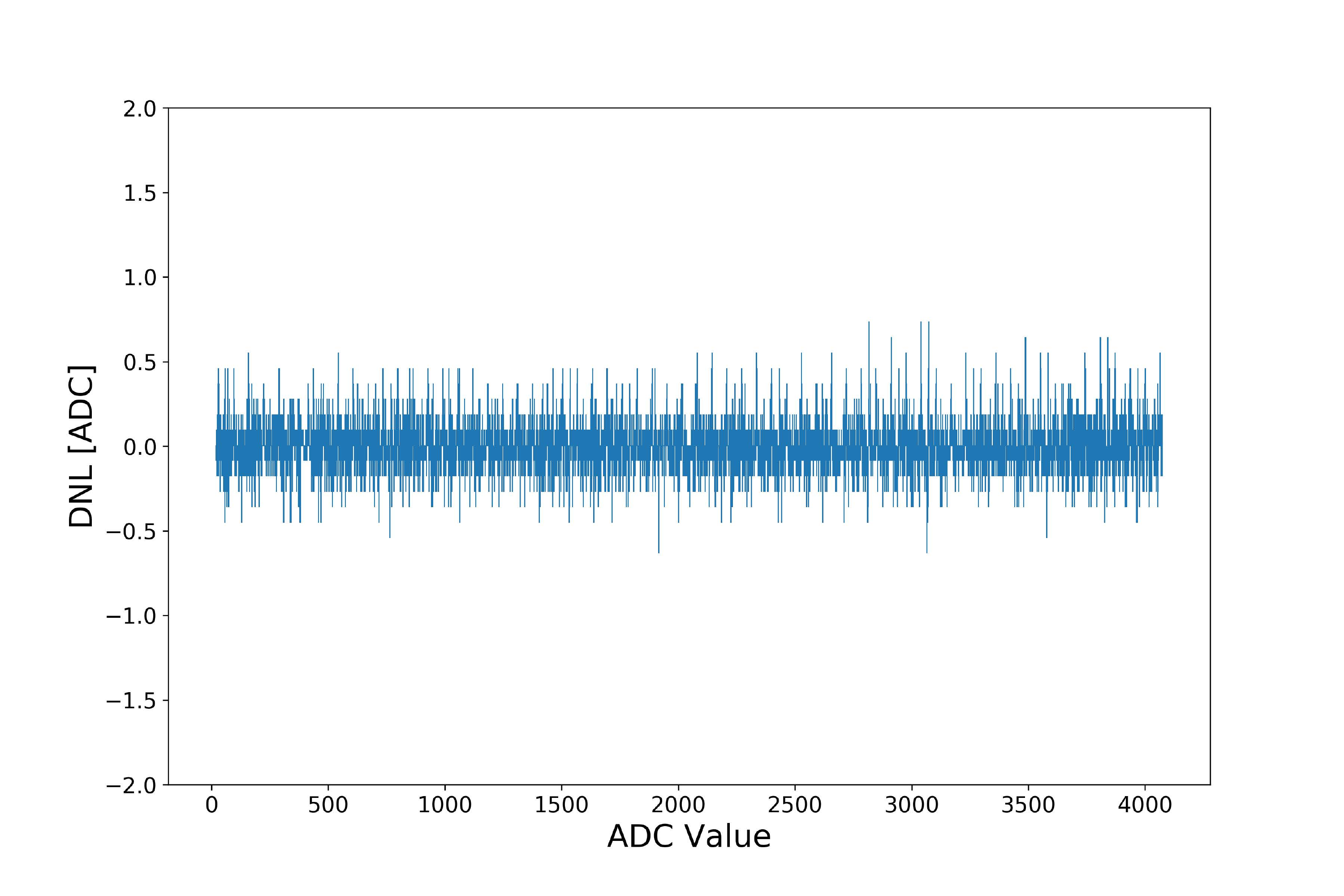}
\includegraphics[width=.495\textwidth]{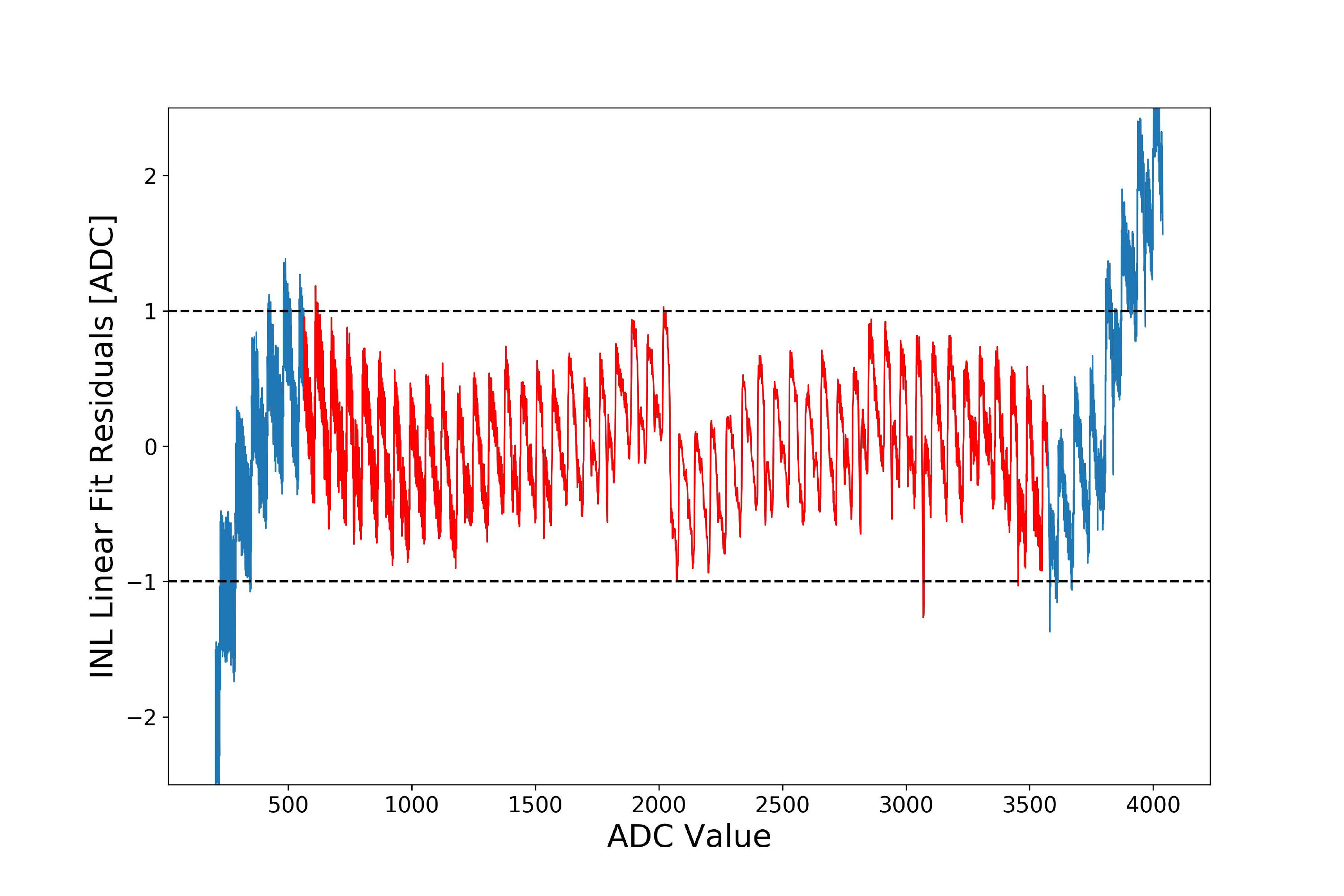}
\end{dunefigure}

\subsubsection{Procedure and Timeline for \dshort{asic} Selection}
\label{sec:fdsp-tpcelec-design-femb-selection}

We are currently pursuing two different \dword{asic} designs and 
planning on qualifying the \dword{cots} \dword{adc} solution that 
will be used for the \dword{sbnd} experiment. 
We plan to continue developing both the three-\dword{asic} 
solution and the \dword{cryo} \dword{asic} for at least a second 
iteration before deciding which \dword{asic} solution to 
implement in the \dword{dune} \dword{spmod}. This plan requires
that multiple versions of the \dword{femb} are also designed 
and tested. The \dwords{femb} populated with the first set of prototypes of 
the two kinds of \dwords{asic} will be available in spring 2020  
and are expected to perform similarly to the
boards used for \dword{pdsp}. We plan to review the results 
of the system tests and of the component lifetimes discussed 
in Section~\ref{sec:fdsp-tpcelec-qa} in early 2020.
In that review, we will also decide whether to change anything on 
the list of specifications for the \dwords{asic} and to further develop
the two custom \dword{asic} solutions, including fixing any 
issues found during the tests of the first version of the \dwords{asic}. 
We expect that the subsequent iteration
of the design, fabrication, and testing of the \dwords{asic} and
\dwords{femb} will take an additional twelve months. At the end 
of this process, when results from standalone tests of the
\dwords{asic} and system tests of the \dwords{femb} are
available, we will have all the information required to select the \dword{asic}
solution to be used in \dword{dune}. We are assuming that the second design iteration of
the \dwords{asic} design will meet all the \dword{dune} requirements.
The schedule for the construction of the \dword{dune} \dword{spmod}
currently has between eight and fourteen months of float for the
\dwords{asic} and \dwords{femb}, which would allow for a third design
iteration, if needed, as discussed
in Section~\ref{sec:fdsp-tpcelec-management-planning}. This does
not apply for the second run of \dword{pdsp} (discussed
later in Section~\ref{sec:fdsp-tpcelec-qa-facilities-pdune}).
Ideally, the \dwords{asic} from the engineering run would be used
for the second run of \dword{pdsp}, but this is not compatible
with the currently planned date for the installation of the \dwords{femb}
on the \dword{apa}s. In order to meet the current goal for the starting
date of the second run of \dword{pdsp}, \dwords{asic} from the
second round of prototyping would have to be used. In case a
third round of prototypes is necessary, the second run of 
\dword{pdsp} would have to be delayed by one year.

The selection of the \dword{asic}(s) to be used for the
construction of the \dword{sp} \dword{detmodule} will be
based on performance, reliability, power density
criteria, as well as consideration of the costs and
resources required during the construction and testing
of the \dwords{femb}. We have not yet decided the weights
to assign to these criteria. Reliability would in principle favor
the single-\dword{asic} solution that requires 
\dwords{femb} with fewer connections, while power 
density considerations could be less favorable to \dword{cryo} option.
We plan to charge a committee to draft a series of recommendations 
on the \dword{asic} selection in spring 2020, at least one year
ahead of the expected decision date. These recommendations could 
also inform the second cycle of design for
\dwords{asic} and \dwords{femb}. Once the second cycle of design
and testing is complete, these recommendations will be used by the
committee charged with the final design review to suggest a
preferred option for the \dword{asic} solution.
The committee's recommendation 
will then be passed to the \dword{dune} \dword{exb}, 
which is tasked with the final \dword{asic} decision.

\subsection{Infrastructure Inside the Cryostat}
\label{sec:fdsp-tpcelec-design-infrastructure}

Each \dword{femb} is enclosed in a mechanical \dword{ce} box 
to provide support, cable strain relief, and control of bubbles of gaseous
argon generated by heat from an \dword{femb} attached to the lower \dword{apa},
which could, in principle, lead to discharge of the \dword{hv} system. The
\dword{ce} box, illustrated in Figure~\ref{fig:ce-box}, is designed to make the 
electrical connection between the \dword{femb} and the \dword{apa} 
frame, as discussed 
in Section~\ref{sec:fdsp-tpcelec-design-grounding}.
Mounting hardware inside the \dword{ce} box connects the ground plane 
of the \dword{femb} to the box casing. If argon bubbles 
form inside
the \dword{ce} box, they must get 
channeled through the two side tubes
of the \dword{apa}'s frame, from where they would reach the top of the cryostat.
As already discussed in Section~\ref{sec:fdsp-tpcelec-overview-requirements},
a test setup has been prepared at \dword{bnl} to measure the
maximum power that can be dissipated in \dword{lar} at a
depth equivalent to that of the \dwords{femb} installed on
the bottom \dword{apa}. Initial measurements indicate that
the \dwords{asic} mounted on the \dwords{femb} are not going
to cause boiling of the \lar  inside the \dword{ce} boxes. 
We have measured the power required to cause boiling at
a pressure equivalent to that of \SI{12}{m} of \lar.
We have also observed that with the current \dword{asic} designs
and power dissipation we have a safety factor of \num{20}
in terms of total power and of at least two in terms of power
density. These measurements will be repeated once prototype
\dwords{femb} with the three-\dword{asic} and 
\dword{cryo} solutions become available.

\begin{dunefigure}
[Prototype CE box used in ProtoDUNE-SP]
{fig:ce-box}
{Prototype \dword{ce} box used in \dword{pdsp}.}
\includegraphics[width=0.45\linewidth]{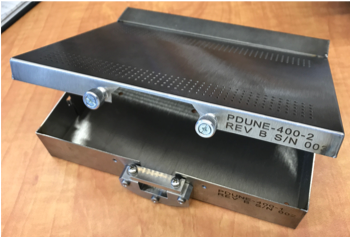}
\end{dunefigure}

The \dword{ce} box casing is electrically connected to the 
\dword{apa} frame via the metal mounting hardware called the 
``Omega bracket'' (not shown in Figure~\ref{fig:ce-box}). The 
input amplifier circuits  are connected to the \dword{cr} board  
and terminate to ground at the \dword{apa} frame, as 
shown in Figure~\ref{fig:CR-board}.  As a backup solution, the casing is 
also connected to the \dword{apa} frame via a wire.

In addition to the \dword{ce} box and mounting hardware, cable trays 
for support and routing the cold cables will be installed in the 
cryostat. One set of cable trays, shown in Figure~\ref{fig:trays} 
(left column), will be attached to the upper \dword{apa}s
to hold the \dword{ce} and \dword{pd} cables. A different cable 
tray design, also shown in Figure~\ref{fig:trays} (right column), 
will support the \dword{ce} cables underneath the 
lower hanging \dword{apa}s. A final set of cable trays will be 
installed inside the cryostat after the \dword{apa}s are 
fixed in their final location to support the cables as they are 
routed to the \dword{ce} and \dword{pd} feedthroughs.

\begin{dunefigure}
[Views of various cable and CE \coldbox{}es supports]
{fig:trays}
{Side and end views of mechanical supports for the \dword{ce} 
boxes on the upper (left column) and lower (right column) 
\dword{apa}s. Shown are the \dword{apa} cable trays in green and pink, 
the \dword{ce} boxes in dark gray, and the Omega brackets and mounting 
hardware between the \dword{ce} boxes and \dword{apa} frame in light gray.  
The \dword{ce} cables are shown in blue; the \dword{pd} cables are not shown.}
\includegraphics[width=0.38\linewidth]{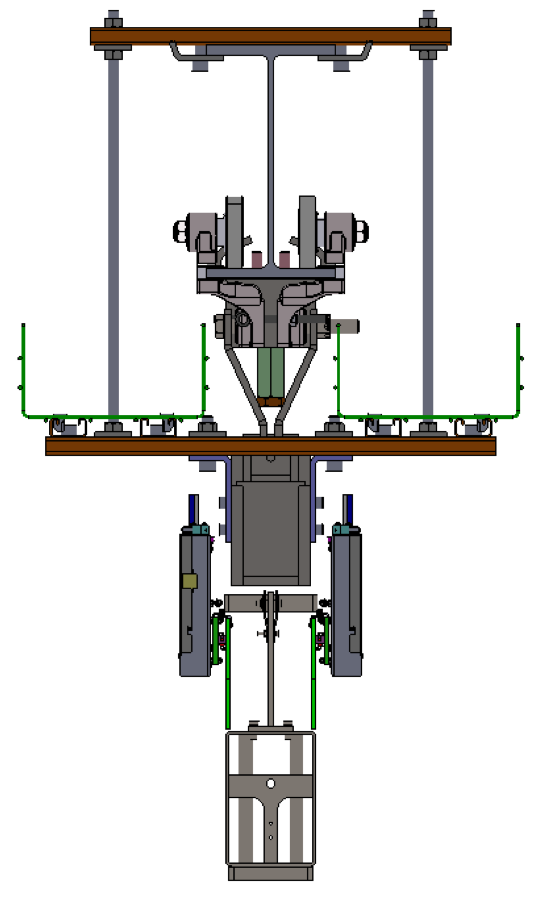}
\hspace{5mm}
\includegraphics[width=0.25\linewidth]{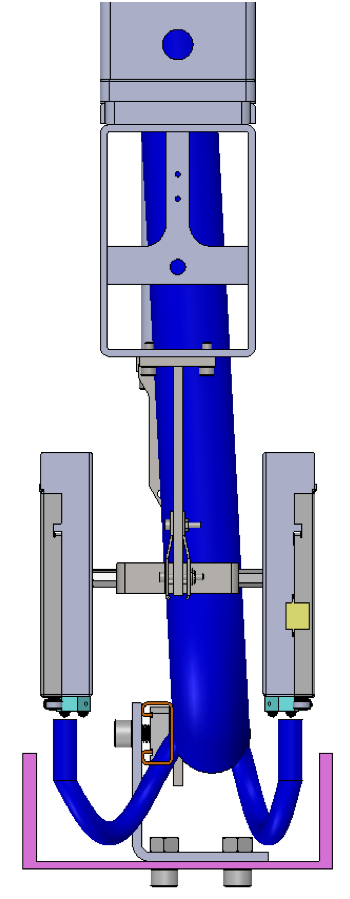} \\
\includegraphics[width=0.32\linewidth]{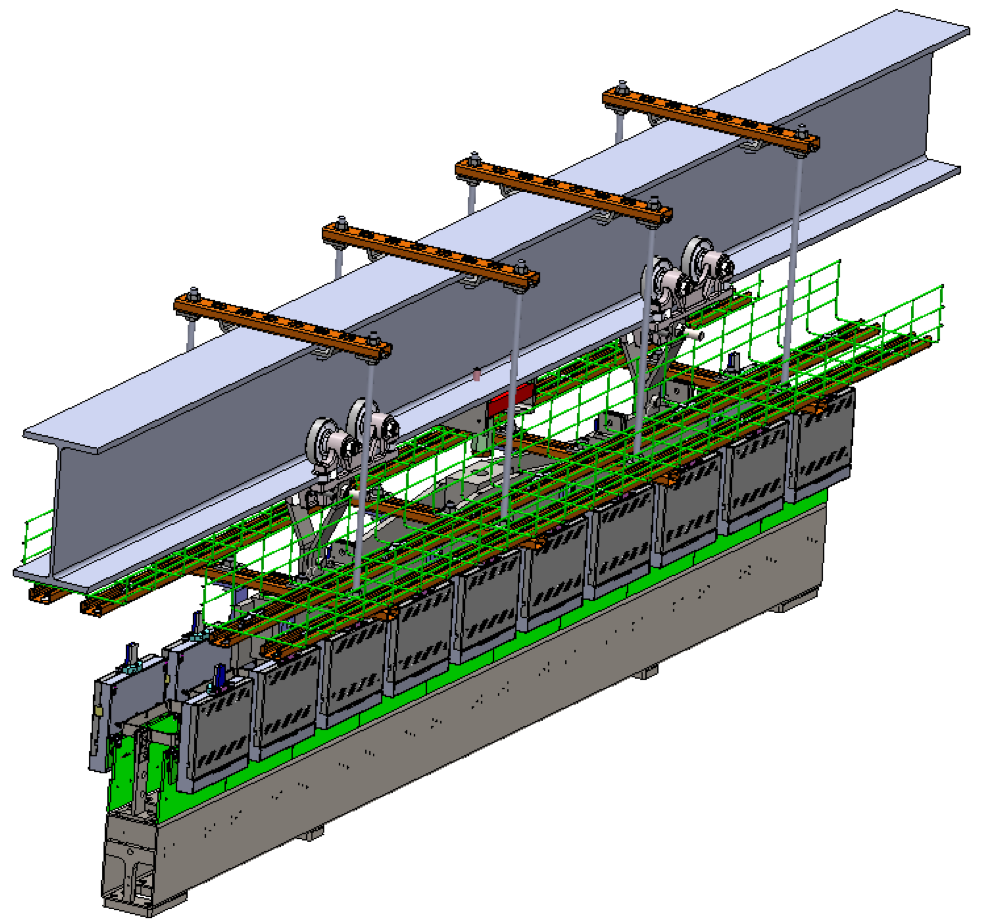}
\hspace{5mm}
\includegraphics[width=0.4\linewidth]{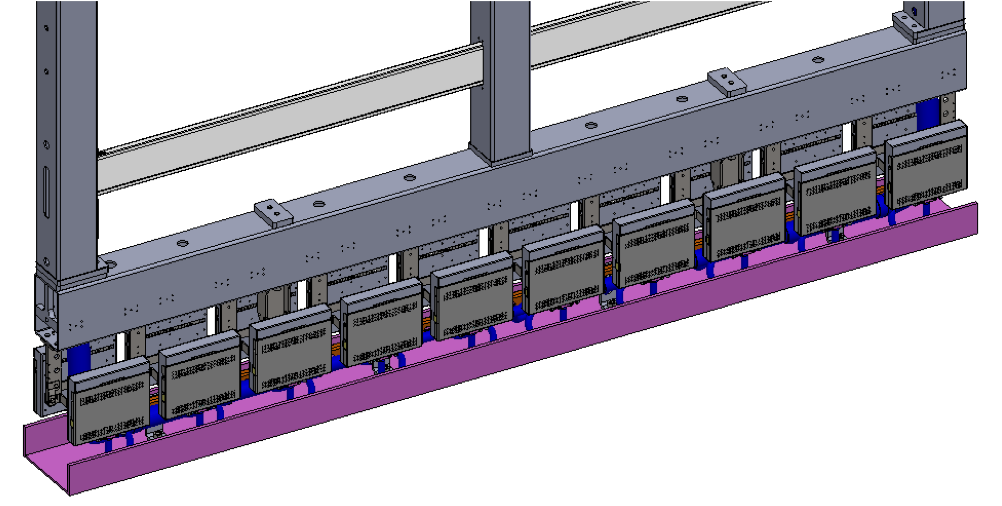}
\end{dunefigure}

\subsection{Cold Cables and Cold Electronics Feedthroughs}
\label{sec:fdsp-tpcelec-design-ft}

All cold cables originating inside the cryostat connect to the outside 
warm electronics through \dword{pcb} \fdth{}s installed in the signal 
flanges that are located on the cryostat roof. The data rate from each
\dword{femb} with four cables is sufficiently low ($\sim\SI{1}{Gbps}$)
that \dword{lvds} signals can easily be driven over more than \SI{22}{m}
of twin-axial transmission line. Additional transmission lines
are available to distribute \dword{lvds} clock and control signals,
which are transmitted at a lower bit rate. The connections 
between the \dwords{wib} on the signal flanges and the 
\dword{daq} (see Chapter~\ref{ch:daq}) and slow 
control systems (see Chapter~\ref{ch:sp-cisc}) are made
using optical fibers.

\begin{dunefigure}
[TPC cold eletronics \fdth]
{fig:tpcelec-signal-ft}
{TPC \dword{ce} \fdth. The \dwords{wib} are seen edge-on in the left 
panel and in an oblique side-view in the right panel, which also shows 
the warm crate for an \dword{spmod} in a cutaway view.}
\includegraphics[width=0.9\linewidth]{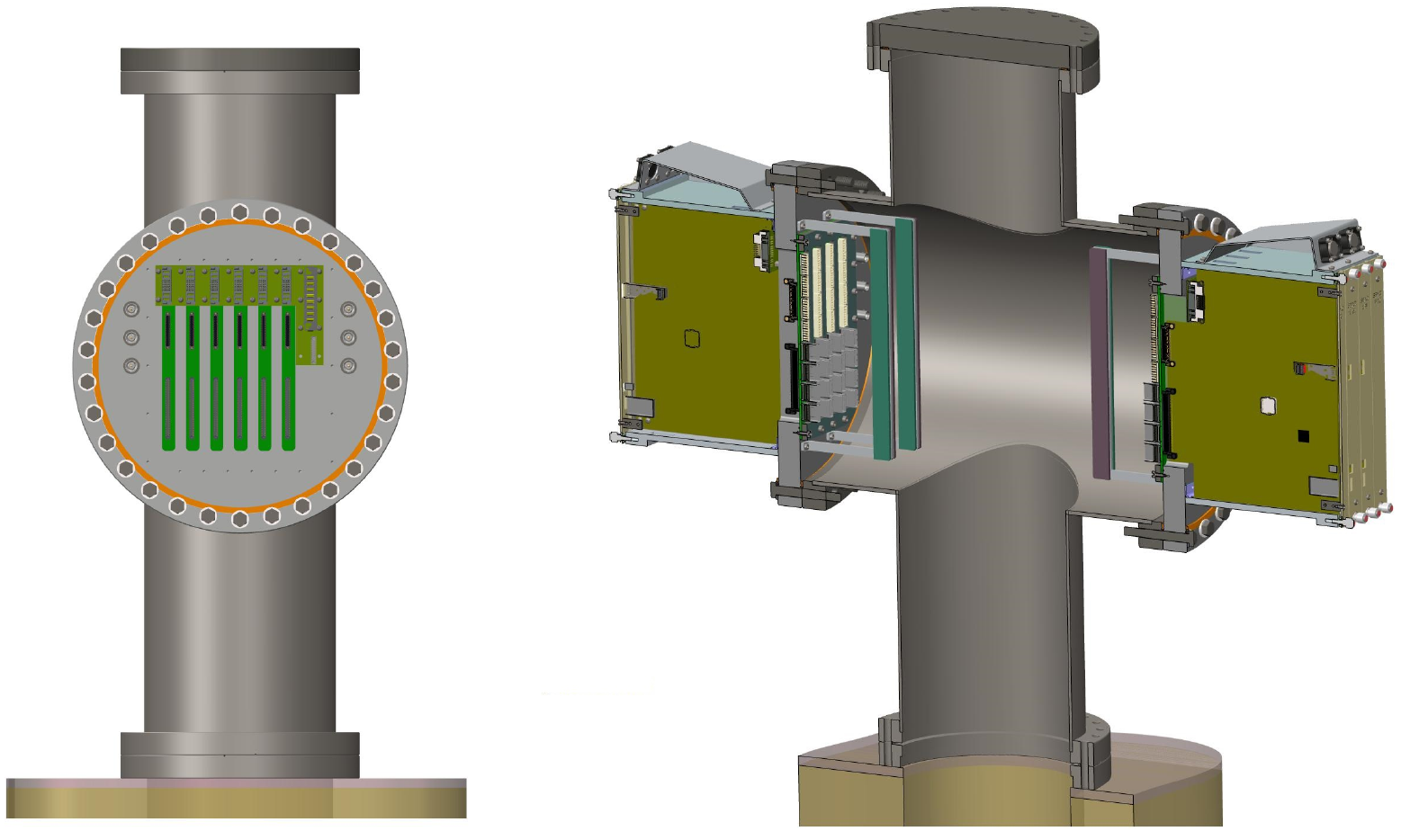}
\end{dunefigure}

The design of the signal flange includes a four-way cross spool 
piece, separate \dword{pcb} \fdth{}s for the \dword{ce} and 
\dword{pds} cables, and an attached crate for the \dword{tpc} 
warm electronics, as shown in Figure~\ref{fig:tpcelec-signal-ft}.
The wire bias voltage cables connect to standard \dword{shv} 
connectors machined directly into the \dword{ce} \fdth{}, ensuring 
no electrical connection between the wire bias voltages and other 
signals passing through the signal flange. Each \dword{ce} \fdth 
serves the bias voltage, power, and digital I/O needs of one \dword{apa}.  

Data and control cable bundles send system clock and control signals 
from the signal flange to the \dword{femb} and stream the $\sim$\SI{1}{Gbps} 
high-speed data from the \dword{femb} to the signal flange. Each 
\dword{femb} connects to a signal flange via one data cable bundle, 
leading to \num{20} bundles between one \dword{apa} and one flange. 
For the reference \dwords{asic} configuration, ten low-skew shielded 
twin-axial cables are required to transmit the following differential signals
between the \dword{wib} and the \dword{femb}:
\begin{itemize}
\item four \SI{1.28}{Gbps} data lines (two from each \dword{coldata});
\item two \SI{64}{MHz} clock signals (one input to each \dword{coldata});
\item one fast command line (shared between the two \dword{coldata} \dwords{asic}); and
\item three \dword{i2c}-like control lines (clock, data-in, and data-out, also shared between the 
two \dword{coldata} \dwords{asic}).
\end{itemize}
As discussed later, this number of connections is compatible with routing
the cables that bring the power and transmit the data and controls for
the lower \dword{apa} through the \dword{apa} frames. We are making the
assumption that the fast command line can be shared between the two 
\dword{coldata} \dwords{asic}, but will also consider other possibilities,
like sharing the \SI{64}{MHz} clock between the two \dwords{asic} or
increasing the data transmission speed to \SI{2.56}{Gbps}, thereby
reducing the number of data transmission lines to two for each \dword{femb}.
This assumption will be tested as soon as the first prototypes of
\dword{coldata} become available.

The \dword{lv} power is passed from the signal flange to the
\dword{femb} by bundles of \SI{20}{AWG} twisted-pair wires, with half of
the wires serving as power feeds and the other half as returns.
Using the measured power consumption for \dword{larasic} and \dword{coldadc}
and the estimates for \dword{coldata}, the total power required to
operate each \dword{femb} is estimated as \SI{6}{W} (\SI{2.4}{A}
at \SI{2.5}{V}), including the power dissipated in the linear voltage
regulators. This assumes that linear voltage regulators are
used on the \dword{femb} to reduce the \SI{2.5}{V} provided by the
\dword{wib} down to the various voltages required by the three
\dwords{asic}:
\begin{itemize}
\item \SI{1.8}{V} for \dword{larasic}, and
\item \SI{2.25}{V} and \SI{1.1}{V} for \dword{coldadc} and \dword{coldata}.
\end{itemize}
We currently assume that only \SI{2.5}{V} will be provided by the 
\dword{wib}, since the largest fraction of the power required by
the \dword{femb} is at \SI{2.25}{V}. We are currently planning on using 
a total of eight \SI{20}{AWG} twisted-pair wires, seven of which will be used
for bringing the \SI{2.25}{V} to the \dword{femb}, with the eighth
one reserved for the \SI{5}{V} bias for the linear voltage regulators
(this connection carries a very low current). With this cable plant, the
resistance of the cable bundle is $\SI{41}{\milli\ohm}$ for the upper 
\dword{apa}s (\SI{9}{m} cable length) and $\SI{101}{\milli\ohm}$ for
the lower \dword{apa}s (\SI{22}{m} cable length). To account for the
voltage drop along the wires and the returns in the case of operation
at room temperature, prior to filling the cryostat, the \dword{wib} needs to
provide \SI{2.7}{V} and \SI{3.0}{V} for the upper
and lower \dword{apa}s, respectively. For one \dword{femb}, the power
dissipated in the cables is \SI{0.5}{W} and \SI{1.2}{W} for the upper
and lower \dword{apa}s, respectively. The values for the power 
dissipated in the cables are reduced by a factor of three for operation
in \dword{lar}, which allows for a reduction of the voltage 
provided by the \dword{wib}. The voltage drop and power dissipation
values are summarized in Table~\ref{tab:SPCE:cablesvdrop}.
We will also consider the possibility of using one pair of wires to
deliver a separate voltage to \dword{larasic} (\SI{2.0}{V} that will be reduced to
the required \SI{1.8}{V} on the \dword{femb}). This solution may
provide a better overall noise performance for the readout electronics,
but will have a slightly larger voltage drop on the cold cables.
The size of the cable bundles planned
for \dword{dune} represents a small reduction compared to that used
for \dword{pdsp}, where bundles of nine \SI{20}{AWG} twisted-pair wires
were used. Overall, the total resistance of the power return wires
are \SI{2}{\milli\ohm} and \SI{5}{\milli\ohm} for the upper and
lower \dword{apa}s, respectively, numbers that are reduced by 
a factor of three for operation in \dword{lar}. For each \dword{apa} pair, the total power 
dissipated inside the power cables ($\sim$\SI{11}{W} at \dword{lar} temperature)
is small compared to the total power dissipated in the \dwords{femb},
\SI{240}{W}.

\begin{dunetable}
[Voltage drop and power dissipation in the FEMBs power cables inside the cryostat]
{p{0.35\textwidth}p{0.15\textwidth}p{0.20\textwidth}p{0.20\textwidth}}
{tab:SPCE:cablesvdrop}
{Voltage drop and power dissipation in the cables bringing power to the \dwords{femb}
at room and at \dword{lar} temperature for the cable lengths corresponding to the upper (\SI{9}{m})
and the lower (\SI{22}{m}) \dword{apa}s. The \dwords{femb} require \SI{2.4}{A} at \SI{2.5}{V} to operate.
At room temperature, the resistances of the seven \SI{20}{AWG} twisted-pair wires are
$\SI{41}{\milli\ohm}$ and $\SI{101}{\milli\ohm}$ for the upper and the lower 
\dword{apa}s, respectively. These resistances are reduced to $\SI{14}{\milli\ohm}$ 
and $\SI{34}{\milli\ohm}$ inside the \dword{lar}.}
 & Voltage & Voltage drop &  Power dissipation  \\
WIB output (room temperature) & \SI{2.7}{V} / \SI{3.0}{V} & \SI{0.2}{V} / \SI{0.5}{V} & \SI{0.5}{W} / \SI{1.2}{W} \\ \colhline
WIB output (\dword{lar} temperature) & \SI{2.6}{V} / \SI{2.7}{V} & \SI{0.1}{V} / \SI{0.2}{V} & \SI{0.25}{W} / \SI{0.5}{W} \\
\end{dunetable}

The cable plant for one \dword{apa} in the \dword{lar} also includes
the cables that provide the bias voltages applied to the $X$-, $U$-, and $G$-plane
wire layers, three \dword{fc} terminations, and an electron diverter,
as shown in Figure~\ref{fig:CR-board}. The voltages are supplied
through eight \dword{shv} connectors mounted on the signal flange.
RG-316 coaxial cables carry the voltages from the signal flange to
a patch panel \dword{pcb} mounted on the top of the \dword{apa} that 
includes noise filtering. From there, wire bias voltages are carried by single wires to
various points on the \dword{apa} frame, including the \dword{cr}
boards, a small \dword{pcb} mounted on or near the patch panel that
houses a noise filter and termination circuits for the \dword{fc}
voltages, and a small board mounted near the electron diverter
that also houses the wire bias voltage filter described
in Section~\ref{sec:fdsp-tpcelec-design-bias}.

In Sections~\ref{sec:fdsp-apa-intfc-cables} and \ref{sec:fdsp-tpcelec-interfaces-apa}
we discuss the problem of routing the cold cables (data, control, power, and
bias voltages) for the bottom \dword{apa}s through the frames of both
the top and bottom \dword{apa}s. Routing tests were initially performed
with the \dword{pdsp} cable bundles, and even after increasing
the cross section of the side tubes from $\num{7.62}\times\SI{7.62}{cm^2}$ ($3"\times3"$)
to $\num{10.16}\times\SI{10.16}{cm^2}$ ($4"\times4"$), routing was difficult. After understanding that we
could reduce the number of cables, we ran a second set of tests with fewer sets
of cables (nine rather than ten sets of 12 data and control cables, nine rather than
ten sets of nine twisted-pair wires for power, and eight bias voltage cables as before).
This insertion test was successful once
a \SI{6.35}{cm} (2.5'') diameter conduit was inserted inside the
\dword{apa} frame to present a uniform cross section to
the cables and the cables were restrained with a mesh. These tests have been
successfully repeated in October 2019 at \dword{ashriver} using the setup with 
two stacked \dword{apa} frames, described in Section~\ref{sec:fdsp-apa-qa-prototyping}.

The cable plant discussed above for the reference design with the three \dword{asic}s
can be used also for \dwords{femb} populated with the
\dword{cryo} \dword{asic}. In that case, the power requirements are 
reduced (\SI{5}{W} at \SI{2.5}{V}), and the internal \dwords{ldo} do
not require an external bias line. Six of the eight \SI{20}{AWG} twisted-pair
wires could be used to bring the low-voltage power to the \dword{femb},
while the remaining two could be used to sense the voltage on the \dword{femb}.
This configuration would entail a small increase ($\sim14$\%) of the 
total resistance seen on the return wires. The same number of low-skew shielded
twin-axial cables are required to transmit the following differential signals
between the \dword{wib} and the \dword{femb}:
\begin{itemize}
\item four \SI{896}{Mbps} data lines (two from each \dword{cryo} \dword{asic});
\item two \SI{56}{MHz} clock signals (one for each \dword{cryo} \dword{asic}); and
\item four shared \dword{saci} signals.
\end{itemize}
In the current design of \dword{cryo} \dword{asic} a total of five \dword{saci} signals
are required: three of them are shared between the two \dword{asic}s on the
\dword{femb}, and two separate ones are required to send commands to the
two \dword{cryo} \dwords{asic}. We are planning to implement internal 
addresses in a future version of \dword{cryo}, such that the two \dwords{asic}
can share the command line.

The proposed cable plant is also compatible with the use of the \dword{cots}
\dword{adc}. The current design of the \dword{sbnd} \dword{femb} uses 
\num{12} low-skew shielded twin-axial cables, instead of ten, but some of
the signals are not used. The low-voltage power is transmitted with a 
bundle of nine \SI{20}{AWG} twisted-pair wires, but two of them are
used for the bias of the linear voltage regulators, which require very
little current.

In all possible configurations of the \dword{femb}, it is very likely that
the cable plant required to bring the low-voltage power and controls
to the \dwords{femb} and to read out the data from the \dwords{femb} is
compatible with the option of routing the cables through the \dword{apa}
frames. This, however, does not leave much room for building redundancy in
the system. The cable connections need to be extremely reliable because
the loss of one connection could result in an entire \dword{femb} becoming
unresponsive.

\subsection{Warm Interface Electronics}
\label{sec:fdsp-tpcelec-design-warm}

The warm interface electronics provide an interface between the 
\dword{ce}, \dword{daq}, timing, and slow control systems, including 
local power control at the flange and a real-time diagnostic readout. 
They are housed in the \dwords{wiec} attached directly to the \dword{ce} 
flange.  
A \dword{wiec}, shown in Figure~\ref{fig:tpcelec-flange}, 
contains one \dword{ptc}, 
 five 
 \dwords{wib} 
 and a passive \dword{ptb}
that fans out clock signals and \dword{lv} power from the \dword{ptc} to the 
\dwords{wib}. The \dword{wiec} must provide Faraday-shielded housing and 
robust ground connections from the \dwords{wib} to the detector ground 
(Section~\ref{sec:fdsp-tpcelec-design-grounding}). Only optical
connections are used for the communication to the \dword{daq} and the
slow controls, to avoid introducing noise in the \dword{ce} \fdth.

\begin{dunefigure}
[Exploded view of the cold electronics signal flange for ProtoDUNE-SP]
{fig:tpcelec-flange}
{Exploded view of the \dword{ce} signal flange for \dword{pdsp}.  
The design for the \dword{dune} \dword{spmod} \dword{ce} 
signal flange will be very similar (with two \dword{ce} signal flanges per \fdth).}
\includegraphics[width=0.9\linewidth]{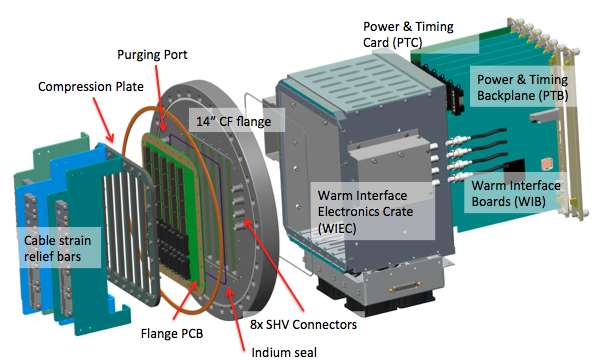}
\end{dunefigure}

The \dword{wib} 
receives the system clock and control signals from the
timing system and provides processing and further distribution of those signals to four
\dwords{femb}. 
It also receives high-speed data signals from the same four 
\dwords{femb} and transmits them to the \dword{daq} system over optical
fibers. The data signals from the \dwords{femb} are recovered on the \dword{wib} with commercial 
equalizers. The \dwords{wib} are attached directly to the \dword{tpc}
\dword{ce} \fdth on the signal flange. The \fdth board is a \dword{pcb} 
with connectors to the cold signal and \dword{lv} power cables fitted
between the compression plate on the cold side and sockets for
the \dword{wib} on the warm side. Cable strain relief for the cold cables is 
provided from the back end of the \fdth.

The \dword{ptc} provides a bidirectional fiber interface to the
timing system. The clock and data streams are separately fanned out to the 
five \dwords{wib} as shown in Figure~\ref{fig:tpcelec-wib-timing}. 
A clock-data separator on the 
\dword{wib} separates the signal received from the timing system into 
clock and data signals.
Timing endpoint firmware for receiving and transmitting 
the clock is integrated into the \dword{wib} \dword{fpga}.
The \dword{spmod} timing system, described in Section~\ref{sec:daq:design-timing}, 
is a further development of the \dword{pdsp} system and is expected to have nearly identical 
functionality at the \dword{wib} endpoint.

\begin{dunefigure}
[PTC and timing distribution to the WIB and FEMBs]
{fig:tpcelec-wib-timing}
{\Dword{ptc} and timing distribution to the \dword{wib} and \dwords{femb} used in \dword{pdsp}.
A similar design will be adopted for the \dword{dune} \dword{spmod}.}
\includegraphics[width=0.75\linewidth]{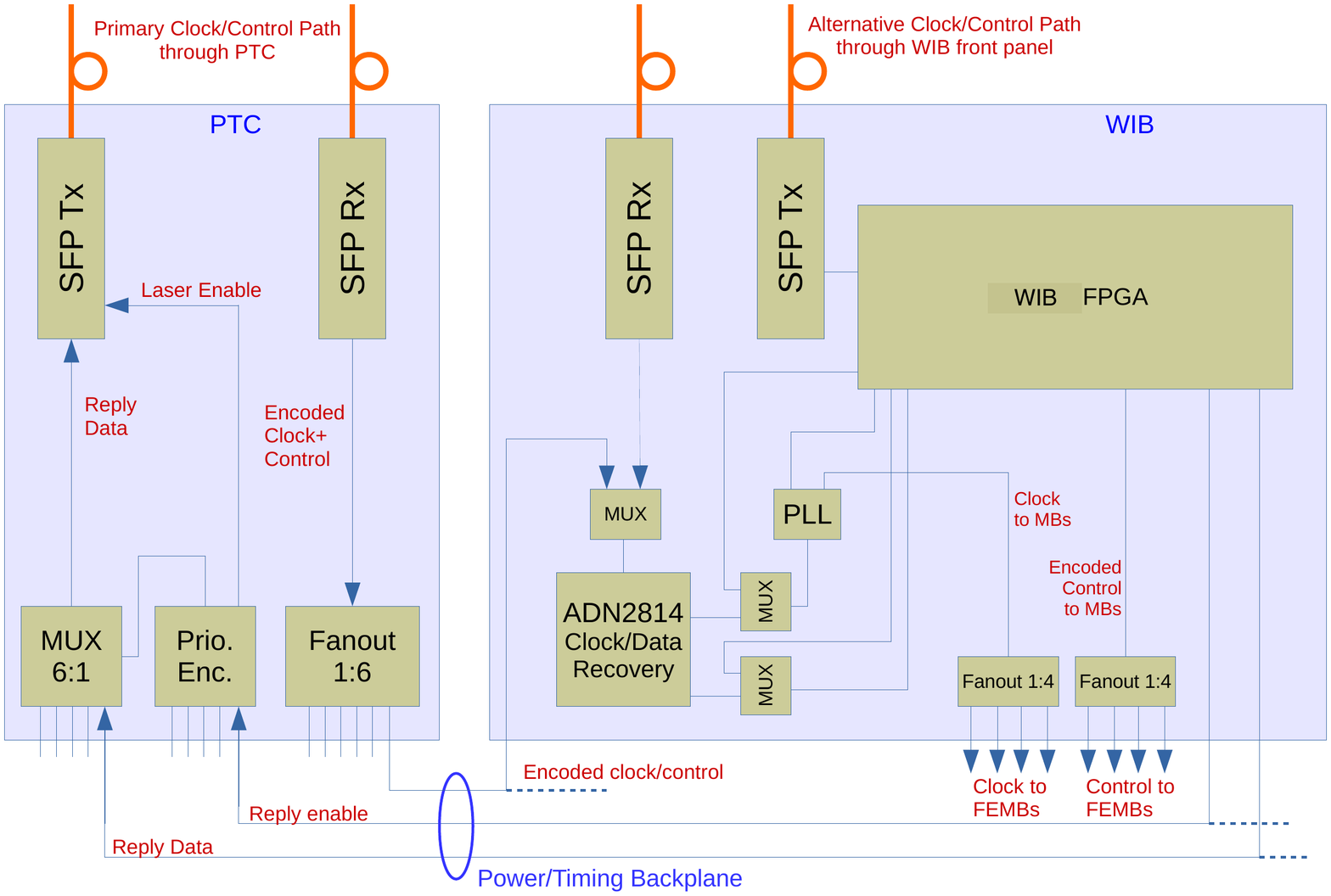}
\end{dunefigure}

The \dword{ptc} receives \SI{48}{V} \dword{lv} power for all \dword{tpc} 
electronics connected through the \dword{tpc} signal flange: 
one \dword{ptc}, five \dwords{wib}, and \num{20}~\dwords{femb}. 
The \dword{lv} power is then stepped down to \SI{12}{V} via 
a \dword{dc}-\dword{dc} converter on the \dword{ptc}. The output 
of the \dword{ptc} converters is filtered with a common-mode choke 
and fanned out on the \dword{ptb} to each \dword{wib}, which provides the 
necessary \SI{12}{V} \dword{dc}-\dword{dc} conversions and fans
the \dword{lv} power out to each of the  \dwords{femb} supplied 
by that \dword{wib}, as shown in Figure~\ref{fig:tpcelec-wib-power}. 
The output of the \dword{wib} converters is also filtered by a 
common-mode choke, and each voltage line provided to the \dwords{femb}
is individually controlled, regulated, and monitored.

\begin{dunefigure}
[Low voltage power distribution to the WIB and FEMBs]
{fig:tpcelec-wib-power}
{\dword{lv} power distribution to the \dword{wib} and \dwords{femb} 
for \dword{dune}. In the current design, up to four separate 
voltages can be provided from the \dword{wib} to each \dword{femb}.
Measurements with prototype \dwords{femb} will inform the final
design of the power distribution, and the number of different
voltages sent to the \dword{wib} will be chosen to reduce the
voltage drops along the cold cables and to minimize the readout
noise.}
\includegraphics[width=0.65\linewidth]{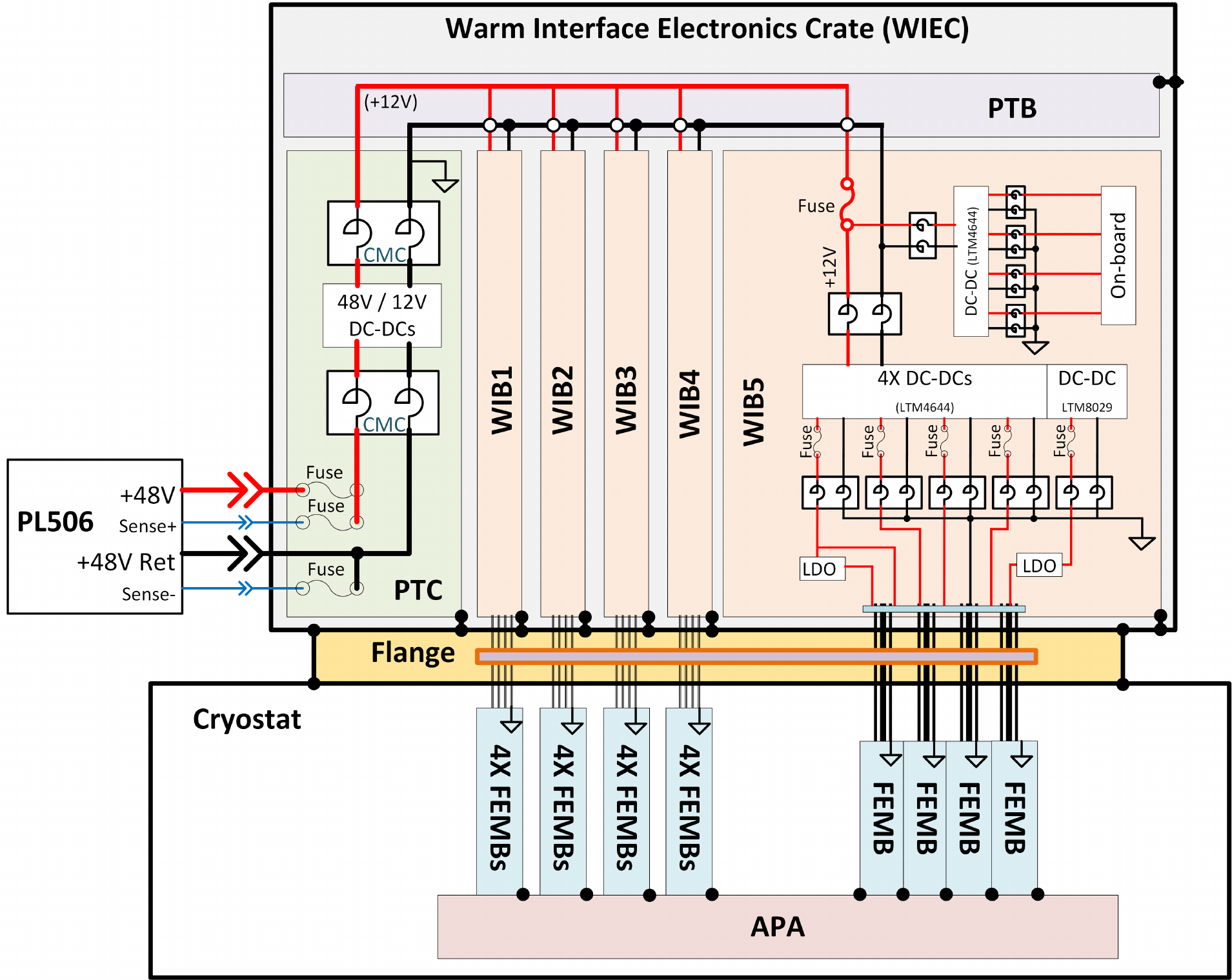}
\end{dunefigure}

Because the \dwords{wib} can provide local power to the \dword{femb} 
and real-time diagnostic readout of all channels, each \dword{tpc} electronics 
system for each \dword{apa} is a complete, stand-alone readout unit. 
The \dwords{femb} and cold cables are shielded inside the cryostat, 
and the \dwords{wib} and \dword{ptc} are shielded inside the Faraday 
cage of the \dword{wiec}, with only shielded power 
cables and optical fibers connecting to external systems.

\begin{dunefigure}
[Warm interface board]
{fig:tpcelec-dune-wib}
{\dword{wib} block diagram including the fiber optic connections to
the \dword{daq} backend, slow controls, and the timing system, as well as the
data readout, clock, and control signals to the \dwords{femb}.}
\includegraphics[width=0.8\linewidth]{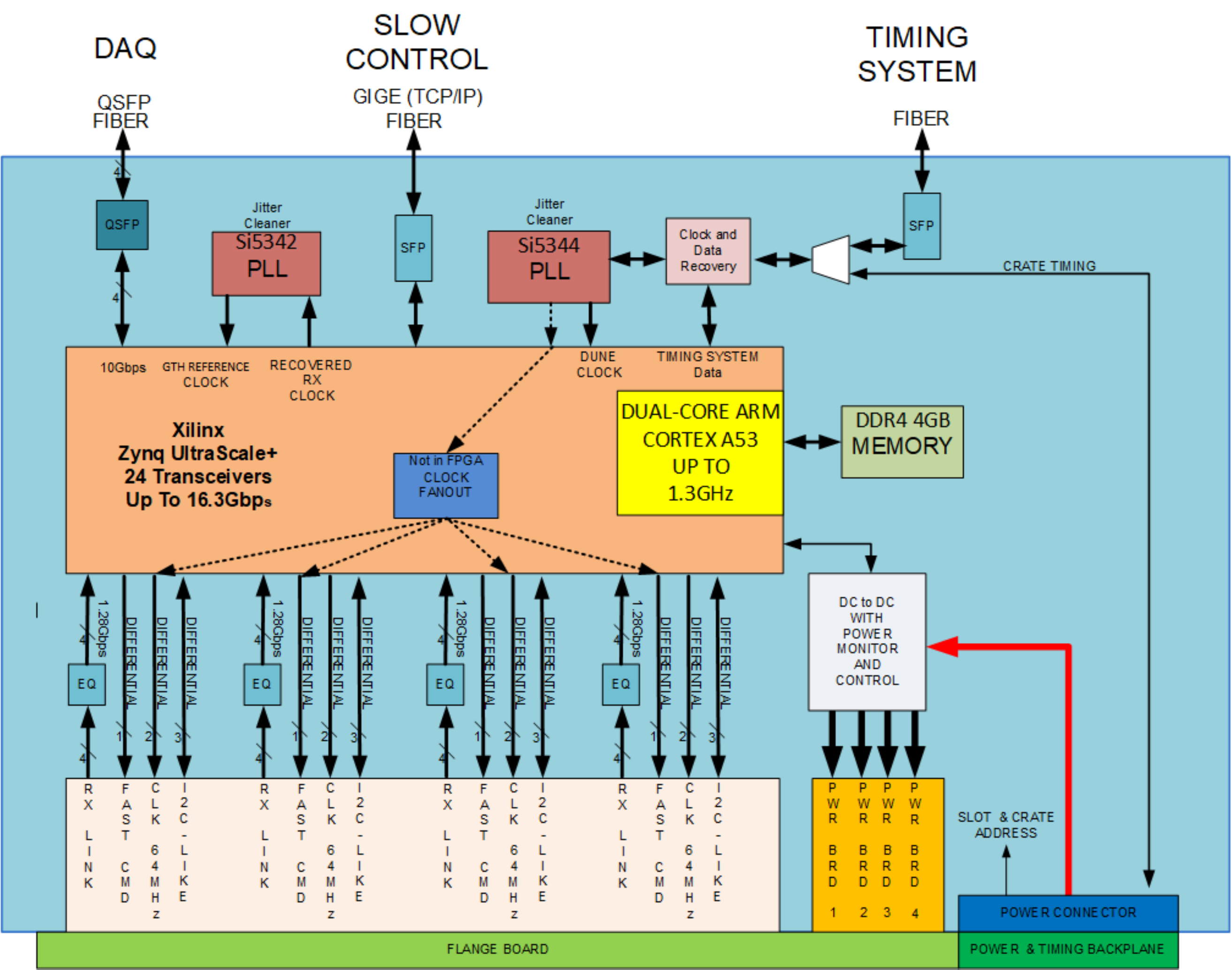}
\end{dunefigure}

As shown in Figure~\ref{fig:tpcelec-dune-wib}, the \dword{wib} can
receive the encoded timing signal over bidirectional optical
fibers on the front panel; it can then process them using either the on-board
\dword{fpga} or clock synthesizer chip to provide the clock required
by the \dword{tpc} electronics.
The reference \dword{asic} design 
currently uses 8b/10b encoding; if the \dword{slac} \dword{cryo} 
\dword{asic} is selected for the \dword{dune} \dword{spmod}, 
12b/14b encoding will be used instead of 8b/10b.

The \dword{fpga} on the \dword{wib} will have transceivers that can 
drive the high-speed data to the \dword{daq} system up to
\SI{10}{Gbps} per link, meaning that all data from
two \dwords{femb} (2$\times\SI{5}{Gbps}$) could be transmitted 
on a single link. The \dword{fpga} will have an additional 
transceiver I/O for an optical \SI{1}{Gbps} Ethernet connection, which 
provides real-time monitoring of the \dword{wib} status to the slow control system.

For system tests, discussed later in Section~\ref{sec:fdsp-tpcelec-qa-facilities},
the \dword{wiec}, \dword{wib}, and \dword{ptc} developed for \dword{pdsp}
are being used. A special version of the \dword{wib} has been developed
for use with \dwords{femb} equipped with the \dword{cryo} \dword{asic},
which require a different power and clock distribution scheme. Plans are
being put in place to redesign the \dword{wib}, and eventually make 
minor changes also to the \dword{ptc} and the \dword{wiec}, to use 
less expensive \dwords{fpga} and to be able to program independently
the voltage rails used to provide power to the \dwords{femb}. While in
the current \dword{wib} it is possible to monitor and turn on and off
these voltage rails independently, the redesign will include the
capability of setting independent voltage and current limits on each
one of them. In addition, it is planned to add the possibility of measuring
the delay of the propagation
of the clock signals between the \dword{wib} and the \dwords{femb}. With
this feature, it will be possible to align the sampling time of different
\dwords{femb} to a precision of a few ns in situ without relying on the 
measurement of the cable lengths, which was necessary
in \dword{pdsp}.

\subsection{Timing Distribution and Synchronization}
\label{sec:fdsp-tpcelec-design-timing}

The charge deposited on each wire of the \dword{apa}s installed 
in the \dword{dune} \dword{spmod} is digitized at a frequency of
$\SI{2}{MHz}$, as discussed in 
Section~\ref{sec:fdsp-tpcelec-overview-requirements}. This requires
that the \dword{tpc} electronics be synchronized to a level of the order of \SI{10}{ns}, which is
much smaller than the time difference between two charge samples.
This level of error in the synchronization between the sampling time of different
\dwords{femb} contributes negligibly to the expected resolution %
of the reconstructed space points measurement,
both in the \dword{apa} plane and along the drift distance.
The timing distribution and synchronization system for the 
\dword{spmod} is described in Section~\ref{sec:daq:design-timing}.
Each \dword{wiec} has a bidirectional optical connection with the
timing system in the \dword{ptc}. Inside the \dword{ptc} the optical
signal from the timing system is converted, as discussed in the previous
section, to an electrical signal and distributed via the backplane to
the \dwords{wib} that constitute an endpoint for the timing distribution
system. Each \dword{wib} contains a standalone jitter-reducing \dword{pll} 
that forwards the clock to all the \dwords{femb}. The
\dword{fpga} contained inside the \dword{wib} implements the 
protocol~\cite{bib:docdb1651,bib:docdb11233} for aligning
the phase of the clock at the endpoint of the distribution tree.

The timing distribution and synchronization system ensures that
all the \dwords{wib} are synchronized to within $\SI{3}{ns}$. 
One possible way of synchronizing the \dwords{femb} is the one
that was used in \dword{pdsp}, which relies on the fact  
that all the cables connecting the \dwords{wib} to the
\dwords{femb} have approximately the same length (a length 
difference of \SI{0.5}{m} corresponds to a difference in the
sampling time of \SI{2.5}{ns}). The same approach could be used
for the \dword{dune} \dword{spmod}, correcting for the top-bottom 
\dword{apa} cable length difference (corresponding to $\sim\SI{65}{ns}$) 
inside the \dword{fpga} of the \dword{wib}. The exact correction
factor could be obtained by measuring the time propagation 
difference for a sample of short and long cables prior to the 
installation of the \dwords{femb} on the \dword{apa}s. 
Synchronizing the \dword{ce} with the \dword{pds} requires one
additional time constants that correspond to the transit 
time of the fast command sent from the \dword{wib} to
\dword{coldata} and from there to the \dword{coldadc},
which includes the propagation time along the cables
(\SI{45}{ns} for the \SI{9}{m} long cables to the top \dword{apa}s
and \SI{110}{ns} for the \SI{22}{m} long cables to the bottom \dword{apa}s) 
plus the propagation time inside the \dwords{asic}. This overall 
time constant can be obtained offline from the data, but it represents 
at most a correction of $\mathcal{O}(\SI{150}{\mu m})$ on the 
position of a track along the drift distance.
Instead of relying on cable measurements, we are also considering
the addition of a timer inside the \dword{wib}'s \dword{fpga}
to measure the transit time of a command
sent to the \dword{femb} and its corresponding return
message. This study will help us understand whether the 
relative phases of the \dword{femb} and the \dword{wib} 
can be aligned more precisely.

The communication between the \dword{wib} and the \dword{daq} 
backend is asynchronous and the \num{64}-bit time-stamp, which
is used to indicate the time at which the signal waveform was
sampled~\ref{sec:daq:design-timing}, is inserted in the data frame
in the \dword{wib}'s \dword{fpga}. 
For the three-\dword{asic} solution, the communication
from the \dword{femb} to the \dword{wib} is also
asynchronous and a \num{8}-bit time stamp is sent to
count the number of \dword{adc} samples (triggered by
the \dword{femb} with a fast command) from the last
``Sync'' signal, as discussed in Section~\ref{sec:fdsp-tpcelec-design-femb-coldata}.
This \num{8}-bit time stamp is used only to ensure
that the \dword{femb} and the \dword{wib} are
still synchronized. 

In contrast to the three-\dword{asic} solution, 
in the \dword{cryo} solution 
the communication between the \dword{femb} and 
\dword{wib} is entirely synchronous, and instead uses a \SI{56}{MHz} 
clock. In order to properly align the phase of the
\dword{adc} sampling for the top and bottom \dword{apa}s,
appropriate delays must be added to the sampling
command in the \dword{wib}'s \dword{fpga}. 
The requirements listed above for synchronizing 
the \dword{ce} relative to the \dword{pds} remain
valid for this case. 

\subsection{Services on Top of the Cryostat}
\label{sec:fdsp-tpcelec-design-services}

Table~\ref{tab:fdsp-tpcelec-power} summarizes the power requirements of the
\dwords{femb}, \dwords{wib}, and \dwords{wiec}, which were discussed in
Sections~\ref{sec:fdsp-tpcelec-design-ft} and~\ref{sec:fdsp-tpcelec-design-warm}. As
shown in Figure~\ref{fig:tpcelec-wib-power}, each \dword{ptc} receives \SI{48}{V}
from a power supply installed on the top of the cryostat; this voltage is stepped down
via voltage regulators to  \SI{12}{V}, which is distributed to each \dword{wib}.
Inside each \dword{wib} the \SI{12}{V} is further reduced to the \num{2.7} or \SI{3.0}{V}
that is  used to power the \dwords{femb}, as discussed in Section~\ref{sec:fdsp-tpcelec-design-ft}.
For these estimates, an efficiency of \num{80}\% is assumed for each voltage
regulation step, while the power requirements for the \dword{fpga} and the
optical components on the \dwords{wib} are based on the measurements from \dword{pdsp}.

The overall power required for each \dword{wiec} is in the range \SIrange{335}{360}{W},
corresponding to the range \SIrange{7}{7.5}{A} at \SI{48}{V}. The \dword{lv} power is delivered
to the \dword{ptc} using a power mainframe that can operate in the \SIrange{30}{60}V range,
providing a maximum of \SI{13.5}{A} and \SI{650}{W} to each \dword{apa}. Using a \SI{10}{AWG}
cable, and assuming a distance of \SI{20}{m} between a \dword{lv} power supply
 on the detector mezzanine and the most distant cryostat penetration
for a row of \dword{apa}s, no voltage drop over \SI{1}{V} should occur along
the cable. At most $\sim\,\SI{150}{W}$ is dissipated inside the cryostat, and another
$\sim\,\SI{200}{W}$ is dissipated inside the (air-cooled) \dword{wiec};
only a few watts are dissipated in the warm cables located below the
false flooring on top of the cryostat. Multiplying by the total number of \dwords{wiec},
less than \SI{1}{kW} of power is dissipated in the cabling system over the entire
surface of the cryostat.

\begin{dunetable}
[Power requirements of FEMBs, WIBs, and WIECs]
{p{0.30\textwidth}p{0.15\textwidth}p{0.15\textwidth}p{0.15\textwidth}p{0.15\textwidth}}
{tab:fdsp-tpcelec-power}
{Power requirements for the \dwords{femb}, \dwords{wib}, and \dwords{ptc}.}
Component & \multicolumn{3}{c}{Current} & Power \\ \colhline
\multicolumn{5}{c}{FEMB (assume \num{80}\% efficiency in the \SI{12}{V} $\rightarrow$ \SI{2.7 / 3.0}{V} conversion)} \\ \colhline
Lower APA & \SI{2.4}{A} at \SI{2.7}{V} & \SI{0.68}{A} at \SI{12}{V} & & \\ \colhline
Upper APA & \SI{2.4}{A} at \SI{3.0}{V} & \SI{0.75}{A} at \SI{12}{V} & & \\ \colhline
\multicolumn{5}{c}{WIB (\num{4} FEMB $+$ FPGA, assume \num{80}\% efficiency in the \SI{48}{V} $\rightarrow$ \SI{12}{V} conversion)} \\ \colhline
\num{4} FEMBs lower APA & & \SI{2.7}{A} at \SI{12}{V} & & \\ \colhline
\num{4} FEMBs upper APA & & \SI{3.0}{A} at \SI{12}{V} & & \\ \colhline
FPGA and optical components & & \SI{1.7}{A} at \SI{12}{V} & & \\ \colhline
Total lower APA & & \SI{4.4}{A} at \SI{12}{V} & \SI{1.4}{A} at \SI{48}{V} & \SI{67}{W} \\ \colhline
Total upper APA & & \SI{4.7}{A} at \SI{12}{V} & \SI{1.5}{A} at \SI{48}{V} & \SI{72}{W} \\ \colhline
\multicolumn{5}{c}{WIEC (\num{5} WIBs $+$ PTC)} \\ \colhline
Total lower APA & & & \SI{7}{A} at \SI{48}{V} & \SI{335}{W} \\ \colhline
Total upper APA & & & \SI{7.5}{A} at \SI{48}{V} & \SI{360}{W} \\ 
\end{dunetable}

Four wires are used for each \dword{ptc} module; two \SI{10}{AWG}, shielded, twisted-pair 
cables for the power and return; and two \SI{20}{AWG}, shielded, twisted-pair 
cables for the sense. The primary protection is the over-current 
protection circuit in the \dword{lv} supply modules, which is set 
higher than the $\sim\SI{8}{A}$ current draw of the \dword{wiec}. 
Secondary sense line fusing is provided on the \dword{ptc}.  
Tests are being performed in \dword{pdsp} to check which is the
best scheme for connecting the shields of the power cables. In
\dword{pdsp} the shield of the warm power cables is connected to
ground on both the power supply side and on the \dword{ce} flange.
Other shield connection schemes are being investigated in \dword{pdsp}
and the connection scheme yielding the lowest readout noise will be
used for \dword{dune}.

Switching power supplies controlled by the slow controls system provide
power to the heaters (\SI{12}{V}) and the fans (\SI{24}{V}) that are 
installed on the \dword{ce} flanges. Temperature sensors mounted on the
flanges, and power consumption and speed controls from the fans are 
connected to the interlock system that is part of the \dword{ddss}, in
addition to being monitored by the slow controls system.

Bias voltages for the \dword{apa} wire planes, the electron diverters,
discussed in Section~\ref{sec:fdsp-apa-intfc-apa}, 
and the last \dword{fc} electrodes are generated by supplies that are 
the responsibility of the \dword{tpc} electronics consortium.  The 
current from each of these supplies should be very close to zero in 
normal operation. However, the ripple voltage must be carefully 
controlled to avoid injecting noise into the \dword{fe} electronics.  
RG-58 coaxial cables connect the wire bias voltages from the bias voltage
supply to the standard \dword{shv} connectors that are machined directly 
into the \dword{ce} \fdth and insulated from the low voltage and 
data connectors.

Optical fibers are used for all connections between the \dwords{wiec} 
and the \dword{daq} and slow 
control systems.  The \dword{wib} reports its temperature 
and the current draw from each \dword{femb} to the slow control system, 
while the current draw for each \dword{apa} is monitored at the 
mainframe itself.

To support the electronics, fan, and heater power cables, as well 
as optical fibers on top of the cryostat, cable trays are installed
below the false flooring on top of the cryostat. These cable trays
run perpendicular to the main axis of the cryostat and connect the
three cryostat penetrations for one row of \dword{apa}s to the detector
mezzanine near the cryostat roof, as shown in Figure~\ref{fig:cryostat-roof}.
All the necessary \dword{lv} supplies and 
the bias
voltage supplies are installed in these racks. Patch panels for
the optical fiber plant used for the control and readout of the
detector are also installed on the detector mezzanine.

\begin{dunefigure}
[Services on top of the cryostat]
{fig:cryostat-roof}
{Services on top of the cryostat. The racks for the \dword{lv} power supplies are shown in blue.}
\includegraphics[width=0.9\linewidth]{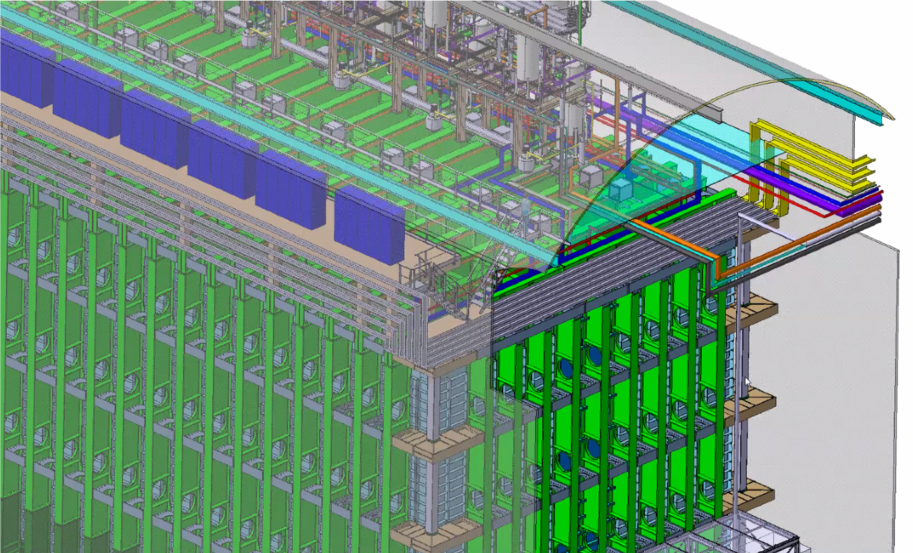}
\end{dunefigure}

\subsection{ProtoDUNE-SP Results}
\label{sec:fdsp-tpcelec-overview-pdune}

The \dword{pdsp} detector features a \dword{lartpc} with 15,360 sense wires
with a fiducial mass of roughly \num{700} tons of \lar. 
The system was deployed in a hadrons and electrons beamline at the CERN Neutrino Platform 
in 2018 and continues to take cosmic event data. 
The goal of 
the \dword{pdsp} \dword{tpc} readout was to validate the concept 
and the design of the integrated \dword{apa}+\dword{ce} readout 
and measure the performance of the \dword{tpc} electronics system with components 
as close as possible in design to those in the final \dword{dune} \dword{tpc} readout.
In the case of the \dword{tpc} electronics, most of the detector components 
used in \dword{pdsp} are prototypes of the \dword{dune} \dword{fd} ones discussed in the 
previous sections. The major difference is the \dword{femb} (and associated
\dwords{asic}), where an
early version of \dword{larasic} (P2) is used for the \dword{fe} \dword{asic}, followed
by the first prototype (P1) of a different \dword{adc}, using the ``domino'' architecture 
and implemented in the \SI{180}{nm} technology, and finally by
an \dword{fpga} that provided the data serialization functionality.

Each of the six \dword{pdsp} \dword{apa}+\dword{ce}
readout units consists of 2,560 sense wires, of which 960 are \SI{6}{m} 
long collection wires and 1,600 are \SI{7.4}{m} long induction wires. 
Five of the six \dword{apa}s were tested in a full-scale \coldbox in 
cold gaseous nitrogen (GN$_2$) with a complete \dword{tpc} electronics readout system,  
identical to the one 
deployed in \dword{pdsp}, before installation in the cryostat;
the sixth was installed without first going through the cold
box testing. Figure~\ref{fig:apa2-cycle} shows the measured noise level, 
represented by the \dword{enc} in units of 
electrons, for the collection ($X$) plane and the two induction ($V$, $U$) 
planes as well as the \dword{femb} temperature in the \coldbox as a 
function of the cold cycle time. At a stable temperature of 
\SI{160}{K}, the \dword{enc} for all three wire planes is less than 500~e$^-$.

\begin{dunefigure}
[ProtoDUNE-SP APA \#2 ENC levels measured in GN$_2$ in the CERN \coldbox]
{fig:apa2-cycle}
{Left $y$ axis: \dword{enc} (in electrons) for $U$, $V$, and $X$ (red, blue, and green 
curves) sense wire planes as a function of time (hours) for the APA 2 cold 
cycle in GN$_2$ in the CERN \coldbox; right $y$ axis: temperature 
(orange curve) measured at the level of the \dword{fe} electronics.}
\includegraphics[width=0.95\linewidth]{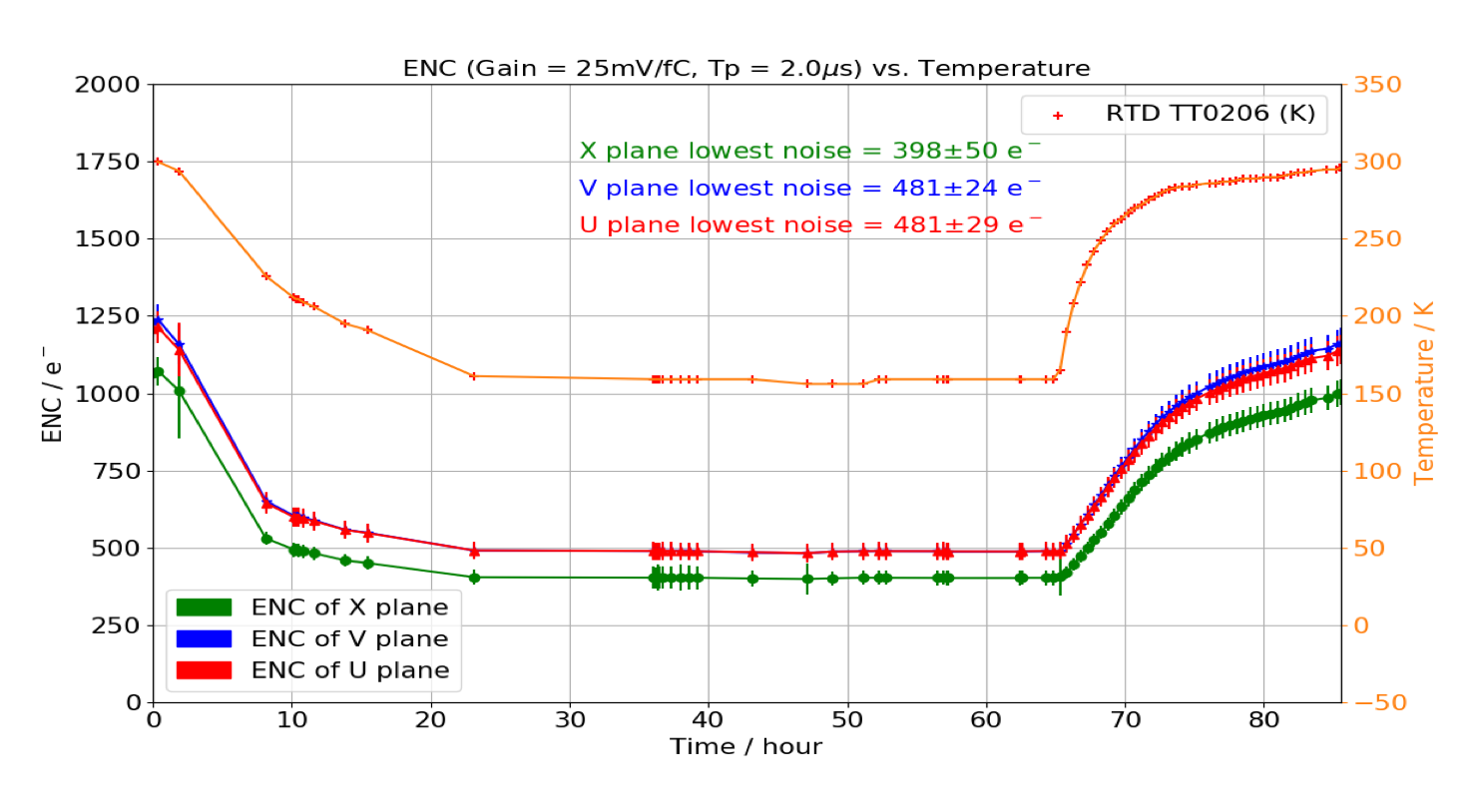}
\end{dunefigure}
After the cryostat was filled with \dword{lar} and the drift and wire 
bias voltages were set to their nominal values,
99.7\% of the \dword{tpc} readout channels were found to be functioning properly.
A total of 42 channels were found to be unresponsive. Of these:
\begin{itemize}
\item{14 channels were identified based on tests performed prior
to the insertion of the \dword{apa} into the cryostat as having
no capacitive load on the \dword{fe} electronics, suggesting an open 
connection somewhere upstream  of the \dword{ce} system;}
\item{24 additional channels showed the same problem after the
cryostat was filled with \dword{lar} (three of these, all on the
first APA, were already observed during testing in the \coldbox); and }
\item{four channels were associated with the \dword{fe} electronics
not functioning properly: two of these
channels appeared in tests performed after the cathode high
voltage was raised to \SI{120}{kV}, and two more
appeared when the high voltage reached \SI{160}{kV}.}
\end{itemize}

As discussed in Section~\ref{sec:fdsp-apa-qa-protodune-ops-dead-channels},
the number of disconnected channels 
due to mechanical failures in the connection between
the \dword{apa} wire and the \dword{fe} electronics has
changed with time, with some channels becoming again active,
and others becoming inactive. Only one additional dead channel
is caused by a permanent failure of the \dword{fe} electronics.
If these numbers are indicative
of the normal rate of channel loss, it would imply that
over the \dunelifetime of \dword{dune} operations at
most \num{0.5}\% of the readout channels would fail. A
similar upper limit can be obtained from the operational
experience of \dword{microboone}, considering also an
additional scale factor for the additional \dwords{asic}
immersed in \dword{lar} (in \dword{microboone} only
the \dword{fe} amplifier is in the liquid). Further 
operation of \dword{pdsp}, as well as operation of
\dword{sbnd} in the coming years, will provide additional
information on the long term stability of the active electronics
components immersed in \dword{lar}.

With the detector operating under nominal
conditions, the 
measured \dword{enc} by the online monitoring program was approximately $\SI{550}{e^-}$ 
on the collection wires and approximately $\SI{650}{e^-}$ on the induction
wires, averaged over all operational channels. The noise increased
relative to the tests performed inside the \coldbox due to the 
larger dielectric constant of \dword{lar} relative to GN$_2$. 
These noise measurements 
are consistent with the 
ratio of the corresponding capacitances of the \dword{apa} wires. 
Figure~\ref{fig:apa3-noise} 
shows the \dword{enc} (in electrons) for all channels of one 
\dword{apa}+\dword{ce} readout unit. 
The collection channels with \dword{enc} larger than \SI{1500}{e$^-$} had a problem 
in the P1-\dword{adc} \dword{asic}; this problem had already been identified 
prior to their
installation on the \dwords{femb}. The channels on all three planes 
with \dword{enc} smaller than \SI{300}{e$^-$} have an open connection somewhere in 
front of the \dword{ce} system. Figure~\ref{fig:ENC-all} summarizes
\dword{enc} levels in the entire \dword{pdsp} detector both before and
after the application of a simple offline common-mode noise filter similar
to the one used in \dword{microboone}~\cite{Acciarri:2017sde};
an improvement of roughly 100~e$^-$ is seen on all planes. 

\begin{dunefigure}
[TPC ENC levels measured at ProtoDUNE-SP after \lar fill]
{fig:apa3-noise}
{\dword{enc} (in electrons) for all $U$, $V$, and $X$ (red, blue, and green curves) sense 
wire planes for one \dword{pdsp} \dword{apa} 
under nominal operating 
conditions.}
\includegraphics[width=0.9\linewidth]{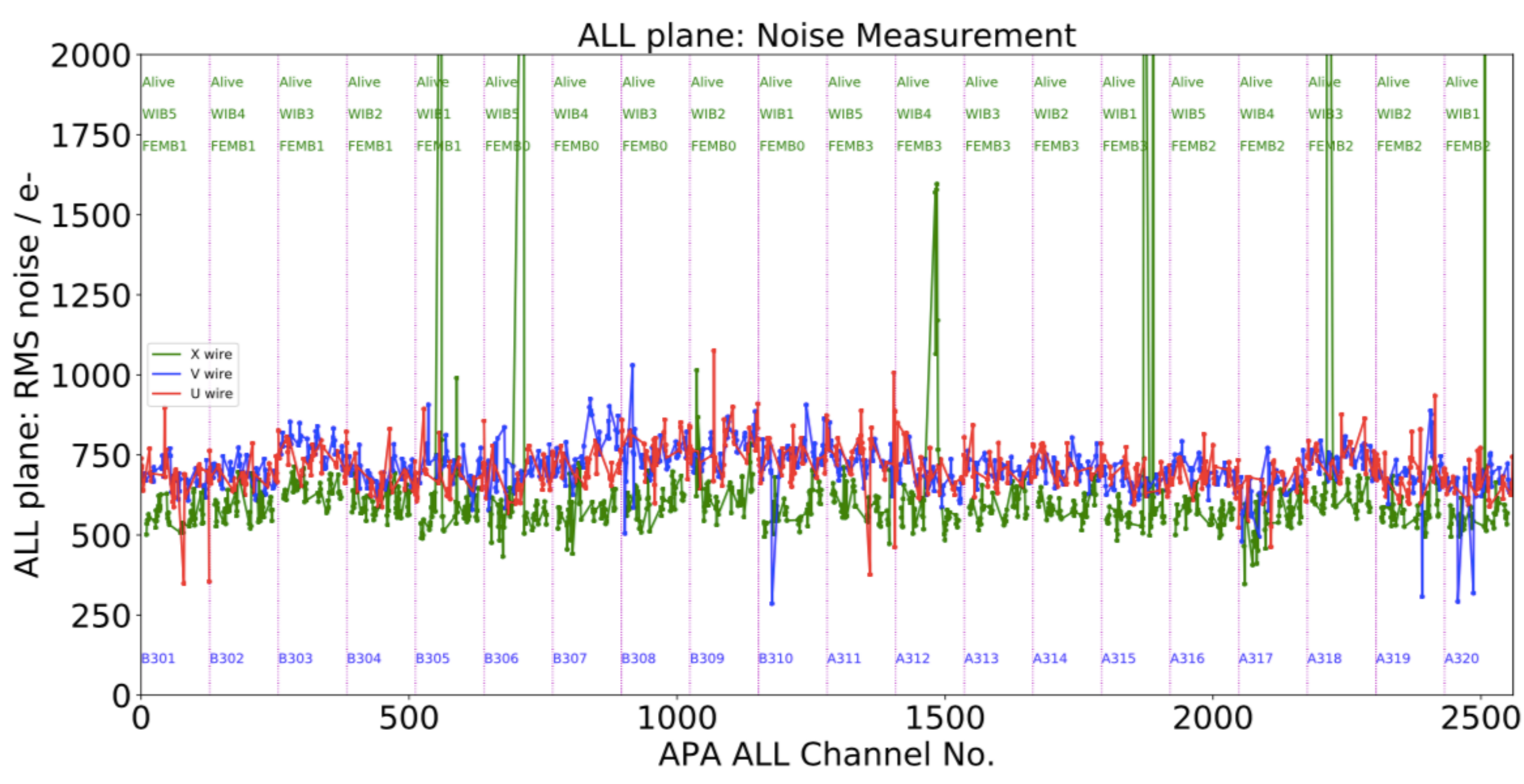}
\end{dunefigure}

\begin{dunefigure}
[TPC ENC levels for all channels of the ProtoDUNE-SP detector]
{fig:ENC-all}
{\dword{enc} levels (in electrons) for all channels of the \dword{pdsp} detector, both
before and after the application of a simple offline common-mode filter.}
\includegraphics[width=0.99\linewidth]{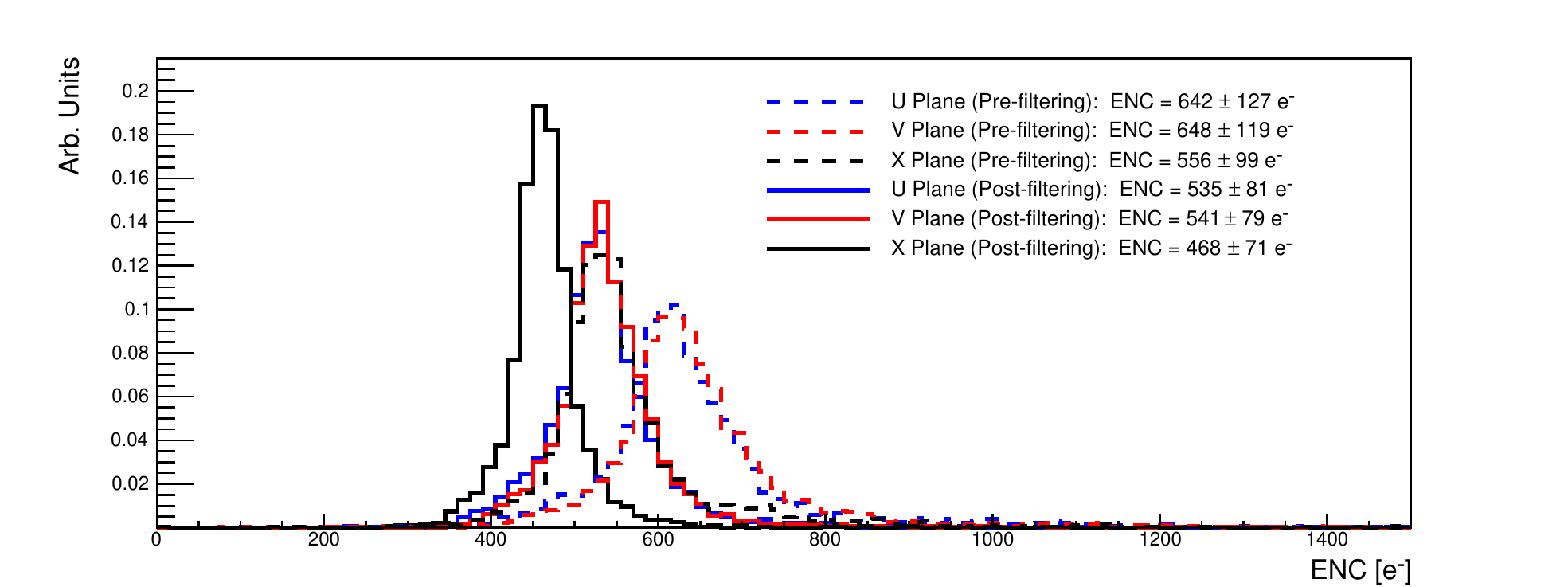}
\end{dunefigure}

The overall performance of the \dword{ce} system in 
\dword{pdsp} satisfies the 
\dword{ce} noise specification 
for the \dword{spmod} 
listed in Section~\ref{sec:fdsp-tpcelec-overview-requirements}. 
A comparison of the raw data from 
a \dword{pdsp} event (Figure~\ref{fig:pdsp-display}) to 
that from a \dword{microboone} event (Figure~\ref{fig:microboone-display}~\cite{Acciarri:2017sde}) 
demonstrates the improvements achieved in \dword{lartpc} performance. 
The \dword{pdsp} event was collected very early in the data taking
period when the charge collection efficiency was still limited
by the amount of impurities in the \dword{lar}; it shows very little
noise and appears to be of the same quality as the \dword{microboone}
event display after offline noise removal. 

\begin{dunefigure}
[Raw data from a ProtoDUNE-SP event]
{fig:pdsp-display}
{Display of the charge deposited on the collection wires (wire number on the $x$-axis) as
a function of the drift time (on the $y$-axis) for a \dword{pdsp} event 
that includes two electromagnetic showers and a four tracks in the
final state of the interaction.
The color associated with each time sample on the \dword{apa}
wires gives a measurement of the charge measured by the \dword{ce}
readout, with blue representing the smallest charge values and
red representing the largest charge values.}
\includegraphics[width=1.0\linewidth]{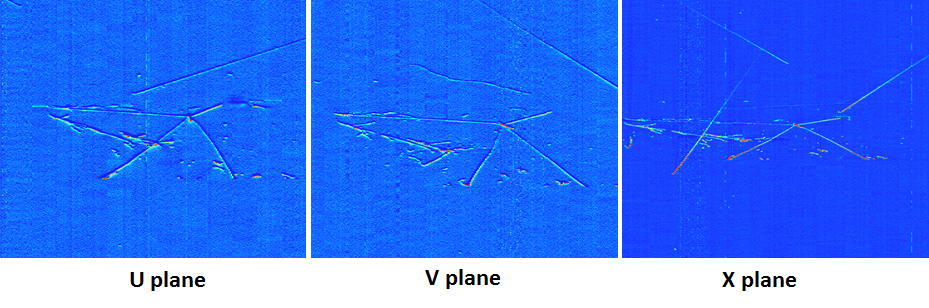}
\end{dunefigure}

\begin{dunefigure}
[Raw data from a MicroBooNE event]
{fig:microboone-display}
{\dword{microboone} \twod event display of the V plane from run 3493 
event 41075 showing the raw signal (a) before and (b) after offline 
noise filtering. A clean event signature is recovered once all of the 
identified noise sources are subtracted~\cite{Acciarri:2017sde}.}
\includegraphics[width=0.9\linewidth]{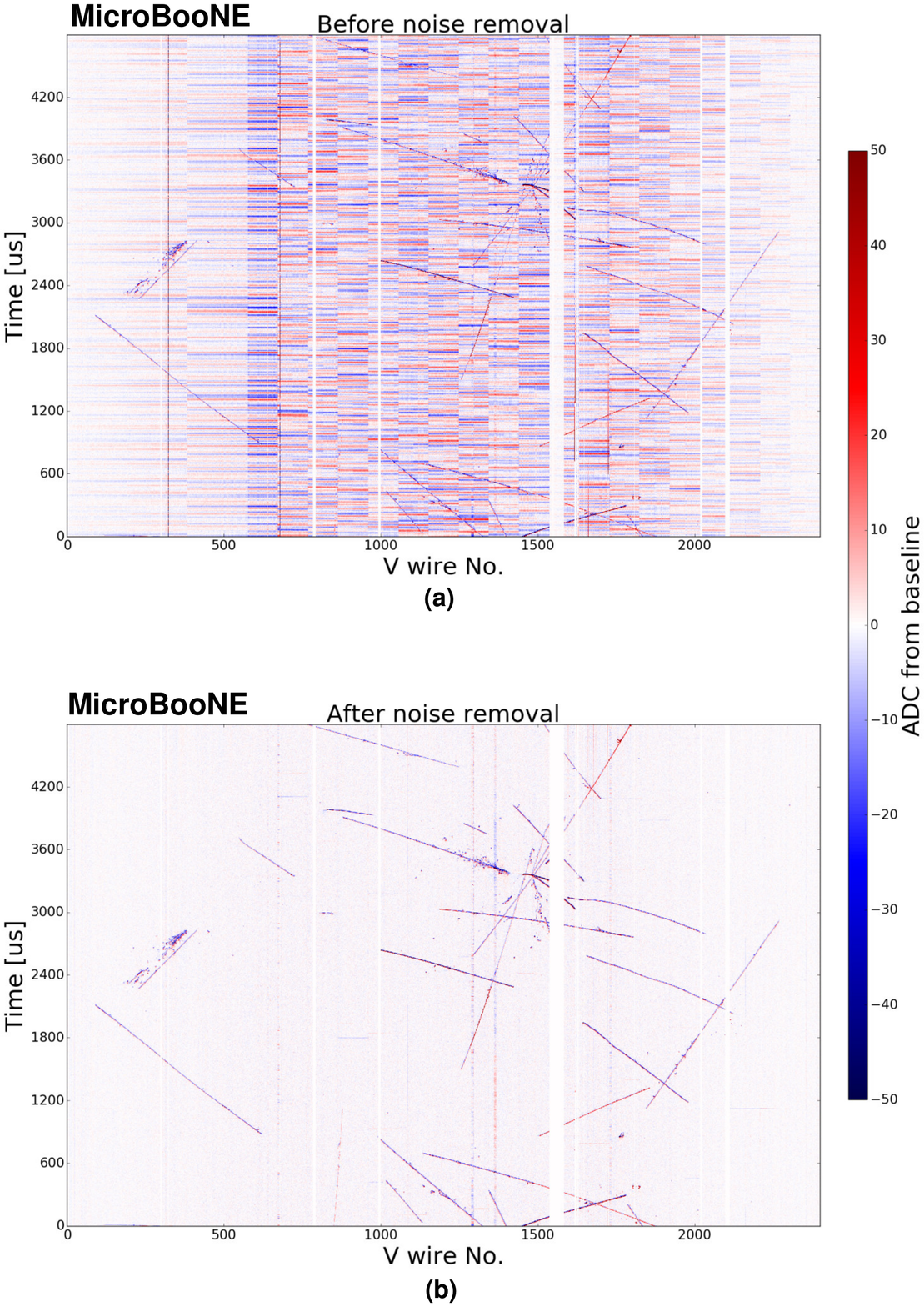}
\end{dunefigure}

\begin{dunefigure}
[Signal over noise for cosmic muon tracks reconstructed in ProtoDUNE-SP]
{fig:pdsp-signalovernoise}
{Angle-corrected peak signal-to-noise ratio for reconstructed cosmic muon tracks in
 \dword{pdsp}, both before and after noise filtering is applied~\cite{pend_PDSP_PerfPaper}.}
\includegraphics[width=0.85\linewidth]{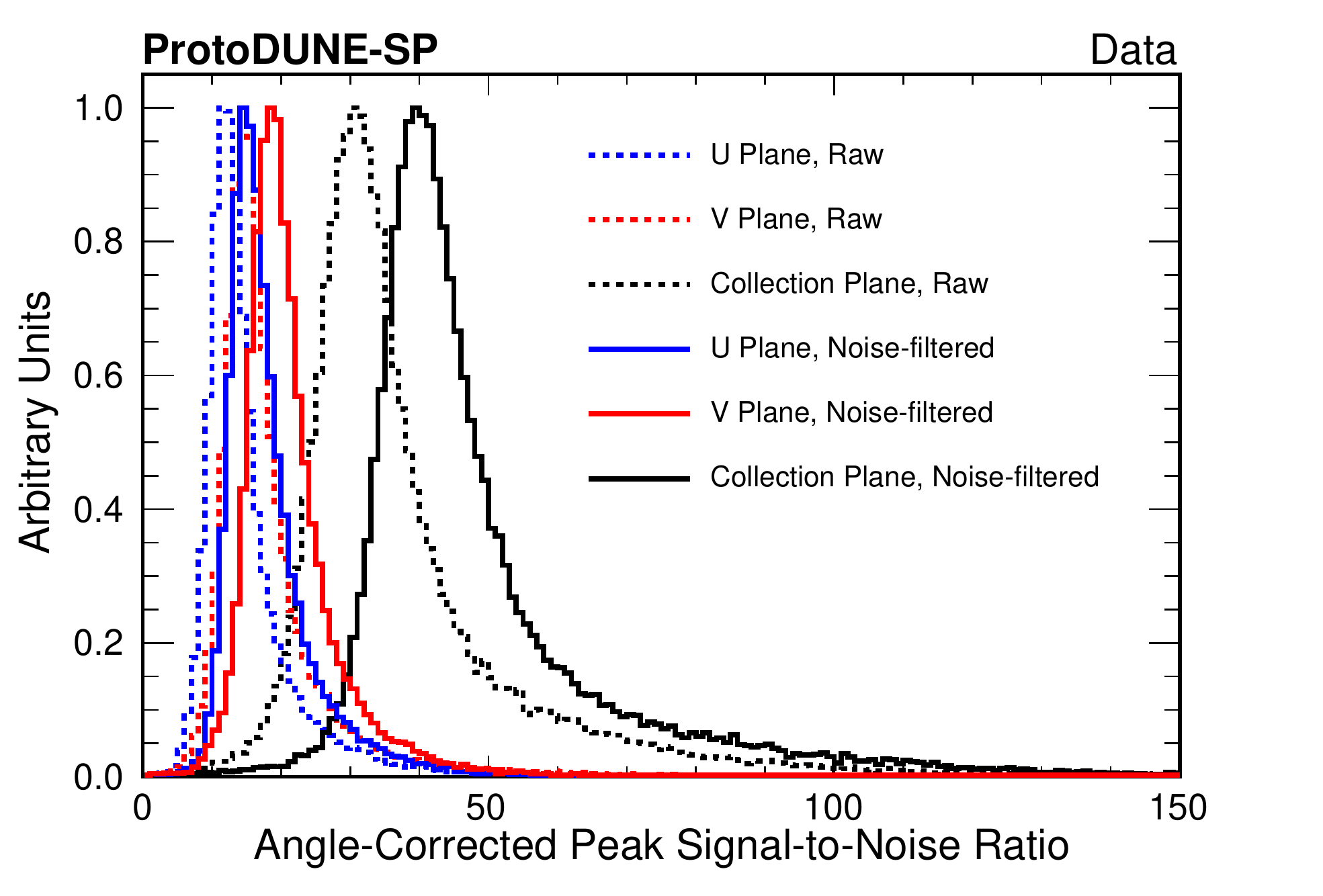}
\end{dunefigure}

The \dword{snr} has been evaluated using a selected cosmic muon sample, with tracks
crossing the \dword{lar} volume at shallow angle with respect to the anode plane and
large angle with respect to the direction of the wires in each plane considered
for \dword{snr} characterization. The charge deposited on each wire in a given plane is evaluated
using the pulse height (peak) of the hit found in the raw waveform.
A correction taking into account different relative angle between track and wire
direction has been applied to normalize the hit response. The noise value is extracted
from the width of the Gaussian fitted on the pedestal distribution of the waveform baseline.
The electric field in the \dword{tpc} volume was at the nominal level of \SI{500}{V/cm},
and the \dword{lar} purity for the runs considered in this analysis was about \SI{5.5}{ms}
as measured by the purity monitors, corresponding to $\sim35$\% charge loss due to
attachment for tracks close to the cathode. This measurement ignores the effect of the
space charge, which introduces distortions of the electric field in the \dword{tpc} volume that
may locally change the recombination factor and therefore affect the \dword{snr} value. Finally,
the measurement is made both before and after the application of a simple offline common-mode noise filter.
The distribution of the \dword{snr} for all the wires in the sample of muon tracks considered is
shown in Figure~\ref{fig:pdsp-signalovernoise}~\cite{pend_PDSP_PerfPaper}. Looking before (after)
the application of the common-mode noise filter, for the collection plane the mean value of the
\dword{snr} distribution is 38 (49), for the first induction plane it is 16 (18), and for the
second induction plane it is 19 (21). 

\subsection{ProtoDUNE-SP Lessons Learned}
\label{sec:fdsp-tpcelec-overview-lessons}

As discussed in Section~\ref{sec:fdsp-tpcelec-overview-pdune}, the initial 
data from \dword{pdsp} show that the \dword{spmod} can meet the noise specification. 
The experience with the \dword{tpc} electronics in \dword{pdsp} nonetheless motivates 
several improvements to the \dword{tpc} electronics system design, some of which
have already been implemented and discussed in the previous sections.
A complete list of the lessons learned from the construction, testing, integration,
installation, commissioning of the \dword{tpc} electronics detector components
is available~\cite{bib:docdb12367}. This reference also discusses
the plans and timeline for addressing the issues observed in \dword{pdsp}. 
This \dword{tdr} section and the following cover only the main issues 
and the plans for their resolution and implementation in the \dword{spmod}. 

During the commissioning of \dword{pdsp}, violations of the
grounding rules described in Section~\ref{sec:fdsp-tpcelec-design-grounding}
have been observed with one of the readout
boards for the \dword{pds} and with the cameras immersed
inside the \lar. The power supply used to provide the \dword{hv} to the 
cathode plane has also been observed
to cause noise inside the detector and has been replaced.
The overall success of \dword{pdsp} owes much to the fact that the
grounding rules were properly implemented
and that any violation was discovered and addressed during
the detector commissioning.

The main problem with the \dword{pdsp} \dword{tpc} electronics readout is 
the poor performance of the 
P1-\dword{adc} \dwords{asic}. This problem was observed as early as 2017 
while these \dwords{asic} were being tested prior to their installation
on the \dwords{femb}. The ``domino'' architecture~\cite{dominoADC} used in this design relies on
excellent transistor matching, which unfortunately 
is worse at \dword{lar} temperature. In \dword{pdsp}, this problem
results in a fraction (about \num{3.2}\%) of the readout channels having a fixed value for some of
the \dword{adc} bits, independent of the input voltage. In a majority
of the cases an approximate value for the charge can be obtained via
interpolation. For about \num{0.9}\% of the channels, the problem is
so severe that the only solution is to remove the channels from the
analysis, resulting in a loss of efficiency. This problem prompted us to abandon
this design and to develop the completely new \dword{coldadc}, to
adapt the \dword{cryo} \dword{asic} for use in \dword{dune}, and to
follow the approach of the \dword{sbnd} collaboration and consider
the \dword{cots} \dword{adc} option as well.

Initial analysis of the \dword{pdsp} data has uncovered a new problem with \dword{larasic} 
that occurs when more than \SI{50}{fC} is collected over a period of \SIrange{10}{50}{$\mu$s}
and the baseline configuration of the amplifier for the collection wires is used. 
The feedback mechanism of the \dword{fe} amplifier stops working for several 
hundred $\mu$s. During this period, the readout does not function and signals 
following the large charge deposited can be completely lost. A ledge is observed 
in the output of the \dword{fe} amplifier, followed by a slow decay 
and a sudden turn-on of the amplifier.
Figure~\ref{fig:pdsp-ledge} shows an example of this behavior.

\begin{dunefigure}
[Pulse shape on a ProtoDUNE-SP wire showing the ledge effect]
{fig:pdsp-ledge}
{Waveform of a channel showing a ledge following significant charge
deposition on the wire, followed by a discharge and a subsequent jump 
to the normal baseline \dword{adc} value.}
\includegraphics[width=1.0\linewidth]{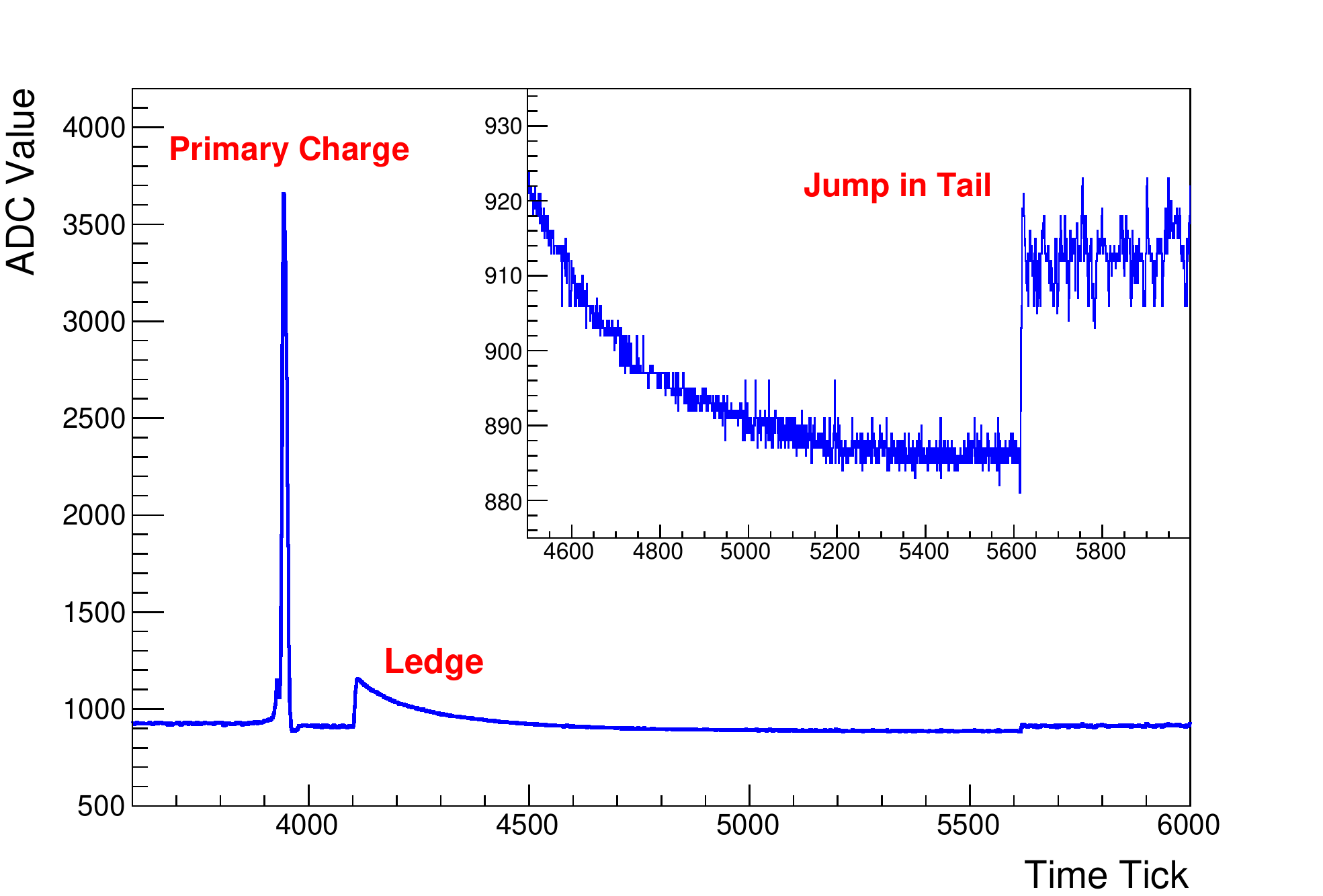}
\end{dunefigure}

This problem has been reproduced in the laboratory and is being actively 
studied. It affects all versions of \dword{larasic} fabricated after 
the one used for the \dword{microboone} experiment. The problem occurs 
when the threshold on the injected charge is small and therefore affects
with larger probability the collection wires, where the \SI{200}{mV} baseline 
is used, compared to the induction wires, which have a \SI{900}{mV} baseline.
After the problem and this difference between the two baselines were 
observed, the decision was taken to operate the \dword{ce} in \dword{pdsp} 
using the \SI{900}{mV} baseline for the collection wires as well, sacrificing
the dynamic range. Data from the wires where the problem occurs can
be masked in analysis, resulting in a loss of efficiency. This problem 
affected a very small fraction of the events: with the \SI{200}{mV}
baseline about \num{0.1}\% of the waveforms were affected, and this
number became almost completely negligible after switching to the 
\SI{900}{mV} baseline. It should be
noted that the problem occurs more often in \dword{pdsp} 
than is expected in the \dword{dune} \dword{spmod} due to the presence of 
cosmic rays traveling parallel to the \dword{apa} wires.
The problem could, however, affect the \dword{spmod}'s ability to detect
electromagnetic showers -- one of the main physics signals.
Section~\ref{sec:fdsp-tpcelec-overview-remaining} discusses the plans and timeline 
for addressing this issue in a new \dword{larasic} prototype.

\begin{dunefigure}
[Image of a connector for the cold cables lifted from the FEMB]
{fig:pdsp-femb-connector}
{Image of a connector for the cold readout and signal cables, which has been lifted from
the \dword{femb} due to the presence of excess epoxy on the 
connection between the cold cables and the printed circuit board
that acts as the ``male'' part of the connector.}
\includegraphics[width=0.85\linewidth]{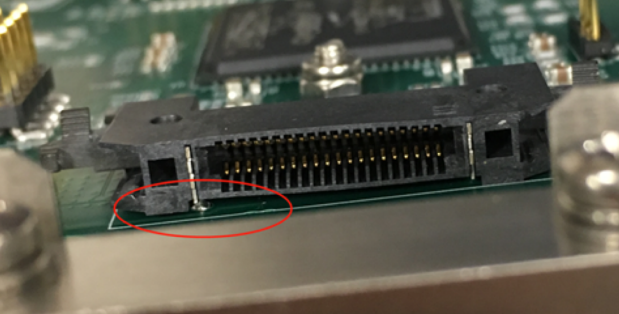}
\end{dunefigure}

During the integration of the \dwords{femb} onto the \dword{apa}s 
and the cold tests that preceded the \dword{apa} installation
inside the \dword{pdsp} cryostat,  multiple connectors  
detached from
the \dwords{femb}, causing a loss of communication.  
We replaced the \dwords{femb} 
on all of the \dword{apa}s that had been tested in the \coldbox. 
One additional \dword{femb} was 
replaced on the \dword{apa} that had been installed without undergoing 
the test in the \coldbox. This 
detachment may also be the cause
of the loss of the external clock signal on one of the \dwords{femb}
that was observed after \cooldown. 
The problem 
with the connector has been traced to a mechanical interference between 
the \dword{pcb} of the \dword{femb} and the epoxy deposited as a protective measure on the small printed circuit
board to which the cold cables are soldered and which forms the 
male part of the connector. The height of the epoxy can cause 
the female part of the connector to lift from the \dword{pcb},
as shown in 
Figure~\ref{fig:pdsp-femb-connector}. 
Section~\ref{sec:fdsp-tpcelec-design-femb} discusses the redesign of the connection to address this problem.

\begin{dunefigure}
[Spectrum of the noise on the ProtoDUNE-SP APA wires]
{fig:pdsp-noise}
{Spectrum of the noise on the first induction plane of the \dword{pdsp} \dword{apa}s before
and after applying a simple offline common-mode filter and partially mitigating \dword{adc} issues in software~\cite{pend_PDSP_PerfPaper}.}
\includegraphics[width=0.7\linewidth]{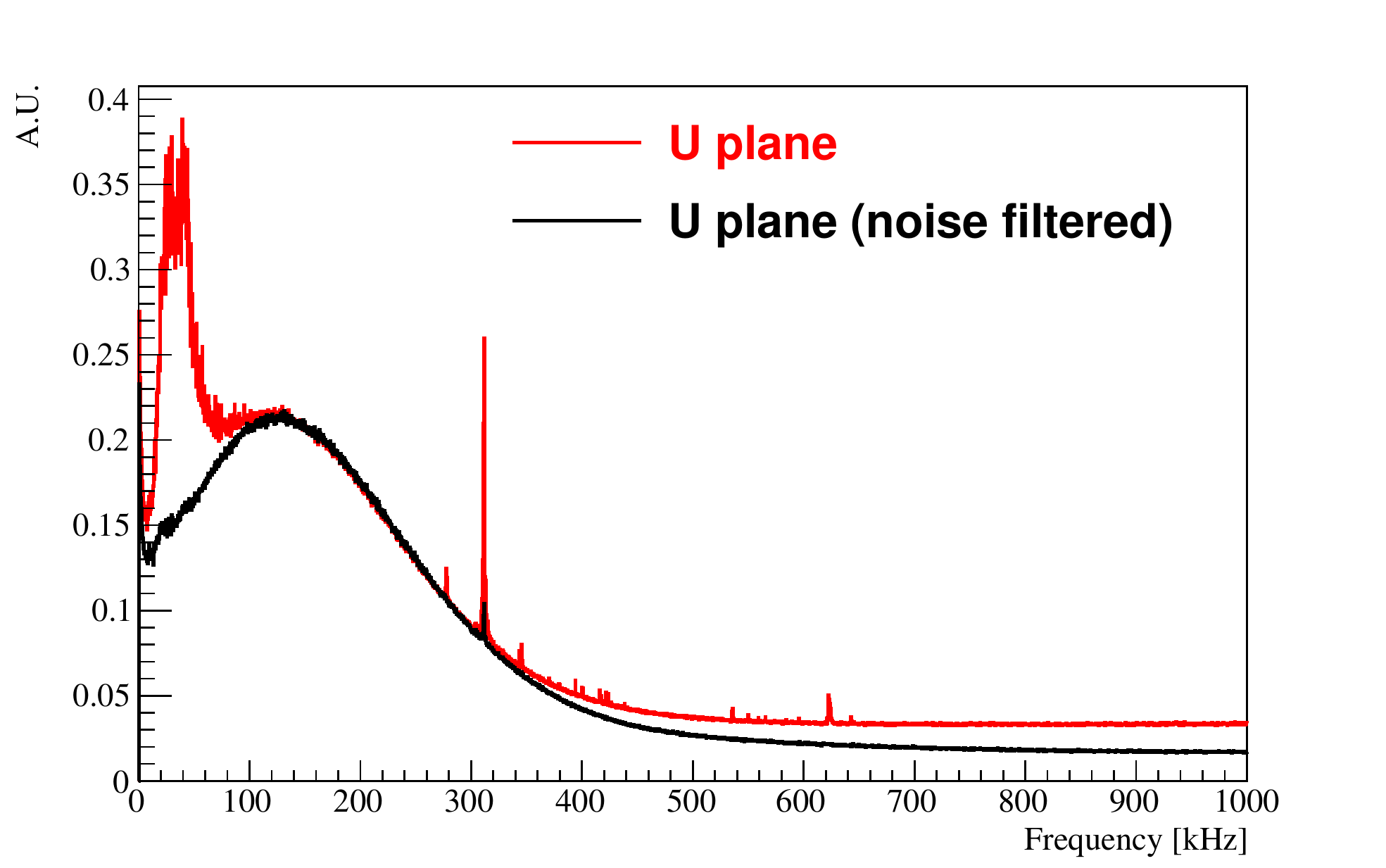}
\end{dunefigure}

Analysis of the \dword{pdsp} data has made significant progress throughout 2019,
leading to many insights on the detector behavior and on the interactions
between its different components. The level of noise mentioned in Section~\ref{sec:fdsp-tpcelec-overview-pdune}
(approximately $\SI{550}{e^-}$ on the collection wires,
and approximately $\SI{650}{e^-}$ on the induction wires) 
was measured with raw data, without any
filtering or selection applied to the pulses on the \dword{apa}
wires. A simple offline common-mode filter can significantly reduce the 
noise, particularly at low frequencies, as shown in Figure~\ref{fig:pdsp-noise},  
which compares the noise spectrum of the first induction plane of the \dword{apa}
before and after the filtering~\cite{pend_PDSP_PerfPaper}. The spectrum prior to the
filtering shows a significant increase at frequencies smaller than \SI{60}{kHz} that in
\dword{microboone} had been associated with the low-voltage regulators that are installed
on the \dwords{femb}~\cite{Acciarri:2017sde}. This contribution to the noise
has been significantly reduced compared to initial observations at 
\dword{microboone} by means of \dword{r-c} filters that have been added
on the \dword{pdsp} \dwords{femb}. Further work is required to
understand why these \dword{r-c} filters do not completely suppress this
specific noise source, as indicated by tests performed in other setups.
The noise spectrum prior to the filtering also shows spikes at multiple 
discrete frequencies, and in some cases the associated noise sources have been 
identified: for example, the operation of cameras inside the cryostat
contributes to the peaks at \SI{310}{kHz} and \SI{630}{kHz};
malfunctioning bias voltage supplies also contribute to the noise. Finally,
the reduced level of white noise in the spectrum at higher frequencies is
due to the application of an algorithm to partially recover from the \dword{adc}
problems described above~\cite{pend_PDSP_PerfPaper}. We
expect that further data analysis and tests with \dword{pdsp}
will result in improvements to the \dword{tpc} electronics design that
already demonstrates excellent performance.

\subsection{Remaining Design and Prototyping Tasks}
\label{sec:fdsp-tpcelec-overview-remaining}

 \dword{pdsp} was 
 built with multiple
goals, one of which was to demonstrate that the specifications
for \dword{dune} could be met with a design that would only require a simple
scale up of the detector size. The data collected with \dword{pdsp}
in fall 2018 has demonstrated that noise levels well below the target
of \SI{1000}{e$^-$} can be achieved in \lar, validating the
detector system design approach planned for the \dword{dune} \dword{spmod}.

Still, additional design and prototyping work is required in several areas before 
the start of \dword{spmod} construction, with differing levels of risk and engineering work, as estimated in Table~\ref{tab:SPCE:designstatus}. 

For example, changing the number and arrangement of cryostat penetrations to accommodate 
 two \dword{ce} flanges in addition to the \dword{pds}
flange can be considered a relatively minimal modification. It may require some structural 
reinforcement and additional \dword{fea} simulations to estimate the proper flow of
argon in order to avoid any back-diffusion of oxygen into the cryostat (in case
of leaks on the flanges), as well as to ensure an acceptable temperature gradient in the \lar.
On the other hand, changes in the design of the \dwords{asic}, with the
development of \dword{coldadc} and \dword{coldata}, are more involved. 

\begin{dunetable}
[Status of the design of the different CE detector components]
{p{0.27\textwidth}p{0.15\textwidth}p{0.50\textwidth}}
{tab:SPCE:designstatus}
{Status of the design of the different CE detector components as well as the expected
amount of engineering and prototyping required prior to construction.}
Component & Status & Expected work \\ \toprowrule
\dword{larasic} & Advanced & Fix issues observed in \dword{pdsp}, port differential output from \dword{coldadc} design \\ \colhline
\dword{cots} \dword{adc} & Complete & None \\ \colhline
\dword{coldadc} & \multicolumn{2}{l}{See text for details} \\ \colhline
\dword{coldata} & \multicolumn{2}{l}{See text for details} \\ \colhline
\dword{cryo} & \multicolumn{2}{l}{See text for details} \\ \colhline
\dword{femb} & Advanced & Experience with multiple prototypes, final design will follow the \dword{asic} selection \\ \colhline
Cold cables & Very advanced & Minor modifications, additional vendor qualification \\ \colhline
Cryostat penetrations & Advanced & Add \dword{ce} flange for bottom \dword{apa} \\ \colhline
\dword{wiec} & Very advanced & Add air filters and hardware interlock system \\ \colhline
\dword{wib} & Advanced & Update design to use cheaper FPGA, modify \dword{femb} power, new firmware \\ \colhline
\dword{ptc} & Very advanced & Add interface to interlock system \\ \colhline
Power supplies & Very advanced & Investigate possible additional vendors, rack arrangement \\ \colhline
Warm cables & Very advanced & Finalize cable layout, identify vendors \\ \colhline
Readout and control fiber plant & Very advanced & Finalize plant layout \\
\end{dunetable}

The area that requires most work is that of the \dword{asic}s that are mounted
on the \dwords{femb}. \dword{larasic} has already gone through eight design
iterations, the last three directly targeted for \dword{dune}, and has already been used 
(in two of its earlier versions) for \dword{microboone} and for \dword{pdsp}, where it has reached the 
noise levels specified for the \dword{dune} \dword{spmod}. At least one additional design iteration is
required to address the issues observed during \dword{pdsp} operations and
to implement a single-ended to differential converter to improve the interface
with the newly developed \dword{coldadc}. To ensure the success of the next 
design iteration, we are investing in the development of appropriate transistor
models for the \SI{180}{nm} \dword{cmos} technology for operation in \lar, such that the saturation effect observed in 
\dword{pdsp} can be properly addressed first in simulation and then with improvements 
in design. It should be noted that, so far, approximate models that were originally developed for 
the same \SI{180}{nm} technology (but with different design rules) have been used for 
the \dword{larasic} development, and therefore it should not be a surprise
that \dword{larasic} may have limitations in certain cases.
The circuitry for the single-ended to differential converter has already been 
developed in the \SI{65}{nm} technology and needs to be ported to the \SI{180}{nm}
technology used for \dword{larasic}. Various measures have been
put in place to minimize the risk associated with the need of a further
prototyping iteration; nevertheless, in Section~\ref{sec:fdsp-tpcelec-risks-design}
we consider a generic risk for a delay in the availability of \dword{asic}s and
argue that this delay would not have an impact on the beginning of \dword{dune} \dword{fd} operations.

It should be noted that even if the reference design for the \dword{spmod} makes use
of custom \dwords{asic} for the \dword{adc} and the data serialization functionality, 
a solution based on commercial components is available and has been 
demonstrated to work by the \dword{sbnd} collaboration. This solution is based on
the use of a \dword{cots} \dword{adc} and an \dword{fpga} for the data serialization,
and further validation is planned for spring 2020. We consider this solution as
a fall-back solution for \dword{dune}. Custom solutions for the \dwords{asic} are 
being developed to simplify the \dword{femb} assembly and reduce the power dissipated by the
electronics in the \lar.

The reference solution for the \dword{adc} and the data serialization is based on two new 
\dword{asic}s, \dword{coldadc} and \dword{coldata}.
The first iteration of the \dword{coldadc} was submitted for fabrication
at the end of October 2018, and the chips were delivered in January 2019.
Initial results from the tests of the \dword{coldadc} prototypes have
been discussed in Section~\ref{sec:fdsp-tpcelec-design-femb-adc}. The
results obtained so far are encouraging, despite the fact that some
flaws have been identified in the design. Even if one additional design
iteration is required, we think that the status of \dword{coldadc} can
be characterized as having reached the ``Advanced'' status.

We are also considering an alternative solution for the readout, where
the three \dword{asic}s are replaced with a single one, the \dword{cryo}
chip that has a development 
timeline similar to that of \dword{coldadc}. Also in this case the
chips from the first submission have been delivered in January 2019;
initial results from the tests of the \dword{cryo} prototypes have
been discussed in Section~\ref{sec:fdsp-tpcelec-design-femb-alt-cryo}.
As in the case of \dword{coldadc}, the results obtained so far are
encouraging. As soon as the noise issue observed in the first prototype
is understood, \dword{cryo} should also be characterized as 
having reached the ``Advanced'' status.

The first complete prototype of \dword{coldata} was submitted in 
April 2019 and the chips have been delivered in July. As discussed
in Section~\ref{sec:fdsp-tpcelec-design-femb-coldata}, all test
results for \dword{coldata} have been positive, and so in this
case it can also be claimed that the design of the \dword{asic} has
reached the ``Advanced'' design status.

There have already been multiple iterations of \dwords{femb} that
have been fabricated and tested and used for data taking in 
\dword{microboone} and in \dword{pdsp}. The \dword{sbnd} collaboration
is starting the production of \dwords{femb} based on the \dword{cots} \dword{adc} and
\dword{fpga} solution. The design of the \dword{femb} needs to be adapted
for the different \dword{asic} solutions that are being considered
for \dword{dune}. This development is already ongoing, as system tests 
where the \dwords{femb} are connected to an \dword{apa} are part
of the qualification tests. The design status for the \dword{femb}
is already at the ``Advanced'' level, and it will reach the 
``Very advanced'' level at the time of the \dword{asic}
selection. At that point, only minor modifications may be
required. 

The only other \dword{tpc} electronics detector components that do not yet
reach the ``Very advanced'' level are the cryostat penetrations, as
discussed above, and the \dword{wib}, where small
design changes will be done prior to production in order to use a more
modern and cheaper \dword{fpga}. Additional changes to
the power distribution scheme will be required as the number of power lines
(and the corresponding voltages) will be reduced compared to
\dword{pdsp}. The transition to a more modern \dword{fpga} will allow 
more extensive data monitoring inside the \dword{wib}, but may
also require developing new software and porting the firmware
from one family of \dword{fpga}s to another. 

For all other detector components, the estimate of the design
maturity is considered ``Very advanced'' based on the experience
gained with commissioning and operation of \dword{pdsp}. The 
cold signal cables will be modified to reduce the number of
connections and to address the issues observed with the connector
on the \dword{femb}. The design of the \dword{wiec} needs to
be modified to include air filters to minimize the possible
damage from dust and/or chemical residues from  explosives 
during the lifetime of the experiment at \dword{surf}.
The \dword{ptc} is going to be modified to add an interface to
the hardware interlocks of the detector safety system. For
cables and fibers on the top of the cryostat, the only work that
remains to be done is the design of the actual cable plant, 
which will then determine the length of the cables. The arrangement
of power supplies in the racks on top of the cryostat is the
only other remaining design task. For many components, the
qualification of additional vendors could also be considered
as part of value engineering; this will reduce the risk of vendor
lock-in and help minimize costs.

\section{Quality Assurance}
\label{sec:fdsp-tpcelec-qa}

The \dword{tpc} electronics consortium is developing a \dword{qa} plan consistent
with the principles discussed in \tcchqa.
The goal of the \dword{qa} plan is to maximize the number of functioning
readout channels in the detector that achieve the performance specifications
for the detector discussed in Section~\ref{sec:fdsp-tpcelec-overview-design},
particularly on noise. Minimizing the noise levels in the detector requires that all
system aspects are considered starting from the design phase, and in this
respect, the experience gained with the \dword{pdsp} prototype is extremely
valuable as it informs necessary design changes in the detector 
components. The lessons learned during the construction of \dword{pdsp},
the commissioning of the detector, and the initial data taking period have
already been discussed in Section~\ref{sec:fdsp-tpcelec-design}. Further
operation of \dword{pdsp} in 2019 has provided information on the
long term stability of the detector components.

Apart from the number of channels, the most important difference
between \dword{pdsp} and \dword{dune} is the projected lifetime of the detector. This
is relevant because a significant fraction of the detector components provided 
by the \dword{tpc} electronics consortium are installed inside the cryostat and cannot 
be accessed or repaired during the operational lifetime of the detector. The 
graded approach to \dword{qa} indicates that particular care must be used for
the \dword{ce} components that will be installed inside the cryostat.

A complete \dword{qa} plan starts with ensuring that the designs of all
detector components fulfill the specification criteria, considering
also system aspects, i.e. how the various detector components interact
among themselves and with the detector components provided by other 
consortia. We discuss validating the design in 
Section~\ref{sec:fdsp-tpcelec-qa-initial} and the facilities that we use
to investigate the interactions among different detector components
in Section~\ref{sec:fdsp-tpcelec-qa-facilities}. 

The other aspects of the \dword{qa} plan involve documenting the 
assembly and testing processes, storing and analyzing the information
collected during the \dword{qc} process, training and qualifying 
personnel from the consortium, monitoring procurement of 
components from external vendors, and assessing whether the
\dword{qc} procedures are applied uniformly across
the various sites involved in detector construction, integration,
and installation. The \dword{tpc} electronics consortium plan involves
having multiple sites using the same \dword{qc} procedures,
many of which will be developed as part of system design tests during the \dword{qa} phase,
with the possibility of a significant turnover in the personnel
performing these tasks. To avoid problems during most of the
production phase, we plan to emphasize training as well as documentation
of the \dword{qa} plan. Reference parts will be tested at
several sites to ensure consistent results. At
a single site, some parts will be tested repeatedly to ensure
that the response of the apparatus does not change and
that new personnel involved in testing detector components are 
as proficient as more experienced personnel. 

All data from
the \dword{qc} process will be stored in a common database, and
the yields of the production will be centrally monitored and 
compared among different sites. The procedures adopted
for detector construction will evolve from the experience
gained with \dword{pdsp}. A first version of the testing procedures
will be put in place in 2020, while the final designs of
the detector components are completed and new prototypes are
tested. The \dword{qc} procedures will then be reviewed
during the engineering design review that precedes pre-production. Lessons learned during pre-production
will be analyzed, and a final and improved \dword{qc} process will be 
developed before the production readiness review that triggers
the beginning of production. During production, the results
of the \dword{qc} process will be reviewed at regular intervals
in production progress reviews. In case of problems, production
will be stopped and the problematic issues assessed, 
followed by changes in the procedures if necessary.

\subsection{Initial Design Validation}
\label{sec:fdsp-tpcelec-qa-initial}

As described in Section~\ref{sec:fdsp-tpcelec-design}, four \dword{asic} designs
are being developed for the \dword{dune} \dword{fd} single-phase \dword{tpc} readout 
(\dword{larasic}, \dword{coldadc}, \dword{coldata}, and \dword{cryo}). 
When a new prototype \dword{asic} is produced, the groups responsible for the \dword{asic} design will perform the first tests of 
\dword{asic} functionality and performance. These tests may use either 
packaged parts or dice mounted directly on a printed circuit board 
and wire bonded to the board.  The goal of these tests is to determine 
the extent to which the \dword{asic} functions as intended, both at room 
temperature and at \lntwo temperature.  For all chips, these tests 
include exercising digital control logic and all modes of operation. Tests 
of \dword{fe} \dwords{asic} include measurements of noise levels as a function 
of input capacitance, baseline recovery from large pulses, cross-talk, linearity, 
and dynamic range. Tests of \dwords{adc} include measurements of the effective noise levels and 
of differential as well as integral non-linearity. Tests of the \dword{coldata} and \dword{cryo} \dwords{asic}
include verification of both the control and high-speed data output links using 
cables with lengths of \SI{9}{m} and \SI{22}{m} as required for the \dword{dune} \dword{fd}.
After the initial functionality
tests by the groups that designed the \dwords{asic}, further
tests will be performed by other independent groups; then the \dwords{asic}
will be mounted on \dwords{femb} so noise measurements can be repeated
with real \dword{apa}s attached to the readout chain.

Tests of \dwords{asic} and \dwords{femb} in a cryogenic environment
are performed in \lntwo instead of \dword{lar} for cost reasons, ignoring
the small temperature difference. These tests can be performed immersing
the detector components in a dewar containing \lntwo for the duration
of the tests. Condensation of water from air can interfere with
the tests or damage the detector components or the test equipment,
particularly during their extraction from the \lntwo. A test dewar
design developed by Michigan State University, referred to as the
\dword{cts}, has been developed to avoid
this problem and to automate the immersion and the retrieval of 
the components being tested. Several \dword{cts} units
were deployed at \dword{bnl} during the \dword{pdsp} construction
and used for the \dword{qc} on the \dwords{asic} and \dwords{femb}
for \dword{pdsp}. Later they were also used to perform similar tasks
during the construction of the electronics for \dword{sbnd}.
Several other \dword{cts} units have been deployed to institutions involved in
developing \dwords{asic} to test the first prototypes of \dwords{asic}
and \dwords{femb} for the \dword{dune} \dword{fd}. Two \dword{cts} units 
in operation at \dword{bnl} are shown in Figure~\ref{fig:CTS}.

\begin{dunefigure}
[The Cryogenic Test System]
{fig:CTS}
{Cryogenic Test System (CTS): an insulated box is mounted on top of a commercial \lntwo dewar.  Simple controls allow the box to be purged with nitrogen gas and \lntwo to be moved from the dewar to the box and back to the dewar.}
\includegraphics[width=0.4\linewidth]{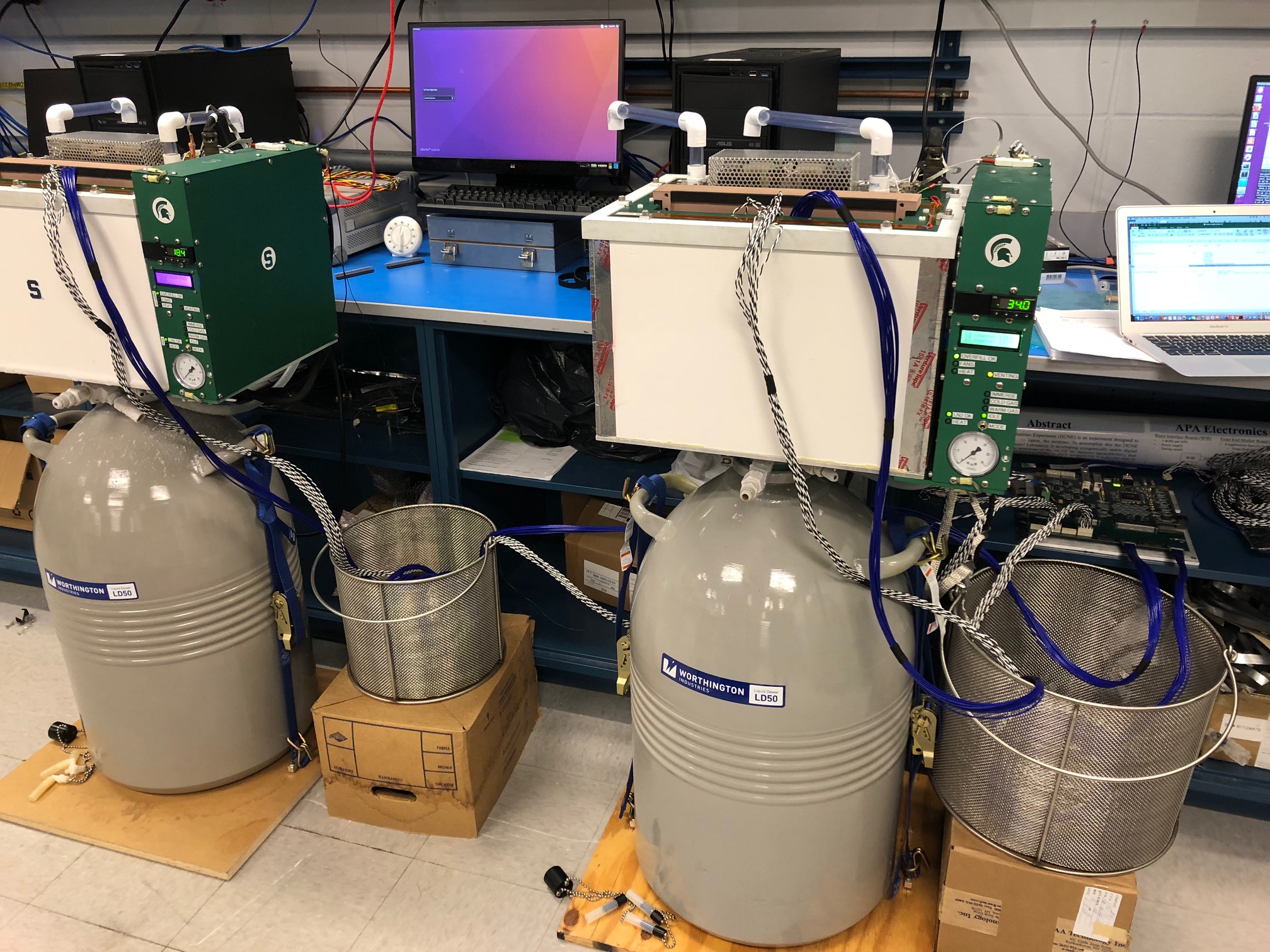}
\end{dunefigure}

\subsection{Integrated Test Facilities}
\label{sec:fdsp-tpcelec-qa-facilities}

The investigation of the system issues that can arise from the interaction 
of different detector components requires that a full system test of a slice
of the entire detector is performed. These tests are performed with \dwords{femb}
attached to \dword{apa}s enclosed in a structure that provides the same
grounding environment planned for the final \dword{dune} \dword{fd}. 
Power, control, and signal readout connections will be provided using cryostat 
penetrations similar to those planned for use in the \dword{dune} \dword{fd}.
Prototypes of the final \dword{dune} \dword{fd} \dword{daq} will be used for readout and
control of the detector, and if possible the \dword{pds} will also be included.
We have identified three such system test stands
that we can use for system tests: the \dword{pdsp} facility at dword{cern}, the
\dword{iceberg} facility at \dword{fnal}, and the \num{40}\,\% \dword{apa} at \dword{bnl}.
We discuss these three setups in this section.

\subsubsection{ProtoDUNE-SP and Cold Box at CERN}
\label{sec:fdsp-tpcelec-qa-facilities-pdune}

\dword{pdsp} is designed as a full slice of the \dword{dune} \dword{spmod}  
using components with a design as close as possible to the one that will
be used in production. It contains six full-size 
\dword{dune} \dword{apa}s instrumented with \num{20} \dwords{femb} each for a 
total readout channel count of \num{15360} digitized sense wires. Critically, 
the wires on each \dword{apa} are read out via a full \dword{tpc} electronics readout 
system, including a \dword{ce} flange and \dword{wiec} with five \dwords{wib} 
and one \dword{ptc}. Each combined \dword{apa} and \dword{ce} readout unit follows 
the grounding guidelines described in Section~\ref{sec:fdsp-tpcelec-design-grounding} 
to operate in a fully-isolated way with respect to the rest of the detector.

\dword{pdsp} took beam data in the \dword{cern} Neutrino Platform in 2018 and will continue
to take cosmic data throughout spring 2020. As described in Section~\ref{sec:fdsp-tpcelec-overview-pdune},
the live channel count (99.7\%) and average noise levels on the collection and induction wires 
(ENC~$\sim$\SI{550}{e$^-$} and $\sim$\SI{650}{e$^-$}, respectively) satisfy the \dword{dune}-\dword{sp} requirements described 
in Section~\ref{sec:fdsp-tpcelec-overview-requirements}. Several lessons learned from the production 
and testing of the \dword{tpc} electronics and the \dword{pdsp} beam data run will be incorporated into the 
next iteration of the system design for the \dword{dune} \dword{spmod}, as discussed in 
Section~\ref{sec:fdsp-tpcelec-overview-lessons}.

Five of the six \dword{apa}s were tested in the \dword{pdsp} \coldbox
before they were installed in the cryostat. These tests were critical
in identifying issues with \dword{ce} components after installation
on the \dword{apa}. Therefore, a very similar set of \coldbox tests are planned at \dword{surf} 
with the fully-instrumented \dword{dune} \dword{apa}s. A seventh \dword{apa} was delivered
to \dword{cern} in March 2019 and will be first equipped with \dword{pdsp} \dwords{femb}. This
\dword{apa} was characterized in October 2019 in the \coldbox like the \dword{apa}s installed
in \dword{pdsp}, establishing a reference point for further tests that will be
performed after replacing half of the \dwords{femb} with new prototypes \dwords{femb} (equipped with prototypes of the
new \dword{asic} designs). Tests will be performed in 
early 2020 using \dwords{femb} equipped with the \dword{cots} \dword{adc}, \dwords{femb}
with the \dword{cryo} \dword{asic}, and \dwords{femb} with the new \dword{coldadc} and
\dword{coldata} \dwords{asic}. The \coldbox will also be used to study the effect of
low temperature on the low-voltage power and bias-voltage power cables as they are 
routed through the \dword{apa} frame.

The \dword{dune} \dword{apa}s and the readout electronics will differ from the ones used 
in \dword{pdsp}. For this reason, we are planning to re-open the \dword{pdsp} cryostat 
and replace three of the six \dword{apa}s with final \dword{dune} \dword{fd} prototypes that also include 
the most recent prototypes of the \dwords{femb} built using the chosen \dword{asic} solution.
If possible, \dwords{asic} from the engineering 
run will be used to populate the \dwords{femb} instead of using prototypes from a multi-purpose
wafer fabrication run. A total of \num{60} \dwords{femb} are required to populate the three 
final \dword{dune} \dword{apa} prototypes to be installed in \dword{pdsp}. A second period of data taking 
with this new configuration of \dword{pdsp} is planned for 2021-2022. This will also 
allow another opportunity to check for interference between the readout of the \dword{apa} 
wires and the \dword{pds} or other cryogenic instrumentation. 

\subsubsection{Small Test TPC (ICEBERG)}
\label{sec:fdsp-tpcelec-qa-facilities-testtpc}

While the \coldbox test at \dword{cern} and \dword{pdsp} operations provide important 
validation of the \dword{tpc} electronics for DUNE, a new cryostat (\dword{iceberg}) 
has been built to test multiple \dword{ce} prototypes in a \dword{lartpc} environment.
\dword{iceberg} will be used for 
\dword{lar} detector R\&D and for system tests of the \dword{ce} prototypes. 
The \dword{iceberg} cryostat allows for rapid turn-around in testing new configurations
of the \dword{ce}. One cycle, including installing new \dwords{femb}, filling the cryostat,
performing measurements, and finally emptying the cryostat, can be completed in less than
one month. While this is slower than the turn-around that can be achieved with the
\coldbox at \dword{cern}, the advantage of \dword{iceberg} is that it houses a small \dword{tpc} which
allows measurements with ionization tracks, which is not possible when performing
tests in the \coldbox. In addition, \dword{iceberg} enables system-wide studies with new
prototypes of the \dword{pds} because the scaled-down \dword{apa} is mechanically compatible with the 
new design of the \dword{pds}, which is not the case for the seventh \dword{pdsp} \dword{apa}.  

The \dword{iceberg} cryostat, shown in Figure~\ref{fig:ICEBERG-cryotee}, is installed
at the Proton Assembly Building at \dword{fnal}. It has an inner diameter of \SI{152}{cm}
and can hold about 35,000 liters of \dword{lar}, sufficient to house a
\dword{tpc} with dimensions \SI{115}{cm}~$\times$~\SI{100}{cm}~$\times$~\SI{60}{cm}. For
\dword{dune} purposes, this cryostat will house a 1,280-channel \dword{tpc}, shown in
Figure~\ref{fig:ICEBERG-tpcdaq}, with an \dword{apa} and two \dwords{fc} that
together enclose two sensitive ionization drift volumes. Each drift volume has 
a maximum drift distance of \SI{30}{cm}. The \dword{apa} 
has been built using wire boards and anchoring elements identical to
those of \dword{pdsp}, as 
described in Section~\ref{sec:fdsp-apa-boards}. It has dimensions of 
1/10$^{\mathrm{th}}$ of a \dword{dune} \dword{apa}. The \dword{apa} mechanics are designed to
accommodate two half-length \dword{pdsp} \dwords{pd} with dimensions and connectors that already 
include the design modifications planned for the \dword{dune} \dword{fd}.

\begin{dunefigure}
[ICEBERG cryostat and top plate spool piece]
{fig:ICEBERG-cryotee}
{\dword{iceberg} cryostat (left) and top plate spool piece (right).}
\includegraphics[width=0.26\linewidth]{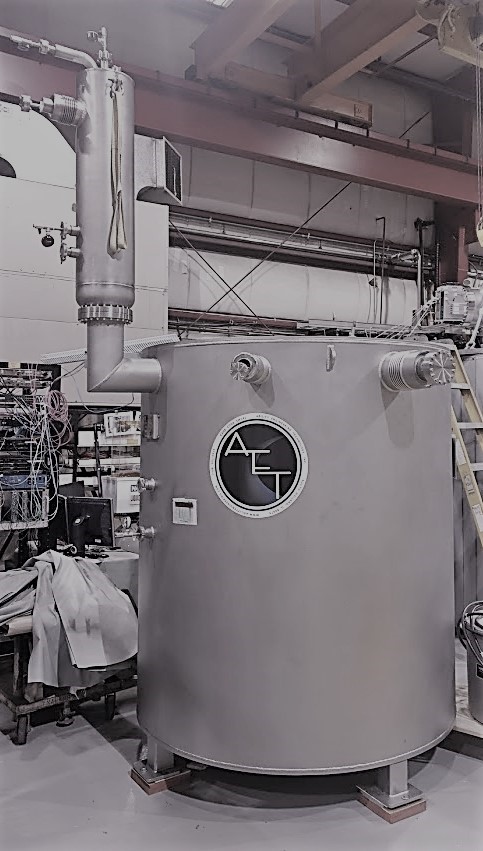}
\includegraphics[width=0.61\linewidth]{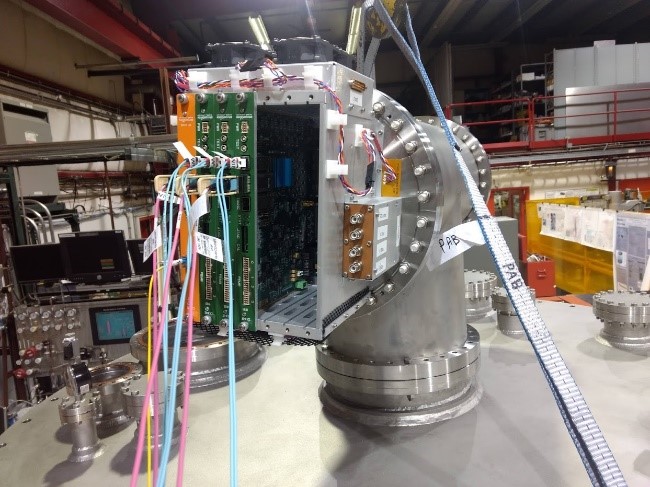}
\end{dunefigure}

\begin{dunefigure}
[ICEBERG TPC and DAQ rack]
{fig:ICEBERG-tpcdaq}
{\dword{iceberg} TPC (left) and DAQ system (right).}
\includegraphics[angle=270,width=0.43\linewidth]{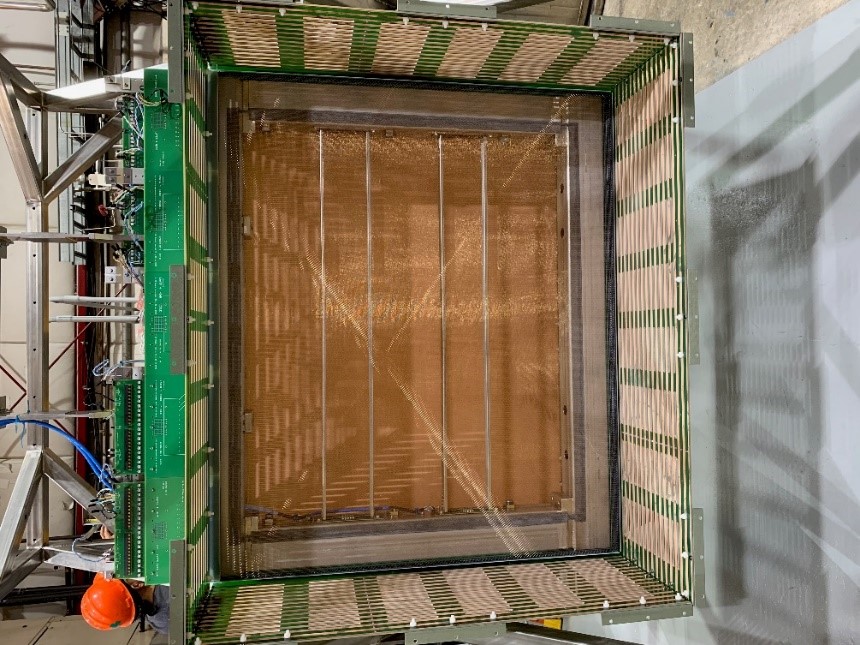}
\includegraphics[angle=270,width=0.43\linewidth]{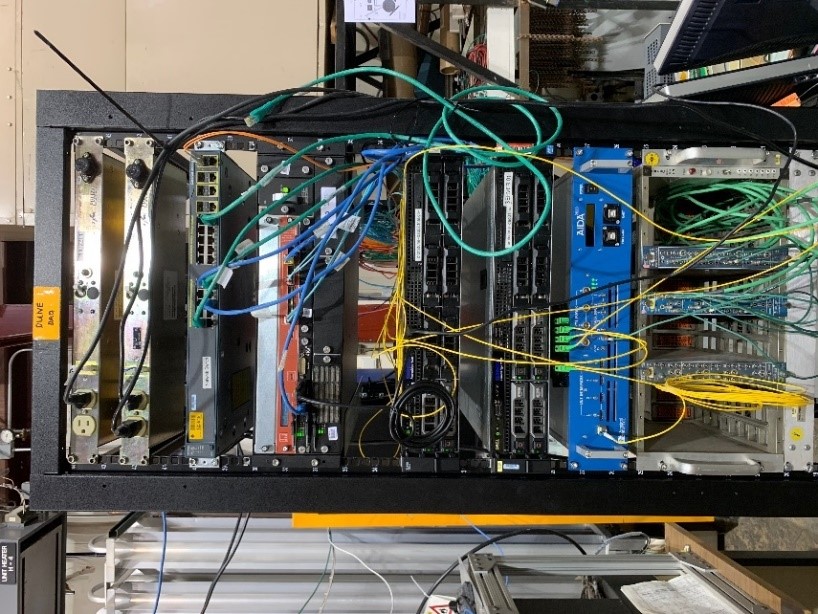}
\end{dunefigure}

Power, readout, and controls use equipment identical to those used for \dword{pdsp}. The
interface between the \dwords{femb} and the \dword{apa} wires uses the same \dword{cr} boards
used for \dword{pdsp} and described in Section~\ref{sec:fdsp-apa-boards}. The \dword{tpc} is
read out via a \dword{daq} system (also shown in Figure~\ref{fig:ICEBERG-tpcdaq})
identical to that of \dword{pdsp}. The power and signal cables for the detector 
are routed through a spool piece installed on the center port of a movable flange on the 
top of the cryostat, which is also used to support the \dword{tpc}. The movable 
flange contains fourteen additional ports that are available for different utilities, 
including \dword{hv}, purity monitoring, cryogenic controls, and visual inspection. 
A condenser as well as \dword{lar} fill and vacuum ports are on the side of the cryostat, 
providing easy access to the detector.

The \dword{fc} for the \dword{tpc} is constructed using printed circuit boards and 
designed to provide  up to \SI{30}{cm} of drift length on both sides of the \dword{apa}. The cathode 
plane is made of a printed circuit board coated with copper and is powered with 
\SI{-15}{kV} DC power. A \SI{1}{G$\Omega$} resistance between the strips of the \dword{fc}
creates a gradient field changing from \SI{-15}{kV} at the cathode to \SI{-1}{kV} near the 
\dword{apa}. In the initial configuration, the sides of the \dwords{fc} are terminated on the \dword{apa}
ground with \SI{156}{M$\Omega$} resistors, which is different from that which has been
planned for the \dword{dune} \dword{fd}, where the last electrode of the \dwords{fc} is also connected to
a separate bias voltage supply. 

The \dword{iceberg} power system that provides power to the detector, electronics, 
\dword{daq}, and cryogenics controls was designed with extreme care to 
isolate the detector and building grounds, following the same principles adopted
for \dword{pdsp} and including a new \SI{480}{V} transformer. The impedance between the detector
and building grounds is continuously monitored. The distribution panel, which is 
at detector ground, provides both \SI{208}{V} and \SI{120}{V} lines for the \dword{tpc} electronics rack, 
providing both the low-voltage power to the \dword{tpc} electronics components and the bias 
voltage to the \dword{apa} wire planes through the \dword{wiec} and \dword{shv}
connectors located on the cryostat penetration. A single WIENER MPOD provides 
\SI{-665}{V}, \SI{-370}{V}, \SI{0}{V}, and \SI{820}{V} to the $G$, $U$, $V$, and 
$X$ planes of the \dword{apa}, respectively. It is also used to provide the 
\SI{-15}{kV} to the cathode plane. A WIENER PL506 provides the low-voltage
power to the \dword{ptc}, to the fans, and to the heaters located on the 
flange that is mounted on the spool piece at the top of the cryostat. 

The \dword{daq} for \dword{iceberg} is a copy of the system used for the readout
of five of the \dword{pdsp} \dword{apa}s at \dword{cern}. The core of the \dword{daq} system 
consists of two Linux PCs that communicate over 10 Gbps optical fibers
with processing units called \dwords{rce}, which are \dwords{fpga} that are
housed on industry-standard \dword{atca} shelves on \dword{cob} motherboards.
The \dwords{rce} can perform data compression and zero suppression. They also buffer
the data while waiting for a trigger and then send it to the Linux PCs where the data can
be analyzed using the \dword{artdaq} framework. A pair of scintillators at the top and
bottom of the cryostat generates a cosmic trigger for the \dword{daq}.
The system is modular and could be upgraded to follow the overall \dword{dune} \dword{fd} \dword{daq} 
development. 

The \dword{iceberg} cryostat was filled for the first time with \dword{lar}
in March 2018, with \dword{pdsp} \dwords{femb} installed on the \dword{tpc}. The
initial data taking run uncovered some issues with the pressure regulation system of
the cryostat and with the field cage. Once these problems were addressed, a second
data taking period started in June 2019, demonstrating stable operations of the
cryostat and \dword{tpc}. The baseline performance of the \dword{iceberg} 
\dword{tpc} has been established using \dword{pdsp} \dwords{femb}. 
As shown in Figure~\ref{fig:ENC-all-ICEBERG}, noise levels of 
$\sim\SI{300}{e^-}$ have been measured on the collection and induction wires,
which are of similar length, in line with expectations given the input capacitance
to the \dword{fe} electronics. A small number of channels have larger noise levels as a
result of the underperforming \dword{adc} electronics, as in the case of \dword{pdsp}.
\dword{tpc} noise levels remained constant while the \dword{pds} was
being operated, demonstrating that there is no interference between that system
and the \dword{tpc} electronics. Moving forward, \dword{iceberg} will be used to
test new \dword{femb} prototypes, equipped with the new \dwords{asic} under
development. 

\begin{dunefigure}
[ENC levels for all channels of the ICEBERG TPC]
{fig:ENC-all-ICEBERG}
{\dword{enc} levels (in electrons) for all channels of the \dword{iceberg} \dword{tpc},
both before and after the application of a simple offline common-mode noise filter.}
\includegraphics[width=0.99\linewidth]{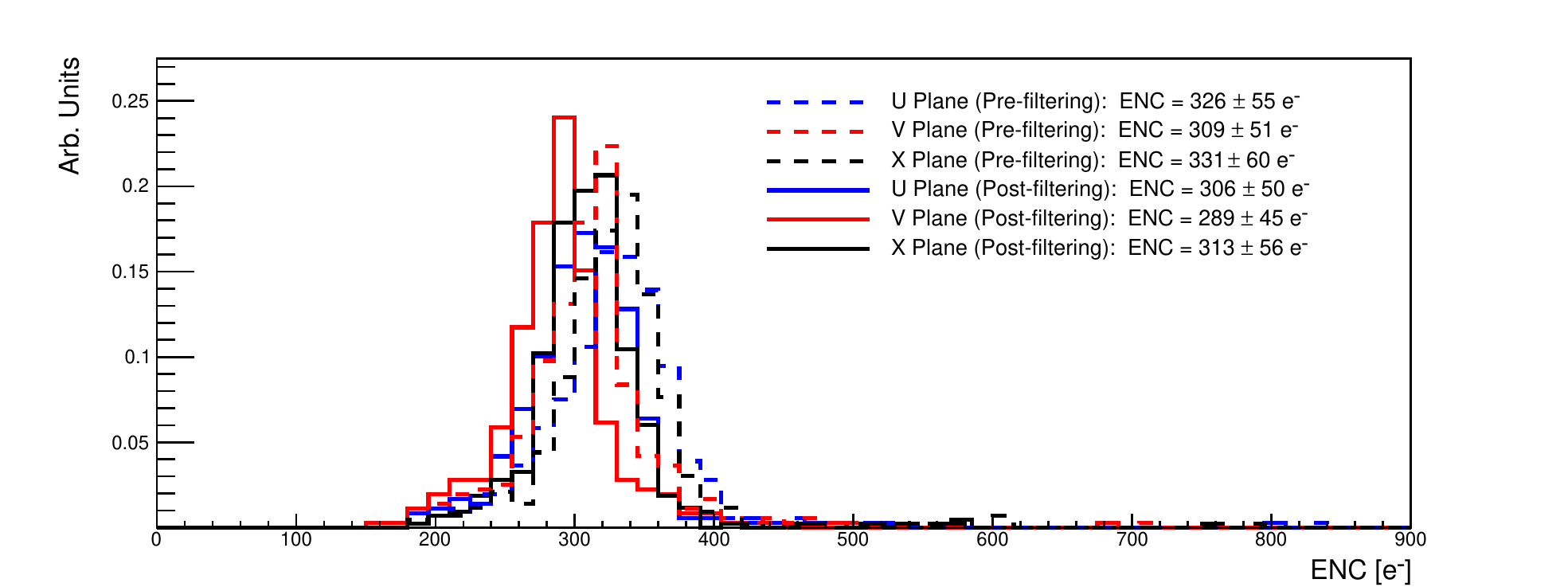}
\end{dunefigure}

\subsubsection{40\% APA at \dword{bnl}}
\label{sec:fdsp-tpcelec-qa-facilities-fortypercent}

One additional facility where the \dword{femb} prototypes can be connected to
an \dword{apa} inside a shielded environment is the \num{40}\,\% \dword{apa} 
test stand at \dword{bnl}. The \num{40}\,\% \dword{apa} at \dword{bnl} is a \SI{2.8}{m}~$\times$~\SI{1.0}{m} 
three-plane \dword{apa} with two layers of \num{576} wrapped ($U$ and $V$) wires 
and one layer of \num{448} straight ($X$) wires. It is read out by up to eight 
\dwords{femb} with the full length (\SI{7}{m}) \dword{pdsp} data and \dword{lv} power 
cables. The readout uses the full \dword{tpc} electronics 
system, including the \dword{ce} flange and \dword{wiec}, as shown in Figure~\ref{fig:tpcelec_40apa}. 
Detailed integration tests of the \dword{pdsp} \dword{ce} readout performance were done 
at the \num{40}\,\% \dword{apa}. During these tests the \dword{dune} grounding and shielding guidelines
were strictly followed.  This system was also used for initial studies of the \dword{cots} \dword{adc}
option that is described in Section~\ref{sec:fdsp-tpcelec-design-femb-alt-cots} and will
be used again for new \dword{femb} prototypes.

\begin{dunefigure}
[One side of the \num{40}\,\% APA with four FEMBs and the \dshort{ce} flange]
{fig:tpcelec_40apa}
{Left: one side of the \num{40}\,\% \dword{apa} with four \dwords{femb}.  Right: the full \dword{ce} \fdth and flange.}
\includegraphics[width=0.72\linewidth]{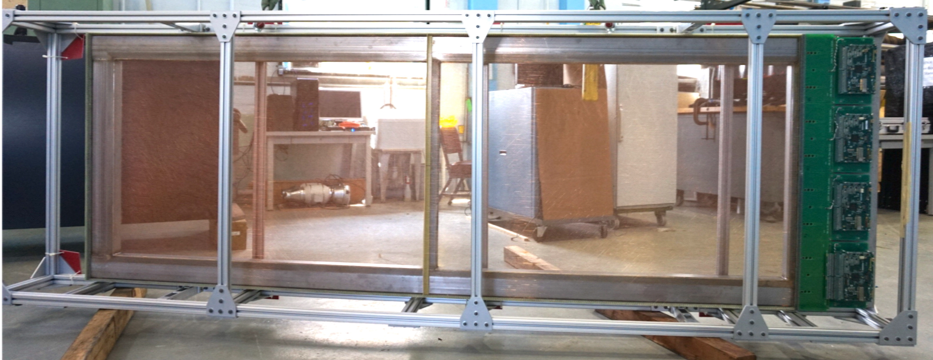}
\hspace{3mm}
\includegraphics[width=0.2\linewidth]{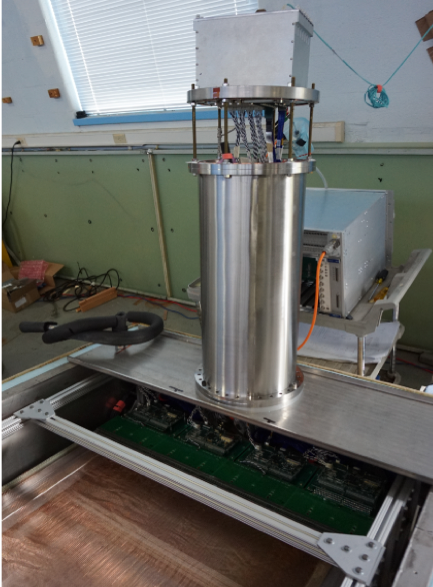}
\end{dunefigure}

Each of the three setups (\dword{apa} in the \coldbox at \dword{cern}, \dword{iceberg} \dword{tpc} at \dword{fnal},
and \num{40}\,\% \dword{apa} test stand at \dword{bnl}) that can be used for system tests has advantages
and disadvantages. Only the \dword{iceberg} \dword{tpc} can be used to 
perform measurements with tracks, but the \dword{apa} is 
much smaller than the \dword{dune} \dword{fd} \dword{apa} (which is also an advantage because it
allows us to determine the ultimate performance of the electronics because
the detector capacitance is reduced). The \dword{iceberg} \dword{tpc}
is for the moment the only setup compatible with the new \dword{pds}
design. Tests performed in the \coldbox at \dword{cern} and with the \num{40}\,\% \dword{apa} 
at \dword{bnl} are limited to noise measurements. These tests are not
performed at \dword{lar} temperature in the \dword{cern} setup. The advantage of
of the \coldbox at \dword{cern} and of tests performed in \dword{pdsp}
is that the \dword{apa} size is the one used in the \dword{dune} \dword{fd},
while the \dword{iceberg} detector is much smaller. We plan to continue
using all these setups for testing during the development of 
new \dwords{asic} and \dwords{femb} designs.

\subsection{Reliability Studies}
\label{sec:fdsp-tpcelec-qa-reliability}

The \dword{tpc} \dword{ce} system of the \dword{dune} \dword{sp} \dword{fd} must meet 
stringent requirements, including a very small number of failures ($<\num{1}$\% of the total number of
channels) for components installed 
on the detector inside the cryostat during the \dunelifetime of 
detector operation. Initial studies of the impact of dead channels indicate
that there is minimal impact on physics measurements even for a large ($\sim5$\%) number of 
channel failures randomly distributed in the detector, given the very high granularity of the \dword{tpc}. 
Further studies are ongoing to understand the impact of failures 
affecting groups of neighboring channels, which could arise from the failure of \dwords{asic} 
(16-64 channels) or \dwords{femb} (128 channels). Reliability must be incorporated in the 
design of all components, and a dedicated analysis of the physics impact of all 
possible failure modes, including a consideration of the number of readout channels
affected, is required before finalizing the design of all \dwords{asic}, printed circuit 
boards, cables, connectors, and their supports, all of which are housed inside the 
\dword{dune} \dword{fd} cryostat. 

A few \dword{hep} detectors have operated without intervention for a 
prolonged period, with few readout channel losses, in extreme 
conditions that are similar to those in the \dword{dune} \dword{fd} cryostats:
\begin{itemize}
\item The NA48/NA62 liquid krypton (LKr) calorimeter has 13,212 channels 
of JFET pre-amplifiers installed on the detector. It has been kept at LKr temperature 
since 1998. The total fraction of failed channels is $<$ 0.2\% in more than \num{20} years of operation.
\item The ATLAS \dword{lar} accordion electromagnetic barrel calorimeter has 
approximately 110,000 readout signal channels, with up to seven connections and different 
circuit boards populated with resistors and diodes inside the cryostat. This
calorimeter has been cold since 2004, for a total of \num{15} years of operation. So far, the
number of readout channels that have failed is approximately 0.02\% of the total channel count.
\item The ATLAS \dword{lar} hadronic endcap calorimeter has approximately 35,000 GaAs
pre-amplifers summed into 5,600 readout channels that are mounted on cold pre-amplifier
and summing boards. The ATLAS \dword{lar} hadronic endcap 
calorimeter \dword{ce} have been in cold since 2004, with
0.37\% of the channels failing during 15 years of operation. 
\end{itemize}
Since neither NA48/NA62 nor the ATLAS \dword{lar} hadronic endcap calorimeter
use \dword{cmos} electronics, the procedures used in the construction and \dword{qc} of \dwords{pcb} and 
for the selection and \dword{qc} of connectors and discrete components mounted
on the \dwords{pcb} represent the most relevant aspect for the \dword{dune} \dword{fd}.

In addition, FERMI/GLAST is an example of a joint project between \dword{nasa} and \dword{hep} groups 
with a minimum mission requirement of five years, and is on its way to achieving a 
stretch goal of ten years of operations in space. Although the requirements are somewhat 
different, examining and understanding the various strategies for a space 
flight project can inform the \dword{dune} project. 

A preliminary list of reliability topics to be studied for the \dword{tpc} electronics operated 
in \dword{lar} environment are:
\begin{itemize}
\item The custom \dwords{asic} proposed for use in \dword{dune} (\dword{larasic}, 
\dword{coldadc}, \dword{coldata}, and \dword{cryo}) incorporate design rules 
intended to minimize the hot-carrier effect~\cite{Li:CELAr,Hoff:2015hax}, 
which is recognized as the main failure mechanism for integrated circuits 
operating at \dword{lar} temperature.
\item For \dword{cots} components, accelerated lifetime testing, a methodology 
developed by \dword{nasa}~\cite{nasa_nepp} will be used to verify the expected
lifetime of operation at cryogenic temperatures. A \dword{cots} \dword{adc} has undergone
this procedure and has been qualified as a solution for the SBND experiment~\cite{Chen:2018zic}.
\item Printed circuit board assemblies are designed and fabricated to survive 
repeated immersions in \lntwo.
\item A study will be undertaken to give guidance on how much components (capacitors,
resistors, etc.) should be de-rated for power dissipation, operating voltage, etc.~in
order to achieve the desired reliability.
\item Similarly, connectors and cables, usually major sources of detector channel failures,
will require a separate study to identify optimal choices.
\item In addition to the \dword{qa} studies noted above, a very detailed
and formal set of \dword{qc} checks of the production pieces will be required in order to ensure
a reliable detector. The \dword{qc} plans for the \dword{tpc} electronics
detector components are discussed in Section~\ref{sec:fdsp-tpcelec-production-qc}.
\end{itemize}
The \dword{tpc} electronics consortium has formed a working group tasked with studying the reliability  
of these components, which is preparing recommendations for the choice of \dwords{asic}, 
the design of printed circuit boards, and testing procedures. This working group will review the 
segmentation of the \dword{ce} to understand which failures will most affect data taking, revisit recommendations for the \dword{asic} design, 
beyond those aimed at minimizing the hot-carrier effect, revisit the industry and 
\dword{nasa} standards for the design and fabrication of printed circuit boards, connectors, 
and cables, and recommend \dword{qc} procedures to be adopted during 
fabrication of the \dword{ce} components. The working group will also review
system aspects, to understand where it is desirable, necessary, and feasible to implement 
redundancy in the system in order to minimize data losses due to single component failures. 

\section{Production and Assembly}
\label{sec:fdsp-tpcelec-production}

In this section, we discuss the production and assembly plans,
including the plans for the spares required during the detector
construction and for operations, for
procurement, assembly, and quality control.


\subsection{Spares Plan}
\label{sec:fdsp-tpcelec-production-spares}

The \dword{apa} consortium plans on building 152 \dword{apa}s
for the first \dword{sp} detector. This means that at least
3,040 \dwords{femb} with the corresponding bundles of cold
cables will be required for the integration (3,040 power cables, 3,040 data cables,
and 1,216 bias voltage cables; half the cables will be long enough for 
integration on the top \dword{apa}s, while the other half will
be compatible with the bottom \dword{apa}s). To have spare \dwords{femb}, the \dword{tpc} electronics consortium plans to
build at least 3,200 \dwords{femb}, 5\% more than necessary. If more spares are needed during the
\dword{qc} process or during integration, additional 
\dwords{femb} can be produced quickly as long as any components that have 
long lead times are on hand. For these components, we plan to keep on hand a
larger number of spares. The \dwords{asic} require a long lead time; a plan for those spares is
discussed below. For other discrete components
(capacitors, resistors, connectors, voltage regulators, oscillators),
plans will be put in place once the final design of the \dword{femb}
is available and vendors contacted. 

For the \dwords{asic}, the number of spare chips is driven by the fact that fabrication
requires batches of 25 wafers at a time. Given the dimensions of the current prototype \dwords{asic}, the
expected number of chips per wafer is about 700 for \dword{larasic}, 930 for \dword{coldadc},
230 for \dword{coldata}, and 220 for \dword{cryo}. These numbers are based on the
assumption that \dword{coldadc} and \dword{coldata} are fabricated on the same wafer. To
estimate the number of usable chips for installation on the \dwords{femb}, we assume that
10\% of the chips will fail during the \dword{qc} process described later in this section,
and an additional 5\% of the chips will fail during dicing and packaging. With these
assumptions, one would need at least 43 \dword{larasic} wafers, 33 \dword{coldadc} and 
\dword{coldata} wafers, and 35 \dword{cryo} wafers for one \dword{sp} \dword{fd} module. Wafers
must be ordered in batches of 25 which implies that we will have a significant number of
spares, meaning that additional batches of wafers would be needed only if the overall
yield of \dword{larasic} falls below 75\% or if the overall yield of the other
\dwords{asic} falls below 60\%. The number of spare chips available can be reduced if wafers are purchased for two
\dword{sp} detectors at a time; however, the wafers are relatively inexpensive and the chosen
processes may not be available after a few years so generous spares of these custom devices
are likely advisable. 

In general, for other components, we plan to procure between 5 and
10\% additional components for spares for the construction of the first \dword{sp}
\dword{fd} module. We will need more spares for components that have
a larger risk of damage during integration and 
installation. For example, for cold cables, we
plan for 10\% additional spare cables for the bottom \dword{apa} because
they must be routed through the \dword{apa} frames, but
for the top \dword{apa}, we foresee needing only 5\% additional spare cables.
Assuming we will have unused spares from the first detector, we will reduce the number of
spares for the second \dword{sp} \dword{fd} module.

The components on top of the cryostat (power supplies, bias
voltage supplies, cables, \dwords{wiec} with their \dword{wib}
and \dword{ptc} boards) can be replaced while the
detector is in operation. For these components, additional spares may be required
during the \dunelifetime operation period of the \dword{dune} \dword{fd}.
The initial plan is to purchase 10\% additional components for spares for the first
\dword{sp} \dword{fd} module and use them for the second  \dword{sp} \dword{fd} as well
(i.e. effectively having 5\% additional components for spares). Once the design of
the \dwords{wib} is finalized, we will decide if 
extra spares should be purchased for \dwords{fpga} and optical
transmitters and receivers. These are commercial components 
that may no longer be available after a certain number of 
years of operation, which could prevent the \dword{tpc} electronics consortium from
fabricating additional spare \dwords{wib} if required. This
risk is discussed in Section~\ref{sec:fdsp-tpcelec-risks-operations}, one that
could be alleviated by placing commercial components
on mezzanine cards to minimize any necessary redesign of
boards if these components are no longer available.
We can also stock additional components
if market trends show that the components will  
become harder or impossible to find in the future.

\subsection{Procurement of Parts}
\label{sec:fdsp-tpcelec-production-procurement}

The construction of the detector components for \dword{dune} requires many large procurements that 
must be carefully planned to avoid delays. For the \dwords{asic}, the 
choice of vendor(s) is made at the time the technology used in designing 
the chips is chosen. For almost all other components, several vendors 
will bid on the same package. Depending on the requirements of the funding
agency and of the responsible institution, this may require a lengthy
selection process. The cold cables used to transmit data from the
\dwords{femb} to the \dwords{wib} represent a critical case. In this case
a technical qualification, including tests of the entire cold chain (from the \dword{femb}
to the receiver on the \dword{wib}) is required. Another problem is the 
large numbers of components required. In some cases, the number of components 
of a given type (resistors, capacitors) may far exceed the number of components
that the usual resellers keep in stock. This will 
require careful planning to avoid stopping the assembly chain for
the \dwords{femb}, for example, because 
one kind of component runs short. Figures~\ref{fig:sp-tpcelec-partsflow1}
and \ref{fig:sp-tpcelec-partsflow2} show the flow of the \dword{tpc}
electronics detector components through procurement, assembly,
\dword{qc} testing, and finally integration and installation at 
\dword{surf} using a color code to indicate the activities that
are performed by external vendors, those that take place at one of
the consortium institutions, and those that take place at \dword{surf}.

\begin{dunefigure}
[Parts flow for the detector components inside the cryostat]
{fig:sp-tpcelec-partsflow1}
{Parts flow for the TPC electronics detector components installed inside the cryostat.}
\includegraphics[width=0.7\linewidth]{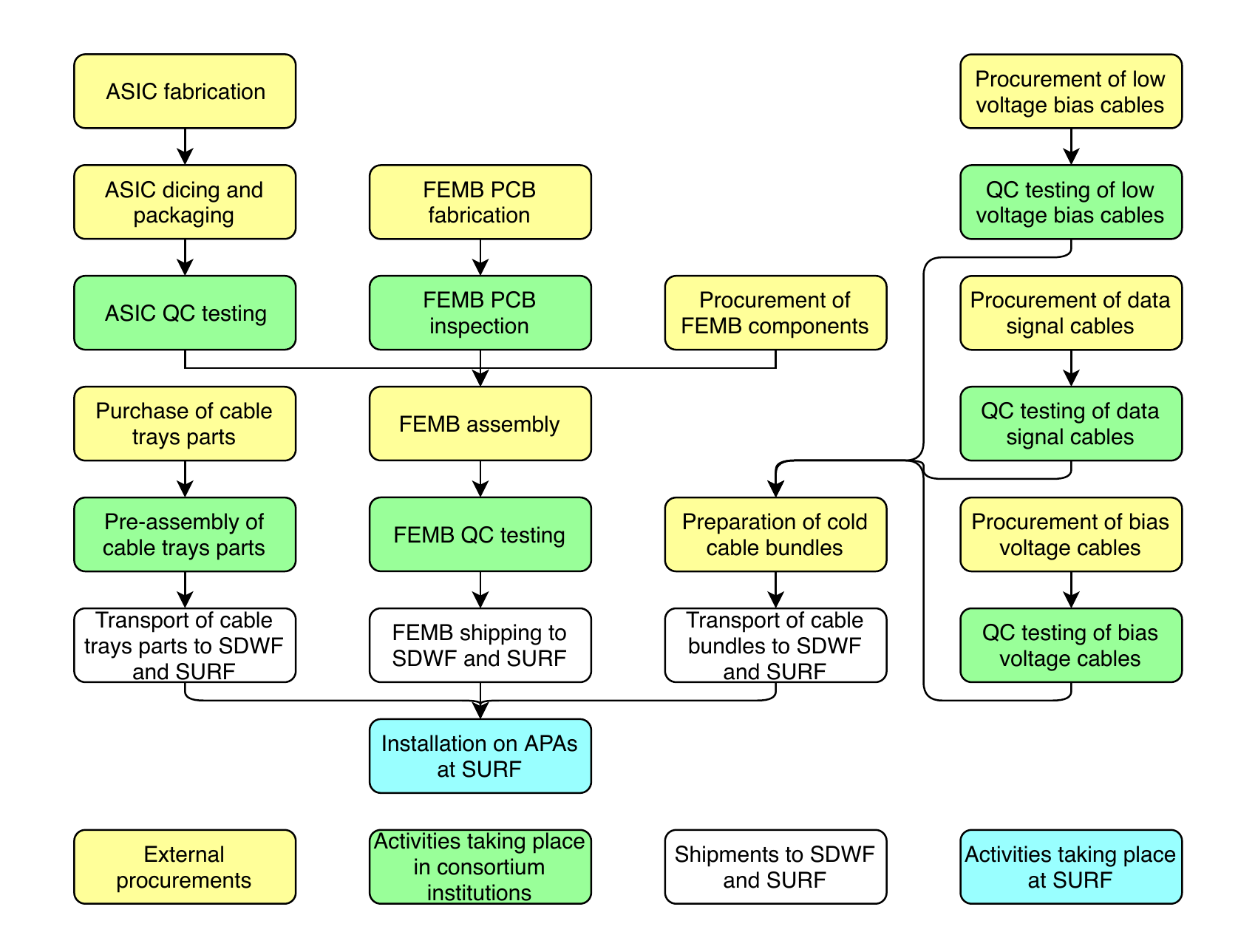}
\end{dunefigure}

\subsection{Assembly}
\label{sec:fdsp-tpcelec-production-assembly}

The \dword{tpc} electronics consortium plans to minimize
the amount of assembly work at any one of the participating
institutions. When assembly work is required, it will be performed
by external companies; examples are the installation of surface 
mount components, \dwords{asic}, \dwords{fpga} on the printed 
circuit boards for the \dwords{femb} and the \dwords {wib}, and
the assembly of the crossing tube cable supports. One of the few
exceptions is the assembly of the \dwords{wiec} that involves
mechanical and electrical connections at the backplane and crate supports.
Other activities that require 
work performed at one of the consortium institutions are the
assembly of the plugs attached to the cold cables, which are used to protect
the \dwords{femb} from \dword{esd} damage, and the preparation of
the bundles of low-voltage power, clock signal, trigger, data readout,
and bias voltage cables. During the engineering
phase and for components fabricated in small quantities, 
like boards used for testing other components, the plan is to
have one of the consortium's institutions assemble the components.
After assembly and testing, discussed below in Section~\ref{sec:fdsp-tpcelec-production-qc},
all detector components are shipped to the \dword{sdwf} and
later to \dword{surf}, where the final detector assembly
takes place as discussed in Chapter~\ref{ch:sp-install}.

\subsection{Quality Control}
\label{sec:fdsp-tpcelec-production-qc}

Once the \dword{apa}s are installed inside the cryostat, only
limited access to the detector components will be available to the \dword{tpc} electronics
consortium. After the \dword{tco} is closed, no access to detector
components will be available; therefore, they should be constructed to
last the entire lifetime of the experiment (\dunelifetime). This
puts very stringent requirements on the reliability of these
components, which has been already addressed in part through 
the \dword{qa} program discussed in Section~\ref{sec:fdsp-tpcelec-qa}. The
next step is to carefully apply stringent \dword{qc} procedures for  
detector parts to be installed in the detector.
All detector components installed inside the cryostat will
be tested and sorted before they are prepared for integration
with other detector components prior to installation. The full
details of the \dword{qc} plan have not been put in place
yet, and the specific selection criteria for the components will
be defined only after the current design and
prototyping phase is completed. For each detector component, a preliminary
version of the \dword{qc} program will be developed before the corresponding 
engineering design review. The program will then be used for
qualification of components fabricated during 
pre-production. It will be modified as needed before the production
readiness review that triggers the start of production of detector components
used for assembling the detector. In most cases the \dword{qc} program
will be informed by the experience gained with the tests of the corresponding
parts fabricated for \dword{pdsp}. Yields from the testing of \dword{larasic}
and of other discrete components mounted on the \dwords{femb} are discussed 
below. 

Some of the requirements for the \dword{qc} plans can
be laid out now based on the lessons learned
from constructing and commissioning the \dword{pdsp}
detector. Experience with \dword{pdsp} shows that a small fraction
(roughly \num{4}\%) of the \dword{larasic} chips that pass the
qualification criteria at room temperature fail the tests
when immersed in \lntwo. Therefore, we plan to test all \dwords{asic} in \lntwo
before they are mounted on the \dwords{femb}; 
cryogenic testing of the \dwords{femb} is also planned. The goal of testing 
the \dwords{asic} in \lntwo is to minimize the need
to rework the \dwords{femb}. This is more important if 
the three \dwords{asic} solution is chosen for the
\dword{femb}. Since in this case there are 18 \dwords{asic} on the
\dword{femb}, an upper limit of 2\% on the fraction of
\dwords{femb} that require reworking translates into a requirement 
of less than 0.1\% of the \dwords{asic} failing during immersion in \lntwo.
If the \dword{cryo} solution is chosen for the \dwords{asic}
to be used on the \dwords{femb}, the 2\% requirement for the
number of \dwords{femb} to be reworked changes to a 
maximum failure rate of 1\%, given that there are only two
\dwords{asic} on the \dwords{femb}. Based on experiences at \dwords{pdsp},
discrete components like resistors and capacitors
need not undergo cryogenic testing before they are installed
on the \dwords{femb}. Capacitors and resistors are commonly
sold in reels of a few thousand components, which
should be typically sufficient for the fabrication of ten
\dwords{femb}. For these components, we are planning to
perform cryogenic tests on samples of a few components
from each reel prior to using the reel in the assembly of
\dwords{femb}. Some other components installed on
the \dwords{femb}, like voltage regulators and crystal oscillators, 
will have to be qualified like the \dwords{asic}
in \lntwo before being mounted on the \dwords{femb}. In the
case of the voltage regulators, it was found that the number
of failures were negligible and that cryogenic testing was
not necessary. One component used for \dword{pdsp}
that we are not planning to use for the \dword{dune} \dword{fd} \dwords{femb},
the memory card used to store the \dword{fpga} programming,
had instead a very high failure rate ($>50$\%).

\begin{dunefigure}
[Parts flow for the detector components on top of the cryostat]
{fig:sp-tpcelec-partsflow2}
{Parts flow for the TPC electronics detector components installed on top of the cryostat.}
\includegraphics[width=0.7\linewidth]{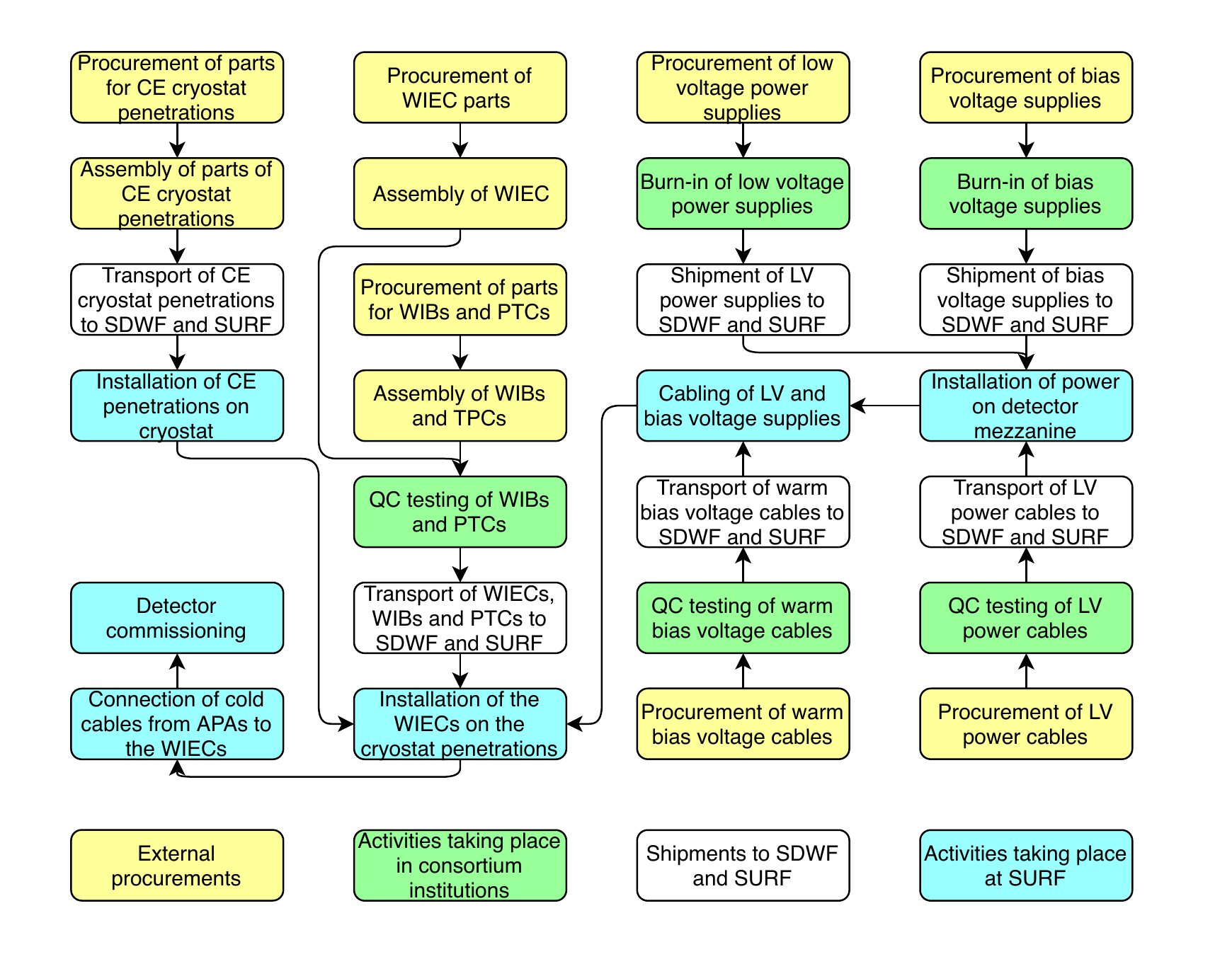}
\end{dunefigure}

\dword{asic} testing is performed with dedicated test boards that
allow tests of the functionality of the chips and are also used
to determine the initial calibration constants that are stored in
a database for later use. The dedicated test boards reproduce the
entire readout chain where the input to the \dword{fe} amplifier
or to the \dword{adc} is replaced by an appropriate signal generator,
and some parts of the backend may be replaced by a simple \dword{fpga}
that is directly connected to a computer. Tests of the \dwords{femb}
can be performed by connecting them directly to a standalone \dword{wib},
as discussed in Section~\ref{sec:fdsp-tpcelec-design-warm}. Given
the large number of \dwords{asic} and \dwords{femb} required for
one \dword{dune} \dword{fd} \dword{sp} detector, we plan to distribute the 
corresponding \dword{qc} activities among multiple institutions
belonging to the \dword{tpc} electronics consortium. Up to six
test sites are needed for the \dwords{asic} plus an additional
five sites for the \dwords{femb}, with each test 
equipped with a cryogenic system such as the \dword{cts}. All tests
will be performed following a common set of instructions.

The choice of distributing the testing activities among multiple
institutions has been made based in part on the experience gained
with \dword{pdsp}, where all associated testing activities were concentrated
at \dword{bnl}. While this approach had some advantages, like the direct availability
of the engineers that had designed the components, a strict conformance
to the testing rules, and a fast turn-around
time for repairs, it also required a very large commitment of
personnel from a single institution. Personnel from other institutions
interested in the \dword{tpc} electronics participated in the test
activities but could not commit for long periods of time. For
this reason, we are planning to distribute the \dword{qc} testing
activities for \dwords{asic} and \dwords{femb} among multiple
institutions belonging to the \dword{tpc} electronics consortium.
It should be noted that this approach is used in the \dword{lhc}
experiments for detector components like the silicon tracker 
modules where both the assembly and \dword{qc} activities take
place in parallel at multiple (of the order of ten) institutions.
To ensure that all sites produce similar
results, we will emphasize training experienced personnel
that will overview the testing activities at each site,
and we will have a reference set of \dwords{asic} and \dwords{femb}
that will be initially used to cross-calibrate the
test procedures among sites and then to check the
stability of the test equipment at each site. 
All testing activities for \dwords{asic} and \dwords{femb}
will be monitored by a member of the management of the \dword{tpc}
electronics consortium who will also have the responsibility of training the
personnel at all sites and conducting site
inspections to ensure that all safety and testing rules
and procedures are applied uniformly. Test results will
be stored in a database, and criteria will be developed
for the acceptance of \dwords{asic} and \dwords{femb}. The acceptance rate
will be monitored, and in case of problems,
the failures will be analyzed and root cause analyses
will be performed. If necessary, the test program will be stopped
at all sites while issues are being investigated. In the case of \dword{cryo},
since each \dword{femb} will only have two chips, it may be possible to bypass
chip-level testing altogether. If the chip-level failure rate is low enough,
it may be sufficient to simply test assembled \dwords{femb} and reject or rework
those that fail the tests.

For the large numbers of
\dwords{asic} required for one \dword{dune} \dword{fd} \dword{sp} detector
(6,000 or 54,000 chips depending on the \dword{asic} solution
chosen), manual testing of the chips requires excessive 
amounts of resources and, based on the lessons learned from  
constructing  \dword{pdsp}, would lead to unacceptable rejection
factors. Ideally, the entire testing
process would be performed using a robotic system, where 
a robotic arm picks up the \dword{asic} from a tray, places
it on a test board, and holds it in place while the test
is performed, followed by sorting it into a second tray
based upon the test result. The requirement that the test
be performed in \lntwo prevents us from using this
scheme. Two of the biggest problems observed during the
construction of \dword{pdsp} were related to the \dword{qc}
of \dwords{asic} in \lntwo. The first one, related to
condensation on the test boards, has been addressed with
the development of the \dword{cts}, discussed in Section~\ref{sec:fdsp-tpcelec-qa-initial}.
The second one is related to placement of the chips into the
sockets on the test boards, and leads to test failures
and in some cases damage to the chips and/or the sockets.
To overcome problems with the manual placement of the chips into
the sockets, we plan to develop a robotic system
to perform this operation. Once the \dwords{asic} are 
placed on test boards, they will be moved manually into 
upgraded versions of the current \dword{cts} that can
house multiple test boards. At the end of the testing
procedure, the robotic system will then remove
the chips from the test boards and sort them according to
the test results. Based on the experience with the tests of
the \dword{pdsp} \dwords{asic}, as well as from other experiments,
we plan to have the sockets on the test boards cleaned on a regular 
basis and then replaced after a certain number of
testing cycles. 

Before assembly, the printed circuit boards for the
\dwords{femb} will be tested by the production vendor for electrical
continuity and shorts. The usual approach for particle physics
experiments is to perform a visual inspection of the boards
before installing the discrete components and 
the \dwords{asic}. This inspection will be repeated after 
installation and before the functionality test, which for \dword{dune} will
be performed in \lntwo. The specifications on vias and pads for the printed
circuit boards for the \dwords{femb} are not outside the 
industrial vendors' capabilities, and therefore we do not
expect these inspections to be absolutely necessary. We will
perform visual inspections on a sample of 
production units, with a higher rate of sampling at the beginning
of the production. We will also investigate the possibility
of using other, possibly automatic, inspection methods for
the bulk of the production. After assembly, 
each \dword{femb} will be tested in \lntwo using
the current \dword{cts} design. Nine \dwords{cts} have already been
fabricated and are being distributed among the institutions
in the consortium. 

The test procedures are likely to be
very similar to the ones adopted for \dword{pdsp}, with the main
difference that the tests will not be performed with
the final cables to be used in the experiment but 
with a set of temporary cables. The final cables will be tested
separately as described below. The tests of the \dwords{femb}
are performed using the \dword{cts}, which allows a turnaround
time of about one hour per \dword{femb}. In the
test, the \dword{femb} is connected to a capacitive load that
simulates the presence of \dword{apa} wires. This allows
connectivity checks for each channel as well as measurements of
the waveform baseline and of the channel noise level. Calibration 
pulses will be injected in the front-end amplifier, digitized,
and read out. These injected pulses will also be used
to determine the calibration constants of the \dword{adc}. 
The test setup requires one \dword{wib} and
a printed-circuit board similar to those used on the cryostat
penetration, allowing simultaneous testing of four \dwords{femb}.
A standalone \SI{12}{V} power supply is required, and the readout
of the \dword{wib} uses a direct Gb Ethernet connection to
a PC. The setup used for \dword{asic} testing is similar.
In both cases, the data can be processed locally on the PC,
and the results from the tests and calibrations are then stored 
in a database. The plan is to have the capability to retrieve  
these test and calibration results throughout the entire life
of the experiment. As in the case of \dword{asic} testing,
we will monitor the test results to ensure that all
sites have similar test capabilities and yields and to
identify possible problems during production.
Further tests will be performed on the \dwords{femb}
before and after their installation on the \dword{apa}s, as
discussed in Section~\ref{sec:fdsp-tc-inst}.

The final component provided by the \dword{tpc} electronics consortium
and installed inside the cryostat is the ensemble of cold
cables: the cables carrying the bias voltage for the \dword{apa}
wires and the field cage termination electrodes;
the cables carrying the low-voltage power to the \dwords{femb};
and the data cables that carry the clock and control signals
to the \dwords{femb} that are also used for signal readout. It is neither
feasible nor necessary to test these cables in \lntwo
because they will usually perform better at cold temperatures than 
room temperature. We will perform checks on all cables 
during production at room temperature before they are installed and 
connected to the \dwords{femb}. These tests will involve continuity checks
and resistance measurements on the low-voltage power and the bias voltage
cables, and also bit-error rate measurements on the clock/control and 
data readout cables. Connectors will be visually inspected to
ensure that they show no sign of damage.  Further tests will take place
when the \dword{apa}s are tested in the cold boxes at \dword{surf}
prior to installation inside the cryostat.

Stringent requirements must be applied to the cryostat
penetrations in order to avoid argon leaks. The cryostat penetrations 
have two parts: the first is the crossing tube with its spool pieces,
and the second one is the three flanges used for
connecting the power, control, and readout electronics with the
\dword{ce} and \dword{pds} components inside the
cryostat. On each cryostat penetration there are two flanges for
the \dword{ce} and one for the \dword{pds}. The crossing
tubes with their spool pieces are fabricated by industrial vendors and pressure-tested
and tested for leaks by other vendors. The flanges are assembled
by institution that are members of the \dword{tpc} electronics and \dword{pds} consortia; the
flanges must undergo both electrical and mechanical tests to ensure their
functionality. Electrical tests comprise checking all of the
signals and voltages to ensure they are passed properly between the two sides of the
flange and that there are no shorts. Mechanical tests involve 
pressure-testing the flange itself, including checking for leaks. Further
leak tests are performed after the cryostat penetrations are installed
on the cryostat and later after the \dword{tpc} electronics and \dword{pds}
cables are attached to the flanges. These leak tests are
performed by releasing helium gas in the cryostat penetration and
checking for the presence of helium on top of the cryostat. Similar
tests were performed during the \dword{pdsp} installation.

All other detector components that are a responsibility of
the \dword{tpc} electronics consortium can be replaced, if necessary,
even while the detector is in operation. Regardless, every component
will be tested before it is installed in \dword{surf} to ensure
smooth commissioning of the detector. The \dwords{wiec} will be
assembled and tested with all of the \dwords{wib} and \dword{ptc}
installed. Testing requires a slice of the \dword{daq} back-end,
power supplies, and at least one \dword{femb} to check all 
connections. All cables between the bias voltage supplies and
the end flange, as well as all of the cables between the low-voltage power
supplies and the \dwords{ptc} will be tested for electrical
continuity and for shorts. All power supplies will undergo a
period of burn-in with appropriate loads before being installed
in the cavern. Optical fibers will be tested by measuring the
eye diagram for data transmission at the required speed. All
test equipment used for qualifying the components to be installed
in the detector will be either transported to \dword{surf} or duplicated
at \dword{surf} in order to be used as diagnostic tools during operations.

After the completion of the initial \dword{qc} testing, all detector 
parts are transported first to the \dword{sdwf} and later to \dword{surf}, 
where all the integration activities will take place as discussed in
Section~\ref{sec:fdsp-tpcelec-integration} and in Chapter~\ref{ch:sp-install}.

\subsection{Test Facilities}
\label{sec:fdsp-tpcelec-production-facilities}

The \dword{qc} plan described in the previous section requires
multiple test stands that must be put in place and used 
before the beginning of production. For the testing of \dwords{asic},
a setup similar to the \dword{cts} (see
Section~\ref{sec:fdsp-tpcelec-qa-initial}) will
be used. In this setup, several test boards housing up to 24
chips will be immersed in \lntwo before running the
electronics tests; the test boards will later be warmed to room temperature in 
a nitrogen atmosphere to avoid condensation on the chips and
boards. As mentioned previously, placing the chips on the test 
boards will be performed using a robotic system. The test setups,
one for each kind of \dword{asic}, will be an evolution of those
used initially for characterizing the \dword{asic} and similar
to the setups used for qualifying the chips used in the \dword{pdsp}
construction. The tests of the \dwords{femb} will be performed with
setups that include using \dwords{cts} units for the cryogenic part
but are otherwise simple modifications (with newer \dwords{wib})
of the setups used to characterize the \dwords{femb}
for \dword{pdsp}. Cold and warm power and bias voltage cables will
be characterized with test stations using appropriate
power supplies and some cable testing equipment; these cables will most
likely be \dword{cots} components.
For the test of the data cables, we will probably rely on a setup
using waveform generators and a high end oscilloscope that 
can handle 2.56 Gbps signals and measure eye diagrams. 
Burn-in stations, with custom-designed loads, may be required for 
the commercial low-voltage power and bias voltage supplies.
A test setup to check \dwords{wiec} with their \dwords{wib}
and \dwords{ptc} installed will require a minimal \dword{daq} back-end that the
\dword{daq} consortium should provide.

Given the delay between the beginning of the \dword{dune} \dword{fd}
\dword{apa}s production and the production of the \dword{tpc} electronics components,
it is desirable to integrate the \dwords{femb} on some of the \dword{apa}s 
and perform tests in cold boxes. For \dword{apa}s fabricated in the UK,
these tests will be performed at CERN using the \dword{pdsp} cold box.
A similar setup needs to be put in place in the US (most probably at 
the University of Wisconsin) to perform these tests ahead of the shipment of the \dword{apa}s to 
\dword{sdwf} and \dword{surf}. Both the setup at \dword{cern} and
the one in the US will require a full power, control, and readout system, similar
to the one described in Section~\ref{sec:fdsp-tc-inst-qaqc}.

\section{Integration, Installation, and Commissioning}
\label{sec:fdsp-tpcelec-integration}

Chapter~\ref{ch:sp-install} provides a complete discussion of the plans for 
integrating, installing, and commissioning the detector.
Here, we briefly discuss the responsibilities of the \dword{tpc} electronics
consortium for the activities taking place 
at \dword{surf}, with the exception of the \dword{qc} process that
is discussed in detail in Section~\ref{sec:fdsp-tc-inst-qaqc}. We also discuss the timeline
and the resources for the integration and installation activities.
Finally, we conclude with a discussion of the commissioning
of the \dword{tpc} electronics detector components that take place while
the cryostat is being filled and immediately after the fill is completed.

\subsection{Timeline and Resources}
\label{sec:fdsp-tpcelec-integration-timeline}

The current \dword{tpc} electronics consortium plan is to receive all detector 
components at the \dword{sdwf}, where they are stored temporarily
prior to being transported to \dword{surf} for integration and 
installation. Only the \dwords{femb} will undergo a reception test,
either in a laboratory on the surface at \dword{surf} or in a nearby
institution, prior to integration with the other \dword{dune} \dword{sp}
detector components. All other integration takes place at \dword{surf} in the clean room
in front of the detector cryostat. After a pair of \dword{apa}s are 
connected and moved inside the clean room, the \dword{ce} cables
for the bottom \dword{apa} are routed through the \dword{apa} frames.
The cables are then connected to the \dwords{femb}, and the bundles
of cables are placed in the trays at the top of the \dword{apa} pair.
At this point, the pair of \dword{apa}s is moved into one of the cold
boxes, and the cables are connected to a patch panel inside the cold box
to save the time that would be required for routing the cables through the cryostat
penetration of the cold box and connecting them to the end flange.
The \dword{ce} is then tested at both room temperature
and at a temperature close to that of \lntwo, much like
what was done for the \dword{apa}s installed in the \dword{pdsp} detector.

Later, the pair of \dword{apa}s is moved to its final position 
inside the cryostat. The \dword{ce}
and \dword{pds} cables are routed through the cryostat penetration and
connected to the corresponding warm flanges, and final leak tests are performed
on the cryostat penetration. At this point, the \dword{wiec} is attached
to the warm flange and all of the cables and fibers required to provide 
power and control signals to the \dword{tpc} electronics and for data
readout are connected. This permits additional testing with the full 
\dword{daq} readout chain and the final power and controls signals distribution
system. Once initial tests are completed successfully, more \dword{apa}s
can be installed, and the \dword{apa}s and their \dwords{femb} 
can remain accessible until the \dword{fc} are deployed.

This installation sequence assumes that all the \dword{tpc} electronics detector 
components required for readout of a pair of \dword{apa}s  on top of the cryostat 
are installed before \dword{apa}s are inserted into the cryostat. This 
includes the \dwords{wiec} with their boards, the power supplies in 
the racks, and all cables and fibers required to distribute power and
control signals as well as for detector readout. Installation should occur at least two weeks before
the \dword{apa}s are inserted into the cryostat to allow time for final checks. 

One exception
is installing the cryostat penetrations with the warm flanges
for both the \dword{ce} and the \dword{pds}. The cryostat
penetrations should be installed, at the latest,
at the same time the detector support structure is installed
inside the cryostat. This ensures the cryostat is almost 
completely sealed to minimize the amount of dust 
entering the cryostat. During the routing of the \dword{ce} and
\dword{pds} cables through the cryostat penetrations, dust entering the cryostat will be minimized by having a small
over-pressure inside the cryostat and by isolating each penetration
from the cavern using a tent mounted over 
the work area.

The schedule of activities at \dword{surf} is designed so all 
\dword{apa}s can be installed in the cryostat on a timescale of eight
months, proceeding at a rate of one row of six \dword{apa}s per week and
allowing for a ramp-up period at the beginning of the process. This
requires that personnel from the \dword{tpc} electronics consortium be available
for two \num{10}-hours shifts per day at \dword{surf} at all times, including weekends. A
total of 30 FTEs/week will be needed to install and test the \dword{apa}s,
under the assumption that there will be a maximum of four shifts per person
per week. 

The installation of all other \dword{tpc} electronics detector components
takes place on the top of the cryostat. The cryostat penetrations are installed
ahead of the installation of the \dword{apa}s inside the cryostat, ideally as
soon as the welding of the cold membrane is completed in part of the cryostat. 
This activity requires at team of eight FTEs, split over two shifts per day, for a 
period of one month. A similar amount of time and personnel is also required
to install the power supplies in the racks, attaching the \dwords{wiec} to
the \dword{ce} flanges, and routing and connecting the warm cables.

\subsection{Internal Calibration and Initial Commissioning}
\label{sec:fdsp-tpcelec-integration-calib}

While the cryostat is being closed (and any time there is welding 
on the cryostat), the electronics should be turned off and all 
cables between the detector racks, including the low-voltage power
and bias voltage, fans, and heater power, should be disconnected 
from the \dwords{wiec}. Once the cryostat is closed, the waveform baseline 
and noise level of all channels should be measured. Dead electronics channels 
should be identified by measuring the response of all channels to 
the internal electronics calibration pulser at a nominal setting, 
such as $\pm\SI{600}{mV}$, which distinguishes between induction 
and collection channels. The noise levels should be measured with the wire bias 
voltages fully enabled on the $G$, $U$, and $X$ planes of the \dword{apa}s. 
It should also be measured with the cathode high voltage on at a very 
low value, e.g. \SI{50}{V}. The non-responsive channels, identified
as having very low noise levels, and the channels that have noise levels that 
significantly exceed the average value should be flagged and recorded.
Sources of excess noise should be identified
and, if possible, fixed. Any warm electronics components with
issues should be replaced with spares.

Once the cryostat is filled with gaseous 
argon, the waveform baseline and noise level of all of the channels will be measured
again, and any new non-responsive channels in the electronics 
should be identified by injecting $\pm\SI{600}{mV}$ with the 
internal calibration pulser. As the cryostat is cooled down, the 
temperature at the electronics and the noise level of all channels should 
be monitored periodically. Any new non-responsive channels should 
be flagged and excess noise sources that are exposed as the 
electronics cools down should be identified and, if possible, fixed.

Once the electronics is fully submerged in \dword{lar}, a full 
set of electronics diagnostic tests should be run, including waveform
baseline and noise level measurements as well as a full gain calibration
on all channels using the internal calibration pulser at settings 
up to the saturation of the \dword{fe} inputs. The shaping time 
should be measured on all channels by injecting the $\pm\SI{600}{mV}$
internal pulser at each of the four settings and fitting the 
pulse shape. Any new non-responsive channels during the pulser 
runs should be flagged. Any new disconnected channels should be 
flagged and excess noise sources should be identified. These tests 
can be performed on the electronics installed on the bottom
\dword{apa}s even while the corresponding wires are in 
the gaseous argon.

\section{Interfaces}
\label{sec:fdsp-tpcelec-interfaces}

Table~\ref{tab:CEinterfaces} contains a brief summary of all of the interfaces
between the \dword{tpc} electronics consortium and other consortia or groups,
with references to the current version of the interface documents. 
In some cases, the interface documents involve more than one 
consortium (one example is the bias voltage distribution system where
the interface involves both the \dword{apa} and the \dword{hv} consortia).
In such cases, the goal is to have all of the corresponding interface documents 
consistent. At this stage, most of the interface documents are
not yet complete; drawings of the mechanical interfaces and diagrams
of the electrical interfaces are still under development. The interface 
documents should be further refined during the 
first half of 2020 before the engineering design
reviews of the detector. All interface documents specify the responsibilities
of different consortia or groups during all phases of the experiment,
including design and prototyping, integration, installation,
and commissioning. In the remainder of this section, the most
important interfaces, specifically those with the \dword{apa}, \dword{daq}, and \dword{hv}
consortia, as well as the interface with technical coordination, are discussed in detail.
Finally, a brief overview of the remaining interfaces is also presented.

\begin{dunetable}
[TPC electronics system interfaces]
{p{0.25\textwidth}p{0.5\textwidth}l}
{tab:CEinterfaces}
{\dword{tpc} electronics system interfaces. }
Interfacing System & Description & Linked Reference 
\\ \toprowrule
\dword{apa} & Mechanical (cable trays, cable routing, connections of CE boxes and 
frames) and electrical (bias voltage, \dword{femb}--\dword{cr} boards connection, grounding 
scheme) & \cite{bib:docdb6670}
\\ \colhline
\dword{daq} & Data output from the \dword{wib} to the \dword{daq} back-end, clock signal distribution,
controls and data monitoring responsibilities & \cite{bib:docdb6742}
\\ \colhline
\dshort{cisc} & Rack layout, controls and data monitoring & \cite{bib:docdb6745}
\\ \colhline
\dword{hv} & Grounding, bias voltage distribution, installation and testing & \cite{bib:docdb6739}
\\ \colhline
\dword{pds} & Electrical (cable routing and installation), cold flange & \cite{bib:docdb6718}
\\ \colhline
Facility & Cable trays inside the cryostat, cryostat penetrations, rack layout and
power distribution on the detector mezzanine, cable and optical fiber trays on top of the
cryostat& \cite{bib:docdb6973}
\\ \colhline
Installation Team & Integration and installation activities at \surf,
equipment required for  \dword{tpc} electronics consortium activities, 
cold boxes for \dword{apa} tests, material handling  & \cite{bib:docdb7000}
\\ \colhline
Physics & Responsibilities for simulation and reconstruction software & \cite{bib:docdb7081}
\\ \colhline
Software \& Computing & Database needs for storing detector calibration and configurations, computing needs for calibration and monitoring & \cite{bib:docdb7108}
\\ \colhline
Calibration & Access to low-level electronics calibration for high-level physics calibration & \cite{bib:docdb7054}
\\ 
\end{dunetable}

\subsection{APA}
\label{sec:fdsp-tpcelec-interfaces-apa}

The most important interface is between the \dword{tpc} electronics
and the \dword{apa} consortia. The design of the \dwords{femb}
and of the \dword{apa}s are intertwined, both from the
mechanical and electrical points of view. The \dword{ce}
boxes, which house the \dwords{femb}, are supported by the \dword{apa}
and are attached to the \dword{cr} boards of the \dword{apa}
through a connector that passes all signals from the wires to
the \dword{fe} amplifiers. The cable trays that house both the
\dword{ce} and the \dword{pds} cold cables are initially
attached to the yoke of the upper \dword{apa}. The \dword{ce}
cables for the lower \dword{apa} must be routed through the 
frames of both the bottom and top \dword{apa}s. The \dword{tpc} electronics
consortium provides the bias voltage for the \dword{apa}
wires as well as for the electron diverters and the \dword{fc} termination electrodes (the latter are a responsibility of
the \dword{hv} consortium) using the \dword{shv} boards mounted
on the \dword{apa}s. The grounding requirements discussed in
Section~\ref{sec:fdsp-tpcelec-design-grounding} inform the
design of all mechanical and electrical interfaces between
the \dword{ce} components and \dword{apa}s as well as the
design of the connections between the top and bottom \dword{apa}s
and between the top \dword{apa} and the \dword{dss}. All 
integration and installation activities at \dword{surf}
must be carefully coordinated by the two consortia and
where appropriate also with the \dword{pds} consortium 
and \dword{tc}.

\subsection{DAQ}
\label{sec:fdsp-tpcelec-interfaces-daq}

The \dword{daq} system receives the data produced by the
\dword{tpc} electronics detector components, further processes this data to
form trigger decisions, and finally transfers the data to 
permanent storage for analysis. The 
\dword{daq} system is also responsible for delivering the clock and control
signals to the \dwords{wiec}. The interfaces are realized 
through optical fibers, ensuring that no electrical noise is fed into
the \dwords{wiec}. One fiber per \dword{wiec} delivers the
clock and control signals to the \dword{ptc}, which then
rebroadcasts the information to the \dwords{wib} in that 
crate. Each \dword{wib} reads out the data from four \dwords{femb}
and transmits the data through two \SI{10}{Gbps} links to the \dword{daq} back-end.

The data signals from the \dwords{wiec} to the \dword{daq} system are carried on
multi-mode optical fibers, compatible with either the \dword{om3} or \dword{om4} standards. Individual
fibers will be merged into bundles of 12 fibers with MTP connectors near the 
\dword{ce} cryostat penetrations. These bundles will be merged in trunks of 144 
fibers on the detector mezzanine, which will then be fanned out to individual
optical fibers inside the \dword{cuc}. The
feasibility of this data transmission scheme using optical fibers with a length up
to \SI{300}{m} has been demonstrated at \dword{bnl} in summer 2018 using 
\dword{pdsp} components. The data format used for 
\dword{dune} will be a modification of the one adopted for
\dword{pdsp}, taking into account the need for an 
extended address space to accommodate the larger number of
\dwords{femb} in the detector. The \dword{daq} consortium
is also responsible for providing the software environment
used for downloading the detector configuration.

\subsection{HV}
\label{sec:fdsp-tpcelec-interfaces-hv}

The \dword{hv} consortium interface 
is driven by the fact that the \dword{ce} flange provides the return path for
the small current that flows from the high-voltage power 
supply through the cathode panels, the \dword{fc}, and finally
the termination electrodes. The hardware interface uses
the \dword{shv} boards mounted on the \dword{apa}s, which
are  the responsibility of the \dword{apa} consortium. The
\dword{shv} boards also distribute the bias voltage to the 
\dword{fc} termination electrodes. The \dword{tpc} electronics 
consortium is responsible for bringing the bias voltage for the 
\dword{fc} termination electrodes to the \dword{shv} boards. Appropriate
rules for avoiding ground loops are also included in the 
interface document.

\subsection{Technical Coordination}
\label{sec:fdsp-tpcelec-interfaces-tc}

In this section, we consider the interfaces with \dword{lbnf} 
and with the \dword{dune} \dword{tc}, including the
interfaces with the \dword{jpo} that oversees the integration
and installation activities that take place at \dword{surf}.
The \dword{tpc} electronics consortium has several
interfaces with the facility, namely the cable trays inside the
cryostat, the cryostat penetrations used by the \dword{tpc} electronics
and \dword{pds} consortia, and the racks and trays on top
of the cryostat. The \dword{tpc} electronics consortium is responsible
for the design, procurement, and installation of the cable trays
inside the cryostat and the cryostat penetrations. The \dword{dune}
\dword{tc} is responsible for providing the racks
where the low-voltage power supplies and the bias voltage supplies are installed,
including their power, cooling, and monitoring systems, as well as the hardware interlock system.
\Dword{tc} is also responsible for the trays connecting these racks 
to the corresponding \dword{wiec} and for the network switches that
connect the controls for the \dword{tpc} electronics to the
\dword{daq} and slow controls back-ends. Finally, \dword{tc} will
provide the \dword{ddss} that will protect
the \dword{tpc} electronics detector components. The \dword{tpc} electronics
consortium will work with \dword{tc} to establish
the action matrix for the \dword{ddss} and the hardware
interlocks, and also to resolve any electronics noise problems
that may be caused by improper grounding of detector components.

The \dword{tpc} electronics consortium will work with the teams responsible
for the underground integration and installation in order
to plan all activities that take place at \dword{surf}.
This includes developing plans to outline the responsibilities
of the consortium and those of \dword{tc} personnel
for the activities at \dword{surf}, including all 
shipments, transport, and logistics, as well as all
integration and installation. Cold boxes for testing
the \dword{apa}s after integration with the \dwords{femb} will be 
provided by \dword{tc} at \dword{surf}.
All other testing equipment will be provided either by the \dword{tpc} electronics
consortium or by other consortia. Equipment required to
minimize risk of \dword{esd} damage to the detector components
will  be provided by the \dword{tpc} electronics consortium. 

\subsection{Other Interfaces}
\label{sec:fdsp-tpcelec-interfaces-other}

The interface with the \dword{pds} consortium is relatively simple.
The \dword{pds} detector component should be isolated from the \dword{ce}
detector component other than sharing a common reference 
voltage point (detector ground) at the chimneys. Inside the cryostat, the 
\dword{pds} and \dword{ce} cables will be housed together in
cable trays that are the responsibility of the \dword{tpc} electronics
consortium. The \dword{tpc} electronics consortium will also assume
the responsibility for routing the \dword{pds} cables through the
cryostat penetration and for connecting them to the \dword{pds}
flange. The \dword{tpc} electronics consortium will also connect
the cables that run from the flange
to the mini-racks housing the \dword{pds} warm electronics. 
The flange itself will be designed and built by the \dword{pds}
consortium, but its integration on the spool piece
of the cryostat penetration will be the responsibility of the \dword{tpc} electronics
consortium.

The \dword{cisc} consortium provides the software infrastructure for the slow
control and monitoring of the status of the \dword{tpc} electronics components.
The \dword{cisc} and \dword{tpc} electronics consortia may also have hardware
interfaces because they may share the same racks on top of the
cryostat. The most important aspect of the interface between these
two consortia is the requirement from the \dword{tpc} electronics consortium
to have all relevant parts of the slow control and monitoring
equipment functional at the beginning
of the installation for \dword{surf}. 

The \dword{tpc} electronics consortium is responsible for many parts of
the \dword{dune} simulation and reconstruction software, which constitute
the interface with the Physics group. These include the
simulation of the material of the detector components (\dwords{femb}, cables, and
cable trays) inside the \dword{lar}, the simulation
of the response of the electronics, from the signal formation to its 
digitization, and finally the methods for the deconvolution of the electronics
response to be used in the event reconstruction software (the latter responsibility is shared 
with the \dword{apa} consortium). Many of these
software tools  already exist and have been used for the simulation
and the reconstruction of \dword{pdsp}. In the coming years they will be
updated to reflect the changes to the detector design relative to 
\dword{pdsp}. Similarly, the \dword{tpc}
electronics consortium has an interface with the Software and Computing
consortium that covers mostly the need for data storage and data access.
This includes both database needs (for storing detector calibration constraints
and detector configurations) and disk space needs (for storing the data 
from special calibration runs that are used to obtain the calibration
data). The calibration data on the response shape of the electronics and
on its gain are also relevant to the Calibration consortium,
which is tasked with obtaining high-level calibrations of the detector
response. When doing this, the most up-to-date electronics calibrations,
or in some cases special electronics calibrations, will be required.

\section{Safety}
\label{sec:fdsp-tpcelec-safety}

Personnel safety during construction, testing, integration,
and installation of the \dword{tpc} electronics components for the \dword{dune}
\dword{sp} \dword{fd} is crucial for the success
of the project. The members of the \dword{tpc} electronics consortium will
respect the safety rules of the institutions where the work is
performed, which may be one of the national laboratories, \dword{surf},
or one of the universities participating in the project. A
preliminary analysis of the risks involved in the design,
construction, integration, and installation of the detector
components provided by the \dword{tpc} electronics consortium has been
performed using the approach discussed in~\tcchesh.

\subsection{Personnel Safety during Construction}
\label{sec:fdsp-tpcelec-safety-personnel}

The main risks for consortium personnel are exposure to
the \lntwo used for cooling down components during testing (risk HA-8
in Table 11.25 in~\tcchappx), falls from heights (risk HA-1), 
electrical shocks (risk HA-6), and oxygen deficiency hazards, possibly 
caused by leaks of either \lntwo or \dword{lar} from test setups (risk HA-8).
The leadership of the
\dword{tpc} electronics consortium will work with the \dword{lbnf-dune}
\dword{esh} manager and other relevant responsible personnel at the
various institutions to ensure all the members of the
consortium, including students and postdocs, receive the appropriate training for the work they
are performing and that all preventive measures to minimize
the risk of accidents are in place. Where appropriate,
we will adopt the strictest standard and requirements among
those of different institutions. Hazard analyses will be performed,
and the level of \dword{ppe} will be determined
appropriately for each task. \Dword{ppe} includes 
appropriate gloves for handling \lntwo dewars, fall
protection equipment for work at height, and steel-toed shoes and
hard hats for integration work with the \dword{apa}. Oxygen
monitors should be used in areas with large potential concentrations of
cryogenic gases.

\dword{esh} plans for the activities to be performed in various
locations, including all universities, national laboratories,
and \dword{surf}, will be discussed in the 
various reviews (Preliminary Design, Engineering Design, Production 
Readiness, Production Progress) that will take place during the construction of the detector.
This will include inspections by a \dword{lbnf}/\dword{dune} \dword{esh}
representative at all the sites where detector components will be 
assembled or tested to ensure conformance to all of the safety
rules.

\subsection{Detector Safety during Construction}
\label{sec:fdsp-tpcelec-safety-detcon}

In addition to personnel safety during detector
construction, including all testing, 
integration, installation, and commissioning, we have also
considered how to protect the detector
components and minimize any chances of damaging
them during handling. We identified two main risks 
to the safety of the detector during construction and one risk during
operation. The most important risk during construction is damage 
induced by \dword{esd} in the 
electronics components. The second risk is mechanical damage to 
parts during transport and handling. For operation risks, we
must consider the risk of damage to the electronics 
caused by accumulated dust inside the components
installed on the top of the detector. In this section, 
we discuss these three risks and ways to minimize their possible 
effect. In the following section, we discuss how to prevent
damage during operation to the \dword{tpc} electronics components 
by using the interlocks of the detector safety system.

\dword{esd} can damage any of the electronics
components mounted on the \dwords{femb}, \dwords{wiec},
the bias voltage supplies, or the power supplies. If 
the damage occurs early in construction, 
the outcome is a reduction
of the yield for some of components, which must be
addressed by keeping a sufficient number of spares on hand to prevent
schedule delays associated with procuring new parts. \dword{esd}
damage on the \dwords{femb} after the \dword{apa}s have been
installed inside the cryostat could result in a permanent
reduction of the fraction of operating channels in the
detector. Even if most components, including the custom 
\dwords{asic} designed for use in the \dwords{femb}, contain 
some level of protection  against \dword{esd}, the possibility of this kind of damage cannot be ignored, and appropriate 
preventive measures must be taken during  
assembly, testing, installation, and shipping of all of the detector 
components provided by the \dword{tpc} electronics consortium. These 
measures include using appropriate \dword{esd}-safe packing materials, 
appropriate clothing and gloves, wearing conducting wrist or foot straps 
to prevent charges from accumulating 
on workers' bodies, anti-static mats to conduct harmful electric 
charges away from the work area, and humidity control. All laboratories with detector components provided by the \dword{tpc} electronics consortium will implement these
measures, including \dword{surf}. 

Additional measures
include using custom-made terminations for 
power, control, and readout cold cables when these
are being routed through the \dword{apa} frames or through the
cryostat penetrations. Storage cabinets where \dwords{asic} and
\dwords{femb} are stored should have \dword{esd} mats
on the shelves and humidity control. Most importantly, all personnel must be trained to take the appropriate preventive measures. We 
will require that all the personnel working on the \dword{tpc} electronics 
consortium components take a training class originally developed 
at \dword{fnal} for handling the charge-coupled devices of the Dark Energy Survey (DES) experiment 
(the material for this training class can also be used at remote 
sites). The scientists in charge of the \dword{tpc} electronics activities at each site involved in the project will be responsible for monitoring the training of personnel at universities, national laboratories,
and \dword{surf}.

Most of the damage to detector components happens during 
transport among the various sites where assembly, testing, integration,
and installation take place. When appropriate, measures to prevent
\dword{esd} damage must also be taken for shipments. Appropriate 
packaging will be used to ensure that parts are not damaged
during transport. We will perform tests upon receiving  
\dwords{femb} as well as integration tests for cold cables as
part of the quality control process discussed in 
Section~\ref{sec:fdsp-tpcelec-production-qc} in order to ensure the 
full functionality of these parts, which are very hard to replace 
after detector integration and installation. For  
components on the top of the cryostat that
can be replaced if damaged during transport, we will 
perform integration tests after installation.

In addition to damage during shipping, we must also consider the
possibility of damage caused by handling of the detector parts.
Additional precautions are being considered for operations where
the risk of damage to \dword{tpc} electronics detector components is
high. For testing \dwords{asic} and \dwords{femb} 
in \lntwo, this has resulted in developing the
\dword{cts} discussed in Section~\ref{sec:fdsp-tpcelec-qa-initial}
to prevent condensation on components after they are extracted from
the \lntwo at the end of a test. For the cold cables, this 
includes modifying the size of the tubes used for the \dword{apa} frames,
adding a conduit inside the frames, and placing a mesh around 
the cables. Special tooling will be designed for arranging the
cold cables on the spools used when cables are routed through
the \dword{apa} frames. Similarly, tooling will be developed to 
support the cold cables while they are being routed through the 
cryostat penetrations. All cables and optical fibers will
be installed in cable trays on top of the cryostat, and for the fibers, additional protection in the form of sleeves or tubes 
may also be used. 

To ensure that the \dword{dune} detector will be operational for a long
time, we also will attempt to minimize 
damage that could happen to detector components inside the 
experimental cavern, which can come from two sources: incorrect
operation of the detector and environmental conditions. We
will discuss the former in the next section. Once the cryostat is filled with \dword{lar}, the environmental
conditions inside the cryostat are extremely stable. Experience from 
previous experiments using electronics inside 
\dword{lar} indicates that, apart from initial problems, little loss of readout channels occurred over long periods. 

Therefore, the main safety concern is related to the electronics 
installed on top of the cryostat. The main problem in this case
is dust accumulation on detector components. Over the long
term, dust could damage the cooling fans used
in the \dwords{wiec} and the \dword{tpc} electronics racks, cause break-down 
on the surface of diodes used in bias voltage supplies,
and, if the dust contains any chemical residue that may
create deposits on printed circuit boards, create leakage paths
between traces on printed circuit boards. While the
experimental cavern should be a very dry environment,
protections should still be in place to prevent water from 
dripping on the \dwords{wiec} and on the racks containing the
\dword{tpc} electronics power and bias voltage supplies. Appropriate filters will be added to the
air supply used to cool the \dwords{wiec} and the \dword{tpc} electronics power and bias voltage supplies, thereby minimizing
the accumulation of dust. The air humidity in the cavern 
will be controlled to prevent condensation.

\subsection{Detector Safety during Operation}
\label{sec:fdsp-tpcelec-safety-detops}

In this section, we discuss where we will use
the detector safety system described in \tcchesh.
To avoid unsafe conditions for the \dword{tpc} electronics detector 
during operations, hardware interlocks will be put in place
in test setups at \dword{surf} to prevent operating or even powering up
detector components 
unless conditions are safe
both for the detector and for personnel. Interlocks will be
used on all low-voltage power and on bias voltage 
supplies, including inputs from environmental monitors both
inside and outside the cryostat. Examples of these interlocks include turning off the power to the \dwords{wiec}
if the corresponding cooling fans are not operational or
if the temperature inside the crates exceeds a preset value.
Similar interlocks will be used for low-voltage power
and bias voltage supplies in \dword{tpc} electronics racks.
Interlocks may be needed to connect the value of the 
bias voltage on the \dword{fc} termination electrodes to the
high voltage applied on the \dword{tpc} cathode. Interlocks will turn 
off transmitters on the \dwords{wiec} if the readout fibers'
bundles are cut. One problem we must address is 
the connection between the \dword{plc} used by the detector 
safety system and the \dwords{wiec} to avoid introducing noise 
inside the detector. We can easily decouple the environmental 
sensors required by the detector safety system inside the 
\dwords{wiec} by following the appropriate grounding rules. 
The connection used to provide the enable/disable signals 
from the \dword{plc} to the \dwords{wiec} will require optical 
fibers to avoid possible ground loops. Interlocks connected
to the detector safety system will also be used during tests 
of the \dword{apa}s in the cold boxes at \dword{cern}, \dword{fnal},
and \dword{surf}. The \dword{cts} has its own interlock system to
prevent condensation from forming on the \dwords{femb} once
they have warmed to room temperature. We cannot exclude the possibility that for some of the smaller test stands we will 
have to rely on software interlocks for detector safety,
but this should be kept to a minimum, and no software
interlock should be used for the cold boxes at
\dword{cern}, \dword{fnal},
and \dword{surf}.

\section{Risks}
\label{sec:fdsp-tpcelec-risks}

In this section, we discuss the risks that could be encountered during design
and construction of the \dword{dune} \dword{sp} \dword{fd}, as well as the
risks that can be encountered later during commissioning and operation.
For every risk, we will describe the mitigating actions 
being put in place even now at the design stage of the experiment, and the 
possible responses that we will take should a specific risk be realized. 
Table~\ref{tab:risks:SP-FD-TPCELEC} contains a list of all risks that we are 
considering. For each risk, we assess a probability for a risk to be
realized (P), as well as cost (C) and schedule (S) impacts, after the 
mitigation activities discussed in the text are put in place. It should
be noted that in the case of poor lifetime of the 
components installed inside the cryostat, there is no cost or schedule
impact, as they cannot be accessed and replaced.
All of these risks are discussed in detail in the remainder of this
section.

\begin{footnotesize}
\begin{longtable}{P{0.18\textwidth}P{0.20\textwidth}P{0.32\textwidth}P{0.02\textwidth}P{0.02\textwidth}P{0.02\textwidth}} 
\caption[TPC electronics risks]{TPC electronics risks (P=probability, C=cost, S=schedule) The risk probability, after taking into account the planned mitigation activities, is ranked as 
L (low $<\,$\SI{10}{\%}), 
M (medium \SIrange{10}{25}{\%}), or 
H (high $>\,$\SI{25}{\%}). 
The cost and schedule impacts are ranked as 
L (cost increase $<\,$\SI{5}{\%}, schedule delay $<\,$\num{2} months), 
M (\SIrange{5}{25}{\%} and 2--6 months, respectively) and 
H ($>\,$\SI{20}{\%} and $>\,$2 months, respectively). \fixmehl{ref \texttt{tab:risks:SP-FD-TPCELEC}}} \\
\rowcolor{dunesky}
ID & Risk & Mitigation & P & C & S  \\  \colhline
RT-SP-TPC-001 & Cold ASIC(s) not meeting specifications & Multiple designs, use of appropriate design rules for operation in LAr & H  & M & L \\  \colhline
RT-SP-TPC-002 & Delay in the availability of ASICs and FEMBs & Increase pool of spares for long lead items, multiple QC sites for ASICs, appropriate measures against ESD, monitoring of yields & M & L & L \\  \colhline
RT-SP-TPC-003 & Damage to the FEMBs / cold cables during or after integration with the APAs & Redesign of the FEMB/cable connection, use of CE boxes, ESD protections, early integration tests & M & L & L \\  \colhline
RT-SP-TPC-004 & Cold cables cannot be run through the APAs frames & Redesign of APA frames, integration tests at Ash River and at CERN, further reduction of cable plant & L & L & L \\  \colhline
RT-SP-TPC-005 & Delay and/or damage to the TPC electronics components on the top of the cryostat & Sufficient spares, early production and installation, ESD protection measures & L & L & L \\  \colhline
RT-SP-TPC-006 & Interfaces between TPC electronics and other consortia not adequately defined & Early integration tests, second run of ProtoDUNE-SP with pre-production components & M & L & L \\  \colhline
RT-SP-TPC-007 & Insufficient number of spares & Early start of production, close monitoring of usage of components, larger stocks of components with long lead times & M & L & L \\  \colhline
RT-SP-TPC-008 & Loss of key personnel & Distributed development of ASICs, increase involved of university groups, training of younger personnel & H & L & M \\  \colhline
RT-SP-TPC-009 & Excessive noise observed during detector commissioning & Enforce grounding rules, early integration tests, second run of ProtoDUNE-SP with pre-production components, cold box testing at SURF & L & L & M \\  \colhline
RT-SP-TPC-010 & Lifetime of components in the LAr & Design rules for cryogenic operation of ASICs, measurement of lifetime of components, reliability studies & L & n/a & n/a \\  \colhline
RT-SP-TPC-011 & Lifetime of components on the top of the cryostat & Use of filters on power supplies, stockpiling of components that may become obsolete, design rules to minimize parts that need to be redesigned / refabricated & L & M & L \\  \colhline

\label{tab:risks:SP-FD-TPCELEC}
\end{longtable}
\end{footnotesize}

\subsection{Design and Construction Risks}
\label{sec:fdsp-tpcelec-risks-design}

Despite the successful operation of \dword{pdsp}, we cannot build and operate
the \dword{dune} \dword{sp} \dword{fd} using the same \dwords{asic}; we are
improving the design of the \dword{fe} amplifier, redesigning
the \dword{adc}, and replacing the \dword{fpga} on the \dword{femb} with a dedicated data
serialization and transmission \dword{asic} (\dword{coldata}). The project schedule has
sufficient time for a second submission of all of these \dwords{asic} in the current 
development cycle. We  nevertheless must consider the risk (RT-SP-TPC-001
in Table~\ref{tab:risks:SP-FD-TPCELEC}) that, even after the second iteration, 
we may not have a set of \dwords{asic} that meets all the \dword{dune} requirements.

To reduce the probability of this happening, we are pursuing the development of the \dword{cryo} \dword{asic}, and we
will also perform system tests of the \dword{cots} solution for the \dword{adc} that is planned 
for use in the \dword{sbnd} experiment. Should this risk become reality, an
additional development cycle would be required, which would delay the availability
of \dwords{femb} for approximately 12 months. Based on the current schedule, this would
not be a problem for the detector integration and installation.
More importantly, such a delay would require
that, during the first part of the \dword{apa} production, a sufficient number
of \dwords{femb} with non-final electronics be available for integration tests
on the \dword{apa}s. These boards would then have to be replaced later
by the final boards.

The second risk (RT-SP-TPC-002) is a general delay in the availability of the \dwords{asic}
and/or \dwords{femb}, which would then not be available for integration
on the \dword{apa}s. This risk has several possible triggers: a lower fabrication yield 
than expected for \dwords{asic} and \dwords{femb},
significant downtime at one of the \dword{qc} sites, or losses during 
handling and transport, in addition to the issues with the design already 
covered in the previous risk. Our planning includes several ways to mitigate this risk. By
procuring spares for the \dwords{asic} and the discrete components to be mounted on the
\dwords{femb}, we can continue integration with the \dword{apa}s;
we are also splitting the \dword{qc} process among various sites. We will also emphasize,
as discussed in Section~\ref{sec:fdsp-tpcelec-safety-detcon}, the use of appropriate
measures that minimize the probability of any \dword{esd} damage. We will then monitor the use
of spare components and, if needed, fabricate additional parts. Appropriate
monitoring of the production yield and spares should minimize
delays in \dword{apa} integration. As in the case of the first risk, should this
second risk become reality, we do not expect to have delays in the installation of the 
detector inside the cryostat. 
 
The third risk (RT-SP-TPC-003) is the possibility of damage to the 
\dwords{femb} during or after integration with the \dword{apa}s. This could
happen while the \dwords{femb} are being installed or, more likely, during the
installation of the cold cables, something that has already taken place during
the installation of \dword{pdsp}. This damage could also happen when the \dword{apa}s are
moved into the cryostat or during the final cabling of the detector. The damage
to the \dwords{femb} and cold cables could be either mechanical damage during 
handling or \dword{esd} damage. We are redesigning the connection between the \dwords{femb}
and the cold cables to minimize the probability of the former and putting procedures in place
to minimize the probability of the latter. The risk of mechanical damage
is further reduced by housing the \dwords{femb} inside the \dword{ce} boxes and by having 
appropriate strain relief on the cold cables. To minimize the possibility of
\dword{esd} damage, we will follow all appropriate procedures, and in addition,
we will use plugs on all the cold cables while they are routed through the
\dword{apa} frames and cryostat penetrations in order to avoid injecting charge on
the \dwords{femb} that could cause \dword{esd} damage. Finally, we are planning for
extensive testing of the \dword{ce} several times during integration
and installation that would allow us to replace the \dwords{femb}
or the cold cables if necessary. This includes significant time for testing the entire readout 
chain after the \dword{apa}s are placed in their final position inside the cryostat,
when repairs are still possible.

We still consider risk (RT-SP-TPC-004) a possibility: that the cold cables cannot
be routed through the frames of the \dword{apa}s. In that case, the cables for the
\dwords{femb} attached to the bottom \dword{apa} would have to be routed along the
walls of the cryostat, requiring a significant redesign of the entire detector. 
To minimize this risk, we have significantly redesigned the \dword{apa} frame 
to use larger tubes, and many studies have been performed in recent months.
These studies are based on the assumption that there can be a small reduction in
the cable plant size compared to \dword{pdsp}. The probability of this risk
being realized has been significantly reduced following the cable insertion
tests performed at Ash River using a stacked pair of \dword{apa}s, discussed
in Sections~\ref{sec:fdsp-apa-qa-prototyping} and~\ref{sec:fdsp-tpcelec-design-ft}.
This risk has not yet been retired, since the expected reduction of the cable plant 
needs to be demonstrated with the design and test of new \dwords{femb}. Should
this risk be realized, we will consider other ways for reducing the cable 
plant, instead of the considering the option of routing the cables along the 
walls of the cryostat. For example, the control signals could be shared between 
multiple \dwords{femb}.

The next risk (RT-SP-TPC-005) involves delays in the availability of or
damage to the \dword{tpc} electronics components installed on top of the
cryostat. As discussed in Section~\ref{sec:fdsp-tpcelec-integration-timeline},
we plan to have all \dword{tpc} electronics detector components on top of the cryostat, including those 
required to power, control, and readout one pair of \dword{apa}s, installed
and available before inserting the \dword{apa}s into the cryostat.
This allows extensive testing of \dword{apa}s to mitigate
the risk of damage to the readout chain. To mitigate the risk associated with
delays in installing the \dword{tpc} electronics components on top of the cryostat,
we plan to have sufficient spares and to use appropriate \dword{esd} 
prevention measures. If only a subset of all components is available, cables 
and fibers on the top of the cryostat would have to be re-routed to allow  
integrating and installing the \dword{apa}s to continue without delays, 
and tests will have to be repeated when all the components become available 
and are installed. The worst possible consequence is a delay in closing 
the cryostat and beginning operation. 

Another risk (RT-SP-TPC-006) is that incompatibilities 
between various components of the \dword{dune} \dword{fd} go undetected until
these components are integrated during prototyping, during integration at 
the \dword{surf}, or during installation. These incompatibilities could 
result in reworking or redesigning some of the components and therefore in
delays of the project. This risk clearly
diminishes as long as integration tests, including mock-ups and prototypes,
come early in design and construction. The schedule for the 
design and construction of the \dword{tpc} electronics detector components foresees many
integration tests to reduce this risk as much as possible. These tests include 
integration tests with components provided by the \dword{apa}, \dword{pds}, 
and \dword{daq} consortia, as well as integration tests of cable routing in 
the cable trays and through the cryostat penetrations. The second run of 
\dword{pdsp}, using pre-production detector components, will further help
in mitigating this risk. At that point, any design change or any deviation from
established procedures will need to go through very extensive vetting to
avoid the introduction of new incompatibilities between detector components.

Another issue that can arise during the detector integration and installation
is a delay caused by the excessive usage of spare detector parts (RT-SP-TPC-007).
For \dwords{asic} and \dwords{femb}, this kind of risk has already been considered
(RT-SP-TPC-002). For the other \dword{tpc} electronics detector components, the
risk should be considered separately, since the other components are needed 
at an earlier point in time and have completely different fabrication and
testing schedules. Some of the actions required to mitigate this risk are
similar, including an early start to the production, careful monitoring of yields
during the \dword{qc} process, and a larger number of spares in the case of
long lead items. 

The final risk we consider for the construction of the \dword{tpc} electronics
detector components is the loss of key personnel (RT-SP-TPC-008). The number of
scientists and engineers that have become involved with the \dword{tpc}
electronics has significantly increased since the construction of \dword{pdsp},
and in some sense this has already contributed to reducing significantly the
probability and possible impacts of this risk. In some areas, like \dword{asic}
design, the addition of large teams of engineers involved in the design of
the new \dwords{asic} means that the probability of this risk is now negligible.
There are areas where the experience from the construction and operation
of \dword{pdsp} resides with a few expert scientists and engineers, and areas
where only a single engineer is responsible for the design of a set of detector
components. The mitigation of this risk involves enlarging the team(s) that
are responsible for the design and prototyping of the detector components.
This has already been done for the \dwords{asic} and plans are already 
in place to involve university groups in the design of the \dwords{femb}
and of the \dword{wiec}. Succession plans for the consortium leadership
need to be put in place, including training younger personnel. 

\subsection{Risks during Commissioning}
\label{sec:fdsp-tpcelec-risks-commissioning}

The biggest risk during the commissioning phase
(RT-SP-TPC009 in Table~\ref{tab:risks:SP-FD-TPCELEC}) is excessive noise 
caused by some detector component not respecting the \dword{dune} grounding
rules. This risk was realized at least twice during the 
integration and commissioning of the \dword{pdsp} detector. During the 
integration of the first \dword{apa}, a source of noise was discovered 
in the electronics used for the readout of the photon  detector, which 
required a simple fix on all of the readout boards. Later, a large noise source was 
discovered in the temperature monitors. The overall noise in \dword{pdsp} 
was reduced compared to previous \dword{lar} experiments or prototypes, 
such as \dword{microboone} or the \dword{35t}. Even if some unresolved source of noise 
is still apparent in the \dword{pdsp} data, this should not preclude using the
collected data for calibration and for physics analyses. Further studies are
planned for 2020 to investigate the remaining sources of noise.

The main problem in going from \dword{pdsp} to the \dword{dune} \dword{fd} is one of scale.
Even if the detector design addresses all possible noise sources, the simple
fact that the detector is 25 times larger and has a correspondingly larger
number of cryostat penetrations requires much more attention to detail
during installation and commissioning. Observations of excessive 
noise in \dword{dune} would result in a delay in commissioning and  
data taking until the source of the noise is found and mitigated. To minimize excessive 
electronic noise, we plan to enforce the grounding  rules throughout the 
design phase, based on the lessons learned from the operation of the 
\dword{pdsp} detector. We also plan to perform  integrated tests to discover 
possible problems as early as possible. This includes system tests at the \dword{iceberg} test stand at \dword{fnal} for each generation 
of the \dwords{femb} and photon detectors. We plan to perform noise 
measurements in the cold boxes at \dword{surf}, and later 
during the insertion of the \dword{apa}s inside the cryostat before the 
\dword{tco} closure, including repeating the measurements directly before the \dword{lar} fill. 
We expect that the extensive testing will allow a quick transition to 
detector operations, first with cosmics and later with beam, as soon as the 
cryostat has been completely filled. 

\subsection{Risks during Operation}
\label{sec:fdsp-tpcelec-risks-operations}

The expectation for the \dword{dune} \dword{fd} is that data taking will continue
for at least two decades. Assuming that the detector operates as designed after commissioning,
two additional risks must still be considered.
These are related to the lifetime of the \dword{ce} components
installed inside (RT-SP-TPC-010 in Table~\ref{tab:risks:SP-FD-TPCELEC}) and on top (RT-SP-TPC-011)
of the cryostat. The components inside the cryostat are not replaceable,
and therefore any malfunction of a detector component will result in
a loss of sensitive volume. The components on top of the cryostat
(with the exception of the flange at the transition from the cold to
the warm volume) can be replaced, and as long as we have sufficient
spares, this will not result in any loss beyond the amount 
of time required for replacing the component. The risk of losing components installed inside the cryostat has been considered
from the earliest stage of the design of \dwords{asic}. As discussed
in Section~\ref{sec:fdsp-tpcelec-qa-reliability}, we have formed a 
reliability committee to ensure that all appropriate measures
are considered in the design and that our \dword{qa} process includes
the relevant tests of component lifetime. These lifetime measurements
should make sure that we will see only minimal losses of sensitivity in the 
detector during operations. 

As discussed in Section~\ref{sec:fdsp-tpcelec-safety-detops}, we
are taking measures (like adding air filters to the
\dwords{wiec} and bias voltage and low-voltage power supplies) to
minimize damage from environmental conditions to the 
detector components on top of the cryostat. We have discussed in
Section~\ref{sec:fdsp-tpcelec-production-spares} our plan for 
spare detector components. We cannot exclude the possibility that we will not have enough
spares, which we plan to build during construction, for the lifetime of the experiment. In this
case, it may become necessary to fabricate new boards or procure new
supplies during operations. One possible issue related to this is the 
continued availability of certain components, in particular
\dwords{fpga} and optical transmitters and receivers, which may 
become obsolete and no longer be available when we need to fabricate new
parts. While it will always be 
possible to design new boards using more modern components, we
wish to keep the maintenance costs for the detector to a minimum,
and this may involve following the technology evolution and stockpiling components that may become obsolete
and/or hard to procure. We are also considering placing the
\dwords{fpga} and the optical components on mezzanine cards
to minimize redesign and procurement costs should these 
components become unavailable during the lifetime of the experiment.

\section{Organization and Management}
\label{sec:fdsp-tpcelec-management}

In this section we first discuss the organization of the \dword{tpc} electronics
consortium that at the moment consists entirely of US institutions: 
fifteen university groups plus groups from four DOE national 
laboratories. Table~\ref{tab:SPCE:institutions} provides a list 
of the participating institutions.
Later we discuss the assumptions that have been made in developing
the construction plan for the detector components that will be
provided by the \dword{tpc} electronics consortium, including the responsibilities
of the different institutions that are part of the consortium. Finally,
we present a schedule for the construction, integration, and installation
of the \dword{tpc} electronics into the detector. 

\begin{dunetable}
[TPC electronics consortium intitutions ]
{ll}
{tab:SPCE:institutions}
{Institutions participating in the \dword{tpc} electronics consortium (all from the US).}
Institution  \\ \toprowrule
Boston University \\ \colhline
Brookhaven National Laboratory \\ \colhline
University of Cincinnati \\ \colhline
Colorado State University  \\ \colhline
University of California, Davis \\ \colhline
\dword{fnal} \\ \colhline
University of Florida \\ \colhline
University of Hawaii \\ \colhline
Iowa State University \\ \colhline
University of California, Irvine \\ \colhline
Lawrence Berkeley National Laboratory \\ \colhline
Louisiana State University \\ \colhline
Michigan State University \\ \colhline
University of Pennsylvania \\ \colhline
University of Pittsburgh \\ \colhline
SLAC National Accelerator Laboratory \\ \colhline
Stony Brook University \\
\end{dunetable}

\subsection{Consortium Organization}
\label{sec:fdsp-tpcelec-management-consort}

The present consortium organization
structure includes a consortium leader and a technical lead (both currently
from \dword{fnal}), with personnel from \dword{bnl} helping with system design. A working group structure has been recently 
provided with subgroups responsible for the detector 
components inside the cryostat (\dwords{asic}, \dwords{femb}, and
cold cables), outside the cryostat (mostly the \dwords{wiec} with 
their boards), and all equipment
used for testing the detector components. When appropriate, a subgroup
responsible for integration and installation activities at
\dword{surf} will be formed (for now, the technical lead oversees 
these activities). In addition,
another subgroup will be in charge of software and physics
preparation activities, including calibrations and simulations.
Mechanical design activities span detector components both inside
and outside the cryostat, requiring strong contacts between 
\dword{tc} and other consortia (mainly the \dword{apa}
consortium). The lead engineer (from \dword{bnl}) on mechanical aspects of the cold
electronics (mechanical interfaces with the \dword{apa}s, cabling, including 
cable trays, and cryostat penetrations) works directly with
the \dword{tc} team. The \dword{tpc} electronics consortium will also have 
contact people for the overall \dword{lbnf-dune} management for
\dword{esh} and for \dword{qa}/\dword{qc}. For the moment, the technical lead oversees these activities, although oversight
will be transferred in part to each subgroup leader for testing
activities. The leadership positions in the consortium 
are listed in Table~\ref{tab:SPCE:leadership} and a diagram of
the organization of the consortium and of its relationships
with other groups in the \dword{lbnf} and \dword{dune} organizations
is shown in Figure~\ref{fig:sp-tpcelec-orgchart}.

\begin{dunetable}
[Leadership positions in the TPC electronics consortium]
{p{0.35\textwidth}}
{tab:SPCE:leadership}
{Current leadership positions in the TPC electronics consortium.}\
Position \\ \toprowrule
Consortium Leader \\ \colhline 
Technical Lead \\ \colhline 
System Aspects \\ \colhline 
Cold Components \\ \colhline 
Warm Components \\ \colhline 
Test Setups \\ \colhline 
Integration and Installation \\ \colhline
Mechanical Design \\ \colhline 
Reliability Task Force \\ \colhline 
TDR Editor \\ \colhline 
\dword{esh} contact \\ \colhline 
\dword{qa}/\dword{qc} contact \\
\end{dunetable}

\begin{dunefigure}
[Organization chart of the TPC electronics consortium]
{fig:sp-tpcelec-orgchart}
{Organization chart of the TPC electronics consortium and relations to
other groups in the \dword{lbnf} and \dword{dune} organizations (shown in green).}
\includegraphics[width=\linewidth]{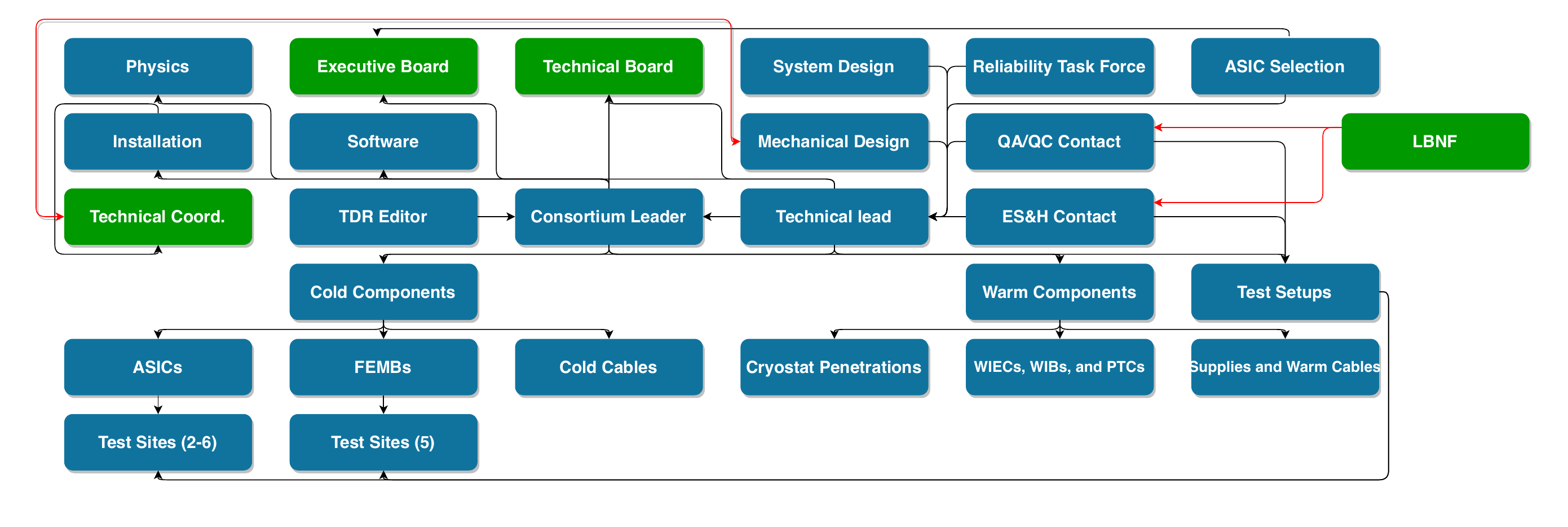}
\end{dunefigure}

In addition to the working groups, task forces for specific issues 
will be formed as necessary. A first example is the task force
charged with studying reliability issues in the \dword{tpc} electronics 
components and preparing recommendations for the choice of 
\dwords{asic}, the design of printed circuit boards, and testing; this
was discussed in Section~\ref{sec:fdsp-tpcelec-qa-reliability}. 
Later on, this task force will help in developing the \dword{qc} 
program for the \dword{tpc} electronics detector components in collaboration
with the testing group leadership, with the \dword{pdsp} 
experience as a starting point. Later, a second task force, with possible
personnel overlap with the first task force and possibly including experts
from outside of the \dword{dune} Collaboration, will be tasked with establishing the
criteria for the \dword{asic} selection. This task force will
be also asked to propose a recommendation that will then go to
the \dword{dune} Executive Board for the final approval, as discussed
in Section~\ref{sec:fdsp-tpcelec-design-femb-selection}.

\subsection{Planning Assumptions}
\label{sec:fdsp-tpcelec-management-planning}

In Section~\ref{sec:fdsp-tpcelec-management-cost}, we describe
the current schedule for the construction, integration, and installation
of the \dword{tpc} electronics detector components for the first \dword{sp} \dword{tpc}
\dword{fd} module. This schedule, as well as the costs associated
with detector construction, are based on experience with constructing
and commissioning the \dword{pdsp}
detector in addition to assumptions for the remaining R\&D program
and the production planning discussed here. 
Section~\ref{sec:fdsp-tpcelec-production-spares} details
the number of spare parts that we plan to fabricate in order to
account for known yield issues during detector construction
and to address possible problems. Table~\ref{tab:SPCE:components}
gives a summary of all fabricated components required for the first \dword{sp}
\dword{fd} module. To develop a schedule for detector construction, 
we must consider the current state of development of 
the \dwords{asic}, \dwords{femb}, and all other components in order to 
estimate the time required for the final production.

A second detector module may be built using the \dword{sp} technology, 
and in that case the construction of the \dword{tpc} electronics components 
for the second module would immediately follow construction of 
the first one. The total number of components for the second 
module will be less than the amount for the first detector, assuming that 
for components inside the cryostat the spare parts from the first 
module can be used for the second. Similarly, for the 
components on the top of the cryostat, the pool of spares from the 
first module should be sufficient to cover the second.
Given the timeline for the construction of the detector components
and their integration on the \dword{apa}s or their installation at
\dword{surf}, an insufficient number of spares presents a risk only
for the second detector module, as discussed in Section~\ref{sec:fdsp-tpcelec-risks-design}.

The critical path for the construction of the first \dword{sp}
\dword{tpc} \dword{fd} module is driven by the availability of \dwords{femb},
which in turn depends on completing the design of the
\dwords{asic}. Thus, construction
of the \dword{apa}s could start as early as spring 2020, while the
decision on the \dwords{asic} to be used on the \dwords{femb}
may come as late as January 2021. The schedule for this decision
depends on the \dword{tpc} electronics consortium, which is
planning a second design iteration on all \dwords{asic}
followed by system tests of various flavors of \dwords{femb}.
The first version of all \dwords{asic} underwent
standalone tests in spring / summer 2019 and will go through the 
sequence of system tests (with the 40\% \dword{apa} prototype at \dword{bnl},
the seventh \dword{pdsp} \dword{apa} in the \coldbox at \dword{cern},
and the \dword{tpc} in the \dword{iceberg} cryostat at \dword{fnal})
in early 2020. At the same time, lifetime tests will be performed 
on all \dwords{asic}. Designs of the \dwords{asic} and \dwords{femb}, including results
from the system test stands and from lifetime tests, will be reviewed
in early 2020. This will trigger any further design
changes on the \dwords{asic}, to be followed by a second round of prototyping
and testing. This gets us to the January 2021 date for the final
\dword{asic} decision. Consequently, the initial
tests of the \dword{dune} prototype \dword{apa}s must be performed
with preliminary versions of \dwords{femb}, which will still be using
the first generation of \dwords{asic} (although the final 
mechanical and electrical connections to the \dword{apa}s will be available and used).

\begin{dunetable}[Number of TPC electronics components for one detector module]
{p{0.5\textwidth}p{0.5\textwidth}}
{tab:SPCE:components}
{Number of TPC electronics components required for a full \dword{spmod} (accounting for spares and yields during \dword{qc}).}
Detector component & Number required \\ \toprowrule
\dword{larasic} & 30,000 chips (at least 43 wafers) \\ \colhline
\dword{coldadc} and \dword{coldata} & 30,000 and 7,500 chips (at least 33 wafers) \\ \colhline
\dword{cryo} & 7,500 chips (at least 35 wafers) \\ \colhline
\dword{femb} & 3,200 \\ \colhline
Cold signal cables & 1,650 and 1,575 (bottom and top \dword{apa}) \\ \colhline
Cold power cables & 1,650 and 1,575 (bottom and top \dword{apa}) \\ \colhline
Cold bias voltage cables & 660 and 630 (bottom and top \dword{apa}) \\ \colhline
Cryostat penetrations & 80 \\ \colhline
\dword{ce} flanges & 160 \\ \colhline
\dword{wiec} & 155 \\ \colhline
\dword{wib} & 775 \\ \colhline
\dword{ptc} & 155 \\ \colhline
Warm power cables & 165 (three different lengths) \\ \colhline
Warm bias voltage cables & 1,320 (three different lengths) \\ \colhline
Wiener PL506 power box & 30 \\ \colhline
Wiener MPOD crate & 30 \\ \colhline
Wiener MPOD modules with 8 HV channels & 180 \\ \colhline
Power supplies and cables for heaters and fans & 30 \\
\end{dunetable}

After the January 2021 review, an additional six months may be 
required for a final iteration of the \dword{femb} design
and a final round of system tests before beginning fabrication
of the \dwords{asic} and \dwords{femb}. Engineering runs
for the \dwords{asic} and \dwords{femb} should then take place in the
second half of 2021, with most production starting in
spring 2022. Production and testing of all chips required
for constructing the first \dword{sp} \dword{tpc} \dword{fd} module
would be completed by August 2023. The first batch of
production \dwords{femb} would then be available for installation on
the \dword{apa}s in February 2023, and production would be
completed in January 2024, roughly eight months before the 
beginning of the integration of the \dwords{femb} on the
\dword{apa}s at \dword{surf}. The final \dwords{femb} for the 
first \dword{spmod} are expected to be available at the \dword{sdwf} 
fourteen months ahead of their for installation on the 
last \dword{apa} to be installed inside the cryostat.
There are therefore between eight and fourteen months of float
in the schedule for the \dwords{asic} and \dwords{femb}, which 
would allow for a third iteration of design and prototyping
if needed. It should be noted that, for the detector components
to be installed on the top of the cryostat, there are between
ten and fourteen months of float.

The schedule presented above assumes that during the
construction of the \dword{apa}s,
integration tests will be performed using preliminary
versions of the \dwords{femb} that must then be replaced
with final versions at a later date. It also assumes that for the
second run of \dword{pdsp}, expected to take place in the
second half of 2021, prototype \dwords{femb} using the
second iteration of prototype \dwords{asic} will be used.
Using the \dwords{asic} from the pre-production, fabricated using
the masks that will be used during the production phase, 
would require a delay of one year in the second run of
\dword{pdsp}. The difference in the masks used for the
fabrication of the \dwords{asic} (multipurpose wafer run instead 
of a dedicated run) does not negate the usefulness of
the second run of \dword{pdsp} as a final validation of
the \dword{dune} \dword{spmod} design.

If a second \dword{sp} \dword{tpc} \dword{fd} module is built, the critical
path will transition from the \dwords{femb}
to the \dword{apa}s, assuming that construction
of both \dword{apa}s and \dwords{femb} will continue 
without interruption after construction of the first module. 
This is because constructing one \dword{apa} requires 
more time than constructing the corresponding \dwords{femb}. Therefore,
toward the end of the constructing this possible second module,
we expect that all required \dwords{femb} will be at the
\dword{sdwf} waiting for the delivery of the \dword{apa}s before
integration can take place.

All other detector components that are the responsibility of the
\dword{tpc} electronics consortium can be produced relatively quickly
in less than two years. The procurement, assembly, and
testing of these components can be scheduled so that
we have sufficient time in the schedule to address possible 
problems during production. Changes in all of these
components, unlike those used in the \dword{pdsp} detector,
are less important than those affecting the \dwords{asic}
and \dwords{femb}. We are assuming that final
designs for the rest of the detector components will be 
available in the second half of 2020 and that the corresponding 
production readiness reviews will occur, at the latest, six months 
before the \dword{asic} design choice.
The first components to be installed on the detector are the cryostat
penetrations, which must be installed before the 
detector support structure inside the cryostat is completed.
In this way, the cryostat can be completely sealed, other than
the manholes used to feed clean air into the cryostat
and the \dword{tco} that is used as an exhaust portal 
and as an entry point for the detector components. The rest of the
\dword{tpc} electronics components, which are installed on top of the 
cryostat (\dwords{wiec} with all of their boards, supplies, cables,
and fibers), should be installed before installing
the corresponding rows of \dword{apa}s and properly connecting
the cables linking the \dword{apa}s to power, control, and readout. 
The \dword{apa}s should be tested as soon as they are installed. 

\subsection{Institutional Responsibilities}
\label{sec:fdsp-tpcelec-management-resp}

Design and prototyping for the \dword{sp} \dword{dune} \dword{fd} have been 
concentrated so far at the DOE national laboratories, mostly because the 
focus has been on designing the new generation of \dwords{asic}. The design 
of the \dword{larasic} was done at \dword{bnl}, the \dword{cryo} 
chip was done at \dword{slac}, and the new \dword{coldadc} was a joint effort of 
\dword{bnl}, \dword{fnal}, and \dword{lbnl}. The \dword{coldata} 
\dword{asic} was designed at \dword{fnal}, with some components 
provided by engineers from the Electrical Engineering Department
at Southern Methodist University (not a consortium member).
The \dword{cts} was designed at Michigan State University.
Most of the design and construction work for the \dword{pdsp} detector 
was done at \dword{bnl}, with other institutions contributing to 
testing, installation, and commissioning. Given the extent of the project, 
particularly testing, additional institutions have begun to contribute
to all of the activities for constructing the \dword{dune} \dword{sp} \dword{fd}, which began 
in the middle of 2018. Almost all the engineering of detector components, 
except for the boards to be installed on the \dwords{wiec} and most boards 
and setups for testing, will remain a responsibility
of the DOE national laboratories. Testing of 
\dwords{asic}, \dwords{femb}, cables, power and bias voltage supplies,
and \dwords{wiec} with their boards will be done at various
universities that are members of the consortium. All institutions
are expected to contribute to the integration and installation activities at
at \dword{surf}, which is very
demanding in terms of personnel. A detailed list of the 
institutions contributing to the development, production, and
testing of the various detector components is given in Figure~\ref{fig:sp-tpcelec-responsibilities}.

\begin{dunefigure}
[Construction responsibilities for the TPC electronics consortium]
{fig:sp-tpcelec-responsibilities}
{Responsibilities of the institutions in the \dword{tpc} electronics 
consortium, matched to the organization chart of the consortium.}
\includegraphics[width=0.8\linewidth]{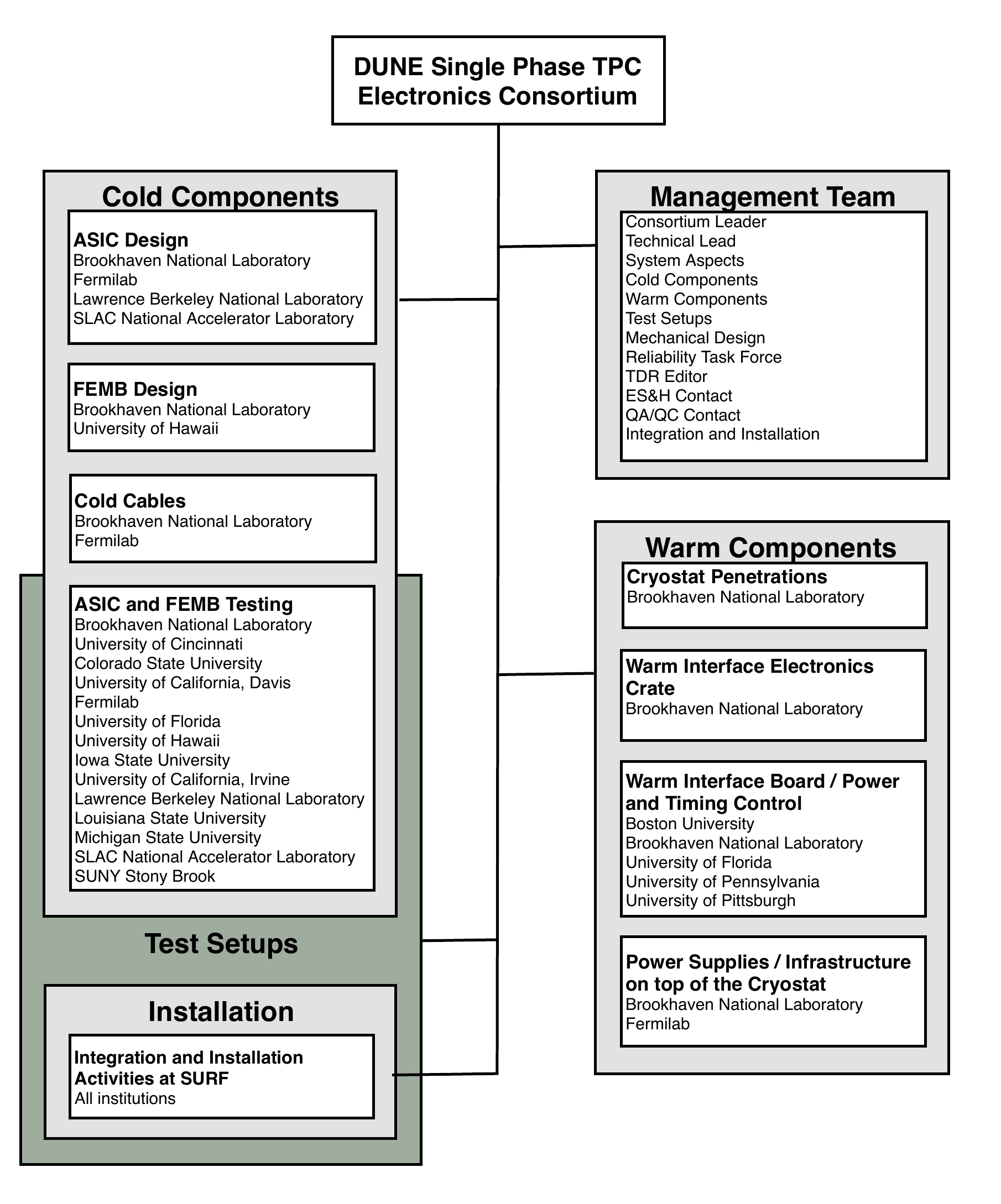}
\end{dunefigure}

\subsection{High-Level Cost and Schedule}
\label{sec:fdsp-tpcelec-management-cost}

\begin{dunetable}
[TPC electronics consortium schedule]
{p{0.75\textwidth}p{0.20\textwidth}}
{tab:SPCE:timeline}
{Milestones of the TPC electronics consortium.}
Milestone & Date \\ \toprowrule
Complete the submission of the first generation of \dwords{asic} & April 2019\\ \colhline
Complete the standalone testing of the first generation of \dwords{asic} & October 2019\\ \colhline
Complete system and lifetime tests on the first generation of \dwords{asic} and \dwords{femb} & May 2020\\ \colhline
Submission of second generation of \dwords{asic} & June 2020 \\ \colhline
Complete the standalone testing of the second generation of \dwords{asic} & September 2020 \\ \colhline
Complete system and lifetime tests on the second generation of \dwords{asic} and \dwords{femb} & January 2021\\ \colhline
Decision on the \dword{asic}(s) to be used for construction &  January 2021\\ \colhline
\rowcolor{dunepeach} Start of \dword{pdsp}-II installation& \startpduneiispinstall      \\ \colhline
Complete characterization of final prototypes of \dwords{asic} and \dwords{femb} including system tests & July 2021\\  \colhline
Complete Engineering Design Reviews and launch pre-production of detector components & September 2021 \\ \colhline
Start of procurement of cold cables & December 2021 \\ \colhline
Start of production of cryostat penetrations & December 2021 \\ \colhline
Start of production of \dwords{wiec}, \dwords{wib}, and \dwords{ptc} & December 2021 \\ \colhline
Start of procurement of power supplies and warm cables & December 2021 \\ \colhline
Complete testing of pre-production of all \dwords{asic} & January 2022\\ \colhline
Complete testing of pre-production \dwords{femb} & April 2022 \\ \colhline
\rowcolor{dunepeach}South Dakota Logistics Warehouse available& \sdlwavailable      \\ \colhline
Complete testing of prototypes and Production Readiness Reviews & May 2022\\ \colhline
Start of \dwords{asic} production & May 2022 \\ \colhline
Start of \dwords{femb} production & May 2022 \\ \colhline
\rowcolor{dunepeach}Beneficial occupancy of cavern 1 and \dword{cuc}& \cucbenocc      \\ \colhline
Completion of cryostat penetrations procurement & February 2023 \\ \colhline
Completion of the procurement and \dword{qc} of cold cables & March 2023 \\ \colhline
\rowcolor{dunepeach} \dword{cuc} counting room accessible& \accesscuccountrm      \\ \colhline
Completion of the \dwords{wiec}, \dwords{wib}, and \dwords{ptc} production and \dword{qc} & May 2023 \\ \colhline
Completion of the procurement of power supplies and warm cables & June 2023 \\ \colhline
Completion of \dwords{asic} production and \dword{qc} & August 2023 \\ \colhline
Completion of \dwords{femb} production and \dword{qc} & January 2024 \\ \colhline
Begin installation of the \dword{tpc} electronics components on top of the cryostat & April 2024 \\ \colhline
Complete installation of the \dword{tpc} electronics components on top of the cryostat & July 2024 \\ \colhline
\rowcolor{dunepeach}Start of \dword{detmodule} \#1 TPC installation& \startfirsttpcinstall      \\ \colhline
Begin integration of the \dwords{femb} on the \dword{apa}s at \dword{surf} & August 2024 \\ \colhline
Complete integration of the \dwords{femb} on the \dword{apa}s at \dword{surf} &  March 2025\\ \colhline
\rowcolor{dunepeach}End of \dword{detmodule} \#1 TPC installation& \firsttpcinstallend \\
\end{dunetable}

In Section~\ref{sec:fdsp-tpcelec-management-planning}, we  
discussed how the project will evolve from the current design
and prototyping phase to production for the \dwords{asic}
and \dwords{femb} by spring 2022. During the same  
period, the engineering of all other detector components will
be completed and prototypes fabricated. The procurement
and qualification of cold cables, cryostat penetrations, \dwords{wiec},
and power and bias voltage supplies can then begin in spring 2021.
Integrating \dwords{femb} on the \dword{apa}s would then
take place over 18 months beginning in early 2023. At
this moment, the installation and testing of \dword{apa}s in 
the cryostat and corresponding activities of the \dword{tpc} electronics consortium, including installing all 
detector components on top of the cryostat, are scheduled to
take four months starting in April 2024. Table~\ref{tab:SPCE:timeline} shows a preliminary list of 
milestones, including the current plan to complete  
the design, R\&D, and engineering phases; and then later  
the production setup, production, integration,
and installation activities. The schedule for the 
completion of the design and prototyping, the pre-production,
the construction of the detector components (including the
\dword{qc} process), and their integration and installation at
\dword{surf} is displayed in Figure~\ref{fig:sp-tpcelec-schedule}.

\begin{dunefigure}
[Schedule for the construction of the TPC electronics detector components]
{fig:sp-tpcelec-schedule}
{Schedule for the completion of the design and prototyping, the pre-production,
the construction of the \dword{tpc} electronics detector components (including the
\dword{qc} process), and their integration and installation at \dword{surf}.}
\includegraphics[width=\linewidth]{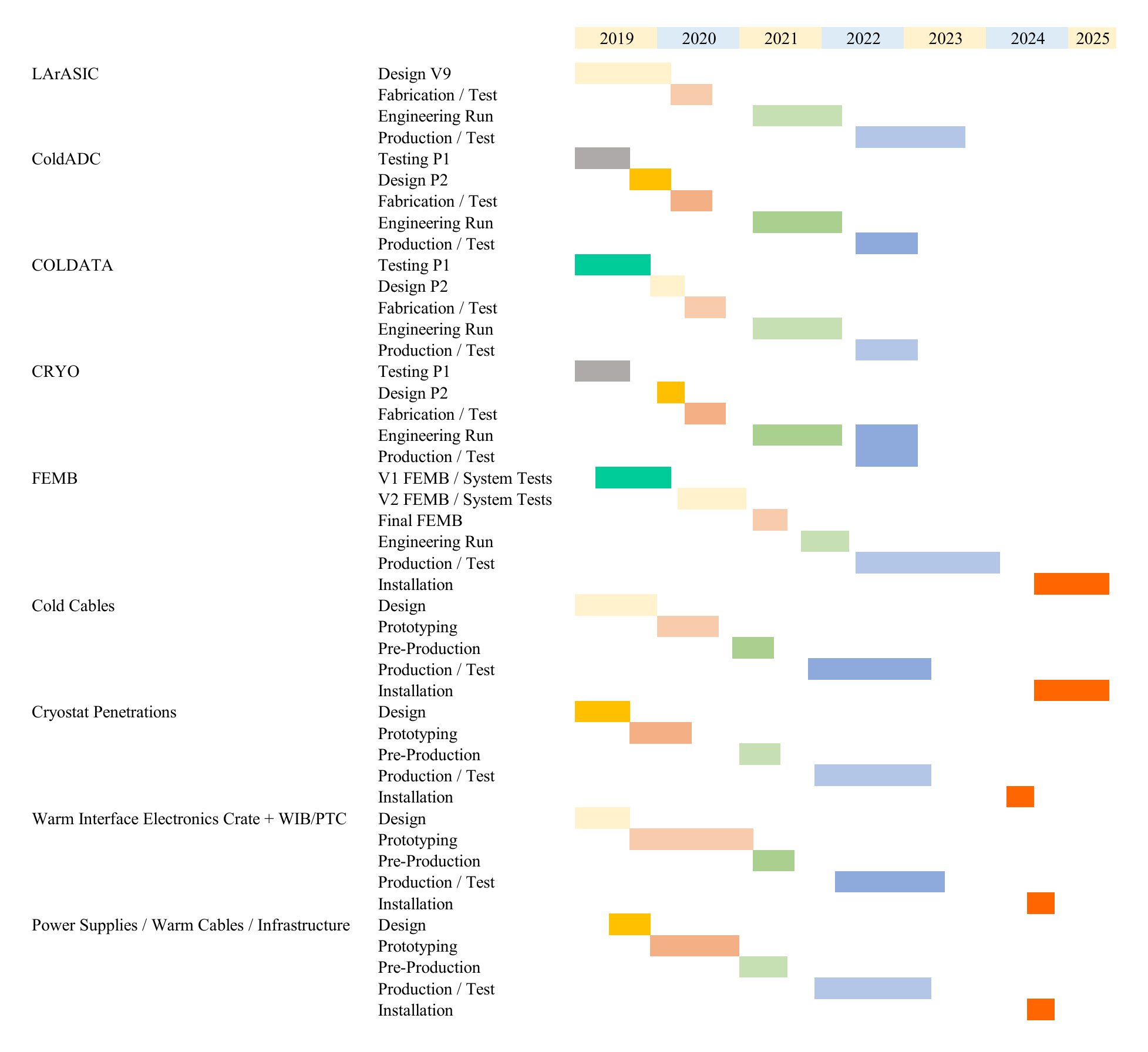}
\end{dunefigure}

\cleardoublepage

\chapter{Photon Detection System}
\label{ch:fdsp-pd}
%

\newcommand{\tzero}{\ensuremath{t_0}\xspace}

\section{Introduction} 
\label{sec:fdsp-pd-intro}

The \dword{dune} \dword{fd} consists of detector systems for charge and light produced by an ionization event in the \dword{lartpc}.  The charge detection system permits both calorimetry and position determination, with two of the three spatial coordinates ($y$ and $z$)  established by the position of the \dword{apa} wires receiving the charge and the third ($x$) by the arrival time of the charge.  Locating the $x$ position requires independently determining the time of the ionization event, a clock start time.  Two systems provide this in \dword{dune}:  the \dshort{fnal} accelerator system for neutrino beam related events and the \dword{pds}.  

Neutrino \dword{cpv} and other elements of the \dword{dune} long-baseline neutrino program are possible without data from the \dword{pds}.  The neutrino beam timing allows full functionality of the \dword{apa}s, and the deep underground location reduces the possibility of in-time background events from cosmic rays and other sources to a negligible level.  
Similarly, \dword{dune} can detect \dwords{snb} originating within the galaxy without the \dword{pds} because the presence of thousands of low-energy neutrino events, even if they consist only of few millimeter long tracks, provides an unambiguous signal in the \dword{tpc}.
By contrast, \dword{dune}'s nucleon decay physics cannot be executed without the \dword{pds}.  The inability to establish a clock start time (\tzero) makes it impossible to determine whether a candidate proton decay event was fully contained in the detector volume or associated with objects entering the detector from the outside.  Determining \tzero also allows the energy reconstructed by the \dword{tpc} to be corrected for charge lost due to electron capture and other transport effects in the \dword{tpc}. This physics sets the requirement on minimum light yield in the dimmest regions of the detector far from the photon detectors 
and timing resolution, 
described further in Appendix Section~\ref{subsec:fdsp-pd-simphys-ndk}.

While only absolutely required for proton decay searches, the \dword{pds} directly enhances physics capabilities for all three \dword{dune} physics drivers, opens up prospects for further physics explorations, and contributes to a more robust set of operating points of the detector that help all physics.   For \dword{snb} neutrino events, the \dword{pds} allows proper location of the event vertex, and improves energy resolution by allowing position-dependent energy corrections and complementary direct calorimetric measurements, 
improving energy resolution and possibly sensitivity to underlying supernova dynamical models.  The \dword{pds} also enables a complementary triggering scheme for the burst itself, increasing 
reliability, reducing dead time, and extending the sensitivity further out to nearby dwarf galaxies (see Appendix Section~\ref{subsec:fdsp-pd-simphys-snb}). Applications to supernova physics set the average light yield requirement. 

The \dword{pds} can also measure energy calorimetrically for all classes of events, working as a crosscheck of the energy measured by the \dword{tpc} or improving the resolution when both measurements are used together (see Appendix Section~\ref{subsec:fdsp-pd-simphys-beam}).  In the event that the \dword{dune} \dword{tpc} cannot operate at its goal electric field of \SI{500}{V/cm}, the \dword{pds} energy measurements could compensate for reduced charge detector performance because light production increases relative to the free charge for lower \efield{}.

The \dword{pds} could open new areas of investigation.  The few-MeV scale solar neutrino interactions occur as isolated events in time and space.  Suppressing radiological and noise related backgrounds to pull out a signal for these events likely requires redundant measurements with charge and light.  The \dword{pds} may also provide a means for identifying events with Michel electrons produced from the decay of a stopped muon. Tagging these electrons can be used to estimate the antineutrino content of the beam flux or further reduce nucleon decay backgrounds (see Appendix Section~\ref{subsec:fdsp-pd-simphys-beam}).
 

Volume~\volnumberphysics{}, \voltitlephysics{},  of this \dword{tdr} describes the detailed physics simulations of the main \dword{dune} physics drivers.  
The \dword{pds} performance specifications have been established and validated in part by simulation.  Details of this simulation, which includes non-uniformity in light yield due to the optical properties of the argon, electronics response, and realistic reconstruction, are presented in Appendix Section~\ref{sec:fdsp-pd-simphys}.

\section{Design Specifications and Scope}
\label{sec:pds:des-specs}
\subsection{Specifications}
Based on the physics drivers and additional simulation studies described in Appendix Section~\ref{sec:fdsp-pd-simphys}, Table~\ref{tab:specs:SP-PDS} summarizes the \dword{pds} specifications necessary to achieve the \dword{dune} science objectives. 
In the remainder of this chapter, we present a design that meets or exceeds the specifications. Section~\ref{sec:fdsp-pd-validation} summarizes an extensive set of prototypes that validate the assumptions used in the design.

\cleardoublepage
\begin{footnotesize}
\begin{longtable}{p{0.12\textwidth}p{0.18\textwidth}p{0.17\textwidth}p{0.25\textwidth}p{0.16\textwidth}}
\caption{PDS specifications \fixmehl{ref \texttt{tab:spec:SP-PDS}}} \\
  \rowcolor{dunesky}
       Label & Description  & Specification \newline (Goal) & Rationale & Validation \\  \colhline

  \newtag{SP-FD-3}{ spec:light-yield }  & Light yield  &  $>\,\SI{20}{PE/MeV}$ (avg), $>\,\SI{0.5}{PE/MeV}$ (min) &  Gives PDS energy resolution comparable to that of the TPC for 5-7 MeV SN $\nu$s, and allows tagging of $>\,\SI{99}{\%}$ of nucleon decay backgrounds with light at all points in detector. &  Supernova and nucleon decay events in the FD with full simulation and reconstruction. \\ \colhline
    
   \newtag{SP-FD-4}{ spec:time-resolution-pds }  & Time resolution  &  $<\,\SI{1}{\micro\second}$ \newline ($<\,\SI{100}{\nano\second}$) &  Enables \SI{1}{mm} position resolution for \SI{10}{MeV} SNB candidate events for instantaneous rate $<\,\SI{1}{m^{-3}ms^{-1}}$. &   \\ \colhline

  \newtag{SP-FD-15}{ spec:lar-n-contamination }  & LAr nitrogen contamination  &  $<\,\SI{25}{ppm}$ &  Maintain \SI{0.5}{PE/MeV} PDS sensitivity required for triggering proton decay near cathode. &  In situ measurement \\ \colhline

  \newtag{SP-PDS-1}{ spec:pds-cleanroom-assbly }  & Clean assembly area  &  Class \num{100000} clean assembly area &  Demonstrated as satisfactory in \dshort{pdsp}, and is the \dshort{dune} assembly area standard. &  \dshort{pdsp} and in \fnal materials test stand \\ \colhline

  \newtag{SP-PDS-2}{ spec:spatial-localization }  & Spatial localization in $y$-$z$ plane  &  $<\,\SI{2.5}{\meter}$ &  Enables accurate matching of \dshort{pd} and \dshort{tpc} signals. &  \dshort{snb} neutrino and \dshort{ndk} simulation in the \dshort{fd} \\ \colhline

  \newtag{SP-PDS-3}{ spec:env-light-exposure }  & Environmental light exposure  &  No exposure to sunlight. All other unfiltered sources: $<\,\num{30}$ minutes integrated across all exposures &  Shown to prevent damage to \dshort{wls} coatings due to UV. &  Studies in \dshort{pdsp}, and at IU \\ \colhline

  \newtag{SP-PDS-4}{ spec:env-humidity-limit }  & Environmental humidity limit  &  $<\,\SI{50}{\%}$ RH at \SI{70}{\degree F} &  Demonstrated to prevent damage to \dshort{wls} coatings due to humidity. &  \dshort{pd} optical coating studies \\ \colhline

  \newtag{SP-PDS-5}{ spec:light-tightness }  & Light-tight cryostat  &  Cryostat light leaks responsible for $<\,\SI{10}{\%}$  of data transferred from PDS to DAQ &  Minimizing false triggers due to cryostat light leaks helps limit the data transfer rate to  \dshort{daq}. &  \dshort{pdsp} and \dshort{iceberg} \\ \colhline

  \newtag{SP-PDS-7}{ spec:mech-deflection }  & Mechanical deflection (static)  &  $<\,\SI{5}{\milli\meter}$ &  Minimize motion of \dshort{pd} modules inside the \dshort{apa} (due to static and dynamic loads) to avoid damaging \dshort{apa}. &  \dshort{pd} \dshort{fea}, \dshort{pdsp}, \dshort{iceberg}; Ash River integration tests and CERN pre-production integration tests pending \\ \colhline

  \newtag{SP-PDS-8}{ spec:apa-install }  & Clearance for installation through APA side tubes  &  $>\,\SI{1}{\milli\meter}$ &  Maintain required clearance to allow \dshort{pd} insertion into \dshort{apa} following wire wrapping. &  \dshort{pd} \dshort{fea}, \dshort{pdsp}, \dshort{iceberg};  Ash River integration tests and CERN pre-production integration tests pending \\ \colhline

  \newtag{SP-PDS-9}{ spec:mech-compatibility }  & No mechanical interference with APA, SP-CE and SP-HV detector elements (clearance)  &  $>\,\SI{1}{\milli\meter}$ &  \dshort{pd} mounting and securing element tolerances must prevent interference with \dshort{apa} and \dshort{ce} cable bundles. &  \dshort{iceberg}, Ash River integration tests, and the CERN pre-production integration tests \\ \colhline

  \newtag{SP-PDS-10}{ spec:pds-cable }  & APA intrusion limit for PD cable routing   &  $<\,\SI{6}{\milli\meter}$ &  \dshort{pd} modules must install into \dshort{apa} frames following wire wrapping.  \dshort{pd} modules must not occlude \dshort{apa} side tubes. &  \dshort{iceberg}, Ash River integration  tests, and the CERN pre-production integration tests \\ \colhline

  \newtag{SP-PDS-11}{ spec:pds-cablemate }  & PD cabling cannot limit upper-lower APA junction gap  &  \SI{0}{\milli\meter} separation mechanically allowed &  \dshort{pd} cable connections must not limit the minimum upper and lower \dshort{apa} separation. &  \dshort{iceberg}, Ash River integration  tests, and the CERN pre-production integration tests \\ \colhline

  \newtag{SP-PDS-12}{ spec:pds-clearance }  & Maintain PD-APA clearance at LAr temperature  &  $>\,\SI{0.5}{\milli\meter}$ &  \dshort{pd} mounting frame and cable harness must accommodate thermal contraction of itself and \dshort{apa} frame. &  Thermal modeling, \dshort{protodune}, \dshort{iceberg}, CERN pre-production integration tests \\ \colhline

  \newtag{SP-PDS-13}{ spec:pds-datarate }  & Data transfer rate from SP-PD to DAQ  &  $<\,\SI{8}{Gbps}$ &  \dshort{pd} data transfer must not exceed \dshort{daq} data throughput capability. &  Maximum bandwidth out of the \dshort{pd} electronics is \SI{80}{Mbps} \\ \colhline

  \newtag{SP-PDS-14}{ spec:pds-signaltonoise }  & Signal-to-noise in SP-PD  &  $>\,\num{4}$ &  Keep data rate within electronics bandwidth limits. &  \dshort{pdsp}, \dshort{iceberg} and \dshort{pdsp2} \\ \colhline

  \newtag{SP-PDS-15}{ spec:pds-darkrate }  & Dark noise rate in SP-PD  &  $<\,\SI{1}{kHz}$ &  Keep data rate within electronics bandwidth limits. &  Pre-production photosensor testing, \dshort{pdsp}, \dshort{iceberg} and \dshort{pdsp2} \\ \colhline

  \newtag{SP-PDS-16}{ spec:pds-dynamicrange }  & Dynamic Range in SP-PD  &  $<\,\SI{20}{\%}$ &  Keep the rate of saturating channels low enough for effective mitigation. &  Pre-production photosensor testing, \dshort{pdsp}, \dshort{iceberg} and \dshort{pdsp2} \\ \colhline

\label{tab:specs:SP-PDS}
\end{longtable}
\end{footnotesize}

\subsection{Scope}
The scope of the \dword{sp} \dword{pds}, provided by the \dword{sp} \dword{pd} consortium, includes selecting and procuring materials for, and the fabrication, testing, delivery and installation of light collectors (\dword{xarapu}), 
photosensors (\dwords{sipm}), electronics, and a calibration and monitoring system. This \dword{tdr} chapter will describe the design, validation, assembly, and \dword{qa}/\dword{qc} testing of the \dword{pds} for a single \nominalmodsize \dword{dune} \dword{spmod}. 
The baseline components for a \dword{spmod} are listed in Table~\ref{tab:pds-config-scope}.

\begin{dunetable}
[PDS baseline configuration]
{p{0.15\textwidth}p{0.35\textwidth}p{0.45\textwidth}}
{tab:pds-config-scope}
{\dword{pds} baseline configuration}
Component  				& Description 						& Quantity		\\ \toprowrule
Light collector 		& \dword{xarapu}							& 10 modules per \dword{apa}; \num{1500} total (\num{1000} single-sided; \num{500} double-sided)\\ \colhline
Photosensor 			& Hamamatsu \dword{mppc} \SI{6}{mm}$\times$\SI{6}{mm}	& 192 \dwords{sipm} per module; \num{288000} total	\\ \colhline
\dword{sipm} signal summing		& 6 passive $\times$ 8 active				& 4 circuits per module; \num{6000}  total	\\ \colhline
Readout electronics		& Based on commercial ultrasound chip& 4 channels/module; \num{6000}  total	\\ \colhline
Calibration and monitoring	& Pulsed UV via cathode-mounted diffusers & 45 diffusers/CPA side; 180 diffusers for 4 CPA sides		\\
\end{dunetable}

Although the configuration of the \dword{sp} and \dwords{dpmod} led to significantly different solutions for the \dword{pds}, a number of scientific and technical issues affect them in a similar way, and the consortia for these two systems cooperate closely on these. See \dpchpds.


\section{Photon Detector System Overview}
\label{sec:fdsp-pd-overview}

\subsection{Principle of Operation}
Liquid argon (\dshort{lar}) is  an abundant scintillator and emits about \SI{40}{photons/keV} when excited by minimum ionizing particles~\cite{Doke:1990rza} in the absence of external \efield{}s.
An external \efield{} suppresses the electron recombination that leads to the excimers responsible for most of the \dword{vuv} luminescence in \dword{lar} and hence reduces the photon yield; for the nominal \dword{dune} \dword{spmod} field of \SI{500}{V/cm}, the yield is approximately \SI{24}{photons/keV}~\cite{PhysRevB.20.3486}. 
As depicted in Figure~\ref{fig:scintAr}, the passage of ionizing radiation in \dword{lar} produces excitations and ionization of the argon atoms that ultimately result in the formation of the excited dimer Ar$^*_2$.  
Photon emission proceeds through the de-excitation of the lowest lying singlet and triplet excited states, $^{1}\Sigma$ and 
$^{3}\Sigma$, to the dissociative ground state. The de-excitation from the $^{1}\Sigma$ state is very fast and has a characteristic time of the order of $\tau_{fast}$ $\simeq$ \SI{6}{ns}. The de-excitation from the $^{3}\Sigma$, state is much slower because it is forbidden by the selection rules; it has a characteristic time of $\tau_{slow}$ $\simeq$ \SI{1.5}{$\mu$sec}. 
In both decays, photons are emitted in a \SI{10}{nm} band centered around \SI{127}{nm}, which is in the \dword{vuv} region of the electromagnetic spectrum~\cite{Heindl:2010zz}.
The relative intensity of the fast versus the slow component is related to the ionization density of \dword{lar} and depends on the ionizing particle: \num{0.3} for electrons, \num{1.3} for alpha particles and \num{3} for neutrons~\cite{PhysRevB.27.5279}. 
This phenomenon is the basis for the particle discrimination capabilities of \dword{lar} exploited by experiments that can separate the two components, but its utility in a large detector is effectively restricted to events with single charged particles. This 
limits its effectiveness in \dword{dune}, where most events in which such \dword{pid} would be beneficial are multi-particle, but it could be a powerful supplement to the charge measurement in some cases.

\begin{dunefigure}[Schematic of scintillation light production in argon]{fig:scintAr}
{Schematic of scintillation light production in argon.}
\includegraphics[width=0.6\columnwidth]{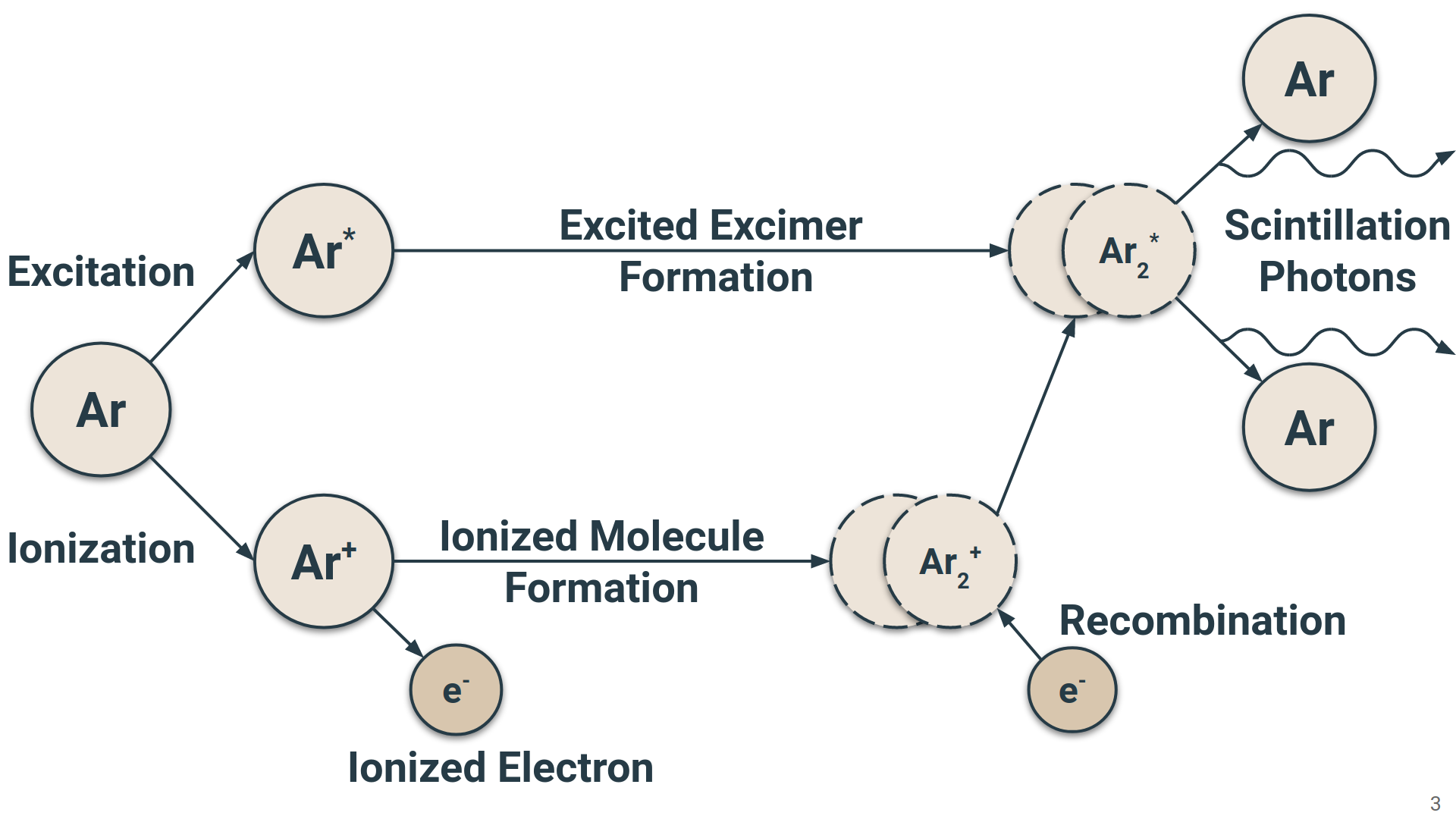}
\end{dunefigure}

\subsection{Design Considerations}
\label{sec:fdsp-pd-des-consid}

The principal task of the \dword{sp} \dword{pds} is to measure the \dword{vuv} scintillation light produced by ionizing tracks in the \dword{tpc} within the geometrical constraints of the \dword{apa} structure. The modular arrangement of the \dword{spmod} calls for a configuration across the width of the cryostat starting with an \dword{apa} plane against one cryostat wall, and following with \dword{apa}s and \dword{cpa}s in the order  APA-CPA-APA-CPA-APA.
The structure of the \dword{apa}, along with the imperative to maximize the active volume of \dword{lar}, precludes the use of traditional large area \dwords{pmt}.  

A solution that reduces the impact of the \dword{pds} on the active volume to zero is to place the light collector modules in the inactive space between the innermost wire planes of the \dword{apa}s. To satisfy \dword{apa} fabrication constraints and mechanical integrity, we must install the modules through slots in a (wound) \dword{apa} frame 
(see Chapter~\ref{ch:fdsp-apa}).  
Individual \dword{pd} modules are restricted to a profile of dimensions 
\SI{23}{mm}$\times$\SI{118}{mm}$\times$\SI{2097}{mm}.  There are ten \dword{pd} modules per \dword{apa}, equally-spaced by \SI{592}{mm}, for a total of \num{1500} per \dword{spmod}.  Of these, \num{500} are mounted in central \dword{apa} frames and must collect light from both directions (dual-face), and \num{1000} are mounted in frames  near the vessel walls and collect light from only one direction (single-face).
Figure~\ref{fig:apa-frame-pds} illustrates the baseline configuration of \dword{pd} modules and \dword{apa}s in an \dword{spmod}. 

To detect scintillation light over a large area in a compact space requires a multi-step process.  First, the \dword{vuv} scintillation photons are converted to longer wavelength by chemical wavelength shifters
\footnote{The most widely used wavelength shifter for \lar detectors is  1,1,4,4-Tetraphenyl-1,3-butadiene (\dshort{tpb}), which absorbs \dword{vuv} photons and re-emits them with a spectrum centered around \SI{420}{nm}, close to the wavelength of maximum quantum efficiency for photo-conversion in most commercial photosensors.}.
These photons are then channeled as efficiently as possible toward much smaller photosensors that produce an electrical signal. Because of the severe space constraints, these must be silicon photosensors with dimensions of just a few millimeters,  not a traditional photomultiplier.
Another requirement, distinct from most previous HEP applications of these devices, is that they must operate reliably for many years at \dword{lar} temperatures. 

\begin{dunefigure}[Arrangement of APAs in a \dshort{spmod} and position of PD modules in APA frame]{fig:apa-frame-pds}
{End-on schematic view of the active argon volume showing the four drift regions and anode-cathode plane ordering of the \dword{tpc} inside the \dword{spmod} (top). The three rows of \dword{apa}s across the width of the \dword{spmod} are two frames high and 25 frames deep. Schematic of an \dword{apa} frame (on its side) showing the ten pairs of \dword{pd} module support rails (almost vertical in figure) (bottom). Notice the five slots on the frame's side that the \dword{pd} modules fit through (top of figure). The other five slots are on the frame's opposite side, at the bottom of the figure.}
\includegraphics[width=0.6\textwidth]{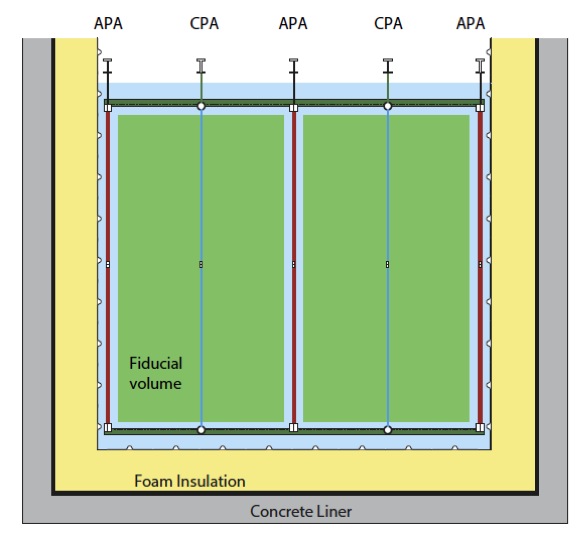}
\includegraphics[width=0.8\textwidth]{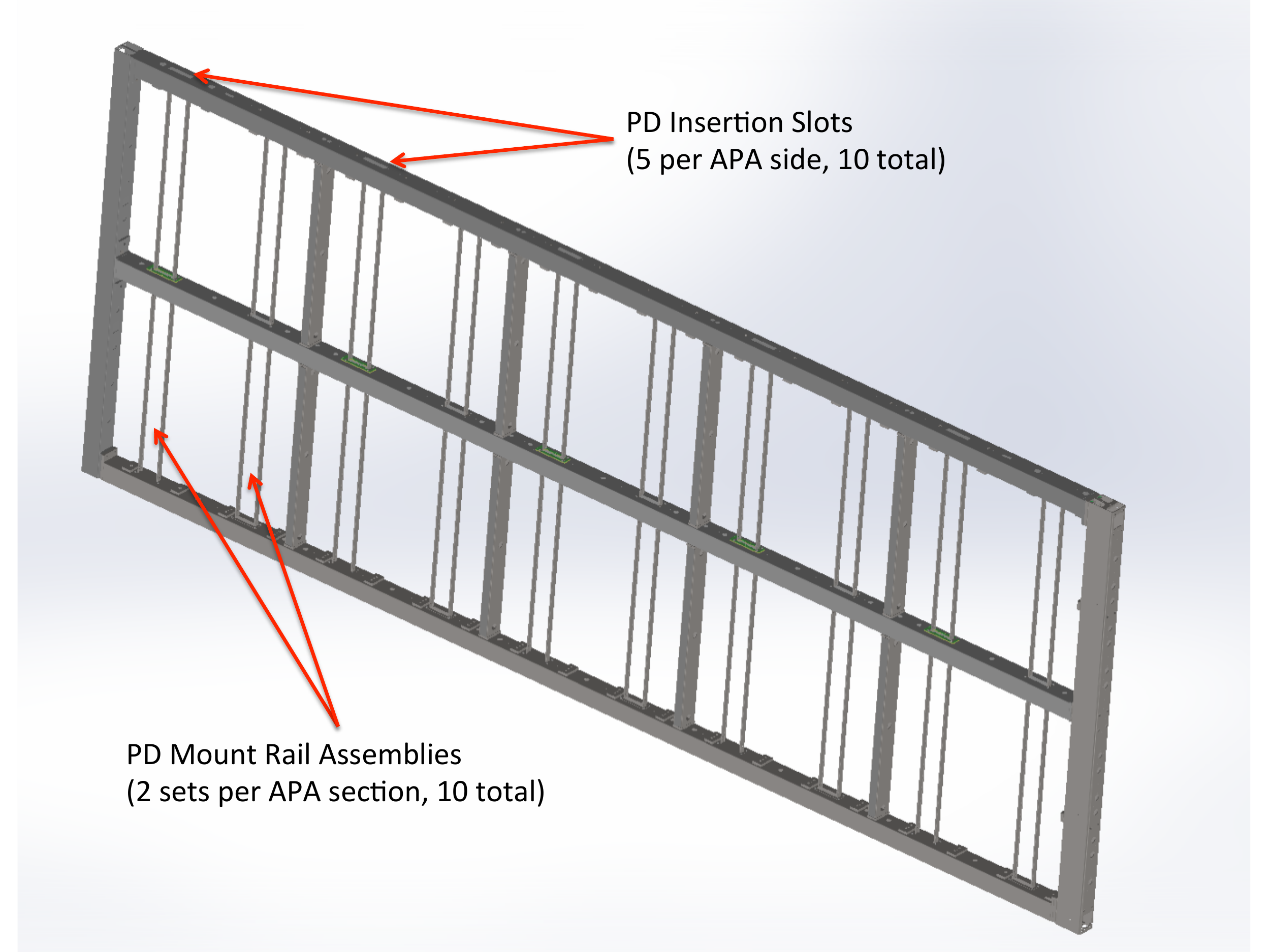}
\end{dunefigure}

An operational consideration for the \dword{pds} design is the presence in the \dword{lartpc} of the long-lived cosmogenic radioisotope \Ar39, which has a specific activity in argon extracted from the atmosphere of approximately \SI{1}{Bq/kg}~\cite{bkds}. The isotope undergoes beta decay at a mean beta energy of \SI{220}{keV} with an endpoint of \SI{565}{keV} and makes up $\sim$70\% of the radiological background signal.
In the \SI{10}{kt} \dword{fd} modules, this leads to a rate of more than \SI{10}{MHz} of very short ($\sim$\SI{1}{mm}) tracks uniformly distributed throughout the module, each of which produces several thousand \dword{vuv} scintillation photons. This continuous background affects the \dword{daq}, trigger, and spatial granularity required of the \dword{pds}. Spatial granularity helps even more with rare but more energetic radiological backgrounds that can produce multi-photon signals, but only a single detector. Low energy neutrinos, on the other hand, will produce coincident  signals on multiple channels, allowing them to be easily identified.

\subsection{Design Overview}
\label{sec:pds:des-ov}

The  large-area light collectors are the core modular elements of the \dword{pds}.  They convert incident \SI{127}{nm} scintillation photons into photons in the visible range (>\SI{400}{nm}) that compact \dword{sipm} photosensors, in turn,  convert to an electrical signal. The light collector design must optimize the costs of various components of the system while meeting the performance requirements.  Even though production cost and key performance parameters of \dwords{sipm} have improved significantly in recent years, covering the light detector surfaces with enough of them to meet the physics requirements of the \dword{pds} would be cost-prohibitive. 
The light collector design should maximize the active \dword{vuv}-sensitive area of the \dword{pds} while minimizing the necessary photocathode (\dword{sipm}) coverage. This is detailed in
Section~\ref{sec:fdsp-pd-lc}.

\subsubsection{Light Collectors} 
\label{sssec:photoncollectors}

\dword{dune} investigated many \dword{pd} light collector module options before forming the \dword{sp} \dword{pd} consortium; we selected four for further development. 
Two designs, \dword{sarapu}\footnote{\textit{Arapuca} is the name of a simple trap for catching birds originally used by the Guarani people of Brazil.} 
and \dword{xarapu}, 
use a relatively new scalable concept designed to provide  
significantly better performance than the other approaches. Functionally, \dword{arapuca} is a light trap that captures wavelength-shifted photons inside boxes with highly reflective internal surfaces until they are eventually detected by \dwords{sipm} or are lost.  The two other designs are based on the use of wavelength-shifters and long plastic light guides coupled to \dwords{sipm} at the ends. Their performance could meet the basic physics requirements but with only a small safety margin, and their performance is not easily scalable within the geometric constraints of the \dword{spmod}. 

The performance of the light collector is characterized by the \emph{collection efficiency} of the device, which is defined as the ratio of the number of detected photons and the number of \SI{127}{nm} scintillation photons incident on the light collector window.  For the \dword{arapuca}, this depends on three distinct aspects of the design:
\begin{itemize}
    \item The efficiency of the conversion of incident \dword{vuv} photons to photons trapped inside the cavity. This depends primarily on the wavelength shifter(s) efficiency and the fraction of converted photons that enter the cavity.
    \item The efficiency for the captured photons to eventually fall on the photosensor. This depends primarily on the reflectivity of the surfaces of the cavity, the geometry of the cavity, and the ratio of the photosensitive area to the light collector window area.
    \item The efficiency for the photosensor to convert incident photons to an electronic signal. This depends on the energy of the converted photons in the cavity and properties of the commercial sensor.
\end{itemize}
The \emph{effective area} of a \dword{pd} module is another useful figure-of-merit that is defined to be the photon collection efficiency multiplied by the photon collecting area of a \dword{pd} module. 

\textit{\dword{sarapu}:} In an \dword{sarapu} cell, enhanced photon trapping is attained when using the wavelength--shifting plates and the technology of the dichroic short-pass optical filter. These commercially available interference filters use multi-layer thin films highly transparent to photons with a wavelength below a tunable cutoff, 
with transmission typically more than 95\%, yet almost perfectly reflective to photons with a wavelength above the cutoff.  Such a filter forms the entrance window to a cell whose internal surfaces are covered by highly reflective acrylic foils
except for a small fraction occupied by \dwords{sipm}.

For the collector to act as a photon trap, the external face of the dichroic filter is coated with a wavelength shifting coating with an emission wavelength less than the cutoff wavelength of the filter. 
The transmitted photons pass through the filter where they encounter a second wavelength-shifter coated on either the inside surface of the filter plate or on the rear surface of the box.
This second wavelength-shifter has emission spectra which exceed the cutoff wavelength, thus trapping the photon inside the box.
Trapped photons reflect off the inner walls and the filter surface(s) (of reflectivity typically greater than \SI{98}{\%}) 
and have a high probability of impinging on a \dword{sipm} before being lost to absorption. 

Several iterations of the \dword{sarapu} design were tested in small cryostats (Section~\ref{sec:valid-initial}) and \dword{pdsp} (Section~\ref{sec:valid-pdsp}), establishing the viability of the concept for \dword{dune}.

\textit{\dword{xarapu}:} The \dword{xarapu}, adopted as the baseline design and detailed in 
Section~\ref{sec:fdsp-pd-lc}, is an evolution of the first generation \dword{sarapu}.  In the \dword{xarapu}, the secondary \dword{wls} layer of the \dword{sarapu} (a vacuum-deposited layer of \dword{wls} applied to the inner surfaces of the cell) is replaced by a \dword{wls} plate with an emission wavelength higher than the filter plate transmission frequency.  Wavelength shifted photons from this plate have two mechanisms for transport to the photosensors inside the cell: either they are transported along the \dword{wls} plate to the photosensors via total internal reflection, or those escaping the plate are captured due to reflection from the dichroic filter by the standard \dword{arapuca} effect.
The concept is illustrated in Figure~\ref{fig:arapuca}.
Validation of the \dword{xarapu} design is described in Sections~\ref{sec:xarapuca-unicamp} and \ref{sec:iceberg-teststand}.

\begin{dunefigure}[Schematic representation of the \dshort{xarapu} operating principle]{fig:arapuca}
{Schematic representation of a single-sided readout  \dword{xarapu} operating principle.  This example assumes a filter cutoff of \SI{400}{nm}. (Note: In the original \dword{arapuca} concept, the second wavelength-shifter was coated on the inner surface of the filter and the \dword{wls} plate shown in the figure was absent.)}               
\includegraphics[height=7cm]{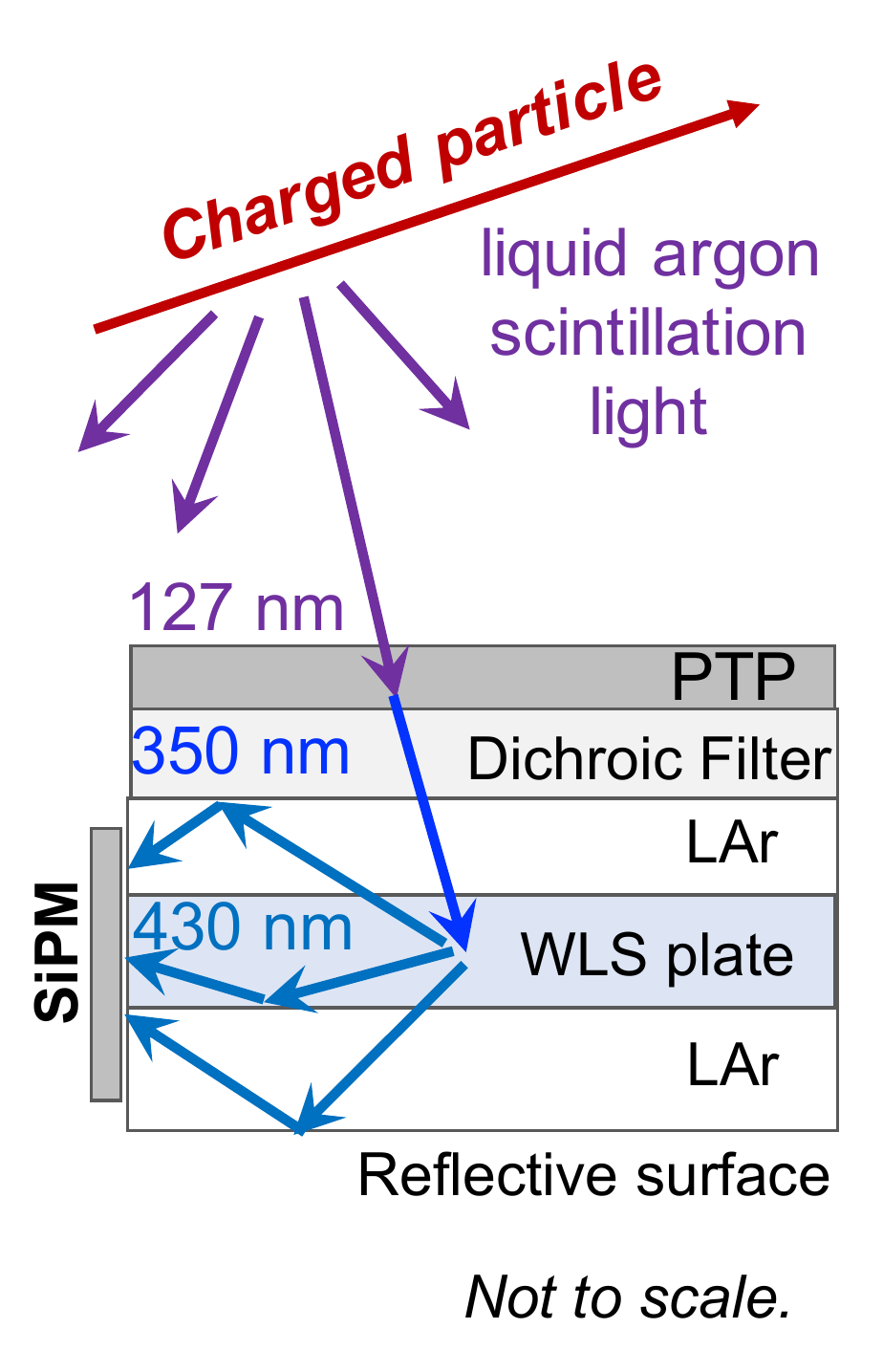}   
\end{dunefigure}

While the \dword{sarapu} modules deployed in \dword{pdsp} collect light from only one direction, the next generation \dword{xarapu} can be deployed as either single-face or dual-face readout by using either an opaque reflector plate (single) or a second dichroic filter window (dual) on the second face. 
Figure~\ref{fig:3dtpc-pd} shows how a light-collector module is incorporated into an \dword{apa}. One module spans the width of an \dword{apa}. Figure~\ref{fig:pds-pd-full-module} (left) shows a detail of the module where the 24 \dword{xarapu} cells are visible on either side of a signal summing and interface board that becomes enclosed by the hollow central beam of the \dword{apa} frame. Figure~\ref{fig:pds-pd-full-module} (right) illustrates how a module is inserted into an \dword{apa} frame.

\begin{dunefigure}[\threed model of \dshorts{pd} in the \dshort{apa}]{fig:3dtpc-pd}
{\threed model of \dwords{pd} in the \dword{apa}. The model on the left shows the full width of a one \dword{apa} deep slice of the \dword{tpc} illustrating the APA-CPA-APA-CPA-APA system configuration. The figure on the right shows a detail of the top far side of the \dword{tpc} where three photon collector modules are visible.}
\includegraphics[height=6.5cm]{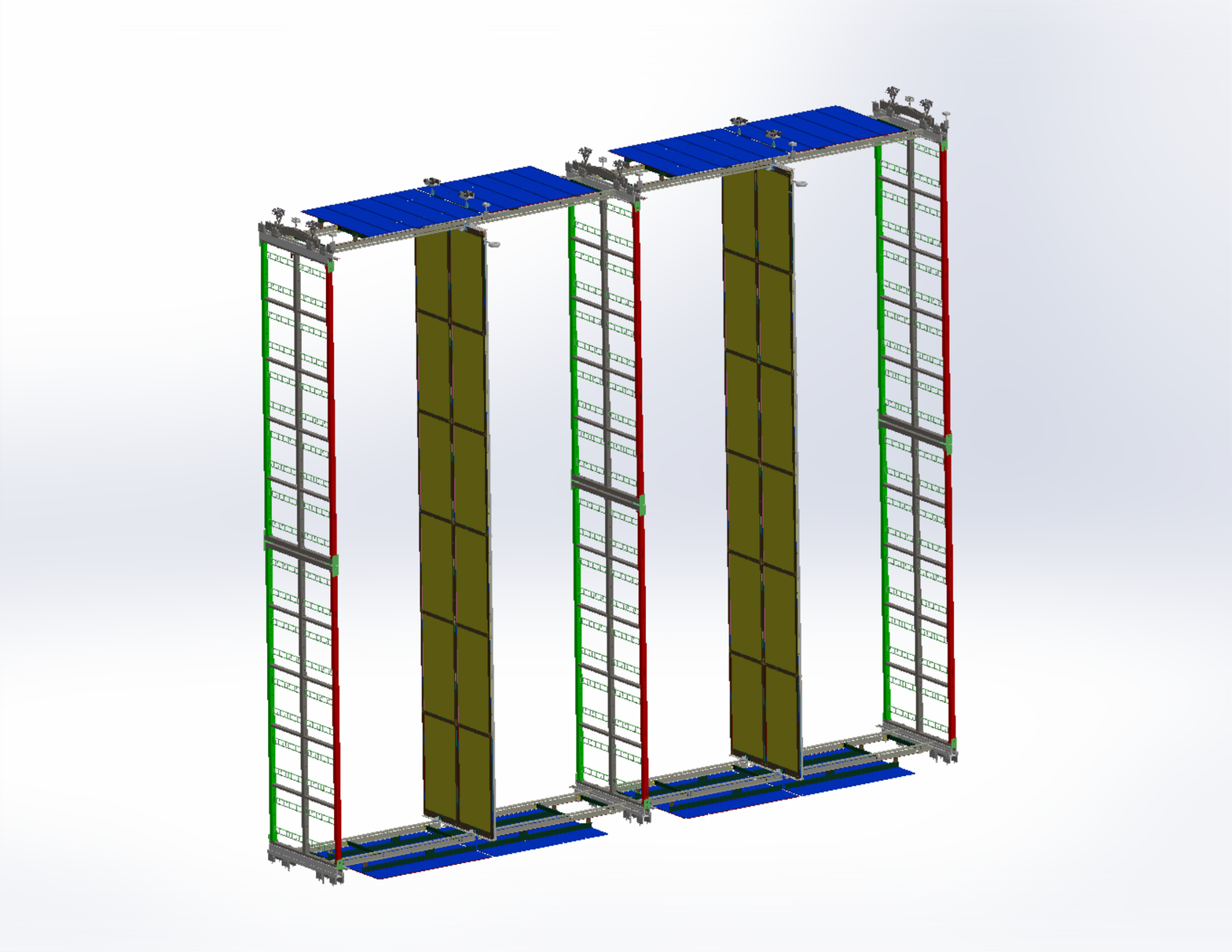}
\includegraphics[height=6.5cm]{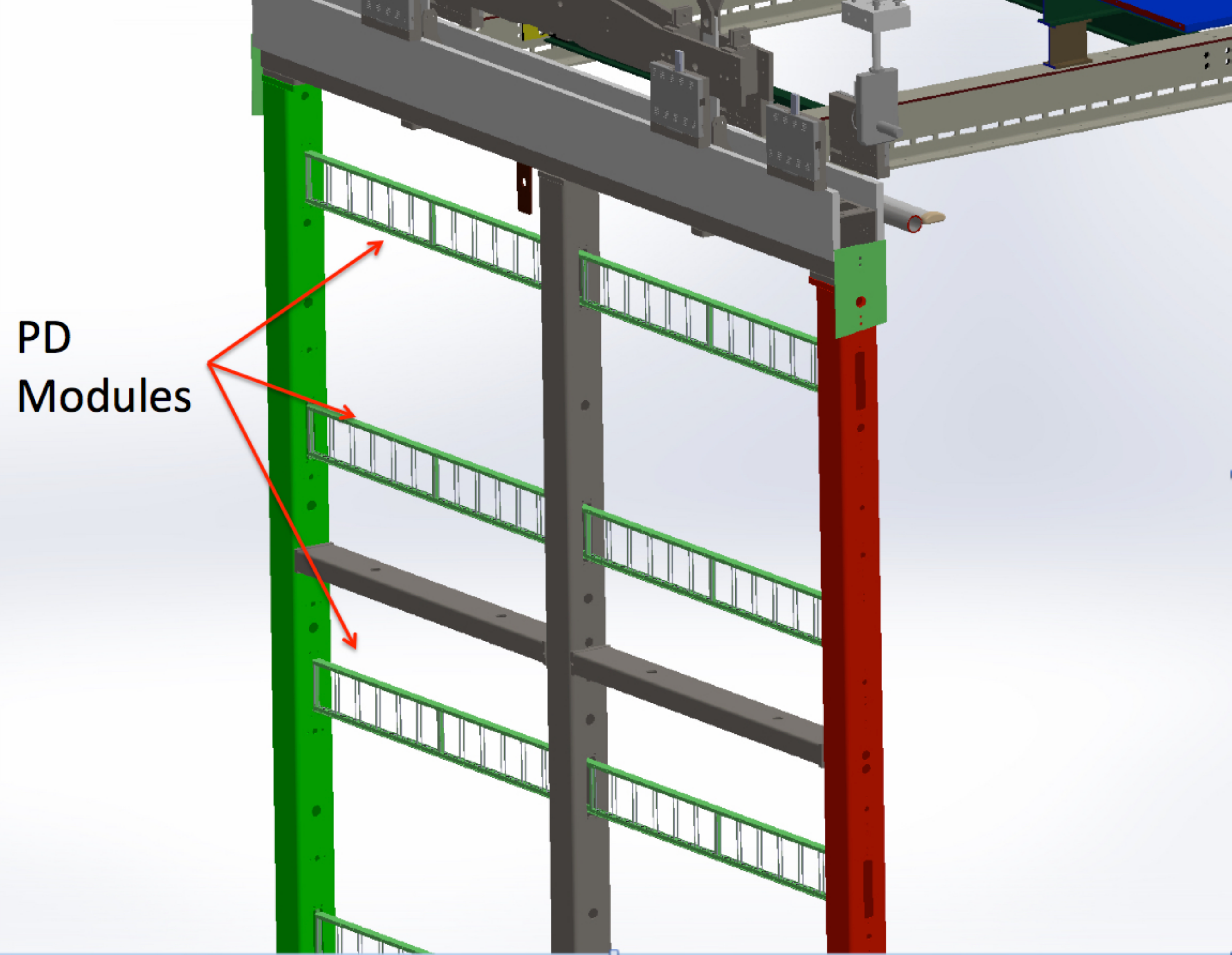}
\end{dunefigure}

\begin{dunefigure}[PD Module being inserted into an APA frame]{fig:pds-pd-full-module}
{Solid model of a \dword{pd} module being inserted into an \dword{apa} frame (wires not shown), which is done after the APA assembly is completed.}
\includegraphics[height=6.5cm]{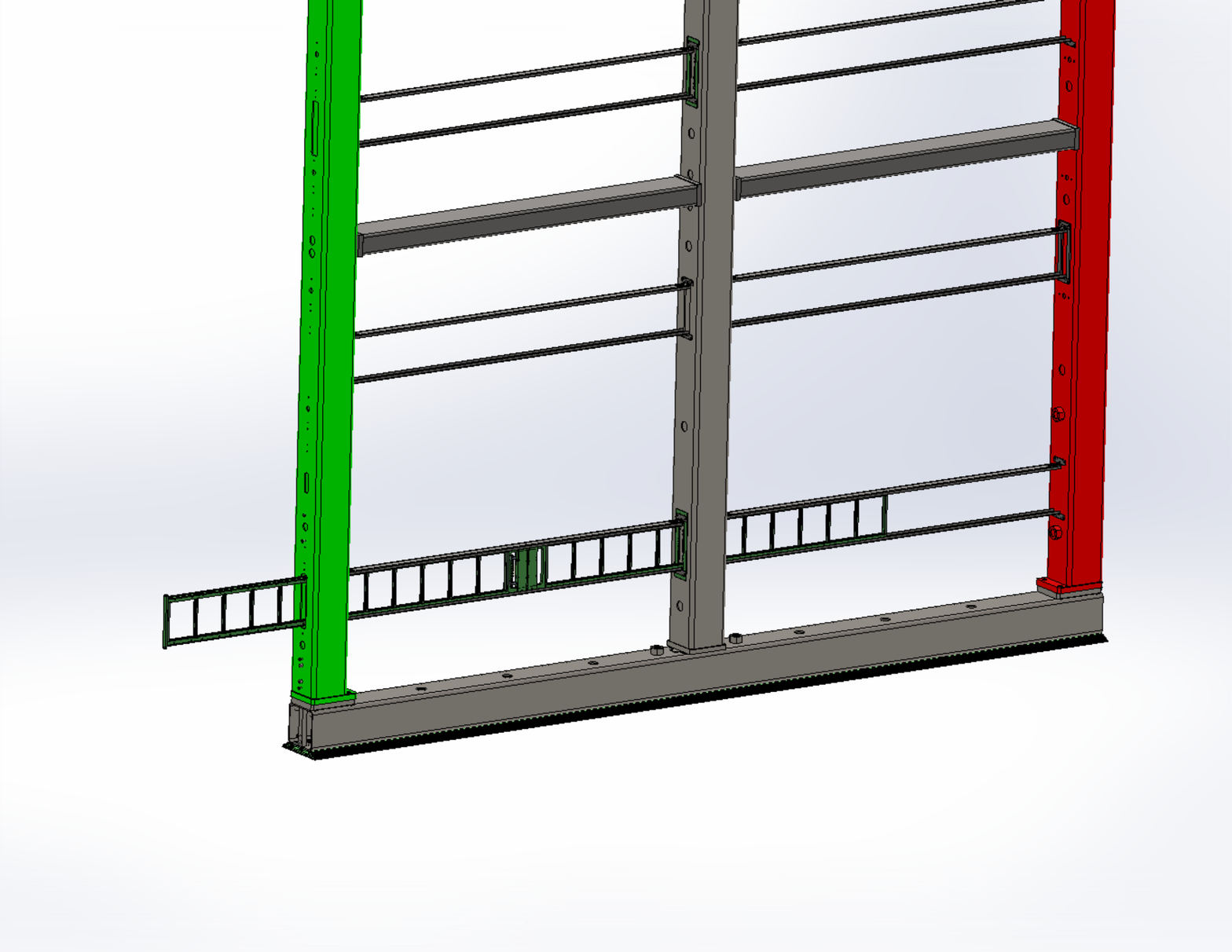}
\end{dunefigure}

The \dword{xarapu} light collector design has the flexibility to accommodate greater demands, such as might be desired for \dword{sbn} physics, without major changes. One example of this flexibility is the ability to increase the number of \dwords{sipm} to increase light yield, which could be incorporated quite late in the final design stages because it would not involve significant mechanical changes.

\subsubsection{Silicon Photosensors} 
\label{sssec:photosensors}

The \dword{sp} \dword{pds} uses a multi-step approach to scintillation light detection with the final stage of conversion into electrical charge performed by \dwords{sipm}. Robust photon conversion efficiency, low operating voltages, small size, and ruggedness make their use attractive in the \dword{sp} design where the \dwords{pd} must fit inside the \dword{apa} frames. 

Based on extensive testing and experience with the vendor, we have selected a \SI{6}{mm}$\times$\SI{6}{mm} \dword{mppc} 
produced by Hamamatsu\footnote{Hamamatsu\texttrademark{} Photonics K.K., \url{http://www.hamamatsu.com/}.} (Japan) as the baseline \dword{sipm} device. 
We are also vigorously pursuing an alternative based on the design of a device developed for operation in \dword{lar} by the DarkSide experiment collaboration and Fondazione Bruno Kessler (FBK)\footnote{Fondazione Bruno Kessler\texttrademark{}, \url{https://www.fbk.eu}.} (Italy).

The baseline \dword{pds} design has \num{192} 
\dwords{mppc} per \dword{pd} module with groups of \num{48} \dwords{mppc} electrically ganged into four electronics readout channels, which provides some spatial granularity within a module and helps to reduce the impact of radiological noise. This configuration has a total of \num{288000} \dwords{mppc} per \dword{spmod}. 


\subsubsection{Readout Electronics} 

The \dword{pds} design requires an electronics readout system that collects and processes electric signals from photosensors in \dword{lar} to (1) provide the interface to trigger and timing systems,
and (2) enable data transfer to an offline storage system for physics analysis. The quantitative requirements for the system are driven by many \dword{fd} level specifications that affect signal size sensitivity, \dword{s/n}, timing resolution, event size and data transfer limits from the \dword{daq}, power needs and dissipation limits, channel density and channel count, and cost. 

As described in Section~\ref{sssec:photosensors}, each electronics signal from a \dword{pd} module is formed from an ensemble of 48 Hamamatsu \dwords{mppc} summed into a single channel by a combination of passive and active ganging.  A cold amplifier adjusts the \dword{mppc} output signal level before transmitting the signal over $\sim\,$\SI{20}{m} long\footnote{Cable lengths are not uniform across all \dwords{pd}, ranging from \SI{15}{m} at the shortest to \SI{27.25}{m} at the longest, averaging \SI{20}{m}.} twisted-pair cables to the input of \dword{fe} \dwords{adc} outside the cryostat.  The twisted pair cable is impedance-matched to  the receiver amplifiers for the \dwords{adc} to optimize common-mode noise rejection at the input of the front-end digitizer. 

The digitizer is a low-cost solution based on commercial ultrasound \dword{asic} chips rather than digitizers based on flash \dwords{adc} used in \dword{pdsp}. Inspiration for this \dword{fe} comes from the system developed for the \dword{mu2e} experiment \dword{crt} readout system as described in Section~\ref{sec:electronics}.

The \dword{fe} will continuously digitize the input signals for each channel and store waveforms alongside event metadata that meet the trigger conditions. In an externally triggered waveform mode, the configured waveform window at the time of the external trigger is also stored. These internally or externally triggered waveforms are transmitted to the \dword{daq} board reader processes for storage. The \dword{daq} system and data storage limitations impose constraints on the data bandwidth, readout rates, and zero suppression. 

Single \phel signals from \Ar39 and radiological backgrounds will dictate threshold level adjustments.
We will configure the photon readout to trigger on signals from the central trigger and timing systems for a variety of configurable events, e.g., beam events, cosmic
muons, periodic triggering, random triggering, or any combination of
these. A special trigger condition needed for \dword{snb} observation will enable readout of all digitized data over predefined periods.
The interface design will define the power, grounding, and rack schemes.

\subsection{Options to Improve Uniformity of Response} 

Because the \dword{pd} modules are installed only in the \dword{apa}, light collection is not uniform over the entire active volume of the \dword{tpc}. 
Though not necessary to meet the basic \dword{dune} performance specifications, improving the uniformity of the response would increase the trigger efficiency, simplify the analysis for \dword{snb} neutrinos and increase the light yield of the detector, which could enable enhanced calorimetric measurements based on light emitted by the ionizing particles.

The primary source of non-uniformity of response is that the Rayleigh scattering length for \SI{127}{nm} scintillation photons is relatively short compared to the size of the \dword{tpc} active volume.   
In parallel to the baseline design, we are pursuing two options that convert \SI{127}{nm} scintillation photons to longer wavelength photons that have a longer Rayleigh scattering length, significantly improving light collection uniformity:
\begin{itemize}
\item Use of a wavelength--shifter--coated cathode plane; see Appendix Section~\ref{sec:fdsp-pd-enh-cathode}.
\item Use of trace amount of xenon in the \dword{lar}; see Appendix Section~\ref{sec:fdsp-pd-enh-xenon}.
\end{itemize}

\subsection{Overview Summary} 
\label{sec:fdsp-pd-ov-summ}

As described in Sections~\ref{subsec:fdsp-pd-simphys-ndk} and~\ref{subsec:fdsp-pd-simphys-snb}, the performance required for the \dword{pds} to achieve 99\% for tagging nucleon decay events is a light yield of \SI{0.5}{PE/MeV} at the furthest point (near the \dword{cpa}), while the requirement to enable a calorimetric energy measurement with the \dword{pds} for low-energy events like \dwords{snb} is \SI{20}{PE/MeV} averaged over the active volume (FD-SP-3 in Table~\ref{tab:specs:SP-PDS}). The relationship between these two different light yields and the collection efficiency of the \dword{pds} depends on the assumed Rayleigh scattering length. Conservatively assuming that this length is \SI{60}{cm}, the \SI{0.5}{PE/MeV} at the \dword{cpa} corresponds to a collection efficiency of 1.3\% and the \SI{20}{PE/MeV} averaged over the active volume corresponds to an efficiency of 2.6\%.

Although full validation of the \dword{sp} \dword{pds} light collection system is still in progress, initial results on the \dword{xarapu} prototype are very encouraging -- the single cell prototype (Section~\ref{sec:xarapuca-unicamp}) has achieved a collection efficiency of $\SI{3.5}{\%}$.  This is significantly higher than the requirement.

A measured collection efficiency in excess of the specification ensures a safety margin against degradation in performance of the optical components over time and failures of a fraction of inaccessible active components during long term operation of the detector. It also opens broader opportunities for the photon detection system to extend the physics reach of the experiment, perhaps in unanticipated ways.



\section{Light Collectors}
\label{sec:fdsp-pd-lc}

The \dword{xarapu}, adopted as the baseline design, is an evolution of the \dword{arapuca} concept that further improves the collection efficiency, while retaining the same working principle, mechanical form factor and active photosensitive coverage. 
 In the original \dword{arapuca} concept, two wavelength-shifters were coated on either side of the dichroic filter window. 
 In contrast, the \dword{xarapu} replaces the inner surface coating with a wavelength-shifter-doped 
 polystyrene light guide\footnote{Eljen EJ-286\texttrademark{}.} occupying a portion of the cell volume, with the silicon photosensor readout mounted along the narrow sides of the cell, as illustrated in Figure~\ref{fig:pds-x-arapuca-cell}. The model shown is a single cell design used for prototypes that allows for photons to enter from either face, however one window can be replaced with an opaque reflecting surface for sensitivity through just one face.

Photons entering the light guide plate are absorbed and wavelength-shifted with high efficiency, and some fraction (those incident on the plate surface at greater than the critical angle) are transported to the readout via total internal reflection. The \dword{lar} gaps between the plate and the surfaces of the cavity ensure the discontinuity of the refractive index that contributes to effective trapping of the photons (n$_{plate}$=1.58 and n$_{LAr}$=1.24 for the wavelengths emitted by the plate).
Those exiting the plate reflect off the filter or other highly reflecting surfaces of the cell, with some fraction eventually incident on a \dword{sipm}, as in a standard \dword{arapuca} cell.
\dword{xarapu} is thus effectively a hybrid solution between the \dword{sarapu} and the \dword{wls} light guide concepts implemented in \dword{pdsp}.

This solution minimizes the number of reflections on the internal surfaces of the cell and thus minimizes the probability of photon loss. 
We have performed a full numerical description of the \dword{xarapu} using the Geant4 framework, following previous studies done for the \dword{sarapu} device~\cite{Marinho:2018doi}.  The comparison between the two kinds of devices is dependent on the value of absorption length of the bar, which was not known precisely, so the gain in efficiency for the \dword{xarapu} with the dimensions tested at \dword{unicamp} is estimated to be between 15 and 40\% when compared to the \dword{sarapu} with same dimensions and number of \dwords{sipm}.
Results from prototype measurements are presented in Sections~\ref{sec:xarapuca-unicamp} and \ref{sec:iceberg-teststand} and are consistent with the simulations.

 \begin{dunefigure}[\dshort{xarapu} conceptual model]{fig:pds-x-arapuca-cell}
{Simplified conceptual model depicting a \dword{xarapu} cell design sensitive to light from both sides: assembled cell (left),  exploded view (right). The yellow plates represent the dichroic filters (coated on their outside surfaces with \dword{ptp} \dword{wls}), the pale blue plate represents the wavelength shifting plate, and the photosensors are visible on the right side of the cell. The size and aspect ratio of the cells can be adjusted to match the spatial granularity required for a \dword{pd} module. .
} 
  \includegraphics[height=.25\textheight]{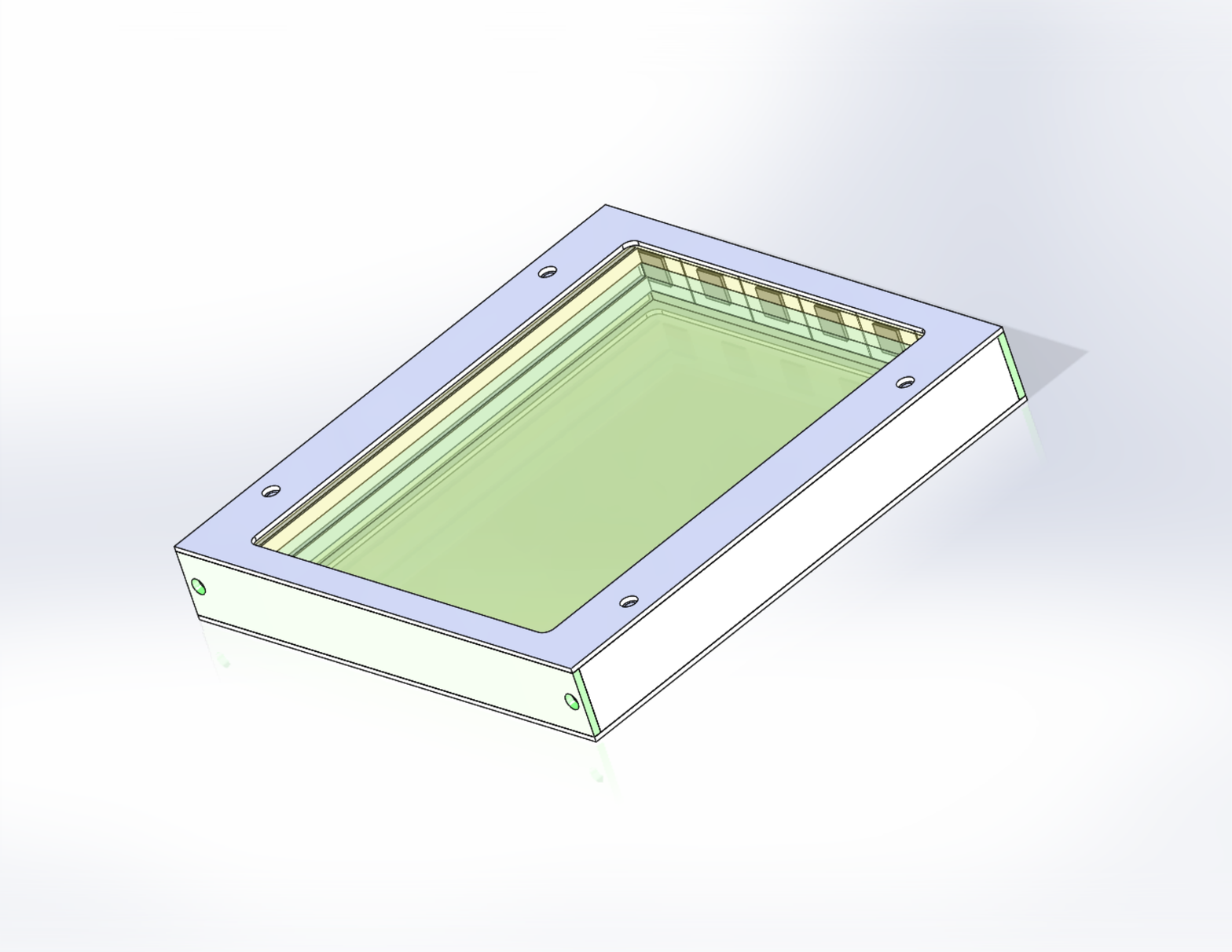}
  \includegraphics[height=.25\textheight]{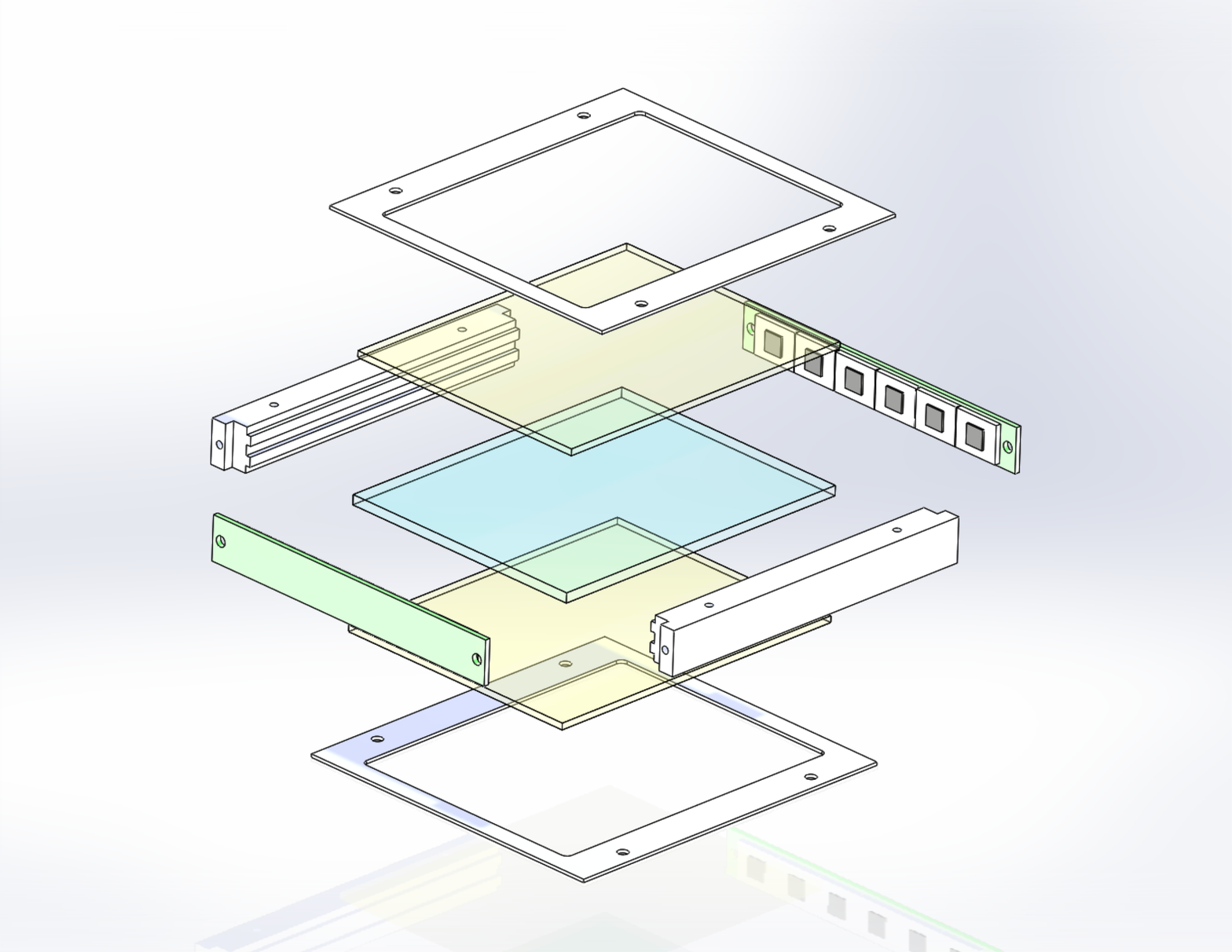}
\end{dunefigure}

The  \dword{pd} module designed for the \dword{dune} \dword{spmod}, illustrated in Figure~\ref{fig:pds-x-arapuca-full-module}, consists of four supercells, each containing a rectangular 
light guide inside the cell positioned behind an array of six dichroic filters that form the entrance window.  
This design is easily configurable to detect light from just one side, as required for the side \dword{apa}s, or from both sides for the central \dword{apa}s. 

For dual-sided \dword{xarapu} modules, dichroic filters are placed on both sides of the cell facing the drift volumes.  In the case of the single-sided device, the back side of the cell has a layer of highly reflective Vikuiti\footnote{3M Vikuiti\texttrademark\  ESR 
}
 to act as a  reflector.  In both cases, the \dword{sipm} arrays are installed on two of the narrow sides of the cell perpendicular to the windows, parallel to and up against the light guide thin ends. Half of the \dword{sipm} active detection area collects photons from the light guide, a quarter of the area on either side of the guide is free to collect the fraction of photons reflected off the cell walls and windows. 
This fraction of photosensor coverage for photons emerging from the light guide ends is a result of using a standard \num{6}$\times$\SI{6}{mm$^2$} \dword{sipm} placed symmetrically with respect to the mid-plane of the bar. Simulation of two additional \dword{sipm} geometries with the same active area (\num{4}$\times$\SI{9}{mm$^2$} and \num{3}$\times$\SI{12}{mm$^2$}) showed no substantial difference in the detection efficiency that would justify a custom geometry for the \dword{sipm}.

The basic mechanical design of the \dword{xarapu}-based \dword{pd} modules is similar to that of the two prototypes produced for \dword{pdsp}. 
The prototype design was modified to include mechanical changes to allow both single-sided and dual-sided readout; an increase in the light collection area made possible by larger slots in the \dword{apa}; and a modified cabling and connector plan necessary to move the \dword{pd} cables out of the \dword{apa} side tubes while reducing cable requirements to one Cat-6 cable per \dword{pd} module.

\begin{dunefigure}[\dshort{xarapu} module indicating four supercells]
{fig:pds-x-arapuca-full-module}
{\dword{xarapu} module overview. A module, which spans the width of an \dword{apa}, includes 24
 \dword{xarapu} cells, grouped into a set of four supercells of six cells each. In the center, active ganging \dwords{pcb} collect the signals and mechanically connect the supercells.}
   \includegraphics[width=17cm]{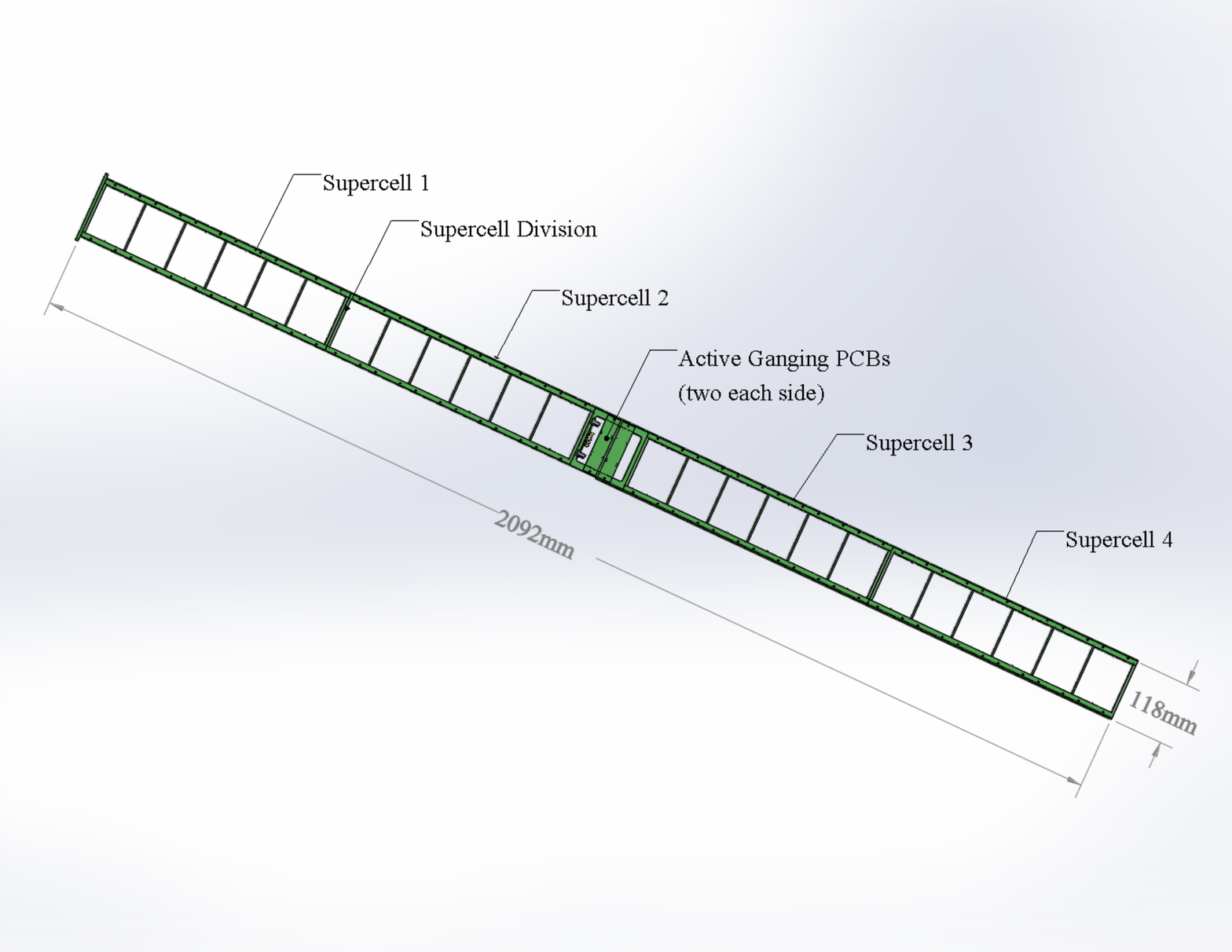}
\end{dunefigure}

An \dword{xarapu} module is assembled in a bar-like configuration with external dimensions inside the \dword{apa} frame of \SI{2092}{mm}$\times$\SI{118}{mm}$\times$\SI{23}{mm},  allowing insertion between the wire planes through each of the ten slots (five on each side) in an \dword{apa}. In addition, there is a header block \SI{5}{mm}(long)$\times$\SI{135}{mm}(wide) at the insertion side of the module used to fix the module inside the \dword{apa} frame, bringing the maximum length to \SI{2097}{mm} and the maximum width to \SI{135}{mm}.
The module contains four \dword{xarapu} supercells, each with six dichroic filter-based optical windows (for the single-sided readout) or twelve windows (double-sided readout) with an exposed area of \SI{78}{mm}$\times$\SI{93}{mm}.  
The total window area for each (single-sided) supercell \dword{xarapu} is \SI{43524}{mm$^2$}.
The internal dimensions of a supercell are approximately \SI{488}{mm}$\times$\SI{100}{mm}$\times$\SI{8}{mm}. A \dword{wls} plate (Eljen EJ-286) of dimensions \SI{487}{mm}$\times$\SI{93}{mm}$\times$\SI{3.5}{mm} is centered in the supercell midway between the dichroic windows. 

The thickness of \SI{3.5}{mm} for the plate is chosen to allow the almost complete absorption of the photons wavelength-shifted by the \dword{ptp} and to ensure the nominal conversion efficiency. This thickness allows a \SI{2}{mm} \dword{lar} gap on both sides of the plate, which prevents any physical contact of the surfaces even considering the tolerances on material thicknesses and plate flatness.   

\begin{dunefigure}[Exploded \dshort{xarapu} supercell]{fig:pds-x-arapuca-exploded-Detail}
{Detailed exploded view of \dword{xarapu} supercell. Note that components are designed to be cut from \frfour G-10 sheets to simplify fabrication.}
   \includegraphics[height=.55\textheight]{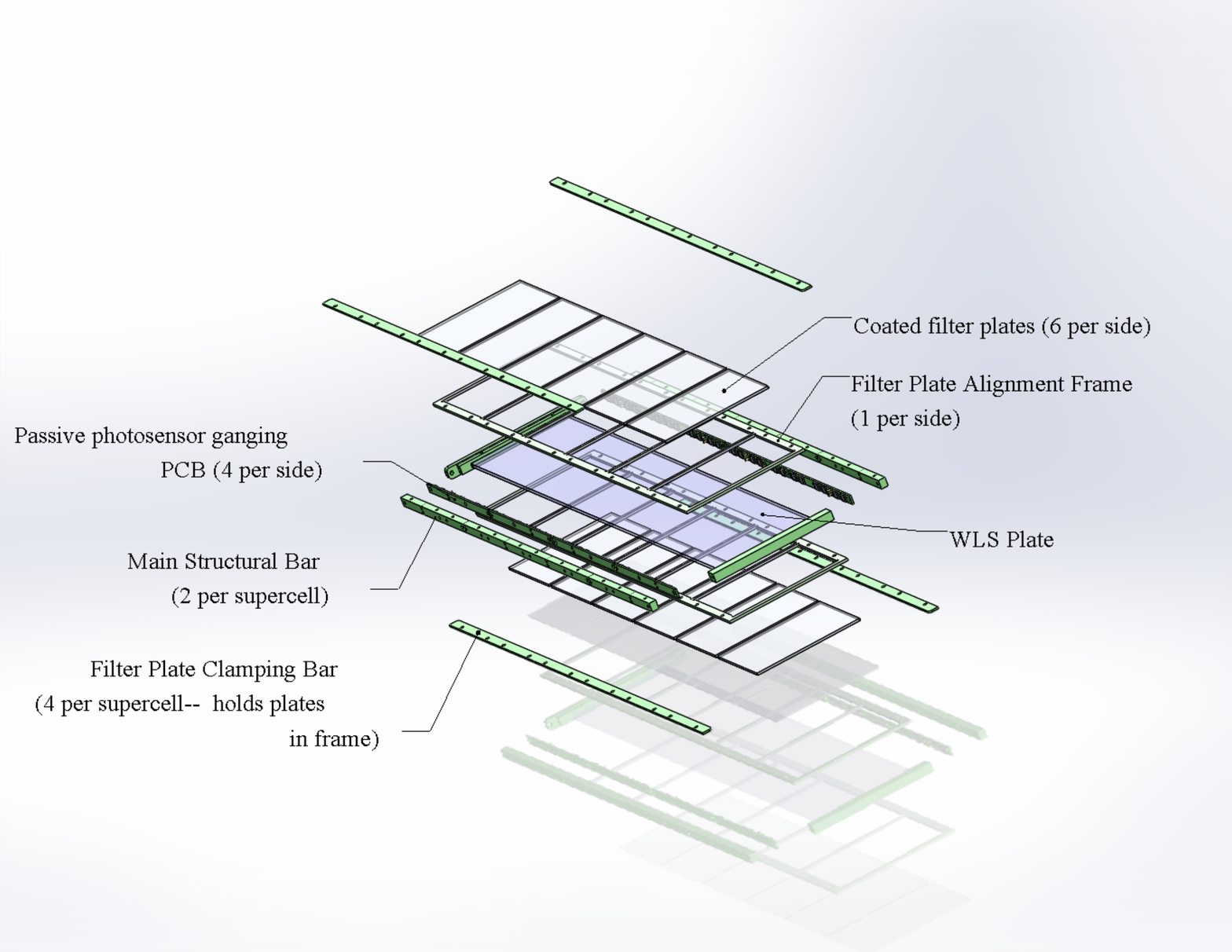}
\end{dunefigure}

To reduce production costs and simplify fabrication, most of the \dword{pd} components are designed to be water-jet cut from sheets of \frfour G-10 material, with minimal post-cutting machining required (mostly the tapping of pre-cut holes).  The current design contains many small fasteners; we will investigate replacing some of the fasteners with epoxy lamination of cut sheets where appropriate and cost effective.

The \dwords{sipm} are mounted to \dwords{pcb} called ``photosensor mounting boards'' that  are positioned on the long sides of the supercell.  
Six \dwords{sipm} are mounted to a single photosensor mounting board.  The mounting boards incorporate spacers that position the face of the photosensors a nominal \SI{0.5}{mm} from the face of the \dword{wls} plate.  All six are electrically connected in parallel (``passively ganged'').

 Before mounting the boards into the \dword{xarapu} module, the boards are tested at room and \dword{ln} temperatures. 
 Each supercell uses eight photosensor mounting boards, each with six \dwords{sipm} (Figure~\ref{fig:mounting-board-routing-board}~(top)), 
 to accommodate the 48 \dwords{sipm}.  The ganged signal outputs from these boards are connected to traces in signal routing boards at the edge of the \dword{pd} module. These signal routing boards also act as mechanical elements in the design, mechanically joining the supercells and providing for rigidity.  The routing boards \dwords{pcb} are four-layer boards, \SI{1046}{mm}$\times$\SI{23}{mm}$\times$\SI{1.5}{mm}.

\begin{dunefigure}[\dshort{xarapu} SiPM mounting and signal routing boards]
 {fig:mounting-board-routing-board}
{Model of photosensor mounting board (top) and signal routing \dword{pcb} (bottom) for \dword{xarapu} module.  Six Hamamatsu \dwords{mppc} are passively ganged and the ganged signals transmitted along the routing board to the active ganging circuits in the center of the module.}
\includegraphics[height=7cm]{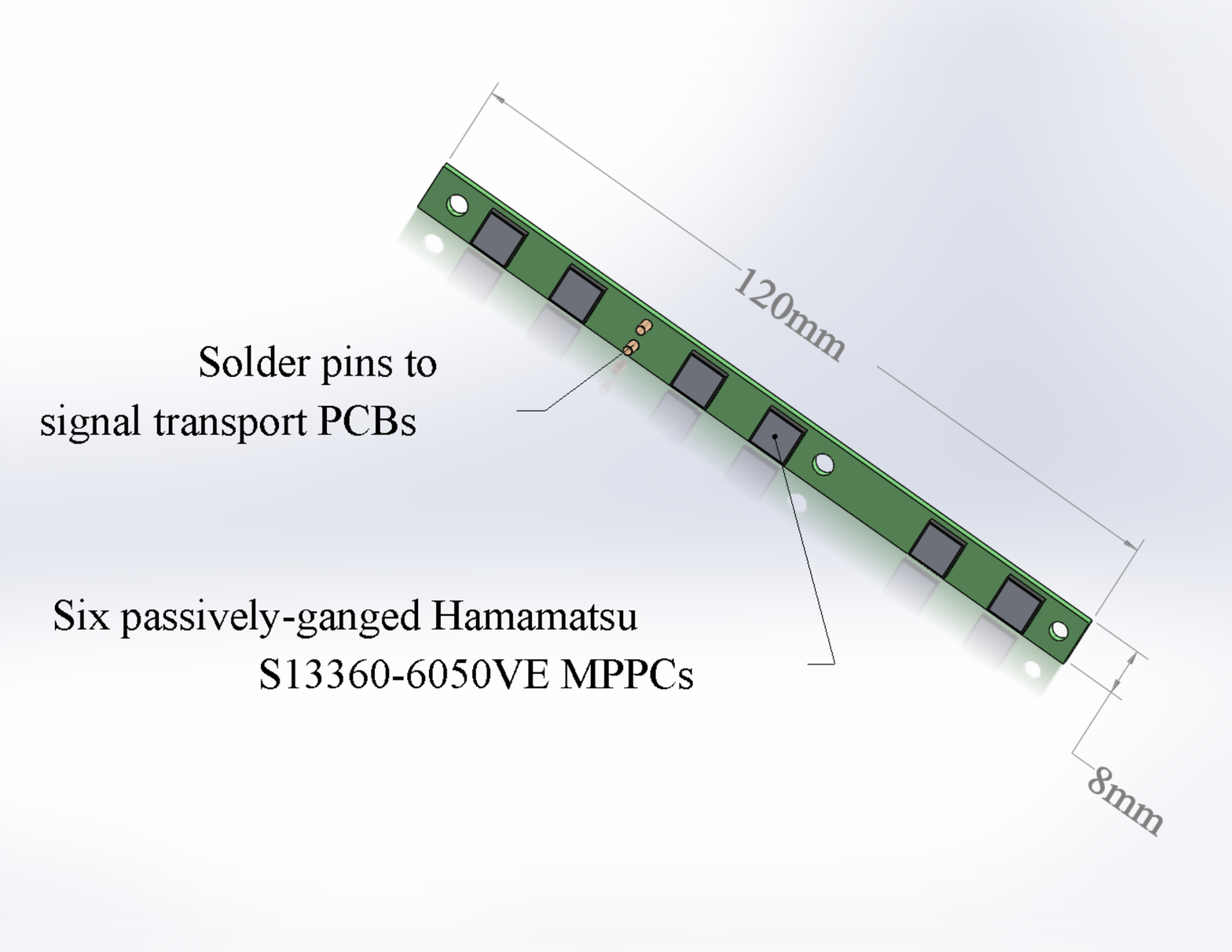}
\includegraphics[height=7cm]{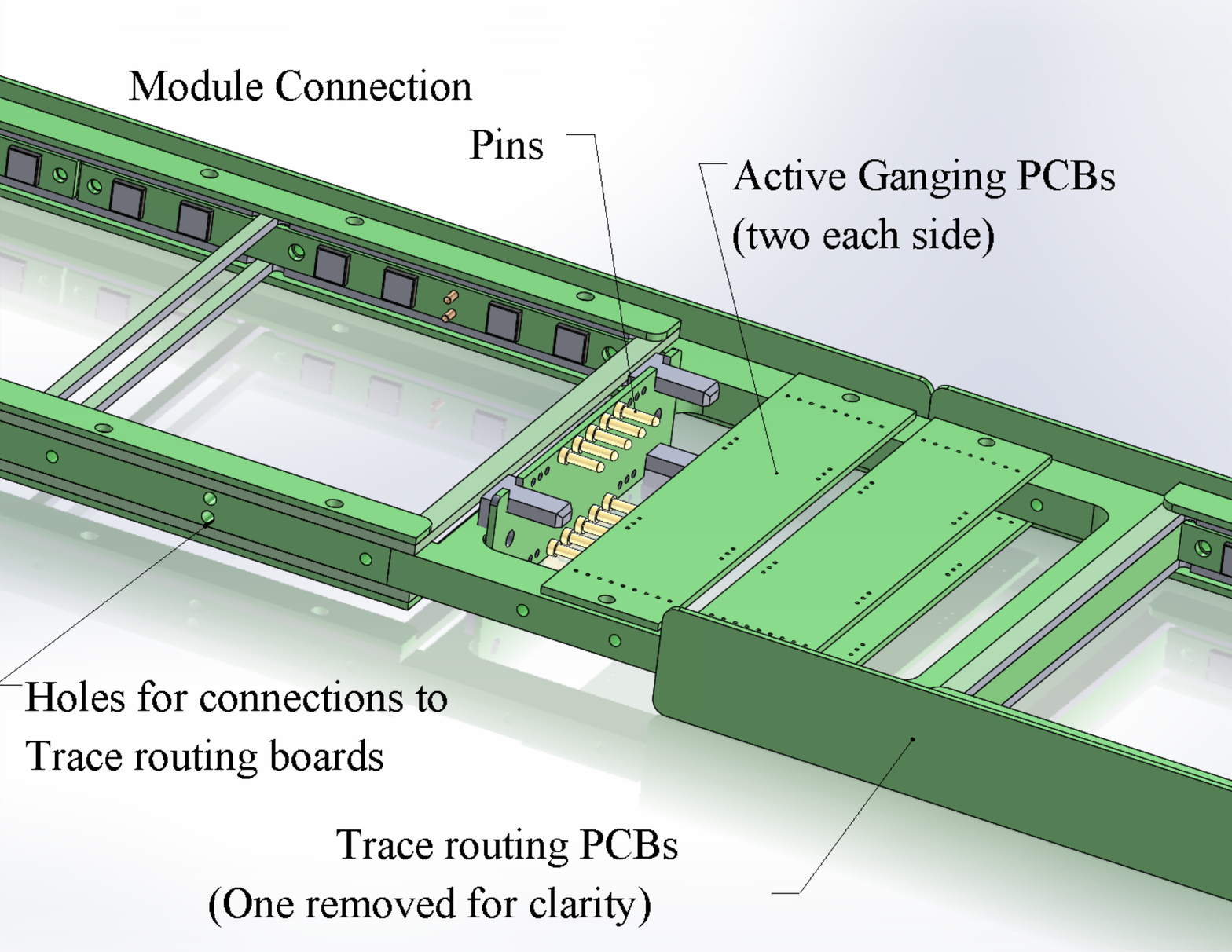}
\end{dunefigure}
The passively ganged signals are then routed through these boards to an active-ganging \dword{pcb} at the center of the module, where all eight passively ganged signals from a single supercell are actively ganged into one output channel (Figure~\ref{fig:mounting-board-routing-board}~(bottom)). This summed output from a single supercell is then connected to a single twisted pair in the Cat-6 readout cable for the module.  The active ganging \dwords{pcb} (one per supercell, four per module) are positioned in the module so that they are located inside the central \dword{apa} mechanical support tube when fully installed.

The  internal surface on the lateral sides of the cell are lined with the Vikuiti\texttrademark\ adhesive-backed dielectric mirror foil
that has been laser cut with openings at the locations of the \dwords{sipm} (i.e., the \dword{pcb} surfaces surrounding the \dwords{sipm} visible in Figure~\ref{fig:mounting-board-routing-board} will be highly-reflective).  In the case of the single-sided readout, the dichroic filter windows on the non-active side of the cell are replaced by a blank \frfour G-10 sheet lined in the cell interior with a Vikuiti\texttrademark\ reflector foil.
These types of foils have been used extensively in the WArP\footnote{Wimp ARgon Program at Gran Sasso: \url{http://warp.lngs.infn.it/}} and \dword{lariat}\footnote{Liquid Argon Time Projection Chamber at \dword{fnal}: \url{https://lariat.fnal.gov/}} experiments, where they performed well optically; no issues were reported related to adhesion of the film or dissolution of the wavelength shifter in \dword{lar}.

In \dword{dune}, we have demonstrated that the foils adhere very strongly to \frfour G10 surfaces cleaned following the \dword{pd} standard cleaning procedures.  Tests by the \dword{pd} group have demonstrated adhesion is maintained through multiple cryogenic (\dword{ln})/warm thermal cycles.  The mechanical design provides additional mechanical constraints on the Vikuiti\texttrademark\ sheets after module assembly, so the foils will be held in place mechanically even if the adhesive fails.  Samples of the adhesive have been used in other experiments with no negative impact on the \dword{lar} purity observed.  Samples will be tested in the \dword{fnal} materials test stand, and in \dword{iceberg}, \dword{sbnd}, and \dword{pdsp2} to confirm that the adhesive does not negatively impact \dword{lar} purity or detector performance.

To allow for air to vent out of the cell and \dword{lar} to completely fill the cell during the detector fill, holes are provided at the end of each supercell (four holes total, top and bottom of the cell when mounted in the \dword{apa}s).

The optical window(s) of each supercell are dichroic filters with a cut-off at \SI{400}{nm}. While the filters used for the \dword{pdsp} prototypes have been acquired from Omega Optical Inc.\footnote{Omega Optical Inc., Brattleboro, VT USA: \url{http://www.omegafilters.com/}}, Opto Eletronica S.A.\footnote{Opto Electronica S.A.: \url{http://www.opto.com.br/}} (in Brazil) is our current primary candidate vendor for \dword{dune} production filters.  Opto is a well-established company with a long history of involvement in research optical components for harsh environments and large thermal gradients (including camera optics for satellite photography).  We plan an extensive suite of testing of their filters at \dword{unicamp}, in \dword{iceberg}, and in \dword{pdsp2}. Other vendors are also being investigated\footnote{ASHAI -Japan, Andover-USA, and Edmunds Optics-USA}.

The filters are coated on the external side facing the \dword{lar} active volume with \dword{ptp}\footnote{p-TerPhenyl, supplier: Sigma-Aldrich\textregistered.}.  The coatings for the \dword{pdsp} modules have been made at the thin film facility facility at \dword{fnal} using a vacuum evaporator. Each coated filter was dipped in \dword{ln} to check the stability of the evaporated coating at cryogenic temperature. 

For the \dword{fd}, filter coatings will be done by the vacuum deposition facility at \dword{unicamp} (see Section~\ref{sec:xarapuca-unicamp}).

\section{Silicon Photosensors} 
\label{sec:fdsp-pd-ps}

The physics goals, the design of the light collectors, and the trigger and \dword{daq} system constraints determine the suite of specifications for the silicon photosensors such as the number of devices, spectral sensitivity, dynamic range, triggering threshold and rate, and zero-suppression threshold. An initial survey of commercial products and a 12-month period of R\&D indicated that the performance characteristics of devices from several vendors effectively meet the \dword{pds} needs. 
However, a key additional requirement is to ensure the mechanical and electrical integrity of these devices in a cryogenic environment. Catalog devices for most vendors are certified for operation only down to \num{-40}$^\circ$C and though one candidate device performed well initially, after an unadvertised production process change a large fraction cracked when submerged in \dword{lar}\footnote{SensL MicroFC-60035C-SMT}. This highlighted the need to be in close communication with vendors in the 
\dword{sipm} design, fabrication, and packaging certification stages to ensure 
robust and reliable long-term operation in a cryogenic environment. 

Nearly one thousand \dwords{sipm}, of several types, are used in \dword{pdsp}'s \dword{pds}\footnote{\dword{pdsp} uses 516 SensL MicroFC-60035C-SMT, 288 Hamamatsu MPPC 13360-6050CQ-SMD with cryogenic packaging, and 180 Hamamatsu MPPC 13360-6050VE.}, providing an excellent test bed for evaluation and monitoring of \dword{sipm} performance in a realistic environment over a period of months. Results from \dword{pdsp} are summarized in Section~\ref{sec:fdsp-pd-validation}.

The \dword{spmod} baseline \dword{pds} design has \num{192} \num{6}$\times$\SI{6}{mm$^2$} \dwords{mppc} per \dword{pd} module with groups of \num{48} \dwords{mppc} electrically ganged into four electronics readout channels. This leads to a total of \num{288000} \dwords{mppc}. 

Two entities have expressed interest to engage with the consortium with an explicit intent to provide a product specifically for cryogenic operation: (1) Hamamatsu Photonics K.K., a large well-known commercial vendor in Japan, and (2) \dword{fbk}\footnote{Hamamatsu Photonics K.K.: \url{https://www.fbk.eu/en/}}
in Italy. 
\dword{fbk} is an experienced developer of solid state photosensors that typically licenses its technology; it is partnering with the DarkSide\footnote{Darkside project: \url{http://darkside.lngs.infn.it/}} collaboration to develop devices with specifications very similar to \dword{dune}'s.  Table~\ref{tab:photosensors} summarizes the key characteristics of the baseline device, Hamamatsu S13360, and two other devices from Hamamatsu and \dword{fbk} that are under consideration.

While the devices from Hamamatsu have been tested extensively by the consortium, those from \dword{fbk} are relatively new to us. The technologies they have developed that are suitable for the needs of \dword{dune} are the NUV-HD-SF (standard field) and NUV-HD-LF (low field)~\cite{Gola:2019idb}. In particular, the LF technology (see Table~\ref{tab:photosensors}) offers the lowest dark current rate and has been successfully employed for the DarkSide experiment. NUV-HV-SF sensors developed by \dword{fbk} specifically for \dword{dune} have been tested in Milano (Italy), CSU (CO, US), and NIU (IL, US). The sensors were characterized both at room and cryogenic temperatures (\SI{77}{K}) and underwent more than \num{50} thermal cycles. The tests confirmed the nominal performance of the photosensors and proved the reliability of the sensors at low temperature. Extensive thermal tests and characterization of sensors in the NUV-HD-LF technology are in progress.   

The milestone for photosensor selection for the first \dword{spmod} is early 2021.  Though a baseline photosensor that meets the requirements has been identified, the addition of experienced INFN groups to the \dword{pds} effort has enabled us to pursue the promising \dword{fbk} option in a way that was not possible previously.  We are carrying out targeted investigations on the performance, cost, and production capability to establish the viability of the alternatives for all or part of the sensors required for either the first or subsequent \dwords{spmod}. Two photosensor types (one from each vendor) will be selected in early 2020 to be used in \Dword{pdsp2}.

As described in Section~\ref{sec:fdsp-pd-lc}, the size and sensitivity of currently available \dwords{sipm} requires that multiple devices are needed for each \dword{xarapu} cell. The spatial granularity of each device is much smaller than required for \dword{dune} so,
along with limitations on the number of readout channels, it is required that the signal output of the \dwords{sipm} must be electrically ganged. The terminal capacitance of the sensors strongly affects the \dword{s/n} when devices are ganged in parallel, which led to a design that passively gangs several sets of \dwords{sipm} in parallel, which are then summed with active components, as described in Section~\ref{sec:pds-design-ganging}.

\begin{dunetable}[Candidate photosensors characteristics]
{p{0.26\textwidth}p{0.26\textwidth}p{0.18\textwidth}p{0.18\textwidth}}
{tab:photosensors}
{Candidate Photosensors Characteristics.}
	                      &Hamamatsu (Baseline)   & Hamamatsu-2    & FBK                 \\ \toprowrule
Series part \#            & S13360                &     S14160         & NUV-HD-LF         \\ \colhline
V$_{\rm br}$ (typical)    & 50 V to 52 V          &   36 V to 38 V & 31 V to 33 V                \\ \colhline
V$_{\rm op}$ (typical)    & V$_{\rm br}$+\SI{3}{V}             &   V$_{\rm br}$+\SI{2.5}{V} & V$_{\rm br}$+\SI{3}{V}                \\ \colhline
Temperature dependence of V$_{\rm br}$  & \SI{54}{mV/K}&  \SI{35}{mV/K}& \SI{25}{mV/K}   \\ \colhline
Gain~at~V$_{\rm op}$(typical)   & $1.7\times10^6$     &      $2.5\times10^6$ &  $0.75\times10^6$          \\ \colhline
Pixel size                & 50 $\mu$m             &       50 $\mu$m    & 25 $\mu$m            \\ \colhline
Size                      & 6 mm x 6 mm           &     6 mm x 6 mm    & 4 mm x 4 mm            \\ \colhline
Wavelength                & 320 to 900 nm         &     280 to 900 nm  & 280 to 700 nm            \\ \colhline
PDE peak wavelength       & \SI{450}{nm}         &      \SI{450}{nm}     & \SI{450}{nm}           \\ \colhline
PDE at peak                & 40\%                  &        50\%        & 50\%            \\ \colhline
DCR at \si{0.5}{PE}               & < \SI{50}{\kilo\hertz\per\square\milli\meter}      & < \SI{100}{\kilo\hertz\per\square\milli\meter}   & < \SI{25}{\kilo\hertz\per\square\milli\meter}               \\ \colhline
Crosstalk                 & <~3\%				  &      <~7\%          & <~3\%             \\ \colhline
Terminal capacitance      & \SI{35}{\pico\farad\per\square\milli\meter}          &   \SI{55}{\pico\farad\per\square\milli\meter}     &      \SI{50}{\pico\farad\per\square\milli\meter}         \\ \colhline
Lab experience            & Mu2e and DUNE prototypes      &                &     Darkside  \\         
\end{dunetable}


\section{Electronics}
\label{sec:fdsp-pd-pde}

The electronic readout system for the \dword{pds} must (1) collect and process electrical signals from \dword{sipm}s reading out the light collected by the \dwords{xarapu}, (2) provide an interface with the trigger and timing systems supporting data reduction and classification, and (3) transfer data to offline storage for physics analysis. Figure~\ref{fig:pds-electronics_signalpath} provides a simple overview of the signal path and key elements. 
\begin{dunefigure}[Overview of the \dshort{pds} signal path]
 {fig:pds-electronics_signalpath}
 {Overview of the \dword{pds} signal path.}
\includegraphics[width=15cm]{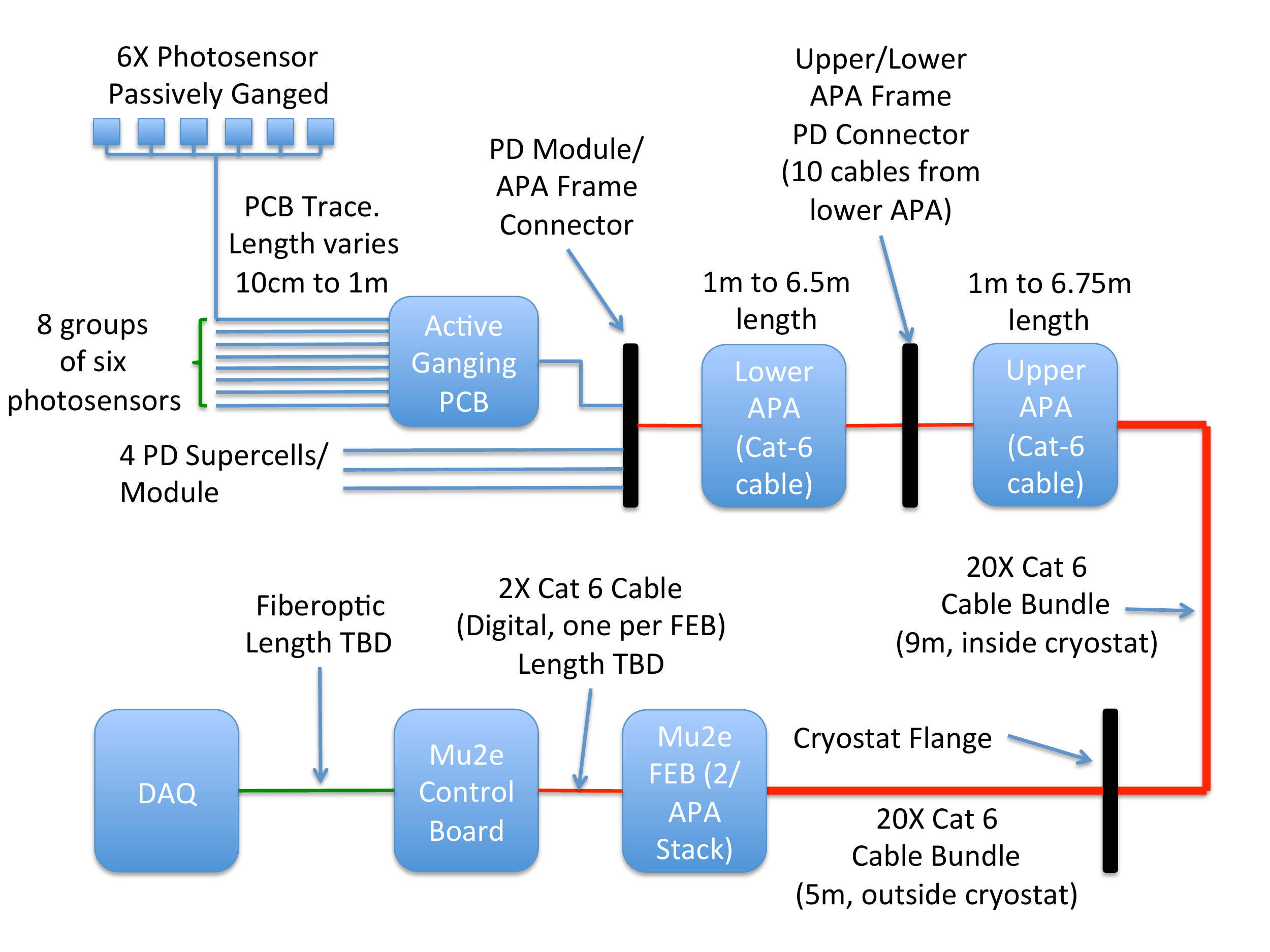}
\end{dunefigure}

As specified in the requirements Table~\ref{tab:specs:SP-PDS}, the readout system must enable the $t_0$ measurement of non-beam events; this capability will also enhance beam physics by recording interaction time of events within 
beam spill more precisely to help separate against potential cosmic background interactions. A highly capable readout system was developed for use with \dword{pdsp} and prototype development as described in Section~\ref{sec:valid-pdsp}. However, a more cost-effective waveform digitization system developed for the \dword{mu2e} experiment has been identified and selected as the baseline choice for the \dword{pd} system.

\subsection{SiPM Signal Ganging}
\label{sec:pds-design-ganging}

The ganging of electrical signals from \dword{sipm} arrays is implemented to minimize the electronics channel count while maintaining adequate redundancy and granularity, as well as to improve the readout system performance.  
Technical factors that affect performance of the ganging system are the characteristic capacitance of the \dword{sipm} and the number of \dwords{sipm} connected together, which together dictate the \dword{s/n} and affect the system performance and design considerations.

We have demonstrated a feasible purely passive summing scheme with twelve Hamamatsu \dword{mppc} sensors now operational in \dword{pdsp}. For optimal performance in \dword{dune}, we have shown that an ensemble of 48 Hamamatsu \SI{6}{mm}$\times$\SI{6}{mm} \dwords{mppc} can be summed into a single channel by a combination of passive and active ganging (see Section~\ref{sec:pds-valid-ganging}).  In this scheme, an amplifier is used to adjust the \dword{mppc} output signal level to the input of an \dword{adc}; the active summing is realized with an OpAmp THS4131. This combination of passive and active ganging with cold signal summing and amplification, illustrated in Figure~\ref{fig:fig-pds-6x8gang}, is the baseline for the \dword{pds}.

\begin{dunefigure}[SiPM signal summing board circuit]
 {fig:fig-pds-6x8gang}
 {\dword{sipm} signal summing board circuit: 6 passive  x 8 active, 48 \dwords{sipm} total.}
\includegraphics[height=14cm]{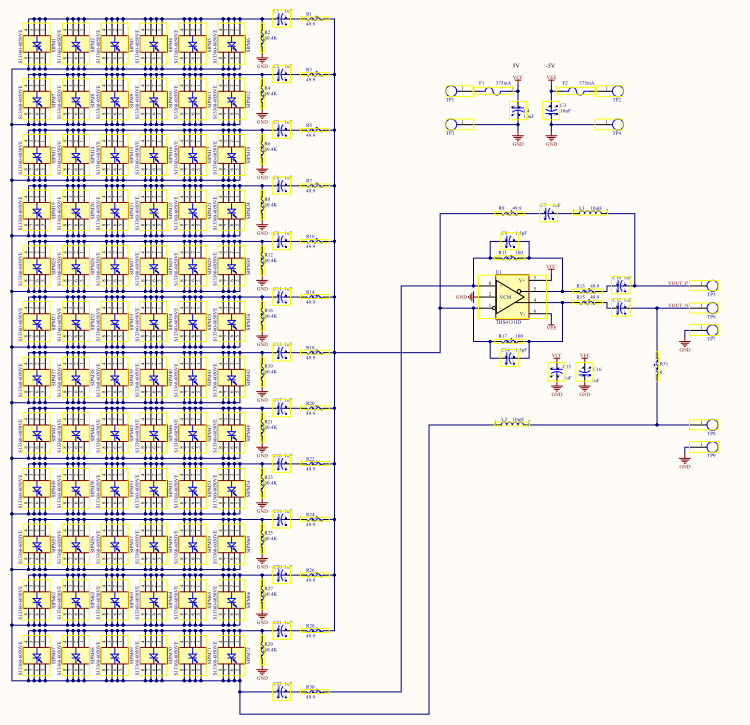}
\end{dunefigure}


\subsection{Front-end Electronics Baseline Design}
\label{sec:electronics}

The \dword{fe} electronics for the development and prototype stages of the  \dword{pds}, including \dword{pdsp}, was provided by a custom-designed \dword{sipm} Signal Processor (\dword{ssp}, see \citedocdb{3126}). This system was highly configurable and provided detailed information on the photosensor signal, which allowed a thorough understanding of the photon system performance.
For the much larger \dword{spmod}, a system is required that meets the performance requirements yet optimizes the cost.
To this end, we have developed a solution based on lower-sampling-rate commercial ultrasound \dword{asic} chips rather than digitizers based on flash \dwords{adc} used in the \dword{ssp}. Inspiration for this cost-effective \dword{fe} comes from the system developed for the \dword{mu2e} experiment cosmic ray tagger readout system.
Both \dword{ssp} and the new design are used in the \dword{pd} validation process summarized in Section~\ref{sec:fdsp-pd-validation} to allow direct comparison.

Development of the readout electronics to date has been primarily by US groups. 
However, since fabrication of the \dword{dune} readout electronics will be conducted by a collaboration of Latin American institutions (including groups in Peru, Colombia, Paraguay, and Brazil), further development is being performed by these groups, with support from the US groups.  
The engineers met and worked together at \dword{fnal} in summer 2019, and that collaboration is continuing.  We expect the first \dword{daphne} prototypes to be complete and tested in \dword{iceberg} in April/May 2020.  
Pre-production \dword{mu2e}-based electronics readout will also be tested in the \dword{pdsp2} run at \dword{cern} in 2021-2022.

\subsubsection{Front-end Board and Controller}

The readout and digitization of the signals from the active summing board described in Section~\ref{sec:pds-design-ganging} will rely on a set of \dword{fe} board (\dword{feb}) readout electronics boards and controller boards, originally designed for the \dword{mu2e} experiment~\cite{bib:mu2e_tdr}
at \dword{fnal}. As discussed in Section~\ref{sec:pds-valid-ganging}, preliminary results indicate that the active-summing board and \dword{mu2e} electronics \dword{feb} combination will perform well together and, in general, meet the readout requirements for the experiment. Figure~\ref{fig:feb} shows the 64~channel \dword{feb} design carried over from \dword{mu2e}. The board has a number of notable features, discussed below. Most importantly,  
the board is designed to utilize commercial, off-the-shelf parts only, and is therefore quite inexpensive compared to other designs. In particular, the digitization implements the low-noise, high-gain, and high-dynamic-range commercial \dwords{adc} used in ultrasonic transducers. 

The \dword{feb} is the centerpiece of the baseline readout electronics system.  
The current 64~channel\footnote{This text assumes 64 channels/\dword{feb} when presenting the \dword{feb} and controller. However, we envision 40~channels/\dword{feb} in the final design, corresponding to a single \dword{apa} as described in Section~\ref{subsec:pds-fe-next}.} \dword{feb} relies on commercial ultrasound chips\footnote{Texas Instruments\texttrademark{} 12~bit, 80~megasamples/s (MS/s); AFE5807.},
with programmable anti-alias filters and gain stages, to read out the \dword{mppc} signals from the active ganging boards inside the \dword{pd} modules. The board currently takes HDMI inputs, with four channels per input.  Each of the eight ultrasound chips on an \dword{feb} handles eight channels (\SI{120}{mW} per channel) of data using a low-noise preamplifier, a programmable gain amplifier, and a programmable low-pass filter. The information is buffered with a total of \SI{1}{GB} DDR SDRAM, divided in four places, and a set of Spartan~6\texttrademark~\dwords{fpga} are used for parallelizing the serial \dword{adc} data, zero suppression, and timing. Each of the four \dwords{fpga} on a board, corresponding to 16 channels, handles two \dword{adc} chips with an available \SI{256}{MB} DDR SDRAM. 

After digitization, the data from each \dword{feb}, in the form of pulses (time-stamp and pulse height), is sent via Ethernet to a master controller that aggregates the signals from 24 \dword{feb}s, or $64\times24=1536$ channels. The 24 \dword{feb}s corresponding to a single controller will come in sets of 12, with each set of 12 \dword{feb}s referenced to a single chassis as shown in 
Figure~\ref{fig:controller}. A trigger decision (e.g., accelerator timing signal) can be produced and/or received by the controller and, depending on the decision, each event's digital information is sent to the controller and then to \dword{daq} computers for processing and storage. The controller-to-\dword{daq} connection will rely on a fiber connection, although an Ethernet-based controller output option is available.

\subsubsection{Bandwidth, readout rates, and zero suppression}

\dword{daq} system and data storage limitations impose constraints on the  data flow from the \dword{fe} electronics system. 
For example, if it were necessary to read out a \SI{5.5}{\micro\second} waveform in order to include more of the longer time constant scintillation light component, the \SI{80}{megasamples/s} (MS/s), 12-bit \dword{adc} device would produce a \SI{5.3}{kbit} waveform. For an envisioned dark count (DC) rate of 250~Hz/channel, this corresponds to a data transfer rate of \SI{53}{MB/s}/\dword{apa} (1~\dword{apa}=40 channels) or \SI{6.6}{MB/s} \dword{feb}-to-controller DC rate. This rate approaches the crucial bottleneck in the electronics readout system with a maximum rate of \SI{10}{MB/s} (per \dword{feb}). However, zero-suppression techniques and multi-channel coincidence/threshold requirements at the \dword{feb} firmware level can be used to significantly mitigate this issue, noting that each on-\dword{feb} \dword{fpga} handles 16 channels. 

The design is flexible enough to accommodate modest changes in system requirements, such as the suppression factor determined by parameters like the readout window length and limits on the overall trigger rate. 
Firmware and zero-suppression technique development is in progress and can easily adapt to the physics and calibration requirements of the \dword{pd}.
In addition to its bandwidth and DC rate readout capabilities, the system can also  manage a highly coincident event in which a large number (or all) channels fire at once. 
For example, the controllers' 24-board write speed of \SI{150}{MB/s} could handle even the unlikely all-detector event featuring 6000~channels firing at once (corresponding to \SI{4}{MB} event size). 

The baseline electronics readout system performance is consistent with the \dword{daq} interface specification of \SI{8}{Gb/s} per connection, given that
each \dword{feb} signal corresponds to a maximum of \SI{10}{MB/s} (\SI{240}{MB/s} total).  

\subsubsection{Power, grounding, and rack schemes} 

Figure~\ref{fig:grounding_power} shows the grounding, power, and data link schemes for the system. The \dword{feb}s are powered via power-over-Ethernet (\SI{600}{mA}, \SI{48}{V} supply) from the controller. One Cat-6 cable from the controller to each \dword{feb} handles the signal and power simultaneously. The reference planes of the controller and \dword{feb} are isolated on both sides. The grounding scheme calls for each set of twelve \dword{feb}s referenced to a single chassis, with each chassis and corresponding controller on detector ground and the \dword{daq}, connected to each controller via fiber, on building ground. 
 
The rack space and power consumption required by the system assume
a total of 6000 channels with 40 channels/\dword{feb}. This system requires 13 chassis (12 \dword{feb}/chassis) at 6U each and seven controllers (controlling 24 \dword{feb} each) at 1U each; these can be accommodated in just over two 42U capacity racks. The power supply on a controller is \SI{700}{W}, with each \dword{feb} taking \SI{20}{W}.

\begin{dunefigure}[\dshort{pds} 64-channel front-end board]
 {fig:feb}
 {Photograph of the 64-channel \dword{pds} \dword{fe} board (\SI{80}{MS/s}, \SI{12}{bit} ADC) (left); schematic of the \dword{fe} board (right).}
\includegraphics[height=5.2in]{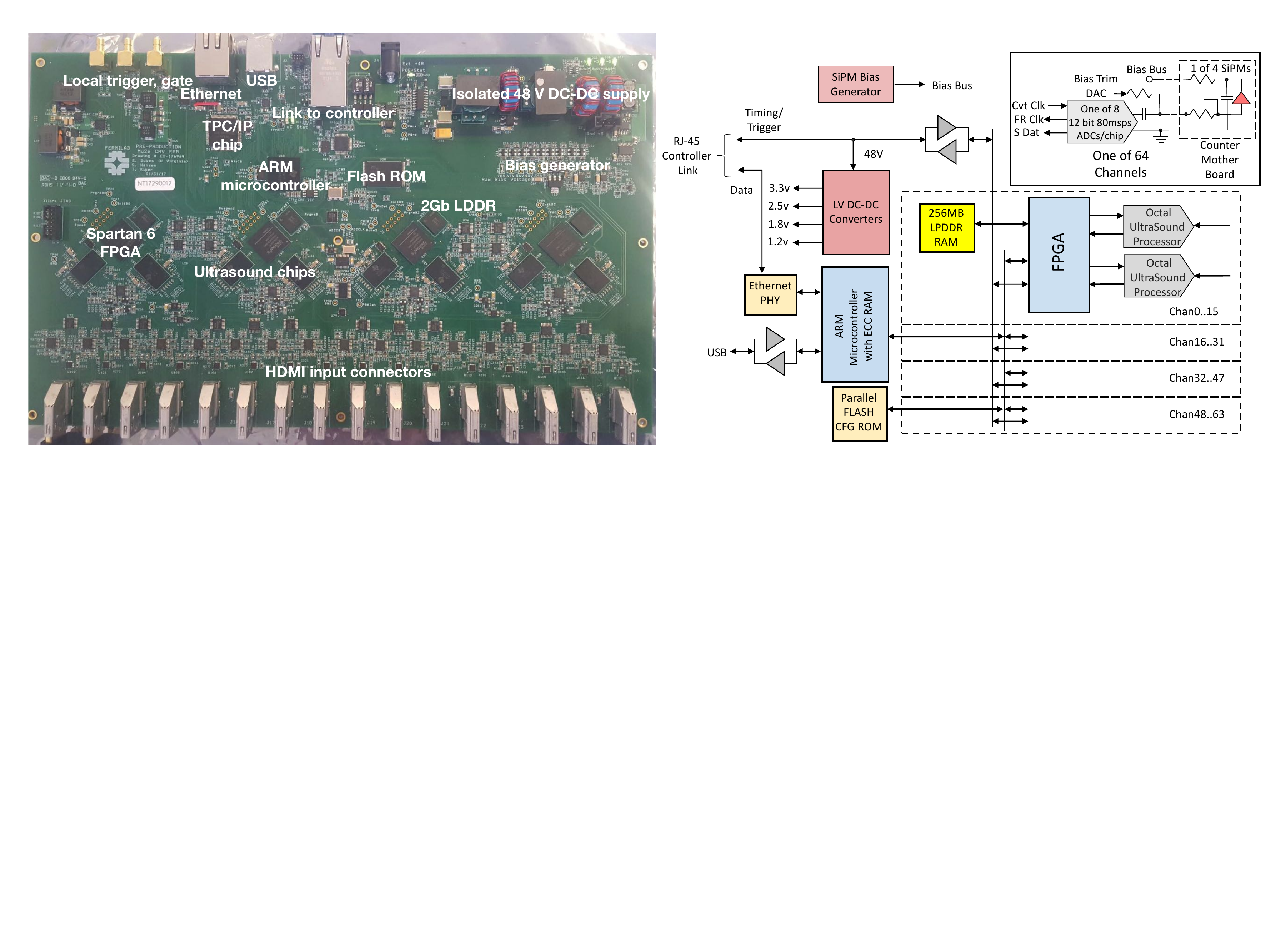} 
\vspace{-6.3cm}
\end{dunefigure}

\begin{dunefigure}[\dshort{pds} front-end electronics controller module]
 {fig:controller}
 {The front (left-bottom) and back views of the controller module (left-top); it is capable of accepting signals from 24 \dword{feb}s; schematic of the controller (right).}
\includegraphics[height=5.1in]{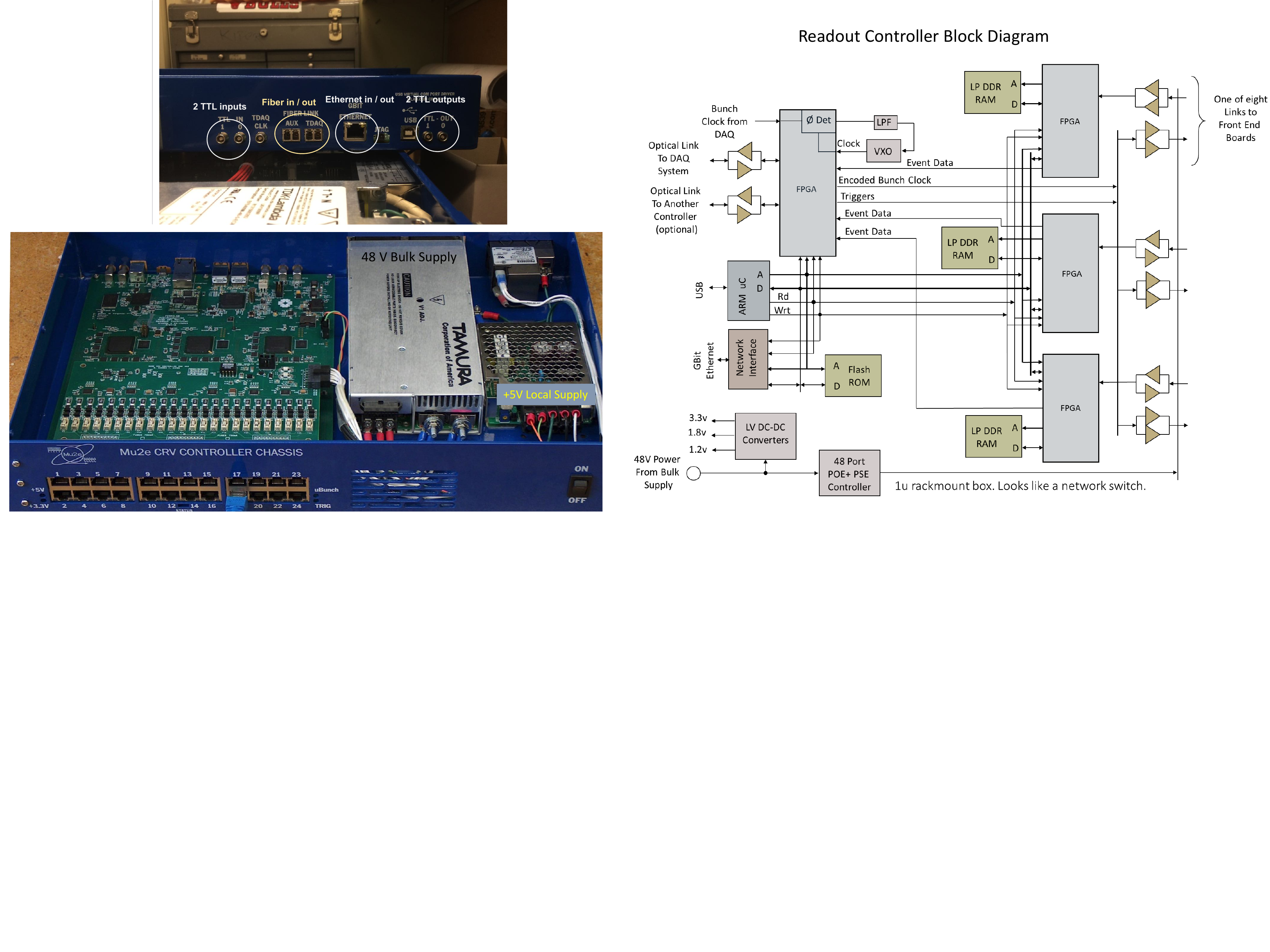} 
\vspace{-5.5cm}
\end{dunefigure}

\begin{dunefigure}[\dshort{pds} front-end electronics grounding scheme]
 {fig:grounding_power}
 {Grounding scheme with 1 chassis, containing 12 \dword{feb}s, a controller module, and a \dword{daq} PC, as an example (left); power and data link arrangement of the \dword{feb} and controller (right).}
\includegraphics[height=5.0in]{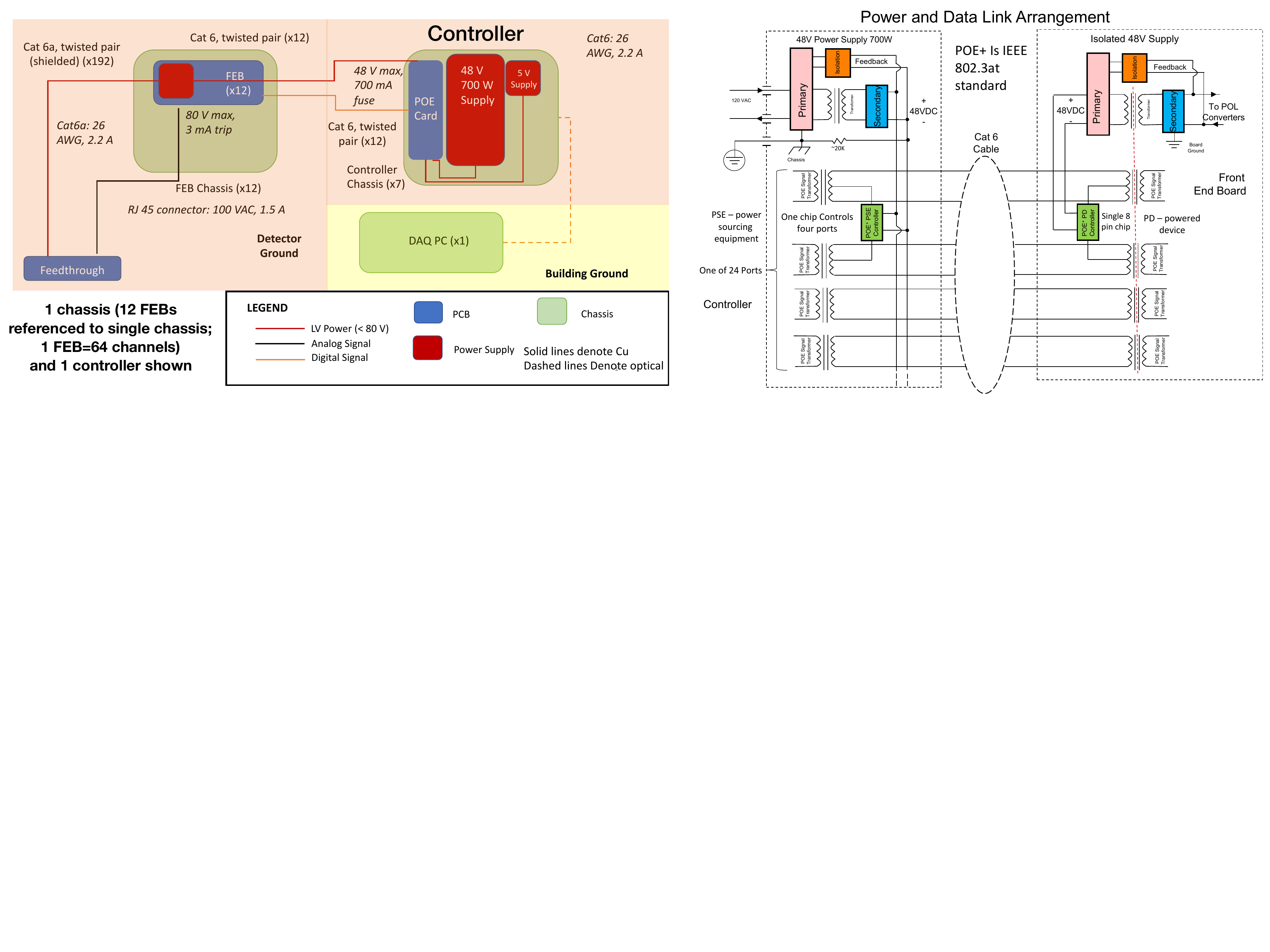} 
\vspace{-7.1cm}
\end{dunefigure}

\subsection{Electronics Next Steps}
\label{subsec:pds-fe-next}

The \dword{feb}s, developed for the \dword{mu2e} cosmic ray veto and proposed for use in \dword{dune}, can read out an array of \dwords{mppc} with an adequate \dword{s/n} ration to be sensitive to single photons. However, we want to optimize and further develop them by pursuing the following tasks:

\begin{enumerate}
\item To better match the 40 \dword{pdsp} channels per \dword{apa}, the system presented here assumes that only 40 out of the 64 channels on the existing  \dword{mu2e} \dword{feb} are populated with active electronics.  A prototype board will test this configuration and validate the associated cost model.

\item The  \dword{mu2e} warm readout electronics use last-generation (Xilinx\texttrademark{} Spartan-6) \dwords{fpga} and other components that have since been superseded by newer devices.  Design and prototyping work will 
incorporate newer \dwords{fpga} (Xilinx Spartan-7 or Artix-7) into the electronics,
improving their performance and maintainability over the lifetime of the \dword{dune} experiment. The Artix-7 \dwords{fpga} have been implemented in the \dword{ssp} readout system used in \dword{pdsp}, and therefore the expertise with these system components has been established. 

\item Results from the  \dword{iceberg} test stand can determine whether there are sufficient logic resources in the \dwords{fpga} to meet a broad range of possible \dword{daq} requirements expected from the warm readout electronics. To that end, the low-cost \dword{fe} solution will be compared to existing \SI{14}{bit}, \SI{150}{MS/s} \dword{ssp} readout.  Straightforward zero suppression schemes that can be implemented on the Mu2e board with the current Spartan-6 \dword{fpga} will be tested with respect to potential \dword{daq} data rate limitations.  However, increases in the number of logic cells can be accommodated by switching to more capable, but still pinout-compatible, devices within the same Xilinx \dword{fpga} family as discussed above.

\item It may be desirable to increase the dynamic range of the \dwords{adc} used on the \dword{feb}s in order to achieve desired physics goals related to the energy resolution of beam neutrino events.  To this end, we plan to investigate replacing the TI AFE5807 ultrasound chip with the TI AFE5808 ultrasound chip, which is pinout-compatible but incorporates a 14-bit \dword{adc}.  Ultimately, a prototype board will incorporate all relevant optimizations and improvements.

\end{enumerate}

The final requirements for the system will be informed by analysis of the data from the readout system implemented in \dword{pdsp} and subsequent data from operating ICEBERG from August 2019 through the end of the year.

Additional testing of the system will continue through \dword{pdsp2} operations. The specifications for the readout electronics system will be reconsidered based on that experience and established before the \dword{pd} final design review (see the high-level schedule in Section~\ref{sec:fdsp-pd-org-cs}).

\section{Calibration and Monitoring}
\label{sec:fdsp-pd-CandM}

Calibration and monitoring is an essential component of the \dword{pds}.
The primary system is a pulsed UV-light source that will allow calibration of the \dword{pd} gain, linearity, and timing resolution and monitoring the stability of the photon response of the system over time.
In many experiments, a pulsed light system is a valuable well-defined, controllable light source for monitoring but for (near) surface detectors, often a supplement to using tracked cosmic ray muons, which provide a much closer replica of the signal from events of interest. 
However, at \dword{dune}, the muon rate per individual photon detector will be very low and insufficient to monitor changes in the system response. In this situation, the pulsed system will play an essential role in achieving and maintaining the \dword{pd} performance required for neutrino calorimetry. 
This system will also be a valuable detector commissioning tool prior to sealing the cryostat, in the cool-down phase, and during the \dword{lar} fill.
Other complementary calibration systems, such as radioactive sources, are described in Chapter~\ref{ch:sp-calib}. 

The system design is almost identical to that deployed in \dword{pdsp}, as described in Section~\ref{sec:fdsp-pd-validation-candm}; the primary differences are the number of diffusers, the lengths of the optical fibers, and the addition of a monitoring diode.

The system hardware consists of both warm and cold components but has no active components within the cryostat. The active component consists of a 1U rack mount light calibration module (\dword{lcm}) located outside the cryostat. The \dword{lcm} generates UV (\SIrange{245}{280}{nm}) pulses that propagate through a quartz fiber-optic cable to diffusers at the \dword{cpa} that distribute the light uniformly across the \dwords{pd} mounted within the \dword{apa}. 
It consists of an \dword{fpga}-based control logic unit coupled to an internal \dword{led} pulser module (\dword{lpm}) and an additional bulk power supply. 
The \dword{lpm} has multiple digital outputs from the control board to control the pulse amplitude, pulse multiplicity, repetition rates, and pulse duration; programmable \dwords{dac} control the \dword{lpm} pulse amplitude. \dword{adc} channels internal to the \dword{lcm} are used to read out a reference photodiode used for pulse-by-pulse monitoring of the \dword{led} light output. The output of the monitoring diode is available for normalizing the response of the \dwords{sipm} in the detector to the monitoring pulse.

\begin{dunefigure}[Calibration system diffuser locations on the SP CPA]
 {fig:pds-calmon-cpa-diffusers}
 {Schematic of a complete SP cathode plane ($\SI{60}{m}\times\SI{12}{m}$) showing the locations of the calibration and monitoring system diffusers. Each diffuser illuminates a region of about $\SI{4}{m}\times\SI{4}{m}$ (indicated by the squares) on \dword{apa}s \SI{3.6}{m} away.}
\includegraphics[angle=0,width=0.9\textwidth]{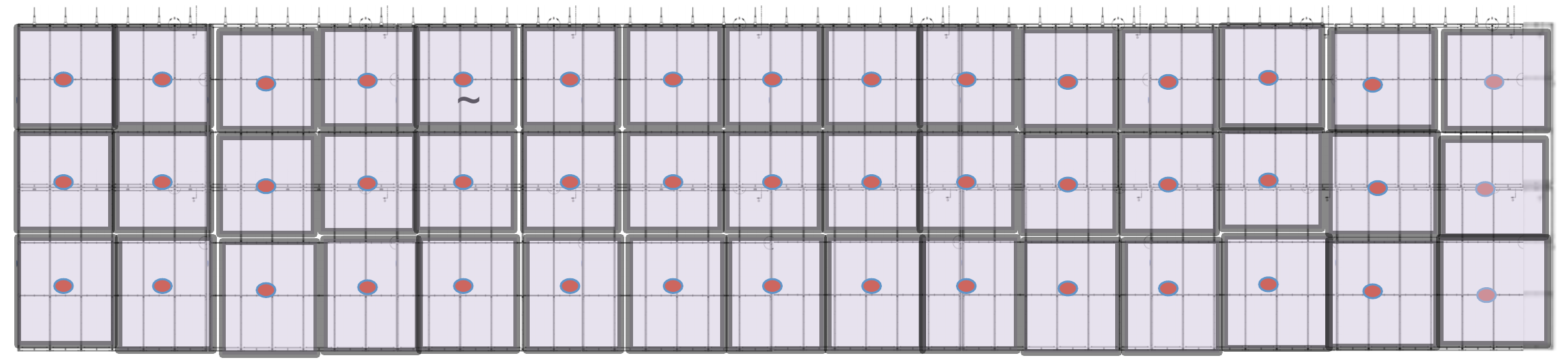}
\end{dunefigure}

Quartz fibers, \SIrange{10}{30}{m} long, are used to transport light from the optical feedthrough (at the cryostat top) through the \dword{fc} \dword{gp}, and through \dword{fc} strips to the \dword{cpa} top frame. 
These fibers are then optically connected to diffusers located on the \dword{cpa} panels using fibers that are \SIrange{2}{10}{m} long. 
The diffusers, \SI{2.5}{cm} in diameter, are mounted on the cathode plane panels acting as light sources to illuminate \dwords{pd} embedded in the \dword{apa}s. There will be \num{45} diffusers uniformly distributed across each of the \dword{spmod} cathode planes facing \dword{apa}s, as indicated in 
Figure~\ref{fig:pds-calmon-cpa-diffusers}. Each diffuser will illuminate an area of approximately $\SI{4}{m}\times\SI{4}{m}$ on the \dword{apa}s that are \SI{3.6}{m} away. 

The diffusers reside at the \dword{cpa} potential, so the \dword{hv} system places a requirement on the fiber electrical resistance to protect the cathode from experiencing electrical breakdown along this path. This requirement is easily met by the fibers. 

As demonstrated in \dword{pdsp}, the system performs the required calibration and monitoring tasks with minimal impact on the \dshort{tpc} design and function. 

\section{Design Validation}
\label{sec:fdsp-pd-validation}

This section summarizes the most important sets of measurements, completed, ongoing, and planned, that validate the \dword{sp} \dword{pds} design.

\subsection{Photosensors and Active Ganging}
\label{sec:pds-valid-ganging}

As described in Section~\ref{sec:pds-design-ganging}, the active ganging of \dwords{sipm} aims to increase the active photo-detecting area while keeping the number of readout channels at a reasonable number. 
Several active ganging detectors were designed and tested during 2017-2018. 
The systems were based on an active summing node mounted near the photosensors in the \dword{lar}. Several incarnations of the cold summing node were designed and tested using SensL and Hamamatsu \dwords{sipm}, 
as were several types of operational amplifiers.
Some of these designs were tested and validated in the \dword{sarapu} prototype measurements.
We describe here only the most recent design that demonstrated that 48 Hamamatsu \dwords{mppc} in the baseline design can be ganged together on a single differential output with excellent signal performance, low noise, and low power dissipation.

Figure~\ref{fig:fig-pds-gang-1} (left) shows a matrix array of 72 \dwords{mppc} organized as 12 rows of six  13360-6050VE \dwords{mppc} each. 
The six \dwords{mppc} per row are connected in parallel, giving a total output capacitance of \SI{7.8}{nF}. The 12 rows are connected to the summing node of an operational amplifier, THS4131, as illustrated in Figure~\ref{fig:fig-pds-gang-1} (right). 
Since the \dword{dune} baseline design is based on 48 \dwords{mppc}/\dword{xarapu} module, only eight rows of six \dwords{mppc} were used for the tests. 
The performance of the cold summing electronics was done by illuminating the \dword{mppc} array with an \dword{led} and digitizing the output with a high-speed oscilloscope and with the \dword{ssp} readout electronics (see 
Section~\ref{sec:valid-pdsp}).

As shown in Figure~\ref{fig:fig-pds-gang-2-3}~(left), the mean signal has a rise time of \SI{60}{ns} and a recovery time of \SI{660}{ns}, well within the \dword{dune} \dword{pd} specifications.

\begin{dunefigure}[Photosensors signal ganging scheme]
 {fig:fig-pds-gang-1}
 {Summing board with a total of 72 \dwords{mppc} used to demonstrate the optimal combination of passive and active ganging with 48 Hamamatsu \SI{6}{mm}$\times$\SI{6}{mm} \dwords{mppc} (left).  Schematic of the summing circuit with a THS4131 operational amplifier (right).}
\includegraphics[height=6cm]{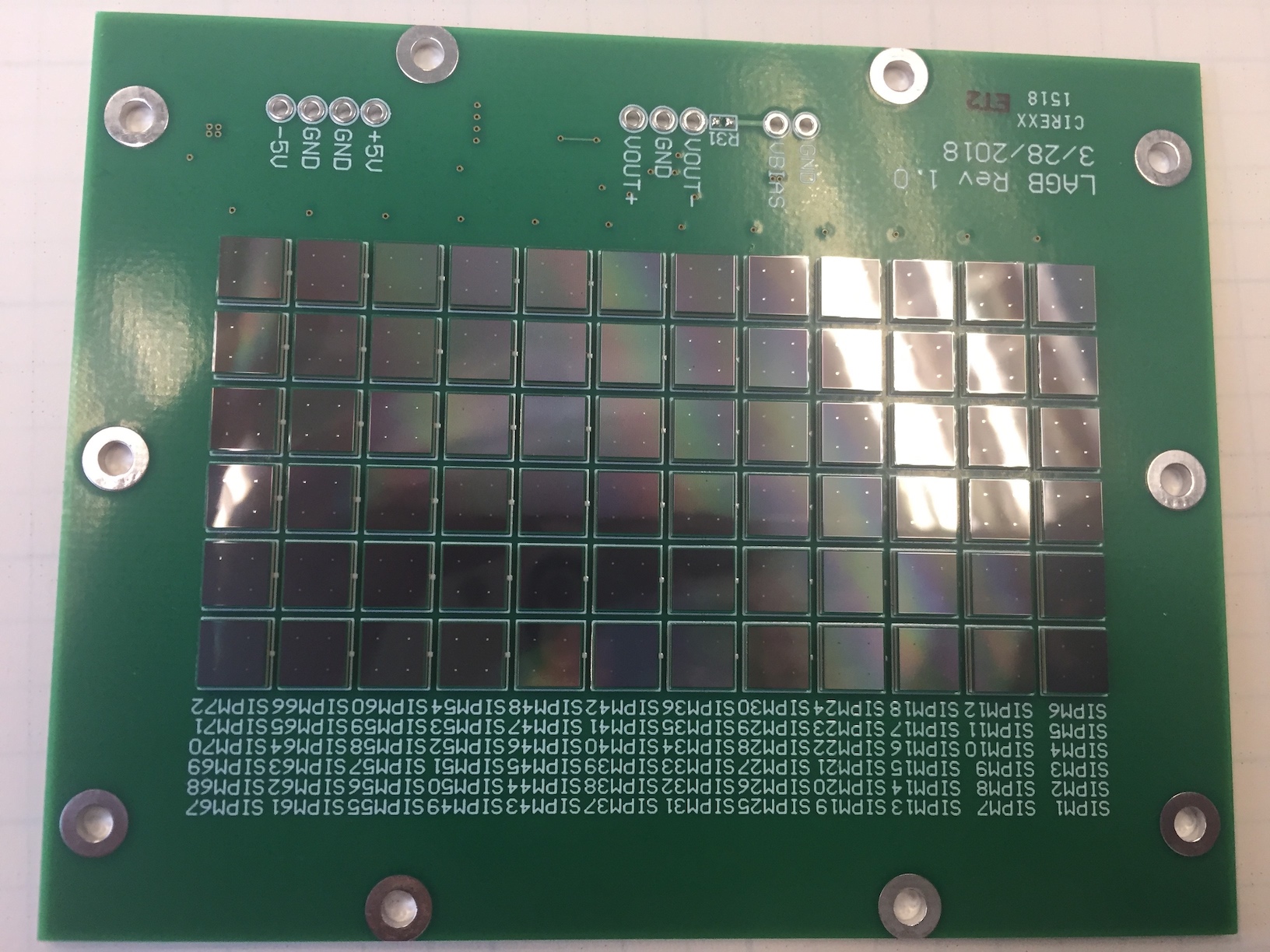}
\includegraphics[height=6cm]{graphics/pds_gang_fig2.png}
\end{dunefigure}

\begin{dunefigure}[Photosensors signal ganging with 48 \dshorts{sipm}]
 {fig:fig-pds-gang-2-3}
 {Waveform signal from 48 \dwords{mppc}/ARAPUCA module, summed with the THS4131 operational amplifier and digitized with the \dword{ssp} \dword{feb} (left); histogram of signals with a \SI{47}{V} bias illustrating the first \phel peak well-separated from the pedestal with an \dword{s/n} = 9.5 (right).}
\includegraphics[height=5.5cm]{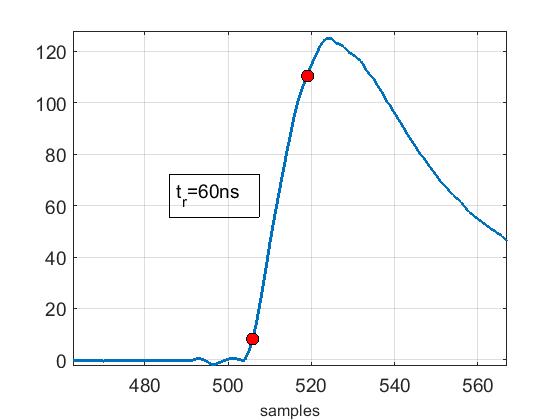}
\includegraphics[height=5.5cm]{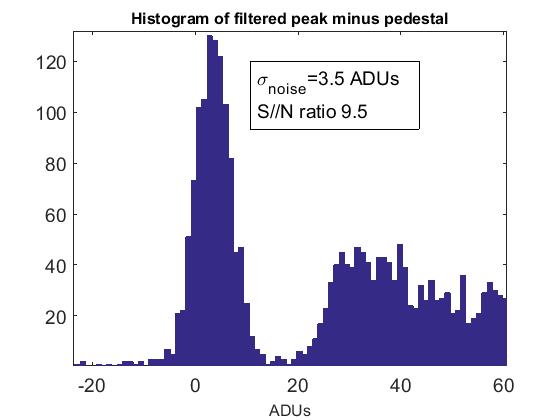}
\end{dunefigure}

Figure~\ref{fig:fig-pds-gang-2-3}~(right) shows a histogram of light collection signals from the array for a bias voltage equivalent to \SI{2}{volts} above the mean breakdown voltage. Since there are 48 \dwords{mppc} in the array, and a single common bias, there is a spread in the gains. Even when the Hamamatsu \dwords{mppc} have a small spread in breakdown voltages, it is enough to smear the peaks in the histogram. It is worth distinguishing the difference between noise and gain spread. The circuit noise can be measured as \dword{fwhm} or \dword{rms} around the 0 \phel signal (\SI{3.5}{\dword{adc} counts} in the figure); the first \phel peak is at \SI{33}{\dword{adc} counts}, resulting in an optimal signal to noise ratio of about ten.

Since the breakdown voltage of the \dwords{mppc} is provided by the manufacturer for each device, the gain spread can be reduced by picking groups of 48 \dwords{mppc} with similar breakdown values for each module. The differential output of the \dword{ce} impedance is matched to the readout electronics and able to reject more than \SI{60}{dB} of common mode noise. This is particularly important since the \dwords{mppc} and output wiring are inside a high voltage \dword{tpc}. The timing properties of the 48 ganged electronics were also measured in \dword{lar} using a $^{241}$Am alpha source. 
Figure~\ref{fig:fig-pds-gang-4} shows the time walk for a constant discrimination threshold which, as expected, is not a linear function of the signal height. The error distribution, which is not Gaussian, has a \dword{fwhm} of \SI{80}{ns}. This value is well within the \dword{dune} specification (Table~\ref{tab:specs:SP-PDS}).

\begin{dunefigure}[Time walk from 48 ganged MPPCs]
 {fig:fig-pds-gang-4}
 {Oscilloscope trace (left) and histogram (right) illustrating time walk from 48 ganged \dwords{mppc} measured with the constant discrimination threshold on the SSP board.}
\includegraphics[height=5.5cm]{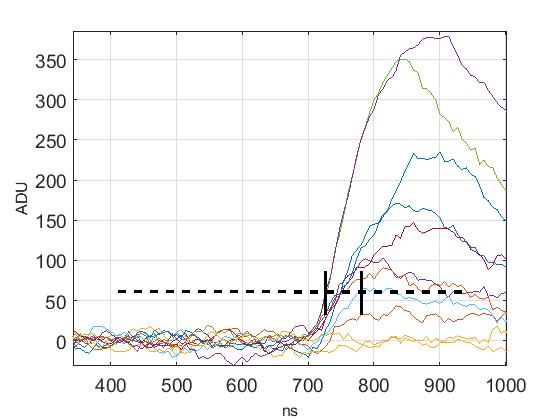}
\includegraphics[height=5.5cm]{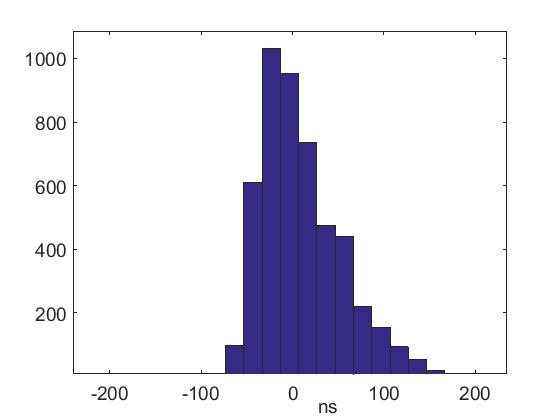}
\end{dunefigure}

\textit{\dword{fbk} Sensors} 
\dword{fbk} has published detailed measurements on photosensors developed in collaboration with the DarkSide cryogenic experiment~\cite{Gola:2019idb}. Figure~\ref{fig:fbk-pde-dcr}~(left) shows that the photon detection efficiency of candidates devices as function of wavelength is very well matched to the needs of the \dword{xarapu}.

An extensive program of evaluation of the key performance characteristics is underway by the \dword{dune} \dword{pds} team.
In the first phase, a sample of (\SI{4}{mm}$\times$\SI{4}{mm}, \SI{40}{$\mu$m} cell-pitch) devices has been tested at INFN-Milano in a dedicated setup optimized for the measurement of very low dark currents. The sensors were operated at \SI{77}{K} and can be biased from \SI{21}{V} (breakdown voltage) up to \SI{31}{V} (maximum overvoltage range at cryogenic temperature is +\SI{10}{V}). The dark count rate at \SI{+4}{V} overvoltage is $\sim$\SI{0.2}{Hz/mm$^2$}, which meets the \dword{dune} requirements (see Figure~\ref{fig:fbk-pde-dcr}~(right)). 
These devices have undergone numerous temperature cycles during the testing with no deterioration in characteristics. Another sample at CSU has undergone more than 50 thermal cycles with no evidence of mechanical failures.  

\begin{dunefigure}[Performance of candidate FBK \dshorts{sipm}]
 {fig:fbk-pde-dcr}
 {PDE measured at \SI{293}{K} for \dword{fbk} NUV-HD-Cryo and NUV-HD \dwords{sipm} (left); dark count rate for NUV-HD-SF \dwords{sipm} at \SI{77}{K} (right).}
\includegraphics[height=5.5cm]{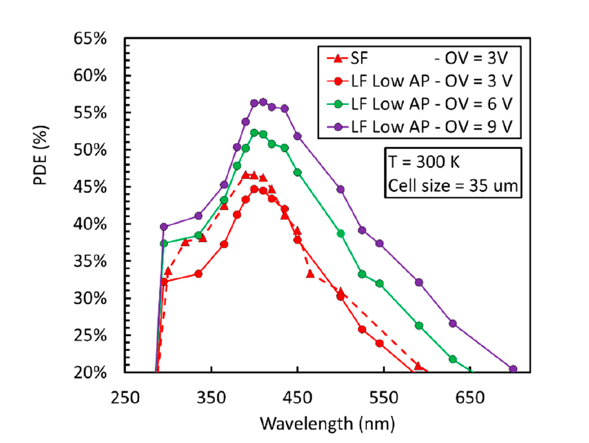}
\includegraphics[height=4.5cm]{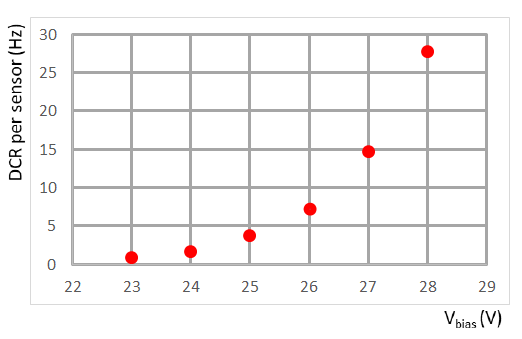}
\end{dunefigure}

\textit{Mu2e Electronics}
The \dword{mu2e} electronics have undergone a series of end-to-end warm and cold tests to demonstrate single-photon sensitivity in various parallel/series ganging and \dword{sipm}/\dword{mppc} configurations. Here we summarize the results with the 72-\dword{mppc} active ganging array described in Section~\ref{sec:pds-valid-ganging}
 at \dword{ln} temperatures. 
A balun\footnote{A transformer used to convert differential (BALanced) signals to single-ended (UNbalanced) ground referenced signal.} is used to convert from the differential actively-ganged \dword{mppc} array output to the single-ended \dword{feb}.

A trigger allowed data to be collected in time with an \dword{led} flasher, with samples taken every 12.5~ns for the length of the readout window ($\sim$3~$\mu$s, in this case). Figure~\ref{fig:pds-board-balun-adc} shows the system used and a histogram of the maximum \dword{adc}  value during each trigger window. The first peak above zero corresponds to the electronic noise and the second peak corresponds to a one-PE signal. The signal to noise from these tests was measured to be 4, calculated from the ratio of the single photon peak (20~\dword{adc} , after subtracting the noise peak) to the spread in the noise ($\sigma_{noise}$ = 5~\dword{adc} ); this is similar to the value found when using the \dwords{ssp} for readout (\dword{s/n} ~=~5).

\begin{dunefigure}[Readout of 72-\dshort{mppc} active ganging array with the \dshort{mu2e} electronics readout board]
{fig:pds-board-balun-adc}
{The \dword{mu2e} electronics readout board was used to read out a 72-\dword{mppc} active ganging array (V$_b$ = \SI{47.2}{V}) (left). The maximum \dword{adc} results are shown, with the first and second peaks representing 0 and 1~\phel signals (right).}
\includegraphics[height=5.2in]{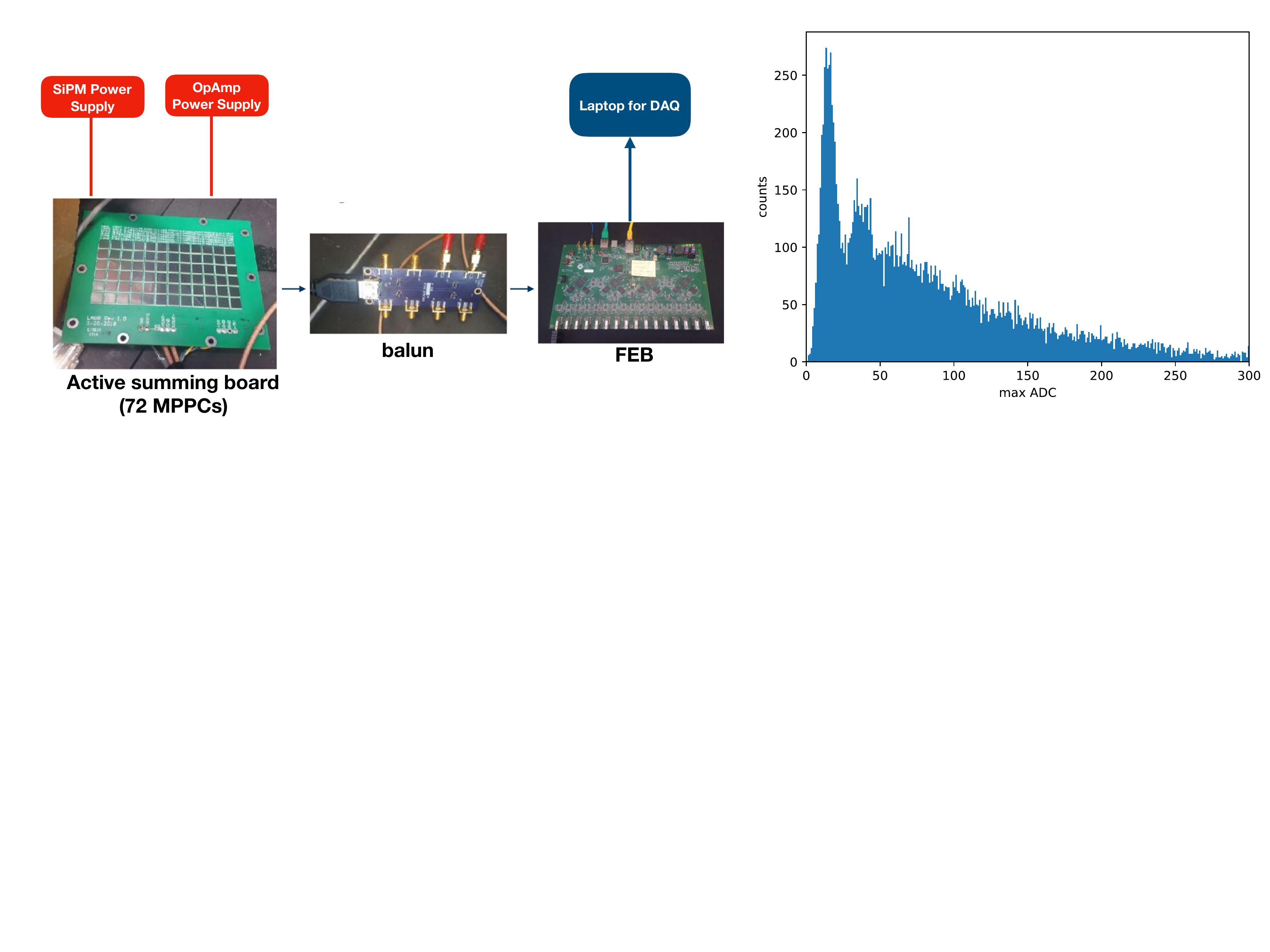} 
\vspace{-7.0cm}
\end{dunefigure}

\subsection{Standard ARAPUCA (S-ARAPUCA)}
\label{sec:sarapuca-prototypes}

\subsubsection{Development Prototypes}
\label{sec:valid-initial}

As outlined in Section~\ref{sssec:photoncollectors}, the design for the \dword{sarapu} features a dichroic filter window coated with a wavelength-shifter on the \dword{lar} active volume face and a second wavelength shifter coated onto the dichroic filter on the surface inside the cell.  
The proof-of-concept measurements of this design were performed on a small cell with internal dimensions of \SI{3.5}{cm} $\times$ \SI{2.5}{cm} $\times$ \SI{0.6}{cm}, with a window formed from a dichroic filter of  dimensions \SI{3.5}{cm} $\times$ \SI{2.5}{cm} and a wavelength cut-off at \SI{400}{nm}. The external side was coated with \dword{ptp} and the internal side was coated with \dword{tpb}. 
The trapped light was detected by a single \SI{6}{mm} $\times$ \SI{6}{mm} SensL MicroFC-60035C-SMT \dword{sipm}\footnote{SensL \dword{sipm}: \url{http://sensl.com/products/c-series/}}. The cell was exposed to scintillation light produced in pure \dword{lar} by an alpha source\footnote{A $^{238}$U-Al alloy in the form of a metallic foil.} that emits three alpha lines with energies of  \SI{4.187}{MeV}, \SI{4.464}{MeV}, and \SI{4.759}{MeV} with relative abundances of 48.9\%, 2.2\%, and 48.9\%. 
The observed spectrum was fit using the predicted photon yield from the three alpha lines to extract the overall collection efficiency for this configuration of 1.10\% $\pm$ 0.15\%~\cite{Segreto:2018jdx} for this configuration, consistent with \dword{mc} expectations~\cite{Marinho:2018doi}. This corresponds to a gain in the effective photosensors area of approximately a factor of \num{3.7}. 

A series of subsequent prototypes with filters from different manufacturers, different reflectors, and different dimensions were evaluated with similar results. 

The final set of prototypes prior to \dword{pdsp} were tested in the \dword{tallbo} facility using an external set of cosmic ray counters as a readout trigger. These consisted of an array of eight \dword{sarapu} cells each with a photon collection area of \SI{80}{cm$^2$}, but the SensL \dwords{sipm} used in previous prototypes were replaced with four \SI{6}{mm} $\times$ \SI{6}{mm} Hamamatsu S13360-6050VE \dwords{mppc}. 
Two double-shift light guide modules were also included in the test and served as a reference for the \dword{sarapu} results.

The measured collection efficiency range for the eight \dword{arapuca} cells was 0.72\% to 0.80\%, with an effective \dword{sarapu} gain of about 4.5 times the photosensor area. These tests demonstrate that the effective area gain is maintained when the area for light collection of the cell is scaled up by almost an order of magnitude.

\subsubsection{\dword{pdsp}}
\label{sec:valid-pdsp}

The most comprehensive set of data on the \dword{sarapu} will come from the fully instrumented modules in the \dword{pdsp} experiment~\cite{Abi:2017aow} that completed first beam running in November \num{2018}. 
Since \dword{pdsp} will remain filled with \dword{lar} for much of the \dword{cern} long shutdown, it will provide a long-term cold test of full-scale \dword{pd} modules for the first time, so it may be possible to quantify any deterioration in their performance.

Three prototype photon collector designs are present in \dword{pdsp}: \num{29} double-shift guides, \num{29} dip-coated guides, and two \dword{sarapu} arrays.
The \dword{tpc} provides precise reconstruction in \threed of the track of any ionizing event inside the active volume, and matching the track with the associated light signal will enable an accurate comparison of the relative photon collection efficiencies of the different \dword{pd} modules. 
The large number of modules and independent channels that record each event can be used to constrain the parameters of the \dword{lar} that regulate \dword{vuv} light propagation in the simulation and are poorly determined in the literature. 
In principle, absolute calculations of of the relative and absolute detection efficiencies are possible using \dword{mc} simulations.
The precision of this approach may be limited by the precision of the constraints on the parameters but in any case will result in a consistent simulation constrained by measurements.

\begin{dunefigure}[Event display from \dshort{pdsp} showing the location of the \dshort{pd} modules]{fig:evtdisplay-pd-protodune}
{Event display from \dshort{pdsp} showing the location of the \dword{pd} modules on the beam entry side of the \dword{tpc}. Reconstructed \dword{tpc} hits from a test beam electron are visible at approximately the same height in the \dword{tpc} as the \dword{sarapu} module mounted in \dword{apa}~3.} 
\includegraphics[height=11.cm]{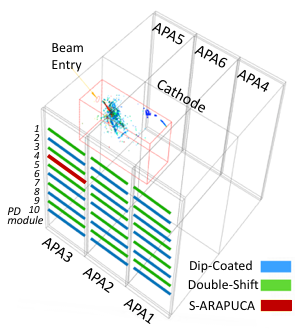}
\end{dunefigure}

\begin{dunefigure}[Full-scale \dshort{sarapu} array installed in \dshort{pdsp} APA~3]{fig:arapuca-protodune}
{Visible in the center of the photograph of \dword{apa}~3 is the 16-cell \dword{sarapu} array installed in \dword{pdsp}.} 
\includegraphics[height=8.cm]{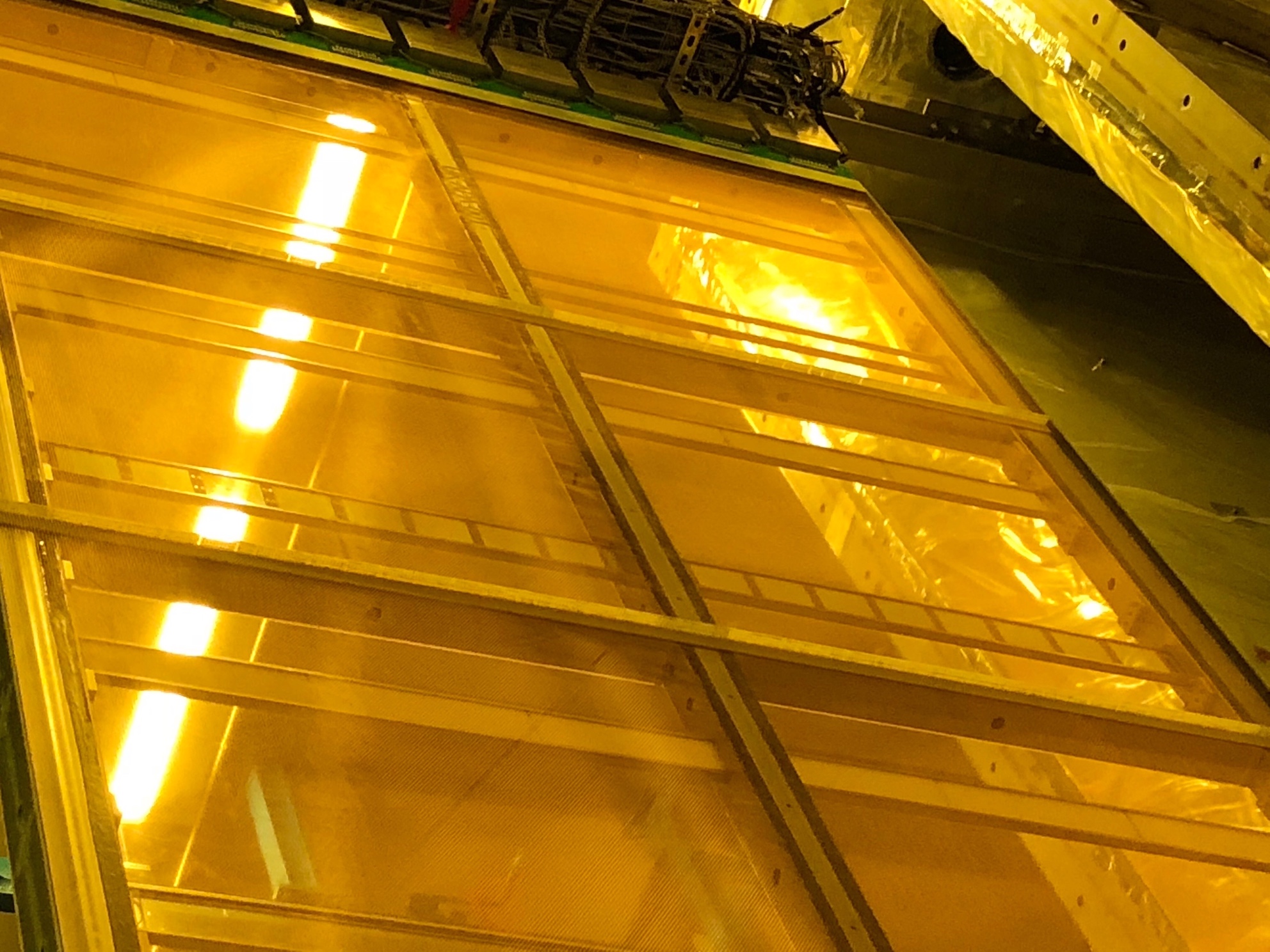} 
\end{dunefigure}

Figure~\ref{fig:evtdisplay-pd-protodune} shows an event display from \dword{pdsp} overlaid with colored bars indicating the positions of \dword{pd} modules that are on the beam entry side of the \dword{tpc}.
Of the two \dword{sarapu} arrays in \dword{pdsp}, the first is installed in \dword{apa}~3 in the fourth position from the top, near the level at which the beam particles enter this drift volume. This module and the surrounding light guide modules are illuminated with a significant amount of light from each beam particle interaction. This module is visible near the center of the photograph of \dword{apa}~3 in Figure~\ref{fig:arapuca-protodune}.
The second is installed in  \dword{apa}~6, in the 6th position from the top, in the drift region on the opposite side of the opaque cathode plane, which does not see entering beam particles; this module does not see significant light from beam events (only from showering particles that pass through the cathode), but it observes photons from a large collection of triggered cosmic rays.
 
Each \dword{pdsp} \dword{sarapu} module array is composed of sixteen cells, where each cell is an \dword{sarapu} box with window dimensions of \SI{7.8}{cm} $\times$ \SI{9.8}{cm}; half of the cells have twelve \dwords{mppc} installed on the bottom side of the cell and  half have six \dwords{mppc}. 
The \dwords{mppc} used are the Hamamatsu model 13360-6050CQ-SMD, which are functionally the same as the 13360-6050VE used for the ganging tests (Section~\ref{sec:pds-valid-ganging}) and also on some of the light guide bars in \dword{pdsp}, but this model incorporates a package specifically designed for cryogenic operation\footnote{A thin glass window mounted in front of uncoated silicon photosensitive surface, as opposed to the thin coating directly on the silicon for the 13360-6050VE.}. 
The \dwords{mppc} have active dimensions \SI{0.6}{cm} $\times$ \SI{0.6}{cm} and account for 5.6\% (\num{12} \dwords{mppc}) or \num{2.8}\% (\num{6} \dwords{mppc}) of the area of the window.
The \dwords{mppc}  are passively ganged together, so that only one readout channel is needed for each \dword{sarapu} grouping of \num{12} \dwords{mppc} (the boxes with six \dwords{mppc} are ganged together to form \num{12}-\dword{mppc} units), so a total of \num{12} channels is required per \dword{pd} module. 
The total width of a module is \SI{9.6}{cm}, while the active width of an \dword{sarapu} is \SI{7.8}{cm}, the length is the same as the light guide modules ($\sim$\SI{210}{cm})\footnote{Since \dword{pdsp} was constructed, the slot opening in the \dword{apa} opening for \dword{pd} module installation has been enlarged allowing for a module with larger collection area.}.

An \dword{sarapu} array during assembly is shown in Figure~\ref{fig:sarapuca_array_prod}; the array installed in \dword{pdsp} is shown in Figure~\ref{fig:arapuca-protodune}. 
A simulation of the \dword{sarapu} cells, using code that was validated by the earlier prototype measurements (Section~\ref{sec:sarapuca-prototypes}), predicts a photon collection efficiency for the module on the beam side of the cathode of 1.5\%, and for the module on the non-beam side with an optimized configuration, (\num{12} \dwords{sipm} and \dword{ptp} coated on the filter substrate), it could be as high as 3.0\%. A full \dword{sarapu} module with the optimized configuration would have an effective area equivalent to a detector with \SI{36}{cm$^2$} active area with 100\% collection efficiency and produce an average light yield across the \dword{tpc} of \SI{20}{PE/MeV}.

\begin{dunefigure}[\dshort{sarapu} module prototype assembly]{fig:sarapuca_array_prod}
{\Dword{pdsp} \dword{sarapu} module being assembled in a class 100,000 clean area.  Front face of assembled module (left) shows the 16 coated dichroic filter plates.  Assembly photos show the reflective rear side (top right) and inner coated surface (right bottom) of Vikuiti reflective foils.  Note the cutouts in foil for \dword{mppc} active area.}
	\includegraphics[angle=90,width=0.6\columnwidth]{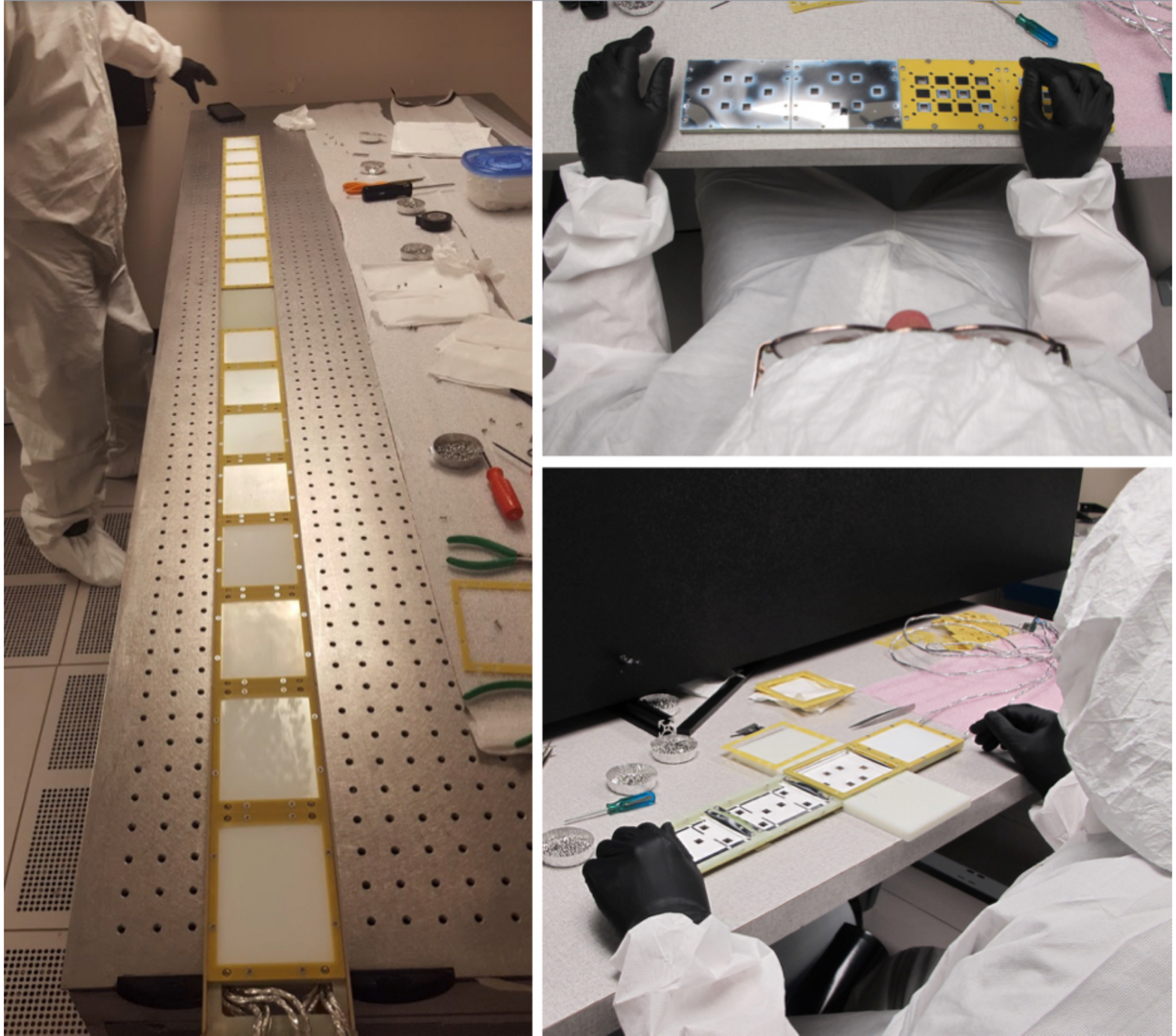}
\end{dunefigure}

As described above, for the \dword{sarapu} modules, the \dwords{sipm} are passively summed in groups of 12 to produce 12 signal channels per module. The 58 light guide-style light collectors each have 12 \dwords{sipm}, which are passively summed in groups of three such that each light guide has four signal channels. 
The unamplified summed analog signals from the \dwords{sipm} are transmitted directly to outside the cryostat for processing and digitization by a
\dword{ssp}.

The \dword{ssp} consists of 12 readout channels packaged in  a self-contained 1U module, where each channel contains a fully-differential voltage amplifier and a \num{14}-bit, \num{150}-MSPS \dword{adc} that digitizes the \dword{sipm} signal waveforms. The \dword{ssp} also provides a programmable bias voltage to the sensors.
The entire set of photon collector arrays are read out by 24 \dword{ssp} units (a total of 288 channels).

\textit{\dshort{pdsp} \dshort{pds} Measurements}
\label{sec:protodune-results}

The \dword{pdsp} beam run provides several distinct sets of data for understanding \dword{pds} performance: beam data sets with triggers determined by the beam instrumentation; cosmic ray data sets from triggers randomly or in coincidence with the \dword{crt} modules; and calibration module data sets, with triggers in coincidence or free running with a programmed light pulse. 
The single avalanche response and gain for all \num{256} 
readout channels has been extracted from \dword{pdsp} data, including runs using the pulsed UV-light calibration system.

The analysis is ongoing, but here we summarize some initial results that illustrate the performance and stability of the \dword{pds}:

\begin{itemize}
   \item Figure \ref{fig:protodune-allmodules-7gev} shows the response\footnote{In units of photoelectrons, not corrected for \dword{sipm} after-pulsing and crosstalk.} to tagged electrons (left) and muons (right) with a momentum of \SI{7}{GeV/c} for each of the \dword{pd} modules on the beam side of the cathode; the \dword{sarapu} module is in position \dword{apa}~3, \dword{pd} module~4. The observed energy is mostly contained within the first \dword{apa} width of \dword{lar} for electrons but is distributed through the whole width of the \dword{tpc} for the minimum ionizing muons. Although a detailed simulation is not completed, the ratio of the average signal (in photoelectrons) in the \dword{sarapu} module to the adjacent double-shifter bars, is approximately a factor of five for both the electron and muon samples. This is consistent with the detection efficiency ratio measured in earlier prototypes.
   
    \item Figure \ref{fig:protodune-tpcpdstime} provides two examples of the timing capability of the \dword{pds}. The left plot shows the excellent correlation between the \dword{tpc} and \dword{pds} track time. 
    The \dword{tpc} track time is the track \tzero time and the \dword{pds} time is the matched flash time for a 4500 track sample. The right plot shows the measured time difference in the \dword{pds}  response to two consecutive flashes from the calibration system, demonstrating timing resolution of \SI{14}{ns}, well below the \SIrange{0.1}{1}{$\mu$s} physics requirement.
 
    \item Figure \ref{fig:protodune-pds-energyresponse} shows the response\footnote{In units of detected photons, corrected for \dword{sipm} after-pulsing and crosstalk.} of the \dword{sarapu} in \dword{apa}~3 to the tagged electron beam as a function of incident electron kinetic energy. 
    The observed energy is mostly contained within the first \dword{apa} (\dword{apa}~3) width of \dword{lar}. The observed number of photons have not been corrected for geometry, attenuation, or scattering effects but nonetheless shows a linear response over the \SIrange{0.3}{7.0}{GeV} beam energy range.   
    
    \item Figure \ref{fig:protodune-pds-stability} shows the stability of the measured light yield in the \dword{sarapu} on the non-beam drift side of the \dword{tpc} (\dword{apa}~6) using the calibration system (left) and a triggered cosmic ray muon sample (right). The left plot shows the response to the calibration flashes over time period spanning November 2018 - June 2019 normalized to the average response over the period; different colors correspond to different readout channels (error bars not shown to increase visibility of points; average of the errors is 1.4\% with a maximum of 4\%). The right plot shows the summed \dword{pd} light yield (all modules of the same type) as a response to cosmic-ray muon samples partitioned by collector and sensor technology over the same period.

    The \dword{sarapu} variation across the entire time period is less than 2\%, but statistical uncertainties are larger than those of the other technologies because there is only a single \dword{sarapu} module with a much smaller light collection area than the combination of approximately 15 times more similar-sized modules for each of the other technologies. 
\end{itemize}

The \dword{pdsp} \dword{pd} modules have been operated for more than six months.  In this time, no failures of \dword{sipm} readout channels have been detected beyond the few seen immediately after installation; none of those failures were in the \dword{mppc} readout channels, which is the baseline photosensor. 

\begin{dunefigure}[\dshort{pds} response to \SI{7}{GeV/c} momentum electrons and muons in \dshort{pdsp}]{fig:protodune-allmodules-7gev}
{\dword{pds} response (in \phel) to \SI{7}{GeV/c} momentum electrons (left) and muons (right) in \dword{pdsp}.}
\includegraphics[angle=0,width=0.48\columnwidth]{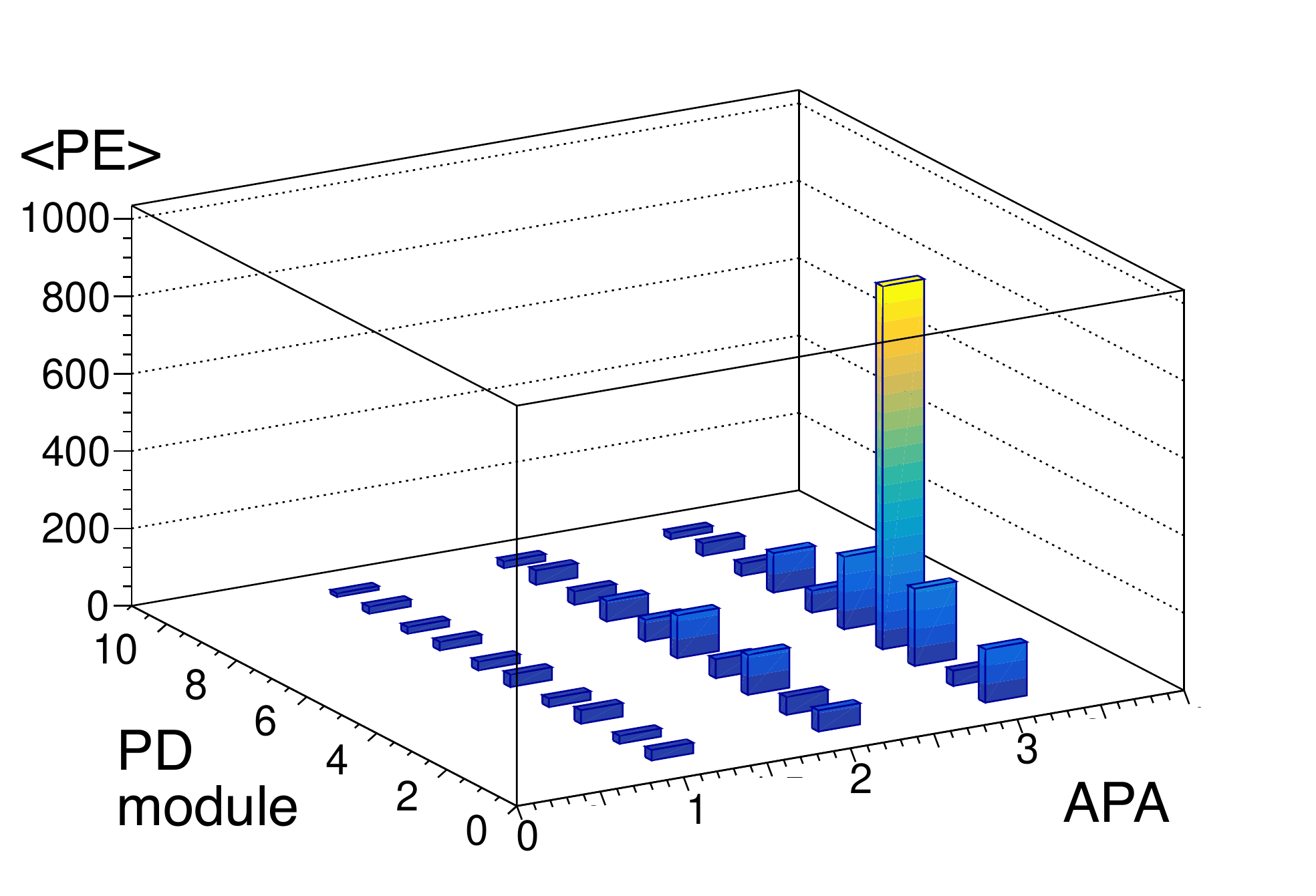}
\includegraphics[angle=0,width=0.48\columnwidth]{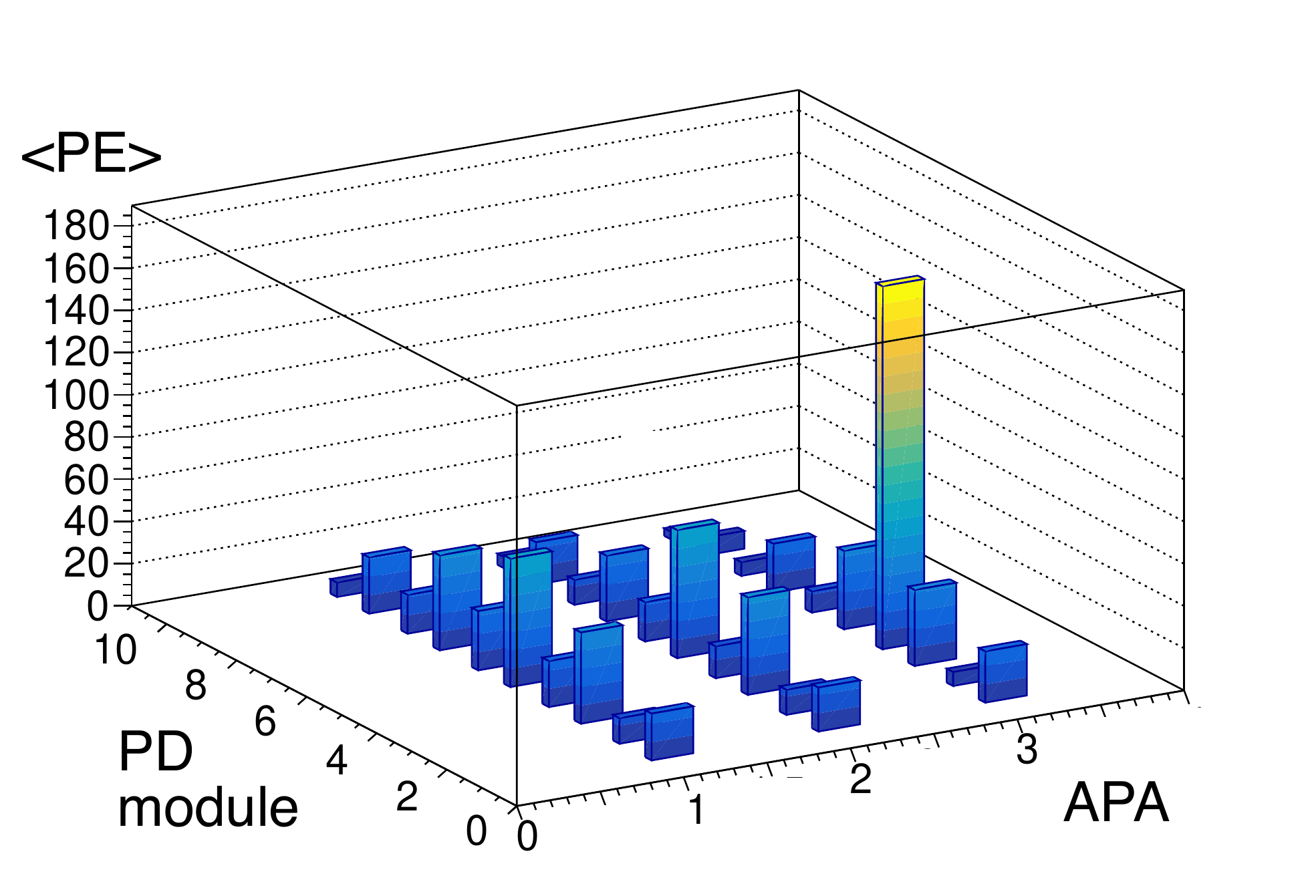}
\end{dunefigure}

\begin{dunefigure}[\dshort{pds} timing measurements with\dshort{pdsp}]{fig:protodune-tpcpdstime}
{\dword{pds} timing measurements: Correlation between the \dword{tpc} and the \dword{pds} track time (left); time difference between two consecutive calibration flashes, demonstrating a resolution of \SI{14}{ns} (right).}

\includegraphics[height=5.5cm,width=0.6\textwidth]{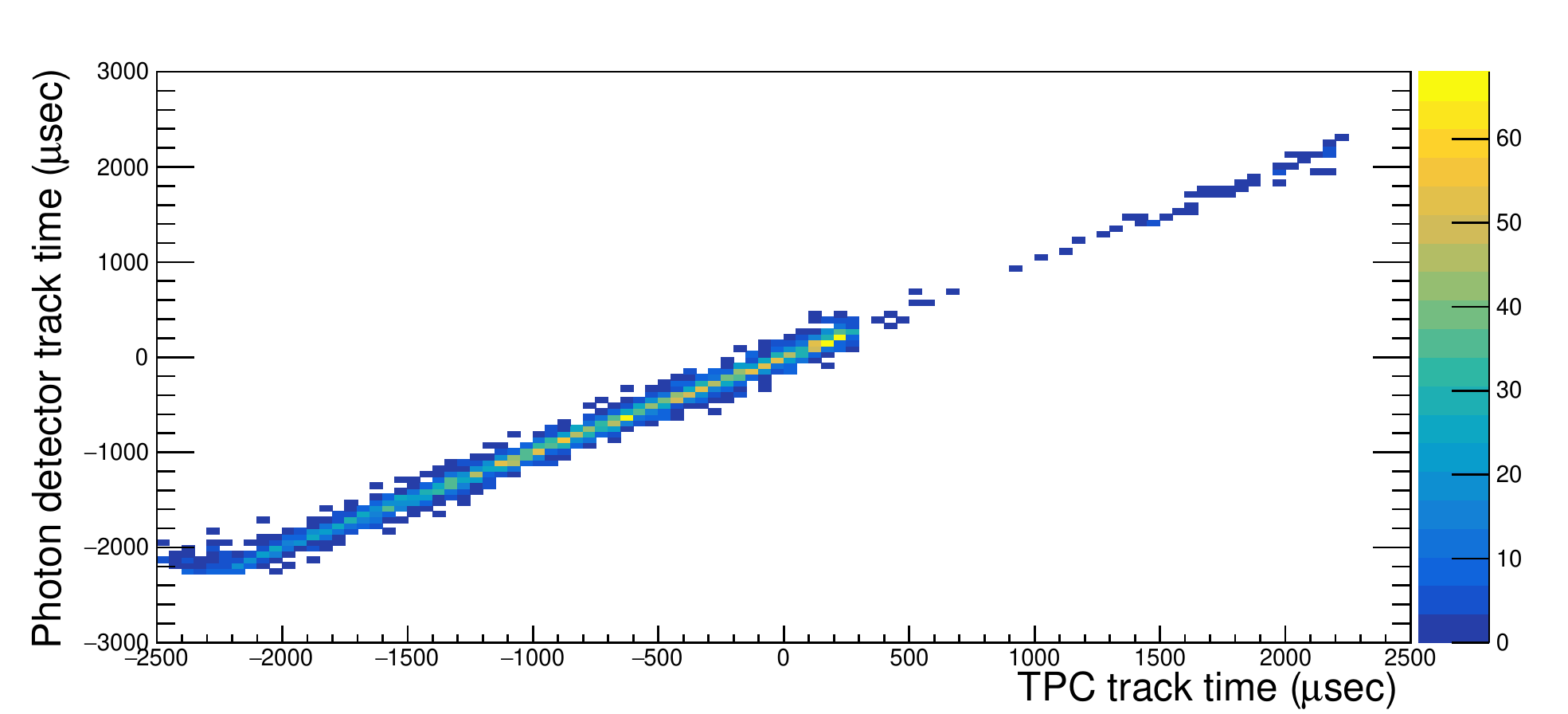}
\includegraphics[height=5.5cm,width=0.3\textwidth]{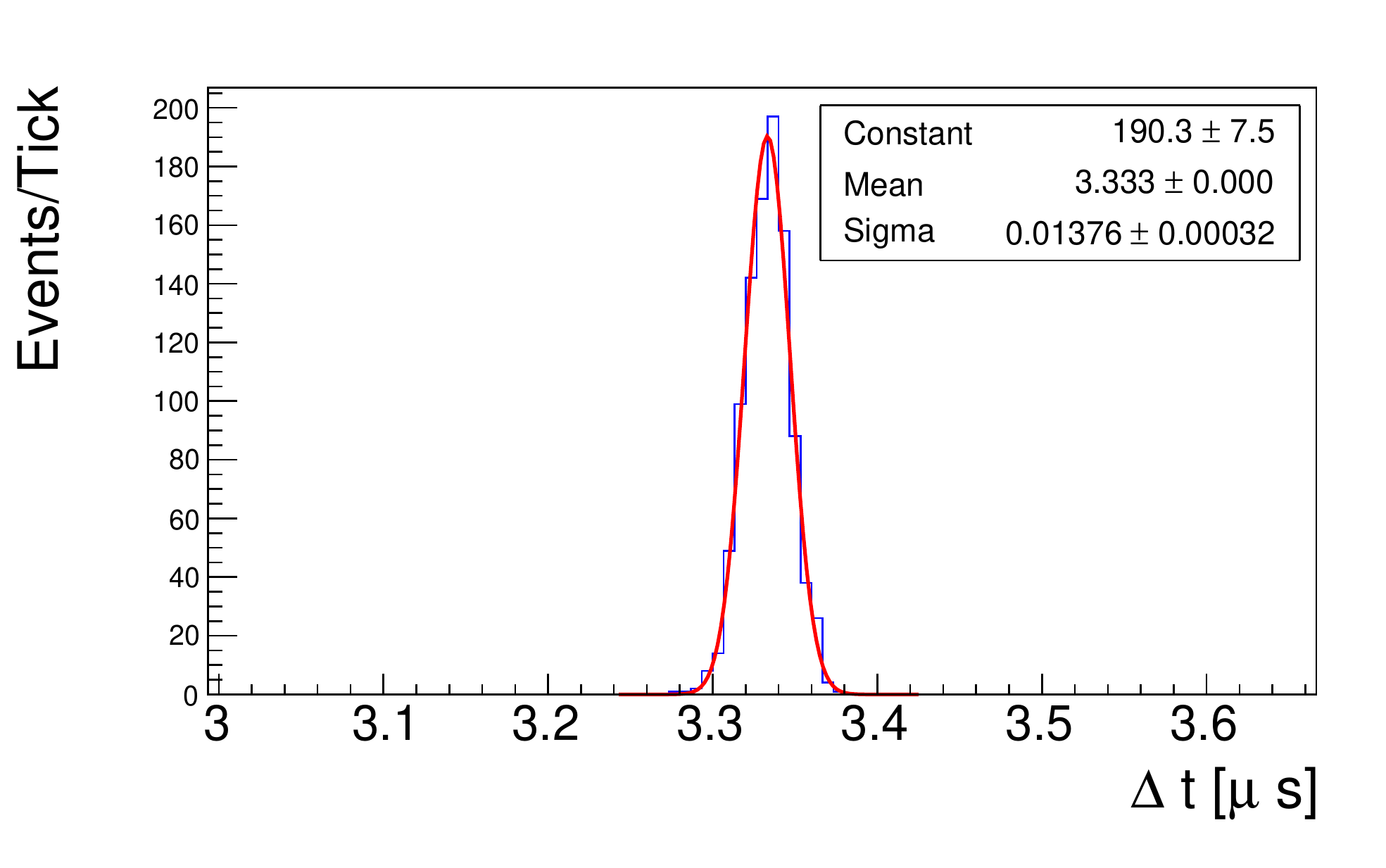}
\end{dunefigure}

\begin{dunefigure}[\dshort{pds} energy response to \dshort{pdsp} electron beam]{fig:protodune-pds-energyresponse}
{Mean number of collected photons as a function of incident electron kinetic energy (left);  photon counting resolution of the \dword{sarapu} array as response to test beam electrons (right).}
\includegraphics[width=0.45\linewidth]{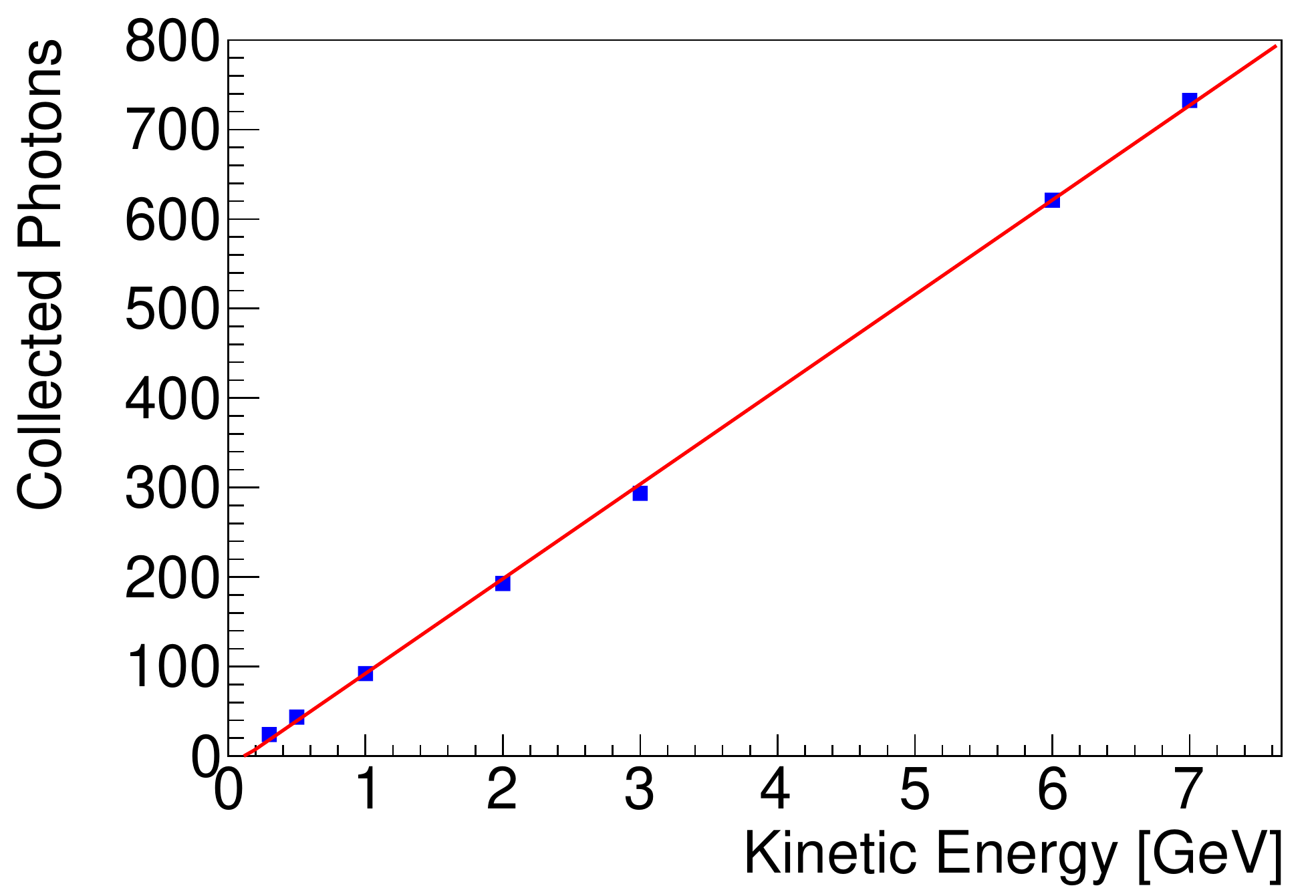}
\includegraphics[width=0.45\linewidth]{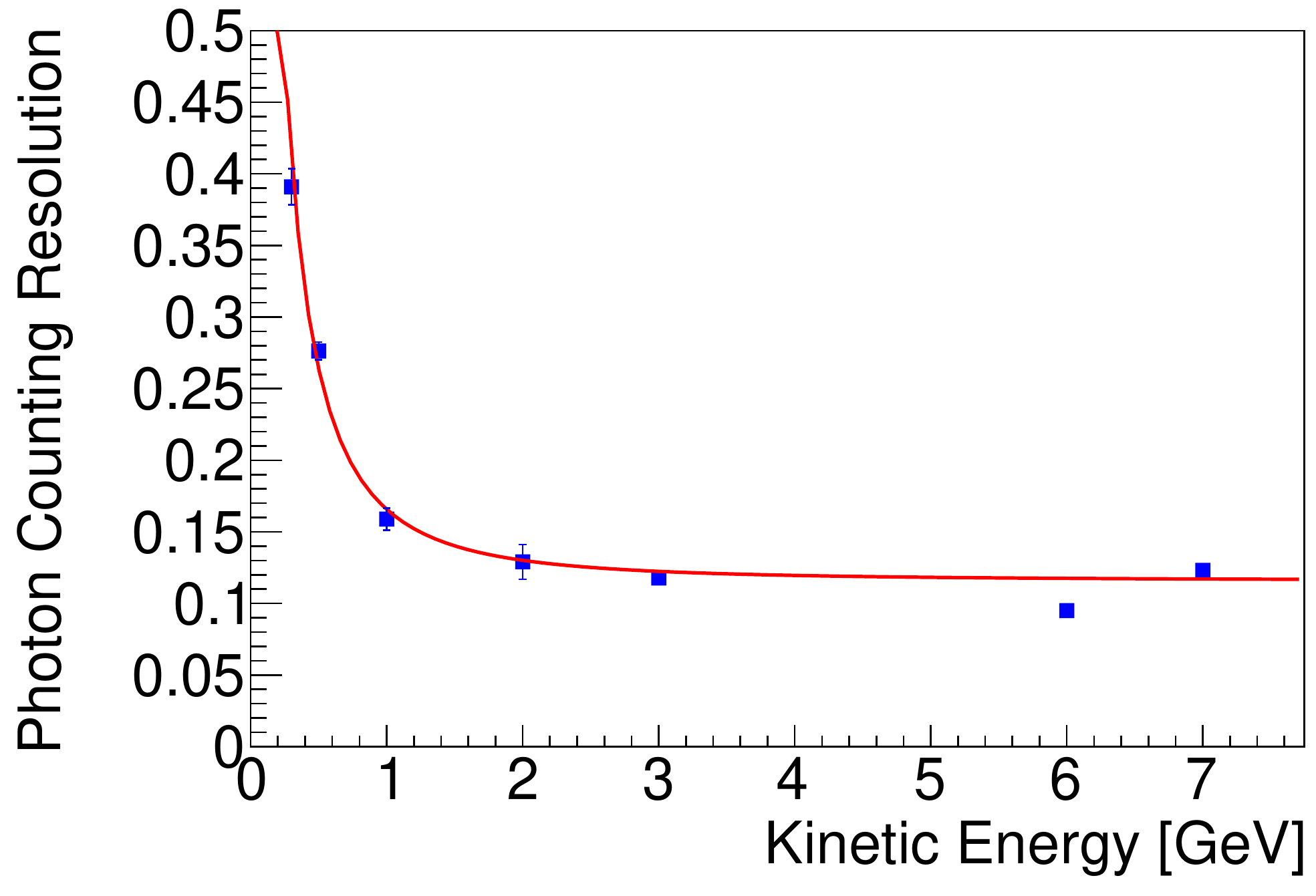}

\end{dunefigure}

\begin{dunefigure}[Stability of the \dshort{pds} during \dshort{pdsp} running]{fig:protodune-pds-stability}
{Stability of the \dword{sarapu} response in \dword{apa}~6 measured with the UV-light calibration system (left) and the stability measurements of all \dword{pds} channels in \dword{apa}s~4-6 with cosmic-ray muons tagged by the \dword{crt} (right).}
\includegraphics[width=0.5\linewidth]{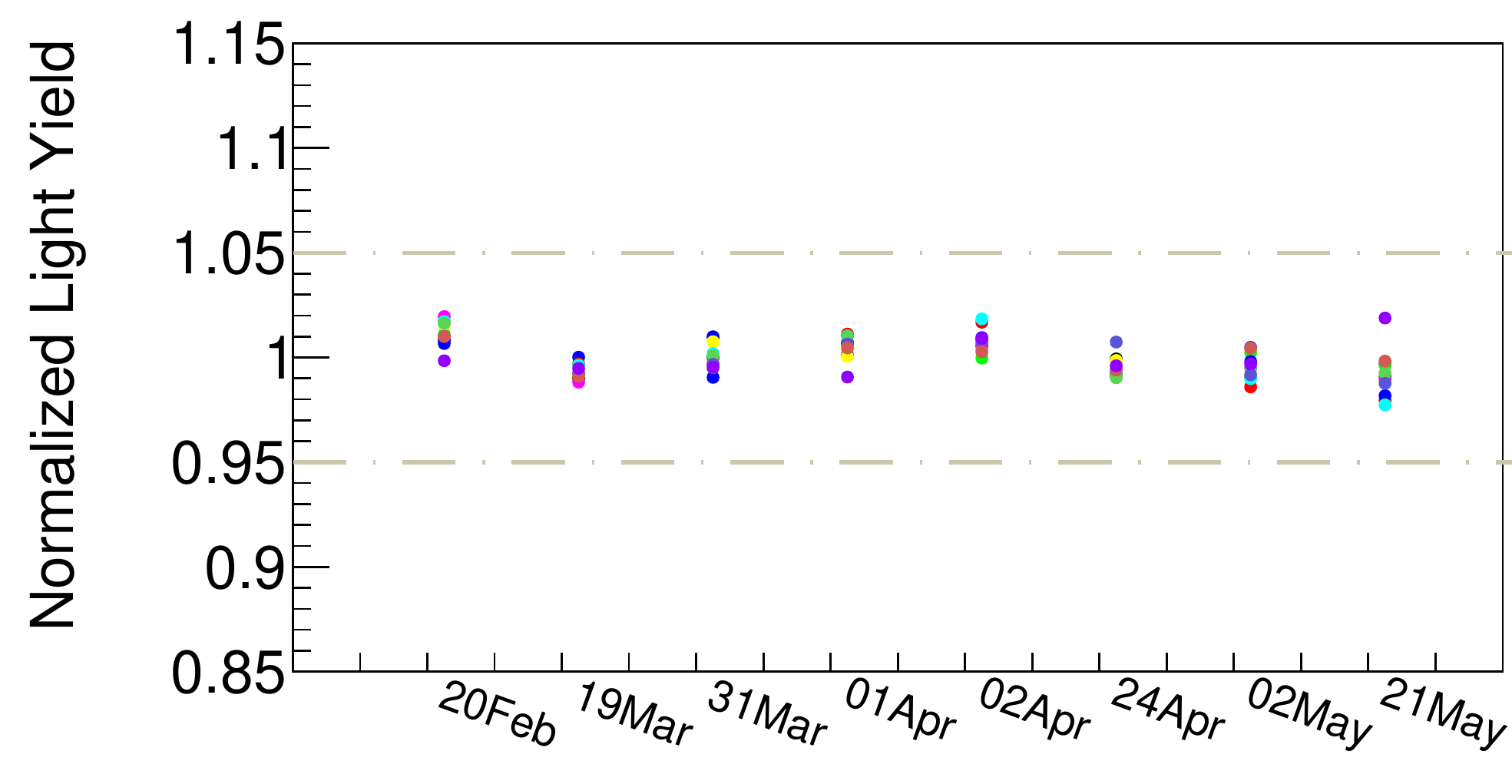}
\includegraphics[width=0.45\linewidth]{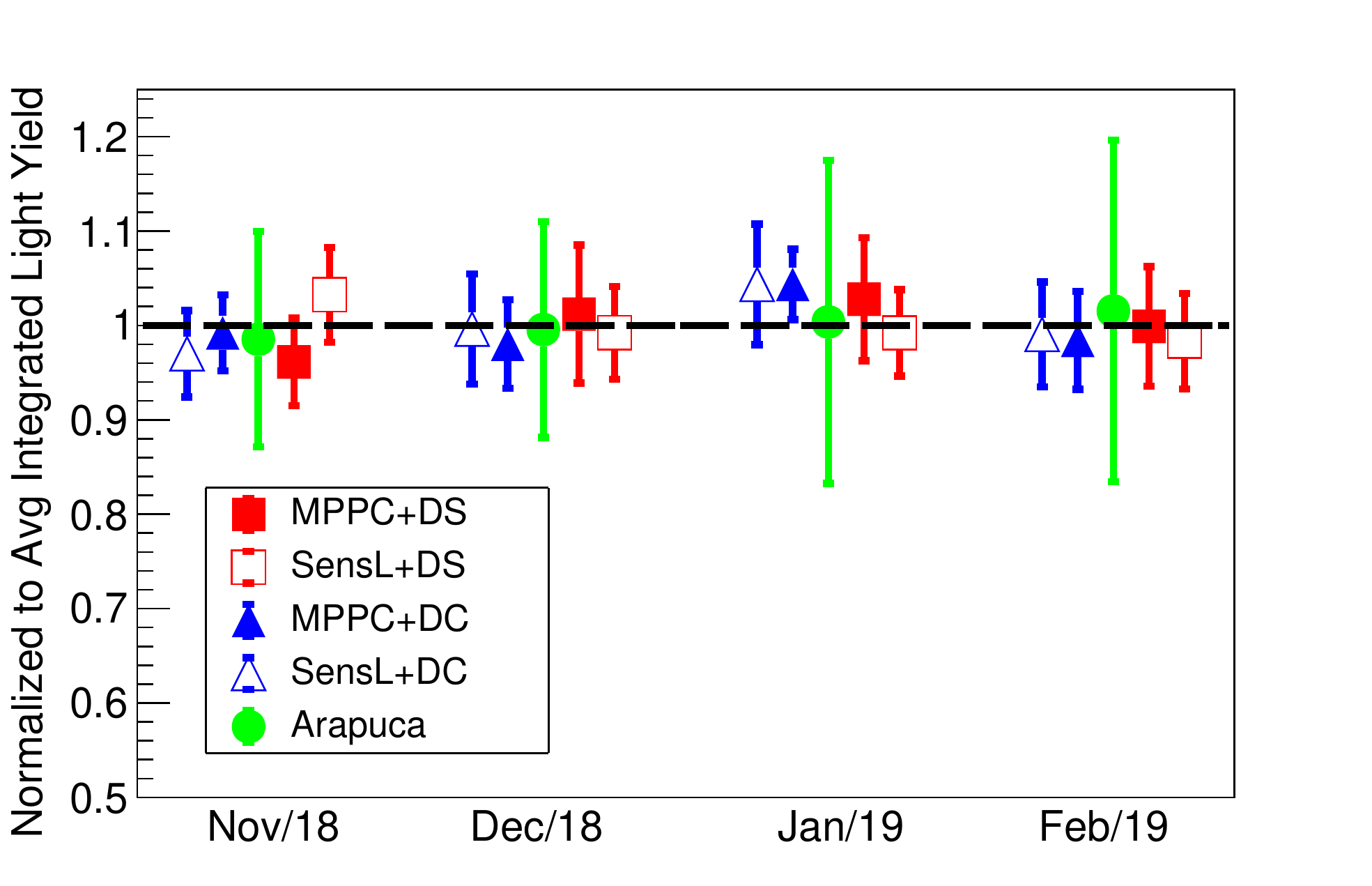}
\end{dunefigure}

\subsection{Extended ARAPUCA (X-ARAPUCA)}
\label{sec:xarapuca-valid}

The \dword{xarapu} is an evolution of the \dword{sarapu} concept that moves the second wavelength-shifter from the inner surface of the dichroic filter to a wavelength-shifter-doped plate that acts as a light guide. This design change was motivated by simulations that indicated a significant increase in collection efficiency for this configuration. 

This section describes in detail the first measurements that demonstrate the validity of the design, followed by a description of ongoing efforts to validate the final design.

\subsubsection{Single Cell \dword{xarapu} Measurements}
\label{sec:xarapuca-unicamp}

The first tests of an \dword{xarapu} cell were made at \dword{unicamp}, Brazil, at the end of November 2018. 
The structure of the cell allowed it to operate as either an \dword{sarapu} or an \dword{xarapu}, with both single- or double-sided readout in both.  This flexibility will allow relative and absolute measurements of performance in the same cryostat and so provide a crucial step to validating the baseline design.

Building on the experience with the \dword{pdsp} prototypes, the frames for the test cell were fabricated from \frfour G-10 in a configuration very similar to that planned for the \dword{spmod} but with some small modifications necessitated by the requirements for holding a single window. The overall dimensions of the cell are \SI{12.3}{cm} $\times$ \SI{10.0}{cm} $\times$ \SI{1.56}{cm}. Figure~\ref{fig:xarapuca-cell} shows an exploded design drawing and the completed cell. 

\begin{dunefigure}[\dshort{xarapu} test cell]{fig:xarapuca-cell}
{\dword{xarapu} test cell:  Assembled cell (left); exploded model (right).  Note that exploded components can be duplicated on the back side (not shown) for double-sided test cell.} 
	\includegraphics[angle=270, origin=c, height=6.5cm]{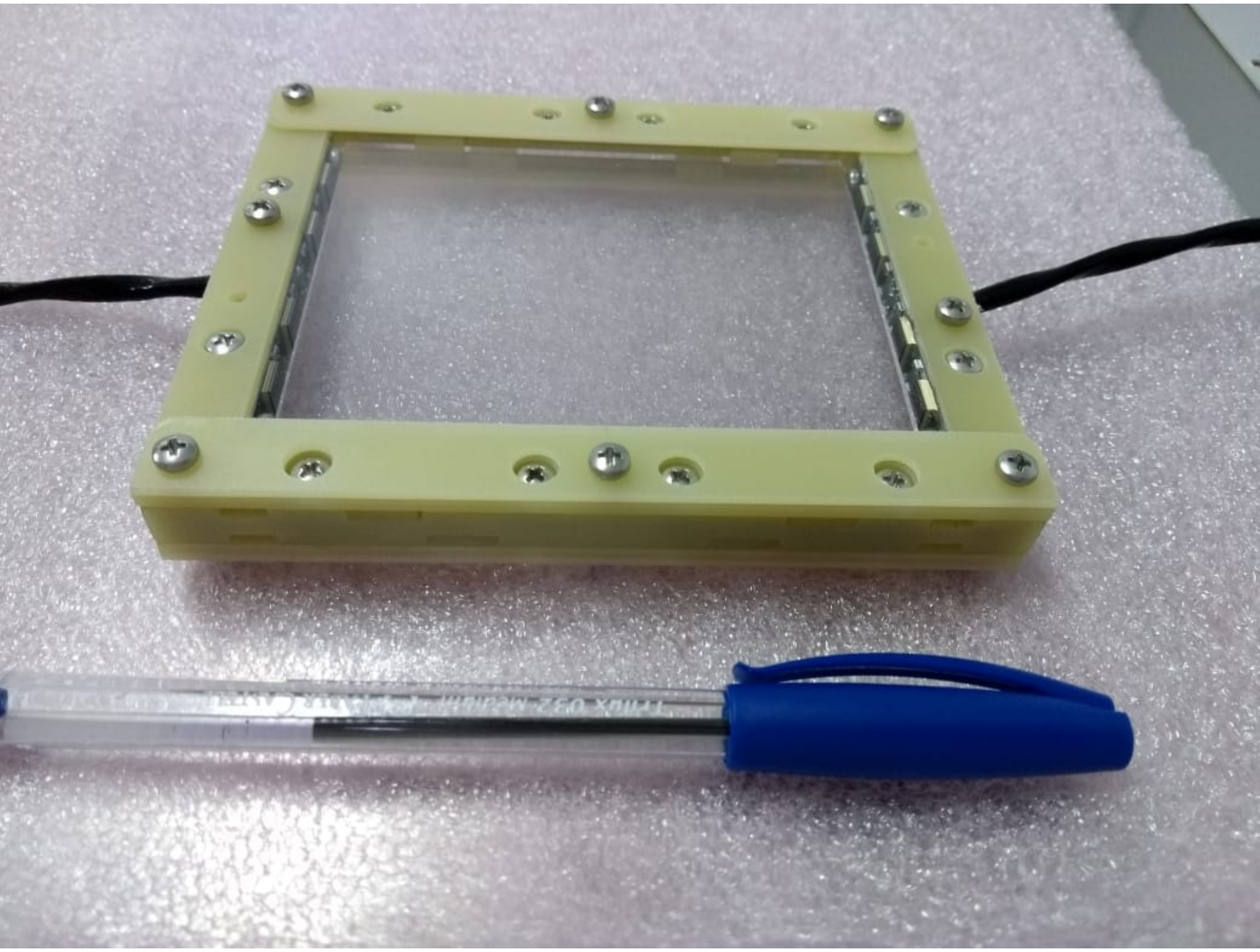}
	    \includegraphics[height=6.5cm]{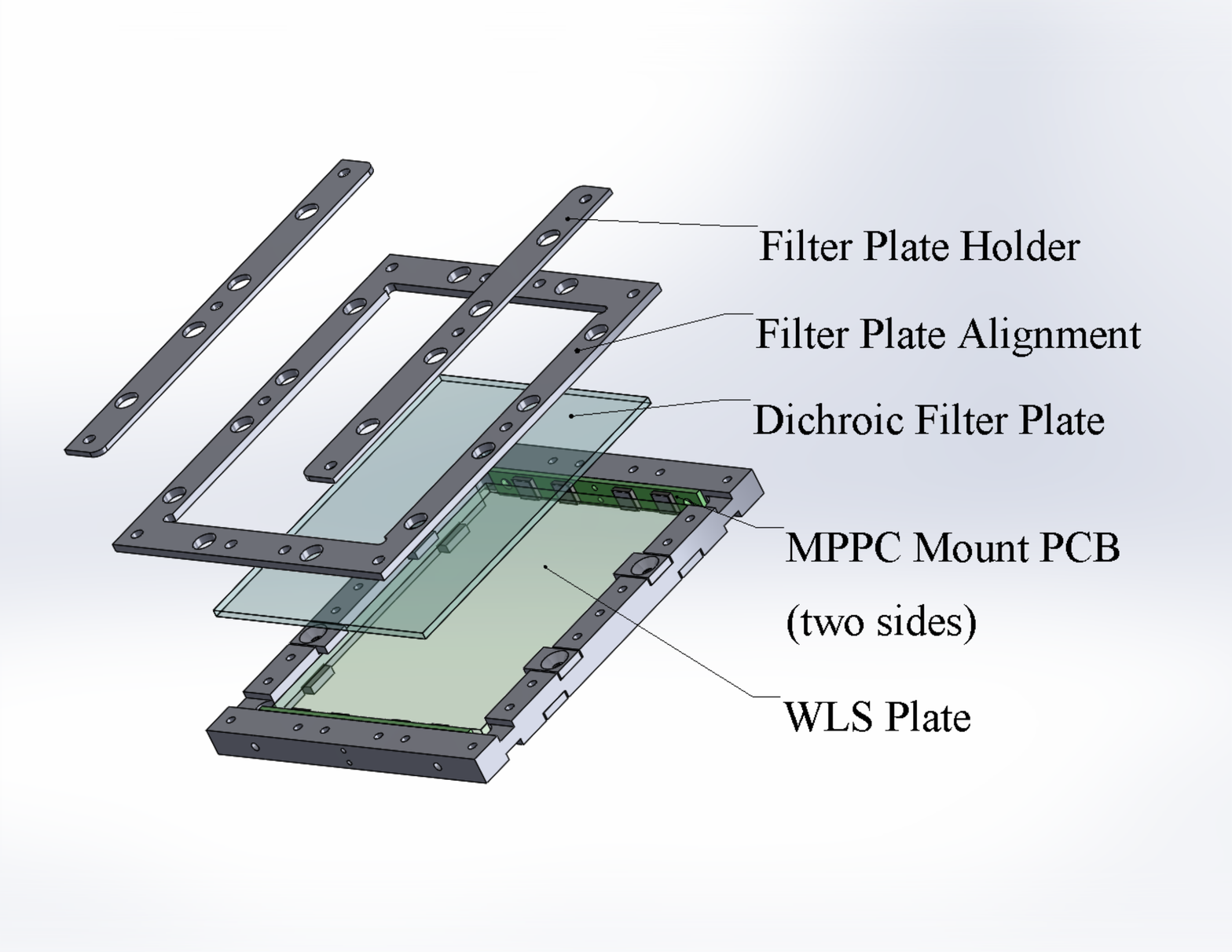}
\end{dunefigure}

The dichroic windows for the prototype are the same size as one of the six windows in a \dword{fd}-design \dword{xarapu} supercell: \SI{10.0}{cm} $\times$ \SI{7.8}{cm}. The filter plate was coated with \dword{ptp} by vacuum evaporation (film thickness $\sim$ \SI{400}{${\mu}$g/cm$^2$})  using an in-house deposition system at \dword{unicamp} (see Figure \ref{fig:xarapuca-plates}). 
The thickness of the film needs to have a minimal value in order to ensure that the \dword{vuv} light is fully absorbed. This minimum value is in the range of \SI{100}{${\mu}$g/cm$^2$} to \SI{200}{${\mu}$g/cm$^2$}. The  \SI{400}{${\mu}$g/cm$^2$} is chosen in order to ensure that the minimum thickness is reached everywhere on the filter even in presence of exceptional fluctuations (measured \dword{rms} of the order of \SI{30}{${\mu}$g/cm$^2$}). When the minimal thickness is reached the maximum conversion efficiency is obtained.

Adhesion was tested by submerging the coated filter in \dword{ln} several times. The coating was visually inspected after each submersion, and no visible effect was observed. At the end of the test, the coated filter was weighed with a precision balance, and no loss of material was measured. The film was also analyzed with an optical microscope, and no signs of degradation could be observed. 

The wavelength shifting plate in the \dword{xarapu} configuration is made from Eljen EJ-286 blue \dword{wls} plate, with dimensions \SI{9.3}{cm} $\times$ \SI{7.8}{cm} $\times$ \SI{0.35}{cm}.  

The \dword{wls} plate thickness of \SI{0.35}{cm} was initially selected to optimize collection of light both by total internal reflection within the \dword{wls} plate and light trapped between the filter plates and the \dword{wls} plate.  Simulation has demonstrated that the detection efficiency performance of the detector reaches a shallow maximum at approximately this value.

The side walls of the test cell are lined with Vikuiti reflector, with cutouts at the positions of the photosensors.

\begin{dunefigure}[Coated dichroic filters and vacuum coating system]{fig:xarapuca-plates}
{Coated dichroic filter plates (left) and \dword{unicamp} thin film coating facility (right).} 
	\includegraphics[angle=90, height=6cm]{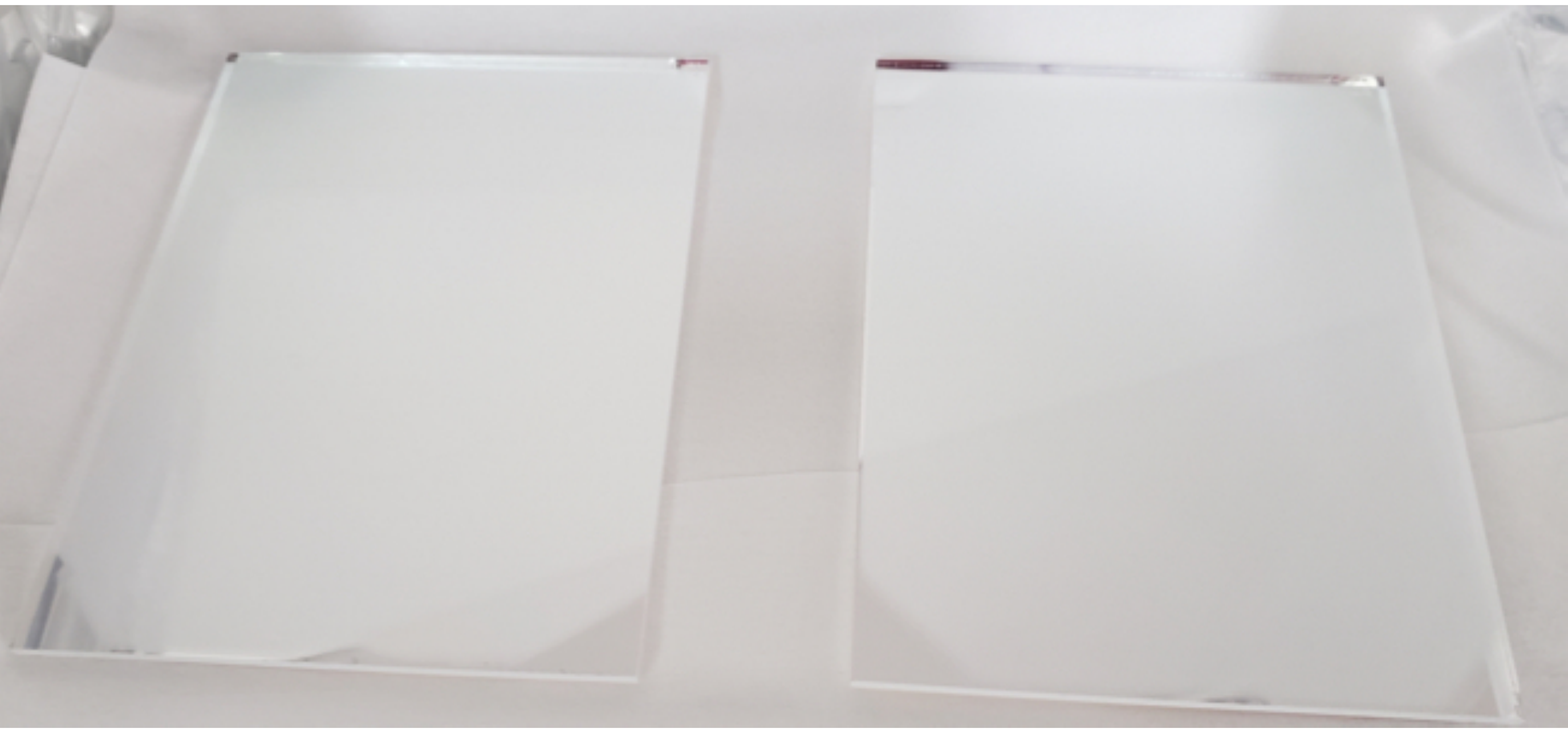}\quad
	\includegraphics[height=8cm]{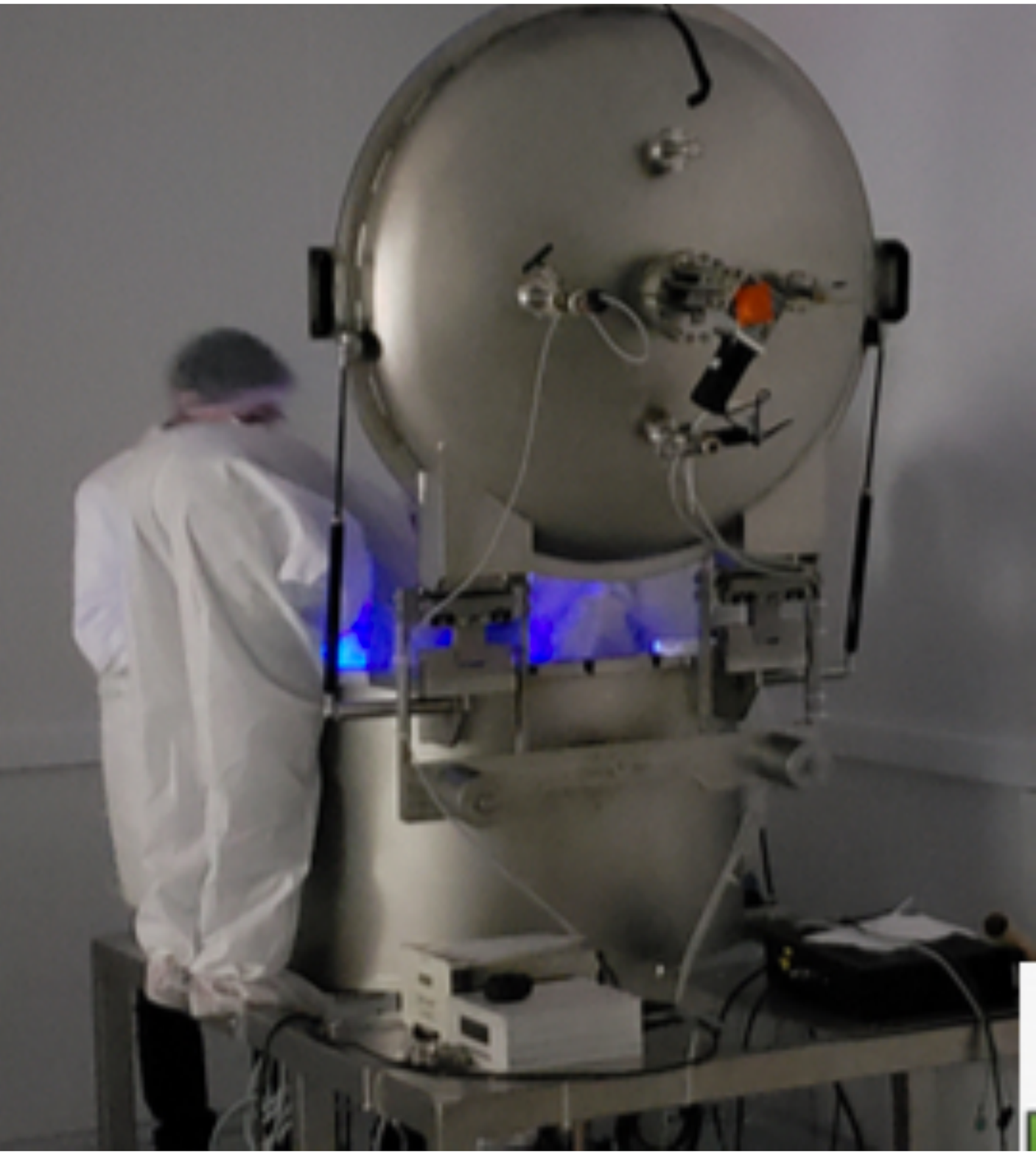}
\end{dunefigure}

The photosensors in the test cell are of the baseline type:  \SI{0.6}{cm} $\times$ \SI{0.6}{cm} Hamamatsu S13360-6050VE \dwords{mppc}.  The photosensors are arranged in the same configuration as in the baseline design, with four \dwords{mppc} (passively ganged) mounted to two sides of the test cell, with positioning relative to the \dword{wls} plates and dichroic filters identical to the baseline design.  In a departure from the baseline design, the two passively ganged groups of four \dwords{mppc} are read out separately; no active ganging circuit is implemented for these tests. 

The test cell is installed at the bottom of a vacuum-tight stainless-steel cylinder (height $\sim$\SI{30}{cm}) closed by two CF160 flanges and exposed to an alpha source\footnote{The same as used in the \dword{sarapu} proof-of-principle tests described in Section~\ref{sec:sarapuca-prototypes}.} 
placed at \SI{3}{cm} from the center of the dichroic window. Figure~\ref{fig:xarapu-teststand} shows photographs of the test cryostat and the test cell in the support structure; the alpha source holder is visible through the windows.
The stainless-steel cylinder  is deployed in a small open Dewar that is filled with commercial-grade \dword{lar} to act as a thermal bath.

\begin{dunefigure}[\dshort{xarapu} test stand]{fig:xarapu-teststand}
{Test cryostat (left) and \dword{xarapu} test cell mounting structure (right).  Note the alpha test source in holder.} 
	\includegraphics[height=6cm]{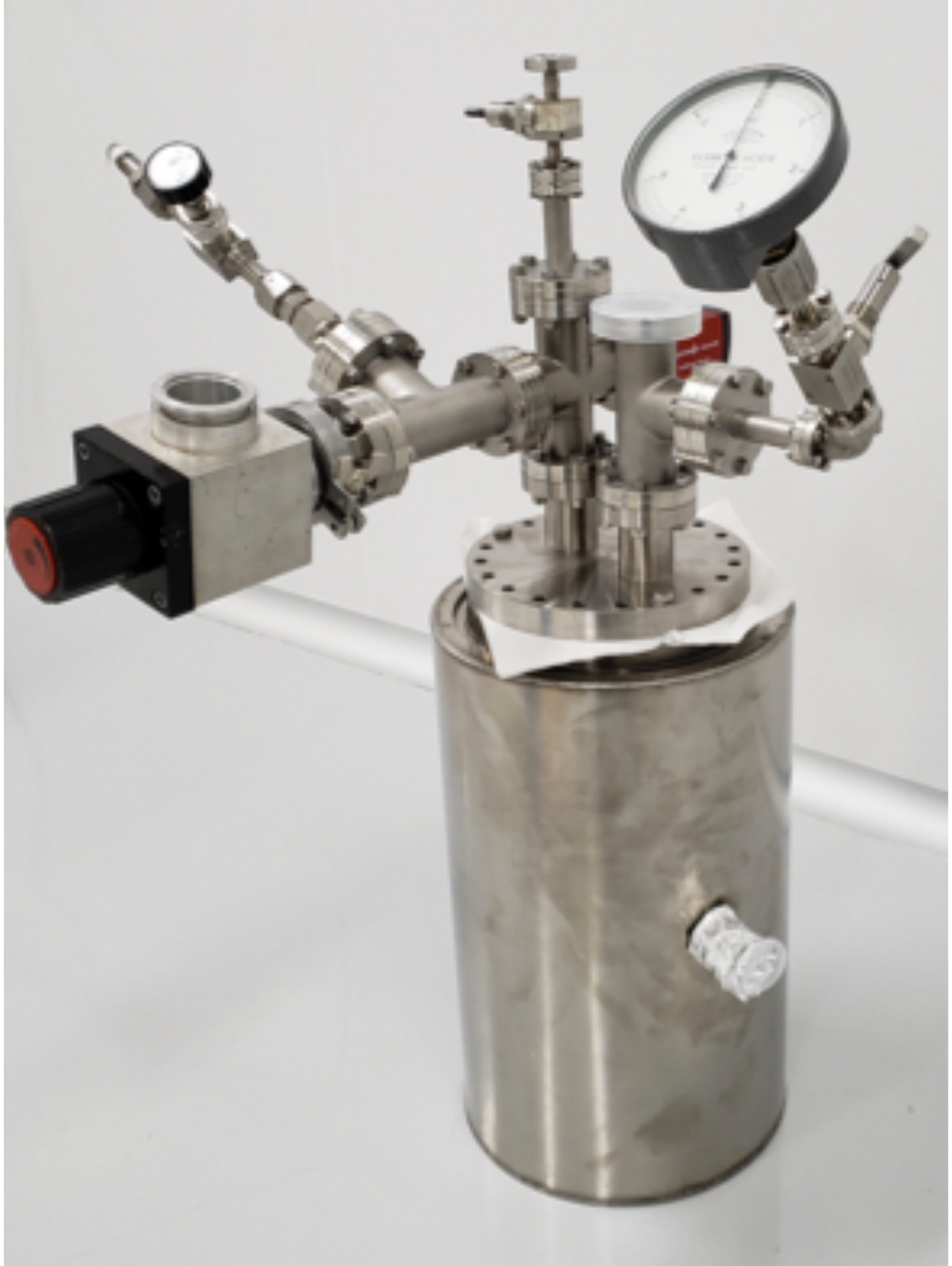} \quad
	\includegraphics[height=7cm]{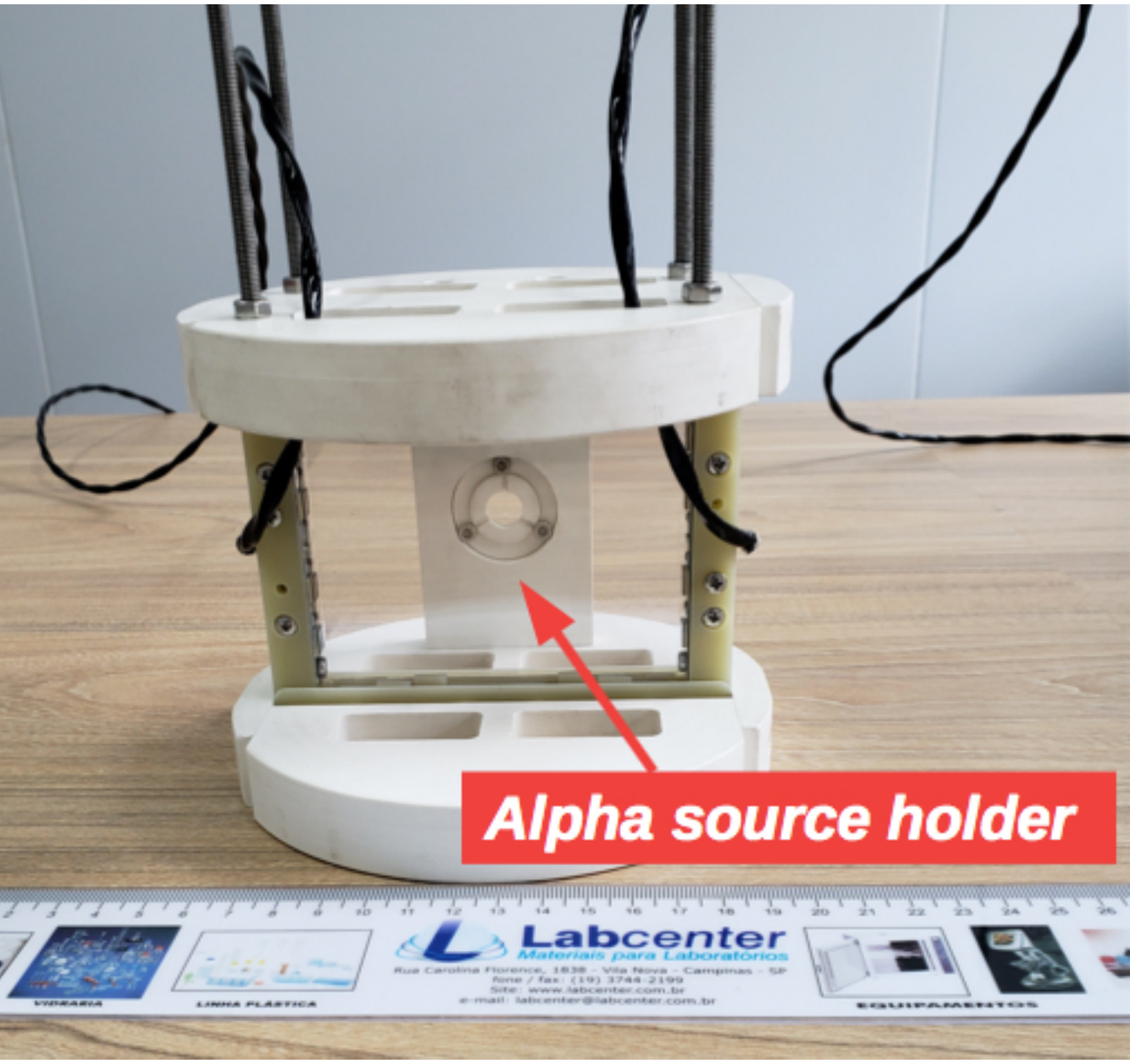}
\end{dunefigure}  

The spectrum of the detected number of photons is shown in Figure~\ref{fig:xarapu-results} with a black line. The same fit procedure as in Section~\ref{sec:sarapuca-prototypes} allows an estimate of the number of detected photons for each alpha line. The result of the fit is shown with a red line in Figure~\ref{fig:xarapu-results}. Comparing the number of detected photons with the number of photons impinging on the \dword{xarapu} provides an estimate the global photon collection efficiency of the device as 3.5 $\pm$ 0.5\%.
This result includes the correction for the crosstalk and after-pulsing of the two arrays of \dword{mppc} at their operating voltage.
The efficiency was stable for the duration of several days of this measurement.

The same test was repeated with the double-sided version of the \dword{xarapu}. Exactly the same set-up and the same device were used with the exception that the back plane coated with Vikuiti was replaced with a dichroic filter coated with \dword{ptp}. The same testing and analysis procedures were followed, and the global detection efficiency was found to be only 10\% less than the single sided version.   

\begin{dunefigure}[Alpha particle energy spectrum in an \dshort{xarapu} test cell]{fig:xarapu-results}
{Spectrum of the total number of \phel collected with the alpha source (blue line) fitted with the Monte Carlo prediction for the single-sided (left) and double-sided (right) \dword{xarapu}. Higher background activity at low energy was found for the second case.}

\includegraphics[height=6cm]{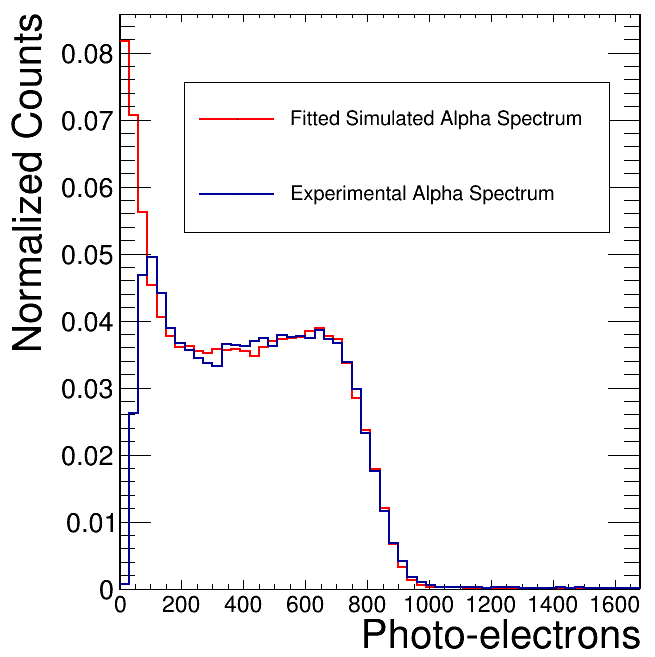}
\includegraphics[height=6cm]{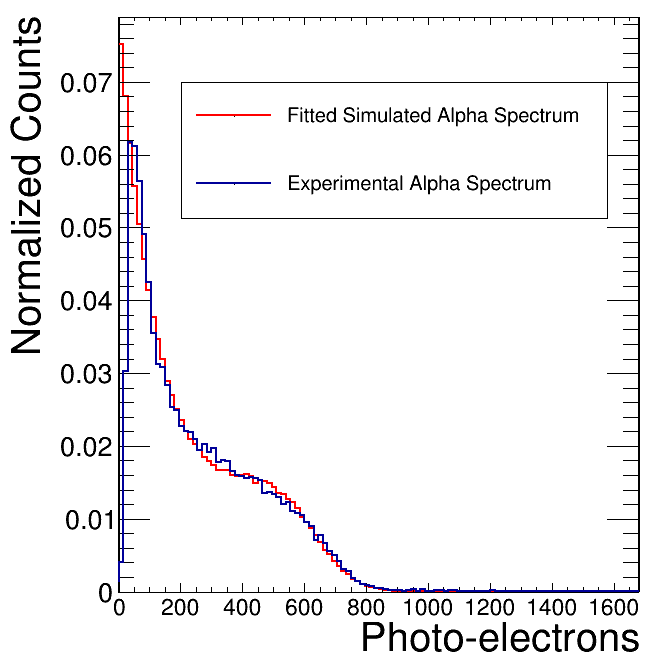}
\end{dunefigure}  

The measured global collection efficiencies translate into an equivalent surface area\footnote{The equivalent surface area is defined as the product of the physical acceptance window of the device multiplied by its global collection efficiency.} of \SI{70}{cm$^2$} for a single sided \dword{xarapu} module and of \SI{63}{cm$^2$} for the double-sided, which exceed the specifications for our system by a substantial factor (Section~\ref{sec:fdsp-pd-simphys}). 

\subsubsection{ICEBERG Test Stand}
\label{sec:iceberg-teststand}

The \dword{iceberg} test-stand is a small-scale \dword{tpc}, using smaller \dword{fd} \dword{apa} and cathode designs, constructed primarily to provide a platform for \dword{dune} \dword{ce} testing at \dword{fnal}. 
The test stand consists of a \SI{94.7}{cm} $\times$ \SI{79.9}{cm} \dword{apa}, with an approximately \SI{30}{cm} drift length to a cathode plane on each side (Figure~\ref{fig:fig-pds-iceberg-tpc}).  
It can accommodate up to two almost 1/2-length \dword{pd} modules\footnote{To enable the use of existing components for the \dword{apa} frame, the \dword{pd} modules are \SI{50}{mm} shorter than final modules, which required a slight modification to the \dword{pd} module design.} in a mounting structure nearly identical to the final \dword{dune} \dword{fd} configuration, allowing for testing of \dword{pd} prototype performance, electrical connections, and interfaces with the \dword{ce} and \dword{apa} systems (Figure~\ref{fig:fig-pds-iceberg-supercell}). 
In addition, the test stand will be used to allow comparisons between \dword{mu2e}-based warm electronics and \dword{pdsp} \dword{ssp} system, as well as testing newer versions of photosensors active ganging circuit designs.

The \dword{iceberg} facility will enable the primary validation of the \dword{xarapu} design prior to a full-scale test envisioned at a future \dword{pdsp} run in late 2021. 

At least three test campaigns are planned for the \dword{iceberg} \dword{tpc} with \dword{pd} modules:  

\begin{enumerate}
    \item The initial photon detector configuration consisted of one full-length \dword{sarapu} supercell and one full-length \dword{xarapu} supercell.  Both of these supercells have single-side windows to allow comparisons of measurements with \dword{sarapu} and \dword{xarapu} prototypes in summer and fall 2018 \dword{pdsp}.  
    The first test run occurred in February-March 2019. This run demonstrated that both module prototypes and the cryogenic active ganging circuitry were operational and saw signals from both modules using a \dword{mu2e} front end electronics system modified for use by the \dword{dune} \dword{pd} (in a separate stream from the \dword{tpc} \dword{daq}). Significantly, no crosstalk between the \dword{pd} and \dword{ce} readout electronics was observed.  Unfortunately, difficulties with the \dword{tpc} \dword{ce} and \dword{hv} systems prevented readout of ionization tracks required to allow direct comparisons between the \dword{pd} modules and required the test to end before significant progress was made.

    \item A second campaign took place late July through December 2019. It used the same \dword{pd} modules as the first campaign. 
    The data will facilitate comparisons of \dword{ssp} and \dword{mu2e} readout systems for similar events as indicated by hodoscope events.  
    Two short runs (\dword{iceberg} 2A and 2B) occurred in August and September 2019, during which the \dword{daphne} and \dword{ssp} were commissioned and initial photosensor bias voltage studies were conducted.  These runs were cut short due to problems with the cryogenic filtering system for the \dword{iceberg} cryostat.

    \item A third campaign is planned for spring of 2020.  This run (\dword{iceberg} run 3) will incorporate four \dword{xarapu} supercells, though it is partially occluded in the frame due to the limitations in \dword{apa} size mentioned above.  Two supercells will be single-sided and two double-sided, allowing for additional comparisons of \dword{pd} technologies.  This campaign will also demonstrate readout of \dword{mu2e} electronics by the \dword{iceberg} \dword{daq}.  In addition, we plan to incorporate a prototype of the \dword{dune} \dword{sp} monitoring system.
    
\end{enumerate}

The test stand will provide testing and validation of the \dword{pds} \dword{mu2e}-based electronics system, including a side-by-side comparison with the \dword{pdsp} \dword{ssp} electronics readout. In addition, concurrent data taking with the \dword{tpc} and light collection system will allow us to study \dword{tpc}-induced noise on the \dword{pd}, \dword{pd}-induced noise on the \dword{tpc}, grounding scheme configuration, controller-\dword{daq} and controller-\dword{feb} interfaces, bandwidth and rates issues, online and offline \dword{pd}-\dword{tpc} interfaces, zero-suppression techniques, firmware development, accepting and producing triggers, and, in general, will inform possible upgrade paths for the system.

\begin{dunefigure}[ICEBERG TPC model and assembled \dshort{apa}]
 {fig:fig-pds-iceberg-tpc}
 {Solid model of \dword{iceberg} \dword{tpc} (left), and assembled \dword{iceberg} \dword{apa} (right).  Note the two sets of \dword{pd} module mounting rails, which are vertical in this image but horizontal during operation. The centrally-mounted \dword{apa} allows for testing of double-sided readout photon detector modules.}
\includegraphics[angle=0,height=6.5cm]{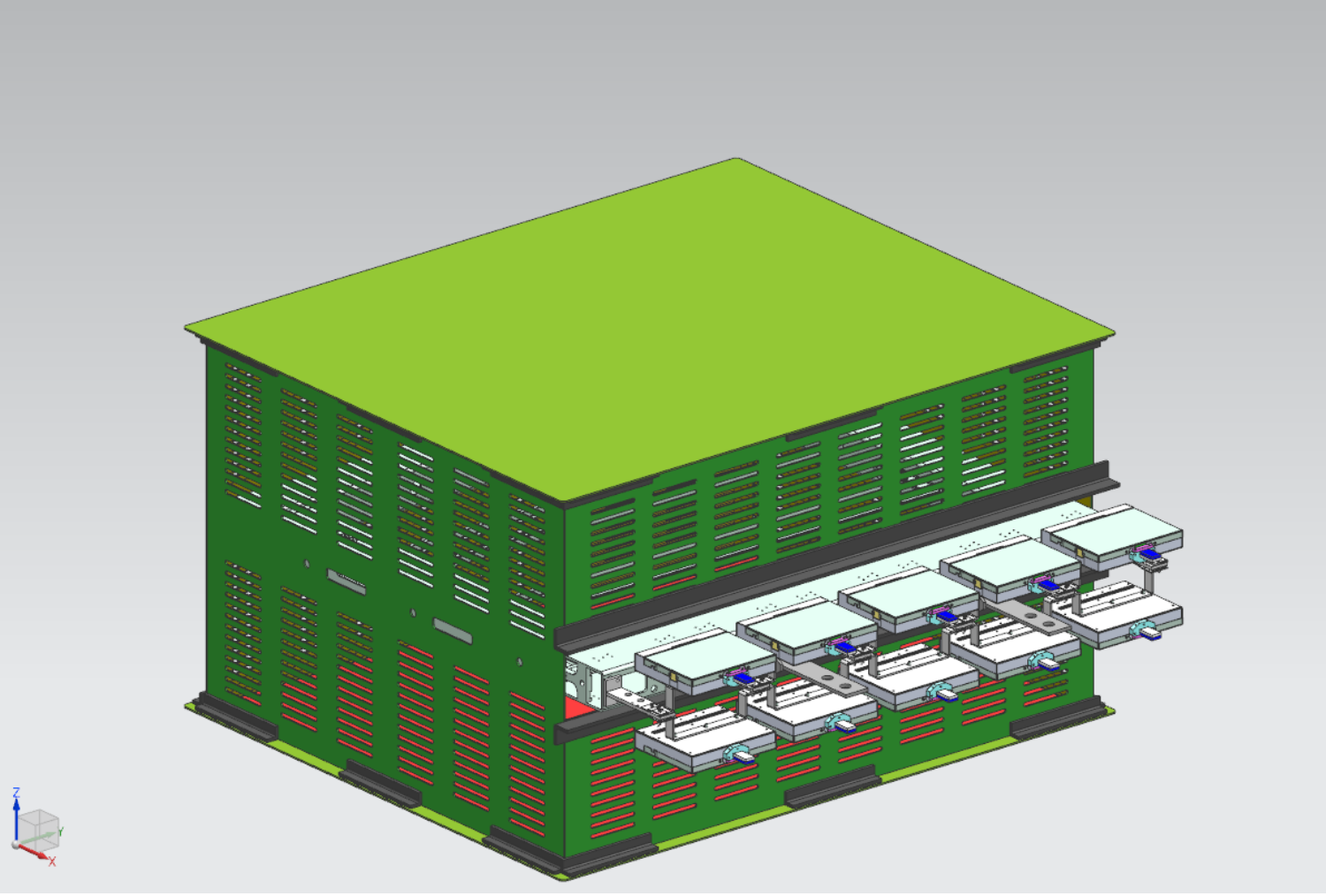}
\includegraphics[angle=0,height=6.5cm]{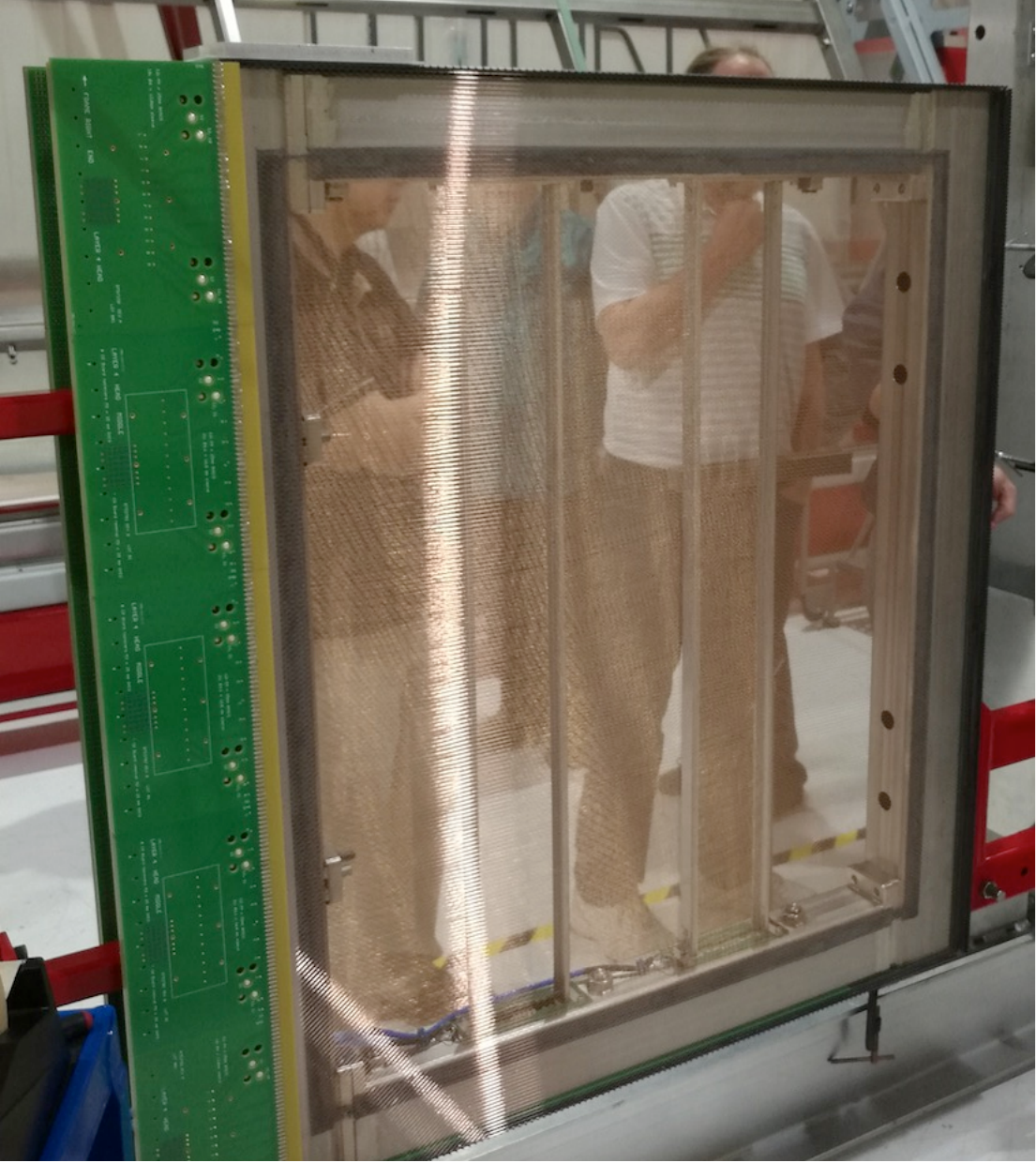}
\end{dunefigure}

\begin{dunefigure}[Single supercell ICEBERG PD module]
 {fig:fig-pds-iceberg-supercell}
 {Software solid model of a single supercell \dword{iceberg} \dword{pd} module (left) and fabricated components during assembly (right).  The connector board (green) in the right photo is mounted to the \dword{apa} frame prior to wire wrapping.}
\includegraphics[angle=0,height=6.5cm]{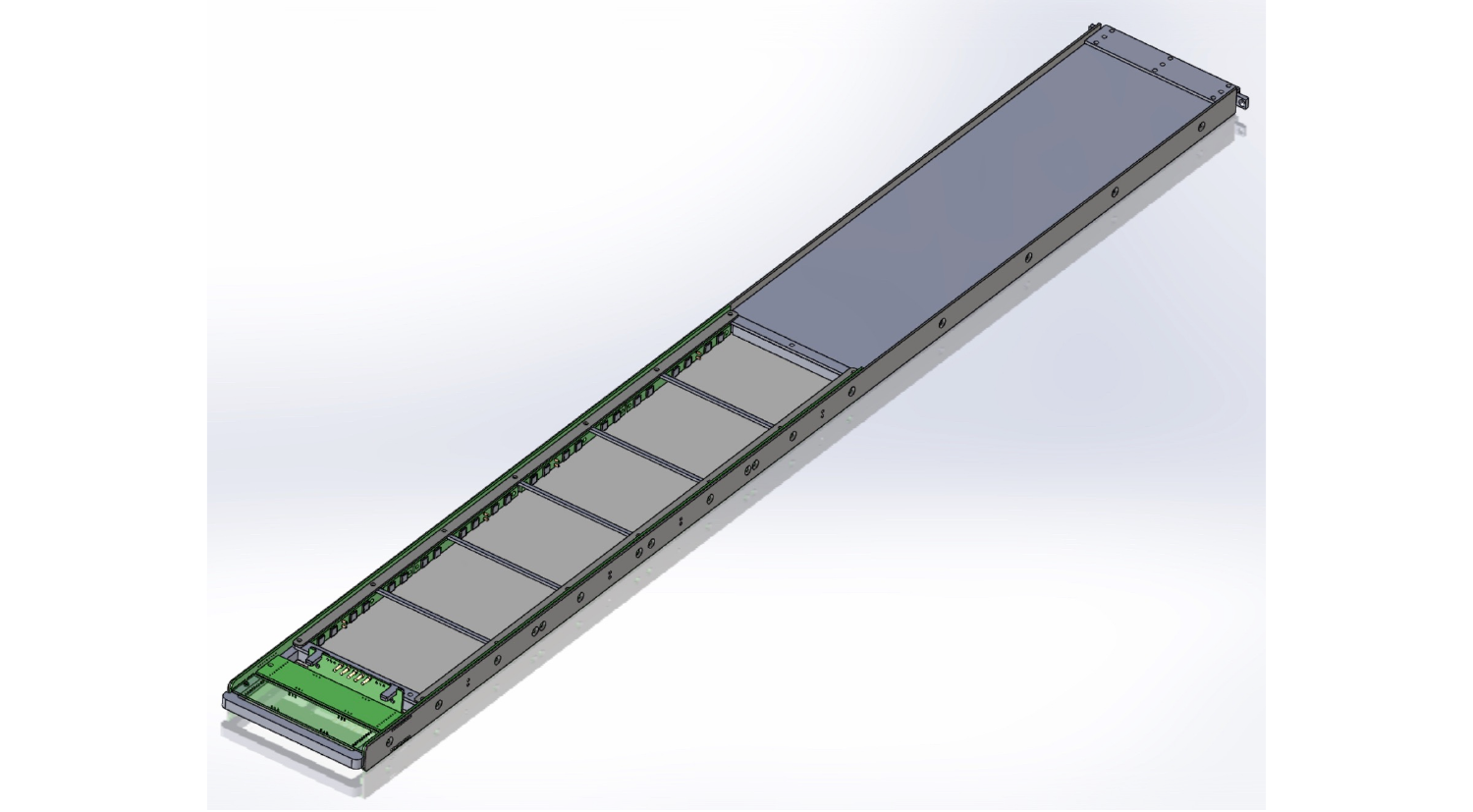}
\includegraphics[angle=0,height=6.5cm]{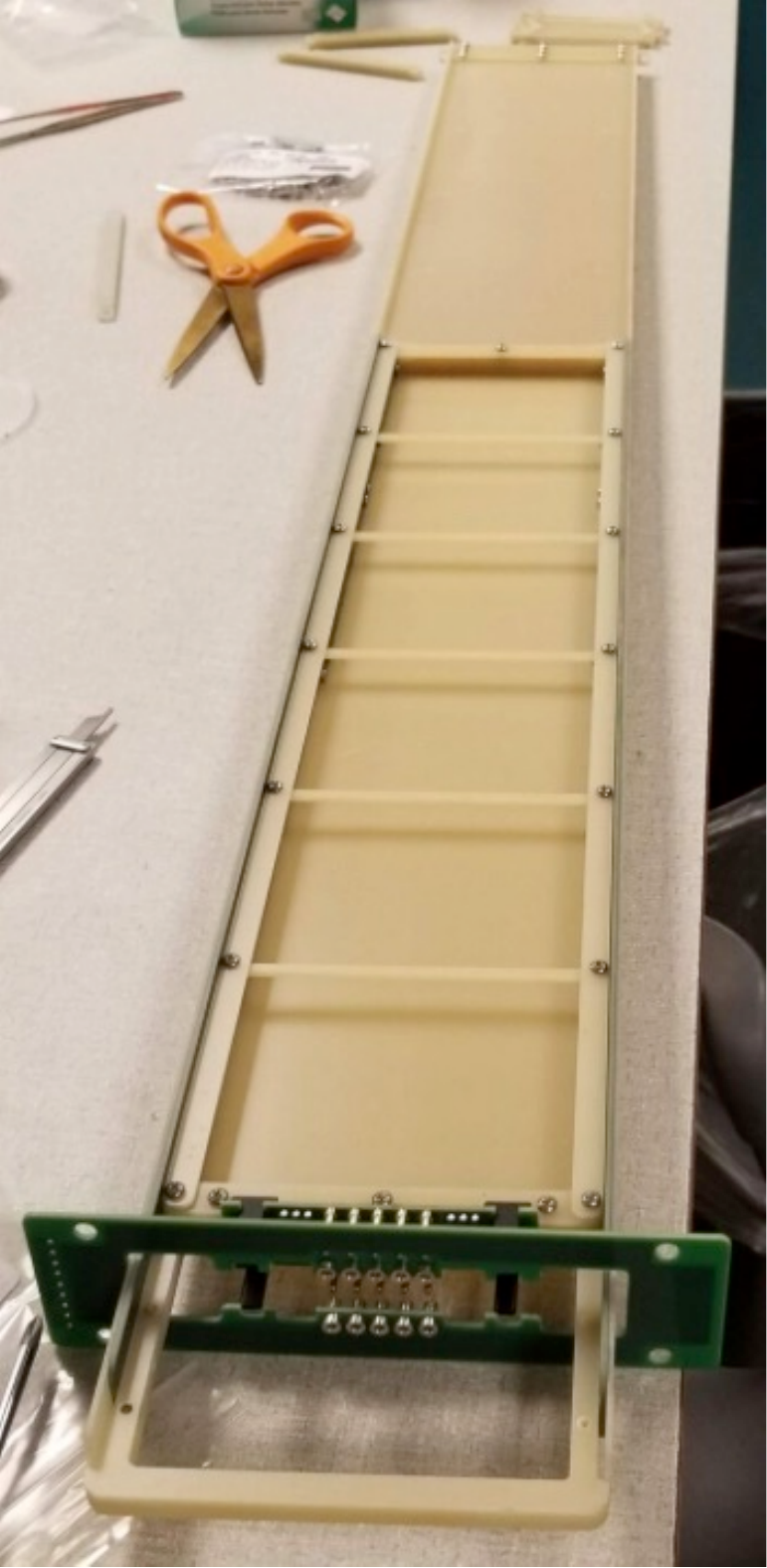}
\end{dunefigure}

Delays in the \dword{iceberg} commissioning schedule unrelated to the \dword{pdsp} system prevented having significant results available in time for this \dword{tdr}. However, test stand data are still expected to provide critical input for the 60\% design review scheduled for May of 2020. Additional runs in 2020 will assist in preparing for the final design review and \dword{pdsp2} module designs.

\subsubsection{SBND}
\label{sec:valid-sbnd}

The baseline \dword{pd} system for the \dword{sbnd} experiment includes three types of detector:  \dword{tpb}-coated cryogenic photomultipliers, an array of dipped light-collector bars similar to those used in \dword{pdsp}, and a small array of \dword{xarapu} modules.  Eight modified \dword{xarapu} modules will provide an opportunity for a long-term system test of a significant number of key components prior to the \dword{sbnd} \dword{pds} \dwords{prr}.  Each \dword{sbnd} \dword{xarapu} will consist of two dichroic filter plates with dimensions \SI{100}{mm} $\times$ \SI{78}{mm} $\times$ \SI{1.5}{mm}
 that are coated with \dword{ptp} at the \dword{unicamp} vacuum deposition facility.  These filters will be produced by Opto Electronics S.A. in Brazil, the leading vendor candidate for the \dword{dune} \dword{pd} modules.  Also, each \dword{sbnd} module will contain an Eljen \dword{wls} plate  
\SI{200}{mm}$\times$\SI{78}{mm}$\times$\SI{4}{mm}. \frfour G-10 frame components will be fabricated at local vendors, representing candidates for eventual \dword{dune} fabrication.  A software solid model of the design is shown in Figure~\ref{fig:fig-SBND-modules}.

The eight \dword{sbnd} \dword{xarapu} modules will be assembled at \dword{unicamp} using the \dword{dune} \dword{pd} consortium assembly plan,
which will provide valuable experience with fabrication of multiple modules at that site.

In the summer of 2019, the \dword{sbnd} collaboration re-opened the question of the composition of their light collection system, eliminating the dipped bar modules in favor of \num{200} additional \dword{xarapu} modules.  These modified modules will use \dword{wls} plates and coated filter plates identical to those proposed for \dword{dune}, and frame components very similar to those in the final \dword{dune} \dword{pd} system.  The modification to the \dword{sbnd} system will provide a larger scale, long-term test of critical \dword{pd} components and will significantly enhance the value of the test as a development run for the \dword{unicamp} facility.

We expect that module assembly for \dword{sbnd} will be complete 
with installation in the detector beginning in spring of 2020.  
Filling with \dword{lar} and operation will occur in summer 2020, and we expect initial results from the \dword{pd} system in fall 2020.  \dword{sbnd} will be the first large-scale operational testing for \dword{xarapu} modules very similar to those to be used in \dword{dune}.
\dword{sbnd} will also use coated reflector foils, which will provide additional valuable information on that \dword{dune} \dword{pds} option.  

While not part of the \dword{dune} project, and not part of the validation schedule for the \dword{pdsp}, \dword{sbnd} results will inform our preparations for the final design review of these components and the fabrication of modules for \dword{pdsp2}.  

\begin{dunefigure}[SBND \dshort{xarapu} modules]
 {fig:fig-SBND-modules}
 {Software solid model of two-cell \dword{xarapu} modules for \dword{sbnd}, exploded (left) and assembled (right).}
\includegraphics[angle=0,width=8.4cm,height=5.5cm]{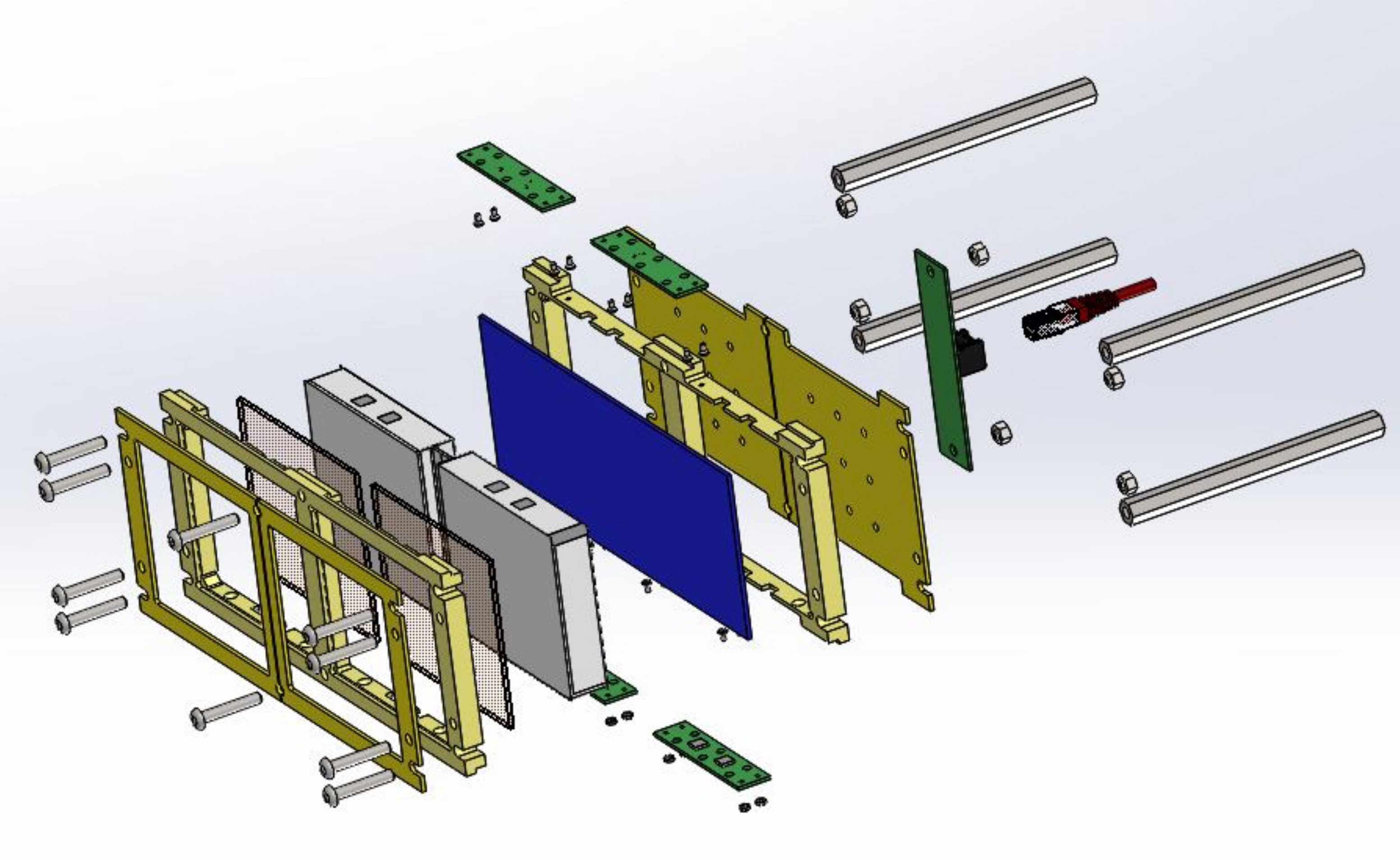}
\includegraphics[angle=0,width=8.4cm,height=5.5cm]{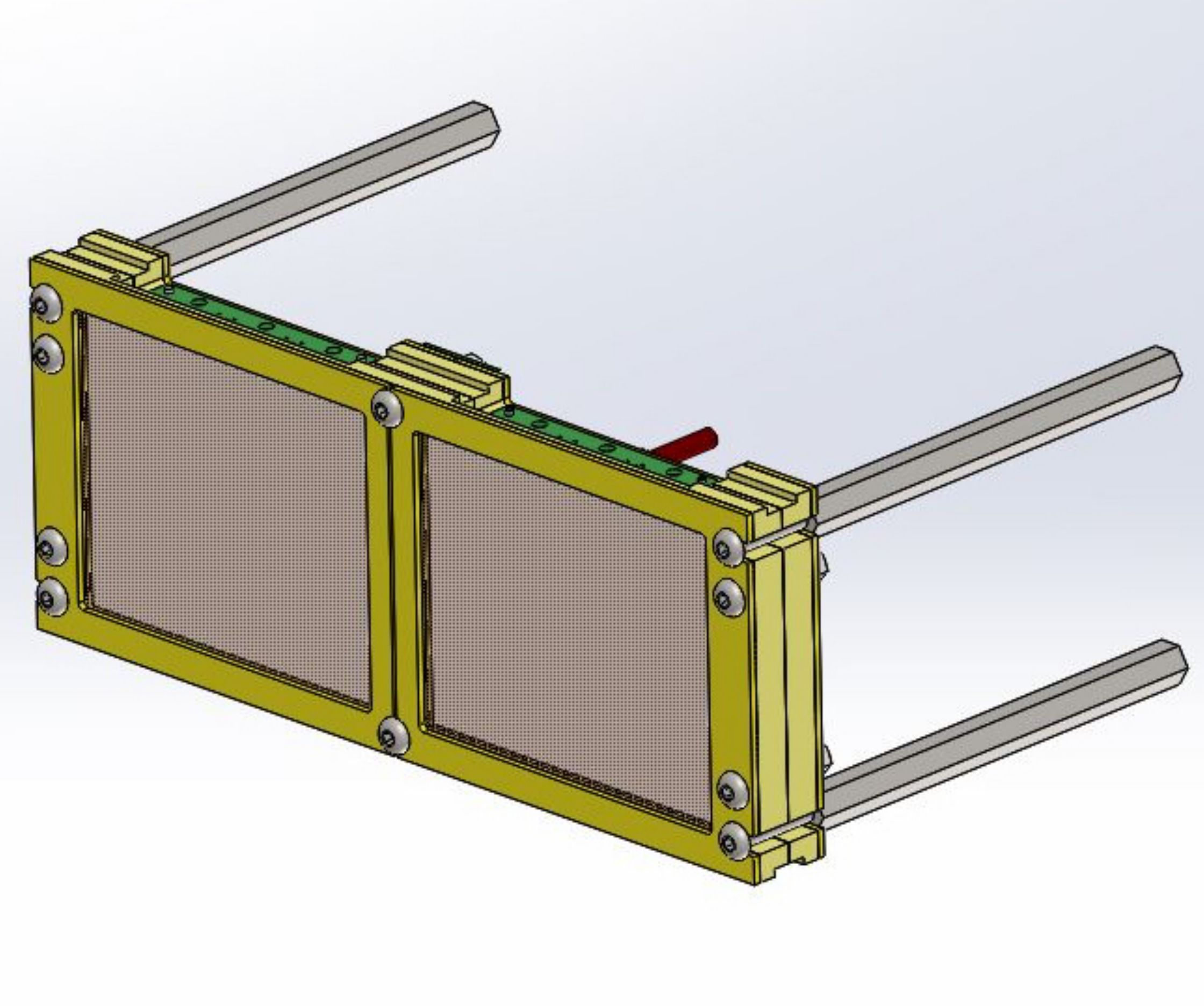}
\end{dunefigure}

\subsubsection{\dword{pdsp2}}
\label{sec:valid-pdune2}

Following completion of the initial run of \dword{pdsp}, a second test run called \dword{pdsp2} is planned in the same cryostat.  This test will serve as a final validation of all pre-production \dword{spmod} detector designs, verifying their performance and ensuring they perform in concert with no interference.  We intend to replace three complete \dword{apa} assemblies with pre-production modules to allow testing \num{40} final-design \dword{pd} modules.  

\dword {pdsp2} will allow for the first end-to-end test of the final \dword{pds}, with significantly redesigned elements:
\begin{itemize}
    \item full-size \dword{xarapu} modules read out in conjunction with a \dword{tpc};
    \item \num{48}-channel photosensor active ganging;
    \item final design electrical connectors for \dword{pd} modules mounted inside full-scale \dword{apa}s;
    \item pre-installed cable harnesses inside \dword{apa}s including final module supports;
    \item readout of full-scale \dword{xarapu} modules using modified \dword{mu2e} electronics, including integrating \dword{tpc} and \dword{pd} event matching into the \dword{daq} system.
\end{itemize}

Two candidate photosensors will be tested (\num{20} modules built with each sensor type), and the experience gained while fabricating \dword{pdsp2} will be an important factor for selecting 
between the candidate sensors or deciding to incorporate both in the \dword{pds} final design.  All other components will be final design components, so at least half of the \dword{pd} modules in \dword{pdsp2} will represent the ``Module 0'' level of design. 

While all of these elements will have been tested previously individually and/or at smaller scale, \dword{pdsp2} will represent the final pre-production testing of all the final design components as an integrated system.

The schedule for \dword{pdsp2} calls for \dword{pd} modules to be installed into \dword{apa} modules at \dword{cern} at the end of summer 2021. However, some components, including module support rails, electrical connectors, and cable harness components, must be mounted inside \dword{apa} frames prior to wire wrapping and so must be available by mid-2020. 
Re-filling of the \dword{pdsp} cryostat will begin in the winter of 2022, with operations beginning in late 2023.  Operation of \dword{pdsp2} will continue for at least one year.  This schedule allows for initial operation of the complete system prior to the \dword{pd} \dword{prr} and the beginning of mass-production of \dword{sp} modules in summer 2022, although some components (including dichroic filter plates and photosensors) will have begun procurement by that time.  These components are physically smaller and more amenable to testing in smaller cryostats, reducing the exposure due to this delay.  These scheduling issues will be addressed in more detail in \ref{sec:fdsp-pd-org-cs}.

If the decision has been taken to proceed with our performance enhancement alternates (xenon doping or \dword{cpa}-mounted reflector foils, see Section\ref{sec:fdsp-pd-enh}), they will be tested in \dword{pdsp2} as well.

\subsubsection{Long Term Cryogenic Aging}
\label{sec:valid-longtermaging}

It is difficult to accelerate aging effects due to long-term immersion of components in cryogenic liquids.  While mechanical aging due to thermal expansion/contraction can be readily accelerated by cycling the components to be tested through multiple cycles, long term aging not related to rapid thermal stresses are not amenable to easy acceleration.  An example of such a process might be dissolution of \dword{ptp} coatings over time.

Mitigation of these risks involves some level of long-term exposure to liquid cryogen with extrapolation to the \dword{dune} timescales.  Several such tests are planned for the \dword{pds}:

\begin{itemize}

    \item \dword{pdsp} will provide a long-term test of two \dword{sarapu} modules for a period of up to one year at the time of draining in the Winter of 2020.  The system will be continuously monitored for gain and response of the detectors, and will provide information regarding aging of \frfour G-10 structures, photosensors, and coated filter plates.  Other \dword{pdsp} detectors (Double-shift bar designs) will give some indication of aging of similar \dword{wls} plates to those used in the \dword{xarapu} modules.

    \item \dword{sbnd} will provide a multi-year test of many \dword{xarapu} components, such as coated filter plates, \dword{wls} plates, and photosensors.  \dword{sbnd} will run for at least three years, starting in late 2020.  As part of a running experiment, the system will be continuously monitored for gain and response of the detectors and will provide information regarding aging of \frfour G-10 structures, photosensors, and coated filter plates as well as \dword{wls} plates to be used in the \dword{xarapu} modules.
    
    In addition, TPB-coated reflector foils will be tested in \dword{sbnd}. While coated reflector foils are not part of the baseline \dword{pds}, this will provide validation for the concept we currently present as an option (Section~\ref{sec:fdsp-pd-enh-cathode}).
    
    \item \dword{pdsp2} will provide a long term test of full-scale \dword{xarapu} modules in the final \dword{dune} configuration. While this test will begin shortly before \dword{dune} \dword{pd} module production fabrication, it will provide long-term validation of all \dword{xarapu} components during module production prior to integration into the \dword{apa} frames, allowing for possible insights and improvements into the \dword{xarapu} design.

\end{itemize}

\subsection{Materials Selection, Testing and Validation}

\subsubsection{\dword{pd} Module Mechanical Frame}

The \dword{apa} mechanical frame components are fabricated from \frfour G-10 (Garolite\textregistered), a glass-epoxy laminate commonly used in printed circuit boards and other mechanical applications where an electrically insulating component with low thermal expansion coefficient is required.  G-10 is widely used in cryogenic applications, including most of the other \dword{dune} subsystems (See \citedocdb{10452} for an extensive discussion in the context of the \dword{hv} system cathode planes). 
\frfour has been certified at the \dword{fnal} materials test stand as an acceptable material for use in \dword{dune} from the standpoint of \dword{lar} contamination.

Thermal contraction of \frfour is similar to that of stainless steel, simplifying design of the module interface with the \dword{apa} frame.
It allows us to use long printed circuit boards for routing photosensor electrical signals along the detector sides without incurring thermal expansion issues.  As an excellent insulator, it simplifies electrically isolating the \dword{pds} from the \dword{apa} frame, as required by the \dword{dune} grounding scheme.
However, selecting \frfour as our main module structural material comes at the cost of some additional difficulty machining components.

All fasteners used in the \dword{pd} are stainless steel alloy 304, widely used in cryogenic applications.  This alloy has also been certified at the \dword{fnal} materials test stand as an acceptable material for use in \dword{dune} from the standpoint of \dword{lar} contamination.

\subsubsection{\dword{pd} Module-\dword{apa} Frame Mechanical Support Structure}

All \dword{pd} mechanical supports (including rails, brackets and fasteners are to be manufactured from stainless steel alloy 304.  This has the benefit of matching thermal contraction coefficients with the \dword{apa} frame and being approved for use in \dword{lar} by the materials test stand.

\subsubsection{Dichroic Filter/Filter coating}

The dichroic filters used in \dword{xarapu} consist of a fused silica substrate, coated on one face to provide the required dichroic properties and on the opposite face with a thin evaporated layer of \dword{ptp}.
Fused silica was selected in part due to its excellent low-temperature properties.  It is widely used as an optical window in low temperature applications, due to its stability and low coefficient of thermal contraction; fused silica dichroic filters have performed well in many \dword{sarapu} validation tests.

Stability of the \dword{ptp} coating is of greater concern.  Initial validation of the \dword{sarapu} in \dword{tallbo}, \dword{pdsp}, and in repeated cryogenic immersion tests at \dword{fnal} and other facilities has demonstrated that while it is possible to generate highly-reliable \dword{ptp} coatings on fused silica substrates, careful surface preparation and deposition procedures are required to prevent failure of the coating.  
Dissolution of similar wavelength-shifting coatings into \dword{lar} has been reported but in a technology-dependent fashion~\cite{Asaadi:2018ixs}, so continued investigation of the design specific to \dword{dune} is necessary to confirm robustness.

A test stand has been developed at Syracuse University to investigate the long-term stability of \dword{xarapu} optical coatings in \dword{lar} that will
subject coated materials to a continuous flow. This will stress the adhesion of the coating to the plates to simulate the convective flow of \dword{lar} in the \dword{spmod}.
 
The test stand consists of a \SI{74}{L} \dword{lar} cryostat with a frame suspending coated filters beneath an impeller driving a continuous flow of \dword{lar} across them. Filters will be inspected monthly for degradation in their opacity, transparency, and wavelength-shifting response. A dark box containing a visible-light and near-UV scanning bed will measure wavelength shifting performance of the tested elements before and after suspension within the argon flow.

Testing began in late 2019, when coated filter plates become available, and is expected to run through late 2022. 

\begin{dunefigure}[\dshort{pds} coating test stand]
 {fig:pds-CoatingTestStand}
 {Two components of the \dword{pd} coating test stand. (1) VUV monochromator (foreground) and 2-axis scanning chamber (background) 
 currently undergoing recommissioning (left); and (2) solid model of \SI{74}{L} \dword{lar} cryostat for quality control studies and future detector development (right).}
\includegraphics[angle=0,height=6.cm]{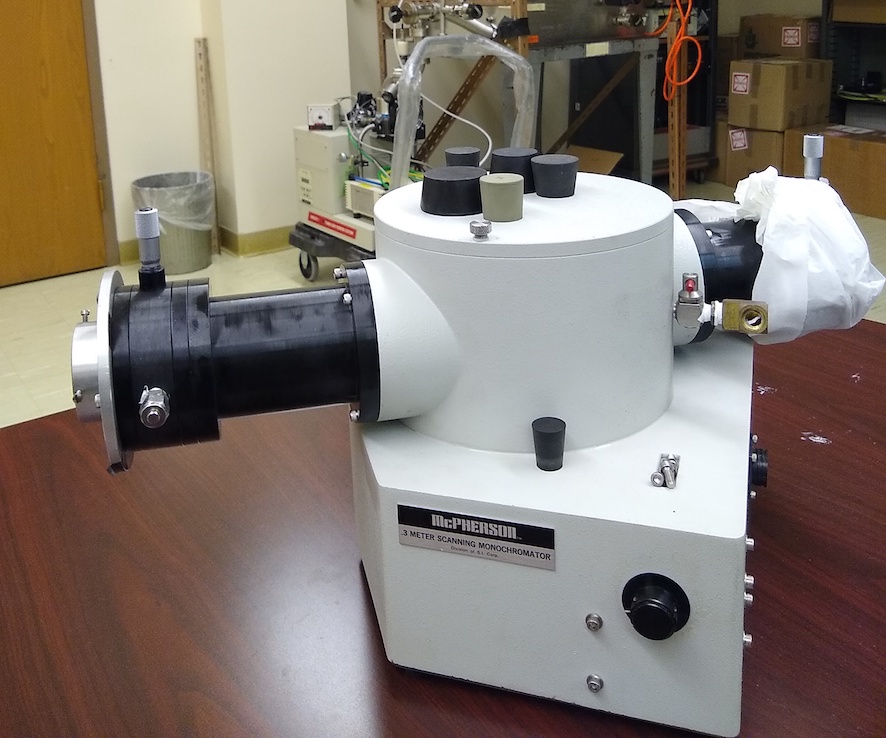}
\hspace{0.02\textwidth}
\includegraphics[angle=0,height=6.cm,trim={2.2in 0.75 2.2in 0.75},clip]{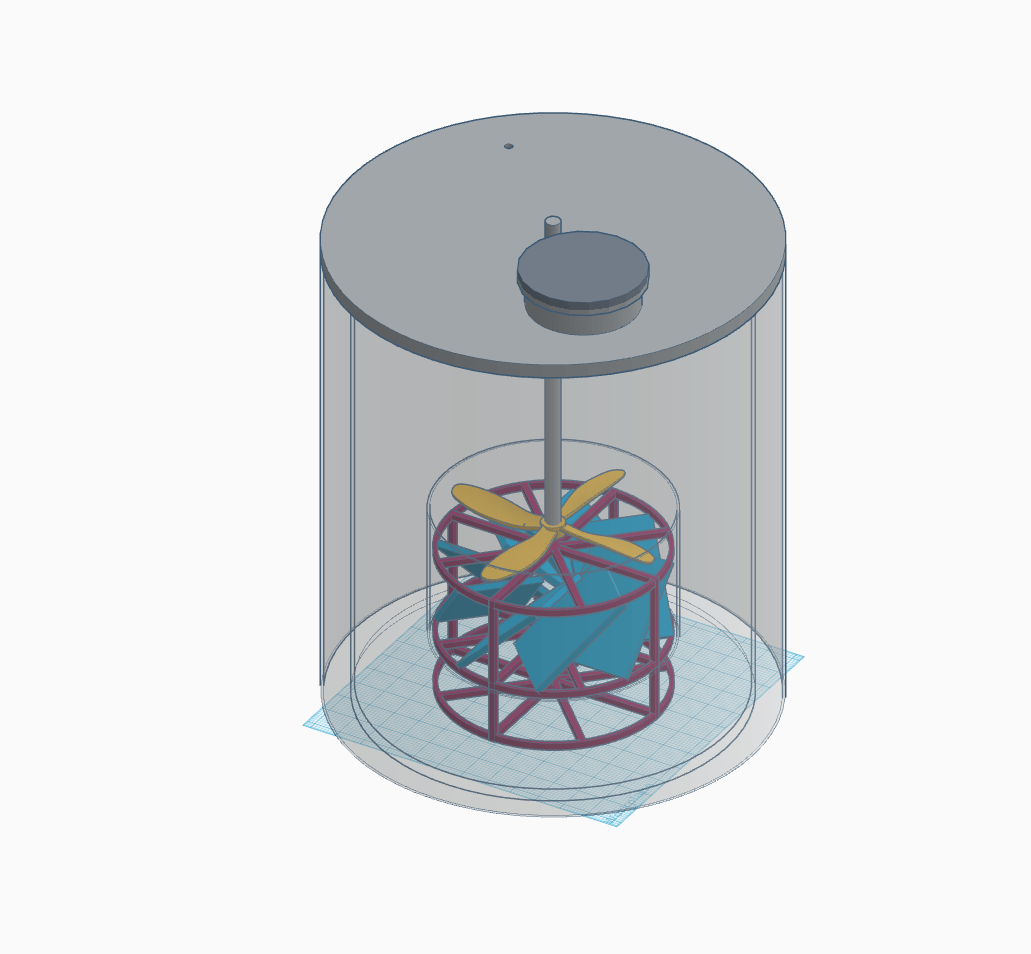}
\end{dunefigure}

\subsubsection{\dword{wls} plates}

The \dword{xarapu} wavelength shifting plates are fabricated by the same vendor as the light guide bars utilized in the double-shift \dword{pdsp} modules.  The plates are made with the same transparent matrix material, but have a different wavelength shifting dopant chosen to provide a better match to the spectral sensitivity of the \dword{pd} \dword{sipm} (around \SI{430}{nm}). It also has an emission spectrum very similar to \dword{tpb}, used in the \dword{sarapu}, which ensures the same performance of the dichroic filter and of the reflective coatings.

While it is possible that the cryogenic properties of this modified \dword{wls} material may be altered by the change in doping agent, it is expected that the tests done for the double-shift bar prototypes are a valuable guide for their expected performance.  As part of the design verification, samples of these bars were manually thermocycled to verify they didn't craze. In addition, we used a \dword{lar} test stand 
with an alpha source behind a small sample of the \dword{wls} plate to scan the attenuation length of a short sample. Finally, we built a darkbox with a $\sim\,$\SI{420}{nm} \dword{led} scanning down the length of a full bar to verify the attenuation length and to compare the results to the \dword{lar} data. (This formed the basis of the threshold requirements on in-air attenuation length measurement for the \dword{pdsp} batch.) 
This same sequence will be undertaken for the light guides selected for \dword{xarapu}. 

As with most other components, it is not possible to simulate long-term exposure to \dword{lar} of a length similar to that expected in \dword{dune} operation, but we will substitute continuous long term exposure by running samples through repeated thermal cycling to maximize thermal stresses in the material.  Finally, samples of the \dword{wls} plates will be certified for use in \dword{dune} in the materials test stand at \dword{fnal}.

\subsection{Calibration and Monitoring}
\label{sec:fdsp-pd-validation-candm}

All major components of the \dword{spmod} \dword{pds} calibration and monitoring system have been designed, fabricated, tested, and operated in \dword{pdsp}. Figure~\ref{fig:pds_calmon_hw_photo} shows the hardware components of the system.
Although at a longer wavelength (\SIrange{245}{280}{nm}) than \dword{lar} scintillation light (\SI{127}{nm}), the UV light from the calibration system exercises the full chain of measurement steps initiated by a physics event in the \dword{detmodule}, starting from the wavelength conversion, photon capture in the \dword{sarapu}, photon detection, and the \dword{fe} electronics readout.

A substantial \dword{pdsp} data set has been collected and the data analysis is underway. 
Goals of the analysis are to verify that the \dword{cpa} includes an optimal distribution of light diffusers for the \dword{spmod};
to demonstrate capability of the system evaluate gain and timing resolution; to perform relative comparisons of photon channels;
and to characterize and monitor stability of the \dword{pds} over the duration of \dword{pdsp}. Here we present preliminary results that demonstrate the timing performance of the system, the stability of the two types of \dword{sipm}, and the photon detection rate over several months.


\begin{dunefigure}[\dshort{pdsp} UV calibration and monitoring system]
 {fig:pds_calmon_hw_photo}
 {The photographs show the hardware components of the \dword{pdsp} calibration and monitoring system.}
 \includegraphics[angle=0, height=9cm]{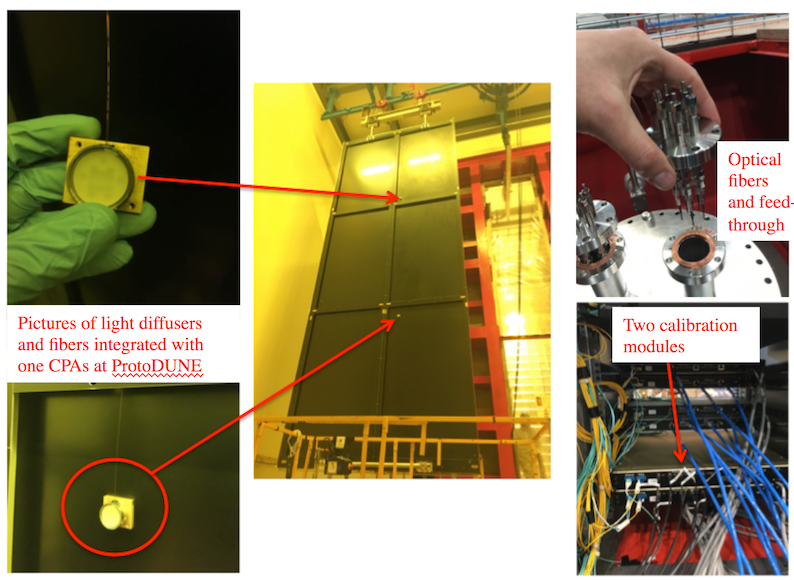}
\end{dunefigure}

Figure~\ref{fig:pds_calmon_timing} (left) shows a typical double waveform recorded by an \dword{pdsp} \dword{ssp} module as a response to calibration system
light pulses illuminating an \dword{sarapu} channel; the figure on the right demonstrates that the calibration system has the precision and stability to meet the system requirements.

Figure \ref{fig:pds-calmon-charge-avalanche} shows the \dword{sipm} gain (charge per \phel-induced avalanche) extracted from the calibration data normalized to the average gain during the period. The left figure shows the results over a period of three months for the \dwords{mppc} mounted on the \dword{sarapu} modules; the right figure shows the results over a period of six months for the SensL \dwords{sipm} that are mounted on a set of double-shift and dip-coated light collector bars. The colors correspond to different readout channels for the left figure and to the average of the sensors in \dword{pd} modules for the right figure.
All sensors are stable at the level of a few percent, with no significant systematic decline.

Figure \ref{fig:pds-calmon-photons-apa6} shows the measured signal (average number of photons, normalized to the average signal over the three-month period) from the double-shift and dip-coated light collector bars in \dword{apa}~6 that are read out with \dword{mppc} \dwords{sipm} (left), and those in \dword{apa}~4 that are read out with SensL \dwords{sipm} (right), in response to the calibration flashes. The colors correspond to the average of the sensors in \dword{pd} modules. The measured signal is sensitive to stability in the intensity of the calibration system light and the response of the light collectors (including effects such as changes in wavelength shifter properties and \dword{sipm} response). The ratio is stable at the few percent level.

These results verify operation and performance of both the \dword{pds} and the UV-light calibration system. This monitoring will continue for the duration of the \dword{pdsp} operation.

\begin{dunefigure}[\dshort{pdsp} \dshort{sarapu} response to UV calibration and monitoring system]
 {fig:pds_calmon_timing}
 {Double waveforms recorded by \dword{pdsp} \dword{ssp} as a response to calibration system light pulses collected by an \dword{sarapu} channel (left). Distribution of measured times of the first light pulse in the two-pulse waveform for 1000 pulse pairs (right).}
  \includegraphics[height=5.8cm,width=0.45\linewidth]{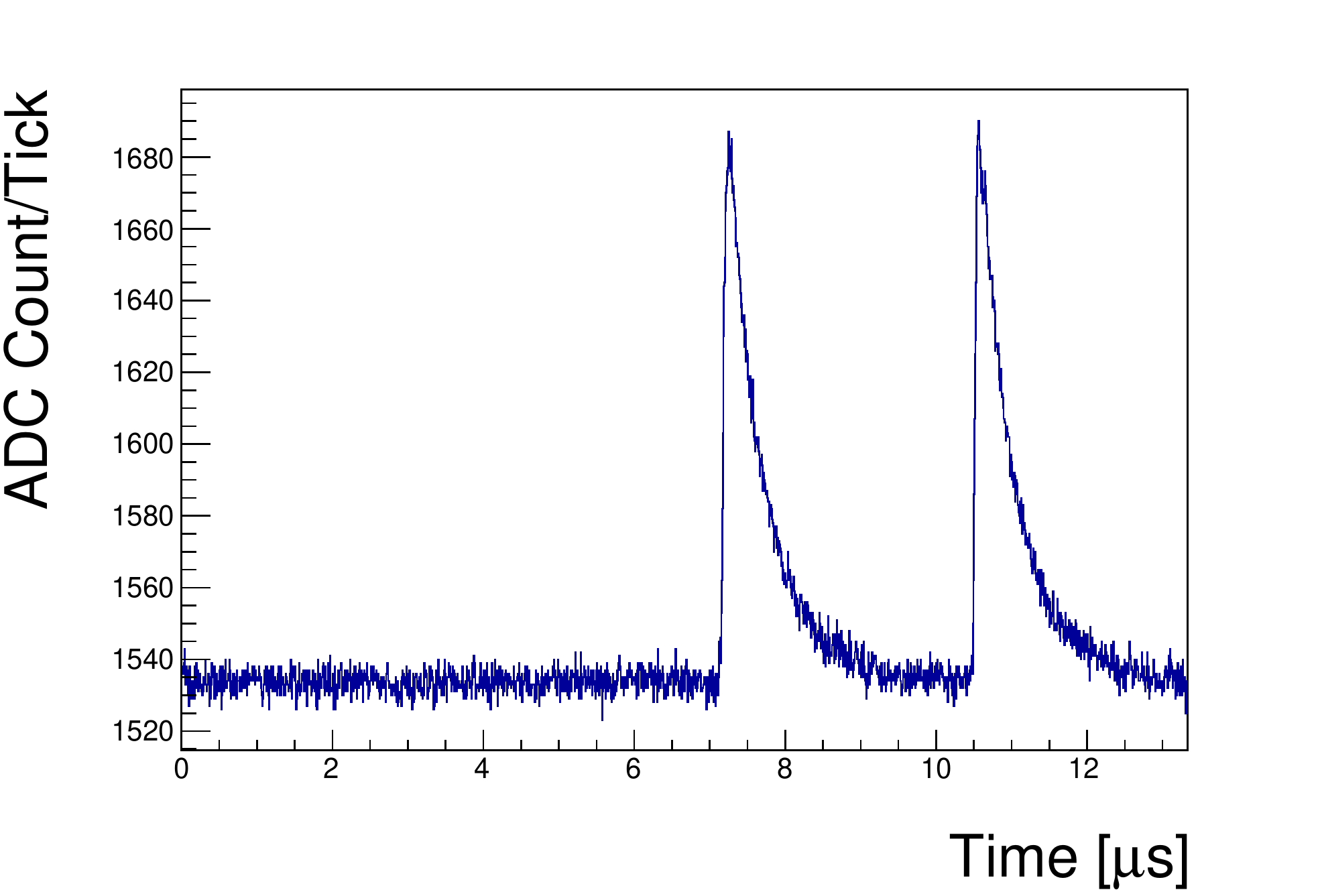}
 \includegraphics[height=6cm,width=0.45\textwidth]{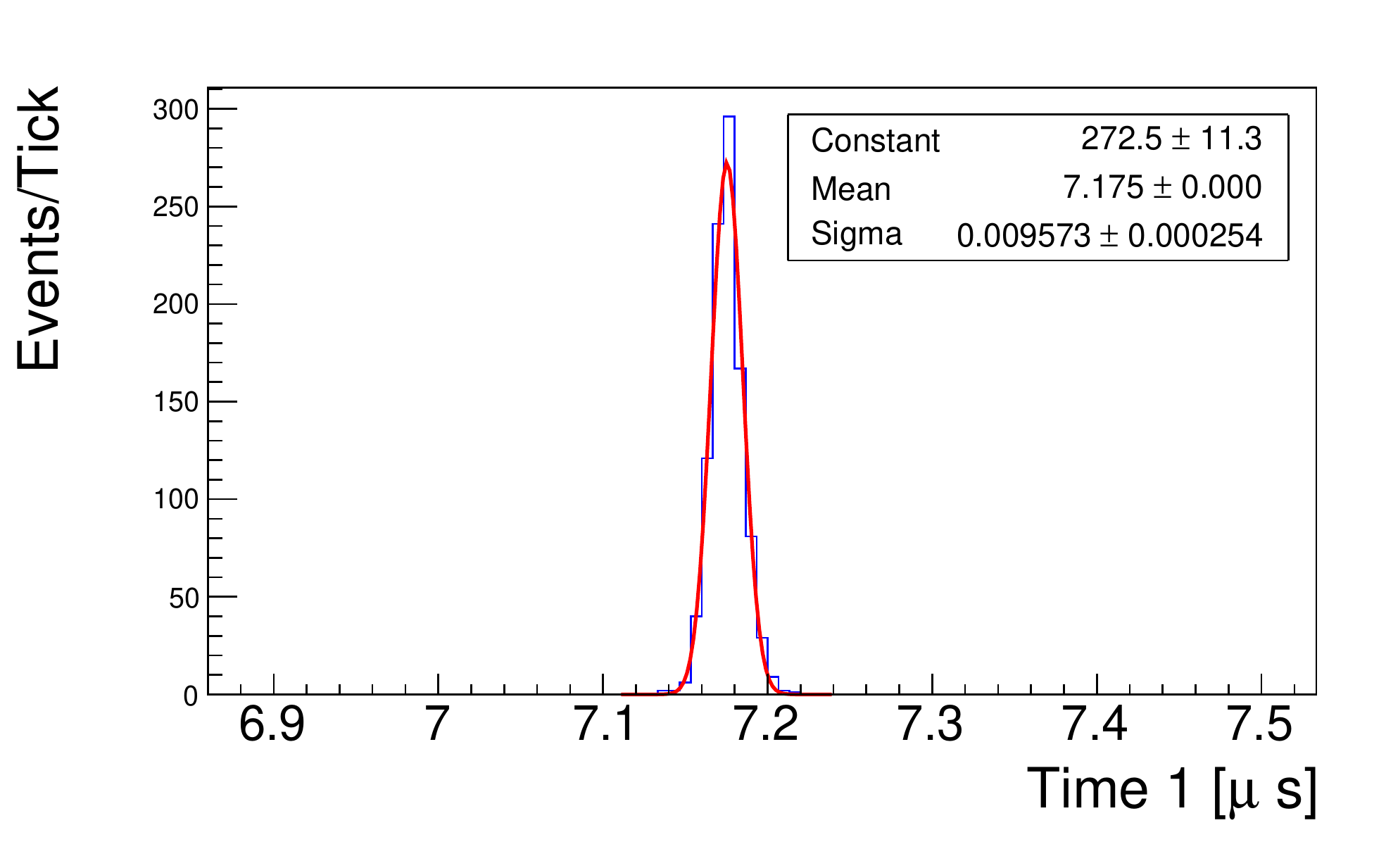}  \end{dunefigure}

\begin{dunefigure}[\dshort{pdsp} gain monitoring with UV calibration and monitoring system]
 {fig:pds-calmon-charge-avalanche}
 {Normalized gain measurements using the calibration system light pulses: \dword{mppc} \dwords{sipm} on the \dword{sarapu} modules (left); SensL \dwords{sipm} on the dip-coated and double-shift bars in \dword{apa}~3 (right). This demonstrates the stability of the gain for both types of device operating in \dword{lar}.}
  \includegraphics[height=5.5cm,width=0.48\textwidth]{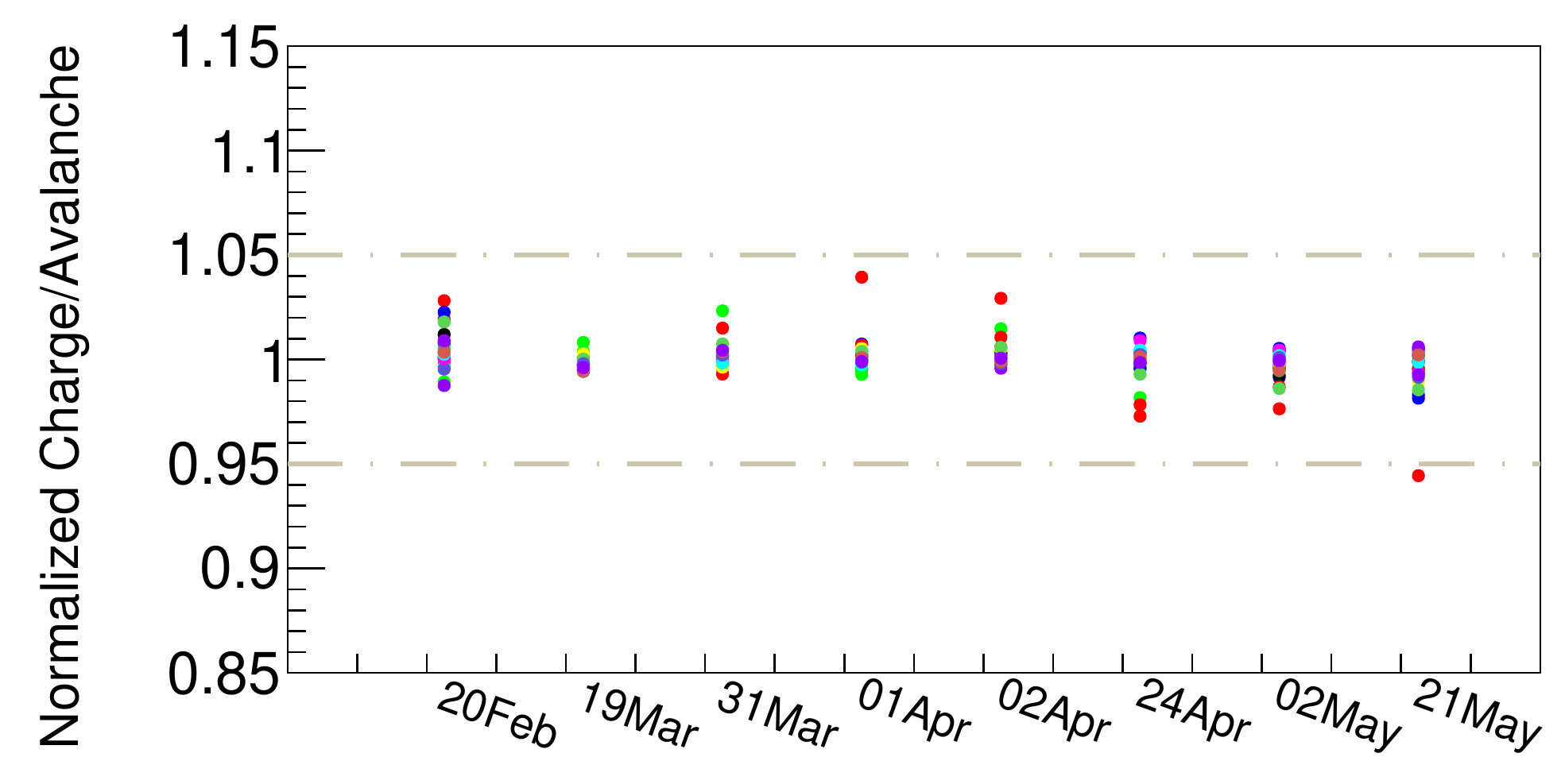}
  \includegraphics[height=5.5cm,width=0.48\textwidth]{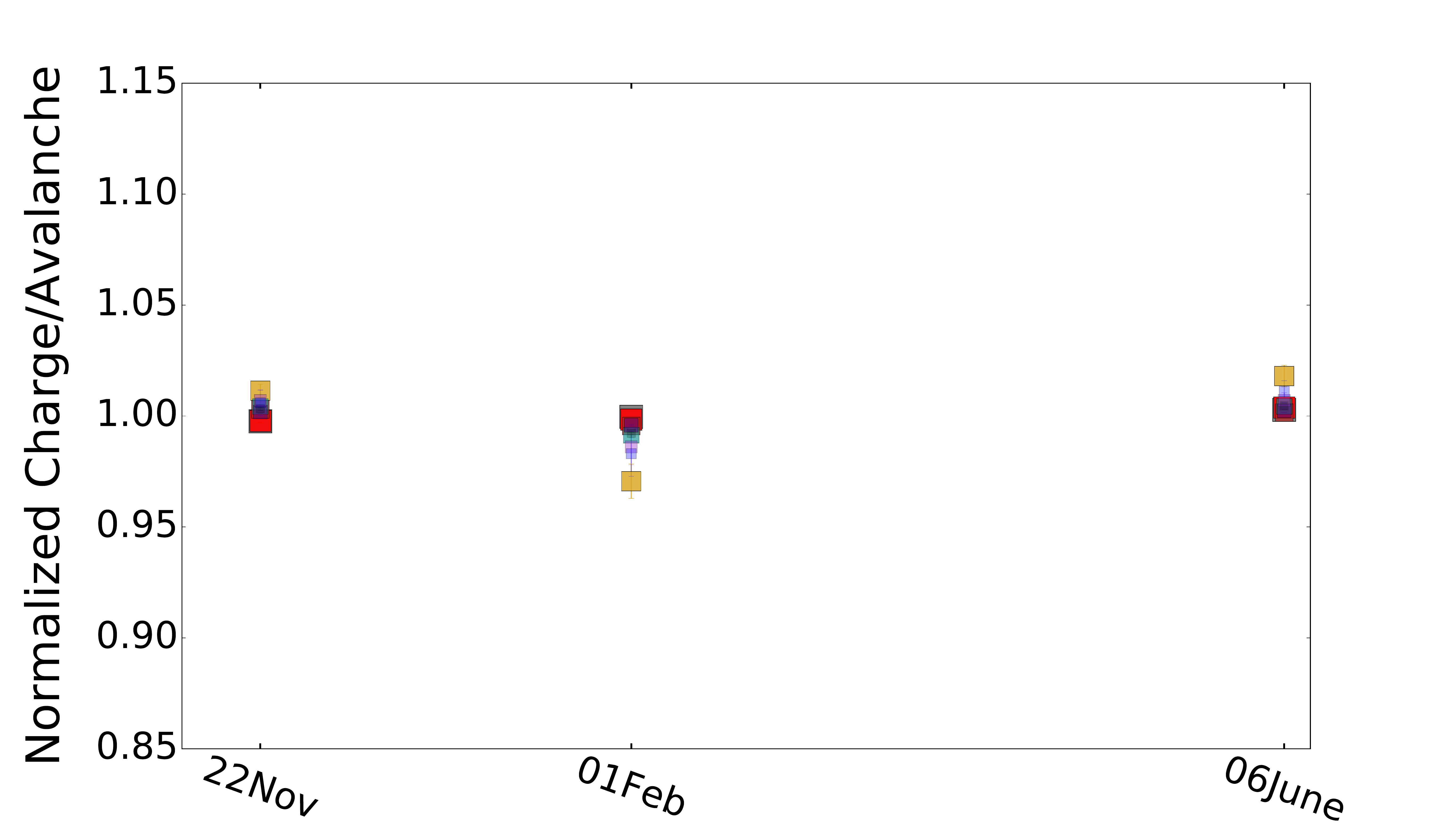}
\end{dunefigure}

\begin{dunefigure}[\dshort{pdsp} UV calibration and monitoring system stability]
 {fig:pds-calmon-photons-apa6}
 {Measurement of the signal from the calibration system using the dip-coated and double-shift bars in \dword{apa}~6 that have \dword{mppc} \dwords{sipm} (left) and in \dword{apa}~4 that have SensL \dwords{sipm}. The signal is normalized to the average response over the entire period.
 }
  \includegraphics[height=5.5cm,width=0.48\textwidth]{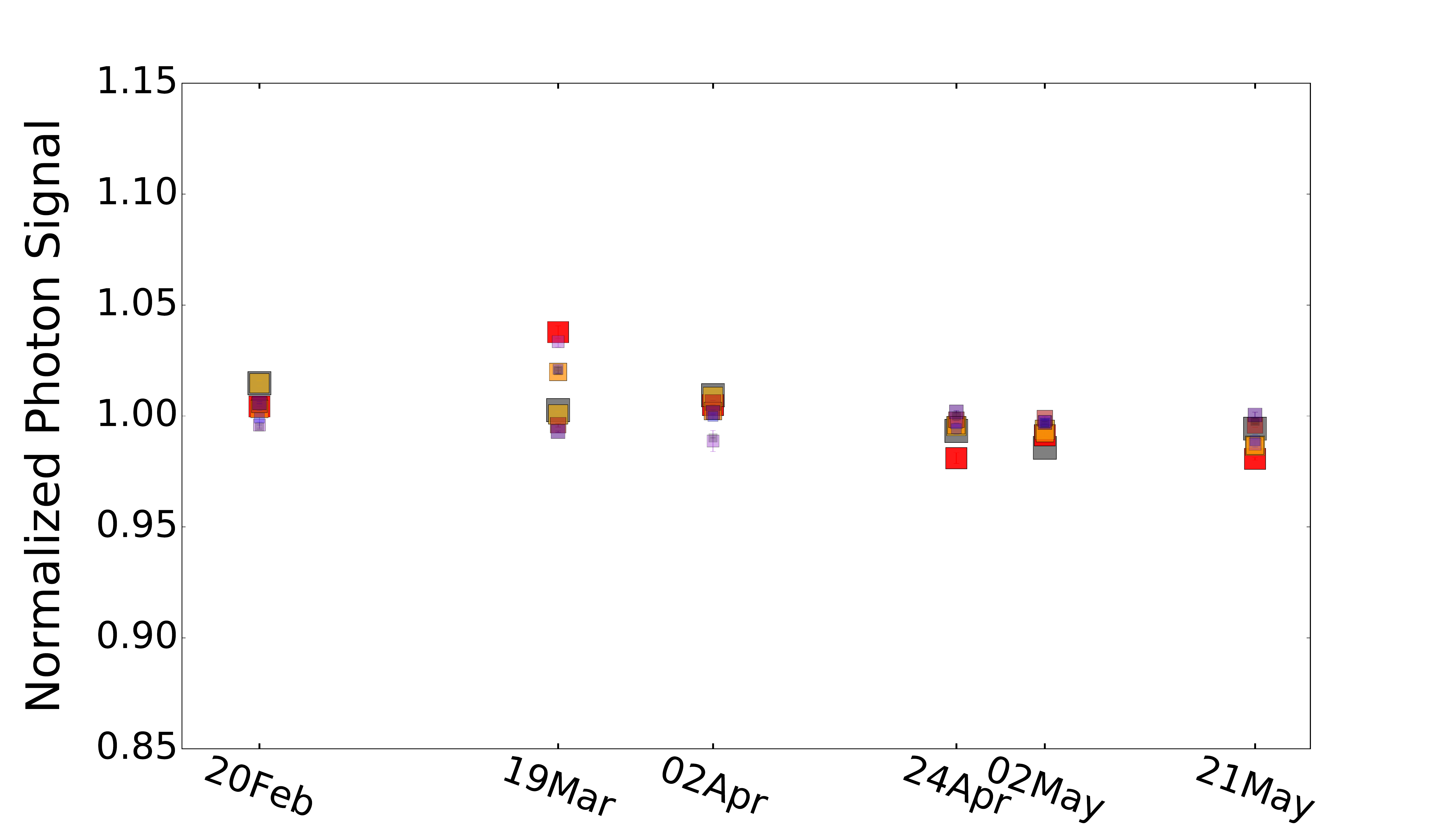}
  \includegraphics[height=5.5cm,width=0.48\textwidth]{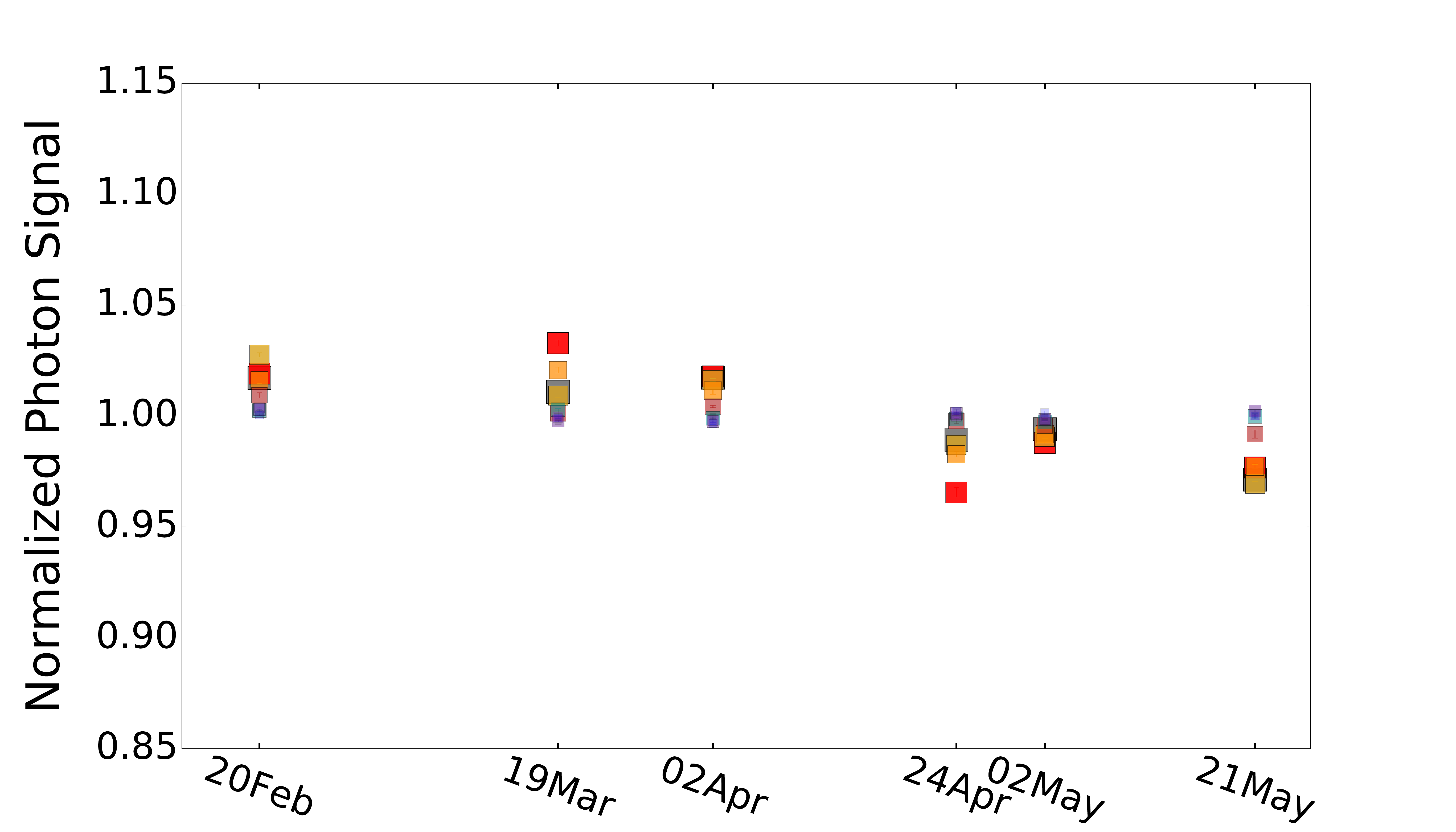}

\end{dunefigure}
\section{Production and Assembly}
\label{sec:fdsp-pd-prod-assy}

The \single \dword{pds} consortium is a geographically diverse group of institutions, collaborating across three continents to fabricate a single integrated system.  As such, careful planning and control of component fabrication, assembly and testing must be maintained.
This section describes the planning for fabrication, assembly, and testing, focusing primarily on the \dword{pd} light collector modules, photosensors and photosensor modules, and electronics. It also covers planning for  calibration and monitoring.
 
This section first describes the fabrication procedure for each of the major \dword{pd} system components.  It concludes (in Section~\ref{sec:fdsp-pd-assy-Assby-plan}), by outlining the work flow and responsible institutions for the assembly plan.

\subsection{Light Collector Module Component Fabrication}

The \dword{pd} light collector modules were designed with ease of fabrication in mind.  The module components can be fabricated and \dword{qc} tested at physically separated facilities, later to be collected and assembled at one or more assembly facilities. 

Each \dword{spmod} detector requires the fabrication of \num{1500} photon detector modules, a significant production effort.  Many of the components for these modules are commercially available or relatively easy to fabricate, but some, such as photosensors (\num{288000} required), filter plates (\num{48000} required), and \dword{wls} plates (\num{6000} required), are specialized items requiring close interfaces with industrial partners.  These issues will be discussed in the relevant sub-sections below.   

\subsubsection{Dichroic Filter and Reflector Foils Fabrication and Coating}

The baseline design for dichroic filters is 
a fused silica plate,  \SI{10}{cm} $\times$ \SI{7.8}{cm} $\times$ \SI{0.2}{cm}, commercially coated (as described in Section~\ref{sec:fdsp-pd-lc}) to provide the dichroic properties of the filter.  
The filter plates will be purchased from a commercial vendor (certified by the vendor for performance), and the performance of a representative sample will be tested at a collaboration institution as part of our \dword{qc} program.  

Prior to coating, the filters are cleaned using the procedure outlined in Section~\ref{sssec:cleaning}.
For \dword{pd} production, the evaporation process will be performed
in Brazil, where a large vacuum evaporator with an internal diameter of one meter is now available. The conversion efficiency of the film deposited on the filters will be measured for a representative sample, with a dedicated set-up that will use the \SI{127}{nm} light produced by a \dword{vuv} monochromator.

Vikuiti reflector foils required for the rear reflector surface of single-sided \dword{xarapu} supercells and the sides of all modules will be purchased and laser-cut to the form factor required by a vendor.  Mechanical and optical \dword{qc} tests will be performed on a representative sample upon receipt.

Coated filter plates and reflector foils represent one of the more challenging fabrication tasks for the consortium.  A total of \num{48000} filter plates will be required, and fabrication will need to occur at a rate of approximately \num{1200} per month.  Dichroic filters will be purchased as part of the Brazilian effort.  A Brazilian candidate vendor for the filters (Opto Eletronica S.A\footnote{www.opto.com.br}), has been selected for the filter manufacturing.  Preliminary contact has been made, a budgetary estimate received, and initial discussions suggest that they will be capable of meeting our production schedule.  Coating of filter plates will be conducted at \dword{unicamp}.  Prototype studies suggest a coating cycle time of less than two hours per \num{24} filter plates, which meets the needs of the project.

\subsubsection{Wavelength Shifting Plates}

The baseline design for the wavelength-shifting plates are Eljen EJ-286 plates of dimensions \SI{48.7}{cm} $\times$ \SI{9.3}{cm} $\times$ \SI{0.35}{cm}.  The edges of a plate will be simultaneously cut and polished with a diamond-edged cutter to increase internal reflectivity, following a proprietary process developed by Eljen.  Plates will be delivered to the consortium institution responsible for this component, where \dword{qc} testing of a representative sample will be performed.

Eljen, Inc. has been involved in \dword{pd} module development for many years and has proven a reliable partner.  We have a budgetary estimate for the plates, indicating a total production time of approximately \num{18} months for all \num{6000} wavelength-shifting plates required.

\subsubsection{Mechanical components}

The mechanical components of the \dword{pd} module frames are fabricated from
\frfour G-10. This material will mitigate thermal expansion issues (see thermal expansion discussion in Section~\ref{sssec:pds-thermal-load}),
but is abrasive and somewhat difficult and expensive to work with using traditional machining processes.

To mitigate these difficulties, most of the \dwords{pd} frame components were designed so that they can be fabricated using water-jet cutting technology.  In some cases, post-cutting fabrication is required, e.g., tapping of pre-cut holes, or (rarely) drilling and tapping holes into the sides of the components where the water jet could not pre-cut pilot holes.

Water jet cutting of \frfour G-10 components will be conducted in-house at \dword{unicamp} to allow for improved quality and schedule control.  A dedicated water-jet cutting machine is being purchased now, and fabrication processes will be validated during \dword{sbnd} and \dword{pdsp2} fabrication.  Secondary machining operations (hole tapping, etc.) will also be conducted at \dword{unicamp}.  \dword{qc} tests will be conducted on a representative sample of finished components.

\label{sec:fdsp-pd-prod-pc}

\subsection{Photon Detector Module Assembly}

\dword{spmod} \dword{pd} module assembly will occur at a \dword{pd} assembly facility at \dword{unicamp}.  Assembly procedures are described below.  Final assembly planning for \dword{pd} modules is guided by the experience gained during the assembly of \num{60} \dword{pdsp} \dword{pd} modules.

\subsubsection{Incoming Materials Control}

Each \dword{pd} sub-component assembly site will have a quality control supervisor, who will be responsible for overseeing all quality-related activities at that site, maintaining all production records and assembly travelers, and uploading them to the production database as appropriate.   This local supervisor will report directly to the \dword{pd} consortium lead.

All materials for \dword{pd} module assembly will be delivered to the \dword{unicamp} module assembly facility with a \dword{qc} traveler (in the case of materials custom fabricated for \dword{dune}) generated prior to arrival at the assembly site or will have an incoming materials traveler generated immediately upon receipt of the component (for commercial components).  These travelers will be scanned (or uploaded if in electronic format) upon receipt at the assembly facility, and the data stored in the \dword{dune} \dword{qc} database.  Materials will either arrive with a pre-existing \dword{dune} inventory control batch/lot number, or will have one assigned prior to entering the assembly area.  Bar code labels attached to storage containers for all components in the assembly area will facilitate traceability throughout the assembly process.

Immediately upon receipt, all materials will undergo an incoming-materials inspection, including confirmation of key dimensional tolerances as specified on their incoming materials documentation.  
The results of these inspections will be included on the traveler for that batch/lot and entered into the database.

In the case of deviations from specification noted in these inspections, the deviation from nominal will be recorded in an exception section of the traveler, as will the resolution of the discrepancy.

\subsubsection{Assembly Area Requirements}

Assembly will occur in a class \num{100000} or better clean assembly area (see specification SP-PDS-1 in Table \ref{tab:specs:SP-PDS}).  

Photosensitive components (e.g., \dword{tpb}-coated surfaces) are sensitive to near-UV light exposure and will be protected by blue-filtered light in the assembly area (>\SI{400}{nm} or better filters\footnote{For example, GAMTUBE T1510\texttrademark{} from GAM Products, Inc., \url{http://www.gamonline.com/catalog/gamtube/index.php}.}); it has been determined that this level of filtering is sufficient to protect coated surfaces during  exposures of up to several days. For exposures of weeks or months, such as in the \dword{pdsp} cryostat assembly area, a higher cutoff yellow filter is used\footnote{F007-010\texttrademark{} Amber with Adhesive - http://www.epakelectronics.com/uv\_filter\_materials\_flexible.htm.}. 

Exposure of photosensitive components will be strictly controlled, per requirement SP-PDS-3 in Table~\ref{tab:specs:SP-PDS}.  
Work flow restrictions will 
ensure no component exceeds a total exposure of \SI{8}{hours} to filtered assembly area lighting (including testing time).

\subsubsection{Component Cleaning}
\label{sssec:cleaning}
All components will be cleaned 
following manufacturer's specifications and \dword{dune} materials test stand recommendations.  
All incoming materials will have written cleaning procedures, and their travelers will document the completion of these procedures.

\subsubsection{Assembly Procedures}

As was done for \dword{pdsp}, detailed step-by-step written procedure documents will 
guide the assembly for each \dword{pd} module, with a \dword{qc} traveler 
completed and recorded in the database.  Travelers will be based on those used for \dword{pdsp}, modified as need to capture additional data needed for \dword{xarapu} module fabrication.

\dword{pdsp} experience suggests that a two-person assembly team is necessary.
Our current assembly plan envisions a pair of two-person assembly teams working simultaneously, with a fifth person acting as shift leader.  This labor force will allow for production of \num{20} \dword{pd} modules per week, meeting our production requirements.  

The shift leader acts as a \dword{qc} officer responsible primarily for ensuring the distribution of materials to the assembly teams, documenting the batch and lot numbers for each \dword{pd} on the travelers, and ensuring that the teams follow the documented assembly procedures.

Assembly fixtures mounted to \SI{2.4}{m} long flat tables will 
support and align \dword{pd} components during assembly.  All workers handling \dword{pd} components will wear gloves, hair nets, shoe covers, and clean-room disposable laboratory jackets at all times.

\subsubsection{Post-Assembly Quality Control}

Post-assembly \dword{qc} planning is also based on \dword{pdsp} experience, modified as appropriate for larger-scale production.  Each \dword{pd} module goes through a series of go/no-go gauges 
designed to control tolerances of critical interface points.  Following this, each module is inserted into a test \dword{apa} support model, representing the tightest slot allowed by \dword{apa} mechanical tolerances.
It is then scanned at a fixed set of positions with \SI{275}{nm} UV \dwords{led}.  
The \dword{pd} response at each position is measured using \dword{pd} readout electronics and the data compared to reference set of values.
Figure~\ref{fig:pds-pd-scanner} shows 
the scanner used for \dword{pdsp} modules. These performance data will serve as a baseline for the \dword{pd} module, and will be compared against those taken in an identical scanner shortly before installation into an \dword{apa} in the \dword{spmod}, as for \dword{pdsp}. 
All data collected are recorded to the module traveler and to the \dword{dune} \dword{qc} database.
Post-assembly immersion into a \dword{ln} cryostat followed by a repeat scan of each \dword{pd} module (as in \dword{pdsp}) is under consideration as a final \dword{qc} check.

\begin{dunefigure}[\dshort{pd} module scanner]{fig:pds-pd-scanner}
{\dword{pd} module scanner.}
  \includegraphics[width=0.5\columnwidth]{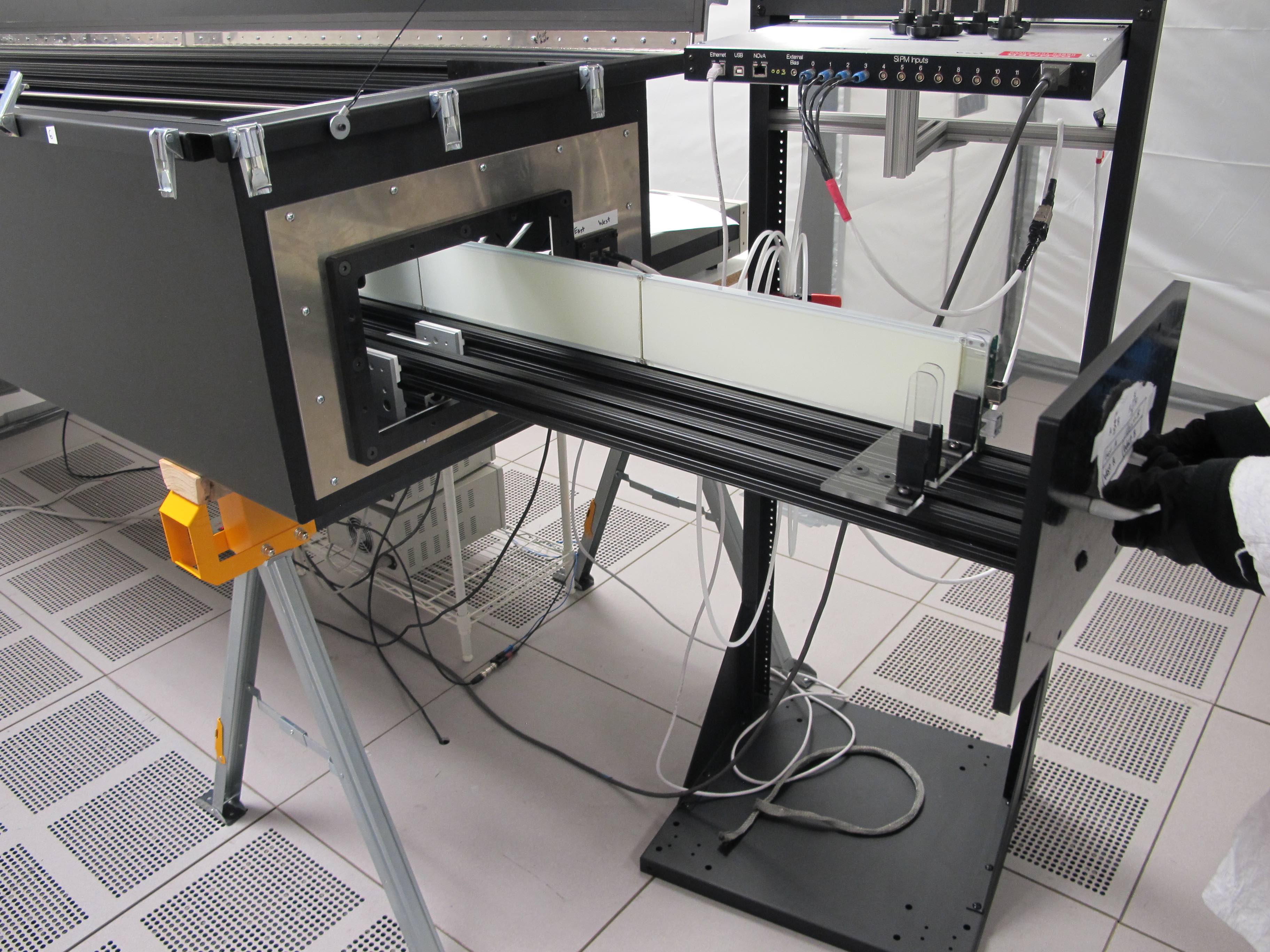}
\end{dunefigure}

\subsection{APA Frame Mounting Structure and Module Securing}	
\label{sec:fdsp-pd-assy-frames}

\Dword{pd} modules are inserted into the \dword{apa} frames through ten slots (five on each side) and are supported inside the frame by stainless steel guide channels.  The slot dimensions for the \dword{spmod} \dword{apa} frames 
are \SI{136.0}{mm} $\times$ \SI{25.0}{mm}\footnote{For \dword{pdsp} they were only \SI{108.0}{mm} $\times$ \SI{19.2}{mm}; the increase allows for larger \dword{pd} modules and an increase in light collection area of nearly 50\% over the \dword{pdsp} design.}   
(see Figure~\ref{fig:pds-pd-mounting}~(left)).
The guide channels are positioned into the \dword{apa} frame prior to application of the wire mesh, and are not accessible following wire wrapping. Following insertion, the \dword{pd} modules are fixed in place using two stainless steel captive screws.

\begin{dunefigure}[\dshort{pd} mounting rails in \dshort{apa} frame]{fig:pds-pd-mounting}
{\dword{pd} mounting in \dword{apa} frame: Fixed end of PD module inside transparent \dword{apa} side tube showing clearance for \dword{ce} cables (left) and showing \dword{pd} mounting rails in an \dword{apa} frame  (right).}
	\includegraphics[height=6.cm]{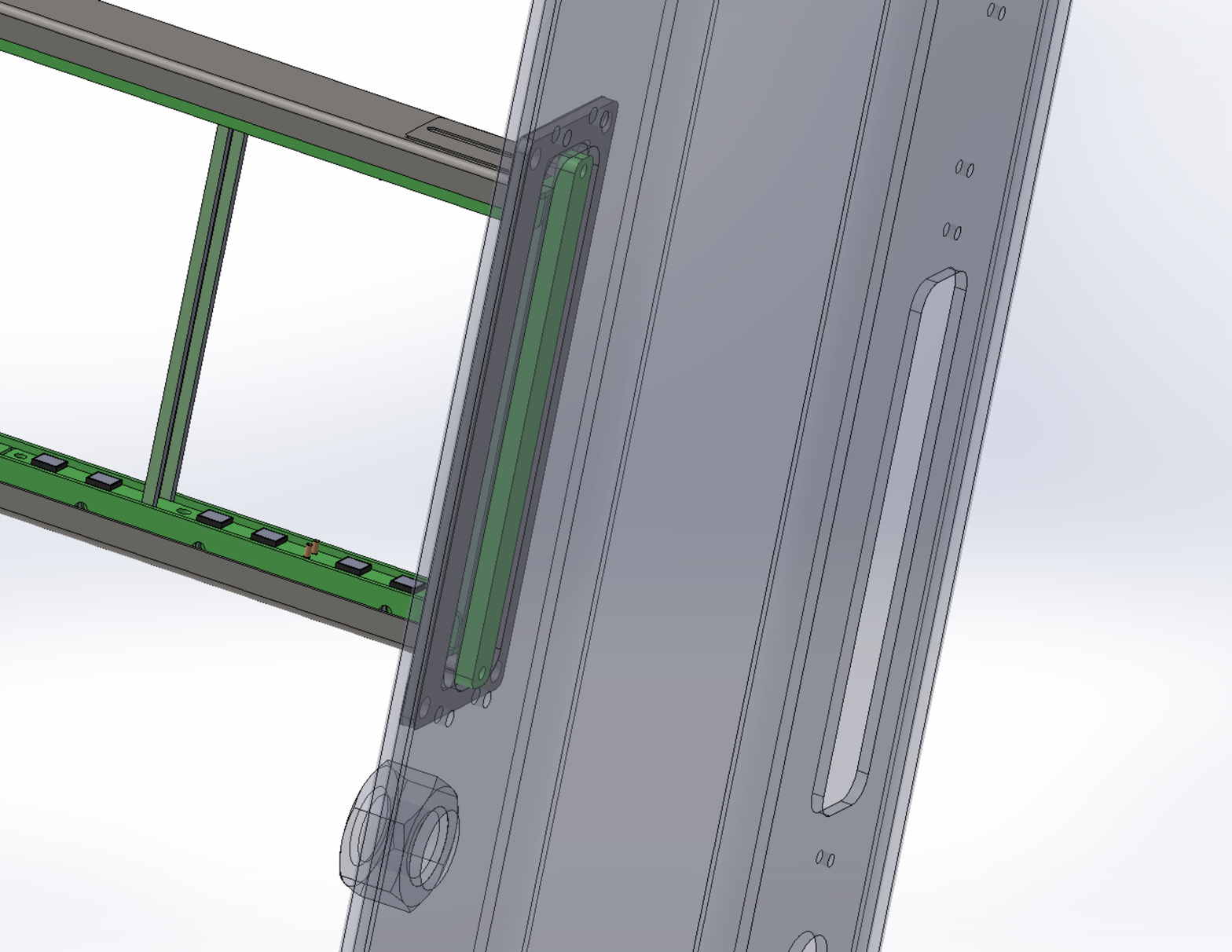}
	\includegraphics[height=6.cm]{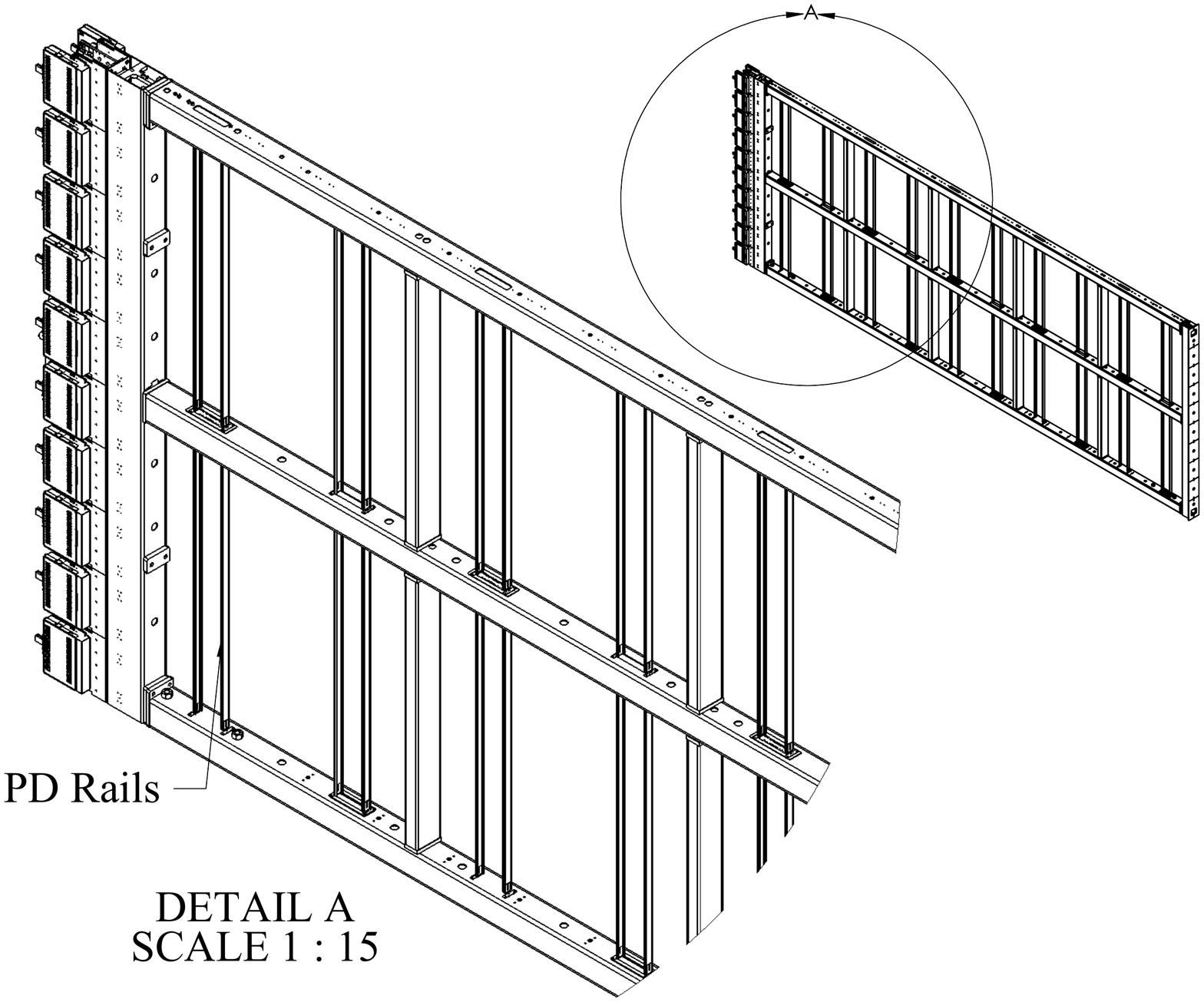}
\end{dunefigure}

\subsubsection{Signal cable and connections}

For the \dword{pdsp}, \dword{pd} cables were run inside the \dword{apa} side tubes, five cables per side.  For the \dword{spmod}, however, this space will be filled by the cable harness for the lower \dword{apa} cold electronics (\dword{ce}) cables.  This change required a revised plan for placing the \dword{pd} cables.  In addition, it was observed during \dword{pdsp} \dword{pds} installation that running the \dword{pd} cables and making electrical connections to the modules during \dword{pd} integration was time-consuming and introduced risk to the process.

For the \dword{spmod}, the \dword{pd} cables will be 
positioned in the \dword{apa} frames prior to installing the 
mesh and wire-wrapping the frame.  
An \dword{apa} in the lower position will house the cables for only the \dwords{pd} in that lower \dword{apa} whereas those in the top position will house the cables for the upper \dword{apa} \dword{pd}s and the pass-through cables from the lower \dword{apa}. The cabling thus requires two different styles of \dword{apa} frame. All cables terminate at the header of the top \dword{apa} after assembly (see Figure~\ref{fig:pd-cable-routing-apa-frames}).


The cable connections between the upper and lower \dword{apa}s are made during \dword{apa} installation into the cryostat, while the \dword{apa} stack is being assembled. The same in-line multi-pin connectors used at the flange penetration in \dword{pdsp}\footnote{Hirose LF 10WBP-12S connectors https://www.hirose.com} are used for this connection.  Superior-Essex \footnote{http://superioressexcommunications.com} Category 6A U/FTP (STP) with FEP jacket (part no. 6S-220-xP) was validated in \dword{pdsp}.  Similar cable will be used in \dword{dune}, but custom-fabricated by the same vendor with two additional twisted pair contained within the external jacket for powering the photosensor active ganging board.

The \dword{pd} signal cables are expected to contract approximately 2\% relative to the \dword{apa} frame during cool-down of the \dword{detmodule} to cryogenic temperatures.  The design accounts for this by leaving cable loops in place between the anchor points to the \dword{apa} frame, allowing for the required relative motion.

To remove interference with the \dword{ce} cables, the electrical connections between the \dword{pd} modules and the \dword{pd} cable harness are moved to the face of the central \dword{apa} tube.  Printed circuit boards with spring-loaded electrical sockets are positioned on the inside face of the tube as part of the \dword{pd} rail installation as shown in Figure~\ref{fig:pd-cable-connectors}~(left).  During \dword{pd} integration into the \dword{apa} frames, a \dword{pcb} with pin contacts mounted to the \dword{pd} module (see Figure~\ref{fig:mounting-board-routing-board}~right) engages into the \dword{pcb} mounted to the \dword{apa} frame, automatically making the electrical connection as shown in 
Figure~\ref{fig:pd-cable-connectors}~(right).

\begin{dunefigure}[\dshort{pd} cable routing in \dshort{apa} frames]
{fig:pd-cable-routing-apa-frames}
{\dword{pd} cable routing in \dword{apa} frames: bottom \dword{apa} (left) and top \dword{apa} (right).}
	\includegraphics[angle=90,height=6.6cm]{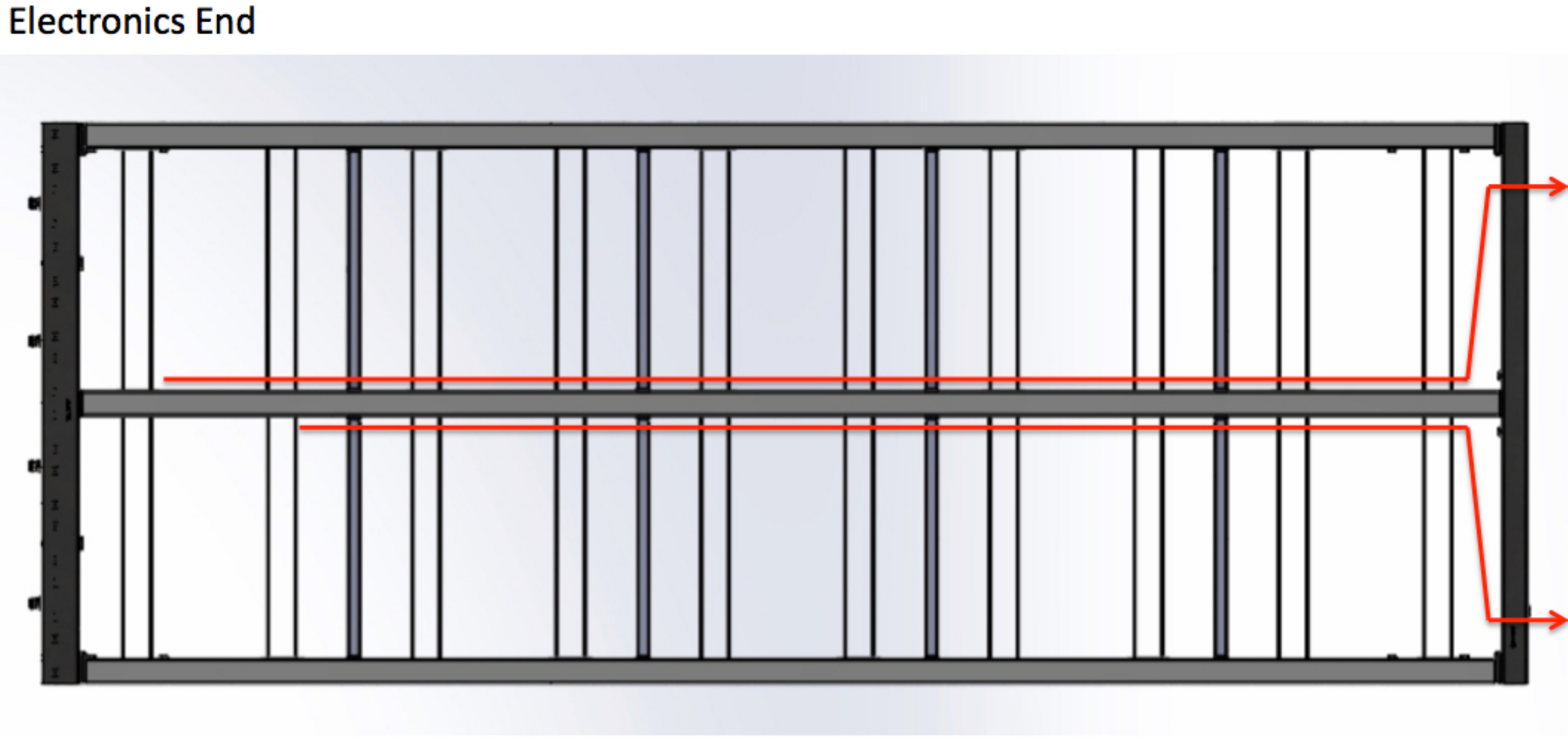}
	\includegraphics[angle=90,height=6.6cm]{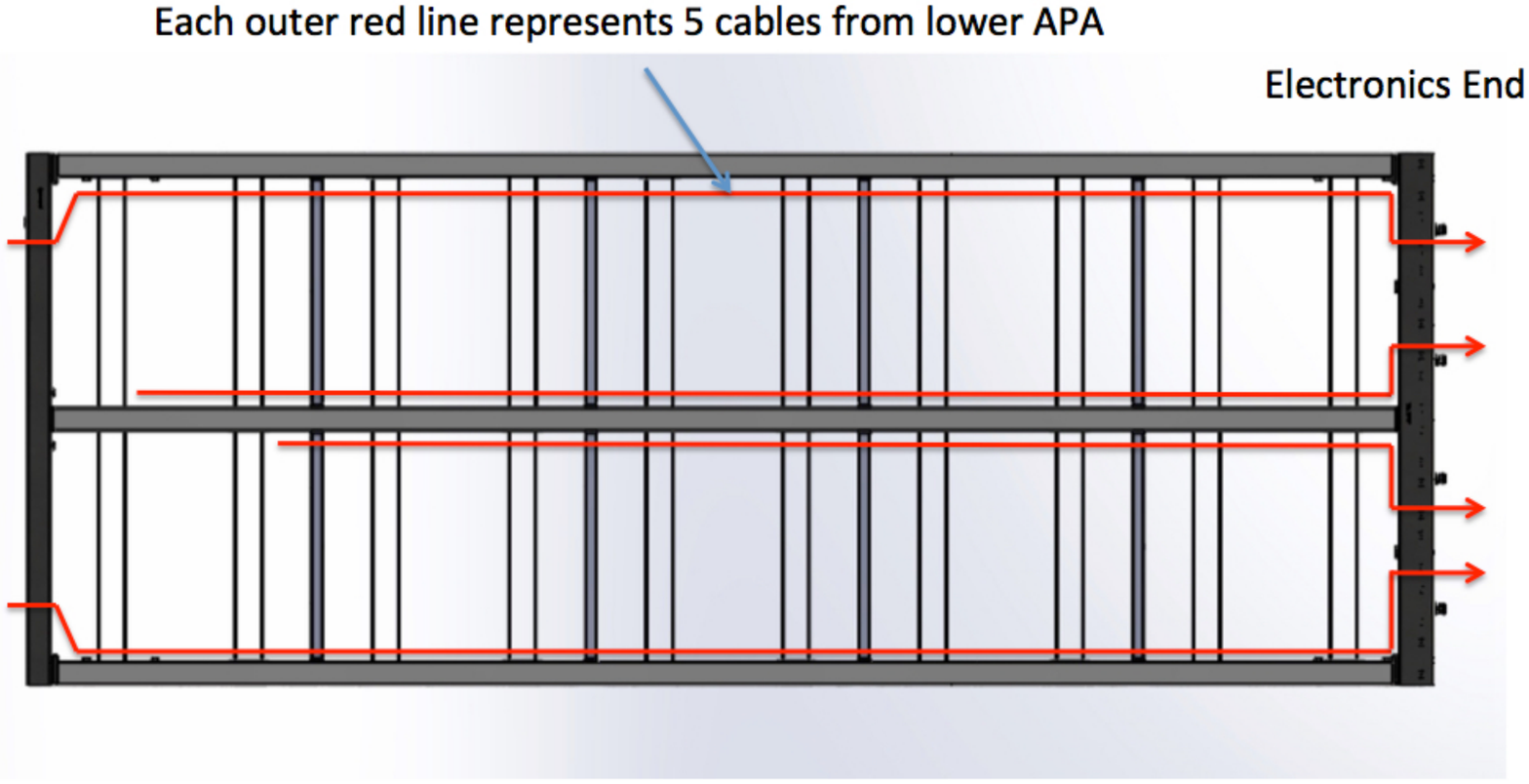}
	\vspace{-1.5cm}
\end{dunefigure}

\begin{dunefigure}[\dshort{pd} cable connectors]{fig:pd-cable-connectors}
{\dword{pd} cable connectors in \dword{apa} frames: \dword{pd} connector plate mounted in \dword{apa} frame (ICEBERG model, left) and a computer model of the mated \dword{pd} and connector assembly in an \dword{apa} (right).  Note that active ganging \dwords{pcb} are buried inside the central tube.}
	\includegraphics[height=6.cm]{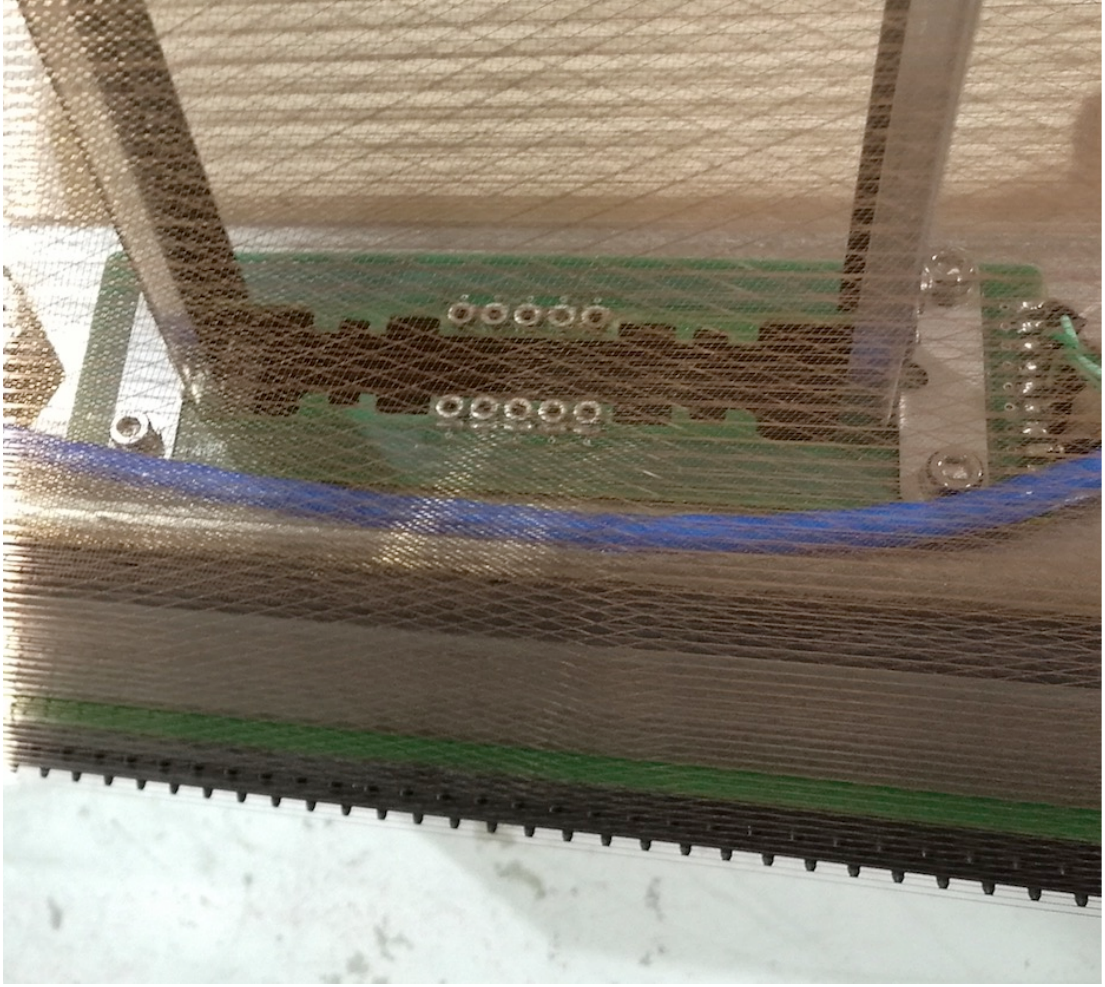}
	\includegraphics[height=6.cm]{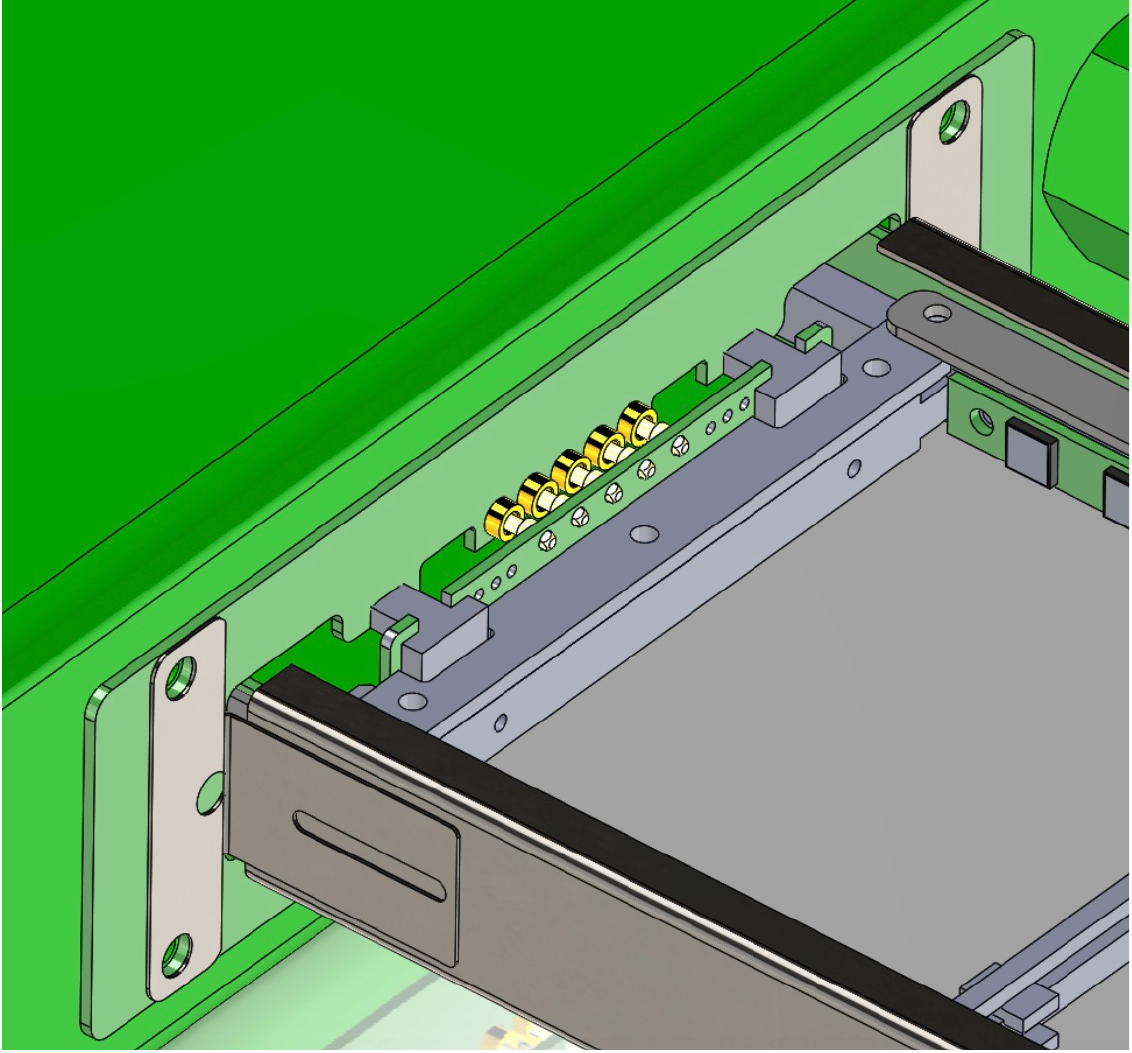}
\end{dunefigure}

\subsubsection{Thermal Contraction and Load Deformation}
\label{sssec:pds-thermal-load}
\textit{\bf Thermal Contraction}

During \cooldown from room 
to \dword{lar} temperatures,  significant relative shrinkage of module components is possible.  Mitigating these effects was a major consideration in the \dword{xarapu} module design.

Thermal expansion coefficients (\dshort{cte}) for the stainless steel \dword{apa} frames and fused-silica filter plates drove the materials selection for the \dword{xarapu} modules.  As shown in Table~\ref{tbl:fdsfpdshrink},  the relative shrinkage of \frfour G-10 and stainless steel are well-matched and fall between the fused silica filter plates and the polystyrene \dword{wls} plates. The frame components are fabricated from \frfour G-10, resulting in a shrinkage of the stainless steel frame structure relative to the frame of approximately \SI{1.2}{mm} along the long ($\sim$\SI{2000}{mm}) axis of the bar, minimizing the motion needed to be accounted for in the electrical connectors to the wiring harness.  The shrinkage of the frame relative to the filter plate is <~\SI{0.2}{mm}.  Both these relative shrinkage factors are accounted for in the dimensions and tolerances of the design.

The largest relative contraction of mechanical components is between the \frfour frame and the polystyrene \dword{wls} plates. The most critical relative shrinkage is between the face of the photosensors and the \dword{wls} plate, where the \SI{92}{mm} width of the plate will shrink significantly more than the \dword{pd} module structure, resulting in more separation (approximately \SI{1.3}{mm}) between the sensor face and the plate.
 Simulation indicates that \dword{xarapu} performance is not strongly affected by this gap size (reducing the gap to zero, direct contact, would be beneficial but would introduce unacceptable risk of damage).  
 The \dword{wls} plate contracts relatively more along the long axis, by \SI{5.8}{mm} for the \SI{487.0}{mm} long plate, but this affects the performance of the detector less; the \dword{wls} bar mounting structure addresses this issue.

Another important potential thermal contraction interference to track in the PD design is the relative contraction of the slots in the \dword{apa} frame and the separation of the photon detector support rails relative to the photon detector cross section.  This requirement is listed in Table~\ref{tab:specs:SP-PDS} as specification SP-PD-12, which requires that a minimum gap between the \dword{pd} module and the \dword{apa} frame of \SI{0.5}{mm} be maintained after cool-down.  Specification SP-PD-08 in the same table that requires a minimum clearance of \SI{1.0}{mm} between the modules and the \dword{apa} frame at room temperature, together with the relative thermal contractions of the stainless steel \dword{apa} frame and G-10 \dword{pd} frames ensures that this specification is met.

\begin{dunetable}[Shrinkage of \dshort{pd} materials]
{lc}
{tbl:fdsfpdshrink}
{Shrinkage of \dword{pd} module materials for a $206^{\circ}$C temperature drop}
Material 			 & Shrinkage Factor (m/m)\\ \toprowrule
Stainless Steel (304) & $2.7\times10^{-3}$\\ \colhline
\frfour G-10 (In-plane) & $2.1\times10^{-3}$\\ \colhline
Fused Silica (Filter Plates) & $1.1\times10^{-4}$\\ \colhline
Polystyrene (WLS Bars) & $1.4\times10^{-2}$\\ 
\end{dunetable}

Mitigation of these contractions is detailed in Table~\ref{tbl:fdsfpdshrinkeffects}.

\begin{dunetable}[Relative shrinkage of \dshort{pd} components and \dshort{apa} frame]
{p{0.2\textwidth}p{0.2\textwidth}p{0.5\textwidth}}
{tbl:fdsfpdshrinkeffects}
{Relative Shrinkage of \dword{pd} components and \dword{apa} frame, and mitigations.}
\textbf{Interface} & \textbf{Relative shrinkage} & \textbf{Mitigation} \\ \toprowrule
\dword{pd} Length to \dword{apa} width & \dword{pd} expands  \SI{1.2}{mm} relative to \dword{apa} frame & \dword{pd} affixed only at one end of \dword{apa} frame, free to expand at other end.  \SI{3}{mm} nominal clearance (beyond tolerance allowance) for expansion in design. \\ \colhline
Width of \dword{pd} in \dword{apa} Guide Rails & \dword{pd} expands \SI{.1}{mm}  relative to slot width & \dword{pd} not constrained in C-channels. C channels and tolerances designed to contain module across thermal contraction range. \\ \colhline
Width of module end mount board to stainless steel frame & Stainless frame shrinks \SI{0.1}{mm}  more than PCB & Diameter of shoulder screws and \frfour board clearance holes selected to allow for motion. \\ \colhline
Length of WLS bar relative to \frfour \dword{pd} frame & WLS bar shrinks \SI{5.8}{mm} relative to \dword{pd} frame & Allowed for in WLS bar mount fixtures. \\ 
\end{dunetable}

\textit{\bf \dword{pd} Mount frame deformation under static \dword{pd} load}

\Dword{fea} modeling of the \dword{pd} support structure was conducted to study static deflection prior to building \dword{pdsp} prototypes.  Modeling was conducted in both the vertical orientation (\dword{apa} upright, as installed in cryostat) and also horizontal orientation.  
Basic assumptions used were fully-supported fixed end conditions for the rails, 
with uniform loading of 3$\times$ \dword{pd} mass (\SI{5}{kg}) along the rails.  
Figure~\ref{fig:pds-rail} illustrates the rail deflection for the \dword{apa} in the horizontal (left) and vertical (right) orientations.
Prototype testing confirmed these calculations.  Similar modeling of final-design \dword{dune} \dword{pd} modules will be completed prior to 60\% design review.


\begin{dunefigure}[\dshort{pd} mechanical support analysis]{fig:pds-rail}
{\dword{pd} mechanical support analysis: Rail deflection for the \dword{apa} in the horizontal (left) and vertical (right) orientations.}
	\includegraphics[height=4.5cm]{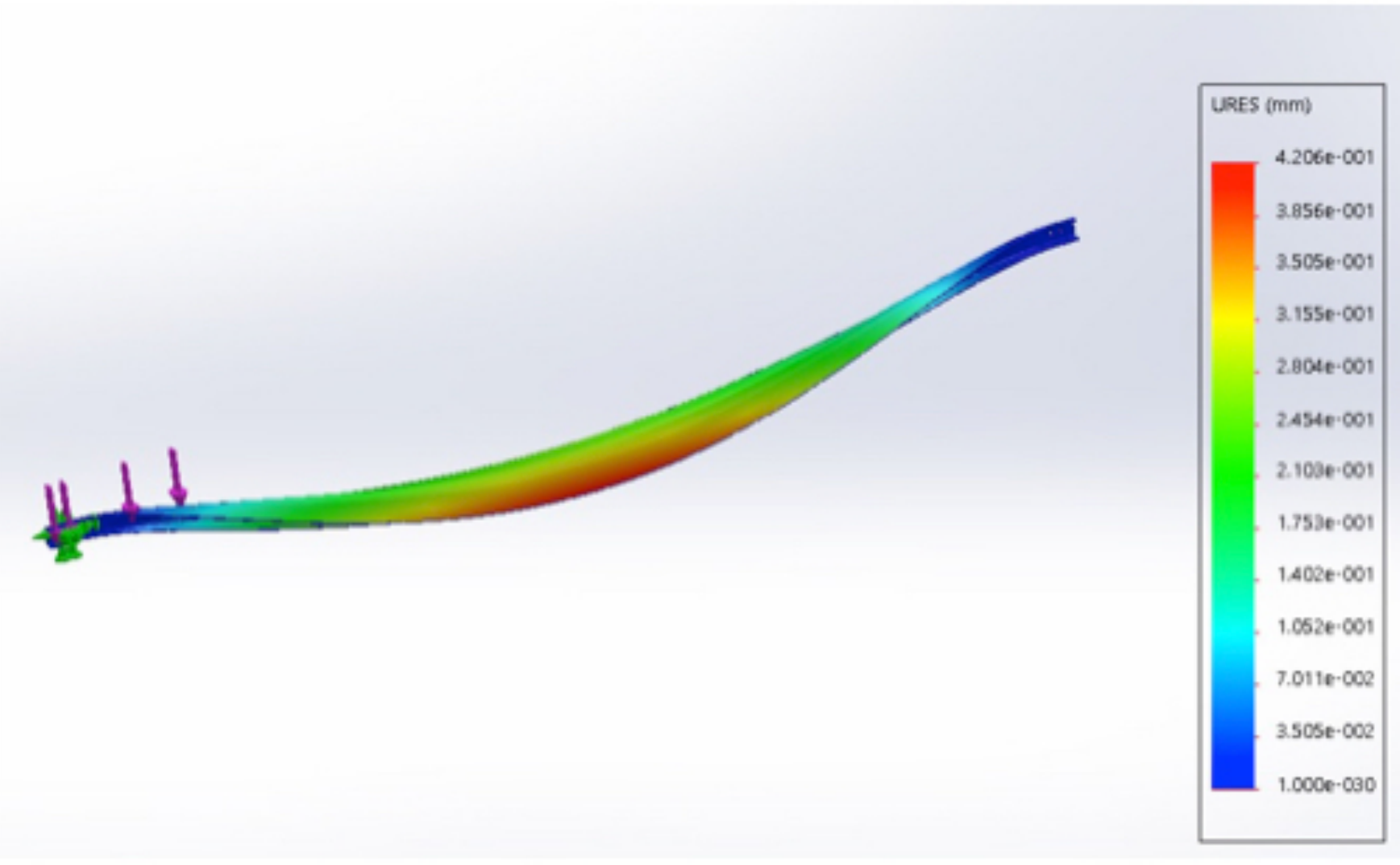} 
	\includegraphics[height=4.5cm]{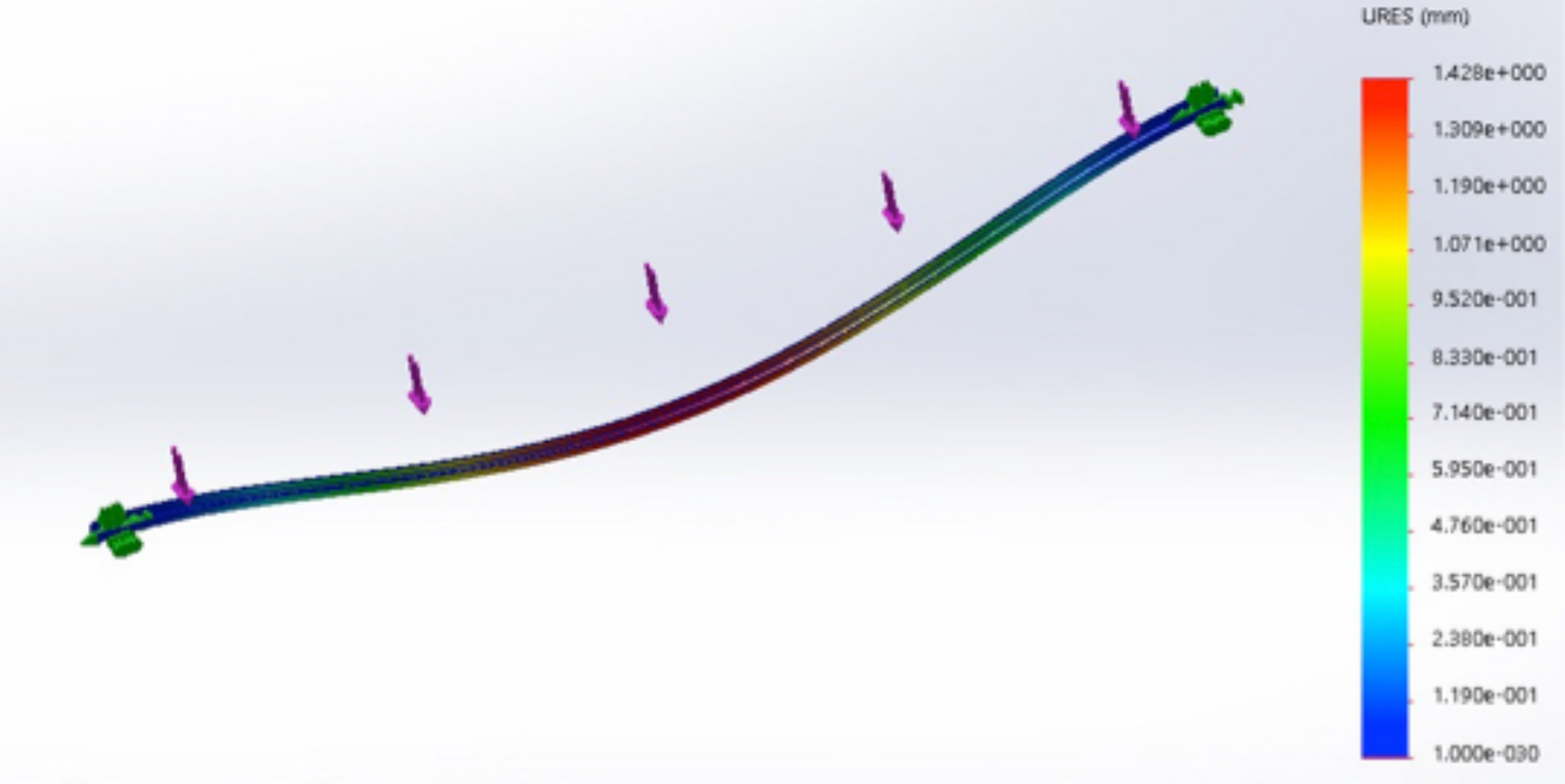}\\
\end{dunefigure}

\subsection{Photosensors and Photosensor Modules}
\label{sec:fdsp-pd-assy-psm}

The use of \dwords{sipm} in noble liquids is relatively new but growing rapidly with experiments such as GERDA, MEG II, DarkSide, and nEXO which are in various stages of preparation. The collaborators of \dword{dune} will learn from these experiments, but in principle, \dword{dune} has a more stringent accessibility and longevity constraints. Risk mitigation through reliability engineering, process control, and vendor and collaboration testing will be a key feature of the \dword{dune} \dword{sipm} production process.

{\textit{Reliability Engineering:}} The primary issue is the change in material properties and thermal stresses induced in the packaging due to differential coefficients of thermal expansion (\dwords{cte}). This is especially critical for interfaces, in particular
die to substrate, substrate to potting mold, potting mold to encapsulation, and solder joints to everything else. Analysis of these interfaces and collaboration with vendors to match \dwords{cte} as much as possible at these interfaces will contribute substantially to the long-term reliability of the photosensors.

{\textit{Process Control:}} Small and seemingly innocuous
changes in the photosensor fabrication process can have a big impact on the robustness of these devices at extreme temperatures. This is one of the reasons why in space applications, for instance, same-day same-batch components are utilized. Given the photosensor quantities involved, this is not feasible for the \dword{dune} \dword{pds}, but
it will be important to establish a \dword{mou} with the vendor regarding 
strict process control once the pre-production batch has been qualified.

{\textit{Procurement:}} Potential vendors have verified that delivery of \num{100000} devices per year is a reasonable expectation so long as the purchase contract is initiated early enough.  Vendor visits to Hamamatsu and \dword{fbk} in June-July 2019 will be used as an opportunity to confirm this guidance.

{\textit{Quality Assurance:}} This will be an essential component in the photosensor risk mitigation strategy consisting of restricting the number of production batches, clearly communicating desired device and packaging parameters to the vendor, and vendor testing to
guarantee device operation down to liquid nitrogen temperatures.
Before shipment to the consortium, the vendor 
must qualify a randomly selected sample of devices from each production batch. The qualification would entail thermally stressing the devices 
with visual and electrical measurements made before and after. 

{\textit{Quality Control:}} The above strategies, while significantly lowering the risk, do not obviate the need for a strict testing regimen. Every sensor will be tested multiple times, at various stages of assembly, before installation in the \dword{spmod}.

\Dwords{sipm} are mounted in groups of six passively-ganged sensors to mounting boards, with eight mounting boards per supercell.  Passive ganging (sensors in parallel) is implemented with traces on the \dword{sipm} mounting board (\dword{pd} module), 
as was done for \dword{pdsp}.  The \dwords{sipm} are mounted using a pick-and-place machine and standard surface-mount device soldering procedures. The outputs from these mounting boards are then routed to active ganging circuits in the center of the \dword{pd} module, where they are collected into a summing amplifier and reduced to a single output channel.

The ganged analog signals exit via long cables (approximately \SI{20}{m}) for digitization outside the cryostat.
\dword{pdsp} has provided essential operational experience with a passive ganging board and signal transport provided by Teflon Ethernet Cat-6 cables, as described in Section~\ref{sec:fdsp-pd-assy-frames}.

\subsection{Electronics}
\label{sec:fdsp-pd-assy-pde}

The \dword{pds} consortium gained extensive experience in manufacturing processes for electronic systems during the development of the \dword{pdsp} \dwords{ssp}.
A general description of the readout system of \dword{pdsp} can be seen in the Section~\ref{sec:fdsp-pd-pde}. Compatibility between elements designed by different institutions is guaranteed when standard procedures are followed, so the circuit design must be done in accordance with mutually agreed-upon specification documents.  A sufficient  number of units must be produced to allow 
for testing both locally and  in the central facility; for example, in \dword{pdsp} five 12-channel  \dwords{ssp} were produced and delivered to \dword{cern} for integration testing. Twenty-four were fabricated for \dword{pdsp} operation. Similar manufacturing test programs are envisioned for \dword{dune}.

The readout electronics of the \dword{pds} will be designed and produced with similar tools and protocols as for  \dword{pdsp}. For example, 
\dword{pcb} layout is performed in accordance with IPC\footnote{IPC\texttrademark{}, Association Connecting Electronics Industries, \url{http://www.ipc.org/}.} specifications. Bare \dword{pcb} manufacturing requirements are embedded within the Gerber file 
fabrication documents (e.g., layers, spacing, impedance, finish, testing, etc.). Components are assembled on circuit boards either by trained \dword{pd} consortium technical staff or by external assembly vendors, based on volume, and in accordance with per-design assembly specification documents. Testing occurs at 
collaboration institutions in accordance with a per-design test procedure that typically includes a mix of manual, semi-automated and automated procedures 
in an engineering test bench followed by overall characterization in a system or subsystem test stand.
Other considerations and practices relevant to readout electronics production and assembly are itemized here:

\begin{itemize}

\item Components: Schematic capture is done using appropriate tools (such as OrCAD 16.6.\footnote{OrCAD\texttrademark{} schematic design tool for \dword{pcb} design http://www.orcad.com} or similar toolset) available within a design facility. Design is hierarchical with common \dword{fe} page referenced multiple times, such as for all input channels. 
The schematic contains the complete bill of materials (BOM) including all mechanical parts. An electronics schematics subversion
repository or similar tool is typically used for version control and backup. Multiple internal design reviews are held before the schematic is released 
for layout. The BOM, stored directly within the schematic, is extracted to a spreadsheet when ordering parts. Every part specifies 
both manufacturer and distributor information. Distributor information may be overridden by a technician at order time due to price or availability. Standard search engines such as Octopart\footnote{Octopart https://octopart.com/}, ECIA\footnote{ ECIA https://www.eciaauthorized.com} and PartMiner\footnote{PartMiner https://www.part-miner.com/} are used to check price or availability across all standard distributors. A parts-availability check 
is performed prior to handoff from schematic to layout since 
obsolete or long lead-time parts 
may have been removed from the design and replaced. BOM information includes dielectric, tolerance, temperature coefficient, voltage rating, and size (footprint) to ensure that all parts are fully described.

\item Boards: Standard tools (such as the Allegro\footnote{Cadence Allegro\textregistered \dword{pcb} design solution https://www.cadence.com} toolset) are available for the \dword{pcb} layout. Conventional \dwords{pcb} are 
controlled-impedance multi-layer boards with many sets of delay-matched nets where necessary.
In usual practice, multiple previously qualified vendors bid competitively. The consortium electronics group provides the complete impedance and delay characteristics  within the layout tool, and the selected vendor cross-checks these values prior to manufacture and performs full electrical and impedance testing.  Multiple internal design reviews are held prior to release of the design.

\item Cable plant: The cabling designed will take into consideration the \dword{apa} space and will be done in close collaboration with the TPC \dword{ce} consortium to avoid crosstalk effects.  
Before making a final decision on cable procurement, we are investigating the possibility of cable manufacturing in a \dword{pds} consortium institution versus the cost of a commercial solution.  

\item Manufacturer list: In addition to the general laboratory procedures for \dword{qa}, the general practice will be to use only \dword{pcb} manufacturers and external assembly vendors whose workmanship and facilities have been personally inspected by experienced production team members. All external assemblers are required to quote in accordance with an assembly specifications document describing the IPC class and specific solder chemistry requirements of the design. The BOM document will show selected and alternate suppliers where available for every component of the \dword{fe} boards.

\item \dword{fe} electronics firmware: This will be specified and updated iteratively in collaboration with other systems. The electronics working group will be responsible for responding to requests for additional firmware development, including for example, modifications to timing interface, modifications to trigger interface, and implemented sensitivity to in-spill versus not-in-spill conditions. Documents describing firmware architecture for each major change will be written and distributed to \dword{pd} and \dword{daq} working groups before implementation. An \dword{fe}  electronics users manual containing all details of new firmware will be distributed with production units when manufactured.

\item Mechanical assembly: With the mechanical assembly of electronics readout boards, it is common practice to use a \threed model generated by the layout software.   
All relevant dimensions of the \dword{pcb} including connector and indicator placement is extracted as a base DXF file from which an overall exploded mechanical diagram of chassis and other mechanical parts is made.  Mechanical items such as shield plates will also be provided. It is assumed that external vendors will make the \dword{fe} chassis 
(one for the chassis, one for front and back panels) from drawings provided by the consortium.

\end{itemize}

\subsection{Calibration and Monitoring}
\label{sec:fdsp-pd-assy-CandM}

The consortium gained extensive experience in manufacturing, testing, and assembly processes 
during the development of the calibration and monitoring system for \dword{pdsp}.
A general description of the proposed calibration and monitoring system can be seen in  Section~\ref{sec:fdsp-pd-CandM}. 

The design and production of the calibration modules including electronics circuitry,  \dword{fpga} implementation for light-source controls, optical timing/trigger and \dword{daq} communication protocols, and UV light sources, closely follows the process described in Section~\ref{sec:fdsp-pd-assy-pde}.

Design and selection of cold diffuser components and selection of cold and warm quartz fiber components follows requirements derived from interface considerations with \dword{hv},  \dword{cpa} and cryostat systems,  
and was tested in \dword{pdsp}.
Installation, \dword{qa} and \dword{qc} of optical fibers is performed during the \dword{cpa} installation process, with diffusers and \dword{cpa} fibers pre-installed on the \dword{cpa}s.

Installation of fibers that connect \dword{cpa}s 
to the optical feedthrough penetrations at the cryostat will be defined with the \dword{dss} and cryostat teams, based on installation experience in \dword{pdsp}. 

\subsection{Outline of PD System Assembly Plan}
\label{sec:fdsp-pd-assy-Assby-plan}

The \dword{sp} \dword{pd} consortium is composed of many institutions in North and South America and Europe and fabrication of the system will occur at many locations. 
Here we present an outline of our assembly 
plan.  The schedule interfaces implicit in this assembly program will be detailed in the overall project schedule.

\begin{itemize}

\item Photosensors, mounting and active ganging:  
Italian groups funded by INFN (and their associated universities) will procure, test, and assemble the active ganging circuits for the \dword{pds}. Photosensor mounting board assembly will likely be outsourced to a yet-to-be selected external firm. 

The groups most involved 
are Bologna, Genova, Milano, Milano-Bicocca, and Laboratori Nazionali del Sud.  Other Italian groups have also expressed interest in joining.  We will allocate tasks among the interested groups prior to the final design review.

Following assembly, these components and their associated \dword{qc} documentation will be shipped to \dword{unicamp} for assembly into \dword{pd} modules.

\item Light collector modules:  Light collector modules will be fabricated primarily in Brazil.  

Dichroic filters and wavelength shifting plates will be procured and received by a combination of CTI Campinas and the National Laboratory of Synchrotron Light (also in Campinas), where they will undergo reception \dword{qc} testing.

Following testing, the dichroic filters will be delivered to \dword{unicamp} for coating with \dword{ptp} in their in-house vacuum deposition system.

The module mechanical components are also the responsibility of \dword{unicamp}.  This includes the \frfour G-10 components (which will be fabricated in-house at \dword{unicamp}), signal routing circuit boards, module electrical connectors, and other miscellaneous components (purchased externally) required to fabricate the modules.

Module assembly and initial \dword{qc} testing will happen at \dword{unicamp}.

Following assembly, the tested modules and all the associated \dword{qc} documentation will be shipped to a reception center in the USA, where they will be retested and stored in the \dword{sdwf} until required for integration into the \dword{apa}s underground.


\item \dword{apa} support rails and electrical connectors:  Stainless steel rails and associated hardware for supporting the \dword{pd} modules inside the \dword{apa} will be fabricated by vendor in the USA.  Components for cable connection between the upper and lower \dword{apa}s, and cable management pieces, will be procured from vendors, assembled, and tested in the USA.  

Following assembly and testing, these components will be shipped to \dword{apa} frame assembly sites for integration into the frames prior to wire wrapping.

\item Readout electronics and \dword{daq} interface:  The \dword{fe} electronics and \dword{daq} interfaces will be built by a collaboration of Latin American countries, particularly Colombia, Peru, and Paraguay, with engineering support from \dword{fnal} and the University of Michigan in the US. 

The \dword{fe} electronics, communications boards, and external cabling between them and the \dword{daq} will be designed by collaboration engineers, fabricated, or purchased from external vendors, and tested at collaboration institutions.  While the exact distribution of effort is still being settled, interested institutions in Colombia include Universidad  Antonio Nari\~no (UAN) and Universidad EIA.  In Peru, they include the Universidad Nacional de Ingenier\'ia de Peru.

\dword{daq}/\dword{pd} interface firmware development will be conducted by Paraguay, particularly the Universidad National de Asuncion (FIUNA), in conjunction with UAN in Colombia.

Following assembly and testing, components will be shipped to a reception center in the US for inspection then stored at the \dword{sdwf} until needed.

\item Cables:  Materials for cables and connectors inside the \dword{apa} frames will be purchased, assembled and tested in the USA.  Cables between the \dword{apa} and the cryostat flange, as well as those between the flange and the \dword{daphne} electronics, will be purchased by \dword{unicamp} and assembled and tested at their facilities.
Following testing, the cables and their associated \dword{qc} documentation will be assembled into groups of 20 cable sets (one \dword{apa} stack) and shipped to the \dword{sdwf} for storage until needed for installation.  Cables intended to be routed inside the \dword{apa} frames will be shipped directly to the \dword{apa} frame assembly facilities.

\item Monitoring system: The monitoring system including \dword{led} drivers, optical fibers, and diffusers will be designed, fabricated and tested in the US by the South Dakota School of Mines and Technology and Argonne National Laboratory. 

Following assembly and testing, the system will be stored at the \dword{sdwf} until needed for installation.

\end{itemize}

\section{System Interfaces}
\label{sec:fdsp-pd-intfc}
\subsection{Overview}
Table~\ref{tbl:SPPDinterfaces} contains a summary and brief description of all the interfaces between the \dword{spmod} \dword{pds} consortium and other consortia, working groups, and task forces, with references to the current version of the interface documents describing those interfaces.  
Drawings of the mechanical interfaces and diagrams of the electrical interfaces are 
included in the interface documents as appropriate.
It is expected that further refinements of the interface documents will take place prior to the final \dword{prr} for the detector. The interface documents specify the responsibility of different consortia or groups during all phases of the experiment including design and prototyping, integration,  installation, and  commissioning.

Additional details describing the interface between the \dword{sp} \dword{pds} and the other consortia, task forces (TF) and subsystems are given below.

\begin{dunetable}
[PDS interfaces]
{p{0.25\textwidth}p{0.5\textwidth}l}
{tbl:SPPDinterfaces}
{Single Phase \dword{pds} interface links.}
Interfacing System & Description & Linked Reference \\ \toprowrule
Detector Subsystems & &\\ \colhline
\dword{apa} & Mechanical support for \dwords{pd}, \dword{pd} installation slots, \dword{pd} cabling support, access slots & \citedocdb{6667} \\ \colhline
\dword{ce} & Electrical signal interference, grounding, cable routing, cryostat flange, installation and testing & \citedocdb{6718} \\ \colhline
\dword{hv} & Mounting of \dword{pd} monitoring system, possible reflector foil support, electrical discharge or corona effect light contamination & \citedocdb{6721} \\ \colhline
\dword{daq} & Data format, data timing, trigger information, timing and synchronization & \citedocdb{6727} \\ \colhline
\dshort{cisc} & Rack layout, flange heaters, power supply selection, power and signal cable selection, monitoring cameras and camera lighting, purity monitor lighting, controls and data monitoring & \citedocdb{6730} \\ \colhline
Technical Coordination & &\\ \colhline
Facility interfaces & Cable trays inside the cryostat, cryostat penetrations, rack layout and power distribution on the detector mezzanine, cable and fiber trays on top of the cryostat & \citedocdb{6970} \\ \colhline
Installation interfaces & Sequence of integration and installation activities at SURF, equipment required for \dword{pd} consortium activities, environmental controls in the cryostat during installation, post-installation testing  & \citedocdb{6997} \\ \colhline
 
Calibration task force interfaces & Interface of \dword{spmod}\ \dword{dp} monitoring system into calibration system. & \citedocdb{7051} \\ \colhline
Physics, Software and Computing interfaces & Covers interfaces between the \dword{pd} group and the joint computing task force, including specifications required for physics, data handling, and computing and storage requirements. & \citedocdb{7105} \\
\end{dunetable}

\subsection{Anode Plane Assembly}

The interface with the Anode Plane Assembly (APA) represents the most significant mechanical interface for the \dword{pds}. Interfaces with the \dword{apa} are involved in meeting specifications SP-PDS-2, SP-PDS-7, SP-PDS-8, SP-PDS-9, SP-PDS-10, SP-PDS-11, and SP-PDS-12 (see Table \ref{tab:specs:SP-PDS}).  The interface document will be written to monitor these specifications.

The \dword{apa} frame is designed to provide:
\begin{itemize}
\item mechanical support and alignment for the \dword{pd} modules, including access slots through the side of the frame for insertion of modules after the \dword{apa}s are wrapped in wire;
\item mounting support for the \dword{pd} electrical connections between the \dword{pd} modules and the cable harness mounted inside the \dword{apa} frame;
\item mechanical support and strain relief for \dword{pd} cables located inside the completed \dword{apa} frame; and
\item provision to connect the \dword{pd} cables from the lower \dword{apa} to the upper \dword{apa} in an assembled \dword{apa} stack and to connect the  
cables from the top of the \dword{apa} stack to the cryostat flange.
\end{itemize}

Work on the two-\dword{apa} connection and inspection in the underground assembly area will be performed by the \dword{apa} group. Work on cabling prior to installation is performed by \dword{pds} and \dword{tpc} electronics groups under supervision of the \dword{apa} installation group. Once the \dword{apa}s are moved inside the cryostat, the \dword{pds} and electronics consortia will be responsible for the routing of the cables in the trays hanging from the top of the cryostat. 

Careful interface control will be required to ensure a successful assembly, which will be guided by the interface control document between the \dword{pd} and \dword{apa} consortia.  

\subsection{TPC Cold Electronics}
\label{sec:fdsp-pd-intfc-ce}

Interfaces with the \dword{tpc} \dword{ce} are involved in meeting specifications SP-FD-2, SP-PDS-8, and SP-PDS-10 (see Table~\ref{tab:specs:SP-PDS}).  The interface between the \dword{pd} and \dword{ce} systems primarily consists of:

\begin{itemize}
    
    \item ensuring no electrical cross-talk between the electronics and cabling harnesses of the \dword{pd} and \dword{ce} systems;
    \item ensuring there be no electrical contact between the \dword{pds} and \dword{ce} components except for sharing a common reference voltage point (ground) at the \fdth{}s;
    \item developing a common cable routing plan allowing the systems to share a common cable tray system on top of the \dword{apa} frame and routing the cables to the cryostat flanges; and 
    \item managing the interface between the \dword{pd} and \dword{ce} flanges in the cryostat cabling tees.
  \end{itemize}  
The \dword{ce} and \dword{pd} use a common cable tray system but separate flanges for the cold-to-warm transition, and each consortium is responsible for the design, procurement, testing, and installation, of their flange on the \fdth{}, together with \dword{lbnf}, which is responsible for the design of the cryostat. 
The installation of the racks on top of the cryostat is a responsibility of the facility, but the exact arrangement of the various crates inside the racks will be reached after common agreement between the \dword{ce}, \dword{pd}, \dword{cisc}, and possibly \dword{daq} consortia. 
The \dword{pd} and \dword{ce} consortia will retain all responsibility for selecting, procuring, testing, and installing their respective racks unless space and cost requires an agreement on shared crates to house the low-voltage or high-bias voltage modules for both systems. 

\subsection{Cathode Plane Assembly and High Voltage System}
\label{sec:fdsp-pd-intfc-le}

Interfaces with the \dword{hv} system should meet mechanical specification SP-PDS-9 (see Table~\ref{tab:specs:SP-PDS}). In addition, light produced in electrical discharges in the \dword{hv} system may increase the \dword{pds} data volume and impact the \dword{daq} system. Communication between the three systems has been established on this issue. 

The primary interface between the \dword{pd} and \dword{hv} systems is summarized as follows:
\begin{itemize}
    \item providing an optical fiber routing path and strain relief system to the cryostat calibration hatch;
    \item mounting the \dword{pd} monitoring system light diffusers to the \dword{cpa} faces; and     
    \item minimizing background light due to electrical discharge (corona effects).
\end{itemize}

This interface has strong overlap with the calibration consortium; this is described in more detail in Section~\ref{sec:fdsp-pd-intfc-calib}.

If the light reflector foil option were to be implemented, production of the \frfour{}+resistive Kapton \dword{cpa} frames will be the responsibility of the \dword{hv} consortium, together with design of the structure for mounting the \dword{pd} reflector foils to the \dword{cpa} structure.  The \dword{hv} consortium will also provide mounting attachment points in the \dword{cpa} frame structure.
 The reflector foils themselves, \dword{tpb} coating of the \dword{wls} foils, and any required hardware for mounting the foils will be the responsibility of the \dword{pd} consortium, with the understanding that all designs and procedures will be approved by the \dword{hv} consortium.

\subsection{Data Acquisition}
\label{sec:fdsp-pd-intfc-daq}

Interfaces with the \dword{daq} system are involved in meeting specifications SP-PDS-5
and SP-PDS-13 (see Table~\ref{tab:specs:SP-PDS}).  The \dword{pds} interfaces with the \dword{daq} system are described in \citedocdb{6727} and include

\begin{itemize}

\item Data physical links: Data are passed from the \dword{pd} to the \dword{daq} on 25 optical links following the 1000Base-SX standard. The links run from the \dword{pd} readout system on the cryostat to the \dword{daq} system in the \dword{cuc}.

\item Data format: Data are encoded using UDP/IP.  The data format consists of a header containing the word count, event time stamp, and channel ID, followed by the digitized waveform in \SI{80}{MHz} samples.
The data format has also been specified to use compression (zero suppression) and custom communication protocol.

\item Data timing: The data must contain enough information to identify the time at which it was taken.

\item Trigger information: The \dword{pd} may provide summary information useful for data selection. If present, this will be passed to the \dword{daq} on the same physical links as the remaining data.

\item Timing and synchronization: Clock and synchronization messages will be propagated from the \dword{daq} to the \dword{pd} using a backwards compatible development of the \dword{pdsp} timing system protocol~\cite{bib:docdb1651}. There will be at least one timing fiber available for each data link coming from the \dword{pds}. 

\item Power-on initialization and start-of-run setup:  The \dword{pd}S may require initialization and setup on power-on and start of run. Power on initialization should not require communication with the \dword{daq}. Start run/stop run and synchronization signals such as accelerator spill information will be passed by the timing system interface.

\end{itemize}

The data format has been determined but it is possible to include additional summary information to the header that depends on the outcome of triggering studies underway. This minor potential modification can be accommodated easily.

Excessive \dword{pd} data may be generated by background effects such as light leaks in the cryostat or light generated due to sporadic short duration current discharge from the HV system (referred to as ``micro-discharges'' or as ``streamers'' in \dword{pdsp}).  The \dword{hv} consortium is trying to reduce the rate at which the discharges happen, but it is not expected to be completely eliminated. 

In the case of light leaks, specification SP-PD-05 limits the acceptable data generated by these leaks to  less than 10\% of the total data transfer rate from the \dword{pd} to the \dword{daq}.
Light flashes due to HV micro-discharges may be harder to mitigate, but the experience of streamers producing light in \dword{pdsp} informs what we are likely to experience in \dword{dune} and indicates that there is low risk that it will be a serious problem:

\begin{itemize}
  
\item They occur at a relatively low rate: once per few hours in \dword{pdsp}, likely much less underground due to the much lower cosmic ray ray that generates charge in the \dword{tpc}.

\item When they occur, they produce a significant amount of light but in a localized region (this was observed in the \dword{pdsp} \dword{pds}).

\end{itemize}

There is an automatic mitigation scheme in the \dword{hv} system slow control that can identify when micro-discharges occur and stop them, but the power supply data is read relatively slowly (a few Hz) compared to the timescale of the \dword{pd}/\dword{daq}. Data corresponding to the \dword{pd} response to the light flash will have already been recorded by the \dword{daq} before the \dword{hv} system can respond, so the mitigation will likely need to be a function of the \dword{daq}.

In summary, since electrical discharges from the \dword{hv} system are not under the control of the \dword{pds} it is not directly a specification for the \dword{pds}.  
Following consultation with the \dword{daq} and \dword{hv} consortia, we determined that it is also not appropriate as a specification on the \dword{hv} system but is better addressed in the \dword{daq}/\dword{pd} interface document.

\subsection{Cryogenics Instrumentation and Slow Control}
\label{sec:fdsp-pd-intfc-xeon}

The primary interactions between the \dword{pd} and the \dword{cisc} include

\begin{itemize}
    \item warm electronics rack controls, power supplies, rack safety equipment;
    \item warm cable and connector selection;
    \item cryogenic camera systems for detector monitoring, including lighting systems;
    \item purity monitor lighting requirements;
    \item cryostat flanges required for \dword{pd} signal cable and monitoring systems; and
    \item \dword{pd} slow control (including bias voltage) and data monitoring.
\end{itemize}

Additional interaction may occur in the case that the xenon doping performance enhancement is selected for inclusion in the detector.  This system requires pre-mixing xenon gas and argon gas to introduce xenon doping into the \lar volume.

Any required hardware for this enhancement will be the responsibility of the \dword{pd} consortium, with the understanding that all designs and procedures will be approved in advance by the cryogenics group.

A proposal is under consideration to mount \dword{cisc} temperature sensors inside the \dword{apa} frames, sharing a readout cable routing inside the \dword{apa} frames and upper-to-lower \dword{apa} connection point with the \dword{pd}.  This decision will be reached prior to the 60\% design review.  In case this plan is adapted, all \dword{pd} cables and connectors will be the responsibility of the \dword{pd} consortium, and all \dword{cisc} components, cables, and connectors will be the responsibility of the \dword{cisc} consortium.  Cable routing plans, junction plates, and cable fixation will be the responsibility of the \dword{pd} consortium.

\subsection{Facility, Integration and Installation Interfaces}

The interface document with the project interface and installation working group covers the interface of the \dword{pd} group with the technical coordination groups who oversee the integration of the \dword{pd} modules and electronics into the \dword{apa} and \dword{daq}. Interfaces with the facility, integration, and installation group are involved in meeting specifications SP-PDS-1, SP-PDS-3, SP-PDS-4, and SP-PDS-5 (see Table~\ref{tab:specs:SP-PDS}).  The interfaces are distributed among the facility, integration, and installation working groups and primarily consist of
\begin{itemize}
    \item electrical racks, cable trays, and cryostat penetrations, and power distribution on the mezzanine;
    \item storage for arriving \dword{pd} modules prior to their integration;
    \item planning of pre-integration tests of \dword{pd} components at the integration area and required equipment/tools;
    \item sequence of integration and installation activities at \dword{surf} (including environmental controls);
    \item quality management testing of \dword{pd} modules during integration and installation;
    \item equipment required for \dword{pd} consortium activities; and
    \item environmental controls in the cryostat during installation, and post-installation testing.
\end{itemize}

The \dword{pd} consortium retains responsibility for providing quality management tooling and test plans at the integration area, as well as specialized labor and supervisory personnel for \dword{pd} module integration and installation. Distribution of these responsibilities is described in~\citedocdb{6970}. 

The installation is described in detail in Chapter~\ref{ch:sp-install}.

\subsection{Calibration and Monitoring}
\label{sec:fdsp-pd-intfc-calib}

This subsection concentrates on the description of the interface between the  \dword{sp}-\dword{pds}, calibration, and \dword{cisc} consortia. 
Main interface items are

\begin{itemize}
    \item cold components: light sources (diffusers and fibers) placed on the cathode planes to illuminate the detectors;
    \item warm components: a controlled pulsed-UV source and warm optics; and 
 
    \item the optical \fdth: used to \dword{pd} bring monitoring system fiber optics through the calibration and monitoring flange.  The flange itself is a shared interface between the \dword{pds}, the calibration task force, and \dword{cisc}.
\end{itemize}

Hardware components required for \dword{pd} monitoring and calibration systems will be designed and fabricated by the \dword{sp}-\dword{pds} consortium. 

Cold components (diffusers and fibers) interface with \dword{hv} and are described in a separate interface document (\citedocdb{6721}). Warm components interface the \dword{pd} calibration and monitoring subsystem with the \dword{cisc}~\citedocdb{6730} and \dword{daq}~\citedocdb{6727} subsystems.

A joint development effort with \dword{hv}/\dword{cpa} groups will define the optimization of materials and location of the photon diffusers, fiber routes, connectors location and also the installation procedure of the diffusers and fiber. The feedthrough ports/locations and fiber routing along \dword{dss} will be determined jointly by \dword{sp}-\dword{pds} and cryostat/\dword{dss} groups. The 
calibration and \dword{pds} 
consortia will share rack spaces. Multi-purpose ports are planned to be shared between various groups, calibration devices such as lasers and cameras will make use of them. 
\dword{sp}-\dword{pds}, calibration, and \dword{cisc} will define the ports for deployment. An interlock system to avoid turning on light sources when the \dword{pds} is in operation will be provided.

\subsection{Physics, Software and Computing}

Interfaces with physics, software, and computing are involved in meeting specifications SP-FD-3, SP-FD-4, SP-PDS-2, SP-PDS-5, 
SP-PDS-14, SP-PDS-15, and SP-PDS-16 (see Table \ref{tab:specs:SP-PDS}). The physics topics covered by the
\dword{snb}/low energy and \dword{ndk}/\dword{hep} working groups are the most closely connected to the \single \dword{pds}. The connection stems from the need for self-triggering for \dword{dune} non-beam physics addressed by these two groups. 
However, there are connections to all physics working groups involving \dword{fd} observables, as scintillation light information will improve event reconstruction/classification beyond what is achievable by \dword{tpc} information only. 

Below is a summary of interfaces between the \dword{sp} \dword{pds} and \dword{fd} and \dword{pdsp} simulation and reconstruction groups:

\begin{itemize}
    \item generating photon libraries, and the tools for doing so;
    \item simulating and evaluating performance of physics events;
    \item \single \dword{pds} reconstruction performance studies;
    \item algorithms for matching flashes to \dword{tpc} tracks; and
    \item analyzing the light produced by various species of charged particles.
\end{itemize}

It is critical that the performance specifications for the \dword{pds} meet the needs of the physics and reconstruction teams, both in terms of detector performance and background (including false triggers from radiologicals and light contamination from cryostat light leaks, \dword{hv} system corona discharges, and calibration system effects such as purity monitors, laser flashers and cameras.  These interfaces will be captured here.
Light contamination of any nature must be studied quantitatively so that the impact on error budget due to misclassification of events can be calculated. Quantitative indicators should be established using \dword{pdsp} data, which should also provide the basis for identification algorithms of spurious signals caused by light leakage. 

The \dword{sp} \dword{pds} shares interfaces with the \dword{dune} core computing systems, primarily with databases. The two databases that will have direct interfaces with the \dword{sp}  \dword{pds} are the hardware/\dword{qc} and calibration databases. All the off-line calibration values will be stored in the \dword{dune} calibration database. Additionally, the system will interface with the \dword{dune} hardware database. During all stages of production/procurement and \dword{qc} evaluations of \dword{pds} components, as well as integration and installation of the system, tracking of the hardware, and test results will be stored in the \dword{dune} hardware/\dword{qc} database. The \dword{sp} \dword{pds} consortium will work with the database group to ensure that all schema, applications, and procedures for the database interfaces are developed. As components of the system will originate at multiple institutions, well defined procedures and management will be required to ensure that all data is archived in the \dword{dune} hardware/\dword{qc} database. 

\section{Risks}
\label{sec:fdsp-pd-risks}

Table~\ref{tab:risks:SP-FD-PD} contains a list of all the
risks that we are currently holding in the \dword{pd} risk register.  Each line includes the official \dword{dune} risk register identification number, a description of the risk, the proposed mitigation for the risk, and finally three columns rating the post-mitigation (P)robability that the risk described comes to pass, the degree of (C)ost risk for that line, and the degree of (S)chedule risk.  Risk levels are defined as (L)ow (<10\% probability of occurring, <5\% cost impact, <2 month schedule impact), (M)edium (10 to 25\% probability of occurring, 5\% to 20\% cost impact, 2 to 6 month schedule impact), or (H)igh (>25\% probability of occurring, >20\% cost impact, >6 month schedule impact).  Most of these risks are reduced to a ``Low'' level following mitigation (as shown in the table), although several of them currently hold a higher risk levels (pre-mitigation), due to the early stage of development of the \dword{pd} system relative to other systems.  

In the following sections, we present a narrative description of each of the risks and the proposed mitigation.

\begin{footnotesize}
\begin{longtable}{P{0.18\textwidth}P{0.20\textwidth}P{0.32\textwidth}P{0.02\textwidth}P{0.02\textwidth}P{0.02\textwidth}} 
\caption[PD system risks]{PD system risks (P=probability, C=cost, S=schedule) The risk probability, after taking into account the planned mitigation activities, is ranked as 
L (low $<\,$\SI{10}{\%}), 
M (medium \SIrange{10}{25}{\%}), or 
H (high $>\,$\SI{25}{\%}). 
The cost and schedule impacts are ranked as 
L (cost increase $<\,$\SI{5}{\%}, schedule delay $<\,$\num{2} months), 
M (\SIrange{5}{25}{\%} and 2--6 months, respectively) and 
H ($>\,$\SI{20}{\%} and $>\,$2 months, respectively). \fixmehl{ref \texttt{tab:risks:SP-FD-PD}}} \\
\rowcolor{dunesky}
ID & Risk & Mitigation & P & C & S  \\  \colhline
RT-SP-PD -01 & Additional photosensors and engineering required to ensure PD modules collect enough light to meet system physics performance specifications. & Extensive validation of \dshort{xarapu} design to demonstrate they meet specification. & L & M & L \\  \colhline
RT-SP-PD-02 & Improvements to active ganging/front end electronics required to meet the specified 1~$\mu$s time resolution. & Extensive validation of photosensor ganging/front end electronics design to demonstrate they meet specification. & L & L & L \\  \colhline
RT-SP-PD-03 & Evolutions in the design of the photon detectors due to validation testing experience require modifications of the TPC elements at a late time. & Extensive validation of \dshort{xarapu} design to demonstrate they meet specification and control of PD/APA interface. & L & L & L \\  \colhline
RT-SP-PD-04 & Cabling for PD and CE within the \dshort{apa} frame or during the 2-APA assembly/installation procedure require additional engineering/development/testing. & Validation of PD/APA/CE cable routing in prototypes at Ash River. & L & L & L \\  \colhline
RT-SP-PD-05 & Experience with validation prototypes shows that the mechanical design of the PD is not adequate to meet system specifications. & Early validation of \dshort{xarapu} prototypes and system interfaces to catch problems ASAP. & L & L & L \\  \colhline
RT-SP-PD-06 & pTB WLS filter coating not sufficiently stable, contaminates \dshort{lar}. & Mechanical acceleration of coating wear.  Long-term tests of coating stability. & L & L & L \\  \colhline
RT-SP-PD-07 & Photosensors fail due to multiple cold cycles or extended cryogen exposure. & Execute testing program for cryogenic operation of photosensors including mutiple cryogenic immersion cycles. & L & L & L \\  \colhline
RT-SP-PD-08 & SiPM active ganging cold amplifiers fail or degrade detector performance. & Validation testing if photosensor ganging in multiple test beds. & L & L & L \\  \colhline
RT-SP-PD-09 & Previously undetected electro-mechanical interference discovered during integration. & Validation of electromechanical designin Ash River tests and at \dshort{pdsp2}. & L & L & L \\  \colhline
RT-SP-PD-10 & Design weaknesses manifest during module logistics-handling. & Validation of shipping packaging and handling prior to shipping.  Inspection of modules shipped to site immediately upon receipt. & L & L & L \\  \colhline
RT-SP-PD-11 & PD/CE signal crosstalk. & Validation in \dshort{pdsp}, \dshort{iceberg} and \dshort{pdsp2}. & L & L & L \\  \colhline
RT-SP-PD-12 & Lifetime of \dshort{pd} components outside cryostat. & Specification of environmental controls to mitigate detector aging. & L & L & L \\  \colhline

\label{tab:risks:SP-FD-PD}
\end{longtable}
\end{footnotesize}

\subsection{Physics Performance Specification Risks}
\label{sec:pds-risks-text}

Risk RT-SP-PD-01 in the Table \ref{tab:risks:SP-FD-PD} addresses the performance specification that the \dword{pd} system detect 0.5 pe/MeV of deposited energy.  The system as designed may not reach this requirement during validation, necessitating additional engineering time and possibly additional system cost.  Current design validation (Section~\ref{sec:fdsp-pd-validation}) 
provides firm indication that this specification will be met by the \dword{xarapu}.  Mitigation of this risk is being achieved by allocating enough development resources to the \dword{pd} to continue developing improved light collection modules; increasing the \dword{apa}  slot size to allow for larger modules; or increasing the number of photosensors per \dword{xarapu} supercell.  
The cost risk is rated M because photosensors are a significant cost driver for the project and increasing their number presents a significant medium level cost risk to the system.

Risk RT-SP-PD-02 addresses the performance specification that the \dword{pds} provide \SI{1}{$\mu$s} time resolution.  While the timing resolution specification has been met by the \dword{pdsp} \dword{ssp}-based \dword{sarapu},  cost-saving modifications to the readout electronics could degrade the performance of the \dword{pds} below the \SI{1}{$\mu$s} requirement.  In addition, the combination of active and passive ganging of 48 photosensors could degrade timing performance.  Current design validation (Section~\ref{sec:fdsp-pd-validation}) provides firm indication that this specification will be met by the \dword{xarapu} and our baseline electronics, so a risk level of L is assigned to this risk.  Mitigation of this risk is being achieved by allocating enough engineering resources to proceed rapidly with the design modifications of our reduced-cost baseline system; extensive testing of passive ganging prototypes, including parallel development of two design options for the active ganging circuit; and testing of timing performance in software simulation and multiple validation test stands.

\subsection{Design Risks}

Risk RT-SP-PD-03 addresses the interface of the \dword{apa} and \dword{pd} designs, and the possibility that in order to meet detector performance or reliability specifications, the \dword{pd} design may evolve in a direction requiring modification of the \dword{apa}.  Our current design validation (Section~\ref{sec:fdsp-pd-validation}) provides firm indication that these specifications will be met by the \dword{xarapu}, but we have not yet completed the validation process.  While the design validation at this point is sufficient to reduce the overall risk to low following validation, this remains one of the principle risks we consider due primarily to the significant potential costs (financial and schedule) associated with such a change following the \dword{tdr}.  Mitigation of this risk involves close interaction between the \dword{apa} and \dword{pd} consortia and assigning significant resources to \dword{pd} validation efforts.

Risk RT-SP-PD-04 covers the plans for running \dword{pd} cables within the \dword{apa} frames.  Lessons learned during the \dword{pdsp} led to the re-design of the \dword{pd} cabling layout, moving the cables inside the \dword{apa} frame where they will be unreachable following installation of the \dword{apa} wires.  Additionally, installation of the \dword{apa}s into the cryostat will require making \dword{pd} cable connections between the upper and lower \dword{apa}s underground.  This risk addresses the concern that difficulties with these \dword{apa}/\dword{pd} interfaces will require changes to the cabling plan.  Mitigation consists of extensive validation tests, including full-scale integration tests at the Ash River installation site.

Risk RT-SP-PD-05 concerns the possibility that continuing validation tests demonstrate that the \dword{pd} mechanical design is in some way not adequate to meet \dword{dune}  specifications.  While validation is ongoing and the possibility or a required design change remains, the impact and cost of such a change is likely relatively low.  Mitigation includes continued design validation testing and sufficient engineering resources.

Risk RT-SP-PD-06 concerns the possibility that continuing validation tests demonstrate that the coatings required on the dichroic filter plates are not sufficiently robust in cryogenic applications and flake or dissolve off the surface and contaminate the \dword{lar}, possibly impacting electron lifetime or optical performance of the detector.  Experience in \dword{pdsp} suggested that coatings of the filters is a delicate operation, and the possibility exists to produce unstable coatings.  Mitigation includes continued validation testing of coated filters and sufficient engineering resources.  This is one of the more significant outstanding risks, due to the possibility of negatively impacting the performance of the \dword{tpc}.

Risk RT-SP-PD-07.  One of the most significant lessons of the \dword{pdsp} for the \dword{pd} system was the failure of a significant number of photosensors during module assembly \dword{qc} due to an unannounced change in the manufacturer's photosensor packaging procedures.  Problems developed with initially reliable photosensors mid-way through fabrication, requiring rapid changes to the \dword{pd} design.  This risk addresses the possibility of a re-occurrence of this or a similar problem.  Mitigation includes (but is not limited to) extensive \dword{qa} testing prior to selecting the final photosensor candidate, careful coordination with photosensor vendor(s), and rigorous \dword{qc} testing procedures (including tracking wafer fabrication and packaging batch information from the vendor) for photosensors.  We are in close contact with both candidate photosensor candidates to develop a \dword{qa}/\dword{qc} plan sufficient to address our concerns.


Risk RT-SP-PD-08 addresses the possibility of a degradation in \dword{pd} performance or outright failure due to the cold amplifiers required by the active ganging circuitry.  In order to reach the baseline design of 48 ganged photosensors per \dword{xarapu} supercell, a mix of active and passive ganging is required.  While initial validation testing is very promising, these circuits remain quite new.  Mitigation of this risk involves additional validation testing in bench-top testing and in the \dword{iceberg} test stand.

\subsection{Risks During Integration}

Risk RT-SP-PD-09 addresses the possibility that a previously undetected flaw in the \dword{pd} module design or the integration plan with the \dword{apa}s manifests itself during the integration process.  Steps taken to mitigate this risk include close coordination between the \dword{pd}, \dword{apa}, \dword{ce}, and the integration task force coordinated by the project, including extensive full-scale testing at Ash River and at other integration test sites.

Risk RT-SP-PD-10 covers risks associated with integration into \dword{dune} detectors.  Fabrication and initial testing of \dwords{pd} will occur in Brazil, follow-on testing will occur at the US reception facility prior to storage at the \dword{sdwf}, and additional logistics and handling will occur prior to the modules arriving at the underground integration facility. This risk addresses the possibility that previously undetected weaknesses will be discovered in \dword{qc} testing following receipt of the modules.  Mitigation of this risk includes careful design engineering and testing of shipping and handling procedures.

\subsection{Risks During Installation/Commissioning/Operations}

The biggest risk that could be realized during the commissioning and operations phase is
the observation of excessive noise caused by failure to follow the \dword{dune}
grounding rules.  Risk RT-SP-PD-11 addresses the possibility of discovering such a failure during installation \dword{qc} testing or commissioning of the detector.  The observation of excessive noise in \dword{dune}
would result in a delay of the commissioning and of data taking until the source of the noise is
found and remedial actions are taken. In order to minimize the probability of observing excessive
electronic noise, we plan to enforce the grounding rules throughout the design phase, based on
the lessons learned from the operation of the \dword{pdsp} detector.  In addition, testing at \dword{iceberg} between \dword{pd} and all generations of \dword{ce} electronics will minimize this risk.  

Risk RT-SP-PD-12 addresses \dword{pd} maintenance during operation.   During operation, most \dword{pd} components are inaccessible due to being submerged in \dword{lar}.  However, some components such as the warm readout electronics remain accessible. It is valuable to assign a risk to the need of their requiring spares beyond those planned for, or replacement due to a previously undetected flaw.  
Mitigation steps include two aspects: (1) designing the warm systems to facilitate repair, and (2) performing a careful mean time between failure analysis to predict failure rates over the lifetime of the experiment that will allow the procurement of sufficient spares in the production phase. 
\section{Transport and Handling}
\label{sec:fdsp-pd-install}

A storage facility near or at the \dword{fd} site (the \dword{sdwf}) will be established to allow storage of materials for detector assembly until needed.  Transport of assembled and tested PD modules, electronics, cabling, and monitoring hardware to the \dword{sdwf} is the responsibility of the \dword{pd} consortium.

Following assembly and quality management testing in Brazil, the \dword{pd} modules  will be packaged and shipped to an intermediate testing facility in the US for post-shipping checkout. Following this, the modules will be stored in their shipping containers in the \dword{sdwf}.  Cables, readout electronics, and monitoring hardware will be shipped directly to the \dword{sdwf} and stored until needed underground for integration.

Packaging plans are informed by the \dword{pdsp} experience.  Each \dword{sp} module will be individually sealed into a light-tight anti-static plastic bag.  Bagged modules will be packaged in groups of ten modules (matching the need for a single \dword{apa} transported in a single shipping box), approximately \SI{20}{cm} $\times$ \SI{20}{cm} $\times$ \SI{250}{cm} long.  These shipping boxes will be gathered into larger crates to facilitate shipping.  The optimal number per shipment is being considered.

Documentation and tracking of all components and \dword{pd} modules will be required during the full logistics process. Well defined procedures are in place to ensure that all components/modules are tested and examined prior to, and after, shipping. Information coming from such testing and examinations will be stored in the \dword{dune} hardware database.  Each \dword{pd} module shipping bag will be labeled with a text and barcode label, referencing the unique ID number for the module contained, and allowing linkage to the hardware database upon unpacking prior to integration into the \dword{apa}s underground.

Tests have been conducted and continue to validate environmental requirements for photon detector handling and shipping. The environmental condition specifications for lighting (SP-PDS-3 in Table~\ref{tab:specs:SP-PDS}), humidity (SP-PDS-4 in Table~\ref{tab:specs:SP-PDS}), and work area cleanliness (SP-PDS-1 in Table ~\ref{tab:specs:SP-PDS})
apply for surface and underground transport, storage and handling, and any exposure during installation and integration underground.

Details of \dword{pd} integration into the \dword{apa} and installation into the cryostat, including quality management testing equipment, tests, and documentation are included in Chapter~\ref{ch:sp-install}.

\section{Quality Assurance and Quality Control}
\label{sec:fdsp-pd-qaqc}

The \dword{qa} and \dword{qc} programs for the \dword{fd} are based on our experience with the \dword{pdsp}.  Our design-phase quality management system is based upon that experience.  Following completion of the 60-percent design review, we will develop a quality final assurance program focused on final specifications and drawings, and developing a formal set of fabrication procedures along with detailed \dword{qc} and test plans.

During fabrication, integration into the detector, and detector installation into the cryostat, our \dword{qc} plan will be carefully followed, including incoming materials and other inspection reports, fabrication travelers, and formal test result reports entered into the \dword{dune} \dword{qa}/\dword{qc} database.

Particular steps in this process are detailed below.

\subsection{Design Quality Assurance}
\label{sec:fdsp-pd-designqa}

\dword{pd} design \dword{qa} focuses on ensuring that the detector modules meet the following goals:
\begin{itemize}
\item physics goals as specified in the \dword{dune} requirements document;
\item interfaces with other detector subsystems as specified by the subsystem interface documents; and
\item materials selection and testing to ensure non-contamination of the \dword{lar} volume.
\end{itemize}

The \dword{pds} consortium will perform the design and fabrication of the components in accordance with the applicable requirements of the \dshort{lbnf}-\dshort{dune} \dword{qa} plan. If the institute (working under the supervision of the consortium) performing the work has a documented \dword{qa} program, the work may be performed in accordance with their own program.

Upon completion of the \dword{pds} design and \dword{qa}/\dword{qc} plan, there will be a pre-production review process, with the reviewers charged to ensure that the design demonstrates compliance with the goals above.

\subsection{Production and Assembly Quality Assurance}
\label{sec:fdsp-pd-prodqa}

The \dword{pds} will undergo a \dword{qa} review for all components prior to completion of the design and development phase of the project.  The \dword{pdsp} test will represent the most significant test of near-final \dword{pd} components in a near-\dword{dune} configuration, but additional tests will also be performed.  The \dword{qa} plan will include, but not be limited to, the following areas:

\begin{itemize}
\item materials certification (in the \dword{fnal} materials test stand and other facilities) to ensure materials compliance with cleanliness requirements;
\item cryogenic testing of all materials to be immersed in \dword{lar}, to ensure satisfactory performance through repeated and long-term exposure to \dword{lar}.  Special attention will be paid to cryogenic behavior of fused silica and plastic materials (such as filter plates and wavelength-shifters), \dwords{sipm}, cables and connectors.  Testing will be conducted both on small-scale test assemblies (such as the small test cryostat at CSU) and full-scale prototypes (such as the full-scale CDDF cryostat at CSU). 
\item mechanical interface testing, beginning with simple mechanical go/no-go gauge tests, followed by installation into the \dword{pdsp2} system, and finally full-scale interface testing of the \dword{pds} into the final pre-production \dword{tpc} system models; and
\item full-system readout tests of the \dword{pd} readout electronics, including trigger generation and timing, including tests for electrical interference between the \dword{tpc} and \dword{pd} signals.
\end{itemize}

Prior to beginning construction, the \dword{pds} will undergo a final design review, where these and other \dword{qa} tests will be reviewed and the system declared ready to move to the pre-production phase.

\subsection{Production and Assembly Quality Control}
\label{sec:fdsp-pd-prodqc}


Prior to the start of fabrication, a manufacturing and \dword{qc} plan will be developed detailing the key manufacturing, inspection, and test steps.  The fabrication, inspection, and testing of the components will be performed in accordance with documented procedures. This work will be documented on travelers and applicable test or inspection reports. Records of the fabrication, inspection and testing will be maintained. When a component has been identified as being in noncompliance to the design, the nonconforming condition shall be documented, evaluated, and dispositioned as: \textit{use-as-is} (does not meet design but can meet functionality as it is), \textit{rework} (bring into compliance with design), \textit{repair} (will be brought to meet functionality but will not meet design), and \textit{scrap}. For products with a disposition of accept, as is, or repair, the nonconformance documentation shall be submitted to the design authority for approval.   

All \dword{qc} data  (from assembly and pre- and post-installation into the \dword{apa}) will be directly stored to the \dword{dune} database for ready access of all \dword{qc} data.  Monthly summaries of key performance metrics (to be defined) will be generated and inspected to check for quality trends.

Based on the \dword{pdsp} model, we expect to conduct the following production testing:

Prior to shipping from assembly site:
\begin{itemize}
\item dimensional checks of critical components and completed assemblies to insure satisfactory system interfaces;
\item post-assembly cryogenic checkouts of \dword{sipm} mounting \dwords{pcb} (prior to assembly into \dword{pd} modules);
\item module dimensional tolerances using go/no-go gauge set; and
\item warm scan of complete module using motor-driven \dword{led} scanner (or UV \dword{led}  array).
\end{itemize}

Following shipping to the US reception and checkout facility but prior to storage at \dword{sdwf}:
\begin{itemize}
\item mechanical inspection;
\item warm scan (using identical scanner to initial scan); and
\item cryogenic testing of completed modules (in CSU CDDF or similar facility).
\end{itemize}

Following delivery to integration clean room underground, prior to and during integration and installation:
\begin{itemize}
\item warm scan (using identical scanner to initial scan);
\item complete visual inspection of module against a standard set of inspection points, with photographic records kept for each module;
\item end-to-end cable continuity and short circuit tests of assembled cables; and
\item an \dword{fe} electronics functionality check.
\end{itemize}

\subsection{Installation Quality Control}
\label{sec:fdsp-pd-installqc}

\dword{pds} pre-installation testing will follow the model established for \dword{pdsp}.  Prior to installation in the \dword{apa}, the \dword{pd} modules will undergo a warm scan in a scanner identical to the one at the \dword{pd} module assembly facility and the results compared.  In addition, the module will undergo a complete visual inspection for defects and a set of photographs of selected critical optical surfaces taken and entered into the \dword{qc} record database.  Following installation into the \dword{apa} and cabling, an immediate check for electrical continuity to the \dwords{sipm} will be conducted.

Following the mounting of the \dword{tpc} \dword{ce} and the \dwords{pd}, the entire \dword{apa} will undergo a cold system test in a gaseous argon cold box, similar to that performed during \dword{pdsp}.  During this test, the \dword{pds} system will undergo a final integrated system check prior to installation, checking dark and \dword{led}-stimulated \dword{sipm} performance for all channels, checking for electrical interference with the cold electronics, and confirming compliance with the detector grounding scheme.
\section{Safety}
\label{sec:fdsp-pd-safety}

Safety management practices will be critical for all phases of the photon system assembly, and testing.  Planning for safety in all phases of the project, including fabrication, testing, and installation will be part of the design process.  The initial safety planning for all phases will be reviewed and approved by safety experts as part of the initial design review.  All component cleaning, assembly, testing,  and installation procedure documentation will include a section on safety concerns relevant to that procedure and will be reviewed during the appropriate pre-production reviews.

Areas of particular importance to the \dword{pds} include
\begin{itemize}
\item Hazardous chemicals (particularly \dword{wls} chemicals such as \dword{ptp} used in filter plate coating) and cleaning compounds:  All potentially hazardous chemicals used will be documented at the consortium management level, with materials data safety sheets (MSDS) and approved handling and disposal plans in place.

\item Liquid and gaseous cryogens used in module testing:  Full hazard analysis plans will be in place at the consortium management level for all module or module component testing involving cryogenic hazards, and these safety plans will be reviewed in the appropriate pre-production and production reviews.

\item High voltage safety:  Some of the candidate \dwords{sipm} require bias voltages above \SI{50}{VDC} during warm testing (although not during cryogenic operation), which may be a regulated voltage as determined by specific laboratories and institutions.  Fabrication and testing plans will demonstrate compliance with local \dword{hv} safety requirements at the particular institution or laboratory where the testing or operation is performed, and this compliance will be reviewed as part of the standard review process.

\item UV and \dword{vuv} light exposure:  Some \dword{qa} and \dword{qc} procedures used for module testing and qualification may require use of UV and/or \dword{vuv} light sources, which can be hazardous  to unprotected operators.  Full safety plans must be in place and reviewed by consortium management prior to beginning such testing.

\item Working at heights, underground:  Some aspects of \dword{pds} module fabrication, testing and installation may require working at heights or deep underground. Personnel safety will be an important factor in the design and planning for these operations, all procedures will be reviewed prior to implementation, and all applicable safety requirements at the relevant institutions will be observed at all times.

\end{itemize}

\section{Organization and Management}
\label{sec:fdsp-pd-org}

The \dword{sp} \dword{pd} consortium benefits from the contributions of many institutions and facilities in Europe and North and South America.  Table~\ref{tab:sp-pds-institutes-i}
lists the member institutions. 

\label{sec:fdsp-pd-org-consortium}

\begin{longtable}
{ll}
\caption{PDS consortium institutions}\\ \colhline
\rowcolor{dunetablecolor} Member Institute  &  Country       \\  \toprowrule
Federal University of ABC & Brazil \\ \colhline
State University of Feira de Santana & Brazil \\ \colhline
Federal University of Alfenas Po\c{c}os de Caldas & Brazil \\ \colhline
Centro Brasileiro de Pesquisas F\'isicas & Brazil \\ \colhline
Federal University of Goi\'as & Brazil \\ \colhline
Brazilian Synchrotron Light Laboratory LNLS/CNPEM & Brazil \\ \colhline
University of Campinas & Brazil \\ \colhline
CTI Renato Archer & Brazil \\ \colhline
Federal Technological University of Paran\'a & Brazil \\ \colhline
Universidad del Atlantico & Colombia \\ \colhline
Universidad Sergia Ablada & Colombia \\ \colhline
University Antonio Nari\~{n}o & Colombia \\ \colhline
Institute of Physics CAS & Czech Republic \\ \colhline
Czech Technical University in Prague & Czech Republic \\ \colhline
Universidad Nacional de Assuncion & Paraguay \\ \colhline
Pontificia Universidad Catolica Per\'{u} & Per\'{u} \\ \colhline
Universidad Nacional de Ingineria & Per\'{u} \\ \colhline
University of Warwick & UK \\ \colhline
University of Sussex & UK \\ \colhline
University of Manchester & UK \\ \colhline
Edinburgh University & UK \\ \colhline
Argonne National Laboratory & USA \\\colhline
Brookhaven National Laboratory & USA \\ \colhline
California Institute of Technology & USA \\ \colhline
Colorado State University   &  USA  \\ \colhline
\dshort{fnal}    &   USA    \\ \colhline
Duke University & USA \\ \colhline
Idaho State University & USA \\ \colhline
Indiana University & USA \\ \colhline
University of Iowa & USA \\ \colhline
Louisiana State University & USA \\ \colhline
Massachusetts Institute of Technology & USA \\ \colhline
University of Michigan & USA \\ \colhline
Northern Illinois University & USA \\ \colhline
South Dakota School of Mines and Technology & USA \\ \colhline
Syracuse University & USA \\ \colhline
University of Bologna and INFN & Italy \\ \colhline
University of Milano Bicocca and INFN & Italy \\ \colhline
University of Genova and INFN & Italy \\ \colhline
University of Catania and INFN & Italy \\ \colhline
Laboratori Nazionali del Sud & Italy \\ \colhline
University of Lecce and INFN & Italy \\ \colhline
INFN Milano & Italy \\ \colhline
INFN Padova & Italy \\  \colhline
\label{tab:sp-pds-institutes-i}
\end{longtable}

The \single \dword{pds} consortium follows the typical organizational structure of \dword{dune} consortia:
\begin{itemize}
\item A consortium lead provides overall leadership for the effort and attends meetings of the \dword{dune} Executive and Technical Boards.
\item A technical lead provides technical support to the consortium lead, attends the Technical Board and other project meetings, oversees the project schedule and \dword{wbs}, and oversees the operation of the project working groups.  
\item A Project Management Board composed by the project leads from the participating countries, the consortium leadership team and few ad hoc members, which maintains tight communication between the countries participating in the consortium construction activity.
\end{itemize}

Below the leadership, the consortium is divided up into six working groups, each led by two or three working group conveners (see Table~\ref{tbl:pds-wgs}).  Each working group is charged with one primary area of responsibility within the consortium, and the conveners report directly to the Technical Lead regarding those responsibilities.

\begin{dunetable}[\dshort{pd} working groups]
{ll}
{tbl:pds-wgs}
{PD working groups and responsibilities}
Working Group			 & Responsibilities\\ \toprowrule
Light Collector WG & Mechanical design, materials selection for PD modules\\ \colhline
Photosensors WG & Selection, validation, procuring of photosensors, cold active ganging\\ \colhline
Readout electronics WG & Warm electronics, cable harness, DAQ interface\\ \colhline
Integration and Installation WG & Internal (inter-WG) and external (inter-consortia) interfaces\\ \colhline
Physics and Simulation WG & Physics and simulations studies to determine \dword{pd} specifications\\ \colhline
ProtoDUNE Analysis WG & Validation of PD system in \dshort{pdsp} and \dshort{pdsp2}\\
\end{dunetable}

The working group conveners are appointed by the \dword{pds} consortium lead and technical lead; the structure may evolve as the consortium matures and additional needs are identified.

\subsection{High-Level Schedule}
\label{sec:fdsp-pd-org-cs}

Table \ref{tab:Xsched} lists key milestones in the design, validation, construction, and installation of the \dword{sp} \dword{pds}.  These milestones include external milestones indicating linkages to the main \dword{dune} schedule (highlighted in color in the table), as well as internal milestones such as design validation and technical reviews.

In general, the flow of the schedule commences with a 60\% design review based on module performance testing at \dword{unicamp} and at \dword{iceberg} and integration testing at Ash River.  Additional similar design validation follows, leading to a final design review (FDR).  Following the FDR, 30 modules and required electronics, cabling, and \dword{pd} monitoring system components for \dword{pdsp2} will be built, installed, and validated during a second \dword{protodune} run at \dword{cern}.  Once the data from this test have undergone initial analysis, production readiness reviews will be conducted and module fabrication will begin.

Some parts of the \dword{pds} system, such as the support rails and electrical connectors required in mid-2020 for \dword{apa} assembly and photosensors and filter plates which have a long procurement cycle, will require an abbreviated design review process as detailed in the narrative earlier in this document and shown in the milestone table.

\begin{longtable}
{p{0.75\textwidth}p{0.25\textwidth}}
\caption{PDS consortium schedule}\\ \colhline
\rowcolor{dunetablecolor}Milestone & Date   \\ \toprowrule
60 percent design validation testing complete & May 2020    \\ \colhline 
60 percent design review & May 2020    \\ \colhline 
\dword{prr} for PD rails, cables. connectors & May 2020\\ \colhline
Final design review for PD rails, cables, connectors & July 2020\\ \colhline
Fabrication of PD rails, cables, connectors begins & August 2020\\ \colhline 
Final design validation testing complete & September 2020    \\ \colhline 
Down selection to two photosensor candidates & September 2020\\ \colhline 
Final design review for remaining PD components & September 2020\\ \colhline 
Start of module 0 component production for \dword{pdsp2} & March 2021\\ \colhline
\rowcolor{dunepeach} Start of \dword{pdsp}-II installation& \startpduneiispinstall      \\ \colhline
End of module 0 component production for \dword{pdsp2} & August 2021\\ \colhline 
End of module 0 installation for \dword{pdsp2} & August 2021\\ \colhline
Start of PD installation in  \dword{pdsp}-II & September 2021     \\ \colhline 
Begin procurement of filter plates  & October 2021\\ \colhline
\dword{pdsp2} initial results available & December 2021\\ \colhline
\dword{prr} for photosensors & March 2022\\ \colhline 
Begin procurement of production photosensors  & April 2022\\ \colhline 
\dword{prr} for remaining PD components & May 2022\\ \colhline 
Begin fabrication/procurement of remaining module components & June 2022\\ \colhline 
Begin assembly of PD monitoring system  & January 2022\\ \colhline
\rowcolor{dunepeach} Start of \dword{pddp}-II installation & \startpduneiidpinstall      \\ \colhline
Begin assembly of front-end electronics modules  & March 2022\\ \colhline
\rowcolor{dunepeach}\dshort{sdwf} available& \sdlwavailable      \\ \colhline
Begin assembly of \dshort{xarapu} modules  & July 2022\\ \colhline
\rowcolor{dunepeach}Beneficial occupancy of cavern 1 and \dword{cuc}& \cucbenocc      \\ \colhline
Initial batch (80 PD modules) assembled  & March 2023\\ \colhline
\rowcolor{dunepeach} \dword{cuc} counting room accessible& \accesscuccountrm      \\ \colhline
Initial batch (80 PD modules) arrive at US PD Reception Facility  & June 2023\\ \colhline
Second batch (160 PD modules) assembled  & July 2023\\ \colhline
Initial batch (80 PD modules) arrive at \dshort{sdwf}  & September 2023\\ \colhline
Second batch (160 PD modules) arrive at US PD Reception Facility  & October 2023\\ \colhline
\dword{pd} monitoring system at \dshort{sdwf}   & October 2023\\ \colhline
Third batch (320 PD modules) assembled  & November 2023\\ \colhline
Second batch (160 PD modules) arrive at \dshort{sdwf}  & December 2023\\ \colhline
\rowcolor{dunepeach}Top of \dword{detmodule} \#1 cryostat accessible& \accesstopfirstcryo      \\ \colhline
Third batch (320 PD modules) arrive at US PD Reception Facility  & January 2024\\ \colhline
Front end electronics modules at \dshort{sdwf}   & February 2024\\ \colhline
Fourth batch (320 PD modules) assembled  & February 2024\\ \colhline
Third batch (320 PD modules) arrive at \dshort{sdwf}  & April 2024\\ \colhline
Fourth batch (320 PD modules) arrive at US PD Reception Facility  & May 2024 \\ \colhline
Fifth batch (320 PD modules) assembled  & June 2024\\ \colhline
\rowcolor{dunepeach}Start of \dword{detmodule} \#1 TPC installation& \startfirsttpcinstall      \\ \colhline
Fourth batch (320 PD modules) arrive at \dshort{sdwf}  & August 2024\\ \colhline
Fifth batch (320 PD modules) arrive at US PD Reception Facility  & September 2024 \\ \colhline
Final batch (300 PD modules) assembled  & December 2024\\ \colhline
Fifth batch (320 PD modules) arrive at \dshort{sdwf}  & December 2024\\ \colhline
Final batch (300 PD modules) arrive at US PD Reception Facility  & February 2025 \\ \colhline
Final batch (300 PD modules) arrive at \dshort{sdwf}  & April 2025\\ \colhline
\rowcolor{dunepeach}End of \dword{detmodule} \#1 TPC installation& \firsttpcinstallend      \\ \colhline
\rowcolor{dunepeach}Top of \dword{detmodule} \#2 accessible& \accesstopsecondcryo      \\ \colhline
 \rowcolor{dunepeach}Start of \dword{detmodule} \#2 TPC installation& \startsecondtpcinstall      \\ \colhline
\rowcolor{dunepeach}End of \dword{detmodule} \#2 TPC installation& \secondtpcinstallend      \\ 
\label{tab:Xsched}
\end{longtable}

\subsection{High-Level Cost Narrative}

In the fall of 2018, we completed an initial cost estimate for fabrication of \dword{pd} modules for one \SI{10}{kt} \dword{dune} module and updated the estimate extensively in March/April of 2019.  The estimates are based on \dword{pdsp} costs, modified as necessary for an \dword{xarapu} design.  Vendor quotations or vendor estimates are used for all the major components. 
For fabrication costs, the biggest uncertainties center around the photosensor fabrication; this constitutes approximately half the total \dword{pds} cost.
We have estimates from Hamamatsu for photosensors which would reduce this line by nearly a factor of two, significantly reducing the system cost.  We also have preliminary indications that similar cost savings may also be available from using \dword{fbk} photosensors.  As noted earlier in this \dword{tdr}, a major focus of our remaining development work is focused on realizing these potential savings.

The dichroic filter procurement and coating represent the other major cost driver for the project.  The costing for the filter plates is based on initial contacts with a Brazilian filter firm.  Initial samples of filter substrates have been received at \dword{unicamp} and have been successfully coated and tested through multiple cryogenic cycles with no indication of failure. Extensive additional validation of the Brazilian filters  will occur during late 2019 as part of the \dword{sbnd} module fabrication.

These filter plates are significantly cheaper than the filters manufactured by Omega, Inc. that were tested in our earlier validation studies.  Until these tests are complete, the filter plates remain a significant cost and schedule risk.

Extensive use of design-for-fabrication techniques throughout the module development phase, as well as multiple rounds of prototype development, have allowed us to minimize the component cost for the remaining components.  In-house fabrication and assembly using university shop facilities and student labor for assembly (particularly at \dword{unicamp}) have also reduced costs.

Modification of an existing and well understood readout electronics system has very significantly reduced initial cost estimates for that portion of the system.

\section{Appendix}
\label{appdx:fdsp-pd}

%
\subsection{Simulation}
\label{sec:fdsp-pd-simphys}

The broad performance specifications for the \dword{pds} are determined by a series of physics deliverables addressing the major physics goals of \dword{dune}: nucleon decay searches, supernova burst neutrinos, and beam neutrinos. Detailed subdetector specifications, such as light yield of the light collectors, are determined using a full simulation, reconstruction, and analysis chain developed for the \dword{larsoft} framework. 

\subsubsection{Simulation and Reconstruction Steps} 
\label{subsec:fdsp-pd-simphys-sim}

The first step in the simulation specific to the \dword{pds} is the simulation of the production of light and its transport within the volume to the \dwords{pd}. Argon is a strong scintillator, producing \SI{24000}{$\gamma$s/MeV} at our nominal drift field. Even accounting for the efficiency of the \dwords{pd}, it is prohibitive to simulate every optical photon with \dword{geant4} in every event. So, prior to the full event simulation, the detector volume is voxelized and many photons are produced in each voxel. The fraction of photons from each voxel reaching each photosensor is called the visibility, and these visibilities are recorded in a 4-dimensional library.
This library includes Rayleigh scattering length ($\lambda_R=$ \SI{60}{cm}~\cite{Grace:2015yta}), absorption length ($\lambda_A=$ \SI{20}{m}), and the measured collection efficiency versus position of the double-shift light-guide bars. There is significant uncertainty on the scattering length in the literature, so the value is conservatively chosen at the low end of those reported. With these optical properties, there is a factor of 20 difference in total amount of light collected between events right in front of the photon detectors and those on the far side of the drift volume \SI{3.6}{m} away. 

When a particle trajectory is simulated, the amount of charge and light it produces is calculated in small steps. The light produced in each step is 
distributed onto the various \dwords{pd} using the photon library as a look-up table, and the 30\% early (\SI{6}{ns}) plus 70\% late (\SI{1.5}{$\mu$s}) scintillation time constants are applied. Transport time of the light through the \lar is not currently simulated but is under development. It is not expected to make a significant difference in the studies presented here.

The second step is the simulation of the sensor and electronics response. For the studies shown here, the SensL \dword{sipm} and \dword{sipm} signal processor (\dword{ssp}) readout electronics used for \dword{pd} development and in \dword{pdsp} is assumed (see Section~\ref{sec:fdsp-pd-pde}). However, a range of \dword{s/n} and dark rates are considered in order to set requirements on the needed performance of the electronics.
Crosstalk (where a second cell avalanches when a neighbor is struck by a photon generated internal to the silicon) is introduced by adding a second \phel \num{16.5}\% of the time when an initial \phel is added to the waveform. Additional uncorrelated random noise is added to the waveform with an \dword{rms} of
\SI{0.1}{\phel{}}. The response of the \dword{ssp} self-triggering algorithm, based on a leading-edge discriminator, is then simulated to determine if and when a \SI{7.8}{$\mu$s} waveform will be read out, or in the case of the simulation, 
stored and passed on for later processing.

The third step is reconstruction, which proceeds in three stages. The first is a ``hit finding'' algorithm that searches for peaks on individual waveforms channel-by-channel, identifying the time (based on the time of the first peak) and the total amount of light collected (based on the integral until the hit goes back below threshold). The second step is a ``flash finding'' algorithm that searches for coincident hits across multiple channels. All the coincident light is collected into a single object that has an associated time (the earliest hit), an amount of light (summed from all the hits), and a position on the plane of the \dword{apa} ($y$-$z$) that is a weighted average of the positions of the photon collectors with hits in the flash.  
The final step is to ``match'' the flash to the original event by taking the largest flash within the allowed drift time that is within \SI{240}{cm} in the $y$-$z$ plane. Since the \dword{tpc} reconstruction is still in active development, especially for low-energy events, we match to the true event 
vertex of the event in the analyses presented here. This is a reasonable approximation since the position resolution of the \dword{tpc} will be significantly better than that of the \dword{pds}. 

These tools (or subsets of them) are then used to evaluate how the performance of the \dword{pds} affects the following set of physics deliverables.

\subsubsection{Nucleon Decay}
\label{subsec:fdsp-pd-simphys-ndk}

Nucleon decays are rare events, so excluding backgrounds is of the utmost importance. Since some backgrounds can be generated by cosmic rays passing outside the active detector area, setting a fiducial volume to exclude such events is critically important.

\textit{Fiducialization with \tzero}\nopagebreak

The physics deliverable: the \dwords{pd} must be able to determine \tzero with approximately \SI{1}{\micro s} resolution (SP-FD-4: time resolution) for events with visible energy greater than \SI{200}{MeV} throughout the active volume and do so with $>99\%$ efficiency (SP-FD-3: light yield), as described in \physchndk{}, Section~6.1.4. 
This energy regime is relevant for nucleon decay and atmospheric neutrinos. The time measurement is needed for event localization for optimal energy resolution and rejection of entering backgrounds. 
This resolution is required for comparable spatial resolution to the \dword{tpc} along the drift direction.

\begin{dunetable}[PD system efficiency for nucleon decay events]
{ccc}
{tab:pds-ndk}
{Efficiency for tagging nucleon decay events with the \dword{pds} at the \dword{cpa}, the dimmest region of the detector, which is \SI{3.6}{m} from the \dwords{pd}, shown for range of light yields (LY) at that position. Also shown is the total PD module collection efficiency required for that light yield with the simulated scattering length, \SI{60}{cm}.}
\dword{cpa} Light yield (PE/MeV) & Collection Efficiency  (\%) & Efficiency at the \dword{cpa} (\%) \\
\toprowrule
0.09 & 0.24   & $93.8 \pm 0.4$ \\ \colhline
0.28 & 0.75  & $97.7 \pm 0.4$ \\ \colhline
0.33 & 0.88  & $98.4 \pm 0.2$ \\ \colhline
0.50 & 1.3  & $98.9 \pm 0.2$ \\ 
\end{dunetable}

The physics here feeds down to a requirement on the minimum light yield (SP-FD-3: light yield), determined by measuring how often the correct flash was not assigned to nucleon decay 
events\footnote{The most relevant sample is actually the \textit{background} to nucleon decay events. However, efficiently simulating background that can mimic nucleon decays is challenging since they can be quite rare topologies. It is therefore easier to simulate the nucleon decay signal that should be representative of the background.} 
in the dimmest region of the detector, near the \dword{cpa}. A minimum light yield of \SI{0.5}{PE/MeV} is required to meet the requirement of 99\% efficiency, as shown in Table~\ref{tab:pds-ndk}. 

A light collector with 1.3\% collection efficiency (defined as the probability that a photon reaching the surface of the light collector will be recorded as a \phel) achieves this light yield with the simulated \SI{60}{cm} scattering length. This efficiency is equivalent to having \SI{23}{cm^2} of active area per module with 100\% efficiency. At this scattering length, there is a factor of 20 difference in light yield between the brightest and dimmest regions of the detector, so techniques to improve light yield uniformity (discussed in Appendix~\ref{sec:fdsp-pd-enh}) would reduce the inefficiency still further and ease understanding the detector systematic uncertainties.

\subsubsection{Supernova Neutrinos}
\label{subsec:fdsp-pd-simphys-snb}

Supernova bursts are also rare events, though here the event is made up of many interactions (spread over several seconds) instead of a single interaction. For distant supernovae (at the far side of the Milky Way or in the Large Magellanic Cloud), the top priority is to ensure that the detector can identify a burst when it happens and trigger the detector readout. For nearby supernovae, triggering will not be a challenge, and instead the goal is to record as much information as possible about the burst.

\textit{Burst Triggering}\nopagebreak

The physics deliverable: the \dword{pds} must be able to trigger on \dwords{snb} which produce 50 neutrino interactions in a \SI{10}{kt} volume\footnote{About the amount expected for a burst at the far side of our galaxy.} with almost 100\% efficiency with a false positive rate of less than one per month. This deliverable is most important for distant supernovae where the most important requirement is that we trigger and record the data. If both the \dword{pds} and \dword{tpc} triggers have good efficiency, they can provide redundancy against one another or be combined to increase efficiency or lower the background rate. The once-per-month false positive rate is determined by limits in data handling.

The \dword{pds} trigger performance was studied for a plausible but challenging signal: a supernova burst in the Large Magellanic Cloud, which we conservatively assumed would produce only 10 signal events in the far detector. The trigger efficiency was studied with variations in light yield, dark rate, and signal-to-noise ratio, keeping the requirement from the \dword{daq} that the fake rate be held to less than one per month. The burst trigger efficiency for 10 supernova neutrino events in one \SI{10}{kt} module (a pessimistic prediction for a supernova in the LMC), was found to be approximately $80\%$, and it is relatively insensitive to all these parameters for average light yield $>$\SI{7}{PE/MeV} (equivalent to 0.9\% collection efficiency with the simulated optical properties), dark rate $<$\SI{1}{kHz}, and signal-to-noise $>3$. The uncorrelated noise from dark rate and low signal-to-noise was easily excluded from trigger primitives by the clustering scheme, and the increased light yield makes both backgrounds and signal brighter together, so performance stays basically constant. Thus this physics deliverable, while important, does not constrain any detector requirements.

\textit{\dshort{tpc} Energy Measurement and Time Resolution with \tzero}\nopagebreak

The physics deliverable: the \dwords{pd} must be able to provide \tzero determination with \SI{1}{\micro s} resolution (SP-FD-4: time resolution) for at least 60\% of the neutrinos in a typical \dword{snb} energy spectrum. The \tzero measurements are used in concert with the \dword{tpc}-reconstructed event in two ways: to correct for the attenuation of the charge signal as a function of how far the charge drifts through the \dword{tpc} and to provide more precise absolute event times for resolving short time features in the \dword{snb} neutrino event rate. This deliverable is important primarily for nearby supernovae where the number of events is large enough that time and energy resolution will be the limiting factors in extracting physics, as described in \physchsnb{}. 

\begin{dunefigure}[Supernova neutrino energy resolution from the \dshort{tpc}]
{fig:pds-snb-driftcor}
{The energy resolution for supernova neutrino events when reconstructed by the \dword{tpc} with the drift distance corrected using three assumptions on the performance of the \dword{pds}. The options considered range from drift correction for no events (black), to 60\% of events (blue), to 100\% of events (red).
}
  \includegraphics[width=0.5\textwidth]{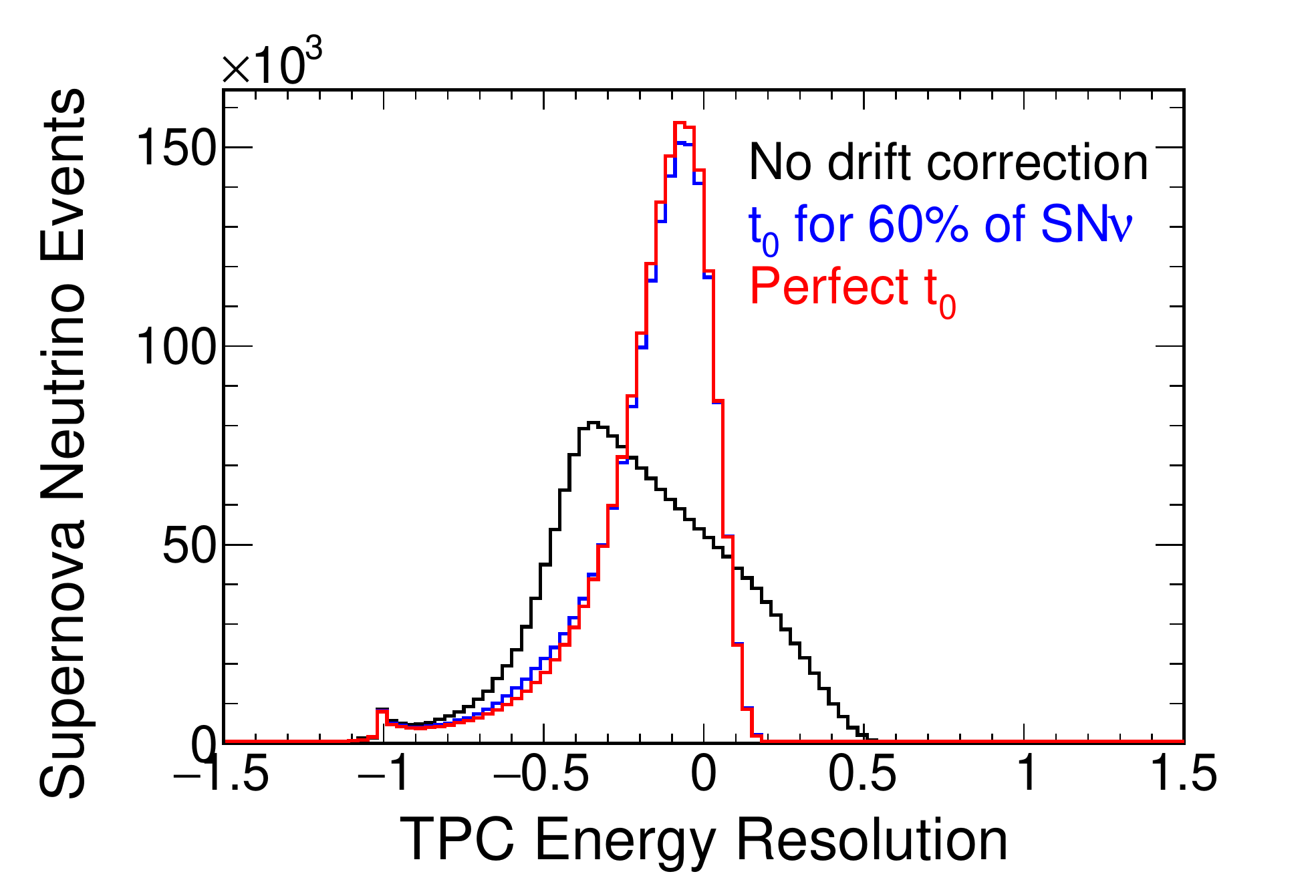}
 \end{dunefigure}

The 60\% \tzero tagging requirement comes from two studies of a typical \dword{snb} neutrino spectrum under varying \dword{pd} performance assumptions: the resolution of the energy reconstructed with the \dword{tpc} and drift-corrected using the time from the \dwords{pd}, and the observability of the in-fall `notch' in the \dword{snb} event time distribution. Both studies show significant improvement when going from no \dwords{pd} to a system that has a collection efficiency of at least 0.25\% (equivalent to \SI{0.5}{PE/MeV} for 60\% of the detector volume), but only marginal improvements past that point, as can be seen in Figure~\ref{fig:pds-snb-driftcor}. The light yield required here is sufficiently low that this deliverable does not set any additional detector requirements.

\textit{Calorimetric Energy}\nopagebreak

Physics deliverable: the \dword{pds} should be able to provide a calorimetric energy measurement for low-energy events, like \dwords{snb}, complementary to the \dword{tpc} energy measurement. 
Improving the energy resolution will enable us to extract the maximum physics from a \dword{snb} (see \physchsnb{}), and with the goal to achieve energy resolution comparable to the \dword{tpc}, we can take full advantage of the anti-correlation between the emission of light and charge signals imposed by the conservation of energy. In addition, this requirement allows the photon detection system to provide redundancy if a supernova occurs during adverse detector conditions. If the argon purification system is offline, the photon signal is significantly less sensitive to electronegative impurities, and if the drift field is low, the reduced charge signal can be partially recovered by increased light.

\begin{dunefigure}[SNB neutrino energy resolution from the PD system]
{fig:pds-snb-calo}
{The energy resolution (determined from the distribution widths of the fraction of difference between reconstructed and true to true neutrino energy for simulated events) for supernova neutrino events when reconstructed directly through \dword{pds} calorimetry for a range of light yields, represented by different colors. The red line labeled \textit{Physics} 
shows the energy smearing inherent to the neutrino interactions and thus serves as a theoretical minimum resolution. The black line shows the energy resolution achieved by the \dword{tpc}, defined in a similar way. The performance improves significantly up until approximately \SI{20}{PE/MeV} where the \dword{pds} and \dword{tpc} give comparable resolution below approximately \SI{7}{MeV}.. 
}
  \includegraphics[width=0.6\columnwidth]{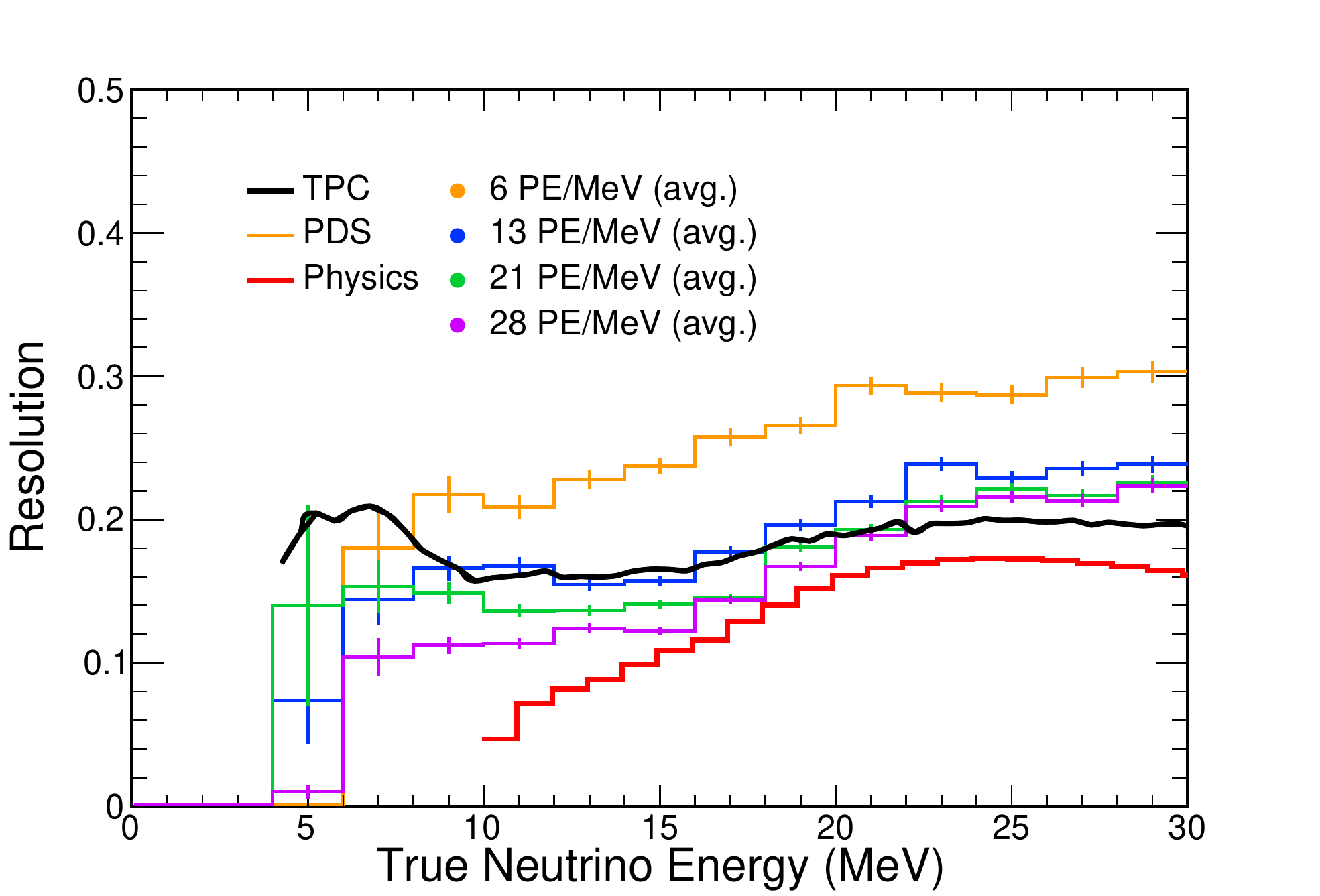}
 \end{dunefigure}

The calorimetric energy performance was studied for supernova burst neutrino events simulated in the far detector for a range of different detector performance assumptions. The energy reconstruction was simple, correcting the total observed amount of photons for the average number of photons expected per MeV as a function of position along the drift direction. Events were required to be well away from the side walls to avoid any possible edge effects. The energy resolution vs. true energy is shown in Figure~\ref{fig:pds-snb-calo}. There is a significant benefit to achieving a photon detector with an average light yield of \SI{20}{PE/MeV}, where the \dword{pds} and \dword{tpc} have comparable resolution for the lowest energy ($<$\SI{7}{MeV}) supernova neutrinos. Past this light yield, the improvement appears to plateau in this analysis. This physics deliverable thus sets a requirement, FD-SP-3: light yield, of \SI{20}{PE/MeV} averaged over the active volume.

While options that can improve the uniformity of the detector are not essential to achieve required resolution, they are likely to improve the calorimetric energy reconstruction above and beyond total light yield. A detector that is more uniform will be easier to calibrate, and the impact of uncertainties on the optical parameters of the liquid argon will be reduced. This effect is potentially important for supernova neutrinos, and certainly more important for the beam neutrino events described in the next section. In addition, for Xe-doping specifically, speeding up the late light will allow for flashes that are narrower in time, reducing the amount of radiological contamination mixed in with the signal, which is of particular importance with these relatively small signals.

\subsubsection{Beam Neutrinos}
\label{subsec:fdsp-pd-simphys-beam}

The \dword{pds} is not required for fiducializing beam neutrino events since the pulsed beam will provide sufficient precision to place the interactions in space. However, the \dwords{pd} can potentially contribute to the energy measurement, and the better timing resolution can help identify Michel electrons from muon and pion decay.

\textit{Calorimetric Energy}\nopagebreak

Physics deliverable: the \dword{pds} should be able to provide a calorimetric energy measurement for high-energy events, like neutrinos from the \dword{lbnf} beam, complementary to the \dword{tpc} energy measurement.
Neutrino energy is an observable critical to the success of the oscillation physics program (see \physchlbl{}),
and a second independent measurement can provide a cross-check that reduces systematic uncertainties or directly improves resolution for some types of events. 

In order to provide a meaningful cross-check, the resolution and uncertainty of the \dword{pds} measurement must be comparable to the calorimetric resolution of the \dword{tpc}. The limit on this measurement will likely come from how well the efficiency of the detector and the optical properties of the argon can be determined (both must be known to approximately 5\% to have a comparable measurement of electron shower energy), which define a program of measurements between now and the operation of the detector rather than requirements on the system itself. The requirement that does flow down from this is that the dynamic range of the system be sufficient to allow for accurate measurement of the amount of light reaching the \dword{pds}. 

Some amount of saturation is tolerable since it can be corrected for using the pulse shape or the neighboring unsaturated channels. However, if the saturation is too large, and too many channels are saturated, the corrections become difficult, so we require that no more than $20\%$ of beam neutrino events have saturating channels (SP-PDS-16: dynamic range), consistent with but looser than the \dword{tpc} requirement of $10\%$.

We studied the likelihood of channels saturating by simulating beam neutrino events in the far detector. The likelihood of saturation depends on the digitization frequency, the dynamic range, and the collection efficiency of the detector design. Assuming the baseline electronics, a 12-bit and \SI{80}{MHz} digitizer, we find the likelihood of saturation vs. average light yield shown in Table~\ref{tab:pds-dynamicrange}. 

\begin{dunetable}[Fraction of beam events with channels that saturate]
{ccc}
{tab:pds-dynamicrange}
{The fraction of beam events which have saturating \dword{pds} channels for different light yields, and the corresponding \dword{pds} collection efficiencies.}
Avg. Light Yield (PE/MeV) & Collection Efficiency (\%) & Saturation Fraction (\%) \\ \toprowrule
6 & 0.88  & 6 \\ \colhline
13 & 1.8   & 13 \\ \colhline
21 & 2.6   & 20 \\ \colhline
28 & 3.5   & 24 \\ 
\end{dunetable}

\textit{\it Michel Electron Tagging}\nopagebreak

Physics deliverable: the \dword{pds} should be able to identify events with Michel electrons from muon and pion decays.
The identification of Michel electrons can improve background rejection for both beam neutrinos and nucleon decay searches. 
Some Michel electrons are difficult to identify with the \dword{tpc} since they appear simultaneous within the time resolution of the \dword{tpc} and colinear with their parent. However, because the \dword{pds} can observe the fine time structure of events in the detector, it can identify Michel electrons that appear separated in time from the main event. While \dword{dune}-specific studies of Michel electron tagging have not been performed, the \dword{lariat} experiment has demonstrated that Michel electrons can be identified and studied using photon signals.


\subsection{Options to Enhance Light Yield Uniformity}
\label{sec:fdsp-pd-enh}

Due to a combination of geometric effects and the impact of Rayleigh scattering, the baseline \dword{sp} \dword{pds} design will result in non-uniformity of light collection along the drift direction. Light emitted from interactions close to the \dword{apa}s has an order of magnitude larger chance of being detected compared to interactions close to the \dword{cpa}.    

Though the designs described in the previous sections will meet the \dword{pd} performance requirements, 
two options for enhancing both the light yield and light yield uniformity are under consideration.  Both approaches mitigate the impact of a short Rayleigh scattering length by converting \SI{127}{nm} scintillation photons to longer wavelength photons with a significantly longer Rayleigh scattering length. 
An increase in uniformity 
will enhance the ability to do calorimetric reconstruction with scintillation light, 
thus enhancing the charge-based energy reconstruction 
and increasing the efficiency of triggering on low energy signals.

These options will be pursued in parallel with the baseline design and may be implemented after appropriate review if resources are available and if they do not interfere with, or produce unacceptable risk for, the baseline design schedule.


\subsubsection{Coated Reflector Foils on the \dshort{tpc} Cathode}
\label{sec:fdsp-pd-enh-cathode}


In this option, scintillation light falling on the cathode plane is converted into the visible wavelengths and reflected.
Installing the foils on the cathode represents the option  with the minimal impact on the current design of the \dword{hv} system (field cage and cathode) and ensures a good uniformity of the light yield across the detector.
This light could then be detected by the \dwords{pd} embedded in the \dword{apa}, improving the overall collection efficiency. This option would require at least a fraction of the light collectors be sensitive to visible light. This sensitivity to visible light can be achieved in two ways: (1) by coating the \dword{xarapu} with \dword{tpb} instead of \dword{ptp}, which results in the same \dword{wls} combination as the double shift bars (whose performance is measured in \dword{pdsp})
and/or (2) by leaving some of the \dword{xarapu} detectors without a  
\dword{wls} coating but with an appropriate dichroic filter. In the former case, the \dwords{pd} are sensitive to both the direct and reflected light, in the latter case only to the reflected light. 

Figure~\ref{fig:ly_with_foils} shows the simulated results of a configuration where 50\% of the \dword{apa} light collectors can 
record both direct scintillation light and the reflected visible light from the \dword{cpa},  and 50\% are left uncoated to maximize uniformity. This results in an enhancement of the total light collection close to the cathode (black points). 

Introducing the foils on the cathode may also enable drift position resolution using only scintillation light. This requires the \dwords{pd} to  
differentiate direct \dword{vuv} light from re-emitted visible light (i.e., requires two 
types of \dword{pd}) and 
sufficient timing of arrival of first light.

Coated reflector foils are manufactured through low-temperature evaporation of \dword{tpb} on dielectric reflectors e.g., 3M DM2000 or Vikuiti ESR. Foils prepared in this manner have been successfully used in dark matter detectors such as WArP~\cite{Acciarri:2008kv}. Recently, they have been shown to work in \dwords{lartpc} at neutrino energies, namely  in the \dword{lariat} test beam detector~\cite{Garcia-Gamez:2017cmu}. In \dword{lariat}, they have been installed on the field cage walls and, during the last run, on the cathode. An alternative solution would be to use Polyethylene Naphthalate (PEN) instead of \dword{tpb}. This wavelength-shifter has a similar emission spectrum to \dword{tpb} \cite{Kuzniak:2018dcf} but is provided in sheets, which could greatly simplify the production and installation. The choice of using PEN depends on demonstrating that its performance holds in \lar; these studies are ongoing. 
The method of foil installation is being developed in collaboration with the \dword{dune} \dword{hv} consortium, with the objective of minimizing the impact on the \dword{cpa} design.

A run has been performed using the \dword{cern} FLIC \SI{50}{l} prototype \dword{tpc}, with
the \dword{dune}-like resistive cathode covered with a non-perforated DM2000
foil evaporated with \dword{tpb}. No obvious \dword{hv} problems were observed, but 
the data is still being analyzed to understand whether any field
distortions were present.  A second run with the cathode coated with PEN
was performed in March 2019. The comparison of effects on the electric field between the two solutions is in progress. Preliminary studies show that PEN seems to work as a wavelength-shifter at liquid argon temperatures but may not be as efficient as \dword{tpb}. A future run will involve running with a perforated \dword{tpb}-coated foil. The presence of the holes maintains the resistive character of the cathode and minimizes the effect of electric field distortions.

\begin{dunefigure}[Predicted light yield with \dshort{wls}-coated reflector foils on the CPA]{fig:ly_with_foils}
{Predicted light yield in the \dword{pds} with \dword{wls}-coated reflector foils on the \dword{cpa}. Blue points represent direct \dword{vuv} light impinging on the \dwords{pd} assuming a 2.5\% photon detection efficiency and 70\% wire mesh transmission and half of the detectors left uncoated; red stars - represent scintillation light that has been wavelength-shifted and reflected on the \dword{cpa} assuming the same photon detection efficiency folded in with an 80\% transmittance of the filters to visible light. Black points show the sum of these two contributions.}
\includegraphics[width=0.5\columnwidth]{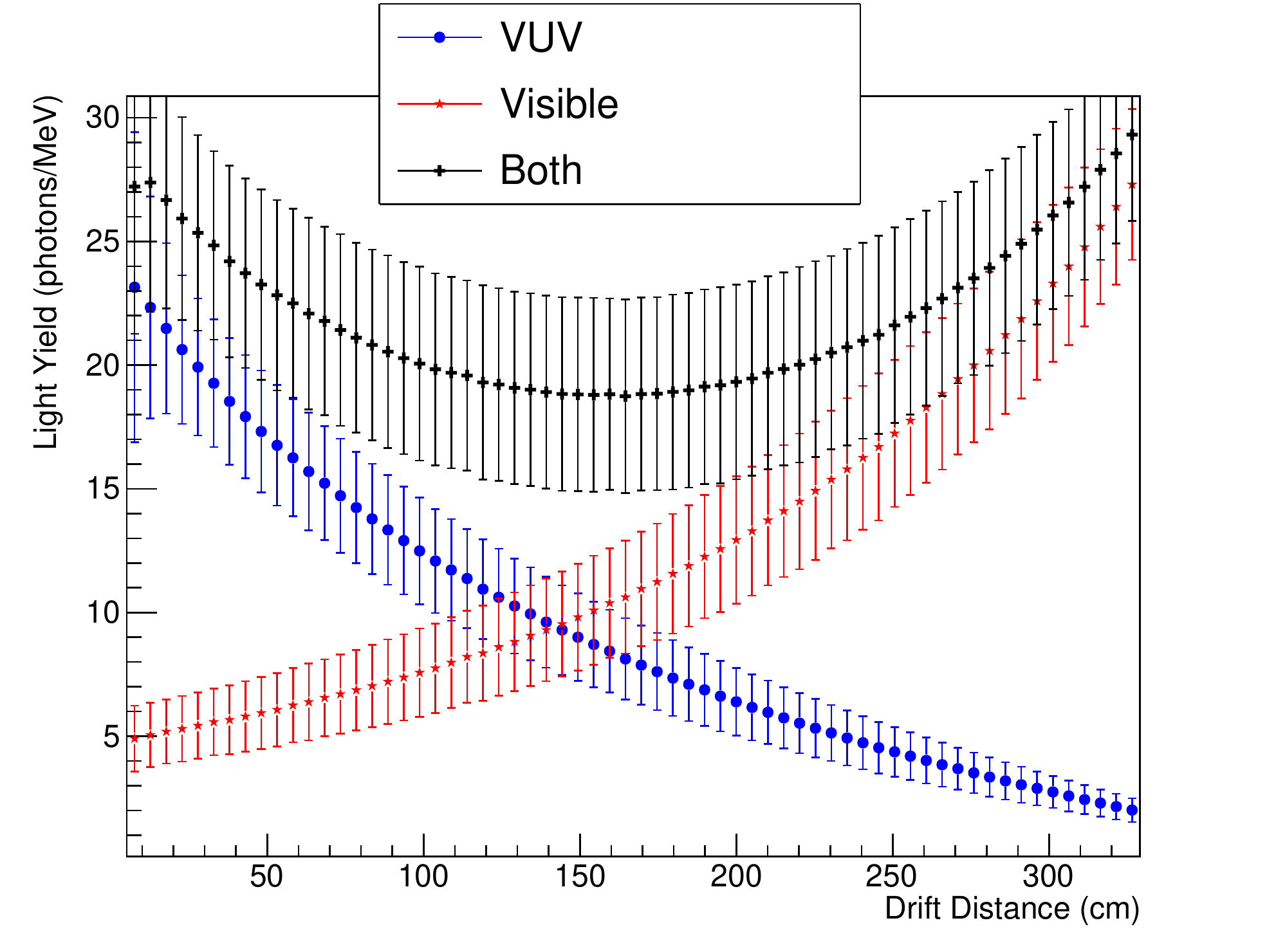}
\end{dunefigure}

\subsubsection{Doping Liquid Argon with Trace Parts of Xenon}
\label{sec:fdsp-pd-enh-xenon}

This option exploits the conversion of the \dword{lar} \SI{127}{nm} light to \SI{175}{nm} by doping the \lar volume with 20-100 ppm of xenon.  While there are indications that the absolute light yield in xenon-doped argon may be higher than in pure argon, in the current estimates, we assume the yields are the same. In this case, the source of the improved performance described here is the much longer Rayleigh scattering length for \SI{175}{nm} light.  The improvement is illustrated in Figure~\ref{fig:visibility_with_xenon} from a \dword{dune} \dword{pd} simulation, assuming an absorption length for the scintillation light of \SI{20}{m}. The gain in average yield for events near the \dword{cpa} is about a factor of five.

\begin{dunefigure}[Simulation of \lar scintillation light with and without Xe doping in a \dshort{spmod}]
{fig:visibility_with_xenon}
{Simulation of visibility of \SI{128}{nm} (\dword{lar} with xenon doping) and \SI{176}{nm} (\dword{lar} scintillation) light in a \dword{spmod}.}
\includegraphics[width=0.5\columnwidth]{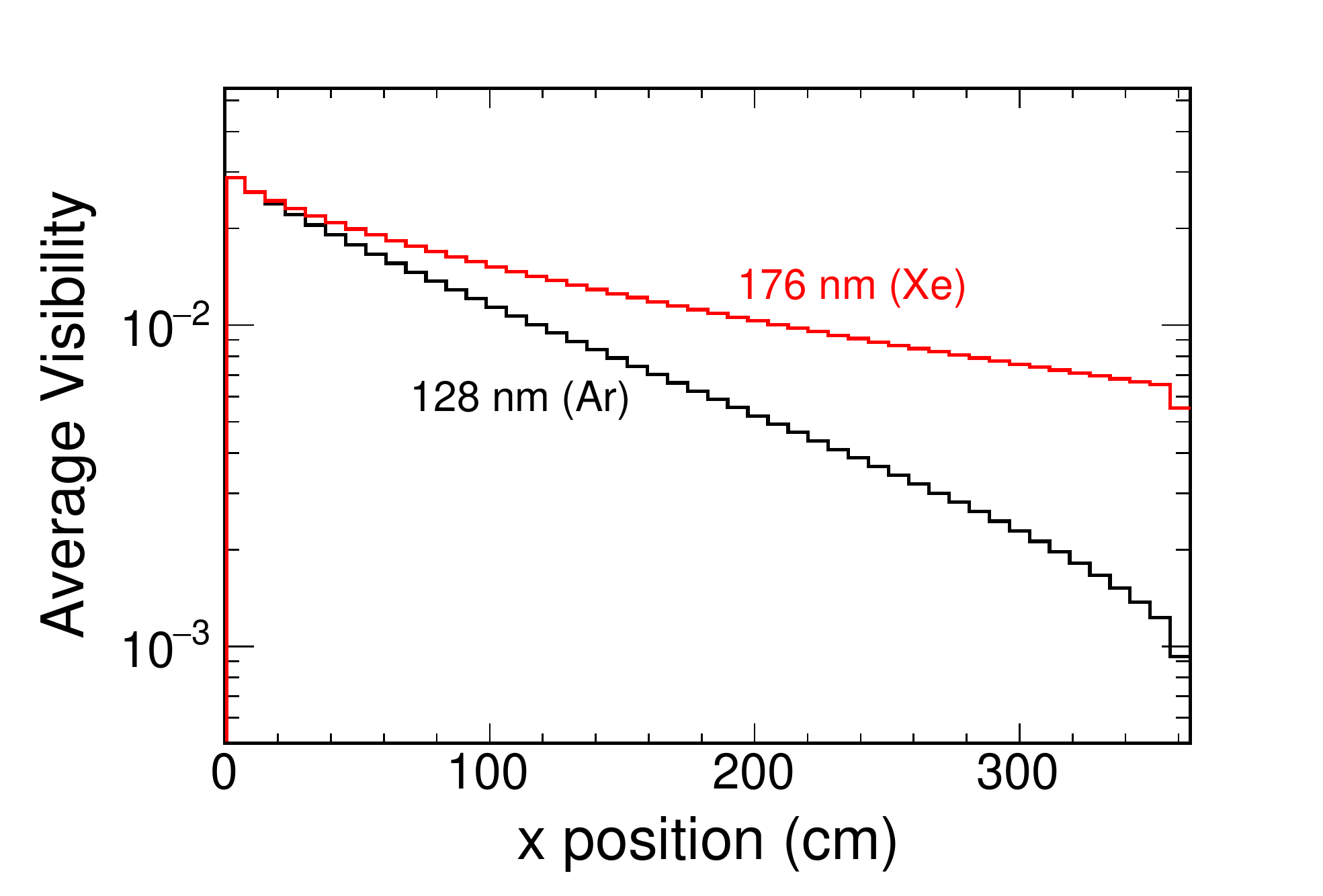}
\end{dunefigure}

Doping with xenon also affects the time structure of the scintillation light and in particular reduces the fraction of late light.  Having a light signal of shorter duration can bring advantages both in physics, such as making it easier to tag Michel electrons from pions and electrons, and in the electronics required. The longer wavelength of the scintillation light resulting from the Xe doping allows the possibility of simplifying the design of the \dword{pds} light collectors (\dword{xarapu}) by dispensing with the use of the outer layer of wavelength shifting material, thereby reducing costs and simplifying the handling of the light collectors during storage and installation. 


Doping the argon with xenon is facilitated by the fact that at the \dword{dune} \dword{fd} the argon is transported from the surface to underground as gas before it is re-condensed for delivery to the cryostats. Xenon and argon can therefore be mixed in gas form before condensation;  in consultation with the cryogenic experts, we have identified locations where this mix could be achieved. Since the operations take place at room temperature, the implementation is relatively straightforward.

{\it\bf Critical Issues and R\&D Work}

Xenon doping must not adversely affect the performance of the \dword{tpc}, and while the doping is expected to be neutral or even beneficial, its effects on charge yield, drift lifetime, and \dword{hv} stability need to be established.  There is experience in xenon doping of argon both at \dword{cern} and \dword{fnal}, and tests for the \dword{tpc} effects are being designed.  
More detailed R\&D is needed to optimize the xenon doping fraction and its interaction with the light-detection system. This can be conducted on a time scale of about a year by a small number of dedicated investigators using resources that, mostly, are expected to be available at \dword{fnal} and \dword{cern}. 


\cleardoublepage

\chapter{Calibration Hardware for Single-Phase}
\label{ch:sp-calib}

\section{Introduction}
\label{sec:sp-calib-intro}


A detailed understanding of the overall detector response is essential for achieving \dword{dune} physics goals. The precision with which each calibration parameter must be measured is spanned by the requirements on the 
systematic uncertainties for the \dword{lbl} 
and \dword{snb} physics programs at \dword{dune}. The calibration program must generally provide measurements at the few-percent-or-better 
level stably across an enormous volume and over a long period and provide sufficient redundancy. This chapter focuses on describing the dedicated calibration hardware systems to be deployed for the \dword{dune} \dword{spmod} that provide necessary information beyond the reach of external measurements and existing sources and monitors. 

A detailed description of the calibration strategy for the \dword{dune} \dword{fd} is provided in \physchtools of this \dword{tdr}. In brief, 
the 
calibration strategy 
uses existing sources of particles, external measurements, and dedicated external calibration hardware systems. Existing calibration sources for \dword{dune} include beam or atmospheric neutrino-induced samples, cosmic rays, argon isotopes, and instrumentation devices such as \dword{lar} purity and temperature monitors. Dedicated calibration hardware systems consist of laser  and neutron source deployment systems.  External measurements by \dword{protodune2} and \dword{sbn} experiments  will validate techniques, tools, and the design of systems applicable to the \dword{dune} calibration program. These sources and systems provide measurements of the detector response model parameters, or provide tests of the response model itself. Calibration measurements can also provide corrections to data, data-driven efficiencies, systematics, and particle responses.

The dedicated calibration hardware systems for the \dword{spmod} include an ionization laser system, a \phel laser system, and a \dlong{pns} system. The possibility of deploying a radioactive source system is also currently being explored. The responsibility of the calibration hardware systems falls under the joint \dword{sp} and \dword{dp} calibration consortium, which was formed in November 2018.

Section~\ref{sec:sp-calib-overview} discusses general aspects driving the calibration program: scope, requirements and data taking strategy.
The baseline calibration hardware designs are described in Section~\ref{sec:sp-calib-systems} and respective subsections. 

Section~\ref{sec:sp-calib-sys-las-ion} describes the baseline design for the ionization laser system that provides an independent, fine-grained measurement of the electric field throughout the detector, which is an essential parameter that affects the spatial and energy resolution of physics signals. 
Volume~\volnumberphysics{}, \voltitlephysics{},  of this \dword{tdr} 
assumes that the \dword{fv} is known to the \SI{1}{\%} level. Through measurements of the spatial distortions and drift velocity map, the laser calibration system mainly helps define the detector \dword{fv}, thus allowing for the correct prediction of the \dword{fd} spectra. The laser system also offers many secondary uses such as alignment checks, stability monitoring, and diagnosing detector performance issues. 
Possible electron lifetime measurements are under study. 
With the goal of knowing precisely the direction of the laser beam tracks, an independent 
\dword{lbls}
is also planned, and is described in Section~\ref{sec:sp-calib-sys-las-loc}.
Alternative designs for the ionization laser system that may improve the physics capability and/or reduce overall cost are also under development and are described in Appendix, Section~\ref{sec:sp-calib-laser-alter}. 
Section~\ref{sec:sp-calib-sys-las-pe} describes the \phel laser system that can be used to rapidly diagnose electronics or \dword{tpc} response issues along with many other useful measurements such as integrated field across drift, drift velocity, and electronics gain. 

Section~\ref{sec:sp-calib-sys-pns} describes the baseline design for the \dword{pns} system, which provides a triggered, well defined, energy deposition from neutron capture in Ar detectable throughout the detector volume. Neutron capture is an important component of signal processes for \dword{snb} and \dword{lbl} physics, enabling direct testing of the detector response spatially and temporally for the low-energy program and the efficiency of the detector in reconstructing the low-energy spectra. A spatially fine-grained measurement of electron lifetime is also planned with this source.
The proposed \dword{rsds} described in the Appendix, Section~\ref{sec:sp-calib-sys-rsds}, 
is in many ways complementary to the \dword{pns}
system, and can provide at known locations inside the detector a source of gamma rays in the same energy range of \dword{snb} and solar neutrino physics. 
The \dword{rsds} is the only calibration system that could probe the detection capability for single isolated solar neutrino events and study how well radiological backgrounds can be suppressed. In contrast, the \dword{pns} is externally triggered and does not provide such a well defined source location for gamma rays inside the detector. On the other hand, the \dword{pns} can probe the uniformity of the full detector, while the \dword{rsds} could only scan the ends of the detector.

For all the calibration hardware systems, the goal is to deploy prototype designs and validate them at \dword{protodune2} during the post long shutdown 2 (LS2) running  at \dword{cern}. The validation plan for calibration systems at \dword{protodune2} and other experiments is described in Section~\ref{sec:sp-calib-val}. 

Section~\ref{sec:sp-calib-intfc} describes interfaces calibration has with other \dword{dune} consortia, especially  with \dword{daq} which are described in more detail in Section~\ref{sec:sp-calib-daqreq}. 

Sections~\ref{sec:sp-calib-const} and \ref{sec:sp-calib-org-manag} conclude the chapter with descriptions of the aspects related to construction and installation of the systems, as well as organizational aspects, including schedule and milestones, discussed in Section~\ref{sec:sp-calib-sched}.

\section{Calibration Overview}
\label{sec:sp-calib-overview}

This section focuses on the general aspects of calibrations in \dword{dune}: the scope of the consortium activities and 
planned systems; the physics and performance requirements driving the design; and the overall strategy for usage of the systems, in combination with natural sources.

\subsection{Scope}
\label{sec:sp-calib-scope}

The scope of the calibration consortium includes a laser ionization system, a \phel laser system, a 
\dlong{lbls}, 
and a \dlong{pns} system. In addition, the consortium is evaluating a \dlong{rsds}.
The calibration consortium is responsible for design through commissioning in the \dword{spmod} for these calibration devices and their associated feedthroughs. Validating the designs of calibration systems at \dword{protodune2} (and other experiments as relevant) is also included under the scope of the consortium. Figure~\ref{fig:calib-scope_chart_v2} shows the subsystems included under the calibration consortium.

Chapters 3, 4, 5, and 8 of Volume~\volnumbersp{}, \voltitlesp{}, of this \dword{tdr} describe other hardware essential for calibration such as \dword{ce} external charge injection systems, \dword{hv} monitoring devices, \dword{pds} stability monitoring devices, and cryogenics instrumentation and detector monitoring devices, respectively. The scope of these systems is described by their respective consortia, and the calibration consortium has substantial interfaces with these consortia. 
The use of other calibration sources such as external measurements and existing sources of particles (e.g., muons, pions) is discussed in the calibration section of \dword{tdr} \physchtools. 
 
We are pursuing the effects of calibration on physics and related studies. 
Calibrations also require simulations (e.g., \efield) to identify desirable locations for calibration devices in the cryostat, away from regions of high \efield, so that their presence does not induce large field distortions. 
The design of the calibration systems and understanding the related physics requires coordination with other consortia and groups. This is discussed in Section~\ref{sec:sp-calib-intfc}.

\begin{dunefigure}[Calibration consortium subsystem chart]{fig:calib-scope_chart_v2}
{Calibration consortium subsystem chart. CTF stands for Calibration Task Force.}
\includegraphics[height=4.0in]{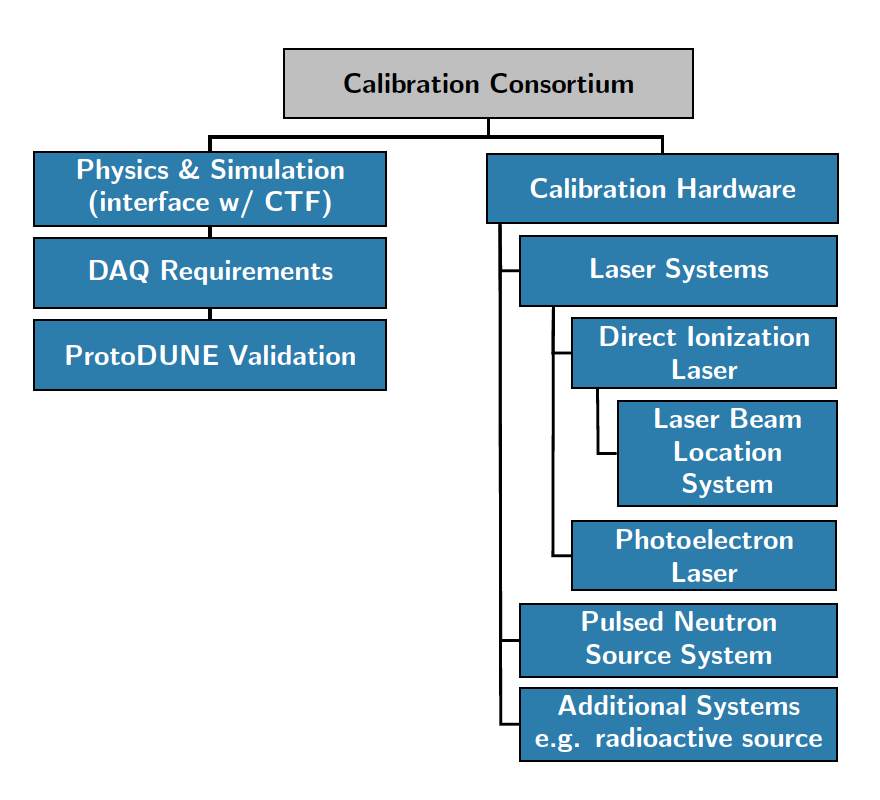}
\end{dunefigure}

\subsection{Design Considerations and Requirements}
\label{sec:sp-calib-requirements}
Some common design considerations for calibration devices include stability, reliability, and longevity, so calibration systems can be operated for the lifetime of the experiment (\dunelifetime). Such longevity is uncommon for any device, so the overall design permits replacing devices where possible, namely the parts that are external to the cryostat. The systems must also adhere to relevant global requirements of the \dword{dune} detector. Table~\ref{tab:specs:SP-CALIB} shows the top-level overall requirements for calibration subsystems along with global \dword{dune} requirements that are relevant for calibration. For example, \dword{dune} requires the \efield  on any instrumentation device inside the cryostat to be less than 30 kV/cm to minimize the risk of dielectric breakdown in \dword{lar}. Another consideration important for event reconstruction is understanding the maximum tolerable level of noise on the readout electronics due to calibration devices and implementing proper grounding schemes to minimize it. 
\dword{pdsp} is evaluating this. In Table~\ref{tab:specs:SP-CALIB}, two values are quoted for most of the parameters: 1) specification, which is the minimum requirement to guarantee baseline performance, and 2) goal, an ideal requirement 
for achieving improved precision.

For the ionization laser system, the energy and position reconstruction requirements for physics measurements lead to requirements for the necessary precision in measuring the \dword{tpc} \efield as well as its spatial coverage and granularity. The precision of the \efield measurement with the laser system must be about \SI{1}{\%} so that the effect from \efield on the collected charge, via the dependence of the recombination factor on \efield, is well below \SI{1}{\%}. This is also motivated by consistency with the high level \dword{dune} specification of \SI{1}{\%} on field uniformity throughout the volume for component alignment and the \dword{hv} system. For laser coverage, to keep the \efield measurement at the $\sim$\SI{1}{\%} level, we are aiming for a coverage of \SI{75}{\%} or more of the total \dword{fv}. The requirement on granularity for the laser is estimated based on the \dword{fv} uncertainty requirements (\SI{1}{\%}) and corresponding uncertainty requirements (\SI{1.5}{\cm}) in each coordinate. A specification is set for a voxel size of \num{30}$\times$\num{30}$\times$\SI{30}{\cubic\cm}, that should be sufficient to satisfy the \dword{fv} uncertainty requirements. A goal is set for \num{10}$\times$\num{10}$\times$\SI{10}{\cubic\cm}, which could allow for a refinement in precision in some detector regions. 

The laser beam location must also meet the level of reconstruction requirement in each coordinate, approximately \SI{5}{\milli\m}. In order to reach that over distances of up to \SI{20}{\m}, where the latter is the maximum distance that any beam needs to travel to cover all detector voxels, this results in a stringent alignment requirement of \ang{0.015} (or \SI{0.25}{\mrad}) on the pointing precision. The laser beam location system is also designed to check the beam location with a precision of \SI{5}{\milli\m} over distances of up \SI{20}{\m}. 
The data volume for the ionization laser system must be 
no more than \num{184}~TB/year/\SI{10}{\kt}, assuming \num{800}k laser pulses, \num{10}$\times$\num{10}$\times$\SI{10}{\cubic\cm} voxel sizes, a \SI{100}{\micro\s} zero suppression window, and two dedicated calibration campaigns per year.

For the \dword{pns} system, the system must provide sufficient neutron event rate to make spatially separated precision measurements across the detector of a comparable size to the voxels probed by the laser (\num{30}$\times$\num{30}$\times$\SI{30}{\cubic\cm}) for most regions of the detector (\SI{75}{\%}). 
For the \dword{snb} program, the sensitivity to distortions of the neutrino energy spectrum depends on the uncertainties in the detection threshold and the reconstructed energy scale and resolution. Studies discussed in the physics \dword{tdr} present target ranges for the uncertainties in these parameters~\cite{bib:docdb14068} as a function of energy. The measurements with the \dword{pns} system aim to provide response corrections and performance estimates, so those uncertainty targets are met throughout the whole volume. This ensures that each voxel has sufficient neutron event rate (percent level statistical uncertainty).

In terms of data volume requirements, the \dword{pns} system requires at least \num{144}~TB/year/\SI{10}{\kt} assuming \num{e5} neutrons/pulse, \num{100} neutron captures/\si{\cubic\m},
and \num{130} observed neutron captures per pulse, and two calibration runs per year. 

Table~\ref{tab:fdgen-calib-all-reqs} shows the full set of requirements related to all calibration subsystems. More details on each of the requirements can be found under dedicated subsections.   


\begin{footnotesize}
\begin{longtable}{p{0.12\textwidth}p{0.18\textwidth}p{0.17\textwidth}p{0.25\textwidth}p{0.16\textwidth}}
\caption{Calibration specifications \fixmehl{ref \texttt{tab:spec:SP-CALIB}}} \\
  \rowcolor{dunesky}
       Label & Description  & Specification \newline (Goal) & Rationale & Validation \\  \colhline

   \newtag{SP-FD-1}{ spec:min-drift-field }  & Minimum drift field  &  $>$\,\SI{250}{ V/cm} \newline ($>\,\SI{500}{ V/cm}$) &  Lessens impacts of $e^-$-Ar recombination, $e^-$ lifetime, $e^-$ diffusion and space charge. &  ProtoDUNE \\ \colhline

  \newtag{SP-FD-2}{ spec:system-noise }  & System noise  &  $<\,\SI{1000}\,e^-$ &  Provides $>$5:1 S/N on induction planes for  pattern recognition and two-track separation. &  ProtoDUNE and simulation \\ \colhline
    
   \newtag{SP-FD-5}{ spec:lar-purity }  & Liquid argon purity  &  $<$\,\SI{100}{ppt} \newline ($<\,\SI{30}{ppt}$) &  Provides $>$5:1 S/N on induction planes for  pattern recognition and two-track separation. &  Purity monitors and cosmic ray tracks \\ \colhline

  \newtag{SP-FD-7}{ spec:misalignment-field-uniformity }  & Drift field uniformity due to component alignment  &  $<\,1\,$\% throughout volume &  Maintains APA, CPA,  FC orientation and shape. &  ProtoDUNE \\ \colhline

  \newtag{SP-FD-9}{ spec:apa-wire-spacing }  & APA wire spacing  &  \SI{4.669}{mm} for U,V; \SI{4.790}{mm} for X,G &  Enables 100\% efficient MIP detection, \SI{1.5}{cm} $yz$ vertex resolution. &  Simulation \\ \colhline

  \newtag{SP-FD-11}{ spec:hvs-field-uniformity }  & Drift field uniformity due to HVS  &  $<\,\SI{1}{\%}$ throughout volume &  High reconstruction efficiency. &  ProtoDUNE and simulation \\ \colhline

  \newtag{SP-FD-13}{ spec:fe-peak-time }  & Front-end peaking time  &  \SI{1}{\micro\second} &  Vertex resolution; optimized for \SI{5}{mm} wire spacing. &  ProtoDUNE and simulation \\ \colhline

  \newtag{SP-FD-22}{ spec:data-rate-to-tape }  & Data rate to tape  &  $<\,\SI{30}{PB/year}$ &  Cost.  Bandwidth. &  ProtoDUNE \\ \colhline

  \newtag{SP-FD-23}{ spec:sn-trigger }  & Supernova trigger  &  $>\,\SI{95}{\%}$ efficiency for a SNB producing at least 60 interactions with a neutrino energy >10 MeV in 12 kt of active detector mass during the first 10 seconds of the burst. &  $>\,$95\% efficiency for SNB within 20 kpc &  Simulation and bench tests \\ \colhline

  \newtag{SP-FD-24}{ spec:local-e-fields }  & Local electric fields  &  $<\,\SI{30}{kV/cm}$ &  Maximize live time; maintain high S/N. &  ProtoDUNE \\ \colhline

  \newtag{SP-FD-25}{ spec:non-fe-noise }  & Non-FE noise contributions  &  $<<\,\SI{1000}\,e^- $ &  High S/N for high reconstruction efficiency. &  Engineering calculation and ProtoDUNE \\ \colhline

  \newtag{SP-FD-26}{ spec:lar-impurity-contrib }  & LAr impurity contributions from components  &  $<<\,\SI{30}{ppt} $ &  Maintain HV operating range for high live time fraction. &  ProtoDUNE \\ \colhline

  \newtag{SP-FD-27}{ spec:radiopurity }  & Introduced radioactivity  &  less than that from $^{39}$Ar &  Maintain low radiological backgrounds for SNB searches. &  ProtoDUNE and assays during construction \\ \colhline
    
   \newtag{SP-FD-29}{ spec:dp-det-uptime }  & Detector uptime  &  $>\,$98\% \newline ($>\,$99\%) &  Meet physics goals in timely fashion. &  ProtoDUNE \\ \colhline
    
   \newtag{SP-FD-30}{ spec:dp-det-mod-uptime }  & Individual detector module uptime  &  $>\,$90\% \newline ($>\,$95\%) &  Meet physics goals in timely fashion. &  ProtoDUNE \\ \colhline

  \newtag{SP-CALIB-1}{ spec:efield-calib-precision }  & Ionization laser \efield measurement precision  &  $\SI{1}{\%}$ &  \efield affects energy and position measurements. &  ProtoDUNE and external experiments. \\ \colhline
    
   \newtag{SP-CALIB-2}{ spec:efield-calib-coverage }  & Ionization laser \efield measurement coverage  &  $>\SI{75}{\%}$ \newline ($\SI{100}{\%}$) &  Allowable size of the uncovered detector regions is set by the highest reasonably expected field distortions, \SI{4}{\%}. &  ProtoDUNE \\ \colhline
    
   \newtag{SP-CALIB-3}{ spec:efield-calib-granularity }  & Ionization laser \efield measurement  granularity  &  $~30\times 30\times 30~$\SI{}{\centi\metre\cubed} \newline ($10\times 10\times 10~$\SI{}{\centi\metre\cubed}) &  Minimum measurable region is set by the maximum expected distortion and position reconstruction requirements. &  ProtoDUNE \\ \colhline
    
   \newtag{SP-CALIB-4}{ spec:laser-location-precision }  & Laser beam location precision  &  $~\SI{0.5}{\milli\radian}$ \newline ($<\SI{0.5}{\milli\radian}$) &  The necessary spatial precision does not need to be smaller than the APA wire gap. &  ProtoDUNE \\ \colhline
    
   \newtag{SP-CALIB-5}{ spec:neutron-source-coverage }  & Neutron source coverage  &  $>\SI{75}{\%}$ \newline ($\SI{100}{\%}$) &  Set by the energy resolution requirements at low energy. &  Simulations \\ \colhline
    
   \newtag{SP-CALIB-6}{ spec:data-volume-laser }  & Ionization laser data volume per year (per 10 kt)  &  $>\SI{184}{TB/yr/10 kt}$ \newline ($>\SI{368}{TB/yr/10 kt}$) &  The laser data volume must allow the needed coverage and granularity. &  ProtoDUNE and simulations \\ \colhline
    
   \newtag{SP-CALIB-7}{ spec:data-volume-pns }  & Neutron source data volume per year (per 10 kt)  &  $>\SI{144}{TB/yr/10 kt}$ \newline ($>\SI{288}{TB/yr/10 kt}$) &  The pulsed neutron system must allow the needed coverage and granularity. &  Simulations \\ \colhline

\label{tab:specs:SP-CALIB}
\end{longtable}
\end{footnotesize}

\begin{dunetable}
[Full specifications for calibration subsystems]
{p{0.45\linewidth}p{0.25\linewidth}p{0.25\linewidth}}
{tab:fdgen-calib-all-reqs}
{Full list of Specifications for the Calibration Subsystems.}   
Quantity/Parameter	& Specification	& Goal		 \\ \toprowrule      

Noise from calibration devices	 & $\ll$ 1000 enc   & \\ \colhline    Max. \efield near calibration devices & < 30 kV/cm & <15 kV/cm \\ \colhline                     

\textbf{Direct Ionization Laser System} &    &   \\ \colhline   
\efield measurement precision & 1\% & <1\% \\ \colhline
\efield measurement coverage & > 75\% & 100\% \\ \colhline
\efield measurement granularity & < \num{30}x\num{30}x\num{30}~cm & \num{10}x\num{10}x\num{10}~cm \\ \colhline
Top field cage penetrations (alternative design) & to achieve desired laser coverage & \\ \colhline
Data volume per 10~kton & 184 TB/year & 368 TB/year \\ \colhline
Longevity, internal parts	& \dunelifetime		& > \dunelifetime   \\    \colhline     
Longevity, external parts	& 5 years			& > \dunelifetime   \\ \colhline 
\textbf{Laser Beam Location System} & & \\ \colhline  
Laser beam location precision & 0.5~mrad & 0.5~mrad \\ \colhline
Longevity	& \dunelifetime		& > \dunelifetime   \\    \colhline     
\textbf{Photoelectron Laser System}	   &   &  \\ \colhline       
Longevity, internal parts	& \dunelifetime		& > \dunelifetime   \\    \colhline 
Longevity, external parts	& 5 years			& > \dunelifetime   \\ \colhline 
\textbf{Pulsed Neutron Source System}	   &   &  \\ \colhline        
Coverage & > 75\% & 100\% \\ \colhline
Data volume per 10~kton & 144~TB/year & 288~TB/year
\\ \colhline 
Longevity	& 3 years			& > \dunelifetime   \\
\end{dunetable}

\subsection{Strategy}
\label{sec:sp-calib-strategy}

Once the far detector is filled and at the desired high voltage, it immediately becomes live for all non-beam physics signals, so it is important to tune the detector response model with calibration data as early as possible. Moreover, since both beam and non-beam physics data will have a fairly uniform rate, regular calibrations in order to monitor space and time dependencies are also needed.

Following those considerations, the strategy for calibration data taking will be organized around three specific periods:
\begin{description}
\item[Commissioning] As soon as the detector is full, with \dword{hv} on and the \dword{daq} operational, it is useful to take laser calibration data. The main goal is to help identify problems in the \dword{apa} wires or the electronics channels, or large cool-down distortions. Depending on how long the ramp-up will take, it could be useful to take data before the \dword{hv} reaches the nominal level, because we can identify problems earlier and possibly learn about dependency of various detector parameters with \efield.
\item[Early data] During the early stages of data-taking, the goal is to do the fullest possible fine-grained laser and neutron calibration (\efield map, lifetime, low energy scale/resolution response) as early as possible, so that all the physics can benefit from a calibrated detector from day 1. 
These results should be combined at a later stage with detector-wide average measurements with cosmics. 
\item[Stable data-taking] The main goal of 
calibrations during stable data-taking is to track possible variations of detector response parameters, and contribute to constraining the 
detector systematics. We expect to combine fine-grained, high statistics scans at regular time intervals -- twice a year for laser, six times for the pulsed neutron source -- with more frequent coarse-grained scans (e.g., photoelectron laser, large voxel ionization laser scan). These, combined with analysis of cosmic ray and radiological backgrounds data, can alert to the need of additional fine scans in particular regions.
\end{description}

\section{Calibration Systems}
\label{sec:sp-calib-systems}
\dword{dune} plans to build two primary systems dedicated to 
calibrate the \dword{spmod} -- a laser system
and a \dlong{pns} system -- both of which require interfaces with the cryostat, that are described in Section~\ref{sec:sp-calib-cryostat}. 

The laser system is aimed at 
determining the essential detector model parameters with high spatial and time granularity. The primary goal is to provide maps of the drift velocity and \efield, following a position-based technique already proven in other \dword{lartpc} experiments. 
Two laser sub-systems are planned. 
With high intensity coherent laser pulses, charge can be created in long straight tracks in the detector by direct ionization of \dword{lar} with the laser beams. This is described in Section~\ref{sec:sp-calib-sys-las-ion}. An auxiliary system aimed at an independent measurement and cross-check of the laser track direction is described in Section~\ref{sec:sp-calib-sys-las-loc}.
On the other hand, laser excitation of targets placed on the cathode creates additional charge from well-defined locations that can be used 
as a general \dword{tpc} monitor and to measure the integrated drift time. This is described in Section~\ref{sec:sp-calib-sys-las-pe}. 

The \dword{pns} system 
provides a ``standard candle'' neutron capture signal (\SI{6.1}{\MeV} multi-gamma cascade) across the entire \dword{dune} volume that is directly relevant to the supernova physics signal characterization thus validating the performance of the detector in the low energy regime. The \dword{pns} system is described in detail in Section~\ref{sec:sp-calib-sys-pns}.  

The physics motivation, requirements and design of these systems are described in the following subsections. Alternative designs
for the ionization laser system, pulsed neutron source system, as well as the proposed radioactive source deployment system, are described in Sections~\ref{sec:sp-calib-laser-alter}, \ref{sec:sp-calib-pns-alt}, and \ref{sec:sp-calib-sys-rsds} of the Appendix, respectively.

\subsection{Cryostat Configuration for Calibration Systems}
\label{sec:sp-calib-cryostat}

Figure~\ref{fig:calib-FTmap} shows the current cryostat design for the 
\dword{spmod} with penetrations for various subsystems. The penetrations dedicated to calibration are the highlighted black circles. 

In addition to these dedicated ports, there are plans to use the \dword{dss} and cryogenics ports (orange and blue dots in Figure~\ref{fig:calib-FTmap}) as needed to route cables for other calibration systems, e.g., fiber optic cables for the \dword{pd} calibration system, which is described in Chapter~\ref{ch:fdsp-pd}. \dword{dss} and cryogenics ports can be accommodated by feedthroughs with a CF63 side flange for this purpose.   

\begin{dunefigure}[Cryostat penetration map with calibration ports]{fig:calib-FTmap}
{Top view of the \spmod 
cryostat showing various penetrations. Circles highlighted in black are multi-purpose calibration penetrations. The green dots are \dword{tpc} signal cable penetrations. The blue ports are cryogenics ports. The orange ports are \dword{dss} penetrations. The larger purple ports at the four corners of the cryostat are human access ports.}
\includegraphics[height=2.0in]{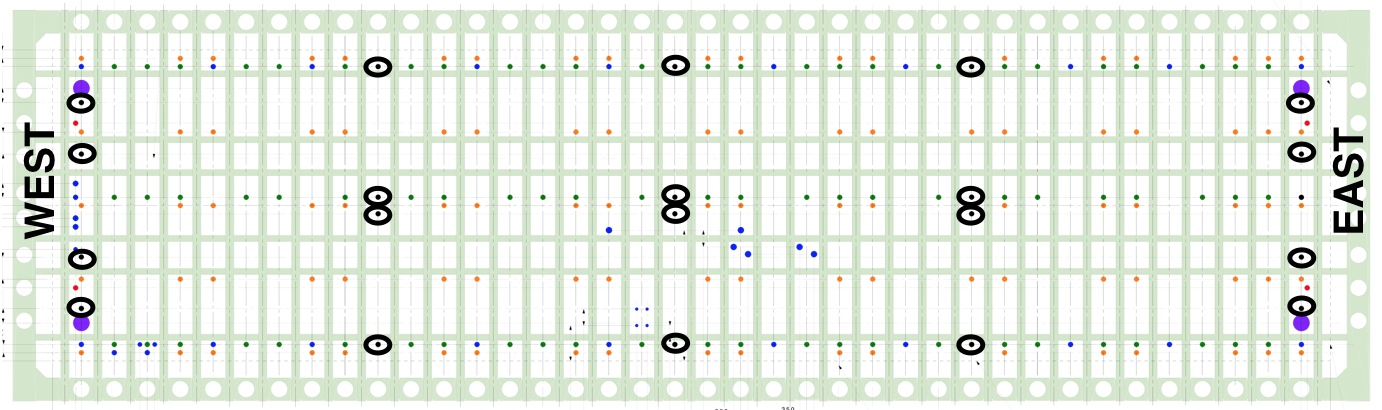}
\end{dunefigure}

The current plan is to use the calibration ports for several different purposes, but their placement is largely driven by requirements for the ionization track laser. 
The ports 
toward the center of the cryostat are placed near the \dword{apa}s, where the \efield is small, 
to minimize any risks due to 
\dword{hv} discharge. \dword{hv} is not an issue for the far east and west ports since they are located outside the \dword{fc} and the penetrations are located 
close to mid-drift (a location favorable for possible source deployment).
Implementing the baseline ionization track laser system as 
described in Section~\ref{sec:sp-calib-sys-las-ion} requires \num{12} feedthroughs, the three central ones in each of the four \dword{tpc} drift volumes; this arrangement allows lasers to be used for full volume calibration of the \efield and associated diagnostics (e.g., \dword{hv}). 

The distance between any two consecutive feedthrough columns shown in Figure~\ref{fig:calib-FTmap} is approximately \SI{15}{\m}. Since the \dword{microboone} laser system has shown that tracks will propagate over that detector's full \SI{10}{\m} length, this distance is considered reasonable. Assuming that the effects of Rayleigh scattering and self-focusing (Kerr effect) do not limit the laser track length, this laser arrangement could illuminate the full volume with crossing tracks in the central region, and single tracks 
in the region closer to the end-walls.  
At this time, the maximum usable track length is unknown, and it may be that the full \SI{60}{\m} \detmodule length could be covered by the laser system after optimization.

Throughout this chapter, the following convention for the coordinate axes will be used: $x$ is parallel to the drift direction, $y$ is the vertical, and $z$ is parallel to the beamline. This is illustrated in Chapter~\ref{ch:sp-cisc} Figure~\ref{fig:cfd-example-geometry}.


\subsection{Laser Calibration: Ionization System}
\label{sec:sp-calib-sys-las-ion}

Through its effect on drift velocity, recombination, and lifetime, the \efield is a critical parameter for physics signals as it ultimately affects the spatial resolution and energy response of the detector. The primary purpose of a laser system is to provide an independent, fine-grained estimate of the \efield in space and time.
It would be extremely valuable to achieve measurements of electron lifetime with the laser system, but the feasibility of that is still under discussion. 
The R\&D plan in \dword{protodune2} will address the feasibility of carrying out charge-based measurements which, if successful, would open up the possibility of using the laser to measure electron lifetime. So, except where specifically indicated, the rest of this section will focus on 
drift velocity and \efield measurement.

\subsubsection{Physics Motivation}
Because it measures spatial distortions of straight tracks, the laser system actually measures the local drift velocity field directly and helps define the detector \dword{fv}, and this in itself is an important input for the \dword{lbl} analysis. 
However, it is still important to use information independent of the charge in order to disentangle effects like lifetime and recombination from \efield distortions. The laser system can do this, by using the position information to derive the \efield from the local velocity map, taking into account the colinearity between both vectors, and the relatively well studied relation between the magnitude of the drift velocity and the \efield, considering a temperature dependence (see~\cite{Li:2015rqa} and references [29, 45-58] therein). A laser system also has the intrinsic advantage of being immune to recombination effects, thus eliminating particle-dependent effects.  

Several sources may distort the \efield temporally  and/or spatially in the detector. Current simulation studies indicate that positive ion accumulation and drift (space charge) due to ionization sources such as cosmic rays or \Ar39 is small in the \dword{dune} \dword{fd}, causing \efield distortions of at most \SI{0.1}{\%}~\cite{bib:mooney2018}.
However, not enough is known yet about the fluid flow pattern in the \dword{fd} to exclude the possibility of stable eddies that may amplify the effect for both \single and \dual modules. 
This effect can be further amplified significantly in the \dword{dpmod} due to accumulation in the liquid of ions created by the electron multiplication process in the gas phase.
Additionally, other sources in the detector (especially detector imperfections) can cause \efield distortions. For example, \dword{fc} resistor failures, non-uniform resistivity in the voltage dividers, \dword{cpa} misalignment, \dword{cpa} structural deformations, and \dword{apa} and \dword{cpa} offsets and  deviations from flatness can create localized \efield distortions. These effects are presented in Figures~\ref{fig:efield_cpa_distortions_boyu2017} and \ref{fig:efield_resistorfailure_mooney2019}, showing the effect of a few \% on the \efield from \SI{2}{\cm} \dword{cpa} position tilts and 
up to \SI{4}{\%} from \dword{fc} single resistor failures.

\begin{dunefigure}[Impact on \efield of \dshort{cpa} position distortions]{fig:efield_cpa_distortions_boyu2017}
{Illustration of a possible distortion of the \dword{cpa} position~\cite{bib:yu2017a}, assuming a \SI{2}{\cm} swing, and its impact on \efield (right).}
\includegraphics[width=0.8\textwidth]{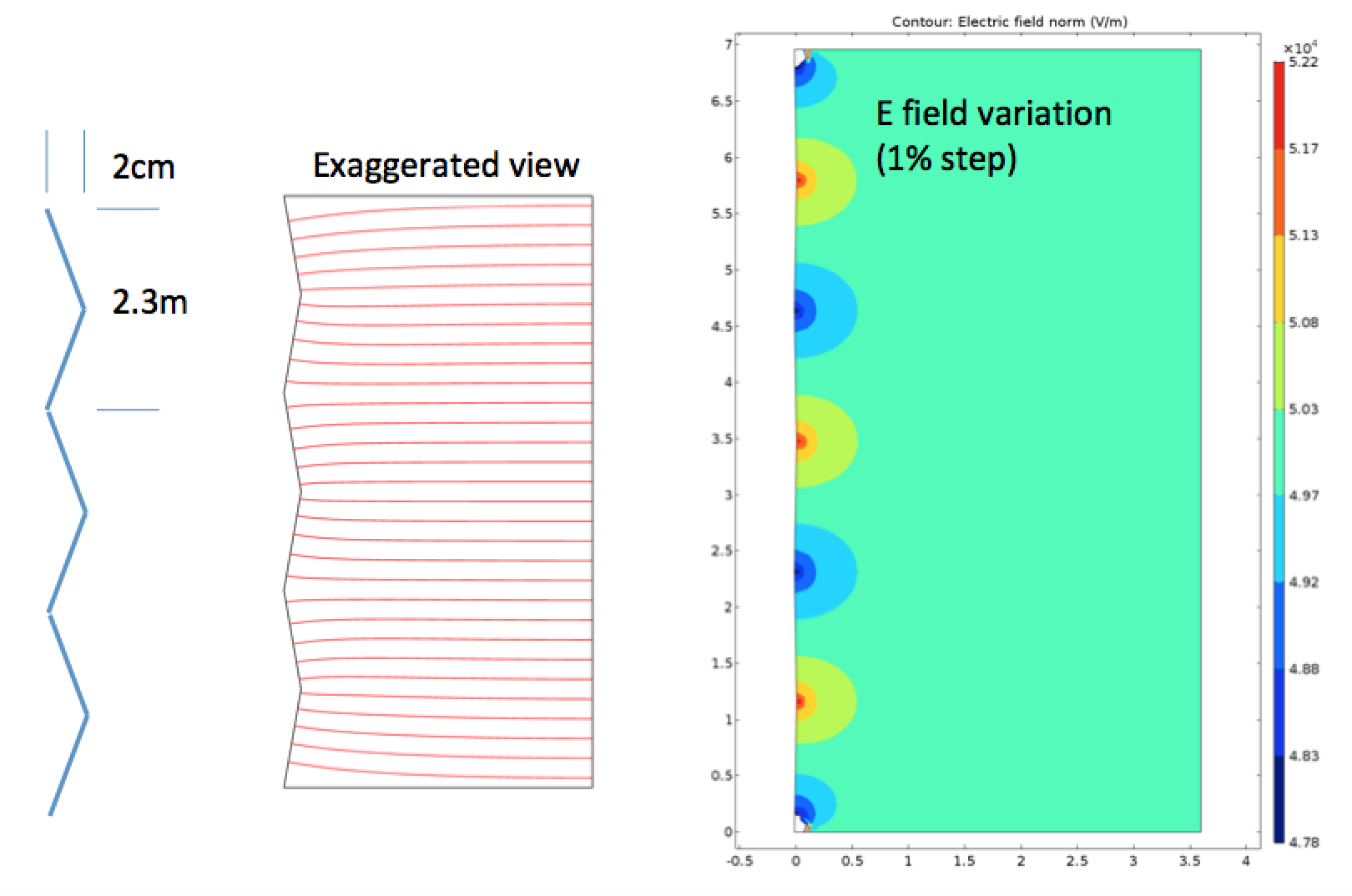}
\end{dunefigure}

\begin{dunefigure}[Impact on \efield of \dshort{fc} resistor failures]{fig:efield_resistorfailure_mooney2019}
{Impact on \efield magnitude distortions of a single \dword{fc} resistor failure~\cite{bib:mooney2019a}.}
\includegraphics[width=0.5\textwidth]{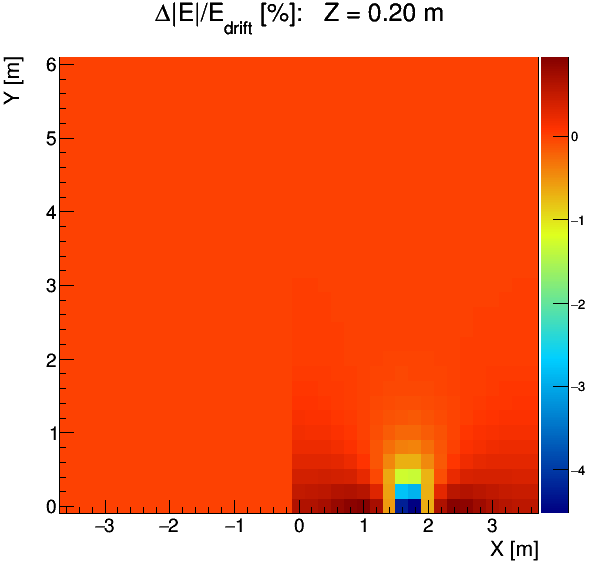}
\end{dunefigure}

In both \single and \dual modules, 
a resistor failure will create significant, local \efield distortions that must be identified. In the \dword{dpmod}, 
four resistors would have to fail to cause a failure across the \dword{fc} gap, but even one failure in the \dword{spmod} can have an effect; this may be partly, but not completely, mitigated by modifying the \dword{hv}. While the resistor failure will be detected temporally, its location in space is not possible to determine from slow controls monitoring data. Misalignments of detector objects or deformations may also create \efield distortions; while individual effects may be small, it is possible to have a combined, significant effect.
Each individual \efield distortion may add in quadrature with other effects, and can aggregate up to \SI{4}{\%} under certain conditions. Understanding all these effects requires in situ  measurement of \efield for proper calibration. 

Useful secondary uses of laser include alignment (especially modes that are weakly constrained by cosmic rays),
stability monitoring, and diagnosing detector performance issues
(e.g., \dword{hv}).  
Misalignment may include physical deformation and/or rotations of objects within the detector. Given the expected low rate of cosmic ray events (about 3500/day/10-kt, inclusive) at the underground location, calibration with cosmic rays is not possible over short time scales. Even over long time scales, certain alignment directions  are difficult to assess with cosmic rays alone, such as distortions of the detector that preserve the gap widths and do not shift the \dword{apa}s in $x$ near the gaps relative to one another.
These distortions include global shifts and rotations in the locations of all detector elements, and crumpling modes where the edges of the \dword{apa}s hold together but angles are slightly different from nominal.   

With respect to electron lifetime, the preliminary results from \dword{pdsp} purity monitors and cosmic ray analyses indicate significant variations with time and space, both between monitors at different vertical coordinates (see Chapter~\ref{ch:sp-cisc}), and between the regions inside and outside the \dword{tpc}. The possibility of carrying out such measurements with the ionization laser is therefore quite interesting. The ArgonTUBE experiment obtained lifetime measurements with laser~\cite{Ereditato:2013xaa} compatible with the cosmic ray ones, but it is not clear yet if this is possible at very large scales, since the modelling of the density of ionization charge created along the tracks presents challenges related to the previously mentioned self-focusing. Therefore the characterization of the ionization charge density from laser tracks will be an important goal of the development plan in \dword{protodune2}.


\subsubsection{Requirements}
\label{sec:sp-calib-laser-req}

The energy and position reconstruction requirements for physics measurements lead to requirements on the necessary precision of the laser 
\efield measurement, its spatial coverage and granularity. The next sections discuss the rationale behind each requirement, which we take as the \dword{dune} specification.

\paragraph{\efield precision:}

In the \dword{lbl} and high-energy range, \physchlbl of this \dword{tdr}
states that the calibration information must provide approximately \num{1} to \SI{2}{\%} understanding of normalization, energy scale and resolution, and position resolution within the detector.
Because a smaller \efield leads to higher electron-ion recombination and therefore a lower collected charge, distortions of the \efield can introduce
energy scale bias. To connect this
to a specification for the necessary precision of the \efield measurement, we note that, via recombination studies~\cite{bib:mooney2018}, we expect a \SI{1}{\%} distortion on \efield to lead to a \SI{0.3}{\%} bias on collected charge.
Because other effects will contribute to the lepton energy scale uncertainty budget, we consider a goal for the 
laser system to measure the \efield to a precision of $\sim$\SI{1}{\%} so that its effect on the collected charge is well below \SI{1}{\%}.
This is also motivated by consistency with the high level DUNE specification on field uniformity throughout the volume due to component alignment and \dword{hv} system, that 
is set at \SI{1}{\%}.
Together with two other high-level \dword{dune} specifications, the \dword{apa} wire spacing (\SI{4.7}{\mm}) and the front end peaking time (\SI{1}{\micro\s}), the effect of this \efield precision requirement on engineering parameters of the calibration laser system is discussed further 
in Section~\ref{sec:sp-calib-sys-las-ion-meas}.

\paragraph{\efield measurement coverage:}

In practice, measuring the \efield  throughout the whole volume of the \dword{tpc} will be difficult, so we must establish a goal for the coverage and granularity of the measurement. 
Until a detailed study of the propagation of the coverage and granularity into a resolution metric is available, a rough estimate of the necessary coverage can be made as follows.

Assuming \SI{4}{\%} as the maximum \efield distortion 
that is 
expected from a compounding of multiple possible effects in the \dword{dune} \dword{fd} 
as described in the previous section,
we can then ask what would be the maximum acceptable size of the spatial region uncovered by the calibration system, if a distortion of that magnitude (systematically biased in the same direction) were present in that region. Our criterion of acceptability is to keep the overall \efield distortion, averaged over the whole detector, at the \SI{1}{\%} level. 
To meet this requirement, the aforementioned spatial region should be no larger than \SI{25}{\%} of the total fiducial volume. Therefore, we aim to have a coverage of \SI{75}{\%} or more.

In addition, we need to consider that the method used to estimate \efield distortions is based on obtaining position displacement maps~\cite{bib:uBlaser2019}, and that the comparison between the reconstructed and true direction of a single track does not 
unambiguously determine a specific displacement map. Having tracks coming from different origins crossing in the same position is a direct way to eliminate that ambiguity, since the displacement vector is given simply by the vector connecting the intersections of the two reconstructed and the two true tracks. A joint iterative analysis of several close-by tracks is the default method for all other positions, but the system design should allow for the maximum possible number of positions 
for crossing tracks from different beams.

\paragraph{\efield measurement granularity:}

Volume~\volnumberphysics~(\voltitlephysics) of this \dword{tdr} states that a \dword{fv} uncertainty of \SI{1}{\%} is required. 
This translates to a position uncertainty of \SI{1.5}{\cm} in each coordinate (see Chapter~\ref{ch:fdsp-apa}). 
In the $y$ and $z$ coordinates, position uncertainty is given mainly by the \dword{apa} wire pitch, and since this is about \SI{4.7}{\mm}, the requirement is met. In the drift ($x$) direction, the position is calculated from timing, and considering the electronics peaking time of \fepeaktime, the uncertainty should be even smaller.

The position uncertainty, however, also depends on the \efield, via the drift velocity. Because the position distortions accumulate over the drift path of the electron, it is not enough to specify an uncertainty on the field. We must accompany it by specifying the size of the spatial region of that distortion. For example, a \SI{10}{\%} distortion would not be relevant if it was confined to a \SI{2}{\cm} region and if the rest of the drift region was at nominal field.
Therefore, what matters is the product of [size of region] $\times$ [distortion]. Moreover, one can distinguish distortions into two types:
\begin{enumerate}
\item Those affecting the magnitude of the field. Then the effect on the drift velocity $v$ is also a change of magnitude. According to the function provided in \cite{Walkowiak:2000wf}, close to \SI{500}{\V\per\cm}, the variation of the velocity with the field is such that a \SI{4}{\%} variation in field $E$ leads to a \SI{1.5}{\%} variation in $v$.
\item Those affecting the direction of the field. Nominally, the field $E$ should be along $x$, so $E = E_L$ (the longitudinal component). If we consider that the distortions introduce a new transverse component $E_T$, in this case, this translates directly into the same effect in the drift velocity, which gains a $v_T$ component, $v_T=v_L  E_T/E_L $, i.e., a \SI{4}{\%} transverse distortion on the field leads to a \SI{4}{\%} transverse distortion on the drift velocity.
\end{enumerate}

Thus, a \SI{1.5}{\cm} shift comes about from a constant \SI{1.5}{\%} distortion in the velocity field over a region of \SI{1}{\m}. In terms of \efield, that could be from a \SI{1.5}{\%} distortion in $E_T$ over \SI{1}{\m} or a \SI{4}{\%} distortion in $E_L$ over the same distance.

\efield distortions can be caused by space-charge effects due to accumulation of positive ions caused by \Ar39 decays (cosmic ray rate is low in \dword{fd}), or detector defects, such as \dword{cpa} misalignments (Figure~\ref{fig:efield_cpa_distortions_boyu2017}), \dword{fc} resistor failures (Figure~\ref{fig:efield_resistorfailure_mooney2019}), resistivity non-uniformities, etc.
These effects added in quadrature can be as high as \SI{4}{\%}. 
The space charge effects due to \Ar39~\cite{bib:mooney2018} can be approximately \SI{0.1}{\%} for the \dlong{sp} (\dshort{sp}), and \SI{1}{\%} for the \dshort{dp} (\dlong{dp}), so in practice these levels of 
distortions must cover several meters to be relevant.
Other effects due to \dword{cpa} or \dword{fc} imperfections can be higher because of space charge, but they are also much more localized. If we assume there are no foreseeable effects that would distort the field more than \SI{4}{\%}, and considering the worst case scenario (transverse distortions), then the smallest region that would produce a \SI{1.5}{\cm} shift is \SI{1.5}{\cm}/\num{0.04}~=~\SI{37.5}{\cm}. This provides a target for the granularity of the measurement of the \efield distortions in $x$ to be smaller than approximately \SI{30}{\cm}, with, of course, a larger region if the distortions are smaller. Given the above considerations, then a voxel size of \num{10}$\times$\num{10}$\times$\SI{10}{\cubic\cm} appears to be enough to measure the \efield with the granularity needed for a good position reconstruction precision. In fact, because the effects that can likely cause bigger \efield distortions are problems or alignments in the \dword{cpa} (or \dword{apa}) or in the \dword{fc}, it is conceivable to have different size voxels for different regions, saving the highest granularity of the probing for the walls/edges of the drift volume.

\subsubsection{Design}
\label{sec:sp-calib-sys-las-ion-des}

The design of the laser calibration system for \dword{dune} is largely based on the design of the system built for \dword{microboone}~\cite{microboone}, which in turn was based on several previous developments~\cite{Rossi:2009im,Zeller:2013sva,Ereditato:2014lra,Ereditato:82014tya}. A similar system was also built for \dword{captain}~\cite{Berns:2013usa} and in the near future, will be built for \dword{sbnd}~\cite{Antonello:2015lea}. Operation of the \dword{microboone} system has already taken place. A preliminary report was given in~\cite{bib:chen2018}, and more details on the data analysis are available in~\cite{bib:uBlaser2019}.

\paragraph{Design overview}
Ionization of \dword{lar} by laser can occur via a multiphoton process in which two-photon absorption~\cite{Badhrees:2010zz} leads the atom to the excited states band, and a third photon subsequently causes ionization. This can only occur with high photon fluxes, and so the lasers must provide pulse energies of \SI{60}{\milli\joule} or more within a few ns. Unlike muons, the laser beams do not suffer multiple scattering and travel along straight lines determined by the steering mirror optics. The basic measurement consists of 
generating laser ionization tracks in the \dword{tpc} and comparing the reconstructed tracks with the direction known from the steering hardware. 
An apparent curvature of the measured track is attributed to drift velocity, and therefore \efield, distortions (either in direction or magnitude).

While the Rayleigh scattering length for \SI{266}{\nano\m}  light is approximately \SI{40}{\m}, additional optics effects may limit the maximum practical range of laser beams of that wavelength to a distance smaller than that. Those can include the Kerr effect  due to the dependency of the refractive index on the \efield. In the presence of an intense field, such as that caused by the laser beam itself, the change in refractive index can lead to lensing, or focusing, that distorts the coherence of the beam\footnote{The Kerr effect is so far believed to be the cause of non-homogeneity of the ionization along the laser beam observed in \dword{microboone}, which prevents the use of the charge information. Its effect on the position measurement and \efield uncertainty has been studied by \dword{microboone}.}. 
Despite this, laser beams with lengths of \SI{10}{\m} in \dword{lar} have been observed in \dword{microboone}, and beams with \SI{20}{\m} lengths (possibly more) can be reasonably expected to obtain with a similar system.
This has determined the choice of locating five calibration ports in the cryostat roof at \SI{15}{\m} intervals along each of the four drift volumes of the \dword{spmod}, for a total of \num{20} ports. In fact, there are four ports just outside each of the \dword{fc} end-walls, and \num{12} ports located over the top \dword{fc}, close to the \dword{apa} of each drift volume, as shown in Figure~\ref{fig:calib-FTmap}. As is discussed further below, the number of ports 
currently assigned for the ionization laser system in the baseline design is \num{12}, a compromise between having the maximum possible coverage with crossing tracks and cost considerations.

\paragraph{Mechanical and optical design for a single port sub-system}

For each of 
the used calibration ports, a laser sub-system can be schematically represented by Figure~\ref{fig:uB_laser_schematic} (left) and consists of the following elements:
\begin{itemize}
    \item a laser box (see Figure \ref{fig:uB_laser_schematic}, right) that provides
    \begin{itemize}
        \item a Nd:YAG laser, with the fourth harmonic option providing \SI{266}{\nano\m} in intense \SI{60}{\milli\joule} pulses with about \SI{5}{\nano\s} width, with a divergence of \SI{0.5}{\milli\radian}. The Surelite SL I-10 laser\footnote{Amplitude Surelite\texttrademark{} https://amplitude-laser.com/wp-content/uploads/2019/01/Surelite-I-II-III.pdf} is a possible choice since it has been successfully used in the past in other experiments.
        \item an attenuator and a collimator to control the intensity and size of the beam;
        \item a photodiode that gives a \dword{tpc}-independent trigger signal;
        \item a low-power red laser, aligned with the UV laser, to facilitate alignment operations; and
        \item a Faraday cage to shield the surrounding electronics from the accompanying electromagnetic pulse. 
    \end{itemize}
    \item a feedthrough (see Figure \ref{fig:uB_laser_ft}, left) into the cryostat that provides
    \begin{itemize}
        \item the optical coupling that allows the UV light to pass through into the cryostat directly into the liquid phase, avoiding distortions due to the gas-liquid interface and the gas itself;
        \item a rotational coupling that allows the whole structure to rotate while maintaining the cryostat seal;
        \item a periscope structure (see Figure~\ref{fig:uB_laser_ft}; Right) mounted under the rotating coupling that supports a mirror within the \dword{lar};
        \item the additional theta rotation of the mirror accomplished by a precision mechanism coupled to an external linear actuator; and
        \item both the rotation and linear movements of the steering mechanism read out by precision encoders.
    \end{itemize}
    
\end{itemize}

\begin{dunefigure}[\dshort{microboone} laser calibration system schematics]{fig:uB_laser_schematic}
{Left: Schematics of the ionization laser system in one port~\cite{Antonello:2015lea}. Right: Schematics of the laser box~\cite{microboone}.}
\includegraphics[width=0.45\linewidth]{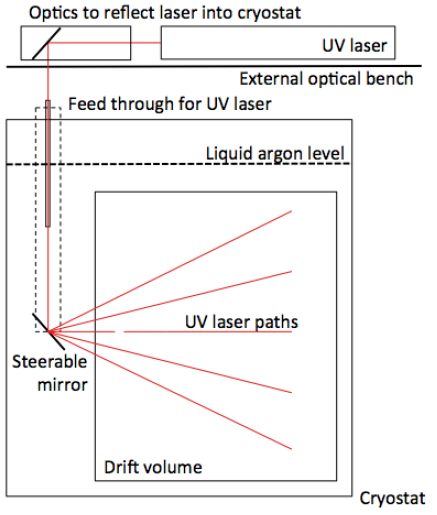}
\includegraphics[width=0.5\linewidth]{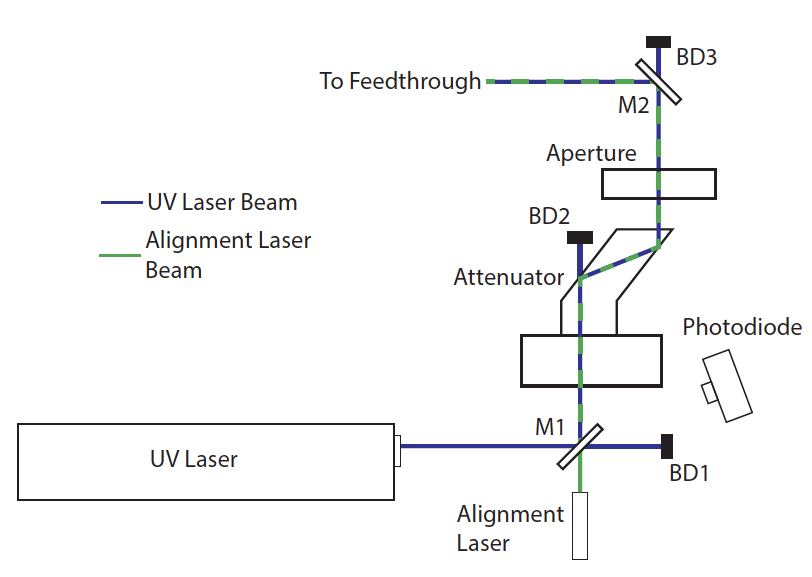}
\end{dunefigure}

\begin{dunefigure}[\dshort{microboone} laser calibration system drawings]{fig:uB_laser_ft}
{CAD drawings of the \dword{microboone} laser calibration system~\cite{microboone}. Left: calibration port feedthrough. Right: laser beam periscope. 
}
\includegraphics[width=0.49\linewidth]{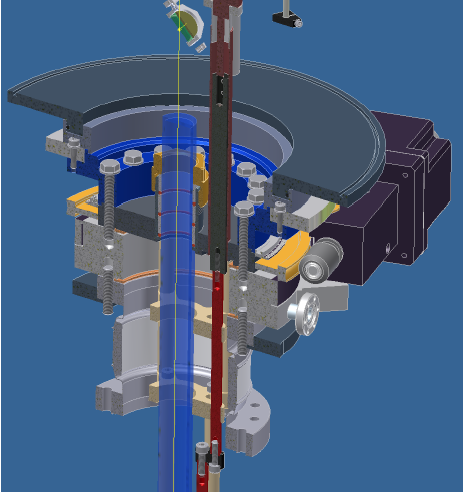}
\includegraphics[width=0.248\linewidth]{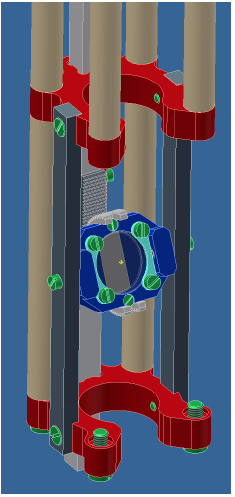}
\end{dunefigure}

The goal of the mechanical design of the system is to achieve a precision close to that of the \dword{tpc} position measurements, so that no single factor dominates 
the overall systematics. The \dword{tpc} precision of about \SI{5}{\milli\m} in the $y$, $z$ coordinates is given primarily by the wire spacing of \uvpitch and \xgpitch. The precision of about \SI{2}{\milli\m} on the $x$ coordinate comes essentially from the \fepeaktime peaking time of the front-end electronics and the typical drift velocity (\driftvelocity).

The starting point of the laser beams is given by the position of the mirror in the periscope, which is known from construction drawings, warm surveys and cool down calculations. The angle of the beam is given by the angles ($\theta$, $\phi$) of the mirror, which are set by the periscope motors and read out by the encoders. 
For \dword{microboone}, reference~\cite{bib:chen2018} quotes a very good \SI{0.05}{\mrad} precision (\SI{0.5}{\milli\m} at \SI{10}{\m}) from the encoders alone, and an overall pointing precision of \SI{2}{\milli\m} at \SI{10}{\m}, driven mostly by beam size and divergence. In fact, with a \SI{0.5}{\mrad} divergence, we expect the beam to be \SI{5}{\milli\m} wide at \SI{10}{\m}.

In \dword{dune}, we aim to reach a similar precision. This will require a number of design and installation considerations: having encoders of similar high accuracy, carrying out surveys in various reference frames, and a capability to do location checks with a precision of about \SI{5}{\milli\m}  at \SI{20}{\m} from the beam origin. Therefore we aim to have a system that can locate the beam end point in few positions and attached to different references, at least one per drift volume and laser beam. 
The independent laser beam location system is described in Section~\ref{sec:sp-calib-sys-las-loc}.

\paragraph{Coverage estimations and top \dword{fc} penetration}
\label{sec:lasercoverage}

A crucial aspect of the design of the full array is the position of the periscope and the cold mirror with respect to the \dword{fc}, since its profiles can induce significant shadows and limit the beam's coverage. In order to address this aspect and motivate the design choices, we carried out a set of shadowing calculation studies.

Given that the \dword{fc} profiles are \SI{4.6}{\cm} wide with only a small \SI{1.4}{\cm} gap between them, the shadows produced if the laser source is located outside the \dword{fc} would be substantial. We estimate that the maximum angle at which beams can go through is about \ang{45}. Given the limitations of the region above the \dword{fc} (shown in Figure~\ref{fig:laser_topfc}, left), especially the geometry of the ground plane, it is likely that the mirror cannot be placed much higher up than \SI{40}{\cm} away from the \dword{fc}. 
With those assumptions, we have carried out a rough estimation of the fraction of voxels that would be crossed by any unblocked track. For simplicity, we are considering only a single vertical plane, so the coverage is actually overestimated since it does not consider the effect of the \dword{fc} I-beams, transverse to the \dword{fc} profiles.
Figure~\ref{fig:laser_topfc} (right) shows an example of those calculations. Assuming \SI{10}{\cm} voxels and no track directed at the \dword{apa}, the coverage is at most \SI{30}{\%}. Assuming \SI{30}{\cm} voxels and allowing all tracks directed at the \dword{apa}, the maximum coverage would be \SI{58}{\%}.

\begin{dunefigure}[View of top field cage and laser coverage estimation]{fig:laser_topfc}
{View of the top field cage (left) and laser 2D voxel coverage estimation for one drift volume (right).}
\includegraphics[width=0.7\linewidth]{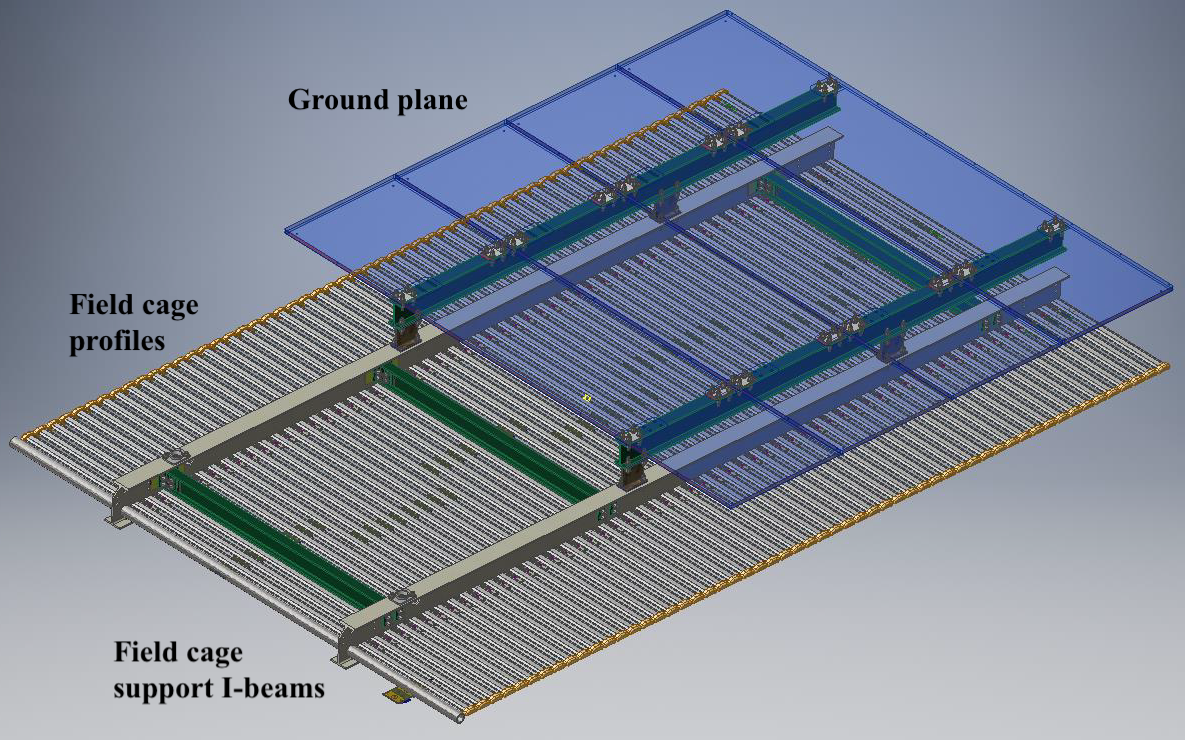}
\includegraphics[width=0.29\linewidth]{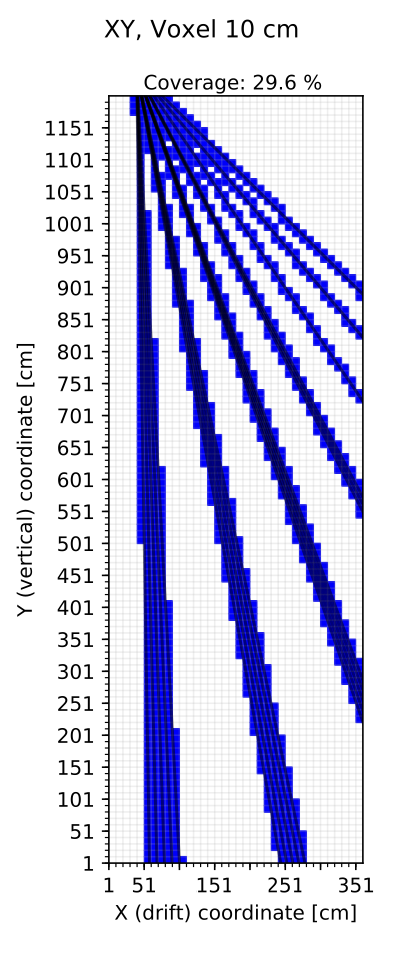}
\end{dunefigure}

Penetration of the \dword{fc} would eliminate most of these shadows and allow for a practically unimpeded coverage. Depending on the depth of the periscope within the \dword{tpc}, some partial shadowing from the field cage support I-beam would still remain.
Figure~\ref{fig:laser_fcpenetration} shows a possible way to accomplish this for the top-of-TPC ports~\cite{bib:yu2019a}. A CAD model of the \dword{sbnd} laser calibration system periscope was used as 
reference design for \dword{dune}. The \dword{sbnd} periscope, when rotating over its axis, requires a \SI{12}{\cm} diameter circular region free of impediments. In order to take into account a tolerance for the estimated \SI{0.3}{\%} shrinkage of the \dword{fc} at cryogenic temperatures, we chose an opening of three profiles, equivalent to \SI{18}{\cm}. 
Still, in order to minimize any risk associated with the presence of material close to the \dword{fc}, ongoing design studies will evaluate the feasibility of implementing vertically retractable periscopes, with a travel range sufficient for them to clear the top of the \dword{fc}. 

\begin{dunefigure}[Laser periscope penetrating the field cage]{fig:laser_fcpenetration}
{CAD drawing of one way the periscope could penetrate the \dword{fc}~\cite{bib:yu2019a}.}
\includegraphics[width=0.6\linewidth]{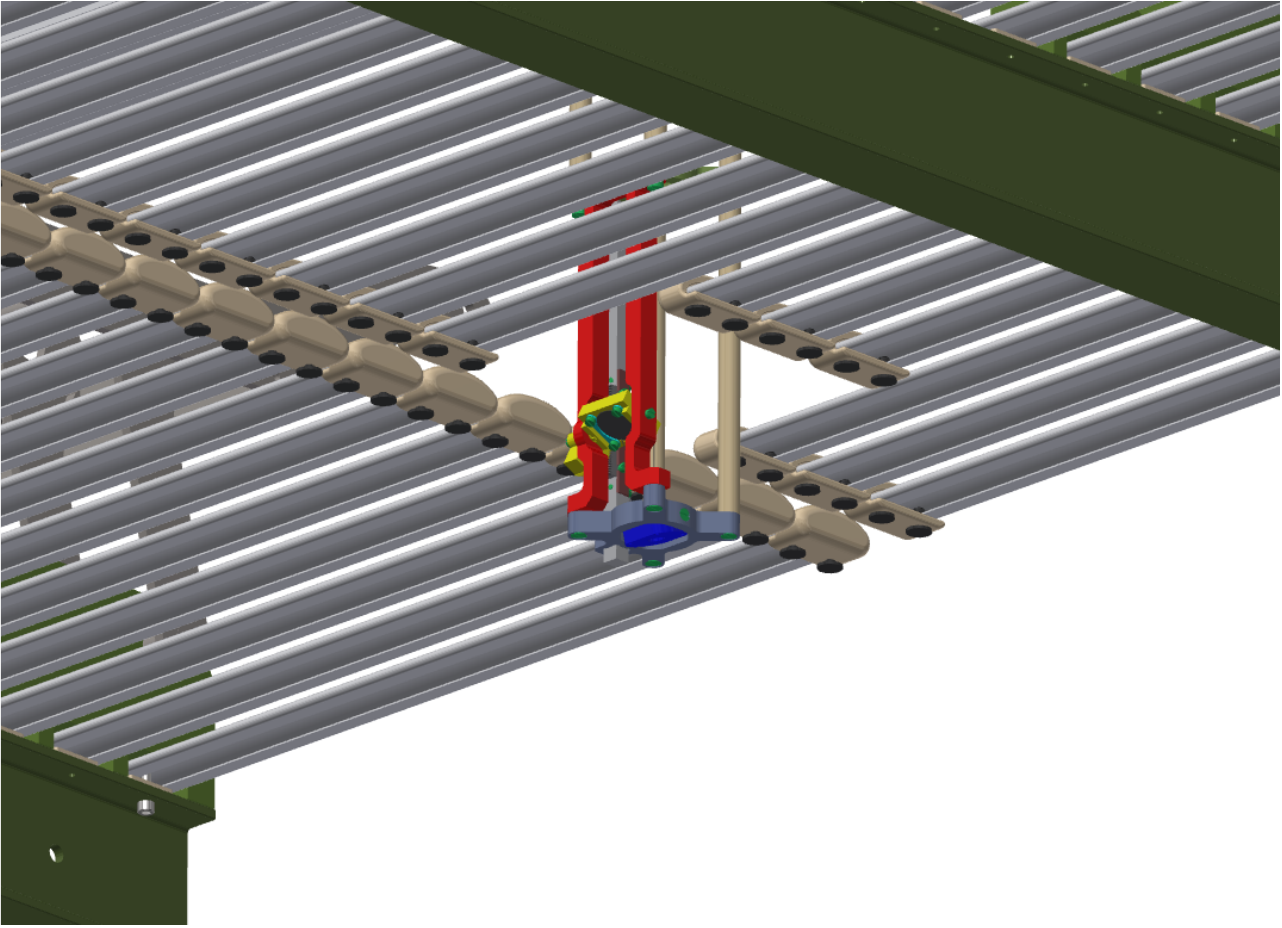}
\end{dunefigure}

Simulations of the effect of \dword{fc} penetrations on the \efield were carried out~\cite{bib:yu2017b}, and are illustrated in Figure~\ref{fig:efield_penetration_boyu2017}. These have shown that the effect of a \SI{12x12}{\cm} opening (equivalent to two profiles), located at \SI{40}{\cm} (along the $x$ direction) from the \dword{apa}, is small and tolerable, with a maximum \SI{10}{\kilo\volt\per\cm} \efield caused by the opening and periscope.
These simulations need to be redone with a larger opening of \SI{18x18}{\cm}  (i.e., three profiles).
Still, if we were to choose, conservatively, to discard from the physics data analysis the volume within the \dword{tpc} determined by the periscope lateral size, a vertical penetration of \SI{10}{\cm}, and the full drift length (\SI{12x10x360}{\cm} = \SI{43}{\litre} for each of the \num{12} periscopes), it would represent only a very small fraction of \num{5e-6} of the full detector volume.

\begin{dunefigure}[Simulation of impact on \efield of \dshort{fc} penetration]{fig:efield_penetration_boyu2017}
{Simulation of the effect on the \efield of a laser periscope penetration of the \dword{fc}. In this case, an opening of only two profiles was considered.}
\includegraphics[width=0.6\linewidth]{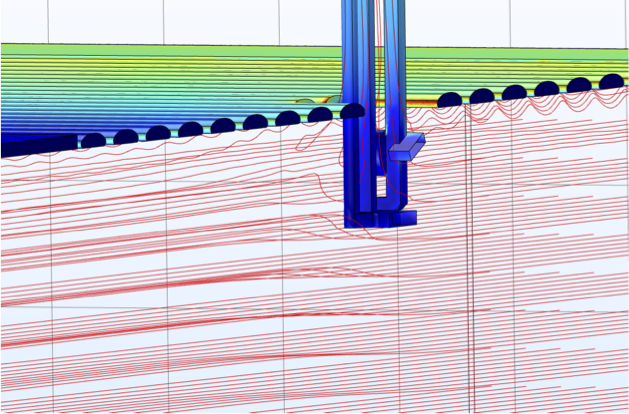}
\end{dunefigure}

\paragraph{Full array scope considerations}

As mentioned earlier (Section~\ref{sec:sp-calib-laser-req}), the system should allow for crossing laser beam tracks wherever possible. In order to 
collect them in the full \dword{spmod} volume, that would require using all the available \num{20} calibration ports. Since it is possible to use an iterative method to obtain displacement maps in regions where no crossing tracks are available, 
to minimize the overall cost of the system, the baseline design will use only the \num{12} central ports, providing crossing tracks in essentially \SI{50}{\%} of the detector volume. 
In addition, for the six most central ports, close to the central \dword{apa}, the distance between them is small enough that we can consider having the same laser box serve two feedthroughs to reduce the costs associated with the laser and its optics. In that case, the total number of lasers needed would be nine.

Usage of the end-wall ports, which are not 
on top of the \dword{tpc}, is therefore not part of the baseline design, and is considered only as an alternative in Section~\ref{sec:sp-calib-laser-alter}. A coverage calculation for possible end-wall periscopes, taking into account the shadowing of both the \dword{fc} profiles and the support beams, gives a maximum of \SI{56}{\%} coverage for \SI{30}{\cm} voxels (allowing all tracks directed at the \dword{apa}). In this case the laser beams would enter the \dword{fc} laterally and \dword{fc} penetration would be harder to consider, so an alternative mechanical design aimed at improving the coverage, is considered in Section~\ref{sec:sp-calib-laser-alter}.

A scan of the full detector using \num{10}$\times$\num{10}$\times$\SI{10}{\cubic\cm}
volume elements would require a number of tracks approximately \num{8e5} 
and can take about three days. Shorter runs could be done to investigate specific regions. The sampling granularity, and therefore the amount of data taken, depends on \dword{daq} requirements. In fact, even to be able to record the desired \num{8e5} tracks, a dedicated data reduction algorithm must be devised, so that only a drift window of about \SI{100}{\micro\s}
of data is recorded, and the position of that window depends on the beam position and direction and which wires are being read out. More details on this are given in Section~\ref{sec:sp-calib-daqreq}.



\subsubsubsection{Measurement Program}
\label{sec:sp-calib-sys-las-ion-meas}

This section describes the methods used to measure 
parameter maps and their expected precision, given the design outlined above.

\paragraph{\efield and drift velocity measurement}
The method for \efield measurement is based on the measurement of apparent position displacements of the straight laser tracks. The laser produces straight tracks with a known starting position and direction. If, when reconstructed under the assumption of uniform and homogeneous drift velocity, any deviations from that are observed, they are attributed to \efield distortions. 

The first step in the analysis~\cite{bib:uBlaser2019} is to obtain a field of position displacements by comparing the known and reconstructed tracks. If two crossing tracks are used, the displacement vector is simply given by the vector connecting the point where the reconstructed tracks cross and the point where the known tracks cross. However, since those displacements can vary both in direction and magnitude, there will be ambiguity in that determination if only one track is used in a given spatial region. An iterative procedure was developed by the \dword{microboone} collaboration~\cite{bib:chen2018,bib:uBlaser2019} to obtain a displacement map from a set of several non-crossing tracks from opposite directions. Following this, a set of drift velocity field lines, which are the same as \efield lines, can be obtained from the displacement map, assuming that all charge deposits along a field line will be collected in the same position. Using the relationship between \efield and drift velocity~\cite{Li:2015rqa,Walkowiak:2000wf}, we can then also obtain the magnitude of the \efield.

Since the observed position distortion in one location depends on \efield distortions in many locations along the drift path, this method of analysis clearly requires the acquisition of data from many different tracks crossing each detector drift volume at many different angles. 

As already indicated in the previous section (section~\ref{sec:sp-calib-sys-las-ion-des}), the pointing precision will be on average \SI{2}{\milli\m} (at average distances of \SI{10}{\m}), and the \dword{tpc} precision is \SI{2}{\milli\m} in $x$ and \SI{5}{\milli\m} in $y$, $z$. Conservatively taking those in quadrature, we get $\sigma_x$ =  \SI{3}{\milli\m} and $\sigma_{yz}$ = \SI{5.4}{\milli\m}.
If we would use only one track per direction, in regions of size $l$ = \SI{300}{\milli\m}, we would therefore be sensitive to drift velocity field distortions of $\sigma /l$, i.e., \SI{1}{\%} in $x$ and \SI{1.8}{\%} in $y$, $z$. 

In order to estimate the \efield precision, we must distinguish between the $x$ and $y$, $z$ coordinates. To first order, distortion in $y$, $z$ do not affect the magnitude of the field, and so the relative distortions on \efield are equal to the relative distortions of the velocity. Along $x$, we must consider the relation between the magnitudes of the drift velocity and \efield. Using the formula from~\cite{Li:2015rqa,Walkowiak:2000wf} we can see that, at \spmaxfield, a \SI{1}{\%} change in \efield leads to a corresponding change of \SI{0.375}{\%} in drift velocity. We therefore reach the values of \SI{2.7}{\%} ($=1./0.375$) in $x$ and \SI{1.8}{\%} in $y$, $z$ for a conservative estimate of \efield precision using a single track per direction. 

This is a conservative estimate because it does not take into account the fact that the centroid of the beam should be known better than its full width, and because it is based on the assumption of a single track per direction.  
As observed in \dword{microboone}~\cite{bib:uBlaser2019}, using several tracks improves the precision, and in most of the volume an accuracy of \SI{1}{\%} was reached so the amount of statistics needed to reach \SI{1}{\%} will be an important question to address in the development plan.

On one side, this gives us an ultimate limit to the \efield precision achievable with the laser system, but on the other side, since these \dword{tpc} precision considerations apply to physics events also, it tells us that an \efield precision much better than \SI{1}{\%} should not have an effect on the physics.

\paragraph{Charge-based measurements}

Electron drift-lifetime~\cite{bib:uBlifetime, Antonello:2014eha} is the parameter that governs the dependence of the amount of collected charge on the drift time. A possible measurement of electron drift-lifetime would therefore require a very good control over the charge profile of the ionization laser tracks. This was achieved in a small scale experiment that measured lifetime with laser beams~\cite{Ereditato:2013xaa}, but is harder with longer distances. The charge produced by the laser tracks along its path depends on distance because the light intensity is reduced due to beam divergence and scattering, as well as non-linear effects such as the self-focusing, or Kerr effect. For this reason, the first steps in any laser-based charge measurement are a fine-tuning of the laser intensity in order to reduce self-focusing to a minimum, and ``charge profile calibration scan'' which consists of acquiring tracks parallel to the \dword{apa}. In order to get good statistical precision, several tracks could be acquired, in the same or different direction, but always parallel to the \dword{apa} in order to factorize out any effect from electron drift-lifetime. This set of data provides a calibrated laser beam charge profile that can then be used to analyse and normalize the measured charge profile from tracks that do have an angle with respect to the \dword{apa} and therefore span different drift times.

As for electron-ion recombination, since the $dE/dx$ for laser beams is much smaller than for charged particles, the effect should also be much smaller. However, that small effect has been observed~\cite{Badhrees:2010zz}, so a similar method than described above could be used to evaluate any dependence of the electron-ion recombination factor on the angle $\phi$ between the track and the electric field, that is predicted in some models~\cite{Acciarri:2013met}. This would entail taking data with tracks as parallel as possible to the \efield, in order to enhance the angular dependence term on the recombination expression (that goes with $1/sin \phi$), and to compensate for the smaller $dE/dx$ for laser beams.

\subsection{Laser Calibration: Beam Location System}
\label{sec:sp-calib-sys-las-loc}
\label{sec:calib-laser-pos}
Because the precision of the \efield measurement relies heavily on a precise knowledge of the laser beam tracks, an independent measurement of their direction for some specific positions is required. The 
\dlong{lbls} (\dshort{lbls})
addresses this requirement. While the direction of the laser beam will be very well known based on the reading from the encoders on the laser beam steering mechanism,  residual uncertainty or unpredictable shift in the pointing direction will remain. 
Keeping in mind the long length of the ionization track of more than \SI{15}{\m}, even a small offset in the pointing direction can lead to vastly different ionization track locations, especially close to the end of the track. Such inaccuracies will directly affect our ability to precisely calibrate any variations in the \efield.

\subsubsection{Design}
The \dword{lbls} is designed to provide precise and accurate knowledge of the laser track coordinates.  
Two complementary systems are planned, one based on PIN diodes and another based on mirrors.

\paragraph{PIN diode system for laser beam location}

The design for the system using PIN diodes is based on the existing system that was built for the mini\dword{captain} experiment~\cite{Berns:2013usa}.

The \dword{lbls} consists of groups of \num{9} PIN diodes, operating in passive, photovoltaic mode. These are GaP diodes with a sensitivity range extending down to  \SI{200}{\nano\m} wavelength; thus, detecting \SI{266}{\nano\m} light is straightforward. 
PIN diodes are placed at the bottom of the cryostat and receive direct laser light\footnote{This is a difference with respect to the miniCAPTAIN system, which does not observe direct light, but detects fluorescence in the \frfour.} passing through the gaps between the \dword{fc} profiles to minimize interference with the \dword{fc}. Drawings of one such group of PIN diodes are shown in Figure~\ref{fig:GaP_assembly}. With the group of \num{9} photodiodes, we can detect not only the beam but also crudely characterize its profile, giving a more precise location of the central beam pulse axis.

\begin{dunefigure}[Cluster assembly of the miniCAPTAIN \dshort{lbls}]{fig:GaP_assembly}
{(Left) \dword{lbls} cluster mounted on the opposite wall from the laser periscope to detect and accurately determine the end point of the laser beam. (Right)
Profile of the \dword{lbls} group mounted on the PCB. GaP diodes come with pins that use pair of twisted wires to transport the signal.}
\includegraphics[width=0.35\linewidth]{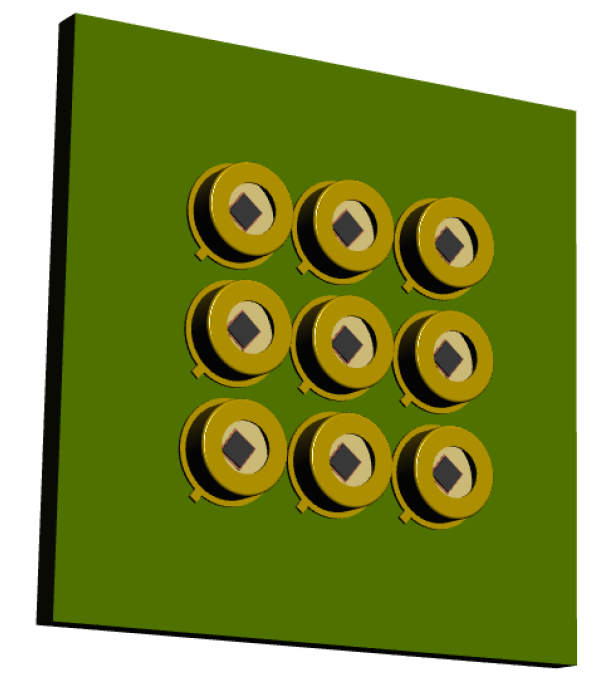} 
\includegraphics[width=0.45\linewidth]{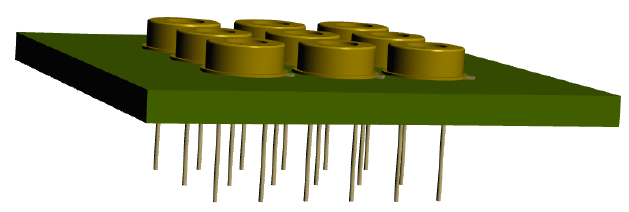} 
\end{dunefigure}

There will be two \dword{lbls} pads per laser, with each pad visible by two different lasers, to maximize precision and ensure sufficient redundancy in the system. There will be a total of 16 locations (4 per volume) for a total of 32 pads. Pads will be placed on the central line of each of the four volumes, in the middle between each pair of adjacent lasers, located under the \dword{fc}. The locations of the pads will be carefully surveyed after installation and prior to closing of the cryostat. The laser should always send the first pulse in the direction of the \dword{lbls} before proceeding into a calibration sequence. In this way, the absolute location of the initial laser track will be determined with high accuracy. The location of the other laser tracks will also be determined with high accuracy with respect to the initial track thanks to the high precision of the rotary encoders.

\paragraph{Mirror-based beam location system:}

In addition to the PIN diode system, we will also have clusters of small mirrors that allow measuring the beam end position via its reflections.

Figure~\ref{fig:laser_mirror_positioning} shows a conceptual sketch with a cluster of \num{6} mirrors located close to each other, but with different angles. When the beam hits one of the mirrors, it will be reflected back into the \dword{tpc}, and the reflection angle unambiguously identifies which mirror was actually hit. With small mirrors, \SI{5}{\milli\m} in diameter, the required positioning precision would be met if these mirrors are placed at distances of more than \SI{10}{\m}. The preferred location is, therefore, at the bottom \dword{fc}. Because the cluster can be small (a few cm), it can fit inside the \dword{fc} profiles. For each drift volume segment seen by two lasers, we plan to install at least two clusters, for redundancy, so the total number of clusters would be \num{32}. 

The simplest solution would be to use polished aluminum as the reflecting surface, so that the cluster could be a single block. 
Tests of the actual reflectivity of the (oxidized) surface will be part of the development plan.  An alternative would be small dielectric mirrors.

\begin{dunefigure}[Mirror-based laser beam location system]{fig:laser_mirror_positioning}
{View of the mirror cluster for the beam location system inserted in the \dword{fc} profiles~\cite{bib:yu2019a}.}
\includegraphics[width=0.5\linewidth]{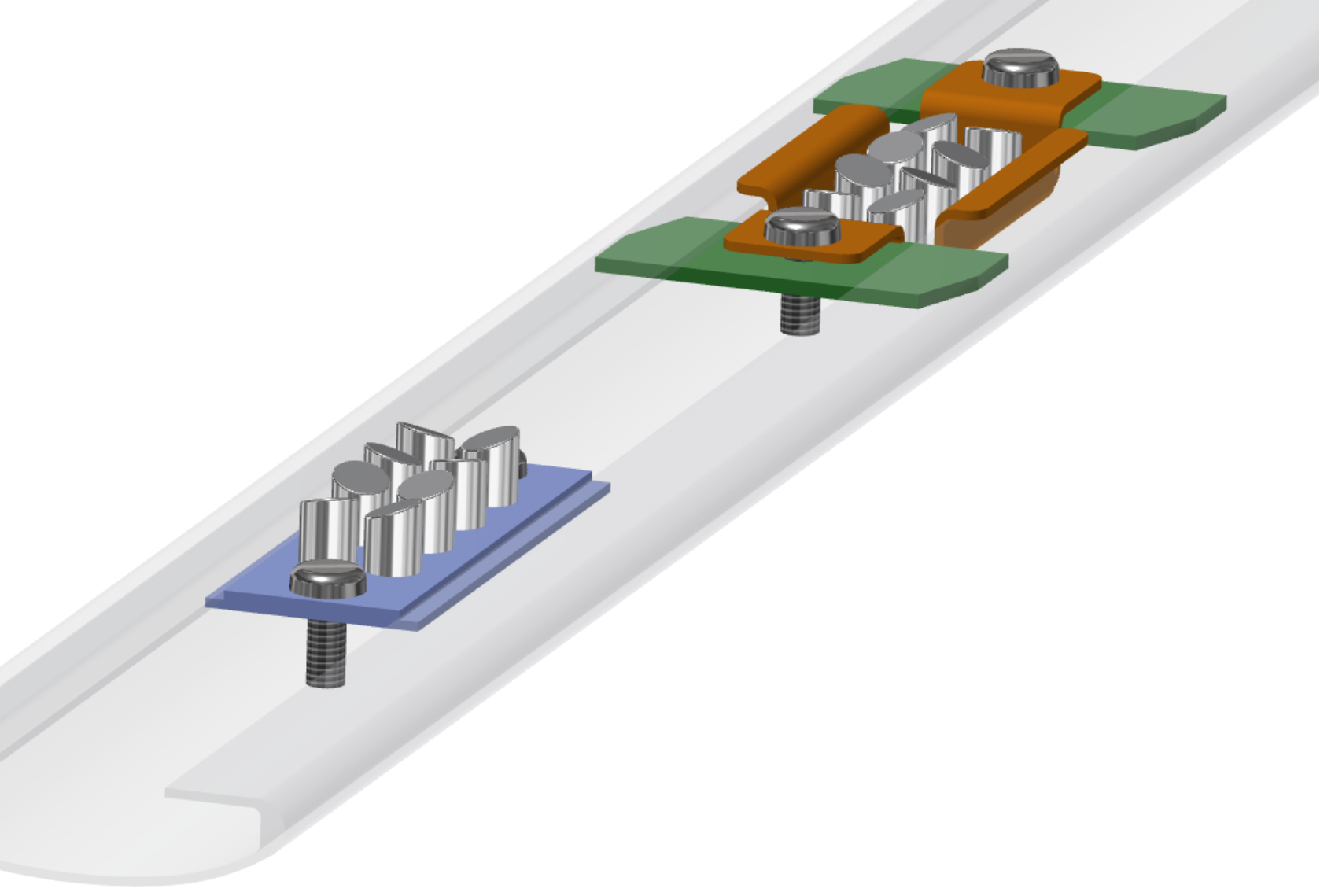}
\end{dunefigure}

\subsubsection{Development plan}
 Further optimization of the PIN diode 
 assembly to reduce electronic noise and cross-talk is required. Also, the size and shape of the cluster that would best collect the light coming through the field cage gaps needs to be optimized.  Another important aspect is durability of the system that will require extensive running in the cryogenic conditions with  a large number of cool-downs to validate GaP for extended use in DUNE. Finally, alternatives to GaP diodes such as SiPMs are under consideration. While SiPMs require power, their sensitivity to single photons makes them a desirable candidate for low light signals and more accurate beam direction reconstruction. 

As for the mirror-based system, the capability of the \dword{tpc} to identify the reflected beam will depend on how diffuse the reflectivity on the aluminum surfaces will be. A full test must be carried out at \dword{pdsp}, including alternative options such as using mirrors. Small dielectric mirrors for \SI{266}{\nano\m} with \SI{6.35}{\milli\m} diameter are commercially available.

\subsection{Laser Calibration:  Photoelectron System}
\label{sec:sp-calib-sys-las-pe}

Well localized electron sources represent excellent calibration tools for the study of electron transport in the \dshort{lartpc}. 
A photoelectron laser system can provide such sources at predetermined locations on the cathode, leading to precise  measurements of total drift time and integrated spatial distortions when the charge is not collected in the expected wires. These are achieved by simply measuring the time difference between the laser pulse trigger time and the time when the electron cloud reaches the 
\dword{apa}.
Such measurement will result in an improved spatial characterization of the \efield, and consequent reduction of detector instrumentation systematic errors.

Being an operationally simpler system compared to the ionization laser system, the photoelectron laser can be used as a ``wake-up'' system to quickly diagnose if the detector is alive, and to provide indications of detector regions that may require a fine-grained check with the ionization system. This is especially important due to the low cosmic ray environment in the detector underground. The photoelectric laser system will utilize the ionization laser for target illumination, thus eliminating the additional cost associated with the laser purchase.

\subsubsection{Design}
\label{sec:sp-calib-sys-las-pe-des}

In order to produce localized clouds of electrons using a photoelectric effect, small metal discs will be placed on the cathode plane assembly and used as targets. Photoelectric laser systems have been successfully used at T2K~\cite{Abgrall:2010hi} and in the Brookhaven National Laboratory (BNL) \dword{lar} test-stand~\cite{Li:2016ods} to generate well-localized electron clouds for \efield calibration.

The baseline material choice for the metal targets is aluminum, while silver is being considered as an alternative. At \SI{266}{\nano\m} (Nd:YAG quadrupled wavelength) the single photon energy of \SI{4.66}{\eV} is sufficient to generate photoelectrons from aluminum and silver. However, aluminum and silver are prone to oxidization.
In the case of aluminum, a thick layer of aluminum oxide forms the surface, but this does not increase the work function of the material. Table~\ref{tab:metalphotoelectric} lists the relevant features of metals under consideration.  

The main factor driving the electron yield from the photoelectric targets is the quantum efficiency of the material. Although electrons will be released from the metal whenever photons hit the metal surface, most of the ejected electrons carry forward momentum and therefore are never released from the metal. Only a small fraction of released electrons back-scatters or knocks another electron out of the surface. The quantum efficiency for various metals is typically between \num{e-5}  and \num{e-6}, thus quite low.  All material candidates will be studied in the lab to verify the electron yield, and tested in \dword{protodune2} in order to verify the quantum efficiency for different materials.

\begin{dunetable}
[Work function and other features of candidate metal targets for laser PE system]
{cccccc}
{tab:metalphotoelectric}
{Work function and other features of candidate metal targets for laser photoelectron system.}
 Target Material & Work function (\SI{}{\eV}) & $\lambda_{max}$ (\SI{}{\nano\m}) & $\lambda_{laser}$ & Oxidizing & Type of \\ 
\rowtitlestyle 
  & & & required (\SI{}{\nano\m}) & in air & oxidization \\ \toprowrule
 Aluminum & 4.06 & 305 & 266 & Yes & Surface layer \\ \colhline
 Silver & 4.26-4.73 & 291 & 266 & Yes & Surface layer \\ 
  & (lattice dependent) & & & & \\ 
\end{dunetable}


Disc targets will be fabricated with two different diameters: \SI{5}{\milli\m} and \SI{10}{\milli\m} to provide a test of the vertex reconstruction precision. In addition to circular targets, metal strips \num{0.5} $\times$ \SI{10}{\cm} are being considered to calibrate the rate of transverse diffusion in \dword{lar}. However, their impact on the cathode field will be carefully studied before being incorporated into target list, to prevent any disruptions to the cathode electric field.  

The targets will be fastened to field shaping strips located on the rim around the resistive panel of the cathode plane assembly. Figure~\ref{fig:calib-PhotoelectricTargets} illustrates locating the photoelectric targets on the rim around the resistive panel. The distance between the dots will be \SI{10}{\cm} with \num{5} targets at each corner, while the strips will be fastened at the center of each long side of the resistive plane. The total number of disc targets per resistive panel is \num{20} and the total number of strips per resistive panel is \num{2} as illustrated in Figure~\ref{fig:calib-TargetsOnCPA}.  Given that there are \num{600} resistive panels per \single module, there will be a total of \num{12000} disc targets and \num{1200} strip targets per module. 
The photoelectric dots and strips layout will be further refined based on the calibration requirements and performance simulation results. It will be essential to conduct a survey of the photocathode disc locations on the cathode after installation and prior to detector closing. In this way, the absolute spatial calibration of the \efield will be achieved. 

\begin{dunefigure}[Placement of phototargets on the cathode plane assembly]{fig:calib-PhotoelectricTargets}
{The best place to place the photo targets without being intrusive for the \efield, is the surface of the field shaping strips around the rim of the resistive panel. Circular targets will be implemented, while the strip targets are still under consideration.}
\includegraphics[width=0.65\linewidth]{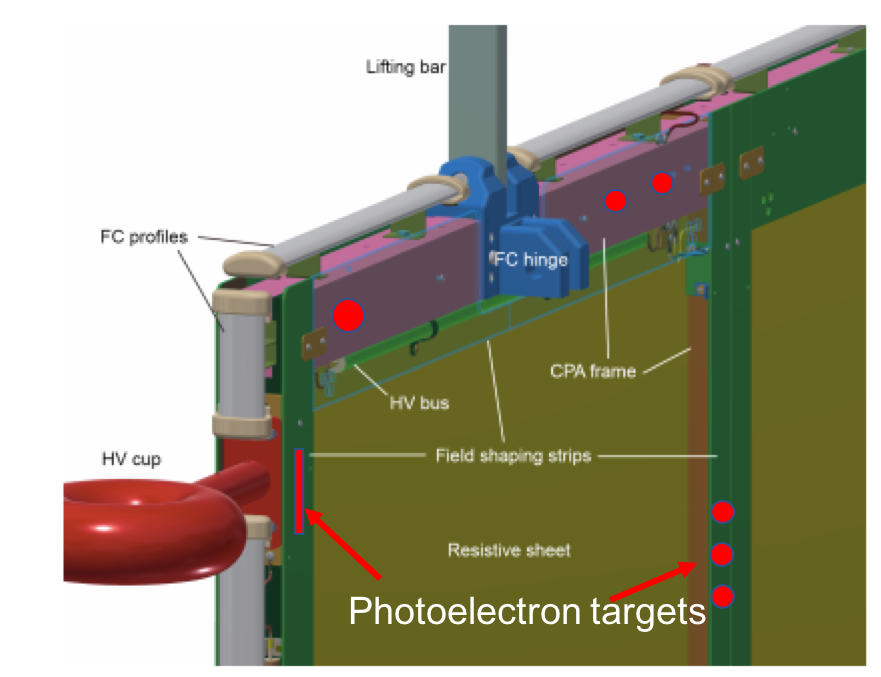} 
\end{dunefigure}

\begin{dunefigure}[Anticipated positions and number of  phototargets on the cathode plane assembly]{fig:calib-TargetsOnCPA}
{There are a total of 5 circular targets in each corner, for a total of 20 circular targets: 12 large and 8 small diameter targets total. In addition, 2 strips at the center along the long sides of the resistive panels may be added if not disruptive to the high voltage on the cathode plane.} 
\includegraphics[width=0.55\linewidth]{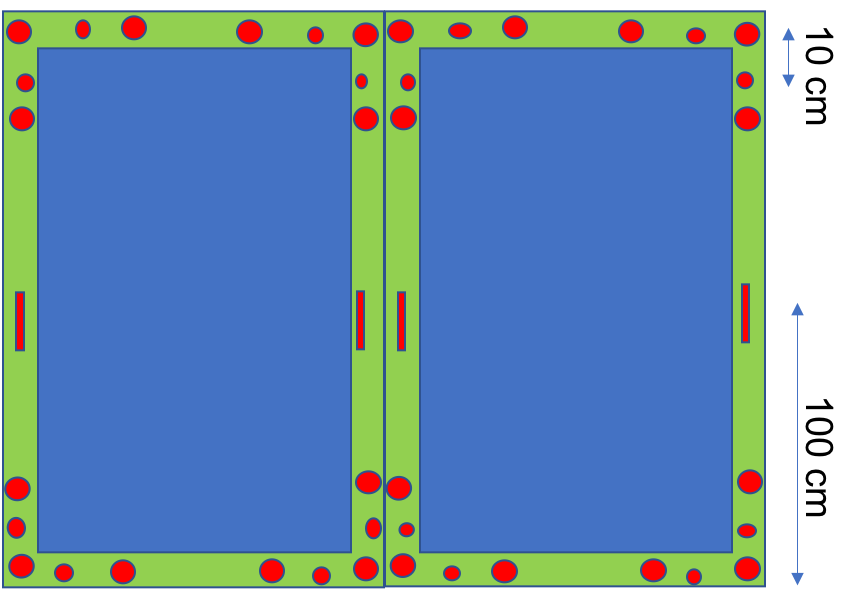} 
\end{dunefigure}

A few thousand electrons are required per spill from each dot to produce the signal above the noise level on the wire and this number will be achieved with high intensity lasers (pulses of the order of \SI{100}{\milli\joule}). The laser beams used to illuminate the targets will be injected into the cryostat via cryogenic optical fibers guided into mounting points in the \dword{apa}, where they are coupled with defocusing elements that will illuminate \SI{10}{\m} diameter surface on the \dword{cpa}  with a single fiber. Fibers will be fastened along the central line of the \dword{apa} in the space between the top and bottom \dword{apa}, on the top of the upper \dword{apa} and on the bottom of the lower \dword{apa}. With the aid of the defocusing elements, the entire single phase module can be illuminated with a total of \num{72} fibers, corresponding to just \num{6} fibers along the central line along with \num{6} fibers on top and bottom for a total of \num{18} fibers per each of the four drift volumes. Figure~\ref{fig:calib-CPAIllumination} shows the conceptual view of the \dword{cpa} illumination.

\begin{dunefigure}[View of the \dshort{cpa} illumination with fibers on the \dshort{apa}]{fig:calib-CPAIllumination}
{Conceptual view of the \dword{cpa} illumination with fibers placed on the top and bottom of the \dword{apa} for better coverage and overlap.} 
\includegraphics[width=0.55\linewidth]{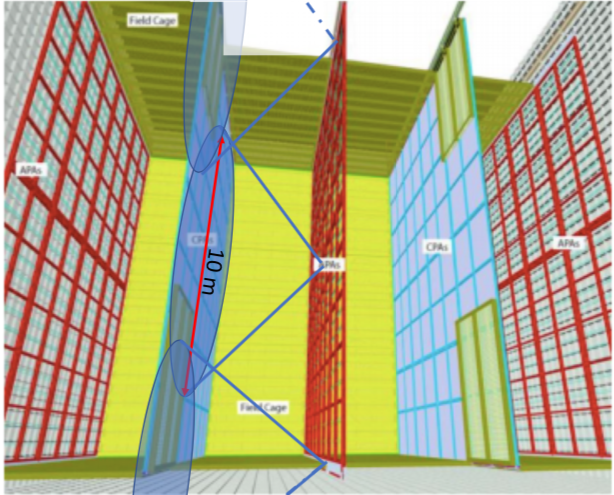} 
\end{dunefigure}

While the current plan aims for illumination of the entire \dword{cpa}, the Kapton material that composes the resistive panels undergoes photoelectric effect, albeit with three orders of magnitude lower quantum efficiency at cryogenic temperatures when compared to phototargets at \SI{266}{\nano\m}. While the noise produced is expected to be tolerable, in case the noise is higher than anticipated, the solution is to illuminate only the areas where phototargets are placed reducing the resistive panel exposure. In this case, instead of defocusing elements, bare fibers will be utilized. The bare fiber opening angle is \ang{10} and \SI{1.3}{\m} diameter exposure.   
Assuming parallel running of lasers, photoelectron targets in 3 out of 4 volumes can be illuminated at once, assuming that laser firing can be coordinated and calibrated with sufficient precision and that volumes 2 and 3 share lasers. Lasers typically operate at 10 Hz frequency. If 10000 pulses per laser are assumed, about 15 minutes of running is needed per laser for a single calibration run
as the photoelectron clouds from different dots are very well localized.
With the help of commercial multiplexers per each volume, 1 hour per volume will be sufficient for a single calibration campaign. If the \dword{daq} or lasers themselves prevent parallel running, the entire calibration campaign will take between 15 minutes or up to 1-5 hours. The calibration run duration will depend on the final calibration scheme.

The photoelectron system will use the same lasers used for argon ionization. Stability of the laser pulses will be monitored  with  a power meter. Dielectric mirrors reflective to \SI{266}{\nano\m} light will guide the laser light to injection points, but a fraction of the light will be transmitted instead of reflected to the power meter behind the mirror. The laser will also send a forced trigger signal to the \dword{daq} based on the photodiode that will be triggered on the fraction of the light passing through the dielectric mirror. 

\subsubsection{Development plan}
The photoelectron system will require the following tasks to complete the design that can be done in \dword{protodune2} or in the lab: 
\begin{itemize}
    \item test the mounting of the targets on the \dword{cpa};
    \item use different target materials to compare their performance;
    \item verify the potential of targets to generate several \SI{1000} electron clouds and their 
    ability to diagnose electric field distortions and vertex reconstruction;
    \item allocate ports to insert laser fibers used for illumination;
    \item validate interface with 
    ionization laser in order to inject UV photons into fibers;
    \item validate efficiency of laser light injection in the optical fiber;
    \item validate light attenuation in fibers;
    \item validate design interface with \dword{apa} and optimized locations of fibers between top and bottom \dword{apa}s;  
    \item validate diffuser design and light losses in the diffuser as well as its ability to illuminate large areas of \dword{cpa}s;
    \item validate bare fiber \dword{cpa} illumination; and 
    \item survey of the dots position to the required level of precision. 

\end{itemize}

\subsubsection{Measurement Program}
\label{sec:sp-calib-sys-las-pe-meas} 

Photoelectron systems have been used in other experiments to diagnose electronics issues by using the known time period between the triggered laser signal and read out times, and to perform rapid checks of the readout of the TPC itself. 

A photoelectron laser is an effective diagnostic and calibration tool, that can quickly and accurately sample the electron drift velocity in the entire detector.
In addition, it can be used to identify electric field distortions due to space charge effects. Exact knowledge of the timing and position of the generated electron clouds is useful for vertex calibration.
In addition to electronics issues, discrepancies between the measured and expected drift time can point to either distortions in the position of the detector elements or to a different drift velocity magnitude. 

Another planned measurement is the comparison between the expected and measured $y$, $z$ position of the collected charge, that can point to transverse distortions of the \efield.

\subsection{Pulsed Neutron Source System}
\label{sec:sp-calib-sys-pns}

The \dword{snb} signal includes low-energy  electrons, gammas and neutrons, 
which capture on argon. Each signal channel will have specific detector threshold effects, energy scale, and energy resolution. 
As noted in~\physchsnb, 
the sensitivity to \dword{snb} physics 
depends on the uncertainties of relevant detector response parameters, and so a calibration method to constrain those uncertainties is needed.
Local detector conditions may change with time due to a variety of 
causes that include electronics noise, misalignments, fluid flow, \dword{lar} purity, electron lifetime and \efield. While these are intended to be characterized from other systems via inputs to the detector model, ``standard candles'' provide a method to assess if our detector model is incomplete or insufficient. An ideal standard candle matches one of the relevant signal processes and will provide spatial and/or temporal information. The \dlong{pns} (\dshort{pns}) system, as described below, will provide a standard candle neutron capture signal (\SI{6.1}{\MeV} multi-gamma cascade) across the entire \dword{dune} volume that is directly relevant to the supernova physics signal characterization. The \dword{pns} is also capable of providing a spatially fine-grained measurement of electron lifetime.

Liquid argon is near transparent to neutrons with an energy near or at \SI{57}{\keV} due to an anti-resonance in the cross-section caused by the destructive interference between two high level states of the \argon40 nucleus (see Figure~\ref{fig:PNS_xsections}). The cross-section at the anti-resonance ``dip'' is about \SI{10}{\keV} wide, and at the bottom the cross section of \SI{1.6e-4}{\barn} implies an elastic scattering length of over \SI{2000}{\m}. 
Natural argon has three major isotopes: \isotope{Ar}{36} (\SI{0.3336}{\%}), \isotope{Ar}{38} (\SI{0.0834}{\%}), and \isotope{Ar}{40} (\SI{99.6035}{\%}) each with a slightly different anti-resonance. The average elastic scattering length of the \SI{57}{\keV} neutrons in natural liquid argon is about \SI{30}{\m}.

\begin{dunefigure}[Cross sections enabling the \dshort{pns} concept]{fig:PNS_xsections}
{Illustration of interference anti-resonance dips in the cross section of \isotope{Fe}{56}, \isotope{Si}{28}, \isotope{S}{32}, and \isotope{Ar}{40}. Elastic scattering cross-section data is obtained from ENDF VIII.0~\cite{Brown:2018jhj}.}
\includegraphics[width=10cm]{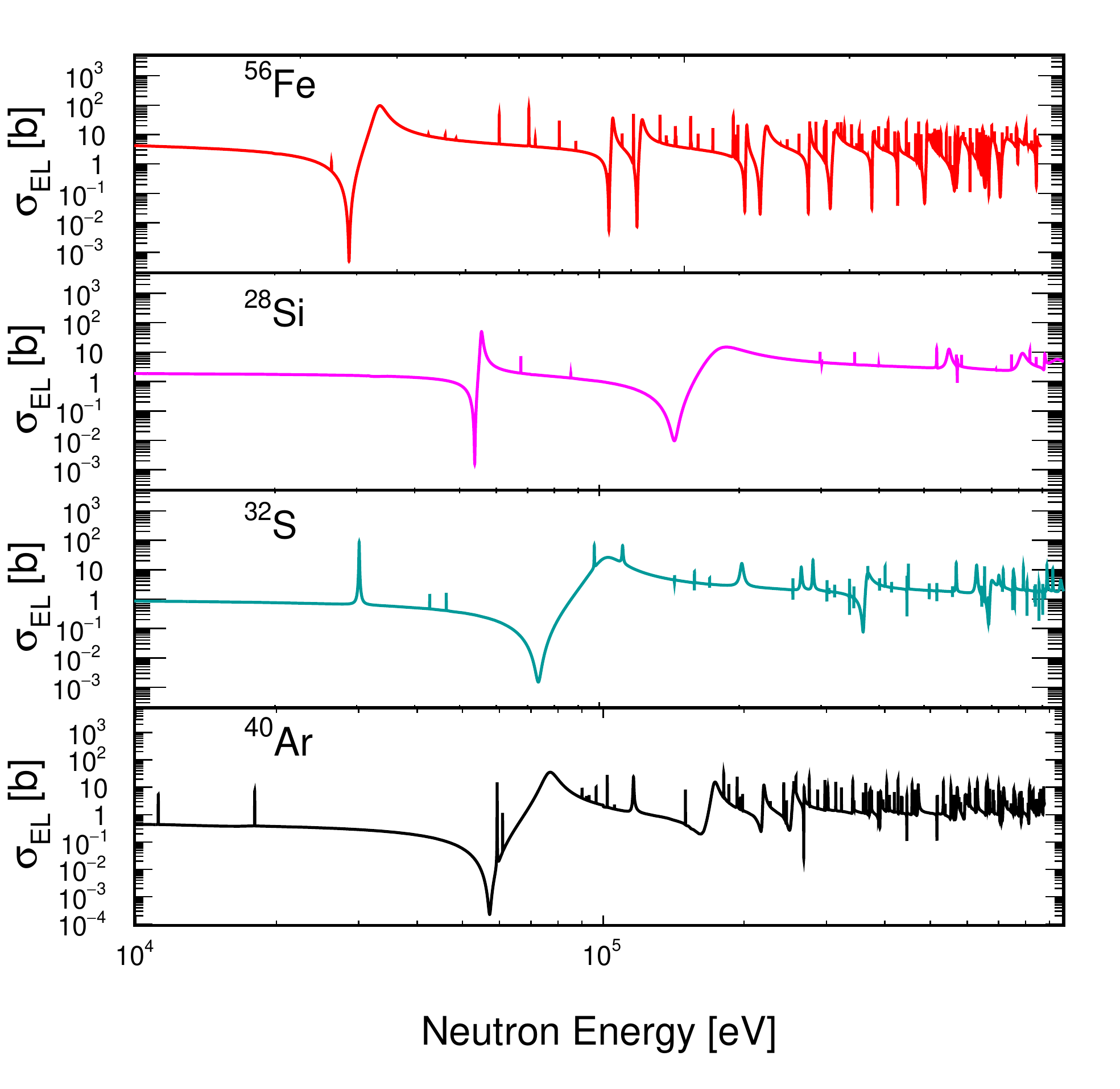}
\end{dunefigure}

The neutrons at the anti-resonance energy could be injected into liquid argon in the \dword{tpc}, provided no materials (e.g., hydrocarbons) block the path. Those that do scatter lose energy, leave the anti-resonance, quickly slow down and are captured. Each capture releases exactly the binding energy difference between \isotope{Ar}{40} and \isotope{Ar}{41}, about \SI{6.1}{\MeV} in the form of $\gamma$ rays.  As will be described below, by using a $DD$ Generator (where $DD$ stands for ``deuterium-deuterium''), a triggered pulse of neutrons can be generated outside the \dword{tpc}, then injected via a dedicated opening in the insulation into the liquid argon, where it spreads through the \SI{58}{\m} volume of the detector to produce \SI{6.1}{\MeV} energy depositions.

One important property of the neutron capture reaction \isotope{Ar}{40}(n,$\gamma$)\isotope{Ar}{41} is that the deexcitation of \isotope{Ar}{41} nucleus produces a cascade of prompt $\gamma$s. Because of the detector threshold effect, the multiplicity and the
total energy of the $\gamma$s within the cascades could be effectively decreased to below the expected values of the neutron capture process. As a consequence, the neutron capture identification and the assessment of neutron tagging efficiency in liquid argon strongly depends on a precise model of the full $\gamma$ energy spectrum from thermal neutron capture reaction. The neutron capture cross-section and the $\gamma$ spectrum have been measured and characterized by the Argon Capture Experiment at DANCE (ACED), where DANCE is the Detector for Advanced Neutron Capture Experiments. Recently, the ACED collaboration performed a neutron capture experiment using DANCE at the Los Alamos Neutron Science Center (LANSCE). The result of neutron capture cross-section was published~\cite{Fischer:2019qfr} and will be used to prepare a database for the neutron capture studies. The data analysis of the energy spectrum of correlated $\gamma$ cascades from neutron captures is underway and will be published soon. 
The $\gamma$ energy spectrum and the branching ratios in the ENDF library will be updated with the ACED result. 

Figure~\ref{fig:PNS_gamma_ACED_v2} shows an example of the energy spectra of individual $\gamma$ clusters measured by ACED~\cite{Fischer:2019qfr}. The most common $\gamma$ cascade emitted from \isotope{Ar}{41} decay has \SI{167}{keV}, \SI{1.2}{MeV} and \SI{4.7}{MeV} $\gamma$s. The peak energy of these $\gamma$s can be clearly seen in the background subtracted data in Ref.~\cite{Fischer:2019qfr}. In liquid argon detectors, the $\gamma$s are detected through calorimetric measurement. Assuming the $\gamma$ cascade from a neutron capture is fully contained in the active volume, it is possible to detect the individual $\gamma$s from the neutron capture. The correlation of the measured $\gamma$ is a strong indication of neutron capture events. Low energy $\gamma$ reconstruction algorithms are being investigated to identify the neutron capture events that could be used for detector response calibration. 

\begin{dunefigure}[Neutron capture gamma spectrum measured by ACED]{fig:PNS_gamma_ACED_v2}
{Energy spectra of individual $\gamma$ clusters measured by ACED. Only events detected in the \SIrange{0.02}{0.04}{eV} neutron energy window are selected.}
\includegraphics[width=10cm]{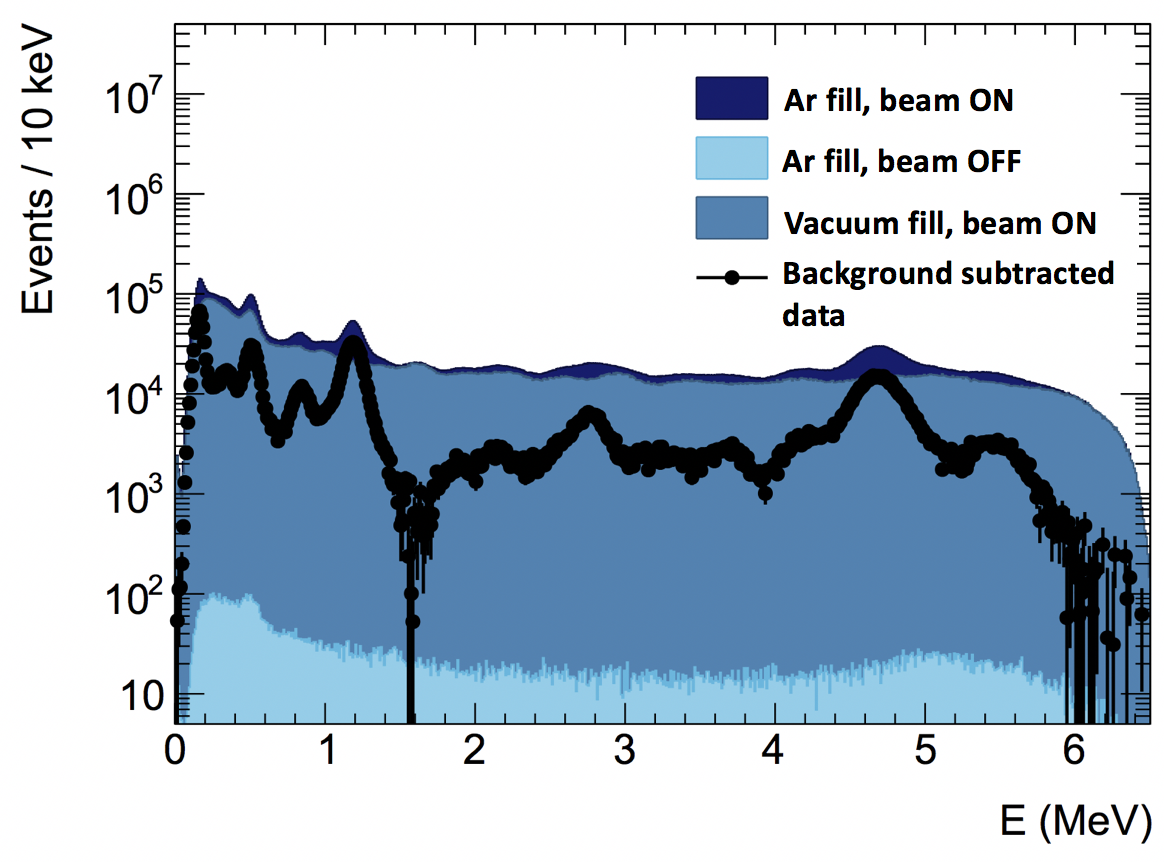}
\end{dunefigure}


\subsubsection{Design}
\label{sec:sp-calib-sys-pns-des}

The basic design concept of sources like the \dlong{pns} are based on successful boron neutron capture therapy~\cite{bib:Koivunoro2004}. The design of the \dword{pns} system used for energy calibration is shown in Figure~\ref{fig:PNS_Moderator}. The system will consist of four main components: a $DD$ generator, an energy moderator reducing the energy of the $DD$ neutrons down to the desired level, shielding materials, and a neutron monitor to confirm neutron flux and safe operation. 

\begin{dunefigure}[Conceptual design of the \dshort{pns}]{fig:PNS_Moderator}
{Conceptual design of the \dlong{pns}. The whole device is placed outside the \dword{tpc} volume on top of the cryostat.}
\includegraphics[width=12cm]{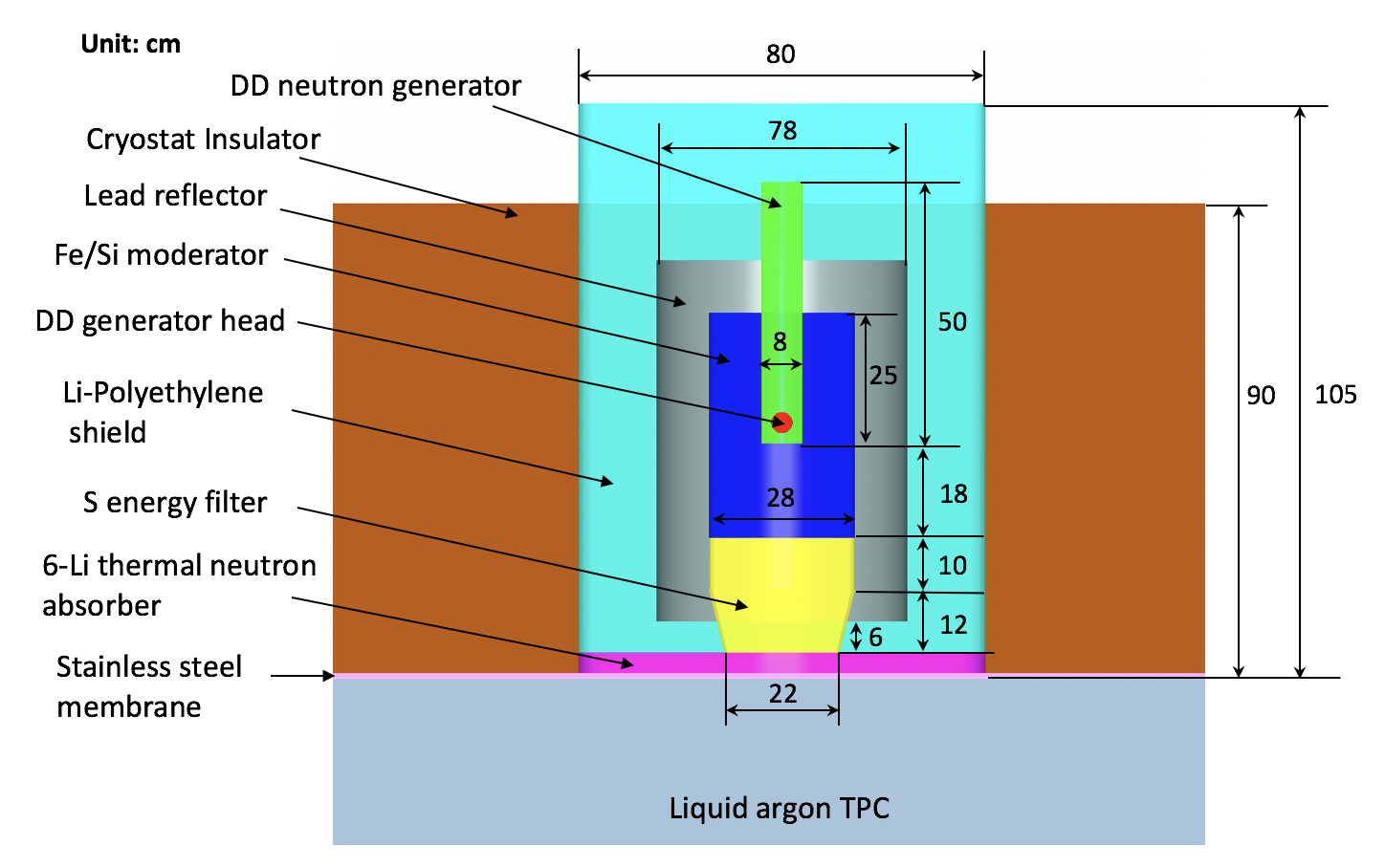}
\end{dunefigure}

{\bf DD generator source:} $DD$ generators are commercial devices that can be readily obtained from several vendors at a cost of about \$\num{125}k each, which includes all control electronics. The pulse width is adjustable and can be delivered from about \SIrange{10}{1000}{\micro\s} (which affects the total neutron output). 

{\bf Moderator:}  A feasible moderator has been designed using a layered moderator~(Fe or Si)-filter~(S)-absorber~(Li) 
configuration. The \SI{2.5}{\MeV} neutrons from the $DD$ generator are slowed to less than \SI{1}{\MeV} by the energy moderator. Natural iron and silicon are found to be efficient moderators for this purpose. Then an energy filter made of sulfur powder is used to further select the neutrons with the desired anti-resonance energy.
The neutron anti-resonance energy in \isotope{S}{32} is \SI{73}{\keV}, right above the \SI{57}{\keV} anti-resonance energy in \argon40. The neutrons at this energy lose about \SI{3.0}{\keV} per elastic scattering length. After a few elastic scattering interactions, most of the \SI{73}{\keV} neutrons selected by the sulfur filter will fall into the \SI{57}{\keV} anti-resonance energy region in \dword{lar}. These materials require no cooling or special handling. Finally, a thermal absorbing volume of lithium is placed at the entry to the argon pool in order to capture any neutrons that may have fallen below the \SI{57}{\keV} threshold. A reflecting volume is added around the $DD$ generator and the neutron moderator to increase downward neutron flux. Figure~\ref{fig:PNS_Energy_Moderator} shows the energy spectrum of the neutrons moderated and injected into the \dword{tpc}.

\begin{dunefigure}[Energy of moderated neutrons produced by the \dshort{pns}]{fig:PNS_Energy_Moderator}
{Energy of moderated neutrons produced by the \dlong{pns}. 
The total number of initial $DD$ generator neutrons is \num{1e6}.}
\includegraphics[width=9cm]{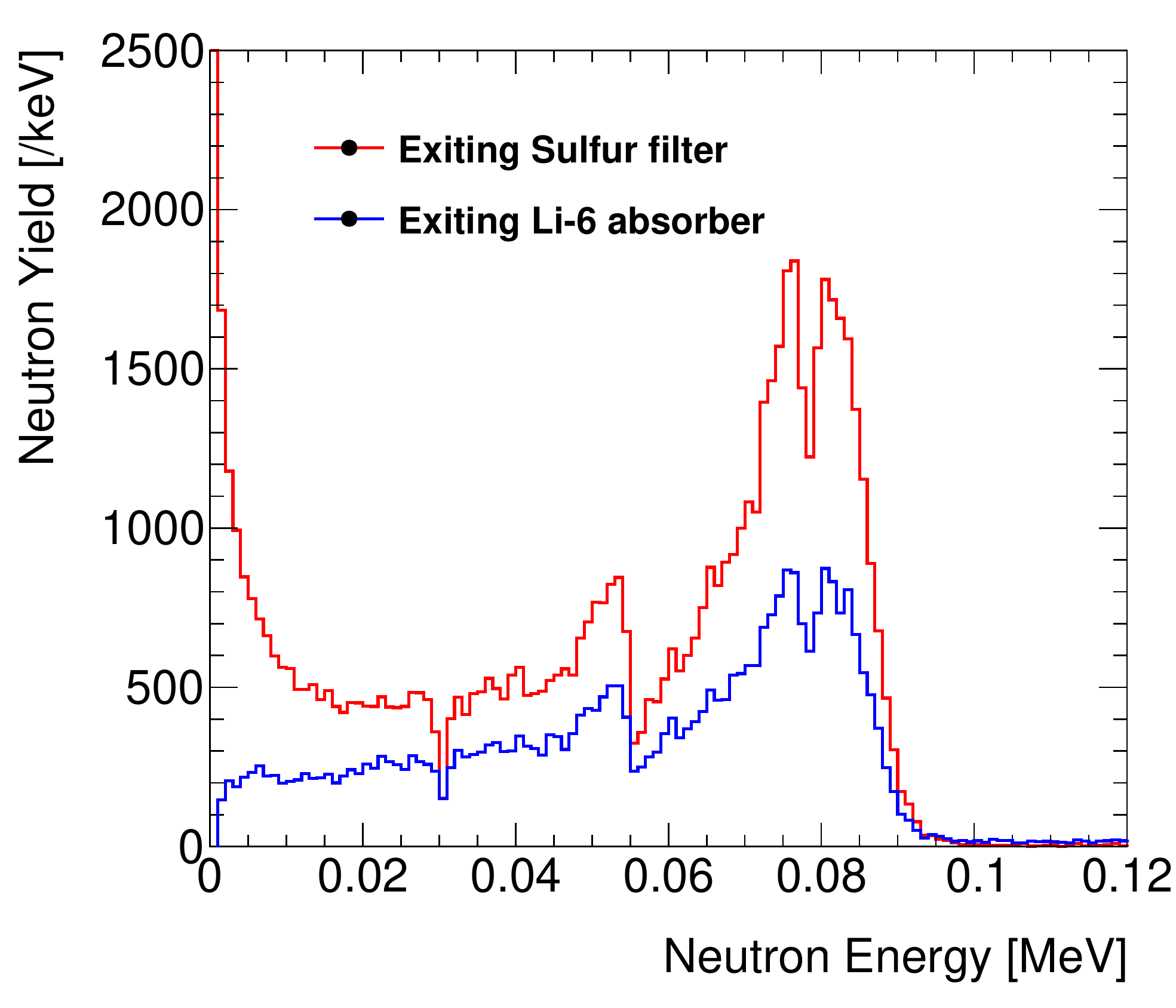}
\end{dunefigure}

{\bf Shielding:} The source will be encased in a shielding volume. The goal of the shield is to block both scattered neutrons and gammas that are produced in the source. Lithium-polyethylene (\SI{7.5}{\%}) is chosen to be the material for the neutron shield because it is rich in hydrogen and lithium atoms which yield a high neutron absorption cross section. Lithium-polyethylene is also light weight, commercially available, and relatively inexpensive. The energy spectrum entering the shield has multiple peaks between \SI{0.5}{\MeV} and \SI{1.5}{\MeV}, and one major spike at \SI{2.2}{\MeV}. The shield can effectively block the lower energy peaks but can only degrade the intensity of the \SI{2.2}{\MeV} because \SI{2.2}{\MeV} gammas are a characteristic signature for neutron captures on hydrogen. A safe thickness of the lithium-polyethylene shield must be found, one that can degrade the dose of \SI{2.2}{\MeV} gammas to safe levels. The dose of radiation from \SI{2.2}{\MeV} gammas was calculated assuming a person standing \SI{1}{\m} away. Simulation indicates that a shield with \SI{12}{\cm} thick lithium-polyethylene satisfies basic safety requirements\footnote{These calculations will be redone assuming a \SI{30}{\cm} personnel safety distance and shielding thickness reestimated to meet \dword{dune} safety requirements.}.

{\bf Neutron Monitor:} The system will need a monitoring system to confirm that the source is operating as expected.  A neutron monitoring detector consisting of an Eljen EJ-420 coupled to an ADIT L51B16S \num{2}-inch \dword{pmt} will be placed just outside of the moderating material surrounding the $DD$ generator and will be read out with a CAEN waveform digitizer with neutron/$\gamma$ pulse-shape discriminating firmware. The monitoring detector will provide relative flux information to the calibration users and will ensure that the intensity of the source is constant, thereby allowing a comparison of data taking at different times.  A small collimator will be placed in front of the neutron detector, and inside the shielding material of the $DD$ source. The collimator dimensions and material specifications (likely a combination of iron, lead, and polyethylene) will be optimized from Monte Carlo simulations.

Based on the general concept described above, Figure~\ref{fig:PNS_Moderator_largeformat} shows a conceptual layout of the neutron injection system. It is referred to as the ``large format moderator'' design. The neutron source is about \SI{0.8}{\m} wide and \SI{1}{\m} high. It would sit above the cryostat insulator. Beneath the neutron source, a cylindrical insulator volume with a diameter of more than \SI{50}{\cm} has to be removed to allow the neutrons enter the cryostat. Such an interface is provided by the human access ports near the endwalls of the detector.
The top flange of the human access port is sealed, and the neutron source sits on top, providing heat insulation. The neutron source weighs about \SI{1.6}{\tonne} and will be supported by the I-beams. 
This design allows a permanent deployment of the neutron source. GEANT4 simulation has shown that \SI{0.13}{\%} of the neutrons generated by the $DD$ generator are expected to be captured inside the \dword{tpc}. It is also possible to place the neutron source inside the human access ports which would allow a factor of \num{6} increase of the neutron flux but will require a modification of the interface flange. This is currently being investigated.

\begin{dunefigure}[\dshort{pns} baseline design]{fig:PNS_Moderator_largeformat}
{Large format neutron source deployed above/inside the human access holes.}
\includegraphics[width=12cm]{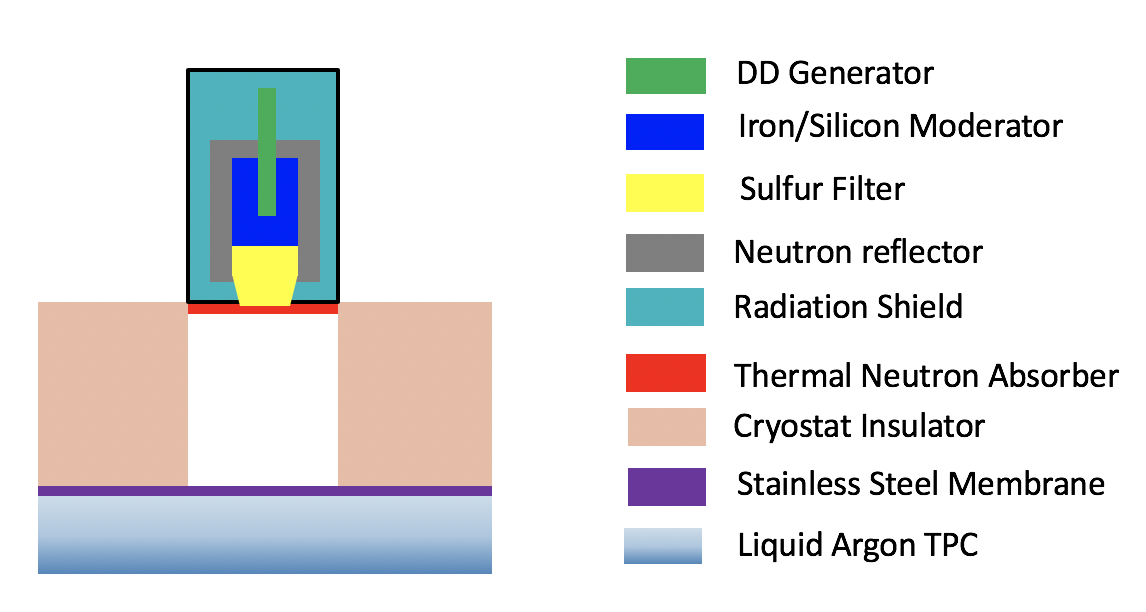}
\end{dunefigure}

Simulation studies were done placing the \dword{pns} system on top of a cryostat with the same size as the \dword{dune} \SI{10}{\kton} \dword{tpc}. Initial simulation results indicate that one \dword{pns} could cover 1/3 of the \dword{tpc} volume, so three identical neutron sources on top of the cryostat would illuminate the whole \dword{tpc} volume of the \dword{dune} \dword{fd}. However, this would require opening three additional neutron injection ports which are not included in the current cryostat design\footnote{Ideally, opening three identical neutron injection ports for each \SI{10}{\kton} TPC would make full use of the neutron source. While this is not possible for the first \dword{fd} module as the cryostat design is frozen, it informs the importance of these ports for subsequent \dword{fd} modules.}. The baseline configuration of the \dword{pns} system consists of two large format neutron sources permanently located at the corner human access ports at the opposite ends on top of the cryostat. 

Figure~\ref{fig:PNS_ncapDistribution_two_sources} shows the position distribution of the neutron captures under baseline configuration. The distribution shows that the baseline deployment can cover a large fraction of the \dword{tpc} volume, but, as evident from the figure, not many neutrons reach the central region of the \dword{tpc}. Neutrons with long scattering lengths can reach the center of the \dword{tpc} but, much longer operation time maybe needed to achieve the required statistics. Assuming a minimum number of \num{100} neutron captures per \si{\cubic\m} in order to carry out a localized energy calibration, and the typical $DD$ generator pulse intensity of \num{e5} neutrons/pulse, the number of pulses needed to calibrate the high rate regions is of the order of \num{1000}, and at least \num{10} times that for the low rate regions. But given the \SI{0.5}{\hertz} \dword{daq} limitation, this would mean calibration runs will increase from 40 minutes to about \num{7} hours to cover the low rate regions. More details on this are given in Section~\ref{sec:sp-calib-daqreq}.
If the neutron capture events at the center of the \dword{tpc} are not sufficient, the detector response calibration would depend on simulations and extrapolation using results from the regions with high neutron coverage. To increase the low coverage at the center of the \dword{tpc}, an alternative deployment strategy is proposed using a small format neutron source design described in Section~\ref{sec:sp-calib-pns-alter}. 

\begin{dunefigure}[Pulsed neutron system neutron capture positions inside a \dshort{dune}-sized \dshort{tpc}]{fig:PNS_ncapDistribution_two_sources}
{Neutron capture positions inside a \dword{dune}-sized \dword{tpc}, assuming baseline configuration with two large format neutron sources located at the corner human access ports at the opposite corners on top of the cryostat. $L$=\SI{60}{\m} (along $Z$ axis, horizontally parallel to the beam direction), $W$=\SI{14.5}{\m} (along $X$ axis, horizontally perpendicular to the beam direction), $H$=\SI{10}{\m} (along $Y$ axis, vertically perpendicular to the beam direction). \num{1.8e7} $DD$ generator neutrons with \SI{2.5}{\MeV} energy were simulated in each moderator and propagated inside the \dword{tpc}. Top (left) and side (right) views of neutron capture positions are shown.}
\includegraphics[width=18cm]{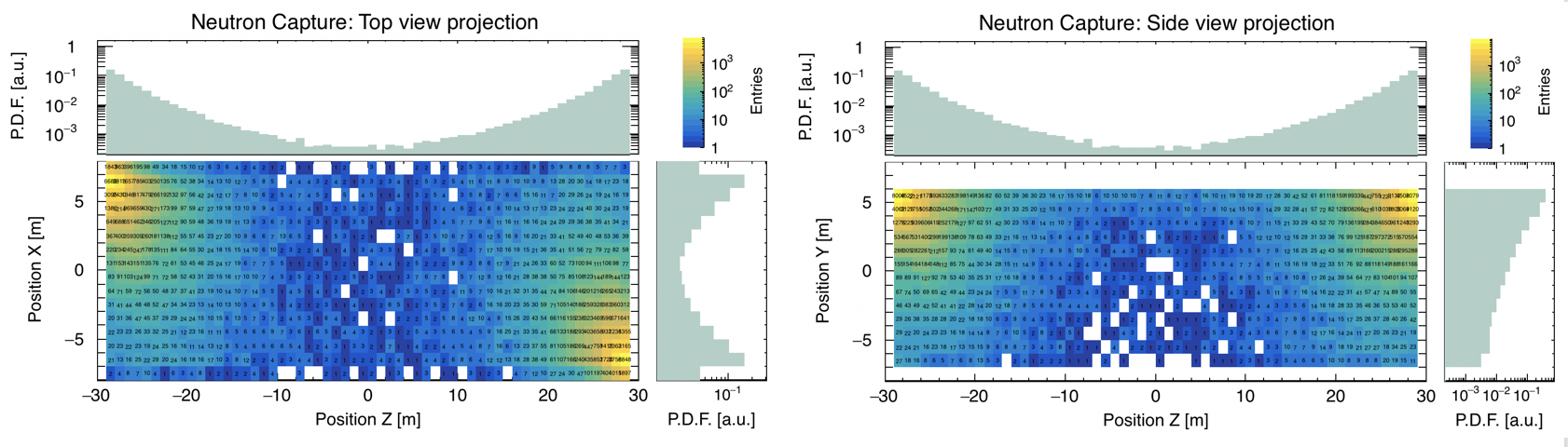}
\end{dunefigure}

The system is expected to have a long lifetime of operation, as the \dword{pns} system sits on top of the cryostat, with no opening to the \dword{lar}, so it is possible to replace the system in case of failure with only crane support.



\subsubsection{Measurement Program}
\label{sec:sp-calib-sys-pns-meas}

The \SI{6.1}{\MeV} $\gamma$ cascade will provide a uniform signal for neutron capture, part of the supernova signal. The source may also be used to determine the relative efficiency across the detector for neutron capture, and provide measurements of energy resolution and energy scale spatially and temporally. Simulation studies are currently underway.


The first goal of the simulation is to provide the expected distribution of signals, with a normalization given by the pulse width of \dword{pns} operation, and neutrons energy and angular correlated distribution, depending on the source filter and shield design.
It is envisaged that the calibration can be done in two modes. First, a short \dword{pns} pulse can provide isolated neutron captures closer to the entrance path; and then a longer 
pulse, for which the same region is saturated, but 
captures 
happen in the full volume.

By using an external trigger coupled to the \dword{pns} operation and running the usual trigger algorithms in parallel, the calibration will provide the efficiency of the trigger and \dword{daq} systems as a function of total fluxes. Changing the pulse width can result in 
higher or lower detector activity. The source will be used for \dword{snb} calibration to test
the capabilities of triggering for low energy signals, but also 
to identify them in different pile-up conditions.
The transmission of the global timing from the external \dword{pns} trigger to the \dword{daq} provides a strong constraint on the initial timing for the \dword{tpc}
as the neutron capture times are of the order of \SI{0.15}{\milli\s}, much lower than typical drift time 
for the TPC. The \dword{pds}, with resolution of \SI{100}{\nano\s}, can discriminate between different neutron captures. The calibration will measure the efficiency of the \dword{pds} response for low energy events, depending on the distance due to Rayleigh scattering in \lar. We will then study the usage of the \dword{pds} time information for improving the position reconstruction of \dword{tpc} signals. In the absence of the \dword{pds} system, the global timing from the \dword{pns} translates to an uncertainty of around \SI{10}{\cm}.

Individual event positions can be translated into response maps of both the photon detectors and the \dword{lartpc} to standard candles of \SI{6.1}{\MeV} electromagnetic depositions. When the cascades can be more precisely reconstructed, individual $\gamma$s within the cascades can be identified, and this provides a lower energy ``standard candle'' close to the solar electron-neutrino threshold. Comparing the collected charge for equal energy signals at different distances from the \dword{tpc} gives a measurement of the electron lifetime, a key detector response parameter. High \dword{pns} flux runs can generate momentary local space charge effects, in the 
upper regions of the detector, that will need to be characterized; low flux runs should be taken before to ensure 
expected space charge distributions.
The global simulation will be tested in the (smaller scale) \dword{protodune} detectors. The neutron mean free path will be larger than the \dword{protodune} size, and so 
external events and interactions with materials of the \dword{pds}, \dword{apa}, and \dword{cpa} systems will be more prominent. These effects must be simulated.

Note that captures of external background neutrons, entering the active volume is a main background for low energy physics; a comparison of simulations of \dword{pns} events and external neutron backgrounds will be interesting, as will a comparison of simulated supernova and solar neutrino signals. For the high energy beam events, the number and energy distributions of neutrons depend on the type of neutrino interaction and are significantly different for neutrinos and anti-neutrinos. Measuring the number and distance of neutron captures around the main hadronic cascades can thus help in identifying which extra proton scattering signals to associate to the hadronic cascades. This can also help make a statistical correction to the energy reconstruction of the neutrino and anti-neutrino events.

\subsection{Validation of Calibration Systems}
\label{sec:sp-calib-val}

All calibration designs presented in the previous section require full system validation before being deployed in the \dword{dune} \dword{fd}. Here, we describe the validation of a complete baseline design and some of the alternative designs described in the Appendix, Section~\ref{appx:calibration}.

Although laser calibration systems are being operated in other \dword{lartpc} experiments (e.g., \dword{microboone}, future \dword{sbnd} runs), they have stringent requirements in terms of mechanical and optical precision , long-term reliability, laser track length, performance of the \dword{lbls}, \dword{daq} interface, and effect on \efield, especially due to the \dword{fc} penetration. 
All of these lead to corresponding goals for a test installation and operation in \dword{pdsp} that could be done in the post-LS2 run. As Figure~\ref{fig:protoDUNESP_topView_marked} shows, \dword{pdsp} has ports of the same size as the \dword{dune} \dword{fd} that could be used for these tests. If a pair of ports can be used, then one could even have crossing tracks within a single drift volume. If one of the ports external to the \dword{tpc} can be used, then we would test the double-rotary alternative system described in Section~\ref{sec:sp-calib-laser-alter} and aimed at improving the coverage from the end-wall locations.

\begin{dunefigure}[Top view of the \dshort{pdsp} cryostat showing various penetrations]{fig:protoDUNESP_topView_marked}
{Top view of the \dword{pdsp} cryostat showing various penetrations. Ports marked in red are free and could be used to test the calibration systems. The four largest ports have the same diameter (\SI{250}{\milli\m}) as the calibration ports of DUNE \dword{fd}, and are located over the \dword{tpc}. The largest ports at the right side corners of the cryostat are the human access ports.}
\includegraphics[height=4.0in]{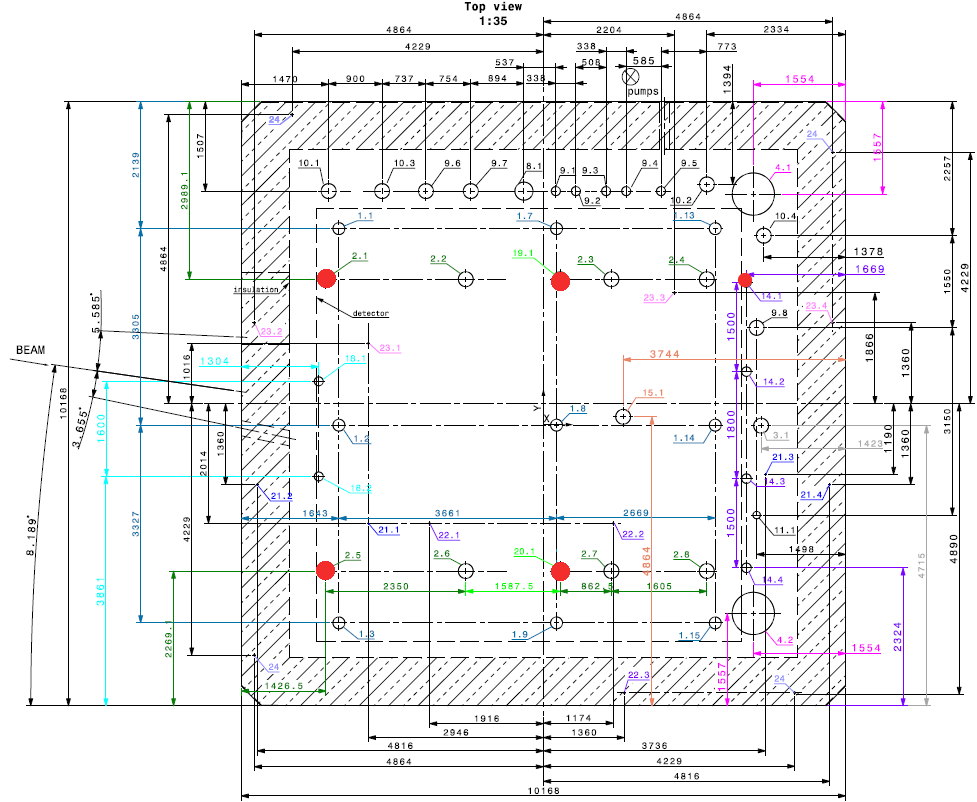}
\end{dunefigure}

The goal for validation would be to test all aspects of the system design, installation, alignment, operation, interfaces with \dword{daq}, and analysis, among others. \dword{pdsp}, because it is located at the surface, could measure the \efield map with cosmic rays to compare with the one from the laser system to improve the analysis methods or identify weak aspects in the design. An important design parameter is the length of a laser track. Our design assumes that \SI{20}{\m} is possible. \dword{microboone} has demonstrated only up to \SI{10}{\m}, but the track could be longer, depending on laser intensity. Measurements are limited by the size of the detector, but one way to gain information on longer tracks is to make a scan with low laser intensities, so that the end of the track is  visible, and register how the maximum obtained track length scales with intensity. An extrapolation to the \dword{dune} \dword{fd} laser intensity would tell us the maximum length possible. Such a measurement could also be done at \dword{microboone} or \dword{sbnd}.

An important aspect of the development plan, to be carried out at \dword{pdsp2}, is the characterization of the charge created by the laser beam ionization as a function of distance travelled in the \dword{lar} and the laser beam intensity. This dependence is thought to be affected by self-focusing effects due to the high light intensity, but it can be studied by measuring the collected charge distribution from a series of tracks close, and parallel, to the \dword{apa}s in order to break any correlations with the electron lifetime. This measured charge function could then be used with tracks in different directions to obtain a measurement of electron lifetime, which would significantly increase the capabilities of the laser system. 

The pulsed neutron source is a new idea never used in other experiments, so a \dword{pdsp2} test is essential. The corner human access ports similar to the ones in the \dword{dune} \dword{fd} could be used for this test.

In addition to dedicated hardware validation runs at \dword{pdsp2}, other \dword{lar} experiments provide ample opportunities to develop and validate calibration tools and techniques, especially those relevant to the hardware being deployed. For example, the \dword{microboone} experiment is currently leading the development of analysis methods using laser data to extract an \efield map. Energy calibration techniques and related software tools are also being developed at various experiments (\dword{microboone}, ICARUS, \dword{lariat}, \dword{protodune}) that involve estimating and propagating uncertainties like \efield distortions, recombination, and other effects into physics signals. Other calibration related developments include \dword{daq} and calibration database design, all of which are being improved at \dword{sbn} and \dword{protodune}.

\section{Interfaces with other Consortia}
\label{sec:sp-calib-intfc}

Interfaces between calibration and other consortia have been identified and appropriate documents have been developed. 
The documents are currently maintained in the \dword{cern} Engineering and Equipment Data Management Service (EDMS) database, with a \dword{tdr} snapshot kept in the \dword{dune} document database (DocDB).
\dword{dune} document database (DocDB). 
A brief summary is provided in this section. Table~\ref{tab:fdgen-calib-interfaces} lists the interfaces and corresponding DocDB document numbers. 
The main systems calibration has interfaces with are \dword{hv}, \dword{pds}, and \dword{daq}, and the important issues that must be considered are listed below.

\begin{description}
    \item[HV] Evaluate the effect of the calibration hardware on the \efield due to laser system periscopes and \dword{fc} penetration. 
    Evaluate the effect of the incident laser beam on the \dword{cpa} material (Kapton); Integrate the hardware of the 
    photoelectron laser system (targets) and the \dword{lbls} (diodes) within the \dword{hv} system components. Ensure that the radioactive source deployment is in a safe field region and cannot do mechanical harm to the \dword{fc}.
    \item[PDS] Evaluate long term effects of laser light, even if just diffuse or reflected, on the scintillating components (\dword{tpb} plates) of the \dword{pds}; establish a laser run plan to avoid direct hits; evaluate the effect of laser light on alternative \dword{pds} ideas, such as having reflectors on the \dwords{cpa}; validate light response model and triggering for low energy signals. 
    \item[DAQ] Evaluate \dword{daq} constraints on the total volume of calibration data that can be acquired; develop strategies to maximize the efficiency of data taking with data reduction methods; study how to implement a way for the calibration systems to receive trigger signals from \dword{daq} to maximize supernova live time. More details on this are presented in Section~\ref{sec:sp-calib-daqreq}.
\end{description}

Integrating and installing calibration devices will interfere with other devices, requiring coordination with the appropriate consortia as needed. Similarly, calibration will have significant interfaces at several levels with cryostat and facilities in coordinating resources for assembly, integration, installation, and commissioning (e.g., networking, cabling, safety). Rack space distribution and interaction between calibration and systems from other consortia will be managed by \dword{tc} in consultation with those consortia.

\begin{dunetable}
[Calibration system interfaces]
{p{0.14\textwidth}p{0.40\textwidth}p{0.14\textwidth}}
{tab:fdgen-calib-interfaces}
{Calibration Consortium Interface Links.}   
\small
Interfacing System & Description & Reference \\ \toprowrule
\dshort{hv}	&
effect of calibration hardware (laser and radioactive source) on \efield and field cage; laser light effect on \dshort{cpa} materials, field cage penetrations; attachment of positioning targets to HV supports 
& \citedocdb{7066} 
\\ \colhline
\dword{pds}	& 
effect of laser light on \dshort{pds}, reflectors on the \dshort{cpa}s (if any); validation of light response and triggering for low energy signals 
& \citedocdb{7051}
\\ \colhline
\dshort{daq}	& 
DAQ constraint on total volume of the calibration data; receiving triggers from DAQ
& \citedocdb{7069}  
\\ \colhline
\dshort{cisc} &
multi-functional \dshort{cisc}/calibration ports; space sharing around ports; fluid flow validation; slow controls and monitoring for calibration quantities 
& \citedocdb{7072} 
\\ \colhline
TPC Electronics	         &  
Noise, electronics calibration
& \citedocdb{7054}  
\\ \colhline
\dshort{apa}	&
\dshort{apa} alignment studies using laser and impact on calibrations
& \citedocdb{7048} 
\\ \colhline
Physics	&
tools to study impact of calibrations on physics
& \citedocdb{6865}  
\\ \colhline
Software and Computing	  &
Calibration database design and maintenance
& \citedocdb{6868} 
\\ \colhline
TC Facility              &   
Significant interfaces at multiple levels   
& \citedocdb{6829}   \\ \colhline
TC Installation     	  &     
Significant interfaces at multiple levels
& \citedocdb{6847}    \\ 

\end{dunetable}

\subsection{Calibration Data Volume Estimates}
\label{sec:sp-calib-daqreq}

The calibration systems must interface with the DUNE \dword{daq} system, discussed in detail in Chapter~\ref{ch:daq}.
Trigger decisions for physics events are made 
hierarchically: \dwords{trigprimitive} are generated from \dword{tpc} and \dword{pds} ``hits'', and these \dwords{trigprimitive} are then used to create \dwords{trigcandidate} which are collections of \dwords{trigprimitive} satisfying selection criteria such as exceeding a threshold number of adjacent collection wire hits, or total collection wire charge recorded, etc. These \dwords{trigcandidate} are passed on to a \dword{mlt} which then makes decisions about whether a given \dword{trigcandidate} is accepted as a detector-wide trigger.  If so, the \dword{mlt} sends trigger commands to the \dword{daqdfo} which in turn passes them to an available \dword{eb} that then requests data from the \dword{fe} readout of the \dword{daq} (servers that host \dword{felix} cards). The management of \dwords{trigdecision}---whether they are generated by candidates from the \dword{tpc}, \dword{pds}, calibrations, or other systems---is done in the \dword{mlt}. 

The trigger commands are in the form of absolute time stamps that are used to extract snapshots of the data stored in the \dword{fe} readout buffers. For physics triggers, all \dword{tpc} information for a snapshot of time (roughly twice the drift time, or \SI{5.4}{\milli\s}) are read out, without any additional zero suppression or localization. For calibration events, this approach would create an unmanageable amount of data and, in any case, is unnecessary because calibration events create interactions or tracks at known positions or times, or both.

To reduce data volume from calibrations, therefore, calibration systems that can be triggered externally are desirable. Like the distribution of trigger commands to the \dword{fe} readout buffers, the external trigger for a calibration system will take the form of an absolute time stamp. The time stamp is generated by the \dword{mlt}, thus ensuring that (for example) a calibration event does not occur during a candidate supernova burst.  The distribution of these time stamps will be done through the \dword{daq}'s timing and synchronization system. Thus triggerable calibration systems (like the laser or \dword{pns}) will have to be synchronized to the rest of the \dword{daq} system and be capable of accepting time stamps. There will be differences in the details of how different calibration systems are handled, as discussed later in this section. 

Table~\ref{tab:calib-daq} shows the estimated data volume needs for various calibration systems assuming each system is run twice per year. For the ionization laser system, as noted earlier, a scan of the full detector can take about three days, resulting in a total of about six days or a week per year per \SI{10}{\kt} module. For the \dword{pns} system, as noted later in this section, a single run can take about seven hours;  doing that twice, or even four times, per year will result in a total of about one day per year per \SI{10}{\kt} module. It is expected that once the detector launches into stable operations, the need for full calibration campaign runs will reduce to 
one nominal run per year. 
We also expect some shorter runs may be needed in smaller, targeted regions of the detector, or for detector diagnostic issues. 
           
\begin{dunetable}
[Calibration DAQ summary]
{p{0.2\textwidth}p{0.15\textwidth}p{0.5\textwidth}}
{tab:calib-daq}
{Estimated data volume needs per year per \nominalmodsize for various calibration systems.}   
System & Data Volume (TB/year) & Assumptions  \\ \toprowrule
Ionization Laser System & \num{184} & \num{800}k laser pulses, \num{10}$\times$\num{10}$\times$\SI{10}{\cubic\cm} voxel sizes, a \SI{100}{\micro\s} zero suppression window (lossy readout), and \num{2} times/year  \\ \colhline
Neutron Source System & \num{144} & \num{e5}~neutrons/pulse, \num{100} neutron captures/m$^{3}$, \num{130} observed neutron captures per pulse, \num{2}~times/year  \\ 
\end{dunetable}           
           
\subsubsection{Laser System}

The \efield vector from ionization laser calibration is determined by looking at the deflection of crossing laser tracks within detector voxels.  Because any given laser track illuminates many such voxels, one laser pulse can be used for several measurements; essentially, what matters is how many voxels it takes to cover three walls of a given drift volume -- \dword{cpa}, bottom and end-wall \dword{fc}, taking into account that we divide that volume by \num{4} because of beam coverage.

Considering a small voxel size of  \num{10}$\times$\num{10}$\times$\SI{10}{\cubic\cm}, the total number of independent track directions is estimated to be \num{800000}: about half the rate of cosmic rays and thus nominally a substantial total data volume. 
However, with the specification voxel size of \num{30}$\times$\num{30}$\times$\SI{30}{\cubic\cm}, that number would be \num{27} times smaller, so that would allow a larger number of tracks per direction. Keeping to the overall estimate of \num{800000} tracks per scan, the choice of voxel granularity and track statistics per direction can be made until the commissioning period.

Fortunately, unlike every other event type in the detector, the laser track has both a reasonably well known position and time; thus the trigger command issued to the \dword{fe} buffers can be much narrower than the window used for physics triggers. A \SI{100}{\micro\s} zero suppression window should be wide enough to avoid windowing problems in the induction plane wire deconvolution process.
To ensure that the interesting part of each waveform is recorded, the \dword{daq} will need to know the current position  of the laser, which will be transmitted from the laser system to the \dword{mlt} via the \dword{daqccm}.

From the standpoint of data volume, therefore, the total assuming the \SI{100}{\micro\s} zero-suppression window is
\begin{equation}
\num{800000}/{\rm scan}/\nominalmodsize \times \SI{100}{\micro\s} \times \num{1.5}{\rm Bytes/sample}\times \SI{2}{\mega\hertz}\times \num{384000}~{\rm channels}   = \num{92}~{\rm TB/scan/\nominalmodsize.}   
\end{equation}
If such a calibration scan were done twice a year, then the total annual data volume for the laser is \num{184}~TB/year/\nominalmodsize and four times a year would result in \num{368}~TB/year/\nominalmodsize

\subsubsection{Pulsed Neutron Source}

The \dlong{pns} (\dshort{pns}) system creates a burst of neutrons that
are captured throughout a large fraction of the total cryostat volume. For triggering and data volume, this is very convenient: the existing scheme of taking \SI{5.4}{\milli\s} of data for each trigger means all these neutrons will be collected in a single \dword{dune} event. Thus, the data volume is simply \num{6.22}~GB times the total number of such pulses, but these are likely to be few: a single burst can produce thousands of neutrons whose $t_0$ is known up to the neutron capture time of \SI{200}{\micro\s} or so.

To trigger the \dword{pns}, the \dword{mlt} will provide a time stamp for the source to fire, and then send a trigger command to the \dword{fe} readout buffers (via the \dword{daqdfo} and \dword{eb}) that will look like a physics trigger command.  The \dword{mlt} itself then tags that trigger command with the expected trigger type (in this case, \dword{pns}).

Typically, a commercial $DD$ neutron generator produces \num{e5} - \num{e8} neutrons/pulse, depending on the adjustable pulse width. 
The current assumption for neutron yield from the $DD$ generator is \num{e5} neutrons per pulse\footnote{Realistic assumption based on commercially available $DD$ generators that produce the most neutron yield with a pulse width less than \SI{100}{\micro\s}. 
$DD$ generators with higher neutron yield are being developed in laboratories; commercial devices may require further development to reach a higher level of performance.}. 
With the current baseline deployment design in Figure~\ref{fig:PNS_Moderator_largeformat}, approximately \num{130} neutron captures per $DD$ generator pulse should be observed inside a \nominalmodsize module. As shown in Figure~\ref{fig:PNS_ncapDistribution_two_sources}, the deployment of two large format neutron sources at the corner human access ports could approximately provide calibration for about half of the total \dword{tpc} volume (\SI{30}{\kt}). 
As the suggested number for localized energy calibration is \num{100} neutron captures per \si{\cubic\m}, a total number of \num{2300} pulses would be needed to calibrate regions under high neutron coverage. Assuming two identical \dlong{pns}s operating in synchronization mode, \num{1150} triggers are needed for each calibration run. Therefore, the total data volume per run would be
\begin{equation}
\num{1150}~{\rm Triggers} \times \num{1.5}~{\rm Bytes}\times
\SI{2}{\mega\hertz}\times \SI{5.4}{\milli\s}\times \num{384000}~{\rm channels} = \num{7.2}~{\rm TB/run}.
\end{equation}
The recommended trigger rate of the \dword{pns} system is \SI{0.5}{\hertz} which is limited by the bandwidth of the DAQ event builder. Assuming that the spatial distribution of the neutron capture is near-uniform for the regions that are covered by the two large format neutrons sources, the operation time per calibration run would be 40 minutes.   
Running the \dword{pns} calibration system twice a year would result in a total data volume of \SI{14.4}{TB} per \nominalmodsize per year. 
For realistic neutron capture distribution that is non-uniform, we expect to operate the \dword{pns} system for a period of 10 times longer than that under the ideal assumption (7.2 TB/run). As a consequence, the data size per calibration run would be 72 TB/run and running the \dword{pns} calibration twice a year would result in a total data size of \num{144}~TB/year/\nominalmodsize and four times a year results in \num{288}~TB/year/\nominalmodsize.


\section{Construction and Installation}
\label{sec:sp-calib-const}

\subsection{Quality Control}
\label{sec:sp-calib-qc}

The manufacturer and the institutions in charge of devices will conduct a series of tests to ensure the equipment can perform its intended function as part of \dword{qc}. \dword{qc} also includes post-fabrication tests and tests run after shipping and installation. The overall strategy for the calibration devices is to test the systems for correct and safe operation in dedicated test stands, then at \dword{pdsp2}, then as appropriate near \dword{surf} at South Dakota School of Mines and Technology (SDSMT), and finally underground. Electronics and racks associated to each full system will be tested before transporting them underground.

\begin{itemize}
    \item {\bf Ionization Laser System:} The first 
    important test is  design validation in \dword{pdsp2}. For assembly and operation of the laser and feedthrough interface, this will be carried out on a mock-up flange for each of the full hardware sets (periscope, feedthrough, laser, power supply, and electronics). All operational parts (UV laser, red alignment laser, trigger photodiode, attenuator, diaphragm, movement motors, and encoders) will be tested for functionality before being transported underground.
    \item {\bf Photoelectron Laser System:} The most important test is to measure the light transmission of all fibers at \SI{266}{\nano\m}. A suitable transmission acceptance threshold will be established based on studies during the development phase. Studies to estimate the number of photoelectrons emitted as a function of intensity (based on distance of fiber output to the metallic tab) will also be undertaken.
    \item {\bf Laser Beam Location System:} For the \dword{lbls}, the main test is checking that the PIN diodes are all functional, and with a light detection efficiency within a specified range, to ensure uniformity across all clusters. For the mirror-based system, the reflectivity of all mirrors will also be tested prior to assembly.
    \item {\bf Pulsed Neutron Source System:} The first test will be safe operation of the system in a member institution radiation-safe facility. Then, the system will be validated at \dword{pdsp2}. The same procedure 
    will be carried out for any subsequent devices before the devices are transported to \dword{surf} and underground. System operation will be tested with shielding assembled to confirm safe operating conditions and sufficient neutron yields using an external dosimeter as well as with the installed neutron monitor. The entire system, once assembled, can be brought down the Ross shaft.
    
\end{itemize}

\subsection{Installation, Integration and Commissioning}
\label{sec:sp-calib-iic}

This section describes the installation plans for calibration systems. Most of the hardware is to be installed outside the cryostat so, space on mezzanine surrounding each calibration port is important for powering and operating the calibration systems. However, some sub-systems have internal components which will be installed 
following a specific installation sequence, coordinated with other consortia.

\subsubsection{Ionization Laser System} 
 
Checking the alignment of the optical components is an essential step of the ionization laser system installation. The system includes a low power visible laser that can be used for the several mirror alignment operations, but before that use, both the UV and the visible lasers in the laser box need to be aligned.
Alignment of the visible and UV (Class 4) lasers requires special safety precautions and must be carried out once for each periscope/laser system before installing further \dword{tpc} components. For that reason, the laser boxes must be installed on the cryostat roof as soon at that area becomes accessible.  
 
The periscopes are the only components of the ionization laser system that will be inside the cryostat, but they will be installed from the top of the cryostat and not from the \dword{tco}, including the alternative options. However, this installation should be done very carefully in the presence of an operator inside the cryostat, to ensure there are no collisions of the long laser periscopes with other detector components, especially \dword{fc} elements and \dword{ce} cable trays.
The periscopes should be installed after the relevant structural elements, especially the top \dword{fc} modules. Installation should proceed in sequence with the assembly of other components, with the furthest from \dword{tco} assembled first.

The relevant \dword{qc} is essentially an alignment test.
The \dword{lbls} can be used to align the periscopes as they are installed, so it is important that the \dword{lbls} is also installed in the same sequence as the periscopes.

A support beam structure closest to the \dword{tco} temporarily blocks the calibration ports, but it is removed after the last \dword{tpc} component is installed. After that, the final calibration components can be installed, including the periscopes on the \dword{tco} end wall. 

\subsubsection{Laser Beam Location System}
This system has several parts that need to be installed inside the \dword{tco}, and some must be integrated with the 
\dword{hv} system during installation underground. 

The PIN diode system uses a set of diodes that fire when the laser beam hits them. Because the laser shoots from above and the diodes must be in a low voltage region, the plan is to place the diodes below the bottom \dword{fc}, facing upward, simply on a tray close to the cryostat membrane.

For the pointing measurement, the beams will pass through the \dword{fc} electrodes and hit the diodes below. There will be \num{32} of these diode clusters to be installed. The installation will consist of positioning the cluster trays in pre-determined locations, and routing the cables to the respective feedthroughs (work is still underway to decide how to route cables and which flanges to use). 

The second \dlong{lbls} consists of a set of \num{32} mirror clusters: a plastic or aluminum piece holding four to six small mirrors \SI{6}{\mm} in diameter, each at a different angle; the ionization laser will point to these mirrors to obtain an absolute pointing reference. These clusters will be attached to the bottom \dword{fc} profiles facing into the \dword{tpc}. 
This attachment/assembly of the mirror clusters on corresponding \dword{fc} profiles will be done during \dword{fc} assembly underground.

\subsubsection{Photoelectron Laser System} 
A large number of photoelectric targets (about \num{4000}) must be 
attached to the cathode. Experience from other experiments indicates that targets can be glued to the cathode surface, which can be done after cathode assembly but before the cathode is installed in the cryostat. 

Once the \dwords{cpa} are in place, the photoelectric target locations will need a high precision survey, which is necessary for the absolute calibration of the electric field with the photoelectron laser. 

The third part of the installation is  quartz optical fibers on the \dword{apa}, needed to illuminate  the photoelectric targets with light from the Nd:YAG laser. 
Fiber tips must be properly fastened and oriented for effective illumination, and fiber bundle routing will bring the fiber bundles to the outside of the cryostat where Nd:YAG laser injection points will be located. 

\subsubsection{Pulsed Neutron Source System} 
The \dword{pns} will be installed after the human access ports are closed because the source sits above the cryostat. Installing the system should take place in two stages. In the first stage, the assembly of the system would be independent of the \dword{tpc} installation. The whole system will be 
assembled on the ground outside the cryostat at a dedicated radiation safe area. Once assembled, the neutron source will be lifted by crane and integrated with the cryostat structure. Final \dword{qc} testing for the system will be operating the source and measuring the flux with integrated monitor and dosimeter.

\subsection{Safety}
\label{sec:sp-calib-safe}

This section discusses risks to personnel safety. Detector safety and risks involving damage to detector components are discussed in Section~\ref{sec:sp-calib-risks}.

Human safety is of critical importance during all phases of the calibration work, including R\&D, laboratory testing, prototyping (including \dword{pdsp} deployment), and integration and commissioning at the \dword{dune} \dword{fd} site. \dword{dune} \dword{esh} personnel review and approve the work planning for all phases of work as part of the initial design review, as well as before implementation. All documentation of component cleaning, assembly, testing, installation, and operation will include hazard analysis and work planning documentation and will be reviewed appropriately before production begins. In addition, in the case of planned \dword{protodune2} tests, the consortium will interface with \dword{cern} safety system to ensure all requirements are met.

Several areas are of particular importance to calibration are
\begin{itemize}
\item {\bf Underground laboratory safety:} All personnel working underground or in other installation facilities must follow appropriate safety training and be provided with the required \dword{ppe}. Risks associated with installing and operating the calibration devices include, among others, working at heights, confined space access, falling objects during overhead operations, and electrical hazards. Appropriate safety procedures will include aerial lift and fall protection training for working at heights. For falling objects, the corresponding safety procedures, including hard hats (brim facing down) and a well restricted safety area, will be part of the safety plan. More details on \dword{ppe} are provided in \dword{tdr} \tcchesh.

\item {\bf Laser safety:} The laser system requires operating a class IV laser~\cite{FNAL:Class4Lasers,CERN:Class4Lasers}. This requires an interlock on the laser box enclosure for normal operation, with only trained and authorized personnel present in the cavern for the one-time alignment of the laser upon installation in the feedthroughs. The trained personnel will be required to wear appropriate laser protective eye wear. A standard operating procedure will be required for the laser which will be reviewed and approved by the \dword{fnal} laser safety officer. 

\item {\bf Radiation safety for \dword{pns}:} A $DD$ neutron generator will be used as a calibration device. The design of safety systems for this system include key control, interlock, moderator, and shielding. Lithium-polyethylene (\SI{7.5}{\%}) is chosen to be the material for the neutron shield which is rich in hydrogen. The gammas from neutron capture on hydrogen in the shielding material could cause potential radiation hazards. The design of the radiation safety systems (custom shielding and moderator) will be designed to meet \dword{fnal} Radiological Control Manual (FRCM) safety requirements and will be reviewed and approved by \dword{fnal} radiological control organization. Material safety data sheets will be submitted to the \dword{dune} \dword{esh} to understand other safety hazards such as fire. Before beginning any operations at \dword{pdsp}, the entire system will be assembled in a neutron shielded room and tested to confirm no leaking of neutrons will occur. The system will also have a neutron monitor that can provide an interlock.

\item {\bf High voltage safety:} Some of the calibration devices will use high voltage. Fabrication and testing plans will show compliance with local \dword{hv} safety requirements at any institution or laboratory that conducts testing or operation, and this compliance will be reviewed as part of the design process.

\item {\bf Hazardous chemicals:} Hazardous chemicals (e.g., epoxy compounds used to attach components of the system) and cleaning compounds will be documented at the consortium management level, with a material safety data sheet as well as approved handling and disposal plans in place.

\item {\bf Liquid and gaseous cryogens:} Cryogens (e.g., liquid nitrogen and \dword{lar}) will most likely be used in testing of calibration devices. Full hazard analysis plans will be in place at the consortium management level for full module or module component testing that involves cryogens. These safety plans will be reviewed appropriately by \dword{dune} \dword{esh} personnel before and during production.

\end{itemize}

\section{Organization and Management}
\label{sec:sp-calib-org-manag}
\subsection{Consortium Organization}
\label{sec:sp-calib-org}

The calibration consortium was formed in November 2018 as a joint single and dual phase consortium, with a consortium leader and a technical leader. Figure~\ref{fig:orgchart} shows the organization of the consortium. The calibration consortium board currently comprises institutional representatives from 11 institutions as shown in Table~\ref{tab:gen-calib-org}. The consortium leader is the spokesperson for the consortium and responsible for the overall scientific program and management of the group. The technical leader of the consortium is responsible for managing the project for the group. 

The consortium's initial mandate is the design and prototyping of a laser calibration system, a neutron generator, and possibly a radioactive source system, so the consortium is organized into three working groups, each dedicated to one system. Each group has a designated working group leader.

\begin{dunefigure}[Organizational chart for the calibration consortium]{fig:orgchart}
{Organizational chart for the calibration consortium.}
\includegraphics[height=3.0in]{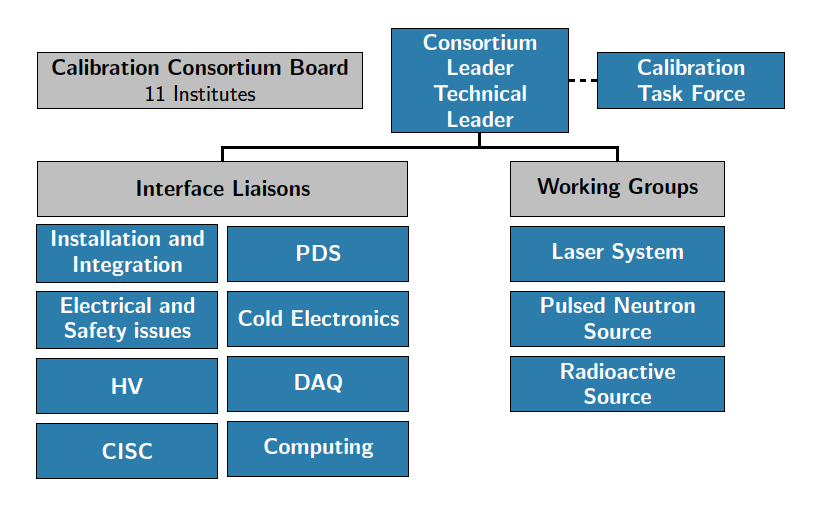}
\end{dunefigure}

\begin{dunetable}
[Calibration consortium institutions]
{lc}
{tab:gen-calib-org}
{Current Calibration Consortium Board Institutional Members and Countries.}
Member Institute     &  Country       \\
LIP & Portugal \\ \colhline
University of Bern (Bern) & Switzerland \\ \colhline
Los Alamos National Lab (LANL) & USA \\ \colhline
Michigan State University (MSU) & USA \\ \colhline
Colorado State University (CSU) & USA \\ \colhline
University of Iowa & USA \\ \colhline
University of Hawaii (Hawaii) & USA \\ \colhline
University of Pittsburgh (Pitt) & USA \\ \colhline
Boston University (BU) & USA \\ \colhline
University of California, Davis (UC Davis)& USA \\ \colhline
South Dakota School of Mines and Technology (SDSMT) & USA \\ 
\end{dunetable}

In addition, Figure~\ref{fig:orgchart} shows several liaison roles currently being established 
to facilitate connections with other groups and activities:
\begin{itemize}
    \item Detector integration and installation,
    \item Electrical and safety issues,
    \item \dword{daq},
    \item Computing,
    \item Cryogenic instrumentation and slow controls (CISC),
    \item Cold electronics,
    \item High voltage,
    \item Photon Detection System.
\end{itemize}


Currently, new institutions are added to the consortium  following an expression of interest from the interested institute and upon obtaining consensus from the current consortium board members.

\subsection{Institutional Responsibilities}
\label{sec:sp-calib-resp}


Calibrations will be a joint effort for \single and \dual. Design validation, testing, calibration, and performance of calibration devices will be evaluated using \dword{protodune} data.

Following the conceptual funding model for the consortium, various responsibilities have been distributed across institutions within the consortium. 
Table~\ref{tab:calib-inst-resp} shows the current institutional responsibilities for primary calibration subsystems. 
For physics and simulations studies and validation with \dword{protodune}, a number of institutions are interested. 

\begin{dunetable}
[Institutional responsibility for calibrations]
{p{0.25\textwidth}p{0.65\textwidth}}
{tab:calib-inst-resp}
{Institutional responsibilities in the Calibration Consortium}  Subsystem & Institutional Responsibility \\ \toprowrule
Ionization Laser System & Bern, LIP, LANL, Hawaii \\ \colhline 
\dlong{lbls} & Hawaii, LIP \\ \colhline 
Photoelectron Laser System & LANL, Hawaii \\ \colhline
Pulsed Neutron Source System & BU, CSU, UC Davis, Iowa, LIP, MSU, LANL, SDSMT \\ \colhline
Proposed Radioactive Source System & SDSMT \\ \colhline
Physics \& Simulation & BU, CSU, Hawaii, LANL, LIP, MSU, SDSMT, UC Davis, Pittsburgh, Iowa \\  
\end{dunetable}


\subsection{Risks}
\label{sec:sp-calib-risks}

Table~\ref{tab:risks:SP-FD-CAL} lists the possible risks identified by the calibration consortium along with corresponding mitigation strategies and impact on probability, cost, and schedule post-mitigation.
The table shows all risks are medium or low level, mitigated with necessary steps and precautions. More discussion on each risk is provided below. 
\begin{itemize}
    \item \textit{Risk 1:} The \dword{pdsp} design tests being inadequate for the \dword{fd} is an important one because this requires early validation from \dword{protodune} data so we can perform R\&D of alternate designs and/or improvements on a reasonable time scale. In addition, the calibration ports will be designed to be multipurpose to enable deployment of new systems if they are developed. Therefore, in general, the calibration systems mitigate risk to the experiment as the systems sit above the cryostat and/or use multipurpose ports and may be removed.
    \item \textit{Risk 2:} This is a medium-level risk where the elements of the calibration system fail engineering requirements, such as laser beam divergence and precision of the mechanical system, in which case the as-built system will not meet the physics requirements. The mitigation strategy for this involves testing the same designs envisioned for the \dword{fd} in dedicated lab tests and \dword{pdsp2}, to identify any issues and address them. The pre-installation \dword{qc} will also allow us to reject parts that do not meet requirements. 
    \item \textit{Risk 3:} If the ionization laser beam directly hits the elements of the \dword{pds} system for an extended time, the scintillation efficiency might be degraded. The mirror movement controller of the laser system must avoid the beam directly hitting the \dword{pds}. An automated system will block or turn off the laser beam in case of saturation at one of the \dword{pds} channels. The laser electrical system must allow the later implementation of a hardware interlock if that is found to be necessary.
    \item \textit{Risk 4:} This is a low level risk, where the laser beam location system fails; this would reduce the precision of the \efield measurement but will not prevent the measurement from being made. Pre-fill \dword{qc} will be carried out to minimize this risk. Additionally, redundancy will be built into the system, with alternative targets, including some passive ones. A possible alternative way to obtain an absolute measurement is to use reflections off of the aluminum \dword{fc} profiles, with a very slow angular scan.
    \item \textit{Risk 5:} This risk relates to the laser beam misalignment. If the laser beam becomes misaligned with the mirror sequence, then that specific ionization laser module becomes unusable for calibration. To mitigate this, the ionization laser system includes a visible (red) laser specifically for the purpose of alignment. If the misalignment is not just with the warm mirrors, but also with the cold ones, cryostat cameras might be needed to check arrival of red light to the \dword{tpc}.
    \item \textit{Risk 6:} If the effective attenuation length of \SI{57}{keV} neutrons in \dword{lar} turns out to be significantly smaller than \SI{30}{m}, then \dword{pns} system will not cover the whole detector, or additional modules will be needed. This will be resolved in the next year by a measurement at the Los Alamos National Lab (LANL); the \dword{protodune} run will also provide a full end-to-end demonstration.
    \item \textit{Risk 7:} If the neutron flux from the $DD$ generator of the \dword{pns} system is enough to activate the moderator and cryostat insulation, then a new source of radiological backgrounds might be created. This can be mitigated by neutron activation studies of insulation material, and \dword{protodune} testing at neutron flux intensities and durations well above the run plan, as well as simulation studies done in collaboration with the \dword{dune} Background Task Force.
    \item \textit{Risk 8:} If the neutron yield from the $DD$ generator is not high enough to provide sufficient neutron captures inside the \dword{tpc}, then either the neutron calibration cannot be done or a higher flux generator must be obtained, or additional sources must be used. Investigation is being done on both commercially available and  custom $DD$ generators. Additionally, operating the $DD$ generator with wider  pulse is under consideration, which would require the \dword{pds} to provide the neutron capture time t$_{0}$. Another possibility is to carry out dedicated runs at higher pulse rate and, to ensure that the \dword{daq} can handle it, one would acquire only the data from the \dword{apa}s farthest from the source. All of this will be tested in the \dword{protodune2} run. Placing the neutron source closer to the \dword{tpc} may increase the neutron yield by a factor of 6. An alternative design (Figure~\ref{fig:PNS_Two_Designs}) with neutron source inside the calibration feedthrough ports (centrally located on the cryostat) is being studied. This compact neutron source would be light enough to be moved across different feedthroughs and will provide additional coverage.
    \item \textit{Risk Opportunity 9:} The ionization laser system assumes that the laser beams will be sufficiently narrow for a measurement up to \SI{20}{m} distances. However, as the Rayleigh scattering is of the order \SI{40}{m}, it is possible the laser may travel further than \SI{20}{m}. This may reduce the number of lasers needed and therefore the overall cost. The maximum laser distance will be assessed in \dword{protodune2}.
\end{itemize}


\begin{footnotesize}
\begin{longtable}{P{0.18\textwidth}P{0.20\textwidth}P{0.32\textwidth}P{0.02\textwidth}P{0.02\textwidth}P{0.02\textwidth}} 
\caption[Calibration risks]{Calibration risks (P=probability, C=cost, S=schedule) The risk probability, after taking into account the planned mitigation activities, is ranked as 
L (low $<\,$\SI{10}{\%}), 
M (medium \SIrange{10}{25}{\%}), or 
H (high $>\,$\SI{25}{\%}). 
The cost and schedule impacts are ranked as 
L (cost increase $<\,$\SI{5}{\%}, schedule delay $<\,$\num{2} months), 
M (\SIrange{5}{25}{\%} and 2--6 months, respectively) and 
H ($>\,$\SI{20}{\%} and $>\,$2 months, respectively).  \fixmehl{ref \texttt{tab:risks:SP-FD-CAL}}} \\
\rowcolor{dunesky}
ID & Risk & Mitigation & P & C & S  \\  \colhline
RT-SP-CAL-01 & Inadequate baseline design & Early detection allows R\&D of alternative designs accommodated through multipurpose ports & L & M & M \\  \colhline
RT-SP-CAL-02 & Inadequate engineering or production quality & Dedicated small scale tests and full prototyping at ProtoDUNE; pre-installation QC & L & M & M \\  \colhline
RT-SP-CAL-03 & Laser impact on PDS & Mirror movement control to avoid direct hits; turn laser off in case of PDS saturation & L & L & L \\  \colhline
RT-SP-CAL-04 & Laser beam location system stops working & QC at installation time, redundancy in available targets, including passive, alternative methods & L & L & L \\  \colhline
RT-SP-CAL-05 & Laser beam misaligned & Additional (visible) laser for alignment purposes & M & L & L \\  \colhline
RT-SP-CAL-06 & The neutron anti-resonance is much less pronounced & Dedicated measurements at LANL and test at ProtoDUNE & L & L & L \\  \colhline
RT-SP-CAL-07 & Neutron activation of the moderator and cryostat & Neutron activation studies and simulations & L & L & L \\  \colhline
RT-SP-CAL-08 & Neutron yield not high enough & Simulations and tests at ProtoDUNE; alternative, movable design & L & M & M \\  \colhline
RO-SP-CAL-09 & Laser beam is stable at longer distances than designed & tests at ProtoDUNE & M & H & L \\  \colhline

\label{tab:risks:SP-FD-CAL}
\end{longtable}
\end{footnotesize}

\subsection{Schedule and Milestones}
\label{sec:sp-calib-sched}

\begin{dunetable}
[Calibration consortium schedule]
{p{0.65\textwidth}p{0.25\textwidth}}
{tab:calib-sched}
{Key calibration construction schedule milestones leading to commissioning the first \dword{fd} module. (*) Schedule items related to the \dlong{rsds} (\dshort{rsds}) are to be considered pending system approval.}   
Milestone & Date (Month YYYY)   \\ \toprowrule
Laser systems design decision (including ionization, beam location and photoelectron systems) & January 2020 \\ \colhline 
Laser systems design review & February 2020 \\ \colhline 
\dword{pns} design decision  & March 2020 \\ \colhline
\dword{pns} design review & April 2020 \\ \colhline
\dword{rsds} design review & May 2020 \\ \colhline
Start of module 0 component production for ProtoDUNE-II & April 2020      \\ \colhline
End of module 0 component production for ProtoDUNE-II &  February 2021    \\ \colhline
\rowcolor{dunepeach} Start of \dword{pdsp}-II installation& \startpduneiispinstall      \\ \colhline
\rowcolor{dunepeach} Start of \dword{pddp}-II installation& \startpduneiidpinstall      \\ \colhline
 \dword{prr} dates &   March 2022   \\ \colhline
\rowcolor{dunepeach}South Dakota Logistics Warehouse available& \sdlwavailable      \\ \colhline
\dword{rsds} demonstration test at ProtoDUNE-SP-II (*) &   April 2022   \\ \colhline
Start of  Laser and \dword{pns} production  &   May 2022   \\ \colhline
\rowcolor{dunepeach}Beneficial occupancy of cavern 1 and \dword{cuc}& \cucbenocc      \\ \colhline
End of \dword{pns} production & March 2023 \\ \colhline 
End of Laser system production  & July 2023     \\ \colhline
End of \dword{rsds} production (*) & August 2023     \\ \colhline
\rowcolor{dunepeach} \dword{cuc} counting room accessible& \accesscuccountrm      \\ \colhline
Start assembly of calibration production units in the cavern & May 2023\\ \colhline
\rowcolor{dunepeach}Top of \dword{detmodule} \#1 cryostat accessible& \accesstopfirstcryo      \\ \colhline
Start installation and alignment of Laser boxes & May 2024 \\ \colhline 
\rowcolor{dunepeach}Start of \dword{detmodule} \#1 TPC installation& \startfirsttpcinstall      \\ \colhline
Start installation of Laser System periscopes & August 2024 \\ \colhline 
Start installation of \dword{rsds} guide system (*) & August 2024 \\ \colhline 
\rowcolor{dunepeach}End of \dword{detmodule} \#1 TPC installation& \firsttpcinstallend      \\ \colhline
Installation of \dword{rsds} purge-boxes (*) & May 2025 \\ \colhline 
Installation of the \dword{pns} main components & June 2025 \\ \colhline
\rowcolor{dunepeach}Top of \dword{detmodule} \#2 accessible& \accesstopsecondcryo      \\ \colhline
 \rowcolor{dunepeach}Start of \dword{detmodule} \#2 TPC installation& \startsecondtpcinstall      \\ \colhline
\rowcolor{dunepeach}End of \dword{detmodule} \#2 TPC installation& \secondtpcinstallend      \\ 
\end{dunetable}

Table~\ref{tab:calib-sched} shows the schedule and key milestones for the calibration consortium that lead to commissioning the first \dword{fd} module. The demonstration of calibration systems design, operation, and performance at the \dword{pdsp}-II running is a key part of calibration schedule; those milestones are also listed in the table. The technology design decisions on calibration subsystems should be made by January 2020 for the laser system and by March 2020 for the neutron source system followed by technical design reviews. The production of design prototypes to be deployed at \dword{pdsp}-II running should be finished by February 2021 followed by assembly and deployment in \dword{pdsp} in March 2021. The \dlong{rsds} (\dshort{rsds})

design will follow a demonstration R\&D program outlined in detail in Table~\ref{tab:calib-rsds-sched} in the Appendix, with major milestones highlighted in this section. The major steps for systems approval are the design review in May 2020 and the deployment test at \dword{pdsp}-II in April 2022. 

Production of calibration systems for the \dword{fd} should start in March 2022, followed by assembly of the systems underground once the detector cavern becomes available in early 2023. Installing the laser system can begin as soon as the cryostat roof is accessible and conclude once the \dword{tpc} is ready to install. If it is approved, the \dword{rsds} guide system can begin installation just before\dword{tpc} is installed. The purge-boxes on top of the cryostat can be done later.
Installing the main components of the \dword{pns} will begin once the human access ports are no longer needed for \dword{tpc} installation in June 2025.

\newpage

\section{Appendix}
\label{appx:calibration}
\subsection{Laser System Alternative Designs}
\label{sec:sp-calib-laser-alter}


\subsubsection{End-wall coverage enhancement}

The eight calibration ports closer to the end-walls (four on each side) are not positioned on top of the \dword{tpc}, but instead located about \SI{40}{\cm} away from the \dword{fc} along the $z$ (beamline) direction. If positioned on top, \dword{fc} penetration would be quite complicated, having to come from the sides. Use of the periscope baseline design for the end-wall periscopes would severely limit the volume coverage, similar to the coverage limitation mentioned in Section~\ref{sec:sp-calib-sys-las-ion-des}.

We describe here an alternative design for the end-wall ports that would improve the laser beam coverage without requiring \dword{fc} penetration.

The periscope is exactly the same as the baseline design but, at the top of the calibration port, is mounted on a flange that has an additional rotation degree of freedom. Figure~\ref{fig:laser_extrarotation} presents a preliminary drawing of the concept. The \SI{250}{\milli\m} diameter calibration port has on top of it the main rotary flange that, itself, has another smaller port off-centered by \SI{40}{\milli\m} with respect to the main one. On this smaller port, a secondary rotary flange is installed and it is this one that holds the laser periscope, including the optical feedthrough and the linear stage for mirror movement. When the main flange rotates, the periscope also moves along a circular (\SI{40}{\milli\m} diameter) trajectory. Consequently, within the cryostat, the relative position between the beam mirror and the \dword{fc} profiles changes as well, and so the shadowed regions also change, by parallax. Using different main rotary flange angles, it should be possible to locate the mirror in enough different positions in order to cover all the previously shadowed angles.

Calculations similar to the ones showed earlier show that, using only \num{3} different positions (separated by \ang{90}), a coverage of \SI{94}{\%} should be possible for \SI{30}{\cm} voxels and allowing all tracks directed at the \dword{apa}.

\begin{dunefigure}[CAD drawing of the double rotary flange for the end-wall laser calibration ports]{fig:laser_extrarotation}
{Exploded CAD drawing (preliminary) of the double rotary flange for the end-wall laser calibration ports. The calibration port is shown in brown at the bottom; the primary and secondary rotary flanges are shown in yellow,  with the (black) motors next to them. The optical feedthrough is shown in the center, in blue. On top, the mirror arrangement allows the laser beam to be aligned with the optical feedthrough no matter the angle of each of the rotary flanges.}
\includegraphics[width=0.4\linewidth]{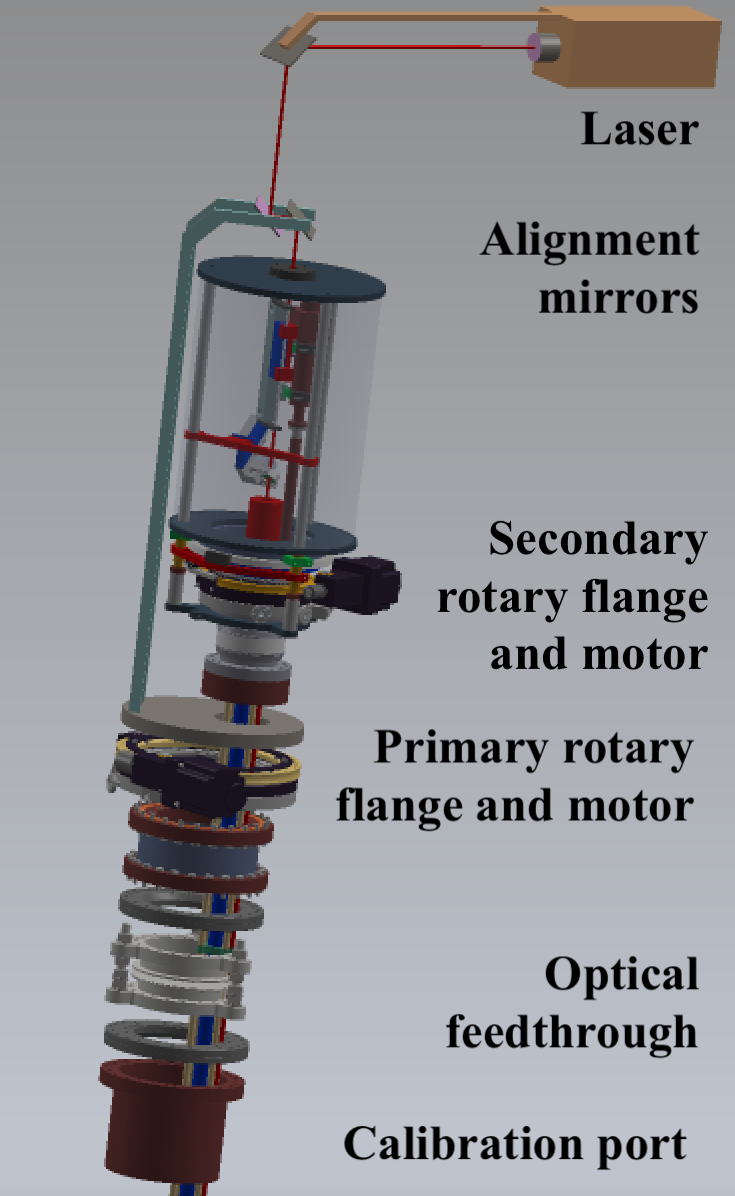}
\end{dunefigure}

\subsection{PNS System Alternative Designs}
\label{sec:sp-calib-pns-alter}
\label{sec:sp-calib-pns-alt}

\subsubsection{Small Format Moderator}
An alternative method for delivering the neutrons is to use the existing calibration feedthroughs. In the current cryostat design, \num{20} calibration feedthroughs with a \SI{25}{\cm} outer diameter will be available on top of the cryostat. One can design the neutron source with an ultra-thin $DD$ generator that fits the size of the feedthrough as shown in Figure~\ref{fig:PNS_Two_Designs} (right). The problem is that there will be no space in the feedthrough for the shielding materials to fit in, so additional shielding will need to be placed around the feedthrough. The weight of this compact neutron source will be about \SI{140}{\kg}, so minimal special mounting is needed. In addition, the source may be moved as well, allowing further flexibility. The effective neutron flux is expected to be similar to that of the baseline deployment. 

\begin{dunefigure}[\dshort{pns} two designs]{fig:PNS_Two_Designs}
{(right) Small format neutron source deployed inside the calibration feedthrough ports. (left) For comparison, large format neutron source deployed above/inside the human access ports is shown on the left.}
\includegraphics[width=16cm]{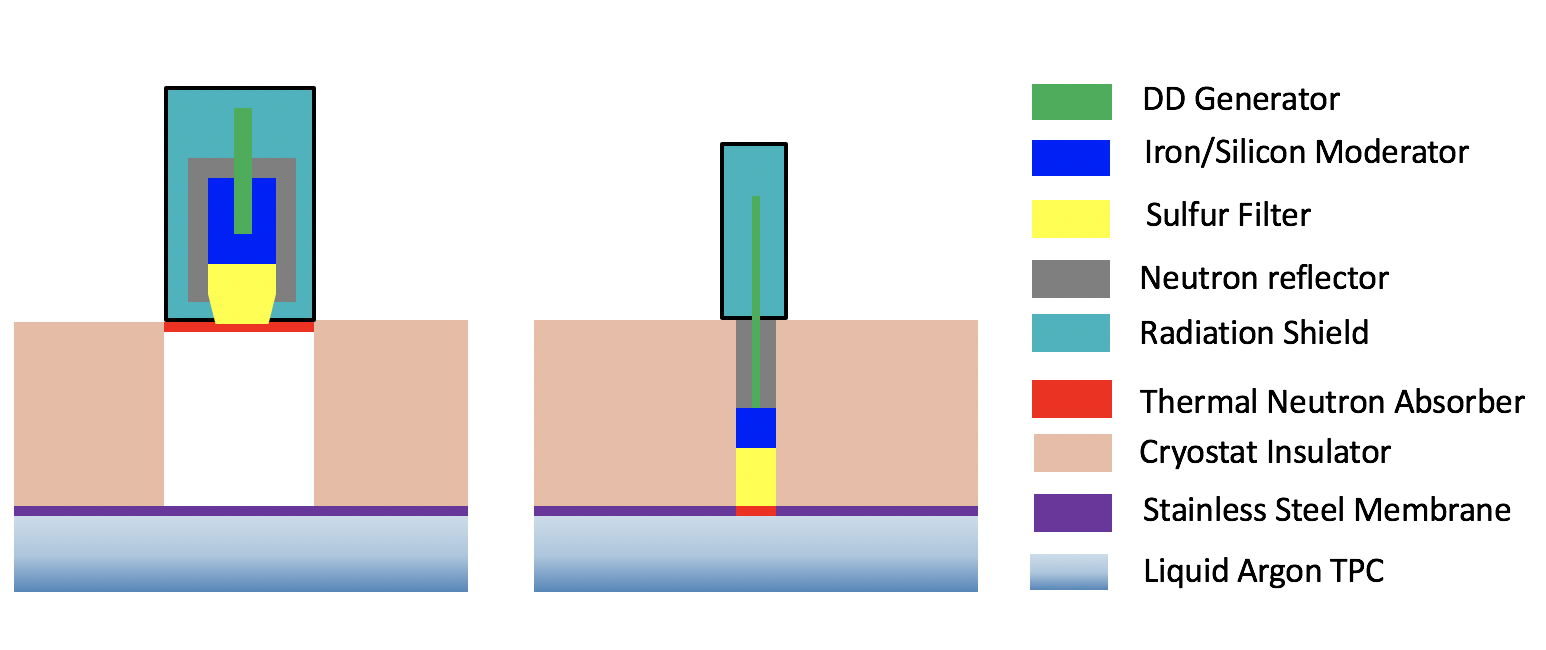}
\end{dunefigure}

The volume coverage at the center of the detector can be significantly increased by using a small format neutron source deployed on top at the center of the cryostat using the multi-purpose feedthroughs.  
Figure~\ref{fig:PNS_alternative_ncap} shows the position distribution of the neutron captures using two large format sources at the corner human access ports and one additional small format source in the middle of the cryostat. The small format source is 
important to complement the coverage at the center of the TPC. The alternative small format neutron source is very compact and lightweight, so further coverage improvement is possible by moving the source to different calibration feedthroughs. The deployment of the small format source would require sharing of the feedthrough ports with other calibration systems, which is currently under investigation.

\begin{dunefigure}[Pulsed neutron system neutron capture positions inside a \dshort{dune}-sized \dshort{tpc}]{fig:PNS_alternative_ncap}
{Neutron capture positions inside a \dword{dune}-sized \dword{tpc}, assuming alternative configuration with two large format neutron sources located at the corner human access ports and one small format neutron source located at the center of the cryostat, which compensates the missing volume coverage of the two large format sources at the center of the detector.  $L$=\SI{60}{\m} (along $Z$ axis, horizontally parallel to the beam direction), $W$=\SI{14.5}{\m} (along $X$ axis, horizontally perpendicular to the beam direction), $H$=\SI{10}{\m} (along $Y$ axis, vertically perpendicular to the beam direction). \num{2.7e7} $DD$ generator neutrons with \SI{2.5}{\MeV} energy were simulated in each moderator and propagated inside the \dword{tpc}. Top (left) and side (right) views of neutron capture positions are shown.}
\includegraphics[width=18cm]{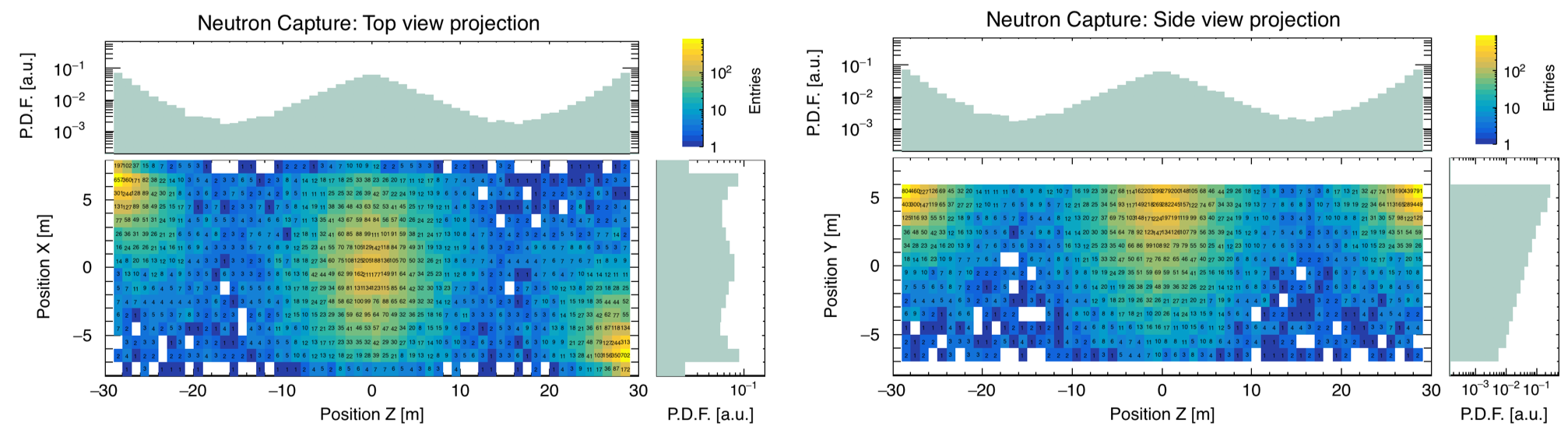}
\end{dunefigure}

In principle, for the baseline deployment plan as shown in Figure~\ref{fig:PNS_ncapDistribution_two_sources}, we can run the neutron source for a longer time to increase the coverage at the central region of the detector. However, this would result in a huge data volume. So, the best way to complement the coverage at the center of the detector is to use the alternate deployment as discussed here. This design would require a total number of \num{4600} pulses to calibrate the entire \nominalmodsize module. Assuming that three neutron sources with identical neutron capture yield are operated in synchronization mode, \num{1500} triggers are needed for each calibration run. Therefore, the total data volume per run would be 

\begin{equation}
\num{1500}~{\rm Triggers} \times \num{1.5}~{\rm Bytes}\times
\SI{2}{\mega\hertz}\times \SI{5.4}{\milli\s}\times \num{384000}~{\rm channels} = \num{9.5}~{\rm TB/run}.
\end{equation}

The recommended trigger rate of the \dword{pns} system is \SI{0.5}{\hertz} which is limited by the bandwidth of the DAQ event builder. Assuming that the spatial distribution of the neutron capture is uniform across the whole detector volume, the operation time per calibration run would be \num{50} minutes.   
Running the \dword{pns} calibration system twice a year would result in a total data volume of \SI{19}{TB} per \nominalmodsize per year. 
For realistic neutron capture distribution that is non-uniform, we expected to operate the \dword{pns} system for a period of 10 times longer than that under the ideal assumption (9.5 TB/run). As a result, the data size per calibration run would be 95 TB/run and running the PNS calibration twice a year would result in a total data size of 190 TB/year and four times a year would result in 380 TB/year.

\subsection{Proposed Radioactive Source Calibration System}
\label{sec:sp-calib-sys-rsds}



Radioactive source deployment provides an in-situ source of physics signals at a known location and with a known activity that can be chosen such that there is only one calibration event per drift time window. The 
primary
source design probes de-excitation products ($\gamma$-rays) which are directly relevant for detection of supernova neutrinos and \isotope{B}{8}/hep solar neutrinos. The \dlong{rsds} (\dshort{rsds}) is the only calibration system that could probe the detection capability for single isolated solar neutrino events and study how well radiological backgrounds can be suppressed. The trigger efficiency could be studied as a function of threshold. 

Other measurements with the primary source include electro-magnetic (EM) shower characterization for long-baseline $\nu_e$ CC events, electron lifetime and electric field as a function of \dword{detmodule} vertical position, individual light detector response, and determination of radiative components of the Michel electron energy spectrum from muon decays. Aside from the 
primary nickel source that produces \SI{9}{\MeV} $\gamma$-rays via the \isotope{Ni}{58}(n,$\gamma$)\isotope{Ni}{59} reaction, other sources could be deployed with the same multi-purpose system, for example an ($\alpha$,$\gamma$)
source, and \isotope{Cf}{252} and/or AmBe neutron sources that probe the impact of various radiological backgrounds, like radon (causing  ($\alpha,\,\gamma$) events) or radiological neutrons, or simply measure the neutron tagging efficiency, useful for improved calorimetry of beam neutrino interactions. In contrast to the primary 
nickel source with \SI{9}{\MeV} gamma-rays, the ($\alpha$,$\gamma$) source producing gamma-ray energies around \SI{15}{\MeV} via the
\isotope{Ar}{40}($\alpha$,$\gamma$)\isotope{Ca}{44}
reaction could even be deployed outside of the cryostat, to probe the upper visible energy range and trigger efficiency for \isotope{B}{8}/hep solar neutrinos. 

Both the \dshort{rsds} and the \dword{pns} systems are needed to address the integrated response of the detector for low energy physics, especially \dword{snb} and $^{8}B$/hep solar neutrinos. 
The \dword{rsds} primarily probes for trigger efficiency, the \dword{pns} tests mostly for uniformity. Response in argon may change rapidly as a function of photon energy due to underlying nuclear physics mechanisms. A combination of \SI{6}{\MeV} (direct neutron capture response), \SI{9}{\MeV} (from the nickel source),
\SI{15}{\MeV} (from the ($\alpha$,$\gamma$) source)
is needed to map the low energy response. In terms of complementarity, radioactive sources provide a known position, known-energy single photon events that could be triggered on, while the pulsed neutron source provides a simple, potentially, non-invasive design with externally triggered multi-photon energy signature which is visible across the entire detector with a known time signature.

\subsubsection{Design Considerations}

A composite source can be used that consists of \isotope{Cf}{252}, a strong neutron emitter, and \isotope{Ni}{58}, which, via the \isotope{Ni}{58}(n,$\gamma$)\isotope{Ni}{59} process, converts one of the 
\isotope{Cf}{252} fission neutrons, suitably moderated, to a monoenergetic \SI{9}{\MeV} photon~\cite{Rogers:1996ks}. 
The source is envisaged to be inside a cylindrical moderator with mass of about \SI{15}{\kg} and a diameter of \SI{20}{\cm} such that it can be deployed via the multipurpose instrumentation ports discussed in Section~\ref{sec:sp-calib-cryostat}. The activity of the radioactive source is chosen such that no more than one \SI{9}{\MeV} capture $\gamma$-event occurs during a single drift period. This forms the main requirement for this system as this allows one to use the arrival time of the measured light as a $t_0$ and then measure the average drift time of the corresponding charge signal(s). Table~\ref{tab:fdgen-calib-all-reqs-rsds} lists the full set of requirements for the \dlong{rsds}.

The sources would be deployed outside the \dword{fc} within the cryostat to avoid regions with a high electric field, about \SI{30}{\cm} from the field cage. The $\gamma$-ray would need to travel about two attenuation lengths (including the \SI{10}{\cm} radius of the source body). Such high $\gamma$-energies are typically only achieved by thermal neutron capture, which invokes a neutron source surrounded by a large amount of moderator, thus driving the size of the source.

\begin{dunetable}
[Full specifications for the radioactive source deployment system]
{p{0.45\linewidth}p{0.25\linewidth}p{0.25\linewidth}}
{tab:fdgen-calib-all-reqs-rsds}
{Full list of Specifications for radioactive source deployment system.}   
Quantity/Parameter	& Specification	& Goal		 \\ \toprowrule   
Distance of the source from the field cage & \SI{30}{\cm} & \\ \colhline
Rate of \SI{9}{\MeV} capture $\gamma$-events inside the source (top-level requirement) & < \SI{1}{\k\hertz} & \\ \colhline 
Data volume per \SI{10}{\kt}  & \SI{50}{\TB\per\year} & \SI{100}{\TB\per\year} \\ \colhline 
Longevity	& \dunelifetime			& > \dunelifetime   \\     

\end{dunetable}

A gamma source based on the \isotope{Ni}{58}(n,$\gamma$)\isotope{Ni}{59} reaction, and triggered by an AmBe neutron source, has been successfully built~\cite{Rogers:1996ks}, yielding high $\gamma$-energies of \SI{9}{\MeV}. DUNE 
proposes to use a \isotope{Cf}{252} (or AmLi as backup) neutron source with lower neutron energies, which requires less than half of the surrounding moderator, and making the \isotope{Ni}{58} (n, $\gamma$) source only \SI{20}{\cm} or less in diameter. The multipurpose instrumentation feedthroughs at either end of the cryostat are sufficient for this, and have an outer diameter of \SI{25}{\cm}.  The moderator material chosen for DUNE is Delrin,\footnote{DuPont\texttrademark Delrin\textregistered, \url{http://www.dupont.com/products-and-services/plastics-polymers-resins/thermoplastics/brands/delrin-acetal-resin.html}.} which has a large enough density to avoid flotation. Further, the end caps of the source body are round to avoid distorting the electric field and to eliminate the risk of the source getting stuck during deployment. 
Figure~\ref{fig:RadioactiveSource_zm40cm_xp220cm} depicts the primary source design of a cylindrical Delrin moderator with a diameter of \SI{20}{\cm}, a height of \SI{40}{\cm} including half-spheres at either end with radius of \SI{10}{\cm}, deployed at $z$=\SI{40}{\cm} leaving a gap of \SI{30}{\cm} towards the \dword{fc} and at a distance to the \dword{apa} of $x$=\SI{220}{\cm}, which is slightly further than mid-drift.

\begin{dunefigure}[Fish-line deployment scheme in \dshort{dune} for an encapsulated radioactive source]{fig:RadioactiveSource_zm40cm_xp220cm}{Fish-line deployment scheme in \dword{dune} for a radioactive source encapsulated inside a cylindrical Delrin moderator body \SI{20}{\cm} in diameter and \SI{40}{\cm} high, including half-spheres with a radius of \SI{10}{\cm} at either end. A \isotope{Cf}{252} neutron source and a natural Ni target are sealed inside at the center. The fish-line is deployed \SI{40}{\cm} outside of the \dword{fc} and \SI{220}{\cm} away from the \dword{apa} (red plane).}
\includegraphics[width=1.0\linewidth]{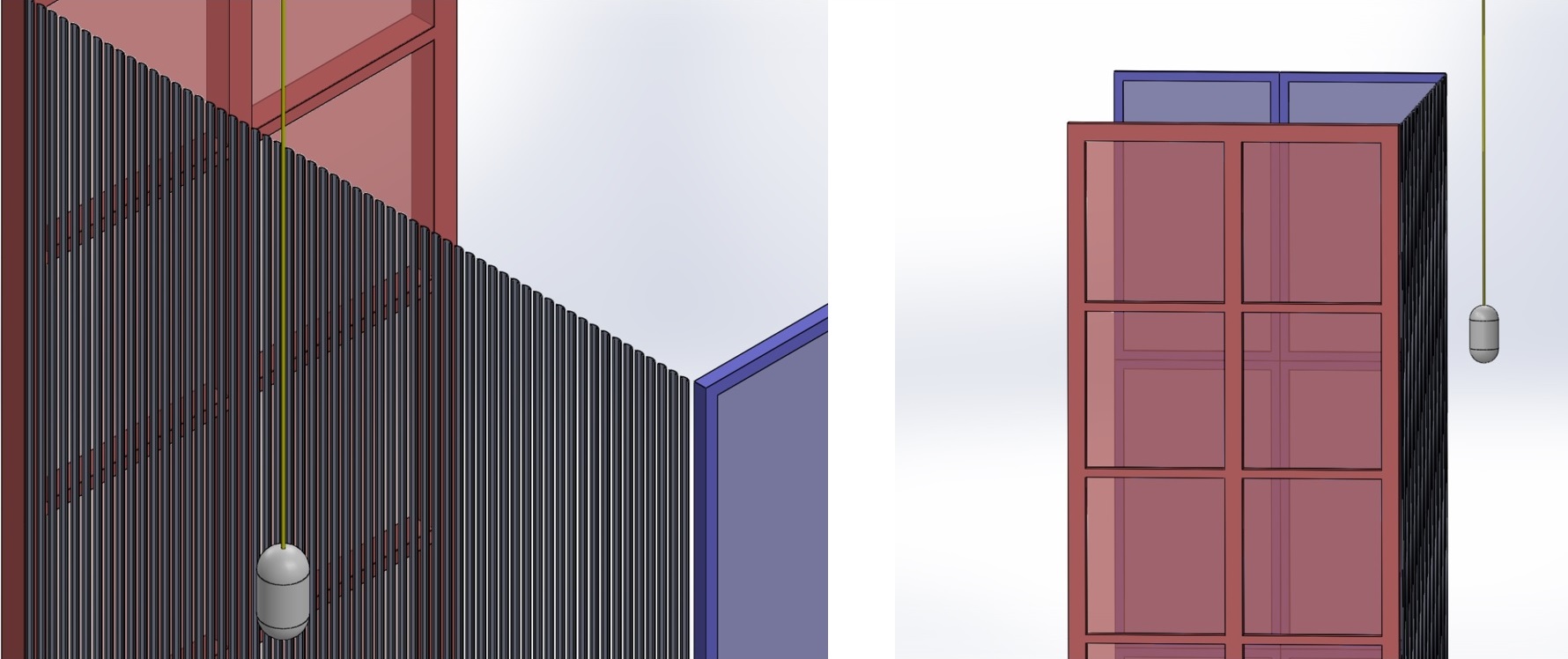}
\end{dunefigure}

A successfully employed multipurpose fish-line calibration
system for the Double Chooz reactor neutrino experiment has become available 
after the decommissioning of Double Chooz in 2018. The system can be easily refitted for use in \dword{dune}. The system will be housed inside a purge-box that is connected via a neck to a multipurpose calibration feedthrough with a closed gate valve on top of the cryostat. Before deployments, the source will be gently cooled-down by blowing liquid argon boil-off onto it inside a sealed purge-box. After the source has reached 
near-\dword{lar} temperatures, the purge-box will be evacuated by a vacuum pump to remove any residual oxygen and nitrogen which is monitored at the ppm level. Then, the entire purge-box interior is purged with boil-off liquid argon, and the pressure equalized with the gas pressure inside the detector, before the gate-valve is opened and deployments can commence. This procedure ensures that no significant impurities are introduced into the detector during a deployment and that no significant amount of liquid argon is boiled-off from the detector.

Deployed near mid-drift (in each TPC module) the \SI{9}{\MeV} $\gamma$-ray source can illuminate the full drift length from \dword{apa} to \dword{cpa}. The sources are retrieved from the detector after each deployment and stored outside the cryostat following approved safety protocols, and the gate-valves are kept closed after deployments. More details on radiation safety and handling procedures are presented in Section~\ref{sec:sp-calib-rsds-safety}.

\subsubsection{Development Plan}
The major development plans for the \dlong{rsds} include the following.
\begin{itemize}
\item Continue development of relevant simulation tools including geometry representation of the source deployment system and impact from various radiological contaminants on detector response. 
\item Conduct studies to suppress radiological backgrounds for the calibration source.
\item Conduct simulation studies to understand data and trigger rates.
\item Study a baseline design source with Delrin moderator, $^{252}$Cf neutron source, and natural nickel target, both sealed inside at the moderator's center.
\item Validate \SI{9}{\MeV} capture $\gamma$-ray yield of source using spectroscopic measurements with the `RABBIT' germanium detector at South Dakota School of Mines and Technology (SDSMT), that has an assay chamber large enough to fit the bulky moderator. 
\item Validate with $^{3}$He based hodoscope at SDSMT to ensure that the flux of neutrons escaping the moderator is not an issue; otherwise use lower energetic AmLi neutron source instead and/or more moderator material, and/or different geometric configuration of nickel target. 
\item Test gentle GAr cooling of source and validate material integrity. Measure tensile strength of braided SS-304 wire-rope at cryogenic temperatures and ensure a safety factor of one order of magnitude by adjusting number of steel braids and their diameters. Validate cryogenic shrinkage of sectional teflon sleeves, that enclose the braided steel wire-rope and electrically insulate it towards the \dword{fc}. 
\item Validate that anticipated fluid flow in \dword{lar} does not cause oscillations of the source; otherwise design vertical guide wires to be pre-installed during detector installation 
which will keep source in stable position during deployment along the vertical axis.
\item Explore other radioactive sources beyond the primary 
\SI{9}{\MeV} $\gamma$-ray nickel source, such as the previously mentioned \SI{15}{\MeV} $\gamma$-ray source based on the  $^{40}$Ar($\alpha,\,\gamma$)$^{44}$Ca process with $^{241}$Am as the alpha emitter. This is currently being assembled at SDSMT.
Furthermore, investigate hybrid neutron sources ($^{252}$Cf and AmBe) that emulate the kinetic neutron energy spectrum of radiological neutrons and probe the neutron tagging efficiency.
\end{itemize}

A successful demonstration of the \dword{rsds} in \dword{protodune2} running is the main priority for this system towards making a decision on deploying this system for the \dword{fd}. A schedule with main steps towards \dword{protodune2} deployment is shown in Table~\ref{tab:calib-rsds-sched}.

\begin{dunetable}
[Key milestones for commissioning RSDS in \dshort{protodune2}]
{p{0.65\textwidth}p{0.25\textwidth}}
{tab:calib-rsds-sched}
{Key milestones towards commissioning the \dlong{rsds} in \dword{protodune2}.}  
Milestone & Date (Month YYYY)   \\ \toprowrule
Baseline \dword{rsds} design validation & January 2020 \\ \colhline 
\dword{rsds} mock-up deployment test at SDSMT & March 2020 \\ \colhline 
\dword{rsds} Design review  & May 2020 \\ \colhline
\dword{rsds} Production readiness review (PRR) & July 2020 \\ \colhline
Start of module 0 \dword{rsds} component production for \dword{protodune2} & September 2020      \\ \colhline
End of module 0 \dword{rsds} component production for \dword{protodune2} &  February 2021    \\ \colhline
\textbf{Start of \dword{protodune2} (\single) installation} & \textbf{March 2021} \\ \colhline
Start of \dword{rsds} installation &  April 2021    \\ \colhline
\dword{rsds} demonstration test at \dword{protodune2}  & April 2022\\ 
\end{dunetable}

\subsubsection{Measurement Program}
\label{sec:sp-calib-sys-src-dep-meas}

The proposed primary \SI{9}{\MeV} single $\gamma$ source may also be used to test the $\gamma$ component of the \dword{snb} and \isotope{B}{8}/hep solar neutrino signal along the full drift but only in the endwall regions of the detector. The source may also be used to determine the relative charge and light extraction efficiency in the vertical direction for measurements of energy resolution and energy scale. 

Figure~\ref{fig:rsds-fig1}
depicts in a top view of the detector the simulated charge extraction efficiency for the 9~MeV $\gamma$-ray source deployed \SI{40}{\cm} outside of the \dword{fc}, near mid-drift i.e., \SI{220}{\cm} away from the \dword{apa} in the $x$ direction, in the presence of expected background before (a) and after (b) applying selection cuts. The selection cuts
are based on the amplitude and location of wire hits, and require a coincidence with a suitable signal in the \dword{pds}.
Figure~\ref{fig:rsds-fig1}(b)
shows that the selection cuts can reject radiological backgrounds almost entirely, and that
the \dword{rsds} should allow the study of the trigger efficiency for isolated solar neutrino events, and its threshold dependence.

Figure~\ref{fig:rsds-fig2}
shows exemplary simulated 
\dword{rsds} measurements of the \efield strength (a) and of the electron lifetime (b), each for three different scenarios. The analysis is based on fitting the measured distribution of drift-time, i.e., the time difference between the \dword{pds} signal
and the recorded hit times on collection wires, passing the selection cuts. 
Figure~\ref{fig:rsds-fig2}(a)
illustrates that with this method the \efield strength could be measured at $\sim$ \SI{1}{\%} precision at each vertical deployment position at the endwalls. Likewise, Figure~\ref{fig:rsds-fig2}(b)
illustrates that the electron lifetime could be measured at about $\sim$ \SI{10}{\%} precision (possibly better at higher lifetimes) at each vertical deployment position at the endwalls. 

Figure~\ref{fig:rsds-fig2}(c)
illustrates that it is not convincingly possible to unambiguously measure both the electron lifetime and the electric field strength with a recorded charge spectrum (after selection cuts) alone, since both parameters simply shift the upper falling edge of the charge spectrum up or down. However, when combined with the drift-time measurement, the charge measurement would provide an additional constrain that could possibly break correlations.

Aside from the primary
9 MeV $\gamma$-ray nickel source, other sources could be deployed with the same multi-purpose system, for example a $^{40}$Ar($\alpha,\,\gamma$)$^{44}$Ca gamma-ray source, a $^{252}$Cf and/or AmBe neutron source that probe the impact of various radiological backgrounds, like radon ($\alpha,\,\gamma$) or radiological neutrons, or simply measure the neutron tagging efficiency, useful for improved calorimetry of beam neutrino interactions. 
In contrast to the nickel source, the \SI{15}{\MeV} ($\alpha,\,\gamma$) could be deployed outside of the cryostat.

An external \dword{protodune2} deployment can demonstrate the feasibility of the non-invasive \SI{15}{MeV} \argon40$(\alpha,\,\gamma)^{44}$Ca $\gamma$-ray source despite the lack of overburden to shield cosmic rays. In contrast to cosmic muons, \SI{15}{MeV} $\gamma$-ray induced hit clusters will start inside the detector volume, and are not tracks that begin at the detector edges. Thus, the \dword{rsds} calibration events could therefore be easily selected and the detected charge can be analyzed. The detected light, however, will be obscured from the high light level in each drift period from cosmic muons hitting \dword{protodune}. 

\begin{dunefigure}[Detected charge from 9 MeV gamma-ray source with radiological backgrounds]
{fig:rsds-fig1}
{
Detected charge (a) without cuts and (b) with selection cuts for a simulated 9~MeV $\gamma$-ray source deployed at $z=$\SI{-40}{\cm} outside of the \dword{fc}, $x=$\SI{220}{\cm} away from the \dword{apa}, and $y=$\SI{300}{\cm} half-height of an upper endwall \dword{apa} with simulated expected radiological background, that gets almost eliminated by selection cuts.}
\centering
   (a)
   \includegraphics[width=0.76\linewidth]{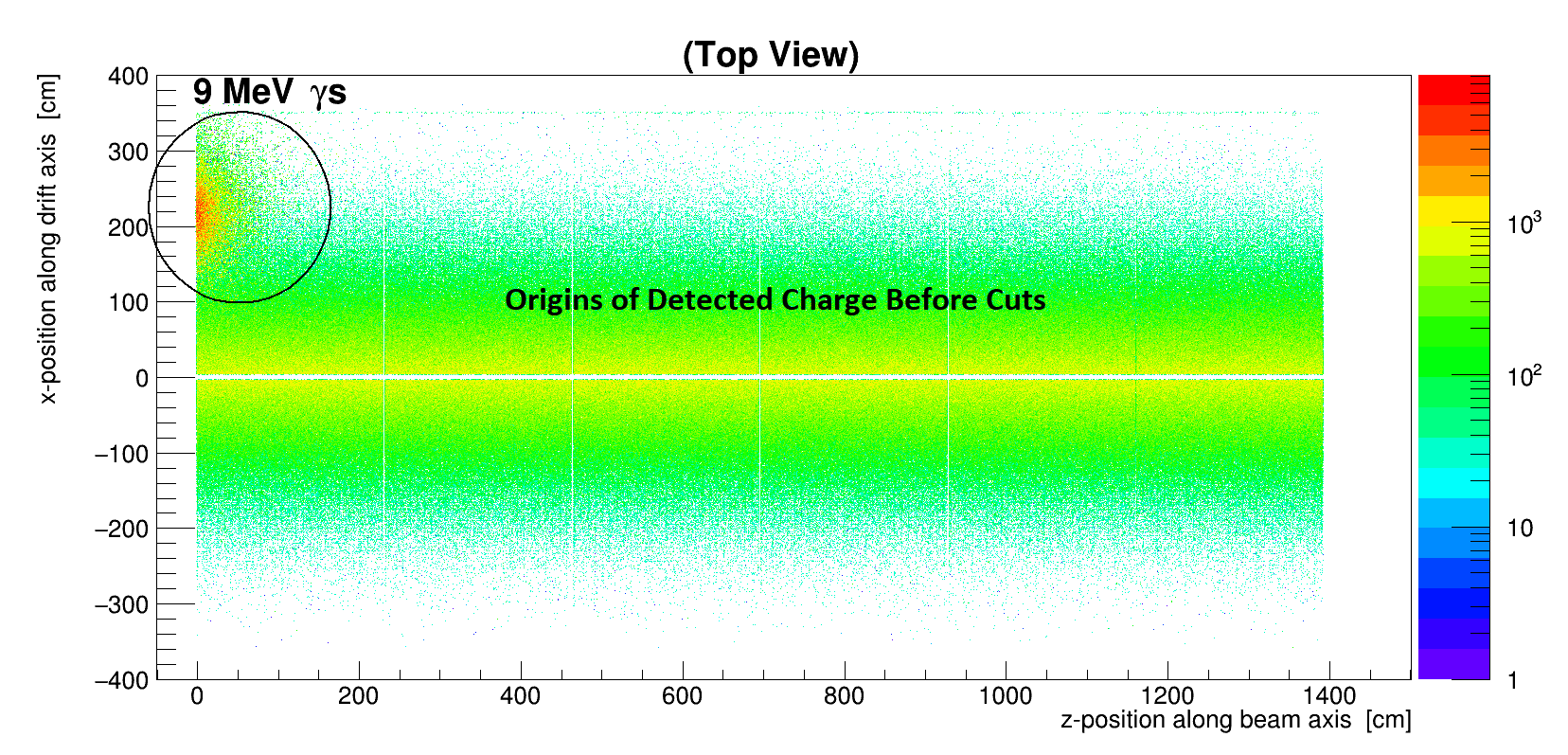}
   (b)
   \includegraphics[width=0.75\linewidth]{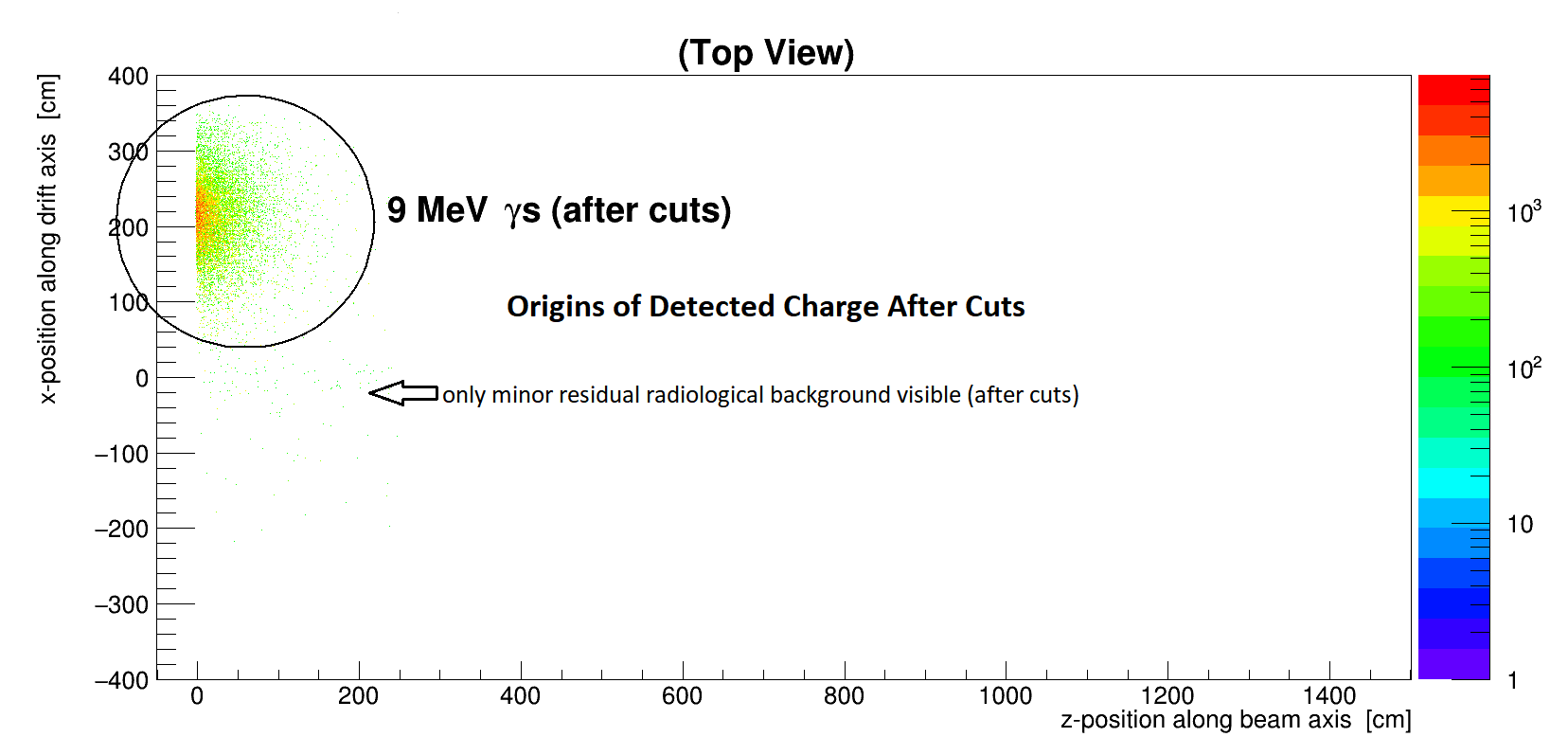}

\end{dunefigure}

\begin{dunefigure}[Measured \efield and e$^-$ lifetime from \SI{9}{MeV} gamma; with radiological backgrounds]
{fig:rsds-fig2}
{Simulated measurements of (a) \efield strength from drift-time distribution, (b) electron lifetime from drift-time distribution, and (c) electron lifetime from charge distribution when electric field is unambiguously known from drift-time distribution. All spectra were created with applied selection cuts for a simulated \SI{9}{\MeV} $\gamma$-ray source with radiological backgrounds deployed at $z=$\SI{-40}{\cm} outside of the \dword{fc}, $x=$\SI{220}{\cm} away from the \dword{apa}, and $y=$\SI{300}{\cm} half-height of an upper endwall \dword{apa}. (Colors of histograms are matching colors of corresponding labels in each histogram.)}
\centering
\includegraphics[width=0.9\linewidth]{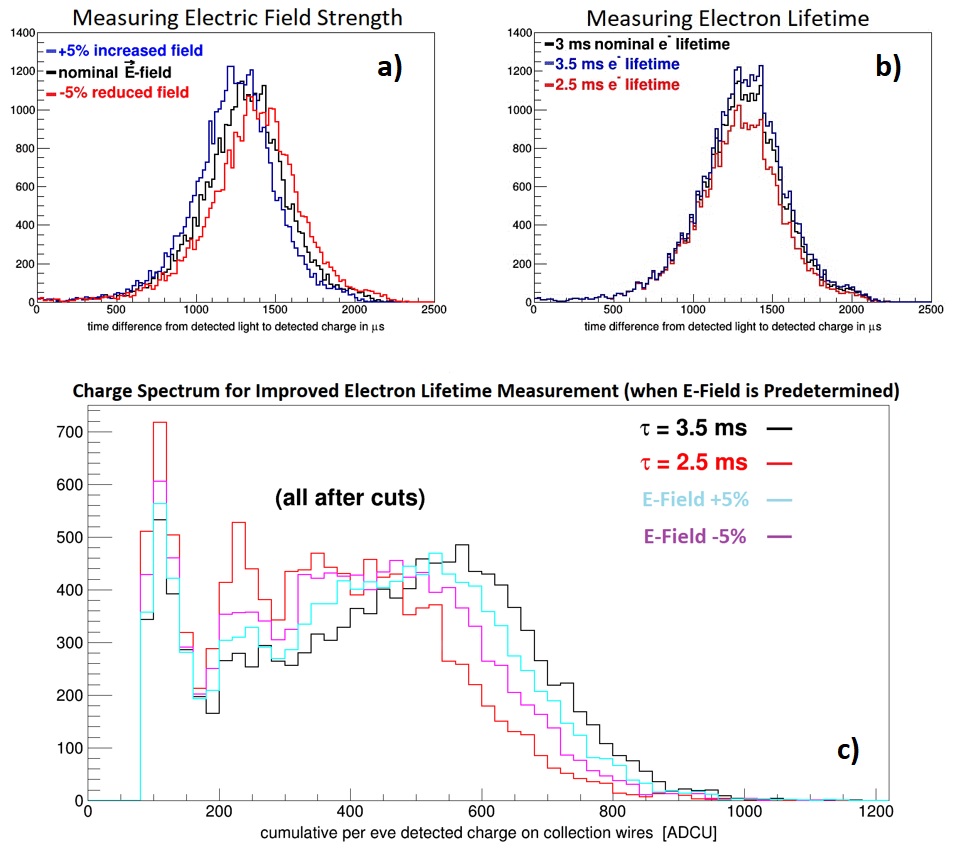}    
\end{dunefigure}

\subsubsection{\dword{rsds} Design Validation}
The cosmic induced background rate at \dword{protodune} is too high at the surface to detect responses to the \dword{dune} $\gamma$-ray source; a higher intensity source could be deployed to test the detector response and analysis method. However, tests of functionality,  reliability, and safety of the mechanical deployment system are essential to show the source can be deployed and retrieved with no issues, so these will be the main goals of the \dword{protodune2} deployment. As mentioned earlier, tests of the source design itself, in terms of $\gamma$ activity, will be done at SDSMT. 

\subsubsection{DAQ Requirements}
Section~\ref{sec:sp-calib-daqreq} provides an overall discussion of the Calibration and \dword{daq} interface. Here, the \dword{daq} requirements for the \dlong{rsds} are discussed. The radioactive source will not be triggerable by the \dword{mlt}.  Rather, it will deliver a tag to the \dword{mlt} and that tag will include a time stamp that can be used by the \dword{mlt} to issue a trigger command to the \dword{fe} readout.  The trigger command will have a standard readout window size of \SI{5.4}{\milli\s}, but to keep data rates manageable, the command will only be send to \dword{fe} readout buffers that are expected to be illuminated by the source. The localization of trigger commands thus reduces the data volume by \num{150}, if only one \dword{apa} is read out.

Nevertheless, if the rate of such a source is anywhere close to one per \SI{5.4}{\milli\s}, the detector would be running  continuously in the current scheme. Therefore we assume that the interaction rate in the detector is \SI{10}{\hertz} or less. The tag from the source will likely be much higher than this, because not all $\gamma$s interact in the active \dword{tpc} volume. Thus the radioactive source trigger will be a coincidence in the Module-Level Trigger between a low-energy trigger candidate from the illuminated \dword{apa}, and a source tag with a relevant time stamp.  With this rate, and with localization of events to one \dword{apa}, the total data volume would be
\begin{equation}
\num{8}~{\rm hours} \times \num{4}~{\rm FTs} \times \SI{10}{\hertz} \times \num{1.5}~{\rm Bytes}\times \SI{2}{\mega\hertz}\times \SI{5.4}{\milli\s}\times \num{2560}~{\rm channels} = \SI{50}{\TB}/scan.
\end{equation}
Running this calibration four times/year would yield \SI{200}{\TB} of data in \SI{10}{\kt} per year. Table~\ref{tab:calib-daq-rsds} summarizes the data volume requirements for \dword{rsds}.

\begin{dunetable}
[Calibration \dshort{daq} summary for \dshort{rsds}]
{p{0.2\textwidth}p{0.15\textwidth}p{0.5\textwidth}}
{tab:calib-daq-rsds}
{Estimated data volume per year per \SI{10}{\kt} for the radioactive source system.}   
System & Data Volume (\SI{}{\TB\per\year}) & Assumptions  \\ \toprowrule
Proposed Radioactive Source System & \num{200} & Source rate < \SI{10}{\hertz}; single \dword{apa} readout,  lossless readout; \num{4} times/year   \\ 
\end{dunetable}           

\subsubsection{Risks}
The risks associated with the radioactive source system are described in Table~\ref{tab:risks:SP-FD-CAL-RSDS} along with appropriate mitigation strategies and the impact (low, medium or high risk levels) on probability, cost, and schedule post-mitigation. There are three residual medium-level risks in the table, more discussion on them is provided below:
\begin{itemize}
    \item \textit{Radioactivity leak:} If radioactivity leaks into the detector during a deployment, radiological backgrounds in the detector might increase. Rigorous source certification under high pressure and cryogenic temperatures mitigates this risk.
    \item \textit{Source stuck or lost:} If the source gets stuck or is lost in the detector, then it becomes a permanent localized radiological background source. Fish-line an order of magnitude stronger than needed to hold the weight, round edges of the moderator and a torque limit of the stepper motor will mitigate this risk.
    \item \textit{Oxygen and nitrogen contamination:} If the purge-box has a small leak, oxygen and nitrogen could get into the \dword{lar}. Leak checks before deployments will mitigate this risk.
\end{itemize}

\begin{footnotesize}
\begin{longtable}{P{0.18\textwidth}P{0.20\textwidth}P{0.32\textwidth}P{0.02\textwidth}P{0.02\textwidth}P{0.02\textwidth}} 
\caption[Radioactive source calibration system risks]{Radioactive source calibration system risks (P=probability, C=cost, S=schedule) The risk probability, after taking into account the planned mitigation activities, is ranked as 
L (low $<\,$\SI{10}{\%}), 
M (medium \SIrange{10}{25}{\%}), or 
H (high $>\,$\SI{25}{\%}). 
The cost and schedule impacts are ranked as 
L (cost increase $<\,$\SI{5}{\%}, schedule delay $<\,$\num{2} months), 
M (\SIrange{5}{25}{\%} and 2--6 months, respectively) and 
H ($>\,$\SI{20}{\%} and $>\,$2 months, respectively).  \fixmehl{ref \texttt{tab:risks:SP-FD-CAL-RSDS}}} \\
\rowcolor{dunesky}
ID & Risk & Mitigation & P & C & S  \\  \colhline
RT-SP-CAL-10 & Radioactive source swings into detector elements & Constrain the system with guide-wires & L & L & L \\  \colhline
RT-SP-CAL-11 & Radioactivity leak & Obtain rigorous source certification under high pressure and cryogenic temperatures & L & L & M \\  \colhline
RT-SP-CAL-12 & Source stuck or lost & Safe engineering margins, stronger fish-line and a torque limit in deployment system & L & M & L \\  \colhline
RT-SP-CAL-13 & Oxygen and nitrogen contamination & Leak checks before deployments & L & M & M \\  \colhline
RT-SP-CAL-14 & Light leak into the detector through purge-box & Light-tight purge box with an infrared camera for visual checks & L & L & L \\  \colhline
RT-SP-CAL-15 & Activation of the cryostat insulation & Activation studies and simulations & L & L & L \\  \colhline

\label{tab:risks:SP-FD-CAL-RSDS}
\end{longtable}
\end{footnotesize}

\subsubsection{Installation, Integration, and Commissioning}
The first elements of the radioactive source guide system are installed before the \dword{tpc} elements on the end wall farthest from the \dword{tco} and as the last system, concurrent and coordinated with the alternative laser system (if any deployed), once the \dword{tpc} is installed before closing the \dword{tco}. The radioactive source deployment system is installed at the top of the cryostat and can be installed when \dword{dune} becomes operational.

The commissioning plan for the source deployment system will include a dummy source deployment (within \num{2} months of the commissioning) followed by first real source deployment (within \num{3} to \num{4} months of the commissioning) and a second real source deployment (within \num{6} months of the commissioning). Assuming stable detector conditions, the radioactive source will be deployed every half a year. Ideally, a deployment before and after a run period are desired so at least two data points are available for calibration. This also provides a check if the state of the system
has changed before and after the physics data run. 
It is estimated that it will take a few hours (e.g. \num{8}~hours) to deploy the system at one feedthrough location and a full radioactive source calibration campaign might take 
a week.

\subsubsection{Quality Control}
A mechanical test of the Double Chooz fish-line deployment system with a \dword{lar} mock-up column will be done in the high bay laboratory at SDSMT. The ultimate test of the system will be done at \dword{protodune}. Safety checks will also be done for the source and for appropriate storage on the surface and underground. 

\subsubsection{Safety}
\label{sec:sp-calib-rsds-safety}
A composite source is used for the radioactive source system that consists  of \isotope{Cf}{252}, a strong neutron emitter, and \isotope{Ni}{58}, which, via the \isotope{Ni}{58}(n,$\gamma$)\isotope{Ni}{59} process, converts one of the \isotope{Cf}{252} fission neutrons, suitably moderated, to a monoenergetic \SI{9}{\MeV} gamma. This system also poses a radiation risk, which will be mitigated with a purge-box for handling, and a shielded storage box and an area with lockout-tagout procedures, also applied to the gate-valve on top of the cryostat. Material safety data sheets will be submitted to DUNE ES\&H and specific procedures will be developed for storage and handling of sources to meet Fermilab Radiological Control Manual (FRCM) requirements. These procedures will be reviewed and approved by \dword{surf} and \fnal radiation safety officers. Sources that get deployed will be checked monthly to ensure they are not leaking. A designated shielded storage area will be assigned for sources and proper handling procedures will be reviewed periodically. A custodian will be assigned to each shielded source.


\cleardoublepage

\glsresetall

\chapter{Data Acquisition}
\label{ch:daq}

\section{Introduction}
\label{sec:daq:introduction}

The \dword{fd} \dword{daq} system receives,
processes, and records data from the \dword{dune} \dword{fd}.
It provides
timing and synchronization for all \dwords{detmodule} and
subdetectors; receives, synchronizes, compresses, and buffers data
streaming from the subdetectors; extracts information from the data at a
local level to subsequently make local, module, and cross-module data
selection decisions; builds event records
from selected space-time data volumes 
and relays them to permanent storage; and carries out subsequent data
reduction and filtering as needed.

This chapter provides a description of the design of the \dword{dune}
\dword{fd} \dword{daq} system developed by the \dword{dune} \dword{fd}
\dword{daq} consortium. 
This consortium brings together resources and expertise from \dword{cern},
Colombia, Czech Republic, France, Italy, Japan, the Netherlands, the UK, and the USA. 
Its members bring considerable experience from \dword{icarus}, \dword{microboone},
\dword{sbnd}, and the
\dword{dune} prototype \dwords{lartpc}, as well as from \dword{atlas} at the \dword{lhc} and other major
\dword{hep} experiments across the world.

The system is designed to service all \dword{fd} \dword{detmodule} designs
interchangeably.  However, some aspects of the \dword{daq} design described in
this chapter are tailored to meet the specific needs of the 
\dword{sp}
detector module technology.  Adaptations to detector technology are
implemented in the upstream part of the \dword{daq}, leaving the remainder generic. The
individual detector modules are serviced by the \dword{daq} independently and the two modules are only loosely
coupled through a cross-module triggering mechanism.

The chapter begins with an overview of the \dword{daq} design
(Section~\ref{sec:daq:overview}), including requirements that the design
must meet and specifications for interfaces between the \dword{daq}  and other
\dword{dune} \dword{fd} systems.  Subsequently,
Section~\ref{sec:daq:design}, which comprises the bulk of this chapter,
describes the design of the \dword{fd} \dword{daq} in greater detail.
Section~\ref{sec:daq:validation} describes design validation efforts to
date, as well as future design development and validation plans. At the center of
these efforts is the  \dword{protodune} \dword{daq} system (described in
Section~\ref{sec:daq:protodune}), which has demonstrated
several key aspects of the  \dword{dune} \dword{fd} \dword{daq}  design and continues
to serve as a platform for further developing and validating 
the final design.  The chapter finishes with two sections
(sections~\ref{sec:daq:production} and \ref{sec:daq:organization}), which
detail the management of the \dword{daq} project, including the
schedule for completing the design,  production, and installation of the
system, as well as safety considerations.

\section{Design Overview}
\label{sec:daq:overview}

Figure~\ref{fig:daq:layout} provides an overview of the \dword{dune} \dword{fd} \dword{daq} system 
servicing a single \dword{fd}
\dword{detmodule}. The system is
physically located at the \dword{fd} site, split between the
underground \dword{dune} caverns and the surface level at \dword{surf}. Specifically, \dword{daq} uses space and
power both in the underground \dword{cuc} and the above-ground \dword{mcr}.
The upstream part of the system, responsible for
raw detector data reception, buffering, and pre-processing, resides in the \dword{cuc}.
The \dword{daqbes},
which is responsible for
event-building, run control, and monitoring, resides on the
surface.
Data flows through the \dword{daq} from 
upstream to the back-end parts of the \dword{daq} and then offline. Most raw data is processed and buffered underground, 
thus controlling consumption of available data bandwidth  to the surface. 

A
hierarchical \dword{daqdss} consumes minimally-processed
information from the upstream \dword{daq} and, through further data processing, 
carries out a module-level \dword{trigdecision} leading to a \dword{trigcommand}.
The command is subsequently executed by a \dword{daqdfo} residing in the \dword{daqbes}
by retrieving the required data from memory buffers maintained by the upstream \dword{daq}.
The results are aggregated across the detector module into a cohesive record and saved to non-volatile storage.
During or after aggregation, an optional down-selection of the data is possible via
high level filtering.
Finally, the data is  transferred offsite and archived by the \dword{dune} offline group.
All
detector modules and their subcomponents are synchronized and timed against a global,
common clock, provided by the timing and synchronization
subsystem. Cross-module communication and communication
to the outside world for data selection (trigger) purposes is facilitated
through an \dword{daqeti}, which is part of the \dword{daqdss}. The
specifics of design implementation and data flow are described in Section~\ref{sec:daq:design}.

\begin{dunefigure}[DAQ conceptual design overview for one
  module]{fig:daq:layout}{\dword{daq} conceptual design overview focusing
    on a single \nominalmodsize module. Included are the upstream
    \dword{daq} subsystem in orange, the \dword{daqdss} in
    blue, and the \dword{daqbes} subsystem in yellow,
    which includes the \dword{daqdfo}, \dword{eb}, and storage buffer. 
    Also shown, in brown, is the subsystem for timing and
    synchronization, in gray, and the subsystem for control,
    configuration, and management.
  }
  \includegraphics[width=0.8\textwidth,trim=4cm 3cm 4cm 3cm]{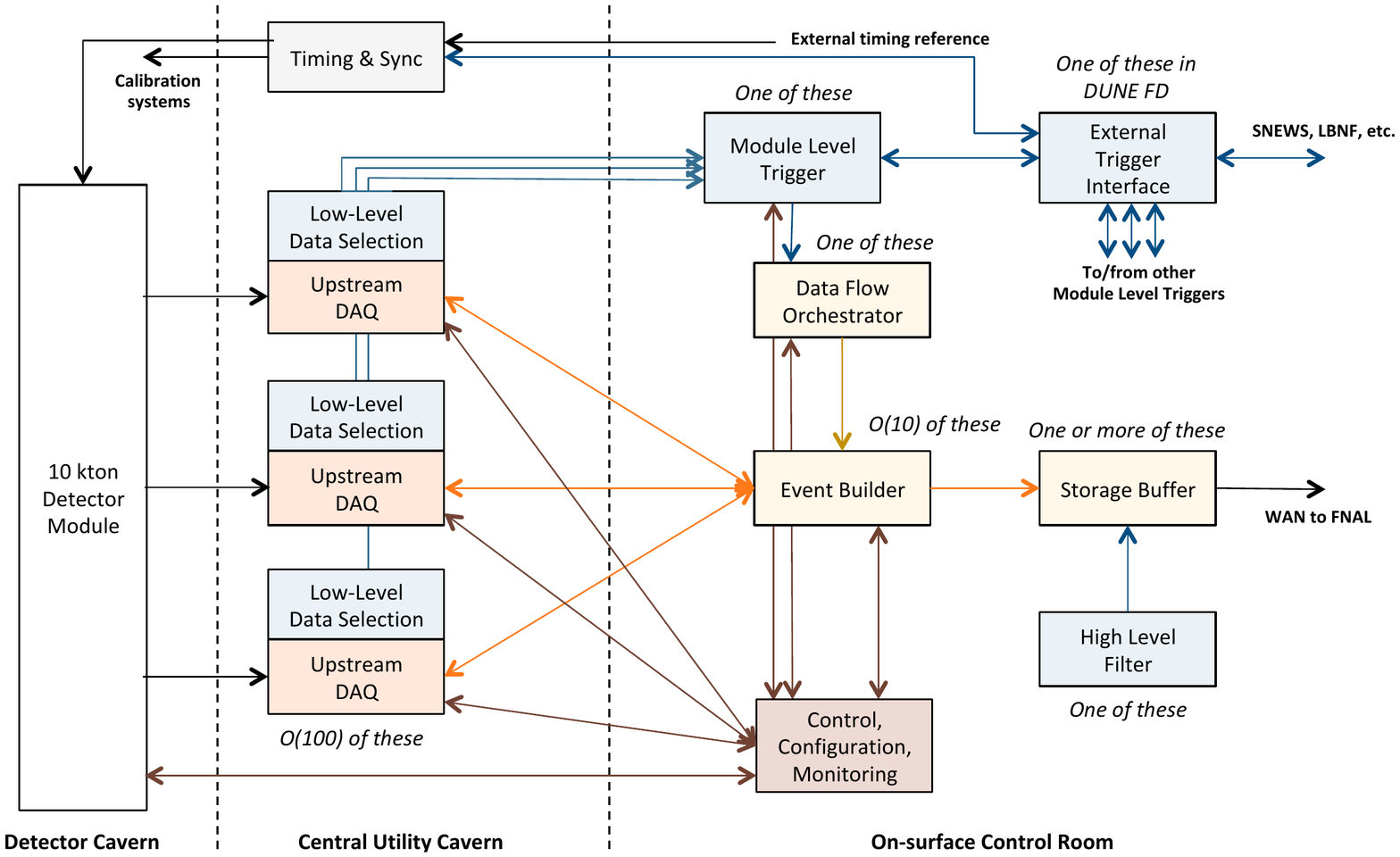}
\end{dunefigure}

\subsection{Requirements and Specifications}
\label{sec:daq:requirements}

The \dword{dune} \dword{fd} \dword{daq} system is designed to meet the
\dword{dune} top-level as well as \dword{daq}-level requirements
summarized in 
Table~\ref{tab:specs:SP-DAQ}. 
The \dword{daq}-level requirements ensure that the 
system
can record all necessary information for offline 
analysis of data associated with on- and off-beam physics events, as directed
by the \dword{dune} physics mission, with minimal compromise to
\dword{dune}'s physics sensitivity. The requirements must be met following the 
specifications provided in the same table.
Those specifications are
associated with trigger functionality, readout,
and operations and are described further in the following subsections.

\subsubsection{How DUNE's Physics Mission Drives the DAQ Design}

The \dword{dune} \dword{fd} has three main physics drivers: measuring neutrino \dword{cpv} and related
long baseline oscillation using the high intensity beam provided
by \fnal; measuring off-beam atmospheric neutrinos and searches
for rare processes such baryon-number-violating decays;
and detecting neutrinos from a nearby \dword{snb}. The
\dword{dune} \dword{fd} \dword{daq} system must facilitate data
readout to deliver on these main physics drivers while keeping
within physical (space, power) and resource constraints for
the system. In particular, the off-beam measurements require continuous readout of the detector, 
and the lack of external triggers for such
events requires real-time or online data processing and
self-triggering capabilities. Because the
continuous raw data rate of the far detector module, as received by
the \dword{daq} system, reaches multiple
terabits per second, significant data buffering and processing
resources are needed as part of the design, as specified in
later sections of this chapter.

The \dword{dune} \dword{fd} modules use two active detector components from which the 
\dword{daq} system must acquire data: the \dword{tpc} and the \dword{pds}. 
The two components access the physics by sensing and collecting signals associated with very different sensing time scales.

Ionization charge measurement by the \dword{tpc} for any given activity in the detector 
requires a nominal recording of data over a time window of approximately \SIrange{1}{10}{\milli\second}. 
This time scale is determined by the ionization electron drift speed in \dword{lar} and the detector 
dimension along the drift direction, nominally set to 
\spreadout, 
corresponding to 2.4$\times$\SI{2.25}{\milli\second}.
The latter (\SI{2.25}{\milli\second}) assumes a drift electric field of \mindriftfieldgoal.
The 2.4 factor ensures capturing ionization information from at
least a full drift before and after the trigger time associated with the
activity. 
Early commissioning data will be used to evaluate and optimize this nominal readout time.

On the other hand, the \dword{pds} measures argon scintillation light emission, which
occurs and is detected over a timescale of multiple \si{\nano\second} to
\si{\micro\second} for
any given event and/or subsequent subevent process. Unlike the \dword{tpc},
the \dword{pds} data is zero-suppressed in
the \dword{pds} electronics (see Chapter~\ref{ch:fdsp-pd}). Although
the \dword{pds} system readout sampling frequency is higher than the \dword{tpc}, the combination of zero-suppression and expected activity
levels should have significantly lower data rates than the \dword{tpc}. Therefore, the total raw data volume received by
the \dword{daq} system should be dominated by
the \dword{tpc} data, which is sent out from the \dword{tpc} electronics as a continuous stream.
 
Figure \ref{fig:daq-rates} provides the expected activity rates in a
single far detector module as a function of true energy associated
with given types of signal.
At low energy ($<$\SI{10}{\MeV}), activity is dominated by radiological backgrounds
intrinsic to the detector and
low-energy solar neutrino interactions. Supernova burst neutrinos,
expected to arrive at a galactic \dword{snb} rate of once per century, 
would span the \SIrange{10}{30}{\MeV} range. At higher energies (generally
more than \SI{100}{\MeV}), rates are dominated by cosmic rays, beam neutrino interactions,
and atmospheric neutrino interactions. With the exception of supernova
burst neutrinos, the activity associated with any of these physics
signals is localized in space and particularly in time. Supernova burst
activity, on the other hand, is characteristically distinct, because it can
manifest as up to several thousands of low-energy neutrino interactions
arriving over multiple seconds. Supernova burst neutrinos
are thus associated with activity that extends over the entirety of the
detector and over a relatively long time.

The nature and rates of these signatures necessitate a \dword{daqdsn} strategy that handles two
distinct cases: a localized high-energy activity trigger, prompting an event record readout for
activity associated with a minimum of \SI{100}{\MeV} of deposited energy; and an extended low-energy
activity trigger, prompting an event record readout when multiple localized low-energy activity
candidates with low deposited energy each (approximately \SI{10}{\MeV}) are found over a short (less than
\SI{10}{\second}) time and over the entirety of a \nominalmodsize  module. Because of the high
granularity of the detector readout elements, a hierarchical \dword{daqdss} is used to
provide data processing and triggering and to facilitate optional data reduction and filtering. 

The \dword{daq}  system must have $>$99\% efficiency for particles
depositing $>$\SI{100}{\MeV} of energy in the detector for localized
high-energy triggers. The system's architecture must also provide a 
mechanism for triggering on galactic supernova bursts with $>$95\%
efficiency for a supernova burst producing at least 60 
interactions with a neutrino energy $>$\SI{10}{\MeV} in 12~kt of active
detector mass, during the first \SI{10}{\second} of the burst, per
\dword{dune} requirements. This requirement ensures sensitivity to the
great majority of \dwords{snb} in our galaxy as well as some bursts in small
nearby galaxies, as described in this document's \dword{snb} physics requirements
section. The \dword{daq} architecture must also provide a mechanism for recording
neutrino interactions associated with those bursts over a \SI{30}{\second}
period, with a goal of \SI{100}{\second}. During this period, the full raw
data information must be stored. The rationale for the latter is that most
models of \dwords{snb} show structure in the neutrino flux for up to
\SI{30}{\second}, and there is potential for interesting measurements to be made
up to \SI{100}{\second}.

Offline considerations require the \dword{daq} to reduce the
full \dword{fd} data volume for offline permanent storage to  \offsitepbpy.
An \dword{fd} composed of four single-phase modules using a
strategy by which the entire \dword{fd} is read out for \spreadout, 
given the presence of a localized high energy trigger, will be limited
by this offline permanent storage constraint to an average readout rate of \SI{0.3}{\hertz}. 
Strategies will be developed and validated during commissioning and early running that will limit the readout to some subset of the detector module, which should allow an increase in this rate limit by about an order of magnitude.
The instantaneous readout rate can be much higher, for example to accommodate calibrations.

For planning, the \dword{daq} will allot an average \dword{snb} trigger rate of one per month.
Given current understanding of \dword{snb} rates and the $>$95\%
expected efficiency for a \dword{snb} with at least 60 interactions
each of minimum \SI{10}{\MeV} in true neutrino energy, most such triggers will be due to fluctuations
of low energy radiological backgrounds and, potentially, excess
noise. 
Such triggers will prompt \snbtime of data from the entire module to be read out.
At this average rate and if saved to offline storage, the \dword{snb}
triggers will produce \SI{1.8}{\peta\byte/\year} uncompressed from one
single-phase module. There is, however, no requirement to
permanently store \dword{snb}
data that is deemed, after further offline analysis, to be due to fake triggers.

The capability of recording data losslessly is built into the design
as a conservative measure; a particular concern is charge reconstruction
efficiency and resolution in the case of zero suppression, in particular for \dword{tpc}
induction wire readout channels.
\dword{microboone} is currently investigating the effect of zero suppression
on reconstruction efficiency and energy resolution for low-energy
events~\cite{Crespo-Anadon:2019lht}. 
Expected data rates from physics signals of interest that fit the
requirement of \offsitepbpy sent to permanent storage are summarized in Table~\ref{tab:daq:rates} and detailed in~\citedocdb{9240}. Potential bottlenecks are analyzed in~\citedocdb{11461}.

\begin{dunefigure}[Expected physics-related activity
    rates in one FD module]{fig:daq-rates}{Expected physics-related activity
    rates in a single \nominalmodsize module.
}
  \includegraphics[width=0.7\textwidth,clip,trim=6cm 6cm 10cm 2cm]{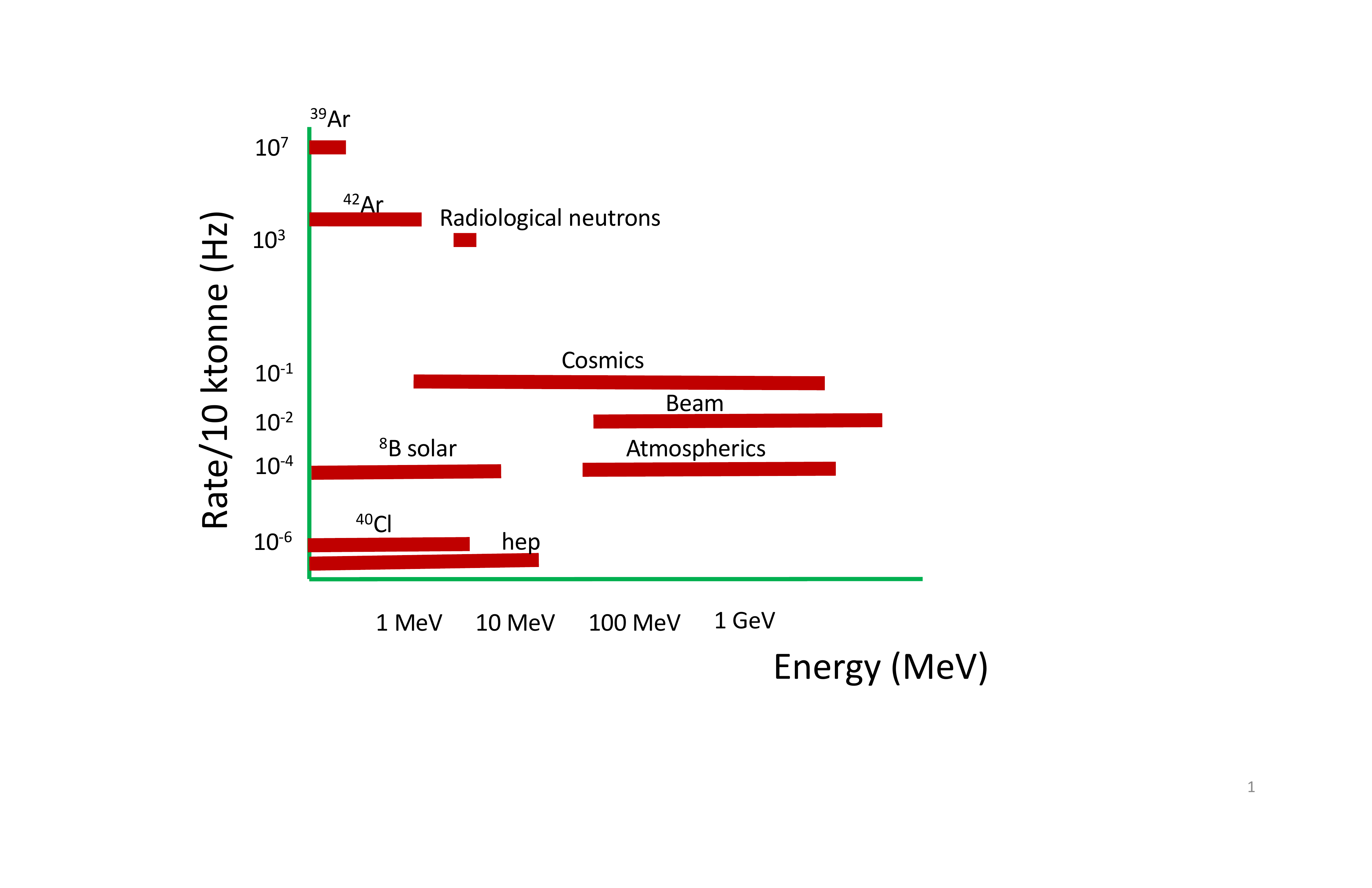}
\end{dunefigure}

\begin{dunetable} [Expected DAQ yearly produced TPC data volumes] {p{0.3\textwidth}p{0.1\textwidth}p{0.5\textwidth}} {tab:daq:rates}{Summary of expected data volumes produced yearly for initial single-module running.
    The numbers assume \dword{tpc} only rates with no compression and are given for a single
    \SI{10}{\kilo\tonne} module, assuming the parameters listed in Table~\ref{tab:daq:parameters}. (A 2-4x lossless
    compression factor is expected.)
    Trigger primitives (see Section~\ref{sec:daq:design-data-selection}), fake \dword{snb} data as well as additional data recorded for detector performance studies and debugging need not be stored offline permanently.}
Source                                              & Annual Data Volume & Assumptions \\\toprowrule
Beam interactions                                   & \SI{27}{\TB}       & \SI{10}{\MeV} threshold in coincidence with beam
time, including cosmic coincidence; \spreadout readout \\\colhline
Cosmics and atmospheric neutrinos                   & \SI{10}{\PB}       & \spreadout readout \\\colhline
Radiological backgrounds                            & $<\SI{2}{\PB}$     & $<1$ per month fake rate for SNB
trigger; \SI{100}{\second} readout\\\colhline
Cold electronics calibration                        & \SI{4}{\TB}       & scaled from \dword{pdsp} experience \\\colhline
Radioactive source calibration                      & \SI{100}{\TB}     & $<10$ Hz source rate; single
APA readout; \spreadout readout \\\colhline
Laser calibration                                   & \SI{200}{\TB}     & 10$^6$ total laser pulses; half the
TPC channels illuminated per pulse; lossy
compression (zero-suppression) on all channels\\\colhline
Random triggers                                     & \SI{60}{\TB} & 45 per day; \spreadout readout \\\colhline
  Trigger primitives and detector performance studies & $<\SI{15}{\PB}$  & $^{39}$Ar dominated\\
\end{dunetable}

Self-triggering on \dword{snb} activity is a unique challenge for the
\dword{dune} \dword{fd}, and an aspect of the design that has never been demonstrated
in a \dword{lartpc}. The challenge of \dword{snb} triggering is two-fold. 
First, the activity of the individual \dword{snb} neutrino interactions
should be relatively low energy (\SIrange{5}{30}{\mega\electronvolt}),
often indistinguishable from pile up of radiological background activity in the
detector.  Triggering on an ensemble of \bigo{100} events expected on
average in the case of a galactic supernova burst is, therefore,
advantageous; however, this ensemble of events will likely be rare over the
entire detector and over an extended period of \bigo{10}\si{s}, so
sufficient buffering capability must be designed into the system to
capture the corresponding signals. 
Furthermore, to ensure high efficiency in collecting \dword{snb} interactions
that, individually, are below low-energy activity threshold, data from
all channels in the detector will be recorded over an extended and
contiguous period,  which is specified to \SIrange{30}{100}{\second}, around every \dword{snb}
trigger. This time has been defined in consultation with the \dword{dune}
physics groups.

\begin{footnotesize}
\begin{longtable}{p{0.12\textwidth}p{0.18\textwidth}p{0.17\textwidth}p{0.25\textwidth}p{0.16\textwidth}}
\caption{DAQ specifications \fixmehl{ref \texttt{tab:spec:SP-DAQ}}} \\
  \rowcolor{dunesky}
       Label & Description  & Specification \newline (Goal) & Rationale & Validation \\  \colhline

  \newtag{SP-FD-22}{ spec:data-rate-to-tape }  & Data rate to tape  &  $<\,\SI{30}{PB/year}$ &  Cost.  Bandwidth. &  ProtoDUNE \\ \colhline

  \newtag{SP-FD-23}{ spec:sn-trigger }  & Supernova trigger  &  $>\,\SI{95}{\%}$ efficiency for a SNB producing at least 60 interactions with a neutrino energy >10 MeV in 12 kt of active detector mass during the first 10 seconds of the burst. &  $>\,$95\% efficiency for SNB within 20 kpc &  Simulation and bench tests \\ \colhline

  \newtag{SP-DAQ-1}{ spec:DAQ-readout }  & DAQ readout throughput: The DAQ shall be able to accept the continuous data stream from the TPC and Photon detectors.  &  1.5 TB/s per single phase detector module &  Specification from TPC and PDS electronics &  Modular test on ProtoDUNE; overall throughput scales linearly with number of APAs \\ \colhline

  \newtag{SP-DAQ-2}{ spec:DAQ-throughput }  & DAQ storage throughput: The DAQ shall be able to store selected data at an average throughput of 10 Gb/s, with temporary peak throughput of 100 Gb/s.  &  10 Gb/s average storage throughput; 100 Gb/s peak temporary storage throughput per single phase detector module &  Average throughput estimated from physics and calibration requirements; peak throughput allowing for fast storage of SNB data ($\sim 10^4$ seconds to store 120 TB of data).  &  ProtoDUNE demonstrated steady storage at $\sim$ 40 Gb/s for a storage volume of 700 TB. Laboratory tests will allow to demonstrate the performance reach. \\ \colhline

  \newtag{SP-DAQ-3}{ spec:DAQ-readout-window }  & DAQ readout window: The DAQ shall support storing triggered data of one or more APAs with a variable size readout window, from few $\mu$s (calibration) to 100 s (SNB), with a typical readout window for triggered interactions of 5.4 ms.  &  10 $\mu$s < readout window < 100 s &  Storage of the complete dataset for up to 100 s is required by the SNB physics studies; the typical readout window of 5.4 ms is defined by the drift time in the detector; calibration triggers can be configured to readout data much shorter time intervals. &  Implementation techniques to be validated on the ProtoDUNE setup and in test labs. \\ \colhline

  \newtag{SP-DAQ-4}{ spec:trigger-calibration }  & Calibration trigger: The DAQ shall provide the means to distribute time-synchronous commands to the calibration systems, in order to fire them, at a configurable rate and sequence and at configurable intervals in time. Those commands may be distributed during physics data taking or during special calibration data taking sessions. The DAQ shall trigger and acquire data at a fixed, configurable interval after the distribution of the commands, in order to capture the response of the detector to calibration stimuli.  &   &  Calibration is essential to attain required detector performance comprehension. &  Techniques for doing this have been run successfully in MicroBooNE and ProtoDUNE.  \\ \colhline

  \newtag{SP-DAQ-5}{ spec:data-record }  & Data record: Corresponding to every trigger, the DAQ shall form a data record to be transferred to offline together with the metadata necessary for validation and processing.  &   &  Needed for offline analysis. &  Common experimental practice. \\ \colhline

  \newtag{SP-DAQ-6}{ spec:data-verification }  & Data verification: The DAQ shall check integrity of data at every data transfer step. It shall only delete data from the local storage after confirmation that data have been correctly recorded to permanent storage.  &   &  Data integrity checking is fundamental to ensure data quality. &   \\ \colhline

  \newtag{SP-DAQ-7}{ spec:trigger-high-energy }  & High-energy Trigger: The DAQ shall trigger and acquire data on visible energy deposition >100 MeV. Data acquisition may be limited to the area in which activity was detected.  &  $>$\SI{100}{\MeV} &  Driven by DUNE physics mission. &  Physics TDR. 100 MeV is an achievable parameter; lower thresholds are possible. \\ \colhline

  \newtag{SP-DAQ-8}{ spec:trigger-low-energy }  & Low-energy Trigger: The DAQ shall trigger and acquire data on visible energy deposition > 10 MeV of single neutrino interactions. Those triggers will normally be fired using a pre-scaling factor, in order to limit the data volume.  &  $>$\SI{10}{\MeV} &  Driven by DUNE physics mission. &  Physics TDR. 10 MeV is an achievable parameter; lower thresholds are possible. \\ \colhline

  \newtag{SP-DAQ-9}{ spec:daq-deadtime }  & DAQ deadtime: While taking data within the agreed conditions, the DAQ shall be able to trigger and acquire data without introducing any deadtime.  &   &  Driven by DUNE physics mission. &  Zero deadtime is an achievable inter-event deadtime but a small deadtime would not significantly compromise physics sensitivity.  \\ \colhline

\label{tab:specs:SP-DAQ}
\end{longtable}
\end{footnotesize} 

\subsubsection{Considerations for Design}
\label{sec:daq:considerations}

The \dword{daq} system is designed as a single, scalable system that can
service all \dword{fd} modules. It is also designed so the
system can record and store full detector data with zero dead
time, applying appropriate data reduction through data selection and
compression. The system should be evolutionary, taking advantage of the
staged construction of the \dword{dune} \dword{fd}, thus beginning very
conservatively for the first \dword{dune} \dword{fd} module, but aggressively reducing
the design conservatism with further experience of detector
operations. At the same time, the system is designed to be able to add capacity as required. Most processing and buffering of raw
detector data is done underground, in the upstream \dword{daq} and low
level data selection parts of the
system (see Figure~\ref{fig:daq:layout}), with only event building and data
storage on surface.

Power, cooling, and space are constrained both in the \dword{cuc} and on the surface, limited to \daqpower and \daqracks (out of \cucracks total) in the \dword{cuc}, and \surfdaqpower and \surfdaqracks on the surface for \dword{daq} for all four \dword{fd} modules.
The underground computing required for the \dword{daq} to service the 
\dword{sp} 
detector module should require less than a quarter of the total power and rack space provided for all four detector modules. 
The hardware for upstream \dword{daq}, low level data selection,
various services, networking, and the timing system, including spares,
all kept in the \dword{cuc}, 
should consume less than \SI{63}{\kilo\watt} and all fill less
than 300U of rack space \cite{bib:docdb15544}.

There are five key challenges for the \dword{dune} \dword{fd} \dword{daq} system: 
\begin{itemize}
\item First, the high overall experiment uptime goal requires \dword{daq} to be stringently designed for reliability, fault tolerance, and redundancy, criteria that aim to reduce overall downtime.

  The \dword{daq} system is fully configurable, controllable, and operable from remote locations, with authentication and authorization implemented to allow exclusive control. The \dword{daq} monitors the quality of the detector data and of its own operational status, as well as automated error detection and recovery capabilities.

\item Second, the system must be able to evolve to accommodate newly commissioned sub-components as they are installed into a detector module that is under construction. 
  The \dword{daq} must also continue to service existing modules that are operational while simultaneously accommodating subsequent detector modules as they are installed and commissioned. 
  To support this ongoing variability, the \dword{daq} will support operating as multiple independent instances or partitions.

  Partitioning will also be supported  within a single detector module for special calibration or debugging runs that are incompatible with physics data taking, while the rest of the detector remains in physics data taking mode. Partitioning, i.e., allowing several instances of the \dword{daq} to operate independently with different configurations on different parts of the detector, will also be important during the installation and commissioning, so experts can work in parallel, e.g., for \dwords{pd} and \dword{tpc}.

\item Third, the \dword{snb} physics requirements require heavy buffering in the upstream \dword{daq}. 

  Implementing a buffer element in the upstream \dword{daq} allows
  the formation and capture of delayed, data-driven data selection
  decisions:
  The trigger accumulates low energy signals over an extended period
  while carrying out the \dword{trigdecision}, thus identifying activity compatible with \dword{snb}.
  The depth of this buffer is determined in
  consultation with physics groups and driven primarily by the need
  to retain all unbiased data while processing up to \SI{10}{s} of
  data for the \dword{trigdecision} preceding a \dword{snb} trigger.
  Collecting data containing information on other types of interactions and decays does not pose additional requirements on the upstream \dword{daq} buffer because the latency required for triggers should be well below \SI{10}{s}.

\item Fourth, the \dword{daq} must support a very wide range of readout windows and trigger rates. This includes acquiring localized events in both time and space up to the very large and rare \dword{snb} detector-wide readouts over \SI{100}{s}.

\item Finally, the \dword{daq} must reduce the volume of data to be permanently stored offline to a maximum of \offsitepbpy.
  The \dword{daq} system should be able to select interesting time windows in which activity was detected, apply lossless compression to data records, and filter records to remove unnecessary data regions.

A programmable trigger priority scheme ensures that the readout for the main physics triggers is never or rarely inhibited, thus making it easy to determine the live-time of these triggers.

\end{itemize}

Table~\ref{tab:daq:parameters} summarizes the important parameters driving the \dword{daq} design. 
These parameters set the scale of data buffering, processing, and transferring resources that must be built 
into each \dword{fd} module. 

\begin{dunetable}
[Key DAQ parameters]
{ll}
{tab:daq:parameters}
{Summary of important parameters driving the \dword{daq} design. The \dword{pds}
  system parameters are under study, but the \dword{pds} raw
  data volume that must be handled by the \dword{daq} should be an order of
  magnitude smaller than the \dword{tpc} raw data volume.}
Parameter                                          & Value \\ \toprowrule
TPC Channel Count per Module                       & \num{384000}\\ \colhline
TPC Collection Channel Count per Subdetector (APA) & \num{960}\\ \colhline
TPC Induction Channel Count per Subdetector (APA)  & \num{1600}\\ \colhline
PDS Channel Count per Module                       & \num{6000}\\ \colhline
PDS Channel Count per Subdetector (PDS per APA)    & \num{40} \\ \colhline
TPC \dword{adc} Sampling Rate                      & \SI{2}{\mega\hertz}\\ \colhline
TPC \dword{adc} Dynamic Range                      & \SI{12}{bits}\\ \colhline
PDS \dword{adc} Sampling Rate                      & Under study \\ \colhline
PDS \dword{adc} Dynamic Range                      & Under study \\ \colhline
PDS \dword{adc} Readout Length                     & Under study \\ \colhline
Localized Event Record Window                      & \spreadout \\  \colhline
Extended Event Record Window                       & \SI{100}{\second}\\  \colhline
Full size of Localized Event Record per Module     & \SI{6.5}{\GB} \\  \colhline
Full size of Extended Event Record per Module      & \SI{120}{\TB}\\  
\end{dunetable}

\subsection{Interfaces}
\label{sec:daq:interfaces}

The \dword{daq} system scope begins at the optical fibers streaming raw digital data from the detector active components
(\dword{tpc} and \dword{pds}) and ends at a wide-area network interface that
distributes the data from on site at \dword{surf} to offline centers off
site. The \dword{daq} also provides common computing and network services for
other \dword{dune} systems, although slow control and safety functions
fall outside \dword{daq}'s scope.

Consequently, the \dword{dune} \dword{fd} \dword{daq} system interfaces with the \dword{tpc} \dword{ce}, \dword{pds}
readout, computing, \dword{cisc}, and calibration systems of the 
\dword{fd}, as well as with facilities and underground installation. The
 interface agreements with the \dword{fd} systems 
are listed in Table~\ref{tab:daq:interfaces} and described
briefly in the following subsections. Interface agreements with
facilities and underground installation are described in Section~\ref{sec:daq:production}.

\begin{dunetable}
[DAQ system interfaces]
{p{0.3\textwidth}p{0.3\textwidth}p{0.2\textwidth}}
{tab:daq:interfaces}
{Data Acquisition System Interface Links. }
Interfacing System & Description & Reference \\ \toprowrule
TPC CE & Data rate and format, number and type of links, timing, inherent noise & \citedocdb{6742}{v6}\\ \colhline
PDS Readout & Data rate and format, number and type of links, timing &  \citedocdb{6727}{v2} \\ \colhline
Computing & Off-site data transfer rates, methods, data file content, disk buffer, software development and maintenance &  \citedocdb{7123} \\ \colhline
CISC & Information exchange, hardware and software for rack and server monitoring & \citedocdb{6790}{v1} \\ \colhline
Calibration & Constraint on total volume of the calibration data; trigger and timing distribution from the \dword{daq} & \citedocdb{7069} \\ \colhline
Timing and Synchronization & Clients, clock frequency, protocols,
transports, accuracy, synchronization precision, monitoring&  \citedocdb{11224} \\ \colhline
Facilities & Detector integration, coordination, cables, racks, safety, conventional facilities, lack of impact on cryo and DSS &  \citedocdb{6988}{v1} \\ \colhline
Installation & Prototyping, planning, transport, underground equipment and activity, safety & \citedocdb{7015} \\ 
\end{dunetable}

\begin{description}
\item[\dword{tpc} \dword{ce}] The \dword{daq} and \dword{tpc} \dword{ce} interface is described in
\citedocdb{6742}. The physical interface is in the \dword{cuc} where optical links from the \dwords{wib} transfer
the raw \dword{tpc} data to the \dword{daq} front-end readout (\dword{felix}; see
Section~\ref{sec:daq:design-upstream}). This ensures the \dword{daq} is electrically decoupled from the detector
cryostat. Ten \SI{10}{\Gbps} links are expected per \dword{apa} and have
been specified as \SI{300}{\meter} OM4 multi-mode fibers from \dword{sfp}+ at the \dword{wib} to
\dword{minipod} on \dword{felix}. (The optical fibers themselves are
the responsibility of the \dword{daq} consortium.) The data format has been specified
as a custom communication protocol with no
compression.

\item[\dword{pds} readout] The \dword{daq} and \dword{pds} readout interface is described in
\citedocdb{6727}. It should
require no more than 150  \SI{10}{\Gbps} OM4 fibers from one \dword{fd} module. 
This
is similar to the interface to the \dword{tpc} \dword{ce}, except the overall
data volume is lower by an order of magnitude. The data format has been specified to use
compression (zero suppression) and a custom communication protocol.

\item[Computing] The \dword{daq} and computing interface is described in \citedocdb{7123}.
 The computing consortium
 is responsible for the online areas of WAN connection between \dword{surf} and
\fnal providing \SI{100}{\Gbps} bandwidth, while the \dword{daq} consortium is responsible for disk buffering
to handle any temporary WAN disconnects and the infrastructure needed
for real-time data quality monitoring.  The computing consortium 
is also
responsible for the offline development and operation of the tools for data
transfers to \fnal. The primary
constraint in defining the \dword{daq} and offline computing interface is the
requirement to produce less than \offsitepbpy 
into final storage at
\fnal. The \dword{daq} and
computing consortia are jointly responsible for data
format definition and data access libraries, as well as real-time data
quality monitoring software. The former is specified in the form of a 
data model documented in \citedocdb{7123}.

\item[CISC] The \dword{daq} and \dword{cisc} interface is described in
\citedocdb{6790}. The \dword{daq} provides a network in the \dword{cuc} for \dword{cisc},  operation information and hardware
monitoring information to \dword{cisc}, and power distribution and
rack status units in \dword{daq} racks. The information from \dword{cisc}
feeds back into the \dword{daq} for run control operations.

\item[Calibration] The \dword{daq} and calibration interface is described in \citedocdb{7069}. 
Two calibration systems are envisioned for the \dword{fd}: a laser calibration system and a neutron generator. 
  Two-way communication between the calibration system and the
  \dword{daq} is needed.  Specifically, the calibration system must
  notify the data selection system, thus informing \dword{trigdecision}
  when activity has been induced in the detector. At the same time, 
the \dword{daq} must provide input to the calibration system, so it can avoid inducing activity in 
the detector during certain periods such as \dword{snb} readout time or during a beam spill.
This second communication will be initiated by the data selection
system and distributed via the \dword{daqeti} and
subsequently through the
\dword{daq} timing system.

\item[Timing and Synchronization] The timing system of the \dword{dune} 
\dword{fd} connects with almost all detector systems and with the calibration 
system.  It has a uniform interface to each of them. 
  A single interface document \citedocdb{11224} describes all timing interfaces. 

Accuracy of timestamps delivered to  detector endpoints will be 
$\pm$\SI{500}{\nano\second} with respect to UTC. 
Synchronization between any two endpoints in the detector will be less than 
\SI{10}{\nano\second}. Between detector modules, synchronization will be less than \SI{25}{\nano\second}.  
\end{description}

\section{Data Acquisition System Design}
\label{sec:daq:design}

This section begins with an overview of the \dword{daq}
design followed by
descriptions of the subsystem designs and implementation specifics.

\subsection{Overview}
\label{sec:daq:design-overview}

The  \dword{daq} system comprises five distinct subsystems:
(1) upstream \dword{daq} (Section~\ref{sec:daq:design-upstream}),
(2) \dword{daqdss} (Section~\ref{sec:daq:design-data-selection}),
(3) \dword{daqbes} (Section~\ref{sec:daq:design-backend}), 
(4) \dword{daqccm} (Section~\ref{sec:daq:design-run-control}), and
(5) \dword{daqtss} (Section~\ref{sec:daq:design-timing}).
Figure~\ref{fig:daq:layout} shows the physical extent of the subsystems: the
upstream \dword{daq} and \dword{daqtss}  live underground in the \dword{cuc}; \dword{daqdss} occupies
both underground and above-ground spaces; \dword{daqbes} is above-ground
and includes data flow orchestration, event building, and buffering before distribution of data
to offline storage; and \dword{daqccm} extends throughout the entire physical layout of the
system, supported on a private network throughout the \dword{daq} system. Each of these subsystems is described in further
detail in the following subsections.

Front-end readout is carried out by the upstream \dword{daq} using
custom data receiver and
co-processing \dword{fpga} and commodity computing hardware, 
all of which is hosted in 80-85 servers in the \dword{cuc}. A
corresponding number of additional servers execute
subsequent software-based low-level processing of trigger
primitives
generated in the upstream \dword{daq} for the purposes of \dword{daqdsn}. 
The trigger candidates constructed from trigger primitives are propagated to a central server responsible
for further processing and module-level triggering. The module level
trigger also
interfaces to a second server that receives and
propagates cross-module and external trigger and timing
information. The module level trigger considers trigger candidates and
external trigger inputs in issuing a \dword{trigcommand} to the \dword{daqbes}
subsystem. The \dword{daqbes} subsystem 
facilitates event building in a few servers and buffering for built
events on non-volatile storage. Upon receiving a \dword{trigcommand}, the \dword{daqbes}
 queries
data from the upstream \dword{daq} buffers and builds that into an event
record, which is temporarily stored in (a number of) files. Event records are optionally processed in a high-level
filter/data reduction stage, which is part of overall data selection,
 before event records are shipped to \dword{dune} offline. Pervasively,
 the  \dfirst{daqccm} orchestrates data taking 
 (Section~\ref{sec:daq:design-run-control}), and the
 \dfirst{daqtss} provides synchronization and synchronous command distribution
 (Section~\ref{sec:daq:design-timing}). Figure~\ref{fig:daq-conceptual-overview}
 provides a conceptual illustration of the overall \dword{daq} system
 functionality.

\begin{dunefigure}[DAQ system overview]{fig:daq-conceptual-overview}{Conceptual
   Overview of \dword{daq} system functionality for a single \nominalmodsize module}
  \includegraphics[width=0.9\textwidth]{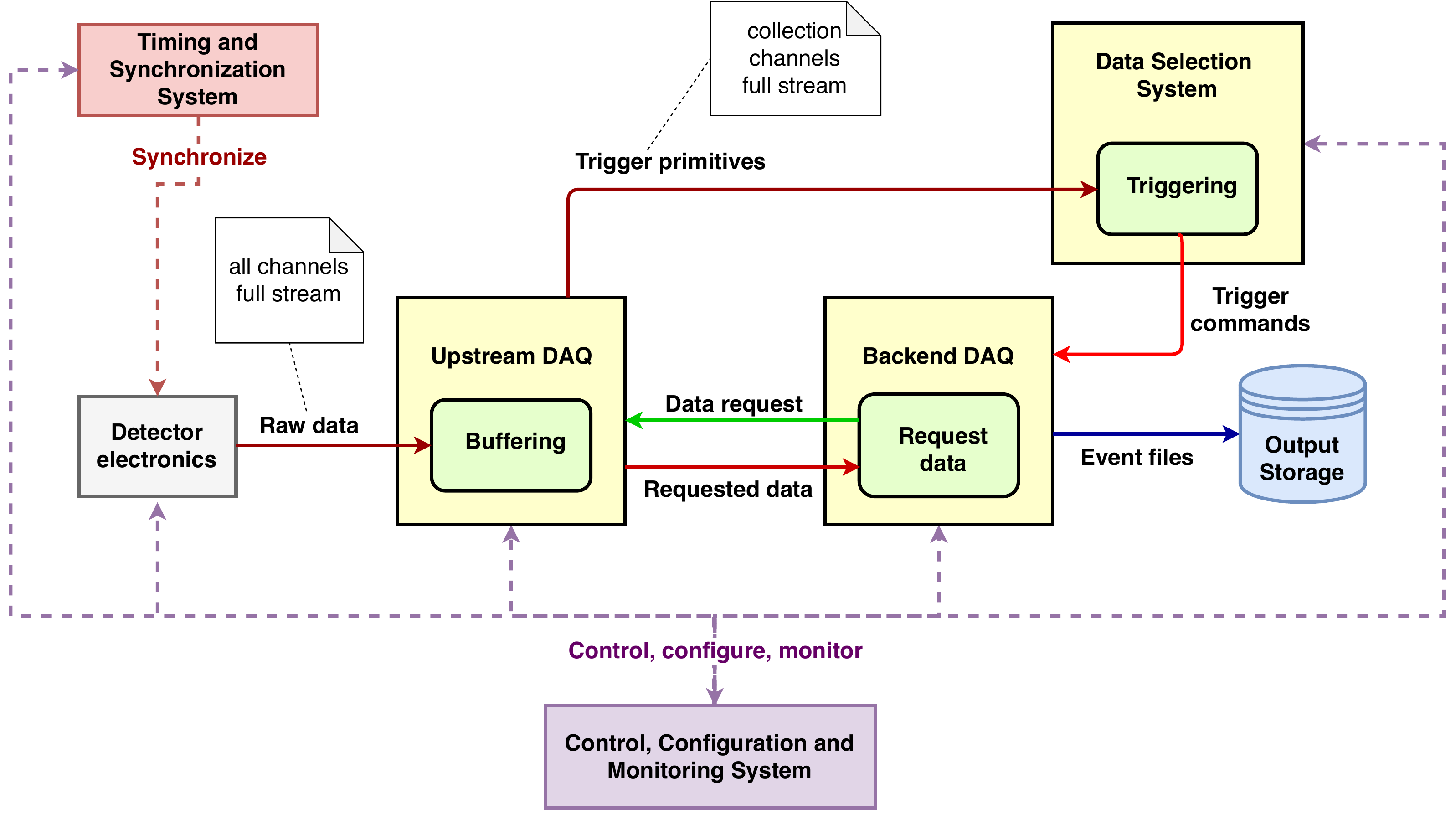}
\end{dunefigure}

Key to implementing the \dword{daq} design is the requirement that
the system can be partitioned. Specifically, the system can operate in
as multiple independent \dword{daq} instances, each 
executed across all \dword{daq} subsystems and uniquely mapped among subsystem components. 
More specifically, a given partition may span the entire 
\dword{detmodule} or some subset of it; its extent is configurable at
run start. This ensures continual readout of most of the detector in normal physics data-taking run mode, while
enabling simultaneous calibration or test runs of small portion of the
detector without interrupting normal data taking. 

\subsection{Upstream DAQ}
\label{sec:daq:design-upstream}

The upstream \dword{daq} provides the first link in the data flow chain of
the \dword{daq} system; it receives raw data from detector electronics.
The upstream \dword{daq} implements a receiver, buffer, and a portion of low-level data
selection (trigger primitive generation; see Section~\ref{sec:daq:design-data-selection}) as detailed in Figure~\ref{fig:daq:readout}.
It is physically connected to the detector electronics via optical
fiber(s) and buffers and serves data to other \dword{daq} subsystems,
namely the \dword{daqdss} and the \dword{daqbes}.

\begin{dunefigure}[DUNE upstream DAQ subsystem and connections]{fig:daq:readout}{\dword{dune} upstream \dword{daq} subsystem and its connections.}
  \includegraphics[width=0.8\textwidth]{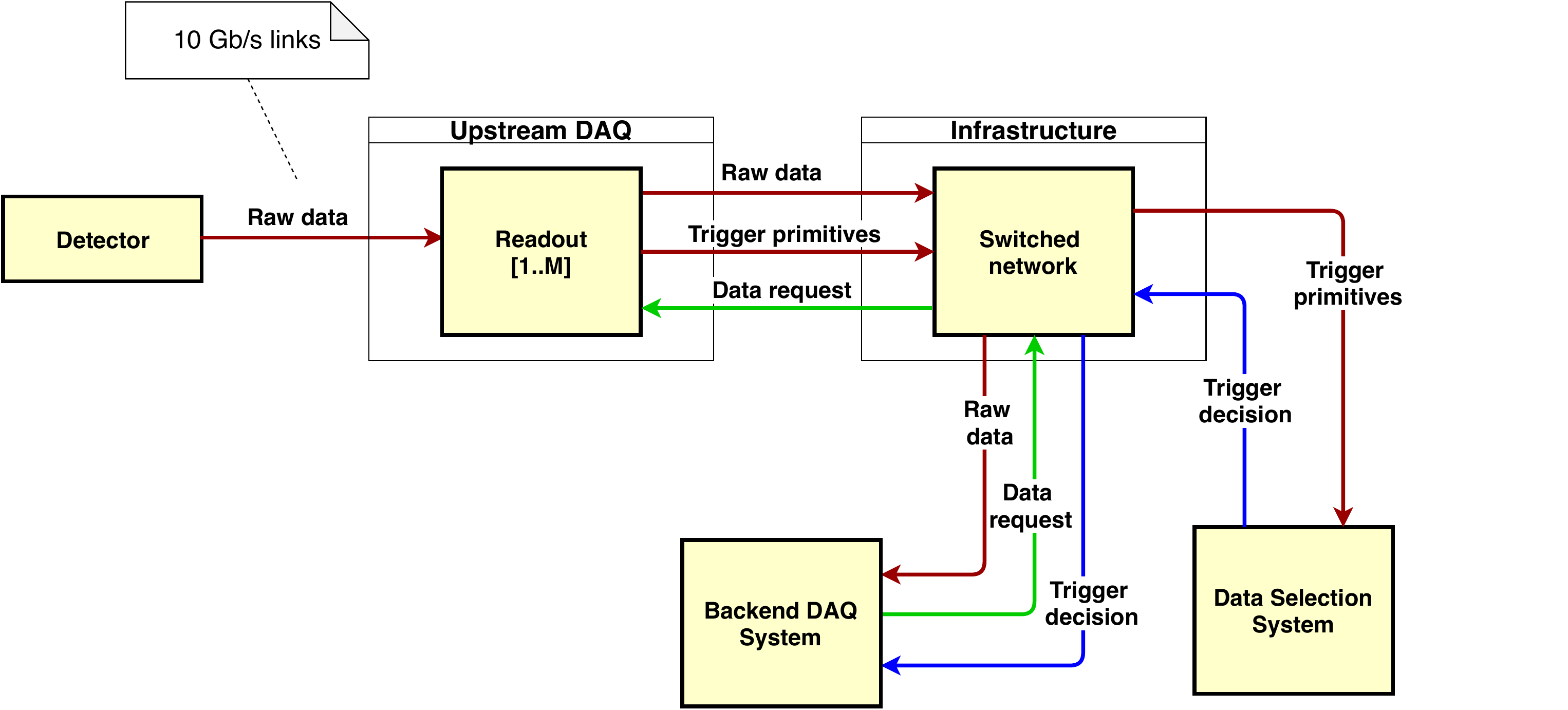}
\end{dunefigure}

\begin{dunefigure}[Flow diagram of the DUNE upstream DAQ subsystem.]{fig:daq:readout-blocks}{\dword{dune} upstream
    \dword{daq} subsystem functional blocks.}
 \includegraphics[width=0.8\textwidth]{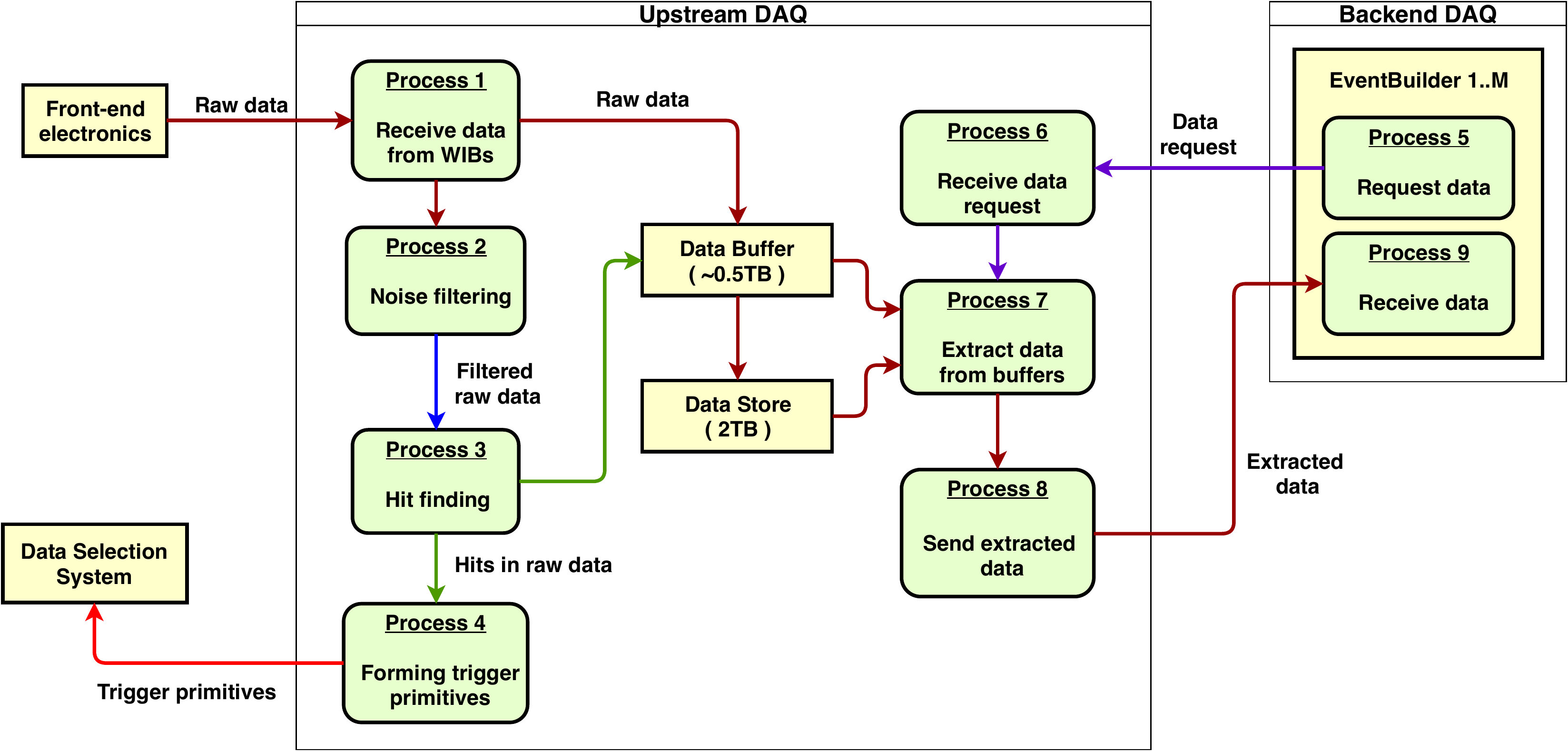}
\end{dunefigure}

The upstream \dword{daq} system comprises many similar \dwords{daqrou}, each
connected to a subset of electronics from a detector module and
interfacing with the \dword{daq} switched network. In the case of the
\dword{tpc}, 75 functionally identical \dword{daqrou} are each responsible for the readout of raw data from two
\dword{apa}s. In the case of the \dword{pds}, up to eight \dword{daqrou} are each responsible for the
readout of raw data from a collection of \dword{pds} subdetectors. 
Each \dword{daqrou} consists of a \dword{fec}, commodity server that hosts a
collection of custom hardware, 
firmware, and software that collectively implement a set of functional blocks.

In the baseline design, used also for system costing, each \dword{daqrou} is composed of
\begin{enumerate}
\item one dual socket multicore 2U server, with two \SI{10}{\Gbps} and two \SI{1}{\Gbps} ethernet ports for redundant data transmission and control, with at least \SI{256}{\GB} of DDR4 RAM, and sufficient PCIe lanes to host \SI{2}{\TB} of SSD disks;
\item two \dword{felix} cards~\cite{atlas-felix}, each with a PCIe 3.0 x16 interface and supporting ten \SI{9.6}{\Gbps} bidirectional serial optical links;
\item only for the \dword{tpc} readout, four custom-designed co-processor cards mounted onto the two \dword{felix} cards for additional processing power and data buffering.

\end{enumerate}

The main functional blocks of the upstream \dword{daq} are described below:

\begin{itemize}
  
  \item The physical interface between the detector electronics and the \dword{daq} are \SI{9.6}{\Gbps}
point-to-point serial optical links running a simple (e.g., 8/10 bit encoded) protocol.  Each
\dword{apa} has ten such links connecting its \dwords{wib} to the \dword{daq}. To reduce space and
power consumption in the \dword{cuc}, high data aggregation is needed. In the baseline design, the
\num{20} fibers from two \dword{apa}s are aggregated into one \dword{fec} with each \dword{apa} connected
to one \dword{felix} FPGA PCIe 3.0 board.  If commodity computing technology advances sufficiently, a PCIe 4.0 version of
\dword{felix} may be produced to increase the data aggregation, so each board would accept \num{20}
links, and a total of \num{40} links per \dword{fec} would be accommodated. Tests were performed to verify
\dword{om3} and \dword{om4} fibers support \SI{10}{\Gbps} links over a run of \SI{300}{\meter}.
Higher-speed fiberoptic links may be used to reduce the number of fibers if run
lengths can be reduced. The \dword{felix} board and firmware were developed initially by and for the
\dword{atlas} experiment and is now proposed or already in use by several other experiments including
\dword{protodune}.
  
  \item The upstream \dword{daq} subsystem provides access to the
\dword{daqdss} and \dword{daqbes} through a commodity switched network as
illustrated in Figure~\ref{fig:daq:readout}. The network communication
protocol is described in Section~\ref{sec:daq:design-ipc}. The data flow
is handled by the \dwords{daqrou} via software. Dedicated hardware or firmware
development is not required.

\item The data processing functional block is tasked with identifying and forming \dwords{trigprimitive} (see
Section~\ref{sec:daq:design-trigger-primitives}), after a stage of data
reorganization and noise filtering for both the \dword{tpc} and \dword{pds}. \Dwords{trigprimitive} summarize time periods on a channel where the digitized waveform is no longer consistent with noise.
These regions of interest are then used as input to the \dword{daqdss} where they begin the process of forming a \dword{trigdecision}.

In the baseline design, this functional block is implemented in \dword{fpga}.
R\&D studies are ongoing to evaluate alternative implementations (in GPUs or CPUs) that may have advantages in flexibility or cost.

\item In \dword{dune}, the upstream \dword{daq} system is in charge of buffering all
detector data until the \dword{daqdss} has issued a \dword{trigcommand}
(see Section~\ref{sec:daq:design-data-selection}) and until the
\dword{daqbes} (Section~\ref{sec:daq:design-backend}) has requested
and received the corresponding selected data. 
In addition, in the case of a \dword{snb} trigger, data received after
the issuing of the trigger must be buffered for longer to
avoid loss of data due to any possible downstream
bottlenecks. Localized and extended trigger activity are associated
with two rather different time scales and data throughput metrics, and
those collectively dictate the temporary storage technology and scale. 

A \dword{trigdecision} based on localized
activity should require buffering the full data stream for no more than one second.
Extended triggers present a far more challenging set of buffering
requirements; 
early activity from a \dword{snb} may occur at a rate near that of radiological activity.
Theoretical estimates indicate \snbpretime of integration time may be needed for the
\dword{snb} interaction rate to be deemed significantly enough above
background rate to form a trigger decision.
In order to locate interactions in this low-rate period, the full data rate must be buffered until an \dword{snb} trigger may be formed.
The throughput and endurance required by this buffer is satisfied by RAM technology like DDR4.

A second challenge in recording data during an \dword{snb} is to assure
essentially 100\% efficiency for collecting the individual, low-energy
interactions during any given \dword{snb} burst. 
To achieve this, the full-stream of data is recorded for a time duration that should cover the time envelope of the burst.
Guided by \dword{snb} models, this duration is set to \snbtime.
This requires extracting as much as \SI{120}{\tera\byte} from the \dword{tpc} upstream \dword{daq} across one 
single-phase detector module.
It is not cost effective to design the \dword{daq} to extract such
extended data record along the same path as nominal readout, so
additional buffering is needed.

The technology and scale of this additional buffering must satisfy several requirements. 
Each \dword{daqrou} must accept the full data rate of the portion of the detector module it services.
The media must have sufficient capacity and allow sufficient extraction throughput to make it unlikely to ever be too full to accept another extended data record: \SI{4}{\tera\bit/\second} guarantee the ability to store two \dword{snb} events simultaneously.
Furthermore, assuming that, on average, an \dword{snb} trigger condition will be
satisfied once per month, the optimal technology is
solid-state \dword{nvme} devices, which, at the scale required to provide suitable input bandwidth, can provide a capacity to write the data from several extended activity triggers. The recorded data will be transferred to the \dword{daqbes} system in less than a day.

For both types of activity, the buffering requirements may be reduced by
using lossless compression to the data before it enters the buffer. A
factor of at least two to four compression is expected, based on
\dword{microboone}~\cite{Crespo-Anadon:2019lht} 
and \dword{protodune} experience using a modified Huffman encoding of
differential \dword{adc} values.  Efforts are currently underway to understand the
costs and technology requirements in exploiting this benefit.

\end{itemize}

Figure~\ref{fig:daq:readout-blocks} shows the flow diagram of data and control messages within the upstream \dword{daq} 
as well as the main interaction of the upstream \dword{daq} with other subsystems.

\subsection{Data Selection}
\label{sec:daq:design-data-selection}

The \dword{daqdss}  is a hierarchical, online, primarily
software-based system. It is responsible for immediate and continuous processing of a substantial fraction of the entire input data stream. 
This includes data from 
\dword{tpc} and \dword{pds} subdetectors.
From that input, as well as external inputs provided by
the accelerator and detector calibration systems, the \dword{daqdss} must form a \dword{trigdecision},
which in turn produces a \dword{trigcommand}.
This command summarizes the observed activity that led to the decision
and provides addresses (in channel-time space) of the data in the
upstream \dword{daq} buffers that capture raw data
corresponding to the activity.
This command is sent to, then consumed, and executed by the \dfirst{daqbes} as described in Section~\ref{sec:daq:design-backend}. 
It may also be propagated to an \dword{etl} stage, and from there, it may be
distributed to other \dwords{detmodule} or other detector systems
(e.g., calibration) for further consideration.

To facilitate partitioning, the \dword{daqdss}  can be instantiated several times, 
and multiple instances can operate in parallel. Within any
given partition, the \dword{daqdss} will also be
informed and aware of current detector configuration and conditions and
apply certain masks and mapping on subdetectors or their fragments in
its decision making. This information is delivered to the
\dword{daqdss} by the \dword{daqccm} system (Section~\ref{sec:daq:design-run-control}).

Following \dword{dune} \dword{fd} and \dword{daq} requirements, the
\dword{daqdss} must select, with sufficiently high ($>$99\%) efficiency, data associated with calibration
signals, as well as beam interactions,
atmospheric neutrinos, rare baryon-number-violating events, and cosmic
ray events that deposit visible energy in excess of \SI{100}{\MeV}. 
It must also select data associated with potential
\dwords{snb} producing 60 neutrino interactions over a span of \SI{10}{\second} in 12~ktons of active
liquid argon mass each with \SI{10}{\MeV} in neutrino energy, with $>$95\% efficiency. 
Furthermore, to meet the requirement that the \dword{dune} \dword{fd} maintain
$<$\offsitepbpy to permanent storage, the \dword{daqdss}
must make \dword{daqdsn} decisions in a way that allows the \dword{daq} 
system to effectively reduce its input data by almost four orders of magnitude,
without compromising the above efficiencies.

To meet its requirements, the \dword{daqdss} design follows a hierarchical data selection strategy, 
where low-level decisions are fed forward into higher-level ones until a module-level trigger is activated. 
The hierarchy is illustrated in
Figure~\ref{fig:daq:data-selection-hierarchy}. 
At the lowest level, trigger primitives are formed on a per-channel basis, and represent, for the baseline design, a hit on a wire/channel activity summary. 
\Dwords{trigprimitive} are aggregated into \dwords{trigcandidate}, which represent information associated with higher-level constructs derived from trigger primitives, for example ``clusters of hits''. 
\Dword{trigcandidate} information is subsequently used to inform a
module-level \dword{trigdecision}, which generates a \dword{trigcommand};
this takes the form of either a localized high energy trigger or an
extended \dword{snb} trigger, and each prompts the corresponding readout of an
event record. Details on the
exact algorithm implementation for \dword{trigprimitive}, \dword{trigcandidate}, and \dword{trigcommand} generation can be found
in~\citedocdb{11215} and~\citedocdb{14522} and references therein. 
After event-building, further data selection is carried out in the form
of down-selection of event records through a high level
filter. Details on a possible post-event-building filtering algorithm
implementation can be found in~\citedocdb{11311}.

This data selection strategy is applicable to both the \dword{pds} and
the \dword{tpc}, operating in parallel up to the module level trigger stage, where \dword{pds} and \dword{tpc} information can be combined to form a module level \dword{trigdecision}. 
Data selection design efforts have taken the approach of validating
and demonstrating a \dword{tpc}-based data selection. Nevertheless,
the data selection design by construction allows an additional
\dword{pds}-based data selection component to be accommodated within
the same design, which will augment data selection capabilities, efficiency, and robustness.

The \dword{daqdss} subsystem structure is illustrated in
Figure~\ref{fig:daq:data-selection}. The structure
reflects the three stages of \dword{daqdsn}: (1) low level trigger, which consists of
\dword{trigprimitive} generation (facilitated in upstream \dword{daq}; see
Section~\ref{sec:daq:design-upstream}) and subsequent
\dword{trigcandidate} generation; (2) module level trigger; and (3)
high level filter. Each stage is described in further detail in subsequent
sections. An additional subsystem component is the \dword{daqeti},
which serves as a common interface for the
module level trigger of each of the \dword{fd} \dwords{detmodule} and between
the module level trigger and other systems (calibration,
accelerator, and timing system) within a single
\dword{detmodule}. An additional responsibility of the
\dword{daqeti} is to send \dword{snb} triggers
to global coincidence trigger recipients like \dword{snews}
\cite{snews} after sufficient confirmation of trigger quality.

\begin{dunefigure}[Data selection strategy and hierarchy]{fig:daq:data-selection-hierarchy}{Data selection
    strategy and hierarchy.}
 \includegraphics[width=0.9\textwidth, trim=0cm 0cm 0cm 0cm]{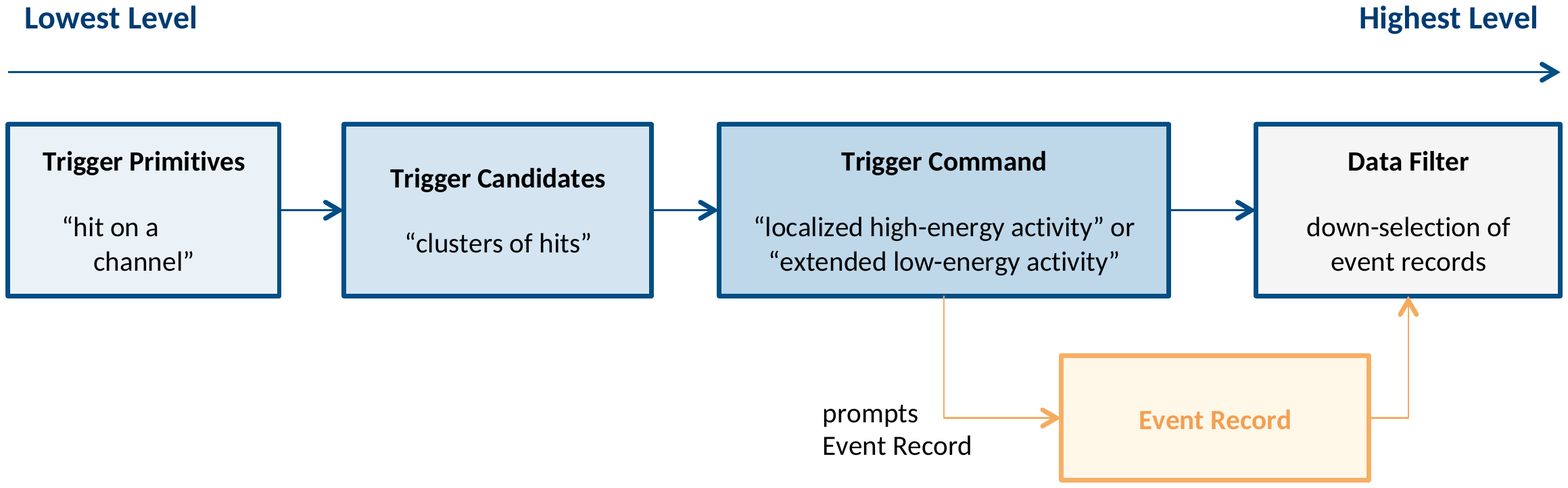}
\end{dunefigure}

\begin{dunefigure}[DUNE DAQ data selection
  subsystem]{fig:daq:data-selection}{Block diagram of the \dword{dune} \dword{daq}
    \dword{daqdss} illustrating hierarchical structure of
    subsystem design, subsystem functionality, and data flow.}
  \includegraphics[width=0.95\textwidth, trim=0 2cm 0 0]{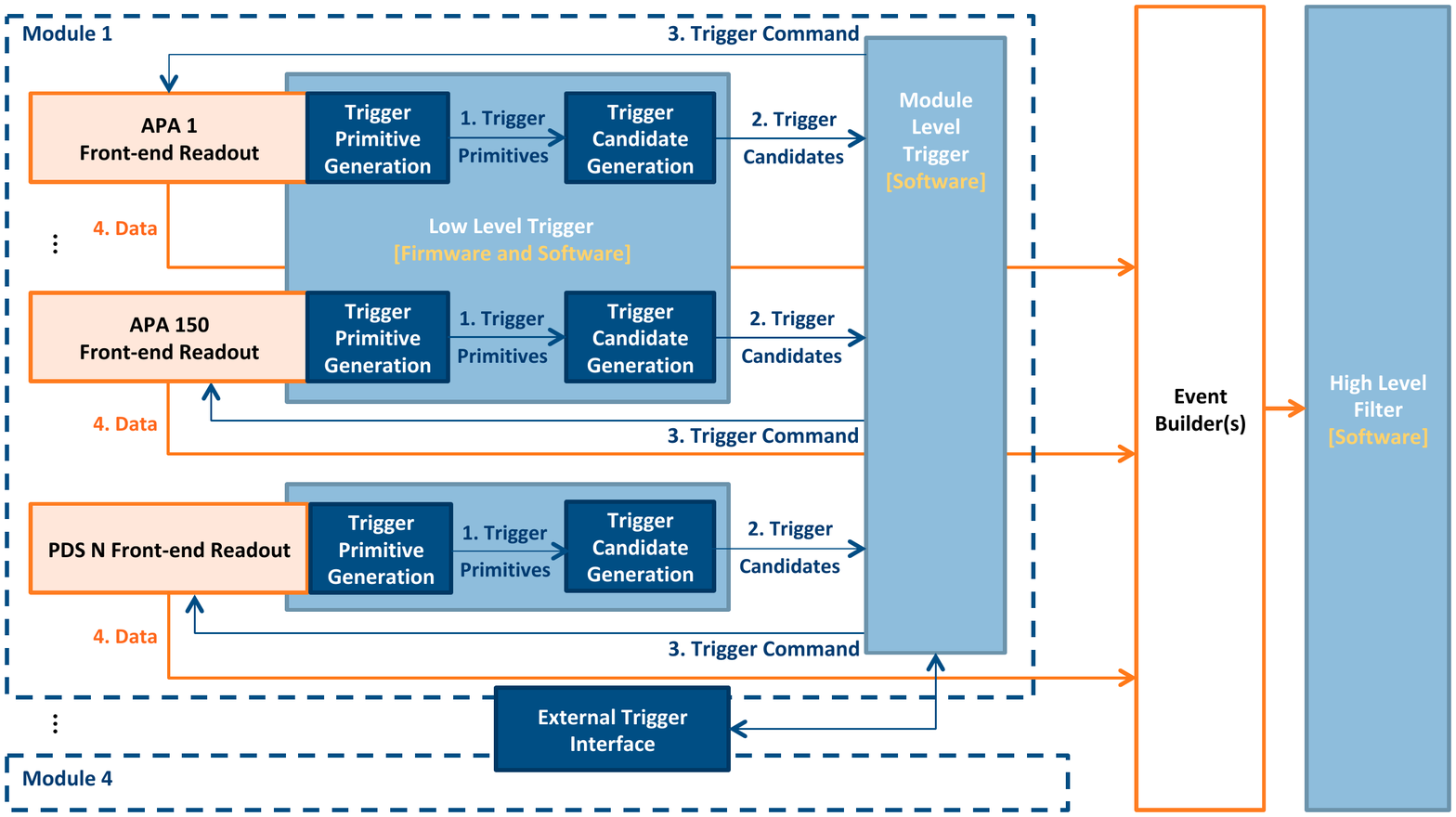}
\end{dunefigure}

The first stage of \dword{dune} \dword{fd} operations will trigger on two general
classes of physics, each handled differently at the trigger level:

\begin{description}
\item[High-energy interactions] High-energy interactions include cosmic muons, neutrino beam interactions, atmospheric neutrinos, and nucleon decays. 
  The trigger efficiency for these interactions must be $>$99\% for any given particle type (electron, muon, photon, etc.) that has a localized (confined to a limited number of neighboring channels) visible energy deposition above \SI{100}{\MeV}.
  To achieve this requirement, algorithms for creating high-energy trigger candidates  target a trigger efficiency of 50\% at \SI{10}{\MeV} visible energy, thus ensuring $>$99\% efficiency or higher at \SI{100}{\MeV}.
  This type of trigger is referred to as localized high energy trigger. 
  Pushing the high-energy threshold down could enable detection of diffuse supernova neutrinos and solar neutrinos if radiological and neutron backgrounds are low enough.

\item[Low-energy interactions] The primary physics target for
  low-energy interactions is a neutrino burst from a nearby supernova. 
  Low-energy trigger candidates (with thresholds at or below
  \SI{10}{\MeV} visible energy) are generated and are input to an
  extended low-energy trigger data selection algorithm that looks for bursts inconsistent with fluctuations in low-energy background events. 
  The time window for detecting such bursts is tuned to ensure
  nearly 100\% efficiency out to the galactic edge, and the pre-burst
  buffers are sized to handle the associated latency for detection.

\end{description}

\noindent Each trigger type prompts readout
of the entire module but over significantly different time
ranges: localized triggers prompt readout of \spreadout event records; extended
triggers prompt readout of \SI{100}{\second} event records. 

Ultimately, each \dword{trigdecision} culminates in a command sent to
the \dword{daqbes} subsystem. 
This command contains all logical detector addresses and time ranges
required, so an \dword{eb} can properly query the upstream \dword{daq}
buffers and finally collect and output the corresponding detector data
and the corresponding trigger data. The details for forming this
command are described next, while the operation of the \dword{daqbes} is
described in Section~\ref{sec:daq:design-backend}.

Viable data selection algorithms for the low level and module level triggers already exist, including
algorithms for a module level \dword{snb} trigger.  Monte Carlo simulations have demonstrated that the resulting
\dword{snb} trigger efficiency reaches $>$99\% for any \dwords{snb}
occurring within our galaxy, and efforts to extend this reach to the
Large Magellanic Cloud using refined algorithms are currently under way \cite{bib:docdb11215,bib:docdb14522}. At the same time, the
pipelines of processing required 
for \dword{daqdsn} can be executed using different firmware and software
implementations. Development is actively ongoing to demonstrate
and compare performance of different implementations. In satisfying
the philosophy and strategies of the \dword{daq} design, there is built-in
flexibility in defining whether each element of a pipeline executes on
\dword{fpga}, CPU, GPU, or, in principle, some other future hardware
architecture. A purely software implementation of data selection
(including \dword{trigprimitive} generation) is being
implemented for demonstration at \dword{protodune}; it will be then
modified to match the baseline design in which \dwords{trigprimitive} are
generated in upstream \dword{daq} \dword{fpga}.

\subsubsection{Low Level Trigger: Trigger Primitive Generation}
\label{sec:daq:design-trigger-primitives}

A \dword{trigprimitive} is defined nominally on a per-channel basis. In the case of
the \dword{spmod} \dword{tpc}, it is identified as a
collection-channel signal rising above a (configurable) noise-driven
threshold for a (configurable) minimum period of time (here called a
hit).
A \dword{trigprimitive} takes the form of an information packet that 
summarizes the above-threshold waveform information in terms of its
threshold crossing times and statistical measures of its \dword{adc} samples. 
In addition, these packets carry a flag indicating the occurrence of any
failures or other exceptional behavior during \dword{trigprimitive} processing.

Algorithms for generating \dwords{trigprimitive} are under development
\cite{bib:docdb11275}.  \Dword{trigprimitive} generation proceeds
by establishing a waveform baseline
for a given channel, subtracting this baseline from each sample, maintaining
a measure of the noise level with respect to the baseline, and searching for the waveform to cross a
threshold defined in terms of the noise level. 
The  \dword{trigprimitive} or hit is said to span the time period when the waveform is above the noise threshold.
Such algorithms ( e.g.,~\cite{bib:docdb11236}) have been validated
using both Monte Carlo simulations and 
real data from \dword{protodune}. 
\Dword{trigprimitive} generation performance is summarized in
Section~\ref{sec:daq:design-validation}.

The format and schema of \dwords{trigprimitive} are subject to further
optimization because they are further tightly coupled with the generation of
\dwords{trigcandidate}, discussed in the following subsection. Nominally,
each \dword{trigprimitive} comprises the channel address (\SI{32}{\bit}), hit
start time (\SI{64}{\bit}), the time over
threshold (\SI{16}{\bit}), the integral \dword{adc} value (\SI{32}{\bit}),
an error flag (\SI{16}{\bit}), and possibly also
the waveform peak (\SI{12}{\bit}) associated with the hit. 
Thus, \SIrange{20}{22}{\byte} provides a generous data
representation of \dword{trigprimitive} information. 
The \dword{trigprimitive} rate will be dominated by the rate of decay of naturally occurring
$^{39}$Ar, which is about \SI{10}{\mega\hertz} per module. Radioactive
isotopes of krypton may also contribute to the trigger primitive rate,
but based on results from dark matter experiments, this rate is much
smaller than the intrinsic $^{39}$Ar rate. 
This leads to a detector module \dword{trigprimitive} aggregate rate of
\SI{200}{\mega\byte/\second}.
The subsequent stage of the \dword{daqdsn} must continuously absorb and process this
rate providing \dwords{trigcandidate} as described next.

\subsubsection{Low Level Trigger: Trigger Candidate Generation}

At the \dword{trigcandidate} generation stage of the low level trigger,
\dwords{trigprimitive} from individual, contiguous fragments of the
detector 
module are cross-channel and time correlated, and further selection
criteria are applied. This may result in the
output of \dwords{trigcandidate}. 
More specifically, once activity is localized in time and channel
space, we can apply a rough energy-based threshold based on the combined
metrics carried by the cross-correlated \dwords{trigprimitive};
satisfying this criteria defines a \dword{trigcandidate}. 

A \dword{trigcandidate} packet carries information about all the \dwords{trigprimitive}  used in its formation. 
In particular, the packet provides a measure of the total activity represented
by these primitives, as well as a measure of their collective time, channel
location, and extent within the module.
These measures are used downstream by the module level trigger, 
as described more in the next section.

While the selection applied in the previous stage (\dword{trigprimitive}
generation) is driven by a measure of noise, at the \dword{trigcandidate}
generation stage, before applying any thresholds, the rate is driven by
background activity.  
In particular, $^{39}$Ar decays would provide \SI{50}{\kilo\hertz} of
\dwords{trigcandidate} per \dword{apa} face if the threshold was set very low, i.e., at
\SI{0.1}{\MeV}.
Next, activity from the $^{42}$Ar decay chain would be substantial for a
threshold below \SI{3.5}{\MeV}.
Nominally, individual candidates, or groups of candidates nearby in
detector space and time, with measures of energy higher than these two
types of decays, will be passed to the module level trigger. 

This stage of data selection is implemented \num{75} (\dword{tpc}) plus eight
(\dword{pds}) CPU servers, which receive the \dword{trigprimitive} stream from the upstream \dword{daq} and distribute \dwords{trigcandidate} to the module level trigger
stage, described next, via the \SI{10}{\Gbps} \dword{daq} network. Studies are underway to demonstrate CPU
resource use and latency, as are efforts to demonstrate online \dword{trigcandidate} generation
at \dword{protodune}. \Dword{trigcandidate} generation performance is summarized in
Section~\ref{sec:daq:design-validation}.

\subsubsection{Module Level Trigger}
\label{sec:daq:mlt}

Data selection is further facilitated as \dwords{trigcandidate} are consumed
by the module level trigger in order to form the ultimate \dword{trigdecision} that prompts the readout of data records from buffers kept by the upstream \dword{daq}. 
The physical (channel and time) location, extent, and energy measure of the
candidates are used at this stage to categorize the activity in terms
of a localized high energy trigger or an extended low energy trigger. 
Specifically, a suitable number of isolated, low energy candidates found in coincidence
over the integration period of up to \SI{10}{\second} across the full \dword{detmodule}
indicate the latter; individual high energy candidates, found
otherwise, indicate the former.

When a particular condition in a category is satisfied, the \dword{trigdecision} is made and a \dword{trigcommand} is formed. 
The \dword{trigcommand} packet includes information about the candidates (and primitives)
that were used to form it. 
The decision also provides direction as to what set of detector subcomponents
should be read out and over what time period (localized or extended as described above). 
The module level trigger publishes a stream of \dwords{trigcommand} and the primary subscriber should be the \dword{daqdfo} of the \dword{daqbes} as described in Section~\ref{sec:daq:design-backend}.

The module level trigger is implemented in \bigo{1} CPU server (with 100\%
redundancy), which
receives the \dword{trigcandidate} stream from the low level trigger stage
of the data selection and distributes \dwords{trigcommand} to the
\dword{daqbes} via the \SI{10}{\Gbps} \dword{daq} network. Studies are
underway to demonstrate CPU resource use and latency, as are
efforts to demonstrate online \dword{trigcommand} generation at \dword{protodune}.
\Dword{trigcommand} generation performance is summarized in
Section~\ref{sec:daq:design-validation}.

\subsubsection{External Trigger Interface}

The \dword{daqeti} provides a loose coupling between the
\dwords{mlt}, sources of external information such as beam spill times
and information to or from components of \dword{dune} \dword{fd}
calibration systems. As an interface between \dwords{mlt}, the \dword{daqeti} receives and distributes information about module-level \dword{snb} \dwords{trigcommand}.
This allows any detector module, which alone may not have satisfied a \dword{snb} trigger requirement, to nonetheless perform an \dword{snb} readout.
The \dword{daqeti} is also responsible for forming a coincidence between module-level \dword{snb} \dwords{trigcommand} and publishing the results, e.g.,~for consumption by the \dword{snews}.

The \dword{daqeti} also receives information about beam spill times from the accelerator.
These times can drive a model of the beam timeline to predict when
beam spills, and consequently beam-related interactions, should occur. 
These predictions can then be sent to the \dwords{mlt}, so they
can either alter \dword{trigdecision} criteria or merely include the
information in contemporaneous \dwords{trigdecision}. The beam time information can also be distributed to components of the calibration system to avoid producing activity in the detector that may interfere with activity from beam neutrinos.

The external trigger interface is implemented in \bigo{1} CPU server (with 100\%
redundancy), with \SI{10}{\Gbps} networking and interfacing hardware
components (to \dword{wr} and \dword{dune} timing system) for timing and external trigger signal I/O.

\subsubsection{High Level Filter}
\label{sec:daq:design-data-reduction}

The last processing stage in the \dword{daqdss} is the
high level filter, which resides in the \dword{daqbes}, and physically,
on the surface at \dword{surf}.
The high level filter acts on triggered, read out, and aggregated data. 
It therefore serves primarily to down-select and thus
limit the total triggered data rate to offline storage, thereby keeping 
efficiency high in collecting information on activities of interest,
while maintaining low selection and content bias, reducing the output data
rate. It may do so using 
further filtering, lossy data reduction, and/or further event
classification.
Because it benefits from operating on relatively low-rate data, it can accommodate a higher level of
sophistication in algorithms for \dword{daqdsn} decisions.

More specifically, the high level filter may further reduce the rate of data output to offline storage by
applying refined selection criteria that may otherwise be impossible
to apply to the pre-trigger data stream.  For example, instrumentally-generated signals (e.g.,~correlated noise)
may produce \dwords{trigcandidate} that cannot be rejected by the module
level trigger and, if left unmitigated, may lead to an undesirably high
output data rate. 
Post processing the triggered data may reduce this unwanted
contamination.
Furthermore, the high level filter can also reduce the triggered data set by further identifying
and localizing interesting activity. A likely candidate hardware
implementation of this level of \dword{daqdsn} is a GPU-based system
residing on surface at \dword{surf}.

To fully understand how much and what type of data reduction may be
beneficial, simulation studies are ongoing \cite{bib:docdb11311},
summarized in Section~\ref{sec:daq:design-validation}, and will
must be
validated with initial data analysis after the
first \dword{dune} \dword{fd} operation. Development efforts are also ongoing to determine the scale of 
processing required by the \dword{fd}.

\subsection{Back-end DAQ}
\label{sec:daq:design-backend}

The  \dword{daqbes} moves data of interest identified by the data-selection system from the readout \dword{daq} buffers, 
serving them to the high level filter and storing the filtered data into the output buffer. From there, 
data will be transferred to permanent storage off-site.

The \dword{daqbes} system accepts \dwords{trigcommand} produced by the \dword{daqdss} as described
in Section~\ref{sec:daq:design-data-selection}.  It queries the upstream \dword{daq} buffers and
accepts returned data as described in Section~\ref{sec:daq:design-upstream}. Finally, it records
\dwords{trigcommand} and the corresponding detector data to the output storage buffer.

The principal components of the \dword{daqbes} are the \dfirst{daqdfo}, \dfirst{eb} and the output
storage buffer (OB) in Figure~\ref{fig:daq:layout}.

\subsubsection{Data Flow Orchestration}

The \dword{eb} stage is implemented as a pool of redundant \dword{eb} processes to maximize the system tolerance to faults and to handle the readout of long SNB events in parallel to nominal readout requests. This asynchronous, parallel readout will be coordinated by a \dword{daqdfo}.  Its operation is illustrated in Figure~\ref{fig:daq:backend} and is discussed here:

\begin{dunefigure}[DUNE DAQ back-end operation]{fig:daq:backend}{Illustration of \dword{dune} \dword{daqbes} operation.}
  \includegraphics[width=\textwidth]{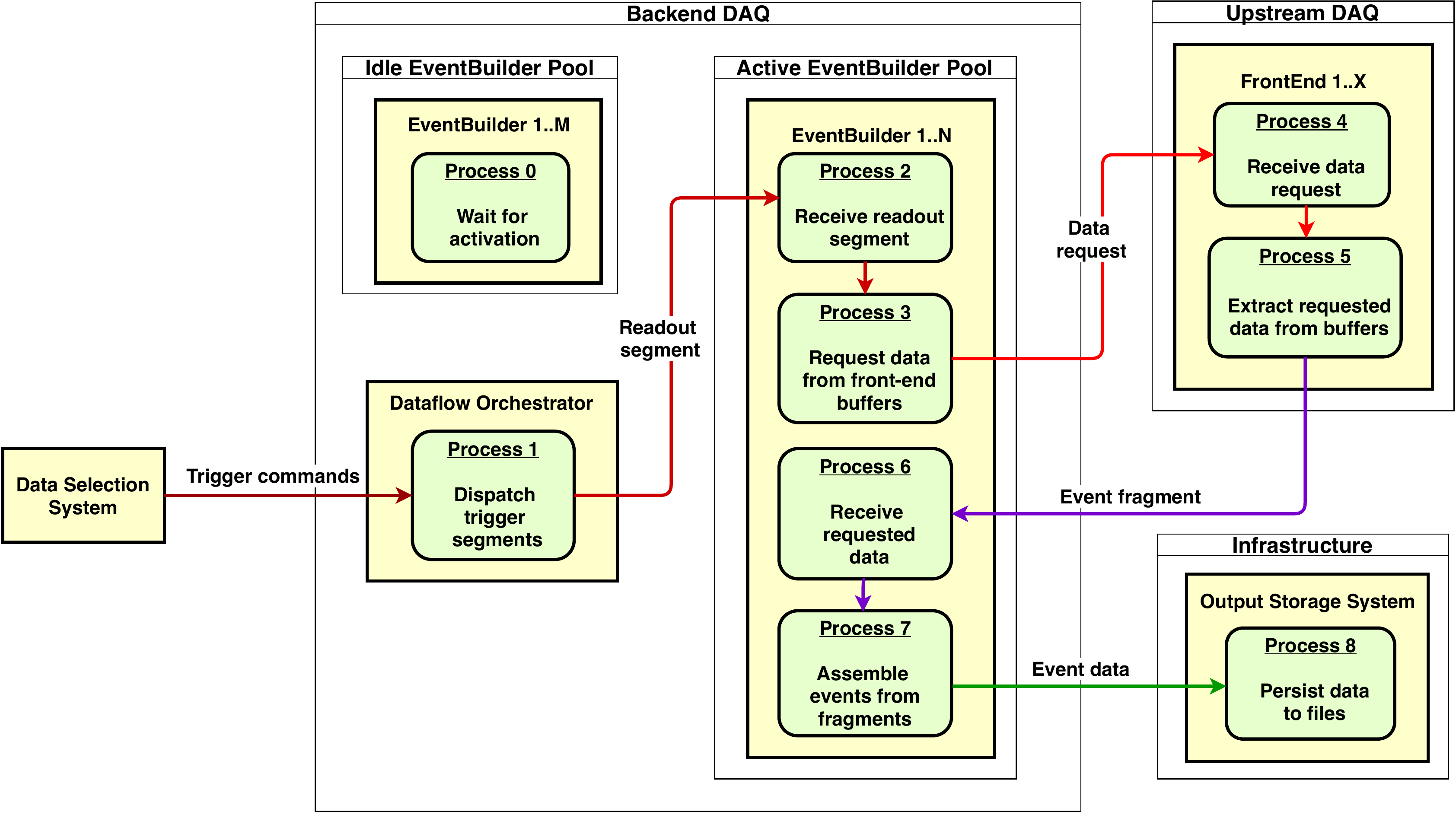}
\end{dunefigure}

\begin{itemize}
\item \dword{daqdfo} accepts a stream of \dwords{trigcommand} and dispatches each to an available \dword{eb} process as described in Section~\ref{sec:daq:design-event-builder} for execution.
\item In atypical situations in which there are insufficient event builder resources available to handle the rate of triggers produced by the data selection subsystem, the DFO will alert the DS subsystem that the rate of triggers needs to be reduced.  When such reductions are requested, the \dword{daqdss} will update the calculation of the module-level \dword{daq} livetime appropriately.
\item The DFO will provide relevant data flow status and statistics information to the monitoring processes that are described in Section~\ref{sec:daq:design:ccm:monitoring}. Given its central role and knowledge of the state of available event builder buffers, it will be able to provide important information about the health and performance of the system.
\end{itemize}

\subsubsection{Event Builder}
\label{sec:daq:design-event-builder}

The \dword{daqbes} will provide the instances of the \dfirst{eb} most likely as
\dword{artdaq}~\cite{artdaq} components of the same name. 
As described above, each EB instance will:

\begin{itemize}
  \item Receive a readout segment for execution. Execution entails interpreting the \dword{trigcommand} segment and querying the appropriate upstream \dword{daq} units to request data from the period of time. 
  \item Requests and their replies may be sent synchronously, and replies are expected even if data has already been purged from the upstream \dword{daq} units. (In that case, and empty fragment will be generated with appropriate error flags set).
  \item The received data then processed and aggregated, is finally saved to one or more files on the output storage system before it is transferred offline.
\end{itemize}

As part of this, the EB subsystem will provide book-keeping functionality for the raw data.  This will include the documenting of simple mappings, such as which trigger is stored in which raw data file, as well as more sophisticated quality checks. For example, it will know which time windows and geographic regions of the detector are requested for each trigger, and in the unlikely event that some fraction of the requested detector data can not be stored in the event record, it will document that mismatch.

\subsubsection{Output Buffer}

The output buffer system is composed by the physical hardware resource
to host the incoming data and by the software services handling the
final processing stages through the High Level Filter and the transfer off-site to permanent storage.

It has two primary purposes.  First, it decouples the production of data from filtering and the
transfer of filtered data offline. It provides the elasticity needed by the \dword{daq} to deal with
perturbations in the flow of data, therefore minimizing the impact of temporary loss in filtering
performance due to hardware or software issues. Second, it provides local storage sufficient for
uninterrupted \dword{daq} operation in the unlikely event that the network connection between the
\dword{fd} and \fnal is lost.  A capacity of at least a few \si{\peta\byte} is envisioned,
sufficient to buffer the nominal output of the entire \dword{fd} for about one week even in the case of SNB events. Based on prior experience of the consortium with unusual losses of  connectivity at
other far detector experiment sites, this is a conservative storage capacity value.

The output buffer system will provide the relevant data flow status and statistics information to the monitoring processes that are described in Section~\ref{sec:daq:design:ccm:monitoring}. The knowledge of the health and performance of the of the buffer system will enable the monitoring system to promptly identify and address developing faults before they can have an impact on data taking.

\subsubsection{Data Network}
Upstream \dword{daq}, \dword{daqdss} and \dword{daqbes} \dword{daq} are interconnected by a \SI{10/100}{\Gbps} ethernet network for data exchange.
In particular the upstream \dword{daq} and \dword{daqdss} servers are connected through redundant \SI{10}{\Gbps} links to top-of-rack switches with \SI{10}{\Gbps} uplinks.
The \dword{eb} and Output buffer hardware will support \SI{100}{\Gbps} directly.
The \dword{daq} data network is connected to the \fnal network via a WAN interface.

\subsubsection{Data Model}
\label{sec:daq:design-data-model}

The data model for the DUNE far detector describes the format and characteristics of the triggered data at each stage in the analysis chain, the grouping of the data into logical units such as runs, and the characteristics of ancillary data such as detector configuration parameters and calibration results.

The requirements that these place on the \dword{daq} are primarily in the areas of flexibility and traceability.  The \dword{daq} will have the flexibility to handle the readout of triggers that have a time window that is on the order of a single TPC drift time (for example a trigger associated with a beam gate window), triggers that have a time window of many seconds (such as for a supernova burst trigger), and windows between those two extremes (for detector and electronics studies).  In the area of traceability, the \dword{daq} system will provide the necessary level of detail regarding the conditions that triggered each event, the expected and actual regions of the detector that contributed raw data to each event, the conditions of the detector and electronics during data taking, the version and configuration of the software components used in the \dword{daq} chain, etc.

\subsection{Control, Configuration, and Monitoring}
\label{sec:daq:design-run-control}

The \dfirst{daqccm}, illustrated in Figure~\ref{fig:daq-ccm-subsys}, consist of the software
subsystems to control, configure, and monitor the \dword{daq} system, as well as the detector components
participating to data taking. It provides a central access point for the highly distributed \dword{daq}
components, allowing them to be treated and managed as a single, coherent system, though their
corresponding subsystem interfaces. It is responsible for error handling and recovery, which is
achieved by designing a robust and autonomous fault-tolerant control system. The main goal is to
maximize system up-time, data-taking efficiency and data quality when the system faces programmatic
(i.e. calibrations) and unforeseen (hardware failures or software faults) change of data-taking
conditions. The \dfirst{daqccm} provides an access point, which is delegating user's actions to the
corresponding interfaces. The detector components and infrastructure elements access the \dfirst{daqccm}
subsystems through their provided interfaces. 

\begin{dunefigure}[DAQ CCM subsystem interaction]{fig:daq-ccm-subsys}{Main interaction among the three \dword{daqccm} subsystems.}
 \includegraphics[width=0.9\textwidth]{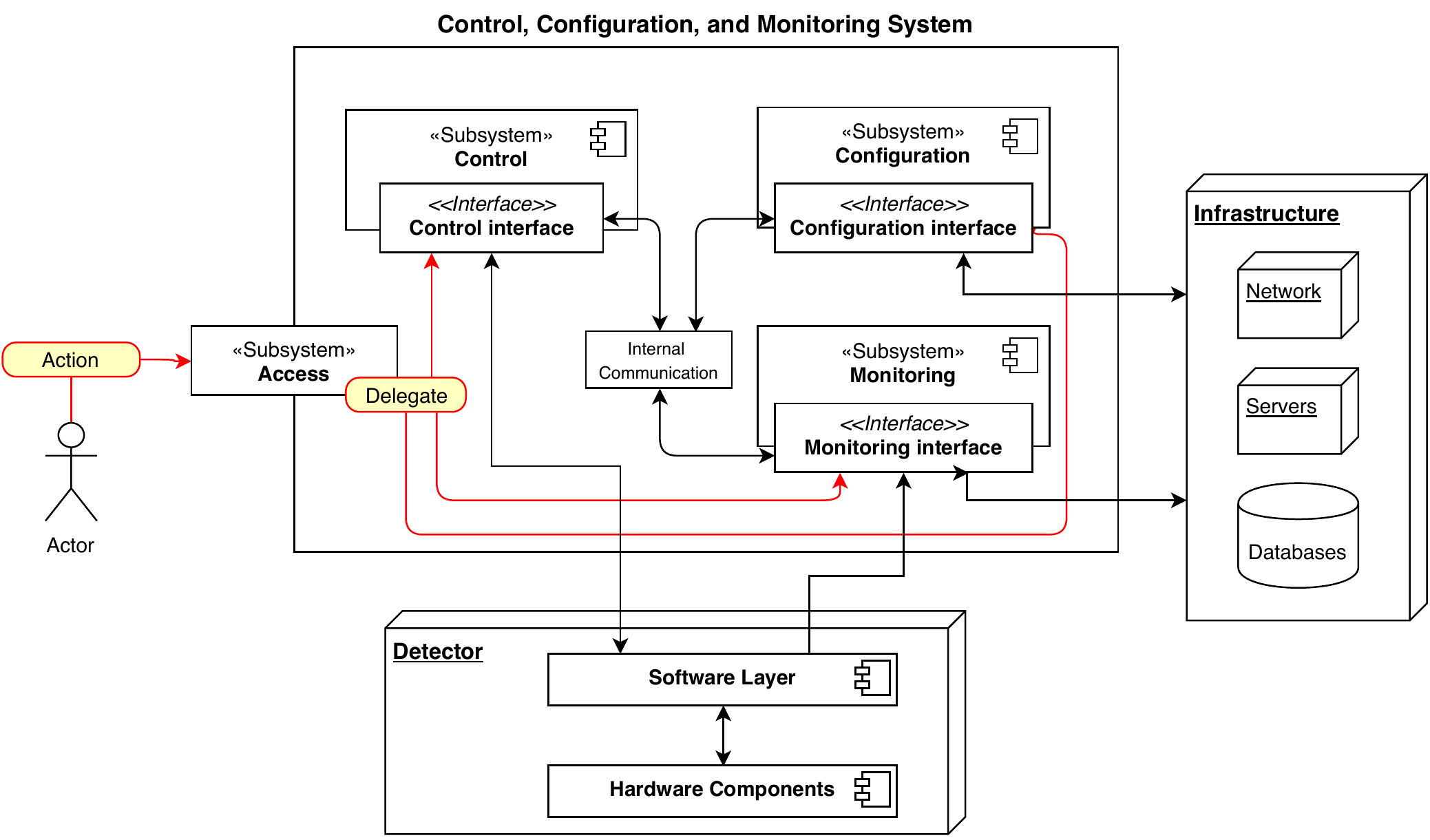}
\end{dunefigure}

The following sections describe each \dfirst{daqccm} subsystem, covering internal functions and dependencies between each other.

\subsubsection{Access}
\label{sec:daq:design:ccm:access}
Actions are defined as any kind of human interaction with the \dfirst{daqccm}. The access subsystem is responsible for the action delegation to internal function calls and procedures. Its implementation is driven by the control, configuration and monitoring interface specifications, and protects the direct access to detector and infrastructural resources. It also controls authentication and authorization, which locks different functionalities to certain actor groups and subsystems. As an example, only the detector experts can modify front-end configuration through the configuration interfaces, or only an expert user can exclude an APA's readout from data taking.

\subsubsection{Control}
\label{sec:daq:design:ccm:control}

The control subsystem consists of several components and utilities, and also has additional subsystems to carry out dedicated roles. It enforces the implementation of required interfaces and actively manages \dword{daq} process lifetimes. It operates in a distributed, fault-tolerant manner due to protocols that will drive the FSM for state sharing. 

\begin{dunefigure}[DAQ control subsystem roles and services]{fig:daq-ccm-control}{Roles and services that compose the \dword{daq} control subsystem.}
  \includegraphics[width=0.8\textwidth]{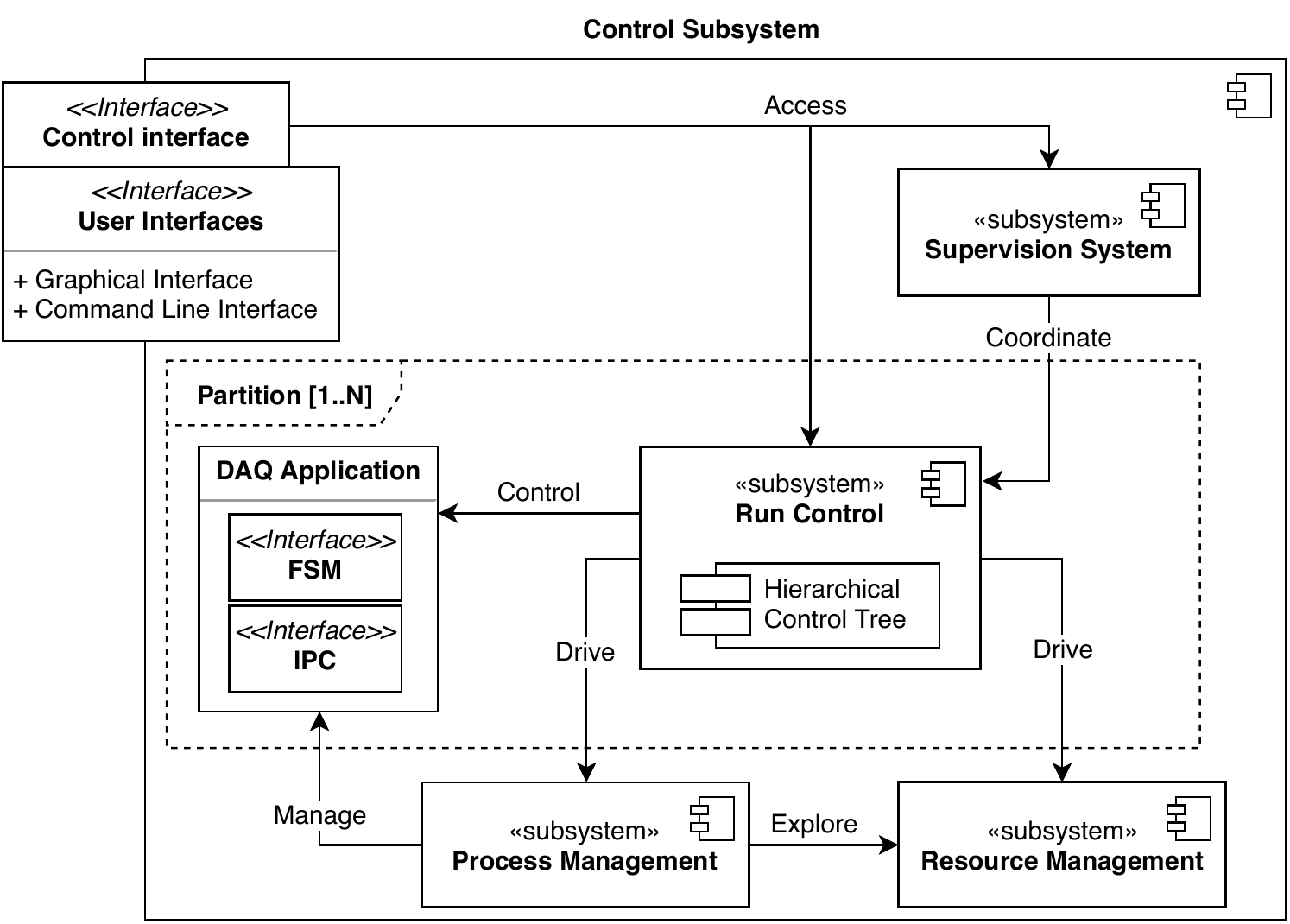}
\end{dunefigure}

It contains the following core components:
\begin{itemize}
\item Supervision System - It is responsible for manual and automated control and supervision of \dword{daq} components at any given time. In autonomous mode, the system makes attempts for fault-recovery, failover to backup instances of subsystems, and isolation of problematic regions of the control tree. This is carried out by a hierarchical rule-based planning or fuzzy logic system.
\item \dword{daq} Application - The CCM provides interfaces in order to communicate with processes of the \dword{daq}, and the ability to control and communication with the CCM. The Inter Process Communication (IPC) supports a mechanism to interact with all actors participating to data taking. The Finite State Machine (FSM) enforces the possible states and transitions that are specific to the experiment's components, and also describes them in a uniform way.
\item Run Control - This part of the control subsystem coherently steers the data taking operations. It interacts with all actors participating to data taking in a given partition. It consists of a hierarchical control tree, which can subdivide the \dword{daq} components into separated regions that may be acted upon independently.
\item Resource Management - It provides a global scope of available resources for the \dword{daq} components. This includes the mapping between the detector front-end readout units, processes, servers where they are spawned and required resources for the processes.
\item Process Management - It is responsible for managing process lifetime.
\end{itemize}

\subsubsection{Configuration}
\label{sec:daq:design:ccm:configuration}

The configuration subsystem provides several key elements for the configuration management of \dword{daq} components and detector front-end electronics. It provides a description of system configurations, the ability to define and modify configurations, and graphical user interfaces for the human user to access the data. Data access libraries will hide the technology used for the databases implementation. The subsystem is also responsible for the serialization, persistency, and bookkeeping of configurations. 

\begin{dunefigure}[DAQ CCM subsystem interaction]{fig:daq-ccm-config}{Main components of the \dword{daqccm} configuration subsystem.}
  \includegraphics[width=0.8\textwidth]{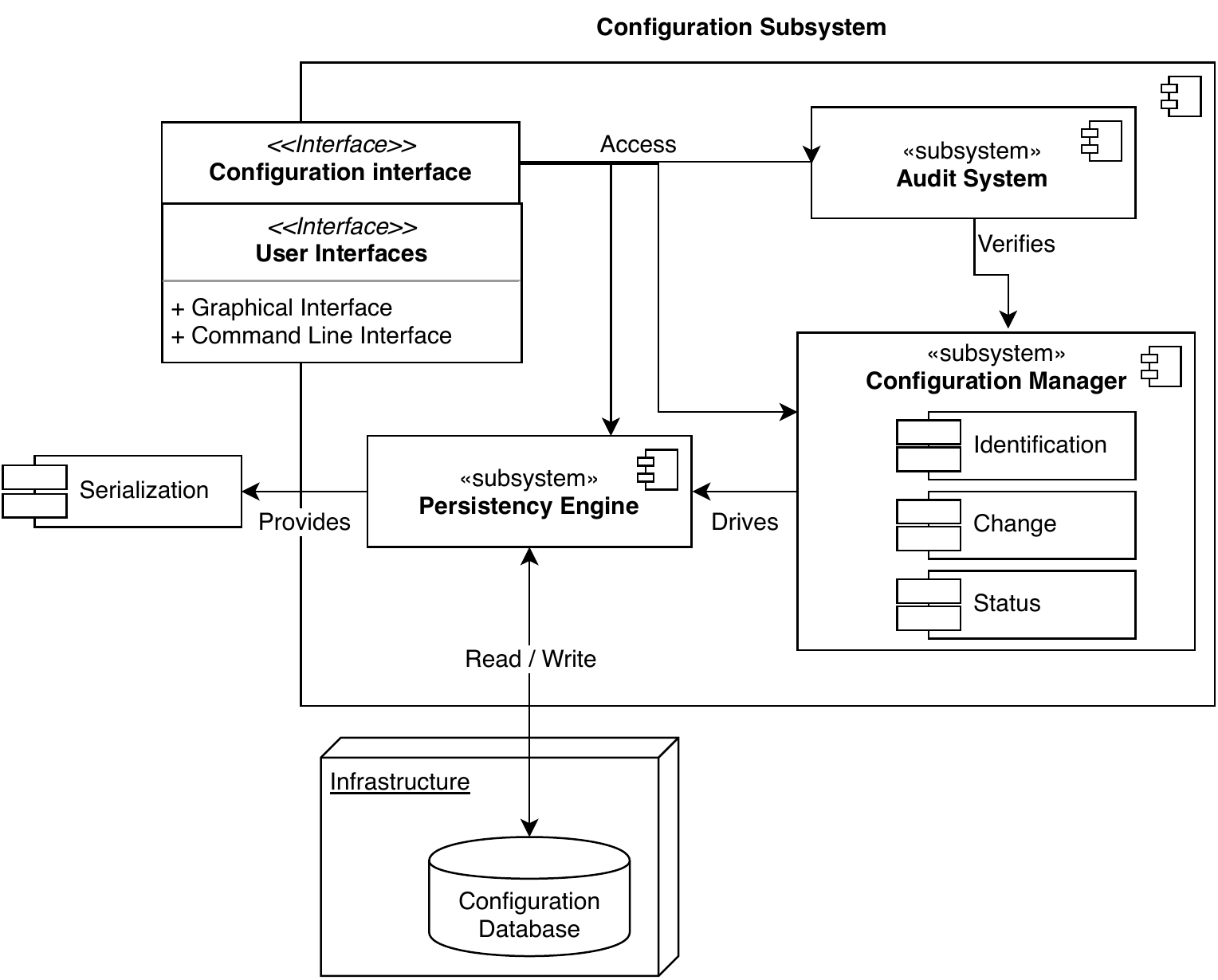}
\end{dunefigure}

The main components of the configuration subsystem are the following:
\begin{itemize}
\item Configuration Manager - It consists of the three main components of configuration management systems. The Identification Engine is a set of functionalities that are responsible for the definition of \dword{daq} components and their corresponding configuration specification. The Change Manager is responsible for providing control over altering the configuration specifications of components. The Status Engine is providing status and information about configuration specifications of individual, or set of \dword{daq} elements.
\item Audit System - This important subsystem is supporting the experts and decision making systems to verify the consistency of configuration specifications against the \dword{daq} and detector components. It provides results on mis-configurations and potential problems on configuration alignment and dependencies between components.
\item Persistency Engine - This component provides a single and uniform serialization module, which is strictly followed by every \dword{daq} component. Also responsible for configuration schema evolution and communication with the configuration database. The storage engine privileges will be only read and write operations, not allowing updates and removal of configurations. It also provides a redundant session layer for high-availability and load distribution.
\end{itemize}

The configuration system will mainly consist of standard configuration management components, with a high emphasis on the audit system, in order to verify that the global configuration of the \dword{daqccm} complies with the detector, physics and operational requirements.

\subsubsection{Monitoring}
\label{sec:daq:design:ccm:monitoring}

\begin{dunefigure}[DAQ monitoring subsystem roles and services]{fig:daq-ccm-monitoring}{Roles and services that compose the \dword{daq} monitoring subsystem.}
  \includegraphics[width=0.8\textwidth]{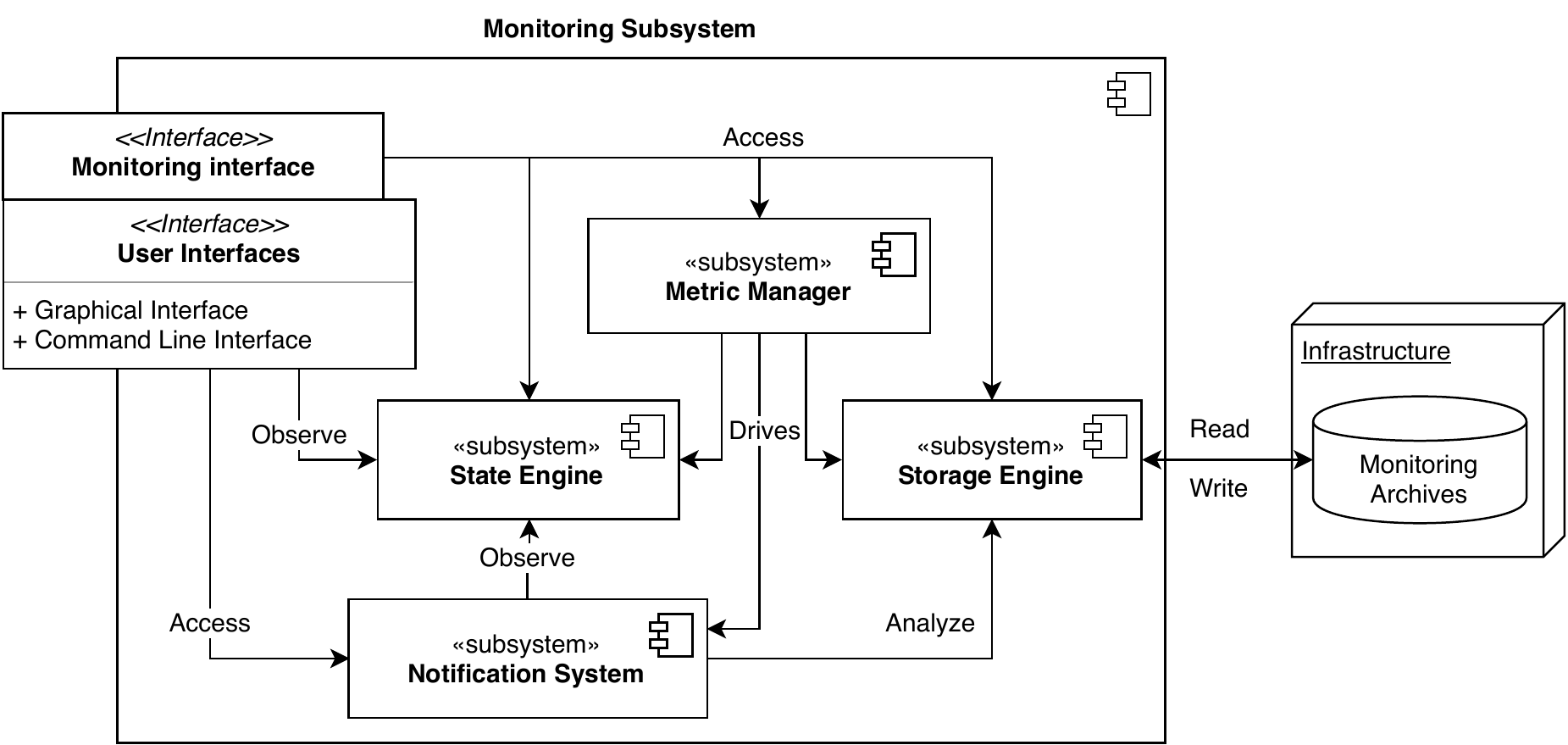}
\end{dunefigure}

Highly-scalable and efficient operational monitoring is essential during data-taking periods. Any malfunctioning component of the experiment must be identified and reported as soon as possible. Therefore, the aim of the monitoring subsystem is probing and notifying the status of \dword{daqccm} components, services, and resources. There is also a requirement of \dword{daqccm} infrastructure monitoring and log aggregation. The types of monitoring information vary greatly depending on operational complexity, which require flexibility from the monitoring infrastructure for seamless additions, modifications and aggregated view on service degradation.
It consists of the following main components, also shown in Figure~\ref{fig:daq-ccm-monitoring}:

\begin{itemize}
  \item Metric Manager - It is responsible for the registration of metric data streams and corresponding aggregator functions. This central element is the collection of features and services that provides support for configurable operational monitoring of \dword{daqccm} services. \dword{daqccm} components and services that are registered via the Metric Manager, are reporting monitoring data to the State and the Storage Engines in a publish-subscribe fashion. 
  \item State Engine - This engine is responsible for providing the global state of the system at the current time. It subscribes to a set of registered metrics in the Metric Manager, and records the actual global state of the set for decision making systems (supervision, notification, and visualization).
  \item Storage Engine - Metrics may have different persistency requirements, for which the engine is responsible for archiving the data with the settings of interval, smoothing, etc. It also provides an implementation for the most common communication protocols for the database back ends (SQL, REST, etc.).
  \item Notification System - It is a rule-based system of scheduled and aimed notifications that occur in the case of state combinations. It defines soft and hard states of events and grace periods of alarms.
  \item User Interfaces - Provides graphical and command line user interfaces for monitoring configuration management and visualization of metric data.
\end{itemize}

Monitoring being a key requirement in the industry for computer clusters and their applications, the proposed solution is the adaptation of mature, robust, and open-source third party tools, with the extension of DUNE \dword{daqccm} specific interface implementations and configurations of these monitoring systems.

\subsubsection{Inter-Process Communication}
\label{sec:daq:design-ipc}

The DUNE FD \dword{daq} is an asynchronous, parallel distributed data processing system. 
It is composed of many independent processes which ingest and produce messages. 
The mechanisms of such message passing are generally called \dword{ipc}. 
Referring to Figure~\ref{fig:daq:layout}, \dword{ipc} is used for both in-band detector data flow between upstream \dword{daq} and back-end \dwords{eb} and for out-of-band messages as part of \dword{daqccm}.  The \dword{ipc} used by the \dword{daqdss} spans both descriptions as it passes derivations of a subset of detector data (trigger primitives, candidates) and culminates in a source of out-of-band message (\dwords{trigcommand}) to direct the readout by \dword{eb} and other components of detector data that is held in the upstream \dword{daq} buffers.

The ZeroMQ~\cite{zeromq} smart socket library is 
the basis of a system being developed and evaluated for parts of both in-band and out-of-band \dword{ipc}. 
As part of the \dword{daqccm}, this includes the issuing of control and reconfiguration commands to and receiving of monitoring messages from essentially all \dword{daq} components. 
As part of \dword{daqdss}, this includes the transfer of \dword{trigprimitive}, candidate and command messages. 
In the upstream \dword{daq} this includes the \dfirst{daqubi} that  provides access to the upstream \dword{daq} primary buffers for queries by \dword{eb} and other components. 
\dword{ipc}  must be implemented broadly across many \dword{daq} systems and ZeroMQ allows their problems to be solved in common, shared software.  As \dword{daqccm} has the most complex \dword{ipc}  needs, this work is organizationally considered part of this system.

As described in~\ref{sec:daq:design-backend}, \dword{artdaq}~\cite{artdaq} utilizes \dword{ipc}  between its back-end components. 
It has been well tested with \dword{protodune} and other experiments. 
\dword{artdaq} may be used for some portions of the \dword{ipc}  described above. 
For example, if the \dword{daqubi} is implemented as an \dword{artdaq} Board Reader it would necessarily use \dword{artdaq} \dword{ipc} . 
This would limit the types of clients that could query for data in the buffers to be \dword{artdaq} modules. 
Understanding how to optimally select an \dword{ipc}  for such parts of the \dword{daq} connection graph is an area of ongoing R\&D effort.

\subsubsection{Hardware}
\label{sec:daq:design:ccm:hardware}

The \dword{daqccm} software suite will run on approximately 15 servers interconnected by a \SI{1}{\Gbps} ethernet network to upstream \dword{daq}, \dword{daqdss}, \dword{daqbes} as well as detector and calibration daq interface elements. While this network has a lower throughput compared to the data network, it has many more endpoints O(2000).

\subsection{Data Quality Monitoring}
\label{sec:daq:design-data-quality}

While the \dword{daqccm} contains an element of monitoring (Section~\ref{sec:daq:design:ccm:monitoring}), here \dfirst{dqm} refers to a subsystem that quickly analyzes the data in order to determine the general quality of the detector and \dword{daq} operation.
This is in order to allow operators to promptly detect and respond to any unexpected changes and assure high exposure times for later physics analyses. 
A \dword{daq} \dfirst{dqm} 
will be developed (including necessary infrastructure, visualization,
and algorithms), which will process a subset of detector data in order
to provide prompt feedback to the detector operators. 
This system will be designed to allow it to evolve as the detector and its data is understood during commissioning and early operation and to cope with any evolution of detector conditions.

\subsection{Timing and Synchronization}
\label{sec:daq:design-timing}

The \dfirst{daqtss} provides synchronous time services to the \dword{daq} and the detector electronics.
All components of the \dword{fd} use clocks derived from a single
\dfirst{gps} disciplined source, and all module components are
synchronized to a common \SI{62.5}{MHz} clock.
This rate is chosen in order for this common clock to satisfy the requirements of the detector electronics of both the single-phase and the dual-phase far detector modules.
To make full use of the information from the \dword{pds}, the common clock must be aligned within a single detector 
module with an accuracy of \bigo{\SI{10}{\nano\second}}. 
For a common trigger for a \dword{snb} between modules, the timing must have an accuracy of order \SI{1}{\milli\second}.
However, a tighter constraint is the need to calibrate the common clock to universal time derived from \dword{gps} so the \dword{daqdsn} algorithm can be adjusted inside an accelerator spill, which again requires an absolute accuracy of order \SI{1}{\micro\second}. The design of the timing system allows for an order-of-magnitude better synchronization precision than these requirements, allowing a substantial margin of safety and the possibility for future upgrades to front-end electronics.

The \dword{dune} \dword{fd} uses a improved version of the \dword{protodune} timing
system, where a design principle is to transmit synchronization messages over
a serial data stream with the clock embedded in the data. The format
is described in \citedocdb{1651}. The timing system design is
described in detail in \citedocdb{11233}.

Central to the timing system are four types of signals:
\begin{itemize}
\item a \SI{10}{\mega\hertz} reference used to discipline a stable master clock,
\item a \dfirst{pps} from the GPS,
\item a \dword{ptproto} signal providing an absolute time for each \dword{pps}, and
\item an \dfirst{irig} time code signal
  used to set the timing system \SI{64}{\bit} time stamp.
\end{itemize}

The timing system synchronization codes are distributed to the \dword{daq} readout components in the \dfirst{cuc} and the readout components on the cryostat via single mode fibers and passive splitters/combiners.
All custom electronic components of the timing system are contained in two \dword{utca} shelves; at any time, one is active while the other serves as a hot spare.
The \SI{10}{MHz} reference clock and the \dword{pps} signal are received through a single-width \dword{amc} at the center of the \dword{utca} shelf.
This master timing \dword{amc} is a custom board and produces the timing system signals, encoding them onto a serial data stream.
This serial data stream is distributed over a backplane to a number of fanout \dwords{amc}.
The fanout \dword{amc} is an off-the-shelf board with two custom \dwords{fmc}.
Each \dword{fmc} has four \dword{sfp} cages where fibers connect the timing system to each detector component (e.g., \dword{apa}) or where direct attach cables connect to other systems in the \dword{cuc}.

To provide redundancy, two independent GPS systems are used,
one with an antenna at the surface at the Ross shaft, and the other
with an antenna at the surface at the Yates shaft. Signals from either
GPS are fed through single-mode optical fibers to the \dword{cuc}, where
either GPS signal can act as a hot spare while the other is active. 
Differential delays between these two paths are resolved by a second pair of 
fibers, one running back from the timing system to each antenna, 
allowing closed-loop delay estimation.

\section{Design Validation and Development Plans}
\label{sec:daq:validation}

The following strategy is being followed in order to validate and
develop the \dword{dune} \dword{fd} \dword{daq}:
\begin{itemize}
\item use \dword{protodune} and other test stands as a basis for prototyping, development and validation of the \dword{daq} design; 
\item use Monte Carlo simulations and emulations in order to augment actual hardware
demonstrations and validate triggering schemes in the FD environment; and 
\item benefit from experience in LArTPC data processing by other experiments.
\end{itemize}

The design and implementation of the \dword{daq} is being carried out using an iterative prototyping model, which is well suited for a system that largely relies on commercial off-the-shelf hardware components, on communication and information technologies that are rapidly evolving, and that mainly requires software and firmware development effort.
The advantage of the prototyping model is also that it facilitates the identification of and collaboration among experts from a large number of institutions, through focussed efforts to achieve the short-term objectives established through each prototyping iteration.

Once the identification of applicable technologies are completed and the overall system requirements are refined, 
the project will switch to an iterative incremental model, ensuring that, step-by-step,
functional and performance requirements will be met by each of the sub-components individually, and by the \dword{daq} system globally.
The overall schedule, summarized in Section~\ref{sec:daq:schedule},
reflects the different development and production time scales that are
envisioned for the various \dword{daq} components.

While \dword{dune} data processing challenges are unique in both form
and scale, other ongoing or planned near-term \dword{lartpc}
experiments, including \dword{microboone}, \dword{sbnd}~\cite{Karagiorgi:2019qzp}, and \dword{icarus}, are exercising
a number of relevant data processing and data reduction techniques, and already providing valuable
inputs to the \dword{dune} \dword{fd} \dword{daq} design. For example,
MicroBooNE has demonstrated successful implementations of lossless Huffman
compression and of continuous readout via use of lossy compression
(dynamic and fixed-baseline zero-suppression)
\cite{Crespo-Anadon:2019lht}; 
this has provided guidance and 
confidence on anticipated achievable data reduction factors and data compression
impact on physics performance.

The following subsections summarize past, ongoing, and planned
development and validation studies and identify how anticipated outcomes
will be used to finalize the \dword{daq} design.

\subsection{ProtoDUNE Test Beam}

\label{sec:daq:protodune}

The \dword{fd} \dword{daq} consortium constructed and operated the \dword{daq} system~\cite{Sipos:2018auh} 
for \dword{protodune}, which included two parallel \dword{daq} readout architectures, one based on \dword{felix}, 
developed by ATLAS \cite{atlas-felix}, and the other on \dword{rce}, developed at SLAC. \dword{daq} design and construction for
\dword{protodune} began in Q3 of 2016, and the system became operational at the start of the beam data run in Q4 of 2018. 
The detector is continuing to run at the time of writing, recording cosmic ray activity, and providing further input for 
\dword{daq} development toward \dword{dune}. 

Figure~\ref{fig:daq-protodune} depicts the \dword{protodune} \dword{daq} system. 
The \dword{daq} is split between the \dword{felix} and \dfirst{rce}
\cite{bib:docdb1881} implementations. The two architectures share the
same back end and timing and trigger systems.  
Neither of these tested architectures explicitly represents the baseline design for the DUNE \dword{fd}. Instead, each roughly maps onto one of two data processing approaches: one in which the data is processed exclusively in custom-designed \dword{fpga} firmware and carrier board, and the other in which the data is processed primarily with custom software run in commodity servers. The baseline design presented here merges elements of the two approaches. Specifically, it uses \dword{felix} as the hardware platform for data receiving and handling, and a co-processor \dword{felix} daughter card (analogous to the \dword{rce} platform used at \dword{protodune}) to provide additional, dedicated data processing resources.

\begin{dunefigure}[ProtoDUNE DAQ system]{fig:daq-protodune}{The \dword{protodune} \dword{daq} system.}
 \includegraphics[width=0.8\textwidth]{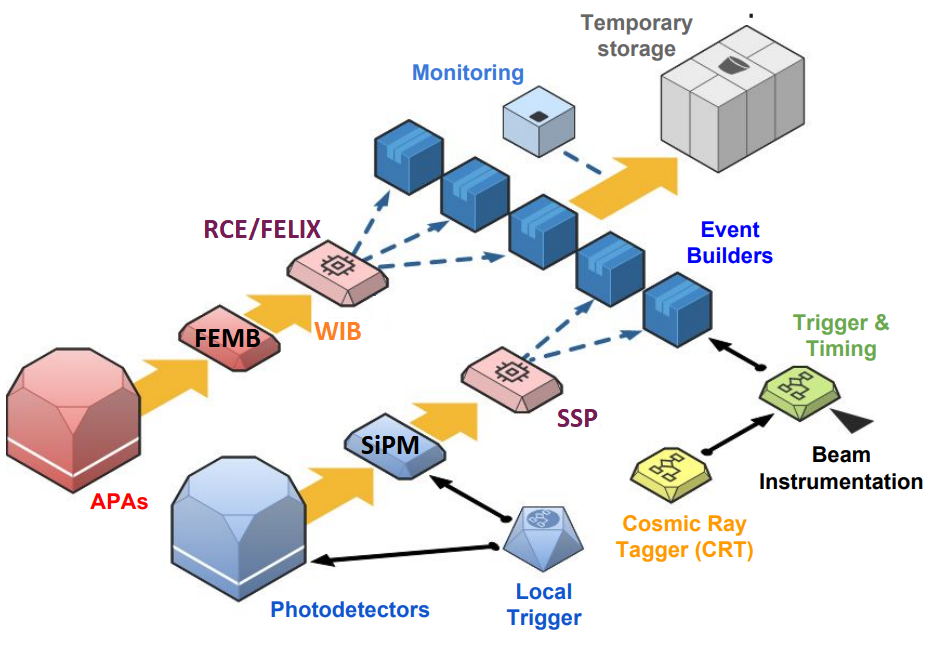}
\end{dunefigure}

Besides overall readout architecture, the \dword{protodune} and \dword{dune} \dword{daq} systems exhibit two key differences. 
First, the \dword{protodune} \dword{daq} is externally triggered (and at a trigger rate over an order of magnitude higher than that anticipated for \dword{dune}). Because of this, the \dword{protodune} \dword{daq} does not facilitate online data processing from the \dword{tpc} or \dword{pds} for self-triggering. 
Second, the \dword{protodune} system sits at the surface with a much higher data occupancy due to cosmic ray activity.
Overcoming the first key difference in order to demonstrate
\dword{daqdsn} capability for the \dword{fd} \dword{daq} design is a
main component of current and future \dword{daq} development plans.

\subsubsection{ProtoDUNE Outcomes}

The \dword{pdsp} \dword{daq} supported a test-beam experiment, and the requirements of the DUNE \dword{daq} are substantially different in scale and performance.
However, the successful operation of the \dword{pdsp} \dword{daq} has provided several key demonstrations for final system, in particular the data flow architecture, run configuration and control, and back-end functionality.

Specifically, \dword{pdsp} has demonstrated: 
\begin{itemize}
\item Upstream \dword{daq}: front-end readout hardware and data flow functionality servicing two out of the six \dword{apa}s.
  The data from each APA was continuously streamed to a single \dword{felix} board hosted in a dedicated computer which then transferred it to two other computers for buffering and readout.
  The baseline upstream \dword{daq} for \dword{dune} FD will retain one APA per \dword{felix} board but place two \dword{felix} boards and the buffering and readout functionality all together in a single \dword{fec}. 
  In addition to data flow functionality, \dword{protodune} front-end readout also demonstrates the interface to the front-end TPC electronics, and scalability to DUNE. It also supports host server requirements and specifications. Finally, it serves as platform for further development involving co-processor implementation and data selection aspects.
\item \dword{daqbes}, \dword{daqccm} and software infrastructure:
 Successful \dword{daqbes} implementation, including event builder
  farm and disk buffering, as well as an initial implementation of \dword{daqccm} functions. This has allowed the
  development and exercising of system partitioning, and provides a
  basis for scalability to DUNE. \dword{protodune} also serves as
  a platform for further system development, in particular in \dword{daqccm} and for the data flow orchestrator part of the
  \dword{daqbes}.
\item Data selection and timing: successful operation of the timing
  distribution system, and external trigger distribution to the
  front-end readout.
\end{itemize}

Besides demonstrating end-to-end data flow, an important outcome of
\dword{protodune} has been the delineation of cross-system
interfaces, i.e.~understanding the exact \dword{daq} scope and the interfaces to \dword{tpc}, \dword{pds}, and offline computing. The use of commodity solutions
where possible, and leverage of professional support from CERN IT 
substantially expedited the development and success of the project, as
did the strong on-site presence of experts from within the consortium during early installation and
commissioning. 
Outcomes specific to \dword{protodune} subsystems are discussed in
greater detail in~\cite{Hennessy:CDRReview}. 

\subsection{Ongoing Development}
\label{sec:daq:design-validation}

\dword{daq} subsystem development is ongoing at \dword{protodune} at the time of
writing. A detailed schedule for 2019 is available
in \cite{bib:docdb14095}. Major development plan milestones, a number
of which have already been achieved at the time of this writing, are:
\begin{itemize}
\item optimization and tuning of the front-end readout;
\item optimization and tuning of the \dword{artdaq} based dataflow software;
\item enhancement of monitoring and troubleshooting capabilities;
\item introduction of CPU-based hit finding (necessary for \dword{pds} readout);
\item introduction of \dword{fpga}-based hit finding (for \dword{tpc} readout);
\item implementation of online software data selection beyond trigger
primitive stage (introduction of trigger candidate generation, 
\dword{trigcommand} generation), and tests on well identified interaction
topologies (e.g,. long horizontal tracks, or Michel electrons from muon decay);
\item integration of online \dword{trigcommand} and modified data flow to event
builder to facilitate self-triggering of detector;
\item implementation of extended \dword{fpga}-based front-end functionality
(e.g., compression); and
\item prototyping of fake \dword{snb} data flow in front end and the back end.
\end{itemize}

Below, we focus on ongoing developments related to upstream \dword{daq},
data selection, and \dword{daqccm} prototyping, which is relevant to all \dword{daq} subsystems. These are the key areas where new technologies, beyond \dword{protodune}, remain to be designed and tested.

\subsubsection{Upstream DAQ Development}

The use of \dword{felix} as the front-end readout technology for DUNE was
successfully prototyped at \dword{protodune}, initially for the readout of one
\dword{apa}. In \dword{protodune}, \dword{felix} allows streaming of data arriving o- multiple \SI{10}{\Gbps} optical point-to-point links into commercial host memory and,
from there, storing, dispatching or processing of the data via
software. 

In \dword{protodune}, a single \dword{apa} \dword{felix}-based readout consists of two servers with a point-to-point \SI{100}{\Gbps} network connection to the \dword{felix} host computer.
The \dword{felix} I/O card interfaces with its host PC through 16-lane PCIe 3.0 (theoretical bandwidth of \SI{16}{GB/s}).
It transfers the incoming WIB data directly into the host PC memory using continuous DMA transfer. The \dword{felix} host PC runs a software process that publishes the data to any client subscribing to it.
Subscriptions may identify from which input optical links the received data stream will be sourced.
The clients consuming these streams are ``BoardReader'' processes implemented in the \dword{artdaq} data acquisition toolkit.
In order to sustain the data rate, modest modifications of the firmware and software were carried out specifically for \dword{protodune}: each \dword{felix} host receives and publishes data at $\sim$\SI{75}{\Gbps}. The BoardReader hosts are equipped with embedded Intel QuickAssist (QAT)~\footnote{Intel\textregistered QuickAssist Technology, \url{https://www.intel.com/content/www/us/en/architecture-and-technology/intel-quick-assist-technology-overview.html}.} technology for hardware accelerated data compression. The \dword{protodune} application of the \dword{felix} front-end readout is shown in Figure~\ref{fig:daq:felix-pd-impl}.

In DUNE only a very small fraction of the data received via the \dword{felix}
system will ever need to leave the host: thus it is not required to
implement very high speed networked data dispatching. On the other
hand it may be interesting to carry out data processing and buffering
on the host. While this is not the baseline design for DUNE, R\&D is
ongoing at \dword{pdsp} to evaluate the feasibility of
implementing hit finding, data buffering, and possibly even local
\dwords{trigcandidate} generation on the \dword{felix} host.

\begin{dunefigure}[Topology of the FELIX-based
    upstream DAQ of ProtoDUNE]{fig:daq:felix-pd-impl}{The topology of the \dword{felix}-based
    upstream \dword{daq} of \dword{protodune} (from~\cite{pdsp-felix}). The \dword{felix} host servers are publishing the data from the \dwords{wib} over \SI{100}{Gb\per\second} network interfaces. The BoardReader hosts are carrying out trigger matching, data compression and forwarding of fragments to the event builder.}
 \includegraphics[width=0.9\textwidth]{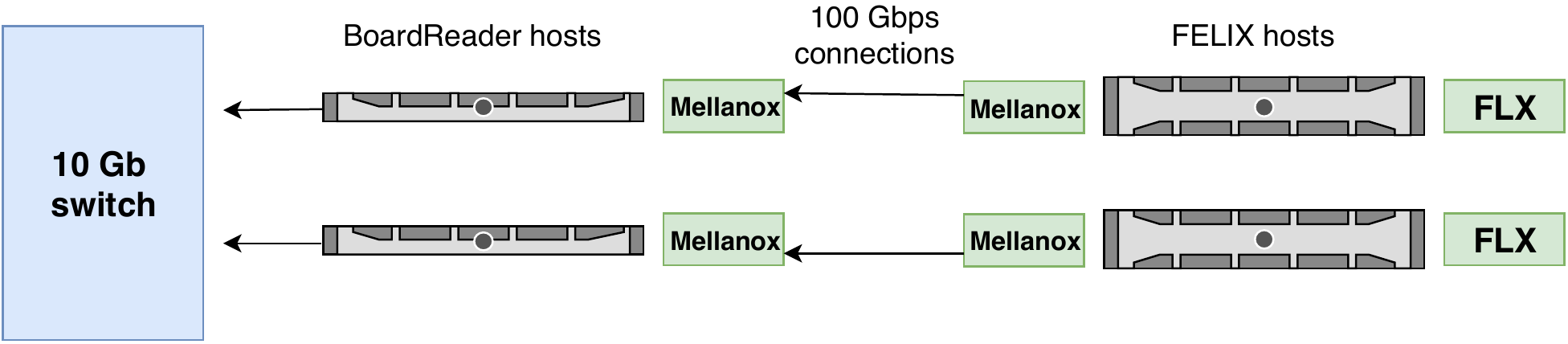}
\end{dunefigure}

The \dword{daq} team is investing substantial effort into the introduction of
a triggering chain based on the \dword{tpc} data into \dword{protodune}, which will
allow to carry out pre-design prototyping studies of the complete flow
of data of the DUNE \dword{daq}.  
The \dword{felix}-based readout system will be adapted to support the
different studies, from co-processor based data handling in firmware
(including trigger primitive generation, compression, and buffering) to software-based processing, on a single
server. Benchmarking and optimization of the \dword{felix} firmware and
software will also continue, with the aim of further compacting the
readout by supporting two \dword{apa}s on a single server. 

\subsubsection{Co-processor Development}

Upstream \dword{daq} development efforts at \dword{protodune} include a parallel test platform for trigger primitive generation, compression, and buffering firmware validation for the \dword{fpga} co-processor board. The platform for these tests will initially be a Xilinx ZCU102 development board. Passive optical splitters will be inserted into the fiber path downstream of the WIBs, providing duplicate data inputs for the test hardware, without disrupting the main readout path of \dword{protodune}. Tests using the development board will first focus on generation of trigger primitives, which will be read out over the network via IPBus\cite{Larrea:2015wra}. The ZCU102 includes 512MBytes of DDR4 RAM connected to the \dword{fpga} programmable logic, as well as a four lane PCIe Gen 2 socket which will host an \dword{nvme} SSD on an adapter. This combination of hardware will allow tests of buffering and compression of readout data, in parallel with trigger primitive generation. The ZCU102 will subsequently be replaced with a ``\dword{felix} demonstrator'' using a Xilinx Virtex-7 Ultrascale+ \dword{fpga}, connected to a custom \dword{fpga} co-processor board via an FMC+ connector. These boards represent the first prototypes for the final system hardware.

Tests using the development board will focus on functionality rather than data throughput. However, the tests will provide estimates of \dword{fpga} logic and memory resources that can be scaled up to the full system. Tests using the \dword{felix} Demonstrator and PBM at \dword{protodune} will focus on scaling the functional tests performed using the ZCU102 to a full demonstration of trigger and readout functionality for a full \dword{apa}. In addition, this platform will facilitate integration with the prototype \dword{daqdss} and \dword{daqbes} subsystems at \dword{protodune}. 

\subsubsection{Data Selection Development}

\begin{dunefigure}[CPU core-time for primitives and CPU utilization in live data]{fig:daq-cpu-hf-speed}{CPU core-time (left) required to find primitives in simulated signal and noise data across 960 collection channels.
    In this test, the data has been pre-formatted to facilitate the use of SIMD hardware accelerated functions (AVX2). 
    Per thread CPU utilization (right) to process live data from a \dword{protodune} \dword{apa}.
    Processing includes data reformatting, trigger primitive generation, and output message formation.
    Each thread, one per \dword{felix} link, requires about 75\% CPU core utilization to keep up.
    Variation in utilization reflects the variations in levels of ionization and noise activity in the input data and is due to output message formation.
    The peak usage at the start arises from input buffering while the process is initializing. 
    Once operational state is achieved, this brief backlog is processed.}
  \includegraphics[height=6cm,clip,trim=4cm 16.5cm 4cm 2cm]{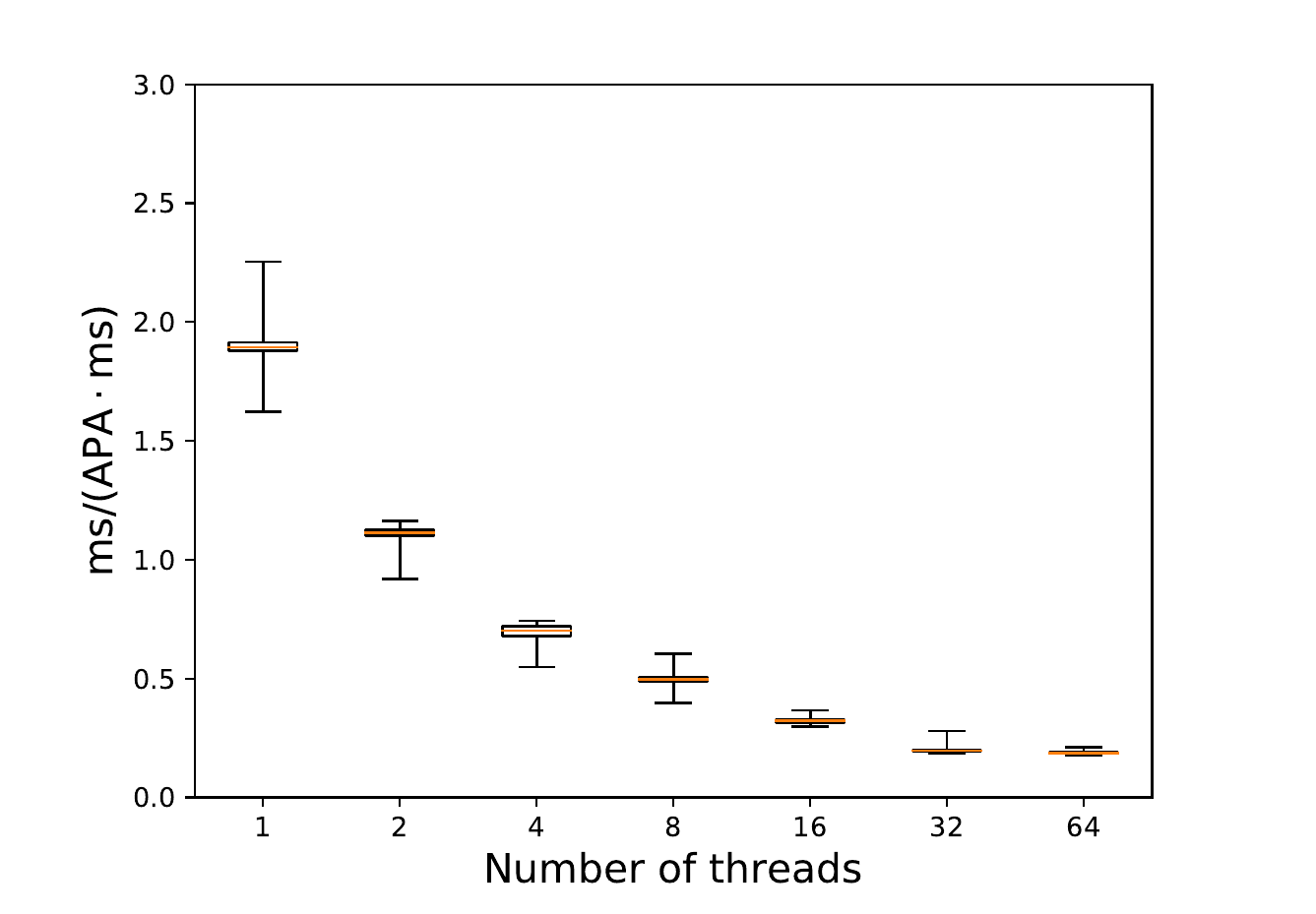}%
 \includegraphics[height=6cm,clip,trim=0cm 0cm 0cm 5mm]{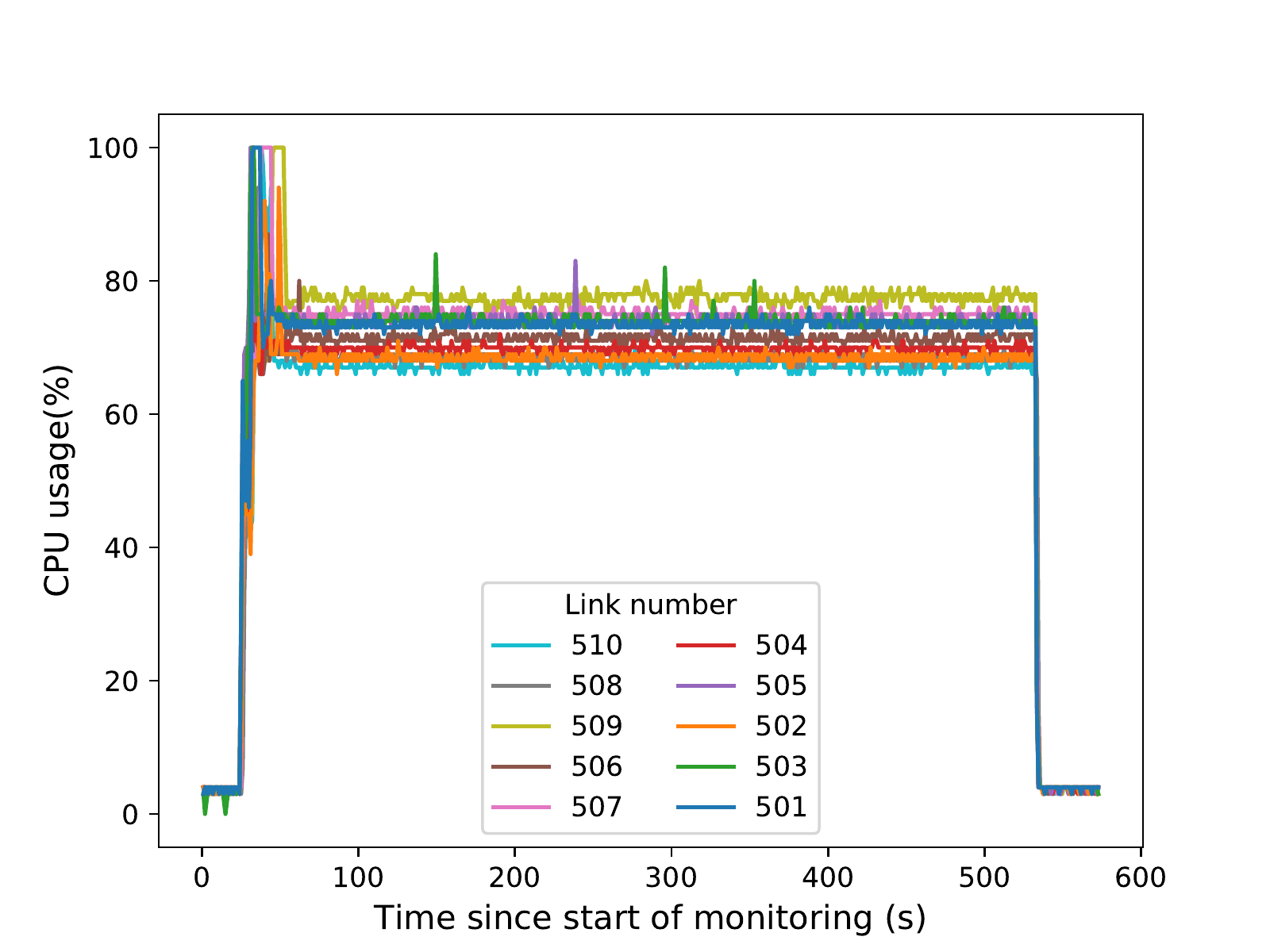}
\end{dunefigure}

\begin{dunefigure}[Trigger primitives in ProtoDUNE-SP data]{fig:daq-hitfinder}{Example result of the trigger-primitive software (``hit finder'') applied to ADC waveform data spanning 96 collection channels that are sensitive to activity in the drift volume of \dword{pdsp}. 
    At left, the figure shows the ADC sample values relative to pedestal which are input to the algorithm. 
    At right, these same data are shown with a green $\times$ marking each trigger primitive found. 
    The inset is a zoomed region showing in detail the alignment of the input waveforms and the derived trigger primitives.
    In this test, the algorithm runs on the continuous stream of ADC waveforms prior to readout while actual readout was prompted by an external trigger.}
  \begin{tikzpicture}
    \node[inner sep=0] (raw) at (0,0) {
     \includegraphics[height=7cm]{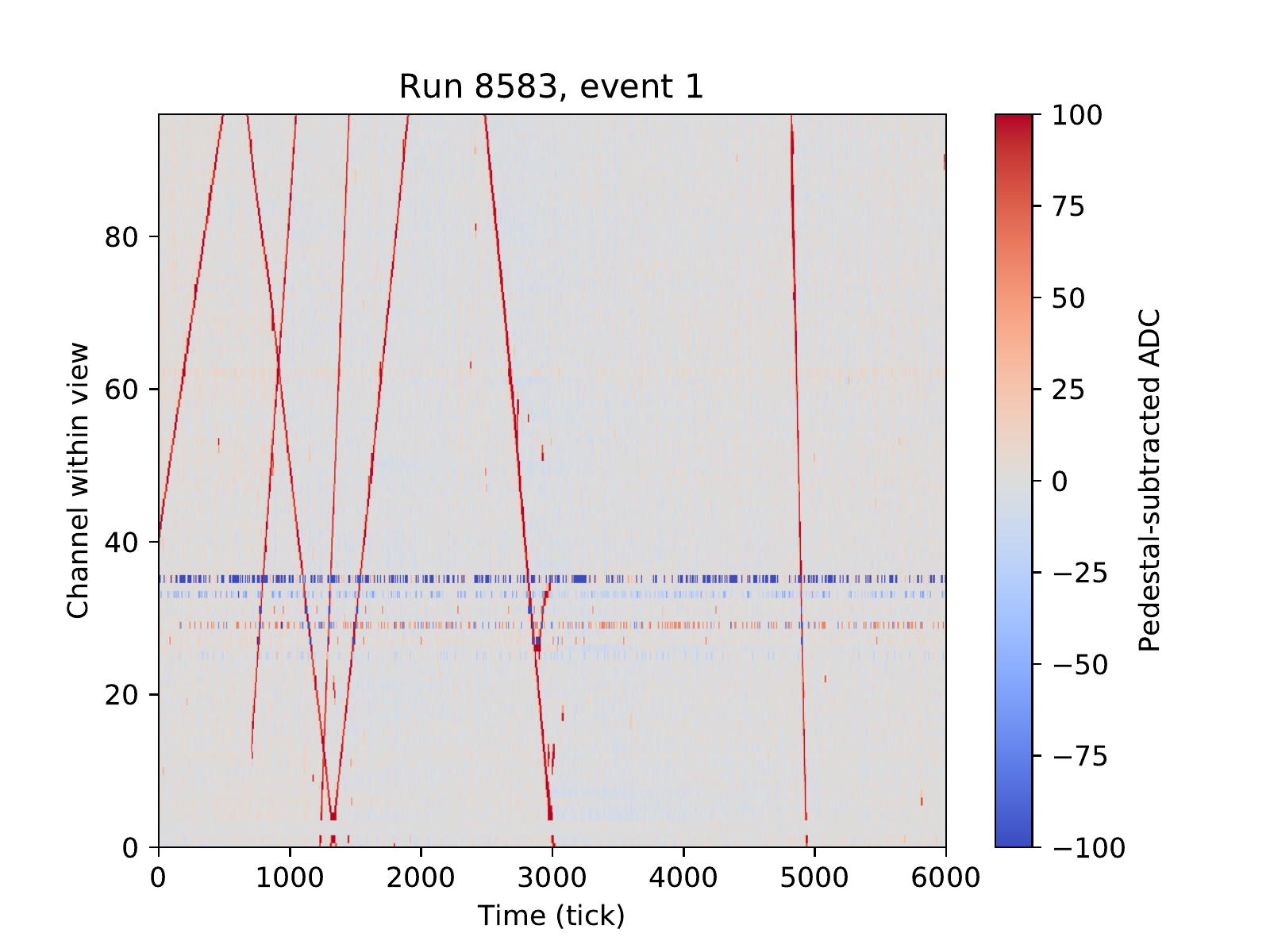}
    };
    \node[inner sep=0, right = 0cm of raw] (hit) at (raw.east) {
     \includegraphics[height=7cm]{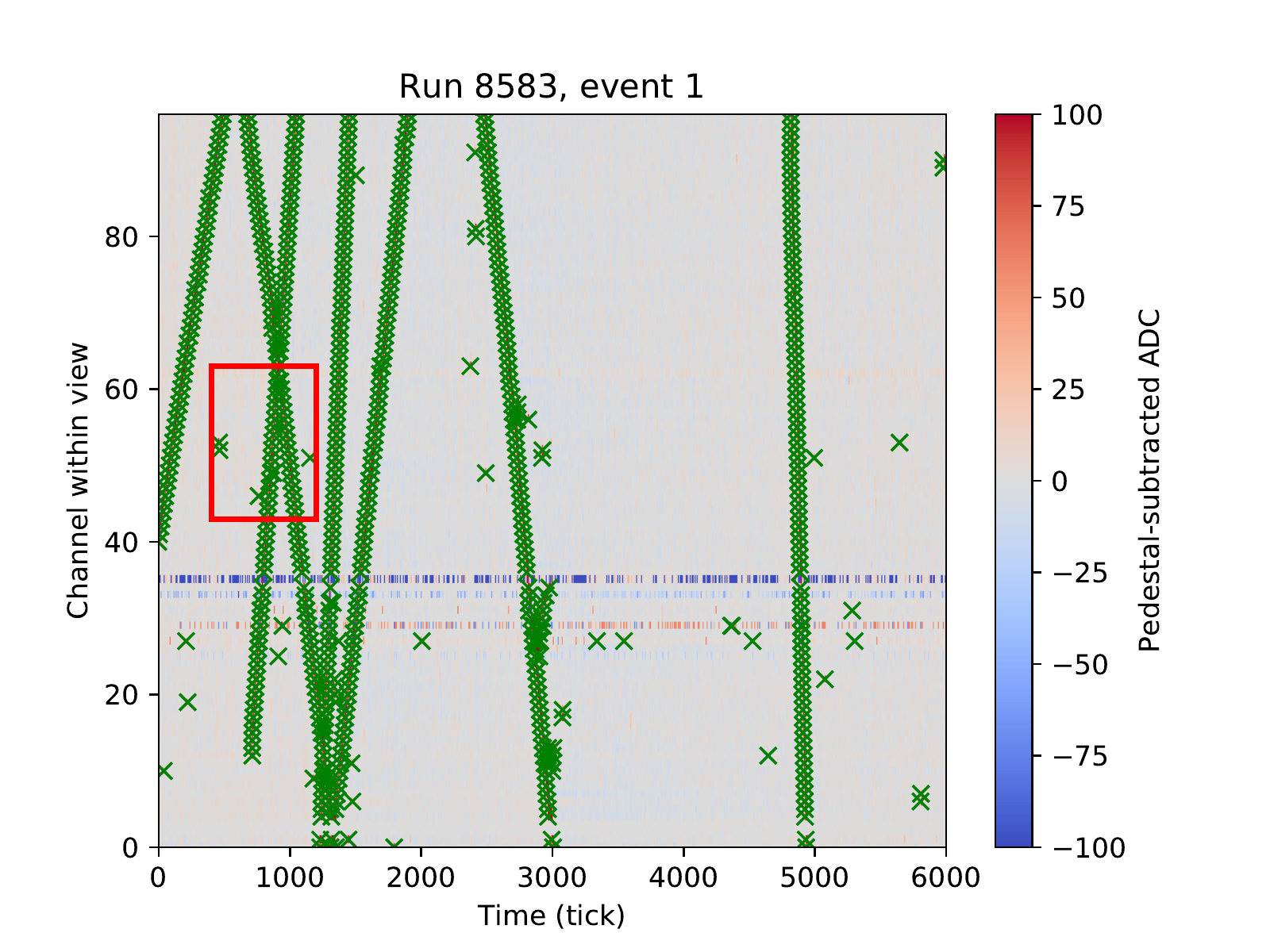}
    };
    \node[inner sep=0] (zoom) at (4,4) {
     \includegraphics[height=5cm]{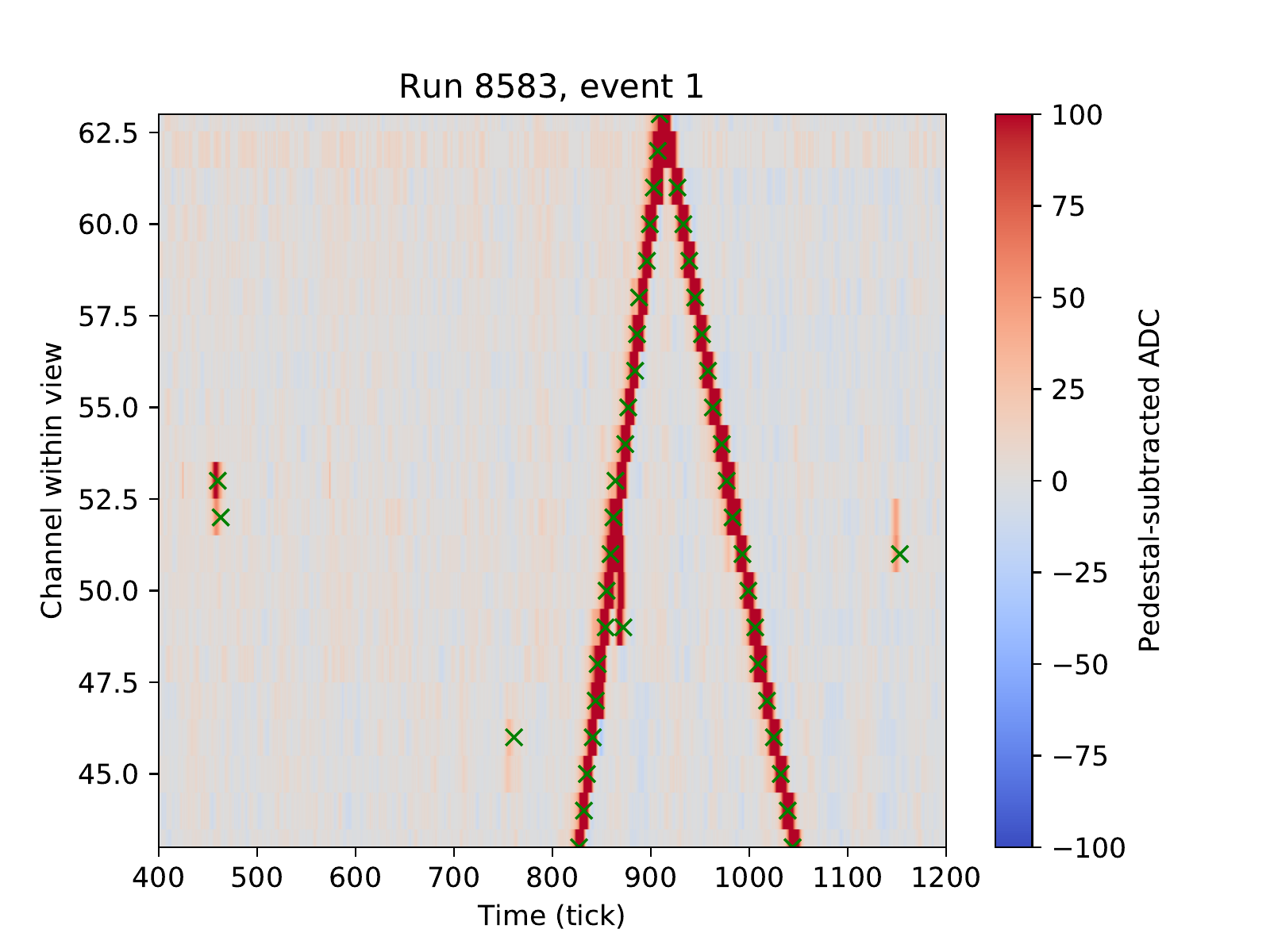}
    };
    \draw[->, line width=0.3mm, red] (6.25, -.32) -- (1.5, 2.05);
    \draw[->, line width=0.3mm, red] (7.01,  0.8) -- (5.65, 5.9);    
  \end{tikzpicture}
\end{dunefigure}

\begin{dunefigure}[Trigger primitive rates in ProtoDUNE-SP, four categories]{fig:daq-tp-rates}{Trigger primitive rates in \dword{pdsp} as a function of threshold in four categories: all data (top left),  after removal of particularly noisy channels (top right), with HV off so no contribution to signal (bottom left) and HV off and noisy channels excluded (bottom right).}
  \begin{minipage}[b]{0.5\linewidth}
    \begin{center}
     \includegraphics[page=1,width=0.8\textwidth,clip,trim=4cm 16cm 4cm 2cm]{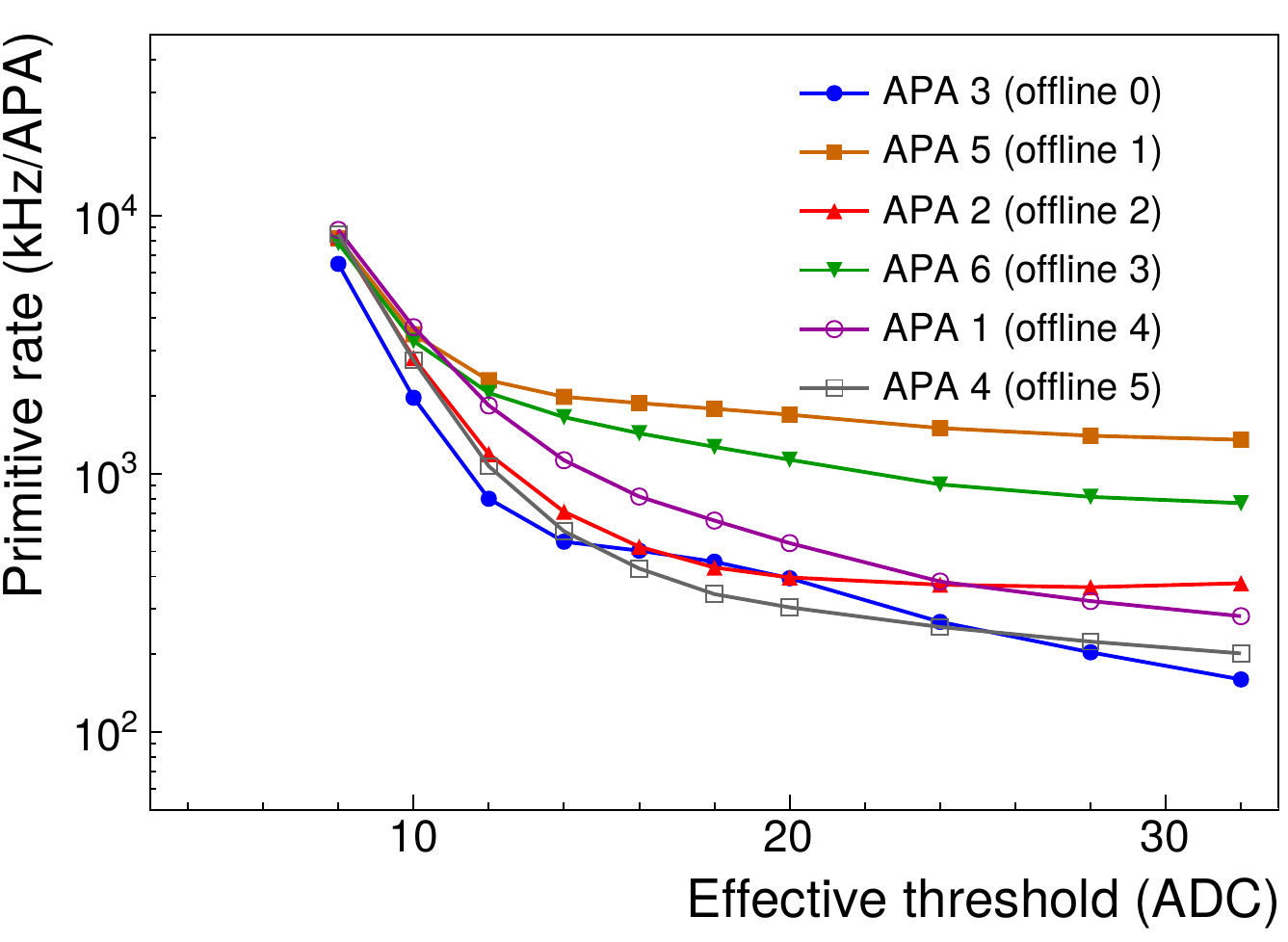}

     \includegraphics[page=2,width=0.8\textwidth,clip,trim=4cm 16cm 4cm 2cm]{daq-primitive-rates-thresholds.pdf}
    \end{center}
  \end{minipage}%
  \begin{minipage}[b]{0.5\linewidth}
    \begin{center}
     \includegraphics[page=3,width=0.8\textwidth,clip,trim=4cm 16cm 4cm 2cm]{daq-primitive-rates-thresholds.pdf}

     \includegraphics[page=4,width=0.8\textwidth,clip,trim=4cm 16cm 4cm 2cm]{daq-primitive-rates-thresholds.pdf}
    \end{center}
  \end{minipage}

\end{dunefigure}

During early stages of design, significant effort has been dedicated to
trigger primitive generation studies through simulations.
Specifically, charge collection efficiency and fake rates due to noise
and radiologicals have been studied as a function of hit threshold,
demonstrating that data rate requirements can be met, given sufficiently low
electronics noise levels and radiological rates~\cite{bib:docdb11236}. 
Ongoing efforts within DUNE's radiologicals task force aim to validate
or provide more accurate background predictions, against which this
performance will be validated.
In addition, offline studies demonstrate the performance of trigger
primitive generation algorithms as a function of the number of CPU cores
used.  
The results are summarized in Figure~\ref{fig:daq-cpu-hf-speed} and show
that four cores are sufficient to keep up with 960 channels.
The test does not include reformatting of the data required to put it in
a form that allows AVX2 hardware SIMD acceleration.
In tests with live \dword{protodune} data, it is found that ten cores at
an average 65\% usage were enough to handle both reformatting and
trigger primitive generation. 
Effort on understanding and removing contribution
from cosmics/cosmogenics and (known) noisy channels is ongoing.
These results are summarized in Figure~\ref{fig:daq-tp-rates}
Additional details may be found in~\citedocdb{14062}.

\begin{dunefigure}[Efficiency for forming trigger candidates as input trigger primitives]{fig:daq-tc-eff-vis}{Efficiency for forming trigger candidates as input trigger primitives from two algorithms, online (blue) and offline (red).}
  \includegraphics[width=0.7\textwidth]{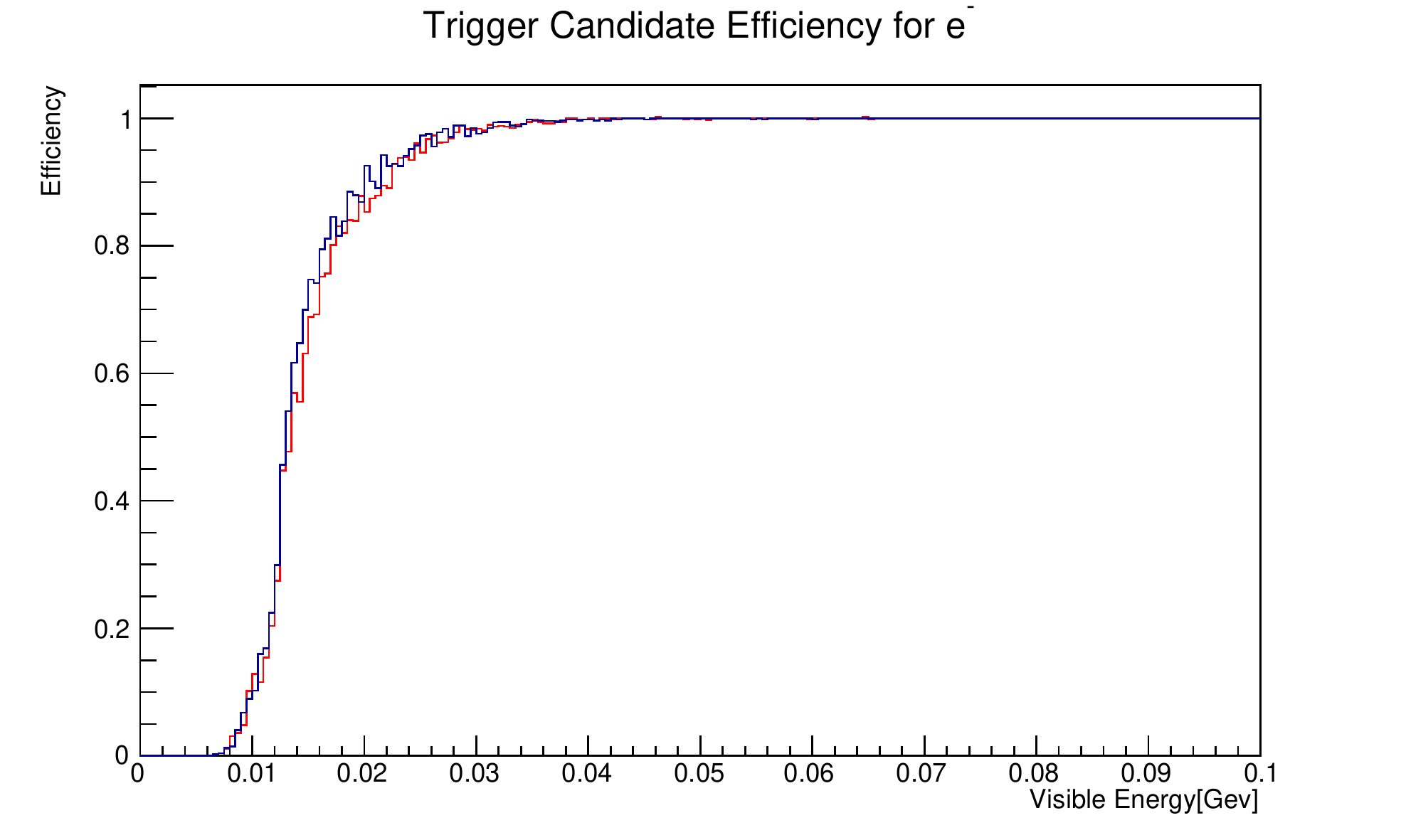}
\end{dunefigure}

\begin{dunefigure}[Efficiency for forming trigger candidates from ionization activity]{fig:daq-tc-eff-allinc}{All-inclusive efficiency for
    forming trigger candidates from ionization activity from beam $\nu_e$
    (left) and beam $\nu_\mu$ (right) interactions at or above a given
    visible energy. 
    At smaller energies, somewhat more events with their visible energy
    dispersed in space and time fail the trigger candidate selection
    criteria. 
    High efficiency is obtained at \SI{100}{\MeV} visible. 
    The trigger candidate algorithm used is the offline version, see
    Figure~\ref{fig:daq-tc-eff-vis} for comparison with online version.}
 \includegraphics[width=0.45\textwidth]{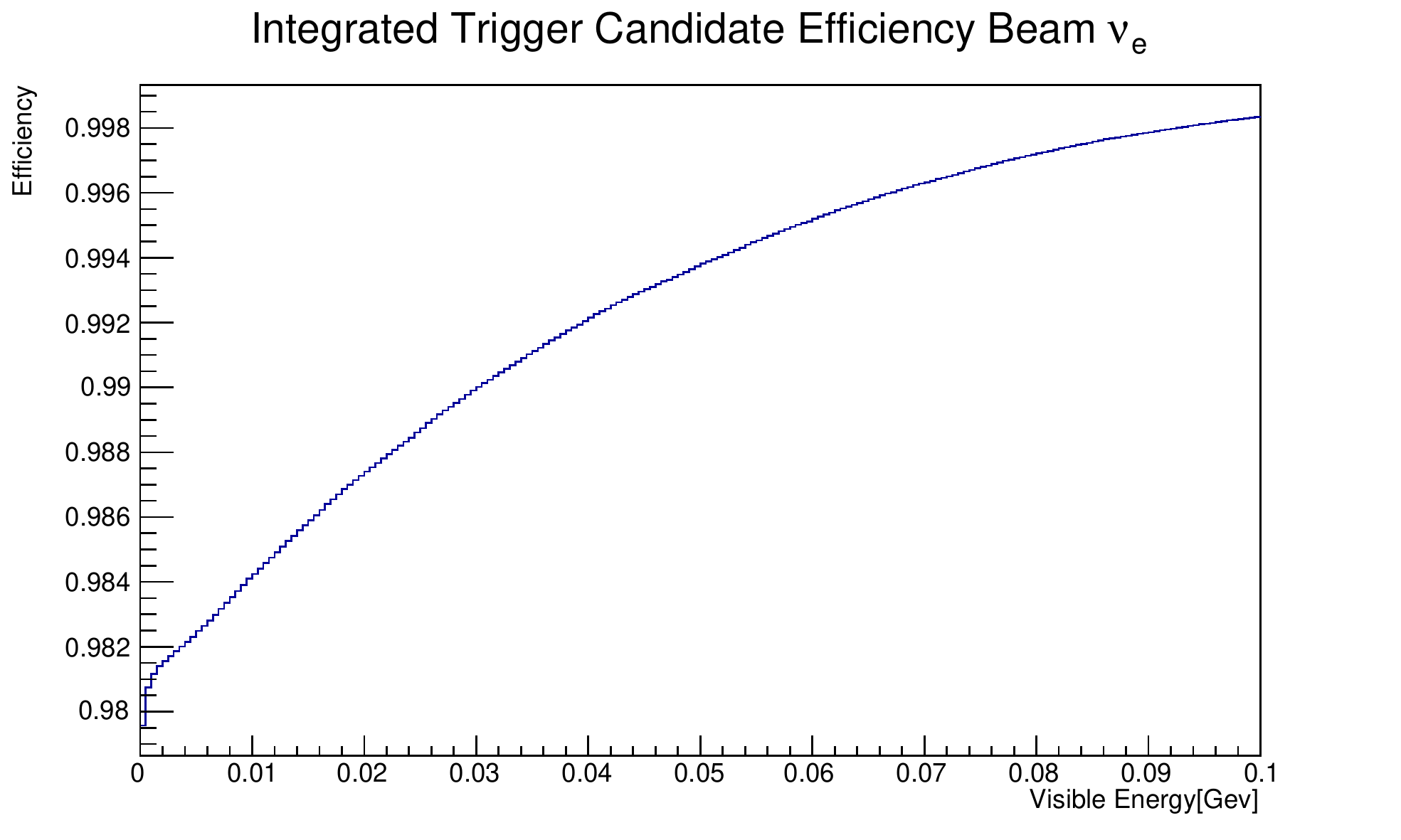}%
 \includegraphics[width=0.45\textwidth]{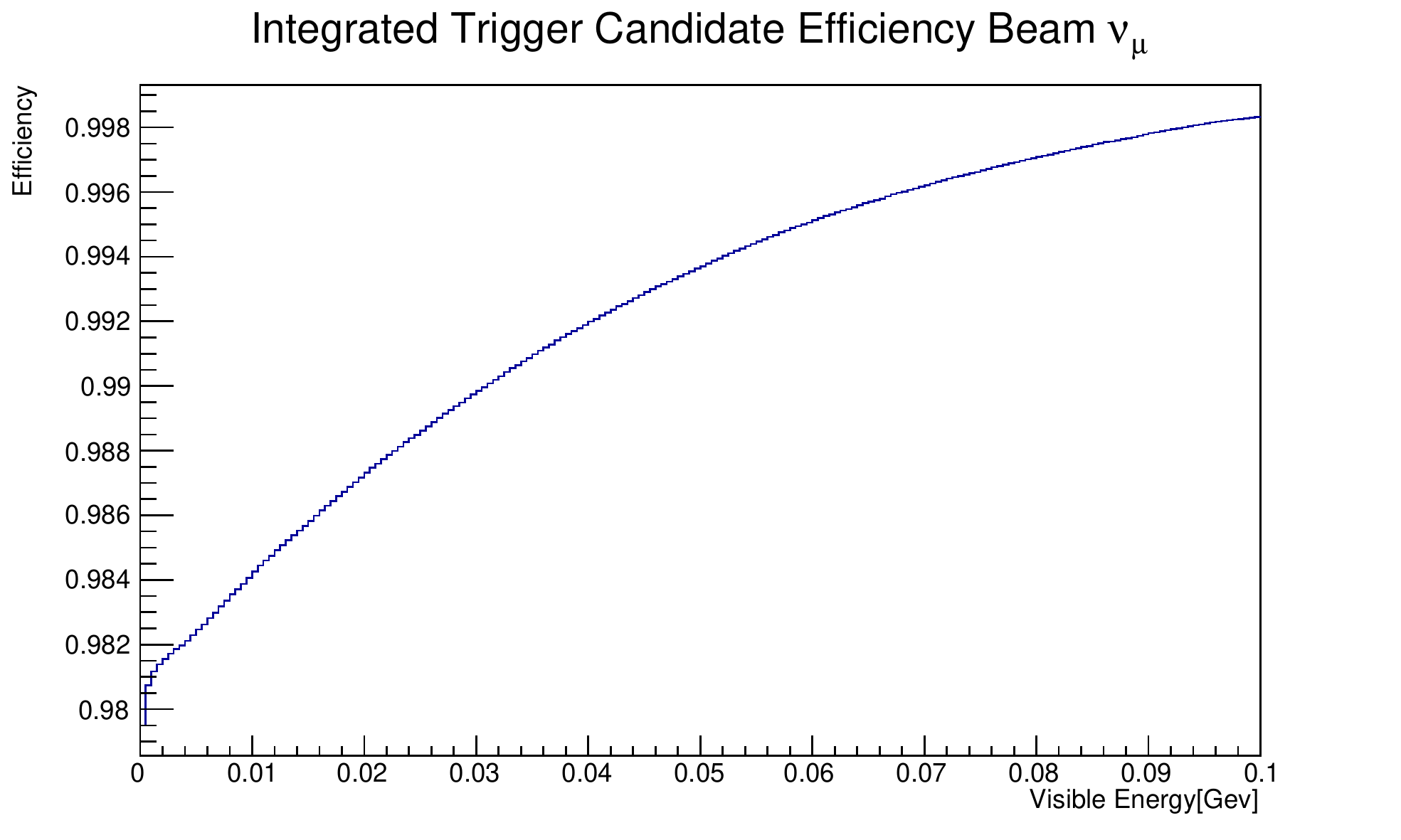}
\end{dunefigure}

\Dword{trigcandidate} generation, building on trigger primitives information
and considering integral \dword{adc} and trigger primitive proximity by channel
and time (in \SI{50}{\micro\second}) space, has also been studied with Monte
Carlo simulations~\cite{bib:docdb11215}.
Trigger candidates with sufficient total integral \dword{adc}  can be accepted to
generate corresponding \dwords{trigcommand} for localized high energy
activity, such as for beam, atmospheric neutrinos, baryon number
violating signatures, and cosmics.
Simulation studies demonstrate that this scheme meets efficiency
requirements for localized high energy triggers.
Specifically, simulations demonstrate that $>99$\% efficiency is
achievable for $>\SI{100}{\MeV}$ visible energy deposited by single particles
(shown in Figure~\ref{fig:daq-tc-eff-vis} for $e^-$), and that the
corresponding effective threshold for localized triggers for the system
is at $\sim$\SI{10}{\MeV}. This translates to all-inclusive efficiencies for
beam $\nu_e$ and $\nu_\mu$ events in excess of 99\% for visible energies
above \SI{100}{\MeV}, as shown in Figure~\ref{fig:daq-tc-eff-allinc}.

Low-energy \dwords{trigcandidate} furthermore can serve as input to the
\dword{snb} trigger. Simulations demonstrate that the trigger candidate
efficiency for any individual \dword{snb} neutrino interaction is on the order
of 20-30\%. However, a multiplicity-based \dword{snb}
\dword{trigdecision} that integrates low-energy trigger candidates over an up to \SI{10}{\second}
integration window yields high trigger efficiency out to the
galactic edge while keeping fake \dword{snb} trigger rates to one per
month. An energy-weighted multiplicity count scheme can be
applied to further increase efficiency and minimize background
\cite{bib:docdb14522}. This is illustrated in
Figure~\ref{fig:daq-snb-eff-Eweight}, demonstrating nearly 100\% efficiency
out to the edge of the galaxy, and 70\% efficiency for a \dword{snb} at
the Large Magellanic Cloud (or for any \dword{snb} creating 10
events). This performance is obtained by considering the sum ADC
distribution of \dwords{trigcandidate} over \SI{10}{\second} and comparing to a background-only
vs.~background plus \dword{snb} hypothesis. The efficiency gain compared to a
simpler, \dword{trigcandidate} counting-based approach is quite significant; using
only counting information, the efficiency for a \dword{snb} at
the Large Magellanic Cloud is 6.5\%. The \dword{daq}
consortium is working on further refining these algorithms to further improve
\dword{snb} trigger efficiency for more distant \dwords{snb}. For
additional efficiency increase, the design provides flexibility
for a slightly higher fake \dword{snb} trigger rate to be handled by
the \dword{daqbes}, combined with more aggressive data reduction
applied at the high level trigger stage so as to respect the data rate to
offline storage requirement.

The dominant contributor to fake \dword{snb} triggers is
radiological backgrounds from neutrons, followed by radon. It is
crucial to continue working closely with the radiological task force
to validate radiological the background assumptions.

\begin{dunefigure}[Supernova burst trigger
  efficiency]{fig:daq-snb-eff-Eweight}{Top: \dword{snb} trigger
    efficiency as a function of the number of supernova neutrino
    interactions in the \SI{10}{\kt} module, for a likelihood trigger
    approach that utilizes sum ADC shape information of
    \dwords{trigcandidate} input into the trigger decision. Bottom:
    For a \dword{snb} at the Large Magellanic Cloud, where 10 neutrino
  interactions are expected, the efficiency gain over a counting-only
  trigger is significantly improved. } 
\includegraphics[width=0.65\textwidth, trim=0cm 9cm 3cm 6cm]{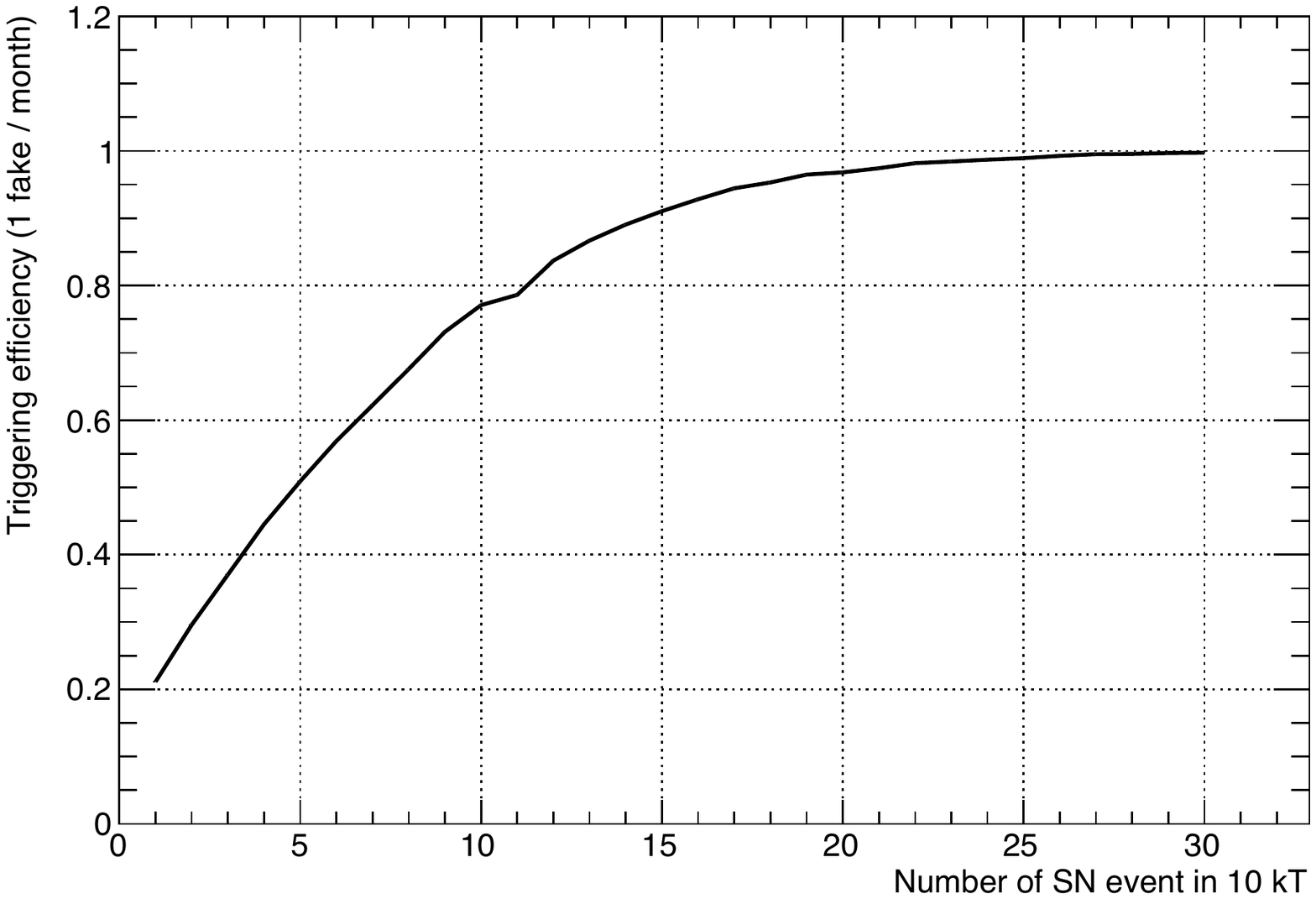}%
\newline
\includegraphics[width=0.65\textwidth, angle=-90,  trim=0cm 2cm 0cm 0cm]{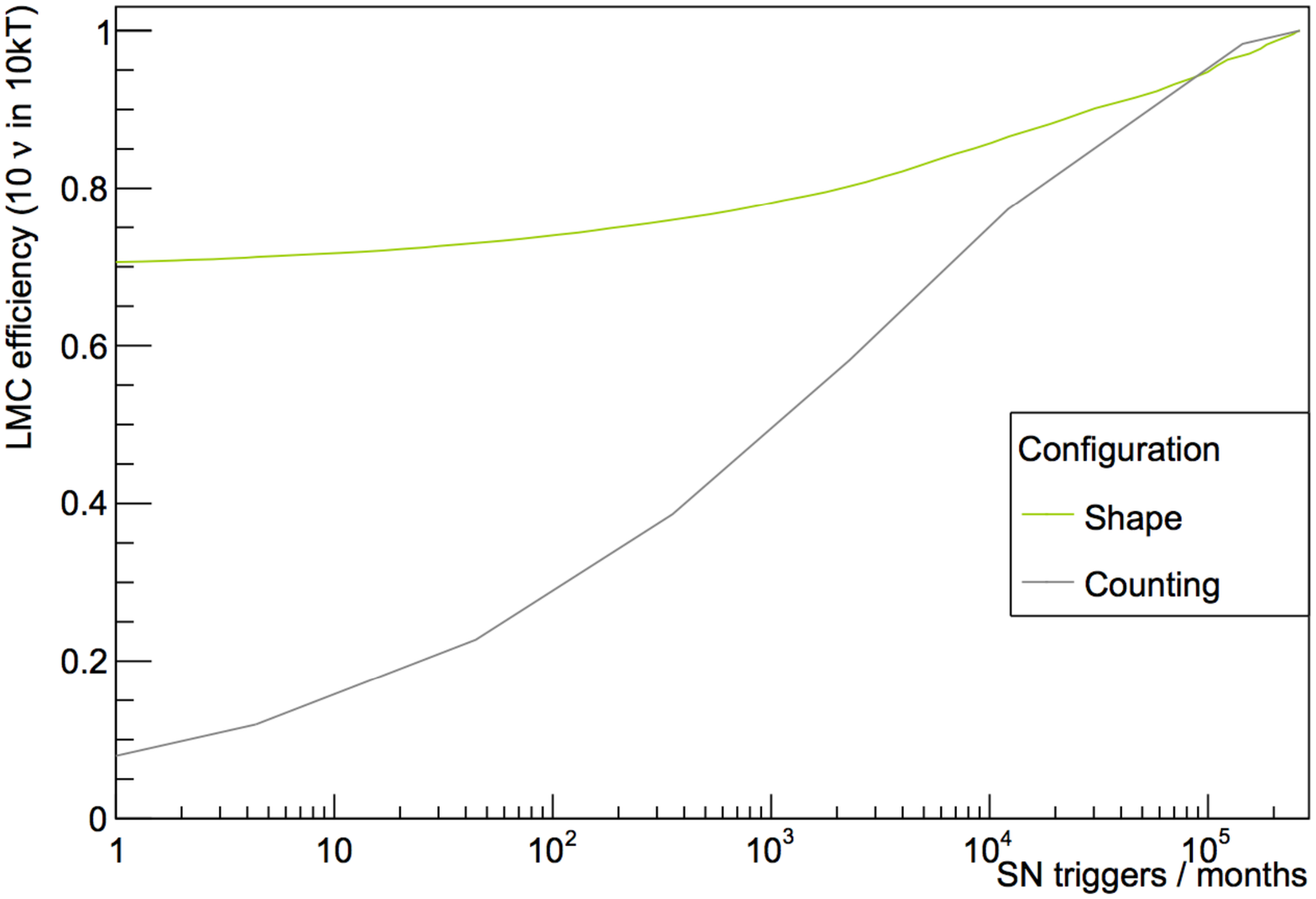}%
\end{dunefigure}

In the case of the high level filter, the consortium is exploring the
use of machine learning techniques, specifically image classification
with the use of \dwords{cnn} on GPUs, as a way to
classify and down-select individual sections of \dword{tpc} channel vs. time
(``frames''), with extent of one \dword{apa}'s worth of collection plane
channels by one drift length (2.25 ms, or, 4500 samples). \dwords{cnn} have
been trained on \dword{mc} simulations of frames with each of
the following off-beam event topologies: 
atmospheric neutrino interactions, baryon number violating
interactions (proton decay or neutron-antineutron oscillation), cosmic
ray interactions, supernova neutrino interactions, or no interactions at all -- all
with radiological and noise background included in the
simulations. Preliminary studies show that a \dword{cnn} can be successfully trained classify any
given input frame as one of three categories: empty, containing a \dword{snb} neutrino
interaction, or containing a high-energy (atmospheric neutrino, baryon
number violating, or cosmic ray) interaction. Specifically, empty frames can be
rejected with an efficiency of $>$99\%, while frames containing or
partially containing 
a supernova neutrino, an atmospheric/cosmic interaction, or a baryon
number-violating interaction can be preferentially selected with efficiency
$>$88\%, $>$92\%, or $>$99\%, respectively. Such a 
filter could potentially be applied to reduce the event record size by
more than two orders of magnitude. Details may be found in~\citedocdb{11311}.
The speed at which machine learning inference may be applied is under study~\cite{jwa2019accelerating}.

\subsubsection{Prototype Trigger Message Passing}

A prototype trigger message passing (PTMP) system using elements of the \dword{ipc} mechanism described in Section~\ref{sec:daq:design-ipc} is currently under development and testing at \dword{protodune}.
The primary goals of this prototype is to add a self-triggering mechanism to the \dword{protodune} detector that includes many of the features needed for the far detector \dword{daq}.
Throughput and latency of the mechanism is being evaluated and optimized.
Message schema and application level protocols have been designed and are being improved.
Future work will include prototyping \dword{daqccm} functionality including \dword{daqdispre}.

PTMP has been successfully exercised to transfer trigger primitives from the software based hit finder. The short-term goal will be to successfully aggregate information from across an APA and feed the result to a trigger candidate finder which identifies horizontal muons. From this output a \dword{trigdecision} can be made.

\subsection{Plan for Future Development}
\label{sec:daq:design-plans}

As mentioned in the introduction of this section, at present, the development model chosen for the \dword{daq} system is the one of iterative prototyping.
This model is widely used in projects in which requirements are still
being refined and particularly for systems relying on rapidly evolving technologies, such as today's information and computing sector. 
This model will be used throughout 2019, making use of the \dword{protodune} setup to explore architectural options, software solutions, etc.
At a later stage, the \dword{daq} development will move to a more streamlined incremental model, ensuring that careful design precedes the final implementation of individual components.
Most of the development will be carried out emulating the data inputs.
On the other hand, the \dword{daq} will be validated regularly via test stand integration with detectors, such as \dword{protodune} or pre-installation sites.
The overall development schedule with the main \dword{daq} milestones is shown in Section~\ref{sec:daq:schedule}.


\section{Production, Assembly, Installation and Integration}
\label{sec:daq:production}

The \dword{daq} system relies largely on commercial off-the-shelf components, with the exception of the timing system and the first stage of the upstream \dword{daq}.
Therefore, the production and assembly aspects are simpler than for other systems, while of course the installation and integration stages are very important and have to be planned carefully, due to the large number of interfaces of the \dword{daq} system with other parts of the experiment.

\subsection{Production and Assembly}
\subsubsection{Timing System}
A prototype of the timing system already exists and has been used at \dword{pdsp}. The final hardware prototype will be used in the second run of \dword{pdsp} in 2021 and production is planned right afterwards, allowing detector communities to have an early integration with the timing hardware and firmware.

\subsubsection{Upstream DAQ}
The upstream \dword{daq} will have FPGA mezzanine cards connecting to the detector electronics readout fibres, and processing and storing data temporarily. Prototype cards implementing parts of the required functionality exist already, but more prototypes are planned before the production readiness review planned in December 2022. While the hardware design will be done at the institutions working in this area, the production of prototypes and final cards will be outsourced to companies, allowing for early identification of those companies that can guarantee a high quality cards production.

\subsubsection{Racks, Servers, Storage and Network}
While commercial devices do not need to be produced or assembled, enough time has to be planned, once the proper devices are identified, for the tendering and procurement procedures. Racks and fibers will be procured in order to be available early in 2023; servers and switches will be purchased in two batches, one to be ready for supporting the installation and commissioning of the detector components and \dword{daq} infrastructure and one to reach nominal performance, in time for the start of data taking.

\subsection{Installation and Integration}

The \dword{daq} will be installed in an enclosure in the west end of the \dword{cuc} (``Data Room'' in Figure~\ref{fig:cavern-layout}.  Roughly
half of this space will be office space (including control
workstations) and the other will be a computer room to hold the \dword{daq}
front-end computing and network equipment (Figure~\ref{fig:install-cuc}). Further details of the interface of \dword{daq} with underground facilities
may be found in \citedocdb{6988}, and the installation interface document for \dword{daq} in \citedocdb{7015}.

\begin{dunefigure}[Layout of the DUNE underground areas]{fig:cavern-layout}
  {Top view of the layout at the \dword{4850l} at \dword{surf}. Shown are the three large excavations and the location of detectors in the north (upper) and south caverns. 
  The \dword{cuc} in the middle houses the \dword{dune} data room where the \dword{daq} will be installed and the underground utilities. }
\includegraphics[width=.9\textwidth]{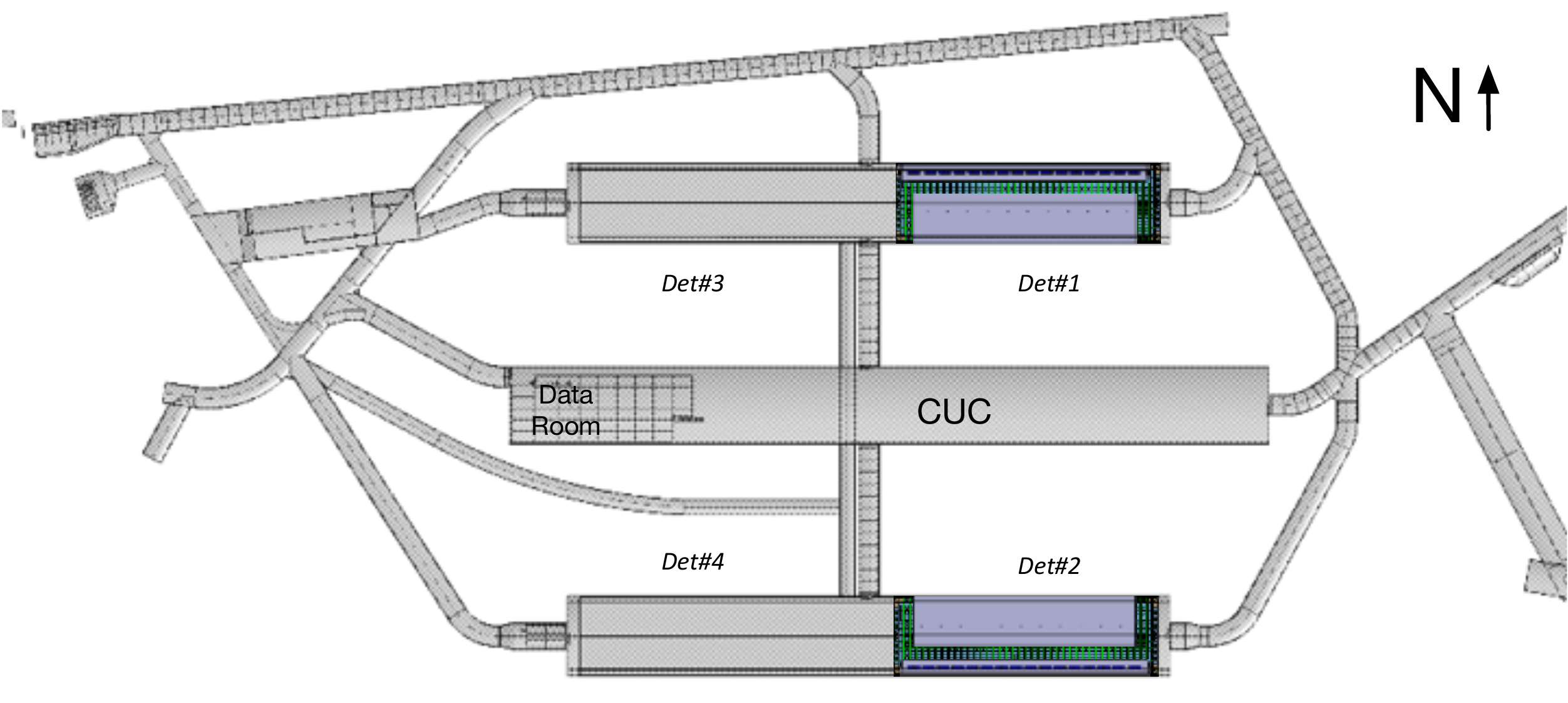}
\end{dunefigure}

\begin{dunefigure}[Layout of the DUNE data room and work area in CUC]{fig:install-cuc}
  {Top: The overall layout of the \dword{dune} spaces in the \dword{cuc}. A110 is the \dword{dune} data room, which houses the underground computing, and A111 is a general-purpose work area (not a control room, as labeled) that we call the experimental work area. Bottom: The first row of ten racks in the data room is shown. The first two represent the \dword{cf} interface racks. The images were taken from the ARUP 90\% design drawings U1-FD-A-108 and U1-FD-T-701~\cite{bib:docdb14242}.}
\includegraphics[width=.85\textwidth]{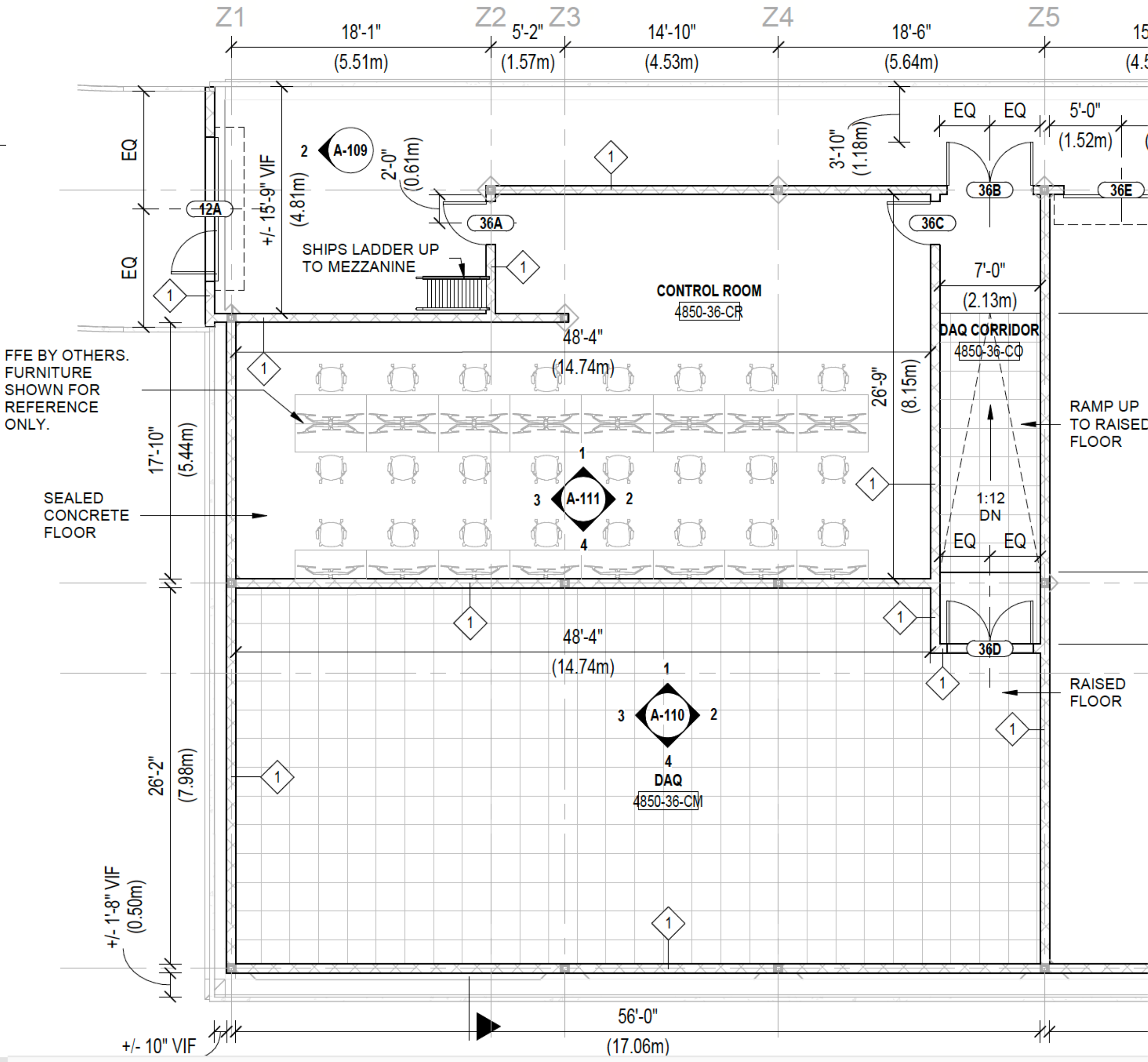}
\vspace{-2pt}
\includegraphics[width=.9\textwidth]{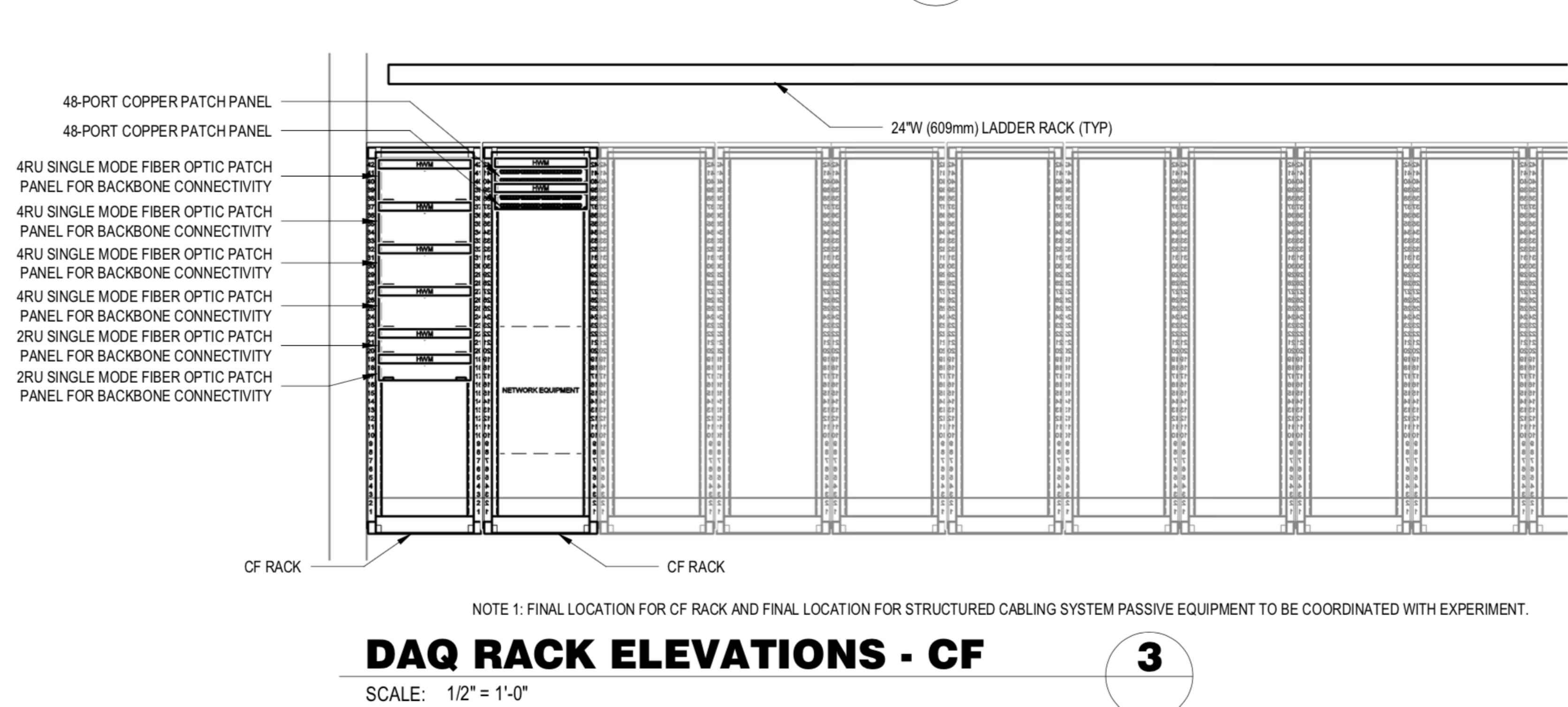}
\end{dunefigure}

Infrastructure in the \dword{cuc} will be installed starting in Q4~2022
(Table~\ref{tab:daq-sched}).  At that point, CF will have
handed the \dword{daq} group an empty room with cooling water and power
connections.  Over the next nine months, racks for the \dword{daq} computing
will be installed, plumbed into cooling water, and connected to power
and networking.  The network connection from this data room to the fiber
trunks going up the shaft will also be made, as will preparations to
receive the multi-mode fiber connections from the \dword{wib} to the
\dword{felix} cards housed in servers in these racks. There is space
for \cucracks racks, with four set aside for other consortia, 12 per
module for upstream \dword{daq} electronics, and the remaining space for
networking and other \dword{daq} computing needs. An initial engineering design of the computer room is shown in Figure~\ref{fig:daq-room}, which meets all requirements for capacity, cooling, safety and installation schedule.

\begin{dunefigure}[DAQ counting room in the CUC]{fig:daq-room}{Initial engineering design for the \dword{daq} counting room in the CUC}
 \includegraphics[width=0.9\textwidth]{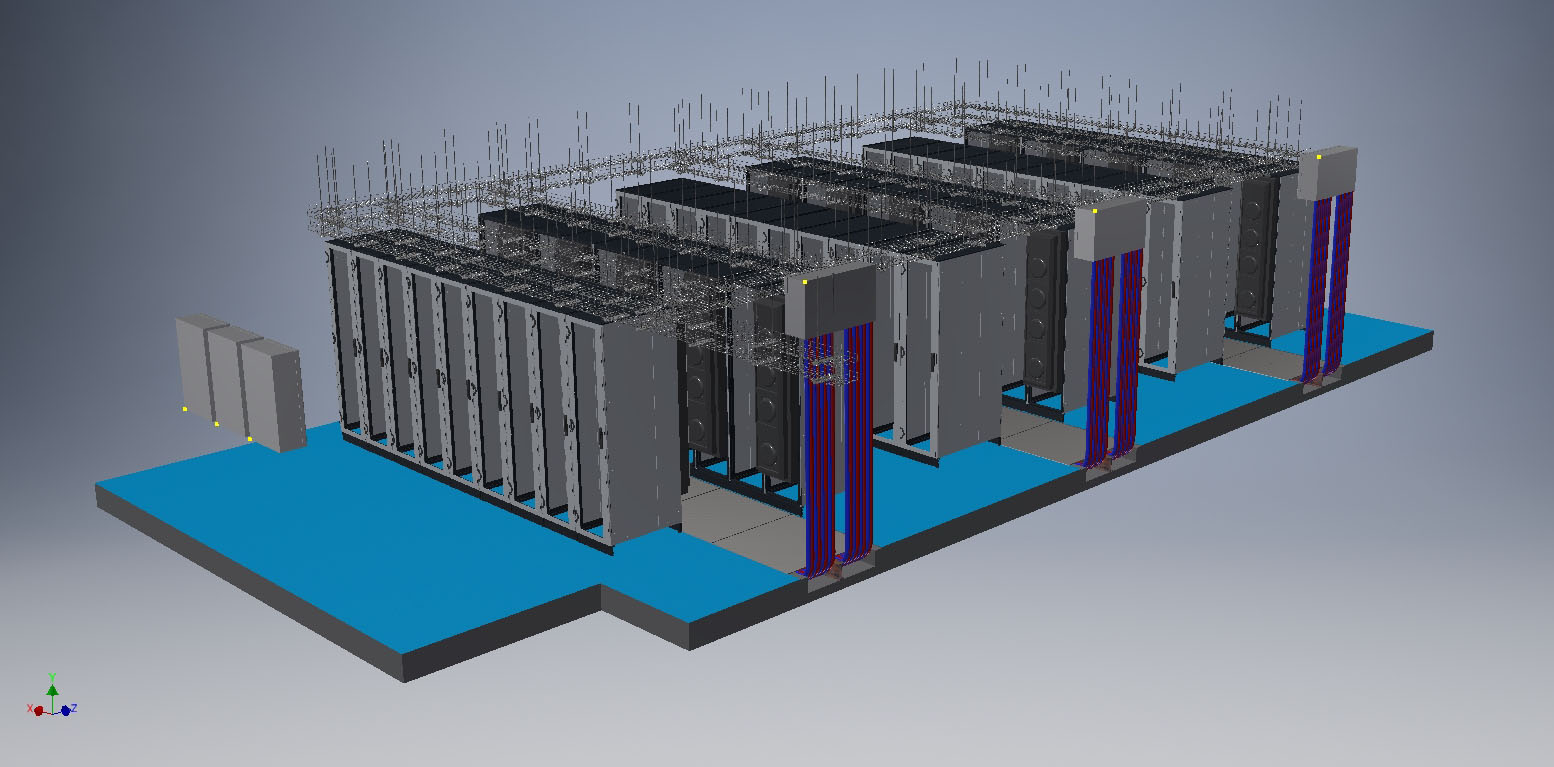}
\end{dunefigure}

Starting in Q3~2023, the data room will be ready for the installation of
the \dword{daq} servers servicing the first module described in
Section~\ref{sec:daq:considerations}.  This will proceed over the
next year, with servers being installed, cabled up, and tested.
As much configuring and commissioning work as possible will be done over
the network from the surface (or home institutions), to limit the number
of people underground.  Note that this data room is sized for all four
modules of \dword{daq} computing, so one quarter will be installed at this
point.  If more computing power is needed for the commissioning of the
first module (for example, to deal with unexpected noise levels), space
and power will be borrowed from the provision for future modules until the problems are
alleviated. 
Additional space for eight racks will be on the surface in the Ross Dry
Room. This will house the back-end computers, high level filter servers, and associated network equipment.

Starting in Q3~2024, the \dword{daq} will thus be ready to connect fibers to the
\dwords{wib} on the detector top as planes are installed, to allow their commissioning.

The underground installation phase of the \dword{daq} system has the largest safety
implications, which are discussed in Section~\ref{sec:daq:safety}.

\section{Organization and Project Management}
\label{sec:daq:organization}

\subsection{Consortium Organization}

The \dword{daq} consortium was formed in 2017 as a joint single and
dual phase consortium, with a consortium leader and a technical
leader. The current organization of the consortium is shown in
Figure~\ref{fig:daq-org}. The \dword{daq} institution board currently comprises
representatives from 34 institutions as shown in Table~\ref{tab:daq-ib}. The consortium leader is the spokesperson for the consortium and responsible for the overall scientific program and management of the group. The technical leader of the consortium is responsible for
managing the project for the group. The leadership is assisted in its duties by the Project Office, populated by the Resource Manager, the Software and Firmware coordinator and the Integration and Support Coordinator, providing support in the corresponding areas. 
The consortium is organized into working groups addressing the design,
R\&D, integration, and, in the future, construction, commissioning and installation of the key \dword{daq} systems. The Physics Performance and Facility working groups are not associated to a specific system but provide oversight of the general \dword{daq} performance in the physics context and the on the interface with the facility infrastructure. The \dword{daq} working group mandates are detailed in~\citedocdb{14938}.

\begin{dunefigure}[DAQ consortium org chart]{fig:daq-org}{Organizational chart for the \dword{daq} Consortium
}
  \includegraphics[width=0.9\textwidth]{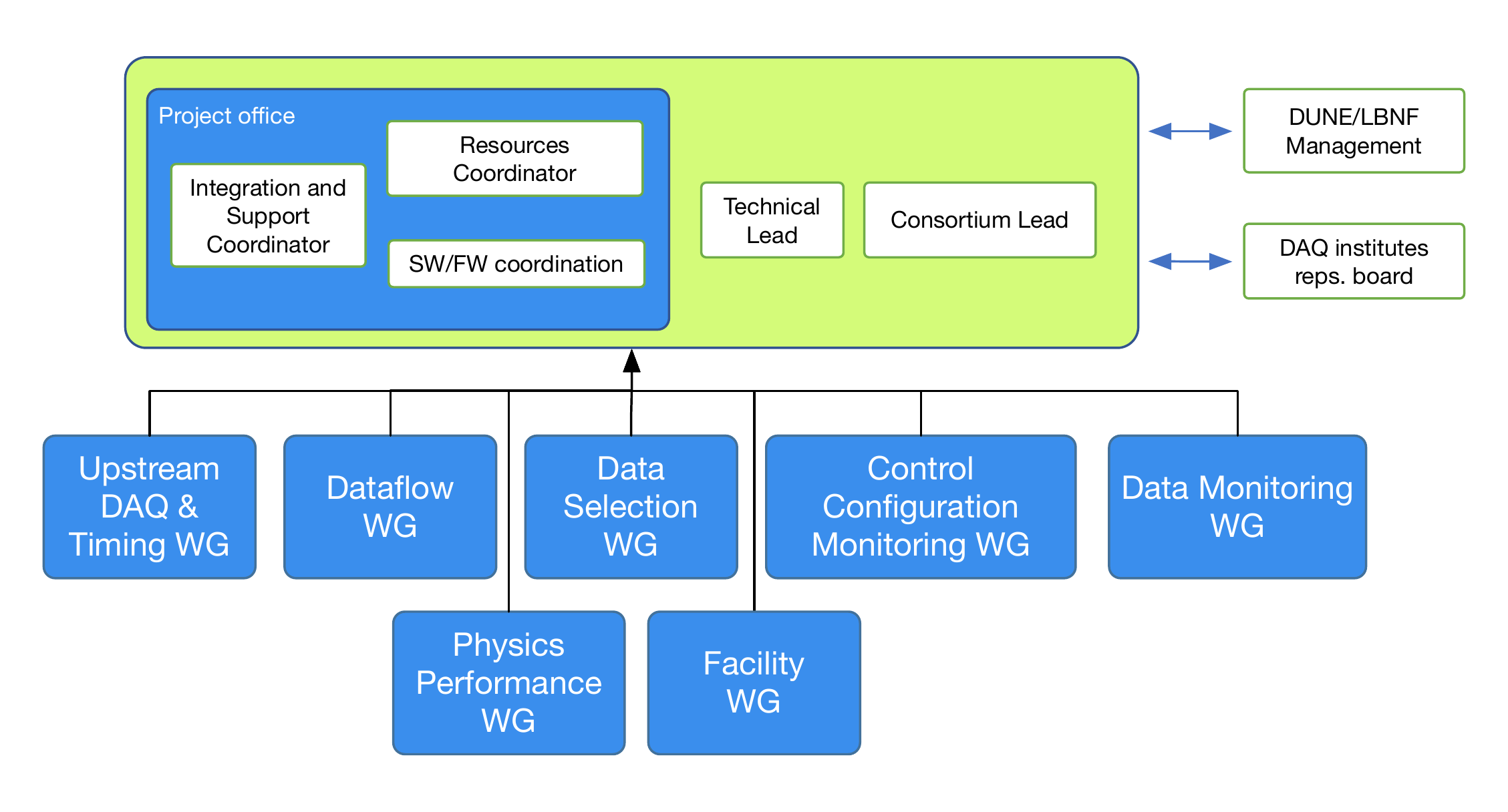}
\end{dunefigure}

\begin{dunetable}
[DAQ consortium institutions]
{p{0.65\textwidth}p{0.25\textwidth}}
{tab:daq-ib}
{DAQ Consortium Board institutional members and countries.}   
Member Institute & Country  \\ \toprowrule
CERN & CERN     \\ \colhline
Universidad Sergio Arboleda (USA) & Colombia     \\ \colhline
Czech Technical University & Czech Republic \\ \colhline
Lyon & France \\ \colhline
INFN Bologna & Italy \\ \colhline
Iwate & Japan     \\ \colhline
KEK & Japan     \\ \colhline
NIT Kure & Japan     \\ \colhline
NIKHEF & Netherlands    \\ \colhline
University of Birmingham & UK     \\ \colhline
Bristol University & UK     \\ \colhline
University of Edinburgh & UK     \\ \colhline
Imperial College London & UK     \\ \colhline
University College London (UCL) & UK     \\ \colhline
University of Liverpool & UK     \\ \colhline
Oxford University & UK     \\ \colhline
Rutherford Appleton Lab (RAL) & UK     \\ \colhline
University of Sussex & UK     \\ \colhline
University of Warwick & UK     \\ \colhline
Brookhaven National Lab (BNL) & USA     \\ \colhline
Colorado State University (CSU) & USA     \\ \colhline
Columbia University  & USA     \\ \colhline
University of California, Davis (UCD) & USA     \\ \colhline
Duke University & USA     \\ \colhline
University of California, Irvine (UCI) & USA     \\ \colhline
Fermi National Accelerator Laboratory (Fermilab) & USA     \\ \colhline
Iowa State University & USA     \\ \colhline
University of Minnesota, Duluth (UMD) & USA     \\ \colhline
University of Notre Dame & USA     \\ \colhline
University of Pennsylvania (Penn) & USA     \\ \colhline
South Dakota School of Mines and Technology (SDSMT) & USA     \\ \colhline
Stanford Linear Accelerator Lab (SLAC) & USA     \\ 
\end{dunetable}

\subsection{Schedule and Milestones}
\label{sec:daq:schedule}

The high-level \dword{daq} milestones are listed in Table~\ref{tab:daq-sched}, interleaved with the top-level DUNE project milestones, and illustrated in Figure~\ref{fig:daq:schedule}. Since the \dword{daq} project is largely based on commercial off-the-shelf components, it can be seen in the overall timeline that many of the components are procured relatively late in the project schedule.

\begin{dunetable}
[DAQ consortium schedule]
{p{0.65\textwidth}p{0.25\textwidth}}
{tab:daq-sched}
{\dword{daq} Consortium Schedule}   
Milestone & Date (Month YYYY)   \\ \toprowrule

Upstream \dword{daq} Architecture Technology Decision & June 2020 \\ \colhline
Engineering Design Review for Timing System &  June 2020   \\ \colhline
\rowcolor{dunepeach} Start of \dword{pdsp}-II installation& \startpduneiispinstall      \\ \colhline
Production Readiness Review for Timing System & June 2021 \\ \colhline
Preliminary Software Design Review & January 2022 \\ \colhline
Engineering Design Review for Hardware/Firmware & March 2022 \\  \colhline
\rowcolor{dunepeach} Start of \dword{pddp}-II installation& \startpduneiidpinstall      \\ \colhline
\rowcolor{dunepeach}South Dakota Logistics Warehouse available& \sdlwavailable      \\ \colhline
Start of Racks Procurement & July 2022  \\ \colhline
Start of \dword{daq} Server Procurement (I) & September 2022  \\ \colhline
\rowcolor{dunepeach}Beneficial occupancy of cavern 1 and \dword{cuc}& \cucbenocc      \\ \colhline 
Production Readiness Review for Readout Hardware/Firmware & December 2022  \\ \colhline
End of Racks Procurement & March 2023  \\ \colhline
Start of  \dword{daq} Custom Hardware Production &  March 2023    \\ \colhline
\rowcolor{dunepeach} \dword{cuc} counting room accessible& \accesscuccountrm      \\ \colhline
End of \dword{daq} Server Procurement (I) & May 2023  \\ \colhline
Start of \dword{daq} Installation & May 2023 \\ \colhline
\dword{daq} Software Final Design Review & June 2023  \\ \colhline

End of  \dword{daq} Custom Hardware Production &  December 2023    \\ \colhline
\rowcolor{dunepeach}Top of \dword{detmodule} \#1 cryostat accessible& \accesstopfirstcryo      \\ \colhline
\rowcolor{dunepeach}Start of \dword{detmodule} \#1 TPC installation& \startfirsttpcinstall      \\ \colhline
Start of \dword{daq} Server Procurement  (II) & September 2024  \\ \colhline
\rowcolor{dunepeach}Top of \dword{detmodule} \#2 accessible& \accesstopsecondcryo      \\ \colhline

End of \dword{daq} Server Procurement  (II) & May 2025  \\ \colhline
End of \dword{daq} installation & May 2025 \\ \colhline
\rowcolor{dunepeach}End of \dword{detmodule} \#1 TPC installation& \firsttpcinstallend      \\ \colhline
 \rowcolor{dunepeach}Start of \dword{detmodule} \#2 TPC installation& \startsecondtpcinstall      \\ \colhline
End of \dword{daq} Standalone Commissioning & December 2025 \\ \colhline
\rowcolor{dunepeach}End of \dword{detmodule} \#2 TPC installation& \secondtpcinstallend      \\ \colhline
\dword{daq} Server Procurement  (III) & July 2026  \\ \colhline
End of \dword{daq} Commissioning & December 2026  \\ 
\end{dunetable}

\begin{dunefigure}[DAQ schedule for first \SI{10}{\kilo\tonne} module]{fig:daq-schedule}{\dword{daq} schedule for first \SI{10}{\kilo\tonne} module. \label{fig:daq:schedule}}
  \includegraphics[width=0.95\textwidth,clip,trim=1cm 2cm 1cm 2cm]{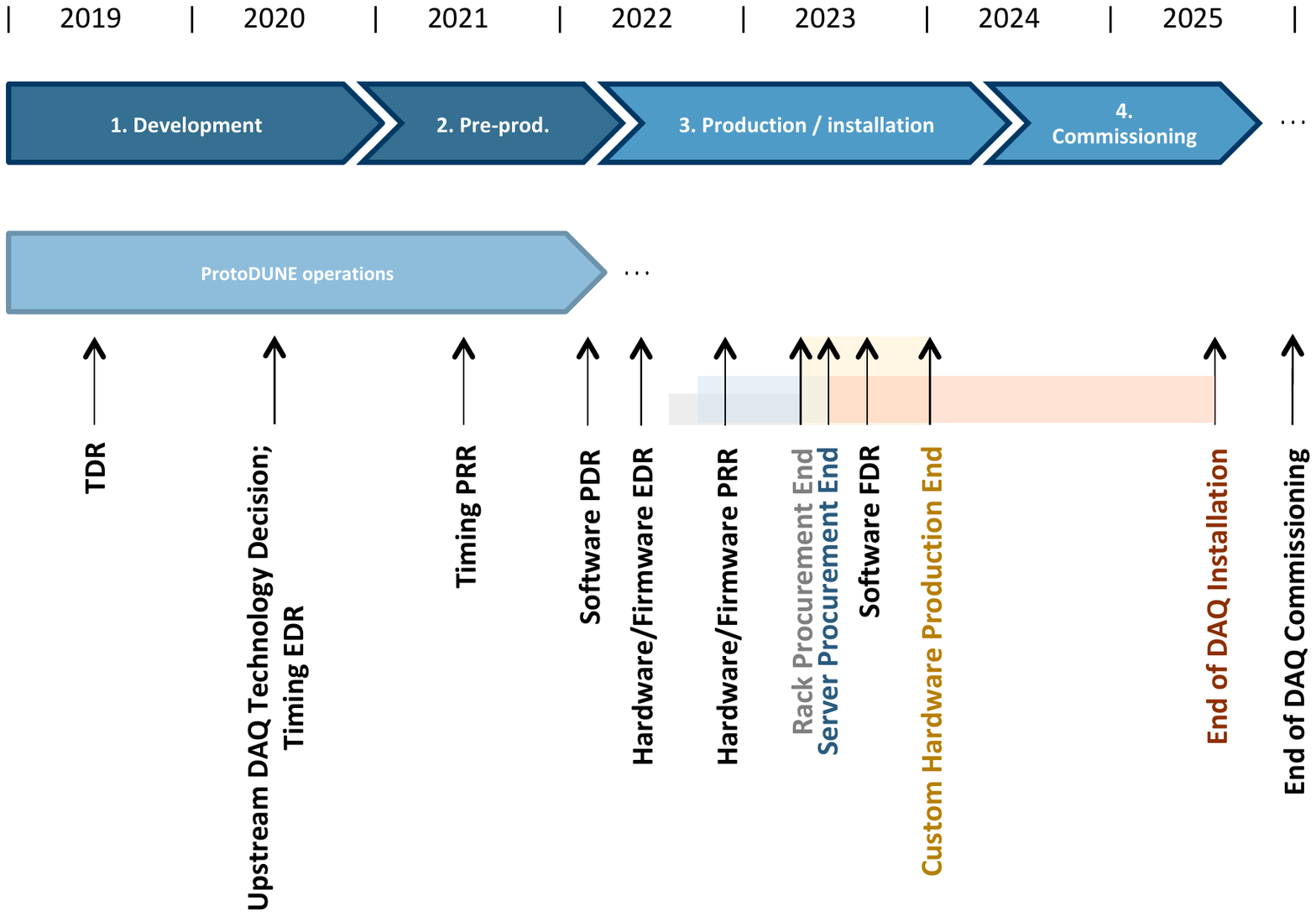}
\end{dunefigure}

\subsection{Safety and Risks}
\label{sec:daq:safety}

Personnel safety during design, construction, testing, installation, and
commissioning of the system is crucial for the success of the
project. Detector safety during installation,
commissioning and operations is also key to project success.
The consortium will strictly follow ES\&H guidelines for the
project as well as follow the safety rules of the institutions where
the work is performed, including national laboratories, SURF, and
participating universities.  

Two overall safety plans will be followed by the FD \dword{daq}. General work underground will comply
with all safety procedures in place for working in the detector
caverns and in the \dword{cuc} underground at
SURF. \dword{daq}-specific procedures for working with racks full of
electronics or computers, as defined 
at Fermilab, will be followed, especially with respect to electrical safety and the fire suppression
system chosen for the counting room. For example, a glass wall between the server room space and
the other areas in the CUC will be necessary to prevent workers in the server room from being
unseen if they are in distress, and an adequate hearing protection
regime must be put in place.

There are no other special safety items for the \dword{daq} system not already covered by the more general safety plans. The long-term emphasis is on remote operations capability from around the world, limiting the need for physical presence at SURF, and with underground access required only for urgent interventions or hardware replacement. 

A set of risks to the successful construction and operation of the \dword{daq} system has been
identified by the consortium, and is provided,
together with mitigation strategies and pre-mitigation risk level, 
in Table~\ref{tab:risks:SP-FD-DAQ}. Post-mitigation risk levels are
currently being re-evaluated. Risk is quantified with respect to
probability, cost impact, and schedule impact. High (H), medium (M), and low (L)
probability is identified as
$>25$\%, $10-25$\%, and $<10$\%, respectively; high (H), medium (M), and low (L)
cost impact is identified as
$>20$\%, $5-20$\%, and $<5$\% cost increase, respectively; and high
(H), medium (M), and low (L) schedule impact is identified as 
$>6$ months, $2-6$ months, and $<2$ months delay, respectively.

\begin{footnotesize}
\begin{longtable}{P{0.18\textwidth}P{0.20\textwidth}P{0.32\textwidth}P{0.02\textwidth}P{0.02\textwidth}P{0.02\textwidth}} 
\caption[DAQ risks]{DAQ risks (P=probability, C=cost, S=schedule) The risk probability, after taking into account the planned mitigation activities, is ranked as 
L (low $<\,$\SI{10}{\%}), 
M (medium \SIrange{10}{25}{\%}), or 
H (high $>\,$\SI{25}{\%}). 
The cost and schedule impacts are ranked as 
L (cost increase $<\,$\SI{5}{\%}, schedule delay $<\,$\num{2} months), 
M (\SIrange{5}{25}{\%} and 2--6 months, respectively) and 
H ($>\,$\SI{20}{\%} and $>\,$2 months, respectively).  \fixmehl{ref \texttt{tab:risks:SP-FD-DAQ}}} \\
\rowcolor{dunesky}
ID & Risk & Mitigation & P & C & S  \\  \colhline
RT-SP-DAQ-01 & Detector noise specs not met & ProtoDUNE experience with noise levels and provisions for data processing redundancy in DAQ system; ensure enough headroom of bandwidth to FNAL. & L & L & L \\  \colhline
RT-SP-DAQ-02 & Externally-driven schedule change & Provisions for standalone testing and commissioning of production DAQ components, and schedule adjustment & L & L & L \\  \colhline
RT-SP-DAQ-03 & Lack of expert personnel & Resource-loaded plan for DAQ backed by institutional commitments, and schedule adjustment using float & L & L & H \\  \colhline
RT-SP-DAQ-04 & Power/space requirements exceed CUC capacity & Sufficient bandwidth to surface and move module 3/4 components to an expanded surface facility & L & L & L \\  \colhline
RT-SP-DAQ-05 & Excess fake trigger rate from instrumental effects & ProtoDUNE performance experience, and provisions for increase in event builder and high level filter capacity, as needed; headroom in data link to FNAL. & L & L & L \\  \colhline
RT-SP-DAQ-06 & Calibration requirements exceed acceptable data rate & Provisions for increase in event builder and high level filter capacity, as neeed; headroom in data link to FNAL. & L & L & L \\  \colhline
RT-SP-DAQ-07 & Cost/performance of hardware/computing excessive & Have prototyping and pre-production phases, reduce performance using margin or identify additional funds & L & L & L \\  \colhline
RT-SP-DAQ-08 & PDTS fails to scale for DUNE requirements & Hardware upgrade & L & L & L \\  \colhline
RT-SP-DAQ-09 & WAN network & Extensive QA and development of failure mode recovery and automation, improved network connectivity, and personnel presence at SURF as last resort. & L & M & M \\  \colhline
RT-SP-DAQ-10 & Infrastructure & Design with redundancy, prior to construction, and improve power/cooling system. & M & M & L \\  \colhline
RT-SP-DAQ-11 & Custom electronics manifacturing issues & Diversify the manifacturers used for production; run an early pre-production and apply stringent QA criteria. & L & M & M \\  \colhline

\label{tab:risks:SP-FD-DAQ}
\end{longtable}
\end{footnotesize}

The following risks and mitigation strategies have been identified:

\begin{description}
\item[Detector noise specs not met] Excessive noise will make it
  impossible for the \dword{daqdss} to meet physics goals while
  generating reasonable data volumes. Prior (to construction) mitigation includes
  studying noise conditions at \dword{protodune}, and leaving provisions in the
  system for additional front-end filtering (in the form of the
  upstream \dword{daq} upgradable processing resources) and/or post-event builder processing (in
  the form of the high level filter). Mitigation (post-construction) includes augmenting
  filtering resources using a larger computing system for the high
  level filter.

\item[Externally-driven schedule change] The \dword{daq} has schedule links during
  testing, construction, and installation phases with most other
  subsystems. Schedule slip elsewhere will potentially cause
  delay to the \dword{daq}. Prior mitigation includes making provisions for
  stand-alone testing and commissioning of \dword{daq} components, in the form
  of vertical and horizontal slice tests, at \dword{protodune} or
  elsewhere. Mitigation includes adjusting schedule for stand-alone
  testing and commissioning phases. 

\item[Lack of expert personnel] A significant number of experts in
  hardware, software, firmware are needed, and must be sustained
  throughout the project. Lack of personnel will increase technical
  risks and cause delay. Prior mitigation includes developing a full
  resource-loaded plan for \dword{daq}, backed by national and institutional
  commitments, and avoiding single points of failure. Mitigation
  includes adaptation of the \dword{daq} schedule, using schedule float.

\item[Power/space requirements exceed CUC capacity] The CUC has fixed
  space and power allocation for \dword{daq} 
that cannot be exceeded.  Prior mitigation includes allowing
sufficient bandwidth up the shafts to move the upstream \dword{daq} components for
subsequent DUNE far detector modules (modules 3 and 4) to the
surface, or moving some of the \dword{daq} components to the detector caverns,
and carrying out a feasibility study for doing so. 
Mitigation includes expending additional resources on an
expanded surface facility.

\item[Excess fake trigger rate from instrumental effects] Instrumental
  effects (beyond excessive noise) can cause fake triggers. Prior
  mitigation includes studying \dword{protodune} performance in detail, and monitoring detector
  performance during installation. Mitigation includes substantially increasing
  data volume, and increasing processing resources in the high level
  filter.

\item[Calibration requirements exceed acceptable data rate]
  Calibration schemes may require substantial data 
volumes, far in excess of triggered data volume, beyond currently
envisioned; e.g., due to offline analysis inefficiencies. Prior
mitigation includes allowing for back-end (\dword{eb}) system expansion to cope
with the increased data rate, and allowing for a high-level filter data
selections stage to carry out online analysis and data
reduction. Mitigation includes increasing the back-end \dword{daq} and high
level filter system capacity. 

\item[Cost/performance of hardware/computing excessive] Costs of
  system-as-designed may exceed available budget, due to the IT technology (FPGA, servers, storage) market evolving in an unfavorable way. Prior mitigation
  includes the planning of prototyping and pre-construction phases to allow
  realistic appraisal of system costs, and applying sufficient margin
  in performance estimates.  Mitigation includes reducing performance
  or identifying additional funds.

\item[\dword{protodune} timing system fails to scale for DUNE requirements] The \dword{protodune} timing system concept may
  not scale to DUNE in scale or performance.  Prior mitigation
  includes testing the system at realistic scale before the final
  design. Mitigation includes replacing the system with upgraded
  hardware.

\item[WAN network] The network connectivity to the experiment from
  remote locations may be proven unstable, making remote control and monitoring
  inefficient. Prior mitigation includes ensuring that minimal human intervention
  is needed on the system for steady data taking and that automated
  error recovery is well developed. Mitigation includes effort to further
  improve automated data taking, and increased cost for improving network
  connectivity. In the worst case, one would foresee presence of personnel at SURF.

\item[Infrastructure] The power/cooling systems on which the \dword{daq}
  relies on cause more frequent than expected downtime. Prior
  mitigation includes designing, wherever possible, independent and redundant
  systems. Mitigation includes adding more uninterruptible power
  supplies, and improving the water cooling system to overcome otherwise
  degraded experiment uptime. 

\item[Custom electronics manufacturing issues] Large-scale production of high-speed custom
  electronics proves challenging, resulting in \dword{daq} installation delays. Prior mitigation includes diversifying manufacturers used for prototype production; assess manufacturer capability to meet specifications.  Mitigation includes running early pre-production with selected manufacturers, applying stringent QA criteria to ensure compliance with specifications.

\end{description}

\cleardoublepage

\chapter{Cryogenics Instrumentation and Slow Controls}
\label{ch:sp-cisc}

\section{Introduction} 

The \dfirst{cisc} consortium provides comprehensive monitoring for all detector  components and for \dword{lar} quality and behavior as well as a control system for many detector components.
The \dword{sp} and \dword{dp} modules both use the same control
system and have nearly identical cryogenics instrumentation except
for differences in location due to the different \dword{tpc}
geometries and the addition of dedicated instrumentation for
monitoring temperature and pressure in the gas phase for the
\dword{dpmod}.  \dpchcisc 
of this
\dword{tdr} is virtually the same as this chapter apart from
those few differences.

The consortium responsibilities are split into 
two main branches: cryogenics instrumentation and slow controls, as illustrated in  Figure~\ref{fig:cisc-subsystem-chart}. 

\begin{dunefigure}[CISC subsystem chart]{fig:cisc-subsystem-chart}
  {\dword{cisc} subsystem chart}
  \includegraphics[width=0.8\textwidth]{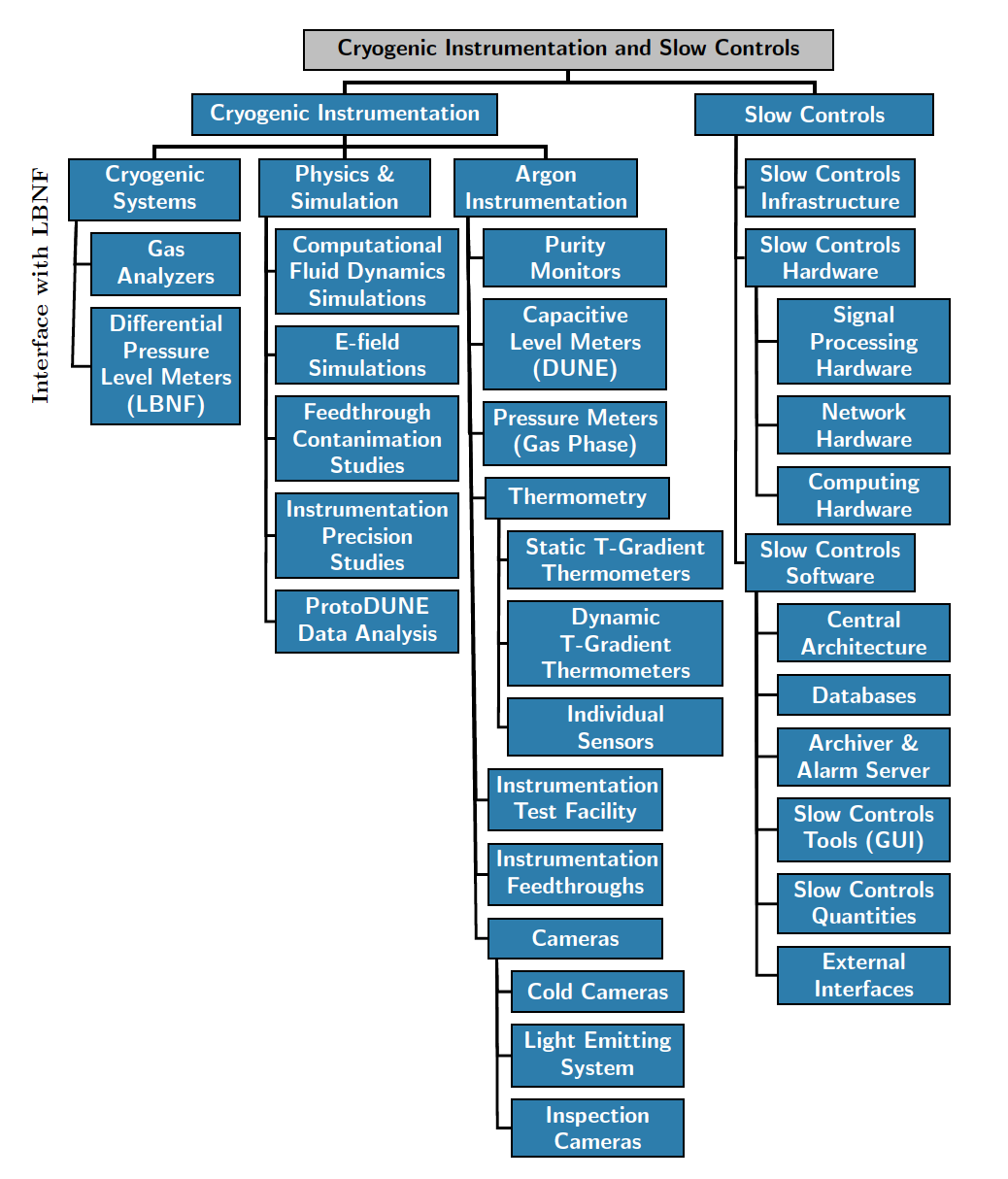}
\end{dunefigure}

Each element of \dword{cisc} contributes to the DUNE physics program primarily through the maintenance of high detector live time.  As described in Volume~\volnumberphysics{}, \voltitlephysics{}, of this \dword{tdr}, neutrino \dword{cpv} and resolution of the neutrino mass hierarchy over the full range of possible neutrino oscillation parameters will require at least a decade of running the \dword{fd}.  Similar requirements apply to searches for nucleon decay and \dword{snb} events from within our galaxy.  Throughout this long run-time the interior of any DUNE cryostat remains completely inaccessible.  No possibility exists for repairs to any components that could be damaged within the \dword{tpc} structure; hence environmental conditions that present risks must be detected and reported quickly and reliably. 
 
Detector damage risks peak during the initial fill of a module with \dword{lar}, as temperature gradients take on their highest values during this phase.  Thermal contractions outside of the range of design expectations could result in broken \dword{apa} wires, \dwords{sipm} in \dwords{pd} that detach from the \dword{xarapu} light detectors, or poor connections at the cathode \dword{hv} feedthrough point that could lead to unstable \efield{}s.  These considerations lead to the need for 
a robust temperature monitoring system for the detector, supplemented with liquid level monitors, and a high-performance camera system to enable visual inspection of the interior of the cryostat 
during the filling process.  These systems are fully described in Section~\ref{sec:fdsp-cryo-therm} of this chapter.
 
Argon purity must be established as early as possible in the filling process, a period in which gas analyzers are most useful, and must maintain an acceptable value, corresponding to a minimum electron drift lifetime of \SI{3}{ms}, throughout the data-taking period.  Dedicated  purity monitors (Section~\ref{sec:fdgen-slow-cryo-purity-mon}) 
provide precise lifetime measurements up to values of \SI{10}{ms}, the range over which electron attenuation most affects \dword{s/n} in the \dword{tpc}.  The purity monitors and gas analyzers remain important even after high lifetime has been achieved as periodic detector ``top-off''  fills occur; the new \dword{lar} must be of very high quality as it is introduced into the cryostat.
 
The \dword{cisc} system must recognize and prevent fault conditions that could develop in the \dword{detmodule} over long periods of running.  For example, the liquid level monitors must register any drop in liquid level; a drop in the level could place top sections of the \dword{fc} or bias \dword{hv} points  for the \dword{apa}s close enough to the gas-liquid boundary to trigger sparking events. 
Very slow-developing outgassing phenomena could conceivably occur, with associated bubble generation creating another source of \dword{hv} breakdown events.  The cold camera system enables detection and identification of bubbling sites, and the development of mitigation strategies such as lower \dword{hv} operation for some period of time.  A more subtle possibility is the formation of quasi-stable eddies in argon fluid flow that could prevent positive argon ions from being cleared from the \dword{tpc} volume, resulting in space charge build up that would not otherwise be expected at the depth of the \dword{fd}.  The space charge could in turn produce distortions in the \dword{tpc} drift field that degrade tracking and calorimetry performance.  The high-performance thermometry  of the DUNE \dword{cisc} system creates input for well developed complex fluid flow models described in Section~\ref{sec:fdgen-cryo-cfd} that should enable detection of  conditions associated with these eddies.
 
Finally, a high detector live-time fraction over multi-year operation cannot be achieved without an extensive system to monitor all aspects of detector performance, report this information in an intelligent fashion to detector operators, and archive the data for deeper offline studies.  Section~\ref{sec:sp-cisc-slowctrl} details the DUNE slow controls system designed for this task.
 
The baseline designs for all the \dword{cisc} systems have been used in \dfirst{pdsp}, 
and most design
parameters are extrapolated from these designs. The \dword{pdsp} data (and in some cases \dword{pddp} data) will therefore be used to validate the instrumentation designs and to understand their performance.

\subsection{Scope}

\subsubsection{Cryogenics Instrumentation}
Cryogenics instrumentation includes purity monitors,  various types of temperature monitors, and cameras with their associated light emitting systems. Also included are 
gas analyzers and \dword{lar} level monitors that are directly related to the external cryogenics system, which have substantial interfaces with the \dword{lbnf}. \dword{lbnf} provides the needed expertise  for these systems and is responsible for the design, installation, and commissioning, while the \dword{cisc} consortium provides the resources and supplements labor as needed. 

A \dword{citf} for the instrumentation devices is also part of the cryogenics instrumentation.
 \dword{cisc} is responsible for design through commissioning in the \dword{spmod} 
of \dword{lar} instrumentation devices: purity monitors, thermometers, capacitive level meters, cameras, and light-emitting system, and their associated feedthroughs.

Cryogenics instrumentation 
requires significant engineering, physics, and
simulation work, such as \efield simulations and cryogenics modeling
studies using \dfirst{cfd}. \efield simulations
identify desirable locations for instrumentation
devices in the cryostat, away from 
regions of high \efield, so that 
their presence does not induce large field distortions. 
\dword{cfd} simulations help identify 
expected temperature, impurity, and velocity flow distributions and guide the placement and distribution of instrumentation devices inside the cryostat.

\subsubsection{Slow Controls}
\label{sec:sp-cisc-slowctrl}
The slow controls portion of \dword{cisc} consists of three main components: 
hardware, infrastructure, and software. The slow controls hardware and infrastructure comprises networking hardware, signal processing hardware, computing hardware, and associated rack infrastructure. The slow controls software provides, for every slow control quantity, the central slow controls processing architecture, databases, alarms, archiving, and control room displays.

\dword{cisc} provides software and infrastructure for controlling and monitoring all detector elements that provide data on the health of the \dword{detmodule} or conditions important to the experiment, as well as  some related hardware. 

Slow controls base software and databases are the central tools needed to develop
control and monitoring for various detector systems and interfaces. These include:
\begin{itemize}
\item base input/output software;
\item alarms, archiving, display panels, and similar operator interface tools; and 
\item slow controls system documentation and operations guidelines.
\end{itemize}

Slow controls for external systems collect data from systems
external to the \dword{detmodule} and provide status monitoring for operators
and archiving. 
They 
collect data on beam status, cryogenics status,
\dword{daq} status, facilities systems status, interlock
status bit monitoring (but not the actual interlock mechanism), ground
impedance monitoring, and possibly building and detector hall
monitoring, as needed.

The \dword{ddss} can provide inputs to \dword{cisc} on safety interlock status, and \dword{cisc} will monitor and make that information available to the experiment operators and experts as needed. However, \dword{ddss} and \dword{cisc} are separate monitors, and the slow controls portion of \dword{cisc} does not provide any inputs to \dword{ddss}. A related question is whether \dword{cisc} can provide software intervention before a hardware safety interlock. In principle such intervention can be implemented in \dword{cisc}, presumably by (or as specified by) the hardware experts. For example, at \dword{pdsp}, the automatic lowering of \dword{hv} to clear streamers was implemented in the software for the \dword{hv} control using \dword{cisc}-level software.

Slow controls 
covers software interfaces for detector hardware devices, including:
\begin{itemize}
\item monitoring and control of all power supplies,
\item full rack monitoring (rack fans, thermometers and rack protection system),
\item instrumentation and calibration device monitoring (and control to the extent needed),
\item power distribution unit and computer hardware monitoring,
\item \dword{hv} system monitoring through cold cameras, and
\item detector components inspection 
using warm cameras.
\end{itemize}
%
\dword{cisc} will develop, install, and commission any hardware related to rack monitoring and control. Most power supplies may only need a cable from the device
to an Ethernet switch, but some power supplies might need special cables (e.g., GPIB or RS232) for communication. The \dword{cisc} consortium is responsible for providing these control cables.

\dword{cisc} 
has additional activities outside the scope of the consortium that require coordination with other groups. This is discussed in Section~\ref{sec:interfaces}.

\subsection{Design Considerations}

Important design considerations for instrumentation devices include stability, reliability, and longevity, so that devices can survive for at least \dunelifetime.
Such longevity is uncommon for any device, so the overall design allows replacement of devices where possible.
Some devices are critical for filling and commissioning but less critical for later operations; for these devices we specify a minimum lifetime of 18 months and 20 years as a desirable goal.
DUNE requires the \efield  on any instrumentation devices inside the cryostat to be less than \localefield to minimize the risk of dielectric breakdown in \dword{lar}. 

A consideration important for event reconstruction is the maximum noise level induced by instrumentation devices that the readout electronics  can tolerate. \dword{pdsp} is evaluating this. 
Table~\ref{tab:specs:SP-CISC} shows the top-level specifications that determine the requirements for \dword{cisc} together with selected high-level specifications for \dword{cisc} subsystems. The physics-driven rationale for each requirement and the proposed validation are also included in the table.
Tables~\ref{tab:fdgen-slow-cryo-requirements-1} and \ref{tab:fdgen-slow-cryo-requirements-2} show the full set of specifications 
for the \dword{cisc} subsystems. In all these tables two values are quoted for most of the design parameters: 
 (1) specification, which is the intended value or limits for the parameter set by physics and engineering needs, and (2) goal, an improved value offering a benefit which the collaboration aims to achieve where it is cost-effective to do so.

Data from purity monitors and different types of thermometers will be used to validate the \dword{lar} fluid flow model. 
A number of requirements drive the design parameters for the precision and granularity of monitor distribution across the cryostat. 
For example, the electron lifetime measurement precision must be \SI{1.4}{\%} to keep the bias on the charge readout in the \dword{tpc} below \SI{0.5}{\%} at \SI{3}{ms} lifetime. For thermometers, the 
parameters are driven by the \dword{cfd} simulations based on \dword{pdsp} design.
The temperature measurement resolution must be less than \SI{2}{mK}, and the relative precision of those measurements must be less than \SI{5}{mK}. The resolution is defined as the temperature \dword{rms}  for individual measurements and is driven by the electronics. The relative precision also includes the effect of reproducibility for successive immersions in \dword{lar}. 
The relative precision is particularly important in order to characterize 
gradients below \SI{20}{mK}. 
As will be described below, the laboratory calibration data and the recent analysis of thermometer instrumentation data from \dword{pdsp} shows that a \SI{2.5}{mK} relative precision is achievable. 

The level meters must have a precision of 0.1\% over \SI{14}{m} (i.e., \SI{14}{mm}) for measurement accuracy during filling. This precision is also sufficient to ensure that the \dword{lar} level stays above the \dwords{gp} of a \dfirst{sp} module. As shown in Table~\ref{tab:fdgen-slow-cryo-requirements-2}, several requirements drive the design of cold and warm cameras and the associated light emitting system. The components of the camera systems must not contaminate the \dword{lar} or produce bubbles 
so as  to avoid  increasing the risk of \dword{hv} discharge. Both cold and warm cameras must provide coverage of at least \SI{80}{\%} of the \dword{tpc} volume 
with a resolution of \SI{1}{cm} for cold cameras and \SI{2}{mm} for warm cameras on the \dword{tpc}.

For the \dword{citf}, a cryostat with a capacity of only \num{0.5} to approximately \SI{3}{m^3} 
will suffice and will keep turn-around times and filling costs 
lower. 
For gas analyzers, the operating range must allow establishment of useful electron lifetimes;
details  are in Table~\ref{tab:fdgen-slow-cryo-requirements-1}.

For slow controls, the system must
be sufficiently robust to monitor 
a minimum of 150,000 variables per \dword{detmodule}, and 
support a broad range of monitoring and archiving rates;
the estimated variable count, data rate, and archive storage needs
are discussed in Section \ref{sec:fdgen-slow-cryo-quant}.
The system must also
interface with a large number of detector subsystems and
establish two-way communication with them for control and monitoring. 
For the alarm rate, 150 alarms/day is used as the specification as it is the maximum to which humans can be expected to respond. The goal for the alarm rate is less than 50 alarms/day. The alarm logic system will need to include features for managing ``alarm storms'' using alarm group acknowledgment, summaries, delays, and other aids.

\begin{footnotesize}
\begin{longtable}{p{0.12\textwidth}p{0.18\textwidth}p{0.17\textwidth}p{0.25\textwidth}p{0.16\textwidth}}
\caption{CISC specifications \fixmehl{ref \texttt{tab:spec:SP-CISC}}} \\
  \rowcolor{dunesky}
       Label & Description  & Specification \newline (Goal) & Rationale & Validation \\  \colhline

   \newtag{SP-FD-5}{ spec:lar-purity }  & Liquid argon purity  &  $<$\,\SI{100}{ppt} \newline ($<\,\SI{30}{ppt}$) &  Provides $>$5:1 S/N on induction planes for  pattern recognition and two-track separation. &  Purity monitors and cosmic ray tracks \\ \colhline

  \newtag{SP-FD-15}{ spec:lar-n-contamination }  & LAr nitrogen contamination  &  $<\,\SI{25}{ppm}$ &  Maintain \SI{0.5}{PE/MeV} PDS sensitivity required for triggering proton decay near cathode. &  In situ measurement \\ \colhline

  \newtag{SP-FD-18}{ spec:cryo-monitor-devices }  & Cryogenic monitoring devices  &   &  Constrain uncertainties on detection efficiency, fiducial volume. &  ProtoDUNE \\ \colhline

  \newtag{SP-FD-25}{ spec:non-fe-noise }  & Non-FE noise contributions  &  $<<\,\SI{1000}\,e^- $ &  High S/N for high reconstruction efficiency. &  Engineering calculation and ProtoDUNE \\ \colhline
    
   \newtag{SP-FD-29}{ spec:dp-det-uptime }  & Detector uptime  &  $>\,$98\% \newline ($>\,$99\%) &  Meet physics goals in timely fashion. &  ProtoDUNE \\ \colhline
    
   \newtag{SP-FD-30}{ spec:dp-det-mod-uptime }  & Individual detector module uptime  &  $>\,$90\% \newline ($>\,$95\%) &  Meet physics goals in timely fashion. &  ProtoDUNE \\ \colhline

  \newtag{SP-CISC-1}{ spec:inst-noise }  & Noise from Instrumentation devices  &  $<<\,\SI{1000}\,e^- $ &  Max noise for 5:1 S/N for a MIP passing near cathode; per SBND and DUNE CE &  ProtoDUNE \\ \colhline
    
   \newtag{SP-CISC-2}{ spec:inst-efield }  & Max. E field near instrumentation devices  &  $<\,\SI{30}{kV/cm}$ \newline ($<\,\SI{15}{kV/cm}$) &  Significantly lower than max field of 30 kV/cm per DUNE HV  &  3D electrostatic simulation \\ \colhline
    
   \newtag{SP-CISC-3}{ spec:elec-lifetime-prec }  & Precision in electron lifetime  &  $<\,$1.4\% \newline ($<\,$1\%) &  Required for accurate charge reconstruction per DUNE-FD Task Force report. &  ProtoDUNE-SP and CITF \\ \colhline

  \newtag{SP-CISC-4}{ spec:elec-lifetime-range }  & Range in electron lifetime  &  \SIrange{0.04}{10}{ms} in cryostat, \SIrange{0.04}{30}{ms} inline &  Slightly beyond best values observed so far in other detectors.  &  ProtoDUNE-SP and CITF \\ \colhline
    
   \newtag{SP-CISC-11}{ spec:temp-repro }  & Precision: temperature reproducibility  &  $<\,\SI{5}{mK}$ \newline (\SI{2}{mK}) &  Enables validation of CFD models, which predicts gradients below 15 mK &  ProtoDUNE-SP and CITF \\ \colhline
    
   \newtag{SP-CISC-14}{ spec:temp-stability }  & Temperature stability  &  $<\,\SI{2}{mK}$ at all places and times \newline (Match precision requirement at all places, at all times) &  Measure the temp map with sufficient precision during the entire duration &  ProtoDUNE-SP \\ \colhline
    
   \newtag{SP-CISC-27}{ spec:camera-cold-coverage }  & Cold camera coverage  &  $>\,$80\% of HV surfaces \newline (\num{100}\%) &  Enable detailed inspection of issues near HV surfaces. &  Calculated from location, validated in prototypes. \\ \colhline
    
   \newtag{SP-CISC-51}{ spec:slowcontrol-alarm-rate }  & Slow control alarm rate  &  $<\,$150/day \newline ($<\,$50/day) &  Alarm rate low enough to allow response to every alarm. &  Detector module; depends on experimental conditions \\ \colhline
    
   \newtag{SP-CISC-52}{ spec:slowcontrol-num-vars }  & Total No. of variables  &  $>\,\num{150000}$ \newline (\SIrange{150000}{200000}{}) &  Scaled from ProtoDUNE-SP &  ProtoDUNE-SP and CITF \\ \colhline
    
   \newtag{SP-CISC-54}{ spec:slowcontrol-archive-rate }  & Archiving rate  &  \SI{0.02}{Hz} \newline (Broad range \SI{1}{Hz} to \num{1} per few min.) &  Archiving rate different for each variable, optimized to store important information  &  ProtoDUNE-SP \\ \colhline

\label{tab:specs:SP-CISC}
\end{longtable}
\end{footnotesize}

\begin{dunetable}
[Specifications for CISC subsystems (1)]
{p{0.45\linewidth}p{0.25\linewidth}p{0.25\linewidth}}
{tab:fdgen-slow-cryo-requirements-1}
{List of specifications for the different CISC subsystems (1).}   
Quantity/Parameter				                             & Specification			                                        & Goal		                                              \\ \toprowrule                     
Noise from Instrumentation devices				             & $\ll$ \elecnoisefe                                      & 
\\ \colhline                     
Max. \efield near instrumentation devices				     & <\localefield			                                                & <15 kV/cm		                                          \\ \colhline                     
\textbf{Purity Monitors}	                                             &                                                                      &                                                         \\ \colhline                      
Precision in electron lifetime				                 & <1.4\% at 3~ms,  <4\% at 9~ms,  relative differences <2.5\%			                                            & < 1\%		                                              \\ \colhline                     
Range in electron lifetime				                     & 0.04 - 10 ms  			                    & (0.04 - 30 ms inline)       
\\ \colhline                         
Longevity				                                     & \dunelifetime			                                                    & > \dunelifetime		                                      \\ \colhline                     
Stability				                                     & Match precision requirement at all places/times			    & 
\\ \colhline  	                   
Reliability				                                     & Daily Measurements			                                        & Measurements 
as needed	  \\ \colhline                         
\textbf{Thermometers}	                                             &                                                                      &                                                         \\ \colhline                      
Vertical density of sensors for T-gradient monitors			 & > 2 sensor/m			                                                & > 4 sensors/m		                                      \\ \colhline                 
2D horizontal density for top/bottom individual sensors		 &  1 sensor/5(10) m 			                                        &  1 sensor/3(5) m 		                          
        \\ \colhline 
Swinging/deflection of T-Gradient monitors                   &  < 5 cm 			                                                    &  < 2 cm 		          
        \\ \colhline                     
Resolution of temperature measurements				         & < 2 mK			                                                    & <0.5 mK		                                          \\ \colhline                         
Precision: temperature reproducibility 				         & < 5 mK			                                                    & 2 mK		                                              \\ \colhline                     
Reliability				                                     & 80\% (in 18 months)			                                        & 50\% (during 20 years)		                              \\ \colhline                     
Longevity				                                     & > 18 months			                                                & > 20 years		                                      \\ \colhline                         
Stability 	  &  < \SI{2}{mK} at all places and times	 &   Match precision requirement at all places/times \\ \colhline                 
Discrepancy between lab and  in situ calibrations for temperature sensors			             & < 5 mK			                                                    & < 3 mK		                                          \\ \colhline                           
Discrepancy between measured temperature map and CFD simulations in \dword{pdsp}	 & < 5 mK & 
\\ \colhline                             
\textbf{Gas Analyzers}	   &   &  \\ \colhline            
Operating Range O2	 & 0.2 (air) to 0.1 ppt  & 
\\ \colhline    
Operating Range H2O				                             & 
Nom. air to sub-ppb; contaminant-dependent & 
\\ \colhline           
Operating Range N2				                             & Nominally Air Nom. air to sub-ppb; contaminant-dependent	& 
\\ \colhline             
Precision: 1 sigma at zero				                     & 
per gas analyzer range
& 
\\ \colhline     
Detection limit: 3 sigma & Different analyzer modules needed to cover entire range	& 
\\ \colhline           
Stability   & <\% of full scale range.		 & 
\\ \colhline         
Longevity		 & >10 years	  & 
\\   \colhline
\textbf{Pressure Meters (GAr)}	          &    &          \\ \colhline            
Relative precision (DUNE side)		   & 0.1~mbar	& 
\\ \colhline  
Absolute precision (DUNE side)		   & <5~mbar	& 
\\  
\end{dunetable}

\begin{dunetable}
[Specifications for CISC subsystems (2)]
{p{0.45\linewidth}p{0.25\linewidth}p{0.25\linewidth}}
{tab:fdgen-slow-cryo-requirements-2}
{List of specifications for the different CISC subsystems (2)}   
Quantity/Parameter		     & Specification	  & Goal   \\ \toprowrule   
\textbf{Level Meters}	          &    &          \\ \colhline            
Precision (LBNF scope)		   & 0.1\% over 14 m (14 mm)			                                    & 
\\ \colhline           
Precision (capacitive level meters, \dword{dune} scope) & 1~cm  &  <5 mm
\\ \colhline         
Longevity (all)		  & 20 years	   & > 20 years		                                                  \\ \colhline     
\textbf{Cold cameras}	                                             &                                                                      &                                                                     \\ \colhline        
Coverage				                                     & 80\% of the exterior of HV surfaces			                        & 100\% 	                                                          \\ \colhline         
Frames per second	   & yet to be defined	  & 
\\ \colhline             
Resolution 	 & 1 cm on the \dword{tpc}	 & yet to be defined
\\ \colhline           
Duty cycle	  & yet to be defined	 & 
\\ \colhline         
longevity			 & > 18 months			                                                & > 20 years		                                                  \\ 
\textbf{Inspection cameras}	     &                                                                      &                                                                     \\ \colhline        
Coverage	 & 80\% of the \dword{tpc}		  & yet to be defined		                                              \\ \colhline         
Frames per second		   & yet to be defined	   & 
\\ \colhline             
Resolution 	  & 2 mm on the \dword{tpc}			                                            & yet to be defined		                                              \\ \colhline           
heat transfer	  & no generation of bubbles			                                & 	
\\ \colhline         
longevity			  & > 18 months			                                                & > 20 years		                                                  \\ \colhline         
\textbf{Light emitting system}	                                     &                                                                      &                                                                     \\ \colhline        
radiant flux     & > 10 mW/sr		 & 100 mW/sr \\ \colhline         
power				     & < 125 mW/LED			                                        & 
\\ \colhline           
wavelength	   & red/green		 & IR/white	   \\ \colhline         
longevity	  & > 18 months (for cold cameras) 			                            & > 20 years		                                              \\ \colhline         
\textbf{\dfirst{citf}}	                 &                                                                      &                                                                     \\ \colhline            
Dimensions		  & 0.5 to 3  cubic meters 			                                    & 
\\ \colhline             
Temperature stability	 & $\pm$1K	 & 
\\ \colhline                                       
Turn-Around time	 & $\sim\,$9 days   & 9 days 	  \\ \colhline                                       
LAr purity		   & O2, H2O: low enough  to measure drifting electrons of devices under test, $\sim\,\SI{0.5}{ms}$.    N2: ppm for scintillation light tests. 	        &  >1.0 ms                                                            \\ \colhline
\textbf{Slow Controls}		                                         &                                                                      &                                                                     \\ \colhline
Alarm rate	  & <150/day			                                                    &  < 50/day                                                           \\ \colhline
Total No. of variables per \dword{detmodule}				                         & 150,000			                                                    &  150,000 - 200,000                                                   \\ \colhline
Server rack space				                             & 2 racks			                                                    &  3 racks                                                            \\ \colhline
Archiving rate 				                                 & 0.02 Hz			                                                    &  Broad range 1 Hz  to 1 per few min.                                \\ \colhline
Near Detector Status & Beam conditions and detector status	                                &  Full beam and detector status                                      \\          
\end{dunetable}                                  

\subsection{Fluid Dynamics Simulation}
\label{sec:fdgen-cryo-cfd}

Proper placement of purity monitors, thermometers, and liquid level monitors within the \dword{detmodule} requires knowing how \lar flows within the cryostat, given its fluid dynamics, heat and mass transfer, and distribution of impurity concentrations. Fluid flow is also important in understanding how the positive and negative ion excess created by various sources (e.g., ionization from cosmic rays and $^{36}$Ar) 
 is distributed across the detector as it affects \efield uniformity. 
Finally, \dword{cfd} simulations are crucial to predict the purity of the argon in regions where experimental data is unavailable. The overall goal of the \dword{cfd} simulations
is to better understand and predict the fluid (in either liquid or vapor state) motions and the implications for detector performance. 

Fluid motion within the cryostat is driven primarily by small changes in density caused by thermal gradients within the fluid although pump flow rates and inlet and outlet locations also contribute. Heat sources include exterior heat from the surroundings, interior heat from electronics, and heat flow through the pump inlet. In principle, purity monitors can be placed throughout the cryostat to determine if the argon is pure enough for experimentation. However, some areas inside the cryostat are off limits for such monitors.

The fluid flow behavior can be determined by simulating \dword{lar} flow within a \dword{detmodule} 
using Siemens Star-CCM$+$\footnote{https://mdx.plm.automation.siemens.com/star-ccm-plus}, a commercially available \dword{cfd} code.  Such a model must properly define the fluid characteristics, solid bodies, and fluid-solid interfaces, as well as provide a way to measure contamination, while still maintaining reasonable computation times. In addition, these fluid dynamics simulations can be compared to available experimental data to assess simulation accuracy and credibility.

Although simulation of the \dword{detmodule} presents challenges, 
acceptable simplifications can 
accurately represent the fluid, the interfacing solid bodies, and variations of contaminant concentrations. Because of the magnitude of thermal variation within the cryostat, modeling of the \dword{lar} is simplified by using constant thermophysical properties, calculating buoyant force with the Boussinesq Model (using a constant density for the fluid with application of a temperature-dependent buoyant force), and a standard shear stress transport turbulence model. Solid bodies that touch the \dword{lar} include the cryostat wall, cathode planes, anode planes, \dword{gp}, and \dword{fc}. As in previous \dword{cfd} models of the \dword{dune} \dword{35t} and \dword{pdsp}
\cite{bib:docdb5915}, the \dword{fc} planes, anode planes, and \dword{gp} can be represented by porous bodies. Because impurity concentration and electron lifetime do not affect fluid flow, these variables can be simulated as passive scalars, as is commonly done for smoke released \cite{cfd-1} 
in air or dyes released in liquids.

Discrepancies between real data and simulations may affect detector performance. 
Simulation results contribute to decisions about where to place sensors and monitors, and to 
the definitions of various calibration quantities. Methods of mitigating such risks include well established convergence criteria, sensitivity studies, and comparison to results of previous \dword{cfd} simulation work. Moreover, the simulation will be improved with input from \lar temperature and purity measurements and validation tests from \dword{pdsp}\footnote{Because \dword{pddp} was not instrumented with high-precision thermometers in the liquid phase and because the cryogenics design is the same for \dword{sp} and \dword{dp} modules of the \dword{dune} \dword{fd}, \dword{pdsp} data will be used to validate the liquid \dword{cfd} model.}.

Taking into account that the \dword{cfd} model can predict both temperature and impurity levels, the procedure for validating and tuning the \dword{cfd} model will be the following: (1) use temperature measurements in numerous locations in the cryostat to constrain temperature predictions and  improve the \dword{cfd} model, (2) use the improved model  to predict the \lar impurity level at the purity monitor locations, and (3) compare the predictions to the actual  purity monitor measurements to further constrain the \dword{cfd} model.  

Figure~\ref{fig:cfd-example} shows an example of the temperature
distribution on a plane intersecting a \dword{lar} inlet and at a
plane halfway between an inlet and an outlet; 
the geometry used for
this simulation is shown in Figure~\ref{fig:cfd-example-geometry}\footnote{The inlet and outlet map has recently changed; it now consists of two rows of 64 inlets each at each longer side of the cryostat and four outlets along the shorter sides (drift direction) of the cryostat.}. Note the plume of higher temperature \dword{lar} between the walls and
the outer \dword{apa} on the inlet plane. The current placement of instrumentation in
the cryostat as shown in Figure~\ref{fig:cisc-tsensor-map} was determined using temperature and impurity distributions from previous simulations.

\begin{dunefigure}[\dshort{cfd} example]{fig:cfd-example}
  {Distribution of temperature on a plane intersecting an inlet (left) and halfway between an inlet and an outlet (right), as predicted by previous \dword{cfd} simulations (from~\cite{bib:docdb5915}). (See Figure~\ref{fig:cfd-example-geometry} for geometry.)}
  \includegraphics[height=0.4\textwidth]{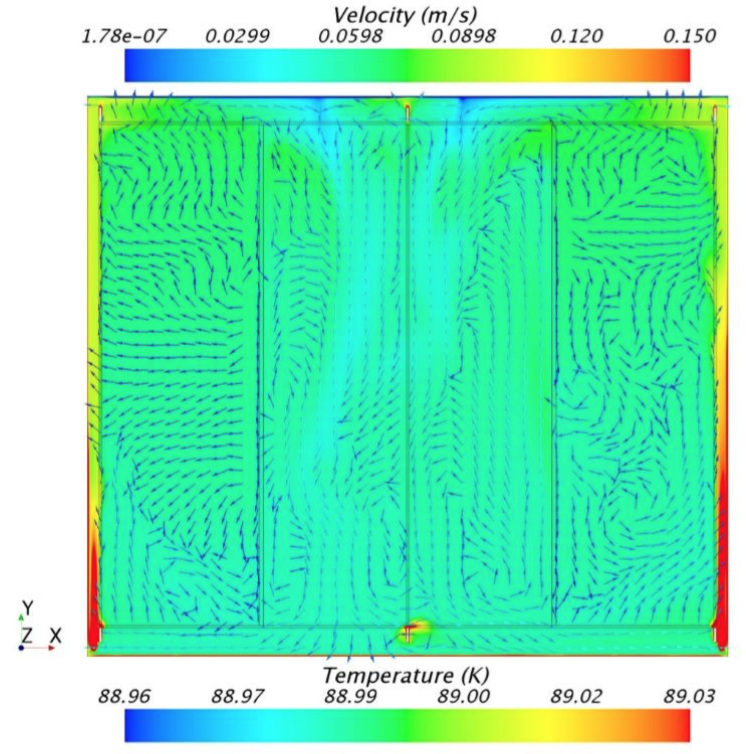}
  \includegraphics[height=0.4\textwidth]{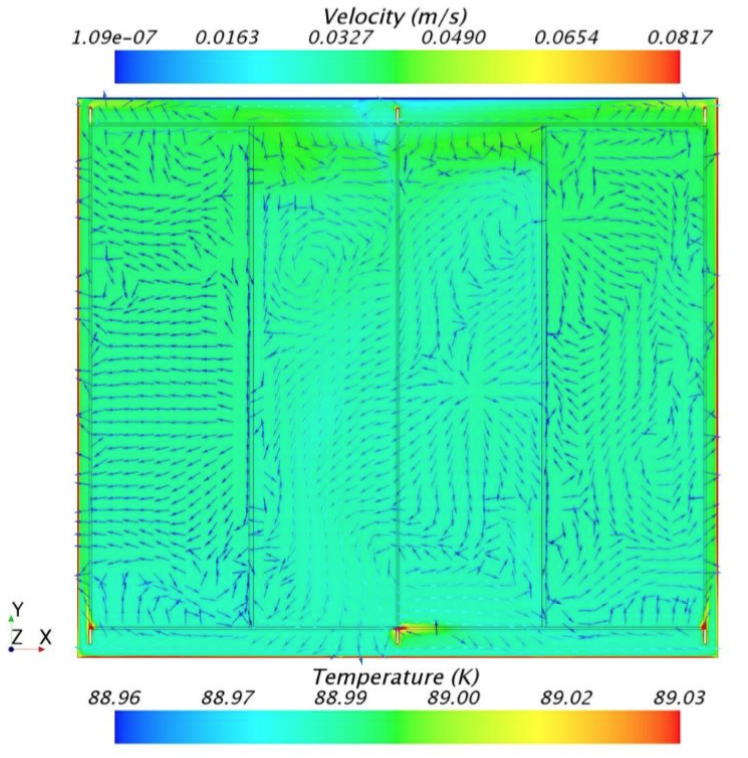}
\end{dunefigure}

\begin{dunefigure}[\single CISC geometry layout]{fig:cfd-example-geometry}
  {Layout of the \dword{spmod} within the cryostat (top) and positions of \dword{lar} inlets and outlets (bottom) as modeled in the \dword{cfd} simulations~\cite{bib:docdb5915}. The $y$ axis is vertical and the $x$ axis is parallel to the \dword{tpc} drift direction. Inlets are shown in green and outlets are shown in red.}
  \includegraphics[width=0.7\textwidth]{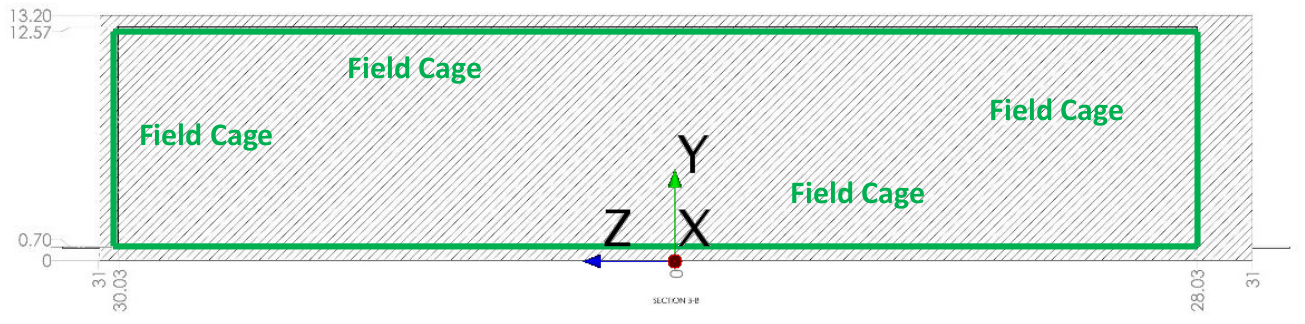}
  \includegraphics[width=0.7\textwidth]{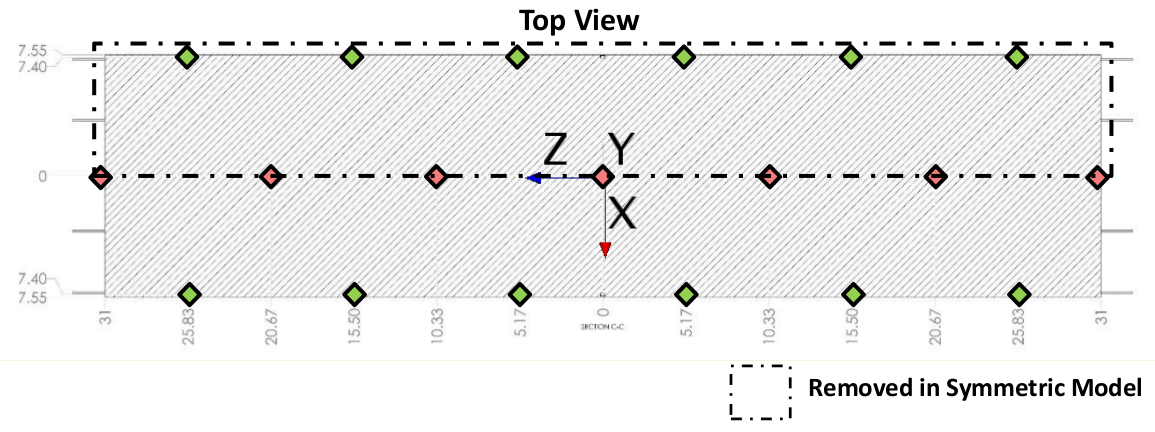}
\end{dunefigure}

The strategy for  future \dword{cfd} simulations begins with understanding the performance of the \dword{pdsp} cryogenics system and modeling the \dwords{detmodule} to derive specifications 
for instrumentation. We are pursuing a prioritized set of studies to help determine the requirements for other systems. We plan to 
\begin{itemize}
\item Review the \dword{dune} \dword{fd} cryogenics system design and verify the current implementation in simulation 
to ensure that the simulation represents the actual design.
\item 
Model the \dword{pdsp} liquid and gas regions with the same precision as the \dword{fd}. Presently, we have only the liquid model, which is needed to interpret the thermometer data. The gas model is needed to see how to place thermometers in the ullage and verify the design of the gaseous argon purge system.
\item Verify the \dword{cfd} model for the \dword{spmod} in a simulation performed by LBNF; this defines the requirements for instrumentation devices (e.g., thermometry).
\end{itemize}

\subsubsection{Validation in ProtoDUNE}
\label{sec:cfdvalid}
\dword{pdsp} has collected data to validate the \dword{cfd} using: 
\begin{itemize}
\item static and dynamic T-gradient thermometers, 
\item individual temperature sensors placed in the return \dword{lar} inlets, 
\item two \twod grids of individual temperature sensors installed below the bottom ground planes and above the top ground planes, 
\item a string of three purity monitors vertically spaced from near the bottom of the cryostat to just below the \dword{lar} surface,
\item two pressure sensors (relative and absolute) in the argon gas,
\item H$_{2}$O, N$_{2}$, and O$_{2}$ gas analyzers, 
\item \dword{lar} level monitors, and
\item standard cryogenic sensors including pressure transducers, individual temperature sensors placed around
the cryostat on the membrane walls, and recirculation flow rates transducers.
\end{itemize}

The data, which have been logged through the \dword{pdsp} slow control system \cite{pdspdcs_proc}, are available for offline analysis. 

In parallel, \dword{cisc} has produced a \dword{pdsp} \dword{cfd} model 
with input from \dword{pdsp} measurements (see Table~\ref{tab:fdgen-cisc-CFDparam}). Streamlines\footnote{In fluid mechanics, a streamline is a line that is everywhere tangent to the local velocity vector. For steady flows, a streamline also represents the path that a single particle of the fluid will take from inlet to exit.} from the current  simulation (Figure~\ref{fig:cisc-cfd-larflow-inlets}) show the flow paths from the four cryostat inlets to the outlet. The validation of this model consists of an iterative process in which several versions of the \dword{cfd} simulation, using different input parameters, eventually 
converge 
to a reasonable agreement with data from instrumentation devices. Those comparisons will be shown in Section~\ref{sec:fdgen-slow-cryo-temp-ana}.

\begin{dunetable}
[CFD parameters for ProtoDUNE]
{p{0.24\textwidth}p{0.17\textwidth}p{0.49\textwidth}}
{tab:fdgen-cisc-CFDparam}
{\dword{cfd} input parameters for \dword{protodune}-SP}

Parameter  &	Value &	Comments \\ \colhline
Cryostat height
&
7.878 m
&
Measured with laser (1 cm error approx.)
\\  \colhline
LAr surface height
&
7.406 m
&
Measured by capacitive level meter ($<1$ cm error)
\\  \colhline	
Ullage pressure		
&
1.045 bar
&
Measured by pressure gauges
\\  \colhline
\lar surface temperature
&
87.596 K
&
Computed using ullage pressure and \cite{larpropertiesbnl}
\\  \colhline
\lar inlet temperature
&
bulk LAr + 0.2 K
&
Estimated from pressure settings in cryo-system
\\  \colhline
\lar flow rate per pipe
&
0.417 kg/s
& Estimated from cryostat filling rate 
\\
\end{dunetable}

\begin{dunefigure}[Streamlines for \lar flow inside  ProtoDUNE-SP]{fig:cisc-cfd-larflow-inlets}
  {Streamlines for \dword{lar} flow inside  \dword{pdsp}}
  \includegraphics[width=0.8\textwidth]{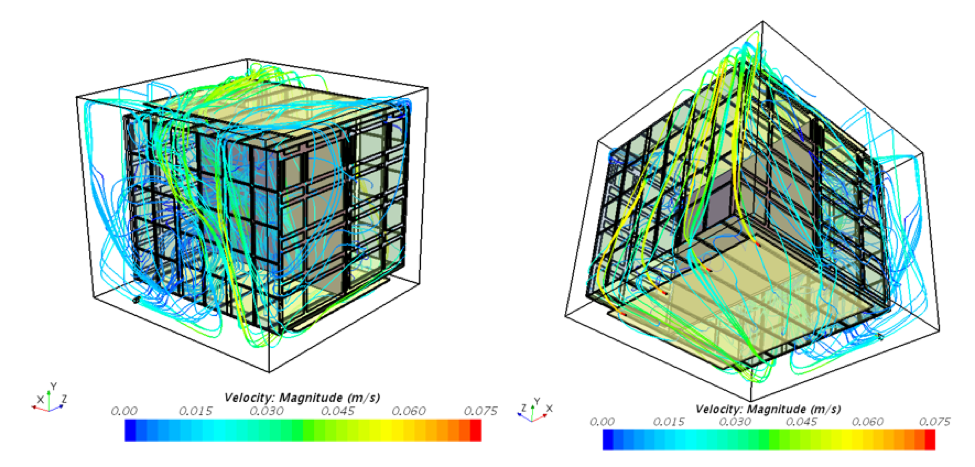}
\end{dunefigure}

Once the \dword{pdsp} \dword{cfd} model 
predicts the fluid temperature in the entire cryostat to a reasonable level under different conditions, we will use it 
to produce maps of impurity levels in the \dword{detmodule}. These can be easily converted into electron lifetime maps, which we will 
compare to the 
\dword{pdsp} purity monitor data.

\section{Cryogenics Instrumentation}
\label{sec:fdgen-cryo-instr}
Instrumentation inside the cryostat must accurately report the condition of the \dword{lar} so that we can ensure that it is adequate to operate the \dword{tpc}.
This instrumentation includes 
purity monitors 
to check the level of impurity in the argon and 
to provide high-precision electron lifetime measurements,
as well as gas analyzers to verify that the levels of atmospheric contamination do not rise above 
certain limits during the cryostat purging, cooling, and filling. 
Temperature sensors deployed in vertical arrays and at the top and bottom of the \dword{detmodule} monitor the cryogenics system operation, providing a 
detailed \threed temperature map that helps predict the \dword{lar} purity across the entire cryostat. The cryogenics instrumentation also includes \dword{lar} level monitors and
a system of internal cameras to help find sparks in the cryostat and 
to monitor the overall cryostat interior. 

The proper placement of purity monitors, thermometers, and liquid-level monitors in the \dword{detmodule} requires 
understanding the \dword{lar} fluid dynamics, heat and mass transfer, and the distribution of impurity concentrations within the cryostat. 
Both this and 
coherent analysis of the instrumentation data require \dword{cfd} simulation results.

\dword{pdsp} is testing the performance of 
purity monitors, thermometers, level monitors and cameras
for the \dword{spmod}, validating the baseline  
design.

\subsection{Thermometers}
\label{sec:fdsp-cryo-therm}
As discussed in Section~\ref{sec:fdgen-cryo-cfd}, a detailed \threed temperature map is important for monitoring 
the cryogenics system for correct functioning and the \dword{lar} for uniformity.
Given the complexity and size of purity monitors, they can only be installed on the sides of the cryostat to provide a local measurement of
\dword{lar} purity.  
A direct measurement of the \dword{lar} purity across the entire cryostat is not feasible, but a sufficiently detailed \threed temperature map based on \dword{cfd} simulations can predict it. The vertical coordinate is especially important because it will relate closely to the 
\dword{lar} recirculation and uniformity. 

The baseline sensor distribution and the cryostat ports used to extract cables (with indication of number of cables per port) are shown in Figure~\ref{fig:cisc-tsensor-map}. The baseline distribution will evolve as more information becomes available (precise \dword{cfd} simulations, better understanding of \dword{dss} ports, installation interfaces with other groups), but the baseline suffices to establish the overall strategy.

\begin{dunefigure}[Distribution of temperature sensors inside the cryostat]{fig:cisc-tsensor-map}
  {Distribution of temperature sensors inside the cryostat}
  \includegraphics[width=0.95\textwidth]{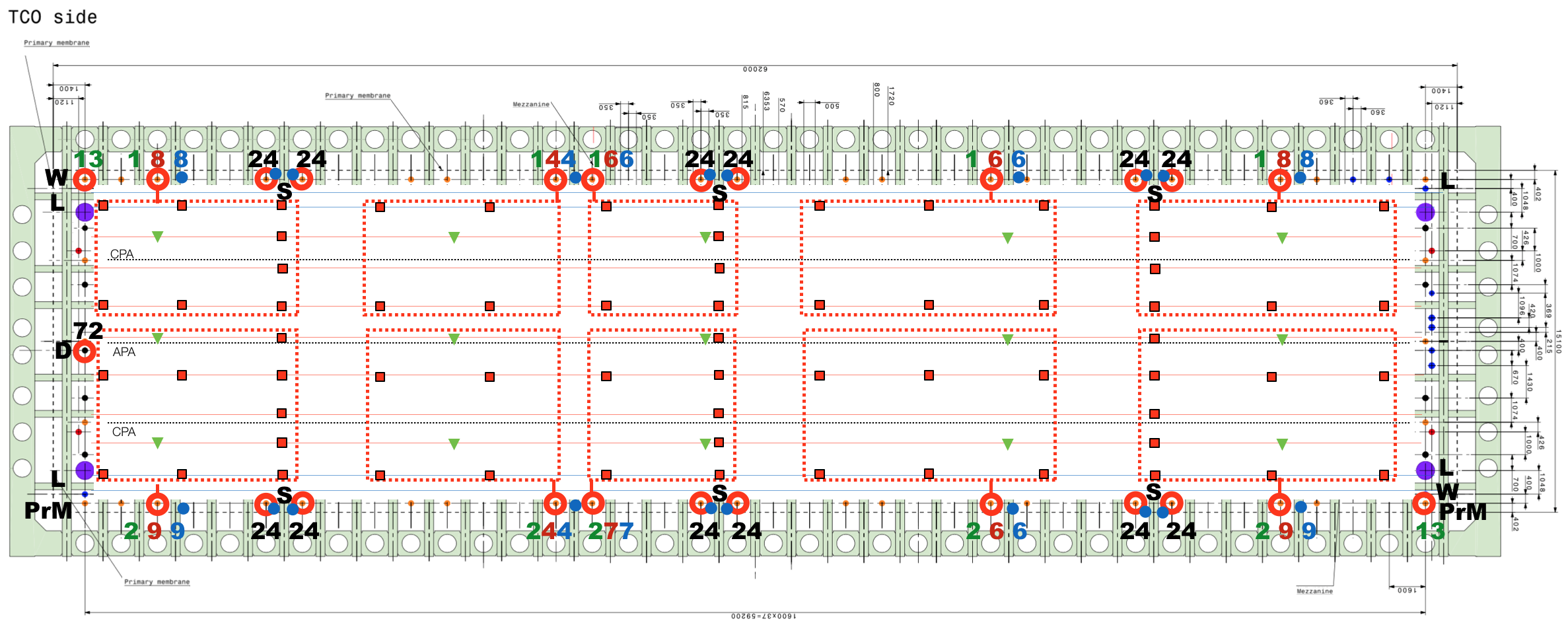}
  \includegraphics[width=0.85\textwidth]{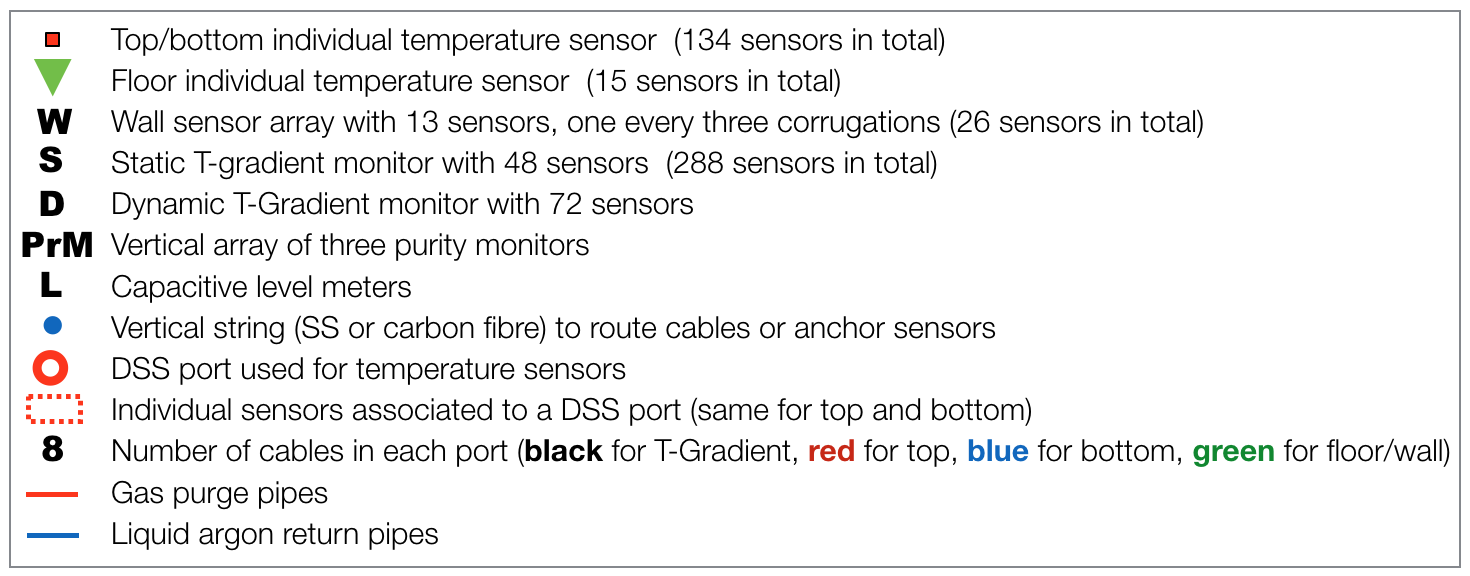}
\end{dunefigure}

High-precision temperature sensors will be distributed near the \dword{tpc} walls in two ways:
(1) forming high density (\(>2\) sensors/\si{m}) vertical arrays 
(called T-gradient monitors) and (2) in coarser ($\sim$ 1 sensor/\SI{5}{m}) 2D arrays 
(called individual sensors) at the top and bottom of the \dword{detmodule}, where it is most crucial to know the temperature.

Expected temperature variations inside the cryostat are very small ($\SI{0.02}{K}$; see Figure~\ref{fig:cfd-example}),
so sensors must be cross-calibrated to better than $\SI{0.005}{K}$. Most sensors will be calibrated in the laboratory before installation
(installation is described in Section \ref{sec:fdgen-slow-cryo-install-th}).
Calibration before installation is the only option for sensors installed on the long sides of the detector and the top and bottom of the cryostat, where space is limited.
Given the precision required and the unknown longevity of the sensors -- possibly requiring another  calibration after some time -- an additional method
will be used for T-gradient monitors installed on the short ends of the detector in the space between the field cage end walls and the cryostat walls. There is sufficient space in this area for a movable system, which can be used to cross calibrate
the temperature sensors in situ, as described in \ref{sec:fdgen-slow-cryo-dynamic-therm}.

The baseline design for all thermometer systems have three elements in
common: sensors, cables, and readout system. We plan to use Lake Shore
PT100-series\footnote{Lake Shore Cryotronics\texttrademark{} platinum RTD series,
  \url{https://www.lakeshore.com/}.} 
platinum sensors with \SI{100}{\ohm} resistance 
because in
the temperature range \SIrange{83}{92}{K} 
they 
show high reproducibility of $\sim\SI{5}{mK}$ and absolute temperature
accuracy of \SI{100}{mK}.  Using a four-wire readout greatly reduces
issues related to lead resistance, any parasitic resistances,
connections through the flange, and general electromagnetic noise
pick-up. Lakeshore PT102 sensors (see
Figure~\ref{fig:sensor-support}, right) were used in the \dword{35t} and \dword{pdsp}, 
giving excellent results. For the inner
readout cables, a custom cable made by Axon\footnote{Axon\texttrademark{} Cable, \url{http://www.axon-cable.com}.}
is the baseline. It
consists of four teflon-jacketed copper wires (\dword{awg} 28), forming two
twisted pairs, with a metallic external shield and an outer teflon
jacket.
The readout system is described in Section \ref{sec:fdgen-slow-cryo-therm-readout}.

Another set of lower-precision sensors epoxied into the bottom membrane of the cryostat will monitor  the cryostat filling in the initial stage.   
Finally, the inner walls and roof of the cryostat will have the same types of sensors to monitor the temperature during \cooldown and filling (``W'' sensors in Figure~\ref{fig:cisc-tsensor-map}).

\subsubsection{Dynamic T-gradient monitors}
\label{sec:fdgen-slow-cryo-dynamic-therm}

 To address concerns about potential differences in sensor readings prior to and after installation in a \dword{detmodule}, and potential drifts over the lifetime of the 
 module that may affect accuracy of the vertical temperature gradient measurement, 
 a dynamic temperature monitor allows cross-calibration of sensor readings 
 in situ.
Namely, this T-gradient monitor is motorized, allowing vertical motion of the temperature sensor array 
in the \dword{detmodule}, 
enabling precise cross-calibration between the sensors, 
as illustrated in Figure~\ref{fig:sensor-cross-calibration}.  

\begin{dunefigure}[Principle of cross-calibration with dynamic T-gradient monitor]{fig:sensor-cross-calibration}
  {In step 1, sensor temperature measurements are taken with the T-gradient monitor in the home position. In step 2, the entire system is moved up \SI{25}{cm} and another set of temperature readings is taken by all sensors. Then, the offsets between pairs of sensors are calculated for each position. In step 3, offsets are linked together, providing cross-calibration of all sensors, to obtain the entire vertical temperature gradient measurement with respect to a single sensor (number 1 in this case). }
  \includegraphics[width=1.0\textwidth]{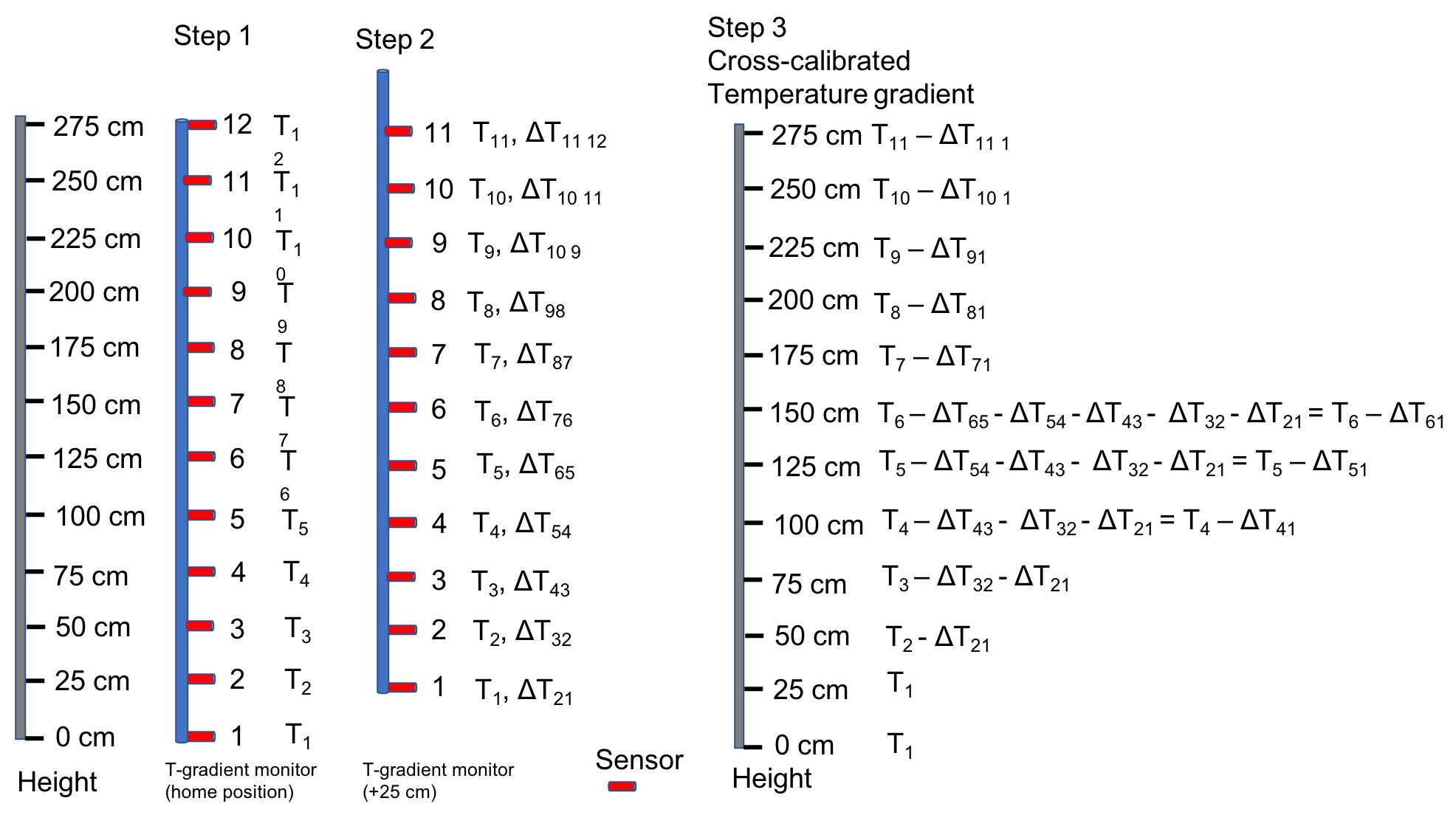}%
\end{dunefigure}

The procedure for cross-calibrations is the following: in step 1, the temperature reading  of all sensors is taken at the home (lowest) position of the carrier rod. In  step 2, the stepper motor moves the carrier rod up \SI{25}{cm}. Since the sensors along the entire  carrier rod are positioned \SI{25}{cm} apart, when the system is moved up \SI{25}{cm}, each sensor is positioned at the height that was occupied by another sensor in step 1. Then a second temperature reading is taken. In this manner, except for the lowest position, two temperature measurements are taken at each location with 
different sensors. Assuming that the temperature at each location is stable over the few minutes required to make the measurements, 
any difference in the temperature readings between the two different sensors is due to their relative measurement offset. This 
difference is then calculated for all locations.  In step 3, readout differences between pairs of sensors at each location are linked to one another, expressing temperature measurements at all heights with respect to a single sensor. In this way, temperature readings from all sensors are cross-calibrated 
in situ, canceling all possible offsets due to electromagnetic noise or any parasitic resistances that may have prevailed despite the four-point connection to the sensors that should cancel most of the offsets. These measurements are taken with a very stable current source, which ensures high precision of repeated temperature measurements over time. The motion of the dynamic T-monitor is stepper-motor operated, delivering measurements with high spatial resolution.

A total of \num{72} sensors will be installed with \SI{25}{cm} spacing, decreased to \SI{10}{cm} spacing for the top and bottom \SI{1}{m} of the carrier rod.  
 The vertical displacement of the system is such that every sensor can be moved to the nominal position of at least five other sensors, minimizing the risks associated with sensor failure and allowing for several points of comparison. The total expected motion range of the carrier rod is \SI{1.35}{m}.

This procedure was tested in \dword{pdsp}, where the system was successfully moved up by a maximum of \SI{51}{cm}, allowing cross-calibration of all sensors (22 sensors with \SI{10.2}{cm} spacing at top and bottom and \SI{51}{cm} in the middle). 

Figure~\ref{fig:dynamic_t_pumps-off} shows the temperature profile after calibration when the recirculation pumps are off. Under these conditions the  temperature should be very homogeneous except near the surface. This is indeed what is observed in that figure, demonstrating the reliability of the method.  

\begin{dunefigure}[Temperature profile for dynamic T-gradient with pumps-off]{fig:dynamic_t_pumps-off}
  {Temperature profile as measured by the dynamic T-gradient monitor after cross-calibration, when the recirculation pumps are off. Temperature variation is of the order of \SI{3}{mK} except close to the top and the gas phase interface, as expected.}
  \includegraphics[width=0.6\textwidth]{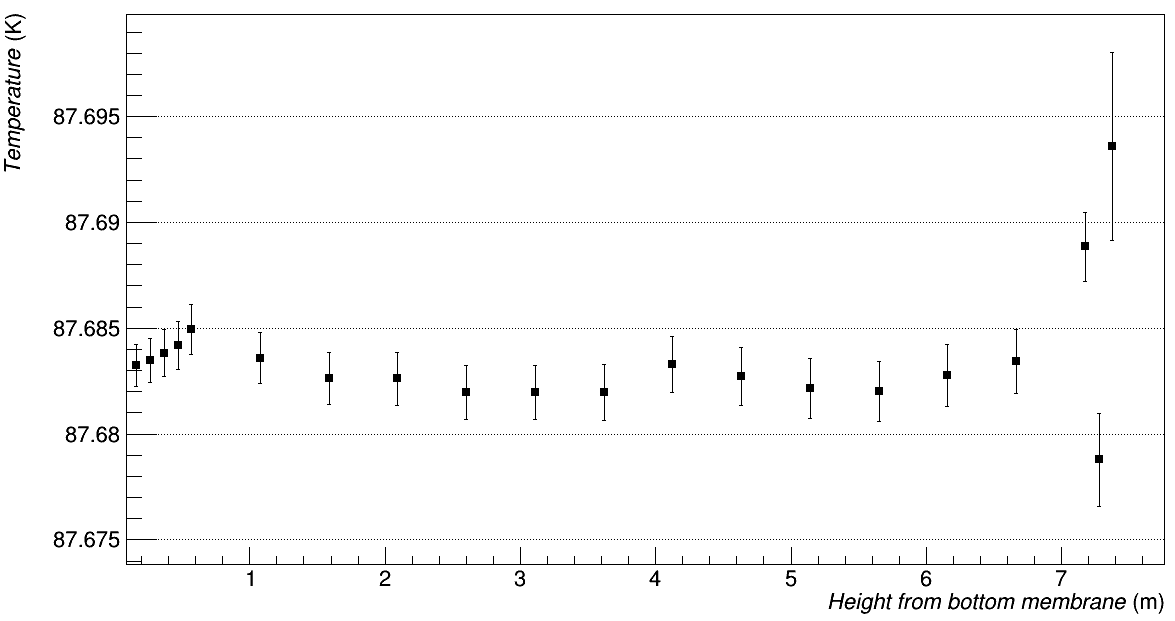}%
\end{dunefigure}

A dynamic T-gradient monitor has three parts: a carrier rod on which sensors are mounted; an enclosure above the cryostat housing space that allows the carrier rod to move vertically  \SI{1.5}{m} over its lowest location; and the motion mechanism. The motion mechanism consists of a stepper motor connected through a ferrofluidic dynamic seal to a gear and pinion motion mechanism. The sensors have two pins soldered to a \dword{pcb}. 
Two wires are individually soldered to the common soldering pad for each pin.  A cutout in the \dword{pcb} around the sensor allows free flow of argon for more accurate temperature readings.  Stepper motors typically have very fine steps that allow highly precise positioning of the sensors.  Figure~\ref{fig:fd-slow-cryo-dt-monitor-overview} shows the overall design of the dynamic T-gradient monitor. 
The enclosure has two parts connected by a six-cross flange. One side of this flange will be used for signal wires, another will be used as a viewing window, and the two other ports will be spares. Figure~\ref{fig:fd-slow-cryo-sensor-mount}, left shows the \dword{pcb} mounted on the carrier rod and the sensor mounted on the \dword{pcb} along with the four point connection to the signal readout wires. 
Figure~\ref{fig:fd-slow-cryo-sensor-mount}, right shows the stepper motor mounted on the side of the rod enclosure. The motor remains outside the enclosure, at room temperature, 
as do its power and control cables. 

\begin{dunefigure}[Dynamic T-gradient monitor overview]{fig:fd-slow-cryo-dt-monitor-overview}
  {
  A schematic of the dynamic T-gradient monitor.}
 \includegraphics[width=0.95\textwidth,angle=0]{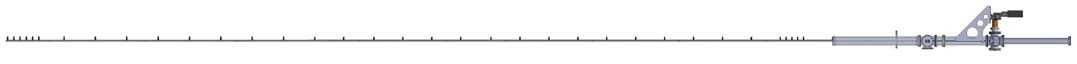}
\end{dunefigure}
\begin{dunefigure}[Sensor-cable assembly for dynamic T-gradient monitor]{fig:fd-slow-cryo-sensor-mount}
  {Left: Sensor mounted on a \dword{pcb} board and \dword{pcb} board mounted on the rod. Right:
    The driving mechanism of the dynamic T-gradient monitor, consisting of a stepper motor driving the pinion and gear linear motion mechanism. }
  \includegraphics[width=0.40\textwidth]{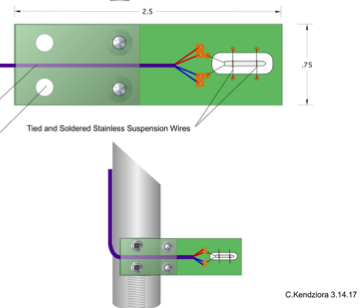}
  \hspace{3cm}%
  \includegraphics[width=0.12\textwidth]{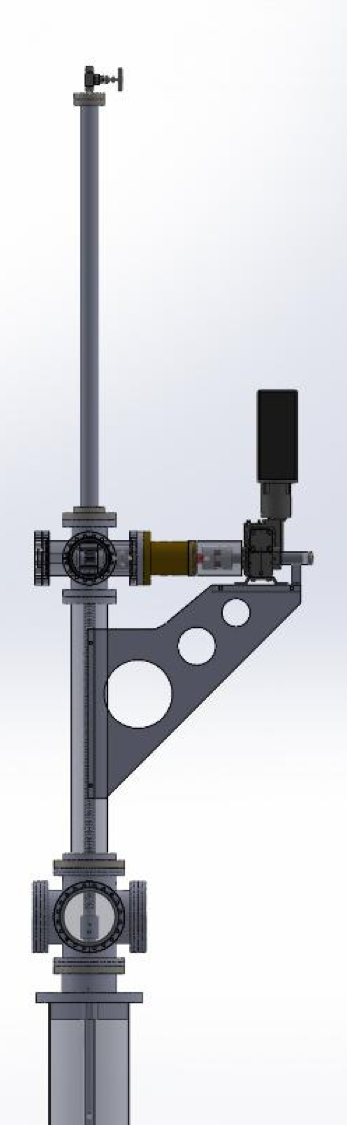}
\end{dunefigure}

\subsubsection{Static T-gradient monitors}
\label{sec:fdgen-slow-cryo-static-therm}

Several vertical arrays of high-precision temperature sensors cross-calibrated in the laboratory will be installed behind the \dword{apa}s.  
The baseline design assumes six arrays with \num{48} sensors each. Spacing between sensors
is \SI{20}{cm} at the top and bottom and \SI{40}{cm} in the middle area. This configuration is similar to the one used in \dword{pdsp} but with nearly double the spacing. 
As shown in Figure~\ref{fig:pd_static_t_results} a configuration with \num{48} sensors was appropriate in \dword{pdsp}, as it should be in the \dword{spmod} where the expected total gradient is no larger than in \dword{pdsp} (see Figure~\ref{fig:cfd-example}). 

\begin{dunefigure}[ProtoDUNE-SP static T-gradient results]{fig:pd_static_t_results}{
 Left: Temperature profile as measured by the static T-gradient monitor for two different calibration methods. Right: Distribution of the difference between both methods.}
  \includegraphics[height=0.34\textwidth]{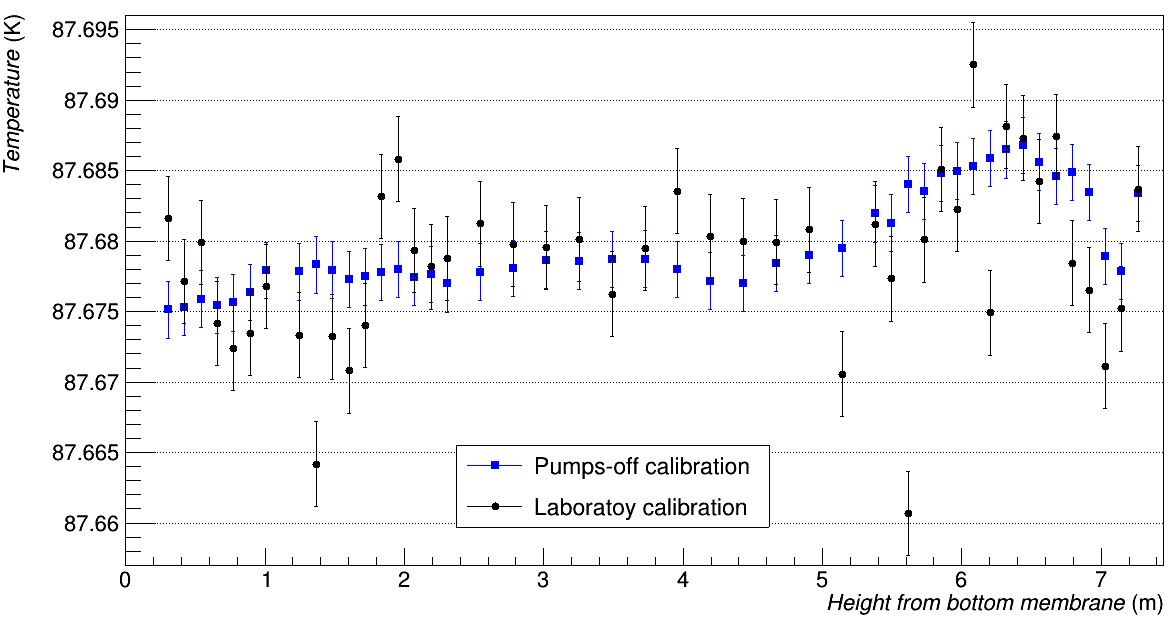}%
  \includegraphics[height=0.34\textwidth]{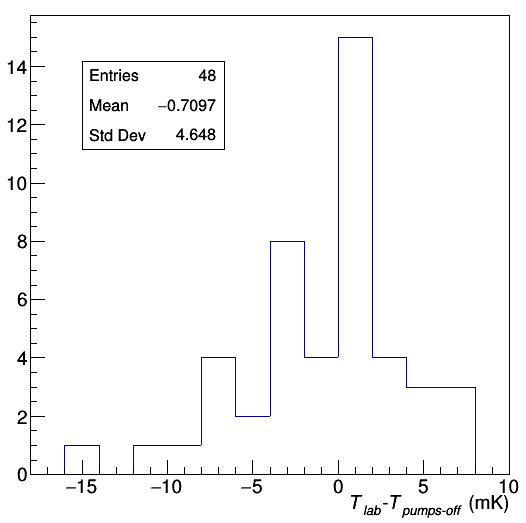}%
\end{dunefigure}

Sensors will be cross-calibrated in the laboratory using a controlled environment and a high-precision readout system, described in Section~\ref{sec:fdgen-slow-cryo-therm-readout}.
The accuracy of the calibration for \dword{pdsp} was estimated to be $\SI{2.6}{mK}$, as shown in Figure~\ref{fig:Trepro}. Preliminary results for the analysis of \dword{pdsp} static T-gradient monitor data are shown in Figure~\ref{fig:pd_static_t_results}. The temperature profile has been computed using both the laboratory calibration and the so-called ``in-situ pump-off calibration,'' which consists 
of estimating the offsets between sensors assuming the temperature of \dword{lar} in the cryostat is homogeneous when the re-circulation pumps are off (the validity of this method is demonstrated in Section~\ref{sec:fdgen-slow-cryo-dynamic-therm}).  
The \dword{rms} of the difference between both methods is $\SI{4.6}{mK}$, slightly larger than the value quoted above for the accuracy of the laboratory calibration, due to the presence of few outliers (under investigation) and to the imperfect assumption of homogeneous temperature when pumps are off (see Figure~\ref{fig:dynamic_t_pumps-off}).

\begin{dunefigure}[Temperature sensor resolution and reproducibility]{fig:Trepro}{
 Left:   Temperature offset between two sensors as a function of time for four independent immersions in \dword{lar}. The reproducibility of those sensors, defined as the RMS of the mean offset in the flat region, is $\sim\,\SI{1}{mK}$,
    The resolution for individual measurements, defined as the RMS of one offset in the flat region, is better than \SI{0.5}{mK}. Right: Difference between the mean offset obtained with two independent calibration methods for the 51 calibrated sensors. The standard deviation of this distribution is interpreted as precision of the calibration.}
  \includegraphics[height=0.44\textwidth]{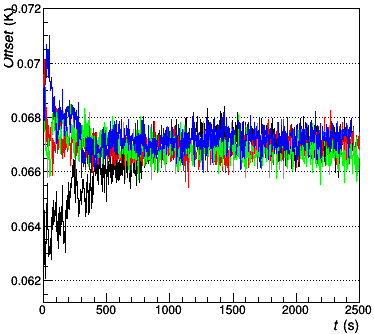}%
  \includegraphics[height=0.45\textwidth]{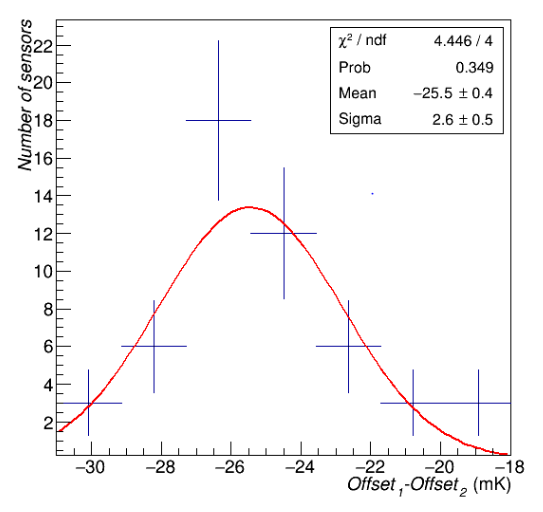}%
\end{dunefigure}

\begin{dunefigure}[Conceptual design of the static T-gradient monitor]
{fig:cisc-static-tgradient}
  {Conceptual design of the static T-gradient monitor.}
  \includegraphics[width=0.8\textwidth]{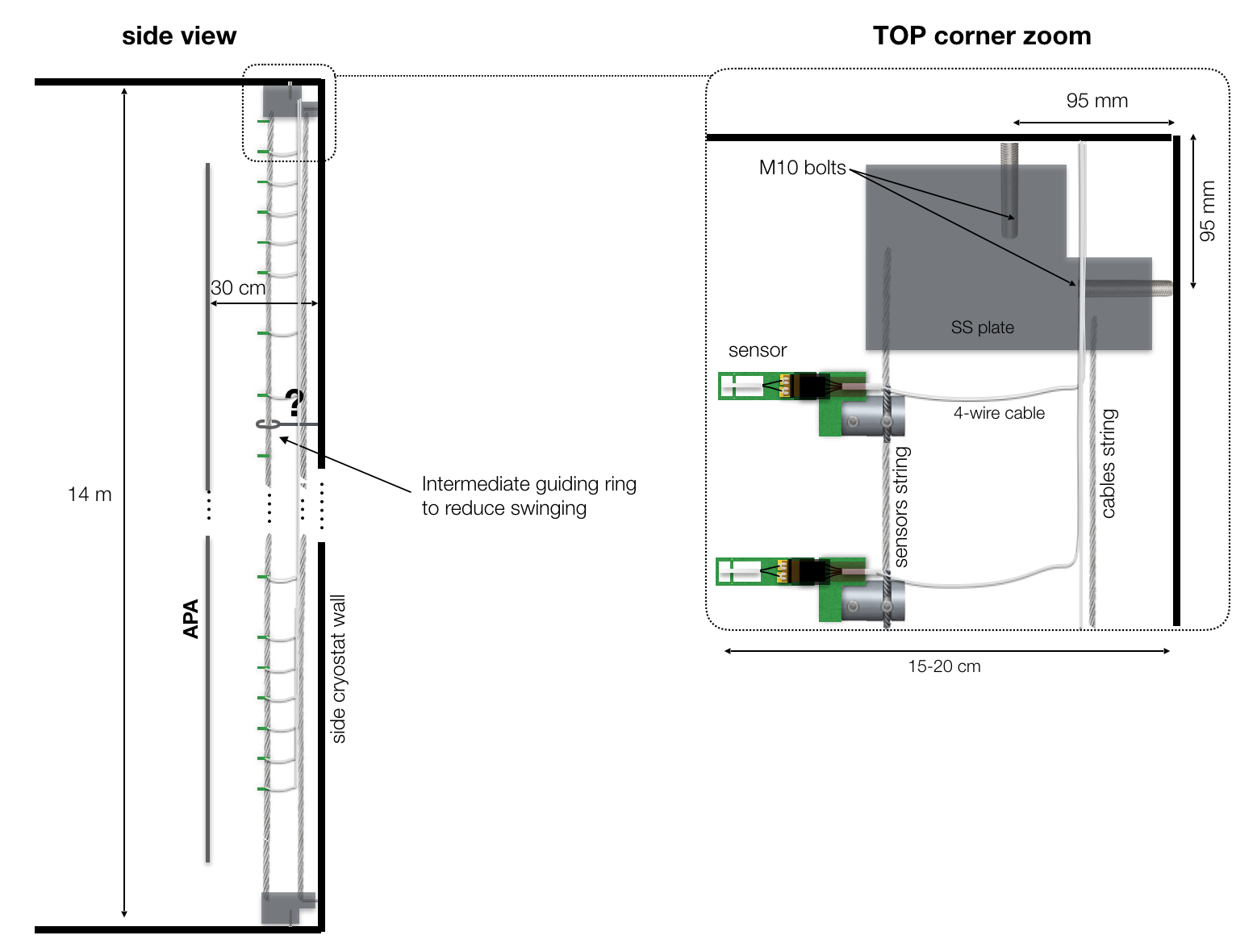}
\end{dunefigure}

Figure~\ref{fig:cisc-static-tgradient} shows the baseline mechanical design of
the static T-gradient monitor. Two strings (stainless steel or carbon fibre) are anchored at the top and bottom corners of the cryostat using the available M10 bolts (see Figure~\ref{fig:sensor-support}, left). One string routes the cables while the other supports the temperature sensors.
Given the height of the cryostat, an intermediate anchoring point to reduce swinging is under consideration. 
A prototype is being built at \dfirst{ific}, Spain, where the full system will be mounted using two dummy cryostat corners.

Figure~\ref{fig:sensor-support} (right) shows the baseline design of the ($52\times \SI{15}{mm^2}$) \dword{pcb} support for temperature sensors with an IDC-4 male connector. A narrower connector (with two rows of two pins each) is being studied. This alternative design would reduce the width of the \dword{pcb} assembly and allow more sensors to be calibrated simultaneously. Each four-wire cable from the sensor to the flange will have an IDC-4 female connector on the sensor end; the flange end of the cable will be soldered or crimped to the appropriate connector, whose type and number of pins  depend on the final design of the \dword{dss} ports that will be used to extract the cables. SUBD-25 connectors were used in \dword{pdsp}.

\begin{dunefigure}[Cryostat bolts and temperature sensor support]{fig:sensor-support}
  {Left: bolts at the bottom corner of the cryostat. Right: Lakeshore PT102 sensor mounted on a \dword{pcb} with an IDC-4 connector.}
  \includegraphics[height=0.2\textwidth]{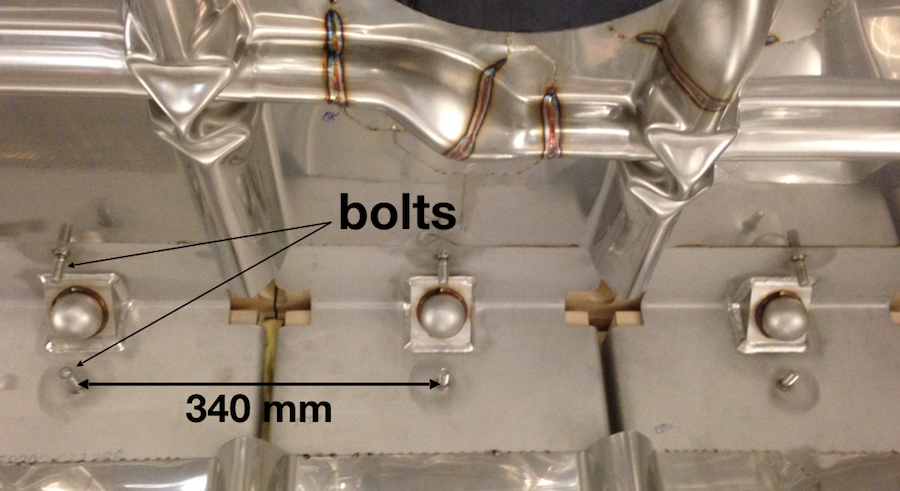}%
    \hspace{1cm}%
  \includegraphics[height=0.2\textwidth]{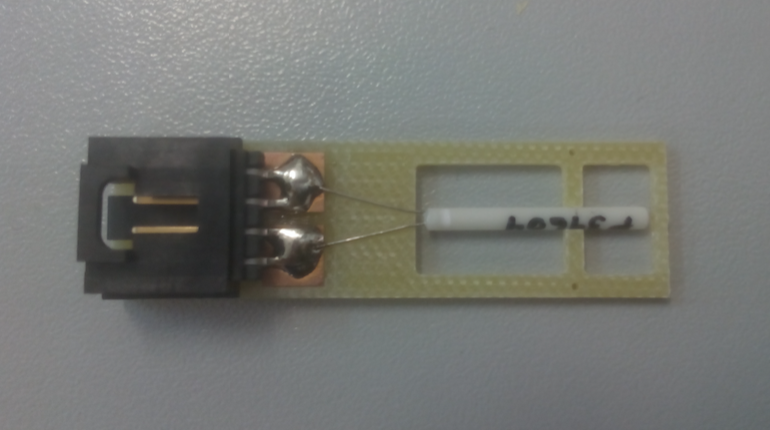}%
\end{dunefigure}

\subsubsection{Individual Temperature Sensors}
\label{sec:fdgen-slow-cryo-individual-therm}

T-gradient monitors will be complemented by a coarser 2D array (every \SI{5}{m}) of precision sensors at the top and bottom of the \dword{detmodule}, as shown in Figure~\ref{fig:cisc-tsensor-map}. Following the \dword{pdsp} design, bottom sensors will use the cryogenic pipes as a support structure, while top sensors will be anchored to the \dwords{gp}. Although sensors at the top will have a similar distribution to those at the bottom, 
suitable anchoring points at the top and bottom will differ. 

As in \dword{pdsp}, another set of standard sensors will be evenly distributed and epoxied to the bottom membrane. They will detect the presence of \dword{lar} when cryostat filling starts. Finally, two vertical arrays of standard sensors will be epoxied to the lateral walls in two opposite vertical corners, with a spacing of \SI{102}{cm} (every three corrugations), to monitor the cryostat membrane temperature during the \cooldown and filling processes. 

Whereas in \dword{pdsp} cables were routed individually (without touching neighboring cables or any metallic elements) to prevent grounding loops in case the outer Teflon jacket broke, such a failure has been proved to be very unlikely. Thus, in the \dwords{detmodule}, cables will be routed in bundles, simplifying the design enormously. As Figure~\ref{fig:cisc-tsensor-map} shows, up to 20 sensors will use the same \dword{dss} port, large enough for a cable bundle \SI{16}{mm} in diameter.

Cable bundles of several sizes will be configured using custom made Teflon 
pieces 
that will be anchored to different cryostat and detector elements to route cables from sensors to \dword{dss} ports. For sensors at the bottom (on pipes and floor), cables will be routed towards the cryostat bottom horizontal corner using stainless steel split clamps anchored to pipes (successfully prototyped in \dword{pdsp}), and from there, to the top of the cryostat using vertical strings (as with static T-gradient monitors). For sensors on the top \dwords{gp}, cables bundles will be routed to the corresponding \dword{dss} port using Teflon supports attached to both the \frfour threaded rods in the union between two \dword{gp} modules and to the \dword{dss} I-beams (both successfully prototyped in \dword{pdsp}). Sensors on the walls will use bolts in the vertical corners for cable routing. 

For all individual sensors, \dword{pcb} sensor support, cables, and connection to the flanges will be the same as for the T-gradient monitors.

\subsubsection{Analysis of temperature data in \dword{pdsp}}
\label{sec:fdgen-slow-cryo-temp-ana}

Temperature data from \dword{pdsp} has been recorded since \dword{lar} filling 
in August 2018. The analysis of this data and the comparison with \dword{cfd} simulations is actively underway, but interesting preliminary results are available and are described below. Figure~\ref{fig:pd_inst_map} shows the distribution of temperature sensors in the \dword{pdsp} cryostat.  

\begin{dunefigure}[ProtoDUNE-SP instrumentation map]{fig:pd_inst_map}{Distribution of temperature sensors in the \dword{pdsp} cryostat. Notice that four of the bottom sensors are located right above the \dword{lar} inlets. Purity monitors and level meters are also indicated. }
  \includegraphics[width=0.7\textwidth]{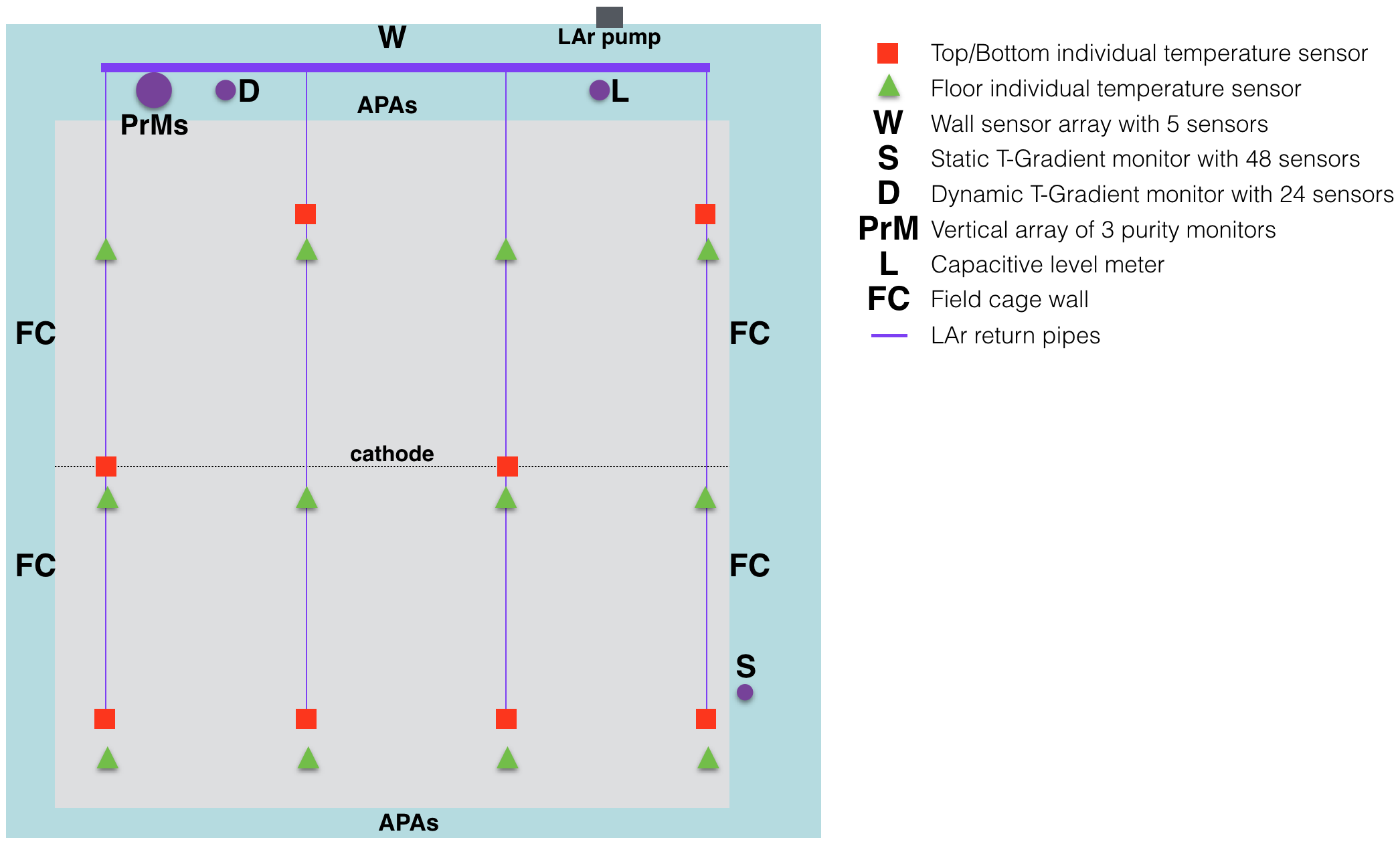}%
\end{dunefigure}

All precision temperature sensors (for the static and dynamic T-gradient monitors,  and the \twod arrays at top and bottom) were calibrated in the laboratory prior to installation, as described in Section \ref{sec:fdgen-slow-cryo-static-therm}.
However, since the calibration method was still under development when those sensors 
were installed, this calibration was only satisfactory (\SI{2.6}{mK} precision) for the static T-gradient monitor, for which the sensors were calibrated last and  plugged in just few days prior to installation in the cryostat. 
In Section~\ref{sec:fdgen-slow-cryo-static-therm}, an additional calibration method, the so-called ``pumps-off calibration,'' is described and the agreement with the laboratory calibration was demonstrated 
(see Figure~\ref{fig:pd_static_t_results}). Since this is the only reliable calibration for individual sensors, this method is used for the data analysis presented in this section, for all sensors except for the dynamic T-gradient monitor, for which the calibration based on the movable system is more precise (see Section~\ref{sec:fdgen-slow-cryo-dynamic-therm}). 

Figure~\ref{fig:pd_tgradient_results} shows the vertical temperature profiles as measured by the dynamic and static T-gradient monitors during a 10 minute period in May 2019. The stability of these profiles has been carefully studied: the relative variation between any two sensors on the same profiler remained below \SI{3}{mK} during the entire data taking period, demonstrating that the shape of the profiles is nearly constant in time. In Figure~\ref{fig:pd_tgradient_results} it is clear  
that the shapes of the two profiles are similar, with a bump at \SI{6.2}{m}, but the magnitude of 
variation of the static profile almost doubles 
compared to the dynamic profile. This effect is attributed to the fact that the dynamic T-gradient monitor is in the path of the \lar flow, which makes the temperature more uniform, while the static profiler is on the side.

\begin{dunefigure}[ProtoDUNE-SP T-gradient results]{fig:pd_tgradient_results}{Temperature profiles measured by the T-gradient monitors and comparison to the CFD model with different boundary conditions. Left: dynamic T-gradient monitor; Right: Static T-gradient monitor.}
  \includegraphics[width=0.5\textwidth]{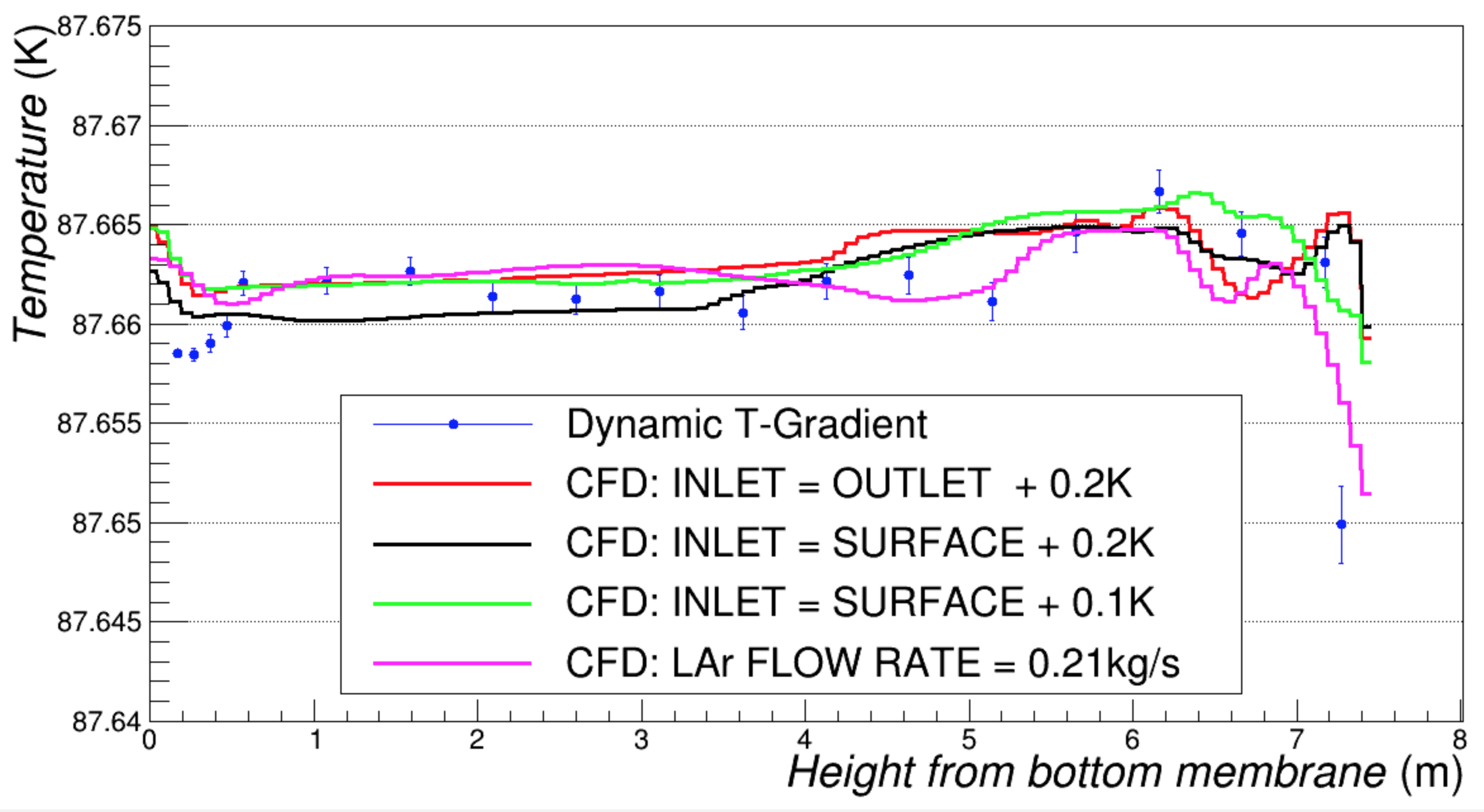}%
  \includegraphics[width=0.5\textwidth]{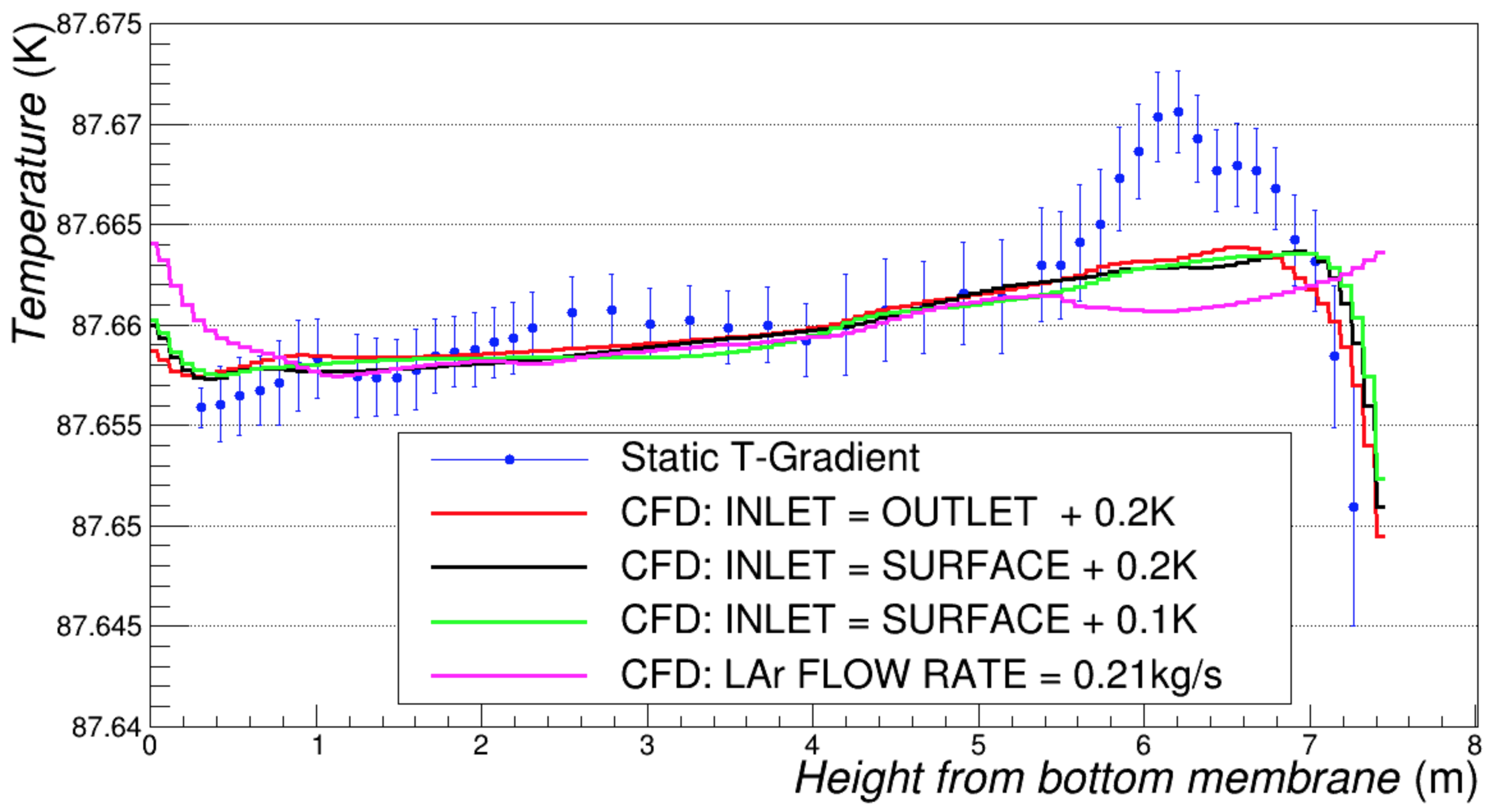}%
\end{dunefigure}

We can use temperature measurements by the bottom sensor grid to connect the two different 
regions covered by the T-gradient monitors. Figure~\ref{fig:pd_t_bottom_results} shows the temperature difference between bottom sensors and the second sensor of the static T-gradient monitor, \SI{40}{cm} from the floor, which is used as a reference. 
Also shown in the figure is the  dynamic T-gradient monitor's third sensor, located at a similar height. Three different periods are shown in the figure: two periods with pumps-on and one period with pumps-off. It is observed that when the pumps are working, the temperature decreases towards the \dword{lar} pump, and is 
higher in the sensors below the cathode. The horizontal gradient observed in this situation is of the order of \SI{20}{mK} -- larger than the vertical gradient. When the pumps are off the horizontal gradient decreases, although a residual gradient of  \SI{5}{mK}  is observed. This gradient is attributed to the inertia of the liquid once the pumps are switched off: it takes more than one day to recover the horizontal homogeneity.    

\begin{dunefigure}[\dshort{pdsp} bottom sensor results ]{fig:pd_t_bottom_results}{Temperature difference between bottom sensors at  \SI{40}{cm} from the floor and static T-gradient sensor at same height. The third dynamic T-gradient sensor, at the same height, is also shown. Two pumps-on periods (left and middle panels) and one pump-off period (right panel) are shown.}
  \includegraphics[width=0.95\textwidth]{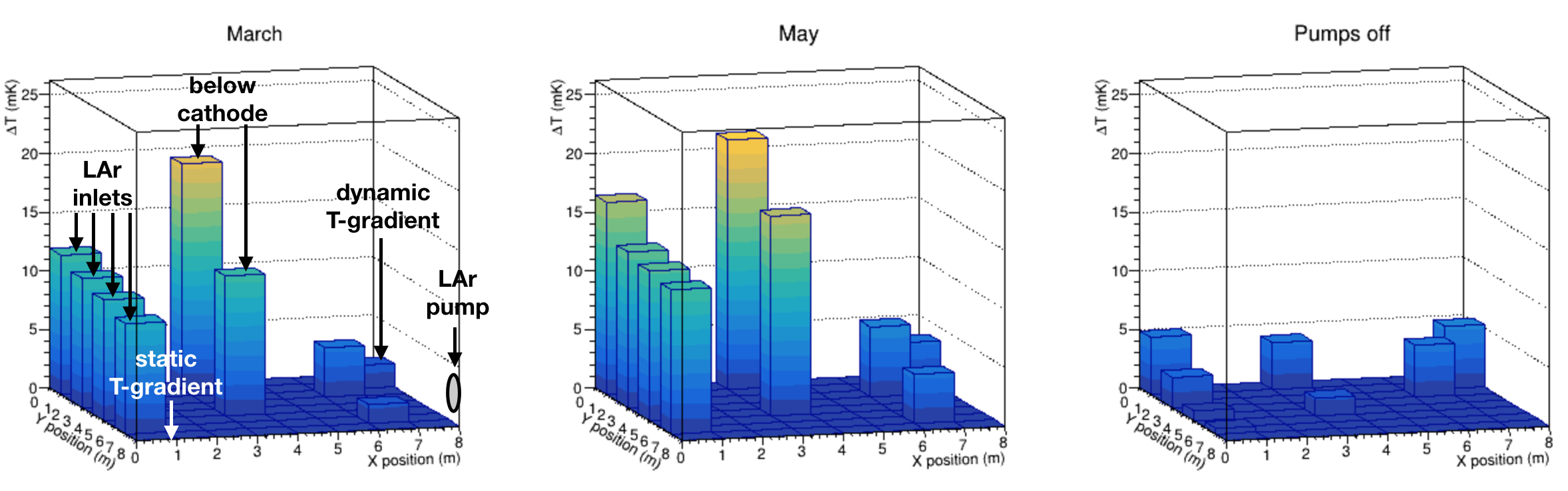}%
\end{dunefigure}

\dword{cfd} simulations have been produced using different inputs. We have identified two  parameters 
as potential drivers of the convection regime: (1) the incoming \dword{lar} flow rate, and (2) the incoming \dword{lar} temperature. Figure~\ref{fig:pd_tgradient_results} shows the result of varying these parameters. 
The \dword{cfd} model reasonably predicts the main features of the data, but some details still need to be understood, in particular the bump at \SI{6.2}{m} and the lower measured temperature at the bottom. It is worth noting that the simulation depends minimally on the \dword{lar} temperature while the flow rate has more impact, especially in the regions where discrepancies are larger. All simulations use the nominal \dword{lar} flow rate, \SI{0.42}{kg/s}, except the one explicitly indicated 
in the plots that uses half that rate. More simulations with other \dword{lar} flow rates 
are still in progress.

The \dword{cfd} reassuringly predicts a reasonable response for more than one set of initial conditions. It is still important to 
measure the 
instrumentation response to help establish the validity of the \dword{cfd} model. We did not run tests with differing initial conditions during the beam run because even controlled changes of the cryostat environment could have undesirable effects. However, recently we ran dedicated tests to validate the \dword{cfd} under various deliberate changes of the cryostat. 
These  additional tests included pump and recirculation manipulations (such as pump on-off, change of pumping speed, and bypassing of filtration), and changing the cryostat pressure set point to a higher (or lower) value\footnote{The HV 
was ramped down for this exercise because dropping the pressure too fast might provoke boiling of the \dword{lar} near the surface.} to induce changes in the pressure for a specified time while monitoring the instrumentation. Any change in pressure could change the temperatures everywhere in the cryostat. Studying the rate of this change, as detected at various heights in the cryostat, might provide interesting constraints on the \dword{cfd} model.

\subsubsubsection{Comparison of calibration methods}

Three different calibration methods have been described above: a laboratory calibration prior to installation, the ``pumps off'' calibration, and the movable system calibration.
The underlying assumption is that reliable temperature monitoring at the few \si{mK} level is desirable during the entire lifetime of the experiment, both to guarantee the correct functioning of the cryogenics system and to compute offline corrections based on temperature measurements and \dword{cfd} simulations. This is only possible if an in situ calibration method is available, since relative calibration between sensors is expected to diverge with time. 
Two in situ methods have been used. The pumps-off calibration method is very powerful since it is the only way of cross-calibrating all sensors in the cryostat at any point in time. However, it relies on the assumption that the temperature is uniform when the recirculation pumps are off. The validity of this assumption has to be bench-marked with real data, and this is done in \dword{pdsp} using the calibration based on the movable system (see Figure~\ref{fig:dynamic_t_pumps-off}). The movable system calibration method
is the most precise and the one that sustains all other methods, providing a reliable reference during the entire lifetime of the experiment. This method is even more crucial for the \dword{fd}. Indeed, recirculation pumps will be located on one side of the cryostat, very far (almost \SI{60}{m}) from some regions of the \dword{lar} volume, where the inertia will be more pronounced and the homogeneous temperature assumption becomes less valid.   

\subsubsection{Readout system for thermometers}
\label{sec:fdgen-slow-cryo-therm-readout}

A 
high-precision and very stable system is required to achieve a readout level of $<\,\SI{5}{mK}$.
The proposed readout system was used in \dword{pdsp} and relies on a variant of an existing mass PT100 temperature readout system developed at
\dword{cern} for an \dword{lhc} experiment; it has already been tested and validated in \dword{pdsp}.
The system has an electronic circuit that includes
\begin{itemize}
\item a precise and accurate current source\footnote{The actual current delivered is monitored with high-precision resistors such that the effect of ambient temperature can be disentangled.} for the excitation of the temperature sensors measured using the four-wires method, 
\item a multiplexing circuit connecting the temperature sensor signals and forwarding the selected signal to a single line, and 
\item a readout system  with a high-accuracy voltage signal readout module\footnote{National Instrument CompactRIO\texttrademark{} device  with a signal readout NI9238\texttrademark{} module.} with 24 bit resolution over a \SI{1}{V} range.
\end{itemize}
This readout system also drives the multiplexing circuit and calculates temperature values. The CompactRIO device is connected to the detector Ethernet network, sending temperature values to the \dword{dcs} software through a standard \dword{opc-ua} driver.

The current mode of operation averages more than \num{2000} samples taken every second for each sensor. 
Figure~\ref{fig:Trepro} shows the system has a resolution higher than 
\SI{1}{mK}, the \dword{rms} of one of the offsets in the stable region.



\subsection{Purity Monitors}
\label{sec:fdgen-slow-cryo-purity-mon}

A fundamental requirement of a \dword{lartpc} is that ionization electrons drift over long distances in the liquid. Part of the charge is inevitably lost due to electronegative impurities in the liquid. To keep this loss to a minimum, monitoring impurities and purifying the \dword{lar} during operation is essential.

A purity monitor is a small ionization chamber used to infer independently  the effective free electron lifetime in the \dword{lartpc}.  
It illuminates a cathode with UV light to generate a known electron current, then collects the drifted current at an anode a known distance away.  Attenuation of the current is related to the electron lifetime.
Electron loss can be parameterized as
\(N(t) = N(0)e^{-t/\tau},\)
where $N(0)$ is the number of electrons generated by ionization, $N(t)$ is the number of electrons after drift time $t$, and $\tau$ is the electron lifetime.

For the \dword{spmod}, the drift distance is \spmaxdrift, and the \efield is \SI{500}{\volt\per\centi\meter}. Given the drift velocity of approximately \SI{1.5}{\milli\meter\per\micro\second} in this field, the time to go from cathode to anode is roughly \SI{\sim2.4}{\milli\second} \cite{Walkowiak:2000wf}.
The \dword{lartpc} signal attenuation, \([N(0)-N(t)]/N(0)\), must remain 
less than \SI{20}{\percent} over the entire drift distance~\cite{bib:docdb3384}.  
 The corresponding electron lifetime is $\SI{2.4}{ms}/[-\ln(0.8)] \simeq \SI{11}{ms}$.

Residual gas analyzers can be used to monitor the gas in the ullage of the tank and would be an obvious choice for analyzing argon gas. 
Unfortunately, suitable and commercially available mass
spectrometers have a detection limit of \SI{\sim 10}{\dword{ppb}},
whereas \dword{dune} requires a sensitivity of \dword{ppt}. Therefore,
specially constructed and distributed purity monitors will measure \dword{lar} purity in all 
phases of operation. 
These measurements,
along with an accurate \dword{cfd} model, enable the
determination of purity at all positions throughout the \dword{detmodule}.

Purity monitors are placed inside the cryostat but outside of the 
\dshort{tpc}. They are also placed  within the recirculation system outside the cryostat, both in front of and behind the filtration system. 
Continuous monitoring of  the \dword{lar} supply lines to the \dword{detmodule} provides a strong line of defense against 
contamination from sources both in the \dword{lar} volume and from components in the recirculation system. 
Similarly, gas analyzers (described in Section~\ref{sec:fdgen-slow-cryo-gas-anlyz}) 
protect against contaminated gas.  

Furthermore, using several purity monitors to measure lifetime with high precision at carefully chosen points provides key input, e.g.,  vertical gradients in impurity concentrations, for \dword{cfd} models of the \dword{detmodule}.

Purity monitors were deployed in previous \dword{lartpc} experiments, e.g., \dword{icarus}, \dword{microboone}, and the \dword{35t}. During the first run of the \dword{35t}, two of the four purity monitors stopped working during \cooldown, and a third operated intermittently. We later found that this was due to poor electrical contacts between the resistor chain and the purity monitor. A new design was implemented and successfully tested in the second run.

\dword{pdsp} and \dword{pddp} 
use purity monitors to 
measure electron lifetime at different heights, and they use a similar design. 

Figure~\ref{fig-pdsp-prmassembly} shows the assembly of the \dword{pdsp} purity monitors. The design reflects improvements to ensure electrical connectivity and improve signals. \dword{pdsp} uses a string of purity monitors connected with stainless steel tubes to protect the optical fibers. 

\begin{dunefigure}[The ProtoDUNE-SP purity monitoring system]{fig-pdsp-prmassembly}
  {The \dword{pdsp} purity monitoring system}
  \includegraphics[width=0.9\textwidth]{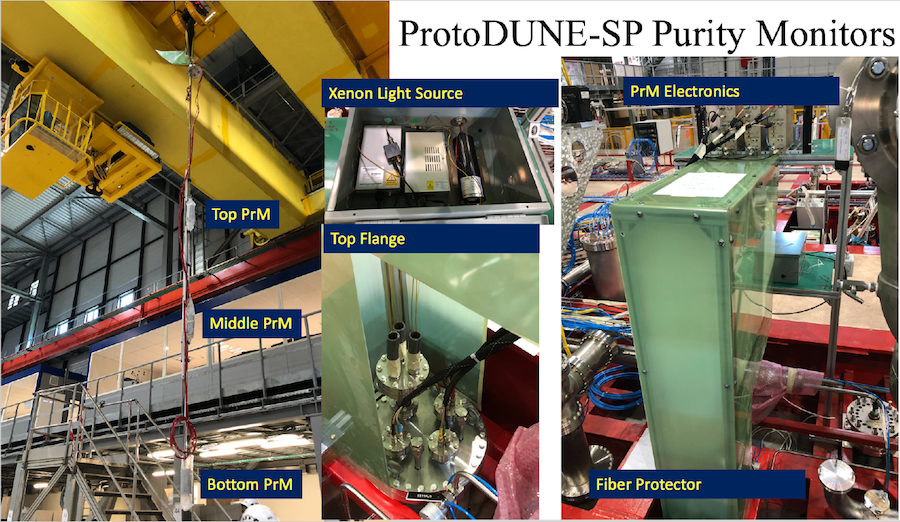}
\end{dunefigure}

\dword{pdsp} implements three purity monitors. The purity monitors continuously monitored \dword{lar} purity during all commissioning, beam test and operation periods of \dword{pdsp}. Figure~\ref{fig-pdsp-prm} shows the \dword{pdsp} data taken using 
its three purity monitors from commissioning of \dword{pdsp} starting in September 2018, through the entire beam running period, to July 2019.

\begin{dunefigure}[Measured electron lifetimes in the 
three purity monitors at ProtoDUNE-SP]{fig-pdsp-prm}
  {The electron lifetimes measured by three purity monitors in \dword{pdsp} as a function of time, September 2018 through July 2019. The purity is low prior to start of circulation in October. Reasons for later dips include recirculation studies and recirculation pump stoppages.}
  \includegraphics[width=0.9\textwidth]{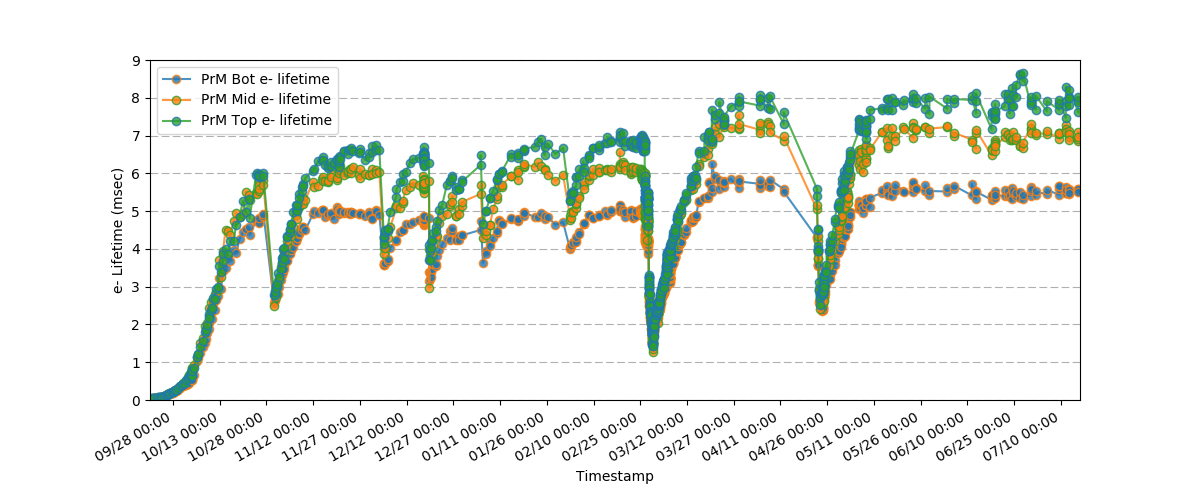}
\end{dunefigure}

Although \dword{pdsp} receives ample cosmic ray data to perform electron lifetime measurements, the purity monitor system was found to be essential for providing quick, reliable real-time monitoring of purity in the detector and to catch purity-related changes in time due to \dword{lar} recirculation issues. 
Each purity monitor electron lifetime measurement is based on purity monitor anode-to-cathode signal ratios from 200 UV flashes within 40 seconds at the same location. The statistical error on this lifetime measurement is less than \SI{0.03}{ms} when the purity is \SI{6}{ms}. 

The purity monitor system at \dword{pdsp} measured electron lifetime every hour during commissioning and daily during the beam test. During 
this time, it alerted the experiment to problems several times. 
The first time was for filter saturation during \dword{lar} filling, and the others were for recirculation pump stoppages, false alarms, and problems from the cryostat-level gauges.  The dips in Figure~\ref{fig-pdsp-prm} show these sudden changes in purity caught by purity monitors. 
Given the high sensitivity, the \dword{pdsp} purity monitors  were immediately able to alert the experiment to the purity drops, 
preventing situations which otherwise would have continued unnoticed for some time, with potentially serious consequences for the ability to take any data. 

During the commissioning and beam test of \dword{pdsp}, the purity monitors operated with different high voltages to change electron drift time, ranging from \SI{150}{\micro\second} to \SI{3}{\milli\second}. This allowed the \dword{pdsp} purity monitors to measure electron lifetime from \SI{35}{\micro\second} to about \SI{10}{\milli\second} with high precision, a dynamic range greater than \num{300}. 
This measurement was also valuable for the \dword{pdsp} lifetime calibration. Because purity monitors have much smaller drift volumes than the \dshort{tpc}, they are less affected by the space charge caused by cosmic rays. 

  A similar 
  operation plan is planned for 
  the \dword{spmod}, with modifications to accommodate the relative positions of the instrumentation port 
  and the purity monitor system, and the 
  different geometric relationships between the \dword{tpc} and cryostat.

\subsubsection{Purity Monitor Design}

The \dword{spmod} baseline purity monitor design follows that of  the \dword{icarus} experiment (Figure~\ref{fig:prm})\cite{Adamowski:2014daa}.  It consists of a double-gridded ion chamber immersed in the \dword{lar} volume with four parallel, circular electrodes, a disk holding a photocathode, two grid rings (anode and cathode), and an anode disk. The cathode grid is held at ground potential. The cathode, anode grid, and anode 
each hold separate bias voltages and are electrically accessible via modified vacuum-grade \dword{hv} \fdth{}s. 
The anode grid and the field-shaping rings are connected to the cathode grid by an internal chain of \SI{50}{\mega\ohm} resistors to ensure the uniformity of the \efield{}s in the drift regions. A stainless mesh cylinder is used as a Faraday cage to isolate the purity monitor from external electrostatic background. 

The purity monitor measures the electron drift lifetime between its anode and cathode. The purity monitor's UV-illuminated 
photocathode generates the electrons via the photoelectric effect. Because the electron lifetime in \dword{lar} is inversely proportional to the electronegative impurity concentration, the fraction of electrons generated at the cathode that arrives at the anode ($Q_A/Q_C$) after the electron drift time $t$ gives a measure of the electron lifetime $\tau$:
\( Q_A/Q_C \sim e^{-t/\tau}.\)

\begin{dunefigure}[Schematic diagram of the 
baseline purity monitor design]{fig:prm}
  {Schematic diagram of the basic purity monitor design \cite{Adamowski:2014daa}.}
  \includegraphics[width=0.9\textwidth]{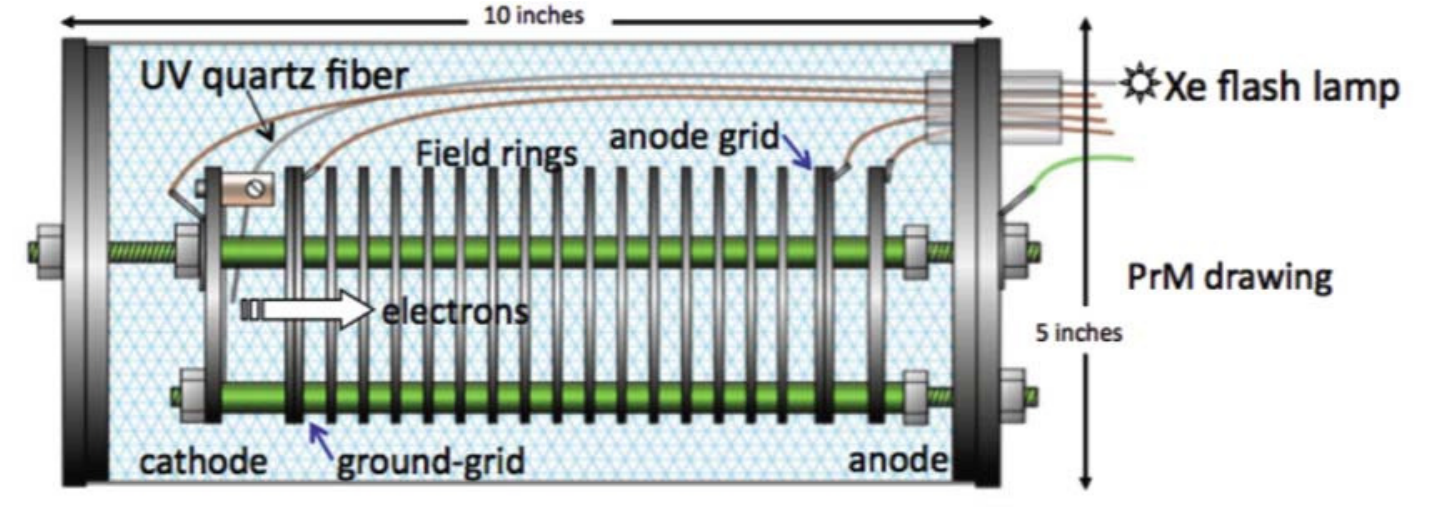}
\end{dunefigure}

Once the electron lifetime becomes much larger than the drift time $t$ the purity monitor reaches its sensitivity limit.  For $\tau >> t$, the anode-to-cathode charge ratio becomes 
$\sim\!1$. Because the drift time is inversely proportional to the \efield, in principle, lowering the 
field should make it possible to measure lifetimes of any length, regardless of the length of the purity monitor.
On the other hand, increasing the voltage will shorten the drift time, allowing measurement of a short lifetime when purity is low. 

The electron lifetime of the purest commercial \dword{lar}, after the first filtering and during the filling process, is typically higher than \SI{40}{\micro\second}. However, when the filter starts to saturate, the lifetime decreases to less than \SI{30}{\micro\second}.  To 
reduce the energy loss due to impurities,  the \dword{spmod} requires an electron lifetime greater than \SI{3}{\milli\second}.

Varying the operational \dword{hv} on the anode from \SI{250}{V} to \SI{4000}{V} in the \dword{pdsp}'s \SI{24}{cm} (9.5 inch) long purity monitor allowed us to make the \SI{35}{\micro\second} to \SI{10}{\milli\second} electron lifetime measurements. 
Purity monitors with different lengths (drift distances) are needed to extend the measurable range to below \SI{35}{\micro\second} and above  \SI{10}{\milli\second}.

The photocathode that produces the \phel{}s is an aluminum plate coated first with \SI{50}{\angstrom} of titanium followed by \SI{1000}{\angstrom} of gold, and is attached to the cathode disk.
A xenon flash lamp is the light source in the baseline design, although 
a more reliable and possibly submersible light source, perhaps \dword{led}-driven, could replace this in the future. The UV output of the lamp is quite good, approximately $\lambda=$ \SI{225}{\nano\meter}, which corresponds closely to the work function of gold (\SIrange{4.9}{5.1}{\eV}). 
Several UV quartz fibers carry the xenon UV light into the cryostat to illuminate the 
photocathode.   Another quartz fiber delivers the light into a properly biased photodiode outside the cryostat to provide the trigger signal when the lamp flashes.

\subsubsection{Electronics, DAQ, and Slow Controls Interfacing}
The purity monitor electronics and \dword{daq} system consist of \dword{fe} electronics, waveform digitizers, and a \dword{daq} PC.  Figure~\ref{fig:cryo-purity-mon-diag} 
illustrates the system.

\begin{dunefigure}[Block diagram of the purity monitor system]{fig:cryo-purity-mon-diag}
  {Block diagram of the purity monitor system.}
  \includegraphics[width=0.95\textwidth]{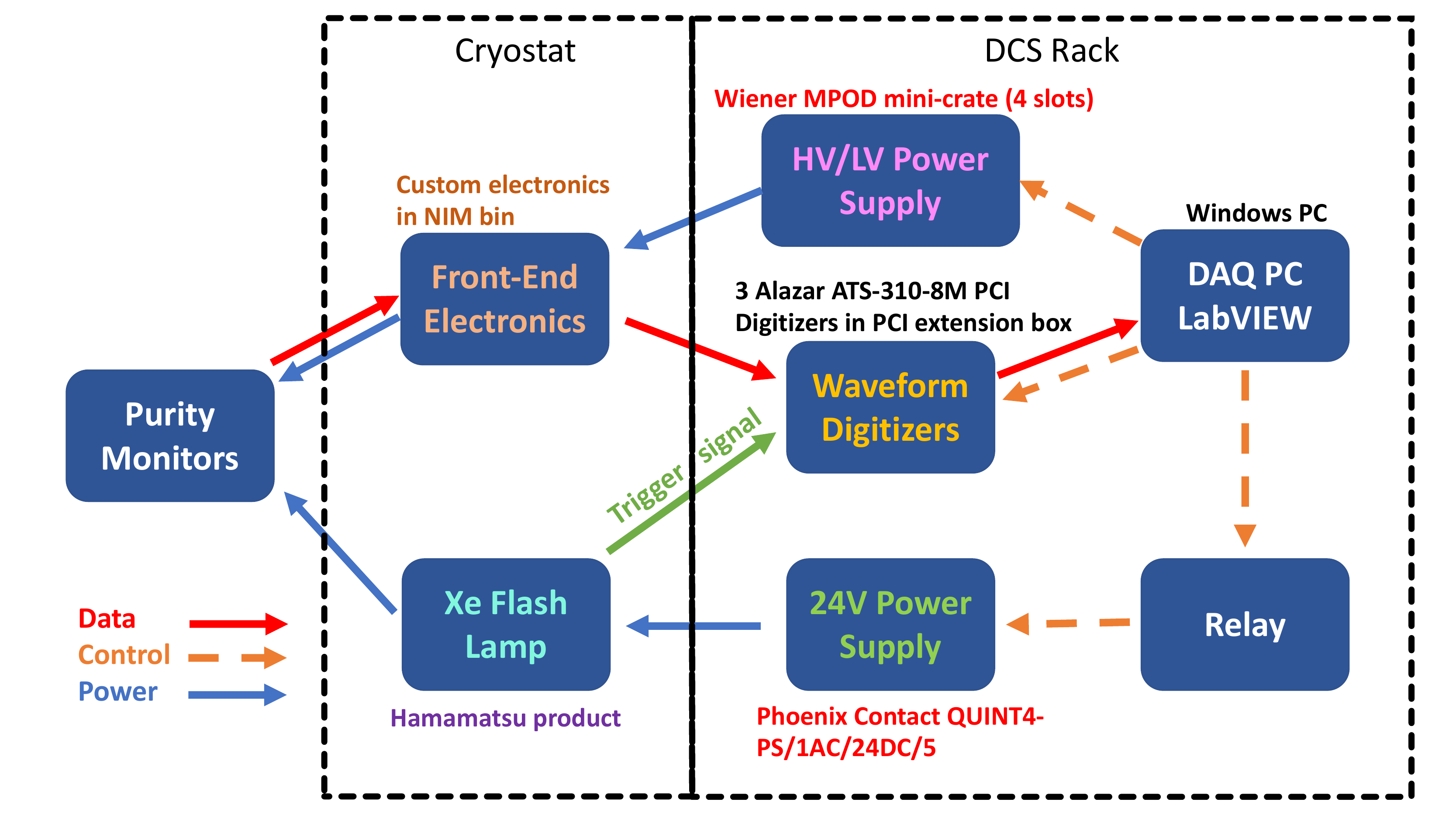}
\end{dunefigure}

The baseline design of the \dword{fe} electronics follows that used in 
the \dword{35t},  \dword{lapd}, and \dword{microboone}. The cathode and anode signals are fed into two charge amplifiers contained within the purity monitor electronics module.
This electronics module includes a \dword{hv} filter circuit and an amplifier circuit, both shielded by copper plates, to allow the signal and \dword{hv} to be carried on the same cable and decoupled inside the purity monitor electronics module. 
A waveform digitizer that interfaces with a \dword{daq} PC records the amplified anode and cathode outputs. 
The signal and \dword{hv} cable shields connect to the grounding points of the cryostat and are separated from the electronic ground with a resistor and a capacitor connected in parallel, mitigating ground loops between the cryostat and the electronics racks. Amplified output is transmitted to an AlazarTech ATS310 waveform digitizer\footnote{AlazarTech ATS310\texttrademark{} - 12 bit, 20 MS/s,  https://www.alazartech.com/Product/ATS310.} containing two input channels, each with 12 bit resolution. Each channel can sample a signal at a rate of \SI{20}{\mega\samples\per\second} to \SI{1}{\kilo\samples\per\second} and store up to \SI{8}{\mega\samples} in memory. One digitizer is used for each purity monitor, and each digitizer interfaces with the \dword{daq} PC across the \dword{pci} bus. 

A custom \dword{labview}\footnote{National Instruments, LabVIEW\texttrademark{}, http://www.ni.com/en-us.html} application running on the \dword{daq} PC has two functions: it controls the waveform digitizers and the power supplies, and it monitors the signals and key parameters. The application configures the digitizers to set the sampling rate, the number of waveforms to be stored in memory, the pre-trigger data, and a trigger mode. A signal from a photodiode triggered by the xenon flash lamp is fed directly into the digitizer as an external trigger to initiate data acquisition.   \dword{labview} automatically turns on the xenon flash lamp by powering a relay when data taking begins and then turns it off when finished.
The waveforms stored in the digitizers are transferred to the \dword{daq} PC and used to obtain averaged waveforms to reduce the electronic noise in them. 

The baseline  is estimated by averaging the waveform samples prior to the trigger. This baseline is subtracted from the waveforms prior to calculating peak voltages of the cathode and anode signals. 
The application performs these processes in real time. 
 The application continuously displays the waveforms and important parameters like measured electron lifetime, peak voltages, and drift time 
in the purity monitors, and shows the variation in these parameters over time, thus pointing out 
effects that might otherwise be missed. 
Instead of storing the measured parameters, the waveforms and the digitizer configurations are recorded in binary form for offline analysis.  \dword{hv} modules\footnote{iseg Spezialelektronik GmbH\texttrademark{} high voltage supply systems, https://iseg-hv.com/en.} in a WIENER MPOD mini crate\footnote{W-IE-NE-R MPOD\texttrademark{} Universal Low and High Voltage Power Supply System, http://www.wiener-d.com/.} supply negative voltages to the cathode and positive voltages to the anode. The  \dword{labview} application controls and monitors the \dword{hv} systems through an Ethernet interface.

The xenon flash lamp and the \dword{fe} electronics are installed close to the purity monitor flange, to reduce light loss through the optical fiber and prevent signal loss. Other pieces of equipment that distribute power to these items and collect data from the electronics are mounted in a rack separate from the cryostat. The slow control system communicates with the purity monitor \dword{daq} software and controls  the \dword{hv} and \dword{lv} power supplies of the purity monitor system. The optical fiber must be placed within  \SI{0.5}{\milli\meter} of the photocathode for efficient \phel extraction, so little interference with the \dword{pds} is expected. The purity monitors could induce noise in the \dword{tpc} electronics, in particular via the current surge through a xenon lamp when it is flashed.  This source of noise can be controlled by placing the lamp inside its own Faraday cage with 
proper grounding and shielding. 
At \dword{pdsp}, after careful checks of the grounding, this noise has remained well below the noise generated by other sources.

In the \dword{spmod} we can make use of triggering to prevent any potential noise from the purity monitor's flash lamp from affecting \dword{tpc} and \dword{pds} signals. The \dword{lartpc} trigger rate is a few hertz, and each trigger window is one or a few milliseconds. 
The pulse from a flash lamp is very short (a microsecond or so, much shorter than the gaps between \dword{lartpc} trigger windows). 
Thus, a \dword{lartpc} trigger signal may be sent to a programmable pulse generator, 
which generates a trigger pulse that does not overlap with \dword{lartpc} trigger windows. This trigger pulse 
is then sent to the external trigger port on the flash lamp \dword{hv} controller so that the lamp flashes between \dword{lartpc} trigger windows. In this way, the electronic and light noises from the flash lamp do 
not affect 
data taking at all.

\subsubsection{Production and Assembly}
\label{sec:PrMon-Production-Assembly}

The \dword{cisc} consortium will produce the individual purity monitors, test them in a test stand, and confirm that each monitor operates at the required level before assembling them into the strings of three monitors each that will be mounted in the \dword{spmod} cryostat using support tubes. The assembly process will follow the methodology developed for \dword{protodune}.

A short version of the 
string with all purity monitors will be tested at the \dword{citf}.
The full string will be assembled and shipped to South Dakota. 
 A vacuum test in a long vacuum tube will be performed on-site before inserting the full assembly into the \dword{spmod} cryostat.

\subsection{Liquid Level Meters}

The goals for the \dword{lar} level monitoring system are basic level sensing when filling, and precise level sensing during static operations. 

Filling the cryostat with \dword{lar} will take several months. During this operation several systems will be use to monitor the \dword{lar} level. 
The first \SI{5.5}{m} will be covered by cameras and by the vertical arrays of \dwords{rtd} at known heights, since temperature will change drastically 
when the cold liquid reaches each \dword{rtd}. Once the liquid reaches the level of the cryogenics pipes going out of the cryostat, 
the differential pressure between that point and the bottom of the cryostat
can be converted to depth using
the known density of \dword{lar}.   Fine tuning of the final \dword{lar} level will be done using several capacitive level meters at the top of the cryostat. 

During operation, liquid level monitoring has two purposes:
the \dword{lbnf} cryogenics system uses monitoring to tune the \dword{lar} flow, and 
DUNE uses monitoring to guarantee that the top \dwords{gp} are always
submerged 
at least \SI{20}{cm} below the \dword{lar} surface to mitigate the risk of dielectric breakdown. This was the value used for the \dword{hv} interlock in \dword{pdsp}. 

The \dword{lar} flow is tuned using two differential pressure level meters, installed as part of the cryogenics system, one on each side of the \dword{detmodule}.  They have a precision of \SI{0.1}{\%}, which corresponds to \SI{14}{mm} at thenominal \dword{lar} surface. Cryogenic pressure sensors will be purchased from commercial sources. Installation methods and positions will be determined as part of the
cryogenics internal piping plan.  

For \dword{hv} integrity, multiple \SI{4}{m} long capacitive level sensors (with a precision of less than \SI{5}{mm}) will be deployed along the top of the fluid 
for use during stable operations, and checked against each other.
One capacitive level sensor at each of the four corners of the cryostat will provide sufficient redundancy to ensure that no single point of failure compromises the 
measurement.

Figure~\ref{fig:cisc_pdsp_level} shows the evolution of the \dword{pdsp} \dword{lar} level over two months as measured by the differential pressure and capacitive level meters. 

\begin{dunefigure}[\lar level measurements]{fig:cisc_pdsp_level}{Evolution of the \dword{pdsp} \dword{lar} level over two months. Left: Measured by the capacitive level meter. Right: Measured by the differential pressure level meter. The units in the vertical axis are percentages of the cryostat height (\SI{7878}{mm}).}
  \includegraphics[width=0.4\textwidth]{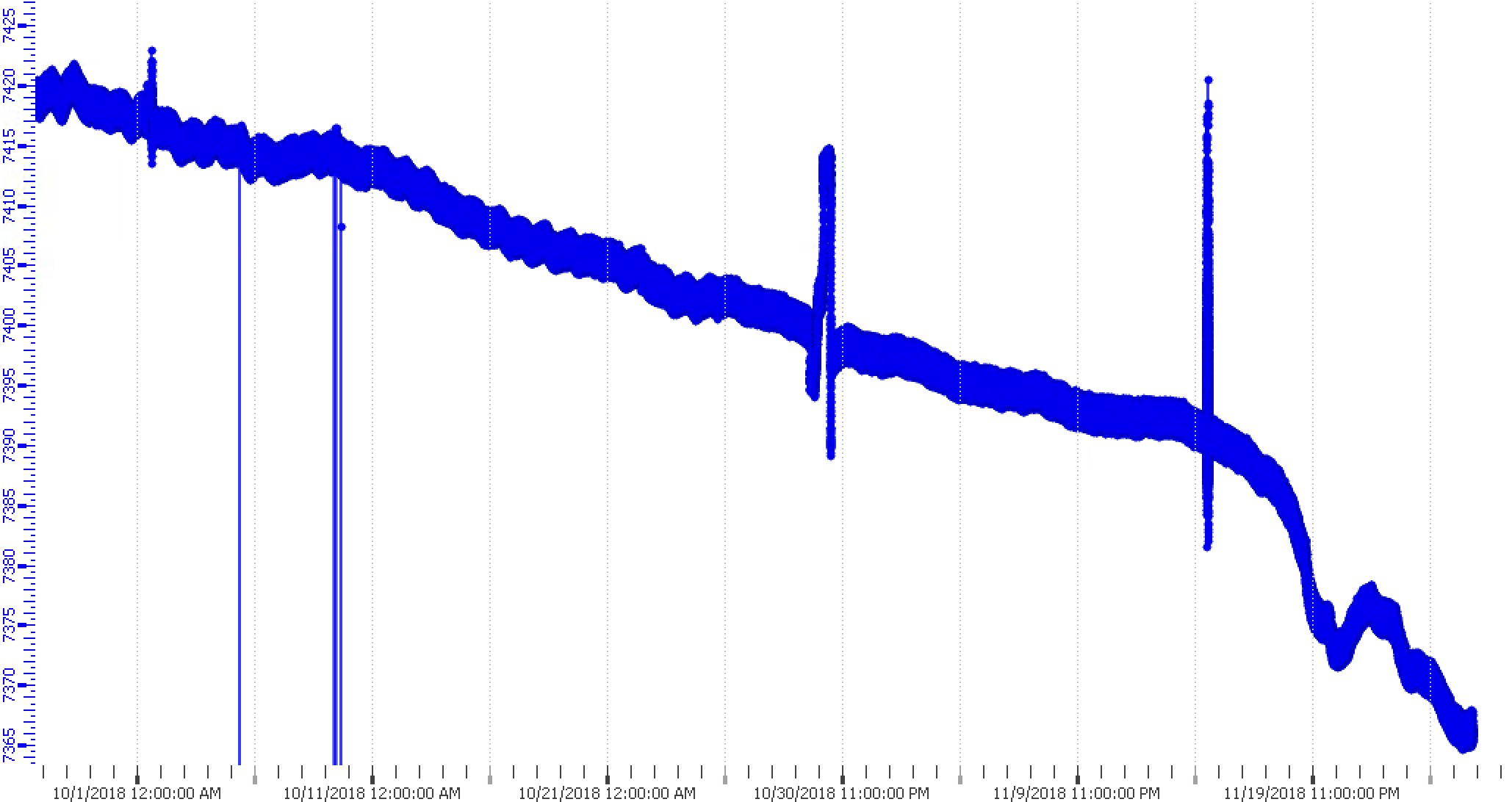}%
  \hspace*{1cm}
  \includegraphics[width=0.4\textwidth]{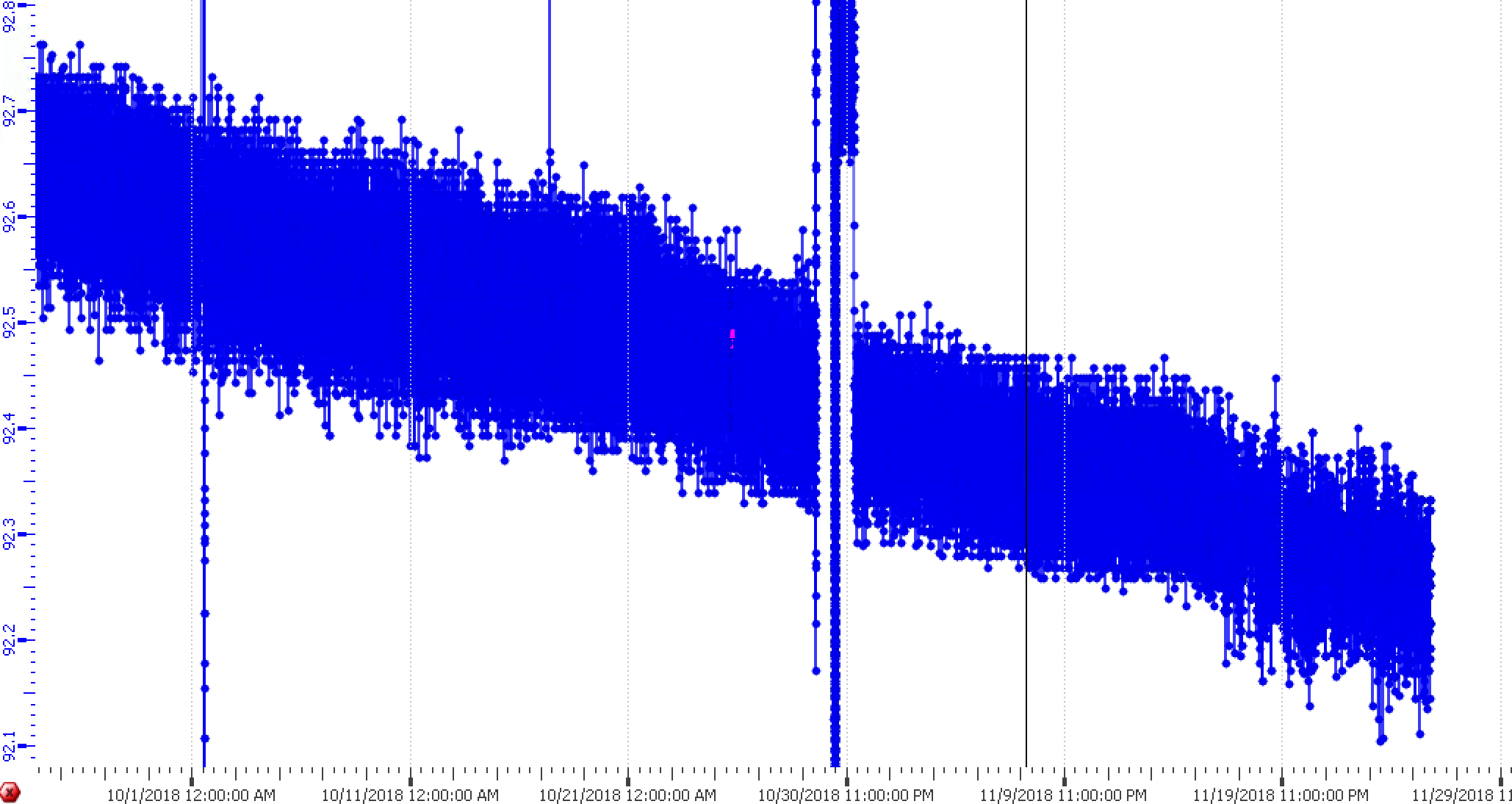}%
\end{dunefigure}

\dword{pdsp}  uses the same design for differential pressure level meters 
as the \dword{spmod}. In the case of capacitive level meters, \dword{pdsp} is using commercially bought \SI{1.5}{m} long level meters while \dword{pddp} is using  \SI{4}{m} long level meters that are custom-built by \dword{cern}. 
We plan to use the longer capacitive level meters custom-built by  \dword{cern} for both \dword{sp} and \dword{dp} modules.

\subsection{Pressure Meters}
\label{sec:fdgen-slow-cryo-press-meter}

The absolute temperature in the liquid varies with the pressure in the argon gas in the ullage of the cryostat, therefore, precise measurements of pressure inside the cryostat allow for a better understanding of temperature gradients and \dword{cfd} simulations. In \dword{pdsp}, pressure values were also used to understand the strain gauge signals installed in the cryostat frame.

Standard industrial pressure sensors can be used to measure the pressure of the argon gas. For the DUNE \dword{fd}, the plan is to follow the same design and configuration used in \dword{pdsp}. \dword{protodune} uses two types of pressure sensors and a pressure switch, 
\begin{itemize}
    \item a relative pressure sensor (range: 0-400~mbar, precision: 0.01~mbar),
    \item an absolute pressure sensor (range: 0-1600~mbar, precision: 0.05~mbar), and
    \item a mechanical relative pressure switch adjustable from 50 to 250~mbar. 
\end{itemize}

Both sensors and the pressure switch are installed in a dedicated flange as shown in Figure~\ref{fig:pdsp-pressure} and are connected directly to a slow controls system \dword{plc} circuit. Dedicated analog inputs are used to read the current values (\SIrange{4}{20}{mA}) which are then converted to pressure according to the sensors range. Given the much larger size of the  \dword{dune} \dwords{detmodule}, the system described above will be doubled for redundancy: two flanges in opposite cryostat sides will be instrumented with three sensors each. 

\begin{dunefigure}[Pressure sensors installed on a flange in ProtoDUNE-SP]{fig:pdsp-pressure}
  {Photograph of the pressure sensors installed on a flange in \dword{pdsp}.}
  \includegraphics[width=0.5\textwidth]{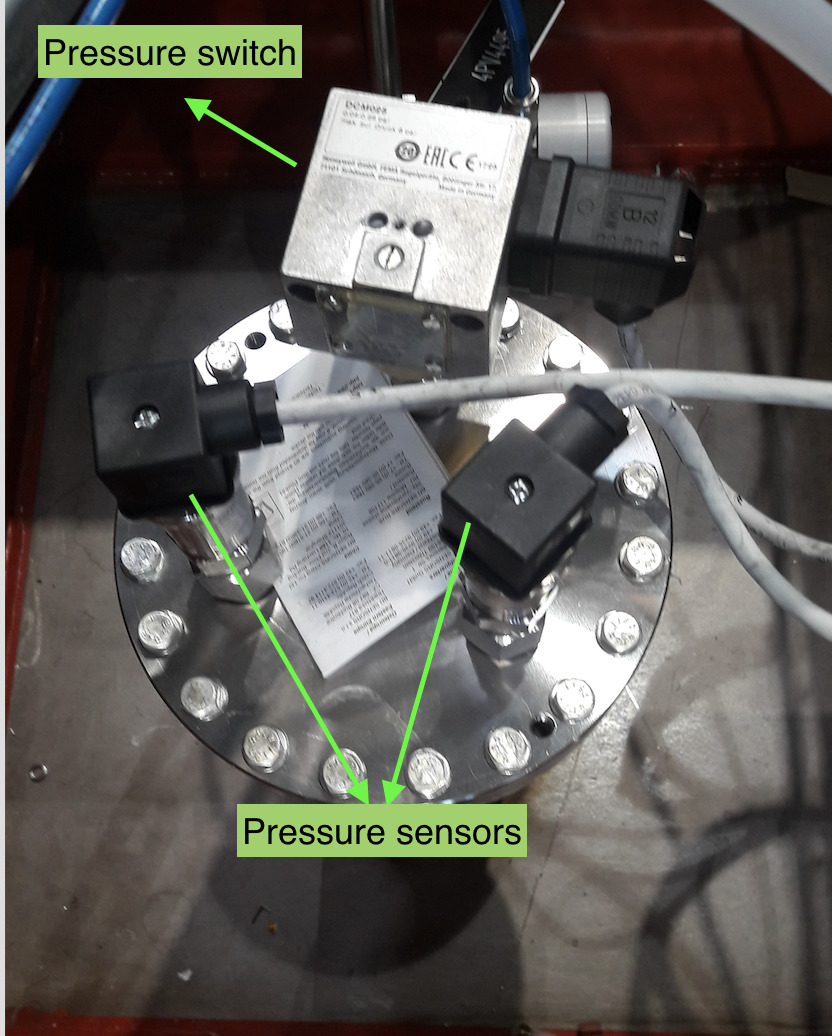}
\end{dunefigure}

Further, relative and absolute pressure sensors (with comparatively lower precision) are installed by \dword{lbnf} that are also recorded by the slow controls system. The availability of two types of sensors from \dword{lbnf} and \dword{cisc} provides redundancy, independent measurements, and cross checks.

\subsection{Gas Analyzers}
\label{sec:fdgen-slow-cryo-gas-anlyz}

 Gas analyzers are commercially produced modules that measure trace quantities of specific gases contained within a stream of carrier gas. The carrier gas for \dword{dune} is argon, and the trace gases of interest are oxygen ($\text{O}_2$), water ($\text{H}_2\text{O}$), and nitrogen ($\text{N}_2$). $\text{O}_2$ and $\text{H}_2\text{O}$ affect the electron lifetime in \dword{lar} and must be kept below \SI{0.1}{ppb} ($\text{O}_2$ equivalent) while $\text{N}_2$ affects the efficiency of scintillation light production at levels higher than \SI{1}{ppm}.
The argon is sampled from either the argon vapor in the ullage or from the \dword{lar} by using small diameter tubing run from the sampling point to the gas analyzer. Typically, the tubing from the sampling points are connected to a switchyard valve assembly used to route the sample points to the desired gas analyzers (see Figure~\ref{fig:GA-switchyard}).

\begin{dunefigure}[A gas analyzer switchyard valve assembly]{fig:GA-switchyard}
  {A gas analyzer switchyard that routes sample points to the different gas analyzers.}
  \includegraphics[width=0.9\textwidth]{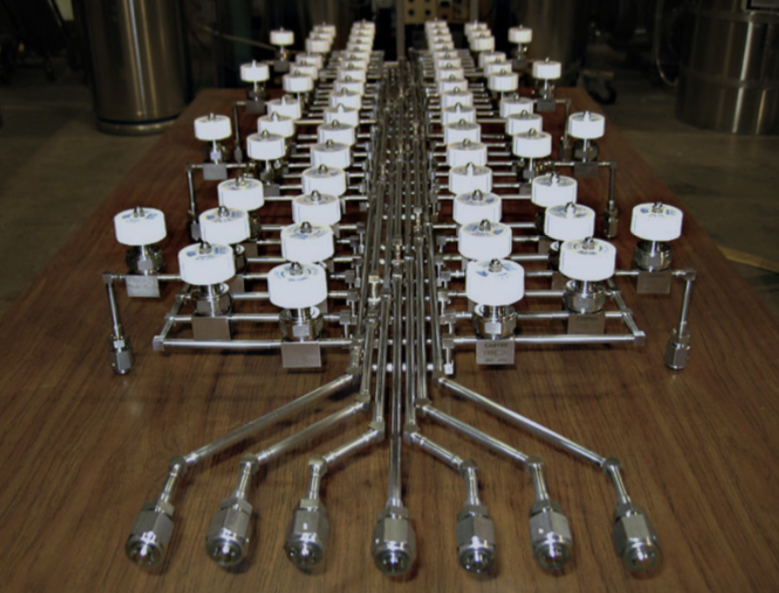}
\end{dunefigure}

The gas analyzer would be predominantly used during three periods:

\begin{enumerate}
\item Once the detector is installed and after the air atmosphere is eliminated from the cryostat to levels low enough to begin \cooldown. This purge and gas recirculation process is detailed in Section~\ref{sec:fdgen-slow-cryo-install-ga}. Figure~\ref{fig:cisc_Phase1_purge_gas_recirculation} shows the evolution of the $\text{N}_2$, $\text{O}_2$, and $\text{H}_2\text{O}$ levels from gas analyzer data taken during the purge and recirculation stages of the \dword{dune} \dword{35t} 
phase 1 run.

\item Before other means of monitoring impurity levels (e.g., purity monitors, or \dshort{tpc} tracks) are sensitive, to track trace $\text{O}_2$ and $\text{H}_2\text{O}$ contaminants from $\>$tens of \dword{ppb} to hundreds of \dword{ppt}.  Figure~\ref{fig:cisc_O2AnalyzerPrM2_HVRun1} shows an example plot of $\text{O}_2$ levels at the beginning of \dword{lar} purification from one of the later \dword{35t} \dword{hv} runs.

\item During cryostat filling to monitor the tanker \dword{lar} delivery purity. This tracks the impurity load on the filtration system and rejects any deliveries that do not meet specifications. 
Specifications for the delivered \dword{lar} are in the \SI{10}{ppm} range per contaminant.

\end{enumerate}

\begin{dunefigure}[Impurity levels during the pre-fill stages for \dshort{35t} phase 1]{fig:cisc_Phase1_purge_gas_recirculation}
  {Plot of the $\text{O}_2$, $\text{H}_2\text{O}$, and $\text{N}_2$ levels during the piston purge and gas recirculation stages of the \dword{35t} Phase 1 run}
  \includegraphics[width=0.7\textwidth]{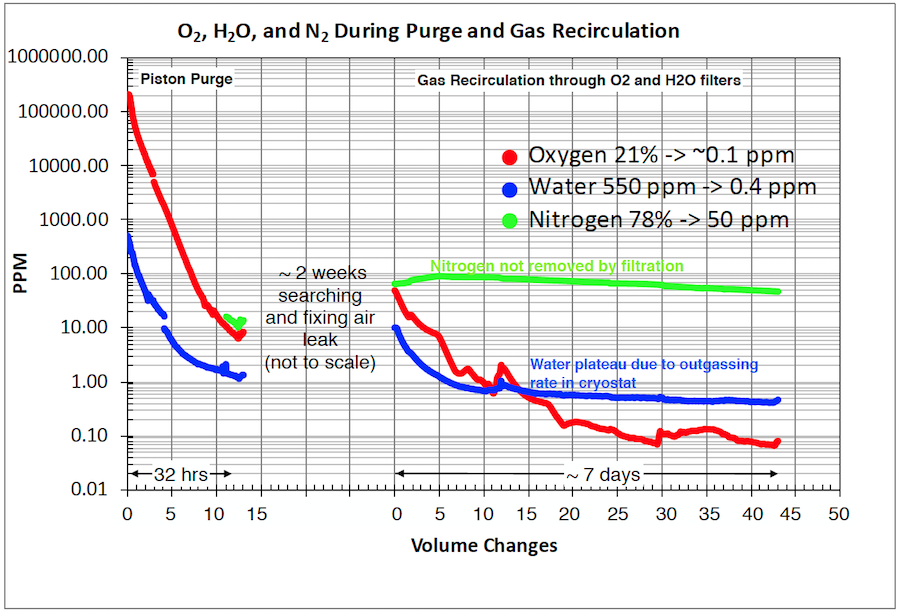}
\end{dunefigure}

\begin{dunefigure}[$\text{O}_2$ just after the \dshort{35t} was filled with \lar]{fig:cisc_O2AnalyzerPrM2_HVRun1}
  {$\text{O}_2$ as measured by a precision $\text{O}_2$ analyzer just after the \dword{35t} cryostat was filled with \dword{lar}, continuing with the \dword{lar} pump start and beginning of \dword{lar} recirculation through the filtration system. As the gas analyzer loses sensitivity, the purity monitor can pick up the impurity measurement. Note that the purity monitor is sensitive to both $\text{O}_2$ and $\text{H}_2\text{O}$ impurities giving rise to its higher levels of impurity.}
  \includegraphics[width=0.7\textwidth]{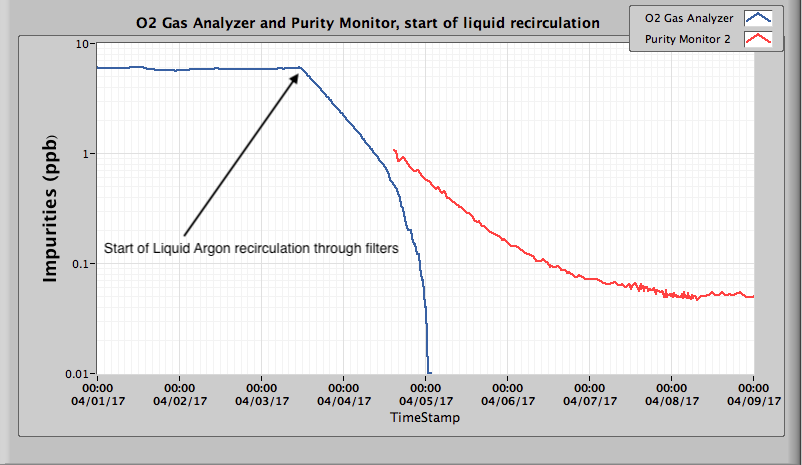}
\end{dunefigure}

Since any one gas analyzer covers only one contaminant species and a range of \numrange{3}{4} orders of magnitude, several units are needed both for the three contaminant gases and to cover the ranges seen between  cryostat closure and the beginning of \dshort{tpc} operations:
\SI{20}{\percent} to $\lesssim 100$~\dword{ppt} for $\text{O}_2$,
\SI{80}{\percent} to $\lesssim 1$~\dword{ppm} for $\text{N}_2$, and
$\sim \SI{1}{\percent}$ to $\lesssim 1$~\dword{ppb} for $\text{H}_2\text{O}$.
Because the total cost of these analyzers exceeds $\SI{100}[\mathdollar]{k}$, we want to be able to  sample more than a single location or cryostat with the same gas analyzers. At the same time, the tubing run lengths from the sample point should be as short as possible to maintain a timely response for the gas analyzer.  This puts some constraints on sharing devices because, for example, argon is delivered at the surface, so a separate gas analyzer may be required there. 

\subsection{Cameras}

Cameras provide direct visual information about the state of the
\dword{detmodule} during critical operations and when damage or unusual
conditions are suspected.  Cameras in the \dword{wa105} showed spray from \cooldown
nozzles and the level and state of the \dword{lar} as it covered the \dword{crp} \cite{Murphy:20170516}.  A camera was
used in the \dword{lapd} 
cryostat\cite{Adamowski:2014daa} to study \dword{hv} discharges in
\dword{lar} and in EXO-100 while a \dword{tpc} was operating
\cite{Delaquis:2013hva}.  Warm cameras viewing \dword{lar} from a distance
have been used to observe \dword{hv} discharges in \dword{lar} in
fine detail \cite{Auger:2015xlo}.  Cameras are commonly used during
calibration source deployment in many experiments (e.g., the
\kamland ultra-clean system \cite{Banks:2014hra}).

In \dword{dune}, cameras will verify the stability, straightness,
and alignment of the hanging \dword{tpc} structures during \cooldown and
filling; ensure that no bubbling occurs near the \dwords{gp}
(\single) or \dwords{crp} (\dual);  inspect the
state of movable parts in the \dword{detmodule} (calibration devices, dynamic
thermometers); and  closely inspect parts of the \dword{tpc} after any seismic activity or other unanticipated
event.  For these functions, a set of fixed
cold cameras are used; they are permanently mounted at fixed points in the cryostat
for use during filling and commissioning, and a movable, replaceable
warm inspection camera can be deployed through any free
instrumentation flange at any time during the life of the
experiment. 

Eleven cameras were deployed in \dword{pdsp} at the locations shown in Figure~\ref{fig:pdsp-camera-locations}. They successfully provided views of the detector during filling and throughout 
its operation. 

\begin{dunefigure}[Camera locations in ProtoDUNE-SP]{fig:pdsp-camera-locations}
  {A \threed view showing the locations of the 11 cameras in \dword{pdsp}.}
  \includegraphics[width=0.6\textwidth]{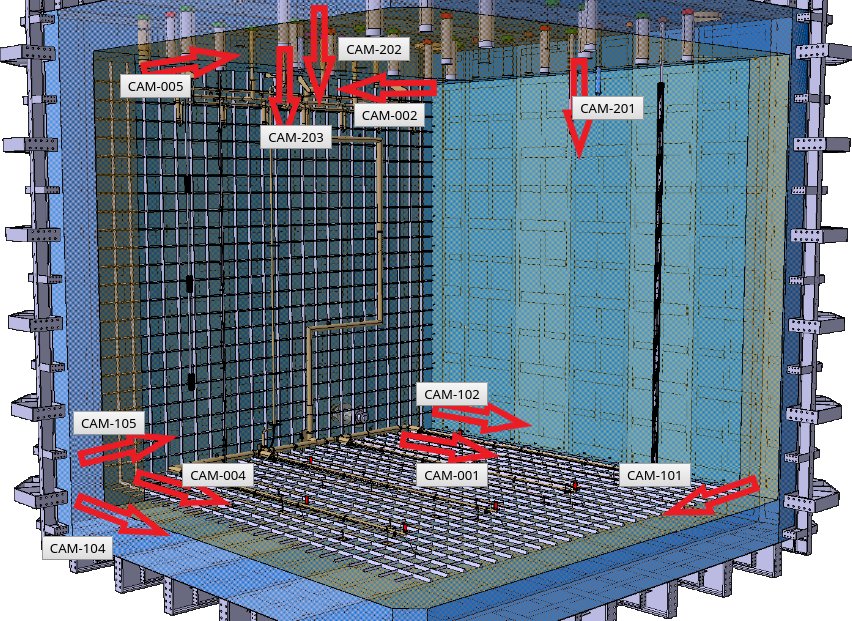}%
\end{dunefigure}

The following sections describe the design considerations for both cold
and warm cameras and the associated lighting system. \dword{pdsp} camera system designs and performance are also discussed.  
The same basic
designs can be used for both the \dword{spmod} and the \dword{dpmod}. 

\subsubsection{Cryogenic Cameras (Cold)}

The fixed cameras
monitor the following items during filling:
\begin{itemize}
\item positions of corners of \dword{apa}, \dword{cpa}, \dwords{fc}, \dwords{gp} (\SI{1}{mm} resolution);
\item relative straightness and alignment of \dword{apa}, \dword{cpa}, and \dword{fc} (\(\lesssim\SI{1}{mm}\));
\item relative positions of profiles and endcaps (\SI{0.5}{mm} resolution); and 
\item the \dword{lar} surface, specifically, the presence of bubbling or debris.
\end{itemize}

One design for the \dword{dune} fixed cameras uses an enclosure similar to
the successful EXO-100 design \cite{Delaquis:2013hva}, which was also
successfully used in the \dword{lapd}
and \dword{pdsp} (see Figure~\ref{fig:gen-fdgen-cameras-enclosure}). Cameras 101, 102, 104, and 105, shown in Figure~\ref{fig:pdsp-camera-locations}, use this enclosure.
A thermocouple in the enclosure allows temperature monitoring, and a heating element provides temperature control.  
SUB-D connectors are used at the cryostat flanges and the camera enclosure for signal, power, and control connections.

\begin{dunefigure}[A camera enclosure]{fig:gen-fdgen-cameras-enclosure}
  {Top left: a CAD exploded view of a vacuum-tight camera enclosure suitable for cryogenic applications \cite{Delaquis:2013hva}.
    (1) quartz window, (2 and 7) copper gasket, (3 and 6) flanges, (4) indium wires, (5) body piece, (8) signal \fdth.
    Top right: two of the \dword{pdsp} cameras using a stainless steel enclosure. 
    Bottom left: one of the \dword{pdsp} cameras using acrylic enclosure.
    Bottom right: a portion of an image taken with \dword{pdsp} camera 105 showing a purity monitor mounted outside the \dword{apa} on the beam left side. This photo was taken with \dword{pdsp} completely filled.
  }
  \includegraphics[width=0.4\textwidth]{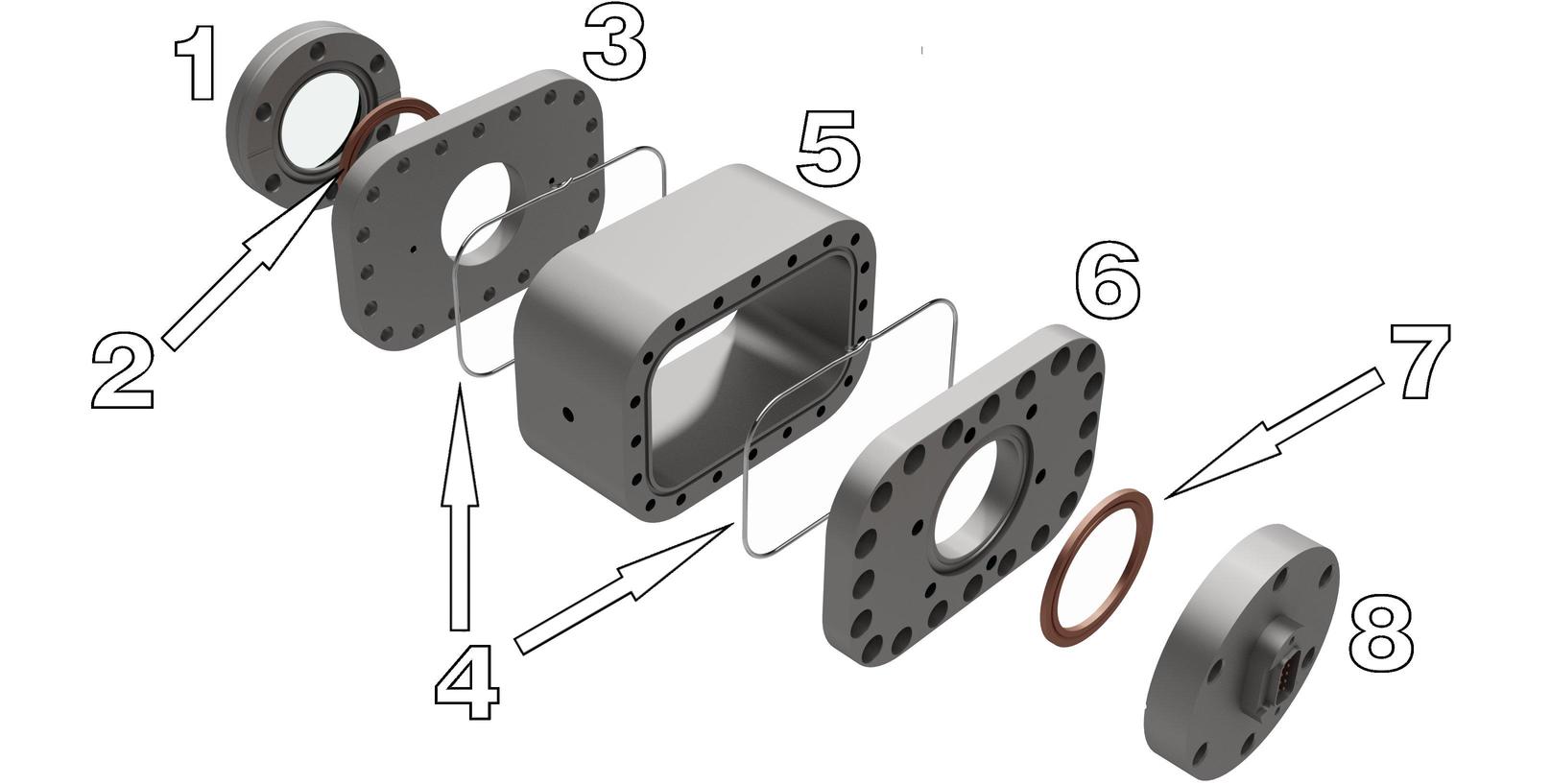}%
  \includegraphics[width=0.4\textwidth]{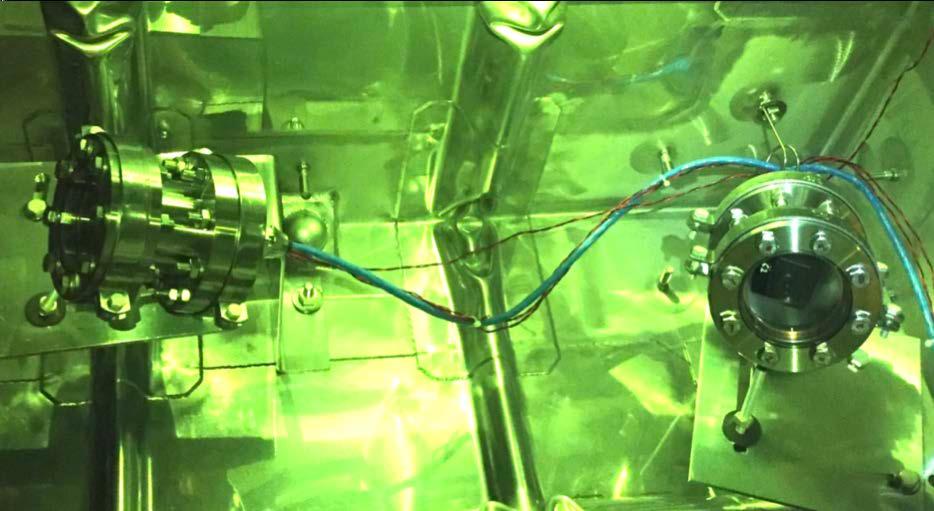}\\
  \hfill \includegraphics[width=0.22\textwidth]{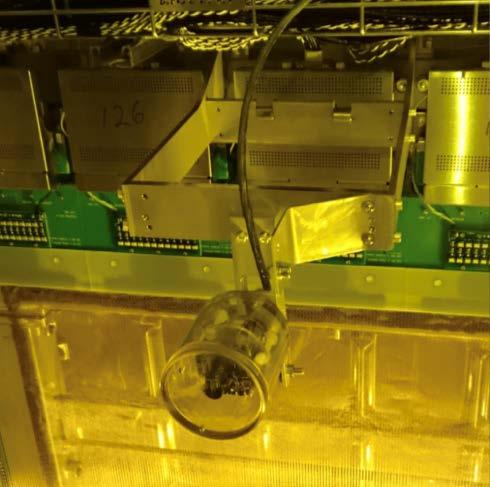}%
  \hfill \includegraphics[width=0.22\textwidth]{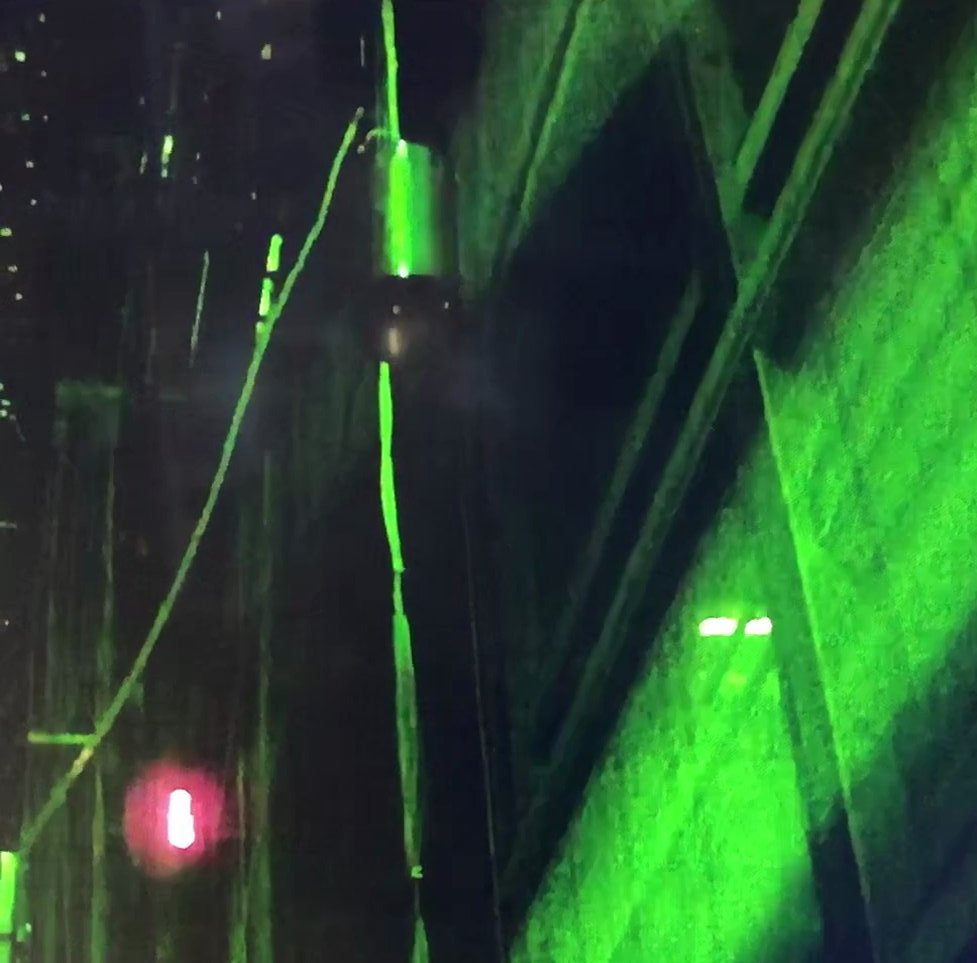}%
  \hfill
\end{dunefigure}

An alternative design uses an acrylic enclosure.
This design was used successfully in \dword{pdsp} (see Figure~\ref{fig:gen-fdgen-cameras-enclosure}, bottom left). Cameras 001, 002, 004, and 005, shown in Figure~\ref{fig:pdsp-camera-locations}, use acrylic enclosures. 
All operate successfully, including those at the bottom of the cryostat. 
The \dword{fd} modules will be twice as deep as \dword{protodune}, and therefore cameras observing the lowest surfaces of the \dword{fc} must withstand twice the pressure.

Improved designs for the cold cameras will be tested in \dword{pddp} and \dword{citf} for improved imaging including focus adjustment, and in \dword{citf} for pressure resistance, during 2020.

\subsubsection{Inspection Cameras (Warm)}

The inspection cameras are intended to be as versatile as possible.
The following 
inspections have been identified as likely uses:
\begin{itemize}
\item status of \dword{hv} \fdth and cup,
\item status of \dword{fc} profiles, endcaps (\SI{0.5}{mm} resolution),
\item vertical deployment of calibration sources,
\item status of thermometers, especially dynamic thermometers,
\item \dword{hv} discharge, corona, or streamers on \dword{hv} \fdth, cup, or \dword{fc},
\item relative straightness and alignment of \dword{apa}, \dword{cpa}, and \dword{fc} (\SI{1}{mm} resolution),
\item gaps between \dword{cpa} frames (\SI{1}{mm} resolution),
\item relative position of profiles and endcaps (\SI{0.5}{mm} resolution), and 
\item sense wires at the top of outer wire planes in \single \dword{apa} (\SI{0.5}{mm} resolution).
\end{itemize}

Unlike the fixed cameras, the inspection cameras must operate only as
long as any inspection; the cameras can be replaced in case of failure.  It
is also more practical to keep the cameras continuously warmer than
 \SI{-150}{\celsius} during deployment; this allows use of  
commercial cameras, 
e.g., cameras of the same model were used successfully to observe discharges
in \dword{lar} from \SI{120}{cm} away \cite{Auger:2015xlo}.

The inspection cameras use the same basic
enclosure design as for cold cameras, but the cameras are mounted on a movable
fork so that each camera can be inserted and removed from the cryostat,
using a design similar to the dynamic temperature probes: see
 Figure~\ref{fig:gen-fdgen-cameras-movable} (left) and
 Figure~\ref{fig:fd-slow-cryo-sensor-mount}.  To avoid contaminating the
\dword{lar} with air, the entire system is sealed, and the
camera can only be deployed through a \fdth equipped with a gate
valve and a purging system, similar to the one used in the vertical axis
calibration system at \kamland~\cite{Banks:2014hra}. The entire system
is  purged with pure argon gas before the gate valve is opened.

\begin{dunefigure}[Inspection camera design]{fig:gen-fdgen-cameras-movable}
  {Left: An overview of the inspection camera design using a sealed deployment system opening directly into the cryostat. Right: A photo of the \dword{pdsp} warm inspection camera acrylic tube, immediately before installation; the acrylic tube is sealed with an acrylic dome at the bottom and can be opened at the top.}
  \includegraphics[height=0.3\textheight]{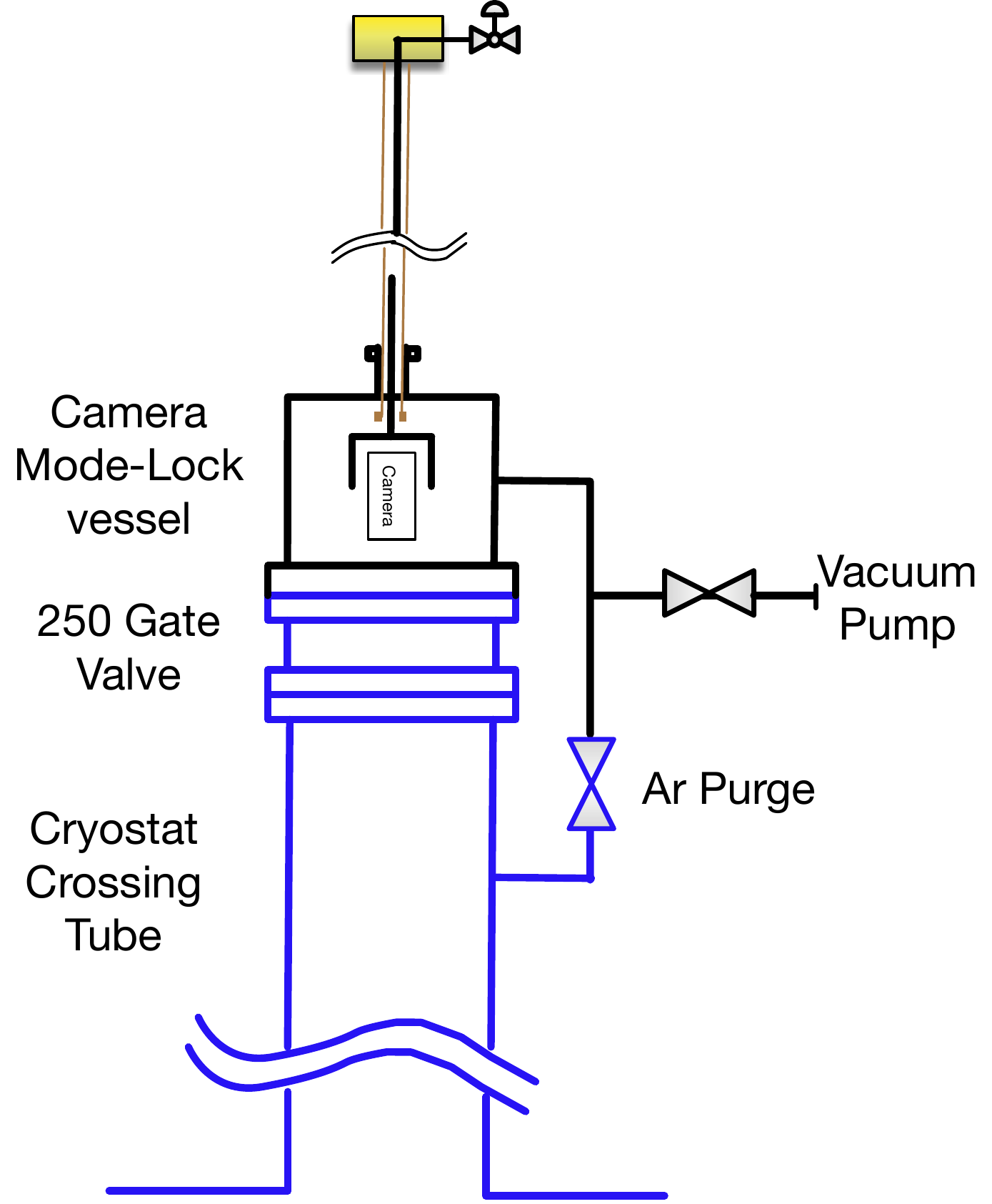}%
  \includegraphics[height=0.3\textheight]{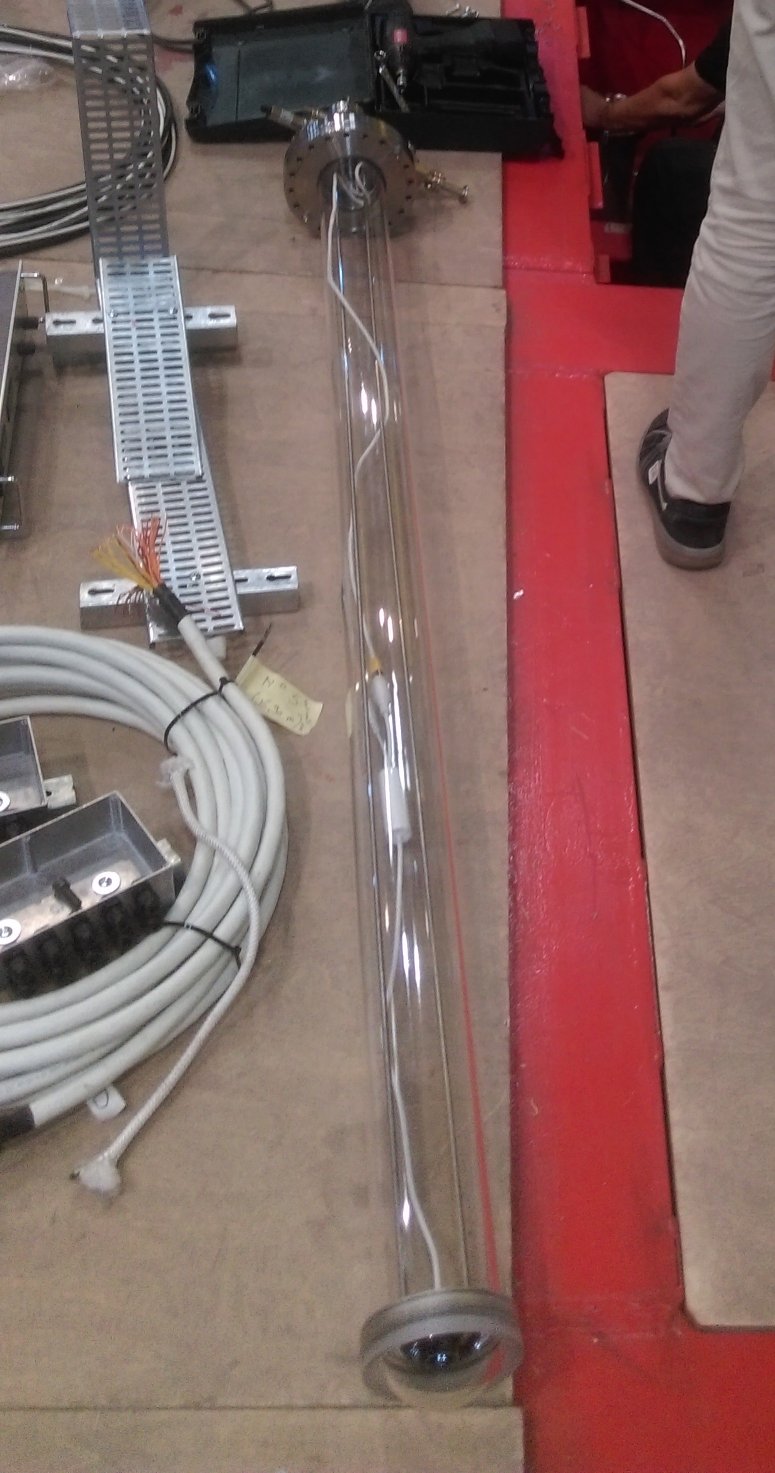}%
\end{dunefigure}

Motors above the flange allow the fork to be rotated and moved vertically to position the camera. 
 A chain drive system with a motor
mounted on the end of the fork allows the camera assembly to tilt, 
creating a point-tilt mount that can be moved vertically.
With the space above the cryostat flanges and the
thickness of the cryostat insulation, cameras can be moved vertically up to
\SI{1}{m} inside the cryostat.
The motors for rotation and vertical motion are outside the sealed
volume, coupled mechanically using ferrofluidic seals, thus reducing any risk of
contamination and allowing manual rotation of the vertical
drive if the motor fails.  

An alternative design was demonstrated in \dword{pdsp}. In this design, the warm camera is contained inside a gas-tight acrylic tube inserted into the feedthrough, so a gate valve or a gas-tight rotatable stage is not needed, and the warm cameras can be removed for servicing or upgrade at any time. Figure~\ref{fig:gen-fdgen-cameras-movable} (right) shows an acrylic tube enclosure and camera immediately before deployment. 
These acrylic tube enclosures for removable cameras were deployed in \dword{pdsp} at the positions marked 201, 202, and 203 in Figure~\ref{fig:pdsp-camera-locations}; they operated successfully. Cameras with fisheye lenses were used in these tubes during initial operation.  One camera was removed without any evidence of contamination of the \dword{lar}. We plan to use other cameras during post-beam running.

Improved designs for the inspection cameras will be tested in the \dword{citf} and \dword{pdsp} during 2020 and 2021, focusing particularly on longevity, camera replaceability, and protection of the \dword{lar}.

\subsubsection{Light-emitting system}
The light-emitting system uses \dwords{led} to illuminate
the parts of the 
\dword{detmodule} in the camera's field of view with selected
wavelengths (IR and visible) that cameras can detect.  Performance criteria for the light-emission system include the efficiency with which the cameras can detect the light and the need to avoid
adding heat to the cryostat. Very high-efficiency
\dwords{led}   
help reduce heat generation; one \SI{750}{nm} \dword{led} \cite{lumileds-DS144-pdf}
has a specification equivalent to
\SI{33}{\%} conversion of electrical input power to light.

While data on how well \dwords{led} perform at cryogenic temperatures
is sparse, some studies of NASA projects~\cite{Carron:2017zzz}
indicate that \dwords{led} are more efficient at low temperatures and
that emitted wavelengths may change, particularly for blue
\dwords{led}.  In \dword{pdsp}, amber \dwords{led} were observed  to
emit green light at \dword{lar} temperature (bottom right photo
in Figure~\ref{fig:gen-fdgen-cameras-enclosure}).  To avoid degradation of
wavelength-shifting materials in the \dword{pds}, short wavelength
\dwords{led} are not used in the \dword{fd}; \dwords{led} will be tested
in \dword{ln} to ensure their wavelength is long enough.

\dwords{led} are placed in a ring around the outside of each
camera, pointing in the same direction as the lens, to 
illuminate nearby parts of the \dword{detmodule} within the camera's field of
view. Commercially available \dwords{led} exist with
a range of angular spreads that can be matched to the needs of the
cameras without additional optics.

Additionally, chains of \dwords{led} connected in series and driven with a
constant-current circuit are used for broad illumination, with each
\dword{led} paired in parallel with an opposite polarity \dword{led} and a resistor
(see Figure~\ref{fig:cisc-LED}).
This allows two different wavelengths of illumination using a single chain simply by changing the direction of the drive current, and allows continued use of an \dword{led} chain even if individual \dwords{led} fail.

\begin{dunefigure}[Example schematic for LED chain]{fig:cisc-LED}
  {Example schematic for LED chain, allowing failure tolerance and two LED illumination spectra.}
  \includegraphics[width=0.6\textwidth]{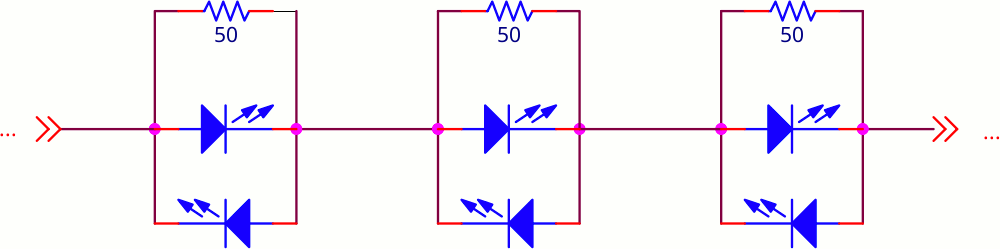}
\end{dunefigure}

\subsection{Cryogenics Instrumentation Test Facility}
The \dword{cisc} consortium plans to build a \dfirst{citf} at \dword{fnal} to facilitate testing of various cryogenics instrumentation devices and small-scale assemblies of \dword{cisc} systems. 
In the past and recent times, various test facilities at \dword{fnal} have provided access to small ($<\,\SI{1}{ton}$) to intermediate ($\sim\,\SI{1}{ton}$) volumes of purified \dword{tpc}-grade \dword{lar}, required for 
any device intended for drifting electrons for millisecond periods.

The \dword{pab} facility at \dword{fnal} houses the \dword{iceberg} \SI {3000} {liter} cryostat, which enables fast turnaround testing for the \dword{dune} \dword{ce}. 

The \dword{pab} facility also includes  \dword{tallbo} (\SI {450} {liter}), Blanche (\SI {500} {liter}), and Luke (\SI {250} {liter}) cryostats. 
In the recent past, Blanche has been used for \dword{hv} studies,  \dword{tallbo} for \dword{pd} studies, and Luke for the material test stand work. These studies have contributed to the design and testing of  \dword{pdsp} components.

\subsection{Validation in ProtoDUNE}
\label{sec:pdsp-cryo-valid}

Design validation and testing of many planned \dword{cisc} systems for
the \dword{spmod} will be done using the data from
\dword{pdsp} and \dword{pddp} as discussed below.

\begin{itemize}
	\item Level Meters: The same differential pressure level meters
	      which are already validated in \dword{pdsp} will be used in the \dword{spmod}. The same capacitive level meters currently used in
	      \dword{pddp} will be used in the \dword{spmod}. These will be
	      validated in the upcoming \dword{pddp} run.
	      
	\item Pressure Meters (GAr): The same high-precision pressure sensors that are already validated in \dword{pdsp} will be used in \dword{sp} \dword{fd}.
	
	\item Gas Analyzers: The same gas analyzers currently used in
	      \dword{pdsp} will be used in the \dword{spmod}, so they have already
	      been validated.
	      
	\item High-precision thermometer arrays in \dword{lar}: The static
	      and dynamic T-gradient thermometers discussed in the previous sections are validated using \dword{pdsp} data.
	      
	\item Purity Monitors: The same purity monitor basic design used in \dword{pdsp} will be used in the \dword{spmod}. \dword{protodune2} at \dword{cern} provides opportunities to
	      test any improvements to the design.
	\item Cameras: various types of cameras are being actively
	      developed in both \dword{pdsp} and \dword{pddp} so these detectors will perform validation of the 
	      designs. Future
	      improvements can be tested in \dword{protodune2} at \dword{cern}. 
\end{itemize}

\section{Slow Controls}

The slow controls system collects, archives, and displays data from
a broad variety of sources and provides real-time status, alarms, and warnings for detector operators. The slow control system also provides control for 
items such as \dword{hv} systems, \dword{tpc} electronics, and \dword{pd} systems. Data is acquired via network interfaces.  Figure~\ref{fig:gen-slow-controls-diagram} shows connections between major parts of the slow controls system and other systems. 

\begin{dunefigure}[Slow controls connections and data]{fig:gen-slow-controls-diagram}
{Typical slow controls system connections and data flow}
\includegraphics[width=0.7\textwidth]{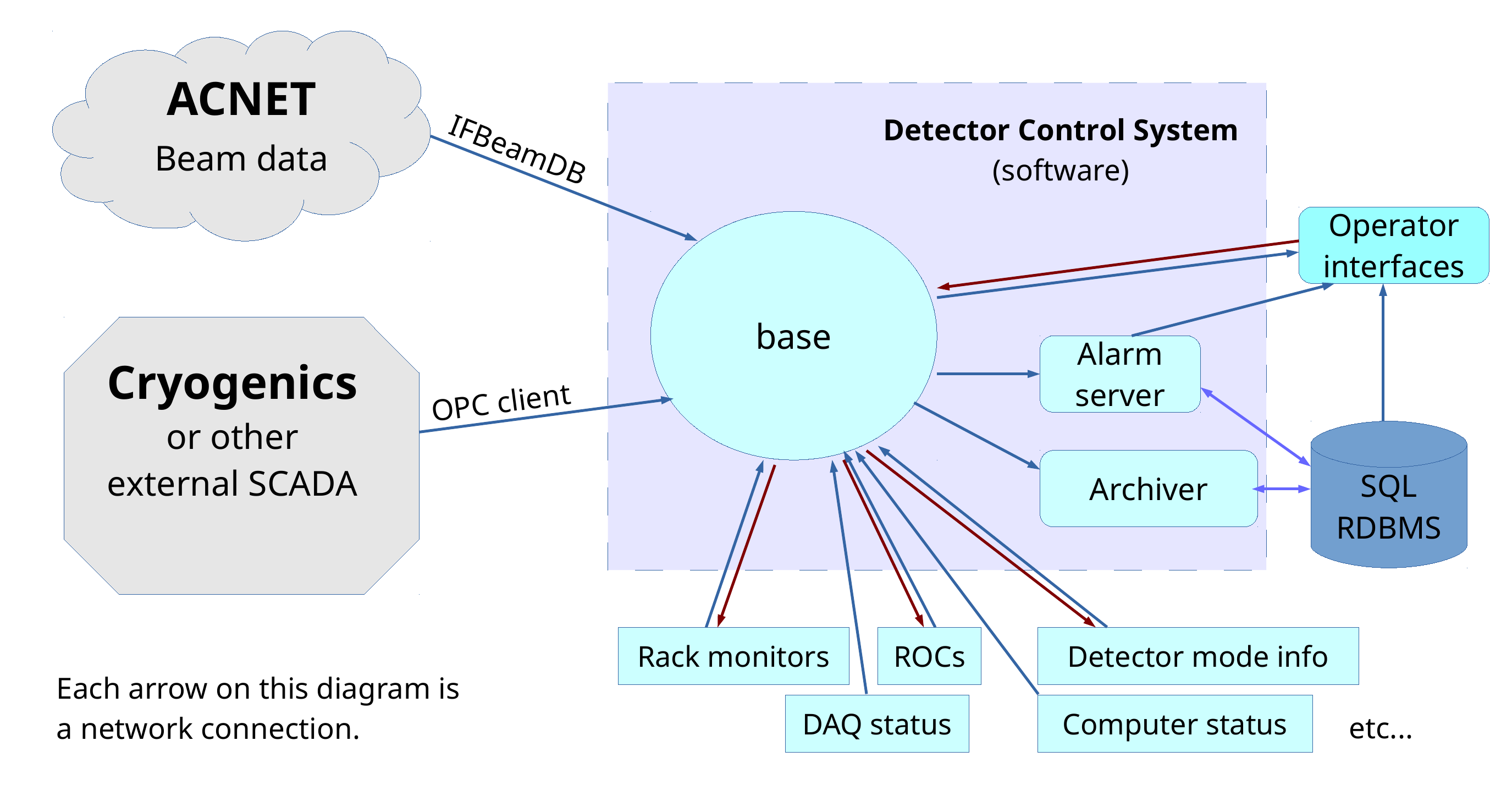}
\end{dunefigure}

The \dword{pdsp} detector control system\cite{pdspdcs_proc} fully met its operational requirements. 
Section~\ref{sec:cisc-slow-control-pdsp} provides a short description of the \dword{pdsp} slow controls and its performance.

 \subsection{Slow Controls Hardware}
\label{sec:fdgen-slow-cryo-hdwr}

Slow controls is expected to need a modest amount of dedicated hardware, largely for rack monitoring,  
and a small amount of dedicated network and
computing hardware. 
Slow controls also relies on common
infrastructure as described in
Section~\ref{sec:fdgen-slow-cryo-slow-infra}.

\subsubsection{Dedicated Monitoring Hardware}

Every rack (including those in the \dword{cuc}) should have dedicated hardware to monitor rack parameters like rack protection system, rack fans, rack air temperatures, thermal interlocks with power supplies, and any interlock bit status monitoring needed for the racks. For the racks in the \dword{cuc} server room, this functionality is built into the proposed water cooled racks, as already in place at \dword{protodune}.  For the racks on the detector itself, the current plan is to design and install a custom-built 1U rack-mount enclosure containing a single-board computer to control and monitor various rack parameters. Such a system has been successfully used in \dword{microboone}. The design is being improved for the \dword{sbnd} experiment (see Figure~\ref{fig:slow-controls-rack-box}). Other slow controls hardware includes interfacing cables like adapters for communication and debugging and other specialized cables like  \dword{gpib} or National Instruments cables. The cable requirements must be determined in consultation with other groups once hardware choices for various systems are finalized.

\begin{dunefigure}[Rack monitoring box prototype for the SBND; based on MicroBooNE design]{fig:slow-controls-rack-box}
{Rack monitoring box prototype in development for the \dshort{sbnd} experiment based on the original design from \dshort{microboone}.}
\includegraphics[width=0.6\textwidth]{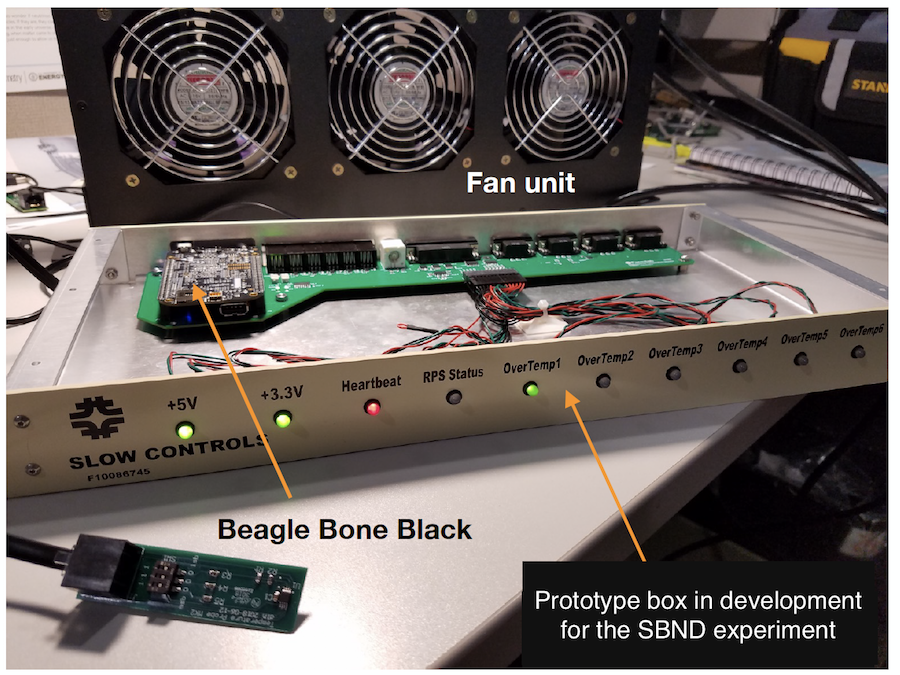}
\end{dunefigure}

\subsubsection{Slow Controls Network Hardware}
\label{sec:fdgen-slow-cryo-slow-network}
The slow controls data originates from the cryogenics instrumentation and from other systems as listed in Table~\ref{tab:gen-slow-quant}. This data is collected by software running on servers
(Section~\ref{sec:fdgen-slow-cryo-slow-compute})
housed in the underground data room in the \dword{cuc},
where data is archived in a central \dword{cisc} database.
The instrumentation data is transported over
conventional network hardware from any sensors located in the cryogenics
plant.  However, the readouts that are in the racks on top of the
cryostats must be cautious about grounding and noise.  Therefore, each
rack on the cryostat has a small network switch that sends
any network traffic from that rack to the \dword{cuc} via a fiber transponder.
This is the only network hardware specific to slow controls and will be provided by 
\dword{surf}'s  
general computing infrastructure.  
The network infrastructure requirements are described in
Section~\ref{sec:fdgen-slow-cryo-slow-infra}.

\subsubsection{Slow Controls Computing Hardware}
\label{sec:fdgen-slow-cryo-slow-compute}
Two servers (a primary server and a replicated backup) suitable for the relational database discussed
in Section~\ref{sec:fdgen-slow-cryo-sw} are located in the \dword{cuc} data
room, with an additional
two servers to service the \dword{fe} monitoring interface.  These additional servers would cover assembling dynamic \dword{cisc} monitoring web pages from adjacent
databases. Yet another server will be needed to run back-end I/O.  Any special purpose software, such as iFix used by the cryogenics experts, would
also run here. One or two additional servers should accommodate these programs.
Replicating this setup on a per-module basis would make commissioning and independent operation easier, accommodate different module
design (and the resulting differences in database tables), and ensure
sufficient capacity.  These four sets of networking hardware would fit tightly into one rack or very comfortably into two. 

\subsection{Slow Controls Infrastructure}
\label{sec:fdgen-slow-cryo-slow-infra}

The data rate will be in the range of tens of kilobytes per second, given the total number of slow controls quantities and the update rate  
(see Section~\ref{sec:fdgen-slow-cryo-quant}), placing minimal demands
on local network infrastructure.
Network traffic out of \dword{surf} to \dword{fnal} will primarily be database calls
to the central \dword{cisc} database, either from monitoring applications or from
database replication to the offline version of the \dword{cisc} database.  This
traffic requires little bandwidth, so the proposed general purpose
links both out of the 
underground area at  \dword{surf} and back to \dword{fnal} can accommodate the traffic.

Up to two racks of space and appropriate power and cooling are
available in the \dword{cuc}'s \dword{daq} server room for \dword{cisc} use. This is ample space as described in Section~\ref{sec:fdgen-slow-cryo-slow-compute}.

\subsection{Slow Controls Software}
\label{sec:fdgen-slow-cryo-sw}

To provide complete monitoring and control of detector subsystems, the slow controls software includes
\begin{itemize}
 \item the control systems base for input and output operations
  and defining processing logic, scan conditions, and alarm conditions;
 \item an alarm server to monitor all channels and send alarm
  messages to operators;
 \item a data archiver for automatic sampling and storing values for history tracking; and 
 \item an integrated operator interface providing display panels for
  controls and monitoring.
\end{itemize}

In addition, the software must be able to 
interface indirectly with external systems (e.g., cryogenics control
system) and databases (e.g., beam database) to export data into
slow controls process variables (or channels) for archiving and status
displays. This allows us to integrate displays and warnings into one
system for the experiment operators and 
provides integrated
archiving for sampled data in the archived database. As one possibility, an input output controller running on a central \dword{daq}
server could provide soft channels for these data.
Figure~\ref{fig:gen-slow-controls-diagram} shows a typical workflow of a
slow controls system.

The key features of the software require highly evolved software designed to manage real-time data exchange, scalable
to hundreds of thousands of channels sampled at intervals of hours to seconds as needed. The software
must be well documented, supported, and reliable. The base
software must also allow easy access to any channel by name. The
archiver software must allow data storage in a database with
adjustable rates and thresholds so data
for any channel can be easily retrieved using channel name and time range. Among other key
features, the alarm server software must remember the state, support an
arbitrary number of clients, and provide logic for delayed alarms and
acknowledging alarms. A standard naming
convention for channels will be part of the software to help handle large
numbers of channels and subsystems.

The \dword{pdsp} detector control system software \cite{pdspdcs_proc} provides a prototype for 
the \dword{fd} slow controls software.
In \dword{pdsp}, the unified control system base is WinCC OA \cite{winccoa}, a
commercial toolkit used extensively at \dword{cern}, with device interfaces
supported using several standardized interface protocols. A more detailed description is in Section~\ref{sec:cisc-slow-control-pdsp} below.
WinCC OA is our baseline for the \dword{fd} slow control software.
EPICS~\cite{epics7} is an alternative controls system which also meets the specifications; it is used in other neutrino experiments including \dword{microboone}~\cite{microboone} and \dword{nova}~\cite{Lukhanin:2012fp}.

\subsection{Slow Controls Quantities}
\label{sec:fdgen-slow-cryo-quant}


The final set of quantities to monitor will ultimately be determined
by the subsystems being monitored, documented in
appropriate  interface control documents (ICDs), and continually revised based on operational
experience.  The total number of quantities to monitor has been roughly estimated by taking the total number monitored
in \dword{pdsp}\cite{pdspdcs_proc}, 7595 as of Nov. 19, 2018, and scaling by the detector length and the number of planes, giving approximately 150,000 per \dword{detmodule}.
Quantities should update on average no more than once per minute.
Transmitting a single update for each channel at that rate translates to a few thousand updates per second, or a few tens of thousands of bytes per second. While this is not a significant load on a network with an efficient slow controls protocol, it would correspond to approximately \SI{1}{TB} per year per \dword{detmodule} if every timestamp and value were stored.
The actual data volume will be lower 
because values are stored only if they vary from previous values by more than an amount that is adjustable channel-by-channel.
Database storage also allows data to be sparsified later.
No slow controls data is planned to be written to the \dword{daq} stream.
With careful management of archiving thresholds for each quantity monitored and yearly reduction of stored sample time density, the retained data volume can be reduced to a few TB over the life of the experiment.

The subsystems
to be monitored include the  
cryogenics instrumentation
described in this chapter, the other detector systems, and relevant
infrastructure and external devices. Table~\ref{tab:gen-slow-quant}
lists the quantities expected from each system.

\begin{dunetable}
[Slow controls quantities]
{p{0.3\textwidth}p{0.6\textwidth}}
{tab:gen-slow-quant}
{Slow controls quantities}
System & Quantities \\ \toprowrule

\multicolumn{2}{l}{\bf Detector cryogenics instrumentation } \\ \toprowrule 
Purity monitors & anode and cathode charge, bias voltage and current, flash lamp status, calculated electron lifetime \\ \colhline
Thermometers & temperature, position of dynamic thermometers \\ \colhline
Liquid level & liquid level \\ \colhline
Gas analyzers & purity level readings \\ \colhline
Pressure meters & pressure readings \\ \colhline
Cameras & camera voltage and current draw, temperature, heater current and voltage, lighting current and voltage \\  \colhline 

\multicolumn{2}{l}{\bf Other detector systems } \\ \toprowrule 
Cryogenic internal piping & \fdth gas purge flow and temperature \\ \colhline
\dword{hv} systems & drift \dword{hv} voltage and current, end-of-field cage current and bias voltage, electron diverter bias, ground plane currents \\ \colhline
\dword{tpc} electronics & voltage and current to electronics \\ \colhline
\dword{pd} & voltage and current for photodetectors and electronics \\ \colhline
\dword{daq} & warm electronics currents and voltages, run status, \dword{daq} buffer sizes, trigger rates, data rates, GPS status, computer and disk health status, other health metrics as defined by \dword{daq} group \\ \colhline
\dword{crp} / \dword{apa} & bias voltages and currents \\  \colhline 

\multicolumn{2}{l}{\bf Infrastructure and external systems } \\ \toprowrule 
Cryogenics (external) & status of pumps, flow rates, inlet and return temperature and pressure (via OPC or similar SCADA interface) \\ \colhline
Beam status & protons on target, rate, target steering, beam pulse timing (via \dword{ifbeam}) \\ \colhline
Near detector & near detector run status (through common slow controls database) \\ \colhline
Rack power and status & power distribution unit current and voltage, air temperature, fan status if applicable, interlock status \\ \colhline

\multicolumn{2}{l}{\bf Detector calibration systems } \\ \toprowrule 
Laser & laser power, temperature, operation modes, other system status as defined by calibration group\\ \colhline
External neutron source  & safety interlock status, power supply monitoring, other system status as defined by calibration group \\ \colhline
External radioactive source & system status as defined by calibration group\\
\end{dunetable}

\subsection{Local Integration}
\label{sec:fdgen-slow-cryo-slow-loc-integ}

The local integration of the slow controls consists entirely of software
and network interfaces with systems that are outside of the scope of the \dword{detmodule}. 
This includes the following:
\begin{itemize}
\item readings from the \dword{lbnf}-managed external cryogenics systems, for status of pumps, flow rates, inlet, and return temperature and pressure, which are implemented via \dword{opc-ua} or a similar \dword{scada} interface; 
\item beam status, such as protons-on-target, rate, target steering, and beam pulse timing, which are retrieved via \dword{ifbeam}; 
and \item near detector status, which can be retrieved from a common slow controls database.
\end{itemize}
Integration occurs after both the slow controls and non-detector
systems are in place.  The \dword{lbnf}-\dword{cisc} interface is managed by the
cryogenics systems working group in \dword{cisc} (see Section~\ref{sec:cisc-slow-controls-org}), which includes members from both \dword{cisc} and \dword{lbnf}. 
The \dword{ifbeam} DB interface for slow controls is an already well established method used in \dword{microboone}, \dword{nova}, and other \dword{fnal} experiments. An internal \dword{nd}/\dword{fd} working group can be established 
to coordinate detector status exchange between the near and far sites.

\subsection{Validation in ProtoDUNE}
\label{sec:cisc-slow-control-pdsp}

The \dword{pdsp} detector control system has met
all requirements for operation of \dword{pdsp}~\cite{pdspdcs_proc} and will be used for \dword{pddp}. The requirements for \dword{protodune} are
nearly identical to those for the \dword{spmod} other than
total channel count. Of particular note, the \dword{protodune} slow control system unified a heterogenous set of devices and data sources
through several protocols into a
single control system, as illustrated in
Figure~\ref{fig:cisc-NP04-DCS-topology}. In addition to what
the figure shows, data were also acquired from external cryogenics and beam
systems.  The topology and data flow of the system matches the general
shape shown in Figure~\ref{fig:gen-slow-controls-diagram}.

\begin{dunefigure}[Diagram of the ProtoDUNE-SP control system topology]{fig:cisc-NP04-DCS-topology}
{Diagram of the \dword{pdsp} control system topology, from \cite{pdspdcs_proc}.}
\includegraphics[height=0.9\textheight,width=0.95\textwidth,keepaspectratio]{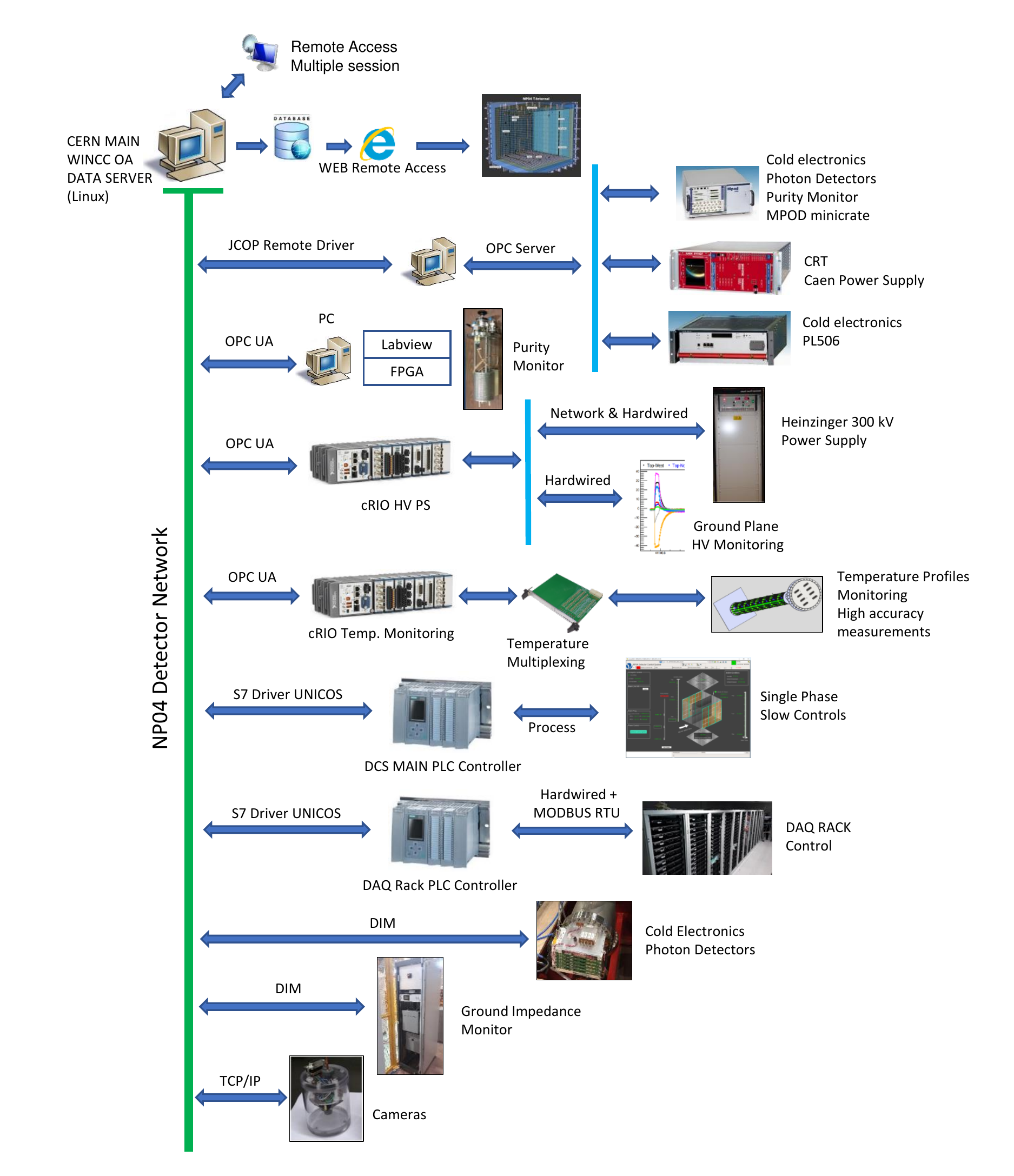}
\end{dunefigure}

In this control system, the unified control system base is WinCC-OA~\cite{winccoa}, a commercial \dword{scada} system for visualizing and operating of processes, production flows, machines, and plants, used
in many businesses. It was chosen at \dword{cern} as a basis for
developing the control systems of the \dword{lhc} experiments, the
accelerators and the laboratory infrastructure for its flexibility and
scalability, as well as for the openness of the architecture, allowing
it to interface with many different types of hardware devices and
communication protocols. Additional software developed at \dword{cern}
is also used, including Joint COntrols Projects~\cite{jcop} and UNified
Industrial COntrol System (UNICOS)~\cite{unicos}. WinCC-OA and the
additional software developed on top of it in the past 20 years, have
grown into a fairly complex ecosystem. While multiple collaboration
members have experience using the \dword{pdsp} control system,
customizing and using WinCC-OA in an effective way for developing the
control system of \dword{dune} requires proper training and a
non-negligible learning effort.

As noted in Sections~\ref{sec:fdgen-slow-cryo-sw} and~\ref{sec:fdgen-slow-cryo-quant},
the slow control archiver will gradually accumulate terabytes of
data, requiring a sizable database to store the value history and
allow efficient data retrieval. Individually adjustable rates and
thresholds for each channel are key to keeping this database
manageable. The \dword{pdsp} operations provided not only a test of
these features as implemented in the \dword{protodune} slow control system, but also insight into
reasonable values for these archiving parameters for each system.

\section{Organization and Management}
\label{sec:cisc-slow-controls-org}

The organization of the \dword{cisc} consortium is shown in
 Figure~\ref{fig:gen-slow-cryo-org}. The \dword{cisc} consortium board currently comprises institutional representatives from 19 institutes as shown in Table~\ref{tab:gen-slow-cryo-org}. The consortium leader is the spokesperson for the consortium and responsible for the overall scientific program and managing the group. The technical leader of the consortium is responsible for managing the project for the group. Currently, the
consortium has five working groups:
\begin{description}
 \item[Cryogenics Systems] gas analyzers and liquid level
  monitors; \dword{cfd} simulations;
  
 \item[Argon Instrumentation] purity monitors, thermometers, pressure meters, capacitive level meters, cameras and light emitting system, and \dword{citf}, also feedthroughs, \efield simulations, instrumentation precision studies, \dword{protodune} data analysis coordination and validation; 
 
 \item [Slow Controls Base Software and Databases]  base I/O software, alarms and archiving databases, and monitoring tools, also 
   variable naming conventions, and slow controls quantities;
 \item [Slow Controls Detector System Interfaces] signal processing software and hardware interfaces (e.g., power supplies), firmware, rack hardware and infrastructure   
 \item [Slow Controls External Interfaces] interfaces with external detector systems (e.g., cryogenics system, beam, facilities, \dword{daq}, and near detector status).
\end{description}

\begin{dunefigure}[CISC consortium organization]{fig:gen-slow-cryo-org}
{CISC Consortium organizational chart}
\includegraphics[width=0.8\textwidth]{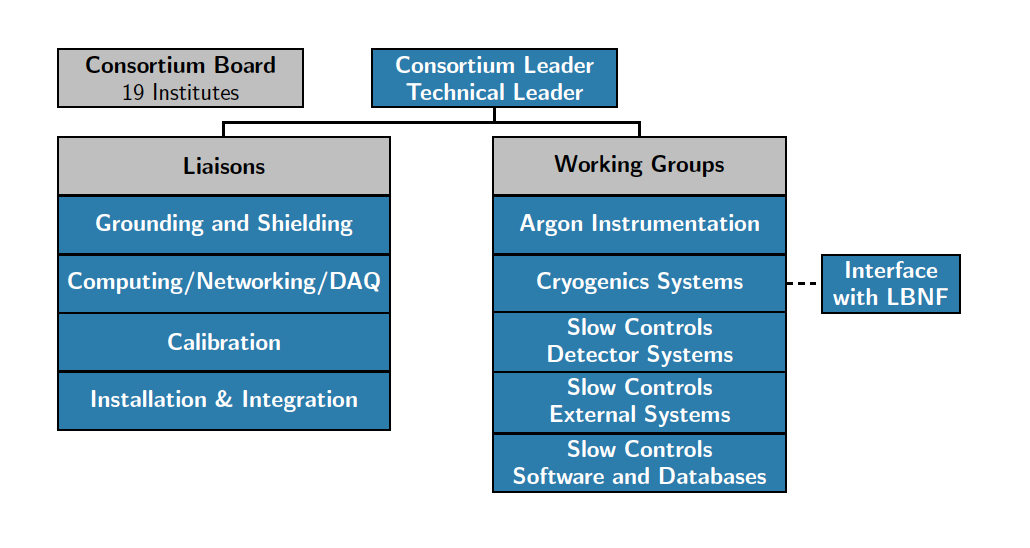}
\end{dunefigure}

\begin{dunetable}
[CISC consortium institutions]
{lc}
{tab:gen-slow-cryo-org}
{Current \dword{cisc} consortium board members and their institutional affiliations}
Member Institute                         &  Country       \\
CIEMAT                                   &  Spain         
\\ \colhline
Instituto de Fisica Corpuscular (IFIC)          &  Spain          
\\ \colhline
University of Warwick                    &  UK 
\\ \colhline
University College London (UCL)             &  UK  
\\ \colhline
Argonne National Lab (ANL)                     &  USA             
\\ \colhline
Brookhaven National Lab (BNL)                  &  USA            
\\ \colhline
University of California, Irvine (UCI)        &  USA            
\\ \colhline
Drexel University                        &  USA           
\\ \colhline
Fermi National Accelerator Lab (\fnal)           &  USA           
\\ \colhline
University of Hawaii                     &  USA            
\\ \colhline
University of Houston                    &  USA           
\\ \colhline
Idaho State University (ISU)                   &  USA           
\\ \colhline
Kansas State University (KSU)                  &  USA            
\\ \colhline
University of Minnesota, Duluth (UMD)         &  USA            
\\ \colhline
Notre Dame University                    &  USA            
\\ \colhline
South Dakota State University (SDSU)           &  USA            
\\ \colhline
University of Tennessee at Knoxville (UTK)     &  USA           
\\ \colhline
Virginia Tech (VT)                            &	USA	           
\\	\colhline
	Boston University (BU)			      & USA 
\\
\end{dunetable}

Moreover, because the \dword{cisc} consortium broadly interacts with other groups, liaisons have been named as shown in Figure~\ref{fig:gen-slow-cryo-org}. 
A short-term task force was recently formed to explore the need for cryogenics modeling for the consortium. A work plan for \dword{cfd} simulations for both \dword{protodune} and \dword{fd} was developed based on input from the task force.

\subsection{Institutional Responsibilities}

The \dword{cisc} consortium  
will be a joint  effort for \dword{sp} and \dword{dp}. 
A single slow controls system will be implemented to serve both the \dword{spmod} and the \dword{dpmod}.

Design and installation of cryogenics systems (e.g., gas analyzers, liquid level monitoring) 
will be coordinated with \dword{lbnf}, with the consortium providing resources and effort, and expertise provided by \dword{lbnf}. \dword{protodune} designs for \dword{lar} instrumentation (e.g., purity monitors, thermometers, cameras) will be the basis for \dword{detmodule} designs. Design validation, testing, calibration, and performance will be evaluated through \dword{protodune} data.

Following the conceptual funding model envisioned for the consortium, various responsibilities have been distributed across institutions within the consortium pending final funding decisions.
Table~\ref{tab:cisc-inst-resp} shows the current institutional responsibilities for primary \dword{cisc} subsystems. Only lead institutes are listed in the table for a given effort. For physics and simulations studies and for validation using \dword{protodune}, a number of institutes are involved. A detailed list of tasks and institutional responsibilities are presented in~\cite{bib:docdb5609}.

\begin{dunetable}
[Institutional responsibilities in the CISC consortium ]
{p{0.4\textwidth}p{0.45\textwidth}}
{tab:cisc-inst-resp}
{Institutional responsibilities in the \dword{cisc} consortium}
CISC Sub-system     &  Institutional Responsibility \\ \toprowrule
Purity Monitors          &  UCI, Houston \\ \colhline
Static T-gradient monitors     &  IFIC \\ \colhline
Dynamic T-gradient monitors & Hawaii \\ \colhline
Individual Sensors & IFIC \\ \colhline
Readout System for Thermometers & IFIC, Hawaii, CIEMAT \\ \colhline
Pressure Meters & UTK \\ \colhline
Cold Cameras & KSU, BNL \\ \colhline
Warm Cameras & KSU, BNL \\ \colhline
Light-emitting System (for cameras) & Drexel \\ \colhline
Gas Analyzers & FNAL, \dshort{lbnf} \\ \colhline
Differential Pressure Level Meters & \dshort{lbnf} \\ \colhline
Capacitive Level Meters & Notre Dame \\ \colhline
\dshort{citf} & FNAL, ANL \\ \colhline
\dshort{cfd} Simulations & SDSU, ANL \\ \colhline
Other Simulation \& Validation Studies & Number of Institutes \\ \colhline
Slow Controls Hardware & UMD, UTK, Drexel\\ \colhline
Slow Controls Infrastructure & UMD, UTK\\ \colhline
Slow Controls Base Software & KSU, UTK, BU, Drexel, Warwick, ANL, IFIC\\ \colhline 
Slow Controls Signal Processing & A number of institutes \\ \colhline
Slow Controls External Interfaces & VT, UTK, UMD \\
\end{dunetable}

\subsection{Schedule}
\label{sec:fdgen-cisc-schedule}

Table \ref{tab:sp-cisc-sched} shows key construction milestones for the \dword{cisc} consortium leading to commissioning of the first \dword{fd} module. \dword{cisc} construction milestones align with the overall construction milestones of the first \dword{fd} module (highlighted in orange in the table). The technology design decisions for \dword{cisc} systems should be made by April 2020 followed by final design reviews in June 2020. Design decisions will largely be based on how a given system performed (technically and physics-wise) in \dword{protodune}. This is currently actively ongoing with the \dword{pdsp} instrumentation data.  
As noted in Section~\ref{sec:pdsp-cryo-valid}, the current plan is to deploy improved designs of static and dynamic T-gradient thermometers, purity monitors, long (\dual-style) level meters and cameras to be validated in \dword{protodune2}. The production of systems aimed for \dword{protodune2} \single should be finished by January 2021 followed by assembly and deployment in March 2021. 

\begin{dunetable}
[SP CISC schedule and milestones]
{p{0.65\textwidth}p{0.25\textwidth}}
{tab:sp-cisc-sched}
{\dword{cisc} construction schedule milestones leading to commissioning of the first FD module. Key DUNE dates and milestones, defined for planning purposes in this TDR, are shown in orange.  Dates will be finalized following establishment of the international project baseline.}   
Milestone & Date (Month YYYY)   \\ \toprowrule
Technology Decision Dates &   April 2020   \\ \colhline
Final Design Review Dates &   June 2020   \\ \colhline
Start of module 0 component production for \dword{protodune2} & August 2020  \\ \colhline
End of module 0 component production for \dword{protodune2} & January 2021  \\ \colhline
\rowcolor{dunepeach} Start of \dword{pdsp}-II installation& \startpduneiispinstall      \\ \colhline
\rowcolor{dunepeach} Start of \dword{pddp}-II installation& \startpduneiidpinstall      \\ \colhline
\rowcolor{dunepeach}South Dakota Logistics Warehouse available& \sdlwavailable      \\ \colhline
 \dword{prr} dates &  September 2022    \\ \colhline
\rowcolor{dunepeach}Beneficial occupancy of cavern 1 and \dword{cuc}& \cucbenocc      \\ \colhline
Start procurement of \dword{cisc} hardware & December 2022 \\ \colhline
\rowcolor{dunepeach} \dword{cuc} counting room accessible& \accesscuccountrm      \\ \colhline
Start of production of \dword{cisc} hardware & April 2023 \\ \colhline
\rowcolor{dunepeach}Top of \dword{detmodule} \#1 cryostat accessible& \accesstopfirstcryo      \\ \colhline
End of \dword{cisc} hardware production  &   April 2024   \\ \colhline

Start integration of \dword{cisc} hardware in the cavern & July 2024   \\ \colhline
\rowcolor{dunepeach}Start of \dword{detmodule} \#1 \dword{tpc} installation& \startfirsttpcinstall      \\ \colhline
Installation of gas analyzers and support structure for all instrumentation devices &  September 2024 \\ \colhline
Installation of individual sensors, static T-gradient thermometers, and level meters & November 2024\\ \colhline
\rowcolor{dunepeach}Top of \dword{detmodule} \#2 cryostat accessible& \accesstopsecondcryo      \\ \colhline
All slow controls hardware, infrastructure, \& networking installed & February 2025\\ \colhline
Slow controls software for I/O, alarms, archiving, displays installed on production systems & May 2025 \\ \colhline
\rowcolor{dunepeach}End of \dword{detmodule} \#1 \dword{tpc} installation& \firsttpcinstallend      \\ \colhline
Install dynamic T-gradient monitors, cameras, purity monitors, and pressure meters & June 2025 \\\colhline
Install all feedthroughs for instrumentation devices & July 2025 \\ \colhline
 \rowcolor{dunepeach}Start of \dword{detmodule} \#2 \dword{tpc} installation& \startsecondtpcinstall      \\ \colhline
Install slow control expert interfaces for all systems in time for testing & September 2025 \\ \colhline
\rowcolor{dunepeach}End of \dword{detmodule} \#2 \dword{tpc} installation& \secondtpcinstallend      \\ 
Full slow controls systems commissioned and integrated into remote operations & July 2026 \\ 
\end{dunetable}

Designs may need review based on performance in \dword{protodune2} and any modifications will be incorporated into the final design before the start of production of \dword{cisc} systems for the \dword{fd} in April 2023. This will be followed by assembly of the systems underground in the detector cavern in July 2024. Installation of instrumentation devices will start in September 2024 following the beneficial occupancy of the interior of the cryostat. Installing gas analyzers, level meters, individual temperature sensors, static T-gradient thermometers, and support structure for all instrumentation devices will be finished before installing \dword{tpc}, but installation of dynamic T-gradient thermometers, purity monitors, pressure meters and cameras will occur afterward. \dword{cisc} will work closely with \dword{lbnf} to coordinate installation of the cryogenics systems and instrumentation devices. For slow controls, the goal is to have the full slow controls system commissioned and integrated into remote operations at least three months before the \dword{spmod} is ready for operations.

\subsection{Risks}

Table~\ref{tab:risks:SP-FD-CISC}  
lists the possible risks identified by the \dword{cisc} consortium along with corresponding mitigation strategies. 
A more detailed list of risks with additional descriptions can be found in \cite{bib:docdb7192}. The table shows 18 risks, all at medium or low level, mitigated with necessary steps and precautions. More discussion on all medium-level risks are provided in the text below. 
\begin{itemize}
    \item Risk 01: The risk associated with \dword{pdsp}-based designs being inadequate for \dword{fd}, is important because this requires early validation from \dword{protodune} data so R\&D of alternate designs can be timely. With \dword{pdsp} data now available, the consortium is focused on validating instrumentation designs. 
    \item Risk 06: Temperature sensors in the dynamic T-gradient monitor are calibrated using two methods: lab calibration to \SI{0.002}{K}  (as in the static T-gradient monitor case) and in situ cross-calibration moving the system vertically. Disagreement between the two methods can occur. In order to mitigate this we need to investigate and improve both methods, specifically the laboratory calibration since this is the only one possible for sensors behind \dword{apa}s, and top/bottom of the detector.  
    \item Risk 10: This risk involves an inability to build a working prototype for cold cameras during R\&D phase that meets all the requirements \& safety, e.g., that cold camera prototypes fail longevity tests or show low performance (e.g. bad resolution). This risk originates from past experience with cold cameras that became non-operational after a period of time in \dword{lar} or showed low performance. In order to address this, we plan to pursue further R\&D to improve thermal insulation and heaters, develop alternative camera models, etc. If problems persist one can use the cameras in the ullage (cold or inspection) with the appropriate field of view and lighting such that elements inside \dword{lar} can be inspected during filling.
    \item Risk 12: Cameras are delicate devices and some of them located near \dword{hv} devices can be destroyed by \dword{hv} discharges. This can be mitigated by ensuring that most important cold cameras have enough redundancy such that the loss of one camera does not compromise the overall performance. In the case of inspection cameras, 
    we can simply replace them.
    \item Risk 17: The gas analyzers and level meters may fail as these are commercial devices purchased at some point in their product cycle and cannot be required to last 20 years. Typical warranties are $\sim$1 year from date of purchase. The active electronics parts of both gas analyzers and level meters are external to the cryostat so they can be replaced. To mitigate this, provisions will be made for future replacement in case of failure or loss of sensitivity. Also, the risk is not high since we have purity monitors in the filtration system that can cover the experiment during the time gas analyzers are being replaced or repaired. 
\end{itemize}

Related to risks 12, 16 and 18, ageing is an important aspect for several monitors, especially for those that are inaccessible. The \dword{protodune} tests demonstrate that the devices survive the commissioning phase, and we continue to learn from \dword{protodune} experience. In addition to \dword{protodune}, other tests are planned. For example, in the case of purity monitors, photocathodes are expected to survive the first five years and if we prevent running them with high frequency at low purity (lifetime~$<$~\SI{3}{\milli\second}), ageing can be prevented for a longer time. To understand long-term aeging, R\&D is planned at \dword{citf} and at member institute sites for many of the devices. 
Systems that are replaceable, such as inline purity monitors, gas analyzers, and inspection cameras, can be replaced when failures occur and maintained for the lifetime of the experiment.

\begin{footnotesize}
\begin{longtable}{P{0.18\textwidth}P{0.20\textwidth}P{0.32\textwidth}P{0.02\textwidth}P{0.02\textwidth}P{0.02\textwidth}} 
\caption[CISC risks]{CISC risks (P=probability, C=cost, S=schedule) The risk probability, after taking into account the planned mitigation activities, is ranked as 
L (low $<\,$\SI{10}{\%}), 
M (medium \SIrange{10}{25}{\%}), or 
H (high $>\,$\SI{25}{\%}). 
The cost and schedule impacts are ranked as 
L (cost increase $<\,$\SI{5}{\%}, schedule delay $<\,$\num{2} months), 
M (\SIrange{5}{25}{\%} and 2--6 months, respectively) and 
H ($>\,$\SI{20}{\%} and $>\,$2 months, respectively).  \fixmehl{ref \texttt{tab:risks:SP-FD-CISC}}} \\
\rowcolor{dunesky}
ID & Risk & Mitigation & P & C & S  \\  \colhline
RT-SP-CISC-01 & Baseline design from ProtoDUNEs for an instrumentation device is not adequate for DUNE far detectors & Focus on early problem discovery in ProtoDUNE so any needed redesigns can start as soon as possible. & L & M & L \\  \colhline
RT-SP-CISC-02 & Swinging of long instrumentation devices (T-gradient monitors or PrM system) & Add additional intermediate constraints to prevent swinging. & L & L & L \\  \colhline
RT-SP-CISC-03 & High E-fields near instrumentation devices cause dielectric breakdowns in \dshort{lar} & CISC systems placed as far from cathode and FC as possible. & L & L & L \\  \colhline
RT-SP-CISC-04 & Light pollution from purity monitors and camera light emitting system & Use PrM lamp and camera lights outside PDS trigger window; cover PrM cathode to reduce light leakage. & L & L & L \\  \colhline
RT-SP-CISC-05 & Temperature sensors can induce noise in cold electronics & Check for noise before filling and remediate, repeat after filling. Filter or ground noisy sensors. & L & L  & L \\  \colhline
RT-SP-CISC-06 & Disagreement between lab and \em{in situ} calibrations for ProtoDUNE-SP dynamic T-gradient monitor & Investigate and improve both methods, particularly laboratory calibration. & M & L & L \\  \colhline
RT-SP-CISC-07 & Purity monitor electronics induce noise in TPC and PDS electronics. & Operate lamp outside TPC+PDS trigger window. Surround and ground light source with Faraday cage. & L & L & L \\  \colhline
RT-SP-CISC-08 & Discrepancies between measured temperature map and CFD simulations in ProtoDUNE-SP & Improve simulations with additional measurements inputs; use fraction of sensors to predict others   & L & L & L \\  \colhline
RT-SP-CISC-09 & Difficulty correlating purity and temperature in ProtoDUNE-SP impairs understanding cryo system. & Identify causes of discrepancy, modify design. Calibrate PrM differences, correlate with RTDs. & L & L & L \\  \colhline
RT-SP-CISC-10 & Cold camera R\&D fails to produce prototype meeting specifications \& safety requirements & Improve insulation and heaters. Use cameras in ullage or inspection cameras instead. & M & M & L \\  \colhline
RT-SP-CISC-11 & HV discharge caused by inspection cameras & Study E-field in and on housing and anchoring system. Test in HV facility. & L & L & L \\  \colhline
RT-SP-CISC-12 & HV discharge destroying the cameras & Ensure sufficient redundancy of cold cameras. Warm cameras are replaceable. & L & M & L \\  \colhline
RT-SP-CISC-13 & Insufficient light for cameras to acquire useful images & Test cameras with illumination similar to actual detector. & L & L & L \\  \colhline
RT-SP-CISC-14 & Cameras may induce noise in cold electronics & Continued R\&D work with grounding and shielding in realistic conditions. & L & L & L \\  \colhline
RT-SP-CISC-15 & Light attenuation in long optic fibers for purity monitors  & Test the max.\ length of usable fiber, optimize the depth of bottom PrM, number of fibers. & L & L & L \\  \colhline
RT-SP-CISC-16 & Longevity of purity monitors & Optimize PrM operation to avoid long running in low purity. Technique to protect/recover cathode. & L & L & L \\  \colhline
RT-SP-CISC-17 & Longevity: Gas analyzers and level meters may fail. & Plan for future replacement in case of failure or loss of sensitivity.  & M & M & L \\  \colhline
RT-SP-CISC-18 & Problems in interfacing  hardware devices (e.g. power supplies) with slow controls & Involve slow control experts in choice of hardware needing control/monitoring.
 & L & L & L \\  \colhline

\label{tab:risks:SP-FD-CISC}
\end{longtable}
\end{footnotesize}

\subsection{Interfaces} 
\label{sec:interfaces}

\dword{cisc} subsystems interface with all other detector subsystems and potentially impact the work of all detector consortia, as well as some
working groups (e.g., physics, software and computing, beam instrumentation), and technical coordination, requiring interactions with all of these entities.  We also interact heavily with \dword{lbnf} beam and cryogenics groups.  
Detailed descriptions of \dword{cisc} interfaces are maintained in the \dword{dune} DocDB. A brief summary is provided in this section. Table~\ref{tab:fdgen-cisc-interfaces} lists the IDs of the different DocDB documents as well as their highlights. Descriptions of the interfaces and interactions that affect many systems are given below. 

\dword{cisc} interacts with the detector consortia because \dword{cisc} will provide status monitoring of all important detector subsystems along with controls for some components of the detector.
\dword{cisc} will also consult on selecting different power supplies to ensure monitoring and control can be established with preferred types of communication. 
Rack space distribution and interaction between slow controls and other modules from other consortia will be managed by \dword{tc} in consultation with those consortia. 

\dword{cisc} will work with \dword{lbnf} to determine whether heaters and \dwords{rtd} are needed on flanges. If so, \dword{cisc} will specify the heaters and  \dwords{rtd}, and will provide the readout and control, while the responsibility for the actual hardware will be discussed with the different groups.

Installing instrumentation devices will interfere with other devices and must be coordinated with the appropriate consortia.  
On the software side, \dword{cisc} must define, in coordination with other consortia/groups, the quantities to be monitored/controlled by slow controls and the corresponding alarms,
archiving, and GUIs.

\begin{dunetable}
[CISC system interfaces]
{p{0.2\textwidth}p{0.55\textwidth}p{0.2\textwidth}}
{tab:fdgen-cisc-interfaces}
{\dword{cisc} system interface links}   
\small
Interfacing System & Description & Linked Reference \\ \toprowrule
\dword{apa}	           &
static T-gradient monitors, cameras, and lights
& \citedocdb{6679} 
\\ \colhline

\dword{pds}	     &

PrMs, light emitting system for cameras
& \citedocdb{6730}  
\\ \colhline

\dword{tpc} Electronics	         &  
noise, power supply monitoring
& \citedocdb{6745}  \\ \colhline

\dword{hv} Systems	           &
shielding, bubble generation by inspection camera, cold camera locations, ground planes
& \citedocdb{6787}   
\\ \colhline

\dword{daq}	                      &
description of \dword{cisc} data storage, 
allowing bi-directional communications between \dword{daq} and \dword{cisc}.      & \citedocdb{6790}
\\ \colhline
Calibration          &
multifunctional \dword{cisc}/\dword{citf} ports; space sharing around ports 
& \citedocdb{7072} 

\\ \colhline
Physics	          &

indirect interfaces through calibration, tools to extract data from the slow controls database 
& \citedocdb{7099} 
\\ \colhline

Software \& Computing	  &

slow controls database design and maintenance
& \citedocdb{7126}  
\\ \colhline

Cryogenics             &  
must be designed and implemented.       
purity monitors, gas analyzers, interlock mechanisms to prevent contamination of LAr
&  -   

\\ \colhline

Beam                      &   
beam status &  -     
\\ \colhline
TC Facility              &   
significant interfaces at multiple levels   
& \citedocdb{6991}   \\ \colhline
TC Installation     	  &     
significant interfaces at multiple levels
& \citedocdb{7018}   \\ \colhline
TC Integration Facility    &    
significant interfaces at multiple levels
& \citedocdb{7045}   \\ 
\end{dunetable}

\subsection{Installation, Integration, and Commissioning}

\subsubsection{Purity Monitors}
\label{sec:fdgen-slow-cryo-install-pm}

The purity monitor system will be built in modules, so it can be assembled outside the 
cryostat  
leaving few steps to complete inside the cryostat.  The assembly itself 
 comes into the cryostat with the individual purity monitors mounted to support tubes, with no \dword{hv} cables or optical fibers yet installed.  The support tube at the top and bottom of the assembly 
 is then mounted to the brackets inside the cryostat, and  
 the brackets attached to the cables trays and/or the detector support structure.  At much the same time, the \dword{fe} electronics and light source can be installed on the top of the cryostat, and the electronics and power supplies can be installed in the electronics rack.  

Integration 
begins by running the  \dword{hv} cables and optical fibers to the purity monitors, through the top of the cryostat.  These cables 
are attached to the  \dword{hv}  feedthroughs with sufficient length to reach each purity monitor inside the cryostat.  
The cables 
are run along cable trays through the port reserved for the purity monitor system. 
Each purity monitor will have three  \dword{hv} cables that connect it to the feedthrough and then further along to the  \dword{fe} electronics.  The optical fibers 
are then run through the special optical fiber feedthrough, into the cryostat, and 
guided to the purity monitor system either using the cables trays or guide tubes, depending on which solution is adopted. 
This should protect fibers from breaking accidentally as the rest of the detector and instrumentation installation continues.  The optical fibers 
are then run inside the purity monitor support tube and to the appropriate purity monitor, terminating the fibers at the photocathode of each monitor while protecting them from breaking near the purity monitor system itself.

Integration 
continues as the  \dword{hv} cables are connected through the feedthrough to the system  \dword{fe} electronics; then optical fibers are connected to the light source.  The cables connecting the \dword{fe} electronics and the light source to the electronics rack 
are also run and connected at this time.  This allows the system to be turned on and the software to begin testing the various components and connections.  Once all connections are confirmed successful, integration with the slow controls system begins, first by establishing communication between the two systems and then transferring data between them to ensure successful exchange of important system parameters and measurements.  

Commissioning the purity monitor system 
begins once the cryostat is purged and a gaseous argon atmosphere is present.  At this time, the \dword{hv} for the purity monitors 
  is ramped up without risk of discharge through the air, and the light source turned on.  Although the drift electron lifetime in the gaseous argon would be very large and therefore not 
  measurable with the purity monitors themselves, the signal strength at both the cathode and anode will 
  give a good indication of how well the light source generates drift electrons from the photocathode.  Comparing the signal strengths at the anode and cathode will indicate whether the electrons successfully drift to the anode.
  Although \dword{qa} and \dword{qc} should make it unlikely for a purity monitor to fail this final test, if that does happen then the electric and optical connections can be fixed before filling.

\subsubsection{Thermometers}
\label{sec:fdgen-slow-cryo-install-th}

Static T-gradient monitors 
must be installed before the outer \dword{apa}s, ideally  
right after the pipes are installed. The profilers 
are preassembled before they are delivered to \dword{surf}. 
Installation will follow these steps:
\begin{enumerate}
\item anchor the stainless steel bottom plates to the four bolts on the bottom corner of the cryostat,
\item anchor the stainless steel support holding the two strings to the four bolts on the top corner of the cryostat,
\item unroll the array with the help of the scissor lift,
\item anchor the strings to the bottom stainless steel support,   
\item check and adjust tension and verticality,
\item review all cable and sensor supports, 
\item route cable from the top anchoring point to the two \dword{dss} ports, and 
\item plan to plug sensors into IDC-4 connectors later, just before moving the corresponding \dword{apa} into its final position. 
\end{enumerate}

Individual temperature sensors on pipes and cryostat floor 
are installed immediately after installing the static T-gradient monitors. First, vertical stainless steel strings for cable routing 
are installed following a procedure similar to the one described above for the static T-gradient monitors. Next, we anchor all cable supports to pipes. Then each cable 
is routed individually starting from the sensor end (with IDC-4 female connector but without the sensor)
to the corresponding cryostat port. Once all cables going through the same port have been routed, we cut the cables to the same length, so they can be properly assembled into the corresponding connector(s). To avoid damaging the sensors, they are installed later (by plugging the IDC-4 male connector on the sensor \dword{pcb} to the IDC-4 female connector on the cable end), just before unfolding the bottom \dwords{gp}.

For the \dword{sp}, individual sensors on the top \dword{gp} must be integrated with the \dwords{gp}. For each \dshort{cpa} (with its corresponding four  \dword{gp} modules)
going inside the cryostat, cable and sensor supports will be anchored to the  \dword{gp} threaded rods as soon as possible.
Once the \dshort{cpa} is moved into its final position and its top \dwords{gp} are ready to be unfolded, sensors on these \dwords{gp} 
are installed. Once unfolded, cables 
exceeding the \dword{gp} limits can be routed to the corresponding cryostat port using either neighboring \dwords{gp} or \dshort{dss} I-beams.

Dynamic T-gradient monitors 
are installed
after the \dword{tpc} components are in place.
Figure~\ref{fig:fd-slow-cryo-dt-monitor-overview} shows the
design of the dynamic T-gradient monitor with its sensor carrier rod,
enclosure above the cryostat, and stepper motor and 
 Figure~\ref{fig:fd-slow-cryo-sensor-mount} shows detailed views of key
components.  Each monitor 
comes in several segments with sensors
and cabling already in place. Additional slack will be provided at
segment joints to make installation easier. Segments of the sensor
carrier rod with preattached sensors 
are fed into the flange one
at a time. Each segment, as it is fed into the 
cryostat, is 
held at the top with a pin that prevents the segment from sliding all
the way in. 
The next segment 
is connected at that
time to the previous segment. Then the pin 
is removed, the first
segment 
is pushed down, and the next segment top 
is held
with the pin at the flange. This process 
is repeated for each
segment 
until the entire sensor carrier rod is in
place.  Next, the enclosure 
is installed on top of the flange,
starting with the six-way cross at the bottom of the enclosure.  (See
 Figure~\ref{fig:fd-slow-cryo-sensor-mount}, right.)  Again, extra cable
slack at the top will be provided to ease connection to the D-sub
flange and to allow the entire system to move vertically.  The wires
are connected to a D-sub connector on the \fdth on one side port
of the cross. Finally, a crane 
positions the remainder
of the enclosure above the top of the cross.  This enclosure includes
the mechanism used to move the sensor rod, which 
is preassembled
with the motor in place on the side of the enclosure, and the pinion
and gear used to move the sensor inside the enclosure.  The pinion
gets connected to the top of the rod. The enclosure is then 
connected to top part of the cross, which finishes the installation of
the dynamic T-gradient monitor.

Commissioning all thermometers will occur in several steps. In the first stage, only cables 
are installed, so
the readout performance and the noise level inside the cryostat 
is tested with precision resistors. Once sensors are installed, the entire chain 
is checked again at room temperature.
Spare cables, connectors and sensors are available for replacement at \dword{surf} if needed. 
The final commissioning phase 
takes place during and after cryostat filling.

\subsubsection{Gas Analyzers}
\label{sec:fdgen-slow-cryo-install-ga}

The gas analyzers 
are installed before the piston purge and gas recirculation phases of the cryostat commissioning. They 
are installed near the tubing switchyard to minimize tubing run length and for convenience when switching the sampling points and gas analyzers. Because each is a standalone module, a single rack with shelves 
is adequate to house the modules.

For integration, the gas analyzers typically have an analog output (\SIrange{4}{20}{mA} or \SIrange{0}{10}{V}), which maps to the input range of the analyzers. They also usually have several relays to indicate the scale they are currently running. These outputs can be connected to the slow controls for readout. However, using a digital readout is preferable because this gives a direct analyzer reading at any scale. Currently, the digital output connections are RS-232, RS-485, USB, and Ethernet. The preferred option is chosen at the time of purchase. 
The readout usually responds to a simple set of text query commands. Because of the natural time scales of the gas analyzers and lags in gas delivery times (which depend on the length of the tubing run), sampling every minute is adequate. Our current plan is to record both analog and digital signals to have both outputs available.  

The analyzers 
must be brought online and calibrated before beginning the gas phase of the cryostat commissioning.  Calibration varies by module because they are different, but calibration often requires using argon gas with zero contaminants, and argon gas with a known level of the contaminant to check the scale. Contaminants are usually removed with a local inline filter for the first gas sample. 
This gas phase usually begins with normal air, with the more sensitive analyzers valved off at the switchyard to prevent overloading their inputs (and potentially saturating their detectors). As the argon purge and gas recirculation progress, the various analyzers 
are valved back in when the contaminant levels reach the upper limits of the analyzer ranges. 

\subsubsection{Liquid Level Monitoring}
\label{sec:fdgen-slow-cryo-install-llm}

Installing differential pressure level meters is the responsibility of \dword{lbnf}, but the capacitive level meters fall within \dword{cisc}'s scope. The exact number of capacitive level meters must still be decided. There will be at least four, located at the four corners of the cryostat. 
They will be attached to the M10 bolts in the cryostat corners after the detector is installed. Cables will be routed to the appropriate \dword{dss} port. If additional capacitive level meters are needed in the central part of the cryostat, those will be installed before the nearby \dword{apa}s. 

\subsubsection{Pressure Meters}
\label{sec:fdgen-slow-cryo-install-press}
Installing pressure meters is the responsibility of \dword{cisc}. A total of six sensors will be mechanically installed in two dedicated flanges (three sensors each) at opposite sides of the cryostat after the detector is installed. Cables will be routed through the same dedicated port assigned for these devices. The pressure signals (absolute and relative) are read and converted to
standard 4--20 mA current loop signals.
A twisted pair shielded cable connects the sensors to the slow controls \dword{plc} controller using software to convert electrical signals to pressure values.

\subsubsection{Cameras and light emitting system}
\label{sec:fdgen-slow-cryo-install-c}

Installing fixed cameras is simple in principle, but involves a
considerable number of interfaces. The enclosure of each camera has
exterior threaded holes to facilitate mounting on the cryostat wall,
cryogenic internal piping, or \dword{dss}. Each
enclosure 
is attached to a gas line to maintain appropriate
underpressure in the fill gas, 
therefore an interface with cryogenic
internal piping will be necessary. Each camera has a cable (coaxial or
optical) for the video signal and a multiconductor cable for power and
control. These 
get run through cable trays to flanges on assigned
instrumentation feedthroughs.

The inspection camera is designed to be inserted and removed on any
instrumentation feedthrough equipped with a gate valve at any time
during operation.  Installing the gate valves and purge system
for instrumentation feedthroughs falls under cryogenic internal
piping.

Installing fixed lighting sources separate from the cameras
requires mounting on cryostat wall, cryogenic internal piping, or
\dword{dss}, and multiconductor cables for power run through cable
trays to flanges on assigned instrumentation feedthroughs.

\subsubsection{Slow Controls Hardware}
\label{sec:fdgen-slow-cryo-install-sc-hard}

Slow controls hardware installation 
includes installing multiple
servers, network cables, any specialized cables needed
for device communication, and possibly some custom-built hardware to monitor racks. The installation sequence will be 
planned with the facilities group and other consortia. The network
cables and rack monitoring hardware will be common across many racks
and will be installed first as part of the basic rack installation, 
to be led by the facilities group. Specialized cables needed for slow controls and servers 
are installed
after the common rack hardware. The selection and installation of these cables will be coordinated
with other consortia, and servers will be coordinated with the \dword{daq} group.

\subsubsection{Transport, Handling, and Storage}
\label{sec:fdgen-slow-cryo-install-transport}

Most instrumentation devices will be shipped in pieces to \dword{surf} via the \dword{sdwf} and mounted on-site. 
Instrumentation devices are in general small, except for  
the support structures for purity monitors and dynamic T-gradient monitors,
which will cover the entire height of the cryostat. The load on those structures is relatively small (\(<\SI{100}{kg}\)), so they can be fabricated in sections of less than \SI{3}{m},
which can be easily transported to \dword{surf}, down the shaft, and through the tunnels.
All instrumention devices except the dynamic T-gradient monitors can be moved into the cryostat without the crane. The dynamic T-gradient monitors, which 
are introduced into the cryostat from above, 
require a crane for the installation of the external enclosure (with preassembled motor, pinion and gear). 

\subsection{Quality Control}
\label{sec:fdsp-slow-cryo-qc}
The manufacturer and the institution in charge of device assembly will conduct a series of tests to ensure the equipment can perform its intended function as part of \dword{qc}. \dword{qc} also includes post-fabrication tests and tests run after shipping and installation. For complex systems, the entire system will be tested before shipping. 
Additional \dword{qc} procedures can be performed 
underground after installation. 

The planned tests for each subsystem are described below.

\subsubsection{Purity Monitors}
\label{sec:fdgen-slow-cryo-qc-pm}

The purity monitor system will undergo a series of tests to ensure the
system performs as intended. These tests 
include electronic tests
with a pulse generator, mechanical and electrical connectivity tests
at cryogenic temperatures in a cryostat, and vacuum tests for short
and full assemblies in a dewar and in a long vacuum tube.

The \dword{qc} tests for \dword{fd} purity monitors begin with testing
individual purity monitors in vacuum after each is fabricated and
assembled.  This test checks the amplitude of the signal generated by
the drift electrons at the cathode and the anode to ensure the
photocathode can provide sufficient numbers of photoelectrons to
measure the signal attenuation
with the required precision, and that the field gradient resistors all work properly to maintain the drift field. 
A smaller version of the assembly with all purity monitors installed will be  tested at the \dword{citf} 
to ensure the full system performs as expected in \dword{lar}.  

Next, 
the entire system 
is assembled on the full-length mounting tubes to check the connections along the way.  Ensuring that all electric and optical connections are operating properly during this test reduces the risk of problems once the full system is assembled and ready for the final test in vacuum.  
The fully assembled system 
is placed in the shipping tube, which 
serves as a vacuum chamber, and tested at \dword{surf} 
 before the system is inserted into the  
 cryostat. During insertion, electrical connections 
 are tested continuously with multimeters and electrometers.

\subsubsection{Thermometers}
\label{sec:fdgen-slow-cryo-qc-th}

\paragraph{Static T-gradient thermometers}
\label{sec:fdgen-slow-cryo-qc-thst}

Static T-gradient monitors undergo three type of tests at the production site before 
shipment to \dword{surf}: a mechanical rigidity test, a calibration of all sensors, and a test of all electrical cables and connectors.
The mechanical rigidity is tested by mounting the static T-gradient monitor between two dummy cryostat corners mounted \SI{15}{m} apart. The tension of the strings is set to match the tension that would occur in a vertical deployment in \dword{lar}, and the deflection of the sensor and electrical cable strings is measured and compared to the expected value; this is to ensure any swinging or deflection of the deployed static T-gradient monitor will be < \SI{5}{cm}, mitigating any risk of touching the \dwords{apa}.
The laboratory calibration of sensors will be performed 
as explained in Section~\ref{sec:fdsp-cryo-therm}. The main concern is reproducibility of results because sensors could change resistance and hence their temperature scale when undergoing successive immersions in \dword{lar}. In this case, the calibration procedure itself provides \dword{qc} because each set of sensors goes through five independent measurements. Sensors with \rms variation outside the requirement (\SI{2}{mK} for \dword{pdsp}) are discarded. This calibration also serves as \dword{qc} for the readout system (similar to the final one) and of the \dword{pcb}-sensor-connector assembly.
Finally, the cable-connector assemblies are tested; sensors must measure the expected values with no additional noise introduced by either cable or connector. 

An integrated system test is conducted at a \dword{lar} test facility at the production site, which has sufficient linear dimension (>\SI{2}{m}) to test a good portion of the system. This 
ensures that  the system
operates in \dword{lar} at the required level of performance.
The laboratory sensor calibration 
is compared with the in situ calibration
of the dynamic T-gradient monitors by operating both dynamic and static T-gradient monitors simultaneously.   

The last phase of \dword{qc} takes place after installation. 
The verticality of each array 
is checked, and the tensions in the stainless steel strings adjusted as necessary.
Before closing the flange, the entire readout chain is 
tested.  
This allows a test of the sensor-connector assembly, the cable-connector assemblies at both ends, and the noise level inside the cryostat.
If any sensor presents a problem, it is replaced. If the problem persists, the cable is checked and replaced as needed.

\paragraph{Dynamic T-gradient thermometers}
\label{sec:fdgen-slow-cryo-qc-thdy}

The dynamic T-gradient monitor consists of an array of high-precision temperature sensors mounted on a vertical rod. The rod can move vertically to cross-calibrate the temperature sensors in situ. 
We will use the following 
tests to ensure that the dynamic T-gradient monitor delivers vertical temperature gradient measurements with a precision of a few \si{mK}.

\begin{itemize}
\item
Before installation, temperature sensors are tested in  \dword{ln}  to verify correct operation and to set the baseline calibration for each sensor with respect to the absolutely calibrated reference sensor. 
\item
Warm and cold temperature readings are taken with each sensor after it is mounted on the \dword{pcb} and the readout cables are soldered. 
\item
The sensor readout is taken for all sensors after the cold cables are connected to electric \fdth{}s on the flange and the warm cables outside of the cryostat are connected to the temperature readout system.
\item 
The stepper motor is tested before and after connecting to the gear and pinion system.
\item
The fully assembled rod is connected to the pinion and gear and moved with the stepper motor on a high platform many times to verify repeatability, possible offsets, and any uncertainty in the positioning. Finally, repeating this test so many times will verify the sturdiness of the system.
\item
The full system is tested after it is installed in the cryostat; both motion and sensor operation are tested by checking 
sensor readout and vertical motion of the system.
\end{itemize} 

\paragraph{Individual Sensors}
\label{sec:fdgen-slow-cryo-qc-is}

To address the quality of individual precision sensors, the same method as for the static T-gradient monitors 
is used.
The \dword{qc} of the sensors is part of the laboratory calibration. After mounting six sensors with their corresponding cables, a
SUBD-25 connector 
is added, and the six sensors tested at room temperature. All sensors 
must 
give values within specifications.  
If any of the sensors present problems, they are replaced.  If the problem persists, the cable is checked and replaced as needed.

For standard \dwords{rtd} to be installed on the cryostat walls, floor, and roof, calibration is not an issue. Any \dword{qc} required for associated cables and connectors 
is performed following the same procedure as for precision sensors. 

\subsubsection{Gas Analyzers}
\label{sec:fdgen-slow-cryo-qc-ga}

The gas analyzers will be guaranteed by the manufacturer. However, once received, the gas analyzer modules 
are checked for both \textit{zero} and the \textit{span} values using a gas-mixing instrument and two gas cylinders, one having a zero level of the gas analyzer contaminant species and the other cylinder with a known percentage of the contaminant gas. This 
 verifies the proper operation of the gas analyzers. When they are installed at \dword{surf}, this process 
 is repeated before commissioning the cryostat. Calibrations will need to be repeated 
 per manufacturer recommendations over the gas analyzer lifetime.

\subsubsection{Liquid Level Monitoring}
\label{sec:fdgen-slow-cryo-qc-llm}

The manufacturer will provide the \dword{qc} for the differential pressure level meters; further \dword{qc} during and after installation 
is the responsibility of \dword{lbnf}.

The capacitive sensors will be tested with a modest sample of \dword{lar} in the laboratory before they are installed. After installation, they are tested in situ 
using a suitable dielectric in contact with the sensor.

\subsubsection{Pressure Meters}
\label{sec:fdgen-slow-cryo-qc-press}
The manufacturer will provide the \dword{qc} for the pressure meters; further \dword{qc} during and after installation is the responsibility of \dword{cisc}.

The pressure sensors will be tested with a modest sample of gaseous argon in the laboratory before they are installed. After installation, they are tested in situ at atmospheric pressure. The whole pressure readout chain, (including slow controls \dword{plc} and software protocol) 
will also be tested and cross-checked with \dword{lbnf} pressure sensors.

\subsubsection{Cameras}
\label{sec:fdgen-slow-cryo-qc-c}

Before 
transport to \dword{surf}, each cryogenic camera unit (comprising the enclosure, camera, and internal thermal control and monitoring) 
is checked for correct operation of all features, for recovery from \SI{87}{K} non-operating mode, for leakage, and for physical defects. Lighting systems 
are similarly checked. Operations tests will verify correct current draw, image quality, and temperature readback and control. The movable inspection camera apparatus 
are inspected for physical defects and checked for proper mechanical operation before shipping. A checklist 
is created for each unit, filed electronically in the \dword{dune} logbook, and a hard copy sent with each unit. 

Before installation, each fixed cryogenic camera unit is inspected for physical damage or defects and checked at the \dword{citf}
for correct operation of all features, for recovery from \SI{87}{K} non-operating mode, and for contamination of the \dword{lar}. Lighting systems are similarly checked. Operations tests verify correct current draw, image quality, and temperature readback and control. After installation and connection of wiring, fixed cameras and lighting are again  checked for operation. The movable inspection camera apparatus is inspected for physical defects and, after integration with a camera unit, tested in the facility for proper mechanical and electronic operation and cleanliness before being installed or stored. A checklist will be completed for each \dword{qc} check and filed electronically in the \dword{dune} logbook. 

\subsubsection{Light-emitting System}
\label{sec:fdgen-slow-cryo-qc-les}

The entire light-emitting system is checked before installation to ensure functionality of light emission. 
Initial testing of the system (see Figure~\ref{fig:cisc-LED}) begins with
measuring the current when low voltage (\SI{1}{V}) is applied, to check
that the resistive \dword{led} failover path is correct. Next, the forward voltage is measured using nominal forward current to
check that it is within \SI{10}{\%} of the nominal forward voltage drop of
the \dword{led}, that all of the \dwords{led} are illuminated, and that each \dword{led} is visible over the nominal angular range. If the \dwords{led} are
infrared, a video camera with the IR filter removed is used for a
visual check. This procedure is then duplicated with the current
reversed for \dwords{led} oriented in the opposite direction. Initial tests are performed at room temperature and then repeated in \dword{ln}. Color shifts in the \dwords{led} are expected and will be noted. A checklist is completed for each \dword{qc} check and filed electronically in the \dword{dune} logbook.

Room temperature tests are repeated during and immediately after installation to ensure that the system has not been damaged during transportation or installation. Functionality checks of the \dwords{led} are repeated after the cameras are installed in the cryostat.

\subsubsection{Slow Controls Hardware}
\label{sec:fdsp-slow-cryo-qc-sc-hard}

Networking and computing systems will be purchased commercially, requiring manufacturer's \dword{qa}. However, the new servers 
are tested after delivery to confirm they suffered no damage during shipping. The new system is allowed to burn in overnight or for a few days, 
running a diagnostics suite on a loop in order to validate 
the manufacturer's \dword{qa} process.

The system 
is shipped directly to \dword{surf} 
where an on-site
expert 
boots the systems and does basic
configuration. 
Specific configuration information 
is pulled over
the network, after which others may log in remotely to do the final
setup, minimizing the number of people underground.

\subsection{Safety}
Safety 
is of critical importance during 
all phases of the \dword{cisc} project, including R\&D, laboratory calibration and testing, mounting tests, and installation. 
Safety experts 
review and approve the initial safety planning for all phases as part of the initial design review, as well as 
before implementation. 
All documentation of component cleaning, assembly, testing, and installation will include a section on relevant safety concerns and will be reviewed during appropriate pre-production reviews.

Several areas are of particular importance to \dword{cisc}.
\begin{itemize}
\item Hazardous chemicals (e.g., epoxy compounds used to attach sensors to cryostat inner membrane) and cleaning compounds:
  All chemicals used will be documented at the consortium management level, with a material safety data sheet and approved handling and disposal plans in place.

\item Liquid and gaseous cryogens used in calibrating and testing: \dword{ln} and \dword{lar} 
are used to calibrate and test instrumentation devices.
  Full hazard analysis plans will be in place at the consortium management level for full module or
  module component testing that involves  
  cryogens. These safety plans will be reviewed in appropriate pre-production and production reviews.

\item High voltage safety:  Purity monitors 
have a voltage of $\sim\,$\SI{2}{kV}. Fabrication and testing plans will show compliance with local
  \dword{hv} safety requirements at 
  any institution or laboratory that conducts testing or operation, and this compliance will be reviewed as part of the standard review process.

\item Working at heights: Some fabrication, testing, and installation of \dword{cisc} devices require working at heights.
  Both T-gradient monitors and purity monitors, which span the height of the detector, require working at heights exceeding \SI{10}{m}.
  Temperature sensors installed near the top cryostat membrane and cable routing for all instrumentation devices
  also require working at heights. 
  The appropriate safety procedures including lift and harness training will be designed and reviewed. 
  
\item Falling objects: all work involving heights have associated risks of falling objects. The corresponding safety procedures, including proper helmet use 
and a well restricted safety area, will be included in the safety plan. 
\end{itemize}

\cleardoublepage

\chapter{Detector Installation}
\label{ch:sp-install}


\section{Introduction}
\label{ch:sp-install-intro}

This chapter covers all the work and infrastructure required to install the \dword{spmod}. 
 
We first provide a reminder of the scale of the task, beginning with the two facts that drive all others: A \dword{dune} \dword{fd} module is enormous, with outer cryostat dimensions  of 
\cryostatlen{}(L) $\times$ \cryostatwdth{}(W) $\times$ \cryostatht{}(H); 
and every piece of a \dword{detmodule} must descend 
\SI{1500}{m} down the Ross Shaft to the \dword{4850l} of \dword{surf} and be transported to a detector cavern.

The \dword{spmod}'s 150 \dwords{apa}, each $6.0$ m high and $2.3$ m wide, and  weighing $600$ kg with $3500$ strung sense and shielding wires, must be taken down the shaft as special ``slung loads'' and moved to the area just outside the \dword{dune} cryostat. 
The \dword{apa}s are moved into a \SI{30x19}{\meter} clean room (a portion of which is \SI{17}{m} high) where they are outfitted  with \dword{pd} units and passed through a series of qualification tests.
Here, two \dword{apa}s are linked into a vertical \SI{12}{m} high double unit and connected to readout electronics. 
They receive a cold-test in place, then move into the cryostat to be connected at the proper location on the previously installed \dword{dss}, and have their cabling connected to \fdth{}s. 
Additional systems are installed in parallel with the \dword{apa}s, e.g., the \dword{fc} and their \dword{hv} connections, elements of the \dword{cisc}, and detector calibration systems. The cathode plane, \dword{fc}, and \dword{apa} together define the \dword{tpc} active volume. 

After twelve months of detector component installation, which follows twelve months of detector infrastructure installation, the cryostat closes (with the last installation steps occurring in a confined space accessed through a narrow human-access port). 
Following leak checks, final electrical connection tests, and installation of the neutron calibration source, the process of filling the cryostat with \SI{17000000}{\kilo\gram} of \dword{lar} begins.

The installation requires meticulous planning and execution of thousands of tasks by well trained teams of technicians, riggers, and detector specialists. 
High-level requirements for these tasks are spelled out in Table~\ref{tab:specs:SP-INST}
\footnote{\dword{apa}s are produced well in advance of their installation date. They are shipped to the storage facility immediately after fabrication and testing in order to control the risk of damage in shipping.} 
and the text that follows it. 
In all the planning and future work, the pre-eminent requirement in the installation process is safety.
\dword{dune}'s goal is zero accidents resulting in personal injury, damage to detector components, or harm to the environment.

\begin{footnotesize}
\begin{longtable}{p{0.12\textwidth}p{0.18\textwidth}p{0.17\textwidth}p{0.25\textwidth}p{0.16\textwidth}}
\caption{Installation specifications \fixmehl{ref \texttt{tab:spec:SP-INST}}} \\
  \rowcolor{dunesky}
       Label & Description  & Specification \newline (Goal) & Rationale & Validation \\  \colhline

  \newtag{SP-INST-1}{ spec:logistics-material-handling }  & Compliance with the SURF Material Handling Specification for all material transported underground  &  SURF Material Handling Specification &  Loads must fit in the shaft be lifted safely. &  Visual and documentation check \\ \colhline

  \newtag{SP-INST-2}{ spec:logistics-shipping-coord }  & Coordination of shipments with CMGC; DUNE to schedule use of Ross Shaft  &  2 wk notice to CMGC &  Both DUNE and CMGC need to use Ross Shaft &  Deliveries will be rejected \\ \colhline

  \newtag{SP-INST-3}{ spec:logistics-materials-buffer }  & Maintain materials buffer at logistics facility in SD   &  $>1$ month &  Prevent schedule delays in case of shipping or customs delays &  Documentatation and progress reporting \\ \colhline

  \newtag{SP-INST-4}{ spec:apa-storage-sd }  & APA stroage at logistics facility in SD  &  700 m$^2$ &  Store APAs during lag between production and installation &  Agree upon space needs \\ \colhline

  \newtag{SP-INST-5}{ spec:cleanroom-specification }  & Installation cleanroom Specificaiton  &  ISO 8 &  Reduce dust (contains U/Th) to prevent induced radiological background in detector &  Monitor air purity \\ \colhline

  \newtag{SP-INST-6}{ spec:cleanroom-uv-filters }  & UV filter in installation cleanrooms for PDS sensor protection  &  filter $<\SI{400}{nm}$ for $>$ 2 wk exp; $<\SI{520}{nm}$ all else &  Prevent damage to PD coatings  &  Visual or spectrographic inspection \\ \colhline

\label{tab:specs:SP-INST}
\end{longtable}
\end{footnotesize}

Installation of the \dword{spmod} presents 
 hazards that include manipulation of heavy loads in the tight spaces at the \dword{4850l} and in the \dword{detmodule},  working at considerable heights above the floor, repeated utilization of large volumes of cryogens, multiple tests with \dword{hv}, commissioning of a Class IV laser system, and deployment of a high-activity neutron source. Mitigation of these hazards begins with the strong professional on-site \dword{esh} teams of the \dword{sdsd} and \dword{surf}.

All installation team members, both at the surface and underground, will undergo rigorous formal safety training. Daily safety meetings will ensure that all workers are aware of the scope of the planned underground work and any related safety considerations. Any team member can stop work at any time for safety purposes. The overall \dword{dune} safety plan is described in   
 \tcchesh{} 
 of this \dword{tdr}.  Individual sections within this chapter provide details on the evolving safety plan for installation. This plan has been informed by the successful safety experience of \dword{surf} with other underground experiments (e.g., \dword{lux}, \dword{mjdemo}, \dword{lz}), \dword{dune} members in executing projects at other underground locations (e.g., \dword{minos} at Soudan, Minnesota, USA), at other locations remote from major international laboratories (e.g., \dword{dayabay}, China and the \dword{nova} far detector (Ash River, Minnesota, USA), and at the home laboratories of both \dword{fnal} and \dword{cern}).

As part of the \dword{dune} design process the detector components and the \dword{tpc} have been prototyped at various stages. \dword{pdsp}, which was assembled from full-scale 
components has 
been completed and has taken data. 
This process has been extremely important in planning the \dword{spmod} 
installation and a detailed list of lessons learned from \dword{pdsp} construction and installation was compiled\cite{bib:docdb8255}. 
These lessons 
and other experience from the team planning the installation were used to develop a list of 
risks for the 
\dword{spmod} installation and to formulate mitigation strategies to reduce the risks.
The highest-impact risks -- those requiring a mitigation strategy -- are listed in Table~\ref{tab:risks:SP-FD-INST}. 
These mitigation strategies and all the lessons learned from \dword{pdsp} will be factored into the detailed installation plan. A description of each of the high level risks follows.

\begin{footnotesize}
\begin{longtable}{P{0.18\textwidth}P{0.20\textwidth}P{0.32\textwidth}P{0.02\textwidth}P{0.02\textwidth}P{0.02\textwidth}} 
\caption[SP module installation risks]{SP module installation risks (P=probability, C=cost, S=schedule) The risk probability, after taking into account the planned mitigation activities, is ranked as 
L (low $<\,$\SI{10}{\%}), 
M (medium \SIrange{10}{25}{\%}), or 
H (high $>\,$\SI{25}{\%}). 
The cost and schedule impacts are ranked as 
L (cost increase $<\,$\SI{5}{\%}, schedule delay $<\,$\num{2} months), 
M (\SIrange{5}{25}{\%} and 2--6 months, respectively) and 
H ($>\,$\SI{20}{\%} and $>\,$2 months, respectively).  \fixmehl{ref \texttt{tab:risks:SP-FD-INST}}} \\
\rowcolor{dunesky}
ID & Risk & Mitigation & P & C & S  \\  \colhline
RT-INST-01 & Personnel injury & Follow established safety plans. & M & L & H \\  \colhline
RT-INST-02 & Shipping delays & Plan one month buffer to store  materials locally. Provide logistics manual. & H & L & L \\  \colhline
RT-INST-03 & Missing components cause delays & Use detailed inventory system to verify availability of  necessary components.  & H & L & L \\  \colhline
RT-INST-04 & Import, export, visa issues  & Dedicated \dshort{fnal} \dshort{sdsd}division will expedite import/export and visa-related issues. & H & M & M \\  \colhline
RT-INST-05 & Lack of available labor  & Hire early and use Ash River setup to train \dshort{jpo} crew. & L & L & L \\  \colhline
RT-INST-06 & Parts do not fit together & Generate \threed model, create interface drawings, and prototype detector assembly. & H & L & L \\  \colhline
RT-INST-07 & Cryostat damage & Use cryostat false floor and temporary protection. & L & L & M \\  \colhline
RT-INST-08 & Weather closes SURF & Plan for \dshort{surf} weather closures & H & L & L \\  \colhline
RT-INST-09 & Detector failure during \cooldown & Cold test individual components then cold test \dshort{apa} assemblies immediately before installation. & L & H & H \\  \colhline

\label{tab:risks:SP-FD-INST}
\end{longtable}
\end{footnotesize}

Personal Injury:
The installation of the \dword{detmodule} requires on the order of fifty person-years of effort. Substantial work at heights, rigging of heavy equipment, use of custom tooling, and some work in confined spaces is necessary. It is critical that all safety measures be implemented and proper oversight be in place. 
\dword{dune} will follow the \dword{fnal} safety program, and if any additional measures are needed to comply with the \dword{surf} program, they will be adopted. However, even with an  excellent safety program, 
 given the large number of hours, the risk of injury remains significant for a project of this scale and needs to be accounted for in the project risk evaluation.

Shipping Delays:
Delays in shipping and availability of components 
presented problems at \dword{pdsp}, and in fact the delays, not technical limitations, ended up driving 
the installation plan 
To avoid this for the \dword{spmod} installation, a one-month buffer of equipment is required from the consortia. The one month period was determined by the maximum delay in customs from a shipment for  \dword{pdsp}, which was three weeks. In addition, a detailed shipping manual will be prepared to provide guidance to collaborators and the LBNF/DUNE logistics manager will be available to provide direct assistance. The residual risk that components are delayed is still considered high, but the total schedule impact is expected to be on the few-week scale.

Missing components cause delay:
Often during \dword{pdsp} installation, parts would arrive at \dword{cern} with small pieces  missing, e.g., brackets or hardware. For the \dword{spmod}, detailed interface drawings will 
define the interfaces clearly and the work packages will 
list all parts. A part-breakdown structure will be defined to clarify 
the ownership of each part 
and track the location of all hardware. 
With these systems in place we expect to minimize the number of instances 
of pieces missing when they are needed. 
The residual risk is considered  highly likely, but with minimal impact.

Import export and VISA issues:
\dword{fnal} has established a new \dword{sdsd} to expedite customs and visa issues. This risk will need re-evaluation after the new division has had time to evaluate the issues.

Lack of available labor:
Unemployment in the Lead area remains low. At the time work is ramping up it may be difficult to find local people with the requisite skills. To mitigate the risk, we plan to hire the core team early and train them at \dword{surf} and \dword{ashriver}.  
In addition, a longer hiring period will be planned (6-12 months) so there will be ample time to hire and train the crew. 
The residual risk is considered low.

Parts do not fit together:
Integration is a critical component of any complex project. \dword{dune} has implemented a process to generate a complete \threed model of the \dword{detmodule} that can be used to detect conflicts. Interface drawings are being generated to clearly define the interfaces between components. Beyond this, an installation prototype of the full assembly is being planned at the \dword{nova} far detector building in \dword{ashriver}. This installation prototype will test the installation of the detector components using full-scale mechanical mockups. For \dword{pdsp}, the \dword{ashriver} test was critical for finding mismatches between components and identifying installation difficulties 
due to limited space. After all the installation steps have been tested it is expected the residual risk will be low. It is highly likely that some small conflict will be found but the impact on the overall schedule will be low.

Cryostat Damage:
The cryostat membrane is a \SI{1.2}{mm} thick stainless steel membrane. A screw driver dropped from \SI{12}{m}, for example, would likely cause damage, and much larger pieces of equipment than that will be used. Equipment must also move within the cryostat. To protect the cryostat, a false floor will be constructed. 
When the false floor is removed, measures will be taken to prevent items from dropping on the membrane. Where possible, all bolts, brackets, and components will be attached to nearby structures so they cannot be dropped. The residual risk of damaging the cryostat is considered small, but in the unlikely event it occurs, it would have a moderate schedule impact.

Weather closes \dword{surf}: 
Weather events in South Dakota that lead to closing \dword{surf} for one or two days occur several times each winter. This risk is accepted, and the average number of snow days is added to the project schedule.

Detector failure during \cooldown: 
As the \dword{detmodule} cools, thermal stresses will develop that the design must be 
sufficiently robust to withstand. 
Risk of breakage due to these stresses is particularly critical as it occurs 
after all components are installed. This risk is minimized by thoroughly testing each component individually in the cold, then cold testing 
each APA-PD-CE assembly 
just prior to moving it into the cryostat. This test of the final assembled components is considered critical in reducing the risk of failure during cool down. The residual risk is classed as low probability, but would have a high impact if it should occur.

The remainder of this chapter is divided into three main sections. 
The first section describes how material will be delivered to the South Dakota region and forwarded to the Ross Headframe on the \dword{surf} site. 
The second section describes the infrastructure needed to install and operate the \dword{spmod}. This includes a cleanroom and its contents, as well as electronics racks, cable trays, storage facilities, and machining facilities. 
The third section describes the installation process itself, which is divided into three phases: the \dword{cuc} setup phase, the installation setup phase, and the detector installation phase. These are summarized in Section~\ref{sec:fdsp-tc-inst}. 

\section{Logistics}
\label{sec:fdsp-tc-log}

Access to the underground installation area for \dword{dune},   \dword{lbnf}, and \dfirst{jpo} personnel, as well as for  \dword{lbnf} and \dword{dune}  materials and equipment, will be provided solely by the \SI{1500}{m}-deep Ross Shaft. Coordinating transport and ensuring on-time delivery of all items are therefore among the more challenging aspects of the \dword{lbnf} and \dword{dune} endeavor. 
The \dword{jpo} (see \tcchjpo 
of this \dword{tdr}) oversees the \dword{sdwf} where deliveries are received before transport to the Ross Headframe. 

Due to the enormous cost of the \dword{lbnf}-\dword{cf} contracts and the risk of increased construction costs due to delays in delivery of materials, the shaft scheduling must be tightly controlled by \dword{lbnf}-\dword{cf} during construction.
The shaft is outfitted with hoists that control the cage and two skips. The cage is used to transport people, equipment and materials, and the skips to bring up muck and transport over-sized equipment and materials. The \dword{lbnf}-\dword{cf} \dword{cmgc} will coordinate overall usage of the Ross Shaft during this period, until the end of the excavation work. At that  time the \dword{jpo} will take over the management of the shaft usage.

To facilitate the flow of non-\dword{cf} \dword{lbnf} and \dword{dune} materials and equipment to the Ross Headframe, the \dword{jpo} will lease a warehouse facility within a maximum one-day roundtrip\footnote{For purposes of warehouse selection ``one-day roundtrip'' is considered three hours of transportation each way and two hours of unloading and loading at the Ross Headframe.} from \dword{surf} by truck. 
It is expected that the lease of this facility, referred to as the \dword{sdwf}, will include warehouse space, personnel, and a \dword{wms} to inventory all incoming materials and equipment. 
A facility has not yet been selected.

Most materials and equipment will be shipped to the \dword{sdwf}; \dword{cf} material, and likely cryogenics equipment, are exceptions and will ship directly to \dword{surf}. 
The \dword{sdsd} logistics  organization will (1) receive and inventory all  goods shipped to the \dword{sdwf}, (2) coordinate with the \dword{cf}-\dword{cmgc}  to transport this material to the Ross Headframe in a just-in-time manner, and (3) transport it underground and into the cavern. 
Figure~\ref{fig:logistics-material-flow} shows a high-level overview of the material flow to the Ross Headframe.

\begin{dunefigure}[Material flow diagram for \dshort{lbnf} and \dshort{dune}]
{fig:logistics-material-flow}
  {Material flow diagram for \dword{lbnf} and \dword{dune}. }
 \includegraphics[width=\textwidth]{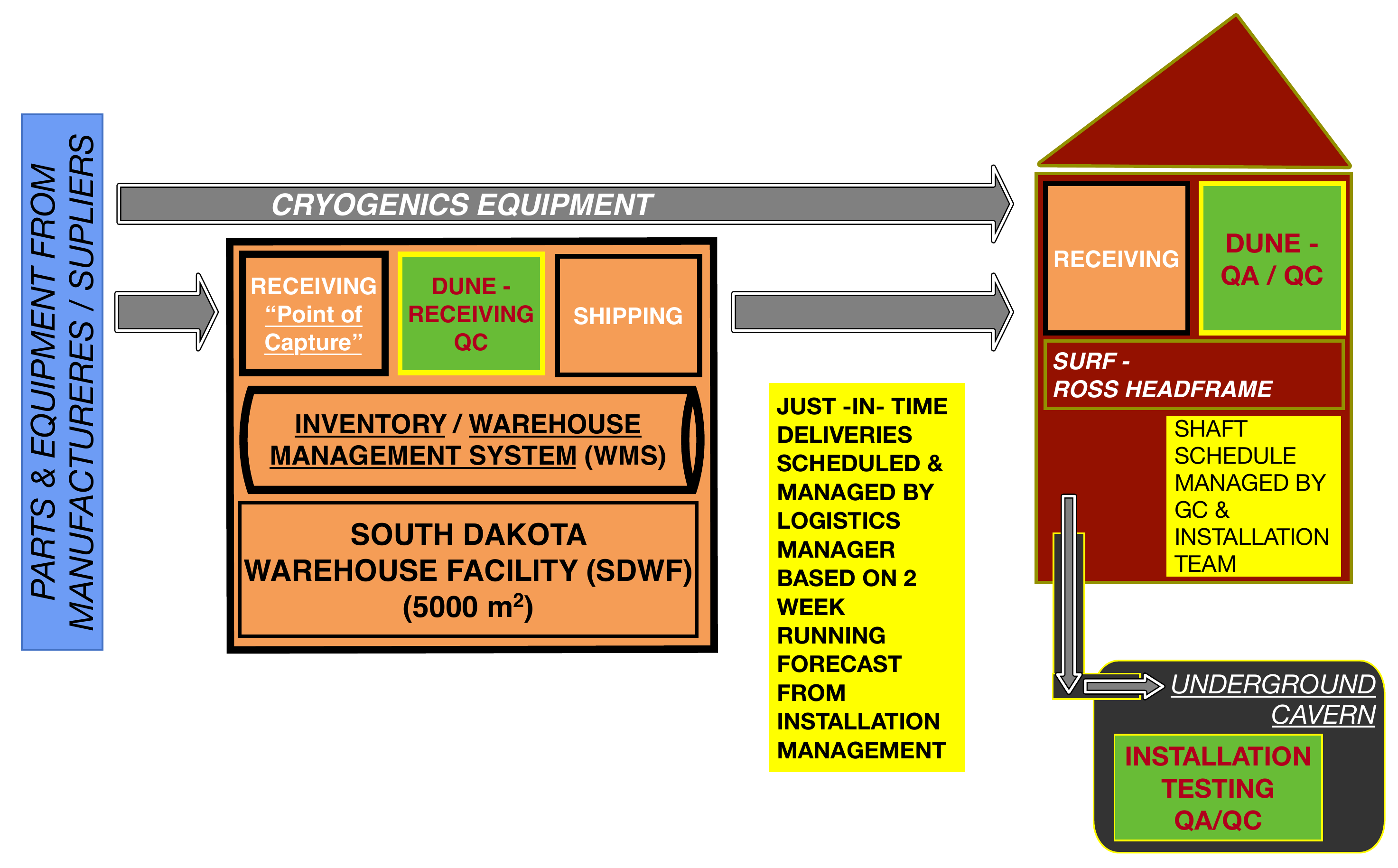}
\end{dunefigure}

\subsection{Logistics Planning}
\label{sec:fdsp-tc-logPln}

The \dword{jpo}/\dword{sdsd} logistics team oversees transportation of the cryostat (steel, foam, and membrane), the cryogenics system, the \dword{dune} detector components, and all related infrastructure not provided by the \dword{cf}. 
\dword{lbnf} specifically oversees the cryostat and cryogenics system, which \dword{lbnf} will discuss in its 
\dword{tdr}. Because \dword{lbnf} materials dominate the logistics, we present a summary of them here, along with an overview of the \dword{dune} materials.  
The steel structure for a single \dword{dune} cryostat requires roughly 1,800 individual steel pieces,  some of which weigh up to \SI{7.5}{t}, as well as \SI{125}{t} of bolts to assemble the steel frame. 
The internal structure for the cryostat, which includes the foam insulation and the thin stainless steel membrane, requires transporting roughly 4,000 boxes of approximate size  \SI{1.5x3.5x1.2}{\meter}. 
 The current plan calls for warehousing all these boxes at the \dword{sdwf} before installation begins. 
The logistics operation will require roughly $\SI{5000}{m^2}$ of area available approximately two years before installation of the first \dword{detmodule} begins, to stage construction of the cryostat, cryogenics system, and detector. 
By the time detector components start arriving, most of the cryostat boxes will have been delivered to \dword{surf}, leaving ample space for the detector and the cryogenics components. 
Additional warehouse space may be required if the boxes for the second cryostat arrive before  \dword{detmodule} \#1 installation is complete; a few buildings of the required size are available in the general area around \dword{surf}. 

\begin{dunefigure}
[Simplified model of the Ross Cage]
{fig:fdsp-tc-Cage}
{Simplified Ross Cage model and specifications.}
\parbox{2.1in}{\includegraphics[width=0.3\textwidth]{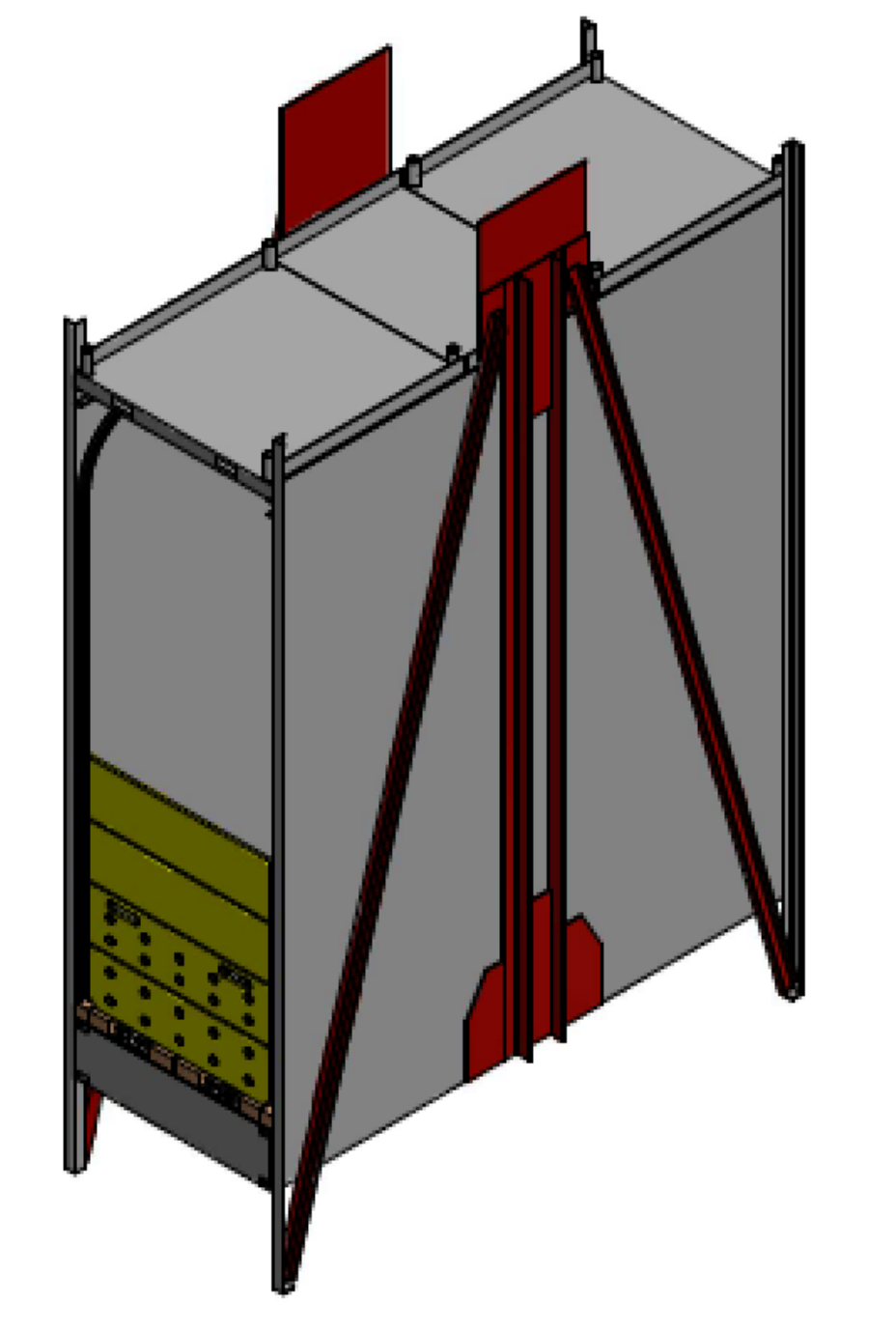}}
\qquad\hspace{10pt}
\begin{minipage}{0.5\textwidth}%
\begin{tabular}{p{3.4cm}p{3.4cm}}        
\multicolumn{2}{c}{Ross Cage Specifications}\\ \toprowrule
Inside height & 3.6 m\\ \colhline
Inside depth  & 3.7 m \\ \colhline
Inside width  & 1.38 m \\ \colhline
Weight limit  &  5,897 kg \\ \colhline
Round trip \newline time & 17 min \newline (incl. unloading) \\ \colhline
\end{tabular}
\end{minipage}
\end{dunefigure}

The \dword{surf} Facility Access Specification~\cite{bib:docdb328} defines the limitations on dimensions and weights for all materials to be transported underground, the most stringent of which are set by the Ross Shaft and Cage. 
It is possible to bring material down the shaft underneath the cage or in the skip compartment  as a slung load, but this is a much slower process and requires careful planning 
and review of detailed procedures for each trip. 
The  \dword{apa}s, for example, require this special handling because they are too tall to fit in the cage. 

Most material will be brought underground inside the cage. Figure~\ref{fig:fdsp-tc-Cage} illustrates the new Ross Cage and summarizes its parameters.  
The roundtrip travel time for the Ross Cage is 17 minutes (actual travel time is \num{3.6} minutes each way), dominated by loading and unloading time.  
Slung loads will require more than an hour round trip.

The Ross Headframe has no loading dock so careful planning of material loading and unloading of shipments is required. 
All materials must arrive at \dword{surf} on a flatbed or curtain-sided chassis, and a forklift will be available for unloading. 
All deliveries, either from the \dword{sdwf} or direct to the Ross Headframe, require (1) coordination with the logistics organization, and (2) minimum two weeks prior notice, per an advance delivery plan.  
 
Logistics will provide to \dword{dune} institutions a shipping manual that  
specifies guidelines on required shipping data and cargo consignment. Adherence to the guidelines will enable the logistics organization to monitor shipping progress and ensure that no delays occur due to incomplete or missing documentation.

In \dword{pdsp}'s experiences with trans-oceanic international shipping highlighted the need to increase delivery schedule duration beyond the shipper-quoted average, which was sometimes exceeded by as much as three weeks. For \dword{lbnf}/\dword{dune} materials, we will plan shipping and transport so that items arrive in South Dakota a minimum of four weeks before they are expected underground. This buffer will allow sufficient advanced planning for the underground work, with confidence that the installation plan can be maintained.

Sufficient space must be made available at the \dword{sdwf} and in the underground area  to house this material.
The \dword{sdwf} staff will deconsolidate or consolidate arriving cargo into appropriately sized boxes and crates, as needed, for delivery to \dword{surf}, to make the most efficient use of available trucks and the Ross Shaft. 

\begin{dunefigure}[Planned usage of underground space during installation setup]{fig:fdsp-tc-setup}
  {CAD image showing the empty half of the north cavern as used during the installation setup phase of the first \dword{detmodule}.  
The entire cavern is \SI{145}{\meter} long, \SI{20}{\meter} wide, and  \SI{28}{\meter} high; the half shown is therefore approximately \SI{73}{\meter} long.  
Half of this empty space will be used for the cryostat work and half for storage of the detector infrastructure. The material shown outside the cavern must be stored in the \dword{sdwf}.}
\includegraphics[width=.9\textwidth]{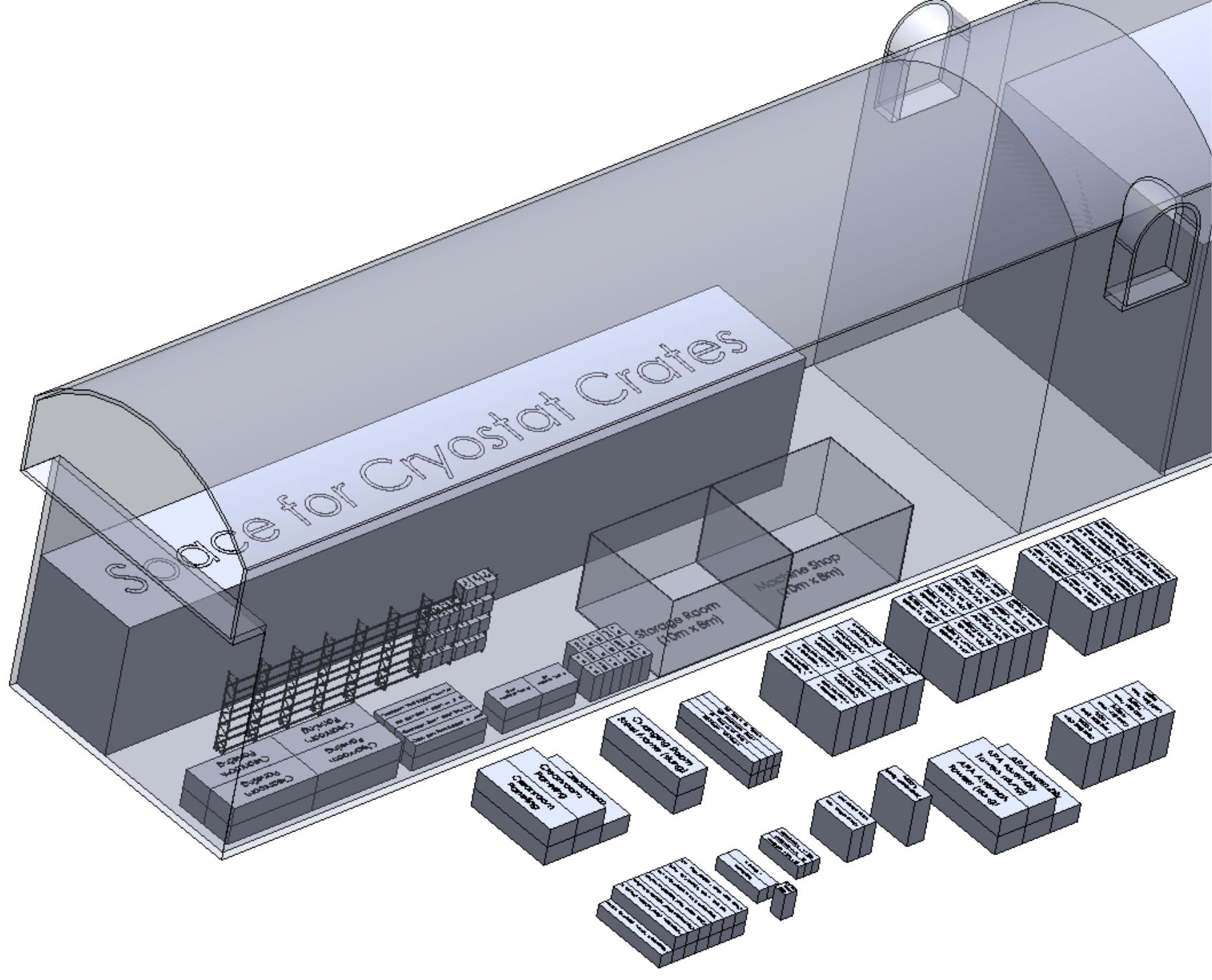}
\end{dunefigure}

To determine the storage space requirements and how much hoist time must be dedicated to \dword{dune}, a detailed inventory of all  \dword{dune} detector equipment and infrastructure is needed. 
A complete list of materials has been solicited from all consortia and technical coordination. 
The entries in the inventory spreadsheet are organized as \textquotedblleft loads\textquotedblright \ for the Ross Shaft where a load is a crate or set of boxes that will be transported underground in one trip, either in the cage or as a slung load~\cite{bib:docdb8426}. 
Information captured in the load spreadsheet includes the number of  
trips, type of trip (slung load or cage), package dimensions, weight, and type of package (crate, pallet, box, or carton). 

The load list at present predicts 1,600 hoist trips and approximately two  months of cage time, most of which is spread over one year. 
Detector installation (see Figure \ref{fig:high-level-schedule}) for the \dword{spmod} will span two years, so we divide the logistics planning into three phases (summarized in Section~\ref{sec:fdsp-tc-inst}: 
 (1) the \dword{cuc} setup phase, (2) the installation setup phase, and (3) the detector installation phase. 
For each phase, a \threed model was generated to show how much material can be stored underground outside the work area and how much material must be stored 
at the \dword{sdwf}, thus setting the surface space requirements. 
The phase with the largest amount of material to transport is the installation setup phase.  
Figure~\ref{fig:fdsp-tc-setup} shows the model of the underground area and the required boxes for surface storage for the first month of this phase. 
The crates outside the cavern were used to estimate \dword{dune}'s storage needs in the surface storage facility. Roughly \SI{1000}{\square\meter} of warehouse space will be needed at this time to buffer \dword{dune} installation equipment.  The \dword{sdwf} will also need space to store up to 150 \dword{apa}s, adding another \SI{700}{\square\meter}. The remaining \SI{3300}{\square\meter} is available for \dword{lbnf} storage. The amount of warehouse space  actually leased can be adjusted to match \dword{lbnf}/\dword{dune} needs, and after the second cryostat construction is complete it will be reduced. 

Managing the hoist and planning all the transport of materials underground is one of the primary responsibilities of the \dword{cmgc}. This task will be challenging and the installation plan with warehouse space on the surface and storage space underground will give critical flexibility in the timing of the delivery of materials. With month-long buffers above and below ground, and a two-week advanced notification of the installation needs, the \dword{cmgc} has freedom to schedule deliveries around the needs of other contractors.

\subsection{Logistics Quality Control}
\label{sec:fdsp-tc-log-qaqc}

The \dword{pdsp} experience offers a couple of significant lessons regarding logistics.

\begin{enumerate}
\item A central inventory system is essential for tracking  shipments.
\item 
It is important to apply realistic shipping durations based on experience into the overall planning so that work can proceed on a predictable schedule. 
 
\end{enumerate}

The central inventory system  implemented at the \dword{sdwf}  and minimum one-month material buffer are the plans we have in place to prevent repetition of the \dword{pdsp} schedule problems. 
The full list of lessons learned from \dword{pdsp} is in~\cite{bib:docdb8255}. 

Component testing at the \dword{sdwf} is presently not planned, however a \dword{dune} \dword{qc} procedure will be followed to detect any damage incurred during transportation and determine remedial action. Any request from the consortia to perform work in the \dword{sdwf} will be addressed on a case-by-case basis.

In critical cases where the shipping dimensions approach the shaft dimensions, a test transportation using a dummy component will be done. At present the \dword{apa} shipping/transport box is planned to be tested in this fashion.
 
The logistics organization in coordination with the  \dword{sdwf} will inventory all received shipments and  ensure that all materials fit in the Ross Cage, or if a slung load is needed, that the necessary procedures are in place and approved before any material is transported to the Ross Headframe.  
\dword{jpo} representatives will verify that no obvious damage occurred in transport. 

The contribution-in-kind model of this project complicates logistics oversight and inventory control, since components will be delivered from many institutions and from different countries. 
Similarly, during production and testing, \dword{qc} information must be gathered from and made accessible to all collaborators. 
Because of the complexity of the project and the different requirements for \dword{qc} and logistics oversight, two different databases will be used. 
A commercial \dword{wms} will control the inventory process at both the \dword{sdwf} (items both received and shipped) and at \dword{surf} (items received at the Ross Headframe). 
A  separate database, the \dword{dcdb}, will store test and other \dword{qc} data, e.g.,  shipping reports and any reported damage. 
The \dword{wms} will need to provide location and  \dword{qc} information to the \dword{dcdb}, which will ultimately archive both sets of data. 
The \dword{dcdb} has not yet been designed.

Until materials arrive at the \dword{sdwf} (or \dword{surf} if directly shipped), the contributors' freight forwarding system will control the logistics supply chain, which will depend on the contractual circumstances and the contributor's choice. 
However, assuming the shipment is consigned as outlined in the  shipping manual (so that the 
logistics organization has access to the shipping data),  the logistics manager will monitor the cargo progress and step in if a problem arises. 
The \dword{qc} and shipping data flow is shown in Figure~\ref{fig:logistics-data-and-mat-flow}.

\begin{dunefigure}[QC and shipping data flow diagram for logistics]{fig:logistics-data-and-mat-flow}
  {\dword{qc} and shipping data flow diagram for the \dword{lbnf} and \dword{dune} logistics.}
 \includegraphics[width=\textwidth]{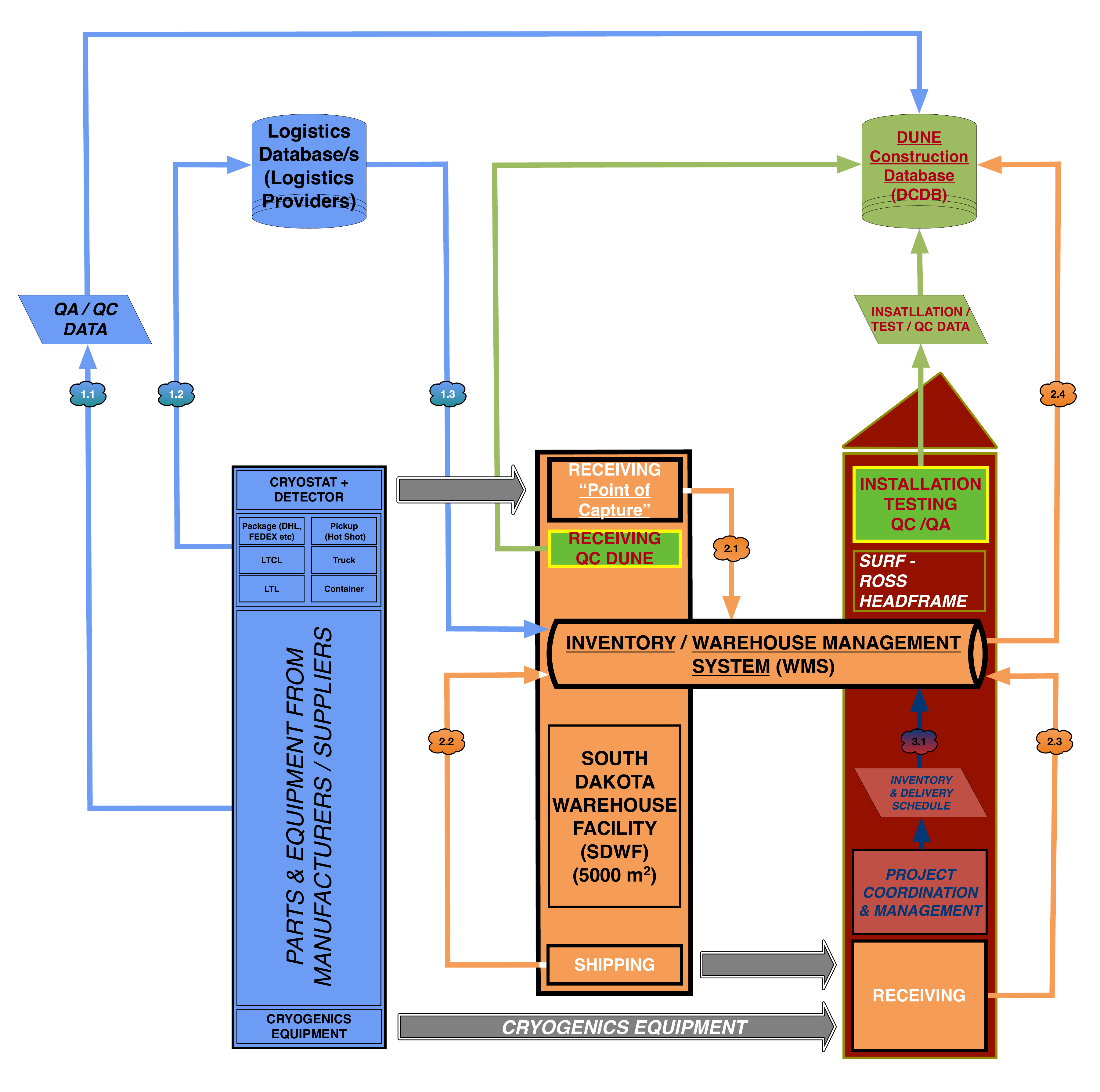}
\end{dunefigure}

The \dword{jpo} installation management team will provide a shipping (supply) report to the \dword{sdsd} logistics organization and \dword{sdwf} for scheduling delivery of parts and equipment two weeks in advance of the required delivery date. 
All deliveries will be inventoried upon receipt at the Ross Headframe in the \dword{wms}.

\subsection{Logistics Safety}
\label{sec:fdsp-tc-log-safety}

The \dword{sdwf} will be managed and operated by an independent contractor under the supervision of the \dword{dune} logistics manager. 

The facility will be operated under the contractor's \dword{esh} program, which must conform to federal regulations and will be reviewed by \dword{fnal}'s \dword{esh} management prior to entering a contractual relationship.

\section{Detector Infrastructure}
\label{sec:fdsp-tc-infr}

The  \dword{jpo} will provide the infrastructure needed to install the \dword{spmod}. The major items, described below, include the \dfirst{dss}, the electronics mezzanine on the cryostat roof (including racks), cable trays, an underground cleanroom with appropriate installation equipment, piping inside the cryostat, and \coldbox{}es with associated cryogenics supply. 

Other items, not described here but also in the \dword{jpo} scope, include a small machine shop, scissor lifts, rigging equipment, hand tools, diagnostic equipment (including oscilloscopes, network analyzers, and leak detectors), local storage with some critical supplies, and \dfirst{ppe}.  

\subsection{Detector Support System}
\label{sec:fdsp-tc-infr-dss}


The \dword{dss} provides the structural support for the detector inside the cryostat.  
It also provides the necessary infrastructure inside the cryostat to move the detector elements into place during assembly. 
The \dword{dss} is a new design, quite different from the \dword{pdsp} \dword{dss}. The detector elements supported by the \dword{dss} include the \dwords{ewfc}, the \dword{apa}s, and the \dwords{cpa} with top and bottom \dword{fc} panels. 
The nominal load of the detector elements both dry (in air) and wet (in \dword{lar})\footnote{The ``wet'' load takes into account the buoyancy of the liquid argon. As G10 is almost neutral buoyant the difference is substantial for some sub-systems.} are shown in Table~\ref{tab:installation-DSS-load}. 
The weights listed are the current design weights.  
The \dword{dss}, however, is designed to accommodate significant design changes --- even if the detector weight were to double the \dword{dss} still meets the design code requirements.  
The \fdth{}s can be adjusted to compensate for deflections due to load.
\begin{dunetable}
[DSS Loads]
{l|c|cc|cc}
{tab:installation-DSS-load}
{The expected dry and wet static loads for the DSS.}
& &  \multicolumn{4}{|c}
{Weight before fill (Dry)}\\ \toprowrule
& & \multicolumn{2}{c|}{Unit Weight} & \multicolumn{2}{c|}{Total Weight}  \\ \colhline

Detector Component &\# Units& (kg)&(lbs) & (kg) &(lbs)\\ \colhline
\dword{dss} & 1 &NA&NA& 12318  & 27100 \\ 
\colhline
\dword{apa} (Installed \dword{apa} pair, no cables)& 75&1184 &2604 &88768  &195290\\ 
\colhline
\dword{cpa} & 100& 233 & 513 & 23331 & 51327 \\ 
\colhline
Top or Bottom \dword{fc} module (FC TB)& 400&149 & 328	 & 59679 & 131294\\ 
\colhline
\dword{tpc} Electronics and Cables &3000& 4.9 & 10.8 & 14700 & 32400\\
\colhline
\dword{ewfc}  & 8	&904 &	1989  & 7234 & 15914\\ 
\colhline
{\bf Total} &  & & & 206,000 &	454,000\\ 

\toprowrule

\rowtitlestyle & &  \multicolumn{4}{c}{Weight after fill (Wet)}\\
\toprowrule
\dword{dss} (not in liquid) & 1 & NA & NA & 12318 & 27100 \\ 
\colhline
\dword{apa} (Installed \dword{apa} pair/No cables)&75&850 &1874 & 64000 &140000\\ 
\colhline
\dword{cpa} & 100& 45 & 99 & 4520 & 9943 \\ 
\colhline
Top or Bottom \dword{fc} module (FC TB)& 400 & 68 & 150	& 27359 & 60191 \\ 
\colhline
\dword{tpc} Electronics and Cables & 3000 &2.9 &6.4 & 8700& 19200 \\
\colhline
\dword{ewfc}  & 8 & 283& 	622& 2263 & 4978\\  
\colhline
{\bf Total} &  & & &110,000	 &242,000 \\ 
\colhline
\end{dunetable}

The \dword{dss} shown in Figure~\ref{fig:DSS} consists of five rows of I-beams inside the detector that support the five rows of \dword{apa}s and \dword{cpa}s. 
The I-beams themselves are supported from the cryostat outer steel structure through a series of vertical supports or mechanical \fdth{}s, also shown in Figure~\ref{fig:DSS}. 
The \dword{dss} constrains the location of the detector inside the cryostat and also accommodates the detector elements' movement and contraction during cooling. The layout of the \dword{dss} sets, in turn, the overall layout of the detector module since the module's elements become a unified mechanical structure only after they are mounted to the \dword{dss} and internally connected.

During installation the detector components are moved along the I-beams using both simple and motorized trolleys. 
The end of the \dword{dss} nearest the \dword{tco} is also designed as a switchyard. An additional set of north-south beams allow a short section of the I-beam rail to be shifted between the five rows of \dword{dss} beams that correspond to the five alternating rows of detector elements  (\dword{apa}-\dword{cpa}-\dword{apa}-\dword{cpa}-\dword{apa}).  
With this the \SI{12}{m} tall detector elements can enter the cryostat on an I-beam through the \dword{tco}, be loaded on the short switchyard beam, moved to the required row of \dword{dss} and then be pushed into position.

\begin{dunefigure}[\threed model of the DSS] {fig:DSS}
  {\threed model of the \dword{dss} showing the entire
  structure on the left along with one \dword{apa} row and one
  \dword{cpa}-\dword{fc} row at each end. The right panel is a zoomed image
  showing the connections between the vertical supports and the
  horizontal I-beams.}
\includegraphics[width=.49\textwidth]{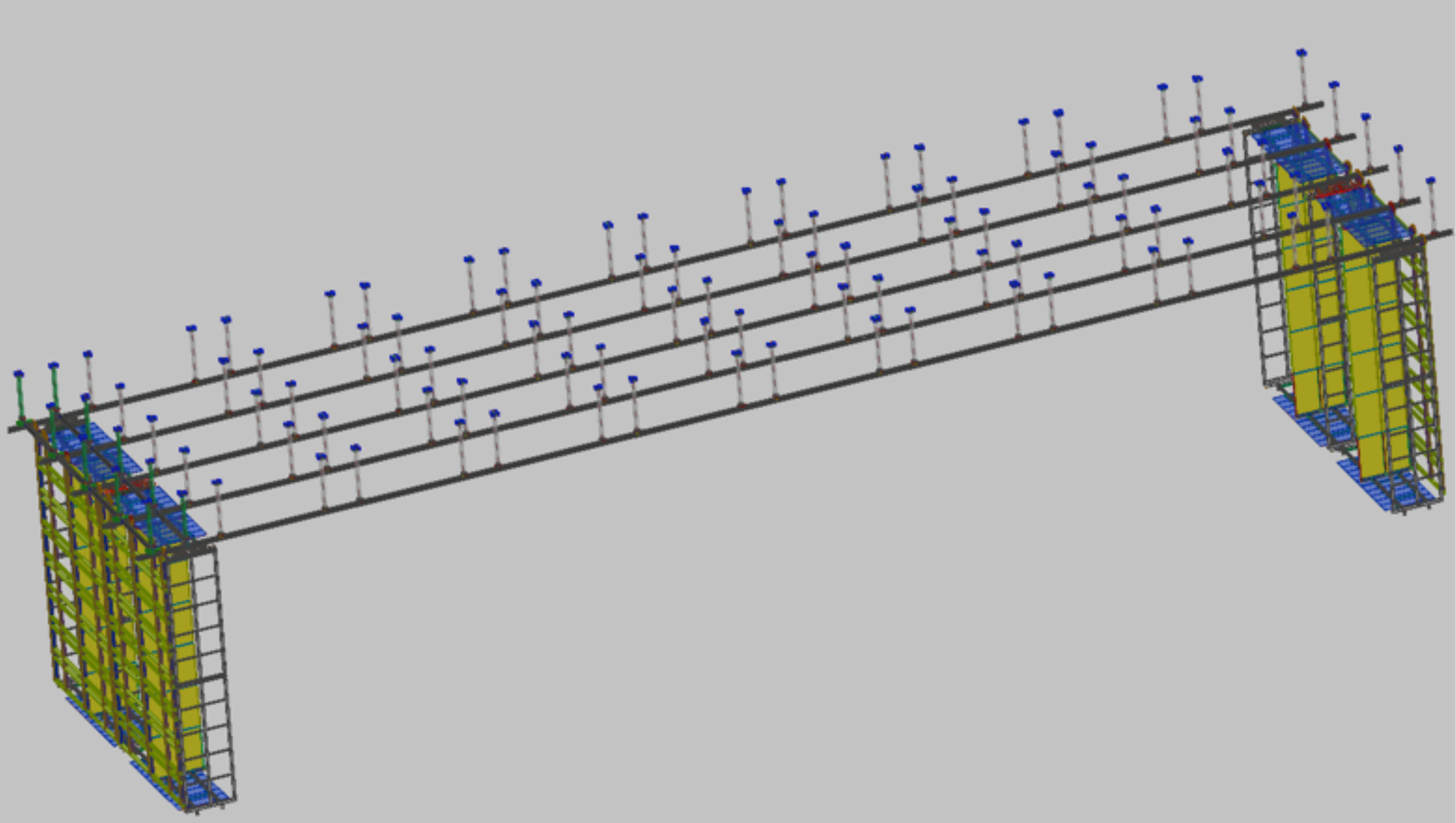}
 \includegraphics[width=.49\textwidth]{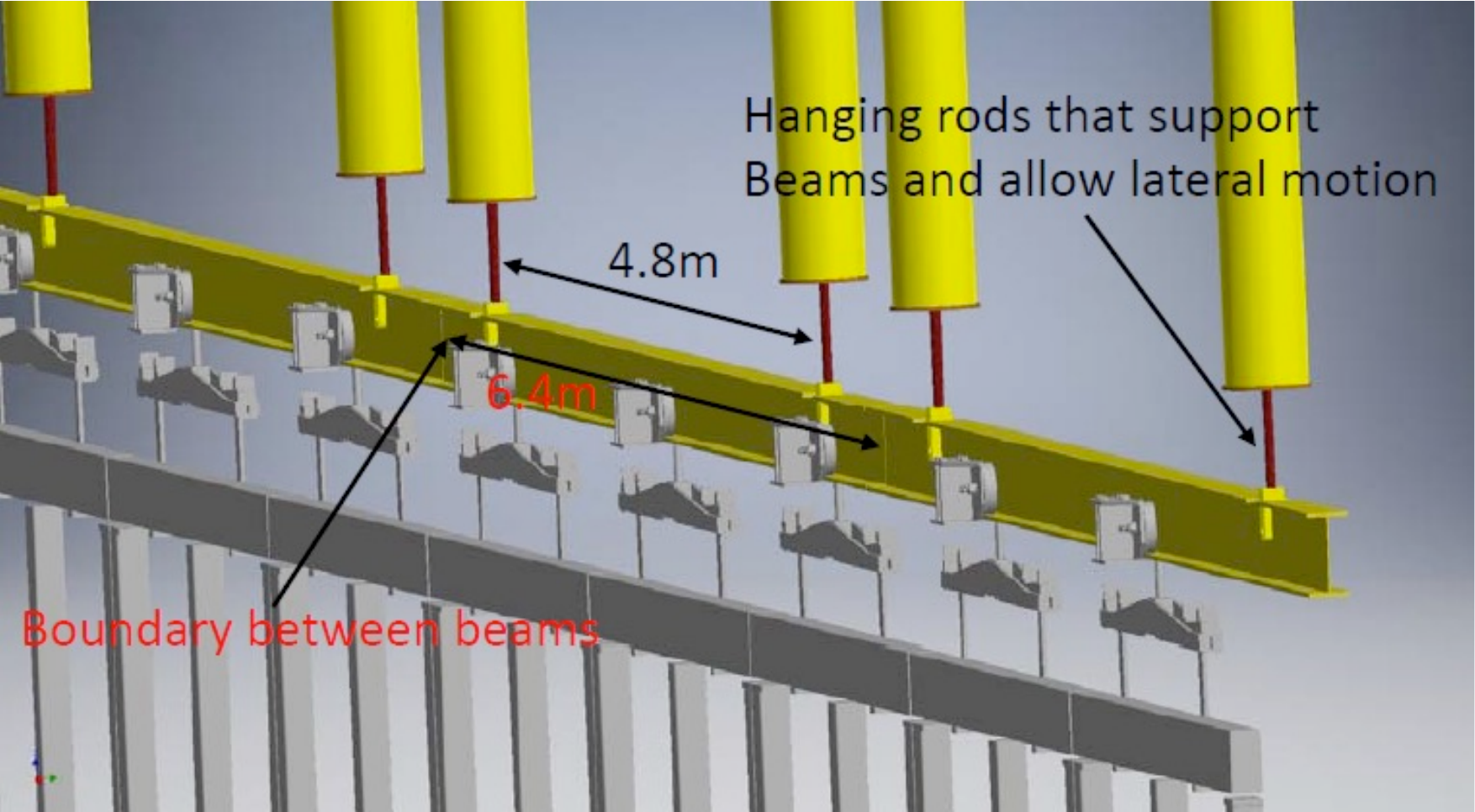}
\end{dunefigure}

The \dword{dss} is designed to meet the following  requirements:
\begin{itemize}
\item support the weight of the detector;
\item accommodate cryostat roof movement during filling, testing, and operation;
\item accommodate variation in \fdth locations and
  variation in the flange angles due to installation tolerances and
  loading on the warm structure;
\item accommodate shrinkage of the detector and \dword{dss} from ambient
  temperature to \dword{lar} temperature;
\item define the positions of the detector components relative to each other; 
\item provide electrical connection to the cryostat ground and remain electrically isolated from the detector;
\item allow support penetrations to be purged with gaseous argon to prevent contaminants from diffusing back into the liquid; 
\item ensure that the instrumentation cabling does not interfere with the \dword{dss};
\item consist entirely of components that can  
be installed through the \dword{tco};
\item meet AISC-360 codes; 
\item meet seismic requirements one mile underground at \dword{surf};
\item consist entirely of materials compatible 
with operation in ultrapure \dword{lar};
\item ensure that the \dword{dss} beams either sit completely submerged in \dword{lar} or sit completely in gas while leaving a \SI{4}-\SI{5}{\%} ullage at the top of the cryostat;  
\item maintain the centerline of the \dword{apa} near the cryostat at \SI{400}{mm} from the membrane flat surface;
\item ensure that the supports do not interfere with the cryostat I-beam structures;
\item ensure that the detector's lower \dword{gp} lies over the cryogenic piping; and
\item include the infrastructure necessary to move the \dword{apa} and \dword{cpa}-\dword{fc} assemblies from outside the cryostat through the \dword{tco} to the correct position.
\end{itemize}

Each row of the \dword{dss} consists of a series of ten  \SI{6.4}{m} long
W10$\times$26 stainless steel I-beam sections, for a total of \num{50} I-beam segments for the five rows. The length of the beam segments was chosen to be a multiple of the \SI{1.6}{m} pitch of the major cryostat beams, which allows the regular placement of the support \fdth across the cryostat roof. With a W10$\times$26 I-beam and \SI{6.4}{m} between the supports,  the beam deflections due to the loads can be kept below \SI{5}{mm}. 
Each I-beam is suspended on both ends by the mechanical \fdth{}s that penetrate the cryostat roof. 
During \cooldown  each I-beam shrinks while the mechanical supports outside the cryostat remain fixed,  causing gaps to form between \dword{apa}s that are adjacent but supported on separate beams.
\dword{apa}s that are supported on the same beam will not have gaps develop because both the beam and \dword{apa} frames are stainless steel and will shrink together.
The gap between two adjacent \dword{dss} beams after \cooldown will be \SI{17}{mm}; this is considered acceptable. 
Increasing the beam length beyond \SI{6.4}{m} was not considered because the deformation of the I-beam under load would increase, as would the gap between \dword{apa}s on adjacent beams and the difficulty of installing the beams.

\begin{dunefigure}[DSS vertical support \fdth]{fig:DSS-Support}
  { Drawing of the \dword{dss} vertical support \fdth. The detector load is carried by the \SI{25}{mm} inner support rod. The outer lateral support tube prevents swinging during installation.  The \fdth mounts to the cryostat crossing tube, which is an integral part of the cryostat. 
  
  Note: A few of the \dword{dss} vertical support \fdth{}s have a short vacuum chamber with side ports inserted between the \dword{dss} support flange and the cryostat crossing tube. These chambers are used to bring the \dword{cisc} cables out of the cryostat and are shown in the \dword{cisc} section in Figure \ref{fig:CISC-feedthru}.}
\includegraphics[width=.85\textwidth]{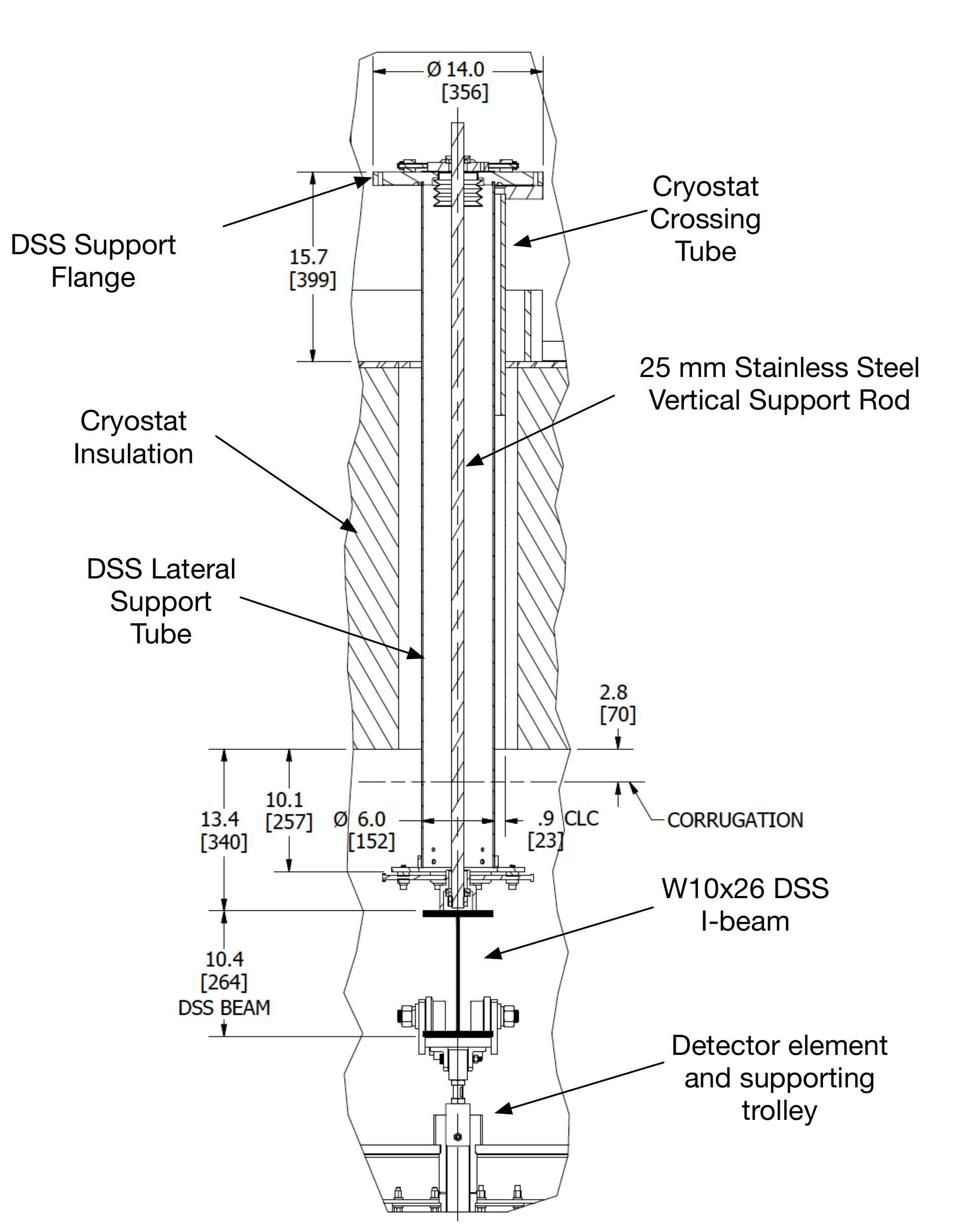}
\end{dunefigure}

The \dword{dss} I-beams are supported on both ends from a vertical support \fdth shown in Figure~\ref{fig:DSS-Support}. A \SI{25}{mm} solid stainless steel rod, which is threaded at both ends, runs down the center of the \fdth and carries the detector load. The support rod connects on the bottom end to a clevis which is then pinned to the \dword{dss} beams shown in Figure \ref{fig:DSS-lateral-support}. At the top the rod bolts to an X-Y table sitting on the top Conflat flange that allows a lateral adjustment of $\pm$\SI{2.5}{cm} (\SI{1}{in}). A swivel washer is used in the bolted connection to the X-Y table to allow the support rod to swing freely. The bolted connection also allows the \dword{dss} I-beams to be adjusted vertically. The vacuum seal is established at the top with a bellows between the rod and the top flange. The top flange of the \dword{dss} support \fdth is a Conflat flange that connects to the cryostat crossing tube's mating flange. The crossing  tube is welded to the cryostat roof and the top flange is mechanically supported from the cryostat's  \SI{1.1}{m} tall support I-beams. The cryostat crossing tubes are shown in Figure~\ref{fig:crossingtube}.

During installation the detector components will be pushed along the \dword{dss} I-beams, placing a lateral load on the \dword{dss}. 
A \SI{15.2}{cm} (\SI{6}{in}) \dword{od}  tube is welded to the top flange of the \dword{dss} \fdth{}. 
This lateral support tube  extends through the cryostat insulation and has a clamping collar at the bottom that is used to fix the I-beam support clevises in position during installation. 
The bottom of the lateral support tube is seen in Figure~\ref{fig:DSS-lateral-support}. 
The long bolts press on the flat sides of the clevis to fix the support rod's location. 
There is a nominal \SI{10}{mm} gap between the \dword{od} of the support tube and the \dword{id} of the clearance tube in the cryostat. 
The clevis can be positioned anywhere inside the \SI{15.2}{cm} tube.

\begin{dunefigure}[DSS support for lateral loads ]{fig:DSS-lateral-support}
  {Left panel shows how the central support rod is locked in position during detector installation. The outer  \SI{15.2}{cm} (\SI{6}{in}) tube is used to fix the support clevis in position. The right panel shows the system as it is connected to the I-Beam.}
\includegraphics[width=.75\textwidth]{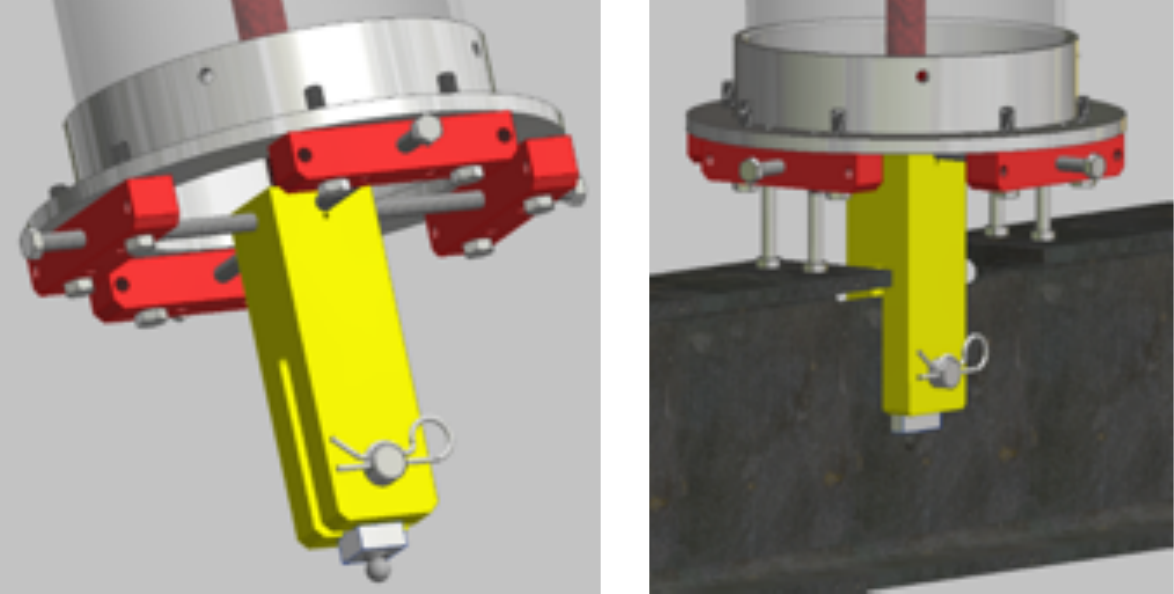}
\end{dunefigure}

After the detector has been installed all restraints on the clevis are released to allow motion as the detector contracts during \cooldown.  The two support rods that support each \dword{dss} beam will contract and move toward each other by \SI{13.1}{mm} along the axis of the detector.  
The drift distance will shrink by \SI{7.4}{mm}  caused by the contraction of the field cage.  The detector is symmetric in the drift direction around the center \dword{apa}.  The drifts on either side of the center \dword{apa} will  shrink toward the center while the center \dword{apa} remains unmoved.  This results in the \dword{cpa}s moving \SI{7.4}{mm}  toward the center and the outer \dword{apa}s moving \SI{14.8}{mm}  (2$\times$\SI{7.4}{mm}) toward the center.  The hanging rod is designed to have a range of motion of \SI{15}{mm}  in the drift direction to accommodate this shrinkage.

Detector components are installed using a shuttle beam system as
illustrated in Figure~\ref{fig:shuttle}.  
The last two columns of \fdth{}s (western-most) support temporary beams that run
north-south, perpendicular to the main \dword{dss} beams.  
A shuttle beam has trolleys mounted to it and traverses 
north-south until it aligns with the required row of \dword{dss} beams.  
The last \dword{apa} or \dword{cpa} in a row is supported by the shuttle beam, which is bolted directly to the \fdth{}s once it is in place.  
As the last \dword{cpa} or \dword{apa} in each row is installed, the north-south beams are removed. This system will be thoroughly tested as part of the Ash River testing program described in \ref{sec:fdsp-tc-inst-qaqc}.

\begin{dunefigure}[\threed models of the shuttle beam end of the DSS]{fig:shuttle}
  {\threed models of the shuttle beam end of the \dword{dss}. The figures show how an \dword{apa}
is translated into position using the north-south beams until it lines up with the correct
row of I-beams.}
\includegraphics[width=.49\textwidth]{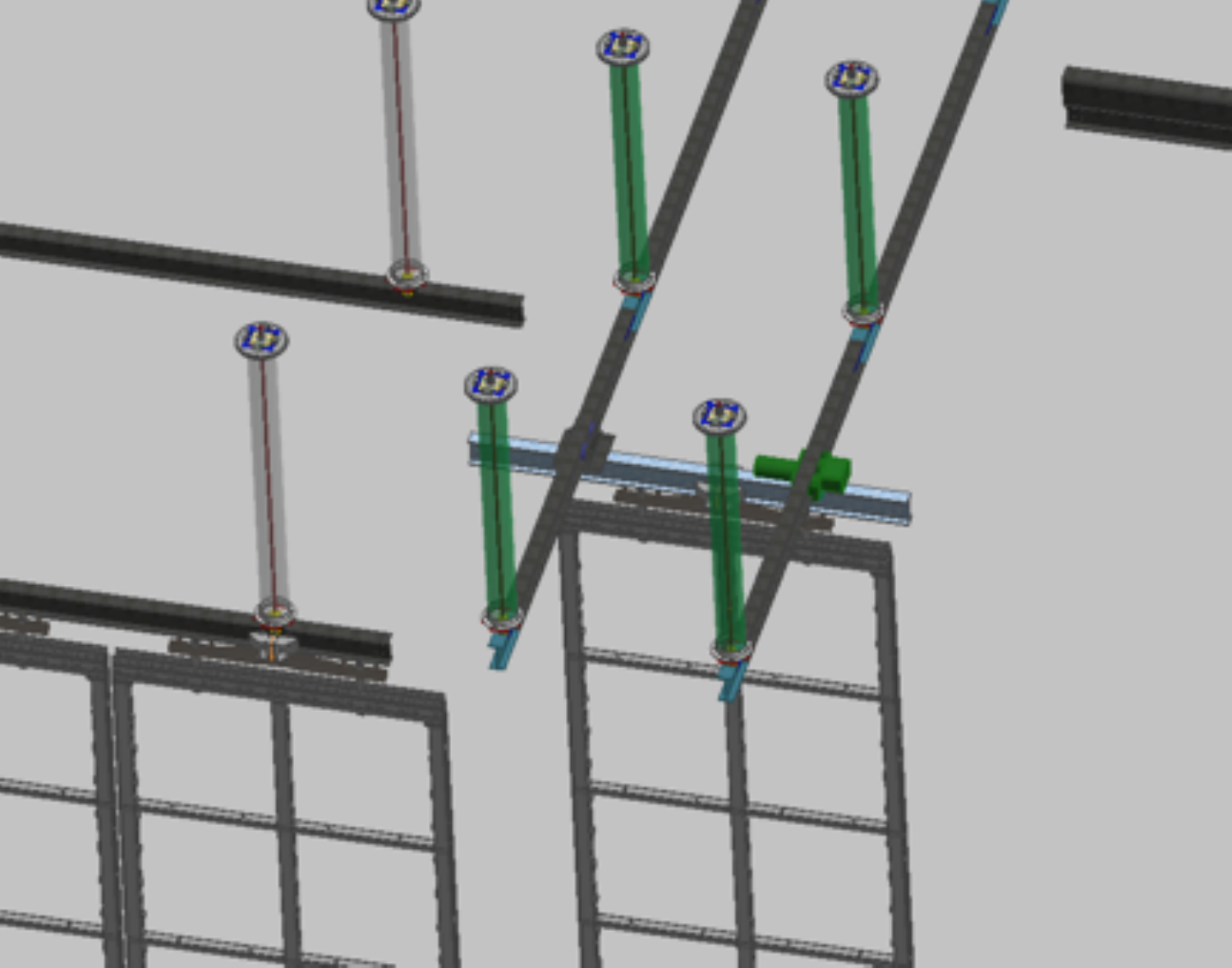}
 \includegraphics[width=.42\textwidth]{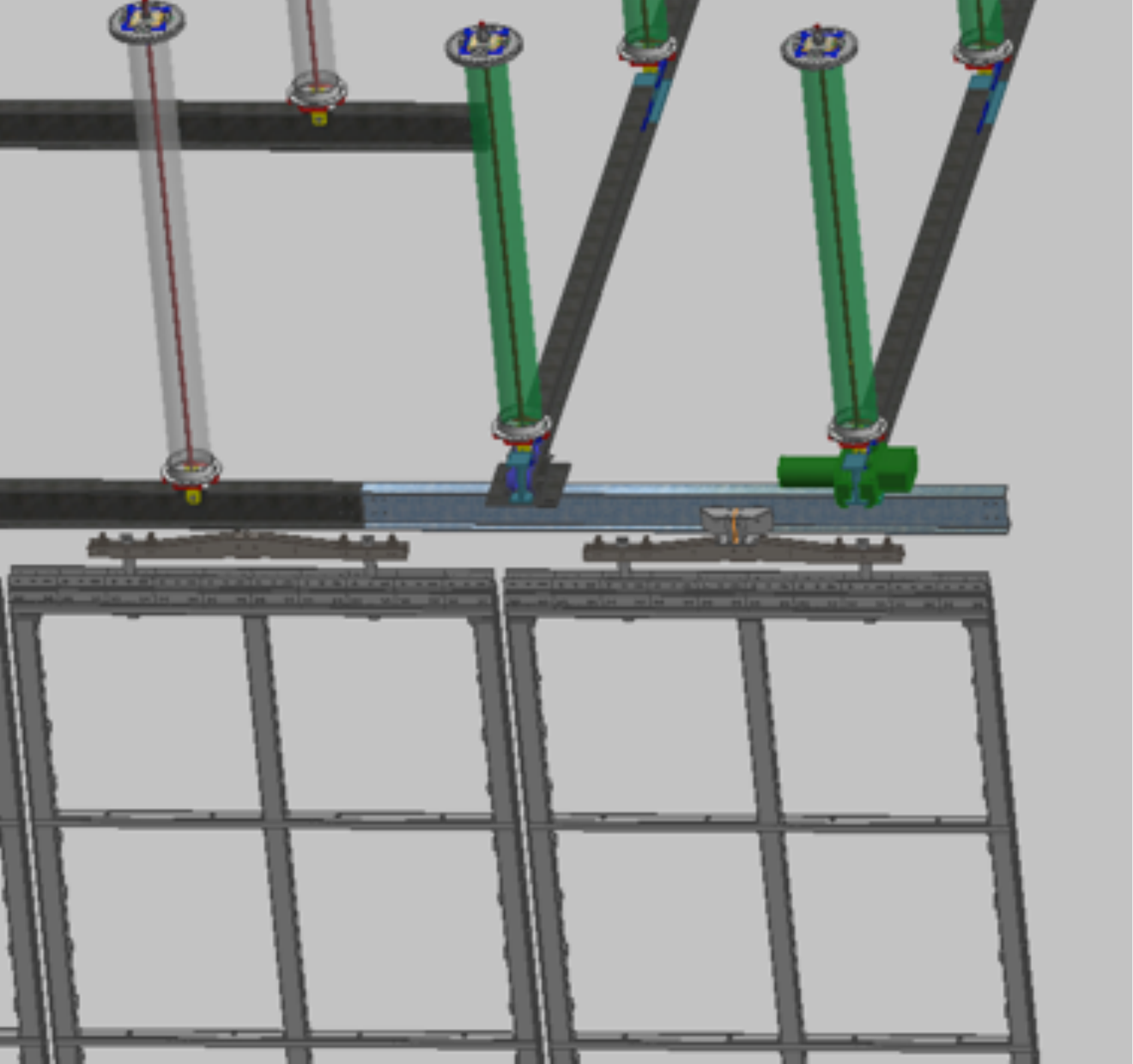}
\end{dunefigure}

The shuttle beam and each detector component are moved using a motorized trolley as seen in Figure \ref{fig:DSS-trolley}.  A commercially available motorized trolley will be modified as needed for the installation. A mechanical stop will prevent the trolley from passing the end of the shuttle beam unless the beam is aligned with a corresponding \dword{dss} beam. A detailed engineering design report for the \dword{dss} is available \cite{bib:docdb6260} and the preliminary design review is complete.

\begin{dunefigure}[Prototype of the motorized DSS trolley ]{fig:DSS-trolley}
  {Prototype of the motorized \dword{dss} trolley that will push the \dword{apa} and \dword{cpa} along the I-beams and through the switchyard.}
\includegraphics[width=.49\textwidth]{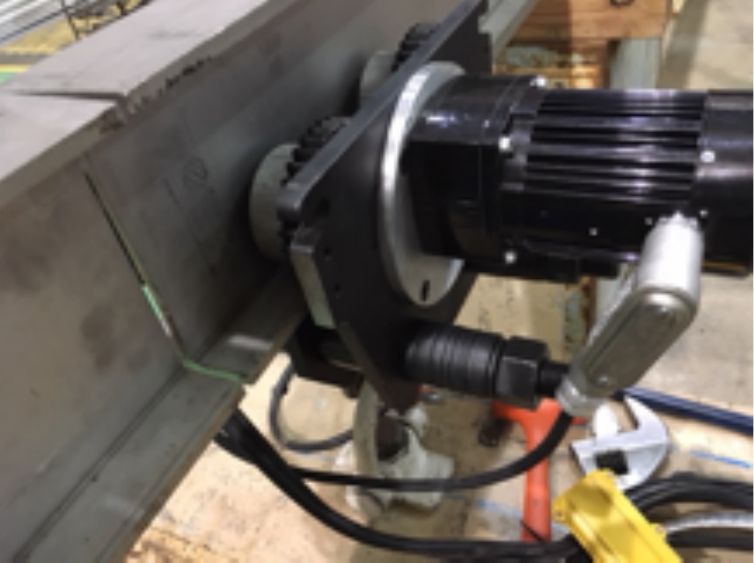}
\end{dunefigure}

A mock-up of the shuttle system will be constructed to test the mechanical interlock and drive systems for the shuttle beam for each \dword{detmodule}.  
Tests will be conducted to evaluate the level of misalignment between beams that can be tolerated and the amount of positional control that can be achieved with the motorized trolley. 
We plan to construct a full scale prototype of a section of the  switchyard and perform tests at floor level. 
Later, the test program will be expanded at Ash River, where a full-scale installation test will be performed; see Section~\ref{sec:fdsp-tc-inst-qaqc}.

\subsection{Cryostat Roof Infrastructure}
\label{sec:fdsp-tc-infr-cryo-roof}

\begin{dunefigure}[Mezzanine and electronics racks]{fig:mezzanine}
  {The electronics racks sit on the \dword{dune} electronics mezzanine. The top image is a view from above the detector looking at the racks from the side. In this view the cavern and cryogenics mezzanine are hidden. The bottom view is from the end of the cryostat looking over the roof. The access stairs to the mezzanine are shown.}
 \includegraphics[width=\textwidth]{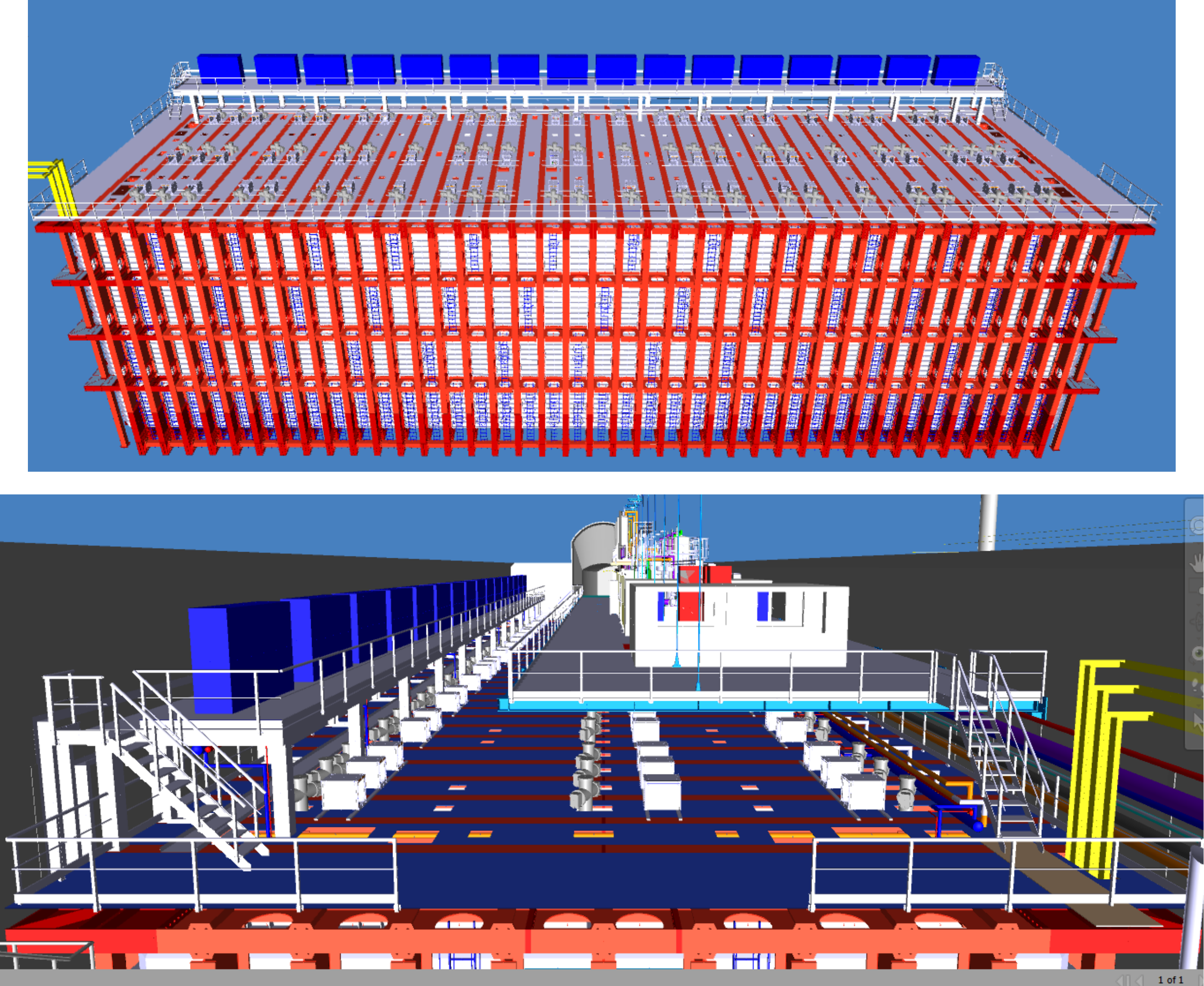}
\end{dunefigure}

\begin{dunefigure}[Electronics rack contents]{fig:rack-build1}
  {The nominal contents of the electronics racks on the mezzanine is shown. Each rack is configured to consume less than 3.5 \si{kW}.  }
 \includegraphics[width=.8\textwidth]{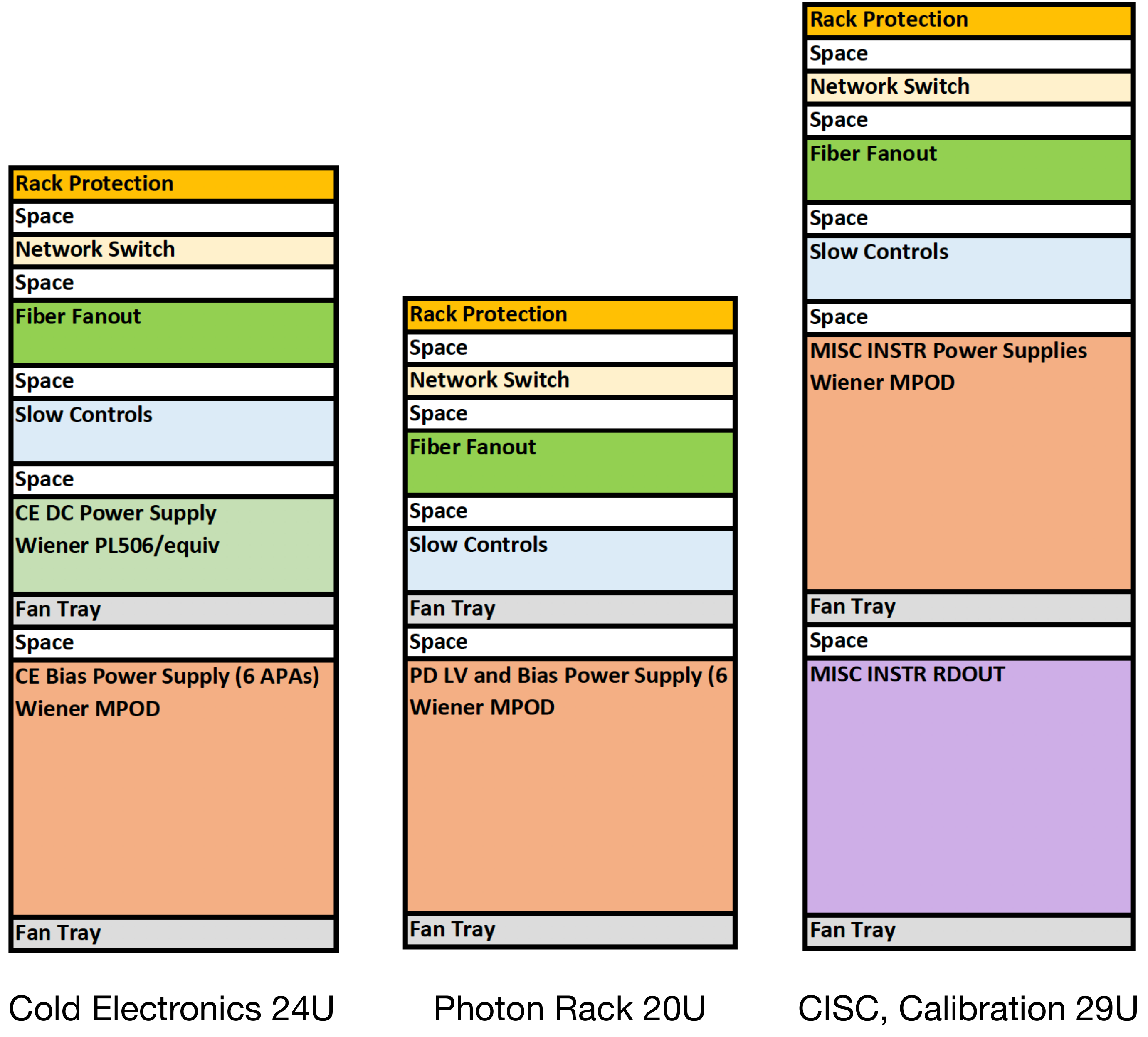} 
\end{dunefigure}

The top image in Figure~\ref{fig:mezzanine} shows the \dword{dune} electronics mezzanine with the 42U tall racks placed on top. 
During the initial design steps, it became clear that the constraints placed on the rack location by the many \dword{dss} support \fdth{}s, the electronics \fdth, and the I-beams themselves make distributing the racks on the roof very challenging. 
By constructing a fixed mezzanine for the electronics above the cryostat at the same height as the cryogenic mezzanine, the electronics \fdth{}s are kept clear. 
This configuration also makes working on the electronics much easier because there are no local obstacles and all the racks are in one place.

Since the electronics modules in the  racks are connected to the detector readout electronics, they are by definition at detector ground. The mezzanine must therefore also be connected to detector ground, which is accomplished by bolting the mezzanine to the cryostat I-beams.

Figure \ref{fig:mezzanine} (top) shows 16 groups of five racks each
on the mezzanine for a total of 80 racks. 
The electronics inside the detector racks will be air-cooled and the heat exhausted into the cavern air. The HVAC system for the detector cavern has a \SI{400}{kW} capacity, which is sufficient for the first \dword{detmodule}.  Note that as the majority of the heat is generated by the \dword{ce} and \dword{pd} electronics located near the cryostat \fdth{}s distributed across the cryostat roof, a water-cooling scheme would be difficult to engineer.
 \dword{cf} will provide sufficient chilled water capacity at the entrance to the north cavern to accommodate the maximum heat load for two \dwords{detmodule}. When detector \#3 is selected and the heat loads are known, the added cooling for this module will be designed.

Of the 80 racks, \dword{ce} \dword{lv} power requires \num{25}, and another \num{25} will be made available collectively for  \dword{apa} wire bias voltage, \dword{pd} power, and miscellaneous additional \dword{ce}, \dword{pds}, and  \dword{apa} electronics modules. 
The remaining 30 will be available for slow control, calibration, and other electrical equipment. 
Small 12U-high mini-racks will  be placed near the electronics \fdth{}s for the \dword{pd} readout electronics and optical patch panels. If this is not enough, additional racks can be placed on the cryostat roof. The present rack configuration for this layout is shown in Figure~\ref{fig:rack-build1}. 
The electronics modules inside the racks are distributed to keep the AC power requirement for each rack below \SI{3.5}{kW}. 
The racks are 42U high, which provides significant extra rack space~\cite{bib:docdb4499}.

The 12U-high mini-racks near the \fdth flanges will be relatively empty because the \dword{pd} readout should need only approximately 2U in height while the \dword{ce} patch panel needs less than 1U. The mini-racks are shown in the lower panel of Figure~\ref{fig:mezzanine}; 
they are the gray rectangles near the electronics crosses.

The north-south cable trays (transverse to the beam) that run from the electronics mezzanine to the electronics \fdth are routed under the floor of the cryostat roof (shown in gray in Figure~\ref{fig:mezzanine}) next to the 
I-beams. 
This keeps the roof reasonably clear, allowing equipment to be transported across it. 
The gap between the web of the I-beams is \SI{1.2}{m} so 
a \SIrange{200}{300}{mm} wide cable tray installed along the beams 
leaves enough space for people to work on the electronics crates while standing directly on the cryostat's outer steel skin (Figure~\ref{fig:install-elect-cross}). 
The cable trays between the \dword{cuc} and the electronics mezzanine will run along the west end of the cryostat under the floor of the cryostat roof. 
We estimate that only half of the \SI{1.6}{m} space is needed, so the cable tray quantity could in principal be doubled, if necessary. 

The flooring material for the top of the \dword{spmod}
will be similar to the \SI{25}{mm} thick plywood used at \dword{pdsp}. 
It is important that it be easy to cut so that it can be fit around many obstacles and pipes on the roof. 
It must be light enough to lift up to allow access under the floor, and it must support the load of a person and a small cart. 
We will investigate fire-retardant options available in the USA and other possible materials, with input from the \dword{fnal} fire life-safety group. 

Air filters for the cleanroom and inside the cryostat will also be placed on the cryostat roof. The present plan is to place fan filter units near the 
access holes on the east end of the cryostat. Initial calculations indicate sufficient airflow is possible to support one air exchange per hour inside the cryostat. The air handling system has yet to be designed in detail.

\begin{dunefigure}[Cryostat crossing tube design]{fig:crossingtube}
  {Draft drawing of the cryostat crossing tubes. The hatched region is the cryostat insulation. Units are mm. Points labeled ``u'' and ``v'' are welds. }
\includegraphics[width=.85\textwidth]{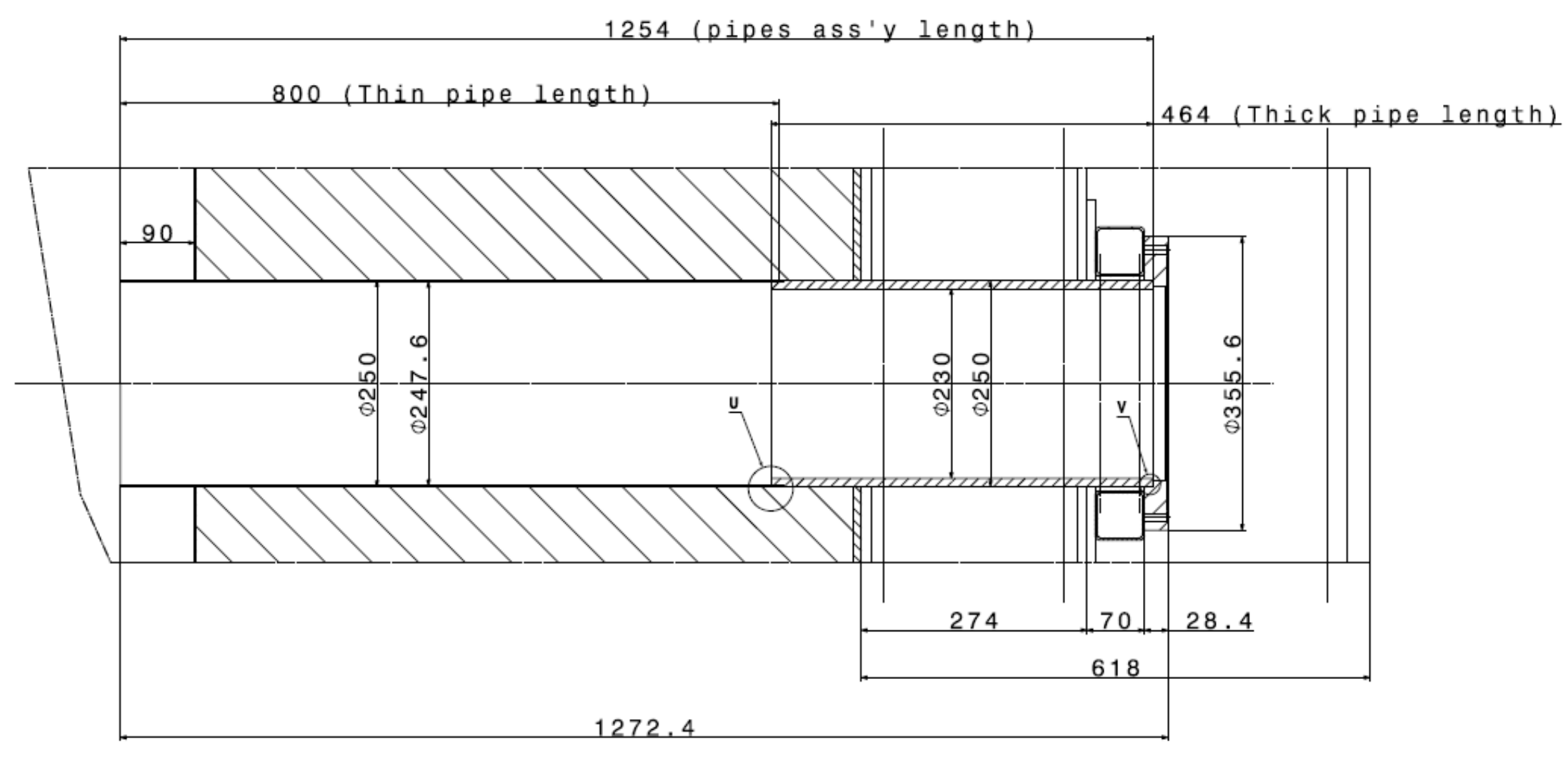}
\end{dunefigure}
 
The cryostat crossing tubes are among the most critical components of the roof infrastructure since they penetrate the cryostat roof and connect to the cold cryostat membrane. The top flange of the crossing tube supports either the electronics \fdth or the detector support \fdth and must be directly tied to the cryostat's steel I-beams for support. 
Accurate placement and true vertical installation of the crossing tubes is important to ensure proper interfacing to the cryostat membrane. 
A draft assembly drawing of the crossing tube is shown in Figure~\ref{fig:crossingtube}. 
The crossing tube consists of a \SI{464}{mm} long stainless steel pipe with a \SI{1}{cm} thick wall. 
One end of the thick-walled section is welded to a \SI{800}{mm} long \SI{250}{mm} diameter thin-walled tube which is also welded to the cryostat membrane.
A custom Conflat flange at the top end of the crossing tube connects to the \fdth. 
The thick tube section is also welded to the steel roof plates (the thin cross-hatched segments in Figure~\ref{fig:crossingtube}).

Each of the 250 crossing tubes has a small side port that connects to the cryogenic gas-handling system through a network of pipes on the cryostat roof. During the initial purge \dword{gar} is withdrawn from each port and analyzed to assess progress and determine when the system is ready to be cooled down. Five \dword{gar} streams, each collecting gas from 50 crossing tubes, are connected independently to the gas analyzers. This provides some redundancy and position-dependant information on the contamination level of the gas at the top of the cryostat during the purge.
 
During filling and normal operation the collection and analysis of the gas from the crossing tubes will continue in order to monitor impurities (mainly water, oxygen and nitrogen) produced by outgassing from the cables in the \fdth{}s and the warmer metal surfaces in the ullage. These impurities can be removed from the \dword{gar} by the cryogenics system.
If the gas analyzers find no significant nitrogen contamination, the \dword{gar} from all or a subset of ports can be sent to the condenser, re-condensed, and purified along with the rest of the \dword{lar}. Simple O$_2$\ sensors monitor the return gas for traces of oxygen, which would indicate development of a leak in the room-temperature \fdth{}s.

A \SI{500}{kVA} transformer provides power to each \dword{detmodule} and the total power budget available for use by detector electronics is derated to \SI{400}{kW} at the power distribution panels.  
The \dword{ce} is the \dword{spmod}'s largest power consumer,  dissipating \SI{306}{W} per \dword{apa}.  
The \dword{lv} power supplies' controller needs about \SI{35}{W} per \dword{apa} and has an efficiency of approximately \SI{85}{\%}. 
This leads to a load of  approximately  \SI{400}{W}  per \dword{apa}, or a total load of  \SI{60}{kW} per \dword{detmodule}.  
The \dword{apa} wire-bias power supplies have a maximum load of  \SI{465}{W} per set of six \dword{apa}s, for a total budget of about   \SI{12}{kW}.   
Cooling fans and heaters near the \fdth{}s will use a nominal amount of power, so the overall power budget for the \dword{ce} and  \dword{apa}s is expected to be less than \SI{75}{kW}.

The \dword{pds} electronics is based on the \dword{mu2e} cosmic ray veto electronics, which reports a power load of approximately  \SI{6}{kW}.  \dword{dune} plans a power budget of  \SI{8}{kW} because of cable drops and  power supply inefficiencies.  

Each of the approximately 80 detector racks will have fan units, Ethernet switches, rack protection, and slow controls modules, adding a load of about \SI{500}{W} per rack, for a total of \SI{40}{kW}.

Twenty-five racks are reserved for cryogenics instrumentation with a per-rack load conservatively estimated at \SI{2}{kW}, for a total of \SI{50}{kW}. 

The \dword{detmodule} will thus use  an estimated \SI{173}{kW} of power.   These numbers provide a safety factor of about two on our power estimates relative to available power.

\subsection{Cryostat Internal Infrastructure}
\label{sec:fdsp-tc-infr-cryo-int}

\label{sec:fdsp-tc-internal-cryo}

The internal cryogenics comprises three sets of pipe distribution networks and two sets of sprayers. All pipes enter the cryostat from the top; some go all the way down to the floor, and others remain in the ceiling. On the floor are:
\begin{itemize}
\item \textbf{Argon gas distribution}: a set of pipes 
for \dword{gar}. These pipes are used only prior to filling to remove air 
in the cryostat. 
They will all have either a longitudinal slit or calibrated holes to distribute \dword{gar} uniformly along the length of the cryostat. 
We have run \dword{cfd} simulations showing that air will be removed from the system as long as \dword{gar} is flowing in at the right speed, calculated and experimentally verified as \SI{1.2}{m/hr} (vertical meters in the cryostat).

\item \textbf{\dword{lar} distribution}: two sets of pipes are required for flowing 
\dword{lar} over a broad  range of flow rates. These pipes are used to fill the cryostat and, during steady state operations, to return the \dword{lar} from the purification system. The pipes have calibrated holes to return the \dword{lar} uniformly throughout the length of the cryostat which is very important to maintain uniform purity. 
Four pumps circulate the \dword{lar} inside the cryostat all of which operate initially to achieve purity. 
Once the target purity is achieved only one or two pumps remain in service. Individual pumps can be isolated for routine maintenance.
\end{itemize}

On the ceiling are:

\begin{itemize}
\item \textbf{Cool down sprayers}: Two sets of cool down sprayers are distributed along the long sides of each cryostat. One set distributes \dword{lar} using liquid sprayers that generate a conical profile of small droplets of liquid. The other set of sprayers distributes \dword{gar} to move the \dword{lar} droplets within the interior and thus cool down the detector and cryostat uniformly. These sprayers are being tested in \dword{pddp}. They are a variation of those implemented in \dword{pdsp}.
\end{itemize}

Figure~\ref{fig:internal-cryo-3D} shows the current layout of the internal cryogenics. 
The \dword{gar} pipes are in red, the \dword{lar} pipes in blue.

\begin{dunefigure}[Layout of the internal cryogenics piping]{fig:internal-cryo-3D}
  {Endview of the inside of the cryostat after the cryogenic piping has been installed. The \dword{gar} pipes used during the piston purge are in red and the pipes which return the purified \dword{lar} to the cryostat are in blue.}
\includegraphics[width=.98\textwidth]{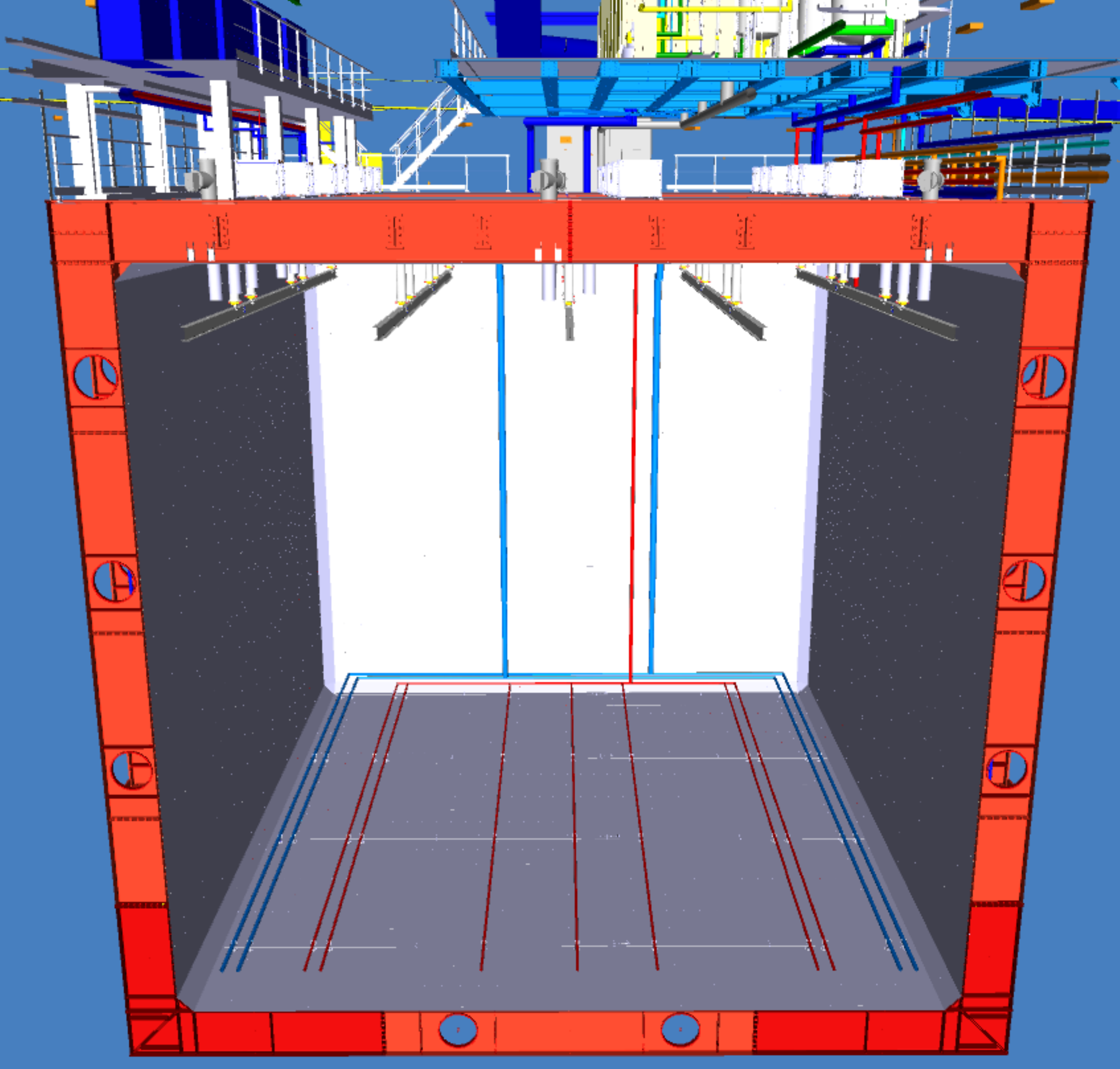}
\end{dunefigure}

Infrastructure inside the cryostat includes the cryostat false floor, the UV-filtered lighting, and the battery-operated scissor lifts. 
The floor must support the load of the scissor lift used to work on the electronic cabling on the inside of the cryostat near the ceiling and allow the scissor lift to get close enough to the \dword{apa}s to work comfortably at the top. 
The floor  must be laid out so that the panels can be removed in sections as equipment is installed. 
This is especially important for the \dword{apa}s since inadequate room exists between the bottom of the \dword{apa}s and the floor to allow removal of panels after installation. 

The cryostat lighting, using UV-filtered\footnote{Light is filtered according to the requirement SP-INST-6} \dword{led} lamps, is expected to be fairly simple. Options for the lighting will be developed during tests at the \dword{ashriver} facility.
Floor-mounted lights with task lighting will be investigated. If needed, lighting can also be mounted to the \dword{dss} and removed as the detector is installed.

We plan to use a commercially available battery-operated scissor lift with a \SI{12}{m} reach. Tests at \dword{ashriver} will verify the stability of the lift at height. If the lift is determined to be suitable, then the remaining issue to resolve is how to install and remove it from the cryostat. 
Commercially available scissor lifts are too wide to fit easily through the \dword{tco} opening where one of the large cryostat support I-beams protrudes above the \dword{tco} floor level,  so 
custom lifting equipment will be needed to insert the lifts into the cryostat from above. 
At the end of the installation process, the last lift may require dismantling before it can be removed from the cryostat.


\subsection{Cleanroom and Cleanroom Infrastructure}
\label{sec:fdsp-tc-infr-comm}

\begin{dunefigure}[Installation Cleanroom layout]{fig:install-cleanroom}
  {Two views of the installation cleanroom.  The top view shows the cleanroom in position in the north cavern. The location of the material airlock and the changing room are indicated. The lower image is a closer view showing the equipment in the cleanroom. The \cryostatht tall cryostat is shown in red.
  } 
\includegraphics[width=1.0\textwidth]{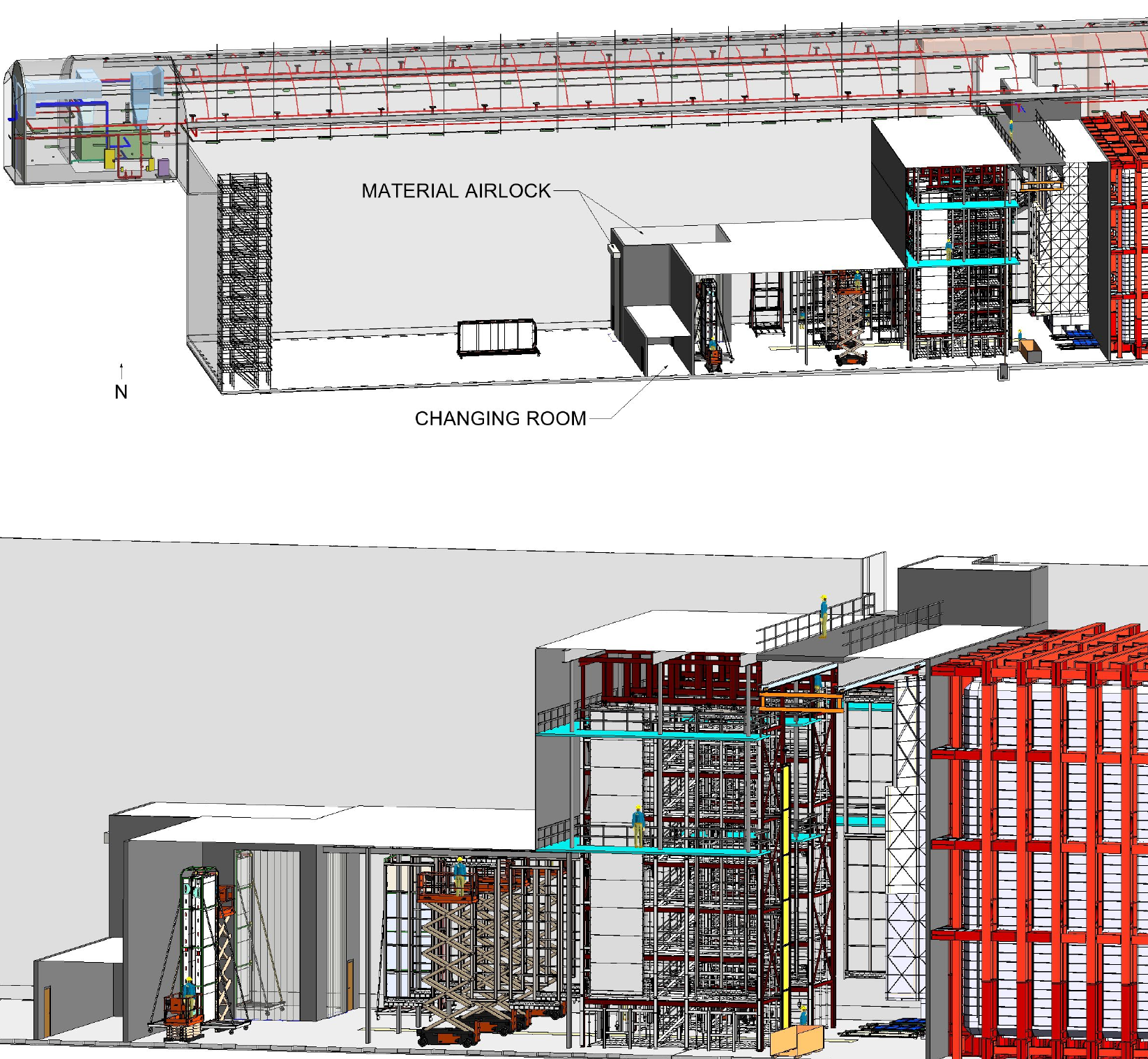}
\end{dunefigure}

Since the \SI{12}{m} tall full \dwords{apa} would be too large and fragile to be brought down the Ross Shaft, they will be assembled in front of the cryostat itself, where there is ample vertical space. A cleanroom meeting the ISO-8 cleanroom standard (3.5$\,$M particles per m$^3$ or 0.1$\,$M per ft$^3$) is required for any work on the detector components in order that the accumulated U/Th contamination due to dust in the detector  produce a background rate lower than the (unavoidable) Ar$^{39}$ decay background. 
We therefore plan to construct a large ISO-8 cleanroom in the assembly area (Figure \ref{fig:install-cleanroom}) to meet the \dword{dune} cleanliness requirements.

Upon arrival in the cavern, the detector elements first pass through a materials airlock (Figure~\ref{fig:install-cleanroom}) before entering the cleanroom. This airlock has large entry doors of dimensions \SI{3}{m} wide and \SI{8}{m} tall, large enough to allow the tallest item -- an \dword{apa} transport box (\SI{2.6}{m} by \SI{6.6}{m}) -- to enter in a vertical orientation while bolted to a custom pallet and be moved with electric pallet jacks. 
The airlock itself is \SI{7}{m} wide, \SI{9}{m} deep, and \SI{9}{m} tall. All materials must be brought through the material airlock and cleaned prior to entering the cleanroom proper. 
The materials airlock, cleanroom, and inside the cryostat will be outfitted with UV-filtered lights to protect the \dwords{pd}. 
Personnel must enter the cleanroom through a changing room. The changing room on the 4910 level is 13 \si{m} wide and 4 \si{m} deep, 
large enough to allow 50 people to gown up for the cleanroom within a reasonable time (roughly 15 people can change at a time). A smaller changing room on the \dword{4850l} allows easier access to the elevated work platform.

The cleanroom infrastructure consists of the cleanroom itself, the \coldbox{}es and cryogenics plant for testing the assembled \dword{apa}s, the assembly towers, rails and a switchyard to allow the \dword{apa}s to move inside the cleanroom, and the \dword{pd} integration infrastructure. 

A combination of contractors, the lead worker, and rigger teams will set up the infrastructure;  they will also assist in detector assembly. 
The attire requirements for work in an ISO-8 cleanroom are a cleanroom lab coat, clean shoes, and nets for hair and beards.  This basic cleanroom attire will be augmented with a clean hard hat and gloves, for safety reasons. 

To keep the cleanroom at least at ISO-8, air is first filtered then forced into the cryostat's east end. 
From there it flows through the cryostat and into and through the cleanroom and the airlocks.

The size of the installation cleanroom depends on the work performed inside it and the required equipment. The dimensions have been defined and are described below, but optimization will continue through fall of 2019.
After the tests at \dword{ashriver} modifications may be necessary. 
Figure \ref{fig:install-cleanroom} illustrates  the conceptual design. The top figure shows the cleanroom situated in the cavern next to the cryostat; the materials airlock and the changing room are on the west end. The bottom image is a closer view showing some of the equipment in the cleanroom. 

The cleanroom proper can be divided into several work areas as follows:
\begin{itemize}
    \item materials and personnel airlocks,
    \item \dword{pd} integration area,
    \item four \dword{apa} assembly lines, where the lower rails are for wire tension measurements and the upper rails are for \dword{apa} assembly and cabling,
    \item the switchyard area used to move the assembled \dword{apa}s around the cleanroom and into the cryostat,
    \item the \coldbox area where the \dword{apa}s are cold tested, and
    \item the \dword{hv} assembly area.
\end{itemize}

The \dwords{pd} are integrated into the \dword{apa}s and the initial \dword{qa} tests upon receipt are performed in the \SI{10}{m} high \dword{pd} integration area at the west end of the cleanroom.

Because \dword{apa} preparation
is time-consuming, four assembly lines will operate in the cleanroom to keep up with the cold tests and installation. 
Three lines will be in continuous usage and the fourth will remain available for repairs or contingency. 
Each assembly line has a lower and upper set of rails for moving the \dword{apa}s. 
The wire tension is measured and the lower \dword{ce} \dwords{femb} are installed at the lower rail section.  The \dword{pd} integration area and the lower rail section of the assembly lines measure \SI{19.5}{m} wide by \SI{18}{m} deep, with a \SI{9}{m} ceiling. A design for this area similar to that used for the \dword{pdsp} cleanroom is under consideration, as the height is similar.  If needed, the rail system inside the integration work area can be used to support the roof.
 
The \dword{apa} assembly and cabling area in the cleanroom is where the top and bottom \dword{apa}s are connected together to form the \SI{12}{m}  doublets, and the \dword{ce} cables are inserted and connected to the \dwords{femb}.

The ceiling in this area is \SI{17.8}{m}, placing the ceiling at the same level as the bridge and the roof of the cryostat. 
A \SI{9.5}{m} by \SI{19.5}{m} area next to the bridge is sufficient to house the two large assembly towers needed to support the assembly lines. 
Above the towers, I-beams running north-south, i.e., transverse to the neutrino beam, are needed to support work platforms that allow access to both faces of the \dword{apa}s. 
These beams can be used to support the cleanroom roof, which can be a light-weight frame with a fire-retardant fabric attached. 
The outer towers' steel structure provides a strong surface to which to attach a polymer sheet intended to serve as the vertical wall connecting the \SI{17.8}{m} area to the \SI{10}{m} tall area.

The switchyard area is the region under the north-south bridge where a bridge crane is mounted. It is used to move the \dword{apa} from the assembly lines to the \coldbox{}es and into the cryostat, and is similar to the shuttle beam system in the \dword{dss} shown in Figure~\ref{fig:shuttle}. The \coldbox{}es and the \dword{hv} assembly area are between the bridge and the cryostat.

The cleanroom spans the width of the cavern excavation. The side walls of the cleanroom will be constructed by hanging reinforced fire-retardant plastic sheets against the walls, providing a low-cost, easy-to-install solution.

\begin{dunefigure}[Installation Integration Lower Rail System]{fig:install-integrate-rail}
  {Two pairs of rails are used to prepare the \dword{apa} for assembly. Each rail holds three \dword{apa}s. This is where the wire tension measurements are performed and the \dword{ce} \dwords{femb} are installed on the lower \dword{apa}s. The lower \dword{ce} is easily reachable from the floor in this arrangement. This view is from the assembly towers looking west along the assembly lines.}
\includegraphics[width=.8\textwidth]{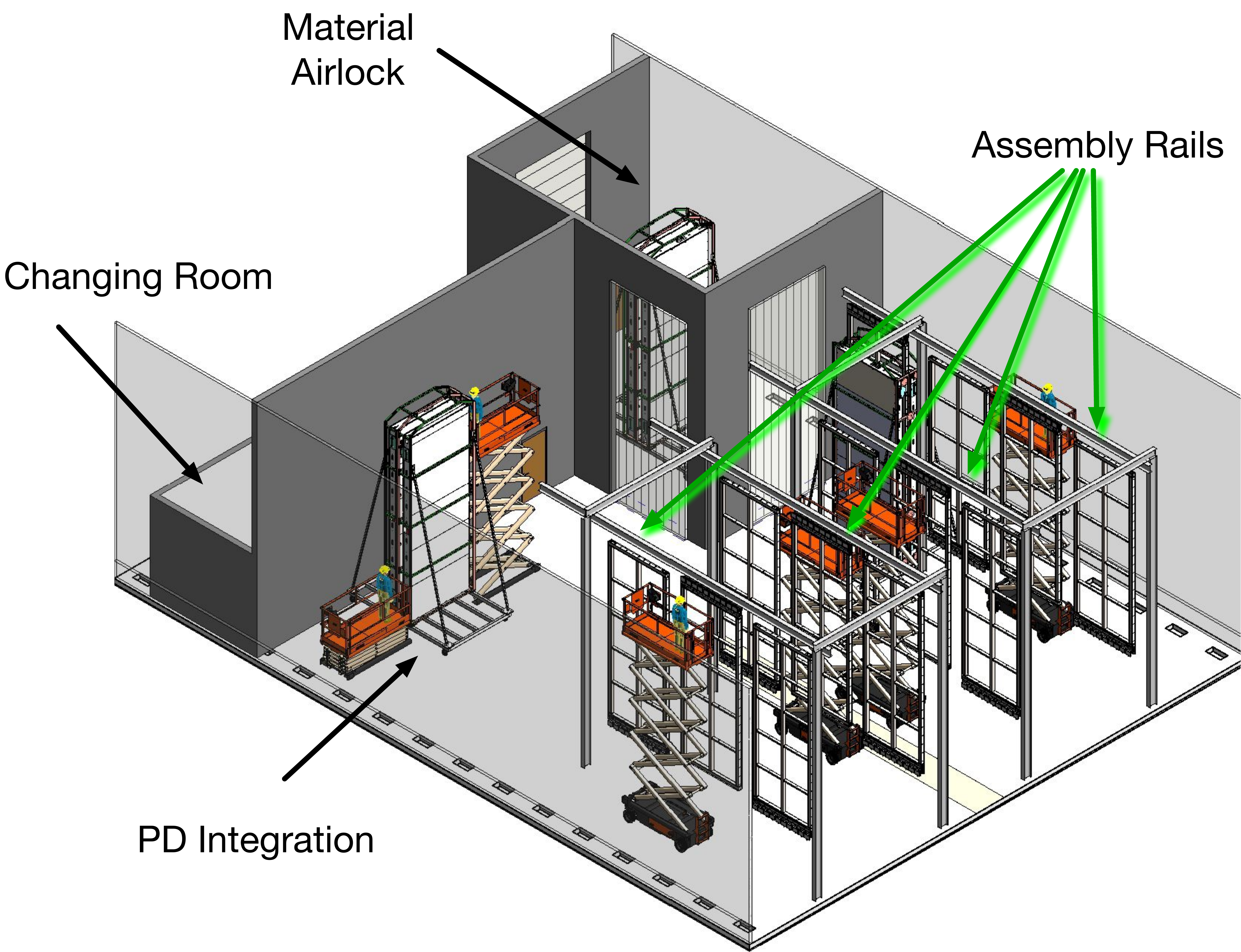}
\end{dunefigure}

Given the substantial size and the significant occupancy of the cleanroom, it will require electrical outlets, Wi-Fi, and fire protection. Monitoring for \dword{odh} will be installed as required by the safety analysis of the \coldbox cryogenics system.

Equipment in the integration work area adjacent to the materials airlock will consist of a station for integrating the \dwords{pd} into the \dword{apa}, two pairs of rails for preparing the \dword{apa} for assembly, and several scissor lifts for working around the \dword{apa}s. 
In the \dword{pd} integration area an \dword{apa} transport box will be positioned between two fixed lifts that will be raised until the \dword{pd} paddles can easily be inserted into the side of the \dword{apa}. 
Figure \ref{fig:install-integrate-rail} illustrates the rail setup in the integration work area. 
The \dword{apa}s are removed from their transport box and mounted to the rails at the far end of the assembly rails near the \dword{pd} integration area and material airlock. 
They then move along the rails using simple trolleys running on the I-beams. 
The rails are long enough to hold three \dword{apa}s at a time. 
This setup is conceptual and the engineering design of the rail supports has not yet started. 
Cross-bracing of the vertical posts will be added during the design stage. 

\begin{dunefigure}[APA cabling tower]{fig:install-assembly-tower}
  {
  Isomtetric view of the \dword{apa} assembly and cabling tower. The steel outer structure is shown in red. The inner scaffolding in gray permits work at different heights.
  The tower is designed to be two \dword{apa}s wide to allow work on two of them side-by-side simultaneously. 
  Both the north and south faces are equipped with assembly rails so that 
   a single tower can support two assembly lines.
  }
\includegraphics[width=.5\textwidth]{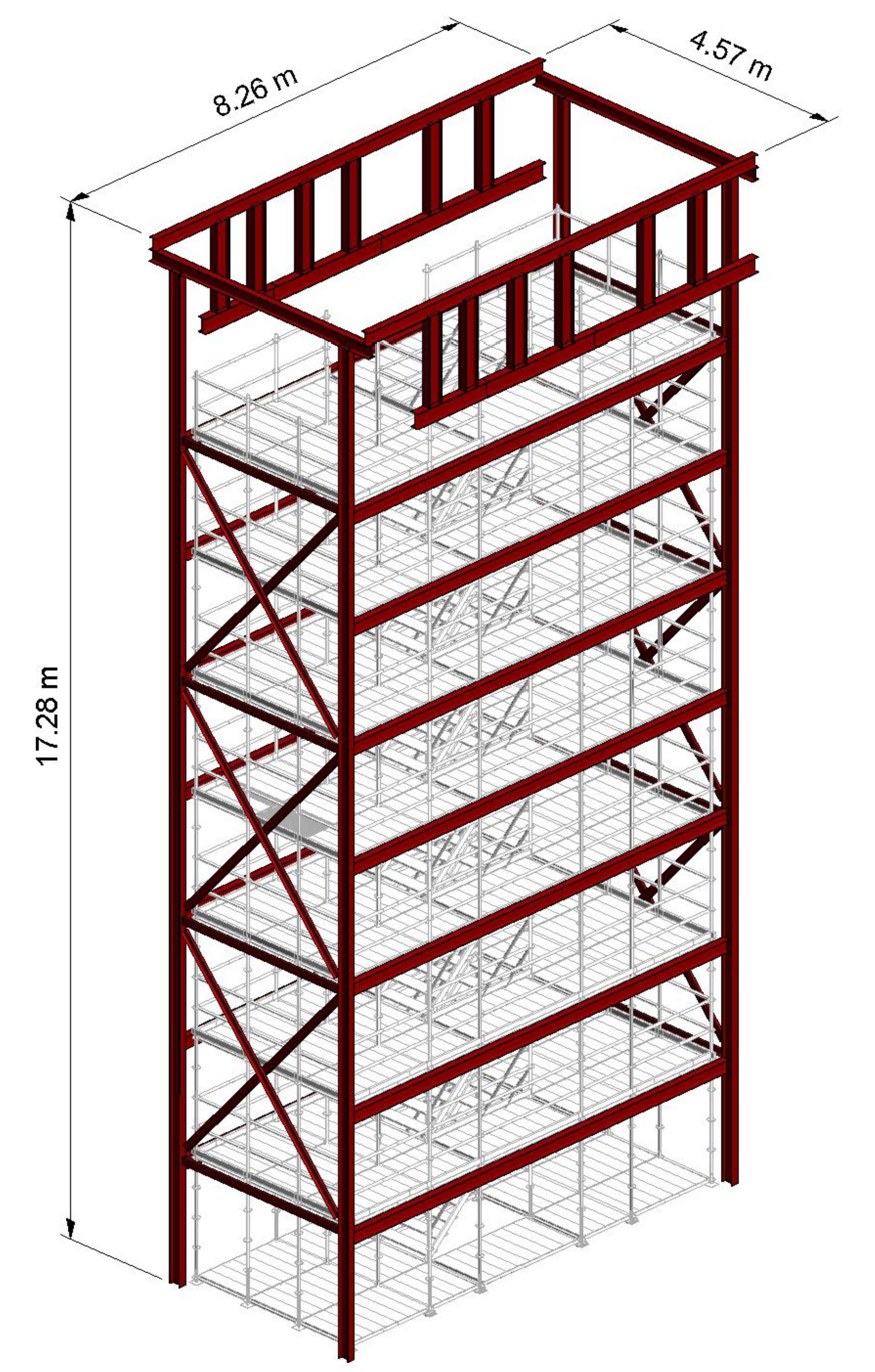}
\end{dunefigure}

In the \SI{17}{m} tall  \dword{apa} assembly and cabling area of the cleanroom two large work towers (shown in Figure~\ref{fig:install-assembly-tower}) support the four assembly lines. 
These towers are designed to be wide enough to hold two \dword{apa}s  side-by-side with enough space between them to walk through or work. 
The tower is seven stories tall with work areas at each landing.
Rails at mid-height and at the top of the towers are used to move the \dword{apa}s to the different locations along the tower. 
The towers also provides support for the tooling needed to hold the upper and lower \dword{apa} during assembly and to bring the two modules together so they can be connected. This tooling is called the \dword{apa} assembly fixture and is provided by the \dword{apa} consortium. 

The  tower is conceived as a steel outer frame that supports the \dword{apa}s and the rails. Inside the steel frame is standard scaffolding that allows workers to access the \dword{apa}s at different heights. 
The scaffolding is wide enough for people to work simultaneously on both sides and it accommodates a stairway in the middle that meets \dword{osha} standards. 
North-south beams spanning the width of the cavern will be placed on top of the towers to support the cleanroom roof and the work platforms shown in Figure~\ref{fig:install-workdeck}.
The image shown in Figure~\ref{fig:install-assembly-tower} is a modified model based on a single-wide \dword{apa} tower that has passed all safety reviews and has already been constructed at \dword{ashriver}. 
The double-wide tower will need to be re-engineered to ensure that all the beam dimensions and bracing are appropriate for the larger spans and loads. 
It will then go through the full safety review and an initial prototype 
will be fabricated for use at  \dword{ashriver}. 
The size and the layout of the top level of the tower will be optimized based on input from the  \dword{ashriver} tests.

\begin{dunefigure}[Cleanroom platforms and material transport system]{fig:install-workdeck}
  {Installation workdeck, assembly towers and  rails, switchyard, \coldbox{}s and HV assembly area in the installation cleanroom. Plan view. 
  }
\includegraphics[width=.75\textwidth]{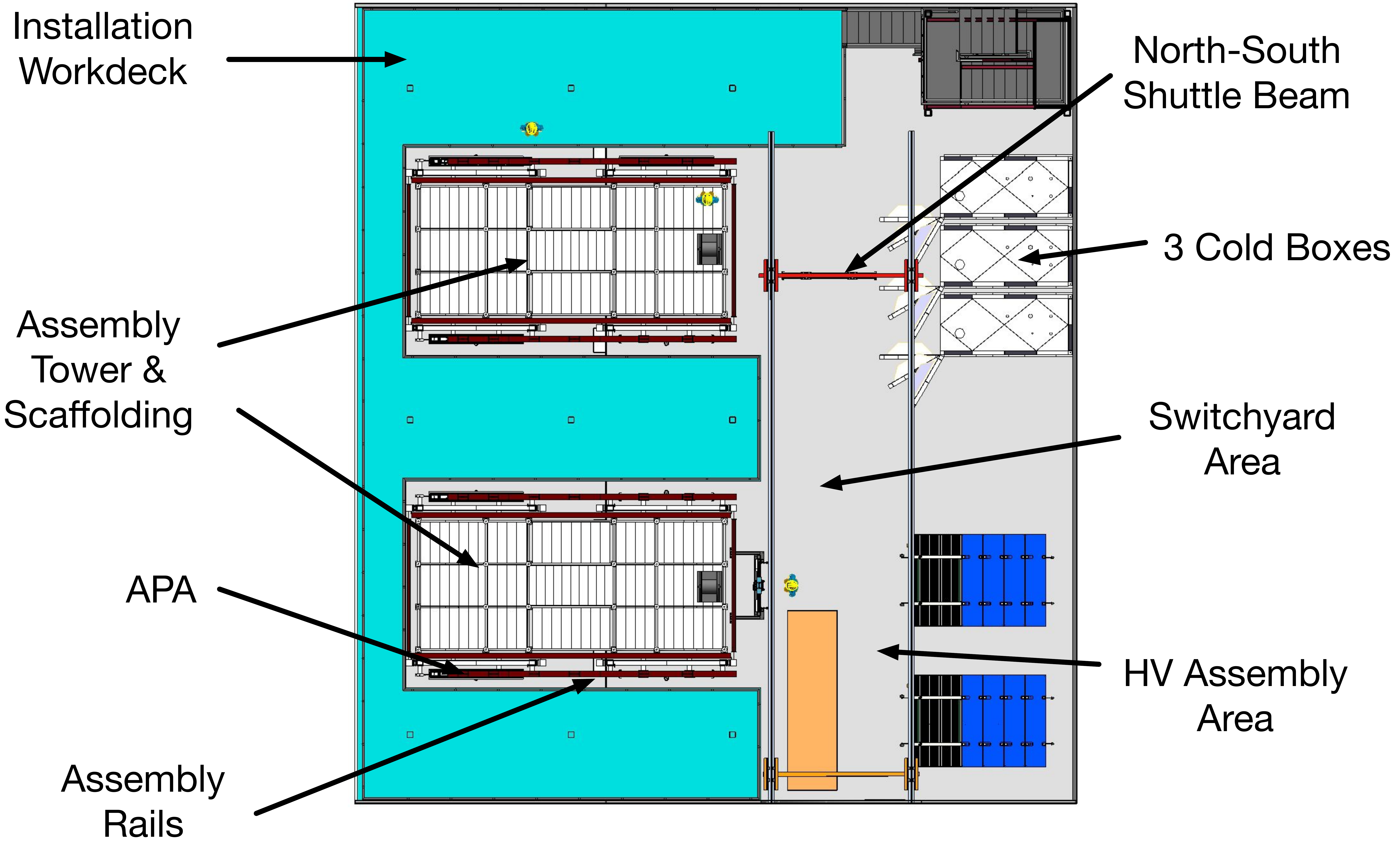}
\end{dunefigure}

Because the cables can only be inserted through the \dword{apa} frames after the top and bottom \dword{apa}s have been assembled, quite a bit 
of work must be performed at the \SI{14.8}{m} height of the \dword{tco} beam. 

The commercial scaffolding inside the \dword{apa} assembly towers 
provide a solid, safe work platform for working on the side of the \dword{apa} facing the tower. However, access to both faces of the \dword{apa} is required to connect the cables to the \dword{femb} and to properly bundle the cable into the cable trays in a way that allows the cables to be installed easily once the assembly is in the cryostat. 
Given the large number of person-hours needed for work at height, all measures will be taken to ensure that this work is safe. 
For this reason a large stable  workdeck will be constructed as shown in Figure~\ref{fig:install-workdeck}.
By running north-south I-beams from the cavern walls across the assembly towers, strong rigid support points are provided. 
Vertical posts down from these beams can then support the fixed workdeck as needed. Access to the workdeck is provided by a walkway along the west end of the platform that connects to both the assembly towers. A second means of egress is provided by a connection to the permanent stairs.

Once a top-bottom \dword{apa} pair is assembled it can be moved onto the switchyard under the bridge. 
This switchyard, illustrated schematically in Figure~\ref{fig:install-workdeck}, is essentially a bridge crane running under the north-south bridge in the cavern.  
Several bridge beams (called shuttle beams) driven by electric trolleys  move on the runway beams of the crane.  
A rail at the bottom of the crane mates with the fixed beams  from the assembly lines, the \coldbox{}es, the \dword{tco} beams and the \dword{hv} assembly area.
By aligning the bridge beams with a set of fixed beams supported from the cleanroom roof, the \dword{apa}s and \dword{cpa}s can be transferred from the fixed beams to the bridge crane and moved to different locations in the cleanroom. 
The \SI{12}{m} tall \dword{cpa} panels will be assembled directly under the bridge crane and transferred directly from the assembly fixture to the switchyard crane.

The division of responsibilities between the installation and the consortia deliverables are defined in interface documents, but they are governed by a simple concept. Any part which bolts, pins or connects to a consortia deliverable is the responsibility of the consortia. General infrastructure, hoists, and cranes are the responsibility of the installation team. Thus, for example, the installation towers are installation's responsibility, while the fixtures that bolt to the towers and to the \dword{apa}s are the responsibility of the \dword{apa} consortium.

\subsection{Cryogenics and \Coldbox{}es}
\label{sec:fdsp-tc-infr-cryo}

After an \dword{apa} pair is fully assembled and cabled but before installation in the detector cryostat, it is thermally cycled in a tall narrow test cryostat, called a \coldbox{}, shown in Figure~\ref{fig:install-coldbox}). 
To test \dword{apa}s at a rate necessary to keep up with the installation plan, we will use three identical \coldbox{}es in the cleanroom. 
The \coldbox{}es require a dedicated cryogenics system that uses a fine mist of cold nitrogen to cool down close to \dword{lar} temperature. This system is designed so that no liquid nitrogen will accumulate.

A \coldbox has external dimensions of 14.0 \si{m} by 3.2 \si{m} by 1.3 \si{m} (H$\times$L$\times$W). With three layers of \SI{100}{mm} thick foam insulation,  
the internal dimensions are 13.4 \si{m} by 2.6 \si{m} by 0.7 \si{m}. A rail section similar to those used elsewhere in the cleanroom will be mounted inside each \coldbox to allow the cleanroom switchyard and trolleys to push an  \dword{apa}  into a \coldbox. The \coldbox{}es will be light-tight when closed to support \dword{pd} testing. A support base under the \coldbox{}es will adjust the height to mate with the cleanroom switchyard.

The \coldbox electronics \fdth{}s  will be  similar to what is used on the top of the \dword{dune} cryostat, except that short cables will be run from the \dword{wiec}  to a patch panel inside the \coldbox. This will allow the cable on the \dword{apa} to connect directly to the test readout without having to remove any cabling. The \coldbox  design is nearly the same as the successful \dword{pdsp} \coldbox. The outer shell is similarly constructed of a stainless steel plate with reinforcing ribs welded on. The height is of course doubled, and a hinged door is planned. Unbolting the door and lifting it off the \dword{pdsp} \coldbox required significant effort, and lacking full crane coverage in this case, doors that can be opened and closed using a scissor lift are necessary. The \dword{dune} \coldbox{}es will collectively need about \SI{11}{t} of stainless steel, according to initial estimates. The finished boxes are too big to fit down the Ross Shaft. As the design continues, 
we will investigate whether the boxes can be brought underground partially assembled, or if they must be fully assembled in place.  

The \coldbox{}es and associated cryogenics system are the responsibility of the \dword{jpo} and the design of the \coldbox{}es will be provided by the \dword{cern} team, which designed the \dword{protodune} test cryostats. The cryogenics system will be designed by the \dword{lbnf} cryogenics team. 

\begin{dunefigure}[Installation \coldbox]{fig:install-coldbox}
  {\Coldbox{}es used to thermally cycle the fully assembled APA pairs. }
\includegraphics[width=.5\textwidth]{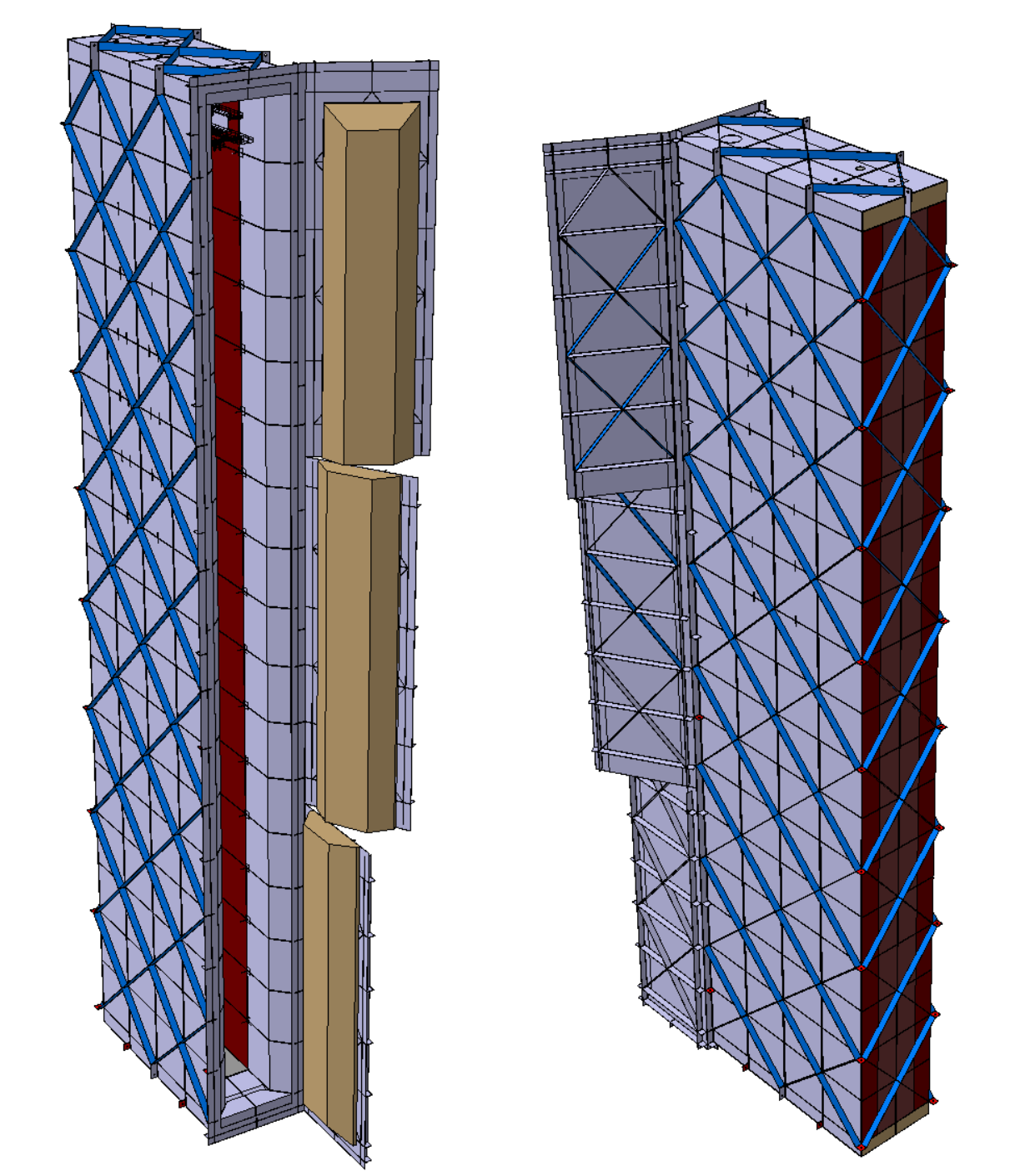}
\end{dunefigure}

\label{sec:fdsp-tc-cryocoldbox}

The  \coldbox{}es will be used to test the \dword{apa}s underground prior to installation.  
The cryogenics supporting the \coldbox{}es  must ensure their reliable and safe operation; to that end, the system must
\begin{itemize}
\item support three \coldbox{}es operating in parallel: 
one in \cooldown mode, two either in steady-state or warm-up modes;
\item allow personnel in the cleanroom during all phases of the purge, \cooldown{}, operation, and warm-up modes; 
\item test the detector modules at near \dword{lar} temperature;
\item operate 24 hours a day;
\item allow remote operations; and 
\item be located in the vicinity of the \dword{tco}, as space is available on top of the cryogenics mezzanine on the roof of the cryostat.
\end{itemize}

It must operate in the following modes: 

\begin{itemize}
\item \textbf{purge}: During this mode, air is removed from the system (\coldbox and cryogenics system) and replaced with dry nitrogen. The concentration of moisture is monitored, and when it no longer decreases, the \cooldown can commence.
\item \textbf{\cooldown}: Cold nitrogen is introduced into the system to cool the inside of the \coldbox and the \dword{apa} inside it. 
This should take 24 hours, during which time the temperature decreases from room temperature to about \SI{90}{K}. 
\item \textbf{steady-state operations}: After reaching approximately \SI{90}{K}, 
the detector is turned on and fully tested. 
This takes about 2 shifts.
\item \textbf{warm-up}: After completing the test, the system is
warmed up to room temperature over a period of 24 hours. 
\end{itemize}

\begin{dunetable}
[\Coldbox  cryogenics system parameters] 
{lc}
{tab:table-cryo-coldboxes}
{Table of parameters for the \coldbox cryogenics system}
Parameter & Value 
\\ \toprowrule
Dual \dword{apa} thermal mass &  1,600 kg\\ \colhline
Temperature uniformity & $+60$ K / $-0$ K \\ \colhline
Electronics load & 300 W \\ \colhline
\Coldbox insulation thickness &  0.3 m \\ \colhline
Target \cooldown temperature &  \SI{90}{K} \\ \colhline
Target \cooldown duration &  24 hr \\ \colhline
Target steady-state duration &  24 hr \\ \colhline
Target warm-up duration &  24 hr \\ \colhline
Maximum cooling power  &  \SI{13}{kW}  \\ \colhline 
Maximum liquid nitrogen consumption  &  \SI{300}{l/hr}  \\ 
\end{dunetable}

The evaporation of liquid nitrogen provides the cooling power for the system. Warm nitrogen and a heater provide the heating power. At peak consumption, the expected maximum heat load is \SI{8.5}{kW}. Assuming a 50\% margin on the refrigeration load, the cryogenics system requires \SI{13}{kW} of net cooling power at peak consumption, which equals about \SI{300}{l/hr} of evaporating liquid nitrogen.

Two layouts are currently under consideration: (1) a closed-loop with mechanical refrigeration, in which liquid nitrogen is generated in situ, circulated, and the spent nitrogen recondensed before being put back into the system; and (2) open-loop, in which liquid nitrogen is transported underground by means of portable dewars, circulated, and the spent nitrogen vented away. For the closed-loop, we would need a mechanical refrigeration capable of supplying \SI{13}{kW} of cooling. For the open-loop, it is possible to use a \SI{2000}{l} dewar, which is commercially available and transportable up and down the Ross Shaft inside the cage. To supply the required amount of nitrogen, four trips per day are needed over the two-year period of detector installation.

The current versions of the closed-loop and open-loop systems are presented in Figures~\ref{fig:mechanical-refrigeration} and~\ref{fig:LN2}, respectively. Both options are viable and a decision will be taken on which to adapt after the analysis is complete.

A full \dword{odh} analysis will be performed once the design has progressed to the point where the process flow and pipe dimensions are fixed (These are a necessary input to the analysis).  
Since no liquid is accumulated and the room volume is large, our initial assessment is that standard \dword{odh} safety measures will be adequate.

\begin{dunefigure}[\Coldbox cryogenics support system based on mechanical refrigeration ]{fig:mechanical-refrigeration}
  {Layout of the cryogenics supporting the \dword{apa} test facility with mechanical refrigeration (closed-loop).}
\includegraphics[width=.98\textwidth]{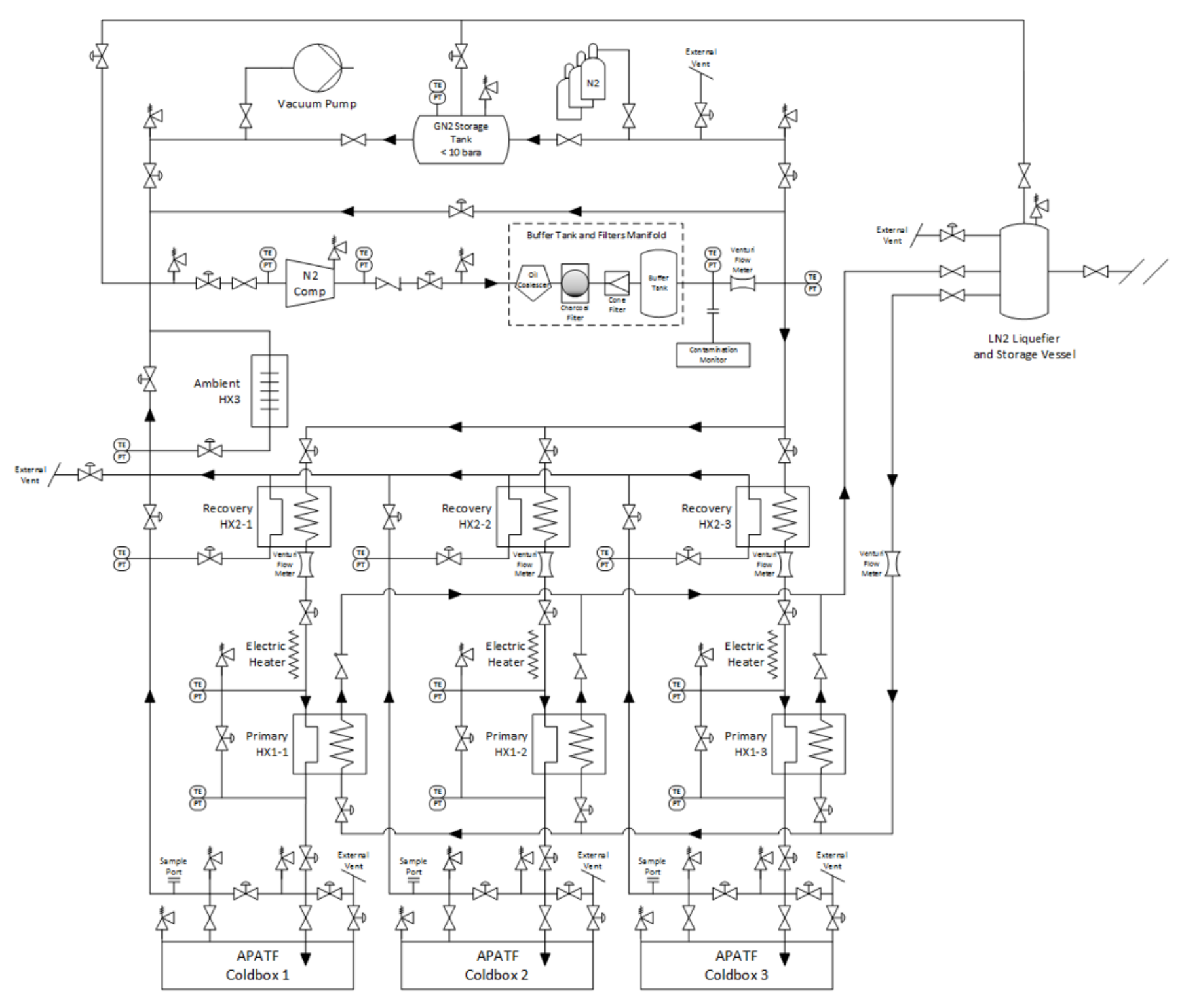}
\end{dunefigure}

\begin{dunefigure}[\Coldbox cryogenics support system based on LN2 ]{fig:LN2}
  {Layout of the cryogenics supporting the \dword{apa} test facility with open-loop refrigeration (open-loop).}
\includegraphics[width=.98\textwidth]{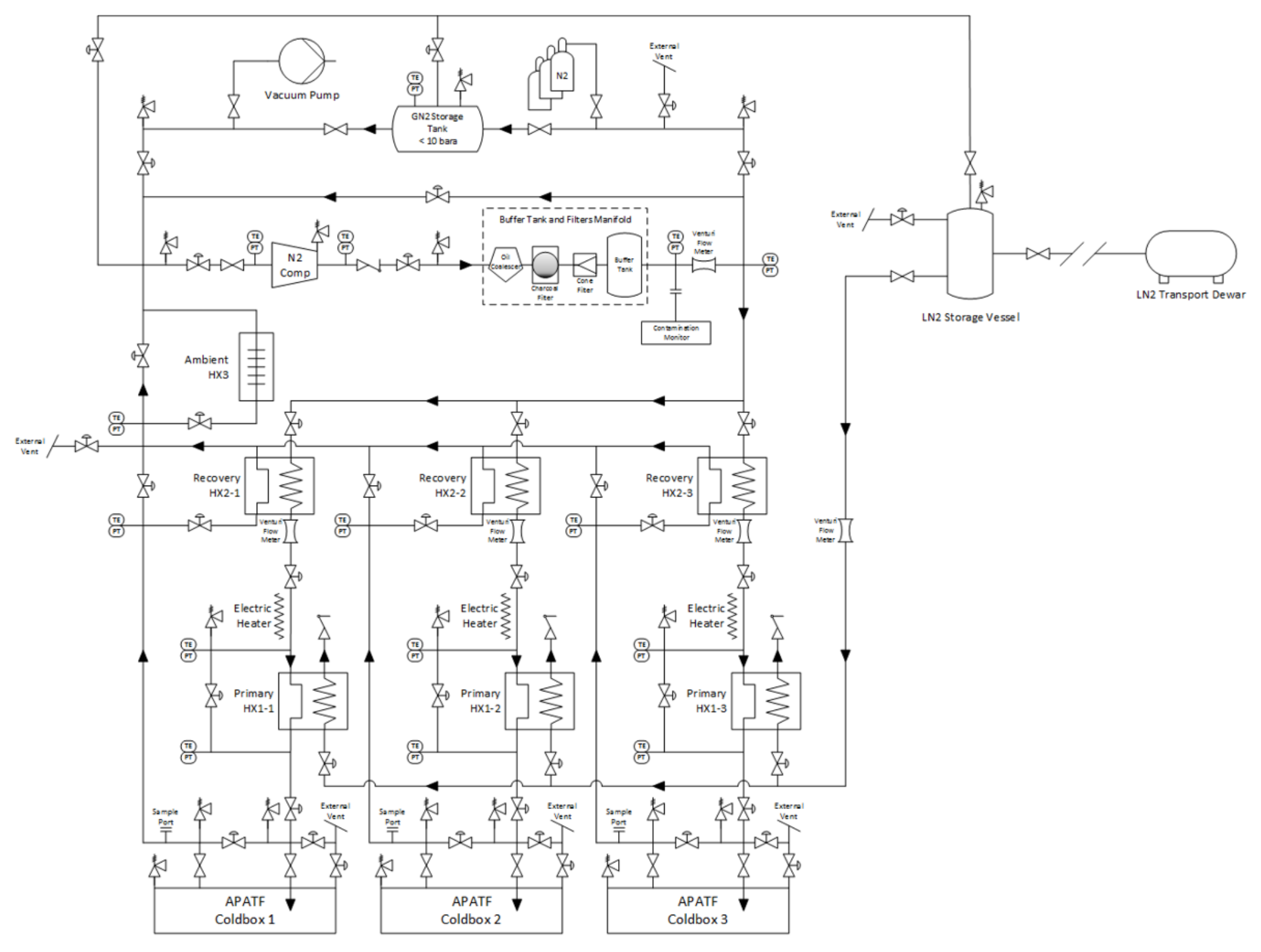}
\end{dunefigure}



\subsection{Prototyping and Testing (QA/QC)}
\label{sec:fdsp-tc-infr-qaqc}

Installing all this new equipment underground during the installation setup phase involves many new techniques  and unique work. While most of the procedures will have been tested during the trial assembly at \dword{ashriver}, everything must be properly approved. The main \dword{apa} and \dword{cpa} towers will already be structurally approved, but all lifting fixtures, shuttle beams, crane tower connections, and \coldbox connections must undergo load tests.

The load test program for the lifting fixtures, shuttle beams, crane tower connections, and \coldbox connections will be documented in test procedures in accordance with the \dword{lbnf} \dword{dune} \dword{qa} program.  
These test procedures will (1) list prerequisites for testing, (2) identify fixtures and test equipment, and (3) provide step-by-step instructions, acceptance criteria, and documentation requirements.
They will be in place prior to the start of testing. 
The test results will be documented and approved by the systems engineering team prior to use of the lifting fixtures, shuttle beams, and crane tower connections. 

The \coldbox{}es and cryogenics system will also be tested, which may require restricting  access to the cleanroom  
for several days for system checks. 
The \coldbox{}es and the associated cryogenic system test program will be similar to the test program that was instituted for \dword{pdsp}. 
This test program will also be documented in procedures in accordance with the \dword{lbnf} \dword{dune} \dword{qa} program. 
These test procedures will (1) list prerequisites for testing, (2) identify test equipment, and (3) provide step-by-step instructions, acceptance criteria, and documentation requirements.
 The test results will be documented and approved by the systems engineering team prior to use of the \coldbox and cryogenics system.


\section{Detector Installation}
\label{sec:fdsp-tc-inst}


\begin{dunefigure}[High level installation schedule]{fig:high-level-schedule}
  {Overview schedule showing the main activities underground.}
\includegraphics[angle=90,height=.99\textheight]{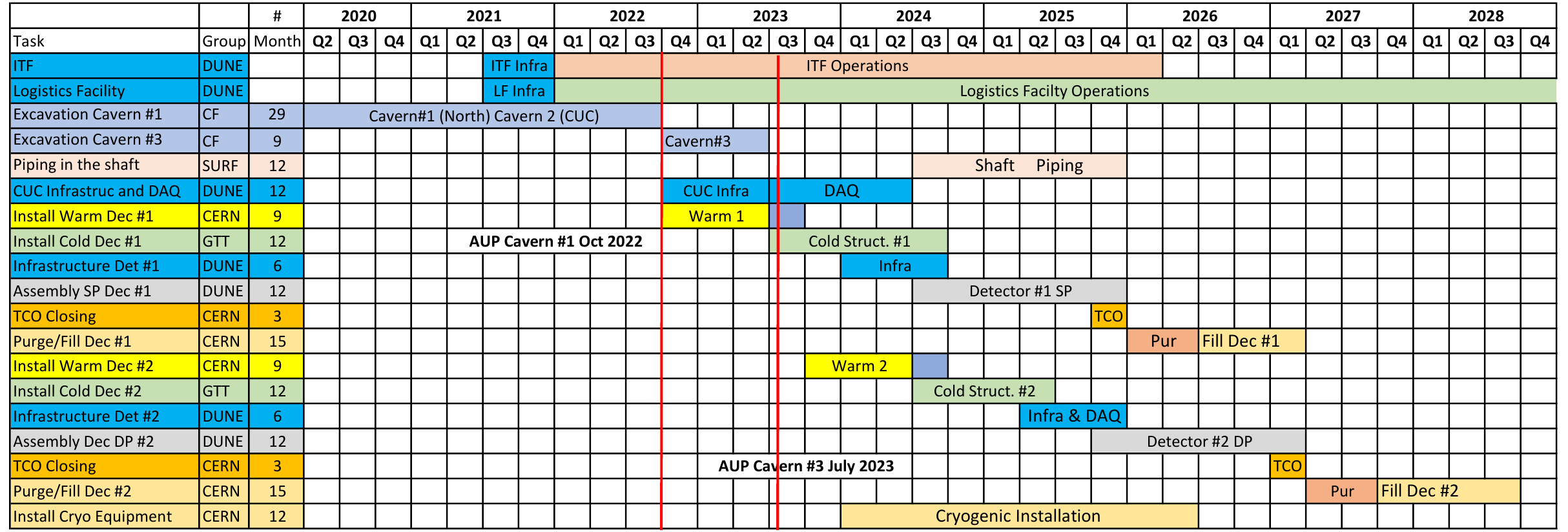}
\end{dunefigure}

As mentioned in Section~\ref{sec:fdsp-tc-log}, the \dword{dune} detector installation will proceed in three phases: \dword{cuc} set up, installation set up, and the detector installation. The schedule in Figure~\ref{fig:high-level-schedule} shows the major underground activities and gives an idea of what work occurs in each phase. 

The first phase, \dword{cuc} set up, begins when the underground area for the north cavern and \dword{cuc} become available to \dword{lbnf} and \dword{dune}. At this time, the \dword{lbnf} cryostat construction begins in the north cavern, and \dword{dune} equipment installation 
begins in the \dword{cuc}, specifically, infrastructure in the \dword{dune} data room. Figure~\ref{fig:cavern-layout2} shows a top view of the underground areas and the location of the dataroom at the west end of the \dword{cuc}. 

The detector installation setup phase (referred to as Infrastructure Det\#1 in the table in Figure~\ref{fig:high-level-schedule}) begins during the cryostat membrane installation period. 
In this phase, the equipment needed for detector installation is erected in the north cavern. This includes the bridge across the cavern, the cleanroom, lifting equipment and work platforms, the \coldbox{}es and cryogenics system for \dword{apa} testing, and the \dword{dss} and switchyard. 
The detector itself is installed in the third phase of the installation. 

\begin{dunefigure}[Layout of the DUNE underground areas]{fig:cavern-layout2}
  {Top view of the layout at the \dword{4850l} at \dword{surf}. Shown are the three large excavations and the location of detectors in the north (upper) and south caverns. 
The detector caverns (north and south) are \SI{145}{\meter} long, \SI{20}{\meter} wide, and  \SI{28}{\meter} high.   The \dword{cuc} in the middle houses the \dword{dune} data room where the DAQ will be installed and the underground utilities. The north, middle and south caverns are also referred to as cavern\#1, cavern\#2 and cavern\#3 in Figure \ref{fig:high-level-schedule}.}
\includegraphics[width=.9\textwidth]{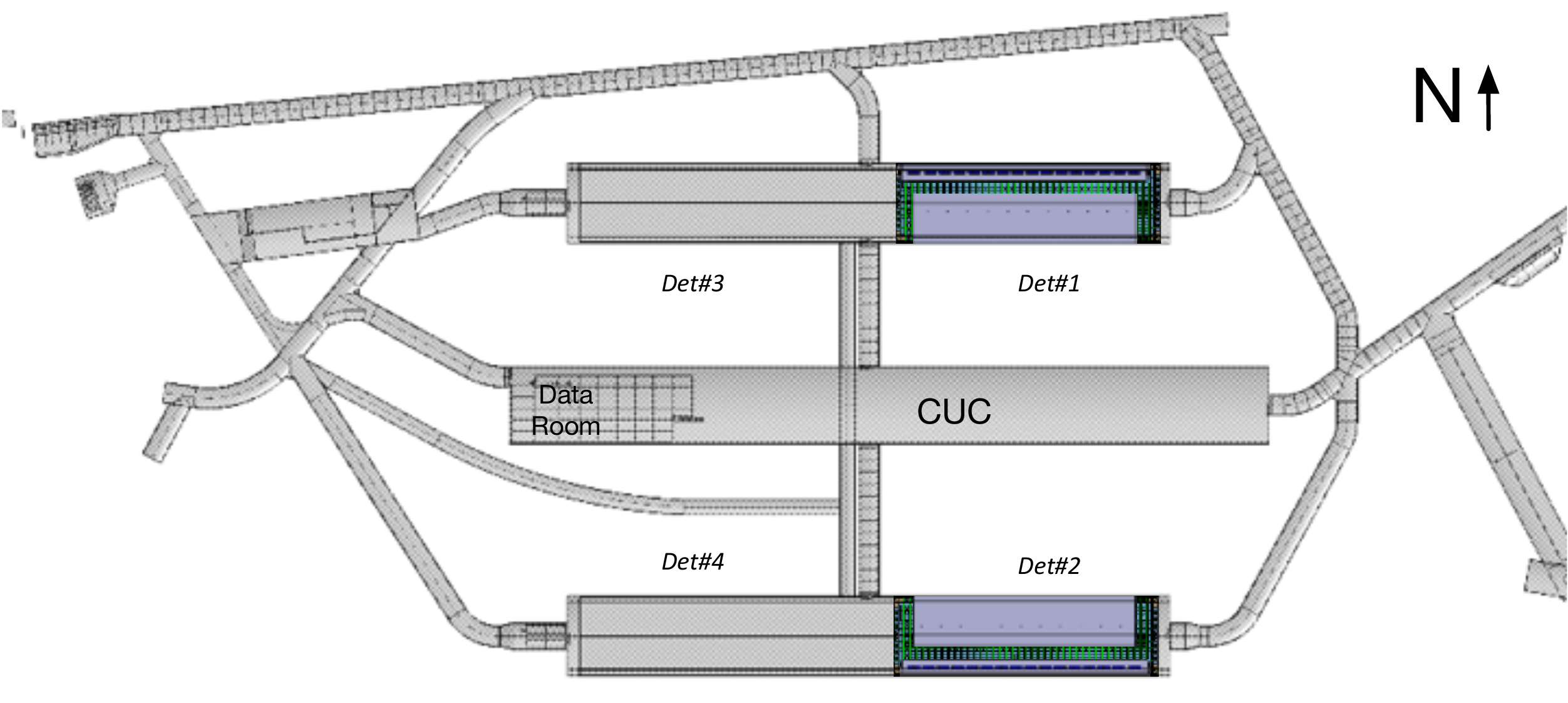}
\end{dunefigure}

\subsection{Installation Process Description}
\label{sec:fdsp-tc-inst-proc}

\subsubsection{CUC Installation Phase}
\label{sec:fdsp-tc-inst-CUC}

\begin{dunefigure}[Layout of the DUNE data room and work area in CUC]{fig:install-cuc2}
  {Top: The overall layout of the \dword{dune} spaces in the \dword{cuc}. A110 is the \dword{dune} data room, which houses the underground computing, and A111 is a general-purpose work area (not a control room, as labeled) that we call the experimental work area. Bottom: The first row of ten racks in the data room is shown. The first two represent the \dword{cf} interface racks. The images were taken from the ARUP 90\% design drawings U1-FD-A-108 and U1-FD-T-701~\cite{bib:docdb14242}.}
\includegraphics[width=.85\textwidth]{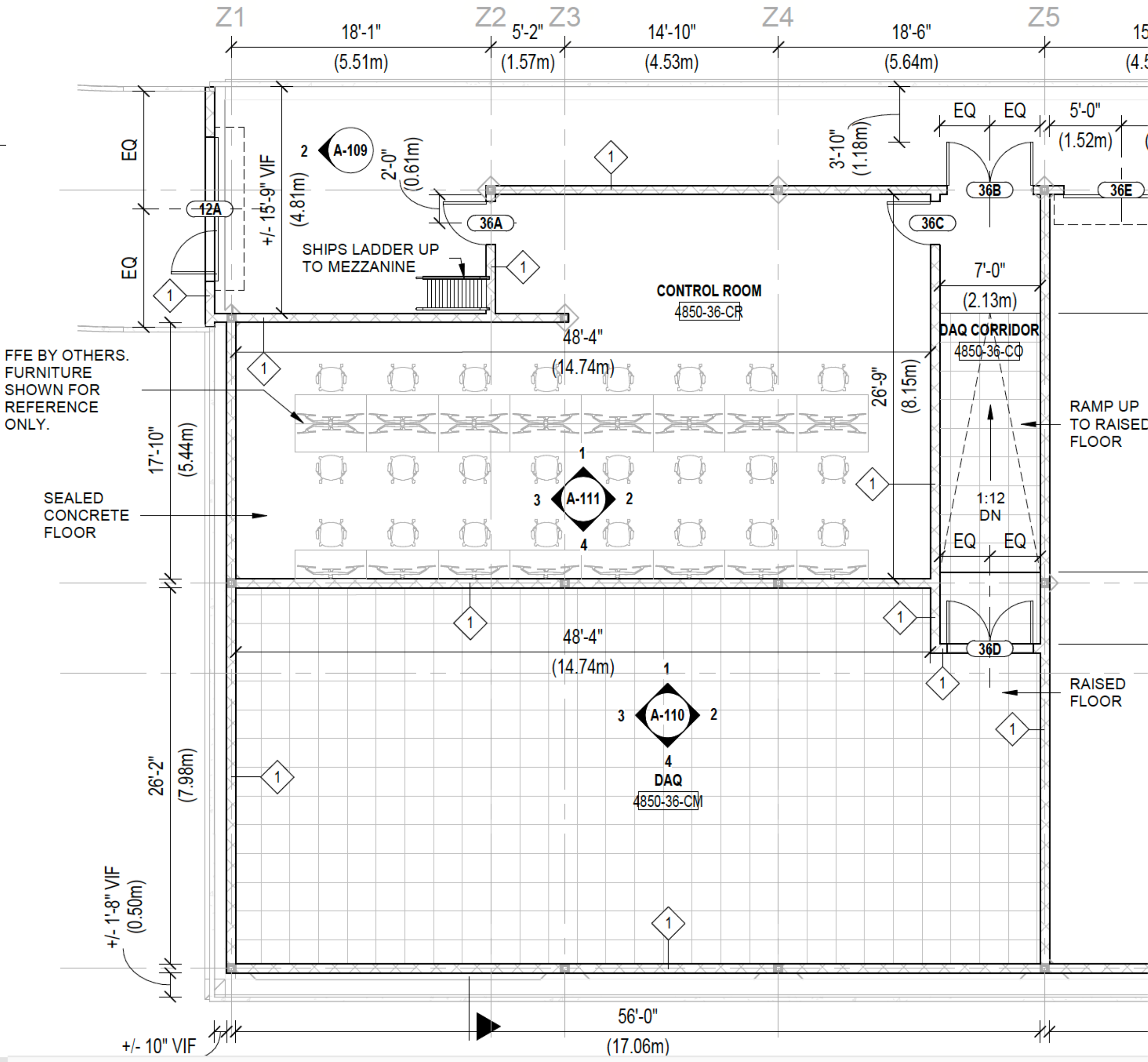}
\vspace{-2pt}
\includegraphics[width=.9\textwidth]{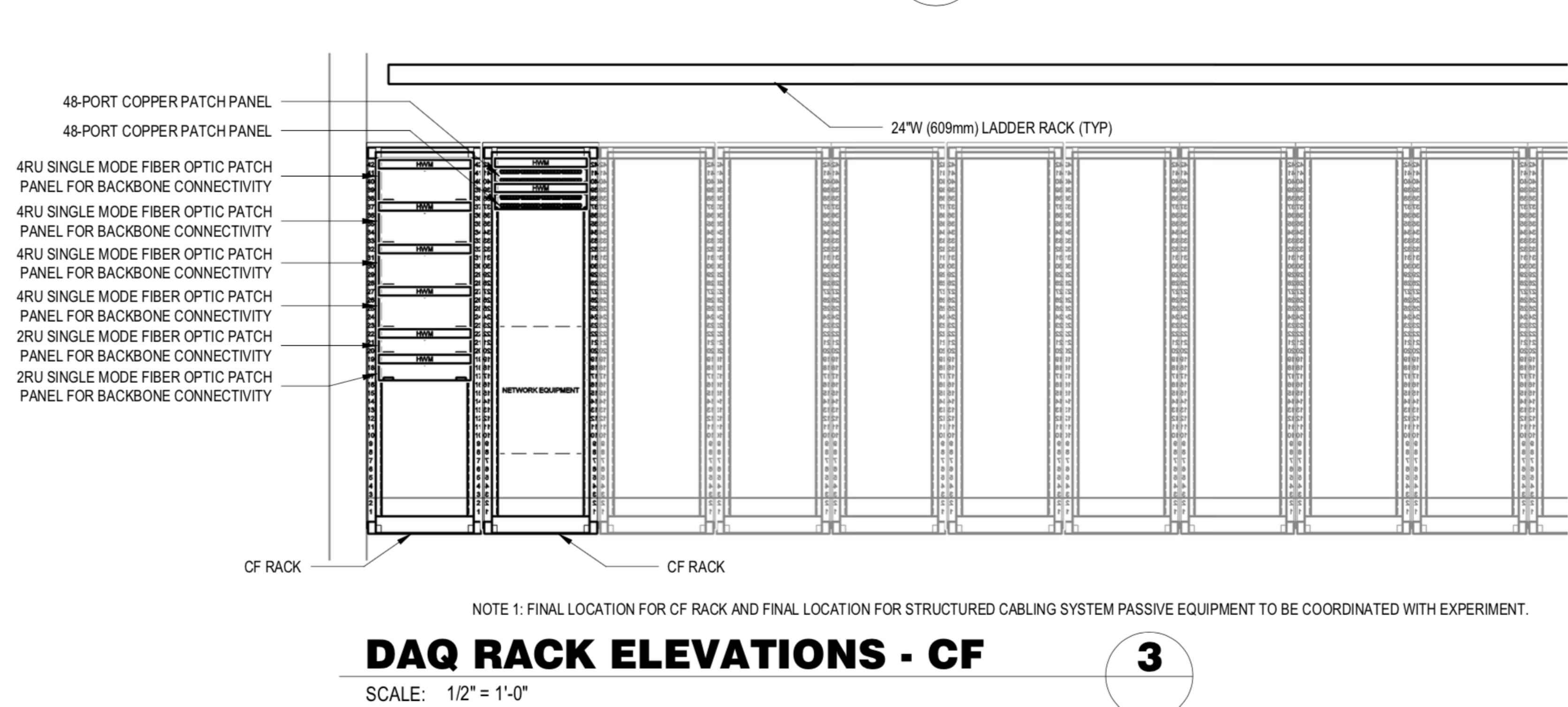}
\end{dunefigure}

Once the \dword{lbnf} \dword{cf} outfitting of the north cavern and the \dword{cuc} is complete,  \dword{lbnf} begins the first cryostat installation in the north cavern and \dword{dune} can begin to install equipment in the data room and work area room in the \dword{cuc}. See Figure~\ref{fig:install-cuc2}.  \dword{dune} will not have access to the north cavern due to the heavy steel work for the cryostat. 
At this point \dword{lbnf}  \dword{cf} will have installed redundant single-mode fiber up the shafts to provide external connectivity, and in the empty data room,  an \SI{18}{in} false floor, a \SI{500}{kVA} power disconnect, and connections for sufficient chilled water to cool the racks. The data room, like the adjacent \dword{cf} electronics room, will be outfitted with a dry fire-extinguishing system.

The water-cooled racks, cable trays, power distribution, and water distribution in the data room are the responsibility of \dword{dune} and will be installed once the space becomes available.  
Installation of the racks must be coordinated with \dword{cf} 
since the first two racks are for \dword{cf} use and must be in place before the first phase of work underground is complete. 
Some small overlap will be needed between \dword{cf} and \dword{dune} at this time. The general-purpose network will be installed by \dword{fnal}'s \dword{sdsd} and connected to the Ross Shaft fiber optics. 
This is required for most subsequent work in the underground area.
The \dword{daq} fiber trunk between the detector cavern and the \dword{cuc} data room will be installed after the cable trays, electronics mezzanine, and racks are available in the north cavern.

Data from the detector electronics will be transmitted over a multimode fiber trunk from the \dwords{wib} on top the \dword{spmod} to the \dword{daq} data room in the \dword{cuc}, shown in Figure~\ref{fig:install-cuc2}.  The data room will contain 60 water-cooled racks, two of which are reserved for \dword{cf} use, two for \dword{cisc} servers, and the rest for \dword{daq} servers and networking. Racks for all four modules will be installed at the beginning of the \dword{cuc} commissioning phase because they must be plumbed into the cooling water below the data room's drop floor and wired into power distribution from the ceiling.  \dword{daq} equipment will populate the racks as needed for servicing the detector commissioning.  For the first \dword{detmodule}, details of this configuration will be informed by \dword{daq} vertical slice tests done at other institutions.  

At the same time, the eight above-ground \dword{daq} racks that receive data from the underground data room and  transmit the data to \dword{fnal} will 
be installed, connected to the network, and connected to the single-mode fiber in the Ross and Yates Shafts.  With this infrastructure in place, the \dword{daq} group can begin constructing and testing the final \dword{dune} \dword{daq}, starting with the timing system. 
Enough \dword{daq} back-end servers to support the first \dword{apa}s will be operational before the \dword{apa}s are installed.  The remainder of the \dword{daq} will grow in parallel with \dword{apa} installation.

The underground experimental work area (shown as ``CONTROL ROOM'' in Figure~\ref{fig:install-cuc2}) must serve a variety of purposes during the \dword{dune} installation. Initially, the area will be outfitted with office equipment for the installation team, workstations for \dword{daq}, and a basic conference area for meetings. The room is \SI{17}{m} wide with portions that are \SI{5.5}{m} and \SI{8}{m} deep.

During this early installation stage, the machine shop and \dword{dune} storage area will be set up in the detector excavation area and shared with the cryostat team. 

\subsubsection{Installation Setup Phase}
\label{sec:fdsp-tc-inst-setup}

\begin{dunefigure}[Top view of the installation area highlighting the infrastructure]{fig:install-infrastructure-top}
  {Top view of the installation area highlighting the infrastructure. The \cryostatwdth wide cryostat is on the right. The cleanroom roof and cavern walls are removed in this view.}
\includegraphics[width=.85\textwidth,trim=0mm 70mm 0mm 70mm,clip]{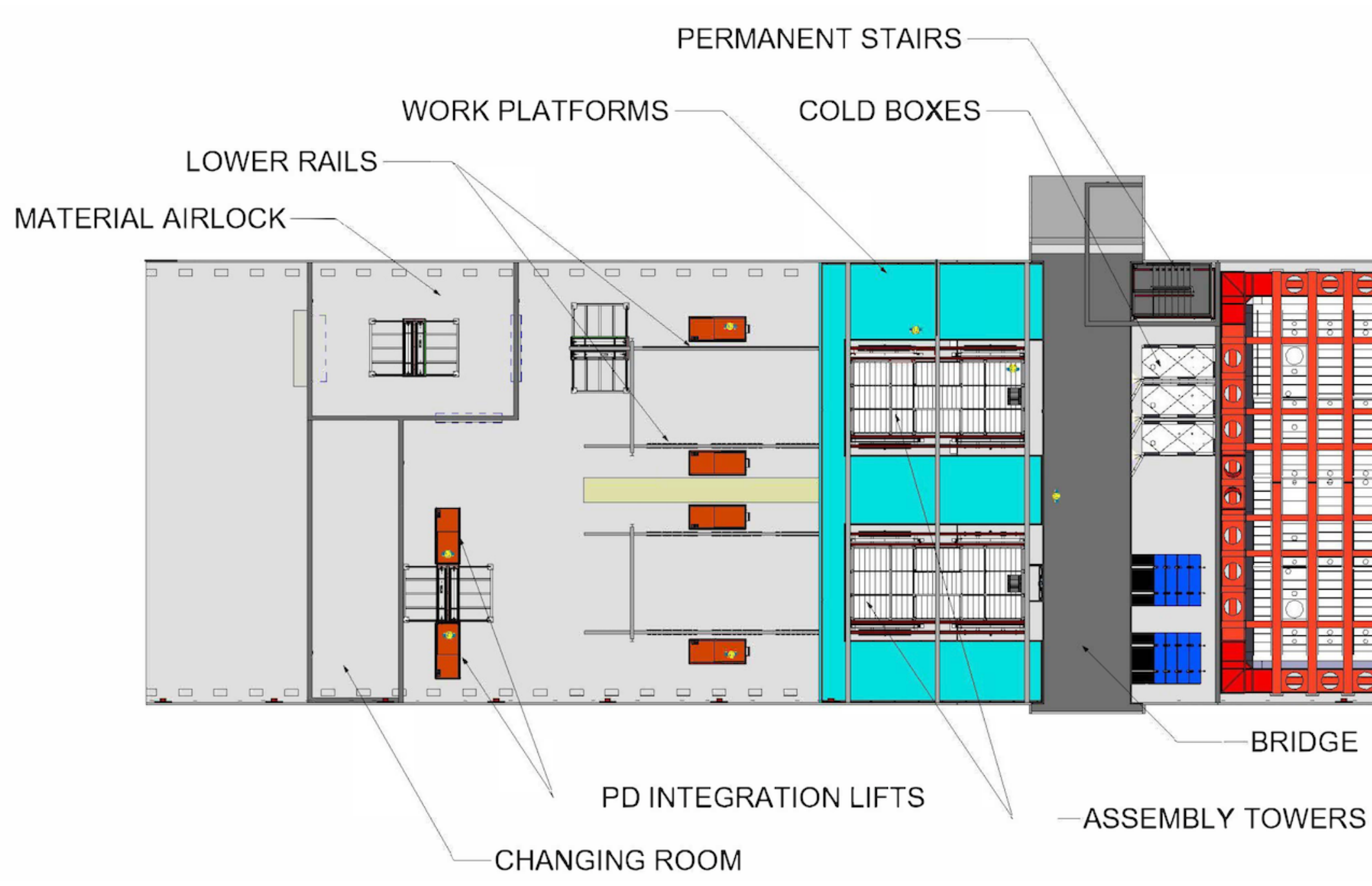} 
\end{dunefigure}

Once the steel structure of the cryostat is complete, the remaining work by the \dword{lbnf} cryostat team will be focused inside the cryostat, installing the insulation and membrane.  
\dword{lbnf} activity outside the cryostat will consist mainly of
transporting the 4,000 crates of foam and other materials from the cavern to inside the cryostat. 
Since the cryostat outer steel structure will be in position, \dword{dune} can 
begin installing the infrastructure needed outside the cryostat to support detector installation. 
Figure~\ref{fig:install-infrastructure-top} shows the major pieces of infrastructure supporting the detector installation.
The first piece of equipment is the bridge between the north and south drifts. 
This will allow the cryogenics equipment to travel from the north drift to the \dword{cuc} and will provide part of the structure for the cleanroom. It also provides an additional means of egress in an emergency. 

As part of the bridge construction, the crane under the bridge will be mounted. This 
will be used to lift crates off the floor and bring them into the cryostat, freeing the cavern crane for work elsewhere. The decking on top of the cryostat is also installed in this early stage to provide a safe work surface. 

The largest and most time-consuming pieces of equipment to  construct in this phase are the three \coldbox{}es and their associated cryogenics system. 
The \coldbox{}es, visible in Figures~\ref{fig:install-coldbox} and \ref{fig:install-infrastructure-top}, will likely need to be constructed in place due to their size (see Section~\ref{sec:fdsp-tc-infr-cryo}) and fabrication will begin as early as possible.  
If it is possible to bring them down the shaft partially assembled, then once the first \dword{detmodule} is complete, we can break them down to their pre-assembled parts and move them to the second \dword{detmodule} area.

After the bridge crane under the north-south bridge is in place the \dword{apa} assembly and cabling towers are installed. 
The two towers have enough space between them, as described in Section~\ref{sec:fdsp-tc-infr-comm}, to walk through or perform work. 
With the towers in place, the north-south support beams and the fixed platforms can be installed. 
This work is done at height so it will only cause temporary interruptions to the material transport along the floor to the cryostat; the lower set of rails and the \dword{cpa} assembly equipment can be installed at the convenience of the cryostat installation crew.

When the cryostat membrane work is complete the cryogenic piping inside the cryostat can be installed, the cryostat cleaned, and the false floor installed. After the cryostat is cleaned the HEPA filters will be installed in air handling units for the cryostat and purified air will begin flowing. A curtain over the \dword{tco} can be used to prevent dust from the cavern from entering the cryostat until the cleanroom walls are constructed.

In parallel to the installation of the cryogenic piping, the walls of the cleanroom can be assembled and AC power and fire suppression installed. Finally, the floor is painted and 
cleaned, making the cleanroom ready for operation.

On the cryostat roof the installation of the cryostat crossing tubes proceeds in parallel with the cryostat membrane assembly sequence. 
The crossing tubes are welded to the \SI{1}{cm} thick steel cryostat roof and cross braced to the large I-beams. 
The thin-walled tubes that penetrate the foam insulation are welded to the cryostat inner membrane.

\begin{dunefigure}[Installation of electronics crosses]{fig:install-elect-cross}
  {Installation of the crosses to which the \dword{tpc} electronics warm readout and the \dword{pd} cables are connected. In this figure the cryostat roof decking is not shown. During the cross installation a section of decking will be removed so people can access the required flanges and work at a comfortable height.  }
 \includegraphics[width=.5\textwidth]{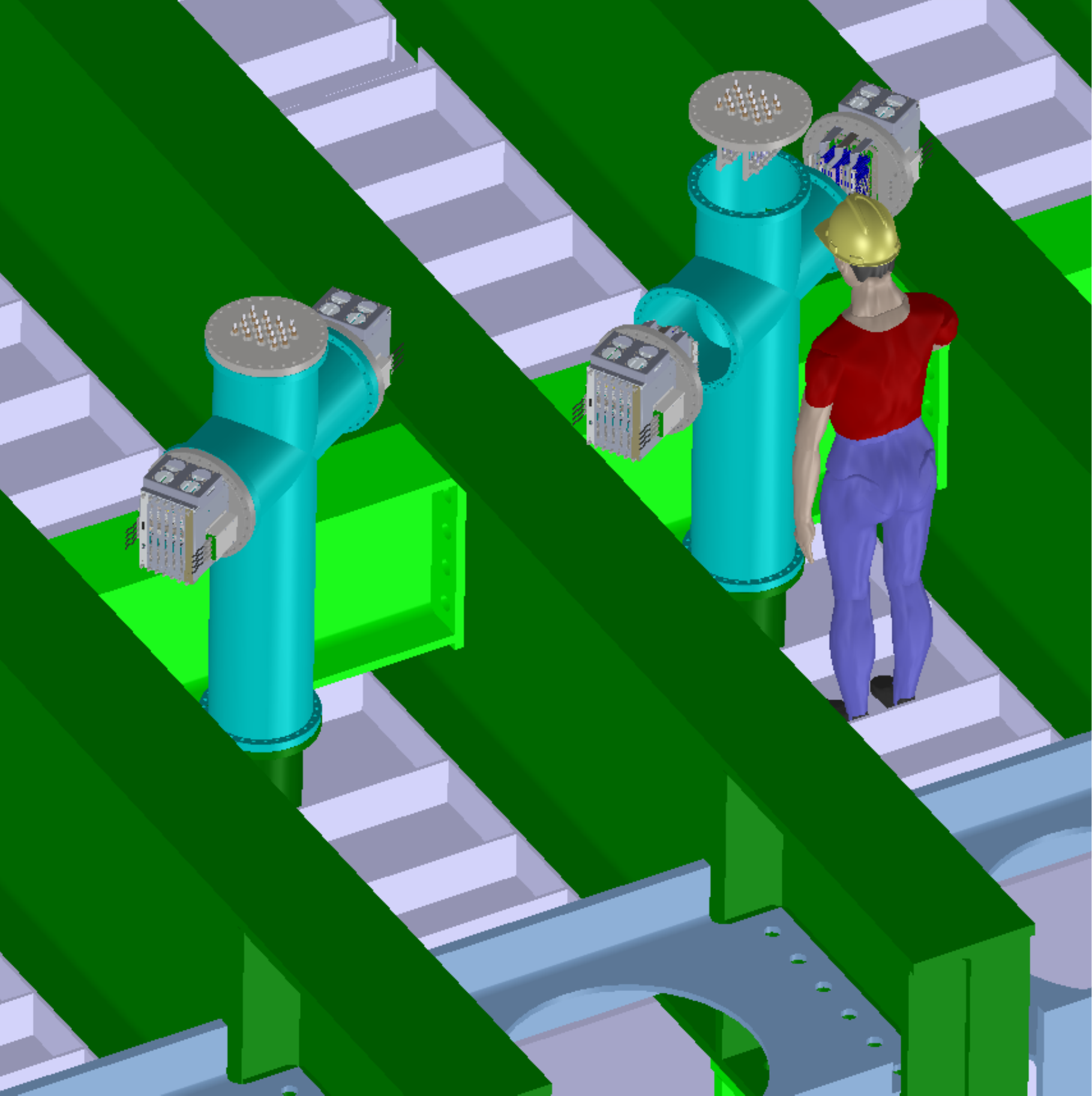}
\end{dunefigure}

Once the crossing tubes are installed and leak-checked, we connect the  \dword{tpc} electronics crosses and mount  the  \fdth flanges for the \dword{ce} and \dwords{pd} onto the crosses. 
The height of the crosses was chosen to allow a person to work comfortably on the \dwords{wiec}  and \dword{pd} flanges while standing on the cryostat roof. 
A fully assembled cross is shown on the left of Figure~\ref{fig:install-elect-cross}, and a cross with a \dword{wiec} extracted in ``assembly position'' is shown on the right. 
The present plan is to install the crosses shortly after the cryostat crossing tubes are installed. 
This allows us to seal the large openings in the cryostat roof to prevent dust from entering.
For this stage, temporary rubber seals are used for the flanges; they must be removed during the cabling process later in the installation. 
When the \dword{wiec} installation is complete the \dword{tpc} electronics is ready for the installation of power and fiber optics for readout. 

\begin{dunefigure}[DSS \fdth  installation]{fig:install-dss-feedthru}
  {The \dword{dss} support \fdth{}s  are installed using a gantry crane running along the roof of the cryostat.
  The cryostat decking is not shown. 
  The gantry can move freely on the cryostat roof decking. The gantry crane is selected to fit under the mezzanine as shown in the right panel.
  }
  \includegraphics[width=.98\textwidth]{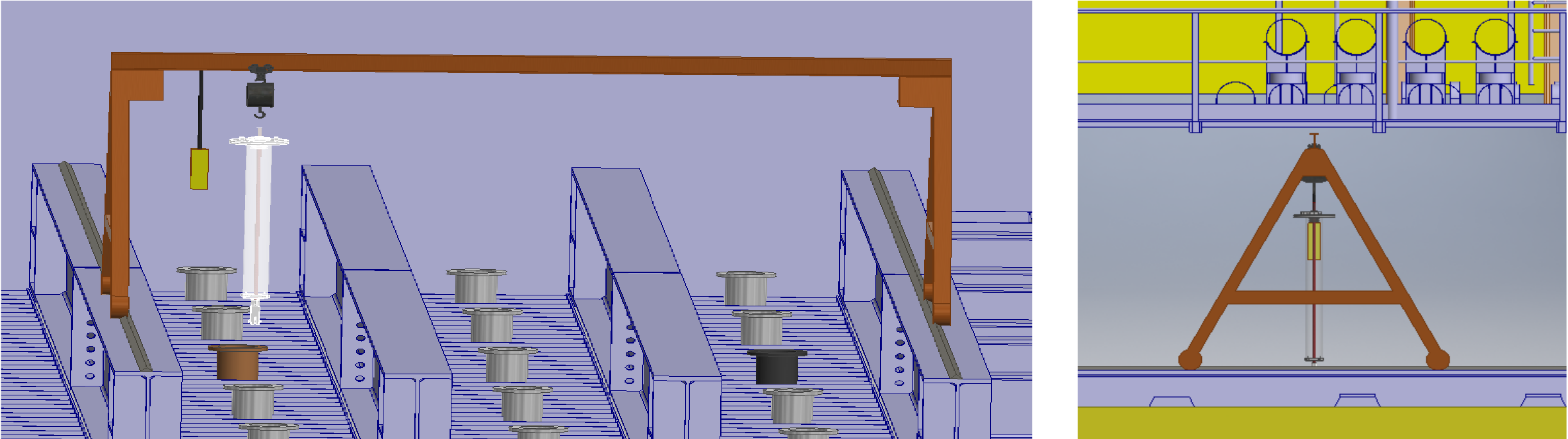}
\end{dunefigure}

The \dword{dss} support \fdth{}s can be installed in parallel to the \dword{tpc} electronics crosses. This is the first step in the \dword{dss} installation. 
A gantry crane on top of the cryostat picks up the \fdth{}s  and lowers them into the cryostat crossing tubes as shown in Figure~\ref{fig:install-dss-feedthru}. 
There are \num{20} \fdth{}s per row and five rows, for a total of \num{100} \fdth{}s.  A fixture with a tooling ball is attached to the clevis of each \fdth{}.  
The horizontal $XY$ and the vertical $Z$ positions of this tooling ball are defined, a survey is performed to determine the location of each tooling ball center, and adjustments are made to get the tooling ball centers to within $\pm$\SI{3}{mm} of the nominal position.  
The \SI{6.4}{m} long I-beams are then raised and pinned to the clevis.  
Each beam weighs roughly \SI{160}{kg} (\SI{350}{lbs}). 
A lifting tripod is placed over each  \fdth{}'s supporting beam, and a \SI{0.64}{cm} (\SI{0.25}{in})  cable is fed through the top flange of the \fdth down \SI{14}{m} to the cryostat floor where it is attached to the I-beam. 
The cable access port and lifting cable are shown in Figure~\ref{fig:dss-beam-lifting}. 
The winches on each tripod raise the beam in unison to position it at the correct height for pinning to the \fdth clevis.  Once the beams are mounted, a final survey of the beams ensures proper placement and alignment. 
 \begin{dunefigure}[DSS I-Beam lifting setup]{fig:dss-beam-lifting}
  {A cable access port is included in the \dword{dss} flange. This is used to feed a cable from the roof through the flange and attach it to the I-beams during \dword{dss} installation.}
 \includegraphics[width=.95\textwidth]{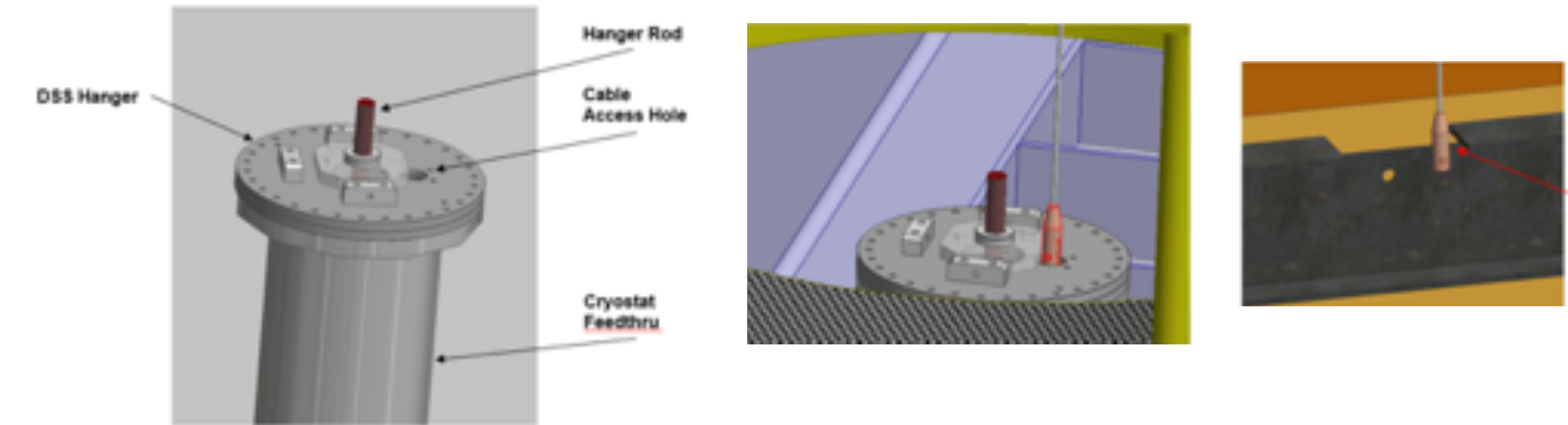}
\end{dunefigure}

Next it is time to install the mezzanines for the cryogenics system and the detector electronics racks, followed by the cable trays,  piping, lighting, and cryostat roof flooring. At this point, the cryostat roof is ready for \dword{daq} and cryogenics system installations to begin; this will proceed in parallel with the detector installation.

\subsubsection{Detector Installation Phase}
\label{sec:fdsp-tc-inst-execute}

At the start of the detector installation phase, the cleanroom and all equipment inside are operational, the \dword{dss} is installed, and the cryostat is clean and ready for installation. 
Figure~\ref{fig:install-infrastructure-top} shows the layout (plan view with the roof removed) of the cleanroom during the detector installation phase; the cryostat is on the right and the open cavern on the left. 
In the north-east corner of the figure (upper right corner), the access  (permanent) stairway is shown. 
This stairway is inside the cleanroom and allows people easy access to both the work platforms and the cleanroom floor. 
The doors to the stairway will never be locked; the stairway is considered a means of egress in emergency, as it leads to an exit through the mucking drift.
A second stairway and an industrial elevator at the west entrance to the cavern provide access to the cavern floor for personnel and equipment. 
The primary changing room is in the southwest corner of the cleanroom and a smaller changing room (shown in Figure \ref{fig:install-cleanroom}) will be situated near the stairs at the \dword{4850l} for people accessing the work platforms.

In the north-west corner of the cleanroom is the material airlock where all materials enter  through tall doors. 
Outfitting the airlock with a removable roof is under consideration. 
It would allow entry of equipment via the cavern crane, which could facilitate the process.
 
Labor for the detector installation phase is split between the \dword{jpo} and the \dword{dune} consortia. 

The  detector installation team includes the  detector installation manager, one installation shift supervisor per shift, the \dword{jpo} technical support team, and \dword{dune} consortia scientific and technical personnel. (See Figure 4.5 in \tcchjpo{}.) The detector installation manager is responsible for communicating with the underground cavern
coordinator for all logistics, shipping, and inventory issues. They organize the daily underground detector installation tasks and lead the detector installation part of the shift-change meeting. The detector installation team is divided into several work crews operating in the cleanroom and cryostat. They are responsible for moving all detector components into the materials airlock, cleanroom, \coldbox  (if needed), and cryostat. Additional activities in the cleanroom include linking the \dword{apa}s together, installing \dwords{pd}, installing and cabling the electronics, and removing  \dword{apa} protective covers. Inside the cryostat, the detector installation team installs the  \dword{tpc} components. 

Each of the \dword{dune} \dword{fd} consortia has specific tasks related to its subsystem. 
The installation activities are planned estimating both the \dword{jpo} and consortia labor contributions.

\begin{dunefigure}[Design of the instrumentation \fdth{}s]{fig:CISC-feedthru}
{Design of the instrumentation \fdth{}s. The signal \fdth is integrated with the \dword{dss} support \fdth{}s. A side port on a short spool piece in the \dword{dss} support structure allows the instrumentation cables to be fed through the cryostat walls where needed.}
\includegraphics[width=0.95\textwidth]{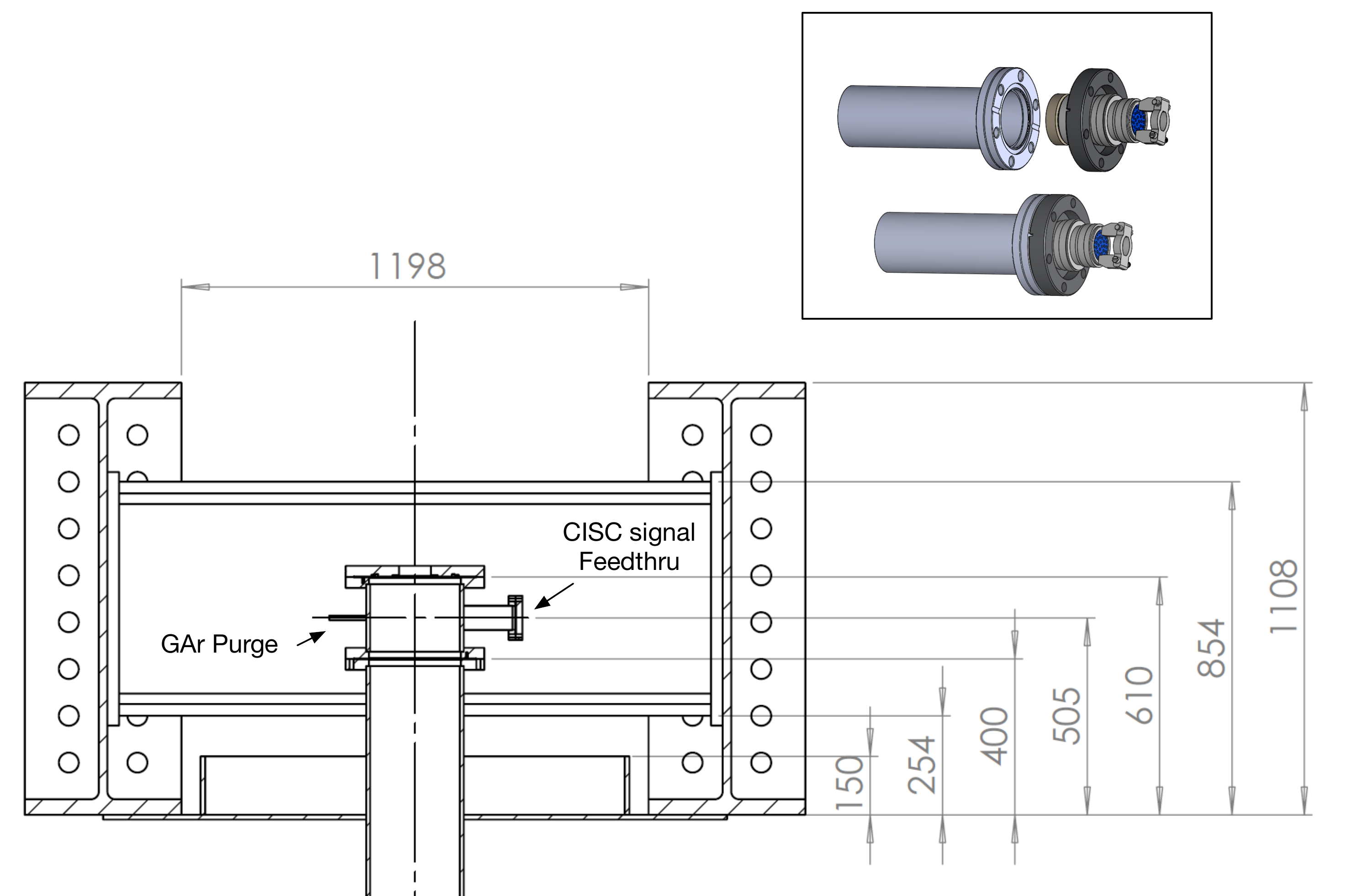}
\end{dunefigure}

The first detector equipment to be installed are \dword{cisc} T-gradient thermometers, an array of purity monitors, and cameras, all at the east end of the cryostat. This equipment will be used to monitor the \cooldown{}, filling, and commissioning of the detector. Some equipment for the laser calibration system is also installed at this time, including some positioning diodes and possibly an optical mirror-based switching system.  The signals exit the cryostat via electrical \fdth{}s distributed across the cryostat roof and integrated with the \dword{dss} support structure, as shown in Figure~\ref{fig:CISC-feedthru}. Because all these components are small, they can be installed using a scissor lift with \SI{12}{m} reach. At present, this is the tallest battery operated (thus cleanroom compatible) scissor lift rated for use in the USA that we have identified.

Cabling for the remaining static T-gradient monitors is also installed before the start of \dword{tpc} installation.  The thermometer cables and mechanical supports are anchored to the cryostat using the bolts running along the cryostat's top and bottom edges and can be installed once the cryostat is clean. 
To avoid  damaging the small and fragile thermometers, they are not plugged into the small IDC-4 connectors until just before the corresponding \dword{apa} is moved into its final position. 
Cables and supports for the thermometers on the pipes below the detector and on the cryostat floor are installed immediately after installing the static T-gradient monitors on the walls.  
Again, the thermal sensors themselves are installed later, just before unfolding the bottom \dwords{gp}, to avoid damage.

Individual sensors on the top \dword{gp} must be integrated with the other \dwords{gp}. For each \dword{cpa} (with its corresponding four \dword{gp} modules), cable and sensor supports will be anchored to the \dword{gp} threaded rods. Once the \dword{cpa} is moved into its final position and its top \dwords{gp} are ready to be unfolded, sensors on these \dwords{gp} are installed.

Installing fixed cameras is, in principle, simple but involves a large number of interfaces. The enclosure for each camera has exterior threaded holes to facilitate its mounting 
on the cryostat wall, the cryogenic internal piping, or the \dword{dss}. Each enclosure is attached to a gas line to maintain appropriate underpressure in the fill gas, requiring an interface with cryogenic internal piping. Camera cables are run through cable trays to flanges on assigned instrumentation \fdth{}s.

A summary of all the cryogenics instrumentation provided by the \dword{cisc} consortium is shown in Figure~\ref{fig:cisc_devices}. 

At this point the quartz optical fibers required for the \dword{pd} monitoring system are run from the optical flange locations (still being finalized) to locations on the \dword{cpa} support beams of the \dword{dss}, to be connected later to the diffusers mounted on the \dword{cpa}s.

The residual gas analyzers that monitor impurities in the \dword{gar} system must be installed before the piston purge and gas recirculation phases of cryostat commissioning.  However the actual installation time depends on the schedule for outfitting the mezzanine and installing the  \dword{gar} purge piping. These  instruments are installed near the tubing switchyard to minimize tubing run length and for convenience when switching the sampling points and gas analyzers. 

\begin{dunefigure}[Distribution of various CISC devices inside the cryostat.]{fig:cisc_devices}
  {Distribution of various \dword{cisc} devices inside the cryostat.
  }
  \includegraphics[width=0.98\textwidth]{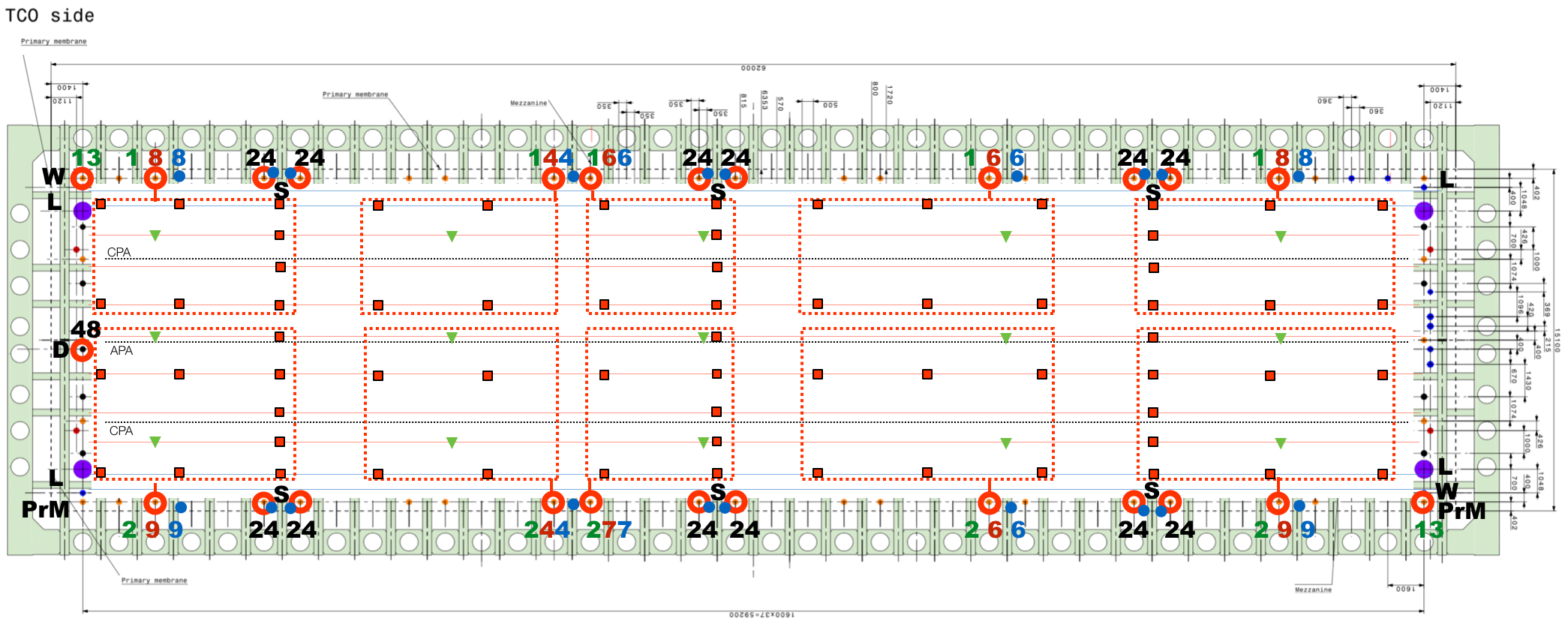}
  \includegraphics[width=0.85\textwidth]{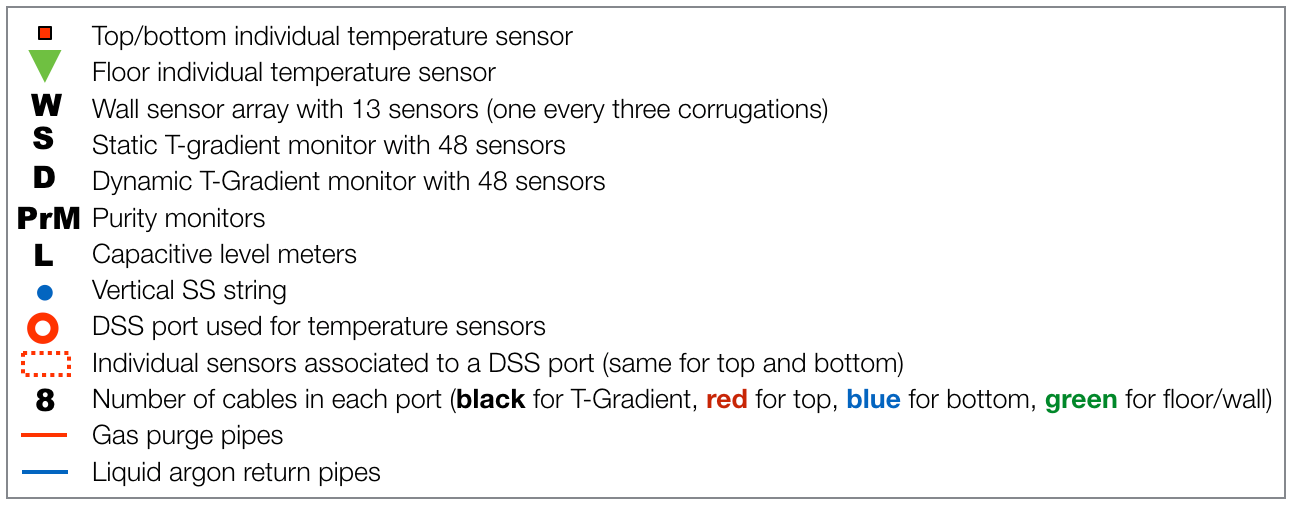}
\end{dunefigure}

\clearpage

Next the east \dword{ewfc} is installed. The endwall planes are brought underground in custom crates. Each of the eight crates holds four endwall modules.  Eight modules are needed to build one complete \SI{12}{m} tall plane.  First a custom hoist is installed on the end of the \dword{dss} beam for lifting and assembling the modules in place, as shown in Figure~\ref{fig:endwall-hoist}. The \dword{ewfc} transport crates are then brought to the material airlock using a forklift where they are cleaned.  Once clean, the crates are moved into the cleanroom and placed next to the \dword{tco}. Then a hoist running on the rails through the \dword{tco} lifts the endwall modules onto the transport cart, which is then rolled into the cryostat. Figure~\ref{fig:endwall-cart} shows an endwall module being transferred to the transport cart. The top endwall module is then attached to the installation hoist and lifted out of the cart. When the module is free of  the cart, the cart is re-positioned so the second module can be attached to the first, and the pair is then lifted. This process is repeated until the full \SI{12}{m} \dword{ewfc} plane is assembled and can be attached to the \dword{dss}. 
Figure~\ref{fig:endwall-hoist} shows an endwall plane being lifted into position.
All the \dword{hv} connections inside the plane can now be tested. The process is then repeated for the remaining three endwall planes comprising the east  \dword{ewfc}. 
 
 \begin{dunefigure}[Endwall hoisting infrastructure]{fig:endwall-hoist}
  {Image showing the hoisting equipment used to lift the endwall into position. The field shaping strips are removed in this image. This shows one of the \SI{3.5}{\meter} endwall planes in place and a second being positioned. }
\includegraphics[width=.5\textwidth]{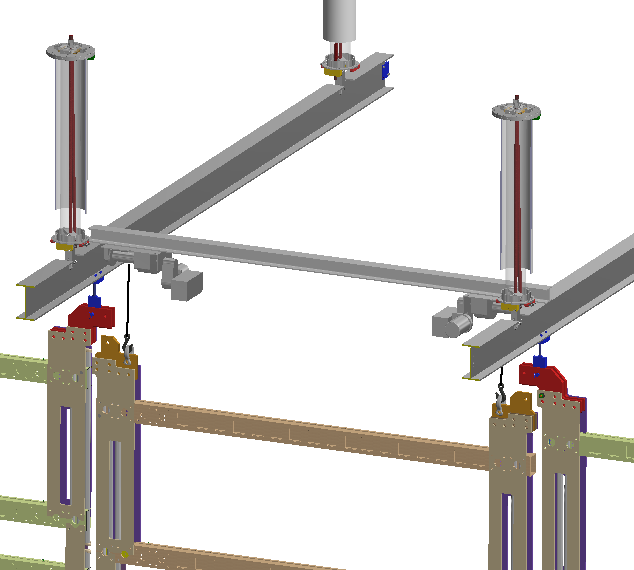}
\end{dunefigure}
 
\begin{dunefigure}[Installation of the first endwall]{fig:endwall-cart}
  {The endwalls are lifted out of the transport crates using the one of the hoists on the installation switchyard. Each module is  placed on a custom cart that is rolled into the cleanroom. The pedestal in front of the TCO is at the height of the cryostat floor so carts can be used to bring material into the cryostat. The guard rails are not shown.}
\includegraphics[width=.5\textwidth]{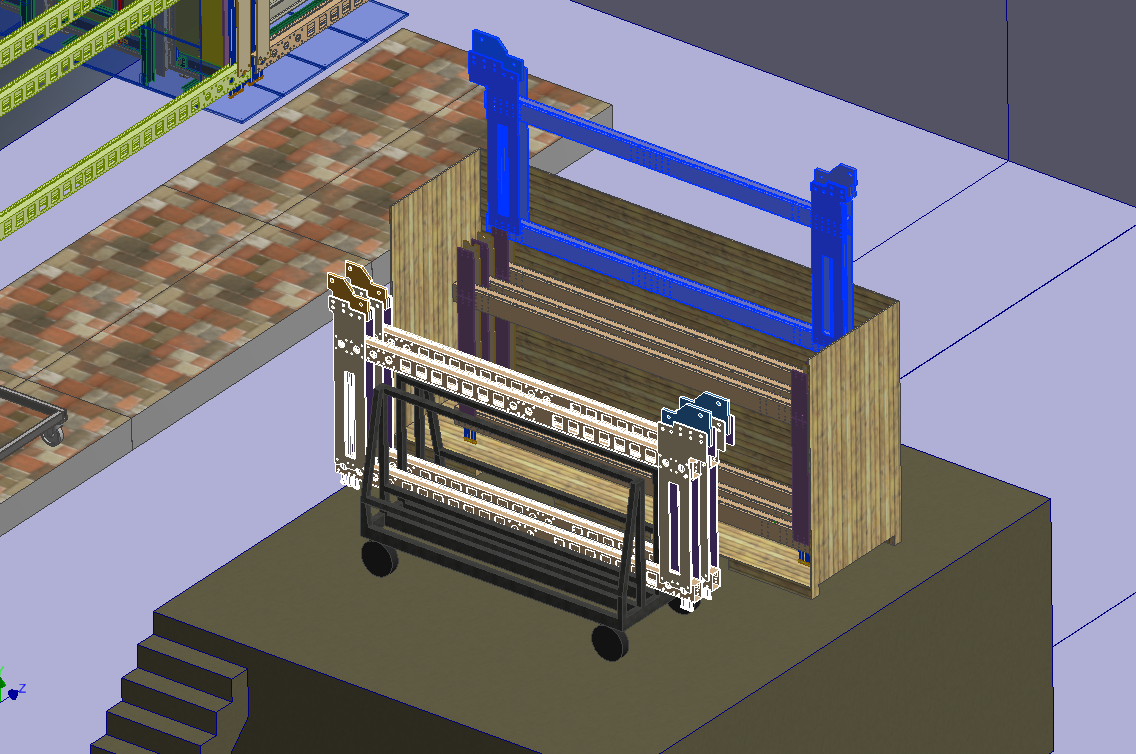}
\end{dunefigure}

The installation of an \dword{apa} and  \dword{cpa} with top and bottom \dword{fc} modules is the most labor-intensive part of the detector installation. Figure~\ref{fig:install-single-row} represents one of the 25 rows of \dword{tpc}.  \dword{dune} aims to perform work in parallel to the extent possible and finish installing one row every week. 
This requires that several separate teams work in the cleanroom, inside the cryostat, and on the cryostat roof simultaneously --  positioning the equipment, integrating \dword{pd} into the \dword{apa}, mounting the \dword{ce} \dword{femb} on the \dword{apa} connecting the cables, cold testing \dword{apa}, installing \dword{apa} in the cryostat, assembling and installing \dword{cpa}-\dword{fc}, and deploying the \dwords{fc}.
Figure~\ref{fig:Single-APA-Schedule} shows the labor breakdown and activities in the airlock,  cleanroom, and cryostat for the \dword{apa} installation. These labor estimates will be refined during time and motion studies at \dword{ashriver}. 
This complicated installation process will be described in steps: first the \dword{apa} assembly work in the cleanroom, followed by the \dword{cpa} assembly, and finally the installation process inside the cryostat.

While the \dword{apa}s, \dword{cpa}, and \dword{fc} are installed, the area outside the cleanroom in the north cavern is available for storage; this area has sufficient capacity to store one full month's worth of equipment. As it is called for, equipment will be brought into the cleanroom's materials airlock through a roll-up or curtain door in the west wall using either an electric forklift or electric pallet jacks.

\begin{dunefigure}[Single row of APA and CPA]{fig:install-single-row}
{One row of the \dword{apa} and \dword{cpa} with associated \dwords{fc}. The \dwords{fc} are shown deployed in the final orientation. The equipment in the figure represents 1/25 of the total \dword{tpc}.}
 \includegraphics[width=0.3\textwidth]{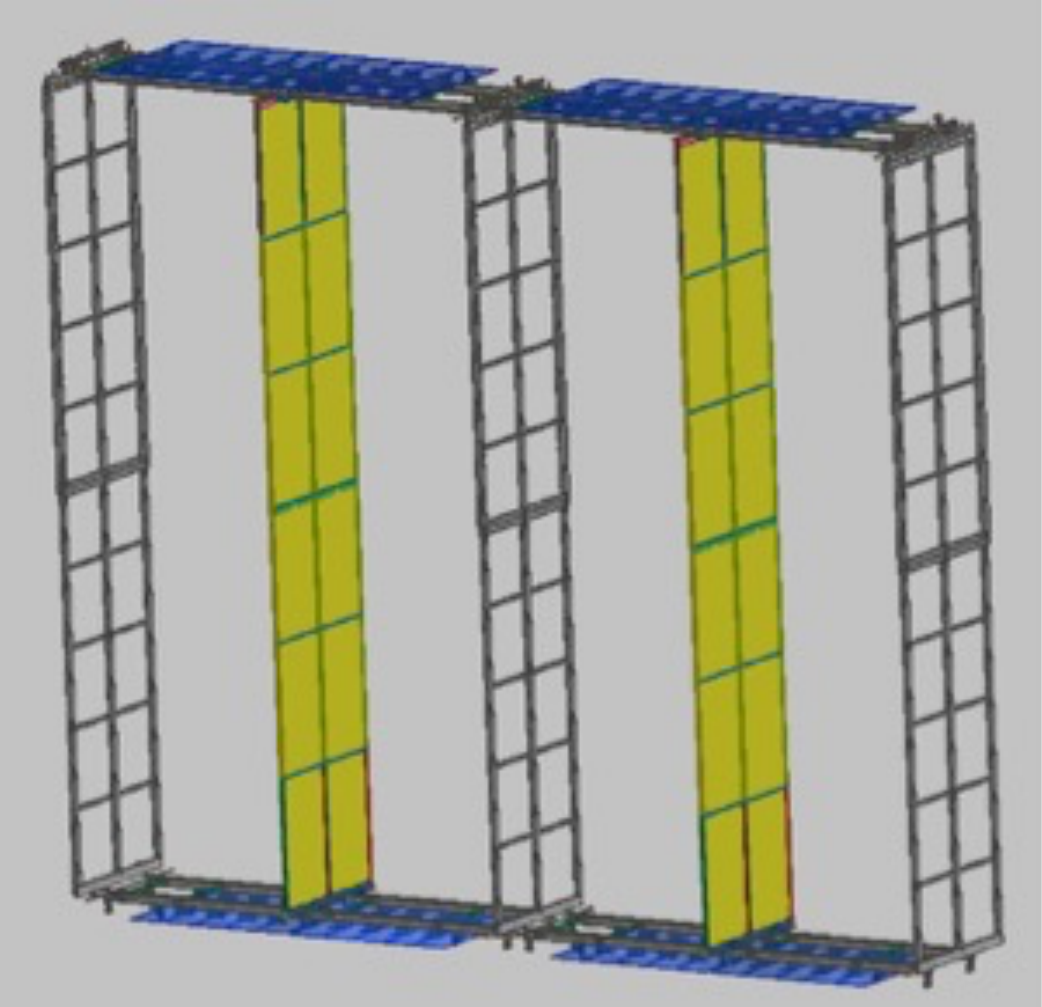}
\end{dunefigure}

\begin{dunefigure}[Typical APA installation schedule]
{fig:Single-APA-Schedule}
{Typical \dword{apa} schedule for \dword{spmod}. As described in Section~\ref{sec:sp-inst-sched}, 
two 10 hour shifts are planned Monday to Thursday and a smaller shift Friday to Sunday for cryogenic and other tests. }
\includegraphics[width=0.95\textwidth]{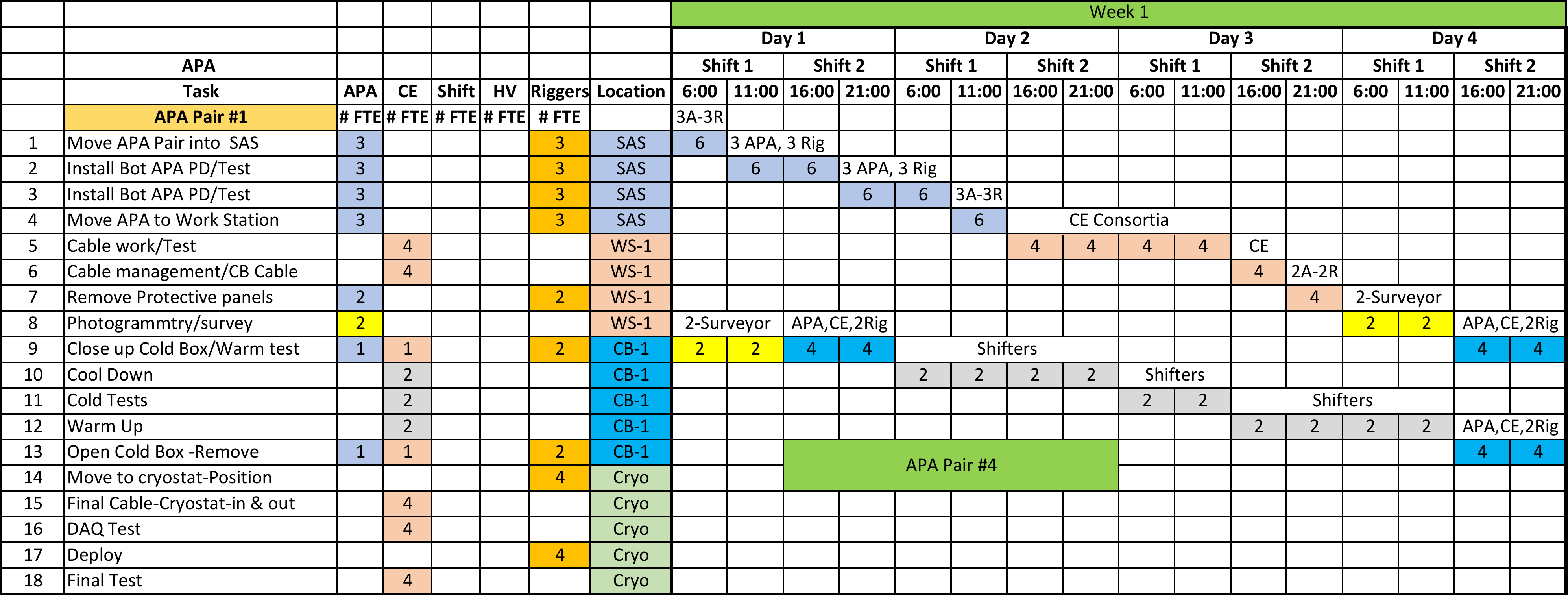}
\end{dunefigure}

An \dword{apa} transport crate that holds two \dword{apa}s  is first rotated to the vertical orientation and bolted to a custom-weighted skid or cart used for moving the crate in the cleanroom. 
Battery powered pallet jacks move the crate into the materials airlock where the outer covers are removed and the outer frame cleaned. 
After the air purity has recovered, the transport box can be brought into the cleanroom proper. 

The \dword{apa}s are first moved to the \dword{pd} integration area where the \dwords{pd} are inserted into the sides of the \dword{apa}s.
The layout of the \dword{pd} integration area is seen in Figure \ref{fig:install-pd-integrate}.
The \dword{apa} transport boxes and \dword{apa} protective covers are designed to keep the slots in the sides for the \dwords{pd} clear so the \dword{pd} modules can be inserted in the sides of the \dword{apa}s without removing them from the transport box. 
The \dword{apa} transport box is placed between two fixed scissor lifts so that two-person teams in the lifts can easily hold a \dword{pd} module on the side of the lift. 
The lift is raised to align the paddle with one of the five slots in the side of the \dword{apa}. The paddle can then slide into the side of the \dword{apa}. 
The guides inside the \dword{apa} frame ensure that the electrical connectors in the middle of the \dword{apa} mate easily. 
The \dwords{pd} are locked into position with two captured screws. 
After each \dword{pd} is installed it can be tested electrically by accessing the connectors at the top using a scissor lift. 
Once the ten \dword{pd} paddles are installed on the first \dword{apa}, the transport crate is shifted slightly and the \dwords{pd} can be inserted into the second \dword{apa} and tested. 
The \dword{apa} transport crate may need to come out from between the lifts to install the lower \dword{pd} modules.

\begin{dunefigure}[Photon detector and APA integration]
{fig:install-pd-integrate}
{Area where the \dwords{pd} are integrated into the \dword{apa} modules. Floor-mounted scissor lifts are used to access the sides of the \dword{apa}.}
\includegraphics[width=0.75\textwidth]{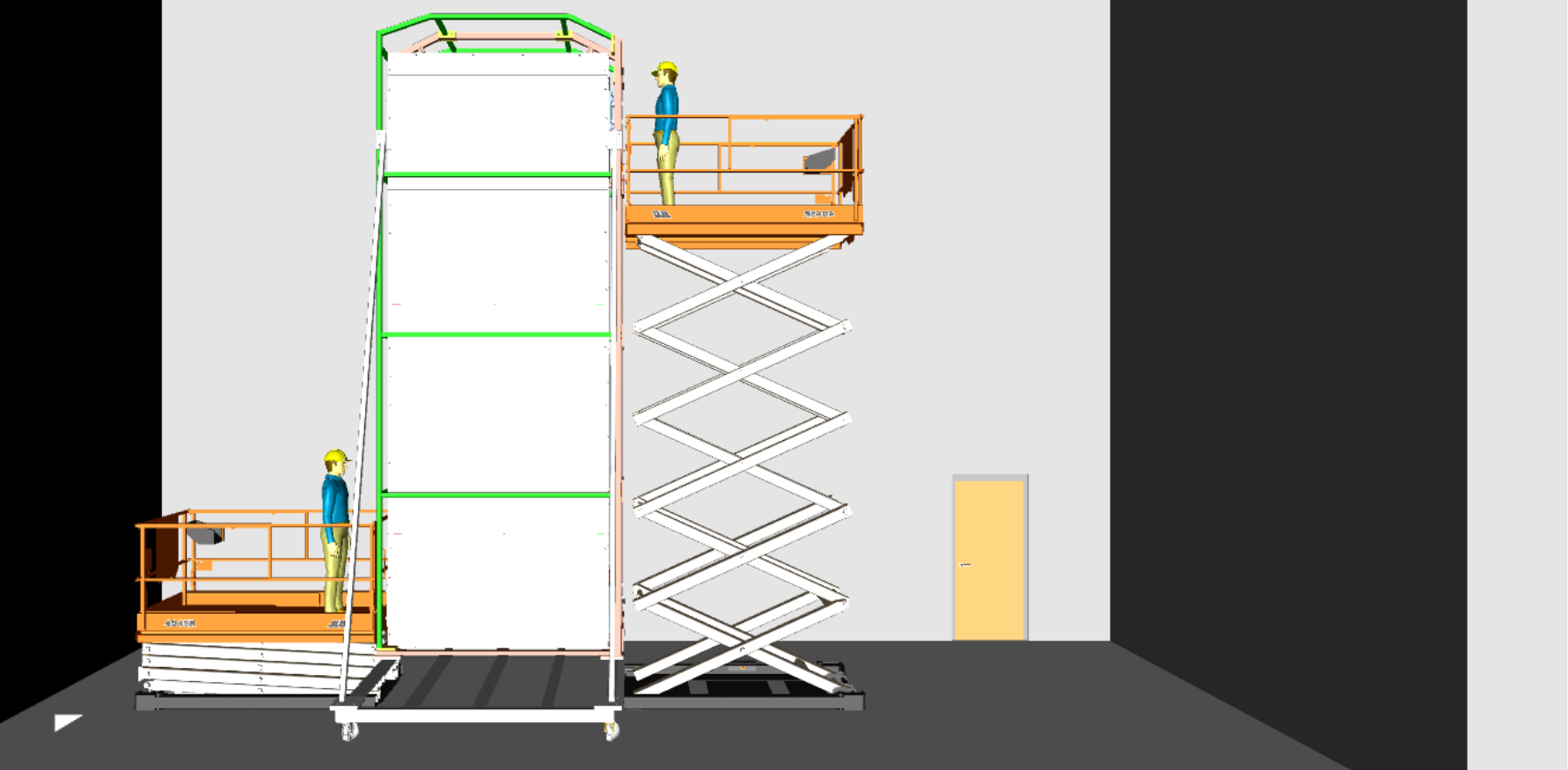}
\end{dunefigure}

After the \dword{pd} integration and testing is complete the transport box with the two \dword{apa}s is moved to the start of one of the four assembly lines (Figure~\ref{fig:install-apa-prep} A). 
The initial time-and-motion studies indicate that three lines are sufficient to keep up with the cold testing and installation in the cryostat; we add a fourth as a spare that can also be used for any needed repair. 

\begin{dunefigure}[Initial APA testing and pair assembly]
{fig:install-apa-prep}
{Initial \dword{apa} testing and assembly into pairs.  Follow row by row from top-left.} 
\includegraphics[width=0.95\textwidth]{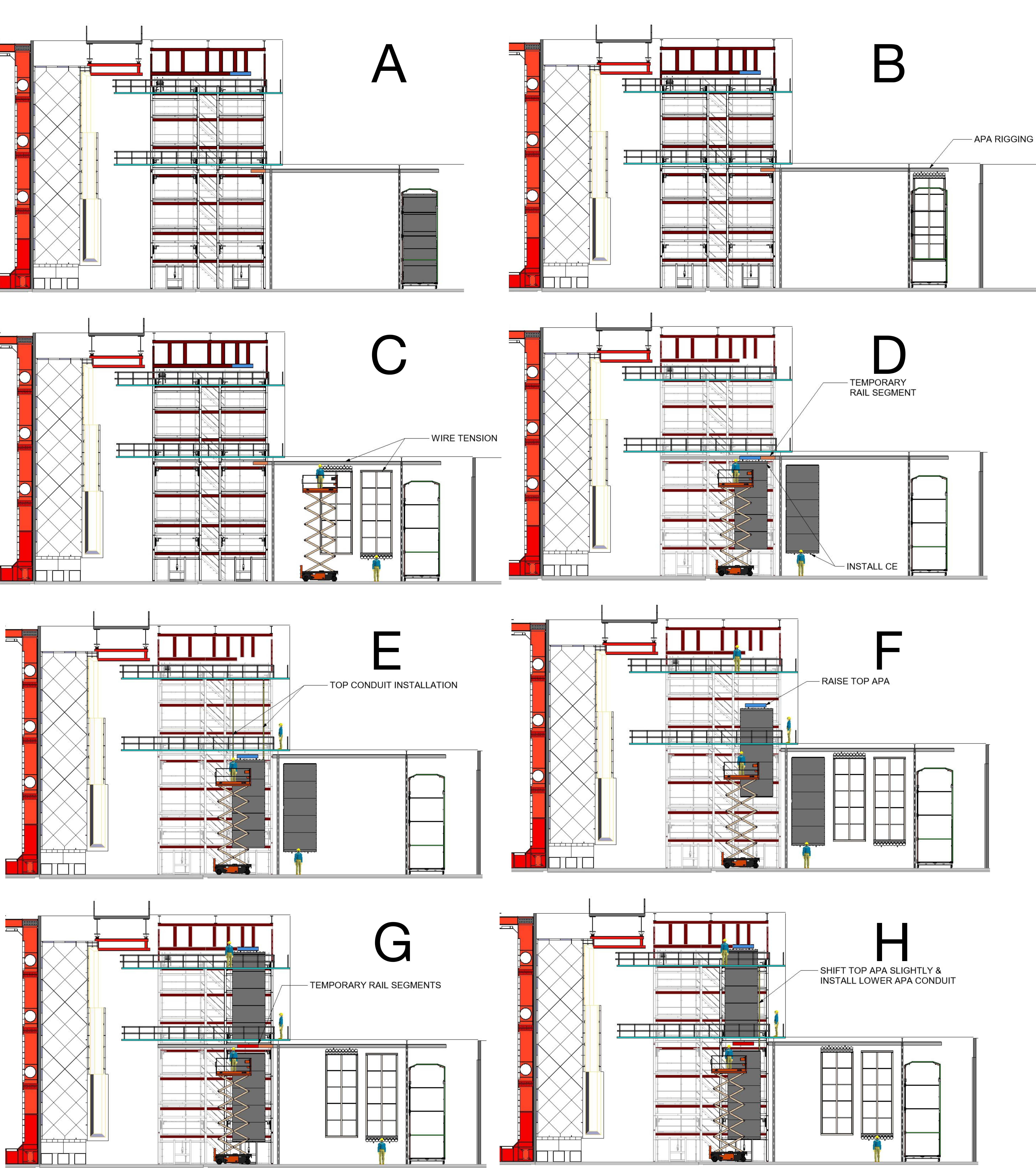}
\end{dunefigure}

First a top \dword{apa} is removed from the transport box and mounted to trolleys on the assembly line rails (Figure~\ref{fig:install-apa-prep} B). 
The \dword{apa} is shifted over to the top \dword{apa} tension measuring station, the protective covers are removed, and a visual inspection performed. 
The bottom of this \dword{apa} cannot support the load of the lower \dword{apa}, so heavy-duty rods are inserted into the sides of the \dword{apa} and bolted to the side tubes using the bolts designed for the linkage connecting the \dword{apa} pair. 
The lower \dword{apa} can then be hoisted out of the transport box and connected to the rail. 
Either the trolleys can be mounted directly to the rods or  a crossbar can be placed between the support rods to hold them.  
Then the lower \dword{apa} is shifted to its tension-measuring location where it is locked in position and its protective covers are removed (Figure~\ref{fig:install-apa-prep} C). 
Wire tension data is collected according to the \dword{qa} plan, similarly to \dword{pdsp}. 
The protective covers are re-installed after the tension measurements to protect the wires during the subsequent assembly steps. 
The top \dword{apa} is moved to the first station on the \dword{apa} assembly tower and attached to a rail section that can be hoisted to the upper level (Figure~\ref{fig:install-apa-prep} D). 
 
The two short sections of the I-beam rail can be removed to the right and left of the beam segment now supporting the top \dword{apa} and the (\SI{6}{m}) cable conduits, needed to install the \dword{ce} cables, are installed (Figure~\ref{fig:install-apa-prep} E). 
Once the conduit is in place, the I-beam segment supporting the \dword{apa} is attached to a hoist, lifted to the upper rails, and attached. 
Locking pins in the \dword{apa} assembly fixture then hold the top \dword{apa} rigidly in position. 
The lower \dword{apa} is then moved into position to be connected to the assembly fixtures. 
At this point the  lower \dword{apa} is supported from the bottom, and guides connected to the sides of the \dword{apa} provide mechanical stability while allowing jacks integrated into the lower support to lift the \dword{apa} and the trolleys and rails it was riding on can be removed.

The cable conduit is installed by freeing the top \dword{apa} and shifting it slightly to allow its insertion from the top through the foot tube, after which  
the \dword{apa} is moved back into position and again locked to the \dword{apa} assembly fixture (Figure~\ref{fig:install-apa-prep} H).
We then test the lower \dword{apa} \dword{pd} paddles to ensure that everything is working. 
At this point, the upper \dword{apa} is supported by the trolleys that move the \dword{apa}s along the upper transport rails, and it is stabilized using the \dword{apa} assembly frame. There is a \SIrange{300}{500}{mm} gap between the upper and lower \dword{apa}, and the 
\dword{pd} cables between upper and lower \dword{apa} can now be connected. 
The connection from the top connectors to the \dwords{sipm} can also be checked. 
To connect the two \dword{apa} modules mechanically, a metal linkage with electrical insulators is inserted into the upper \dword{apa} and bolted into place. Then the lower \dword{apa} is raised until the linkage can be bolted to the lower \dword{apa}.  
At this point the \dword{apa} pair can be released from the assembly tower supports and jacks; it is now supported from the top, where the upper \dword{apa} connects to the transport rail system. The \dword{ce} boxes can be installed at the top and bottom of the \dword{apa} pair and a simple test of the electronics is performed.  
The \dword{apa}s can now be shifted over to the second station on the assembly tower where the cabling is done.

\begin{dunefigure}[APA cabling and cold test]{fig:apa-assembly-v4}
  { Left: \dword{apa} pair  moving on the cleanroom switchyard and cables being inserted on the tower. Right: the \dword{apa}s being inserted into the \coldbox.
  }
\includegraphics[width=.9\textwidth]{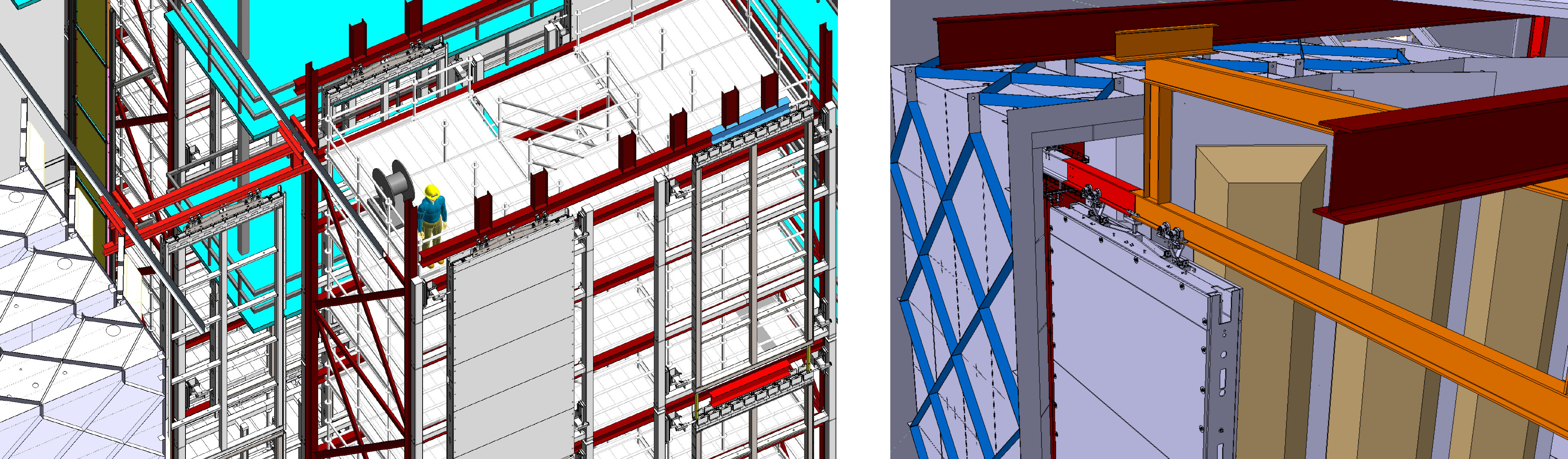}

\end{dunefigure}

The next assembly step is to install and test the electronics cabling.  The left image in Figure~\ref{fig:apa-assembly-v4} shows the \dword{apa} cabling area on the \dword{apa} assembly tower. 
The electronics cables are delivered to the cleanroom on reels,  pre-bundled and tested. 
The switchyard crane lifts the lower \dword{apa} cable reels to the top of the assembly tower, and a cable is spooled over to a motorized deployment spool. 
The cable guide is then attached and fed through a guide sheave and into the conduit on the side of the \dword{apa}. 
The cable bundle is carefully fed through the conduit and anchored in place using a cryogenic-compatible cable grip. 
The cable is connected to the electronics at the bottom and is laid into the cable trays on top. This process is repeated for the second lower \dword{apa} cable bundle. 
Finally, the upper \dword{ce} and \dword{pd} cables can be installed and prepared for transport.

At this time, the functionality of all the electronics is checked. 
After the \dword{apa} electrical test, the \dword{apa} pair is moved onto the switchyard where the protective covers are removed and the assembly is surveyed using photogrammetry. 
The \dword{apa} pair is then transported to a \coldbox where it undergoes a thermal cycle and complete systems test. (The \coldbox{}es were described in Section~\ref{sec:fdsp-tc-infr-cryo}.)
The right image in Figure~\ref{fig:apa-assembly-v4} shows the \dword{apa} being inserted into the \coldbox. The \coldbox is also a Faraday cage, so noise levels can be measured and the \dword{pds} checked for photon sensitivity. 
After the cold test is complete, an \dword{apa} will either move back to a cabling station (if a repair is needed), or into the cryostat for installation. Recall that three assembly lines are needed to keep up with the cold testing but four assembly lines are available. The fourth line will be used for repair and eventually dis-assembly if required. Possible repairs would include repairing electrical connections, replacing electronics modules, replacing photon modules, removing damaged \dword{apa} wires.

\begin{dunefigure}[CPA assembly steps]{fig:install-cpa-assembly}
  {The \dword{cpa} assembly steps are shown. Top row from left:  \dword{cpa}s are delivered to the \dword{cpa} assembly fixture in the cleanroom. The \SI{3}{m} sub-panels are lifted onto the assembly frame and connected. The \dword{cpa} sub-panel is moved in front of the TCO. Bottom row from left: A second sub-panel is assembled. The two sub-panels are combined to make a \dword{cpa} panel. The \dword{fc} modules are attached to the top. The assembly is then moved into the cryostat through the TCO.}
\includegraphics[width=.9\textwidth]{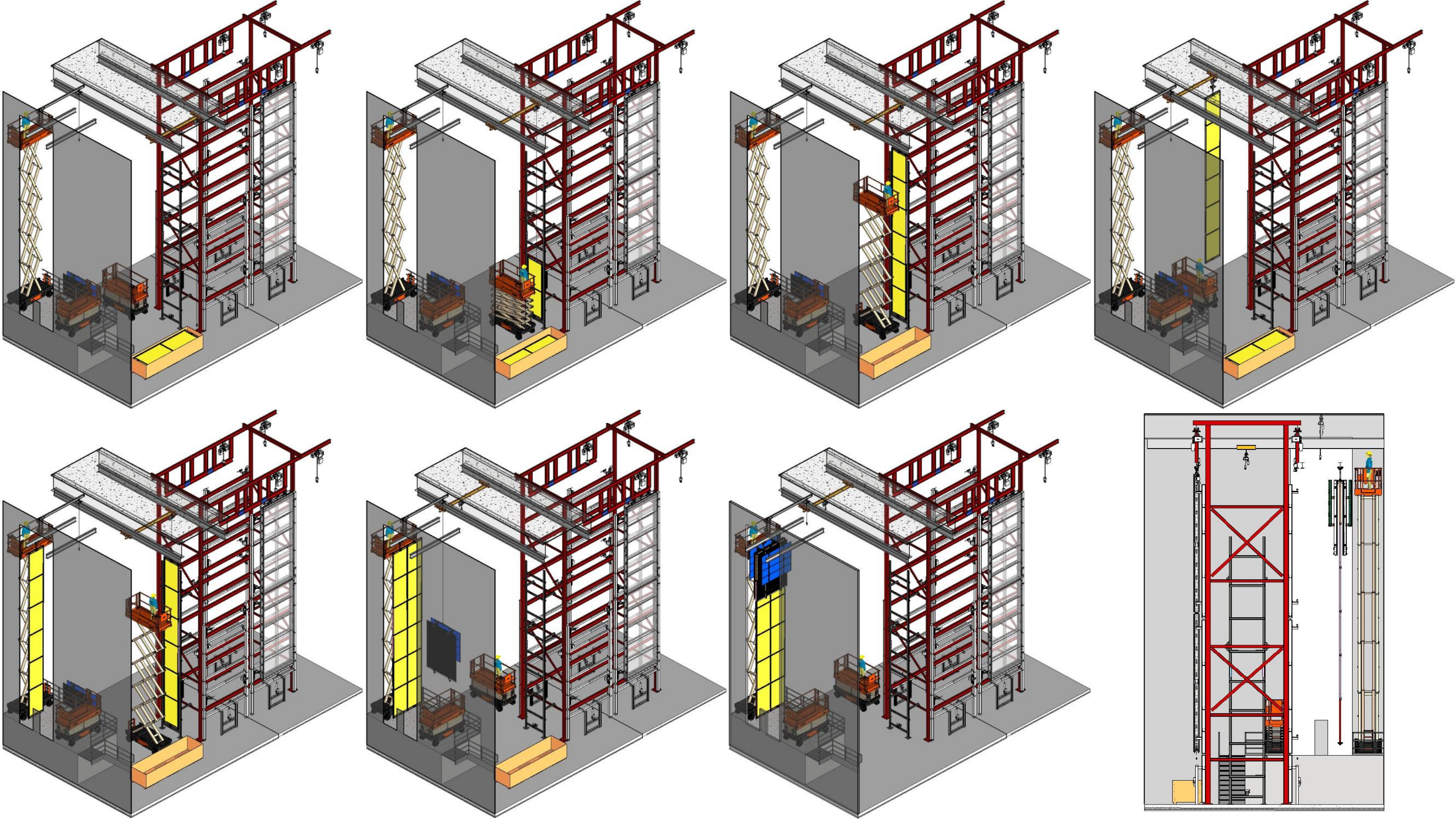}
\end{dunefigure}

The \dword{cpa} and top \dword{fc} modules are assembled in parallel to the \dword{apa} assembly. Figure~\ref{fig:install-cpa-assembly} shows the  assembly sequence. The \dword{cpa} units are delivered to the airlock in crates that hold six \SI{4}{m} long \SI{1.15}{m} wide units. After cleaning, the crates are brought into the cleanroom and opened. The \dword{cpa} units inside are bagged to provide additional dust protection. They are lifted out of the crate and placed on the assembly frame using the cleanroom switchyard hoist. 
The first two of the \SI{4}{m} tall units are assembled together vertically, followed by third one.
The \SI{12}{m} tall \dword{cpa} panel is then lifted, connected to the installation switchyard, and moved to the \dword{tco} beam. The second \SI{12}{m} tall panel is then assembled like the first from the three remaining \dword{cpa} units in the crate. 
The two \SI{1.15}{m} wide panels are then connected to make the \SI{2.3}{m} wide cathode plane.  
A complete set of \dword{qc} measurements is taken of all electrical connections between units and panels.  
The cathode plane  is then moved to a location in the switchyard where the diffuser fibers and top \dword{fc} modules are then attached. 
Finally, the \dword{cpa}-\dword{fc} assembly is moved into the cryostat.
In Figure~\ref{fig:install-cpa-assembly}, the completed assembly is shown outside the cryostat with the lower \dword{fc} modules also attached. 
This is an option, but present planning is to install the lower \dword{fc} modules once it is inside the cryostat.

\dword{pd} monitoring system optical diffusers and short optical fibers must be connected to the \dword{cpa} panels before the panels are installed in the cryostat.  
Discussions are underway about the optimal site for this installation:  at the \dword{cpa} assembly facility before shipping to the site, or as part of the assembly of \dword{cpa} stacks in the underground cleanroom.  
Whichever solution is adopted, quartz optical fibers must be routed from the diffuser to the top of the \dword{cpa} assembly to be connected later to the pre-installed fibers in the cryostat; this connection will occur upon final positioning of the \dword{cpa}.  

Work inside the cryostat proceeds in parallel with the work in the cleanroom. It is critical that the cryostat function as a Faraday cage which shields the \dword{tpc} from external noise sources. A permanent ground monitor will be connected to the cryostat immediately after the construction is complete to monitor if any connections are made between the detector ground and the cavern ground. An acoustical alarm will sound if forbidden connections are detected. This system was used quite successfully at \dword{protodune}.

The large detector components like \dword{apa} pairs and \dword{cpa} modules enter the cryostat using the \dword{tco} rails that connect to the \dword{dss} switchyard. 
Inside the cryostat, the modules are pushed onto one of the switchyard shuttle beams shown in  Figure~\ref{fig:shuttle}. 
The \dword{dss} shuttle beam is then moved to the appropriate row of the \dword{dss}, and  the module is pushed down the length of the cryostat into position. The position of the \dword{dss} beams are well defined and accurately surveyed so that the \dword{apa} and \dword{cpa} modules can be accurately located by precisely positioning them along the \dword{dss} beams. 
A small correction in the height of the modules may be needed to accommodate deflections in the \dword{dss} due to load. Figure~\ref{fig:install-ce-cables} shows the typical situation during the \dword{apa} installation and \dword{ce} cabling. 

\begin{dunefigure}[Cold Electronics cabling inside the cryostat]{fig:install-ce-cables}
  {The installation of the \dword{apa}  and cabling of the cold electronics.
  The left panel shows the \dword{apa}  installation process. 
  One row of \dword{apa}  and \dword{hv} equipment is installed, and a second \dword{apa}  is ready for electrical cabling. 
  The top right image shows the cable trays that will hold the \dword{ce}  cables; one worker is in the scissor lift. 
  The left bottom image shows the work space with the geometry of the \dword{apa}, the cryostat roof, and the cable \fdth .
  }
\includegraphics[width=.95\textwidth]{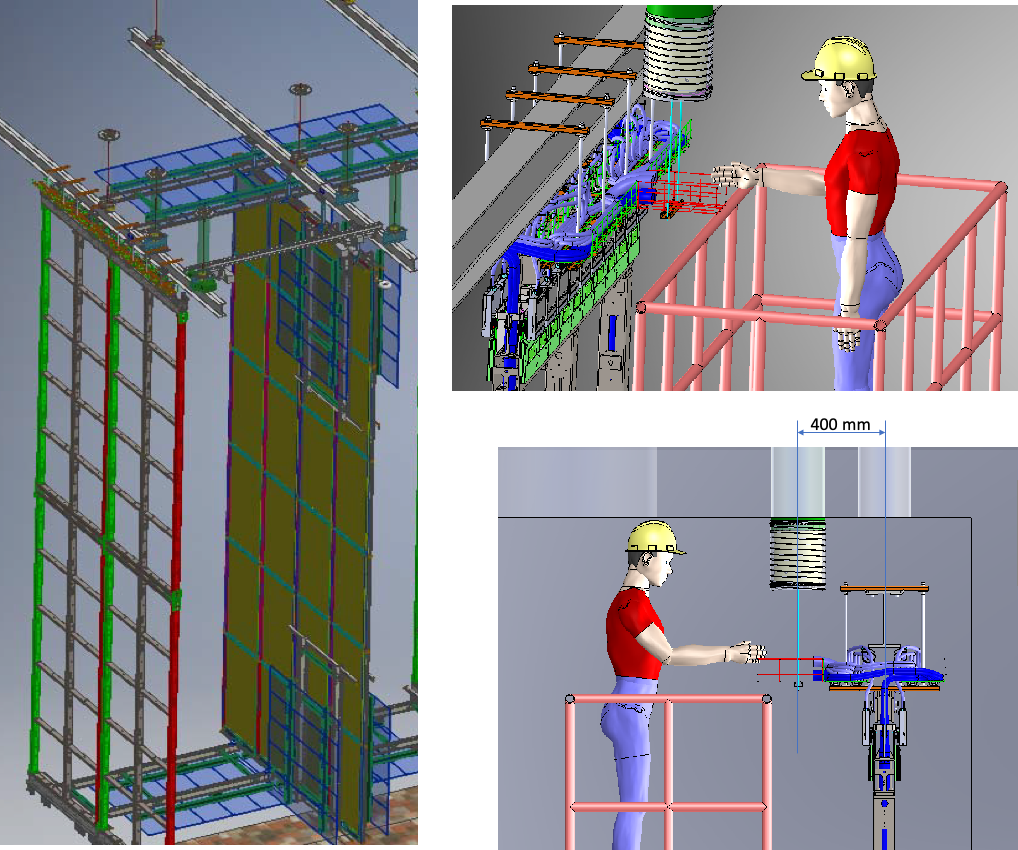}

\end{dunefigure}

After the \dword{apa} is moved into position, the permanent support rod is connected to the \dword{dss} beam, and the trolleys are removed. 
The crawler used to push the \dword{apa} along the rails is then moved back through the shuttle area and can be used for the next module. 
After the \dword{apa} is locked into position, the cable tray \fdth to the \dword{ce} is installed, after which \dword{ce} cabling can start. 
Even when a \dword{cpa} module is in position, more than \SI{3}{m} of space remains free between the \dword{apa} and \dword{cpa}, and a scissor lift can easily be positioned in front of the \dword{apa}. 
The two right images in Figure~\ref{fig:install-ce-cables} show the situation at the top of the cryostat during cabling. The cables are not shown, so the cable trays and their support infrastructure can be seen. 

When cabling begins, all the cables are in the cable trays. 
Two people are in the scissor lift in front of the \dword{apa}, and another two people are on top of the cryostat. 
The \dword{ce} cables from the bottom \dword{apa} emerge from the \dword{apa} side-tube  and are split into two bundles in the cable tray for a total of four cable bundles. 
The top \dword{apa} also has \dword{ce} cables organized into four bundles. 
The photon cables from both the top and bottom \dword{apa}s are bundled into two cable bundles.
During the cabling process, each bundle is partly removed from the cable tray and  fed up through the cryostat crossing tube. 
At the top and bottom of the crossing tube  the cables are strain relieved.
This is repeated for each of the ten cable bundles needed for the \dword{apa} pair. 
When all cables are installed through the cryostat crossing tube, any excess length is returned to the cable tray at the top of the \dword{apa}. 
On the roof, the short individual cables are connected to the \fdth  flange, and all electronics and electrical connections are checked. 
At this point the flange connecting the \dword{wiec} can be sealed to the cryostat \fdth flange, and the cable installation is complete. 

Similarly, the \dword{pd} warm cables are connected to the readout module, and the flange sealed after testing.  
Once testing is complete, the support for the tray holding the excess cabling is transferred to the \dword{dss} beam.  
This minimizes any uneven load on the \dword{apa} pair, so they hang more vertically.   
The electronics for each \dword{apa} is continuously monitored after installation. 

Placement and routing of the cables is complex. 
Figure~\ref{fig:install-cable-routing} shows the working \threed model of the cable routing, showing how the cables will be bundled and placed in the trays. 
A mock up the cabling configuration is planned at \dword{bnl}, and the installation of the cables will be tested as part of the \dword{ashriver} testing program.

\begin{dunefigure}[Model of the electronics and photon detector cabling]{fig:install-cable-routing}
  {Working model of the cable trays and routing of the \dword{pd} cables in the trays. The scale is set by the \SI{2.3}{\meter} \dword{apa} width.}
\includegraphics[width=.95\textwidth]{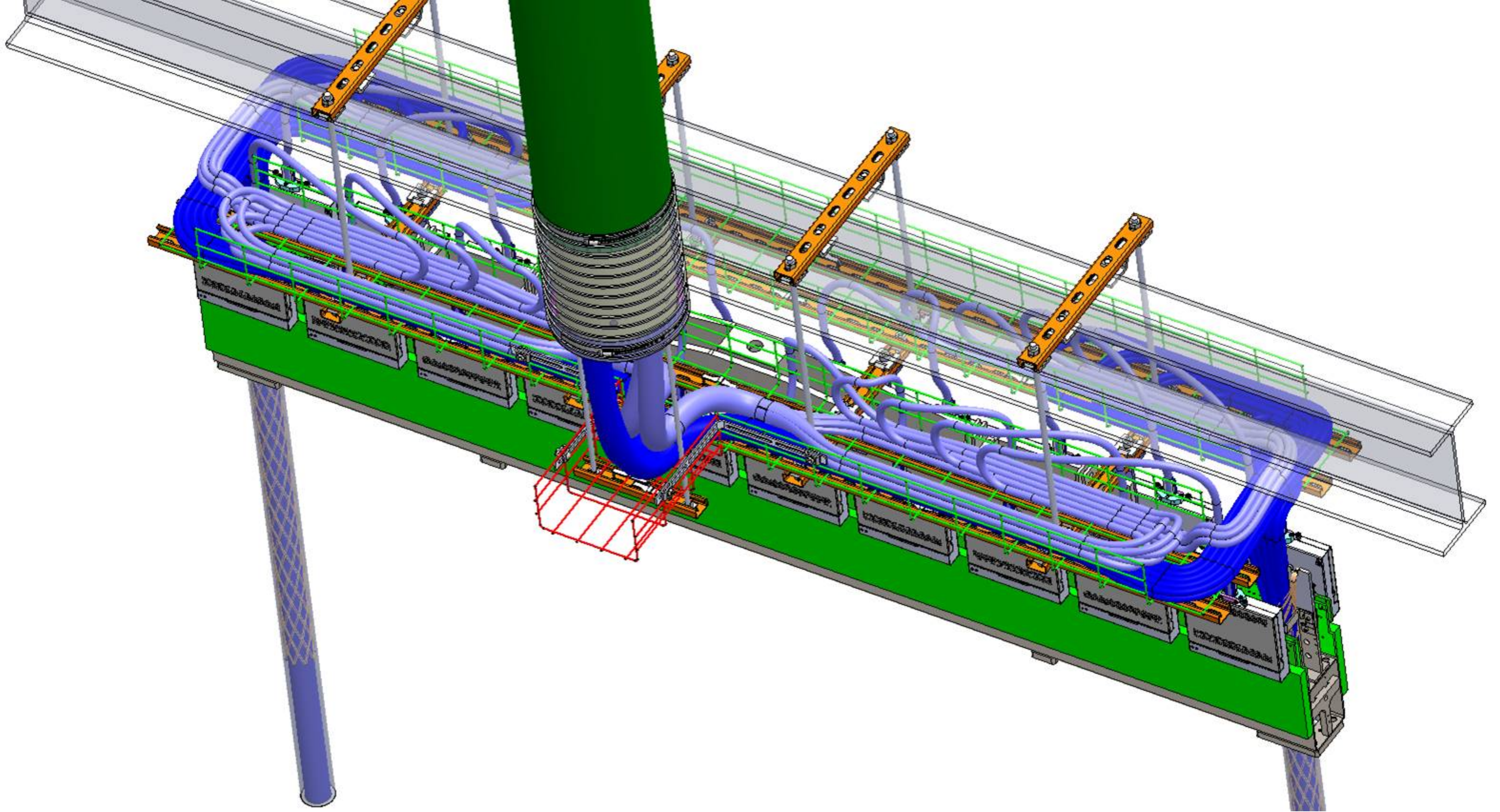}
\end{dunefigure}

\begin{dunefigure}[CPA installation]{fig:install-cpa-fieldcage}
  {The top \dword{fc} assemblies are deployed using a custom tool that mounts to the \dword{dss} beams as seen in the top-left panel. The  \dword{fc} is lifted using the electric winch controlled by an operator in the nearby scissor lift. The lower  \dword{fc} is lowered using a hoist mounted on a wheeled frame. The hoist is on a linear slide to keep it aligned above the connection point.}
\includegraphics[width=.9\textwidth]{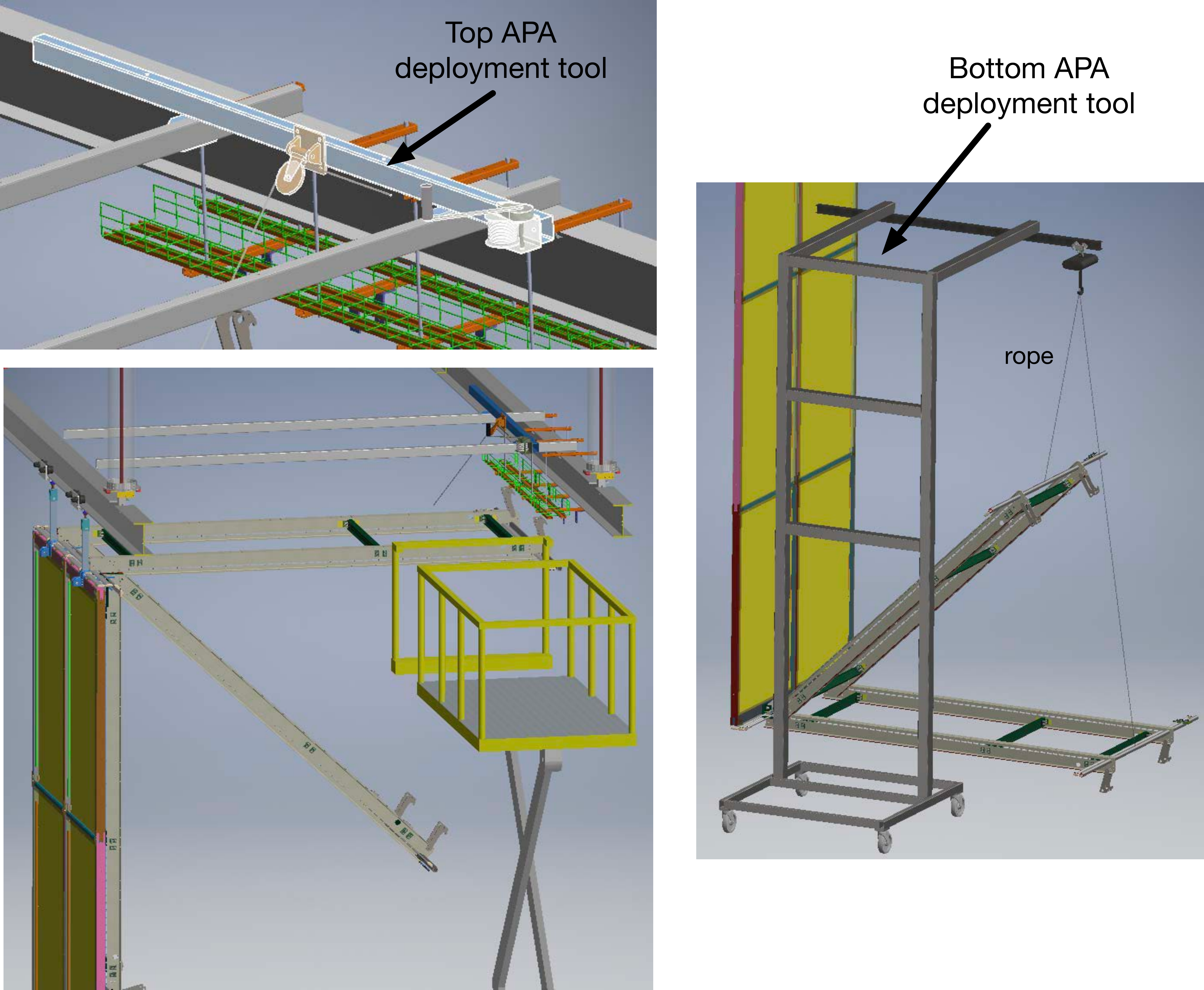}
\end{dunefigure}

The cathode \dword{fc} assemblies are brought into the cryostat like the \dword{apa} pairs, using the overhead rails through the \dword{tco}. 
Inside the cryostat, they are moved into position using the \dword{dss} switchyard and \dword{dss} I-beams. 
Once in position, the load is transferred directly to the \dword{dss} beam, and the trolleys are removed. 
The \dword{cpa} will wait in position until its \dword{apa} pairs are fully tested, after which the \dword{fc} modules could be deployed. 
The deployment sequence of the \dword{fc} has not yet been fixed. 
If the \dword{fc} modules are immediately deployed then the \dword{apa} and \dword{fc} can be tested in the final position. 
If we wait to deploy them, the \dword{ce} can undergo a longer burn-in test and we have an  opportunity to clean the cryostat near the end of the installation. 
The decision on the best time to deploy the \dword{fc} will be made during final design.

Figure~\ref{fig:install-cpa-fieldcage} shows the equipment for deploying the \dwords{fc}. 
The top \dword{fc} modules are raised by connecting a cable to the module and then using a pulley-winch assembly to lift the module, which latches to the \dword{apa} mounts. 
A scissor lift is used to connect the cable to the module and also to control the winch. 
Once the \dword{fc} are latched at the \dword{apa} ends and the \dword{fc} termination connections are made and verified on the \dword{apa}s.

After the module is in place, the deployment tool is moved to the next \dword{apa} and \dword{cpa} sets. 
The lower \dword{fc} is deployed using a custom frame that can be wheeled into position. 
The cable from a small hoist is then attached to the \dword{fc} module, and the module can be lowered. 
The hoist is on a linear slide, so the cable is always directly over the connection point, 
keeping the \dword{cpa} from swinging due to an induced moment. 
When the module is down, it latches to the \dword{apa} frame much like the upper \dword{fc}. 
The electrical connections to the \dword{hv} bus are tested, and deployment is complete. 

In principle, the \dword{cpa}-\dword{fc} assemblies can be constructed faster than the \dword{apa}s and 
the \dword{cpa}-\dword{fc} assembly process could start later than the start of \dword{apa} assembly if the deployment is postponed. The sequence will be baselined prior to completion of preliminary design.

Weekly during the \dword{tpc} installation the \dword{tco} will be optically closed and the cryostat made dark to allow testing of the \dwords{pd} and noise measurements of the \dword{tpc} electronics. 
These tests will be performed on the weekend shift. 

The periscopes for the laser calibration system on the top of the \dword{tpc}  can be installed after the nearby \dword{fc} elements are deployed. The lasers are immediately aligned with the alignment laser system. 
Once for each periscope/laser system, prior to the installation of further \dword{tpc} components, the cavern will need to be cleared for the optical alignment as both UV (Class 4) and visible lasers are used.  
This will require special safety precautions, outlined in \tcchesh 
of this \dword{tdr}.
It may be possible to align all lasers at roughly the same time, to minimize the disruption.

\begin{dunefigure}[Installation of final row of detector components]{fig:install-row25}
  {Detector installation as the last row of detector components is installed. At this time, the switchyard runway beams are removed, and the temporary hoists for the \dwords{ewfc} are installed.}
\includegraphics[width=.9\textwidth]{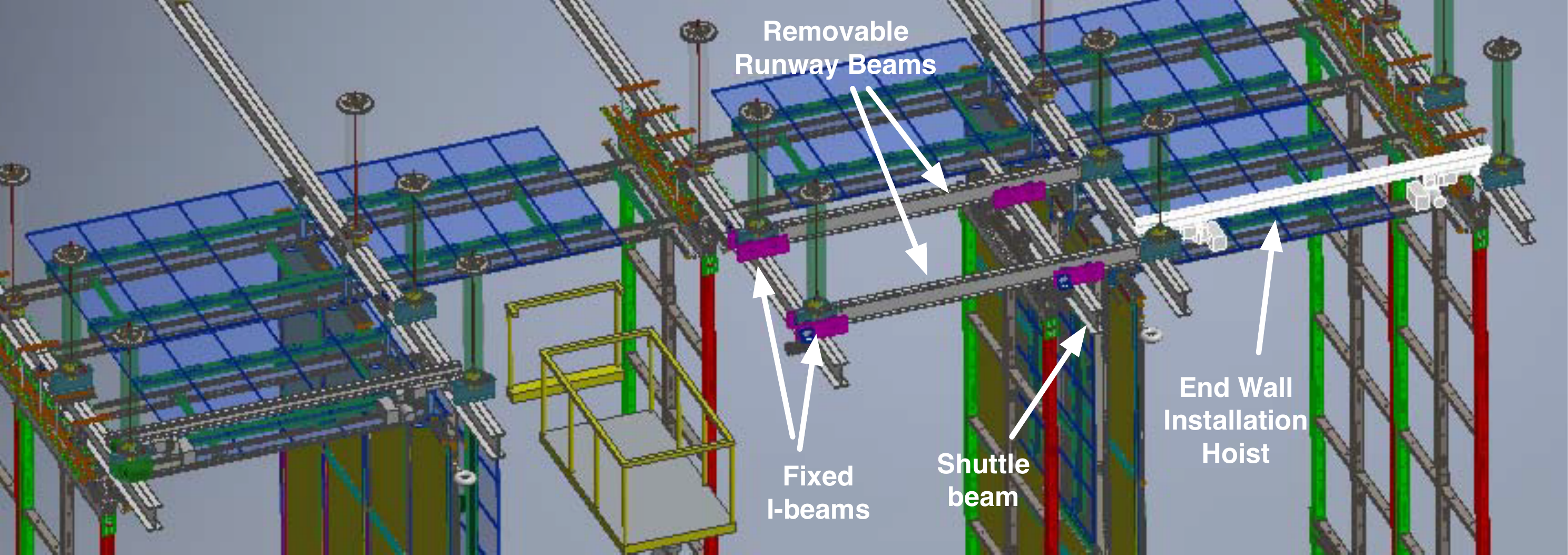}
\end{dunefigure}

The last row of detector elements is installed much like previous rows except that the north-south runway beams in the switchyard need to be removed to allow the last row of the detector to contract as it cools down.  Figure~\ref{fig:install-row25} shows the top of the detector as this last row is installed. When the shuttle beam with the detector component is aligned to the correct \dword{apa} or \dword{cpa} row, it is bolted to a short I-beam section of the runway beam which is in turn permanently fixed to the \dword{dss} support \fdth . When the shuttle beams at both ends of a runway beam section are fixed in position, a section of the runway beam is removed and the \dword{ewfc} insertion hoist mounted. The last \dword{fc} modules could then deployed but they are kept in the folded state to give space for the \dword{ewfc}.

\begin{dunefigure}[Second endwall FC installation]{fig:install-ew2}
  {Installation of the final \dword{ewfc} before closing the \dword{tco}}
\includegraphics[width=.9\textwidth]{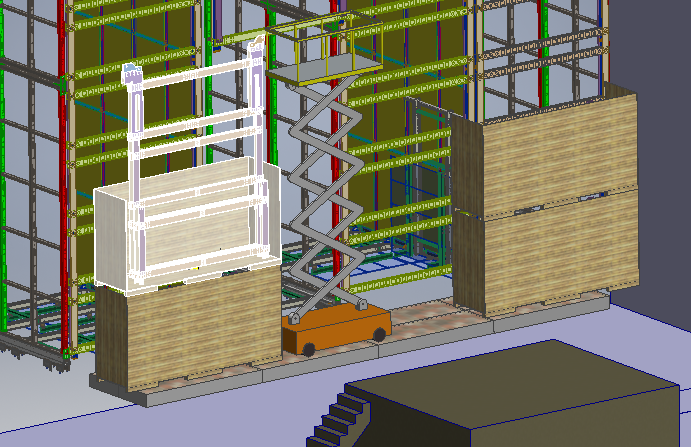}
\end{dunefigure}

The west \dwords{ewfc} are installed like the first \dwords{ewfc} except there is now only \SI{2}{m} of space in which to work. Figure \ref{fig:install-ew2} is an image showing the inside of the cryostat at the time the \dwords{ewfc} is installed. The crates holding the \dwords{ewfc} panels will take much of the available space, but there is room for a person to connect the hoist cables to the \dwords{ewfc} and the hoist can be operated from the scissor lift. After the north and south \dwords{ewfc} panels are installed the center \dword{apa} is rolled into the cryostat, the shuttle beam is bolted to the \dword{dss}, and the runway beams are removed. The two center drift volumes \dwords{fc} are then deployed. The last two \dwords{ewfc} are constructed, and the \dword{tpc} is effectively finished. 

At this point, a frame supported by the shuttle beams is covered with flame-retardant plastic and installed to create a work area for the \dword{tco} closure.  
A scaffold  for egress to the access holes through the roof  must be in place before the \dword{tco} is closed. 
The scissor lift must be removed at this time since the \dword{tco} beams are required to lift it over a piece of the cryostat's structural steel support. 
The \dword{tco} is closed working from the scaffolding inside. 
Once access through the \dword{tco} is blocked, the cryostat is classed as a confined space and the corresponding safety measures are required.
The east end of the cryostat is then cleaned and the plastic sheeting is removed. 

The dynamic T-gradient monitor is installed after the \dword{tco} is closed. 
The monitor comes in several segments with pre-attached sensors and cabling already in place. Each segment is fed into the crossing tube flange one at a time until the entire sensor carrier rod is in place. 
The remainder of the system (the motor system that moves the sensor rod and the sensors) that goes on top of the flange is installed using the bridge crane. 

The purity monitor system will be built in modules so that it can be assembled outside the cryostat, leaving only a few steps to complete inside the cryostat. 
The assembly itself comes into the cryostat with the three individual purity monitors mounted to support tubes that are then mounted to the brackets inside the cryostat. The brackets get attached to the appropriate elements (cables trays, \dword{dss}, and bolts in the cryostat corner are under consideration). Also at this time, the remaining level monitors are installed.

The periscopes at the end of the detector are installed and aligned. 

Once all this work is completed, the scaffolding is taken apart and hoisted out the access holes along with all remaining flooring sections. The area is cleaned, and the last two workers in the cryostat are hoisted out. 

The warm inspection cameras and other possible calibration instruments can be installed from the roof while the \dword{tco} is being closed. 
At this point, the cryostat is classified as a confined space and the corresponding safety measures are required. (These measures include a search of the area before closure, confined-space training for workers, and controlled access into and out of the cryostat. ) After the \dword{tco} is closed the four access holes (two for ventilation and two for personnel access)  can be closed and the pulsed neutron source can be placed in position above two of the access holes. The east end of the cryostat is then cleaned and the plastic sheeting is removed
The pulsed neutron source can be tested to confirm neutron yields with integrated monitors and dosimeters in dedicated runs.

\subsection{Installation Prototyping and Testing (QA/QC)}
\label{sec:fdsp-tc-inst-qaqc}

This section describes the planned \dword{qa} process for developing the installation process and qualifying the installation equipment; the \dword{qc} testing planned during the detector installation follows. 

\subsubsection{Detector Installation Quality Assurance}

An extensive prototyping program designed to develop and test the \dword{pdsp} installation process was executed in the \dword{nova} assembly area at \dword{ashriver} prior to the final design of the \dword{protodune} detector. 
The \dword{nova} Far Detector Laboratory at \dword{ashriver} is owned and operated by the University of Minnesota using grants from the \dword{doe} and \dword{fnal}.
A mechanical mock-up of one sixth of the detector was fabricated from test components where the interface infrastructure was considered final. 
Here all the mounting points, external dimensions, loads, latches and hinges were expected to be exactly as planned for the final \dword{pdsp} detector.
A mock-up cryostat roof and wall were constructed to understand how some tasks could  be performed physically in the space available.
Initially, all the components failed the installation test and had to be modified. A series of hands-on working group meetings with the different consortia were held to resolve installation issues and revise the detector design. 
In some cases, two iterations were required before the components could be assembled together in the space available, and dedicated tooling had to be developed. 
Having the mock-up of the cryostat roof and walls was critical in developing the installation procedures and eventually assembling \dword{pdsp} on schedule; only when handling these objects in the space available did all the constraints become obvious. 
The experience gained during the \dword{protodune} \dword{ashriver}  trial assembly was critical for both verifying the mechanical design and interface, and also for developing the tools and procedures needed for the installation.

The \dword{dune} \dword{tpc} will have half the available work space,  both above and below the \dword{tpc} inside the cryostat, compared to \dword{pdsp}. 
We learned from \dword{pdsp} that access at the top and bottom of the detector was already 
difficult, as was 
properly connecting the bottom latches between the \dword{fc} and the \dword{apa} during the \dword{fc} deployment. 

The process of test-installing the detector mock-up allowed refinement of the hazard analysis and development of the detailed procedure documentation for the assembly process. 
Having complete, well developed procedures prior to the delivery of the components at \dword{cern} allowed the safety approval process to begin early. Along with the test installation, this created a safe work environment.

Mechanical tests at the \dword{dune} trial assembly at \dword{ashriver}  will be key to developing the installation process. We will also perform the time-and-motion studies that are required to develop a reliable schedule.  Other important prototyping tasks performed at  universities, national laboratories, and \dword{cern} will contribute to the installation plan. 
For example, \dword{anl} is testing the \dword{apa} shuttle beam drive system and the \dword{cpa} assembly tower connections before they are shipped to \dword{ashriver}, and \dword{bnl} is planning a test setup to develop the cable management process on top of the detector. 
The \dword{pdsp} experience has led to many small improvements in 
in the assembly process, and the \dword{spmod} prototyping effort will help us develop it further.

\begin{dunefigure}[NOvA Assembly Area at Ash River]
{fig:NOvA-Assembly-Area}
{Top Panel shows the \dword{nova} Assembly Area and the bottom panel shows the \threed model of the installation prototype.}                
\includegraphics[width=0.49\textwidth]{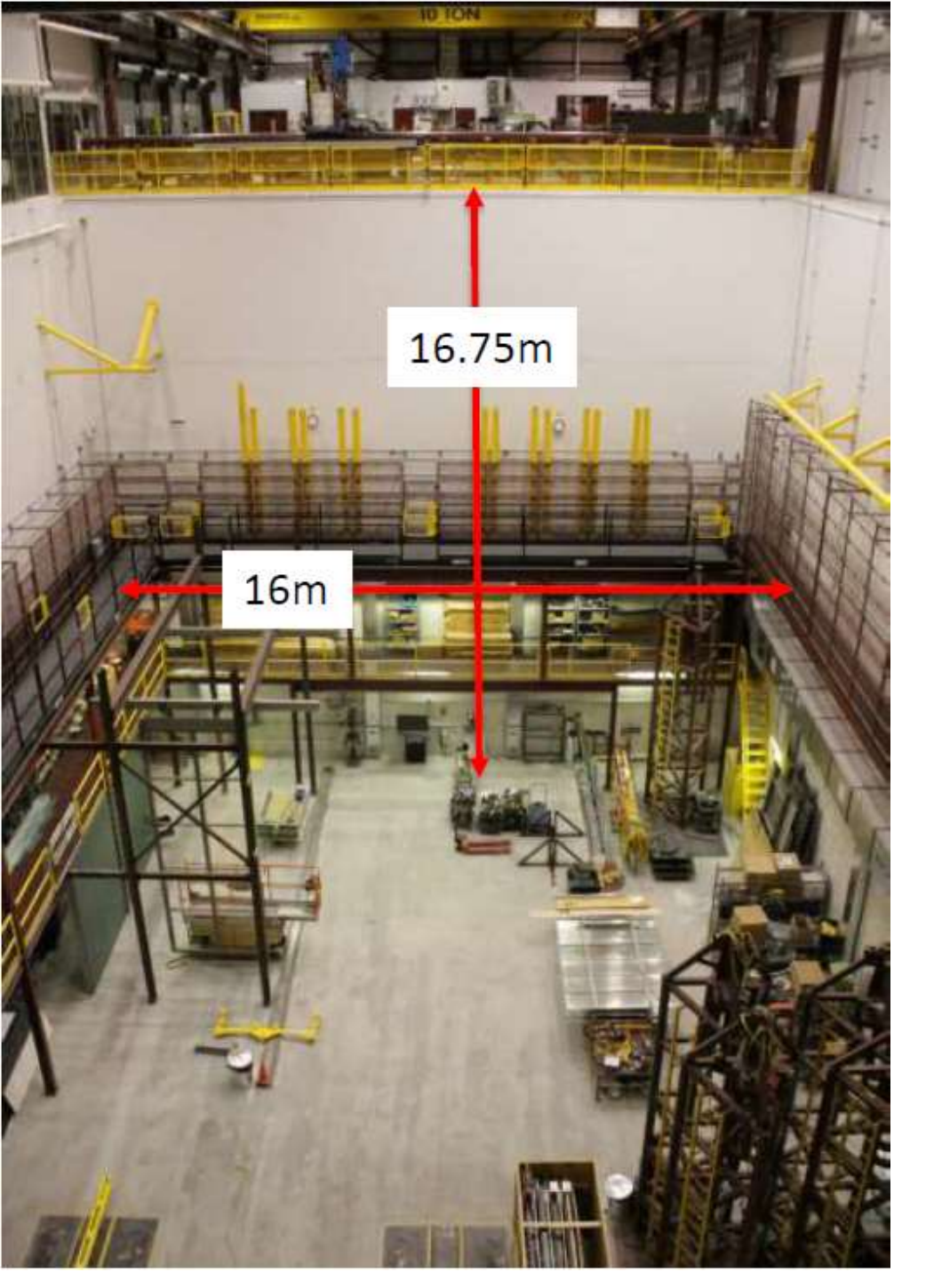}
\vspace{-12pt}
\includegraphics[width=0.7\textwidth,trim=0pt 0pt 0pt 0pt,clip]{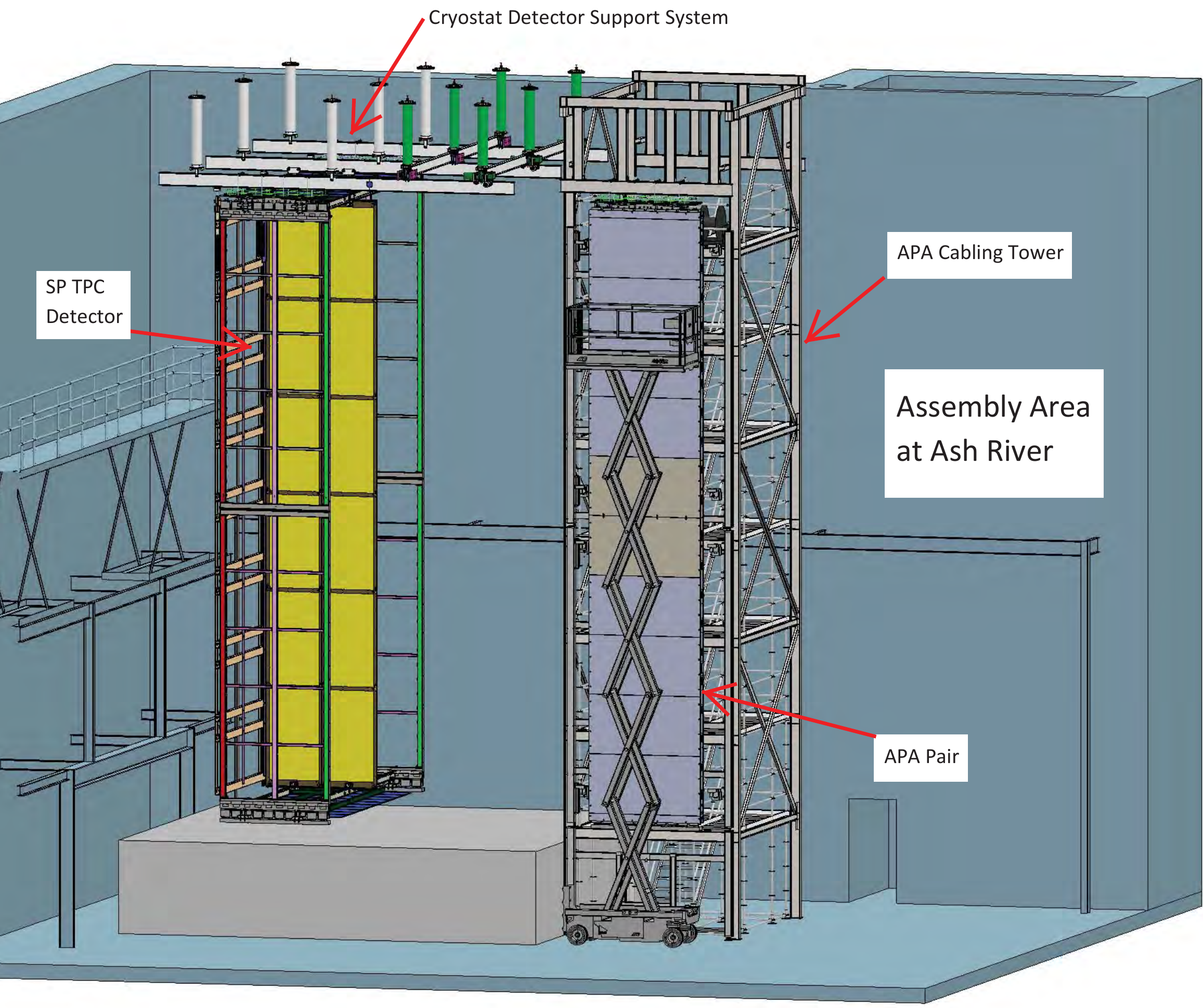}
\end{dunefigure}

Full-scale mechanical testing of the assembly and installation of all the \dword{tpc} components, including the \dword{dss}, will be critical for the success of the \dword{spmod}. 
A prototype of the installation equipment for the \dword{spmod}  will be constructed at the \dword{nova} neutrino experiment \dword{fd} site in \dword{ashriver}, 
and the installation process tested with full-scale mechanical models detector elements.  
This assembly area at \dword{ashriver}, shown in Figure~\ref{fig:NOvA-Assembly-Area}, meets the requirements of both space and available equipment, and experienced technicians that helped construct the \dword{pdsp} detector are available. 
It has both the elevation and floor space available to do a full-scale test of both the component assembly outside the cryostat and a test installation of the \dword{tpc} components  inside the
cryostat. 
Also available are a \SI{75}{ft}$\times$\SI{100}{ft} loading dock and ramp access, two \SI{10}{t} cranes, a machine shop, and a wide assortment of tools. 

While the University of Minnesota has jurisdiction over the safety program at \dword{ashriver}, the facility also follows the \dword{fnal} Safety Program and works together with \dword{fnal} to ensure a safe working environment.  
A key deliverable of the \dword{pdsp} work was a set of documentation including e.g.,   
component design,  hazard analysis, and final assembly procedures, for approval by the \dword{cern} Health Safety and Environment division. 
For \dword{ashriver}  and \dword{dune}, this is all part of the \dword{orc} review process. 
Documentation for both the trial assembly process at \dword{ashriver}  and for \dword{dune} will be stored on the \dword{edms} at \dword{cern}. 
Though many of the \dword{tpc} components are mechanically similar to the \dword{pdsp} components, the access equipment will be different and the need to work at \SI{14}{m} height will make construction of the \dword{spmod} much more challenging.

The \dword{dune} \dword{fd} trial assembly program at \dword{ashriver} has the following goals:

\begin{enumerate}
\item Validate the \dword{apa} design. 
\item Test all full-scale \dword{tpc} components during both the initial assembly stages in the cleanroom outside the cryostat and the deployment stages inside the cryostat:  
\begin{itemize}
    \item \dword{apa}  assembly: manipulation of \dword{apa} shipping frames, joining an \dword{apa} pair together, \dword{ce} cabling, removal and re-installation of the \dword{apa} protection covers, movement on shuttle beam, cryostat cabling, and final deployment in cryostat. 
    \item Integration and installation testing of \dword{pd} components: cable harness routing and cryogenic cable strain relief, module integration into \dword{apa} frames, and electrical connections between upper and lower \dword{apa}s.   Mounting of \dword{pd} monitoring system components and optical fiber routing on the \dword{cpa}. 
    \item \dword{dss} and shuttle beam system, including final detector configuration.
    \item Assembly of \dword{hv} system: construction of an \dword{ewfc}, 
    \dword{cpa} pairs, movement on shuttle beam, and final deployment in cryostat.
\end{itemize}
\item Write full set of hazard analyses and assembly procedure documents; 
gather all component documentation. 
\item Test access equipment (scaffold, scissor lifts, work platforms) and lifting fixtures. 
\item Study assembly time and motion, including labor estimates. 
\item Train lead workers as \dword{dune} begins set up (in the installation setup phase).
\item Test mechanical modifications.
\item Train the installation team prior to the start of \dword{dune} installation (in the installation phase). 
\item Possibly (in future) run assembly tests of the \dword{dpmod} components.
\end{enumerate}

We have developed a staged testing program to meet the above goals.  The initial phase is dedicated to qualifying the \dword{apa} and \dword{fc} designs. The main difference (other than the number of units) between \dword{pdsp} and the \dword{dune} \dword{spmod} is that one \dword{apa} will hang beneath another, doubling the height. The cables from the lower \dword{apa} will need to be routed through the upper \dword{apa}, thus requiring a redesign. 

To date, an \dword{apa} pair has never been assembled and cabled.  It is critical to complete tests of these operations before finalizing the \dword{apa} design. The initial phase of the installation testing is focused on this; it is time critical in order that \dword{apa} production can begin in 2020. The assembly and deployment of the \dword{cpa}s can  be tested in parallel since they will not require large amounts of additional infrastructure.

The second phase of the prototyping program is focused on developing the installation plan and verifying all the detector interfaces. 
We will construct a full-scale model of the major installation equipment in the cleanroom and the inside of the cryostat, and conduct a dry run of the assembly, component transport and deployment in the cryostat. 
This is especially important because the space in the \dword{spmod} cryostat above and below the detector is only half that of \dword{pdsp}, where some of the installation steps were already challenging.

The final prototyping stage includes a mock-up of the top of the cryostat to test the final cabling steps at height and perform  accurate time-and-motion studies to benchmark the installation schedule. Detailed procedures will be drafted and in place before the start of actual installation. The installation team will train to work underground at this time.

Testing the installation process early allows identification of  hazards and remediation measures -- without the time pressure associated with the actual installation. This is critical for reducing risks.  Detector installation is by definition on the critical path, making it vital that the work be performed efficiently and with the lowest possible risk. 

This prototyping program is summarized in Table \ref{tab:AR-test-program} and the \threed model representing the final layout is shown in Figure~\ref{fig:NOvA-Assembly-Area}. 

\begin{dunetable}
[Summary of the tests at Ash River]
{p{.2\textwidth}p{.7\textwidth}} 
{tab:AR-test-program}
{Summary of the tests at Ash River} 
Testing phase & Deliverables\\ \toprowrule
FY-19 20 Phase 0   &  \\ \colhline
 & Build an APA cabling tower for full scale APA pair assembly \\ \colhline
 & Check vertical cabling with a pair of APA side tubes \\ \colhline
 & Test APA shipping frame and underground handling\\ \colhline
 & Build a CPA assembly stand and test assembly process \\ \colhline
 & Test FC deployment and ground plane installation \\ \colhline
  FY-20 21 Phase 1 &  \\ \colhline
  & Build support structure for DSS shuttle, 3 sections of DSS beam \\ \colhline
  &  Test movement of CPA and APA from cleanroom to final destination\\ \colhline
  & Test APA, CPA, endwall and FC deployment in one drift section \\ \colhline
  & Test assembly sequence of final section of TPC \\ \colhline
  & Removal of DSS shuttle beam runway rails \\ \colhline
  & Final deployment after \dword{tco} is closed up \\ \colhline
  FY-21 22 Phase 2&  \\ \colhline
  &  Include the top of the cryostat (no warm structure) with \fdth \\
  \colhline
  & Test DSS installation  \\  \colhline
  &  Test CE cable installation using \fdth \\  \colhline
  & Design \fdth to support Dual Phase installation test \\ \colhline
  & Test shipping and construction using first factory TPC components  \\ \colhline
  & Train lead workers for underground at SURF \\ 
\end{dunetable}

\dword{qc} activities during the integration of the \dword{ce}, \dword{apa}s and \dwords{pd} underground are intended to ensure that the detector is fully functional once the cryostat is filled with \dword{lar}. 
The testing of all detector components will continue throughout the installation of all the elements of the \dword{tpc}, until the cryostat is ready to be filled with \dword{lar}.  All these consortium-provided detector components that arrive underground will have gone through a qualification process to ensure that they are fully functional and that they meet the \dword{dune} specifications. Additional tests and checks will be performed during installation  to ensure that the components have not been damaged during the transport or during the installation itself, and most importantly that all the parts are properly connected.

The individual consortia will retain responsibility for providing quality management, tooling, and test plans at the integration area, as well as specialized labor and supervisory personnel for component integration and installation.

Following the mounting of the \dword{tpc} \dword{ce} and the \dwords{pd}, the entire \dword{apa} will undergo a cold system test in a gaseous argon \coldbox, similar to that performed during \dword{pdsp}. During this test, the system will undergo a final integrated system check prior to installation, checking dark and \dword{led}-stimulated \dword{sipm} performance for all channels, checking for electrical interference with the \dword{ce}, and confirming compliance with the detector grounding scheme.
The \dword{qc} process will be documented through the development of procedures which will define the integration and installation tests required including the appropriate acceptance criteria. 
The integration and installation process will be documented on travelers or manufacturing and inspection plan documents. 
Test results will either appear on these documents or we will have individual test reports including the test data. 
All documents will be retained in the \dword{edms}.


\subsubsection{DAQ QC Testing}

Testing is required at several stages of \dword{daq} installation.  The first is the installation of the data room infrastructure, where, upon installation, professional data center building contractors will test rack airflow, power distribution, and check for cooling-water leaks.

The distance between the detector and the data room is not negligible for multimode fibers and \SI{10}{Gb/s} transmission. 
In order to avoid any issue with signal integrity, path of the detector-to-data room fiber runs will be minimized, high quality multimode OM4 fibers will be purchased, installed professionally and carefully tested in place.
Covered cable trays will protect them after installation.  As \dword{apa}s and servers are commissioned, pre-tested fibers will be connected to the newly installed hardware.

The \dword{daq} servers in the \dword{cuc} data room will be initially received and integrated off site.  Upon installation in the \dword{cuc}, only a simple functionality test is needed.  Sufficient spare capacity will be installed, and the main commissioning work will be software-related, which can be done over the network from the surface or remotely.

\subsubsection{APA QC Testing}

After the \dword{apa} transport boxes are brought into the material airlock the outer covers are removed and a visual inspection can be performed. Next the \dword{apa}s are moved to the \dword{pd} integration area, tests here are described in the \dword{pd}  section. 

When the \dword{pd} integration is complete the \dword{apa} transport box is moved to one of the \dword{apa} assembly lines. Individual \dword{apa}s are removed from the transport frame, mounted to the assembly line rails, and the protective covers are removed. A second detailed visual inspection is performed now that the wires are visible. A spot check of the wire tensions will be done to verify that no change has occurred since the \dword{apa} left the factory.

In the current plan, the tension measurements are performed using a laser focused on individual wires, the same method used at the production site. 
The wire is plucked to induce a vibration, and a photodiode under the wire records the frequency of vibration, which directly translates into the tension value. 
The measured values are stored in the wire \dword{dcdb} database. 
While this method is robust and has been extensively used by \dword{lartpc} experiments, it is very time consuming and thus prohibits measuring every wire. 
Two people over three shifts will be able to measure approximately 350 wires, 10\% of the total. In \dword{pdsp} 10\% of the wires were measured prior to cold testing and no wires were found to be out of tolerance.

An alternative method, using electrical signals, is currently under development and could replace the laser method, potentially allowing measurement of all the wires. 
With this electrical method, where adjacent wires under certain voltages induce the middle wire to vibrate, the resonance frequency vibration of the measured wire correlates directly
to the wire tension. 

The current requirement for tension values are 6$\pm$1 N, however this tolerance is currently under study with \dword{protodune} data.  
If wires are found in \dword{dune} that are outside the tolerance specification then the wire will be removed and a more detailed study of the \dword{apa} performed. 
No more than 25 missing wires will be permitted. 
\dword{apa} wires are also tested for continuity, to make sure they are intact and properly connected to the readout boards.
This test is done as part of the \dword{tpc} electronics testing below. 
Photogrammetry is used to measure the final assembled dimensions of the \dword{apa}, either while the wire tension is being measured or immediately before entering the \coldbox. A measurement of the wire-plane spacing is also performed using a scanning laser combined to a Faro arm (a portable coordinate measuring machine). The exact wire-plane spacing values will be stored in the wire \dword{dcdb} database.  In the case of  
any deviations from the required tolerance of \SI{0.5}{mm}, different bias voltage values may be used to make corrections to the wire plane transparency. 

Once the \dword{apa}s have been installed, the \dword{tpc} electronics will be continuously read out;  this will directly inform wire continuity and the full function of the channels. 

\subsubsection{TPC Electronics QC Testing}

Many of the activities of the \dword{ce} consortium at \dword{surf} aim to ensure the full functionality of the \dword{tpc} once the cryostat is filled with \dword{lar}. 
All the detector components provided by the \dword{ce} consortium that arrive at \dword{surf} will have gone through a qualification process to ensure that they are fully functional and that they meet the \dword{dune} specifications. 
Additional tests and checks are performed at \dword{surf}
to ensure that the components have not
been damaged during the transport or during the installation itself,
and most importantly that all the parts are properly connected.

\dwords{femb} are tested multiple times during this process,  
first after they are received and then 
after 
installation on the \dword{apa}s. Further 
tests are performed before and after the 
\dword{apa}s are installed in the cryostat, using the final cables to connect the \dwords{femb} and the detector flanges. 
Results of these tests at \dword{surf} are compared with the results of the
tests performed during the qualification of \dwords{asic} and
\dwords{femb} to detect possible deviations that could signal 
damage in the boards or problems in the connections. All test 
results will be stored in the same database system used for
results obtained during the qualification of components.

The post-installation \dword{apa} tests, 
performed at room temperature,  
involve connecting up to four \dwords{femb} to a \dword{wib} that is 
connected directly to a laptop computer for readout over 1~Gbps
Ethernet, with power provided by a portable \SI{12}{V} supply. For 
the reception test, the \dwords{femb} are attached to a capacitive 
load to simulate the presence of wires, which allows a test of connectivity, 
and measurement of the baseline and \dword{rms} of the noise for 
each channel. Dead channels are identified using the calibration  pulse internal to the \dword{fe} \dword{asic} as well as the measured
noise level relative to that associated with the temporary capacitive load.
Overall, the reception test and the test performed after attaching the
\dwords{femb} to the \dword{apa}s each require approximately half an hour per \dwords{femb}, 
including  the time for connecting and disconnecting test cables.
The \dword{ce} consortium plans to have a cryogenic test stand available
in a laboratory on the surface at \dword{surf} or at a nearby institution to perform checks at \dword{ln} temperature
of \dwords{femb} that fail the \dword{qc} procedures at \dword{surf},
and eventually for sample checks on the \dwords{femb} as they are received
at \dword{surf}.

Once the pair of \dword{apa}s is in the \coldbox an initial test of the readout is performed at room temperature, to ensure the final cables are
properly connected to the \dwords{femb}. This test is done using elements of the final \dword{daq} system.
Fast Fourier transforms of
the noise measurements made in the closed \coldbox will be inspected for indications
of coherent noise. All \dword{fe} gain and shaping time settings will be exercised,
and the gain will be measured using the integrated pulser circuit in the \dword{fe}
\dword{asic} and/or the \dword{wib}. The connectivity and noise measurements, as well
as the check for dead channels, are repeated later after the \dword{apa} pair cools
down to a temperature close to that of \dword{ln} in the \coldbox. The bias voltage
connections and the \dword{pds} are also checked at this time.

Results of all these tests will be compared with results obtained 
in earlier \dword{qc} tests.  If problems are found, it will be possible 
to fix them by re-seating cables or replacing individual \dwords{femb}.
Noise levels are also monitored during the \cooldown and warm-up 
operations of the \coldbox{}es. These tests also ensure that the power,
control, and readout cables are properly connected
on the \dword{femb} side and that this connection will withstand temperature 
cycles. 
Although the connection between the cables and the \dword{femb}
has been redesigned to address a problem 
seen in \dword{pdsp}, repeating these tests during integration
and installation of the \dword{tpc} is important because a single connection problem would
result in the loss of one entire \dword{femb}. 
In addition, the tests 
performed in the \coldbox at \dword{surf} will demonstrate that the power, control, and
readout cables for the bottom \dword{apa}s are not damaged when they are routed 
through the \dword{apa} frames. 
The \dwords{femb} are tested immediately after installation, after the cables are installed, and during and after thermal cycling. This ensures that the connections are robust before the \dword{apa} enters the cryostat.  Testing at \dword{ln} temperature, done with the final cables attached, indicates clearly the capacitance of the wires and it verifies that the connections to the \dwords{femb} and the cables is maintained during thermal cycling.  After installation in the cryostat the \dwords{femb} are monitored continuously. 
Additional measurements of the noise
level inside the cryostat will be performed regularly by closing 
the \dword{tco} temporarily with an \dword{rf} shield electrically connected 
to the cryostat steel. 

All readout tests are repeated after the \dword{apa}s are put
in their final positions inside the cryostat and after the power, control, and
readout cables are connected to the warm flange attached to the cryostat
penetration. At this point, the connection between the cables and the flange
is validated, and the entire power, control, and readout chain, including the
final \dword{daqbes} used during normal operations, are exercised. The
installation plan for the \dword{tpc} components inside the cryostat (\dword{apa}s,
\dwords{cpa}, and deployment of the \dwords{fc}) allows for minor repairs on some \dwords{femb} without extracting
the \dword{apa}s from the cryostat. The testing of all  components
will continue throughout the installation of 
the \dword{tpc}, 
until the cryostat is ready to be filled with \dword{lar}.  When the 
\dword{apa}s are in their final position, replacing \dwords{femb} 
or cold cables will be more difficult and may require extracting the \dword{apa}s 
from the cryostat. This operation will be performed only if major problems occur with the \dwords{femb}.

In addition to measurements
on the \dword{apa}
readout, 
the \dword{ce} consortium will also
test the bias voltage system together with the \dword{apa}
and \dword{hv} consortia. These tests should show that
the cables providing the bias voltage to the \dword{apa} wires, the
\dword{fc} termination electrodes, and the electron diverters are connected
properly with no short circuits. 
These tests will commence 
as soon as 
the first \dword{apa} pair 
is in its final position, but after connecting the bias voltage cables to the 
\dword{shv} boards on the \dword{apa}. 
The connection will use a resistive load for the 
\dword{fc} termination electrodes and the electron diverters. This ensures
the continuity of the bias voltage distribution system from the bias voltage
supplies to the \dword{apa}s. The test must be repeated for the \dword{fc}
termination electrodes and the electron diverters after 
the \dword{fc} modules are deployed.

Additional tests will be performed on the other components provided by the TPC electronics   consortium prior to insertion into the \dword{apa}s. 
After the cryostat penetrations are put in place, helium leak checks
will be performed. 
These tests will be repeated 
after all cables have been routed through the cryostat penetration.
As soon as the bias voltage and the power supplies are installed on the cryostat mezzanine and cables are put in place between the corresponding racks and
cryostat penetrations, tests will be performed to ensure that the proper 
power and bias voltage can be delivered to the \dwords{wiec} before installing them. 
Even before connecting the \dwords{wiec} to the warm flanges,
tests will be performed to ensure that they can be properly powered up, controlled, and read out by the \dword{dune} \dword{daq}.
Tests will be performed on the readout fiber plant to ensure that all fiber connections are functional and properly mapped. 
Additional tests will be performed on the slow control
system and on the detector safety system several times during the
installation of the detector. These tests will take place before the
corresponding \dword{apa}s are installed, after their installation, after
all the corresponding cables and fibers are connected, and finally
during the integrated tests that take place before
the \dword{tco} is closed and the cryostat is filled. Negative results in any
of these tests will halt integration, installation, and
commissioning activities. The results will then be used in reviews that must
take place before the closure of the \dword{tco} is authorized, the \dword{lar} filling operation takes place, 
and the detector is
commissioned. 

\subsubsection{HV QC Testing}

The \dwords{ewfc} are assembled in eight panel units, four on each end of the \dword{tpc}.  
As each of the eight panels is  removed from the shipping crate and placed on the installation cart, the endwall panel checklist is filled out~\cite{bib:docdb10452}.
This checklist includes a visual inspection of the frames, profiles, and connections, as well as continuity and resistance measurements of the divider boards and their connections.  
After completing an eight-panel endwall in the cryostat, the complete endwall checklist is filled out.  
This includes hanging position, straightness measurements, and continuity checks between panels.

The \dword{cpa} panels are assembled from a set of three units removed from the shipping crates.  
After each unit is removed from its bag, a visual inspection confirms structural integrity and the connections between \dwords{fss}, \dword{hv} bus pieces, and profiles, if present.  
After inspection, the unit is positioned on the \dword{cpa} assembly tower.  
After all three units are connected on the tower, the \dword{cpa} panel checklist is filled out.  
This includes inspecting all mechanical connections, continuity checks of the \dword{fss}, \dword{rp}, profile, and \dword{hv} bus connections, and resistance measurements of the four mini-resistor board connections from the \dword{rp} to the \dword{fss}.  
This is repeated for the second panel in a \dword{cpa} plane.  Then the two panels are paired, each hanging from trolleys on the transport beam.  
Visual inspection of the alignment and hanging straightness is made, and \dword{hv} bus connections at the top and bottom are made and checked for continuity (\dword{cpa} plane checklist).

The \dword{fc} top and bottom units are removed from their crates.  The \dword{fc} unit checklist is filled out with a visual inspection of the frames, profiles, and connections, as well as continuity and resistance measurements of the divider boards and their connections.  
After hanging the top \dword{fc} units on the \dword{cpa} plane, the four jumpers from the first \dword{fc} profile on each side of the \dword{cpa} and the \dword{cpa} \dword{fss} is connected, and the resistance is measured, completing the \dword{cpa}-\dword{fc} top assembly checklist. 
The \dword{fc} bottom units are not attached to the \dword{cpa} but are taken into the cryostat independently after filling out the \dword{fc} unit checklist.

The \dword{cpa}-\dword{fc} top assembly is moved into its position in the cryostat.   
After deploying the \dword{fc} top units, the resistor board/jumper between the \dword{fc} and the \dword{cpa} \dword{fss} are visually inspected.  
Also, the latch connecting the \dword{fc} to the \dword{apa}  is visually inspected.  After deploying the \dword{fc} bottoms, visual inspection verifies the resistor board/jumper connection from the \dword{fc} to the \dword{cpa}.  
Also, the latch connecting the \dword{fc} bottom to the \dword{apa} is visually inspected.  
These are included in the \dword{cpa}-\dword{fc} cryostat checklist.

\subsubsection{CISC QC Testing}

Cryogenics instrumentation systems must undergo a series of tests to guarantee they will perform as expected: 

\begin{itemize}
\item {Purity monitors}: Each of the fully assembled purity monitor arrays is placed in its shipping tube, which serves as a vacuum chamber to test all electric and optical connections at \dword{surf} before the system is inserted into the cryostat. During insertion, electrical connections are tested continuously with multimeters and electrometers.

\item {Static T-gradient thermometers}: Right after each sensor array is installed, its verticality 
is checked, and the tensions in the stainless steel strings adjusted as necessary. Once cables are routed to the corresponding \dword{dss} ports, the entire readout chain is tested. This allows a test of the sensor, the sensor-connector assembly, the cable-connector assemblies at both ends, and the noise level inside the cryostat.
If any sensor presents a problem, it is replaced. If the problem persists, the cable is checked and replaced as needed.

\item {Dynamic T-gradient thermometers}: The full system is tested after it is installed in the cryostat. Two aspects are particularly important: the vertical motion of the system using the step motor, which is controlled through the slow controls system, and the full readout chain, which is tested mainly for failures in sensors, cables, and connectors inside the cryostat. 

\item {Individual sensors}: To address the quality of individual precision sensors, the same method is used as for
the static T-gradient monitors. For standard \dwords{rtd} to be installed on the cryostat walls, floor, and roof, calibration is not an issue. Any \dword{qc} required for associated cables and connectors is performed following the same procedure as for precision sensors.

\item {Gas analyzers}: Once the gas analyzer modules are installed at \dword{surf} and before the cryostat is commissioned, the analyzers 
are checked for both \textit{zero} and the \textit{span} values using a gas-mixing instrument and two gas cylinders, one with a
zero level of the gas analyzer contaminant species and the other with a known percentage of the contaminant gas. This verifies that the gas analyzers are operating properly. 

\item {Liquid level monitoring}: Once installed in the four cryostat corners, the capacitive level meters are tested in situ 
using a suitable dielectric in contact with the sensors.

\item {Cameras}: After installing and connecting the wiring, fixed cameras, movable inspection cameras, and the light-emitting system are checked for operation at room temperature. Good quality images should be obtained of all cryostat and detector areas chosen from the system's design.  
\end{itemize}

\subsubsection{Photon Detector QC Testing}
\dword{pd} modules will arrive underground in custom crates.  
Each crate will contain the ten modules required for one \dword{apa}, each individually packaged in a static-resistant sealed plastic bag filled with clean dry nitrogen. 
Each \dword{pd} module is initially removed from its shipping bag, inspected visually, then (if it passes inspection) loaded into the optical scanner for operational testing shown in Figure \ref{fig:fdsp-tc-pds-scanner}.

The optical scanner tests the operation of the photosensor readout chain to ensure all electrical connections are operational, and measures light-collection performance at several positions along the length of the module.
Identical optical scanners are used at the module assembly facility, to test the module just before shipping so the underground test will detect any changes in performance due to shipping or storage.  
This technique was used successfully in \dword{pdsp}.

The optical scanner consists of a light-tight aluminum box, approximately \SI{2.5}{m} long, with a \SI{0.75}{\square\meter} cross section. 
The box acts as a Faraday cage to minimize electrical interference with measurements. In the \dword{dune} test configuration, two \dword{pd} modules are inserted through slots on the face of the box, guided by support rails of the type used for the \dword{apa}s, which provides a final mechanical check of the \dword{pd} module dimensions.  
The insertion slots are closed and optically sealed.

The \dword{pd} module uses an electrical connector identical to the ones in the \dword{apa} frames, and that important interface is also checked in the scanner.  
Once the scan begins, \dword{dune} \dword{pd} readout electronics is used to bias and read out the photosensors, while a UV \dword{led} is scanned along the length of the modules via an automated stepper-motor-driven translation stage.  Measurements are made at 16 positions along the length (on two sides for double-sided \dword{pd} modules), checking the performance of each  dichroic filter. The response is compared to that measured earlier at the assembly facility. Figure~\ref{fig:fdsp-tc-pds-scanner} shows the scanner used to test the \dword{pdsp} \dwords{pd}.

\begin{dunefigure}[Photon detector scanner used for \dshort{pdsp}]{fig:fdsp-tc-pds-scanner}
{Picture of the \SI{2.5}{m} long scanner used for operational tests of the \dword{pdsp} \dword{pd} modules prior to insertion into an \dword{apa}.} 

\includegraphics[height=.40\textheight, angle=0]{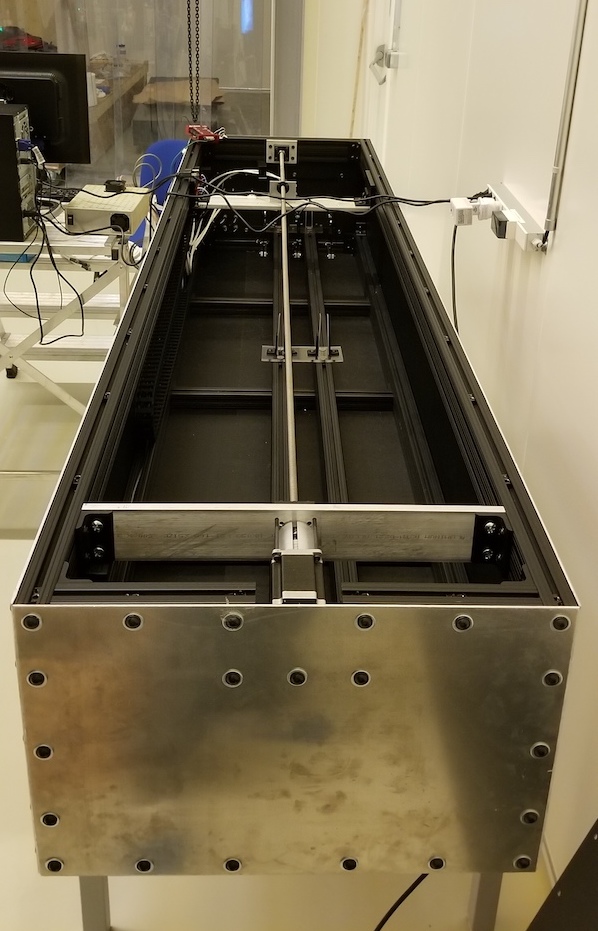}
\end{dunefigure}

Access to the \dwords{pd} inside the \dword{apa}s is severely limited once the \dword{ce} cable conduit is in place, so identifying problems early in the process is necessary to minimize schedule issues caused by required \dword{pd} maintenance or repair due to problems detected during installation.

Following the optical scan, the \dword{pd} modules are inserted into an  \dword{apa} at the \dword{pd} integration area in the cleanroom. The connection to the cable harness, which is pre-installed in the \dword{apa} before wire wrapping, is automatic. An electrical continuity check follows insertion to verify  continuity between the \dword{pd} module and the \dword{pd} cable end connector at the end of the \dword{apa}.

As the upper and lower \dword{apa}s are joined on the assembly tower, \dword{pd} cables from the upper to the lower \dword{apa} will be connected, and at that time, continuity checks will be made.

Once the upper and lower \dword{apa}s are joined, the assembled unit will be moved into a \coldbox in front of the cryostat for final testing.  This is an opportunity to make a final low-temperature check of the complete \dword{pd} \dword{ce} and cabling chain before installation into the cryostat.  \dword{pd} \dword{fe} electronics boards will be used to read out the photon system during the cold test, and results will be compared to previous \dword{qa} test results.

The \dword{apa} stack is rolled into position in the cryostat following the \coldbox test, and the \dword{pd} and \dword{ce} cables connected to the cryostat flange.  At this point, a final continuity check is made from the flange bulkhead to the \dword{pd} module.

Discussions are underway with the installation team to arrange for a one-shift dark test of the installed \dwords{pd} in the cryostat following final installation, verifying end-to-end system operation.

\subsubsection{Calibration System Testing}
The laser system is aligned and tested as the lasers are installed. This requires an initial alignment with the alignment laser followed by testing with the UV laser under controlled conditions. Details of the testing procedures have not yet been developed. The pulsed neutron source will be calibrated offline to verify the shielding design and neutron flux.  The source will be operated in test mode in situ to verify the functionality.

\section{Detector Commissioning}
\label{sec:fdsp-tc-commiss}

Once the \dword{spmod} is installed in the cryostat and the \dword{tco} is closed, a warm commissioning phase can begin in order to test the fully assembled detector.  After completion of the cryogenics installation, cold commissioning of both the cryogenics system and the detector can commence. Cold commissioning of the cryogenics steps through specific operating modes: purge, \cooldown, fill, and circulate. During these steps \dword{dune} conducts its own cold commissioning procedures.

Before the purging starts, a series of tests is performed to verify that the detector is operating nominally. 

\begin{enumerate}

    \item A pedestal and \dword{rms} characterization of all \dword{ce} channels verifies that all \dword{apa} \dword{fe} boards are responding and no dead channel or new noise sources arose following the \dword{tco} closing.
    
    \item A noise scan of all \dword{pd} channels is performed as a last check. 

    \item Each \dword{apa} wireplane is checked to verify it is isolated from the \dword{apa} frame and properly connected to its \dword{hv} power supply through the following steps:
    
\begin{itemize}

    \item The \dword{shv} connector of each wire plane bias channel gets unplugged at the power supply, and both the resistance and capacitance between inner conductor and ground is measured. 
    The resistance should show that the wireplane is electrically isolated from the ground, while the capacitance value should match that of the cold \dword{hv} cable and the capacitance of the circuit on the \dword{apa} top frame.

    \item \SI{50}{V} is applied to each wireplane and the drawn current is checked against the expected value.
    
    \item Nominal voltages are applied to each wireplane, and the  drawn current is checked against the expected value. 
    
\end{itemize}

    \item A low \dword{hv} (i.e., \SIrange{1}{2}{kV}) is applied to the cathode, and the drawn current is checked against the expected value to ensure the integrity of the \dword{hv} line.

\end{enumerate}

Cryogenics plant commissioning begins after installation is complete and the cryogenics system, including cryogenics controls and safety \dword{odh} systems, are approved for operation.
The system first purges the air inside the cryostat  by injecting pure \dword{gar} at the bottom  at a rate that fills the cryostat volume uniformly, but faster than the diffusion rate. This ``piston purge'' process produces a column of \dword{gar}  that rises through the volume and pushes the air up and out through the \dword{gar} purge lines and the \dword{gar} venting lines.  When the piston purge is complete, misting nozzles inject a liquid-gas mix into the cryostat that cools the detector components at a controlled rate.

Once the detector is cold, the filling process begins. \dword{lar} stored at the surface  
is vaporized, brought down the shaft in gaseous form, and re-condensed underground. The \dword{lar} then flows through filters to remove any H$_2$O and O$_2$ before entering the cryostat. Given the volume of the cryostat and the limited cooling power for recondensing, \num{12} months will be required to fill the first \dword{detmodule}. The detector readout electronics will  monitor the status of the detector during the filling period.

A number of the following tests (and likely others) will  take place during the \cooldown and fill phases: 

\begin{enumerate}

    \item Each \dword{apa} wireplane isolation and proper connection to its \dword{hv} power supply will be checked at regular time intervals as was done before sealing the cryostat.
    
    \item \SIrange{1}{2}{kV} will be held on the cathode, and the drawn current will be  monitored constantly to observe the trend in temperature of the total resistance.
    
    \item \dword{ce} noise figures (pedestal, \dword{rms}) will be measured at regular intervals and their trends with temperature recorded.
    
    \item \dword{pd} system noise (pedestal, \dword{rms}) will be measured at regular intervals and its trend with temperature recorded.
    
     \item Values of the temperature sensors deployed in several parts of the cryostat will be monitored constantly to watch the progress of the \cooldown phase and to relate the temperature to the behavior of the other \dword{spmod} subsystems. 
     
\end{enumerate}

Regular monitoring of \dword{ce} and \dword{pd} noise, as well as checks of wire plane isolation and proper connections to the bias supply system will continue throughout the fill period, recording noise variations as a function of the progressively reduced temperature. In addition,

\begin{enumerate}

    \item as each purity monitor is submerged in liquid, it will be turned on every eight hours to check \dword{lar} purity. 
    
    \item as soon as top \dwords{gp} are submerged, \dword{hv} on the cathode will be raised up to \SI{10}{V}-\SI{50}{V} to check that the current drawn by the system agrees with expectations.

\end{enumerate}

Once the \dword{detmodule} is full, the drift \dword{hv} will be carefully ramped up following these steps:

\begin{enumerate}

    \item Evaluate need for a filter regeneration before starting any operation.

    \item Once filter regeneration is completed (if needed), examine the \dword{lar} surface 
    using cameras to verify that the surface is flat, with no bubbles or turbulence;
    
    \item Start \dword{lar} recirculation while monitoring the \dword{lar} surface 
    again to see if activating the recirculation system introduced any turbulence into the liquid.
    
    \item Wait one day after beginning  recirculation to stabilize the \dword{lar} flow inside the \dword{detmodule}, then start the \dword{hv} ramp up.
    
\end{enumerate}

Ramping up of the cathode \dword{hv} represents the final step in the detector commissioning period; the full operating design parameters of the \dword{tpc} detector can be accessed only after the cathode is at full voltage. 
We raise the cathode voltage in steps over three days. 
On the first day, cathode voltage is first raised to \SI{60}{kV}, then to \SI{90}{kV} after waiting two hours, and finally to \SI{120}{kV} after waiting another two hours. It is left at this value overnight.
On the second day, cathode voltage is first raised to \SI{140}{kV}, then to \SI{160}{kV} after waiting four hours, and left at this value overnight. 
On the third day, cathode voltage is first raised to \SI{170}{kV} and then to the nominal operating voltage of \SI{180}{kV} after waiting four hours. 
During each \dword{hv} ramp up, all \dword{ce} current draws are monitored, and the procedure is stopped if any of the current draws go out of the allowed range. 
During each waiting period, regular \dword{daq} runs monitor \dword{ce} and \dword{pd} noise and response, while cathode \dword{hv} and current draw stability are constantly monitored.

In \dword{pdsp}, this process took three days, after which the system was  ready for data-taking. With a detector twenty times larger, the process will take longer, but the turn on time should still be relatively short. 

\section{Schedule}
\label{sec:sp-inst-sched}

The detector installation planning hinges on the date that the \dword{jpo} is permitted to begin work underground. According to the \dword{dune} \dword{cf} schedule, the \dword{jpo} receives the \dword{aup} for  the north cavern and \dword{cuc} in \cucbenocc{}.  The  \dword{sdwf} will be in place approximately six months before the warm structure installation begins, i.e., in spring 2022. Building the schedule for the \dword{detmodule} \#1 installation after \dword{aup} is  complicated and depends on many entities including \dword{cf}, \dword{lbnf}, and \dword{sdsta}.  The maximum number of people allowed underground is 144, which 
is based a one-hour time limit to evacuate all the underground personnel using all available paths.  As a backup, 
the  underground refuge must have capacity to accommodate this number of people. 
This number places a hard bound on how much work can be performed underground at any time and is particularly critical during the excavation of the third cavern when \dword{cf} is still active. Figure~\ref{fig:Overview-of-SinglePhase-Schedule} shows the main activities for the \dword{detmodule} \#1 installation and  the high-level milestones are shown in Table~\ref{tab:sp-iic-sched}.

The cost, schedule, and labor estimates are based on two \SI{10}{hour} shifts per day, four days a week (Monday through Thursday). Work efficiency should be a maximum of \SI{70}{\%}.  The cage ride, shift meetings, lunch, coffee breaks, and cleanroom gowning takes 
up to three hours per day. Some low level of effort is planned on Friday, Saturday, and Sunday to monitor the \coldbox{}es and take data.

\begin{dunetable}
[\dshort{sp} installation, integration, and commissioning milestones]
{p{0.65\textwidth}p{0.25\textwidth}}
{tab:sp-iic-sched}
{\dword{sp} installation, integration, and commissioning schedule}   
Milestone & Date (Month YYYY)   \\ \toprowrule

Ash River phase 0 complete &  March 2020    \\ \colhline
\rowcolor{dunepeach} Start of \dword{pdsp}-II installation& \startpduneiispinstall      \\ \colhline
Ash River phase 1 complete & June 2021     \\ \colhline
Installation Preliminary Design Review & August 2021     \\ \colhline
\rowcolor{dunepeach} Start of \dword{pddp}-II installation& \startpduneiidpinstall      \\ \colhline
\rowcolor{dunepeach}South Dakota Logistics Warehouse available& \sdlwavailable      \\ \colhline
Ash River phase 2 complete &  July 2022    \\ \colhline
Installation \dword{prr} &  August 2022    \\ \colhline
Start production of installation infrastructure for module \#1 & August 2022     \\ \colhline
Installation Final Design Review  & September 2022     \\ \colhline
\rowcolor{dunepeach}Beneficial occupancy of cavern 1 and \dword{cuc}& \cucbenocc      \\ \colhline
Start construction warm structure cryostat \#1   & October 2022     \\ \colhline
Start outfitting of \dword{cuc}  &  October 2022    \\ \colhline
\rowcolor{dunepeach} \dword{cuc} counting room accessible& \accesscuccountrm      \\ \colhline
Start installation of  cold structure cryostat \#1 &  August 2023    \\ \colhline
Start installing Detector\#1 infrastructure  &  August 2023    \\ \colhline
\rowcolor{dunepeach}Top of \dword{detmodule} \#1 cryostat accessible& \accesstopfirstcryo      \\ \colhline
Start installation of \dword{detmodule} \#1 & June 2024   \\ \colhline
\rowcolor{dunepeach}Start of \dword{detmodule} \#1 TPC installation& \startfirsttpcinstall      \\ \colhline
Complete installation of east \dshort{fc} endwall \& first \dword{apa} for module \#1 &   September 2024   \\ \colhline
Complete installation of \dword{apa} \#75 for \dword{detmodule} \#1 &   January 2025   \\  \colhline
\rowcolor{dunepeach}Top of \dword{detmodule} \#2 accessible& \accesstopsecondcryo      \\ \colhline
Complete installation of \dword{apa} \#150 (last) for \dword{detmodule} \#1 &   April 2025   \\  \colhline
\rowcolor{dunepeach}End of \dword{detmodule} \#1 TPC installation& \firsttpcinstallend      \\ \colhline
\dword{tco} of \dword{detmodule} \#1  closed  &  July 2025    \\ \colhline
Start of cryogenic operation for \dword{detmodule} \#1  &  August 2025    \\ \colhline
 \rowcolor{dunepeach}Start of \dword{detmodule} \#2 TPC installation& \startsecondtpcinstall      \\ \colhline
\rowcolor{dunepeach}End of \dword{detmodule} \#2 TPC installation& \secondtpcinstallend      \\ \colhline
Start of \dword{detmodule} \#1  commissioning  & January 2027     \\ 
\end{dunetable}

\begin{dunefigure}[Overview of the single-phase schedule]
{fig:Overview-of-SinglePhase-Schedule}
{Schedule Overview of the \dword{sp} \dword{detmodule} \#1}                
\includegraphics[angle=90, width=0.55\textwidth]{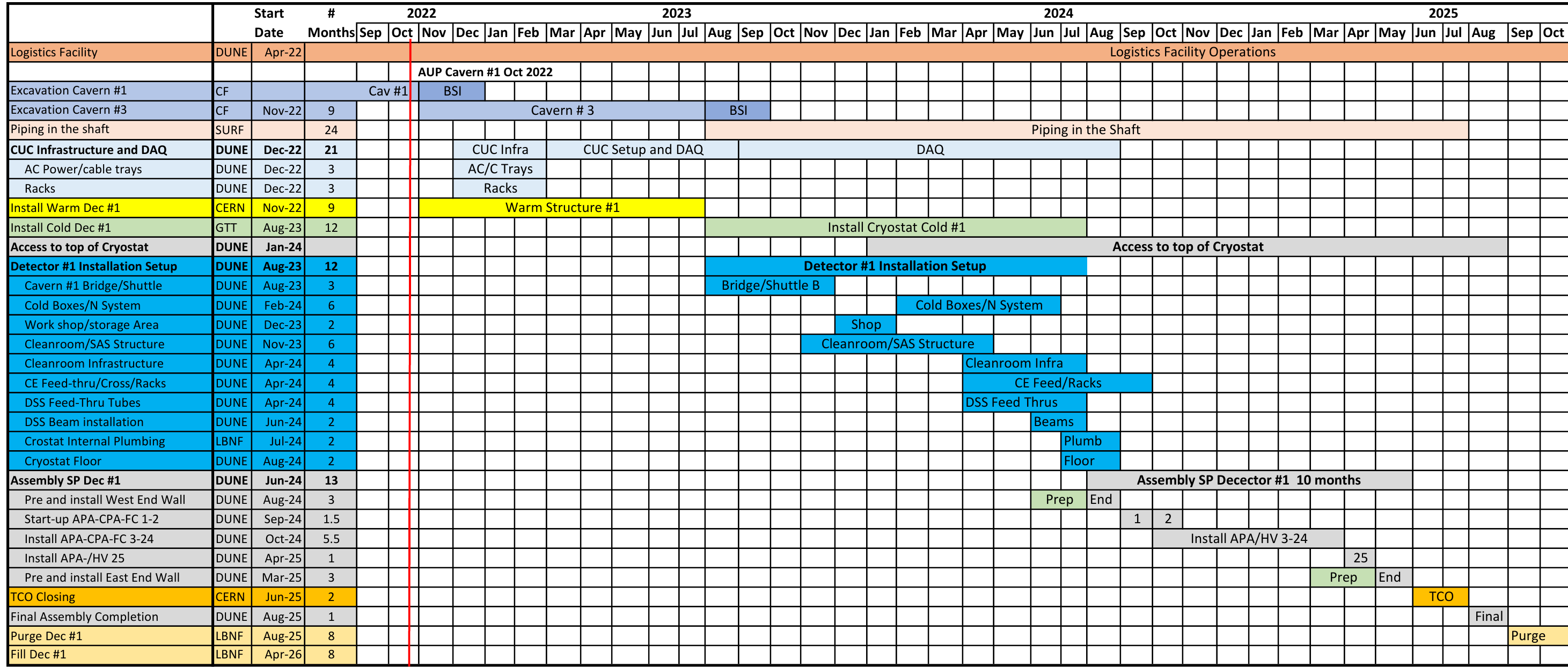}
\end{dunefigure}

We have defined three schedule phases for installation of the first \dword{spmod}:

\begin{itemize}

   \item {\bf \dword{cuc} Installation Phase:}
    This period,described in detail in Section~\ref{sec:fdsp-tc-inst-CUC}, starts once \dword{aup} has been received for the north cavern and the \dword{cuc}. This is the same time that excavation of the south cavern and installation of the warm structure by \dword{lbnf} begins. Since the \dwords{fte} underground is limited to 144 at a time, access will be minimal for \dword{dune} personnel and their work will only take place inside the \dword{cuc} and surface dataroom. Installing the basic rack infrastructure in the dataroom will take an estimated three months. Installation and testing of the \dword{daq} (required at the start of detector installation) will continue over the next 12 month period. Ten \dword{daq} workers are planned on each shift in this period.

    \item {\bf Installation Setup Phase:}  Described in detail in Section~\ref{sec:fdsp-tc-inst-setup}, this period includes installation of  the majority of the infrastructure. The setup phase is a critical training period, so getting lead workers, riggers, and equipment operators familiar with the tasks is a priority, as is  adjusting the crews to ensure balanced teams. The training process will have begun already at \dword{ashriver}. 
    There are many parallel underground activities planned in this phase making it a difficult phase to schedule, and frequent schedule adjustments may be required. Immediately after the cryostat warm structure is complete the north-south bridge is constructed. Following this the bridge crane under the bridge can be installed. A few months after the cryostat warm structure is complete the \dword{cf} work will also complete. Eighty of the 144 underground workers will become available to the \dword{jpo}, and the \dword{uit} team doubles in size. The \dword{jpo} will start two \SI{10}{hour} shifts per day.  Due to space constraints, peripheral work only on the cleanroom structure and assembly towers can begin. Once the  cryostat cold structure is approximately six months into its installation schedule, most of the foam will have been installed and floor space becomes available in the north cavern. The \coldbox construction must begin immediately at this point because the welding takes approximately six months. In parallel, the machine shop area can be set up. As the membrane installation nears completion, the walls of the cleanroom can be installed as can the remaining equipment. 
    
    Installation of the \dword{dss} could begin during the final installation stages of the cryostat cold structure because they both require full-height scaffolding for the welding on the top of the cryostat. The \dword{pdsp} \dword{dss} was installed this way. This requires a crew on top of the cryostat installing the \dword{dss} support feedthroughs from the top, as shown in Figure~\ref{fig:install-dss-feedthru}.  The details have not yet been worked out with the contractor, and work may be done in stages.

    \item {\bf Detector Installation Phase:} The final detector installation phase begins with an \dword{orr} to check that all documentation and procedures are in place. After the east \dwords{ewfc} are installed, a start-up period of 1.5 months begins for the first two rows of \dword{tpc} components.  To meet this schedule, three assembly lines, three \coldbox{}es, and separate crews in the cryostat, all working in parallel, are needed.  It will take 5.5 months to install rows 3 through 24 and about one month for row 25. Closing the \dword{tco} will take approximately two months for the cryostat cold structure contractor; during this time, there is no access to the cryostat.  Once this is completed, we can complete the final instrumentation, and the purge can begin. During this period, up to 50 people will be working in the cleanroom and cryostat.
    
\end{itemize}

The total time to install the detector including the time for the setup phase is two years. Coincidentally, this was roughly the time needed to install \dword{minos} and \dword{nova}.

\section{Environmental, Safety, and Health (ES\&H)}
\label{sec:fdsp-tc-safety}

 \tcchesh of the \dword{dune} \dword{tdr} outlines the requirements and regulations that \dword{dune} work must comply with, whether (1) at \dword{fnal}, (2) in areas  leased by \dword{fnal} or the \dword{doe}, (3) in leased space at \dword{surf}, or (4) at collaborating institutions.
 
\subsection{Documentation Approval Process}

\dword{dune} implements an engineering review and approval process for all required documentation, including structural calculations, assembly drawings, load tests, \dwords{ha}, and procedural documents for a comprehensive set of identified individual tasks. 
As for \dword{pdsp}, all these documents are stored in \dword{edms}. 
For the larger operations and systems like \dword{tpc} component factories, the \dword{dss}, cleanroom, and assembly infrastructure, \dword{dune} safety also reviews the documentation then visits the site to conduct
 an \dword{orr}, which includes a demonstration of the final operations. The \dwords{orr} are listed in project schedule. 
 
 Structural calculations, assembly drawings and proper documentation of  load tests, hazard analyses, and procedures for various items and activities will require review and approval before operational readiness is granted.

\subsection{Support and Responsibilities}

The \dword{esh} coordinator for each shift, who will report to the \dword{dune} project \dword{esh} manager, has overall \dword{esh} oversight responsibility for the \dword{dune} activities at the  \dword{sdwf} and on the \dword{surf} site. 
This person coordinates any \dword{esh} activities and facilitates the resolution of any issues that are subject to the requirements of the \dword{doe} Workers Safety and Health Program, Title 10, Code Federal Regulations (CRF) Part 851 (10 CFR 851) (see Volume~\volnumbertc{}).  The on-site \dword{esh} coordinator facilitates training and runs weekly safety meetings. This  person is also responsible for managing \dword{esh}-related  documentation, including training records, \dword{ha} documents, weekly safety reports, records on materials-handling equipment, near-miss and accident reports, and equipment inspections. 

If the \dword{esh} coordinator is absent, the shift supervisor acts in this capacity.

All workers have work stop authority in support of a safe working environment. 

\subsection{Safety Program}

The on-site \dword{esh} coordinators will guide the \dword{fd} installation safety program, using the following:

\begin{enumerate}
\item	the \dword{feshm};
\item the \dword{dune} Installation \dword{esh} Plan, which includes the fire evacuation plan, fire safety plan, lockdown plans, and the site plan;
\item	work planning and controls documentation which includes both hazard analysis and procedures; 
\item	Safety Data Sheets (SDS); 
\item	the respiratory plan, as required for chemical or \dword{odh} hazards; and 
\item	the training program, which covers required certifications and  training records.
\end{enumerate}

During the installation setup phase, as new equipment is being installed and tested, new employees and collaborators will be trained to access the facility and use the equipment. At the end of this phase, we will require two shifts per day.

The \coldbox and cryogenics system will not be 
tested during the trial assembly work at \dword{ashriver}. 
While the new \coldbox design is very similar to \dword{pdsp}'s, it will be operated under \dword{doe} and \dword{feshm} regulations.  Procedures for operating the \coldbox will be written according to the established requirements.

The \dword{dune} installation team
 will develop an  \dword{esh} plan for detector  installation that defines  
the \dword{esh} requirements and responsibilities for personnel during  assembly, installation, and construction of equipment at \dword{surf}. It will cover at least the following areas:

{Work Planning and \dword{ha}:} The goal of the work planning and \dword{ha} process is to initiate thought about the hazards associated with work activities and plan how to perform the work. Work planning ensures the scope of the job is understood, appropriate materials and tools are available, all hazards are identified, mitigation efforts are established, and all affected employees understand what is expected of them. 
The work planning and \dword{ha} program is documented in Chapter 2060 in the \dword{feshm}.

The shift supervisor and the \dword{esh} coordinator  will lead a work planning meeting at the start of each shift  to (1) coordinate the work activities, (2) notify the workers of potential safety issues, constraints, and hazard mitigations, (3) ensure that employees have the necessary \dword{esh} training and \dword{ppe}, and (4) answer any questions.

{Access and training:}  All \dword{dune} workers requiring access to the \dword{surf} site must (1) register through the \dword{fnal} Users Office to receive the necessary user training and a \dword{fnal} identification number, and (2) they must apply for a \dword{surf} identification badge. 
The workers will be required to complete \dword{surf} surface and underground orientation classes. Workers accessing the underground must also complete \dword{4850l} and 4910L specific unescorted access training, and obtain a \dword{tap} for each trip to the underground area; this is required as part of \dword{surf}'s Site Access Control Program. 
A properly trained guide will be stationed on all working levels. 

{\dword{ppe}:} 
The host laboratory is responsible for supplying appropriate \dword{ppe} to all workers. 

{\dword{ert}:} The \dword{sdsta} will maintain an emergency response incident command system and an \dword{ert}.  The guides on each underground level will be trained as first responders to help in a medical emergency.
  
  Guides: The shift supervisor and lead workers will be trained as guides.
  
  {House cleaning:} All workers are responsible for keeping a clean organized work area. This is particularly important underground. Flammable items must be in proper storage cabinets, and items like empty shipping crates and boxes must be removed and 
transported back to the surface to make space.

{Equipment operation:} All overhead cranes, gantry cranes, fork lifts, motorized equipment, e.g., trains and carts, will be operated only by trained  operators. 
Other equipment, e.g., scissor lifts, pallet jacks, hand tools, and shop equipment, will be operated only by people trained
and certified for the particular piece of equipment. All installation equipment will be electrically powered.

\cleardoublepage


\cleardoublepage
\printglossaries

\cleardoublepage
\cleardoublepage
\renewcommand{\bibname}{References}
\bibliographystyle{utphys} 
\bibliography{common/tdr-citedb}

\providecommand{\href}[2]{#2}\begingroup\raggedright\begin{thebibliography}{100}

\bibitem{Rubbia:1977zz}
C.~Rubbia, ``{The Liquid Argon Time Projection Chamber: A New Concept for
  Neutrino Detectors},'' Tech. Rep. CERN-EP-INT-77-08, 1977.

\bibitem{Icarus-T600}
S.~Amerio {\em et~al.}, ``{Design, construction and tests of the ICARUS T600
  detector},'' {\em Nucl. Instrum. Meth. A} no.~527, 329, (2004) .

\bibitem{Anderson:2012vc}
C.~Anderson, M.~Antonello, B.~Baller, T.~Bolton, C.~Bromberg, {\em et~al.},
  ``{The ArgoNeuT Detector in the NuMI Low-Energy beam line at Fermilab},''
  \href{http://dx.doi.org/10.1088/1748-0221/7/10/P10019}{{\em JINST} {\bfseries
  7} (2012) P10019},
\href{http://arxiv.org/abs/1205.6747}{{\ttfamily arXiv:1205.6747
  [physics.ins-det]}}.

\bibitem{microboone}
R.~Acciarri {\em et~al.}, ``Design and construction of the microboone
  detector,'' {\em Journal of Instrumentation} {\bfseries 12} no.~02, (2017)
  P02017. \url{http://stacks.iop.org/1748-0221/12/i=02/a=P02017}.

\bibitem{Cavanna:2014iqa}
{\bfseries LArIAT} Collaboration, F.~Cavanna, M.~Kordosky, J.~Raaf, and
  B.~Rebel, ``{LArIAT: Liquid Argon In A Testbeam}.'' arXiv:1406.5560,
2014.

\bibitem{Abi:2017aow}
{\bfseries DUNE} Collaboration, B.~Abi {\em et~al.}, ``{The Single-Phase
  ProtoDUNE Technical Design Report},''
\href{http://arxiv.org/abs/1706.07081}{{\ttfamily arXiv:1706.07081
  [physics.ins-det]}}.

\bibitem{bib:mu2e_tdr}
{\bfseries Mu2e} Collaboration, L.~Bartoszek {\em et~al.}, ``{Mu2e Technical
  Design Report},''
\href{http://arxiv.org/abs/1501.05241}{{\ttfamily arXiv:1501.05241
  [physics.ins-det]}}.

\bibitem{bib:cernedms2100877}
F.~Feyzi and D.~Wenman, ``{Structural analysis plans and results},'' CERN EDMS
  2100877, 2019.
\newblock \url{https://edms.cern.ch/document/2100877}.

\bibitem{bib:docdb6670}
D.~Christian, A.~Marchionni, S.~Soldner-Rembold, and M.~Verzocchi, ``{DUNE
  Interface Document: APA / TPC Electronics},'' DUNE doc 6670, DUNE, 2018.
\newblock
  \url{http://docs.dunescience.org/cgi-bin/ShowDocument?docid=6670&asof=2019-11-1}.

\bibitem{bib:docdb6667}
A.~Marchionni, E.~Segreto, C.~Touramanis, and D.~Warner, ``{DUNE Interface
  Document: SP-APA/PDS},'' DUNE doc 6667, DUNE, 2018.
\newblock
  \url{https://docs.dunescience.org/cgi-bin/private/ShowDocument?docid=6667&asof=2019-11-1}.

\bibitem{bib:docdb6673}
A.~Marchionni, F.~Pietropaolo, S.~Soldner-Rembold, and B.~Yu, ``{DUNE Interface
  Document: HVS/APA},'' DUNE doc 6673, DUNE, 2018.
\newblock
  \url{https://docs.dunescience.org/cgi-bin/private/ShowDocument?docid=6673&asof=2019-11-1}.

\bibitem{bib:docdb6676}
G.~Karagiorgi, A.~Marchionni, D.~Newbold, and S.~Soldner-Rembold, ``{DUNE FD
  Interface Document: SP APA to Joint DAQ},'' DUNE doc 6676, DUNE, 2018.
\newblock
  \url{https://docs.dunescience.org/cgi-bin/private/ShowDocument?docid=6676&asof=2019-11-1}.

\bibitem{bib:docdb6679}
A.~C. Villanueva {\em et~al.}, ``{DUNE FD Interface Document for Single
  Phase/Joint: SP APA to Joint CISC},''.
  \url{http://docs.dunescience.org/cgi-bin/ShowDocument?docid=6679}.

\bibitem{bib:docdb7021}
E.~James, S.~Kettell, A.~Marchionni, M.~Marshak, and S.~Soldner-Rembold,
  ``{DUNE FD Interface Document: Integration Facility to SP APA},'' DUNE doc
  7021, DUNE, 2019.
\newblock
  \url{https://docs.dunescience.org/cgi-bin/private/ShowDocument?docid=7021&asof=2019-11-1}.

\bibitem{bib:docdb6967}
E.~James, S.~Kettell, A.~Marchionni, M.~Marshak, and S.~Soldner-Rembold,
  ``{DUNE FD Interface Document: Integration Facility to SP APA},'' DUNE doc
  6967, DUNE, 2019.
\newblock
  \url{https://docs.dunescience.org/cgi-bin/private/ShowDocument?docid=6967&asof=2019-11-1}.

\bibitem{bib:docdb6994}
E.~James, S.~Kettell, A.~Marchionni, M.~Marshak, and S.~Soldner-Rembold,
  ``{DUNE FD Interface Document: Integration Facility to SP APA},'' DUNE doc
  6994, DUNE, 2019.
\newblock
  \url{https://docs.dunescience.org/cgi-bin/private/ShowDocument?docid=6994&asof=2019-11-1}.

\bibitem{bib:docdb7048}
S.~Gollapinni, K.~Mahn, A.~Marchionni, and S.~Soldner-Rembold, ``{DUNE FD
  Interface Document: Calibration to SP APA},'' DUNE doc 7048, DUNE, 2018.
\newblock
  \url{https://docs.dunescience.org/cgi-bin/private/ShowDocument?docid=7048&asof=2019-11-1}.

\bibitem{bib:docdb7102}
A.~Marchionni, A.~Norman, H.~Schellman, and S.~Soldner-Rembold, ``{DUNE FD
  Interface Document: Software and Computing to SP APA},'' DUNE doc 7102, DUNE,
  2018.
\newblock
  \url{https://docs.dunescience.org/cgi-bin/private/ShowDocument?docid=7102&asof=2019-11-1}.

\bibitem{bib:docdb7075}
A.~Marchionni, R.~Patterson, S.~Soldner-Rembold, and E.~Worcester, ``{DUNE FD
  Interface Document: DUNE Physics to SP APA},'' DUNE doc 7075, DUNE, 2018.
\newblock
  \url{https://docs.dunescience.org/cgi-bin/private/ShowDocument?docid=7075&asof=2019-11-1}.

\bibitem{bib:docdb1300}
L.~Greenler, ``{Layer Distortion Summary 6m Frame},'' DUNE doc 1300, DUNE,
  2017.
\newblock
  \url{http://docs.dunescience.org/cgi-bin/ShowDocument?docid=1300&asof=2019-11-1}.

\bibitem{Acciarri:2016ugk}
{\bfseries MicroBooNE} Collaboration, R.~Acciarri {\em et~al.}, ``{Construction
  and Assembly of the Wire Planes for the MicroBooNE Time Projection
  Chamber},'' \href{http://dx.doi.org/10.1088/1748-0221/12/03/T03003}{{\em
  JINST} {\bfseries 12} no.~03, (2017) T03003},
\href{http://arxiv.org/abs/1609.06169}{{\ttfamily arXiv:1609.06169
  [physics.ins-det]}}.

\bibitem{Garcia-Gamez:2018frz}
D.~Garcia-Gamez, V.~Basque, T.~G. Brooks, J.~J. Evans, M.~Perry,
  S.~Soldner-Rembold, F.~Spagliardi, and A.~M. Szelc, ``{A Novel Electrical
  Method to Measure Wire Tensions for Time Projection Chambers},''
  \href{http://dx.doi.org/10.1016/j.nima.2018.09.031}{{\em Nucl. Instrum.
  Meth.} {\bfseries A915} (2019) 75--81},
\href{http://arxiv.org/abs/1804.05941}{{\ttfamily arXiv:1804.05941
  [physics.ins-det]}}.

\bibitem{bib:docdb291}
M.~Andrews, ``{Integrated Environment, Safety and Health Management Plan},''
  DUNE doc 291, 2016.
\newblock
  \url{https://docs.dunescience.org/cgi-bin/private/ShowDocument?docid=291&asof=2019-11-1}.

\bibitem{bib:docdb1320}
{Rescia, S. and Yu, B.}, ``{Cathode HV Discharge Mitigation Design},'' tech.
  rep., 2016.
\newblock
  \url{http://docs.dunescience.org/cgi-bin/ShowDocument?docid=1320&version=1}.

\bibitem{bib:docdb2338}
{Guarino, V. and Horton-Smith, G. and Magill, S. and Yu, B.}, ``{CPA Electrical
  Connections Cold Test},'' tech. rep., 2017.
\newblock
  \url{http://docs.dunescience.org/cgi-bin/ShowDocument?docid=2338&version=2}.

\bibitem{bib:docdb1504}
{Guarino, V.}, ``{CPA and FC Design},'' tech. rep., 2018.
\newblock
  \url{http://docs.dunescience.org/cgi-bin/ShowDocument?docid=1504&version=20}.

\bibitem{bib:docdb1601}
{Nelson, J.}, ``{Mechanical Specifications for ProtoDUNE-SP field cage test in
  35 ton},'' tech. rep., 2016.
\newblock
  \url{http://docs.dunescience.org/cgi-bin/ShowDocument?docid=1601&version=1}.

\bibitem{Blatter:2014wua}
A.~Blatter {\em et~al.}, ``{Experimental study of electric breakdowns in liquid
  argon at centimeter scale},''
  \href{http://dx.doi.org/10.1088/1748-0221/9/04/P04006}{{\em JINST} {\bfseries
  9} (2014) P04006},
\href{http://arxiv.org/abs/1401.6693}{{\ttfamily arXiv:1401.6693
  [physics.ins-det]}}.

\bibitem{bib:docdb16766}
F.~Pietropaolo, ``{DUNE FD Interface Document: DSS to SP HV},'' DUNE doc 16766,
  CERN, 2019.
\newblock
  \url{https://docs.dunescience.org/cgi-bin/private/ShowDocument?docid=16766&asof=2019-11-1}.

\bibitem{bib:docdb6739}
D.~Christian, F.~Pietropaolo, M.~Verzocchi, and B.~Yu, ``{DUNE Interface
  Document: Single Phase TPC Cold Electronics and Joint Far Detector HV
  System},'' DUNE doc 6739, DUNE, 2018.
\newblock
  \url{http://docs.dunescience.org/cgi-bin/ShowDocument?docid=6739\&asof=2019-11-1}.

\bibitem{bib:docdb6721}
D.~Duchesneau, I.~Gil-Botella, E.~Segreto, and D.~Warner, ``{DUNE FD Interface
  Document: SP Photon Detector to joint High Voltage},'' DUNE doc 6721, DUNE,
  2018.
\newblock
  \url{https://docs.dunescience.org/cgi-bin/private/ShowDocument?docid=6721&asof=2019-11-1}.

\bibitem{bib:docdb6985}
F.~Feyzi, E.~James, F.~Pietropaolo, T.~Shaw, and B.~Yu, ``{DUNE FD Interface
  Document: SP TPC Electronics to joint DAQ},'' DUNE doc 6985, DUNE, 2018.
\newblock
  \url{https://docs.dunescience.org/cgi-bin/private/ShowDocument?docid=6985&asof=2019-11-1}.

\bibitem{bib:docdb7066}
S.~Gollapinni, K.~Mahn, F.~Pietropaolo, and B.~Yu, ``{DUNE FD Interface
  Document: Calibration to Joint HV},'' DUNE doc 7066, DUNE, 2018.
\newblock
  \url{https://docs.dunescience.org/cgi-bin/private/ShowDocument?docid=7066&asof=2019-11-1}.

\bibitem{bib:docdb6787}
A.~C. Villanueva {\em et~al.}, ``{DUNE FD Interface Document: High Voltage to
  CISC},''. \url{http://docs.dunescience.org/cgi-bin/ShowDocument?docid=6787}.

\bibitem{bib:docdb7039}
E.~James, M.~Marshak, F.~Pietropaolo, and B.~Yu, ``{DUNE FD Interface Document:
  Integration Facility to Joint HV},'' DUNE doc 7039, DUNE, 2018.
\newblock
  \url{https://docs.dunescience.org/cgi-bin/private/ShowDocument?docid=7039&asof=2019-11-1}.

\bibitem{bib:docdb7093}
R.~Patterson, F.~Pietropaolo, E.~Worcester, and B.~Yu, ``{DUNE FD Interface
  Document: DUNE Physics to Joint HV},'' DUNE doc 7093, DUNE, 2018.
\newblock
  \url{https://docs.dunescience.org/cgi-bin/private/ShowDocument?docid=7093&asof=2019-11-1}.

\bibitem{bib:docdb10452}
V.~Guarino and S.~Magill, ``{FD HV Design},'' DUNE doc 10452, 2019.
\newblock
  \url{http://docs.dunescience.org/cgi-bin/ShowDocument?docid=10452&asof=2019-11-1}.

\bibitem{DeGeronimo:2011zz}
G.~D. Geronimo, A.~D'Andragora, S.~Li, N.~Nambiar, S.~Rescia, E.~Vernon,
  H.~Chen, F.~Lanni, D.~Makowiecki, V.~Radeka, C.~Thorn, and B.~Yu, ``Front-end
  asic for a liquid argon tpc,''
  \href{http://dx.doi.org/10.1109/TNS.2011.2127487}{{\em IEEE Transactions on
  Nuclear Science} {\bfseries 58} no.~3, (June, 2011) 1376--1385}.

\bibitem{bib:sbnddoc1921}
O.~Palamara, M.~Stancari, and D.~Schmitz, ``{Technical Requirements of the SBND
  TPC Front-End Electronics},'' SBND doc 1921, 2017.
\newblock \url{http://sbn-docdb.fnal.gov/cgi-bin/ShowDocument?docid=1921}.

\bibitem{Acciarri:2018myr}
{\bfseries ArgoNeuT} Collaboration, R.~Acciarri {\em et~al.}, ``{Demonstration
  of MeV-Scale Physics in Liquid Argon Time Projection Chambers Using
  ArgoNeuT},'' \href{http://dx.doi.org/10.1103/PhysRevD.99.012002}{{\em Phys.
  Rev.} {\bfseries D99} no.~1, (2019) 012002},
\href{http://arxiv.org/abs/1810.06502}{{\ttfamily arXiv:1810.06502 [hep-ex]}}.

\bibitem{MICROBOONE-NOTE-1050-PUB}
{MicroBooNE Collaboration}, ``{Study of Reconstructed $\mathrm{{}^{39}Ar}$ Beta
  Decays at the MicroBooNE Detector},'' tech. rep., 2018.
\newblock
  \url{{http://microboone.fnal.gov/wp-content/uploads/MICROBOONE-NOTE-1050-PUB.pdf}}.
\newblock MICROBOONE-NOTE-1050-PUB.

\bibitem{Hairapetian1989}
A.~{Hairapetian}, D.~{Gitlin}, and C.~R. {Viswanathan}, ``{Low-temperature
  mobility measurements on CMOS devices},''
  \href{http://dx.doi.org/10.1109/16.30958}{{\em IEEE Transactions on Electron
  Devices} {\bfseries 36} (Aug., 1989) 1448--1455}.

\bibitem{Hot-electron}
C.~Hu, S.~C. Tam, F.-C. Hsu, P.-K. Ko, T.-Y. Chan, and K.~W. Terrill,
  ``Hot-electron-induced mosfet degradation—model, monitor, and
  improvement,'' \href{http://dx.doi.org/10.1109/T-ED.1985.21952}{{\em IEEE
  Transactions on Electron Devices} {\bfseries 32} no.~2, (Feb, 1985)
  375--385}.

\bibitem{radekaNoise}
V.~Radeka, ``{Shielding and grounding in large detectors},'' in {\em
  {Electronics for LHC experiments. Proceedings, 4th Workshop, Rome, Italy,
  September 21-25, 1998}}, pp.~14--18.
\newblock
1998.
\newblock

\bibitem{DUNE:GroundingRules}
{Theresa Marie Shaw}, ``Dune grounding and shielding rules,'' tech. rep.,
  Fermilab, Feb., 2019.
\newblock \url{https://edms.cern.ch/document/2095958}.

\bibitem{nasa_mems}
R.~Patterson and A.~Hammoud, ``Effect of temperature on the performance and
  stability of recent cots silicon oscillators.''
  \url{https://nepp.nasa.gov/files/19908/10_281_GRC_Patterson_Si%20Oscillators%207_12_10.doc},
  2010.

\bibitem{TSMC65}
 \url{http://www.tsmc.com/english/dedicatedFoundry/technology/65nm.htm}.

\bibitem{Nagel:1973qq}
L.~Nagel and D.~Pederson, ``{SPICE (SIMULATION PROGRAM WITH INTEGRATED CIRCUIT
  EMPHASIS)},''.

\bibitem{TSMC130}
 \url{https://www.tsmc.com/english/dedicatedFoundry/technology/0.13um.htm}.

\bibitem{TSMC180}
 \url{https://www.tsmc.com/english/dedicatedFoundry/technology/0.18um.htm}.

\bibitem{PipelinedADC}
D.~W. Cline and P.~R. Gray, ``A power optimized 13-b 5 msamples/s pipelined
  analog-to-digital converter in 1.2 micron cmos,''
  \href{http://dx.doi.org/10.1109/4.494191}{{\em IEEE Journal of Solid-State
  Circuits} {\bfseries 31} no.~3, (Mar, 1996) 294--303}.

\bibitem{CalibrationCorrection}
E.~G. Soenen and R.~L. Geiger, ``An architecture and an algorithm for fully
  digital correction of monolithic pipelined adcs,''
  \href{http://dx.doi.org/10.1109/82.372864}{{\em IEEE Transactions on Circuits
  and Systems II: Analog and Digital Signal Processing} {\bfseries 42} no.~3,
  (Mar, 1995) 143--153}.

\bibitem{bib:I2C}
 \url{http://i2c.info}.

\bibitem{Chen:2018zic}
H.~Chen {\em et~al.}, ``Lifetime study of cots adc for sbnd lartpc readout
  electronics.'' arXiv:1806.09226 [physics.ins-det], 2018.

\bibitem{Li:CELAr}
S.~Li, J.~Ma, G.~D. Geronimo, H.~Chen, and V.~Radeka, ``{LAr TPC electronics
  CMOS lifetime at 300 K and 77 K},''
  \href{http://dx.doi.org/10.1109/TNS.2013.2287156}{{\em IEEE Trans. Nuclear
  Science} {\bfseries 60} (2013) 4737}.

\bibitem{nEXO}
 \url{https://www-project.slac.stanford.edu/exo/about.html}.

\bibitem{5482529}
Y.~{Zhu}, C.~{Chan}, U.~{Chio}, S.~{Sin}, S.~{U}, R.~P. {Martins}, and
  F.~{Maloberti}, ``A 10-bit 100-ms/s reference-free sar adc in 90 nm cmos,''
  \href{http://dx.doi.org/10.1109/JSSC.2010.2048498}{{\em IEEE Journal of
  Solid-State Circuits} {\bfseries 45} no.~6, (June, 2010) 1111--1121}.

\bibitem{yeh}
W.~Yeh {\em et~al.}, ``Efficient suppression of substrate noise coupling in
  cmos technology,'' {\em International Conference on Solid State Devices and
  Materials} (2003) 408--409.

\bibitem{SACI}
 \url{https://confluence.slac.stanford.edu/pages/viewpage.action?pageId=219252577}.

\bibitem{bib:docdb1651}
D.~Newbold, ``{ProtoDUNE-SP Timing System: Interfaces and Protocol},'' tech.
  rep., Bristol, 2016.
\newblock \url{http://docs.dunescience.org/cgi-bin/ShowDocument?docid=1651}.

\bibitem{bib:docdb11233}
D.~G. Cussans and D.~Newbold, ``{DUNE Far Detector Timing and Synchronization
  System},'' DUNE doc 11233, 2018.
\newblock
  \url{https://docs.dunescience.org/cgi-bin/private/ShowDocument?docid=11233&asof=2019-11-1}.

\bibitem{Acciarri:2017sde}
{\bfseries MicroBooNE} Collaboration, R.~Acciarri {\em et~al.}, ``{Noise
  Characterization and Filtering in the MicroBooNE Liquid Argon TPC},''
  \href{http://dx.doi.org/10.1088/1748-0221/12/08/P08003}{{\em JINST}
  {\bfseries 12} no.~08, (2017) P08003},
\href{http://arxiv.org/abs/1705.07341}{{\ttfamily arXiv:1705.07341
  [physics.ins-det]}}.

\bibitem{pend_PDSP_PerfPaper}
{\bfseries DUNE} Collaboration, B.~Abi {\em et~al.}, ``{First results on
  ProtoDUNE-SP LArTPC performance from a test beam run at the CERN Neutrino
  Platform}.'' Paper in preparation, to be submitted to {\it JINST}.

\bibitem{bib:docdb12367}
M.~Verzocchi, ``{Lessons learned from ProtoDUNE, details for the SP-CE
  consortium},'' DUNE doc 12367, DUNE, 2019.
\newblock
  \url{https://docs.dunescience.org/cgi-bin/private/ShowDocument?docid=12367&asof=2019-11-1}.

\bibitem{dominoADC}
M.~Takhti, A.~Sodagar, and R.~Lotfi,
  \href{http://dx.doi.org/10.1109/ISCAS.2010.5537633}{``Domino adc: A novel
  analog-to-digital converter architecture,''} pp.~4057 -- 4060.
\newblock 07, 2010.

\bibitem{Hoff:2015hax}
J.~R. Hoff, G.~W. Deptuch, G.~Wu, and P.~Gui, ``Cryogenic lifetime studies of
  130 nm and 65 nm nmos transistors for high-energy physics experiments,''
  \href{http://dx.doi.org/10.1109/TNS.2015.2433793}{{\em IEEE Transactions on
  Nuclear Science} {\bfseries 62} no.~3, (June, 2015) 1255--1261}.

\bibitem{nasa_nepp}
{Teverovsky, Alexander}, ``Reliability of electronics at cryogenic
  temperatures.''
  \url{https://nepp.nasa.gov/DocUploads/1BCC6326-46F7-4C5E-94D476A8B6176ED1/Approach%20to%20Reliability%20Testing%20at%20cryo%20NEPP.pdf},
  2005.

\bibitem{bib:docdb6742}
D.~Christian, G.~Karagiorgi, D.~Newbold, and M.~Verzocchi, ``{DUNE FD Interface
  Document: SP TPC Electronics to joint DAQ},'' DUNE doc 6742, DUNE, 2018.
\newblock
  \url{http://docs.dunescience.org/cgi-bin/ShowDocument?docid=6742\&asof=2019-11-1}.

\bibitem{bib:docdb6745}
A.~Cervera~Villanueva, D.~Christian, S.~Gollapinni, and M.~Verzocchi, ``{DUNE
  Interface Document: CISC / SP--TPC Cold Electronics},'' DUNE doc 6745, DUNE,
  2018.
\newblock
  \url{http://docs.dunescience.org/cgi-bin/ShowDocument?docid=6745\&asof=2019-11-1}.

\bibitem{bib:docdb6718}
D.~Christian, E.~Segreto, M.~Verzocchi, and D.~Warner, ``{DUNE Interface
  Document: Single Phase TPC Cold Electronics and Single Phase TPC Photon
  System},'' DUNE doc 6718, DUNE, 2018.
\newblock
  \url{http://docs.dunescience.org/cgi-bin/ShowDocument?docid=6718\&asof=2019-11-1}.

\bibitem{bib:docdb6973}
D.~Christian, F.~Feyzi, E.~James, T.~Shaw, and M.~Verzocchi, ``{DUNE Interface
  Document: Single Phase TPC Cold Electronics and Detector and Facilities
  Infrastructure (LBNF) and Technical Coordination (DUNE)},'' DUNE doc 6973,
  DUNE, 2018.
\newblock
  \url{http://docs.dunescience.org/cgi-bin/ShowDocument?docid=6973\&asof=2019-11-1}.

\bibitem{bib:docdb7000}
D.~Christian, E.~James, S.~Kettel, J.~Stewart, and M.~Verzocchi, ``{DUNE
  Interface Document: TPC Electronics / Underground Installation Team},'' DUNE
  doc 7000, DUNE, 2018.
\newblock
  \url{http://docs.dunescience.org/cgi-bin/ShowDocument?docid=7000\&asof=2019-11-1}.

\bibitem{bib:docdb7081}
D.~Christian, R.~Patterson, M.~Verzocchi, and E.~Worcester, ``{DUNE Interface
  Document: DUNE Physics to SP TPC Electronics},'' DUNE doc 7081, DUNE, 2018.
\newblock
  \url{http://docs.dunescience.org/cgi-bin/ShowDocument?docid=7081\&asof=2019-11-1}.

\bibitem{bib:docdb7108}
D.~Christian, M.~Kirby, A.~McNab, H.~Schellman, and M.~Verzocchi, ``{DUNE
  Interface Document: Software and Computing to SP TPC Electronics},'' DUNE doc
  7108, DUNE, 2018.
\newblock
  \url{http://docs.dunescience.org/cgi-bin/ShowDocument?docid=7108\&asof=2019-11-1}.

\bibitem{bib:docdb7054}
D.~Christian, S.~Gollapinni, K.~Mahn, and M.~Verzocchi, ``{DUNE Interface
  Document: Calibration to SP TPC Electronics},'' DUNE doc 7054, DUNE, 2018.
\newblock
  \url{http://docs.dunescience.org/cgi-bin/ShowDocument?docid=7054\&asof=2019-11-1}.

\bibitem{Doke:1990rza}
T.~Doke, K.~Masuda, and E.~Shibamura, ``{Estimation of absolute photon yields
  in liquid argon and xenon for relativistic (1 MeV) electrons},''
\href{http://dx.doi.org/10.1016/0168-9002(90)90011-T}{{\em Nucl. Instrum.
  Meth.} {\bfseries A291} (1990) 617--620}.

\bibitem{PhysRevB.20.3486}
S.~Kubota, M.~Hishida, M.~Suzuki, and J.-z. Ruan(Gen), ``Dynamical behavior of
  free electrons in the recombination process in liquid argon, krypton, and
  xenon,'' \href{http://dx.doi.org/10.1103/PhysRevB.20.3486}{{\em Phys. Rev. B}
  {\bfseries 20} (Oct, 1979) 3486--3496}.
  \url{https://link.aps.org/doi/10.1103/PhysRevB.20.3486}.

\bibitem{Heindl:2010zz}
T.~Heindl, T.~Dandl, M.~Hofmann, R.~Krucken, L.~Oberauer, W.~Potzel, J.~Wieser,
  and A.~Ulrich, ``{The scintillation of liquid argon},''
  \href{http://dx.doi.org/10.1209/0295-5075/91/62002}{{\em EPL} {\bfseries 91}
  no.~6, (2010) 62002},
\href{http://arxiv.org/abs/1511.07718}{{\ttfamily arXiv:1511.07718
  [physics.ins-det]}}.

\bibitem{PhysRevB.27.5279}
A.~Hitachi, T.~Takahashi, N.~Funayama, K.~Masuda, J.~Kikuchi, and T.~Doke,
  ``Effect of ionization density on the time dependence of luminescence from
  liquid argon and xenon,''
  \href{http://dx.doi.org/10.1103/PhysRevB.27.5279}{{\em Phys. Rev. B}
  {\bfseries 27} (May, 1983) 5279--5285}.
  \url{https://link.aps.org/doi/10.1103/PhysRevB.27.5279}.

\bibitem{bkds}
{\bfseries WARP} Collaboration, P.~Benetti {\em et~al.}, ``{Measurement of the
  specific activity of ar-39 in natural argon},''
  \href{http://dx.doi.org/10.1016/j.nima.2007.01.106}{{\em Nucl. Instrum.
  Meth.} {\bfseries A574} (2007) 83--88},
\href{http://arxiv.org/abs/astro-ph/0603131}{{\ttfamily arXiv:astro-ph/0603131
  [astro-ph]}}.

\bibitem{Marinho:2018doi}
F.~Marinho, L.~Paulucci, A.~A. Machado, and E.~Segreto, ``{Numerical
  characterization of the ARAPUCA: a new approach for LAr scintillation light
  detection},'' \href{http://dx.doi.org/10.1088/1742-6596/1056/1/012036}{{\em
  J. Phys. Conf. Ser.} {\bfseries 1056} no.~1, (2018) 012036},
\href{http://arxiv.org/abs/1804.03764}{{\ttfamily arXiv:1804.03764
  [physics.ins-det]}}.

\bibitem{Gola:2019idb}
A.~Gola, F.~Acerbi, M.~Capasso, M.~Marcante, A.~Mazzi, G.~Paternoster,
  C.~Piemonte, V.~Regazzoni, and N.~Zorzi, ``{NUV-Sensitive Silicon
  Photomultiplier Technologies Developed at Fondazione Bruno Kessler},''
\href{http://dx.doi.org/10.3390/s19020308}{{\em Sensors} {\bfseries 19} no.~2,
  (2019) 308}.

\bibitem{bib:docdb3126}
J.~T. Anderson, R.~Dharmapalan, Z.~Djurcic, G.~Drake, A.~Farbin, and
  M.~Oberling, ``{ProtoDUNE ANL Photon Detector Readout PRR Documents},'' DUNE
  doc 3126, DUNE, 2018.
\newblock
  \url{http://docs.dunescience.org/cgi-bin/ShowDocument?docid=3126&asof=2019-11-1}.

\bibitem{Segreto:2018jdx}
E.~Segreto {\em et~al.}, ``{Liquid Argon test of the ARAPUCA device},''
  \href{http://dx.doi.org/10.1088/1748-0221/13/08/P08021}{{\em JINST}
  {\bfseries 13} no.~08, (2018) P08021},
\href{http://arxiv.org/abs/1805.00382}{{\ttfamily arXiv:1805.00382
  [physics.ins-det]}}.

\bibitem{Asaadi:2018ixs}
J.~Asaadi, B.~J.~P. Jones, A.~Tripathi, I.~Parmaksiz, H.~Sullivan, and Z.~G.~R.
  Williams, ``{Tetraphenyl Butadiene Emanation and Bulk Fluorescence from
  Wavelength Shifting Coatings in Liquid Argon},''
\href{http://arxiv.org/abs/1804.00011}{{\ttfamily arXiv:1804.00011
  [physics.ins-det]}}.

\bibitem{bib:docdb6727}
D.~Duchesneau, I.~Gil-Botella, G.~Karagiorgi, and Newbold, ``{DUNE FD Interface
  Document: DP Photon Detector to joint DAQ},'' DUNE doc 6727, DUNE, 2018.
\newblock
  \url{http://docs.dunescience.org/cgi-bin/ShowDocument?docid=6727\&asof=2019-11-1}.

\bibitem{bib:docdb6730}
A.~C. Villanueva {\em et~al.}, ``{DUNE FD Interface Document: SP Photon
  Detector to Joint CISC},'' DUNE doc 6730, DUNE, 2018.
\newblock
  \url{https://docs.dunescience.org/cgi-bin/private/ShowDocument?docid=6730&asof=2019-11-1}.

\bibitem{bib:docdb6970}
F.~Feyzi, E.~James, E.~Segreto, and T.~Shaw, ``{DUNE FD Interface Document:
  Facility Interfaces to SP Photon Detection},'' DUNE doc 6970, DUNE, 2018.
\newblock
  \url{https://docs.dunescience.org/cgi-bin/private/ShowDocument?docid=6970&asof=2019-11-1}.

\bibitem{bib:docdb6997}
E.~James, E.~Segreto, J.~Stewart, and D.~Warner, ``{DUNE FD Interface Document:
  Installation Interfaces to SP Photon Detector},'' DUNE doc 6997, DUNE, 2018.
\newblock
  \url{https://docs.dunescience.org/cgi-bin/private/ShowDocument?docid=6997&asof=2019-11-1}.

\bibitem{bib:docdb7051}
S.~Gollapinni, K.~Mahn, E.~Segreto, and D.~Warner, ``{DUNE FD Interface
  Document: Calibration to SP Photon Detector},'' DUNE doc 7051, DUNE, 2018.
\newblock
  \url{https://docs.dunescience.org/cgi-bin/private/ShowDocument?docid=7051&asof=2019-11-1}.

\bibitem{bib:docdb7105}
A.~Norman, H.~Schellman, E.~Segreto, and D.~Warner, ``{DUNE FD Interface
  Document: Software and Computing to SP Photon Detector},'' DUNE doc 7105,
  DUNE, 2018.
\newblock
  \url{https://docs.dunescience.org/cgi-bin/private/ShowDocument?docid=7105&asof=2019-11-1}.

\bibitem{Grace:2015yta}
E.~Grace and J.~A. Nikkel, ``{Index of refraction, Rayleigh scattering length,
  and Sellmeier coefficients in solid and liquid argon and xenon},''
  \href{http://dx.doi.org/10.1016/j.nima.2017.06.031}{{\em Nucl. Instrum.
  Meth.} {\bfseries A867} (2017) 204--208},
\href{http://arxiv.org/abs/1502.04213}{{\ttfamily arXiv:1502.04213
  [physics.ins-det]}}.

\bibitem{Acciarri:2008kv}
{\bfseries WArP} Collaboration, R.~Acciarri {\em et~al.}, ``{Effects of
  Nitrogen contamination in liquid Argon},''
  \href{http://dx.doi.org/10.1088/1748-0221/5/06/P06003}{{\em JINST} {\bfseries
  5} (2010) P06003},
\href{http://arxiv.org/abs/0804.1217}{{\ttfamily arXiv:0804.1217 [nucl-ex]}}.

\bibitem{Garcia-Gamez:2017cmu}
{\bfseries SBND} Collaboration, D.~Garcia-Gamez, ``{Light Detection System
  simulations for SBND},''
\href{http://dx.doi.org/10.1088/1742-6596/888/1/012094}{{\em J. Phys. Conf.
  Ser.} {\bfseries 888} no.~1, (2017) 012094}.

\bibitem{Kuzniak:2018dcf}
M.~Ku\'{z}niak and B.~Broerman, ``{Polyethylene naphthalate film as a
  wavelength shifter in liquid argon detectors},''
\href{http://arxiv.org/abs/1806.04020}{{\ttfamily arXiv:1806.04020
  [physics.ins-det]}}.

\bibitem{bib:docdb14068}
E.~Conley, ``{Fitting supernova neutrino spectral parameters with the Deep
  Underground Neutrino Experiment},'' DUNE doc 14068, DUNE, 2019.
\newblock
  \url{http://docs.dunescience.org/cgi-bin/ShowDocument?docid=14068&asof=2019-11-1}.

\bibitem{Li:2015rqa}
Y.~Li {\em et~al.}, ``{Measurement of Longitudinal Electron Diffusion in Liquid
  Argon},'' \href{http://dx.doi.org/10.1016/j.nima.2016.01.094}{{\em Nucl.
  Instrum. Meth.} {\bfseries A816} (2016) 160--170},
\href{http://arxiv.org/abs/1508.07059}{{\ttfamily arXiv:1508.07059
  [physics.ins-det]}}.

\bibitem{bib:mooney2018}
M.~Mooney, ``Space charge effects in lartpcs,'' {\em Workshop on Calibration
  and Reconstruction for LArTPC Detectors} (2018) .
  \url{https://indico.fnal.gov/event/18523/session/19/contribution/29/material/slides/0.pdf}.

\bibitem{bib:yu2017a}
B.~Yu, ``Drift field distortions due to cpa/field cage imperfections,'' {\em
  DUNE Calibration Task Force meeting} (2017) .
  \url{https://indico.fnal.gov/event/15245/contribution/1/material/slides/}.

\bibitem{bib:mooney2019a}
M.~Mooney, ``Field cage resistor chain failure study,'' {\em DUNE Calibration
  Task Force meeting} (2019) .
  \url{https://indico.fnal.gov/event/19386/contribution/4/material/slides/0.pdf}.

\bibitem{Ereditato:2013xaa}
A.~Ereditato, C.~C. Hsu, S.~Janos, I.~Kreslo, M.~Messina, C.~Rudolf~von Rohr,
  B.~Rossi, T.~Strauss, M.~S. Weber, and M.~Zeller, ``{Design and operation of
  ARGONTUBE: a 5 m long drift liquid argon TPC},''
  \href{http://dx.doi.org/10.1088/1748-0221/8/07/P07002}{{\em JINST} {\bfseries
  8} (2013) P07002},
\href{http://arxiv.org/abs/1304.6961}{{\ttfamily arXiv:1304.6961
  [physics.ins-det]}}.

\bibitem{bib:uBlaser2019}
{\bfseries MicroBooNE} Collaboration, C.~Adams {\em et~al.}, ``{A Method to
  Determine the Electric Field of Liquid Argon Time Projection Chambers using a
  UV Laser System and its Application in MicroBooNE},'' {\em submitted to
  JINST} (2019) ,
\href{http://arxiv.org/abs/1910.01430}{{\ttfamily arXiv:1910.01430
  [physics.ins-det]}}.

\bibitem{Walkowiak:2000wf}
W.~Walkowiak, ``{Drift velocity of free electrons in liquid argon},''
\href{http://dx.doi.org/10.1016/S0168-9002(99)01301-7}{{\em Nucl. Instrum.
  Meth.} {\bfseries A449} (2000) 288--294}.

\bibitem{Rossi:2009im}
B.~Rossi {\em et~al.}, ``{A Prototype liquid Argon Time Projection Chamber for
  the study of UV laser multi-photonic ionization},''
  \href{http://dx.doi.org/10.1088/1748-0221/4/07/P07011}{{\em JINST} {\bfseries
  4} (2009) P07011},
\href{http://arxiv.org/abs/0906.3437}{{\ttfamily arXiv:0906.3437
  [physics.ins-det]}}.

\bibitem{Zeller:2013sva}
M.~Zeller {\em et~al.}, ``{First measurements with ARGONTUBE, a 5m long drift
  Liquid Argon TPC},''
\href{http://dx.doi.org/10.1016/j.nima.2012.11.181}{{\em Nucl. Instrum. Meth.}
  {\bfseries A718} (2013) 454--458}.

\bibitem{Ereditato:2014lra}
A.~Ereditato, I.~Kreslo, M.~L\"uthi, C.~Rudolf~von Rohr, M.~Schenk, T.~Strauss,
  M.~Weber, and M.~Zeller, ``{A steerable UV laser system for the calibration
  of liquid argon time projection chambers},''
  \href{http://dx.doi.org/10.1088/1748-0221/9/11/T11007}{{\em JINST} {\bfseries
  9} no.~11, (2014) T11007},
\href{http://arxiv.org/abs/1406.6400}{{\ttfamily arXiv:1406.6400
  [physics.ins-det]}}.

\bibitem{Ereditato:82014tya}
A.~Ereditato, D.~Goeldi, S.~Janos, I.~Kreslo, M.~Luethi, C.~Rudolf~von Rohr,
  M.~Schenk, T.~Strauss, M.~S. Weber, and M.~Zeller, ``{Measurement of the
  drift field in the ARGONTUBE LAr TPC with 266 nm pulsed laser beams},''
  \href{http://dx.doi.org/10.1088/1748-0221/9/11/P11010}{{\em JINST} {\bfseries
  9} no.~11, (2014) P11010},
\href{http://arxiv.org/abs/1408.6635}{{\ttfamily arXiv:1408.6635
  [physics.ins-det]}}.

\bibitem{Berns:2013usa}
{\bfseries CAPTAIN} Collaboration, H.~Berns {\em et~al.}, ``{The CAPTAIN
  Detector and Physics Program},'' in {\em {Community Summer Study 2013}:
  {Snowmass on the Mississippi}}.
\newblock 9, 2013.
\newblock \href{http://arxiv.org/abs/1309.1740}{{\ttfamily arXiv:1309.1740
  [physics.ins-det]}}.

\bibitem{Antonello:2015lea}
{\bfseries LAr1-ND, ICARUS-WA104, MicroBooNE} Collaboration, M.~Antonello {\em
  et~al.}, ``{A Proposal for a Three Detector Short-Baseline Neutrino
  Oscillation Program in the Fermilab Booster Neutrino Beam},''
\href{http://arxiv.org/abs/1503.01520}{{\ttfamily arXiv:1503.01520
  [physics.ins-det]}}.

\bibitem{bib:chen2018}
Y.~Chen, ``Laser calibration at lar tpcs,'' {\em Workshop on Calibration and
  Reconstruction for LArTPC Detectors} (2018) .
  \url{https://indico.fnal.gov/event/18523/session/17/contribution/35/material/slides/0.pdf}.

\bibitem{Badhrees:2010zz}
I.~Badhrees {\em et~al.}, ``{Measurement of the two-photon absorption
  cross-section of liquid argon with a time projection chamber},''
  \href{http://dx.doi.org/10.1088/1367-2630/12/11/113024}{{\em New J. Phys.}
  {\bfseries 12} (2010) 113024},
\href{http://arxiv.org/abs/1011.6001}{{\ttfamily arXiv:1011.6001
  [physics.ins-det]}}.

\bibitem{bib:yu2019a}
B.~Yu, ``Fc penetrations and laser interface,'' {\em DUNE Calibration
  Consortium meeting} (2019) .
  \url{https://indico.fnal.gov/event/20390/contribution/0/material/slides/}.

\bibitem{bib:yu2017b}
B.~Yu, ``E field distortion from a laser port opening on the field cage,'' {\em
  DUNE Calibration Task Force meeting} (2017) .
  \url{https://indico.fnal.gov/event/15246/contribution/0/material/slides/}.

\bibitem{bib:uBlifetime}
{\bfseries MicroBooNE} Collaboration, ``{A Measurement of the Attenuation of
  Drifting Electrons in the MicroBooNE LArTPC},'' 2017.
\newblock
  MICROBOONE-NOTE-1026-PUB,\url{http://microboone.fnal.gov/wp-content/uploads/MICROBOONE-NOTE-1026-PUB.pdf}.

\bibitem{Antonello:2014eha}
M.~Antonello, B.~Baibussinov, P.~Benetti, F.~Boffelli, A.~Bubak, {\em et~al.},
  ``{Experimental observation of an extremely high electron lifetime with the
  ICARUS-T600 LAr-TPC},''
  \href{http://dx.doi.org/10.1088/1748-0221/9/12/P12006}{{\em JINST} {\bfseries
  9} no.~12, (2014) P12006},
\href{http://arxiv.org/abs/1409.5592}{{\ttfamily arXiv:1409.5592
  [physics.ins-det]}}.

\bibitem{Acciarri:2013met}
{\bfseries ArgoNeuT} Collaboration, R.~Acciarri {\em et~al.}, ``{A study of
  electron recombination using highly ionizing particles in the ArgoNeuT Liquid
  Argon TPC},'' \href{http://dx.doi.org/10.1088/1748-0221/8/08/P08005}{{\em
  JINST} {\bfseries 8} (2013) P08005},
\href{http://arxiv.org/abs/1306.1712}{{\ttfamily arXiv:1306.1712
  [physics.ins-det]}}.

\bibitem{Abgrall:2010hi}
{\bfseries T2K ND280 TPC} Collaboration, N.~Abgrall {\em et~al.}, ``{Time
  Projection Chambers for the T2K Near Detectors},''
  \href{http://dx.doi.org/10.1016/j.nima.2011.02.036}{{\em Nucl. Instrum.
  Meth.} {\bfseries A637} (2011) 25--46},
\href{http://arxiv.org/abs/1012.0865}{{\ttfamily arXiv:1012.0865
  [physics.ins-det]}}.

\bibitem{Li:2016ods}
Y.~Li {\em et~al.}, ``{A 20-Liter Test Stand with Gas Purification for Liquid
  Argon Research},''
  \href{http://dx.doi.org/10.1088/1748-0221/11/06/T06001}{{\em JINST}
  {\bfseries 11} no.~06, (2016) T06001},
\href{http://arxiv.org/abs/1602.01884}{{\ttfamily arXiv:1602.01884
  [physics.ins-det]}}.

\bibitem{Brown:2018jhj}
D.~A. Brown {\em et~al.}, ``{ENDF/B-VIII.0: The 8th Major Release of the
  Nuclear Reaction Data Library with CIELO-project Cross Sections, New
  Standards and Thermal Scattering Data},''
\href{http://dx.doi.org/10.1016/j.nds.2018.02.001}{{\em Nucl. Data Sheets}
  {\bfseries 148} (2018) 1--142}.

\bibitem{Fischer:2019qfr}
V.~Fischer {\em et~al.}, ``{Measurement of the neutron capture cross-section on
  argon},''
\href{http://arxiv.org/abs/1902.00596}{{\ttfamily arXiv:1902.00596 [nucl-ex]}}.

\bibitem{bib:Koivunoro2004}
H.~Koivunoro, D.~Bleuel, U.~Nastasi, T.~Lou, J.~Reijonen, and K.-N. Leung,
  ``Bnct dose distribution in liver with epithermal d-d and d-t fusion-based
  neutron beams,''
  \href{http://dx.doi.org/https://doi.org/10.1016/j.apradiso.2004.05.043}{{\em
  Applied Radiation and Isotopes} {\bfseries 61} no.~5, (2004) 853 -- 859}.
  \url{http://www.sciencedirect.com/science/article/pii/S0969804304003409}.
  Topics in Neutron Capture Therapy: Proceedings of the Eleventh World Congress
  on Neutron Capture Therapy (ISNCT-11).

\bibitem{bib:docdb7069}
S.~Gollapinni, G.~Karagiorgi, K.~Mahn, and D.~Newbold, ``{DUNE FD Interface
  Document: Calibration to Joint DAQ},'' DUNE doc 7069, DUNE, 2018.
\newblock
  \url{https://docs.dunescience.org/cgi-bin/private/ShowDocument?docid=7069&asof=2019-11-1}.

\bibitem{bib:docdb7072}
S.~Gollapinni {\em et~al.}, ``{DUNE FD Interface Document: Calibration to Joint
  CISC},''. \url{http://docs.dunescience.org/cgi-bin/ShowDocument?docid=7072}.

\bibitem{bib:docdb6865}
S.~Gollapinni, K.~Mahn, R.~Patterson, and E.~Worcester, ``{DUNE FD Interface
  Document: Calibration Task Force to DUNE Physics},'' DUNE doc 6865, DUNE,
  2019.
\newblock
  \url{https://docs.dunescience.org/cgi-bin/private/ShowDocument?docid=6865&asof=2019-11-1}.

\bibitem{bib:docdb6868}
S.~Gollapinni, K.~Mahn, A.~Norman, and H.~Schellman, ``{DUNE FD Interface
  Document: Calibration Task Force to DUNE Software and Computing},'' DUNE doc
  6868, DUNE, 2019.
\newblock
  \url{https://docs.dunescience.org/cgi-bin/private/ShowDocument?docid=6868&asof=2019-11-1}.

\bibitem{bib:docdb6829}
F.~Feyzi, S.~Gollapinni, E.~James, K.~Mahn, and T.~Shaw, ``{DUNE FD Interface
  Document: Facility to Calibration Task Force},'' DUNE doc 6829, DUNE, 2019.
\newblock
  \url{https://docs.dunescience.org/cgi-bin/private/ShowDocument?docid=6829&asof=2019-11-1}.

\bibitem{bib:docdb6847}
S.~Gollapinni, E.~James, K.~Mahn, and J.~Stewart, ``{DUNE FD Interface
  Document: Installation to Calibration Task Force},'' DUNE doc 6847, DUNE,
  2019.
\newblock
  \url{https://docs.dunescience.org/cgi-bin/private/ShowDocument?docid=6847&asof=2019-11-1}.

\bibitem{FNAL:Class4Lasers}
M.~Quinn, ``Lasers,'' FESHM 4260, Fermi National Accelerator Laboratory,
  January, 2019.
\newblock \url{http://esh-docdb.fnal.gov/cgi-bin/RetrieveFile?docid=385}.

\bibitem{CERN:Class4Lasers}
{TIS/RP}, ``Rules for the safe use of lasers at cern,'' Safety Instruction IS
  22, CERN, Jan., 1994.
\newblock \url{https://edms.cern.ch/ui/file/335744/LAST_RELEASED/IS22_E.pdf}.

\bibitem{Rogers:1996ks}
J.~G. Rogers, M.~S. Andreaco, and C.~Moisan, ``{A 7-MeV - 9-MeV isotopic
  gamma-ray source for detector testing},''
{\em Submitted to: Nucl. Instrum. Meth.} (1996) .

\bibitem{Crespo-Anadon:2019lht}
{\bfseries MicroBooNE} Collaboration, J.~Crespo-Anadón, ``{The MicroBooNE
  Continuous Readout Stream for Detection of Supernova Neutrinos},''
  \href{http://dx.doi.org/10.1088/1742-6596/1312/1/012006}{{\em J. Phys. Conf.
  Ser.} {\bfseries 1312} no.~1, (2019) 012006},
  \href{http://arxiv.org/abs/1907.02195}{{\ttfamily arXiv:1907.02195
  [physics.ins-det]}}.

\bibitem{bib:docdb9240}
{Josh Klein}, ``{DUNE Far Detector Data Volumes},'' DUNE doc 9240, Univ. of
  Pennsylvania, 2018.
\newblock
  \url{https://docs.dunescience.org/cgi-bin/private/ShowDocument?docid=9240&asof=2019-11-1}.

\bibitem{bib:docdb11461}
B.~Viren, ``{What are the DUNE FD DAQ Bottlenecks?},'' DUNE doc 11461, 2018.
\newblock
  \url{http://docs.dunescience.org/cgi-bin/ShowDocument?docid=11461&asof=2019-11-1}.

\bibitem{bib:docdb15544}
G.~Lehmann-Miotto and A.~Thea, ``{DUNE FD SP DAQ rack layout and power
  consumption estimate draft},'' DUNE doc 15544, 2019.
\newblock
  \url{https://docs.dunescience.org/cgi-bin/private/ShowDocument?docid=15544}.

\bibitem{bib:docdb7123}
G.~Karagiorgi, D.~Newbold, A.~Norman, and H.~Schellman, ``{DUNE FD Interface
  Document: Software and Computing to Joint DAQ},'' DUNE doc 7123, Columbia,
  Bristol, Fermilab, and Oregon State, 2018.
\newblock
  \url{http://docs.dunescience.org/cgi-bin/ShowDocument?docid=7123&asof=2019-11-1}.

\bibitem{bib:docdb6790}
A.~C. Villanueva {\em et~al.}, ``{DUNE FD Interface Document: DAQ to CISC},''.
  \url{http://docs.dunescience.org/cgi-bin/ShowDocument?docid=6790}.

\bibitem{bib:docdb11224}
D.~G. Cussans, J.~R. Klein, and D.~Newbold, ``{DUNE FD Interface Document:
  Timing/Synchronization},'' DUNE doc 11224, 2018.
\newblock
  \url{https://docs.dunescience.org/cgi-bin/private/ShowDocument?docid=11224&asof=2019-11-1}.

\bibitem{bib:docdb6988}
F.~Feyzi, E.~James, G.~Karagiorgi, D.~Newbold, and T.~Shaw, ``{DUNE FD
  Interface Document: Facility Interfaces to Joint DAQ},'' DUNE doc 6988, DUNE,
  2018.
\newblock
  \url{https://docs.dunescience.org/cgi-bin/private/ShowDocument?docid=6988&asof=2019-11-1}.

\bibitem{bib:docdb7015}
E.~James, G.~Karagiorgi, S.~Kettell, D.~Newbold, and J.~Stewart, ``{DUNE FD
  Interface Document: Installation Interfaces to Joint DAQ},'' DUNE doc 7015,
  DUNE, 2018.
\newblock
  \url{https://docs.dunescience.org/cgi-bin/private/ShowDocument?docid=7015&asof=2019-11-1}.

\bibitem{atlas-felix}
W.~Wu, ``{FELIX: the New Detector Interface for the ATLAS Experiment},'' 2018.
\newblock arXiv:1806.10667.

\bibitem{bib:docdb11215}
D.~Last and D.~Rivera, ``{Data Selection for DUNE Beam and Atmospheric
  Events},'' DUNE doc 11215, 2018.
\newblock
  \url{http://docs.dunescience.org/cgi-bin/ShowDocument?docid=11215&asof=2019-11-1}.

\bibitem{bib:docdb14522}
P.~Lasorak, ``{DUNE Supernovae Triggering},'' DUNE doc 14522, 2019.
\newblock
  \url{https://docs.dunescience.org/cgi-bin/private/ShowDocument?docid=14522&asof=2019-7-23}.

\bibitem{bib:docdb11311}
G.~Ge, Y.~jae Jwa, and G.~Karagiorgi, ``{ML-based triggering for DUNE},'' DUNE
  doc 11311, 2018.
\newblock
  \url{https://docs.dunescience.org/cgi-bin/private/ShowDocument?docid=11311&asof=2019-11-1}.

\bibitem{snews}
P.~Antonioli {\em et~al.}, ``{SNEWS: The SuperNova Early Warning System},''
  \href{http://dx.doi.org/10.1088/1367-2630/6/1/114}{{\em New J. Phys.}
  {\bfseries 6} (2004) 114},
\href{http://arxiv.org/abs/astro-ph/0406214}{{\ttfamily
  arXiv:astro-ph/0406214}}.

\bibitem{bib:docdb11275}
A.~Booth, G.~Karagiorgi, J.~R. Klein, P.~J. Lasorak, D.~Last, D.~Rivera,
  P.~Rodrigues, and B.~Viren, ``{DUNE-SP Data Selection Summary},'' DUNE doc
  11275, DUNE, 2018.
\newblock
  \url{https://docs.dunescience.org/cgi-bin/private/ShowDocument?docid=11275&asof=2019-11-1}.

\bibitem{bib:docdb11236}
P.~Rodrigues, ``{Single phase TPC trigger primitive algorithms for TDR},'' DUNE
  doc 11236, DUNE, 2018.
\newblock
  \url{https://docs.dunescience.org/cgi-bin/private/ShowDocument?docid=11236&asof=2019-11-1}.

\bibitem{artdaq}
K.~Biery, E.~Flumerfelt, J.~Freeman, W.~Ketchum, G.~Lukhanin, and
  R.~Rechenmacher, ``artdaq: {DAQ} software development made simple,''
  \href{http://dx.doi.org/10.1088/1742-6596/898/3/032013}{{\em Journal of
  Physics: Conference Series} {\bfseries 898} (Oct, 2017) 032013}.
  \url{https://doi.org/10.1088%2F1742-6596%2F898%2F3%2F032013}.

\bibitem{zeromq}
P.~Hintjens, {\em {ZeroMQ, Messaging for Many Applications}}.
\newblock O'Reilly Media, 2012.
\newblock \url{http://zguide.zeromq.org/}.

\bibitem{Karagiorgi:2019qzp}
{\bfseries SBND} Collaboration, G.~Karagiorgi, ``{Liquid Argon TPC Trigger
  Development with SBND},'' in {\em {Meeting of the Division of Particles and
  Fields of the American Physical Society (DPF2019) Boston, Massachusetts, July
  29-August 2, 2019}}.
\newblock 2019.
\newblock
\href{http://arxiv.org/abs/1910.08218}{{\ttfamily arXiv:1910.08218
  [physics.ins-det]}}.
\newblock

\bibitem{Sipos:2018auh}
{\bfseries DUNE} Collaboration, R.~Sipos, ``{The DAQ for the Single Phase DUNE
  Prototype at CERN},'' in {\em {21st IEEE Real Time Conference (RT2018)
  Williamsburg, USA, June 11-15, 2018}}.
\newblock 2018.
\newblock
\href{http://arxiv.org/abs/1806.09310}{{\ttfamily arXiv:1806.09310
  [physics.ins-det]}}.
\newblock

\bibitem{bib:docdb1881}
G.~Barr, M.~Graham, R.~Herbst, and L.~Ruckman, ``{ProtoDUNE RCE-Based TPC
  Readout},'' DUNE doc 1881, 2018.
\newblock
  \url{http://docs.dunescience.org/cgi-bin/ShowDocument?docid=1881&asof=2019-7-23}.

\bibitem{Hennessy:CDRReview}
K.~Hennessy, ``Dune cdr review - protodune-sp lessons learned,'' in {\em DUNE
  CDR: DAQ Review}.
\newblock December 3, 2018.
\newblock
  \url{https://indico.fnal.gov/event/18505/contribution/10/material/slides/0.pdf}.

\bibitem{bib:docdb14095}
G.~Lehmann-Miotto, ``{ProtoDUNE DAQ Schedule 2019},'' DUNE doc 14095, DUNE,
  2019.
\newblock
  \url{https://docs.dunescience.org/cgi-bin/private/ShowDocument?docid=14095&asof=2019-11-1}.

\bibitem{pdsp-felix}
A.~Borga, E.~Church, F.~Filthaut, E.~Gamberini, P.~de~Jong, G.~L. Miotto,
  F.~Schreuder, J.~Schumacher, R.~Sipos, M.~Vermeulen, K.~Wierman, and L.~Wood,
  ``{FELIX based readout of the Single-Phase ProtoDUNE detector},'' 2019.
\newblock arXiv:1806.09194.

\bibitem{Larrea:2015wra}
C.~Ghabrous~Larrea, K.~Harder, D.~Newbold, D.~Sankey, A.~Rose, A.~Thea, and
  T.~Williams, ``{IPbus: a flexible Ethernet-based control system for xTCA
  hardware},''
\href{http://dx.doi.org/10.1088/1748-0221/10/02/C02019}{{\em JINST} {\bfseries
  10} no.~02, (2015) C02019}.

\bibitem{bib:docdb14062}
J.~Klein, ``{ProtoDUNE Data Selection Tests},'' DUNE doc 14062, 2018.
\newblock
  \url{https://docs.dunescience.org/cgi-bin/private/ShowDocument?docid=14062&asof=2019-11-1}.

\bibitem{jwa2019accelerating}
Y.-J. Jwa, G.~Di~Guglielmo, L.~P. Carloni, and G.~Karagiorgi, ``{Accelerating
  Deep Neural Networks for Real-time Data Selection for High-resolution Imaging
  Particle Detectors},'' in {\em New York Scientific Data Summit (NYSDS)},
  IEEE.
\newblock 2019.

\bibitem{bib:docdb14242}
J.~Wilhite, ``{LBNF BSI - 90 Pct Final Design},'' LBNF DocDB 14242, LBNF, 2018.
\newblock
  \url{https://docs.dunescience.org/cgi-bin/private/ShowDocument?docid=14242}.

\bibitem{bib:docdb14938}
G.~L. Miotto and A.~Thea, ``{DAQ Consortium Working Group Mandates},'' DUNE doc
  14938, 2019.
\newblock
  \url{https://docs.dunescience.org/cgi-bin/private/ShowDocument?docid=14938&asof=2019-11-1}.

\bibitem{bib:docdb5915}
G.~J. Michna {\em et~al.}, ``{CFD Analysis of Fluid, Heat, and Impurity Flows
  in DUNE FAR Detector to Address Additional Design Considerations},'' DUNE doc
  5915, South Dakota State University, 2017.
\newblock
  \url{http://docs.dunescience.org/cgi-bin/ShowDocument?docid=5915&asof=2019-11-1}.

\bibitem{cfd-1}
{Strons, Ph., and Bailey, J. L.}, ``{Flow visualization methods for field test
  verification of CFD analysis of an ope\ n gloveport},'' 2017.
\newblock \url{https://www.osti.gov/pages/servlets/purl/1402050}.

\bibitem{pdspdcs_proc}
A.~Kehrli, G.~L. Miotto, X.~Pons, S.~Ravat, and M.~J. Rodriguez, ``{The
  protoDUNE Single Phase Detector Control System, in Proceedings of CHEP
  2018},'' tech. rep., 2018.
\newblock
  \url{https://docs.dunescience.org/cgi-bin/private/ShowDocument?docid=11098}.

\bibitem{larpropertiesbnl}
{The Liquid Argon Technology @BNL}, ``Basic properties.''
\newblock \url{https://lar.bnl.gov/properties/basic.html}. Accessed Jan. 15,
  2019.

\bibitem{bib:docdb3384}
L.~Whitehead, ``{DUNE Far Detector Task Force Final Report},'' DUNE doc 3384,
  2018.
\newblock
  \url{http://docs.dunescience.org/cgi-bin/ShowDocument?docid=3384&asof=2019-11-1}.

\bibitem{Adamowski:2014daa}
M.~Adamowski {\em et~al.}, ``{The Liquid Argon Purity Demonstrator},''
  \href{http://dx.doi.org/10.1088/1748-0221/9/07/P07005}{{\em JINST} {\bfseries
  9} (2014) P07005},
\href{http://arxiv.org/abs/1403.7236}{{\ttfamily arXiv:1403.7236
  [physics.ins-det]}}.

\bibitem{Murphy:20170516}
{\bfseries WA105} Collaboration, S.~Murphy, ``{Status of the WA105-3x1x1 m$^3$
  dual phase prototype},'' tech. rep., 2017.
\newblock
  \url{https://indico.fnal.gov/event/12345/session/1/contribution/5/material/slides/0.pdf}.

\bibitem{Delaquis:2013hva}
{Delaquis, S. C. and Gornea, R. and Janos, S. and L\"uthi, M. and Rudolf von
  Rohr, C. and Schenk, M. and Vuilleumier, J. -L.}, ``{Development of a camera
  casing suited for cryogenic and vacuum applications},''
  \href{http://dx.doi.org/10.1088/1748-0221/8/12/T12001}{{\em JINST} {\bfseries
  8} (2013) T12001},
\href{http://arxiv.org/abs/1310.6601}{{\ttfamily arXiv:1310.6601
  [physics.ins-det]}}.

\bibitem{Auger:2015xlo}
{Auger, M. and Blatter, A. and Ereditato, A. and Goeldi, D. and Janos, S. and
  Kreslo, I. and L\"uthi, M. and Rudolf von Rohr, C. and Strauss, T. and Weber,
  M. S.}, ``{On the Electric Breakdown in Liquid Argon at Centimeter Scale},''
  \href{http://dx.doi.org/10.1088/1748-0221/11/03/P03017}{{\em JINST}
  {\bfseries 11} no.~03, (2016) P03017},
\href{http://arxiv.org/abs/1512.05968}{{\ttfamily arXiv:1512.05968
  [physics.ins-det]}}.

\bibitem{Banks:2014hra}
T.~I. Banks {\em et~al.}, ``{A compact ultra-clean system for deploying
  radioactive sources inside the KamLAND detector},''
  \href{http://dx.doi.org/10.1016/j.nima.2014.09.068}{{\em Nucl. Instrum.
  Meth.} {\bfseries A769} (2015) 88--96},
\href{http://arxiv.org/abs/1407.0413}{{\ttfamily arXiv:1407.0413
  [physics.ins-det]}}.

\bibitem{lumileds-DS144-pdf}
``{LUXEON} {C} {Color} {Line} (datasheet).''
\newblock \url{https://www.lumileds.com/uploads/571/DS144-pdf}. Accessed:
  2018-11-19.

\bibitem{Carron:2017zzz}
J.~Carron, A.~Philippon, L.~S. How, A.~Delbergue, S.~Hassanzadeh, D.~Cillierre,
  P.~Danto, and M.~Boutillier, ``{Cryogenic characterization of LEDs for space
  application},'' \href{http://dx.doi.org/10.1117/12.2272409}{{\em Proc. SPIE,
  Sixteenth International Conference on Solid State Lighting and LED-based
  Illumination Systems} {\bfseries 10378} (2017) 20}.

\bibitem{winccoa}
{Siemens AG}, ``{SIMATIC} {WinCC} open architecture.''
\newblock \url{http:///www.siemens.com/wincc-open-architecture}. Accessed:
  2018-11-27.

\bibitem{epics7}
``{EPICS - Experimental Physics and Industrial Control System}.''
\newblock \url{https://epics-controls.org/}. Accessed: July 19, 2019.

\bibitem{Lukhanin:2012fp}
G.~Lukhanin, K.~Biery, S.~Foulkes, M.~Frank, A.~Hatzikoutelis, J.~Kowalkowski,
  M.~Paterno, and R.~Rechenmacher, ``{Application of control system studio for
  the NOvA detector control system},''
\href{http://dx.doi.org/10.1088/1742-6596/396/6/062012}{{\em J. Phys. Conf.
  Ser.} {\bfseries 396} (2012) 062012}.

\bibitem{jcop}
``{About JCOP}.''
\newblock \url{http://jcop.web.cern.ch/}. Accessed: July 19, 2019.

\bibitem{unicos}
``{UNICOS},'' 2015.
\newblock \url{http://unicos.web.cern.ch/}.

\bibitem{bib:docdb5609}
A.~C. Villanueva and S.~Gollapinni, ``{DUNE FD WBS: Slow Control},'' DUNE doc
  5609, DUNE, 2018.
\newblock
  \url{http://docs.dunescience.org/cgi-bin/ShowDocument?docid=5609&asof=2019-11-1}.

\bibitem{bib:docdb7192}
S.~Gollapinni {\em et~al.}, ``{DUNE FD Risks: CISC},''.
  \url{http://docs.dunescience.org/cgi-bin/ShowDocument?docid=7192\&asof=2019-11-1}.

\bibitem{bib:docdb7099}
S.~Gollapinni {\em et~al.}, ``{DUNE FD Interface Document: DUNE Physics to
  Joint CISC},''.
  \url{https://docs.dunescience.org/cgi-bin/private/ShowDocument?docid=7099&version=1}.

\bibitem{bib:docdb7126}
S.~Gollapinni {\em et~al.}, ``{DUNE FD Interface Document: Software and
  Computing to Joint CISC},''.
  \url{http://docs.dunescience.org/cgi-bin/ShowDocument?docid=7126}.

\bibitem{bib:docdb6991}
F.~Feyzi {\em et~al.}, ``{DUNE FD Interface Document: Facility Interfaces to
  Joint CISC},''.
  \url{https://docs.dunescience.org/cgi-bin/private/ShowDocument?docid=6991}.

\bibitem{bib:docdb7018}
S.~Gollapinni {\em et~al.}, ``{DUNE FD Interface Document: Installation
  Interfaces to Joint CISC},''.
  \url{http://docs.dunescience.org/cgi-bin/ShowDocument?docid=7018}.

\bibitem{bib:docdb7045}
S.~Gollapinni {\em et~al.}, ``{DUNE FD Interface Document: Integration Facility
  to Joint CISC},''.
  \url{https://docs.dunescience.org/cgi-bin/private/ShowDocument?docid=7045}.

\bibitem{bib:docdb8255}
R.~Rameika, ``{ProtoDUNE - SP Lessons Learned},'' DUNE doc 8255, DUNE, 2018.
\newblock
  \url{https://docs.dunescience.org/cgi-bin/private/ShowDocument?docid=8255&asof=2019-11-1}.

\bibitem{bib:docdb328}
D.~Montanari, E.~Costopoulos, and J.~Willhite, ``{Far Site Shaft Hoisting
  Assumptions and Parameters},'' DUNE doc 328, DUNE, 2018.
\newblock
  \url{http://docs.dunescience.org/cgi-bin/ShowDocument?docid=328&asof=2019-11-1}.

\bibitem{bib:docdb8426}
J.~Stewart, ``{DUNE Load List for Ross},'' DUNE doc 8426, DUNE, 2018.
\newblock
  \url{https://docs.dunescience.org/cgi-bin/private/ShowDocument?docid=8426&asof=2019-11-1}.

\bibitem{bib:docdb6260}
V.~Guarino, ``{DUNE Installation Integration},'' DUNE doc 6260, DUNE, 2019.
\newblock
  \url{http://docs.dunescience.org/cgi-bin/ShowDocument?docid=6260&asof=2019-11-1}.

\bibitem{bib:docdb4499}
T.~Shaw, ``{Detector Racks -- DUNE working document},'' DUNE doc 4499, 2018.
\newblock
  \url{http://docs.dunescience.org/cgi-bin/ShowDocument?docid=4499&asof=2019-11-1}.

\end{thebibliography}\endgroup

\end{document}